\documentclass[openany,a4paper,11pt]{book}
\usepackage[left=2.5cm,top=3.2cm,right=2.0cm,bottom=2.5cm,asymmetric]{geometry}
\usepackage{amsbsy,epsfig,latexsym,palatino}
\usepackage{mathptmx,mathtools,newtxtext,newtxmath,xfrac,amsmath}
\usepackage{setspace,sectsty,fancyhdr}
\usepackage{rotating}
\usepackage[super,comma,numbers,sort&compress]{natbib}
\usepackage[sectionbib]{chapterbib}
\usepackage{wrapfig}
\usepackage{lscape}
\usepackage{longtable}
\usepackage{dcolumn}
\usepackage{multirow}
\usepackage{makecell}
\usepackage{textcomp}
\usepackage{threeparttable}
\usepackage{caption}
\usepackage[table]{xcolor}
\usepackage{pifont}
\usepackage{diagbox}
\usepackage{siunitx}
\usepackage{appendix}
\usepackage[T1]{fontenc}
\usepackage{titlesec}
\usepackage{graphicx}
\usepackage{subcaption}
\usepackage{soul}
\usepackage{slashbox}
\usepackage{enumitem}
\usepackage[pagewise]{lineno}
\usepackage{tabularx}
\usepackage[table]{xcolor}
\usepackage{colortbl}
\definecolor{green}{HTML}{66FF66}
\definecolor{myGreen}{HTML}{009900}
\usepackage{hyperref}
\hypersetup{linkbordercolor={0 0 1}, citebordercolor={0 0 1},
	colorlinks=true, linkcolor=blue, citecolor=blue}

\usepackage[version=4]{mhchem}

\newcommand*{\citen}[1]{%
	\begingroup
	\romannumeral-`\x 
	\setcitestyle{numbers}%
	\cite{#1}%
	\endgroup
}

\begin{document}

\title{\LARGE The Role of Electron-Nuclear Coupling  on \\Chemical and Physical Processes} 

\vspace{10cm}

\author{Soumya Mukherjee \\
	\textit{Department of Chemistry, Maulana Azad National Institute of Technology, Bhopal, India} \\ \\
	Satrajit Adhikari\\
	\textit{School of Chemical Sciences, Indian Association for the Cultivation of Science, Kolkata, India}}
\date{\today}
\maketitle

%

%
%

\newpage
\tableofcontents


\chapter*{Preface}\label{sec:pref}
\addcontentsline{toc}{chapter}{Preface}

\vspace{-1.0 cm}

\noindent
Over the last few decades, Born-Oppenheimer (BO) and Born-Oppenheimer-Huang (BOH) treatments for electron-nuclear equation of motion have been considered as the cornerstone for qualitative as well as quantitative exploration and understanding of chemical and physical processes. The BO approximation due to the large mass difference between electrons and nuclei as well as the energy separation among the electronic states could  able to predict some of the processes (chemical and physical) satisfactorily at lower energy regime. Such approximation (BO) quite often fails for those cases with the involvement of excited electronic states, in particular, due to the coupling among the states. On the other hand, the BO treatment provides two important quantities, namely, the adiabatic potential energy surfaces (PESs) and the nonadiabatic coupling terms (NACTs) determining the forces, which control the motion of the nuclei in a molecule. The NACTs play a major role in photoexcitation spectra, charge transfer chemical reactions, ferroelectric property and other processes, where significant anomalies are observed between single and multiple PES(s) results. Since the NACTs often exhibit singularity at the close vicinity of degenerate states forming conical intersections (CIs), such coupling terms could break the symmetry of the molecular processes changing the observables qualitatively.\\

\noindent
The BOH treatment brings the electron-nuclear equation of motion from time-independent/time-dependent Schr\"odinger Equation (TISE/TDSE) known as adiabatic representation, where the NACTs could raise the issue of conservation of energy as well as numerical inaccuracies due to their (often) singular nature in the configuration space (CS). While traversing along a contour of nuclear coordinates enclosing CI(s), BO adiabatic eigenfunction undergoes phase change and such multivalued feature of the wavefunction can be taken care by multiplying a complex phase factor known as Longuet-Higgins (LH) phase. Later on, such phase could be  correlated with the topological feature of NACTs originating from the BO treatment. In all practical purposes, such singular terms are of real concern and can be removed by adapting a different representation known as {\it diabatic} SE. In other words, a transition to beyond Born-Oppenheimer (BBO) regime is indeed necessary to provide an equivalent description of the molecular processes, where the equations of motion are theoretically ``exact'' and  numerically ``accurate''.   Such transition can be achieved through a unitary transformation of the adiabatic SE to diabatic one known as {\it adiabatic-to-diabatic transformation} (ADT). In BBO treatment, the kinetically coupled matrix equations are transformed to the potentially coupled ones possessing well-behaved smooth functions of nuclear coordinates. Later on, BBO theory has been generalized for  three or more coupled electronic states with explicit expressions of NACTs, ADT equations, curl-divergence equations and diabatic PESs as functions of nuclear coordinate dependent ADT angles. For a given sub-Hilbert space (SHS), an accurate transformation to the diabatic framework is ensured if the NACT vector fields fulfill \textit{curl condition}. Though the ADT transformation does not provide unique diabatic Hamiltonian (DH) and those DHs depend on the path of the contour integration, calculated observables from DHs for spectroscopy and scattering processes will be unique and can be compared with experimental result as those DHs are related with each other through an orthogonal similarity transformation.  \\

\noindent 
As a complete theory, BBO treatment has been explored for various 
floppy molecules of spectroscopic interest, like Na$_3$, K$_3$, NO$_2$, NO$_3$, C$_6$H$_6^+$, 1,3,5-C$_6$H$_3$F$_3^+$ and C$_4$N$_2$H$_4$; reaction dynamics involving multiple electronic 
states, like H$_3^+$, HeH$_2^+$, F+H$_2$ and H$_3$ as well as solid state systems, like MnO$_6^{9-}$ and TiO$_6^{8-}$ exhibiting phase transition. Such systems are rich in exhibiting
strong Jahn-Teller (JT), Renner-Teller (RT) and pseudo Jahn-Teller (PJT) effects. Since the first principle based BBO theory is expected to generate single-valued, continuous and 
smooth diabatic PESs to perform nuclear dynamics, this approach could able to reproduce experimental spectroscopic (spectra) and scattering results (cross-sections and rate constants).\\

\noindent
The aim of this book is to reach the broad community, who are working in the realm of nonadiabatic dynamics 
connected with the development and application of BBO theory as well as other contemporary approaches over the last few decades. Chapter \ref{sec:intro1} depicts an extensive review in the field of BBO and other theories, triatomic scattering and molecule-surface reaction dynamics. On the other hand, Chapters \ref{sec:math} to \ref{ssec:mol} demonstrate the detailed theoretical development on the formulation of BBO treatment as well as its associated basic mathematical theorems. Chapters \ref{sec:spec} and \ref{sec:scat} present the applications of BBO theory on nonadiabatic chemistry in spectroscopic and reactive scattering, respectively. The NACTs as friction in classical EOM for reaction dynamics is described in Chapter~\ref{sec:traj}. Chapters \ref{sec:phase} and \ref{sec:surface} display the theory and application of Jahn-Teller effect (JTE) in phase transition of solids as well as electron-phonon and electron-hole pair couplings in molecule-surface scattering, respectively.\\

\noindent
Our work on nonadiabatic chemistry and in physical phenomena employing BBO \textit{vis-a-vis} JTE theory are presented as below:

\begin{itemize}
	\item The Period, 2003 - 2011: In this period, my former Ph.D students, Professors Dr. Panchanan Puzari (Tezpur Central University), Dr. Manabendra Sarma (IITG, Guwahati), Dr. Biplab Sarkar (North Eastern Hill University) and Dr. Amit Kumar Paul (Bose Institute, Kolkata) were primarily involved in generalizing the BBO theory for three and four coupled electronic manifold with explicit expressions of NACTs, ADT equations, curl-divergence equations and diabatic PESs as functions of nuclear coordinate dependent ADT angles. In order to explore the workability of such formalism, model diabatic Hamiltonian and non-adiabatic coupling matrix (NACM) were chosen for spectroscopic and scattering systems, like Na$_3$ (Cocchini model), C$_2$H (\textit{ab initio} data from Baer \textit{et. al.}) etc. In addition, Prof. Dr. Panchanan Puzari (Tezpur Central University) and Dr. Subhankar Sardar (Bhatter College, West Bengal) contributed in the development of the program for time-dependent discrete variable representation (TDDVR) dynamics and applied this approach on the vibronic coupling model (VCM) based diabatic Hamiltonian of several molecules, like, pyrazine, butatriene, benzene, allene, monofluoro benzene, isomers of difluoro benzene, trifluoro acetonitrile and many others. On the other hand, Dr. Tapas Sahoo (Haldia Institute of Technology) was engaged in the development of fully closed-coupled (FCC) reactive scattering and molecule-surface scattering formalism. 
	
	\item The Period, 2011 - 2019: During this time, Dr. Saikat Mukherjee (Nicolaus Copernicus University, Poland), Dr. Satyam Ravi (Vellore Institute of Technology, Bhopal), Dr. Srijeeta Talukder (Adamas University, Kolkata), Dr. Sandip Ghosh (Ram Ratan Singh College, Patna), Dr. Bijit Mukherjee (NISER, Bhubaneshwar) and Dr. Soumya Mukherjee (Maulana Azad National Institute of Technology, Bhopal) were involved in calculating {\it ab initio} PESs and NACTs for realistic molecular systems of spectroscopic interest, like Na$_3$, K$_3$ (Dr. Saikat Mukherjee), NO$_2$ (Dr. Saikat Mukherjee and Dr. Subhankar Sardar) and NO$_3$ (Dr. Bijit Mukherjee) as well as triatomic reactive scattering processes involving multiple electronic states, like H$_3^+$ (Dr. Saikat Mukherjee and Dr. Sandip Ghosh) and F+H$_2$ (Dr. Bijit Mukherjee). In case of spectroscopic systems, adiabatic PESs and NACTs were diabatized followed by dynamical calculations using Fast Fourier Transform (FFT) - Lanczos and TDDVR approaches to generate the spectral envelop. For scattering processes, the BBO based diabatic Hamiltonian were constructed to carry out FCC dynamics to compute reaction cross-sections and rate constants. On the other hand, Dr. Sandip Ghosh executed FCC scattering dynamics on D$^+$+H$_2$, O+OH, H+O$_2$ systems to calculate the reaction attributes. Dr. Souvik Mandal (University of Strasbourg, France) applied the molecule-surface scattering formalism on D$_2$-Cu(111) and H$_2$/D$_2$-Ni(111) systems. The JTE theory for the ground and excited state of $d^4/d^9$ systems was reformulated to evaluate the roto-vibrational states as function of barrier height including anharmonicity and the satelite peaks of ellipsometric experiment were reproduced by Dr. Joy Dutta (Durham University, UK). The model Hamiltonian based approach was modified through \textit{ab initio} calculations for the same manganate systems by Dr. Joy Dutta (Durham University, UK) and Dr. Satyam Ravi (Vellore Institute of Technology, Bhopal).

	
	\item The Period, 2019 - 2026: (a) A generalized algorithm and user-friendly 
	OpenMP parallelized program, ``ADT" (see \href{https://pubs.acs.org/doi/10.1021/acs.jctc.9b00948}{doi:10.1021/acs.jctc.9b00948} or 
	\href{https://github.com/AdhikariLAB/ADT-Program}{https://github.com/AdhikariLAB/ADT-Program}) has been developed to perform symbolic manipulation for the formulation of ADT equations and to carry out numerical calculation of various ADT quantities (like ADT equations, ADT matrix elements, diabatic potential energy matrix elements, etc.). This program package can interface various quantum chemistry softwares, like, Molpro, Columbus, Gamess and Gaussian to produce adiabatic PESs and NACTs (input data) and thereafter, to construct diabatic PESs matrices. The development of ``ADT" software is primarily carried out by Dr. Koushik Naskar (NextGen Invent, India), where Dr. Soumya Mukherjee (Maulana Azad National Institute of Technology, Bhopal) and Dr. Satyam Ravi (Vellore Institute of Technology, Bhopal) were involved for smooth implementation of the package. In addition, BBO theory was applied to construct diabatic Hamiltonian of spectroscopic systems, namely, C$_6$H$_6^+$, 1,3,5-C$_6$H$_3$F$_3^+$ (Dr. Soumya Mukherjee), C$_4$N$_2$H$_4$ (Dr. Saikat Hazra) and C$_5$H$_4$N (Dr. Tirthendu Sen and Mr. Avik Guchait), scattering processes, like HeH$_2^+$ and F+H$_2$ (Dr. Koushik Naskar), H$_3^+$ (Dr. Sandip Ghosh and Dr. Saikat Hazra) and H$_3$ (Mr. Amarendra Ghosh and Mr. Ahitagni Roy) as well as unit cell of perovskites, like MnO$_6^{9-}$ (Dr. Joy Dutta) and TiO$_6^{8-}$ (Dr. Mantu Kumar Sah). The implementation of BBO as well as JTE theory on perovskites was primarily carried out by Dr. Mantu Kumar Sah and Dr. Joy Dutta. \\
	
	 (b) The multi-state classical equation of motion considering NACTs as friction has been formulated and investigated to explore the formation of triatomic molecular species and to calculate cross-sections for H$_3^+$ and HeH$_2^+$ systems (Mr. Avik Guchait and Dr. Soumya Mukherjee in collaboration with Professor N. Satyamurthy and Professor M. Baer) in comparison with Trajectory Surface Hopping (TSH) and experimental data. Recently, a mixed Open-MP as well as MPI parallelized program, ``abc$\_$hyp" has been developed to carry out triatomic reactive scattering calculations, which can perform both adiabatic (single) as well as diabatic (multi-state) FCC time-dependent wave packet (TDWP) quantum dynamics in hypersperical coordinates and generate reaction probabilities, state-to-state and integral cross-sections, rate constants as well as differential cross-sections for comparison with other theoretical and experimental data.
	
	\end{itemize}
	
	\noindent
	This book is a followup of our works carried out in collaboration and discussion with national and international colleagues, where other contemporary theoretical developments as well experimental results are discussed in length. In particular, we wish to thank the internationally well-known scientists, Professor Michael Baer, Hebrew University of Jerusalem, Israel; Professor Ant\'onio Joaquim de Campos Varandas, University of Coimbra, Portugal; Professor J\"orn Manz, Freie Universit\"at, Berlin; Professor Kazuo Takatsuka, The University of Tokyo, Japan; Professor Peter Saalfrank, University of Potsdam, Germany; Professor Natalia N. Kovaleva, Lebedev Physical Institute of the Russian Academy of Sciences; Professor Geert-Jan Kroes, Professor Mark Somers and Professor J\"org Meyer, Leiden University; and Professor Jean Christophe Tremblay, Theoretical and Physical Chemistry Lab, Université de Lorraine, France. We also thank Professor N. Sathyamurthy and Dr. K. R. Shamasundar, IISER Mohali, Mohali; Professor Debasis Mukhopadhyay and Professor Pinaki Chowdhuri, University of Calcutta; Dr. Rahul Sharma, St. Xavier's College for collaborative research works. On various occasions, we discuss on critical topics in quantum mechanics with senior colleagues, Professor Debashis Mukherjee, IACS Kolkata; Professor Sankar Prasad Bhattacharya, IACS Kolkata; Professor K. L. Sebastian, IISc Bangalore;  Professor Deb Shankar Ray, IACS Kolkata; Professor Sourav Pal, Ashoka University.

	\vspace{0.3cm}
	
	\noindent
	 I am grateful to extramural funding agencies, namely, Science and Engineering Research Board (SERB), Anusandhan National Research Foundation (ANRF), Board of Research in Nuclear Sciences (BRNS), Council of Scientific and Industrial Research (CSIR) as well as our institute, Indian Association for the Cultivation of Science (IACS), Kolkata for research grants to make a summary of our teaching and research work in the form of a book.\\
	 
	 \noindent
	 Professor Satrajit Adhikari\\
	 IACS, Kolkata, India

\chapter*{Abbreviations}\label{sec:bio}
\addcontentsline{toc}{chapter}{Abbreviations}

	\begin{table}[!htp]
	\centering
	\begin{tabular}{|c|c|}
		\hline 
ADT & Adiabatic-to-Diabatic Transformation \\
AIMS & {\it Ab Initio} Multiple Spawning \\
BBO & Beyond Born-Oppenheimer \\
BO & Born-Oppenheimer \\
CASSCF & Complete active-space self-consistent ﬁeld\\
CI & Conical Intersection \\
CS & Configuration Space \\
DD & Direct Dynamics \\
DD-vMCG & DD variational multi-configurational 
Gaussian \\
DOF & Degree of Freedom \\
EBO & Extended Born-Oppenheimer\\
EF & Exact Factorization \\
FC & Franck-Condon\\
FCC & Fully Close Coupled \\
IREP & Irreducible Representations \\
JT & Jahn-Teller \\
LH & Longuet-Higgins \\
PES & Potential Energy Surface \\
PE & Photoelectron \\
PJT & Pseudo Jahn-Teller\\
MATI & Mass-Analyzed Threshold Ionization\\ 
MCTDH & Multi-Configurational Time-Dependent Hartree \\ 
MRCI & Multi-Reference Configuration Interaction \\
MS & Molecular Symmetry\\
NACT & Non-Adiabatic Coupling Terms \\
RT & Renner-Teller \\
SE & Schr\"odinger Equation\\
SHS & Sub-Hilbert Space\\
TDDVR & Time-Dependent Discrete Variable Representation\\
TSH & Trajectory Surface Hopping \\
VCM & Vibronic Coupling Model\\
\hline
	\end{tabular}

\end{table}

\chapter{Introduction}\label{sec:intro1}

\section{Theory of Electron-Nuclear Couplings}\label{sec:intro1.1}
\noindent
For over a century, Born-Oppenheimer (BO) theory~\cite{BO} and its generalization by Born and Huang~\cite{BH} appear as a cornerstone in Quantum Chemistry for investigating static and dynamic properties of molecular species, chemical reactions and the phenomena in solid state physics. As per the BO treatment,
the electronic motion shows parametric dependence on the nuclei due to the huge mass difference of electrons and nuclei. Such difference introduces the BO approximation, which is expected to be applicable successfully only for few chemical and physical processes occurring at lower energy domain of ground electronic state. On the contrary, a wide variety of molecular phenomena can be observed, which clearly indicates breakdown of BO approximation, specifically at the close vicinity of conical intersection(s) (CIs) / seam(s) arising from degeneracy(ies) between the electronic states. In other words, such failure primarily originates due to the coupling between electronic and nuclear motion leading to nonadiabatic events. Those degeneracies (CIs/seams) are omnipresent in various processes, namely, photosynthesis, vision, charge transfer chemical reactions, solar energy conversion to photochemical reactions
engaging multiple electronic states,~\citep{baer_jcp66,adhikari_jcp111,varandas_jcp112,martinez_chemrev} which cannot be probed with enough ``precision" from single electronic state BO perspective. As a result, the nonadiabatic phenomena and the breakdown of BO approximation are frequently observed in nature, which demands an extensive consideration of electron-nuclear couplings for accurate description of chemical and physical processes.  \\

\noindent
The adiabatic potential energy surfaces (PESs) and the nonadiabatic coupling terms (NACTs) are the two major outcomes of BO treatment, where both the quantities (PESs and NACTs) govern the nuclear motion in a molecular event. In this treatment, the electron Hamiltonian produces adiabatic PESs, which are physically realizable dictating the movement of the nuclei on a specific PES. On the other hand, the momentum couplings between the nuclear and electronic motion represented by NACTs dictate the transition of the nuclei among the PESs and thereby, multiple state processes, can only be understood by incorporating such couplings in the structure and dynamics calculations. Several earlier research works unveiled that calculations on photoexcitation 
spectra,~\citep{bijit_cp515,soumya_bz,soumya_tfb} charge transfer chemical reactions,~\cite{baer_jcp91,
	grimbert_jcp108} scattering cross sections~\cite{baer_jcp105,
	varandas_jcp112,baer_pra62,adhikari_pra62,sarkar_jcp124,sarkar_jpca112,sarkar_ijqc109,akpaul_jcp131} and phase transition of solids~\citep{joy_jpca}
exhibit substantial disparities between single and multiple surface cases along with experimental results, which indicates the necessity of transition from BO approximation to beyond Born-Oppenheimer (BBO) treatment for more accurate
calculations of any molecular process.\\

\noindent
In order to investigate multi-state nuclear dynamics involving electronic transitions from one surface to another, one needs to incorporate the NACTs in the nuclear Schr\"odinger Equation (SE), which often appear as singular functions~\cite{hellmann,feynman_pr56} at the close neighborhood of degenerate point(s)/seam(s). As a matter of fact, 
the major bottleneck of adiabatic framework is the non-conservation of energy~\citep{friction_jpca128} and numerical inaccuracies originating from the singularities of NACTs. On the other hand, one can overcome such issues by transforming the SE from adiabatic to diabatic picture. This can be achieved with the aid of a unitary transformation of the adiabatic SE known as 
adiabatic-to-diabatic transformation (ADT). In the new framework known as diabatic representation of SE, those often singular kinetic couplings transform into smooth, continuous and single-valued potential couplings. The study of CIs vis-\`a-vis NACTs and their effects in polyatomic species on spectroscopic and scattering events are the most important areas of nonadiabatic chemistry, which drew the attention of allied scientific community over decades. \\

\noindent
In the early 1960s, Longuet-Higgins (LH) unveiled that a BO adiabatic eigenfunction  undergoes sign flipping along a 
closed contour of nuclear coordinates encapsulating a CI~\cite{longuet-higgins,longuet-higgins_prsl344} and accumulates a geometric phase. In order to resolve such inconsistency arising from multi-valuedness of the BO eigenfunction, Herzberg and Longuet-Higgins introduced a complex phase factor~\cite{herzberg_dfs35} in the adiabatic eigenfunction~\cite{baer_book} in an ad hoc manner. Alternatively, Mead and Truhlar\citep{mead_jcp77} incorporated a 
vector potential in the nuclear kinetic energy operator to remove the multi-valued property of the adiabatic eigenfunctions, which is a 
reminiscent of LH phase. Varandas {\it et al.}~\cite{varandas_cpl61} first numerically investigated the LH phase for a triatomic 
molecule (LiNaK) by following the dominant coefficient of the electronic eigenfunction while roaming adiabatically around the CI, whereas Adhikari and coworkers~\citep{saikat_jpca117} showed sign inversions of such coefficient of 
adiabatic electronic eigenfunction along a contour encircling symmetry-driven and accidental Jahn-Teller (JT) CI for Na$_3$ cluster.\\

\section{The Methods of Diabatization}\label{sec:intro2}
\noindent
In the adiabatic representation,~\cite{ytlee_jcp82_3045,werner_jcp109} the singular nature of the NACTs
around the CI point(s), in principle, results non-conservation of energy~\citep{friction_jpca128} and numerical inaccuracies in the dynamical calculations. Since the electron-nuclear couplings play a vital role for the electronic
transition and show molecule specific functional features with multiple CIs over configuration space (CS), it is necessary to reformulate 
the nuclear SE to avoid the singular kinetic coupling terms in an alternative way, known as diabatic representation.
Hobey and McLachlan attempted to carry out a transformation to recast the NACTs from the BO close-coupling~\cite{hobey_jcp33} to potentially coupled 
diabatic~\cite{litchen_pr164} SE. Later on, Smith extended such formulation for a diatomic molecule by successfully eliminating NACTs from the adiabatic representation to achieve the corresponding diabatic one~\cite{smith_pr179}. In addition, several other diabatization schemes have been developed for 
realistic systems through out the last three/four decades, namely, vibronic coupling model (VCM), exact factorization (EF) scheme, direct dynamics (DD) approaches, and many others. \\

\noindent
In VCM approach,~\citep{koppel_jcp110,koppel_jcp150,viel_jcp120} diagonal and off-diagonal diabatic matrix elements are expanded with linear, bilinear as well as higher order polynomials of nuclear coordinates with respect to the minima of ground adiabatic PES ($V_0(\textbf{R})$) or more preferably, the CI point ($\textbf{R}_0$) between the first two adiabatic states. The bottleneck of such formalism is its limited applicability around the  
domain of Franck-Condon (FC) region at the close vicinity of reference geometry $\textbf{R}_0$ for the dynamics of spectral features, namely, photoelectron (PE), emission/absorption, Mass-Analyzed Threshold Ionization (MATI) spectra, etc, but for the reactive scattering dynamics, the accuracy of the model is not guaranteed. On the other hand, in the exact factorization method developed by Gross and coworkers,~\citep{gross_prl105,gross_jcp137} the electronic and nuclear SEs are solved simultaneously with the aid of time-dependent PESs, vector potential and a coupling operator, which essentially represent the coupling between electron-nuclear coordinates. In this approach, the asymptotic wavefunctions do not represent any specific adiabatic quantum state and thereby, it is difficult 
to initialize the dynamics of the multi-dimensional system in correlation with experimental conditions. Several direct-dynamics (DD) methods, namely, 
on-the-fly diabatization version of multi-configurational time-dependent Hartree (MCTDH)\citep{worth_irpc27,richings_jcp148} and DD variational multi-configurational 
Gaussian (DD-vMCG)\citep{worth_irpc34,worth_jpca119} approaches were also developed to include the contribution of nonadiabatic couplings in the dynamics. Martinez and co-workers \cite{SPAWN1,SPAWN2} have developed {\it Ab Initio} Multiple Spawning (AIMS) method for nonadiabatic dynamics, where the Gaussian Wave Packets (GWPs) follow classical trajectories, which are spawned during the bifurcation of wavepacket 
in a region with strong nonadiabatic coupling. Nevertheless, among all the proposed theories of diabatization, the first principle based approach developed by M. Baer~\cite{baer_cpl35,baer_jcp66,baer_phys_rep358,baer_book} and later on, generalized and widely applied by Adhikari and coworkers~\citep{sarkar_jcp124,sarkar_jpca112,soumya_jpca121,bijit_fh2,bijit_irpc38,naskar_jctc} has been proven to be theoretically ``exact" and numerically ``accurate" one.
\\

\section{Generalization on First Principle Diabatization}\label{sec:intro3}
\noindent
In order to formulate the nuclear equations of motion for atom-diatom collision process, M. Baer proposed a 
set of differential equations for ADT within a given sub-Hilbert space, where those equations are integrated
along a contour over two-dimensional cross-sections of the multi-dimensional nuclear CS~\cite{baer_cpl35,baer_mp40}. While computing the transformation angle(s), 
the NACTs are required for integrating those differential equations along a closed contour encapsulating the CI point(s) or seam(s)
resulting those angle(s) as $n\pi$ with $n$ being an integer~\cite{baer_phys_rep358,baer_mp75}. On the other hand, Mead, Truhlar and Kendrick
explored the removable and non-removable components of NACTs~\cite{mead_jcp77,kendrick_cp277} depicting only the 
longitudinal (removable) component of NACTs could be wiped out. Furthermore, Yarkony and co-workers unveiled that at the close neighborhood of a CI,   
the removable part of the derivative coupling vector appear as a singular function and 
non-removable component remains insignificant~\cite{yarkony_mp93}. \\

\noindent
Later on, the BBO approach was generalized by Adhikari and co-workers for any three or more coupled
electronic states with extensive formulation of ADT equations, NACT expressions, curl conditions, 
curl-divergence equations and diabatic PESs matrices
in terms of such mixing (ADT) angles~\cite{sarkar_jpca112,akpaul_cpl508,
	akpaul_jcp135,mukherjee_jpca117,mukherjee_cp440,mukherjee_jcp141,sardar_cp416,mukherjee_jcp143,soumya_jpca121,bijit_mp115,bijit_cp515,soumya_jcp150}.
In other words, this generalization is a non-trivial formulation of ADT equations for non-Abelian cases, where curl of the nonadiabatic coupling
elements appears non-zero for more than two-dimensional sub-space of electronic manifold.
Moreover, a generalized algorithm and user-friendly 
	OpenMP parallelized program, ``ADT" (see \href{https://pubs.acs.org/doi/10.1021/acs.jctc.9b00948}{https://
		pubs.acs.org/doi/10.1021/acs.jctc.9b00948} or 
	\href{https://github.com/AdhikariLAB/ADT-Program}{https://github.com/AdhikariLAB/ADT-Program})\citep{naskar_jctc} has been developed to carry out symbolic manipulation for the formulation of ADT equations and perform numerical calculation of various ADT quantities (like ADT equations, 
	ADT matrix elements, diabatic potential energy matrix elements, etc.). The program package can interface with various quantum chemistry softwares, like, Molpro,\citep{MOLPRO_brief} Columbus,\citep{columbus} Gamess\citep{gamess} and Gaussian\citep{g16} to obtain the input data (adiabatic PESs and NACTs) to construct diabatic PESs matrices. For a given sub-Hilbert space (SHS), an accurate transformation
to the diabatic framework can be ensured if and only if a condition known as 
\textit{curl condition} is satisfied by the NACT vector fields~\cite{baer_phys_rep358,baer_book,mantu_ijqc_curl}. 
Later on, Adhikari and coworkers developed a practical way of using curl condition for the formulation of approximated 
Extended Born-Oppenheimer (EBO) equation~\cite{sarkar_jcp124,akpaul_jcp131}, which essentially connect the origin of geometric phase.
In subsequent works, a detailed study of molecular symmetry (MS)
and its relevance for exploring exact sign of NACTs was carried out. In other words, appropriate irreducible representations (IREPs) of the 
MS group are used for correcting the signs of the couplings~\cite{mukherjee_jpca117,mukherjee_cp440} at different geometries over the nuclear CS. \\

\section{Spectroscopy, Reactive Scattering and Phase Transition}\label{sec:intro4}
\noindent
The first principle based BBO formalism has been employed to construct diabatic Hamiltonian  
for spectroscopically interesting molecules~\citep{mukherjee_jpca117,mukherjee_cp440,sardar_cp416,mukherjee_jcp143,
	soumya_jpca121,bijit_mp115,soumya_acs,bijit_cp515,
	soumya_jcp150,soumya_comptc,bijit_irpc38,joy_pccp,soumya_bz,soumya_tfb,saikat_pyra}, triatomic scattering processes\citep{mukherjee_jcp141,jcp-147-074105-SA,bijit_fh2,bijit_irpc38,joy_pccp} as well as solid state perovskites~\citep{joy_jpca} exhibiting a wide variety of nonadiabatic interactions, namely, JT\citep{JT_prs161,bersuker_book,Pablo_prb71} CIs/seams (symmetry driven~\citep{bijit_mp115} degeneracies), Jahn-Teller (JT) type accidental CIs/seams (symmetry allowed~\citep{bijit_mp115} or no symmetry~\citep{bijit_mp115} degeneracies), Renner-Teller (RT)\citep{renner_zp92,Hizhnyakov_jcp147} 
interactions and pseudo Jahn-Teller (PJT)\citep{bersuker_book} interactions (no degeneracy, but non-zero NACTs between the electronic states). It is interesting
to note that for (a) JT or JT type accidental CIs/seams, the ADT angles acquire the magnitude of integer multiple of $\pi$; (b) for RT interactions, those angles attain the value of integer multiple of $2\pi$; and finally, (c) for PJT interactions, such angles attain zero (0) value at the end of closed contour.
It is important to highlight that the theoretically ``exact" and numerically ``accurate"
BBO theory is equally applicable in Franck-Condon (FC) domain as well as in the asymptotic limit. In other words, BBO constructed diabatic PESs and couplings have been used in nuclear dynamics to simulate photoelectron spectra of several molecules and transition metal complexes (Na$_3$,\citep{mukherjee_jpca117} K$_3$,\citep{mukherjee_cp440} NO$_2$,\citep{mukherjee_jcp143} NO$_3$,\citep{bijit_mp116,bijit_cp515} C$_6$H$_6^+$,~\citep{soumya_jpca121,soumya_bz} 1,3,5-C$_6$H$_3$F$_3^+$,~\citep{soumya_jcp150,soumya_tfb} C$_4$N$_2$H$_4$~\citep{saikat_pyra}, MnO$_6^{9-}$~\citep{joy_jpca} and TiO$_6^{8-}$~\citep{mantu_tio6_jcp}) as well as to calculate various reaction attributes (reaction probabilities, integral cross-sections and rate constants) for triatomic reactive systems, namely, H$_3^+$ (three states)\citep{mukherjee_jcp141,jcp-147-074105-SA}, HeH$_2^+$ (four states)~\citep{naskar_jpca_heh2+,koushik_heh2+_dyna} and F+H$_2$ (three states)~\citep{bijit_fh2,koushik_fh2_dyna}.

\section{Phonons and Electron-Hole Pair Effects on Surface Reactions}\label{sec:intro5}

\noindent
Metal surfaces at finite temperatures impart crucial effects on heterogeneous catalyzed reactions, which 
is broadly acknowledged since the days of Haber process, i.e., production of ammonia (NH$_3$) from 
hydrogen (H$_2$) and nitrogen (N$_2$) in presence of iron catalyst at 450 K. The importance of gas-surface 
chemical reactions also ranges from growth of semiconductor devices to hydrogen storage in metals. As a 
result of the wide-ranging relevance, the nature and mechanism of elementary reactions taking place over 
metal surface have been critically investigated both experimentally~\cite{HA2,CTD,CTH,CTR2,CTR,HAM,EXPNEW2,EXPNEW3,Wodtke,Chadwick} and theoretically~\cite{HO1,CDGJK,HA1,SAL,SAG,TSS,TSS1,TSS2,TSSM,FNGJK1,SPO,MWMF,QD-AIMD-H2-D2-Cu-review-2016,H2-Cu-multiple,PSMFS,VV-Hartree-2021,EAM-SCM-QC-Mark-2021,SRP-DF-new-H2-Cu-Kroes-2021,Kroes-Review-2021,SAMARK,Joy_D2_jcp_2022,Mantu_jcp_2024,Bauke_jcp_2024,Mantu_JPCC_2026} for the past few 
decades. In this context, dissociation/scattering of H$_2$/D$_2$ on/from various facets 
of different transition metal surfaces are being considered as the prototypes for studying physisorption,  activated chemisorption, and elastic/inelastic collision phenomena. Traditionally, the approximate adiabatic (BO)-PESs for such dissociative scattering systems are constructed by using density functional theory (DFT), which are then incorporated in time - dependent dynamical [quantum dynamical (QD)~\cite{QD-method-1993,SPO,SRP-DF-24-CH4-step-Ni-2018,SRP-DF-23-CH4-step-Ni-2019,SMIRPC,QD-AIMD-H2-D2-Cu-review-2016,Kroes-Review-2021}, quasi-classical trajectory (QCT)~\cite{SC1,MWMF,SRP-DF-8-H2-D2-Cu-LDFA-ODF-elhp-2018,H2-Cu-multiple,HD-NNP-perspective-2020,EAM-SCM-QC-Mark-2021}] calculations. However, at finite surface temperature situations, effects of surface mode vibrations (phonons) and electron-hole pair (elhp) excitations (non - adiabatic effect) on experimental outcomes become significant. In this regard, there are just few theoretical approaches [e.g., \textit{ab initio} molecular dynamics (AIMD)~\cite{FNGJK,FNGJK1,PBE-19-CH4-on-Ni-AIMD-2014,QD-AIMD-H2-D2-Cu-review-2016,SRP-DF-17-H2-Cu-GLO-AIMDEF-elhp-2017}, lattice reconstruction sudden (LRS) model~\cite{AKT1,AKT}, single/modified oscillator (SO) models~\cite{HA1,HO1,SAL,UZER1}, electron friction (EF)~\cite{JIJ1,SRP-DF-8-H2-D2-Cu-LDFA-ODF-elhp-2018} approach, independent electron surface hopping (IESH) model~\cite{Dynamical-models-review-2018,IESH-2009,JCTully-elhp1}] to take into account the effects of these two channels (phonon modes and elhp excitations). \\

\noindent
Adhikari and coworkers developed a time and temperature dependent effective Hamiltonian (EH)-based model by incorporating molecular degrees of freedom (DOFs)-phonon modes (for linear and quadratic) couplings as well as molecular DOFs-elhp couplings within the framework of mean-field approximation.~\cite{SAG,TSS,TSS1,TSS2,TSSM,SAMARK,SM-JTCC,SM-MOLPHYS,VV-Hartree-2021,Joy_D2_jcp_2022,Mantu_jcp_2024,Mantu_JPCC_2026} In this approach, effective Hartree potentials (EHPs) are constructed by using quantum state distributions [Bose-Einstein statistics (for phonon modes) and Fermi-Dirac statistics (elhps)] of respective dissipation channels (i.e., phonon modes and elhps) at a particular surface temperature situation. During the entire time course of diatom-metal collision, this method considers six (6) DOFs of incoming diatomic projectile as well as phonon modes/elhps, and thereby constituting a seven dimensional (6+1)D quantum scattering problem. An efficient scheme was developed by using MPI-OpenMP parallelization technique (which can solve 10$^9$ number of coupled differential equations) to calculate observables, i.e., reaction probabilities and state-to-state transition probabilities for various collision energies of diatom. In recent time, with the aid of such EH (linear) model and accurate potential, novel vibrational quantum effects on scattering outcomes were revealed. On the other hand, amalgamation of our EH (linear-quadratic) model\cite{VV-Hartree-2021,Joy_D2_jcp_2022,Mantu_jcp_2024,Mantu_JPCC_2026} with a chemically accurate PES\cite{MWMF} is on the process to investigate effects of anharmonic molecular DOFs-surface modes couplings and anharmonic phonon modes on dissociative scattering processes in a more accurate way. \\

\noindent
This book is focused on a concise overview of the theoretical developments vis-\'a-vis 
applications of the BBO approach in conjunction with JT theory, TDDVR approach, Fully Close Coupled (FCC) dynamical approach as well as TSH theory, which is expected
to be beneficial for the associated scientific community.
This book focuses on the theoretical background of the aforementioned theories, followed by
implementation of such approaches in spectroscopic systems, triatomic scattering processes as well as solid state systems exhibiting phase transition phenomenon. 
Finally, the accuracy of the theoretical results is compared with the available experimental as well as other theoretical findings.

\bibliographystyle{aipnum4-1}
\bibliography{ref_book,surface/myrefs_H2.bib,surface/ref_rev.bib, surface/refs_article.bib, surface/refs_new.bib}

\chapter{Quantum Mechanics and Group Theory of Symmetry Breaking}\label{sec:math}

\vspace{0.5 cm}

\section{Degenerate Perturbation Theory}\label{sec:deg}
\noindent
Any non-linear molecular system with degenerate electronic states undergoes distortion to form a system of lower symmetry and energy 
by removing the said degeneracy. Before going into the physical insight of Jahn-Teller (JT)~\citep{JT,longuet-higgins_prsl344} problem, 
we need to discuss the fundamentals of degenerate perturbation theory.\\

\noindent
The unperturbed time-independent Schr\"odinger Equation (TISE) for $m$-fold degenerate states can be written as:
\begin{eqnarray}
	\hat{H}^0\phi_n^{(0)}=E_n^{(0)}\phi_n^{(0)}, \qquad 1\leq n\leq m.
	\label{Eq:unpert}
\end{eqnarray}
On the other hand, the perturbed Hamiltonian and the associated TISE are:
\begin{eqnarray}
	\hat{H}\phi_n &=& E_n\phi_n \\
	\hat{H} &=& \hat{H}^0 + \lambda \hat{V},
	\label{Eq:SE}
\end{eqnarray}
where $\lambda$ represents the extent of perturbation and $\hat{V}$ signifies the perturbation, namely, electric, magnetic, ligand-field, crystal-field, etc.\\

\noindent
The perturbed molecular wave function and total energy can be expanded in terms of $\lambda$:~\citep{klein_jcp61}
\begin{subequations}
	\begin{eqnarray}
		\phi_n &=& \psi_n^{(0)}+\lambda\phi_n^{(1)}+\lambda^2\phi_n^{(2)}+...\\
		E_n &=& E_n^{(0)}+\lambda E_n^{(1)} + \lambda^2 E_n^{(2)} +... \qquad 1\leq n\leq m,
	\end{eqnarray}
	\label{Eq:expand}
\end{subequations}
\noindent
where $\psi_n^{(0)}$ represents the zeroth order wavefunction and can be written as linear combination of the $m$-fold unperturbed eigenfunctions:
\begin{eqnarray}
	\psi_n^{(0)}= \sum_{i=1}^m c_i^n\phi_i^{(0)}
	\label{Eq:zeroth}
\end{eqnarray}
In other words, one can construct $m$-fold orthogonal wavefunctions ($\psi_n^{(0)}$) from a set of $m$-fold linearly independent eigenfunctions ($\phi_i^{(0)}$). Moreover, the higher order corrections on wavefunctions are assumed to be orthogonal with the zeroth order, $\langle\psi_n^{(0)}\vert\psi_n^{(m)}\rangle=\delta_{0m}$.\\

\noindent
Therefore, we can write:
\begin{eqnarray}
	&\left(\hat{H}^0 + \lambda\hat{V}\right)\left(\psi_n^{(0)}+\lambda\phi_n^{(1)}+
	\lambda^2\phi_n^{(2)}+...\right)& \nonumber \\
	&= \left(E_n^{(0)}+\lambda E_n^{(1)} + \lambda^2 E_n^{(2)} +... \right)
	\left(\psi_n^{(0)}+\lambda\phi_n^{(1)}+\lambda^2\phi_n^{(2)}+...\right)&   
\end{eqnarray} 
While equating the first order terms, one can arrive:
\begin{eqnarray}
	\hat{H}^0\phi_n^{(1)} + \hat{V}\psi_n^{(0)} = E_n^{(0)}\phi_n^{(1)} + E_n^{(1)}\psi_n^{(0)}
\end{eqnarray}
Multiplying the above equation by $\phi_i^{(0)\star}$ and subsequently, integrating over electronic space:
\begin{eqnarray}
	&\langle\phi_i^{(0)}\vert \hat{H}^0 \vert \phi_n^{(1)}\rangle + \langle\phi_i^{(0)}\vert \hat{V} 
	\vert \psi_n^{(0)}\rangle = E_n^{(0)} \langle\phi_i^{(0)}\vert\phi_n^{(1)}\rangle + 
	E_n^{(1)}\langle\phi_i^{(0)}\vert\psi_n^{(0)}\rangle, & \nonumber \\ \nonumber \\
	&\Rightarrow E_n^{(1)}\langle\phi_i^{(0)}\vert\psi_n^{(0)}\rangle = \langle\phi_i^{(0)}\vert \hat{V} 
	\vert \psi_n^{(0)}\rangle.&  \nonumber
\end{eqnarray}

\noindent
Using the expansion of $\psi_n^{(0)}$ in terms of the zeroth-order eigenfunctions 
(see Eq. \ref{Eq:zeroth}), the following equation arrives: 
\begin{eqnarray}
	&\qquad& \sum_{j=1}^m \left[\langle\phi_i^{(0)}\vert \hat{V}\vert \phi_j^{(0)}\rangle
	- E_n^{(1)}\langle\phi_i^{(0)}\vert\phi_j^{(0)}\rangle\right]c_j^n = 0 \nonumber \\
	&\Rightarrow& \sum_{j=1}^m \left[V_{ij} - E_n^{(1)}\delta_{ij}\right]c_j^n = 0,  \quad 1\leq i\leq m,
	\quad 1\leq n\leq m.
\end{eqnarray}
\noindent
Thus, $m$ number of linear, homogeneous equations for $c_1^n,c_2^n,...,c_m^n$ become:
\begin{eqnarray}
	\left(V_{11}-E_n^{(1)}\right)c_1^n + V_{12}c_2^n + \ldots + 
	V_{1m}c_m^n &=& 0 \nonumber \\
	V_{21}c_1^n + \left(V_{22}-E_n^{(1)}\right)c_2^n + \ldots +
	V_{2m}c_m^n &=& 0 \nonumber \\
	\vdots \\
	V_{m1}c_1^n + V_{m2}c_2^n + ... + \left(V_{mm} -E_n^{(1)}\right)c_m^n &=& 0 \nonumber
\end{eqnarray}
\noindent
The solution of the above set of equations 
will exist if and only if the determinant constituted by the coefficients does not vanish. Therefore, the secular
determinant is defined as:
\begin{eqnarray}
	\begin{vmatrix}
		V_{11}-E_n^{(1)} & V_{12} & \cdots & V_{1m} \\
		V_{21} & V_{22}-E_n^{(1)} & \cdots & V_{2m} \\
		\vdots  & \vdots  & \ddots & \vdots  \\
		V_{m1} & V_{m2} & \cdots & V_{mm}-E_n^{(1)} 
	\end{vmatrix} = 0
	\label{Eq:secu}
\end{eqnarray}
where the $m$ roots, $E_1^{(1)}, E_2^{(1)},...,E_m^{(1)}$ are the first 
order corrections to the eigenvalues ($E_n^{(0)}$) lifting the degeneracies.\\

\noindent
For a doubly degenerate system, one can write:
\begin{eqnarray}
	\begin{vmatrix}
		V_{11}-E_n^{(1)} & V_{12} \\
		V_{21} & V_{22}-E_n^{(1)} 
	\end{vmatrix} = 0
\end{eqnarray}
where the two roots of the equation will be:
\begin{eqnarray}
	E_{n,\pm}^{(1)} = \dfrac{1}{2}\left[V_{11} + V_{22} \pm\sqrt{(V_{11} - V_{22})^2+4|V_{12}|^2} \right]
\end{eqnarray}
\noindent
The fundamental result of degenerate perturbation theory 
is that the roots correspond to the correction ($E_{n,\pm}^{(1)}$) over the unperturbed energies ($E_{n}^{(0)}$)
and hence, there is a possibility of getting degeneracy lifted.

\section{Degenerate Perturbation Theory: JT Distortion}\label{sec:jt}		
\noindent
If there is coupling between vibrational modes and electronic Hamiltonian (vibronic coupling) of a molecular system, the electron-nuclear Hamiltonian [$\hat{H}^0=
\hat{H}_e (\{Q_k\}= \boldsymbol{0})]$ can be expanded in terms of normal mode coordinates:
\begin{eqnarray}
	\hat{H} = \hat{H}^0 + \sum_k \hat{V}_kQ_k + 
	\dfrac{1}{2}\sum_{kl}\hat{V}_{kl}Q_kQ_l + \cdots,
	\label{Eq:ham_taylor_1}
\end{eqnarray}
\noindent
$\hat{H}^0$ represents electronic Hamiltonian at equilibrium nuclear configuration. On the other hand, $\hat{V}_k \left(=\frac{\partial H}{\partial Q_k}\right)$ and 
$\hat{V}_{kl} \left(=\frac{\partial^2 H}{\partial Q_k \partial Q_l}\right)$ are 
the first and second order perturbations, respectively, due to normal modes, $\{Q_k\}s$.\\

\noindent
In such case, unperturbed TISE is defined as in Eq. \ref{Eq:unpert}:
\begin{eqnarray}
	\hat{H}^0\phi_n^{(0)}=E_n^{(0)}\phi_n^{(0)}, \qquad 1\leq n\leq m, \nonumber
\end{eqnarray}
where $E_n^{(0)}$ is the eigenvalue of $m$-fold degenerate electronic 
states and $\{\phi_n^{(0)}\}$s form a complete set of orthogonal eigenfunctions.	
In a particular nuclear configuration, the $m$-fold degenerate eigenvalue $E_n^{(0)}$ may either split 
into number of different energy levels $E_{n,m}$ or it may get lifted without removing 
the degeneracy.\\

\noindent
The matrix elements for first order energy correction on the $n^{th}$ state ($E_n^{(1)}$) are expressed in the following equation:
\begin{eqnarray}
	\sum_{j=1}^m \left[\langle\phi_i^{(0)}\vert \hat{V}\vert \phi_j^{(0)}\rangle
	- E_n^{(1)}\langle\phi_i^{(0)}\vert\phi_j^{(0)}\rangle\right]c_j^n = 0
\end{eqnarray}

\noindent
When only the linear perturbation w.r.t the normal mode coordinates is considered (see Eq. \ref{Eq:ham_taylor}), the matrix elements for the perturbation can be expressed as:
\begin{eqnarray}
	V_{ij}=\sum_{k} Q_{k} \langle \phi_i^{(0)} \vert \hat{V}_{k}\vert \phi_j^{(0)}\rangle.
\end{eqnarray}
\noindent
If $Q^{0} (\{Q_k\}= \boldsymbol{0})$ coincides with the most stable nuclear configuration of a molecule, the first order perturbation matrix elements should vanish. On the other hand, if the direct product on the irreducible representations (IREPs) of the components of the integrand, $\Gamma(\phi_i^{(0)})\otimes\Gamma(\hat{V}_k)\otimes\Gamma(\phi_j^{(0)})$ does not contain 
the totally symmetric representation ($\Gamma^0$), the perturbation will be zero. 
Since $V_{ij}$ is a number, it is invariant under the transformation of 
all symmetry elements of a group.\\

\noindent
In other words, let (a) $\hat{V}_k \left(=\frac{\partial H}{\partial Q_k}\right)$ and $Q_k$ transform 
according to irreducible IREP $\Gamma^{\beta}$ of the symmetry group at 
configuration $Q^{(0)}$; (b) degenerate wave functions $\phi_i^{(0)}$ transform according to $\Gamma^{\alpha}$,
the necessary condition for JT effect takes the following form:~\citep{JT,longuet-higgins_prsl344,JT-review-Bersuker,JT1}
\begin{eqnarray}
	\Gamma^0 \in \Gamma^{\beta} \otimes \left[ (\Gamma^{\alpha})^2 \right], 
	\label{Eq:JTcond}
\end{eqnarray}
where $\left[(\Gamma^{\alpha})^2 \right]$ represents the symmetrized product of the representation $\Gamma^{\alpha}$.\\

\noindent
Employing Eq. \ref{Eq:JTcond}, one can define four different cases for JT distortion, namely,

\begin{itemize}
	
	\item {\color{black}Electronic state is non-degenerate $\left[\text{i.e.}\,\, (\Gamma^{\alpha})^2=\Gamma^0\right]$ 
		and nuclear displacement ($Q_k$) is totally symmetric $\left[\text{i.e.}\,\, \Gamma^{\beta}=\Gamma^0\right]$. For such case,
		the contribution of perturbation is not necessarily zero but system will readjust 
		until the contribution vanishes, since it involves a symmetric mode and does not 
		destroy the symmetry of the system.}
	
	\item  	{\color{black}Electronic state is non-degenerate $\left[\text{i.e.}\,\, (\Gamma^{\alpha})^2=\Gamma^0\right]$ 
		and nuclear displacement ($Q_k$) is non totally symmetric $\left[\text{i.e.}\,\, \Gamma^{\beta}\neq\Gamma^0\right]$. For such case, the contribution of perturbation is always zero.}
	
	\item  {\color{black}Electronic state is degenerate $\left[(\Gamma^{\alpha})^2=\Gamma^0 + \cdots \right]$ 
		and nuclear displacement ($Q_k$) is totally symmetric $\Gamma^0$. For such case, the contribution of perturbation is not necessarily zero, but situation is analogous to the first case.}	
	
	\item  {\color{black}Electronic state is degenerate $\left[(\Gamma^{\alpha})^2=\Gamma^0 + \cdots + \Gamma^{\beta} + \cdots \right]$ 
		and nuclear displacement ($Q_k$) is non totally symmetric $\left[\text{i.e.}\,\, \Gamma^{\beta}\neq\Gamma^0\right]$. For such case, the contribution of perturbation will be non zero and show the {Jahn-Teller effect}, 
		if the symmetrized product of the states contains the symmetry $\Gamma^{\beta}$.} In the 
	case of linear molecules, the product does not contain the symmetry of the nuclear displacement 
	and hence does not show JT effect.
	
\end{itemize}

\section{JT Condition and Symmetry Breaking}\label{sec:group}

\noindent
While exploring the symmetry breaking of electronic states in a molecule, one can use the concept of direct product between the electronic states and the normal modes, and analyze the survivality of the direct product integral in terms of their irreducible representation. 

\subsection{Symmetry Breaking in H$_2$O Molecule}\label{sec:h2o_group}

\begin{figure}[!htp]
	\centering
	\includegraphics[width= \textwidth, angle=0]{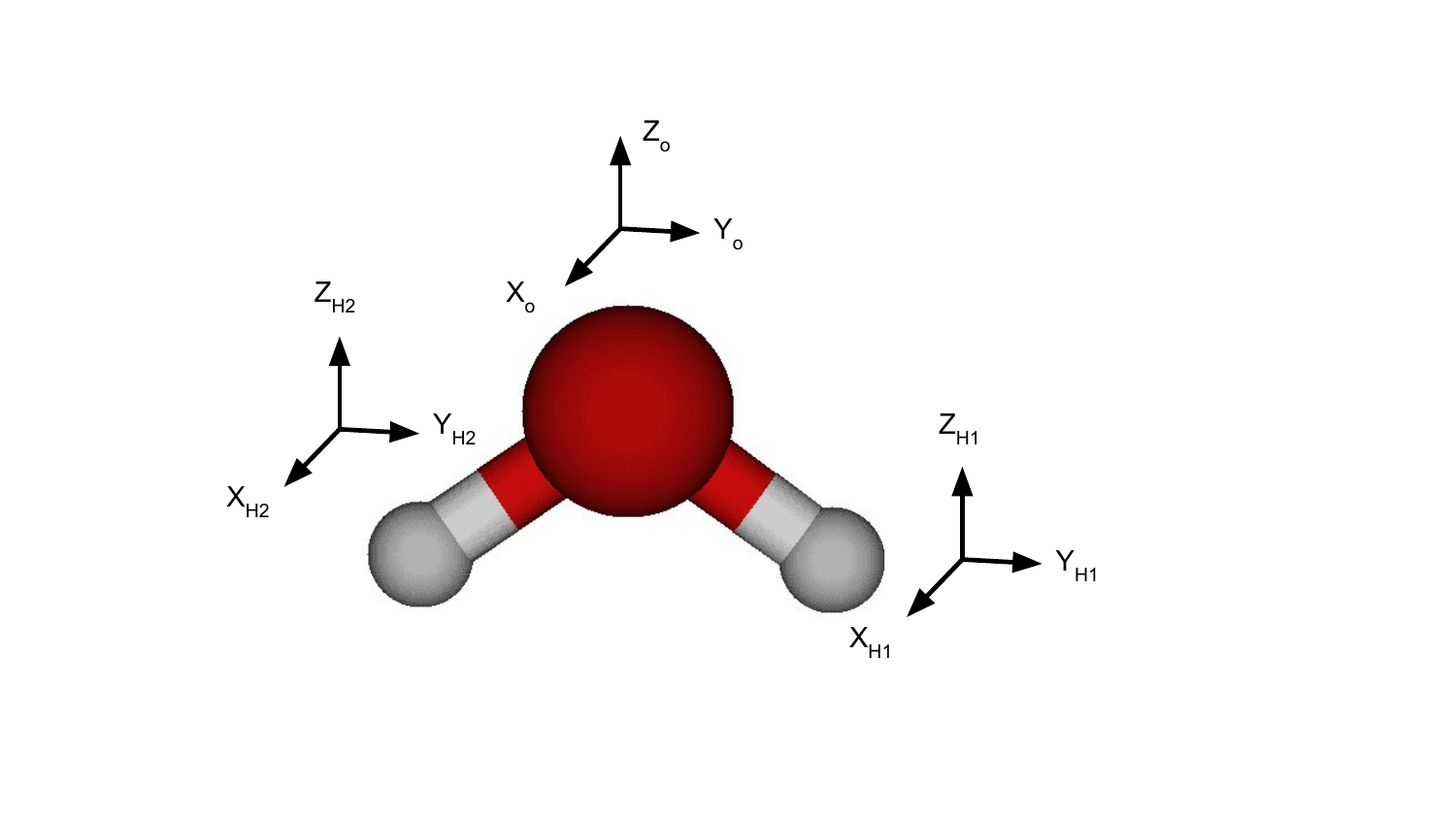}
	\caption{Cartesian coordinates of H$_2$O molecule}
	\label{fig:h2o_cart}
\end{figure}

\begin{figure}[!htp]
	\centering
	\includegraphics[width= \textwidth, angle=0]{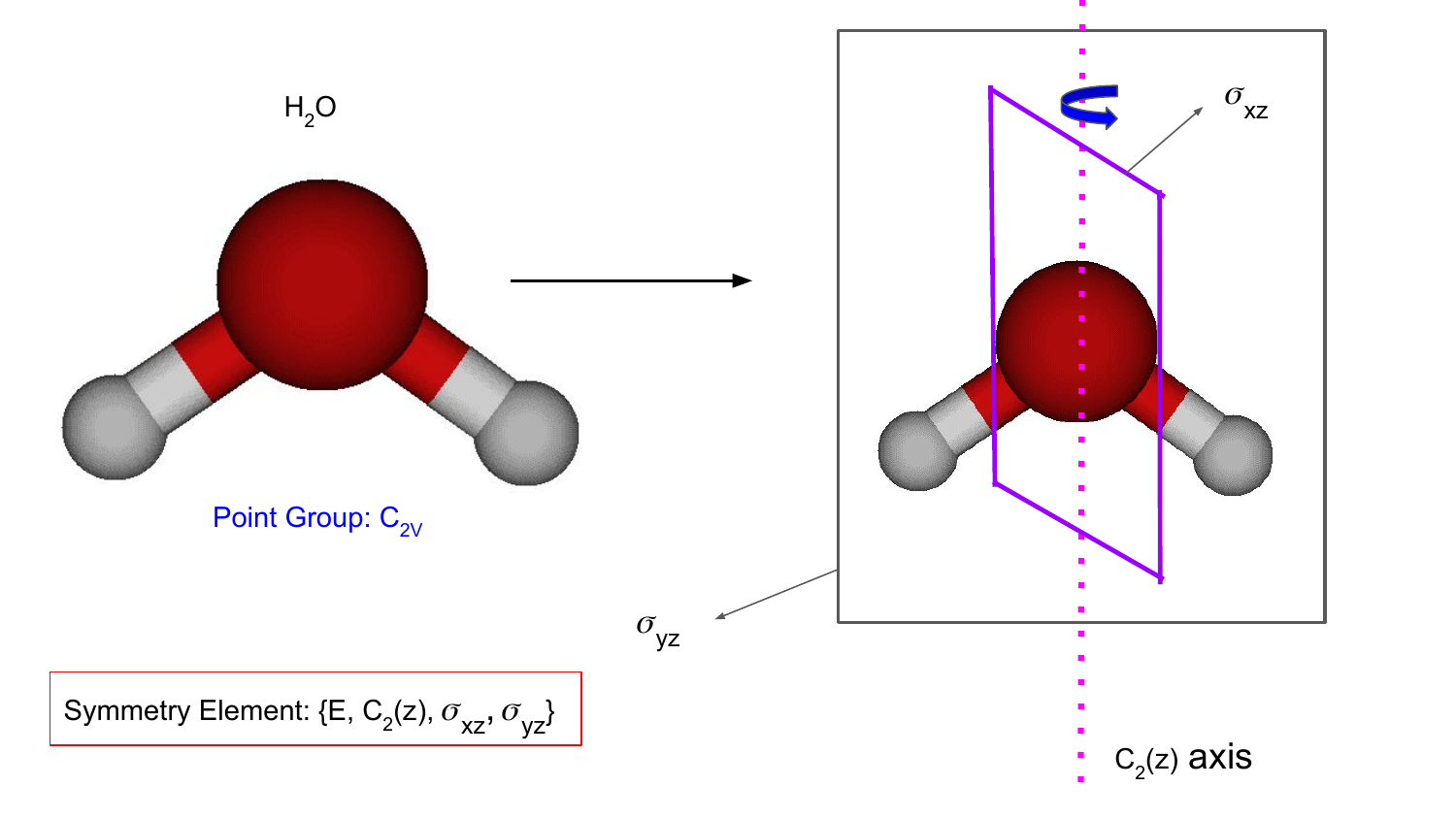}
	\caption{Symmetry elements of H$_2$O molecule}
	\label{fig:h2o_sym}
\end{figure}

\noindent
In case of H$_2$O molecule, Cartesian coordinates of each atom and symmetry elements are depicted in Figures \ref{fig:h2o_cart} and \ref{fig:h2o_sym}, respectively. 

\noindent
While constructing the reducible representation of 3$N$ (= 9 for H$_2$O) coordinates, the transformation matrix for identity operation ($E$) takes
the following form:

\begin{equation*}
	E :
	\begin{pmatrix}
		1 & 0 & 0 & 0 & 0 & 0 & 0 & 0 & 0 \\
		0 & 1 & 0 & 0 & 0 & 0 & 0 & 0 & 0 \\
		0 & 0 & 1 & 0 & 0 & 0 & 0 & 0 & 0 \\
		0 & 0 & 0 & 1 & 0 & 0 & 0 & 0 & 0 \\
		0 & 0 & 0 & 0 & 1 & 0 & 0 & 0 & 0 \\
		0 & 0 & 0 & 0 & 0 & 1 & 0 & 0 & 0 \\
		0 & 0 & 0 & 0 & 0 & 0 & 1 & 0 & 0 \\
		0 & 0 & 0 & 0 & 0 & 0 & 0 & 1 & 0 \\
		0 & 0 & 0 & 0 & 0 & 0 & 0 & 0 & 1 \\
	\end{pmatrix}
	\begin{pmatrix}
		x_{H_{1}} \\
		y_{H_{1}} \\
		z_{H_{1}} \\
		x_{O} \\
		y_{O} \\
		z_{O} \\
		x_{H_{2}} \\
		y_{H_{2}} \\
		z_{H_{2}} \\
	\end{pmatrix}
	= 
	\begin{pmatrix}
		x^{\prime}_{H_{1}} \\
		y^{\prime}_{H_{1}} \\
		z^{\prime}_{H_{1}} \\
		x^{\prime}_{O} \\
		y^{\prime}_{O} \\
		z^{\prime}_{O} \\
		x^{\prime}_{H_{2}} \\
		y^{\prime}_{H_{2}} \\
		z^{\prime}_{H_{2}} \\
	\end{pmatrix}
\end{equation*}

\noindent
Therefore, reducible character for $E$ operation appears as $\chi (\text{E})$ = 9. On the other hand, transformation matrix for $C_2$ operation
becomes: 

\begin{equation*}
	C_{2} :
	\begin{pmatrix}
		0 & 0 & 0 & 0 & 0 & 0 & -1 & 0 & 0 \\
		0 & 0 & 0 & 0 & 0 & 0 & 0 & -1 & 0 \\
		0 & 0 & 0 & 0 & 0 & 0 & 0 & 0 & 1 \\
		0 & 0 & 0 & -1 & 0 & 0 & 0 & 0 & 0 \\
		0 & 0 & 0 & 0 & -1 & 0 & 0 & 0 & 0 \\
		0 & 0 & 0 & 0 & 0 & 1 & 0 & 0 & 0 \\
		-1 & 0 & 0 & 0 & 0 & 0 & 0 & 0 & 0 \\
		0 & -1 & 0 & 0 & 0 & 0 & 0 & 0 & 0 \\
		0 & 0 & 1 & 0 & 0 & 0 & 0 & 0 & 0 \\
	\end{pmatrix}
	\begin{pmatrix}
		x_{H_{1}} \\
		y_{H_{1}} \\
		z_{H_{1}} \\
		x_{O} \\
		y_{O} \\
		z_{O} \\
		x_{H_{2}} \\
		y_{H_{2}} \\
		z_{H_{2}} \\
	\end{pmatrix}
	= 
	\begin{pmatrix}
		x^{\prime}_{H_{1}} \\
		y^{\prime}_{H_{1}} \\
		z^{\prime}_{H_{1}} \\
		x^{\prime}_{O} \\
		y^{\prime}_{O} \\
		z^{\prime}_{O} \\
		x^{\prime}_{H_{2}} \\
		y^{\prime}_{H_{2}} \\
		z^{\prime}_{H_{2}} \\
	\end{pmatrix}
\end{equation*}

\noindent
Hence, reducible character for $C_2$ operation becomes  $\chi (\text{C$_{2}$})$ = -1. In a similar way, $\chi (\sigma_{xz})$ and $\chi (\sigma_{yz})$ appear as 1 and 3, respectively. Finally, the reducible representation by 3$N$ vectors takes the following form (Table \ref{tab:red1}):

\begin{table}[!htp]
	\begin{center}
		\caption{Reducible representation of 3$N$ vectors}
		\label{tab:red1}
		\begin{tabular}{|c|c|c|c|c|}
			\hline 
			$C_{2v}$ & $E$ & $C_2$ & $\sigma_{xz}$ & $\sigma_{yz}$ \\ 
			\hline 
			$\Gamma^{3N}$ & 9 & -1 & 1 & 3 \\ 
			\hline 
		\end{tabular} 
	\end{center}
\end{table}

\noindent
In a reducible representation, the number ($a_{i}$) of IREPs is obtained by the following expression:
\begin{equation}
	a_{i} = \frac{1}{g}\sum_{g} \chi(R) \, \chi_{i} (R),
	\label{eq:got}
\end{equation} 
where $a_{i}$ represents no of a particular $i$th IREP in the reducible representation. On the other hand, $g$ signifies total number of 
symmetry operations in a given point group, whereas $\chi(R) \,\,\text{and} \,\,\chi_{i} (R)$ signify characters of operation $R$ in reducible representation and $i$th IREP, respectively. \\

\noindent
The character table for $C_{2v}$ point group is given in Table \ref{tab:char1}

\begin{table}[!htp]
	\begin{center}
		\caption{Character table for $C_{2v}$ point group}
		\label{tab:char1}
		\begin{tabular}{c|cccc|c}
			$C_{2v}$ & $E$ & $C_2$ & $\sigma_{xz}$ & $\sigma_{yz}$ &  \\ 
			\hline 
			$A_1$ & 1 & 1 & 1 & 1 & $T_z$ \\ 
			\hline 
			$A_2$ & 1 & 1 & -1 & -1 & $R_z$ \\ 
			\hline 
			$B_1$ & 1 & -1 & 1 & -1 & $T_x,R_y$ \\ 
			\hline 
			$B_2$ & 1 & -1 & -1 & 1 & $T_y,R_z$ \\ 
		\end{tabular} 
	\end{center}
\end{table}

\noindent	
Using Eq. \ref{eq:got}, the number of each IREPs is obtained as: 	$n(a_1) = 3$, $n(a_2) = 1$, $n(b_1) = 2$, $n(b_2) = 3$.
Hence, one can write:\\

\noindent
$\Gamma_{H_{2}O}$ = $3 a_1 + a_2 + 2 b_1 + 3 b_2 $ \\
$\Gamma_{Trans}$  = $a_1 + b_1 + b_2$ \\
$\Gamma_{Rot}$ = $a_2 + b_1 + b_2$ \\
$\Gamma_{Vib}$ = $2 a_1 + b_2$\\

\begin{figure}[!htp]
	\centering
	\includegraphics[width= 0.7\textwidth, angle=0]{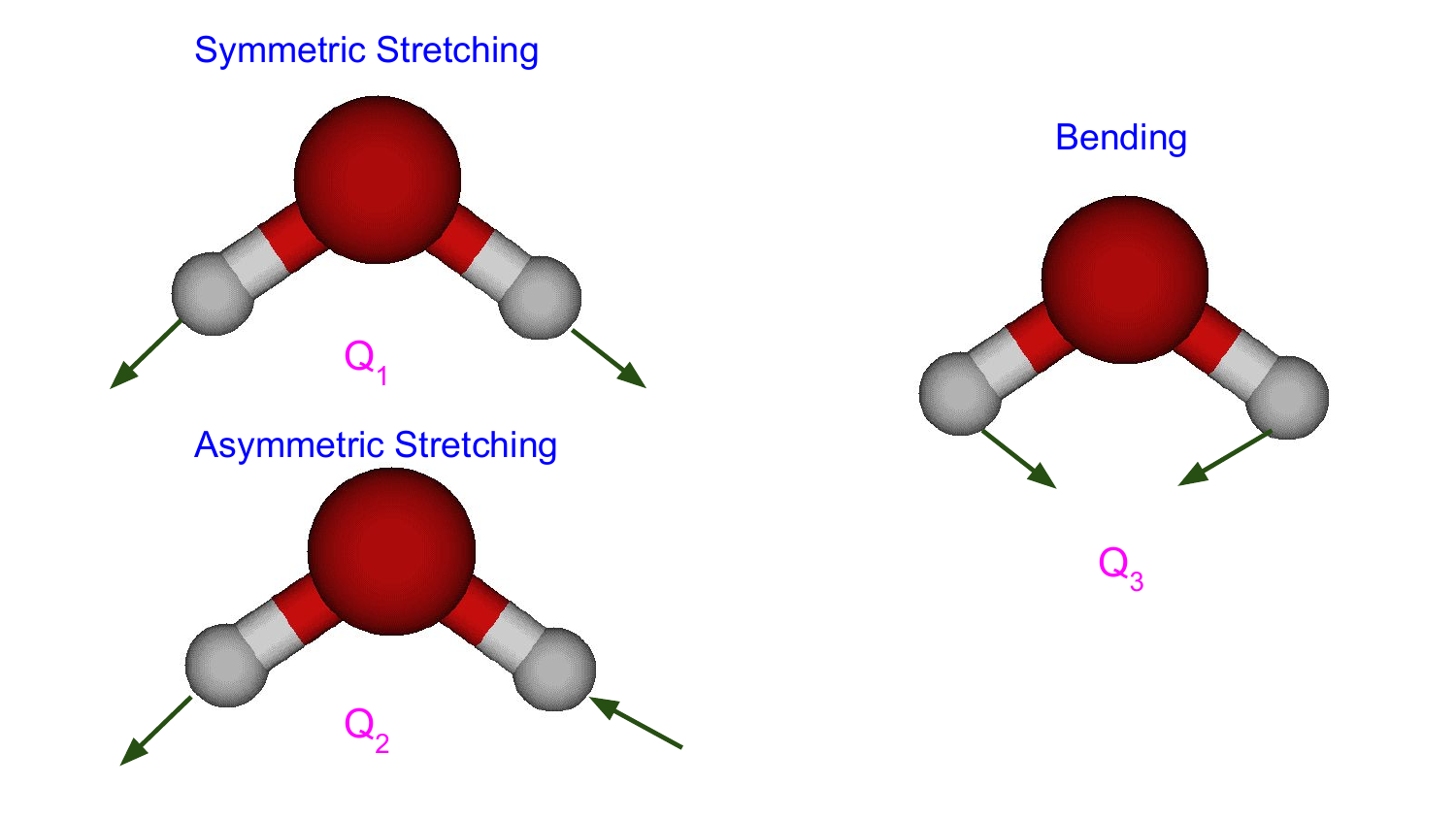}
	\caption{Vibrational modes of H$_2$O molecule with vectorial displacements}
	\label{fig:h2o_sym1}
\end{figure}

\noindent
The H$_2$O molecule has three normal modes, namely, symmetric stretching ($Q_1$), asymmetric stretching ($Q_{2}$) and bending ($Q_{3}$) (see Figure \ref{fig:h2o_sym1} and Table \ref{tab:char2}).\\

\begin{table}
	\begin{center}
		\caption{IREPs of vibrational modes for $C_{2v}$ point group}
		\label{tab:char2}
		\begin{tabular}{c|cccc|c}
			& $E$ & $C_2$ & $\sigma_{xz}$ & $\sigma_{yz}$ &  \\ 
			\hline 
			$Q_1$ & 1 & 1 & 1 & 1 & $A_1$ \\ 
			\hline 
			$Q_2$ & 1 & -1 & -1 & 1 & $B_2$ \\ 
			\hline 
			$Q_3$ & 1 & 1 & 1 & 1 & $A_1$ \\ 
			\hline 
		\end{tabular} 
	\end{center}
\end{table}

\noindent
In order to hold the condition for first order coupling (Eq. \ref{Eq:JTcond}), $\Gamma^{\beta}$ will be $b_2$ symmetry as it can not be totally symmetric IREP ($a_1$). Since H$_2$O has two $a_1$ and one $b_2$ symmetric modes, the different conditions for linear couplings are revisited in the light of C$_{\text{2v}}$ system (H$_2$O). The condition:~\citep{bijit_mp115} 
$$\Gamma^0 \in \Gamma^{\beta} \otimes \left[(\Gamma^{\alpha})^2 \right]$$
\noindent
As per the above condition, if the normal mode is $b_2$ symmetric, JT type distortion can be observed only if one electronic state belongs to $A_1$ symmetry and another one is of $B_2$ symmetry: $$a_1 \in b_2  \otimes [ A_1 \otimes B_2]$$.

\subsection{Symmetry Breaking in NO$_3$ Molecule}\label{sec:no3_group}

\noindent
We consider another non-linear molecule to understand the JT condition in some more details. NO$_3$ radical~\citep{bijit_mp115} is an example of D$_{\text{3h}}$ molecule which has six normal modes, namely, totally symmetric ($Q_1$), umbrella ($Q_2$),
degenerate asymmetric stretching ($Q_{3x}$ and $Q_{3y}$) and degenerate asymmetric bending ($Q_{4x}$ and $Q_{4y}$) modes.
The lowest five electronic states of this doublet system (NO$_3$) are $\tilde{X}^2A^{\prime}_2$, $\tilde{A}^2E^{\prime\prime}$ and $\tilde{B}^2E^{\prime}$ among which $E^{\prime\prime}$ and $E^{\prime}$ states exhibit JT symmetry breaking. The JT active modes lift the degeneracy of the $E^{\prime\prime}$ 
and $E^{\prime}$ states are $Q_{3x}$, $Q_{3y}$, $Q_{4x}$ and $Q_{4y}$.\\

\noindent
While exploring the JT condition (Eq. \ref{Eq:JTcond}), one can find: $\Gamma^0 \in \Gamma^{\beta} \otimes \left[(\Gamma^{\alpha})^2 \right]$, where $\left[(\Gamma^{\alpha})^2 \right]$ represents symmetrized product of the IREP of the degenerate state and $\Gamma^{\beta}$ signifies IREP of the normal mode. In case of $E^{\prime\prime}$ state,  
$\left[(\Gamma^{\alpha})^2 \right] \equiv E^{\prime\prime} \otimes E^{\prime\prime} 
= A^{\prime}_1 + A^{\prime}_2 + E^{\prime}$  and for $E^{\prime}$ state, 
$\left[(\Gamma^{\alpha})^2 \right] \equiv E^{\prime} \otimes E^{\prime} 
= A^{\prime}_1 + A^{\prime}_2 + E^{\prime}$ \\

\noindent
In order to hold JT condition (Eq. \ref{Eq:JTcond}), $\Gamma^{\beta}$ will be either $a^{\prime}_2$ or $e^{\prime}$ mode as it can not be totally symmetric IREP ($a^{\prime}_1$). Since NO$_3$ has four $e^{\prime}$ but no $a^{\prime}_2$ modes, the four $e^{\prime}$ modes, namely, $Q_{3x}$, $Q_{3y}$,
$Q_{4x}$ and $Q_{4y}$ vibrations are JT active ones.\\

\noindent
Therefore, the four different situations of JT condition are revisited in the light of D$_{\text{3h}}$ system (NO$_3$). The JT condition:~\citep{bijit_mp115} 
$$\Gamma^0 \in \Gamma^{\beta} \otimes \left[(\Gamma^{\alpha})^2 \right]$$

\noindent
While applying this condition, the following four situations will arise:	

\begin{itemize} 
	\item[(1)]  Ground state (non-degenerate, $A_2^{\prime}$) couples 
	with the (totally) symmetric stretching mode, $\Gamma^{\beta} = a_1^{\prime}$ and 
	$\left[(\Gamma^{\alpha})^2 \right] = A_2^{\prime} \otimes A_2^{\prime} = A_1^{\prime}$.
	Therefore, the contribution of the perturbation will not be zero. Nevertheless this will not lift 
	the degeneracy of the system and hence, would not show JT effect.
	
	\item[(2)]  Ground state (non-degenerate, $A_2^{\prime}$) 
	couples with a non totally symmetric mode,
	$\Gamma^{\beta} \ne a_1^{\prime}$ and
	$\left[(\Gamma^{\alpha})^2 \right] = A_2^{\prime} \otimes A_2^{\prime} = A_1^{\prime}$.
	Thus, the contribution of the perturbation will be zero.
	
	\item[(3)] A degenerate state ($E^{\prime\prime}$ or $E^{\prime}$) 
	couples with the (totally) symmetric stretching mode, $\Gamma^{\beta} = a_1^{\prime}$  and
	$[(\Gamma^{\alpha})^2] = \quad E^{\prime\prime} \otimes E^{\prime\prime} \quad \text{or} \quad 
	E^{\prime} \otimes E^{\prime} \quad = \quad A_1^{\prime} + A_2^{\prime} + E^{\prime}$. 
	It will be analogous to the first case.
	
	\item[(4)] A degenerate state ($E^{\prime\prime}$ or $E^{\prime}$) 
	couples with a non totally symmetric mode, $\Gamma^{\beta} \ne a_1^{\prime}$ and 
	$[(\Gamma^{\alpha})^2] = \quad E^{\prime\prime} \otimes E^{\prime\prime} \quad \text{or} \quad 
	E^{\prime} \otimes E^{\prime} \quad = \quad A_1^{\prime} + A_2^{\prime} + E^{\prime}$. Such product will survive and 
	leads to symmetry breaking. 
\end{itemize}

\noindent
If the umbrella mode ($a_2^{\prime\prime}$) of NO$_3$ molecule is considered, $\Gamma^{\beta}=a_2^{\prime\prime}$. Since $a_2^{\prime\prime} \otimes (A_1^{\prime} + A_2^{\prime} + E^{\prime})$ 
does not contain $A_1^{\prime}$, $a_2^{\prime\prime}$ mode will not show up JT effect. In case of asymmetric stretching and bending modes, ($e^{\prime}$), $\Gamma^{\beta}=e^{\prime}$. Since $e^{\prime} \otimes (A_1^{\prime} + A_2^{\prime} + E^{\prime})$ contains 
$A_1^{\prime}$, all $e^{\prime}$ modes are JT active.

\begin{center}
	\begin{figure}[!htp]
		\includegraphics[width=0.9\textwidth,angle=0]{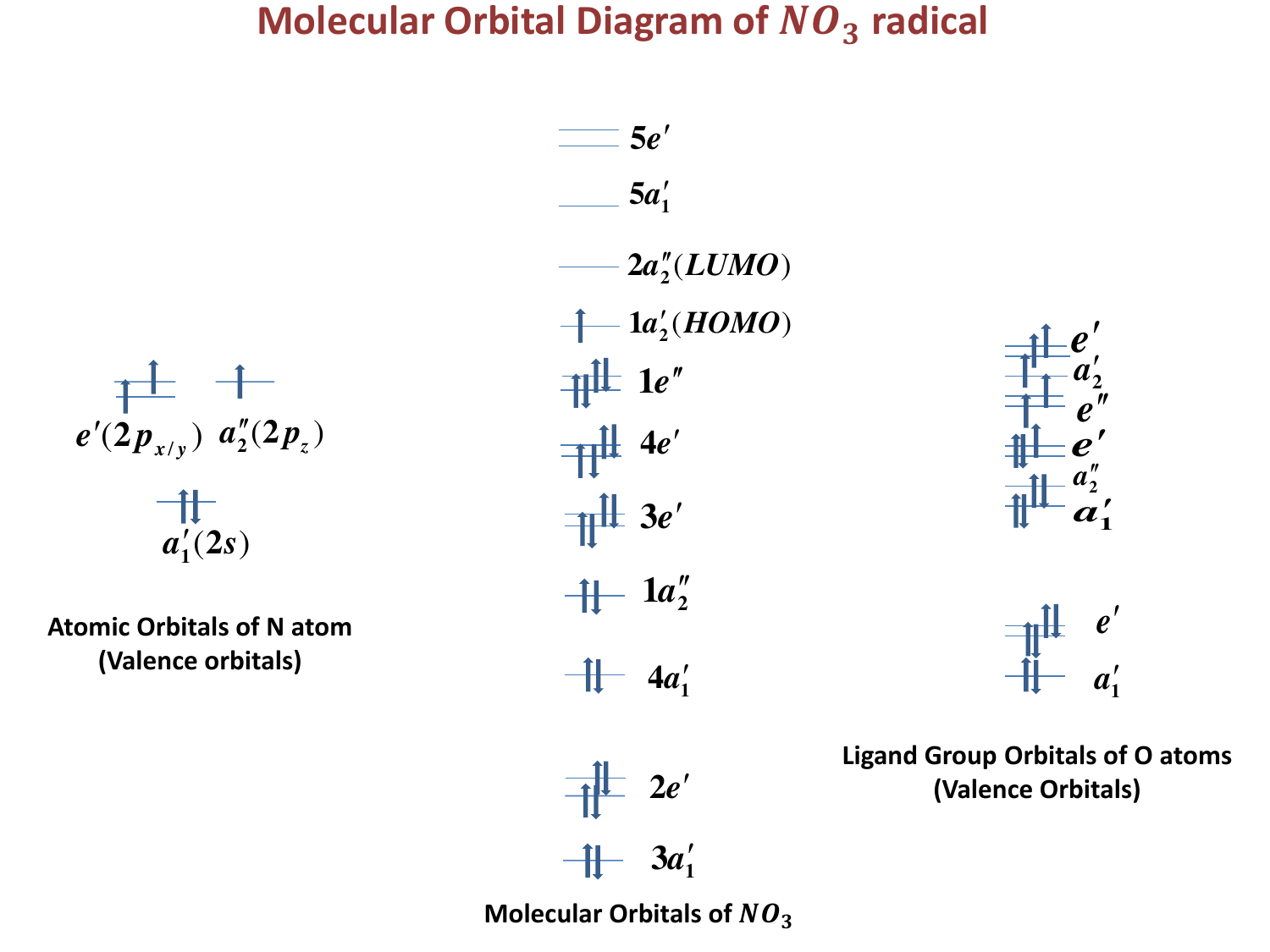}
		\caption{Molecular orbitals of NO$_3$ radical}
	\end{figure}
\end{center}

\begin{center}
	\begin{figure}[!htp]
		\includegraphics[width=0.9\textwidth,angle=0]{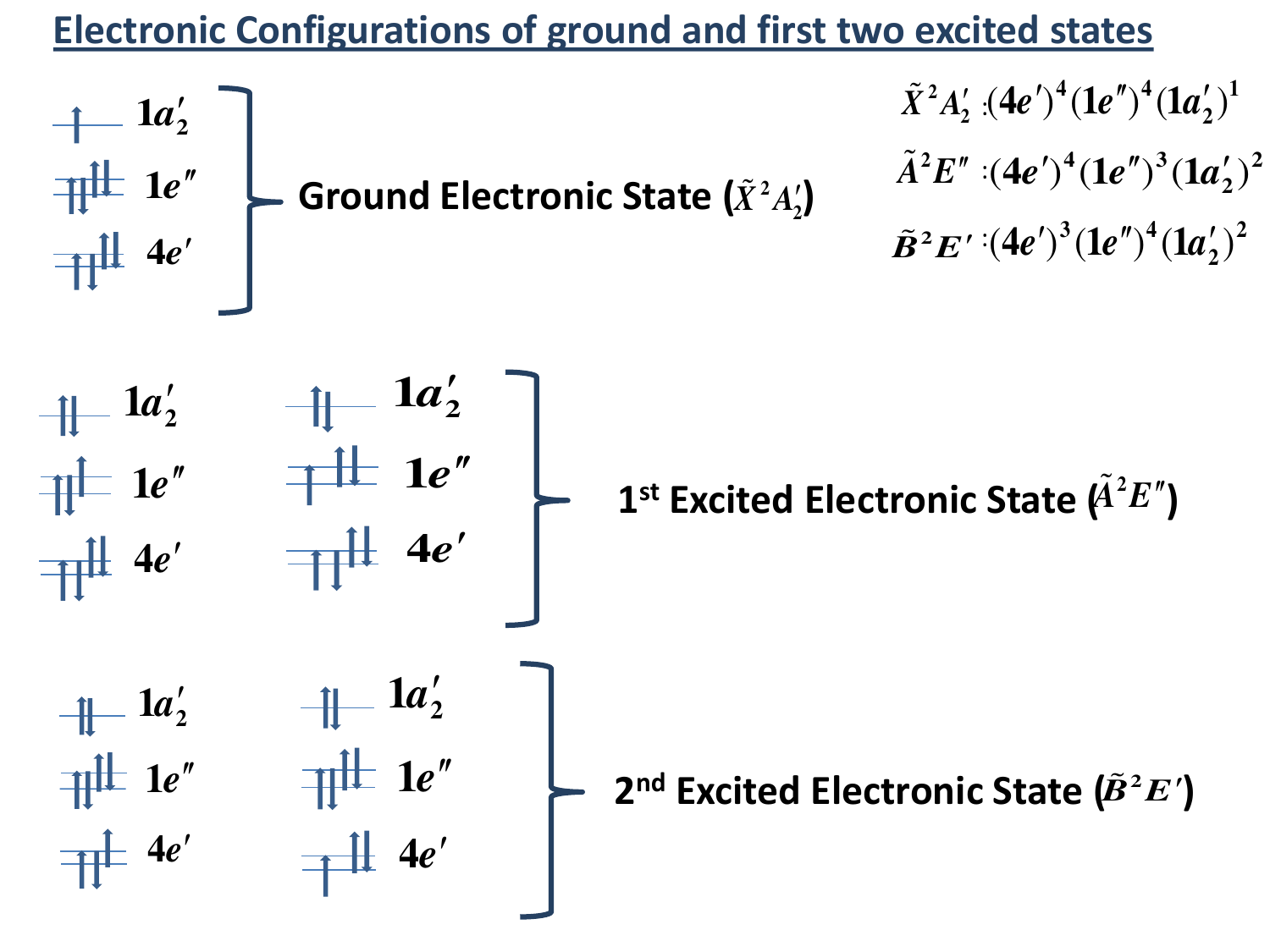}
		\caption{Low lying electronic states of NO$_3$ radical}
	\end{figure}
\end{center}

\begin{center}
	\begin{figure}[!htp]
		\includegraphics[width=0.9\textwidth,angle=0]{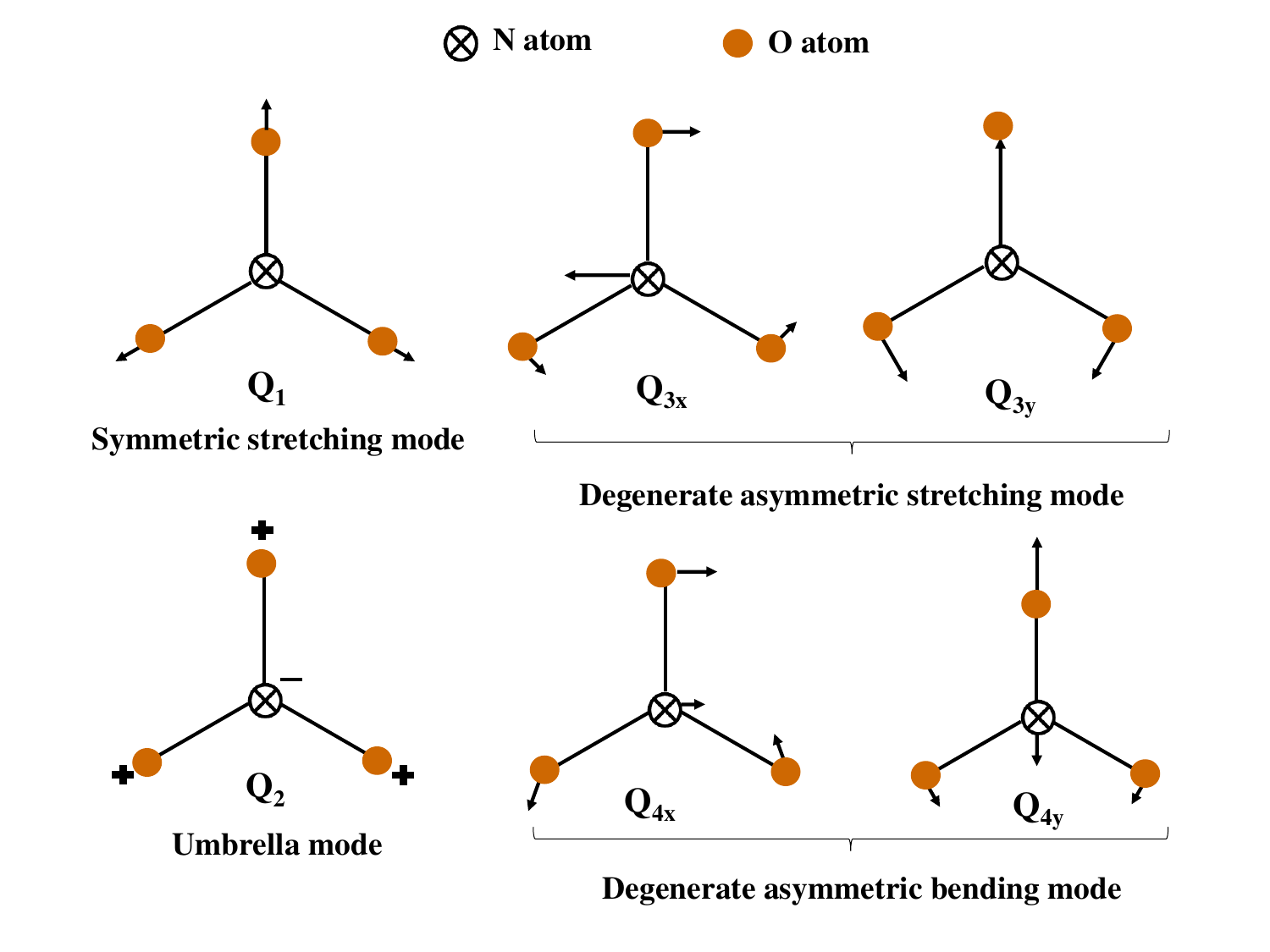}
		\caption{Normal modes of NO$_3$ radical. The results are adopted from Ref. [\protect\citen{bijit_mp115}] and permission has been taken from Taylor \& Francis.}
	\end{figure}
\end{center}

\begin{center}
	\begin{figure}[!htp]
		\includegraphics[width=0.9\textwidth,angle=0]{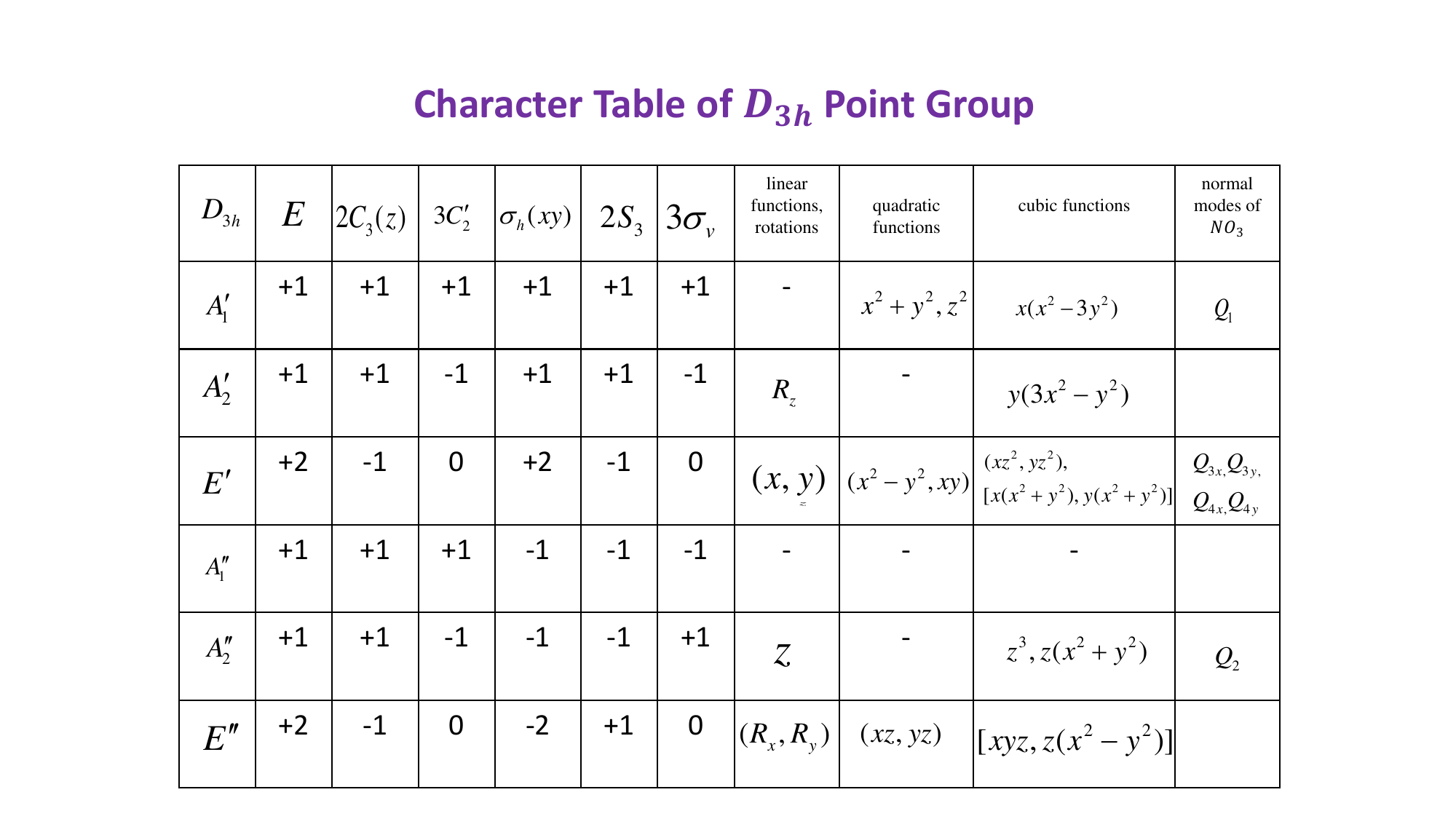}
		\caption{Character table of D$_{3h}$ point group.}
	\end{figure}
\end{center}

\subsection{Symmetry Breaking in CO$_2$ Molecule}\label{sec:co2_group}

\begin{figure}[!htp]
	\centering
	\includegraphics[width= 0.6\textwidth, angle=0]{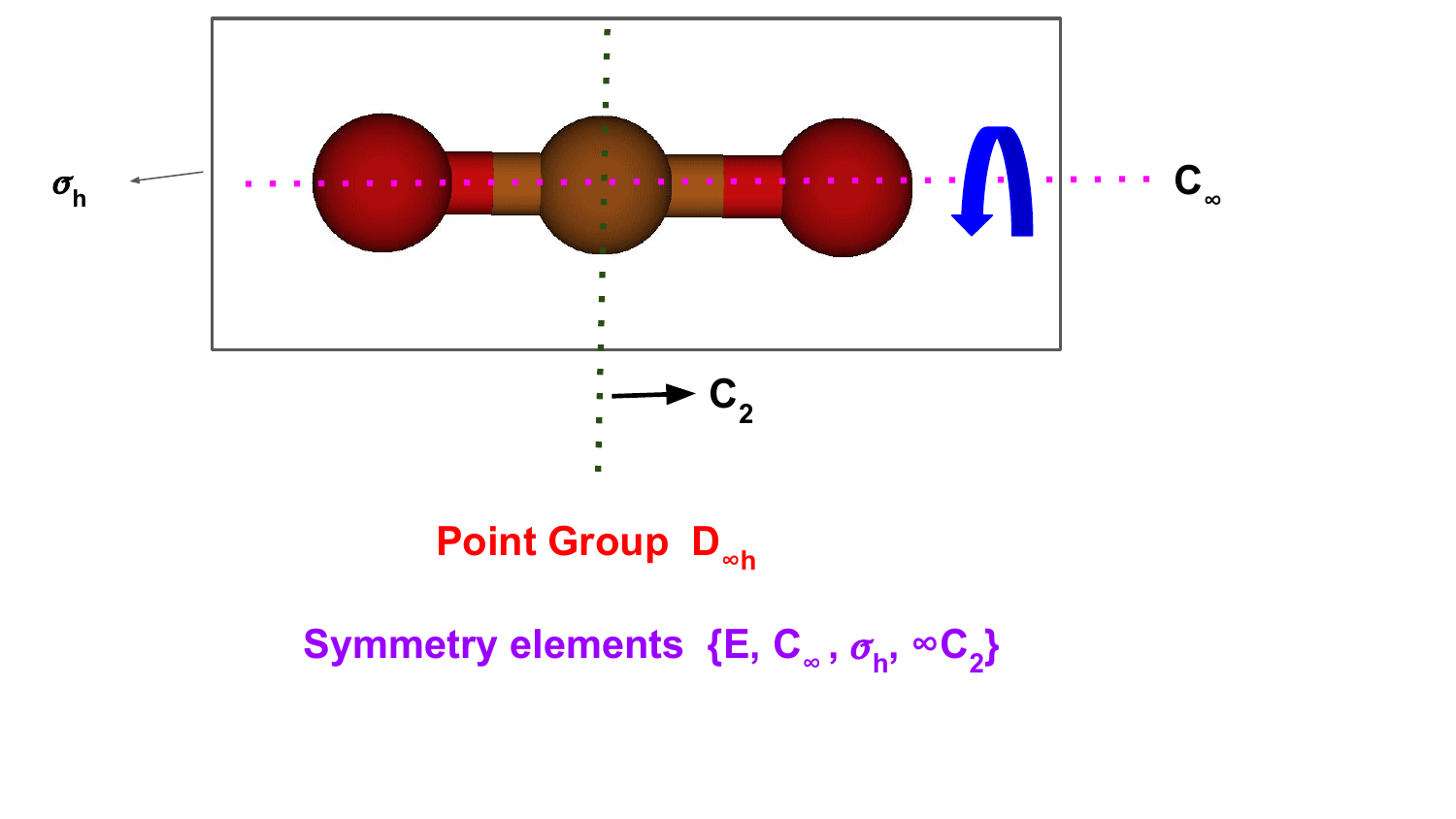}
	\caption{Symmetry elements of CO$_2$ molecule}
	\label{fig:co2_sym}
\end{figure}

\noindent
CO$_2$ is an example of D$_{\infty\text{h}}$ molecule due to linear geometry, which has (3$N$-5 = 4) four normal modes, namely, symmetric stretching ($Q_1$),
anti-symmetric stretching ($Q_{2}$) and degenerate bending ($Q_{3}$ and $Q_{4}$). The symmetry elements are depicted in Figure \ref{fig:co2_sym}.
\noindent
In order to obtain IREPs of normal modes, we explore the symmetry operations in $D_{2h}$ point group, which is a sub-group of
$D_{\infty h}$. For CO$_2$, the reducible representation is depicted in Table \ref{tab:red3}.
\begin{table}[!htp]
	\begin{center}
		\caption{Reducible representation of 3$N$ vectors}
		\label{tab:red3}
		{\begin{tabular} {|c|cccccccc|}
				\hline
				D$_{2h}$ & $E$ & $C_2$ (x) & $C_2$ (y) & $C_2$ (z) & $i$ & $\sigma (xy)$ & $\sigma (xz)$ & $\sigma (yz)$  \\
				\hline
				$\Gamma_{3N}$ & 9 & -3 & -1 & -1 & -3 & 1 & 3 & 3 \\
				\hline
		\end{tabular}}
	\end{center}
\end{table}
\noindent
On the other hand, character table for $D_{2h}$ point group is depicted in Table \ref{tab:char3}.
\begin{table}[!htp]
	\caption{Character table for $D_{2h}$ point group}
	\label{tab:char3}
	\begin{center}
		\scalebox{0.8}{\begin{tabular} {|c|cccccccc|c|}
				\hline
				& $E$ & $C_2$ (x) & $C_2$ (y) & $C_2$ (z) & $i$ & $\sigma (xy)$ & $\sigma (xz)$ & $\sigma (yz)$  & translations,\\
				& & & &  & & & & & rotations \\
				\hline
				$A_{g}$ & 1 & 1 & 1 & 1 & 1 & 1 & 1 & 1 & \\
				$B_{1g}$ & 1 & 1 & -1 & -1 & 1 & 1 & -1 & -1 & $R_{z}$  \\
				$B_{2g}$ & 1 & -1 & 1 & -1 & 1 & -1 & 1 & -1 & $R_{y}$  \\
				$B_{3g}$ & 1 & -1 & -1 & 1 & 1 & -1 & -1 & 1 & $R_{x}$  \\
				$A_{u}$ & 1 & 1 & 1 & 1 & -1 & -1 & -1 & -1 &   \\
				$B_{1u}$ & 1 & 1 & -1 & -1 & -1 & -1 & 1 & 1 & $z$ \\
				$B_{2u}$ & 1 & -1 & 1 & -1 & -1 & 1 & -1 & 1 & $y$ \\
				$B_{3u}$ & 1 & -1 & -1 & 1 & -1 & 1 & 1 & -1 & $x$ \\
				\hline
		\end{tabular}}
	\end{center}
\end{table}
While resolving $\Gamma_{3N}$ into the IREPs by using the character table (see Table \ref{tab:char3}), we find:
\begin{eqnarray}
	\Gamma^{3N} = a_g + b_{2g} + b_{3g} + 2b_{1u} + 2b_{2u} + 2b_{3u}
\end{eqnarray}
On the other hand, the character table shows that:

\begin{eqnarray*}
	\Gamma_{\text{trans}} &=& b_{1u} + b_{2u} + b_{3u}\\
	\Gamma_{\text{rot}} &=& b_{2g} + b_{3g}\\
	\Gamma_{\text{vib}} &=& a_{g} + b_{1u} + b_{2u} + b_{3u}
\end{eqnarray*}

\begin{figure}[!htp]
	\centering
	\includegraphics[width= \textwidth, angle=0]{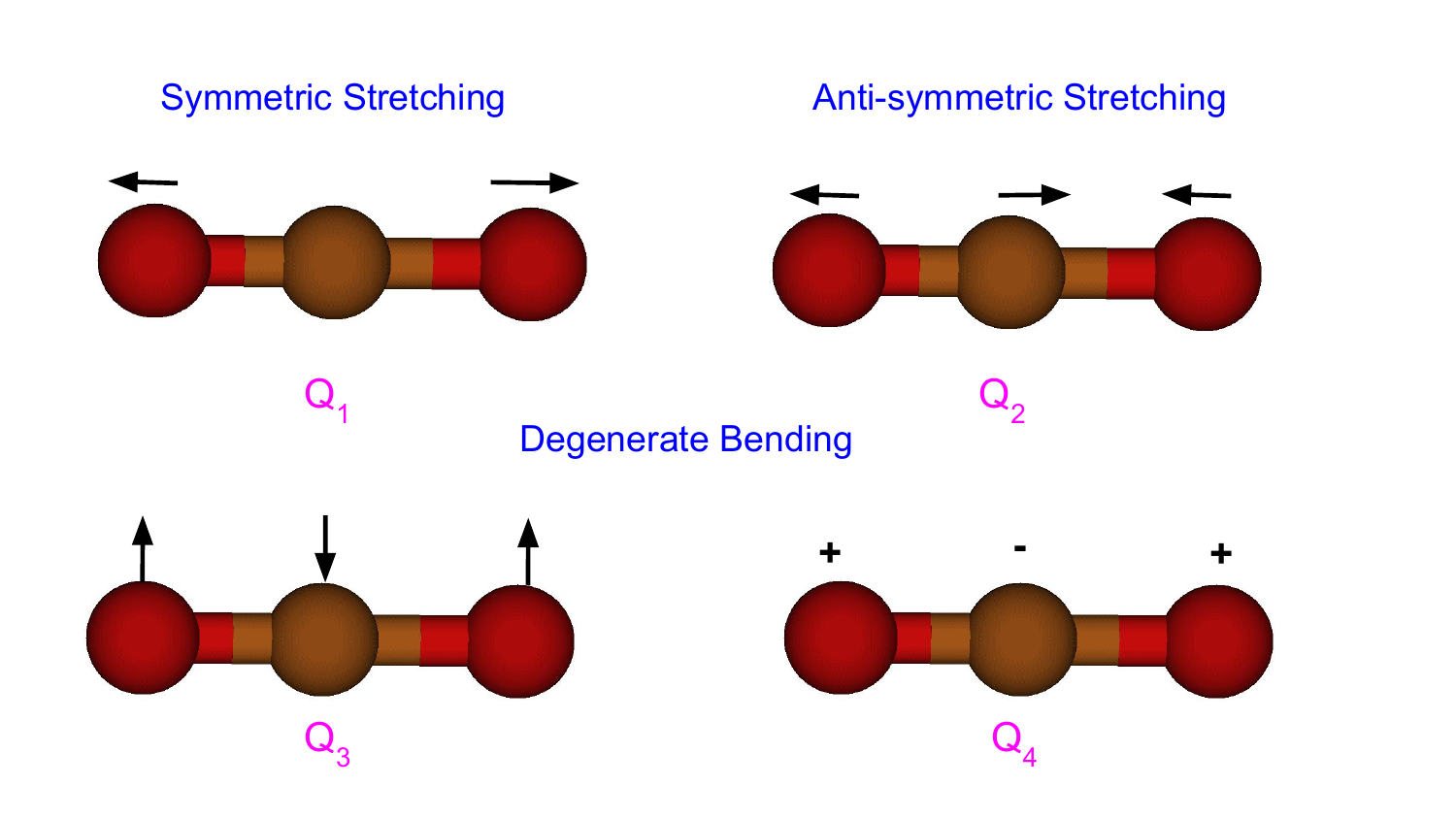}
	\caption{Vibrational modes of CO$_2$ molecule with vectorial displacements}
	\label{fig:vib_co2}
\end{figure}

\noindent
The symmetry of the vibrational modes of CO$_2$ are represented in Figure \ref{fig:vib_co2}, where it is clearly evident: 
\begin{itemize}
	\item $Q_1$ belongs to totally symmetric IREP, $a_g$.
	\item $Q_2$ is symmetric with respect to $E$, $C_2(z)$, $\sigma_{xz}$ and $\sigma_{yz}$ operations, and anti-symmetric with respect to $C_2(y)$, $C_2(x)$,
	$i$ and $\sigma_{xy}$ operations, i.e. $Q_{2}$ belongs to $b_{1u}$ IREP.
	\item $Q_3$ is symmetric with respect to $E$, $C_2(y)$, $\sigma_{xy}$ and $\sigma_{yz}$ operations, and anti-symmetric with respect to $C_2(z)$, $C_2(x)$,
	$i$ and $\sigma_{xz}$ operations, i.e. $Q_{3}$ belongs to $b_{2u}$ IREP.
	\item $Q_4$ is symmetric with respect to $E$, $C_2(x)$, $\sigma_{xy}$ and $\sigma_{xz}$ operations, and anti-symmetric with respect to $C_2(z)$, $C_2(y)$,
	$i$ and $\sigma_{yz}$ operations, i.e. $Q_{4}$ belongs to $b_{3u}$ IREP.
\end{itemize}

\noindent
As $D_{2h}$ point group is a sub-group of $D_{\infty h}$, $a_g$ becomes $\sigma_{g}^{+}$ (non-degenerate), $b_{1u}$ becomes $\sigma_{u}^{+}$ (non-
degenerate) and the remaining two ($b_{2u}$ and $b_{3u}$) turn into $\pi_u$ (degenerate) in $D_{\infty h}$ point group. \\

\noindent
While analysing the JT interaction in $CO_2$ molecule, we need to consider the $D_{2h}$ point group as $D_{\infty h}$ point group contains infinite number of irreducible representations. The character table of $D_{\infty h}$ point group takes following form as depicted in Table \ref{tab:char4}.\\ 

\begin{table}[!htp]
	\begin{center}
		\caption{Character table of $D_{\infty h}$ point group}
		\label{tab:char4}
		\scalebox{0.7}{\begin{tabular} {|c|cccccccc|c|}
				\hline
				& $E$ & $2C_{\infty}$ & $...$ & $\infty\sigma_{v}$ & $i$ & $2S_{\infty}$ & $...$ & ${\infty}C'_{2}$  & translations,\\
				& & & &  & & & & & rotations \\
				\hline
				$A_{1g}=\Sigma_{g}^{+}$ & 1 & 1 & ... & 1 & 1 & 1 & ... & 1 & $x^2+y^2,z^2$ \\
				$A_{2g}=\Sigma_{g}^{-}$ & 1 & 1 & ... & -1 & 1 & 1 & ... & -1 & $R_{z}$  \\
				$E_{1g}=\Pi_{g}$ & 2 & 2cos($\phi$) & ... & 0 & 2 & -2cos($\phi$) & ... & 0 & $(R_{x},R_{y}),(xz,yz)$  \\
				$E_{2g}=\Delta_{g}$ & 2 & 2cos($2\phi$) & ... & 0 & 2 & 2cos($2\phi$) & ... & 0 & $x^2-y^2 ,xy$  \\
				$E_{3g}=\Phi_{g}$ & 2 & 2cos($3\phi$) & ... & 0 & 2 & -2cos($3\phi$) & ... & 0 &   \\
				... & ... & ... & ... & ... & ... & ... & ... & ... &  \\
				$A_{1u}=\Sigma_{u}^{+}$ & 1 & 1 & ... & 1 & -1 & -1 & ... & -1 & $z$ \\
				$A_{2u}=\Sigma_{u}^{-}$ & 1 & 1 & ... & -1 & -1 & -1 & ... & 1 &  \\
				$E_{1u}=\Pi_{u}$ & 2 & 2cos($\phi$) & ... & 0 & -2 & 2cos($\phi$) & ... & 0 & $(x,y)$ \\
				$E_{2u}=\Delta_{u}$ & 2 & 2cos($2\phi$) & ... & 0 & -2 & -2cos($2\phi$) & ... & 0 &  \\
				$E_{3u}=\Phi_{u}$ & 2 & 2cos($3\phi$) & ... & 0 & -2 & 2cos($3\phi$) & ... & 0 &   \\
				... & ... & ... & ... & ... & ... & ... & ... & ... &  \\
				
				\hline
		\end{tabular}}
	\end{center}
\end{table}

\noindent	
On the other hand, the partial correlation table between $D_{\infty h}$ and $D_{2h}$ point groups is presented in Table \ref{tab:char5}.\\

\begin{table}[!htp]
	\caption{Partial correlation table between $D_{\infty h}$ and $D_{2h}$ point groups}
	\label{tab:char5}
	\begin{center}
		\scalebox{0.9}{\begin{tabular} {|c|cccccccc|}
				\hline
				D$_{{\infty}h}$ & $\Sigma_{g}^{+}$ & $\Sigma_{g}^{-}$ & $\Pi_{g}$ & $\Delta_{g}$  & $\Sigma_{u}^{+}$ & $\Sigma_{u}^{-}$ & $\Pi_{u}$ & $\Delta_{u}$  \\
				\hline
				$D_{2h}$ & $A_g$ & $B_{1g}$ & $B_{2g}+B_{3g}$ & $A_{g}+B_{1g}$ & $B_{1u}$ & $A_u$ & $B_{2u}+B_{3u}$ & $A_{u}+B_{1u}$ \\
				\hline
		\end{tabular}}
	\end{center}
\end{table}

\noindent
In $D_{{\infty}h}$ symmetry, $Q_3$ and $Q_4$ normal mode of CO$_2$ molecule are degenerate ($\pi_u$ symmetry). On the other hand, ground ($S_0$), first excited ($S_1$) and second excited (degenerate, $S_2$ and $S_3$) electronic states belong to $^1\Sigma_{g}^{+}$, $^1\Sigma_{u}^{-}$ and $^1\Pi_{g}$ IREPs, respectively. \\

\noindent
In $D_{2h}$ point group, ${\pi_{u}}$ symmetry turns into $b_{2u}+b_{3u}$ and ${\Pi_{g}}^2$ symmetry appears as $2a_g + 2b_{1g}$. While considering the JT active feature of $\pi_u$ ($\Gamma^{\beta}$, see Eq. \ref{Eq:JTcond}) symmetric normal modes for the $^1\Pi_{g}$ states ($\Gamma^{\alpha}$, see Eq. \ref{Eq:JTcond}), the JT condition (Eq. \ref{Eq:JTcond}) takes the following form: 
\begin{eqnarray}
	\Gamma^{\beta} \otimes \left[ (\Gamma^{\alpha})^2 \right] &=& \pi_u \otimes \left[ \Pi_{g}^2 \right] \quad \text{in $D_{\infty h}$ symmetry}\nonumber\\
	&=& (b_{2u} + b_{3u}) \otimes (2A_g + 2B_{1g}) \quad \text{in $D_{2h}$ symmetry} \nonumber
\end{eqnarray}
\noindent
Therefore, $A_g \not\in (b_{2u} + b_{3u}) \otimes (2A_g + 2B_{1g})$, i.e. $\Gamma^0 \not\in \Gamma^{\beta} \otimes \left[ (\Gamma^{\alpha})^2 \right]$ and this implies that normal modes of CO$_2$ are not JT active.

\bibliographystyle{aipnum4-1}
\bibliography{ref_book}

\chapter{Symmetry Breaking and Jahn-Teller Theory}\label{sec:symm}	
\vspace{0.5 cm}

\section{Jahn-Teller Model}\label{sec:symm_jt}	
\begin{figure}[!htp]
		\centering
		\includegraphics[width=0.7\textwidth,angle=0]{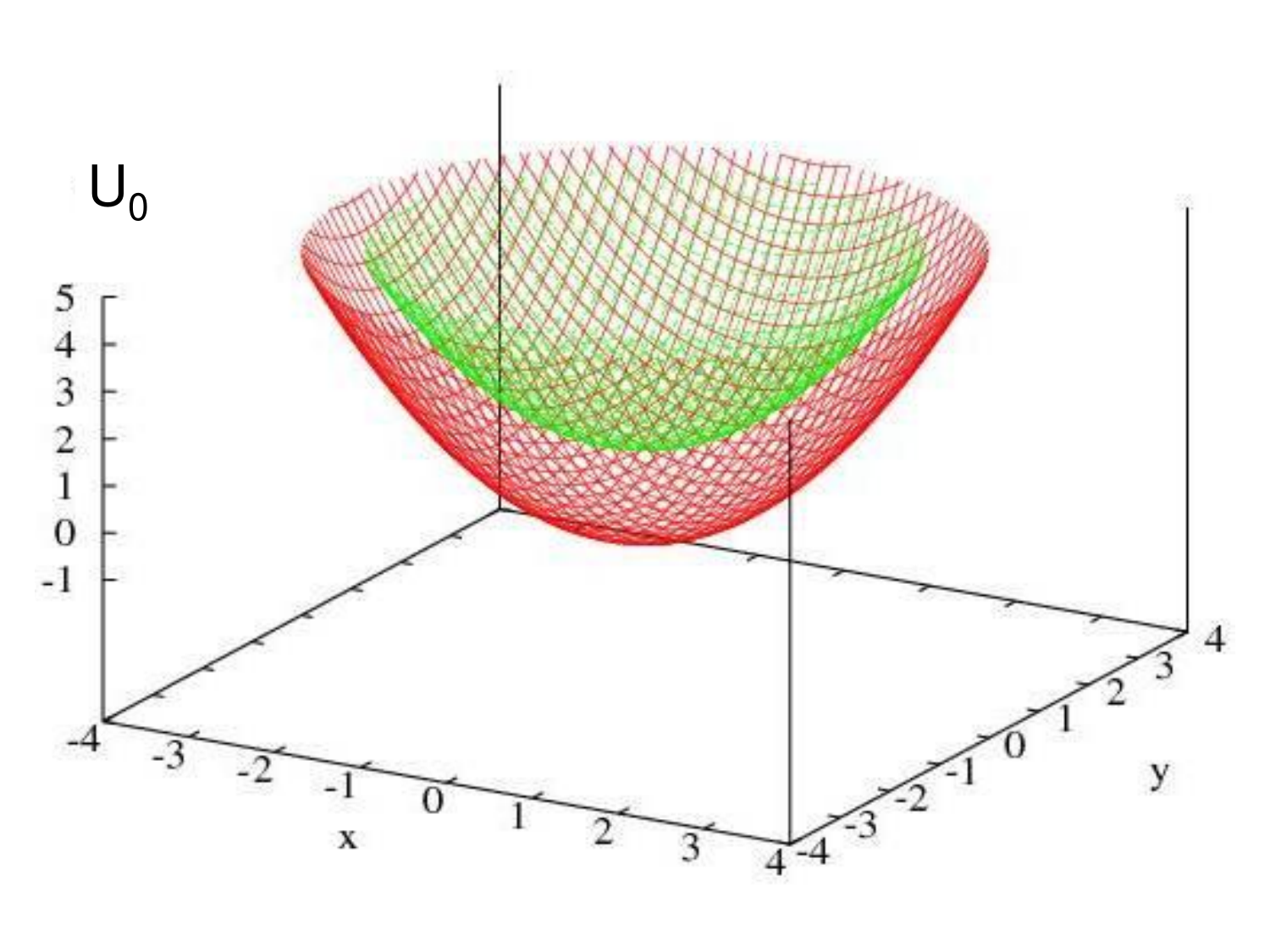}
		\caption{Potential energy surfaces without any JT distortion}
		\label{fig:pes1}
	\end{figure}

\noindent
When any two electronic states of a molecular system are degenerate at some point of the nuclear configuration space (CS), the situation was modeled by two-dimensional (2D) harmonic oscillator for the unperturbed Hamiltonian ($\hat{H_0}$) as defined below:~\citep{JT:PRS} 
\begin{eqnarray}
	\hat{H_0} &=& \hat{T}\cdot I + \hat{U_0}  \\ 
	\hat{U_0} &=& \frac{\omega}{2}\begin{pmatrix}
		x^2+y^2+\epsilon & 0 \\ 
		0 & x^2+y^2-\epsilon \\
	\end{pmatrix} 
\end{eqnarray}
where $\hat{T}$ is the kinetic energy operator, $I$ is the $2\times 2$ unit matrix and the functional forms of $\hat{U_0}$ matrix elements are depicted in Figure \ref{fig:pes1}. If such a Hamiltonian ($\hat{H_0}$) is perturbed due to nuclear degrees of freedom (see Eq. \ref{Eq:ham_taylor_1}), the $2\times 2$ matrix only within the linear perturbations for 2D nuclear CS is modeled by:
\begin{eqnarray}
	\hat{H} &=& \hat{H_0} + \hat{W} \\
	\hat{W} &=& k
	\begin{pmatrix}
		y & x \\ 
		x & -y \\ 
	\end{pmatrix}, \qquad y=q\cos\theta; \quad x=q\sin\theta
\end{eqnarray}

\noindent
On diagonalization of Jahn-Teller (JT) model Hamiltonian ($\hat{W}$), the eigenvalues and eigenfunctions can be written as:
\begin{equation}
	\hat{W}\Phi = \Lambda\Phi, \qquad \qquad \lambda=\pm kq
\end{equation}
\begin{equation}
	\phi_1=\begin{pmatrix}
		\cos{\theta}/{2}  \\
		\sin{\theta}/{2} 
	\end{pmatrix} \qquad \qquad 
	\phi_2=\begin{pmatrix}
		-\sin{\theta}/{2}  \\
		\cos{\theta}/{2} 
	\end{pmatrix} 
\end{equation}
It could be noted that if $\theta$ becomes $\theta + 2\pi$, the adiabatic nuclear wavefunctions ($\phi_1$ and $\phi_2$) change sign as per Figure \ref{fig:ps_inverse}.~\cite{longuet-higgins,longuet-higgins_prsl344}\\

	\begin{figure}[!htp]
		\centering
		\includegraphics[width=0.7\textwidth,angle=0]{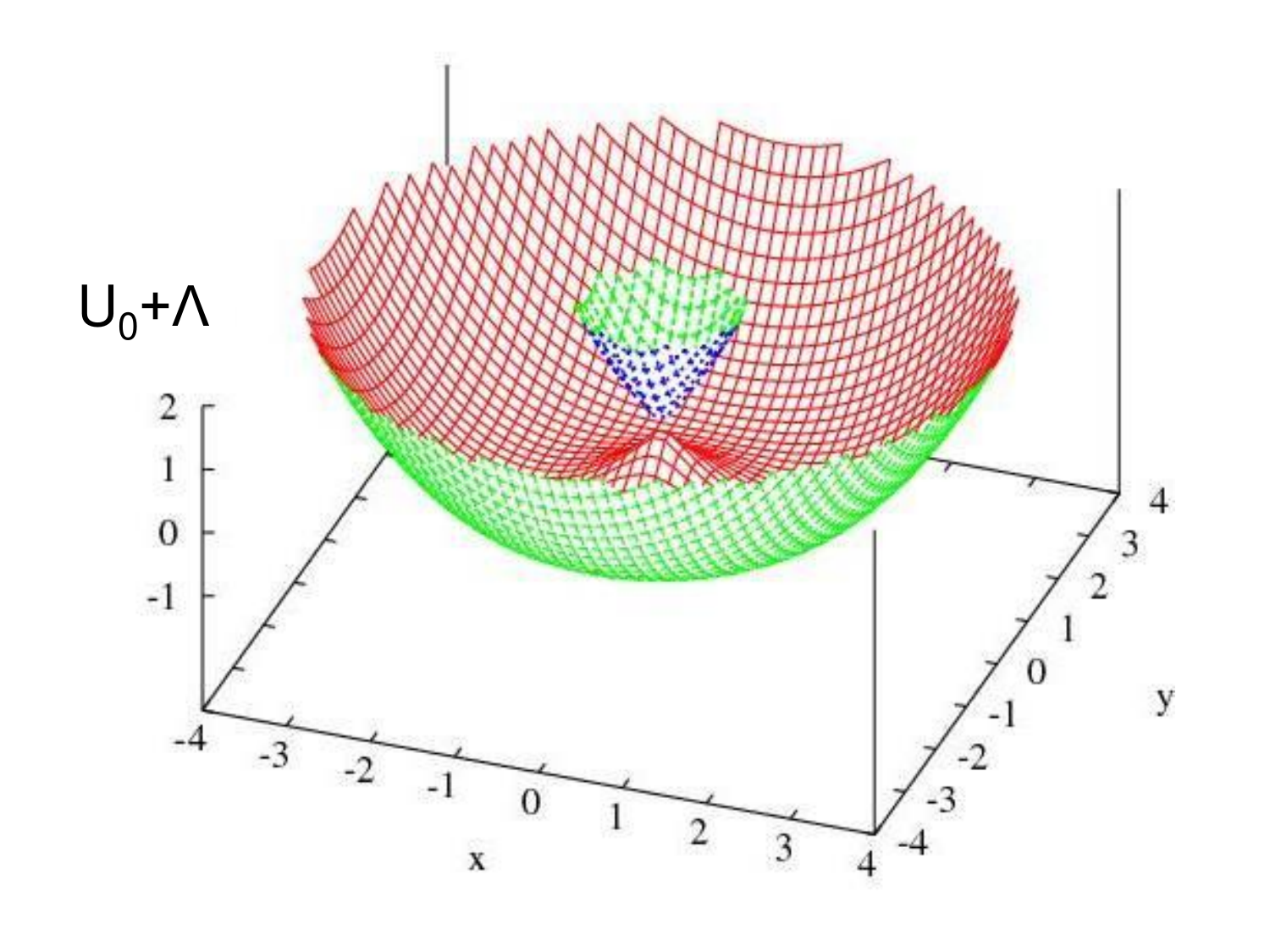}
		\caption{Potential energy surfaces with JT distortion}
	\end{figure}

	\begin{figure}[!htp]
		\centering
		\includegraphics[width=0.5\textwidth,angle=0]{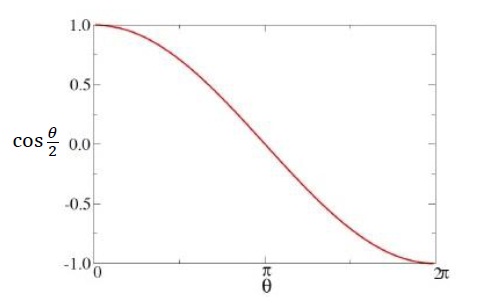}
		\caption{Sign inversion property of adiabatic nuclear eigenfunction along a closed loop encircling degenerate point.}
		\label{fig:ps_inverse}
	\end{figure}

\noindent
While removing the multi-valuedness of the nuclear eigenfunctions, Longuet-Higgins introduced a 
complex phase factor in an \textit{ad-hoc} manner,~\citep{herzberg_dfs35} which takes care of the sign of the
wavefunctions:
\begin{eqnarray}
	\zeta_1 = e^{i\theta/2} \phi_1 = e^{i\theta/2} 
	\begin{pmatrix}
		\cos{\theta}/{2}  \\
		\sin{\theta}/{2} 
	\end{pmatrix} \quad 
	\zeta_2 = e^{i\theta/2} \phi_2 = e^{i\theta/2} 
	\begin{pmatrix}
		-\sin{\theta}/{2}  \\
		\cos{\theta}/{2} 
	\end{pmatrix} \nonumber
\end{eqnarray}
\noindent
Hence, $\zeta_i (\theta + 2\pi) = \zeta_i(\theta)$, $i=1,2$ and the eigenfunctions become single-valued.

\section{Van Vleck's Approach on JT Model for Realistic Systems}
\noindent
To explore the validity of the JT model, an octahedral complex, $XY_6$ is considered having fifteen (15) normal modes. Since such a complex contains a center of symmetry, the modes can be classified into gerade ($g$) and ungerade ($u$) symmetry.~\citep{van-vleck} On using the JT condition (Eq. \ref{Eq:JTcond}), one can write:
\begin{equation}
	\Gamma^0 \in \Gamma^{\beta} \otimes \left[(\Gamma^{\alpha})^2 \right]
\end{equation}	 
Since the symmetrised product of the degenerate electronic states ($d$ orbitals) of the 
central metal atom ($\left[(\Gamma^{\alpha})^2 \right]$) gives $\Gamma^{(g)}$ IREP, JT condition will be true only if: $\Gamma^{\beta} = \Gamma^{(g)}$. \\

\begin{center}
	\begin{figure}[!htp]
		\includegraphics[width=0.9\textwidth,angle=0]{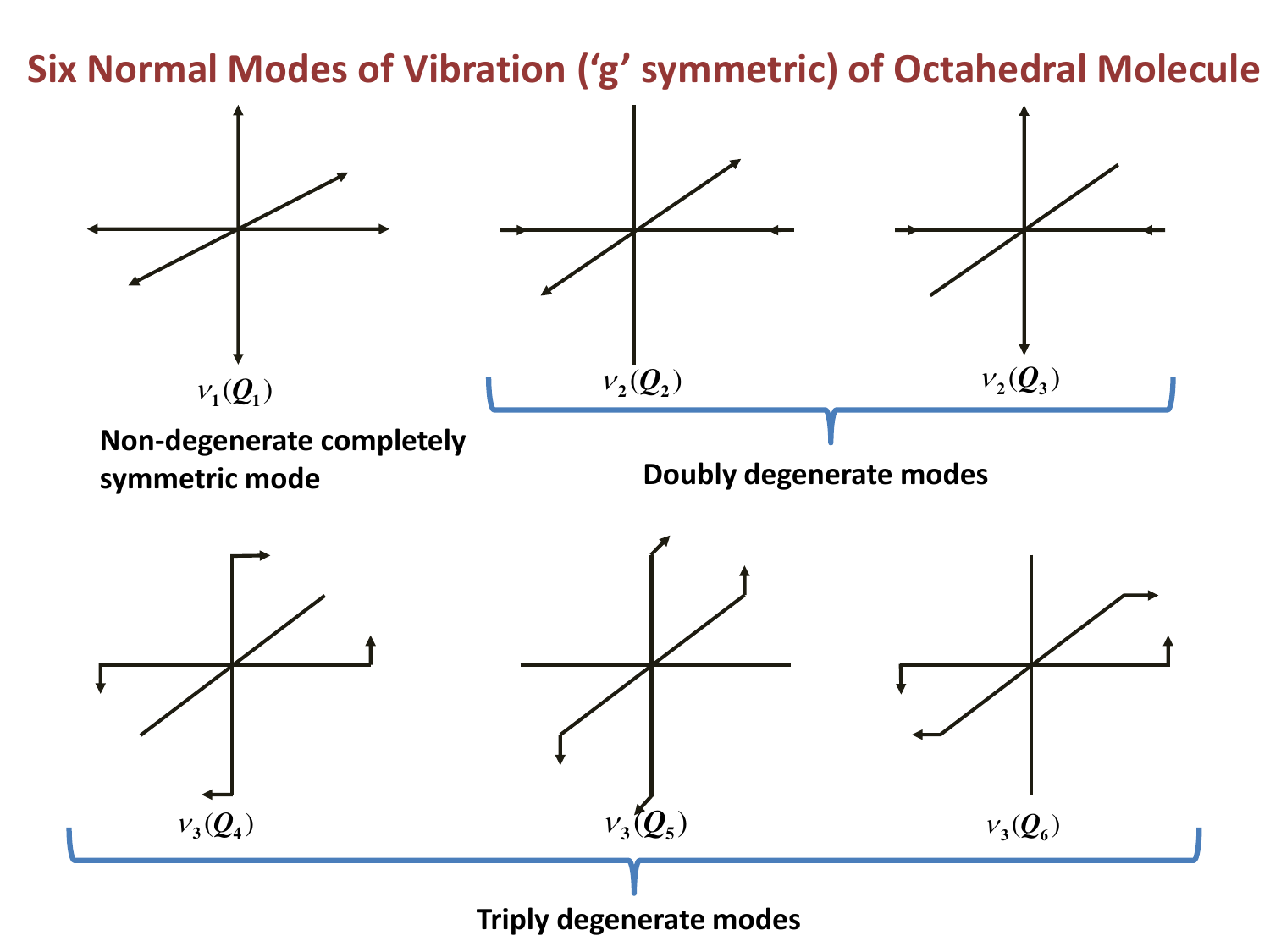}
		\caption{Vectorial displacements of atoms in octahedral molecule for six gerade ($g$) symmetric vibrational modes.}
		\label{fig:oh_JT}
	\end{figure}
\end{center}

\noindent
As mentioned before, in case of linear perturbation term in the Hamiltonian, the gerade modes will be JT active lifting the degeneracy of the unperturbed Hamiltonian, whereas the ungerade ones will not contribute. As a consequence, the following six (6) modes out of fifteen (15) are of even or `$g$' type, where $Q_2 - Q_{6}$ are the five JT active modes. On the contrary, $Q_1$ is not JT active, since it is the totally symmetric one:
\begin{small}
	\begin{eqnarray*}
		Q_1&=&[X_1-X_4+Y_2-Y_5+Z_3-Z_6]/\sqrt{6} \\
		Q_2&=&\frac{1}{2}[X_1-X_4-Y_2+Y_5]       \\
		Q_3&=&[\frac{1}{2}(X_1-X_4+Y_2-Y_5)-Z_3+Z_6]/\sqrt{3} \\ 
		Q_4&=&\frac{1}{2}[Y_1-Y_4+X_2-X_5] \\
		Q_5&=&\frac{1}{2}[Z_1-Z_4+X_3-X_6] \\
		Q_6&=&\frac{1}{2}[Z_2-Z_5+Y_3-Y_6] 
	\end{eqnarray*}
\end{small}

\noindent		
where $X_i, Y_i, Z_i$ represent displacements of $i$th ligand from the ideal octahedral arrangement (see Figure \ref{fig:oh_JT}).
Therefore, the JT distorted Hamiltonian can be written as:~\citep{van-vleck}
\begin{eqnarray*}
	H &=& W_0 + V_{\text{trig}} \nonumber \\ 
	&+& \frac{1}{2}\alpha Q^2_1 +  \frac{1}{2}\beta (Q^2_2+Q^2_3) 
	+ \frac{1}{2}\gamma Q^{\prime2}_4 + \frac{1}{2}\delta (Q^2_5+Q^2_6) \nonumber \\
	&+& V_2Q_2 + V_3Q_3 + V^{\prime}_4Q^{\prime}_4 + V^{\prime}_5Q^{\prime}_5 + V^{\prime}_6Q^{\prime}_6 \nonumber \\ 
	&+& \theta Q_1Q^{\prime}_4 + \kappa(Q_2Q^{\prime}_5 + Q_3Q^{\prime}_6) + \sum_{i=7}^{15} \sum_{j=7}^{15}\xi_{ij}Q_iQ_j, 
\end{eqnarray*}
\noindent
where $Q_i^{\prime} = \sum_{j=4}^6 c_jQ_j, \quad i = 4, 5, 6.$ $W_0$ is the unperturbed Hamiltonian for rotational and translational modes and 
also contains electronic Hamiltonian of the cluster at octahedral geometry of cubic symmetry. $V_{\text{trig}}$ defines the potential where the central atom experiences perturbation due to the bulk crystal atoms. On the other hand, $V_2$ to $V_6$ are the coefficients of the first order perturbation term resulting the JT effect.
\noindent
While applying the degenerate perturbation theory (Eq. \ref{Eq:secu}) and incorporating the above Hamiltonian, the secular equation can be formed and the degeneracy of the $d$ orbitals of the central metal atom will be split exhibiting the JT effect:
\begin{eqnarray}
	& \sum_{j=1}^m \left[\langle\phi_i^{(0)}\vert \hat{V}\vert \phi_j^{(0)}\rangle
	- E_n^{(1)}\langle\phi_i^{(0)}\vert\phi_j^{(0)}\rangle\right]c_j^n = 0&  \\
	& V_{ij}=\sum_{k} Q_{k} \langle \phi_i^{(0)} \vert \hat{V}_{k}\vert \phi_j^{(0)}\rangle.&  \\
	& E_{mn}=E_{n}^0+E_{mn}^{(1)}& 
\end{eqnarray}

\section{Dynamic JT Effect}

\subsection{The distortion phenomenon in the phase transition of rare-earth manganites}

\begin{center}
	\begin{figure}[!htp]
		\includegraphics[width=0.9\textwidth,angle=0]{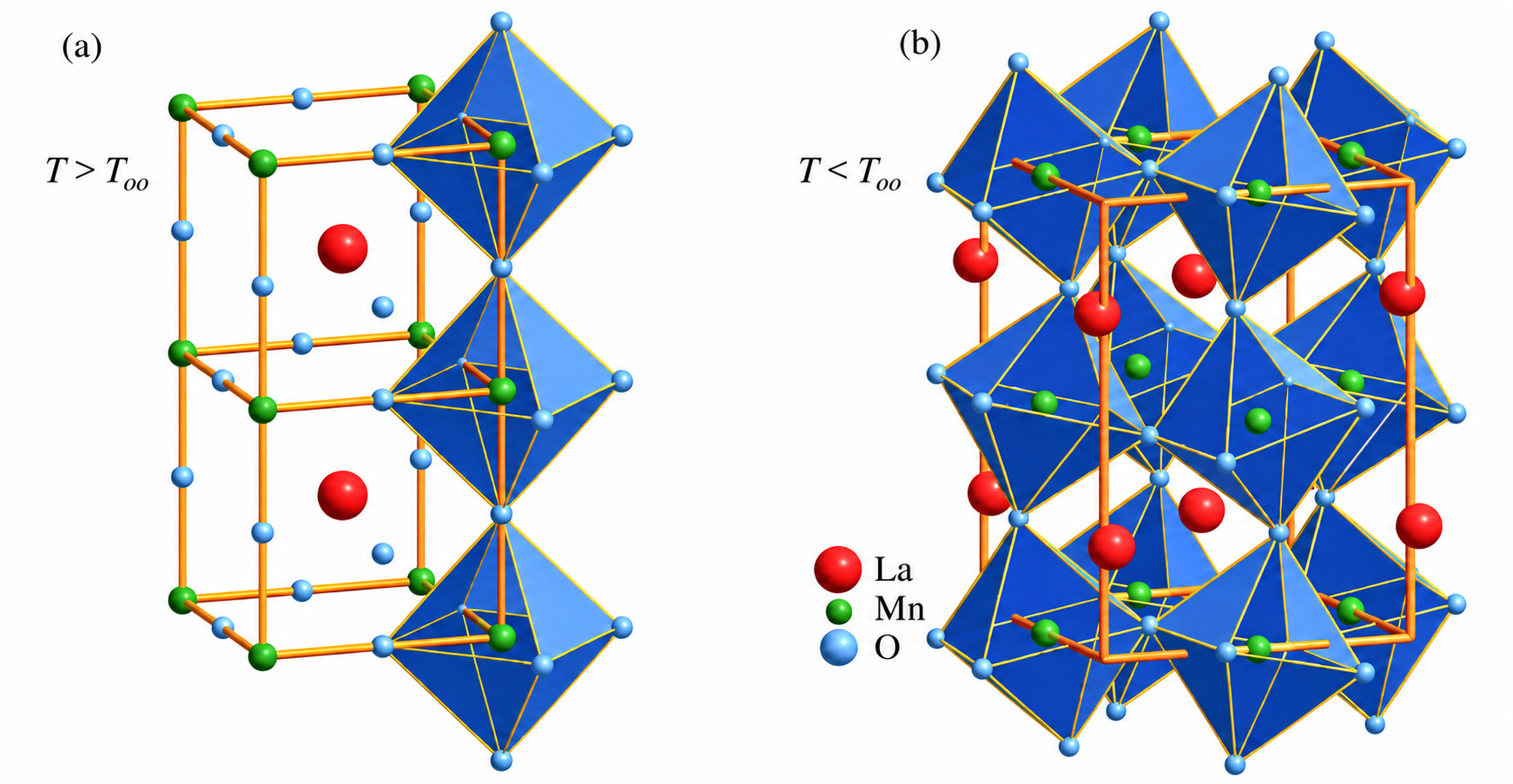}
		\caption{ (a) Structure of the high- temperature pseudo-cubic perovskite phase of rare-earth manganites REMn${O_3}$. (b) Structure of the orthorhombic $Pbnm$ phase of REMn${O_3}$ below the orbital ordering temperature
			${\bf T}_{{\bf oo}}$.\cite{Natalia_JETP}}
		\label{fig:ortho}
	\end{figure}
\end{center}

\noindent
Orthorhombic rare-earth manganites, namely, REMn$O_3$ (RE= La, Pr, Nd, Sm, Eu, Gd, Tb, Dy) are perovskites with M$n^{3+}$ $(d^4)$ ions with $t^3_{2g} e_g^1$ electronic configuration. Its high temperature phase is nearly cubic, which at $T_{oo}$ (orbital ordering temperature), undergoes a transition into the orthorhomic orbitally ordered $Pbnm$ structure with an antiferrodistortive ordering of Mn$O_6$ octahedra as depicted in Figure \ref{fig:ortho}.~\citep{Natalia_JETP}

\noindent
Recent {\it ab initio} studies~\citep{joy_jpca} demonstrate that superexchange (SE)~\citep{Kugel-Khomskii,Oles-Feiner} interaction cannot initiate the transition at $T_{oo}$. SE favors the occupation of the 
$e_g$ orbital with the orbital angle $\theta_{\angle Mn-O-Mn} = 90^{\circ}$, whereas the orbital angle determined from the neutron scattering for LaMn$O_3$ is $\theta_{\angle Mn-O-Mn} =180^{\circ}$ for cubic structure (non-antiferromagnetic) and $\theta_{\angle Mn-O-Mn} \sim 108^{\circ}$ for orthorhombic $Pbnm$ structure (antiferrodistortive). For comparison, the orbital angle of an isolated JT distorted Mn$O_6$ complex is $\theta_{\angle Mn-O-Mn} = 120^{\circ}$.

	\begin{figure}[!htp]
		\centering
		\includegraphics[width=0.7\textwidth,angle=0]{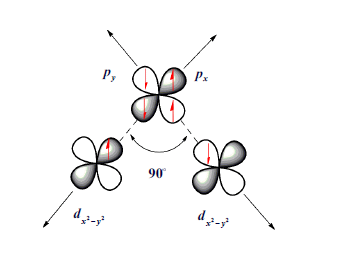}
		\caption{Superexchange antiferromagnetic interaction}
	\end{figure}

\subsection{The Jahn-Teller Effect In Octahedrally Co-ordinated Complex}

\noindent
The phenomenon of JT effect can be modelled~\citep{van-vleck} by the following potential energy operator: 
\begin{eqnarray}
	V= A
	\begin{bmatrix}
		-Q_{3} &  -Q_{2} \\
		-Q_{2} &   Q_{3}
	\end{bmatrix}
	+ \frac{1}{2} M \omega_{b\_vib}^2 \left( Q_2^2 + Q_3^2\right),
\end{eqnarray}
\noindent
which operates on the column matrix $\begin{bmatrix}
	a \\
	b
\end{bmatrix}$ with
$a \vert x^2 - y^2 \rangle + b \vert 3 z^2 - r^2 \rangle $ as the electronic state. The term $A$ is the Jahn-Teller term, which shows linear dependence on the nuclear coordinates, $Q$s. The two new variables $\rho$ and $\phi$ are defined as $Q_3 = \rho \cos \phi,  Q_2 = \rho \sin \phi$. The eigenvalues of the system under the linear Jahn-Teller perturbation are $\pm A \rho$. \\

\begin{center}
	\begin{figure}[!htp]
		\includegraphics[width=0.9\textwidth,angle=0]{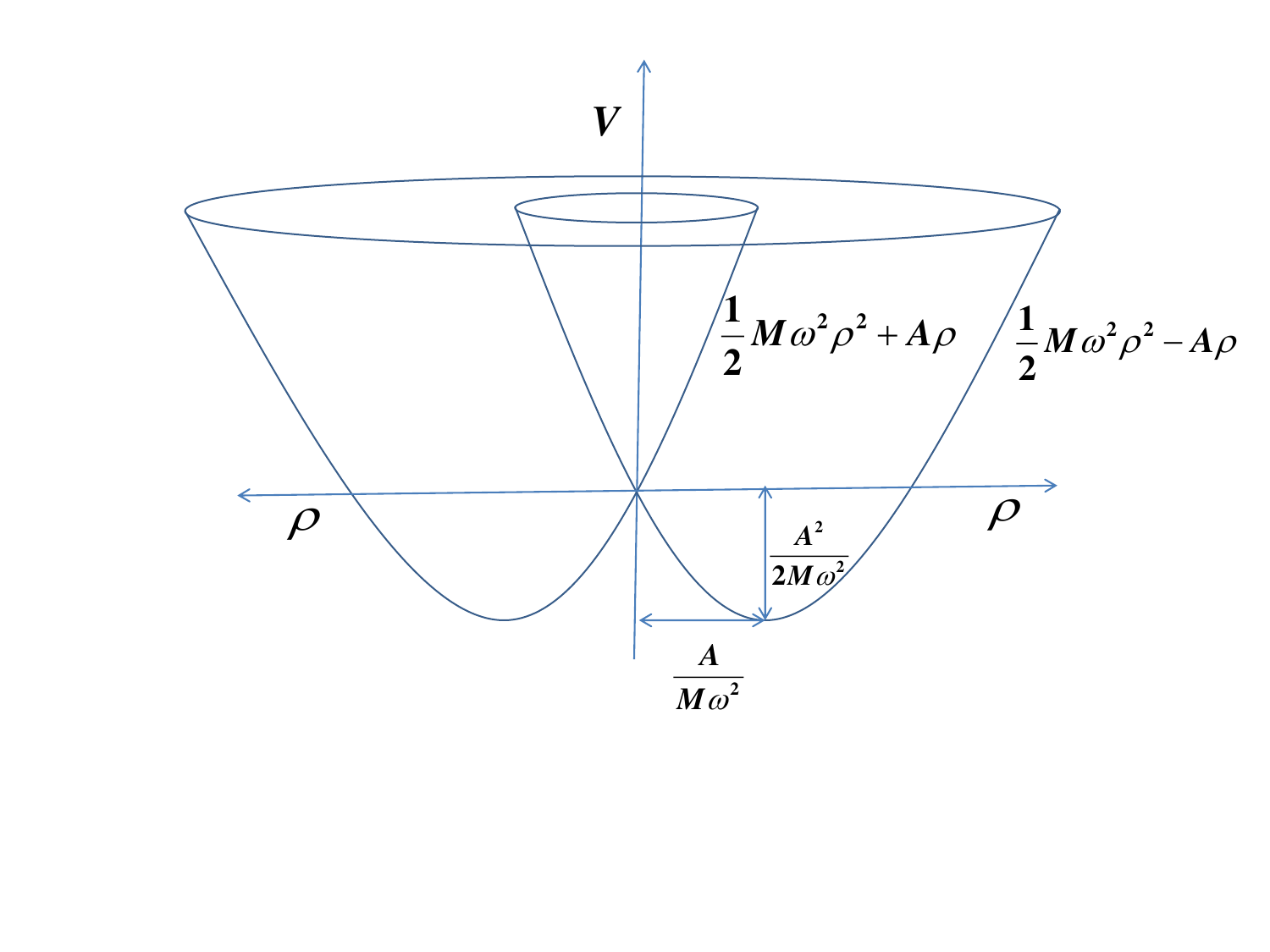}
		\caption{Diagram indicating JT splitting of nuclear potential function into two branches $ \frac{\bf 1}{\bf 2} {\boldsymbol M \boldsymbol \omega_{b \_ vib}^{2} \boldsymbol \rho^2} {\boldsymbol \pm} {\bf A \rho}$ with $E_{JT}=\dfrac{\bf A^2}{\bf 2M \omega_{b \_ vib}^{2}}$. The diagram has been adopted from Ref. [\protect\citen{lh-opik}] and permission has been taken from Royal Society.}
		\label{fig:mex}
	\end{figure}
\end{center}

\noindent
On the other hand, the vibronic potential energy surface appears as a `Mexican Hat' with the following functional form: 				
\begin{eqnarray}
	V_{\pm} = \pm A \rho + \frac{1}{2} M \omega_{b\_vib}^{2} \rho^2
\end{eqnarray}
and can be demonstrated pictorially in Figure \ref{fig:mex}.
\noindent
The JT distortion stabilizes the ground state with the amount of energy, $E_{JT}\,(=\dfrac{\bf A^2}{\bf 2M \omega_{b \_ vib}^{2}})$.~\citep{lh-opik} As a consequence, the molecular species can adapt a new geometry, which may be elongated or compressed. 	

\subsection{A treatment for coupled electron-nuclear motion}

\noindent
The motion of singly occupied electron in E$_g$ electronic state for the d$^{4}$ and d$^{9}$ transition metal ions (Mn$^{3+}$, Cu$^{2+}$) resembles as a particle in a ring motion with electronic angular coordinate, $\theta$. When the circular motion of the electron is elliptically deformed by a doubly degenerate vibrational modes Q$_2$ and Q$_3$ of e$_g$ symmetry, the nature of coupling potential energy between electron and nuclear motion should be an even function of 2$\theta$ - $\phi$.~\citep{lh-opik}

\begin{center}
	\begin{figure}[!htp]
		\includegraphics[width=0.9\textwidth,angle=0]{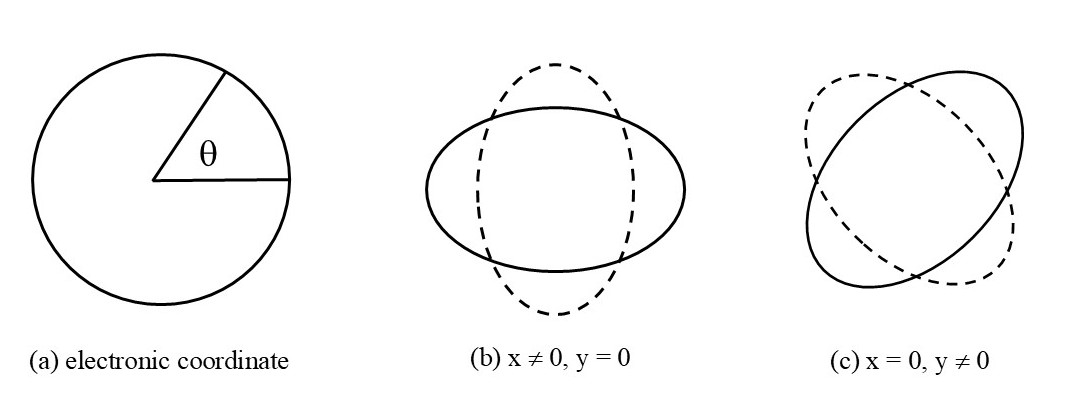}
		\caption{Model system exhibiting the Jahn-Teller effect. The diagrams have been extracted from Ref. [\protect\citen{lh-opik}] and permission has been taken from Royal Society.}
	\end{figure}
\end{center}

\noindent
For such systems, one can write:
\begin{eqnarray}
	H^{\prime} = 2 A \rho \cos(2 \theta - \phi) = A \rho \lbrace e^{i(2\theta-\phi)}+ e^{-i(2\theta-\phi)}\rbrace
	\label{Eq: distortion_coupling_potential}
\end{eqnarray}

\noindent
It is assumed that there is no coupling between the motions of the electron and
the ring. The wave function for the system then takes the following expression:
\begin{eqnarray}
	\Psi = \frac{1}{\sqrt{2 \pi}}\chi e^{i j \theta} \qquad  (j= 0, \pm1, \pm2, ...)
	\label{Eq: wavefunction uncoupled}
\end{eqnarray}
\noindent
where $\chi$ represents wavefunction for an isotropic two dimensional harmonic oscillator and $\frac{1}{\sqrt{2 \pi}} e^{i j \theta}$ defines the quantum mechanical solution of a particle in a ring. In other words, an electron is moving in a circular path with a constant radius, where $j$ is the set of integers representing the rotational quantum number.\\

\noindent
In polar co-ordinate, the unperturbed nuclear Schr\"{o}dinger equation transforms as:
\begin{equation}
	\left[ - \dfrac{\hbar^2}{2 M} \left( \dfrac{\partial^2}{\partial \rho^2}+ \dfrac{1}{\rho} \dfrac{\partial}{\partial \rho} + \dfrac{1}{\rho^2} \dfrac{\partial^2}{\partial \phi^2} \right)+\dfrac{1}{2}M \omega^2 \rho^2\right]\chi = E \chi.
\end{equation}
The solutions of this equation are of the following form:
\begin{eqnarray}
	\chi_{nm} (\rho, \phi) = \tau_{n \mid m \mid } (\rho) e^{i m \phi} , \qquad E_{m n} = n
	\label{Eq: solution_2D_oscillator}
\end{eqnarray}
where $n= 1,2, ..., $ and $ m=n-1, n-3, ..., -n+1 $.

\subsection{Evaluation of non-vanishing matrix element~\citep{lh-opik}}

\noindent
The total wavefunction of the uncoupled system for a particular $j$ value can be written as:~\citep{lh-opik}
\begin{center}
	\begin{eqnarray}
		\Psi^{j}_{n m} = \frac{1}{\sqrt{2 \pi}} \chi_{n m} e^{i j \theta} = \frac{1}{\sqrt{2 \pi}} \tau_{n \mid m \mid} e^{i (m \phi+j \theta)} 
	\end{eqnarray}
\end{center}
where the matrix element is defined as:
\begin{center}
	\begin{eqnarray}
		&\langle \Psi_{n m}^{j} \vert H^{\prime} \vert \Psi_{n^{\prime} m^{\prime}}^{j^\prime} \rangle \nonumber \\
		&= \frac{1}{2 \pi} \tau_{ n \mid m \mid}  \tau_{ n^{\prime} \mid m^{\prime} \mid} A \rho \langle e^{i (m \phi + j \theta)} \vert \lbrace e^{i(2\theta-\phi)}+ e^{-i(2\theta-\phi)} \rbrace \vert e^{i (m^{\prime} \phi + j^{\prime}  \theta)} \rangle 
	\end{eqnarray}
\end{center}

\noindent
Such integral would not vanish, if the following conditions fulfill:
\begin{center}
	$ { \bf j^{\prime} = j - 2 } $ , $ { \bf m^{\prime} = m + 1 }$ \\
	$ { \bf j^{\prime} = j + 2 } $ , $ { \bf m^{\prime} = m - 1 }$
\end{center}
\noindent
Therefore, it can be readily realized that only for the  case, {\color{black}${ \bf \mid j \mid = \mid j^{\prime} \mid =1 }$}, two vibrational levels of the same electronic state interact under the perturbation {\color{black}$H^{\prime}$}.

\begin{center}
	\begin{figure}[!htp]
		\includegraphics[width=0.9\textwidth,angle=0]{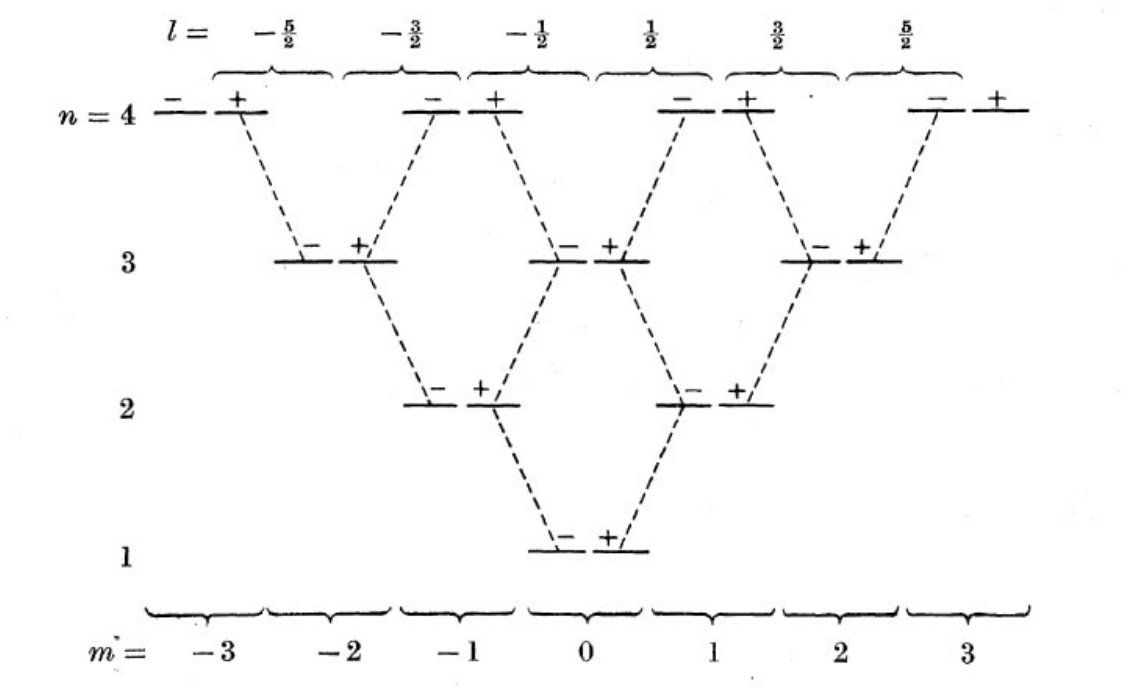}
		\caption{Interaction scheme between the levels of ${\boldsymbol \psi_{\bf n m}}{\boldsymbol (}{\bf \rho}, {\boldsymbol \phi}{\boldsymbol )}$ for ${\boldsymbol \mid} {\bf j} {\boldsymbol \mid}{\boldsymbol =}{\bf 1}$ under ${\boldsymbol H^{\prime}} {\boldsymbol =} {\bf 2} {\bf A \rho} {\boldsymbol \cos} {\boldsymbol (}{\bf 2}{\boldsymbol \theta} {\bf -} {\boldsymbol \phi}{\boldsymbol )}$. The diagram has been adapted from Ref. [\protect\citen{lh-opik}] and permission has been taken from Royal Society.}
	\end{figure}
\end{center}

\noindent
The interesting fact is that levels do not interact together, but only those with the same value of {\color{black}$ l=m+\frac{1}{2}j $. The new quantum number {\color{black}$\boldsymbol l$} defines the electron-vibration coupling. \\
	
	\noindent	
	The perturbation {\color{black}$H^{\prime}$} mixes levels together whose {\color{black}$j$} values are {\color{black}$+1$} and {\color{black}$-1$}. Thereby, the wavefunction for given {\color{black}$l$} is of the following form:	
	\begin{eqnarray}
		\Psi =  \chi_{+} (\rho, \phi) \psi_{+} (\theta) + \chi_{-} (\rho, \phi) \psi_{-} (\theta), 
	\end{eqnarray}
	where $\chi_{\pm}$ and $\psi_{\pm}$ are the nuclear and electronic wavefunctions, respectively, with 
	\begin{eqnarray}
		\chi_{+} = \tau_{l}^{+}(\rho) e^{i(l- \frac{1}{2}j)\phi}, \qquad  \chi_{-} = \tau_{l}^{-}(\rho) e^{i(l+ \frac{1}{2}j)\phi}.
		\label{Eq: Coefficientmixedequation}
	\end{eqnarray}	
	
	\noindent
	Under the linear Jahn-Teller perturbation, ${\color{black}H^{\prime} = A \rho \lbrace e^{i(2 \theta - \phi)} + e^{-i(2 \theta - \phi)} \rbrace}$, the {\color{black}$\chi_{+}$} and {\color{black}$\chi_{-}$} are coupled through the following matrix, which is known as Dynamic Jahn-Teller matrix: 	
	\begin{eqnarray}
		\begin{bmatrix}
			H_{0} - E & A \rho e^{-i \phi} \\
			A \rho e^{i \phi} & H_{0} - E  
		\end{bmatrix}
		\begin{bmatrix}
			\chi_{+} \\
			\chi_{-} 
		\end{bmatrix} 
		= 0 
		\label{Eq: dynamicJahnmatrix}
	\end{eqnarray}	
	leading to
		
	\begin{eqnarray}
			\left[ \frac{\hbar^2}{2M} \lbrace - \frac{d^{2}}{d \rho^{2}} - \frac{1}{\rho} \frac{d}{d \rho} + \frac{(l - \frac{1}{2})^{2}}{\rho^{2}} \rbrace  - E + \frac{1}{2} M \omega_{\text{b}_vib}^{2} \rho^{2} \right] \tau_{l}^{+}(\rho) + A \rho \tau_{l}^{-}(\rho) = 0  \qquad \\
		\text{and}\nonumber\\
		A \rho \tau_{l}^{+}(r) + \left[ \frac{\hbar^2}{2M} \lbrace - \frac{d^{2}}{d \rho^{2}} - \frac{1}{\rho} \frac{d}{d \rho} + \frac{(l + \frac{1}{2})^{2}}{\rho^{2}} \rbrace - E  + \frac{1}{2} M \omega_{\text{b}_vib}^{2} \rho^{2}  \right]\tau_{l}^{-}(\rho) = 0. 
	\end{eqnarray}	
	\noindent
	Subtracting the above two equations and multiplying with $(-1)^{l - \frac{1}{2}}$, we get the following compact form:
	\begin{eqnarray}
		(H_{l} - A \rho - E) u_{g}(\rho) = (-1)^{l- \frac{1}{2}} \left(\frac{\hbar^{2} l}{2 M \rho^2} \right) u_{e}(\rho),
		\label{eq:hamil1} 
	\end{eqnarray}
	where 	${\color{black}u_{e}(\rho) \equiv \tau_{l}^{+} + \tau_{l}^{-}}$ and ${\color{black}u_{g}(\rho) \equiv (-1)^{l- \frac{1}{2}}(\tau_{l}^{+} - \tau_{l}^{-})}$ signify the wavefunctions for upper and lower branches of the PES, respectively. The associated nuclear Hamiltonian for $l^{th}$ quantum number takes the following form:
	\begin{eqnarray}
		H_{l} = \frac{\hbar^2}{2M} \left[- \frac{d^{2}}{d \rho^{2}} -\frac{1}{\rho} \frac{d}{d \rho} + \frac{l^2+\frac{1}{4}}{\rho^2} \right]+ \frac{1}{2} M \omega_{b \_ vib}^{2} \rho^{2}.
	\end{eqnarray}
	
	\noindent
	When {\color{black}$\rho_{min} = \frac{A^2}{M \omega^2}$} is large and {\color{black}$E$} is low, the system is placed on lower Born-Oppenheimer surface. Hence, a good approximation is {\color{black}$u_{e} (\rho) = 0$} and Eq. \ref{eq:hamil1} turns into:
	\begin{eqnarray}
		\left( H_l - A \rho - E^{\textit{g}}\right) u_{g}(\rho) = 0 , 
		\label{Eq: Wave-equation-lowersurf}
	\end{eqnarray}	
	where {\color{black}$E^{\textit{g}}$} corresponds to lower energy surface. While using $u_g(\rho) = \rho^{-1/2} \psi_g(\rho)$ and substituted in above equation, we get the following radial Schr\"odinger equation:
	\begin{eqnarray}
		-\frac{\hbar^2}{2M} \left[\frac{d^2 \psi_g(\rho)}{d \rho^2} - \frac{l^2}{\rho^2} \psi_g(\rho)\right] - E^{g} \psi_g(\rho) + \left( -A \rho + \frac{1}{2} M \omega_{\text{b\_vib}}^2 \rho^2 \right) \psi_g(\rho) = 0.
		\label{Eq: radial_lower_surface_equ_appendix}
	\end{eqnarray}	
	\noindent
	Thus, the total Schr\"odinger equation for the ground state becomes:
	\begin{eqnarray}
		-\frac{\hbar^2}{2 M} \left[\frac{\partial^2 \chi_g}{\partial \rho^2} + \frac{1}{\rho^2} \frac{\partial^2 \chi_g}{\partial \phi^2} \right] - E^{g} \chi_g + \left(-A \rho + \frac{1}{2} M \omega_{\text{b\_vib}}^2 \rho^2\right) \chi_g = 0,
		\label{Eq: Lower-surface-equ-appendix}
	\end{eqnarray}
	with $\chi_g = \psi_g (\rho) e^{i l \phi}$. Moreover, while substituting the expressions of $\chi_{\pm}$, $\psi_g (\rho)$ and $u_g (\rho)$ as shown above, $\chi_g$ can be written in terms of $\chi_+$ and $\chi_-$ as
	\begin{eqnarray}
		\chi_g &=& \sqrt{\rho} \left( -1 \right)^{l - \frac{1}{2}} \lbrace \chi_{+} e^{i \phi/2} - \chi_- e^{-i \phi/2}\rbrace \nonumber \\
		&=& \sqrt{\rho} \left( -1 \right)^{l - \frac{1}{2}} \lbrace \tau_{l}^{+} (\rho) - \tau_{l}^{-} (\rho) \rbrace e^{i l \phi}
	\end{eqnarray}
	
	\noindent
	On the other hand, the Schr\"odinger equation for excited state takes the following form:
	\begin{eqnarray}
		\left( H_{l} + A \rho - E \right) u_{e}(\rho) = (-1)^{-(l - \frac{1}{2})} \left( \frac{\hbar^{2} l}{2 M \rho^{2}}\right) u_{g} (\rho), 
		\label{Eq: higher surface equation}
	\end{eqnarray}
	\noindent
	When the energy ($E$) is substantially high, the Eq. (\ref{Eq: higher surface equation}) turns into the following form:	
	\begin{eqnarray}
		\left( H_{l} + A \rho - E^{e} \right) u_{e}(\rho) = 0.
		\label{Eq: higher surface diff eqn}
	\end{eqnarray}
	\noindent
	Employing $u_e (\rho) = \rho^{-1/2} \psi_e (\rho)$, the following expression of radial nuclear SE for excited state is obtained:
	\begin{eqnarray}
		-\frac{\hbar^2}{2M} \left[\frac{d^2 \psi_e(\rho)}{d \rho^2} - \frac{l^2}{\rho^2} \psi_e(\rho)\right] - E^{e} \psi_e(\rho) + \left( A \rho + \frac{1}{2} M \omega_{\text{b\_vib}}^2 \rho^2 \right) \psi_e(\rho) = 0.
		\label{Eq: radial_excited_surface_equ_appendix}
	\end{eqnarray} 			
	
	\noindent
	As a consequence, the total SE for the excited state can be expressed as:
	\begin{eqnarray}
		-\frac{\hbar^2}{2 M} \left[\frac{\partial^2 \chi_e}{\partial \rho^2} + \frac{1}{\rho^2} \frac{\partial^2 \chi_e}{\partial \phi^2} \right] - E^{e} \chi_e + \left(A \rho + \frac{1}{2} M \omega_{\text{b\_vib}}^2 \rho^2\right) \chi_e = 0,
		\label{Eq: Upper-surface-equ-appendix}
	\end{eqnarray} 
	with $\chi_e = \psi_{e} (\rho) e^{i l \phi}$. Similarly, we have defined $\chi_e$ in terms of $\chi_+$ and $\chi_-$:			
	\begin{eqnarray}
		\chi_e &=& \sqrt{\rho} \lbrace \chi_{+} e^{i \phi/2} + \chi_- e^{-i \phi/2}\rbrace \nonumber \\
		&=& \sqrt{\rho} \lbrace \tau_{l}^{+} (\rho) + \tau_{l}^{-} (\rho) \rbrace e^{i l \phi}
	\end{eqnarray}

	\subsection{Ground state adiabatic Schr\"odinger equation under cubic perturbation}
	\noindent
	The radial Schr\"odinger equation for the ground electronic state is defined as:~\citep{lh-opik}
	\begin{eqnarray}
		-\frac{\hbar^2}{2M} \left[\frac{d^2 \psi_g(\rho)}{d \rho^2} - \frac{l^2}{\rho^2} \psi_g(\rho)\right] - E^{g} \psi_g(\rho) + \left( -A \rho + \frac{1}{2} M \omega_{\text{b\_vib}}^2 \rho^2 \right) \psi_g(\rho) = 0.
		\label{Eq: radial_lower_surface_equ_appendix}
	\end{eqnarray}
	\noindent
	Since the centrifugal term $\frac{\hbar^2 l^2}{2 M \rho^2}$ appears as constant $\frac{\hbar^2 l^2}{2 M \rho_{0}^2}$ at the minima ($\rho_{0}= \frac{A}{M \omega_{\text{b\_vib}}^2}$) of the lower `Mexican hat' PES, the above equation can be transformed with shifted coordinate $s$ ($= \rho - \rho_{0}$) as:
	\begin{eqnarray}
		\left(-\frac{\hbar^2}{2M} \frac{d^2}{d s^2} +  \frac{1}{2} M \omega_{\text{b\_vib}}^{2} s^{2} \right) w_{g}(s) &=& \left( E^{g} +  \frac{A^2}{2 M \omega_{\text{b\_vib}}^2} - \frac{\hbar^2 l^2}{2 M \rho_{0}^{2}}\right)  w_{g}(s),\qquad
	\end{eqnarray}
	\noindent
	where the corresponding expression for the eigenvalues takes the following form:
	\begin{eqnarray}
		E_{p l}^{g} &=& \hbar \omega_{\text{b\_vib}} \left( p + \frac{1}{2} \right) - \frac{A^2}{2 M \omega_{\text{b\_vib}}^2} + \frac{\hbar^2 l^2}{2 M \rho_{0}^{2}},  \quad p = 0, 1, 2, \cdot \cdot \cdot
		\label{Eq: Eigenvalue of lower surface}
	\end{eqnarray}
	Therefore, at the minima of ground state PES, nuclei behave as a rotating oscillator.
	
	\noindent
	On the other hand, the potential energy operator in presence of anharmonicity can be defined as:~\citep{O'Brien}
	\begin{eqnarray}
		V = A
		\begin{bmatrix}
			-Q_{3} &  -Q_{2} \\
			-Q_{2} &   Q_{3}
		\end{bmatrix}
		+ \frac{1}{2} M \omega_{b \_ vib}^2 \left( Q_2^2 + Q_3^2\right) + A_3 Q_3 \left( Q_3^2 - 3 Q_2^2\right).
	\end{eqnarray}
	In polar coordinate, the anharmonic term transforms as:~\citep{O'Brien}
	\begin{eqnarray}
		A_3 Q_3 \left( Q_3^2 - 3 Q_2^2\right) \Longrightarrow A_3 \rho^3 \cos 3 \phi
	\end{eqnarray}
	and the lowest potential surface (see Figure \ref{fig:cubic_ground}) appears as:
	\begin{eqnarray}
		V_{-}= -A \rho + \frac{1}{2} M \omega_{b \_ vib}^2 \rho^2 + A_3 \rho^3 \cos 3 \phi + \dfrac{\hbar^2 l^2}{2 M \rho^2}
	\end{eqnarray}
	Such surfaces possesses three minima as follows:
	\begin{center}
		$\phi = 0, \frac{2 \pi}{3}, \frac{4 \pi}{3}; A_3 < 0$ and $ \phi = \frac{\pi}{3}, \pi, \frac{5 \pi}{3}; A_3 > 0$.
	\end{center}
	
	\begin{center}
		\begin{figure}[!htp]
			\includegraphics[width=0.6 \textwidth, angle=270]{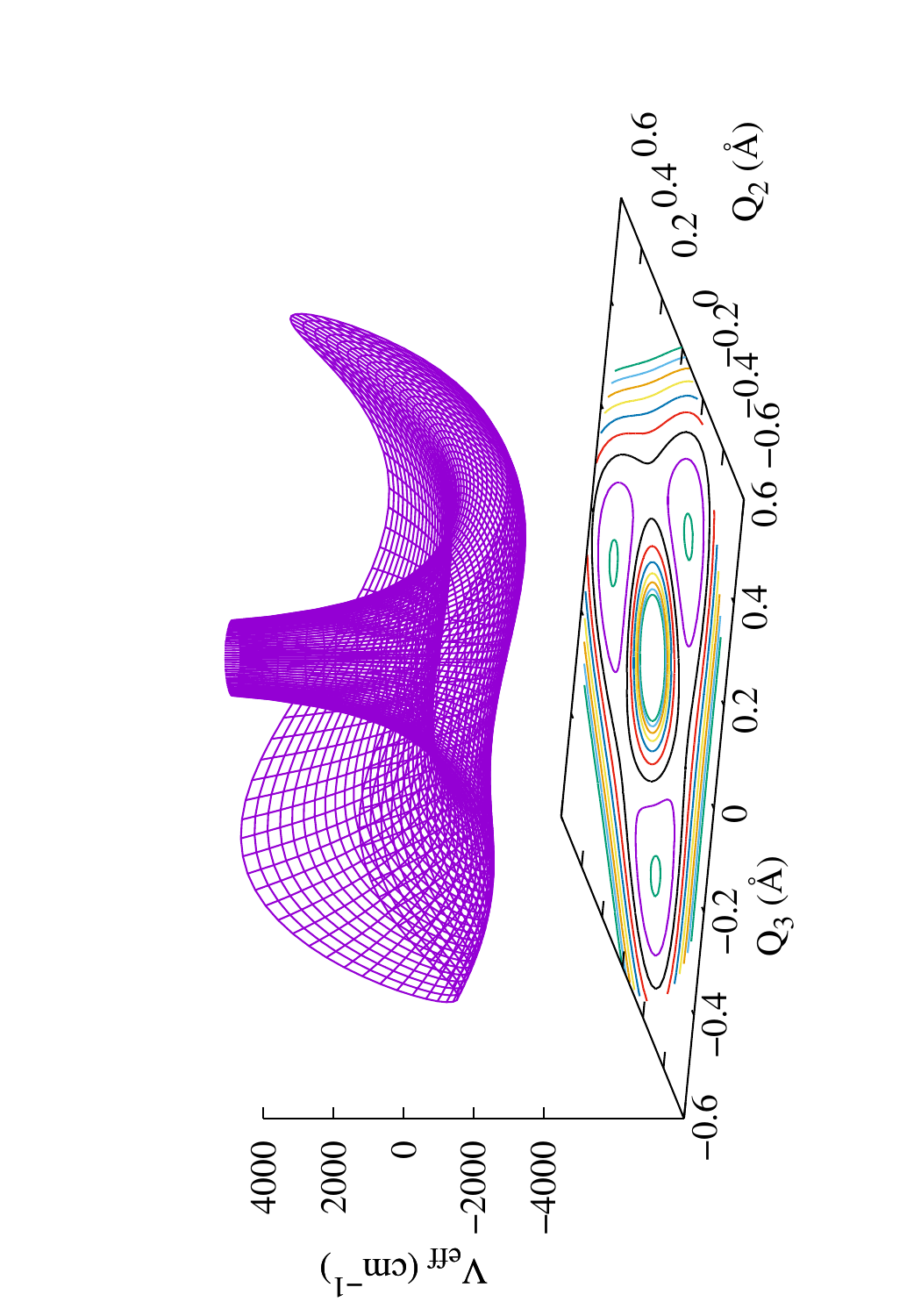}
			\caption{Lower-sheet of the cubic perturbed Mexican hat surface in normal coordinate {\color{black}$Q_{2}$} and {\color{black}$Q_{3}$}, where {\color{black}$V_{-} = -A \rho + \frac{1}{2} M \omega_{b \_ vib}^2 \rho^2 + A_{3} \rho^3 \cos(3 \phi) + \dfrac{\hbar^2 l^2}{2 M \rho^2}$.} Data has been taken from Ref. [\protect\citen{joy_jcp}].			
			}
			\label{fig:cubic_ground}
		\end{figure}
	\end{center}
	
	\noindent
	In the above figure (Figure \ref{fig:cubic_ground}), since the ground adiabatic potential ({\color{black}$V_{-}$}) blows up at the vicinity {\color{black}$r \sim 0$} due to the centrifugal term {\color{black}$\frac{\hbar^2 l^2}{2 M \rho^2}$} though its width is very small, it doesn't affect the three minima away from the center ({\color{black}$\rho \sim 0.3$ \AA})~\citep{Natalia_JETP,joy_jcp} arising at 120$^\circ$({\color{black}$\phi$}) interval. As a consequence, we neglect the centrifugal term and the solution of the {\color{black}{radial part}} would be a {\color{black}{harmonic oscillator}} one.\\
	
	\noindent		
	The Schr\"odinger equation for lower-sheet of the Mexican hat including the cubic perturbation term is of the 
	following form:	
	\begin{eqnarray}
		&&-\frac{\hbar^2}{2M} \left[ \frac{\partial^2}{\partial \rho^2} +  \frac{1}{\rho^2} \frac{\partial^2}{\partial \phi^2} \right] \xi_g (\rho, \phi) - E \xi_g(\rho, \phi) \nonumber \\
		&& + \left( - A \rho + \frac{1}{2} M \omega_{\text{b\_vib}}^2 \rho^2 + A_3 \rho^3 \cos(3 \phi) \right) \xi_g(\rho, \phi) = 0,
		\label{Anharmonic-lower-schroedinger-eqn_1}
	\end{eqnarray}	
	\noindent
	If the cubic term is sufficiently small and the difference of the frequencies of the 
	radial and rotational motion is sufficiently large, an approximation for separation of these variables is possible in the adiabatic approach. \\
	
	\noindent
	The operator 
	\begin{eqnarray}
		-\frac{\hbar^2}{2 M \rho^2} \frac{\partial^2}{\partial \phi^2} + A_3 \rho^3 \cos 3 \phi 
	\end{eqnarray}
	\noindent
	acts as a perturbation within the unperturbed eigenstates, {\color{black}$\chi_g(\rho, \phi)$} (see Eq. \ref{Eq: Lower-surface-equ-appendix}), where the perturbed eigenfunction {\color{black}$\xi_g(\rho, \phi)$ [$= w_g^0 (\rho) \gamma(\phi)$]} is restricted with the lowest radial eigenfunction, {\color{black}$w_g^0(\rho)$}. \\
	
	\noindent
	Plugging the above form of wavefunction {\color{black}$\xi_g(\rho, \phi)$}, in the Eq. (\ref{Anharmonic-lower-schroedinger-eqn_1}), we obtain:			
	\begin{eqnarray}
		&&\frac{\hbar^2}{2 M} \frac{d^2}{d \rho^2} \vert w_{g}^{0}(\rho) \rangle \vert \gamma(\phi) \rangle + \frac{\hbar^2}{2 M \rho^2} \frac{d^2}{d \phi^2} \vert w_{g}^{0}(\rho) \rangle \vert \gamma(\phi) \rangle + E \vert w_{g}^{0}(\rho) \rangle \vert \gamma(\phi) \rangle \nonumber \\
		&& - \left( - A \rho + \frac{1}{2} M \omega_{\text{b\_vib}}^2 \rho^{2} + A_{3} \rho^3 \cos(3 \phi) \right) \vert w_{g}^{0}(\rho) \rangle \vert \gamma(\phi) \rangle = 0 
	\end{eqnarray}		
	
	\noindent
	After averaging over the radial part, the above equation ended up into the following form:
	\begin{eqnarray}
		&&\langle w_{g}^{0}(\rho) \vert \frac{\hbar^2}{2 M \rho^2} \vert w_{g}^{0}(\rho) \rangle \frac{d^2}{d \phi^2} \vert \gamma(\phi) \rangle + \langle w_{g}^{0}(\rho) \vert - A_3 \rho^3 \vert w_{g}^{0}(\rho) \rangle \cos(3 \phi) \vert \gamma(\phi) \rangle \nonumber \\
		&&+ \langle w_{g}^{0}(\rho) \vert \frac{\hbar^2}{2 M} \frac{d^2}{d \rho^2} + \epsilon_{0} - \left( - A \rho + \frac{1}{2} M \omega_{\text{b\_vib}}^2 \rho^2 \right) \vert w_{g}^{0}(\rho) \rangle \vert \gamma(\phi) \rangle \nonumber \\
		&&+ \langle w_{g}^{0}(\rho) \vert \epsilon \vert w_{g}^{0}(\rho) \rangle \vert \gamma(\phi) \rangle = 0
		\label{Eq: gq1} 
	\end{eqnarray}
	
	\noindent
	It is necessary to mention that {\color{black}$E = \epsilon_{0} + \epsilon $}, where {\color{black}$\epsilon_{0}$} defines energy without {\color{black}cubic perturbation} and {\color{black}$\epsilon$} represents energy of {\color{black}cubic perturbation}:
	\begin{eqnarray}
		\frac{\hbar^2}{2 M} \frac{d^2}{d \rho^2} \vert w_{g}^{0}(\rho) \rangle + \epsilon_{0} \vert w_{g}^{0}(\rho) \rangle - \left( -A \rho + \frac{1}{2} M \omega_{\text{b\_vib}}^2 \rho^2\right) \vert w_{g}^{0}(\rho) \rangle = 0,
		\label{Eq: gq2}
	\end{eqnarray}
	
	\noindent
	Using Eq. (\ref{Eq: gq2}), Eq. (\ref{Eq: gq1}) turns into:		
	\begin{eqnarray}
		&&\langle  w_{g}^{0}(\rho) \vert \frac{\hbar^2}{2 M \rho^2} \vert  w_{g}^{0}(\rho) \rangle \frac{d^2}{d \phi^2} \vert \gamma(\phi) \rangle + \langle  w_{g}^{0}(\rho) \vert - A_{3} \rho^{3} \vert  w_{g}^{0}(\rho) \rangle \cos(3 \phi) \vert \gamma(\phi) \rangle \nonumber \\
		&&+ \langle  w_{g}^{0}(\rho) \vert \epsilon \vert  w_{g}^{0}(\rho) \rangle \vert \gamma(\phi) \rangle = 0 \nonumber \\
		&&\Longrightarrow \alpha \frac{d^2}{d \phi^2} \vert \gamma(\phi) \rangle + \beta \cos(3 \phi) \vert \gamma(\phi) \rangle + \epsilon \vert \gamma(\phi) \rangle = 0,
		\label{Eq: ground state Mathieu Eq_1}
	\end{eqnarray}			
	where
	\begin{eqnarray}
		\epsilon&=& \langle w_{g}^{0}(\rho) \vert \epsilon \vert w_{g}^{0}(\rho) \rangle \Longrightarrow \quad \text{Eigenvalue after perturbation} \nonumber \\
		\alpha&=& \langle w_{g}^{0}(\rho) \vert \frac{\hbar^2}{2 M \rho^2} \vert w_{g}^{0}(\rho) \rangle \Longrightarrow \quad \text{Splitting parameter} \nonumber \\
		\beta &=& \langle w_{g}^{0}(\rho) \vert - A_{3} \rho^{3} \vert w_{g}^{0}(\rho) \rangle  \Longrightarrow \quad \text{Barrier parameter}
	\end{eqnarray}		
	
	\noindent
	The solutions of Eq. (\ref{Eq: ground state Mathieu Eq_1}) can conveniently be written as a Fourier series:
	\begin{eqnarray}
		\gamma = \sum_l a_le^{i l \phi} , \\
		l+\frac{1}{2} \equiv \text{integer} \nonumber 
	\end{eqnarray}	
	
	\noindent
	The boundary condition can be defined as:
	\begin{center}
		$\gamma(\phi + 2 \pi )= - \gamma(\phi)$.
	\end{center}
	
	\noindent
	On the other hand, the coefficients satisfy the following recurrence relation			
	\begin{eqnarray}
		a_l \left( \epsilon - \alpha l^2\right) + \frac{1}{2} \beta \left( a_{l-3} + a_{l+3} \right) = 0.
		\label{Eq:determinants-coeff}
	\end{eqnarray}			
	The equation (\ref{Eq:determinants-coeff}) can be written in the following matrix form:				
	\begin{eqnarray}
		\begin{bmatrix}
			(\epsilon - \frac{49 \alpha }{4}) & 0 & 0 & \frac{\beta}{2} & 0 & 0 & 0 & 0 \\ \\
			0 & (\epsilon - \frac{25 \alpha}{4}) & 0 & 0 & \frac{\beta}{2} & 0 & 0 & 0 \\ \\
			0 & 0 & (\epsilon - \frac{9 \alpha}{4}) & 0 & 0 & \frac{\beta}{2} & 0 & 0 \\ \\
			\frac{\beta}{2} & 0 & 0 & (\epsilon - \frac{ \alpha }{4}) & 0 & 0 & \frac{\beta}{2} & 0 \\ \\
			0 & \frac{\beta}{2} & 0 & 0 & (\epsilon - \frac{ \alpha }{4}) & 0 & 0 & \frac{\beta}{2} \\ \\
			0 & 0 & \frac{\beta}{2} & 0 & 0 & (\epsilon - \frac{9 \alpha}{4}) & 0 & 0 \\ \\ 
			0 & 0 & 0 & \frac{\beta}{2} & 0 & 0 & (\epsilon - \frac{25 \alpha}{4}) & 0 \\ \\
			0 & 0 & 0 & 0 & \frac{\beta}{2} & 0 & 0 & (\epsilon - \frac{49 \alpha }{4}) 
		\end{bmatrix}
		\begin{bmatrix}
			\vspace{0.02cm}
			a_{-7/2} \\ \\ 
			a_{-5/2} \\ \\
			a_{-3/2} \\ \\
			a_{-1/2} \\ \\
			a_{1/2} \\ \\ 
			a_{3/2} \\ \\
			a_{5/2} \\ \\
			a_{7/2} \\ \\ 
			\vspace{-0.1 cm}
		\end{bmatrix}
		= 0 \nonumber \\ \nonumber \\ \nonumber
		\label{Eq: Ground-state-Matheiu-determinant}
	\end{eqnarray}	
	\noindent
	Hence, the corresponding determinant can be  diagonalised in order to find the eigenvalues of the Eq. \ref{Eq: Ground-state-Matheiu-determinant}.
	
	\begin{center}
		\begin{figure}[!htp]
			\includegraphics[width=1 \textwidth, angle=0]{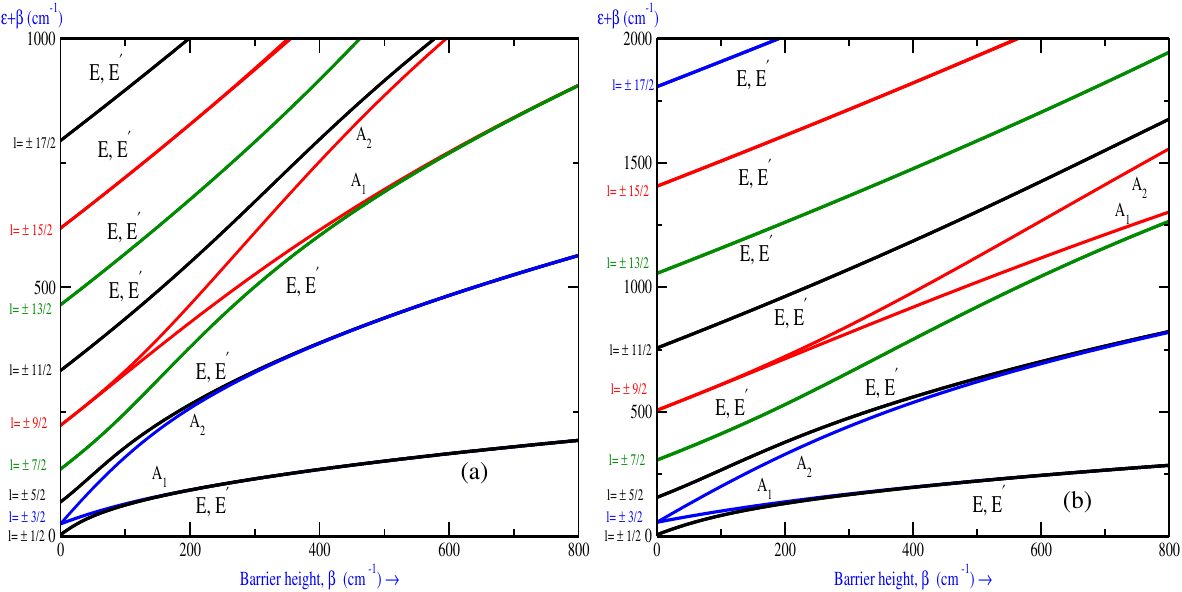}
			\caption{The variation of roto-vibrational energy levels for pseudo rotational quantum number {\color{black}$l= \frac{1}{2}$} to {\color{black}$l = \frac{17}{2}$} with respect to barrier parameter {\color{black}$\beta$} for (a) {\color{black}[Cu(H$_2$O)$_6$]$^{2+}$} and (b) {\color{black}LaMnO$_3$} complexes. In case of {\color{black}[Cu(H$_2$O)$_6$]$^{2+}$} complex, {\color{black}$\alpha$} and {\color{black}$\beta$} values are {\color{black}11 cm$^{-1}$} and {\color{black}544 cm$^{-1}$}, whereas for {\color{black}LaMnO$_3$} complex, {\color{black}$\alpha$} and {\color{black}$\beta$} values are {\color{black}25 cm$^{-1}$} and {\color{black}375 cm$^{-1}$}, respectively.~\citep{Natalia_JETP,joy_jcp} The figures have been taken from Ref. [\protect\citen{joy_jcp}] and permission has been granted from American Institute of Physics (AIP).}  
			\label{Fig: ground state eigenvalue}
		\end{figure}
	\end{center}
	For the above equations, three kinds of solution are possible, namely,
	{\color{black}$ \lbrace l = \pm \frac{3}{2}, \pm \frac{9}{2}, \pm \frac{15}{2},...., \rbrace$}, {\color{black}$ \lbrace l = + \frac{1}{2}, - \frac{5}{2}, + \frac{7}{2}, - \frac{11}{2}....,\rbrace $} and {\color{black}$\lbrace l = -\frac{1}{2}, + \frac{5}{2}, - \frac{7}{2}, +\frac{11}{2}....,\rbrace$}, and the solutions of the last two sets are degenerate pairs. The solution can be classified by representations of the symmetry group {\color{black}$C_{3v}$} to which the equation belongs 
	if {\color{black}$\phi$} supposed to be a rotation about an axis. \\
	
	\noindent
	{\color{black}$l =  \pm \frac{3}{2}, \pm \frac{9}{2}, \pm \frac{15}{2},.... $}  transforms as {\color{black} ${\boldsymbol A_1}$} if {\color{black}$a_l = a_{-l}$} and transforms as {\color{black} ${\boldsymbol A_2}$} if {\color{black}$a_l = -a_{-l}$}. {\color{black}$l = + \frac{1}{2}, - \frac{5}{2}, + \frac{7}{2}, - \frac{11}{2}....,\Longrightarrow$} transforms as  {\color{black} ${\boldsymbol E}$} (labeled as {\color{black} ${\boldsymbol E}$}). {\color{black}$l = -\frac{1}{2}, + \frac{5}{2}, - \frac{7}{2}, +\frac{11}{2}....,\Longrightarrow$} transforms as  {\color{black} ${\boldsymbol E}$} (labeled as {\color{black} ${\boldsymbol E^{\prime}}$}).
	
	\subsection{Excited state adiabatic Schr\"odinger equation under cubic perturbation}
	
	\noindent
	The nuclear SE for radial Hamiltonian of upper sheet of APES can be written as\citep{Slonczewski}:

	\vspace*{-0.6 cm}

	\begin{eqnarray}
		H \psi_e(\rho)= \left(- \frac{\hbar^2}{2M} \frac{\partial^2}{\partial \rho^2} + \frac{\hbar^2 l^2}{2M\rho^2} + A \rho + \frac{1}{2} M \omega^2_{\text{b\_vib}} \rho^2\right)\psi_e(\rho)=E\psi_e(\rho),
		\label{Eq: Radial Hamiltonian}
	\end{eqnarray}
	
	\noindent
	where $A$ and $\omega_{\text{b\_vib}}$ are the linear Jahn-Teller term and bond vibrational frequency, respectively.
	
	\noindent
	Contrary to the ground state with `Mexican hat' PES, effect of centrifugal energy, $
	\frac{\hbar^2 l^2}{2M\rho^2}$ cannot be neglected in excited state, since the linear Jahn-Teller ($A\rho$) term increases with $\rho$ (contrast to the ground state) and gravitates the wavepacket to be localized at the vicinity of $\rho \sim 0 $, where the centrifugal energy $\frac{\hbar^2 l^2}{2M\rho^2}$ is of very large magnitude. In other words, the excited minima ($\lbrace\rho_0^{l}\rbrace$) are mainly formed due to two opposing forces, namely, centrifugal ($F_{centrifugal}= -\frac{\hbar^2 l^2}{M \rho^3}$) and linear Jahn-Teller ($F_{JT}=A$) ones, where the resulted bound states are called centrifugally stabilized vibronic states\citep{Slonczewski}. The expression of minima and eigenvalues of radial Hamiltonian for such centrifugally stabilized states (excluding bond vibration) take the following forms within harmonic approximation\citep{Slonczewski}: \\

	
	
	\begin{eqnarray}
		\frac{\partial V_{+}}{\partial \rho} &=& -\frac{\hbar^2 l^2}{M \rho^3} + A = 0  \nonumber \\
		\rho_0^{l} &=& \left(\frac{\hbar^2 l^2}{A M}\right)^{\frac{1}{3}}
	\end{eqnarray}
	
	\begin{eqnarray}
		E_{s, l} &\cong& \frac{(\hbar A)^{2/3}}{M^{1/3}}  \left[ \frac{3}{2} \mid l \mid^{2/3} + 
		\sqrt{3} \frac{\left( s + \frac{1}{2} \right)}{\mid l \mid^{1/3}} \right],
		\label{Eq: centrifugally_stabilised_eigenvalue}
	\end{eqnarray} 
	
	\noindent
	where $s=0,\ 1,\ 2$, ... and $l=\pm \frac{1}{2},\ \pm \frac{3}{2},\ \pm \frac{5}{2}$, ... are the centrifugally stabilized vibrational and pseudo - rotational quantum numbers, 
	respectively. \\

	\noindent
	Incorporation of harmonic bond vibrational energy into the interaction Hamiltonian would lead to modulate the centrifugally stabilized roto - vibrational states with the following elastically coupled oscillation frequency\citep{joy_jcp}:
	

	\begin{eqnarray}
		\Omega_{l} &=& \left[ \frac{3 A^{4/3}}{(\hbar \mid l \mid M)^{2/3}} +  \omega_{\text{b\_vib}}^2 \right]^{1/2},
		\label{Eq: Modulated_frequency}
	\end{eqnarray}
	
	\noindent
	and the eigenvalues of such entangled states satisfy the following form\citep{joy_jcp}: 
	
	\vspace*{-0.4 cm} 
	
	\begin{small}
		\begin{eqnarray}
			E^{\text{b\_vib}}_{s, l} &=& \frac{3}{2} \frac{(\hbar A \mid l \mid)^{2/3}}{M^{1/3}} + \hbar \left[ \frac{3 A^{4/3}}{(\hbar \mid l \mid M)^{2/3}} + \omega_{\text{b\_vib}}^2 \right]^{1/2} \left( s + \frac{1}{2} \right).
			\label{Eq: centrifugally_stabilised_eigenvalue_b_vib}
		\end{eqnarray}
	\end{small}
	
	\vspace{0.3 cm}

	\noindent
	It is evident from Eqs. (\ref{Eq: centrifugally_stabilised_eigenvalue}), (\ref{Eq: Modulated_frequency}) 
	and (\ref{Eq: centrifugally_stabilised_eigenvalue_b_vib}) that unlike usual oscillators, such 
	centrifugally stabilized vibronic states are very much parametrically dependent on pseudo - rotational 
	quantum number, $l$ and thereby, lead to alteration of energy order of roto - vibrational levels in an 
	interesting way. Moreover, the singularity of 
	energy spectrum located at $l=0$ (see Eq. (\ref{Eq: centrifugally_stabilised_eigenvalue_b_vib})) emerges 
	due to cone PES of the centrifugally stabilized excited state. \\

	\noindent
	The striking difference between ground and excited electronic state is that in the upper sheet of `Mexican 
	hat' APES, centrifugal term contributes substantially, where the minima are determined by such centrifugal 
	and linear Jahn-Teller forces (centrifugally stabilized state). As a result, determination of  minima of 
	cubic perturbed centrifugally stabilized vibronic states requires the entire spectrum of anharmonically 
	perturbed pseudo - rotational energy on each grid point ($\rho_i$) of radial vibrational coordinate ($\rho
	$)[Dutta \textit{et al.}\citep{joy_jcp}]. Consequently, the nuclear SE of cubic perturbed excited state: 
	
	\vspace*{-0.4 cm}
	
	\begin{eqnarray}
		&&-\frac{\hbar^2}{2M} \left[ \frac{\partial^2}{\partial \rho^2} + \frac{1}{\rho^2} \frac{\partial^2}{\partial \phi^2} \right]\xi_e(\rho, \phi)  - E \xi_e(\rho, \phi) \nonumber \\
		&& + \left(A \rho + \frac{1}{2} M \omega_{\text{b\_vib}}^2 \rho^2 + A_3 \rho^3 \cos(3 \phi) \right) \xi_e(\rho, \phi) = 0
		\label{Eq: Hamiltonian of cubic perturbed excited state}
	\end{eqnarray}
	
	\noindent
	has been averaged radially at every grid point ($\lbrace \rho_i \rbrace$) within adiabatic approximation, where $-\frac{\hbar^2}{2 M \rho^2} \frac{\partial^2}{\partial \phi^2} + A_3 \rho^3 \cos 3 \phi$ represents the perturbation and the radial part ($w_{e} (\rho)$) of product type eigenfunction $\xi_e(\rho, \phi) (=w_{e}^{0} (\rho) \gamma (\phi))$ is constrained to be at lowest harmonic vibrational state $w_{e}^{0} (\rho)$. The corresponding angular SE (Mathieu equation) for each $\rho$ grid point turns into\citep{joy_jcp}:
	
	\vspace{-0.4 cm}
	
	\begin{eqnarray}
		\alpha_{i} \frac{\partial^2}{\partial \phi^2} \gamma (\phi) + \beta_{i} \cos (3 \phi) \gamma(\phi) + \epsilon (\rho_i)\gamma(\phi) = 0, 
		\label{Eq: Excited state Mathieu Equation}
	\end{eqnarray}
	
	\noindent
	where both splitting $\alpha_{i}(= \frac{\hbar^2}{2 M \rho_i^2})$ and barrier $\beta_{i}(= - A_3 \rho_i^3)$ parameters become variables unlike for the ground state case ($\alpha(= \frac{\hbar^2}{2 M \rho_0^2}; \rho_0 \equiv  \: \text{ground state minima})$ is constant and $\beta$ is the variable). The spectrum of $\epsilon (\rho)$ is obtained by solving the following secular equation of angular SE (Eq. (\ref{Eq: Excited state Mathieu Equation})) [employing $\gamma(\phi) = \sum_{l} a_{l} \exp( i l \phi)$] at every grid point ($\lbrace \rho_i \rbrace$):
	
	\vspace*{-0.4 cm}

	\begin{eqnarray}
		a_{l} \left( \epsilon(\rho_i) - \alpha_i l^{2} \right) + \frac{1}{2} \beta_i \left( a_{l-3} + a_{l+3} \right) = 0.
		\label{Eq: Recurrence Relation excited state} 
	\end{eqnarray}
	
	\noindent
	Such cubic perturbed roto - vibrational energies ($\epsilon_{l}(\rho)$) of various pseudo - rotational levels ($l$) form bound states with the association of linear Jahn-Teller energy ($A \rho$). Thereby, the expression of cubic perturbed centrifugally stabilized excited state eigenvalues takes the following form within harmonic approximation (expanding $V_{+}$ (=$\epsilon_{l} (\rho) + A \rho$) upto second order)\citep{joy_jcp}:
	
	\vspace*{-0.6 cm}
	
	\begin{eqnarray}
		E_{s, l}^{\text{cp}} = \epsilon_{l} (\rho_{0}^{l}) + A \rho_{0}^{l} + \hbar \left[ \frac{1}{M^{1/2}} \left( \frac{\partial^2 V_{+}}{\partial \rho^2}\right)^{\frac{1}{2}}_{\rho=\rho_{0}^{l}} \right] \left( s + \frac{1}{2} \right),
		\label{Eq: cp_eigenvalue}
	\end{eqnarray}
	
	\noindent
	where $\rho_{0}^{l}$ represents the minimum of cubic perturbed centrifugally stabilized excited state PES for particular pseudo-rotational level, $l$. Such cubic perturbed centrifugally stabilized oscillators are dressed with the bond vibration resulting elastically entangled roto - vibrational levels with the following eigenvalues\citep{joy_jcp}:
	
	\begin{small}
		\begin{eqnarray}
			E_{s, l}^{\text{cp + b\_vib}} = \epsilon_{l} (\rho_{0}^{l}) + A \rho_{0}^{l} + \hbar \left[ 
			\frac{1}{M} \left( \frac{\partial^2 V_{+}}{\partial \rho^2}  \right)_{\rho=\rho_{0}^{l}} + \omega^2_{\text{b\_vib}}  
			\right]^{\frac{1}{2}} \left( s + \frac{1}{2} \right),
			\label{Eq: cp_bond_eigenvalue}
		\end{eqnarray}
	\end{small}
	
	\noindent
	where $s$ and $l$ representing the centrifugally stabilized elastically coupled vibrational and roto -vibrational quantum numbers, respectively.\\

\bibliographystyle{aipnum4-1}
\bibliography{ref_book}

\chapter{Beyond Born-Oppenheimer Theory}\label{sec:bbo}

\section{Born-Oppenheimer Treatment}

\noindent
While advancing for Beyond Born-Oppenheimer (BBO) theory, first we need to formulate the first principle based Born-Oppenheimer (BO) treatment\citep{BO,BH}. Since the total electron - nuclear Hamiltonian governs the molecular properties, such treatment was made with the aid of double perturbative theory on the molecular Hamiltonian. \\

\noindent
Considering $M$, $m$, $X$, $x$, $P$ and $p$ as nuclear mass, electronic mass, nuclear coordinates, electronic coordinates, nuclear momenta and electronic momenta, respectively, the kinetic energy of the nuclei can be written as:
\begin{equation}
	T_{N}=\sum\dfrac{1}{2M}P^{2}=-\sum\dfrac{\hbar^{2}}{2M}\left(\dfrac{\partial^{2}}{\partial X^{2}}\right),
\end{equation}   
where the kinetic energy of the electrons takes the following form:
\begin{equation}
	T_{e}=\sum\dfrac{1}{2m}p^{2}=-\sum\dfrac{\hbar^{2}}{2m}\left(\dfrac{\partial^{2}}{\partial x^{2}}\right).
\end{equation}
The interaction energy between nuclei and electrons is defined by Coulomb potential represented as $U(x,X)$.\\

\noindent
Thus total electronic Hamiltonian with fixed nuclear arrangement can be written as :
\begin{equation}
	T_{e}+U=H_{0}(x,\dfrac{\partial}{\partial x};X)
\end{equation}
\noindent
Since $T_{N}$ is very small due to large nuclear mass, $H_{0}$ can be considered as the zeroth order Hamiltonian,
\begin{equation}
	H=(T_{e}+U)+T_{N}=H_{0}+T_{N},
\end{equation}
and then, $T_{N}$ acts as perturbation. In the above perturbation, expansion parameter is considered as,
\begin{equation}
	\kappa=\left(\dfrac{m}{M_{0}}\right)^{1/4},
\end{equation}
where $M_{0}$ is the mass of any one nucleus or the mean mass. In other words, one can write the following expression,
\begin{equation}
	T_{N}=\kappa^{4}H_{1},
	\label{Eq:nukinetic}
\end{equation}
where
\begin{equation}
	H_{1}=-\sum\left(\dfrac{M_{0}}{M}\right)\left(\dfrac{\hbar^{2}}{2m}\right)
	\left(\dfrac{\partial^{2}}{\partial X^{2}}\right)
\end{equation}
\noindent
Therefore, the total Hamiltonian turns into, 
\begin{equation}
	H=H_{0}+\kappa^{4}H_{1}.
	\label{Eq:totHam}
\end{equation}

\noindent
As a result, molecular Schr\"odinger Equation (SE) can be written as,
\begin{equation}
	(H-E)\psi(x,X)=0,
	\label{Eq:MolSE}
\end{equation}
where the SE of electronic motion for arbitrary fixed nuclear position is as follows:
\begin{equation}
	(H_{0}-E^{0})\phi(x;X)=0,
	\label{Eq:ElecSE}
\end{equation}
with the eigenfunctions depending on nuclear coordinates ($X$) parametrically.\\

\noindent
For $n^{th}$ electronic state, one can write:
\begin{subequations}
	\begin{eqnarray}
		E^{0}&=&\epsilon_{n}(X),\\
		\phi &=& \phi_{n}(x;X).
	\end{eqnarray}
\end{subequations}
\noindent
By considering a nuclear geometry ($X^{0}$) very close to $X$, we can define
\begin{equation}
	X-X^{0}=\kappa\,u,
\end{equation}
$\kappa\,u$ appears as the deviation in nuclear coordinate from reference geometry. \\

\noindent
Therefore, one can write,
\begin{equation}
	\epsilon_{n}(X)=\epsilon_{n}(X^{0}+\kappa u)=\epsilon_{n}^{(0)}+\kappa\epsilon_{n}^{(1)}+\kappa^{2}\epsilon_{n}^{(2)}+....
	\label{Eq:taylor1}
\end{equation}
\begin{equation}
	\phi_{n}(x;X)=\phi_{n}(x;X^{0}+\kappa u)=\phi_{n}^{(0)}+\kappa\phi_{n}^{(1)}+\kappa^{2}\phi_{n}^{(2)}+....
	\label{Eq:taylor2}
\end{equation}
\vspace{0.15 cm}
where
$\lbrace\epsilon_{n}^{(0)},\phi_{n}^{(0)}\rbrace$ are independent of $u$, $\lbrace\epsilon_{n}^{(1)},\phi_{n}^{(1)}\rbrace$ are linearly dependent on $u$ and 
$\lbrace\epsilon_{n}^{(2)},\phi_{n}^{(2)}\rbrace$ are quadratically dependent on $u$. \\

\noindent
On the other hand, $H_{0}\left(x,\dfrac{\partial}{\partial x};X\right)$ can be expanded around the reference geometry, $X^{0}$  as,
\begin{equation}
	H_{0}\left(x,\dfrac{\partial}{\partial x};X\right)=H_{0}(x,\dfrac{\partial}{\partial x};X^{0}+\kappa u)=(
	H_{0}^{(0)}+\kappa H_{0}^{(1)}+\kappa^{2}H_{0}^{(2)}+....).
	\label{Eq:hamil}
\end{equation}
\noindent		
Substituting Eq. \ref{Eq:taylor1}, Eq. \ref{Eq:taylor2} and Eq. \ref{Eq:hamil} in Eq. \ref{Eq:ElecSE} 
[$(H_{0}-E^{0})\phi(x;X)=0$], we arrive,
\begin{eqnarray}
	&&\left[(H_{0}^{(0)}+\kappa 
	H_{0}^{(1)}+\kappa^{2}H_{0}^{(2)}+....)-(\epsilon_{n}^{(0)}+\kappa\epsilon_{n}^{(1)}+\kappa^{2}\epsilon_{n}^{(2)}+....)\right]
	\nonumber\\
	&&\left[\phi_{n}^{(0)}+\kappa\phi_{n}^{(1)}+\kappa^{2}\phi_{n}^{(2)}+....\right]=0.
\end{eqnarray}
\noindent
While comparing zeroth, first and second power of $\kappa$, one can obtain,
\begin{equation}
	(H_{0}^{(0)}-\epsilon_{n}^{(0)})\phi_{n}^{(0)}=0
	\label{Eq:zeroth1}
\end{equation}
\begin{equation}
	(H_{0}^{(0)}-\epsilon_{n}^{(0)})\phi_{n}^{(1)}=-(H_{0}^{(1)}-\epsilon_{n}^{(1)})\phi_{n}^{(0)}
	\label{Eq:first1}
\end{equation}
\begin{equation}
	(H_{0}^{(0)}-\epsilon_{n}^{(0)})\phi_{n}^{(2)}=-(H_{0}^{(1)}-\epsilon_{n}^{(1)})\phi_{n}^{(1)}-
	(H_{0}^{(2)}-\epsilon_{n}^{(2)})\phi_{n}^{(0)}
	\label{Eq:second1}
\end{equation}
\vspace{0.2 cm}

\noindent
If we approach this perturbation originating from nuclear kinetic energy operator on the electronic Hamiltonian in a different way, the above formulation can be carried out as presented below.\\

\noindent
Since the partial derivative takes the following form:
\begin{equation}
	\dfrac{\partial}{\partial X}=\dfrac{1}{\kappa}.\dfrac{\partial}{\partial u},
\end{equation}
Eq. \ref{Eq:nukinetic} turns into the following expression:
\begin{eqnarray}
	T_{N}&=&\kappa^{4}H_{1}\nonumber\\
	&=&\kappa^{4}\left[-\sum\left(\dfrac{M_{0}}{M}\right)\left(\dfrac{\hbar^{2}}{2m}\right)
	\left(\dfrac{\partial^{2}}{\partial X^{2}}\right)\right]\nonumber\\
	&=&\kappa^{4}\left[-\dfrac{1}{\kappa^{2}}\sum\left(\dfrac{M_{0}}{M}\right)\left(\dfrac{\hbar^{2}}{2m}\right)
	\left(\dfrac{\partial^{2}}{\partial u^{2}}\right)\right]\nonumber\\
	&=&\kappa^{2}H_{1}^{(2)}
	\label{Eq:modnuKE}
\end{eqnarray}
where,
\begin{equation}
	H_{1}^{(2)}=-\sum\left(\dfrac{M_{0}}{M}\right)\left(\dfrac{\hbar^{2}}{2m}\right)\left(\dfrac{\partial^{2}}{\partial 
		u^{2}}\right)\nonumber
\end{equation}
\noindent
Putting Eq. \ref{Eq:modnuKE} and Eq. \ref{Eq:hamil} in Eq. \ref{Eq:totHam} [$H=H_{0}+\kappa^{4}H_{1}$], we arrive to the following perturbative form of the molecular Hamiltonian,
\begin{eqnarray}
	H&=&H_{0}+\kappa^{4}H_{1}\nonumber \\
	&=&\left[H_{0}^{(0)}+\kappa H_{0}^{(1)}+\kappa^{2}H_{0}^{(2)}+\kappa^{3}H_{0}^{(3)}+....\right]+\kappa^{4}H_{1}\nonumber \\
	&=&\left[H_{0}^{(0)}+\kappa H_{0}^{(1)}+\kappa^{2}H_{0}^{(2)}+\kappa^{3}H_{0}^{(3)}+....\right]+\kappa^{2}H_{1}^{(2)}
	\nonumber \\
	&=&\left[H_{0}^{(0)}+\kappa H_{0}^{(1)}+\kappa^{2}(H_{0}^{(2)}+H_{1}^{(2)})+\kappa^{3}H_{0}^{(3)}+....\right].
	\label{Eq:hamilpert}
\end{eqnarray}

\noindent
In this perturbative treatment, the electronic Hamiltonian [$H_{0}$] is the zeroth order Hamiltonian of molecular SE 
[$(H-E_n)\psi_n=0$]. In this Taylor series expansion of molecular Hamiltonian, $\kappa H_{0}^{(1)}$ and $\kappa^{2}(H_{0}^{(2)}+H_{1}^{(2)})$ are the first and second order correction on the electronic Hamiltonian ($H_{0}^{(0)}$) at reference geometry ($X_0$). \\

\noindent	
One can write,
\begin{subequations}
	\label{Eq:pert}
	\begin{eqnarray}
		E_n&=&E_{n}^{(0)}+\kappa E_{n}^{(1)}+\kappa^{2}E_{n}^{(2)}+...\\
		\psi_n &=&\psi_{n}^{(0)}+\kappa\psi_{n}^{(1)}+\kappa^{2}\psi_{n}^{(2)}+.....
	\end{eqnarray}
\end{subequations}

\noindent	
Substituting Eq. \ref{Eq:pert}a, Eq. \ref{Eq:pert}b and Eq. \ref{Eq:hamilpert} in [$(H-E_n)\psi_n=0$], we arrive,
\begin{footnotesize}
	\begin{eqnarray}
		&&\left[(H_{0}^{(0)}+\kappa H_{0}^{(1)}+\kappa^{2}(H_{0}^{(2)}+H_{1}^{(2)})+\kappa^{3}H_{0}^{(3)}+....)
		-(E_{n}^{(0)}+\kappa E_{n}^{(1)}+\kappa^{2}E_{n}^{(2)}+...)\right]
		\nonumber\\
		&&\left[\psi_{n}^{(0)}+\kappa\psi_{n}^{(1)}+\kappa^{2}\psi_{n}^{(2)}+...\right]=0.
	\end{eqnarray}
\end{footnotesize}

\noindent
While collecting coefficients of $\kappa_{0}$, $\kappa_{1}$ and $\kappa_{2}$, the following equations are obtained,
\begin{equation}
	(H_{0}^{(0)}-E_{n}^{(0)})\psi_{n}^{(0)}=0,
	\label{Eq:zeroth2}
\end{equation}
\begin{equation}
	(H_{0}^{(0)}-E_{n}^{(0)})\psi_{n}^{(1)}=-(H_{0}^{(1)}-E_{n}^{(1)})\psi_{n}^{(0)},
	\label{Eq:first2}
\end{equation}
\begin{equation}
	(H_{0}^{(0)}-E_{n}^{(0)})\psi_{n}^{(2)}=-(H_{0}^{(1)}-E_{n}^{(1)})\psi_{n}^{(1)}-(H_{0}^{(2)}+H_{1}^{(2)}
	-E_{n}^{(2)})\psi_{n}^{(0)}.
	\label{Eq:second2}
\end{equation}
\vspace{0.2 cm}

\noindent
From Eq. \ref{Eq:zeroth1}, $\phi_{n}^{(0)}$ represents function of $x$ only (at a fixed nuclear position,
$X^0$), whereas from Eq. \ref{Eq:zeroth2}, $\psi_{n}^{(0)}$ appears as function of both $x$ and $u$ (as $\psi_{n}^{(0)}$ is the molecular wavefunction at the zeroth order), but both are eigenfunctions of same Hamiltonian with same eigenvalue $\epsilon_{n}^{(0)}$ [= $E_{n}^{(0)}$].\\ 

\noindent
Therefore, one can write
\begin{equation}
	\psi_{n}^{(0)}(x,u)=\chi_{n}^{(0)}(u)\phi_{n}^{(0)}(x;X^0),
\end{equation}
where $\chi_{n}^{(0)}(u)$ is a multiplier, which is a function of $u$. \\

\noindent			
While projecting Eq. \ref{Eq:first1} and Eq. \ref{Eq:first2} by $\phi_{n}^{(0)}$, the corresponding equations take the following form,
\begin{equation}
	\langle\phi_{n}^{(0)}\vert H_{0}^{(1)}-\epsilon_{n}^{(1)}\vert\phi_{n}^{(0)}\rangle = 0
	\label{Eq:comp1}
\end{equation}
\begin{equation}
	\chi_{n}^{(0)}\langle\phi_{n}^{(0)}\vert H_{0}^{(1)}-E_{n}^{(1)}\vert\phi_{n}^{(0)}\rangle = 0
	\label{Eq:comp2}
\end{equation}
since $\langle\phi_{n}^{(0)}\vert H_{0}^{(0)}-\epsilon_{n}^{(0)}\vert\phi_{n}^{(1)}\rangle=0
\,\,\text{and}\,\,\langle\phi_{n}^{(0)}\vert H_{0}^{(0)}-\epsilon_{n}^{(0)}\vert\psi_{n}^{(1)}\rangle=0$ \\

\noindent
Therefore, comparing Eq. \ref{Eq:comp1} and Eq. \ref{Eq:comp2}, one can obtain:
\begin{equation}
	\epsilon_{n}^{(1)}=E_{n}^{(1)}
\end{equation} 
If the equilibrium geometry is chosen as the reference nuclear geometry, first order correction on energy will be zero: 
\begin{equation}
	\epsilon_{n}^{(1)}=E_{n}^{(1)}=0
\end{equation}
Therefore, Eq. \ref{Eq:first1} and Eq. \ref{Eq:first2} become,
\begin{equation}
	(H_{0}^{(0)}-\epsilon_{n}^{(0)})\phi_{n}^{(1)}=-H_{0}^{(1)}\phi_{n}^{(0)}
	\label{Eq:comp3}
\end{equation} 
\begin{equation}
	(H_{0}^{(0)}-\epsilon_{n}^{(0)})\psi_{n}^{(1)}=-H_{0}^{(1)}\psi_{n}^{(0)}
	\label{Eq:comp4}
\end{equation} 
While multiplying Eq. \ref{Eq:comp3} by $\chi_{n}^{(0)}$ and then, subtracting Eq. \ref{Eq:comp4}, we get,
\begin{equation}
	(H_{0}^{(0)}-\epsilon_{n}^{(0)})(\chi_{n}^{(0)}\phi_{n}^{(1)}-\psi_{n}^{(1)})=0.
	\label{Eq:comp5}
\end{equation} 
From Eq. \ref{Eq:zeroth1}, $\phi_{n}^{(0)}$ appears as function of $x$ only (at a fixed nuclear 
position), whereas from Eq. \ref{Eq:comp5}, it is evident that
$\chi_{n}^{(0)}\phi_{n}^{(1)}-\psi_{n}^{(1)}$ is function of both $x$ and $u$. Nevertheless, both are eigenfunctions of same Hamiltonian ($H_{0}^{(0)}$) with same eigenvalue $\epsilon_{n}^{(0)}$ [= $E_{n}^{(0)}$]. Therefore, we can write
\begin{equation}
	\chi_{n}^{(1)}\phi_{n}^{(0)}=\chi_{n}^{(0)}\phi_{n}^{(1)}-\psi_{n}^{(1)},
\end{equation} 
where $\chi_{n}^{(1)}$ signifies an arbitrary function of $u$.
In other words, the first order correction of electronic eigenfunction takes the following form:
\begin{equation}
	\psi_{n}^{(1)}=\chi_{n}^{(0)}(u)\phi_{n}^{(1)}(x;u)+\chi_{n}^{(1)}(u)\phi_{n}^{(0)}(x;X^0)
\end{equation}
In a similar way, the second order correction takes the following form,
\begin{equation}
	\psi_{n}^{(2)}=\chi_{n}^{(0)}(u)\phi_{n}^{(2)}(x;u)+\chi_{n}^{(1)}(u)\phi_{n}^{(1)}(x;u)+\chi_{n}^{(2)}(u)\phi_{n}^{(0)}(x;X^0)
\end{equation}
Finally, total molecular wavefunction upto second order correction can be written as,
\begin{eqnarray}
	\psi_{n}&=&\psi_{n}^{(0)}+\kappa\psi_{n}^{(1)}+\kappa^{2}\psi_{n}^{(2)}+.....\nonumber \\
	&=& \chi_{n}^{(0)}\lbrace\phi_{n}^{(0)}+\kappa\phi_{n}^{(1)}+\kappa^{2}\phi_{n}^{(2)}+.....\rbrace +\nonumber\\
	&&\kappa\chi_{n}^{(1)}\lbrace\phi_{n}^{(0)}+\kappa\phi_{n}^{(1)}+.....\rbrace +\kappa^{2}\chi_{n}^{(2)}\lbrace\phi_{n}^{(0)}+.....\rbrace+.....\nonumber \\
	&=&\lbrace\chi_{n}^{(0)}+\kappa\chi_{n}^{(1)}+\kappa^{2}\chi_{n}^{(2)}+......\rbrace\nonumber\\
	&&\lbrace\phi_{n}^{(0)}+\kappa\phi_{n}^{(1)}+\kappa^{2}\phi_{n}^{(2)}+.....\rbrace\nonumber\\
	&=&\chi_{n}(u)\,\phi_{n}(x;u)
\end{eqnarray}
In the above expression, the first factor describes nuclear motion and the second one shows the electronic
movement with parametric dependence on nuclear coordinates. Therefore, total wavefunction can be written as a product of nuclear and electronic 
wavefunction. \\

\noindent
Nevertheless, the third order correction is,
\begin{eqnarray}
	\psi_{n}^{(3)}&=&\chi^{(0)}_{n}\phi_{n}^{(3)}+\chi^{(1)}_{n}\phi_{n}^{(2)}+\chi^{(2)}_{n}\phi_{n}^{(1)}\nonumber\\
	&&+\chi^{(3)}_{n}\phi_{n}^{(0)}+F(x,u),
\end{eqnarray}
where the extra term, $F(x,u)$ does not contain $x$ solely through a factor of 
the type $\phi_{n}^{(r)}(x,u)$. Therefore, the molecular wavefunction is corrected up to fourth power of the expansion parameter, $\kappa$ known as 
`Adiabatic Approximation'. On the other hand, as per Born-Huang expansion, one can write the molecular wavefunction by considering the entire Hilbert space,		
\begin{equation}
	\vert\psi(x,X)\rangle = \sum_i\chi_i(X)\vert\phi_i(x;X)\rangle
\end{equation}
so that contribution of higher excited electronic states are incorporated in the molecular 
wavefunction.

\section{Adiabatic Representation}	\label{ssec:adia_th}
\noindent
The Born-Oppenheimer (BO) treatment~\citep{BO,BH} is based on the fact that slow moving nuclei could be distinguished from fast moving electrons 
in molecular species, chemical reactions and solid state processes. This distinction leads to the BO approximation, i.e., the non-adiabatic coupling 
terms (NACTs) are assumed to be negligibly small. On the contrary, the excited electronic states affect the ground and/or any other state
dynamics very strongly due to the so-called NACTs, and the BO approximate 
equation fails to calculate correct transition probabilities.\citep{baer_jcp91,baer_jcp105,grimbert_jcp108} The coupling terms between the electronic states are fundamentally
important, particularly when degenerate states are involved through single 
or multiple conical intersections (CIs) in the nuclear configuration space (CS). The NACTs become a natural requirement in the context of BO treatment,
what we call the Adiabatic picture, to introduce physically
meaningfulness in representing any molecular/chemical process. Since the NACTs play an important role in the description of scattering 
and radiationless processes in molecular systems, the NACTs should be gracefully taken care 
of, rather than neglecting them forcibly. \\

\noindent
The time independent total molecular SE is represented as,
\begin{eqnarray}
	\hat{H}(s_e,s_n)\vert\Psi(s_e,s_n)\rangle = E\vert\Psi(s_e,s_n)\rangle,
	\label{mse_1}
\end{eqnarray}
where $\hat{H}(s_e,s_n)$ and $\vert\Psi(s_e,s_n)\rangle$ represent molecular Hamiltonian and molecular wave function, respectively ($s_e$ and $s_n$ 
signify electronic and nuclear coordinates). On the other hand, molecular Hamiltonian, $\hat{H}(s_e,s_n)$ symbolizes sum of nuclear kinetic energy and
electronic Hamiltonian:
\begin{eqnarray}
	\hat{H}(s_e,s_n) = \hat{T}_n(s_n) + \hat{H}_e(s_e|s_n).
\end{eqnarray}

\noindent
On the other hand, nuclear kinetic energy (KE) in atomic units is given by		
\begin{eqnarray}
	\hat{T}_n(s_n) = -\dfrac{1}{2}\sum_A\left(\dfrac{\nabla^2_{s_n,A}}{M_A}\right),
\end{eqnarray}
and the electronic Hamiltonian is:
\begin{eqnarray}
	\hat{H}_e(s_e|s_n) = -\dfrac{1}{2}\sum_i \nabla^2_{s_e,i} -\sum_i\sum_A\dfrac{Z_A}{r_{iA}} + 
	\sum_i\sum_{j>i}\dfrac{1}{r_{ij}} + \sum_A\sum_{B>A}\dfrac{Z_AZ_B}{R_{AB}},\qquad
\end{eqnarray}
where the first, second, third and fourth terms signify electronic kinetic energy operator, electron-nuclear Coulombic attraction, electron-electron Coulombic repulsion and 
nuclear-nuclear Coulombic repulsion, respectively. \\

\noindent
The electronic Hamiltonian satisfies the time-independent electronic SE:
\begin{eqnarray}
	\hat{H}_e(s_e|s_n)\vert\xi_i(s_e|s_n)\rangle=u_i(s_n)\vert\xi_i(s_e|s_n)\rangle.
	\label{elecse}
\end{eqnarray}
where $\vert\xi_i(s_e|s_n)\rangle$ and $u_i(s_n)$ represent electronic eigenfunctions and adiabatic potential energy surfaces, respectively. It is to 
note that both the quantities depend parametrically on the nuclear coordinates. Since the electronic eigenfunctions form complete set, those will satisfy:
\begin{eqnarray}
	\langle\xi_i\vert\xi_j\rangle = \delta_{ij}; 
	\quad \sum_i\vert\xi_i\rangle\langle\xi_i\vert= \textbf{I}.  
\end{eqnarray}

\noindent		
The total molecular wave function $\vert\Psi(s_e,s_n)\rangle$ can be expanded in terms
of electronic eigenfunctions:
\begin{eqnarray}
	\vert\Psi(s_e,s_n)\rangle = \sum_i \psi_i^{ad}(s_n)\vert\xi_i(s_e|s_n)\rangle,
	\label{boexp}
\end{eqnarray}
which is known as the Born-Oppenheimer~\citep{BO} expansion. In the above equation, $\psi_i(s_n)$ and $\vert\xi_i(s_e|s_n)\rangle$ depict nuclear wave functions
and adiabatic electronic eigenfunctions, respectively. \\

\noindent
One can use the above expansion of $\vert\Psi(s_e,s_n)\rangle$ [Eq. (\ref{boexp})] and substituting it in the molecular Schr\"odinger Equation [Eq. (\ref{mse_1})], we arrive:
\begin{eqnarray}
	(\hat{T}_n + \hat{H}_e)\sum_i \psi_i^{ad}\vert\xi_i\rangle=E\sum_i \psi_i^{ad}\vert\xi_i\rangle.
	\label{molecse11}
\end{eqnarray}
Applying $\langle\xi_j\vert$ from the left on the above equation, the SE takes the following form,
\begin{eqnarray}
	&\quad&\sum_i \langle\xi_j\vert(\hat{T}_n + \hat{H}_e)\psi_i^{ad}\vert\xi_i\rangle 
	= E\sum_i\psi_i^{ad}\langle\xi_j\vert\xi_i\rangle \\
	&\Rightarrow &\sum_i\left( \langle\xi_j\vert \hat{T}_n\psi_i^{ad} \vert\xi_i\rangle 
	+ \delta_{ji}(u_i-E)\psi_i\right)  = 0.
	\label{secontnd}
\end{eqnarray}
where $u_i$ are the adiabatic PESs. \\

\noindent
The nuclear KE term ($\hat{T}_n$) in atomic units:
\begin{eqnarray}
	\hat{T}_n(s_n) = -\dfrac{1}{2}\sum_A\left(\dfrac{\nabla^2_{s_n,A}}{M_A}\right), 
\end{eqnarray}
can be written in terms of mass-weighted coordinates by taking 
a transformation, $\tilde{s}_{n,A} = \sqrt{M_A}s_{n,A}$:	
\begin{eqnarray}
	\hat{T}_n(\tilde{s}_n) = -\dfrac{1}{2}\sum_A\nabla^2_{\tilde{s}_n,A} \equiv -\dfrac{1}{2}\nabla^2_n.
	\label{Eq:nuKE}
\end{eqnarray}

\noindent
Using the form of $T_n$ in Eq. (\ref{Eq:nuKE}), we have,
\begin{eqnarray}
	&\quad&\sum_i\left\lbrace-\dfrac{1}{2}\langle\xi_j(s_e|s_n)\vert\nabla^2_n\psi_i^{ad}(s_n) \vert\xi_i(s_e|s_n)\rangle + 
	\delta_{ji}(u_i-E)\psi_i^{ad}(s_n)\right\rbrace  = 0 \nonumber \\
	&\Rightarrow &\sum_i\left\lbrace-\dfrac{1}{2}\delta_{ji}\nabla^2_n - \langle\xi_j\vert\vec{\nabla}_n\vert\xi_i\rangle.\vec{\nabla}_n 
	-\dfrac{1}{2}\langle\xi_j\vert\nabla^2_n\vert\xi_i\rangle + \delta_{ji}(u_i-E)\right\rbrace\psi_i^{ad} = 0 \nonumber \\
	&\Rightarrow &\sum_i\left\lbrace -\dfrac{1}{2}\delta_{ji}\nabla^2_n - \vec{\tau}^{(1)}_{ji}.\vec{\nabla}_n 
	-\dfrac{1}{2}\tau^{(2)}_{ji} + \delta_{ji}(u_i-E)\right\rbrace\psi_i^{ad} = 0,
	\label{secontnd2}
\end{eqnarray}
where $\langle\xi_j\vert\vec{\nabla}_n\vert\xi_i\rangle=\vec{\tau}^{(1)}_{ji}$ signifies NACTs of first order and 
$\langle\xi_j\vert\nabla_n^2\vert\xi_i\rangle=\tau^{(2)}_{ji}$ represents NACTs of second order.

\subsection{Properties of NACT}

\noindent
Since the first order NACTs, $\vec{\tau}^{(1)}_{ij}$ acts as vector field, one needs to examine the connection between NACTs and standard properties of vector fields. The nature of vector fields are thoroughly discussed in Appendix \ref{sec:app1}. \\

\noindent
On the other hand, while simplifying Eq. \ref{secontnd2}, we explore some crucial 
properties of the NACTs. The elements $\vec{\tau}^{(1)}_{ij}$ constitute a skew-symmetric matrix~\citep{baer_phys_rep358,baer_book} as follows:
\begin{eqnarray}
	\vec{\boldsymbol{\tau}}^{\dagger}=-\vec{\boldsymbol{\tau}}
\end{eqnarray}
which can be shown as follows: 	
\begin{eqnarray}
	&\langle\xi_i\vert\xi_j\rangle = \delta_{ij}& \nonumber \\
	&\Rightarrow \vec{\nabla}_n\langle\xi_i\vert\xi_j\rangle = 0&  \nonumber \\
	&\Rightarrow \langle\vec{\nabla}_n\xi_i\vert\xi_j\rangle + \langle\xi_i\vert\vec{\nabla}_n
	\xi_j\rangle = 0,& \nonumber \\
	&\Rightarrow \vec{\tau}^{(1)}_{ji} = -\vec{\tau}^{(1)}_{ij}.&
	\label{tauskewsym}
\end{eqnarray}

\noindent
On the other hand, second order NACT elements satisfy the following condition:
\begin{eqnarray}
	\boldsymbol{\tau}^{(2)} = \vec{\boldsymbol{\tau}}^{(1)}.\vec{\boldsymbol{\tau}}^{(1)} + \vec{\boldsymbol{\nabla}}_n.
	\vec{\boldsymbol{\tau}}^{(1)}
\end{eqnarray}
which can be established from the first order term through the following way: 
\begin{eqnarray}
	\langle\xi_i\vert\vec{\nabla}_n\vert\xi_j\rangle=\vec{\tau}^{(1)}_{ij},  \nonumber
\end{eqnarray}
and then, applying $\vec{\nabla}_n$ on both sides of the above equation, we obtain:
\begin{eqnarray}
	&\vec{\nabla}_n.\langle\xi_i\vert\vec{\nabla}_n\vert\xi_j\rangle = \vec{\nabla}_n.\vec{\tau}^{(1)}_{ij} &
	\nonumber \\
	&\Rightarrow \langle\vec{\nabla}_n\xi_i\vert\vec{\nabla}_n\xi_j\rangle + \langle\xi_i
	\vert\nabla_n^2 \vert\xi_j\rangle = \vec{\nabla}_n.\vec{\tau}^{(1)}_{ij}& \nonumber \\
	&\Rightarrow \tau^{(2)}_{ij} = \vec{\nabla}_n.\vec{\tau}^{(1)}_{ij} - \langle\vec{\nabla}_n\xi_i\vert\vec{\nabla}_n
	\xi_j\rangle & \nonumber \\
	&\Rightarrow \tau^{(2)}_{ij} = \vec{\nabla}_n.\vec{\tau}^{(1)}_{ij} - \sum_k \langle\vec{\nabla}_n\xi_i\vert
	\xi_k\rangle.\langle\xi_k\vert\vec{\nabla}_n\xi_j\rangle &\nonumber \\
	&\Rightarrow \tau^{(2)}_{ij} = \vec{\nabla}_n.\vec{\tau}^{(1)}_{ij} - \sum_k \vec{\tau}^{(1)}_{ki}.\vec{\tau}^{(1)}_{kj}.&
	\label{eq:nact_2nd}
\end{eqnarray}
Using Eq. \ref{tauskewsym} in Eq. \ref{eq:nact_2nd}, we have
\begin{eqnarray}
	\tau^{(2)}_{ij}=\sum_k \vec{\tau}^{(1)}_{ik}.\vec{\tau}^{(1)}_{kj} + \vec{\nabla}_n.\vec{\tau}^{(1)}_{ij}.
	\label{tauscnd}
\end{eqnarray}
In matrix notation, one can write:
\begin{eqnarray}
	\boldsymbol{\tau}^{(2)} = \vec{\boldsymbol{\tau}}^{(1)}.\vec{\boldsymbol{\tau}}^{(1)} + \vec{\boldsymbol{\nabla}}_n.
	\vec{\boldsymbol{\tau}}^{(1)}.  \nonumber
\end{eqnarray}
On the other hand, the above equation can be rewritten in the following form:
\begin{eqnarray}
	\vec{\boldsymbol{\nabla}}_n.\vec{\boldsymbol{\tau}}^{(1)} =  \boldsymbol{\tau}^{(2)} -
	\vert\vec{\boldsymbol{\tau}}^{(1)}\vert^2, 
	\label{diveq}
\end{eqnarray}
which is known as Divergence Equation~\citep{baer_phys_rep358,baer_book}. \\

\noindent
It needs to mention that the NACT matrix elements satisfy the Curl condition,~\citep{baer_phys_rep358,baer_book} known as Yang-Mill field:
\begin{eqnarray}
	\boldsymbol{\nabla}_q.\boldsymbol{\tau}_p -\boldsymbol{\nabla}_p.\boldsymbol{\tau}_q 
	= \left[\boldsymbol{\tau}_p, \boldsymbol{\tau}_q \right]
	\label{curleq}
\end{eqnarray}
where $p$ and $q$ are any two pair of the nuclear coordinates and the right hand side of the above equation is the commutator of the NACT components along $p$ and $q$. In order to verify such expression, we can consider the $ij^{th}$ component of the left hand side expression as,
\begin{eqnarray}
	(\boldsymbol{\nabla}_q.\boldsymbol{\tau}_p -\boldsymbol{\nabla}_p.\boldsymbol{\tau}_q)_{ij} 
	&=& \nabla_q.\tau^{ij}_p -\nabla_p.\tau^{ij}_q \nonumber \\
	&=& \nabla_q.\langle\xi_i\vert\nabla_p\vert\xi_j\rangle - \nabla_p.\langle\xi_i\vert\nabla_q
	\vert\xi_j\rangle \nonumber \\
	&=& \langle\nabla_q \xi_i \vert \nabla_p \xi_j\rangle + \langle \xi_i \vert \nabla_q.\nabla_p 
	\vert \xi_j \rangle  \nonumber \\
	&-&\langle\nabla_p \xi_i \vert \nabla_q \xi_j\rangle - \langle \xi_i \vert \nabla_p.\nabla_q 
	\vert \xi_j \rangle. \nonumber
\end{eqnarray}

\noindent
Since the $\vert \xi_k\rangle$s are analytic functions, the commutator $(\left[\nabla_p, \nabla_q\right])$ vanishes and the above expression turns into:
\begin{eqnarray}
	(\boldsymbol{\nabla}_q.\boldsymbol{\tau}_p -\boldsymbol{\nabla}_p.
	\boldsymbol{\tau}_q)_{ij} &=&\langle\nabla_q \xi_i \vert \nabla_p 
	\xi_j\rangle - \langle\nabla_p \xi_i \vert \nabla_q \xi_j\rangle 
	\nonumber \\
	&=& \sum_k \left(\langle\nabla_q \xi_i \vert \xi_k\rangle \langle \xi_k\vert 
	\nabla_p \xi_j\rangle - \langle\nabla_p \xi_i \vert \xi_k\rangle \langle 
	\xi_k\vert \nabla_q \xi_j\rangle  \right) \nonumber \\
	&=& - \sum_k \left( \tau_q^{(ik)}.\tau_p^{(kj)} - \tau_p^{(ik)}.
	\tau_q^{(kj)} \right) \nonumber \\
	&=& (\boldsymbol{\tau}_p.\boldsymbol{\tau}_q)^{ij} - (\boldsymbol{\tau}_q.
	\boldsymbol{\tau}_p)^{ij} \nonumber \\
	&=& \left[\boldsymbol{\tau}_p, \boldsymbol{\tau}_q \right]_{ij}. \nonumber
\end{eqnarray}
Eqs. (\ref{diveq}) and (\ref{curleq}) are collectively known as the Curl-Div Equations,~\citep{baer_phys_rep358,baer_book} which establishes the validity of the sub-Hilbert space. \\ 

\noindent
The NACTs satisfy the following definition which is reminiscent of the Hellmann-Feynman Theorem~\citep{hellmann,feynmann}:
\begin{eqnarray}
	\vec{\tau}_{ji} = \dfrac{\langle \xi_j \vert \vec{\nabla}_n \textbf{\^H}_e\vert \xi_i\rangle}{u_i-u_j}
	\label{hftheorem}
\end{eqnarray}
Considering the electronic SE in Eq. (\ref{elecse}),
\begin{eqnarray}
	\hat{H}_e(s_e|s_n)\vert\xi_i(s_e|s_n)\rangle=u_i(s_n)\vert\xi_i(s_e|s_n)\rangle, 
\end{eqnarray}
and	applying $\vec{\nabla}_n$ on both sides of the equation, we obtain the following equation:
\begin{eqnarray}
	\vec{\nabla}_n\hat{H}_e(s_e|s_n)\vert\xi_i(s_e|s_n)\rangle &=& \vec{\nabla}_n u_i(s_n)\vert\xi_i(s_e|s_n)\rangle
	\nonumber \\
	\Rightarrow \left(\vec{\nabla}_n\hat{H}_e\right)\vert\xi_i\rangle + \hat{H}_e\left(\vec{\nabla}_n
	\vert\xi_i\rangle\right) &=& \left(\vec{\nabla}_n u_i\right)\vert\xi_i\rangle + u_i\left(\vec{\nabla}_n\vert\xi_i\rangle 
	\right). \nonumber
\end{eqnarray}
Projecting $\langle \xi_j\vert$ (where, $j \neq i$) on both sides, we will have then:
\begin{eqnarray}
	&\langle \xi_j \vert \vec{\nabla}_n\hat{H}_e \vert\xi_i\rangle + \langle \xi_j \vert\hat{H}_e\vec{\nabla}_n
	\vert\xi_i\rangle = \left(\vec{\nabla}_n u_i\right)\langle \xi_j\vert\xi_i\rangle + u_i\langle \xi_j \vert \vec{\nabla}_n 
	\vert\xi_i\rangle & \nonumber \\
	&\Rightarrow \langle \xi_j \vert \vec{\nabla}_n\hat{H}_e \vert\xi_i\rangle + u_j\langle \xi_j \vert \vec{\nabla}_n 
	\vert\xi_i\rangle = 0 +  u_i\langle \xi_j \vert \vec{\nabla}_n \vert\xi_i\rangle & \nonumber \\
	&\Rightarrow \vec{\tau}_{ji} = \dfrac{\langle \xi_j \vert \vec{\nabla}_n \hat{H}_e\vert \xi_i\rangle}{u_i-u_j}.&
	\label{Eq:helfen}
\end{eqnarray}
The above equation is a reminiscent of the Hellmann-Feynman Theorem, which shows that the NACTs become singular at any degenerate point ($u_i=u_j$) in the nuclear CS.

\section{Diabatic Representation}	
\noindent
When we substitute the expression of $\tau_{ij}^{(2)}$ in terms of $\vec{\tau}_{ij}^{(1)}$, 
Eq. (\ref{secontnd2}) turns into:
\begin{eqnarray}
	&\sum_i\left\lbrace -\dfrac{1}{2}\delta_{ji}\nabla^2_n - \vec{\tau}^{(1)}_{ji}.\vec{\nabla}_n 
	-\dfrac{1}{2}\sum_k\vec{\tau}^{(1)}_{jk}.\vec{\tau}^{(1)}_{ki} - \dfrac{1}{2}\vec{\nabla}_n.
	\vec{\tau}^{(1)}_{ji} + \delta_{ji}(u_i-E)\right\rbrace\psi_i^{ad} = 0 & \nonumber \\
	&\Rightarrow -\dfrac{1}{2}\sum_i\left\lbrace \delta_{ji}\nabla^2_n\psi_i^{ad} + 2\vec{\tau}^{(1)}_{ji}.(\vec{\nabla}_n\psi_i^{ad}) +
	\sum_k\vec{\tau}^{(1)}_{jk}.\vec{\tau}^{(1)}_{ki}\psi_i^{ad} \right. & \nonumber \\
	&\left. +(\vec{\nabla}_n.\vec{\tau}^{(1)}_{ji})\psi_i^{ad} - 2\delta_{ji}(u_i-E)\psi_i^{ad}
	\right\rbrace = 0. & \nonumber
\end{eqnarray}
The aforementioned expression can be further simplified to:
\begin{eqnarray}
	\sum_i \left[-\dfrac{1}{2}\sum_k \left\lbrace(\delta_{jk}\vec{\nabla}_n + \vec{\tau}_{jk}).(\delta_{ki}\vec{\nabla}_n + 
	\vec{\tau}_{ki})\right\rbrace + \delta_{ji}(u_i-E)\right]\psi_i^{ad} = 0,  
	\label{adiaequ}
\end{eqnarray}
In matrix notation, the above equation is represented as~\citep{baer_phys_rep358,baer_book}:
\begin{eqnarray}
	\left[-\dfrac{1}{2}(\vec{\boldsymbol{\nabla}}_n + \vec{\boldsymbol{\tau}})^2 + \textbf{u}-E\right]\boldsymbol{\psi^{ad}}= 0.
	\label{adiase}
\end{eqnarray}
Here, $\boldsymbol{\vec{\tau}}$ is the non-adiabatic coupling matrix (NACM) and $\textbf{u}$ is the adiabatic potential energy matrix. 
When the NACT matrix $(\boldsymbol{\vec{\tau}})$ is neglected in Eq. (\ref{adiase}),
the nuclear SE becomes decoupled (known as the BO approximation):
\begin{eqnarray}
	\left(-\dfrac{1}{2}\boldsymbol{\nabla}_n^2 + \textbf{u}-E\right)\boldsymbol{\psi^{ad}}= 0. 
\end{eqnarray}

\noindent
It is evident from Hellmann-Feynman Theorem [Eq. \ref{Eq:helfen}] that close to any degenerate point or along a seam in the nuclear CS, the NACT takes up substantial magnitudes tending to blow up its magnitude. One can not just neglect those terms from the SE by imposing the BO
approximation since the degeneracies at CI points are ubiquitous rather than exceptions. Within the adiabatic representation, NACTs are not numerically amenable and hence, the pursuit for a different representation is an attempt to avoid 
such singularities. \\

\noindent
It becomes a natural choice to find out a unitary transformation matrix that transforms the adiabatic representation to the diabatic one, where the kinetically coupled SE transforms to potentially coupled Hamiltonian. There are two ways through which one can carry out such a transformation. 
Indeed, it can be shown that the two approaches are equivalent to each other. \\

\noindent
In Method I, choosing an electronic basis set, parameterized by a fixed nuclear coordinate $(s_0)$, the total molecular wave function is written as:
\begin{eqnarray}
	\vert\Psi(s_e,s_n)\rangle = \sum_i \tilde{\phi}_i^{d}(s_n)\vert\xi_i(s_e|s_0)\rangle, 
	\label{totpsidiab}
\end{eqnarray}
where $\tilde{\phi}_i^{d}(s_n) \equiv \tilde{\phi}_i^{d}(s_n|s_0)$, depends parametrically on $s_0$, and
$\vert\xi_i(s_e|s_0)\rangle$'s satisfy the corresponding electronic SE:
\begin{eqnarray}
	\hat{H}_e(s_e|s_0)\vert\xi_i(s_e|s_0)\rangle=u_i(s_0)\vert\xi_i(s_e|s_0)\rangle. \nonumber
\end{eqnarray}
Using the above form of $\vert\Psi(s_e,s_n)\rangle$, Eq. (\ref{molecse11}) becomes:
\begin{eqnarray}
	(\hat{T}_n(s_n) + \hat{H}_e(s_e|s_n))\sum_i \tilde{\phi}_i^{d}(s_n)\vert\xi_i(s_e|s_0)\rangle = 
	E\sum_i \tilde{\phi}_i^{d}(s_n)\vert\xi_i(s_e|s_0)\rangle. \nonumber
\end{eqnarray}
While projecting $\langle\xi_j(s_e|s_0)\vert$ from the left on the above equation, one can obtain:
\begin{eqnarray}
	\sum_i \left[\delta_{ji}(\hat{T}_n(s_n)-E) + \langle \xi_j(s_e|s_0)\vert \hat{H}_e(s_e|s_n) \vert 
	\xi_i(s_e|s_0)\rangle \right]\tilde{\phi}_i^{d}(s_n) = 0  \nonumber \\
	\Rightarrow \sum_i \left[\delta_{ji}(-\dfrac{1}{2}\nabla^2_n-E)+V_{ji}(s_n|s_0)\right]
	\tilde{\phi}_i^{d}(s_n) = 0,
	\label{diabse11}
\end{eqnarray}
where
\begin{eqnarray}
	V_{ji}(s_n|s_0) = \langle \xi_j(s_e|s_0)\vert \hat{H}_e(s_e|s_n) \vert \xi_i(s_e|s_0)\rangle.   
\end{eqnarray}

\noindent
On the other hand, one can write
\begin{eqnarray}
	\hat{H}_e(s_e|s_n) &=& \hat{T}_e(s_e) + \hat{U}(s_e|s_n), \nonumber \\
	\hat{H}_e(s_e|s_0) &=& \hat{T}_e(s_e) + \hat{U}(s_e|s_0), \nonumber \\
	\qquad \hat{H}_e(s_e|s_n) &=& \hat{H}_e(s_e|s_0) + (\hat{U}(s_e|s_n) - \hat{U}(s_e|s_0)).  
\end{eqnarray}
Therefore, the following equation turns into,
\begin{eqnarray}
	V_{ji}(s_n|s_0) = u_i(s_0) + \langle \xi_j(s_e|s_0)\vert (\hat{U}(s_e|s_n) - \hat{U}(s_e|s_0) )\vert 
	\xi_i(s_e|s_0)\rangle.  \nonumber 
\end{eqnarray}
Thus, Eq. (\ref{diabse11}) becomes potentially coupled which in the matrix representation appears as~\citep{baer_phys_rep358,baer_book}:
\begin{eqnarray}
	\left(-\dfrac{1}{2} \nabla_n^2 + \textbf{V} - E\right)\boldsymbol{\tilde{\phi}^d} = 0.
	\label{diabse}
\end{eqnarray}
In this way, one can transform Eq. (\ref{adiase}) into Eq. (\ref{diabse}) where 
the couplings in the kinetic energy term turn into potential energy 
one known as diabatic representation of SE. In this process, the SE appears to be continuous and single-valued, which is expected to be numerically solvable within desired level of accuracy.  \\

\noindent
Using the above formulations, one can observe that,
\begin{eqnarray}
	\vec{\nabla}_n \vert \xi_i(s_e|s_n)\rangle &=& \sum_j \vert \xi_j(s_e|s_n)\rangle \langle\xi_j(s_e|s_n)\vert
	\vec{\nabla}_n \vert \xi_i(s_e|s_n)\rangle   \\
	&=& \sum_j \vec{\tau}_{ji} \vert \xi_j(s_e|s_n)\rangle   \\
	&=& -\sum_j \vec{\tau}_{ij} \vert \xi_j(s_e|s_n)\rangle.  \nonumber
\end{eqnarray}
In matrix notation, it can be expressed as, $\vec{\boldsymbol{\nabla}}_n \vert\boldsymbol{\xi} (s_e|s_n) 
\rangle = -\vec{\boldsymbol{\tau}}\vert \boldsymbol{\xi} (s_e|s_n) \rangle.$ In other words, we have to solve the following differential equation:
\begin{eqnarray}
	\vec{\boldsymbol{\nabla}}_n \vert\boldsymbol{\xi} (s_e|s_n) \rangle + \vec{\boldsymbol{\tau}}\vert \boldsymbol{\xi} (s_e|s_n)
	\rangle = 0,  
	\label{elecwvfndiffeqn}
\end{eqnarray}
to establish the relation between $\vert \xi_i(s_e|s_0)\rangle$ and $\vert \xi_i(s_e|s_n)\rangle$.\\

\noindent
On the other hand, a first order vectorial linear differential equation can be written as:
\begin{eqnarray}
	\vec{\boldsymbol{\nabla}} \boldsymbol{\Omega} + \vec{\boldsymbol{\tau}}\boldsymbol{\Omega} = 0, 
	\label{Eq:veclin}
\end{eqnarray}
\noindent
where $\boldsymbol{\Omega}$ is a scalar matrix. One can show that the approximated solution can be written in the form~\citep{baer_phys_rep358,baer_book}:
\begin{eqnarray}
	\boldsymbol{\Omega} (s) = \zeta \exp \left(-\int_{s_0}^s \vec{\boldsymbol{\tau}} (s^{\prime}) .d\vec{s}^{\prime} \right) 
	\boldsymbol{\Omega} (s_0). 
	\label{vecdiffeqn}
\end{eqnarray}
where $\zeta\rightarrow$ represents ordering operator that defines the order of integration along the given contour
$(s_0 \rightarrow s_1 \rightarrow s_2 ,... \rightarrow s)$. \\

\noindent
Introducing the above form of the solution of the linear differential equation 
[Eq. (\ref{Eq:veclin})], the solution of Eq. (\ref{elecwvfndiffeqn}) can be written as:	
\begin{eqnarray}
	\vert \boldsymbol{\xi} (s_e|s_n) \rangle=\zeta\exp \left(-\int_{s_0}^{s_n} \vec{\boldsymbol{\tau}}(s^{\prime}_n).
	d\vec{s^{\prime}_n}\right)\vert \boldsymbol{\xi} (s_e|s_0)\rangle, \nonumber
\end{eqnarray}
or, alternatively,
\begin{eqnarray}
	\vert \boldsymbol{\xi} (s_e|s_n) \rangle =\tilde{\textbf{A}} (s_n|s_0)\vert \boldsymbol{\xi} (s_e|s_0)\rangle, 
	\label{elecbasischng}
\end{eqnarray}
where,
\begin{eqnarray}
	\tilde{\textbf{A}}(s_n|s_0) = \zeta\exp \left(-\int_{s_0}^{s_n}\vec{\boldsymbol{\tau}}(s^{\prime}_n).d\vec{s}^{\prime}_n \right),
	\label{Amatsol1}
\end{eqnarray}
is the transformation matrix operating on the electronic eigenstate $\vert \boldsymbol{\xi} (s_e|s_0)\rangle$ to
give $\vert \boldsymbol{\xi} (s_e|s_n) \rangle$. Moreover, it can be shown that the matrix $\tilde{\textbf{A}}(s_n|s_0)$
is orthogonal:
\begin{eqnarray}
	\tilde{\textbf{A}}(s_n|s_0)^{\dagger}\tilde{\textbf{A}}(s_n|s_0) &=& \zeta\exp \left(-\int_{s_0}^{s_n}\vec{\boldsymbol{\tau}}
	(s^{\prime}_n).d\vec{s}^{\prime}_n \right)^{\dagger}\zeta\exp \left(-\int_{s_0}^{s_n}\vec{\boldsymbol{\tau}}(s^{\prime}_n).d
	\vec{s}^{\prime}_n\right) \nonumber \\
	&=& \zeta\exp \left(-\int_{s_0}^{s_n}\vec{\boldsymbol{\tau}}^{\dagger}(s^{\prime}_n).d\vec{s}^{\prime}_n \right)\zeta\exp 
	\left(-\int_{s_0}^{s_n} \vec{\boldsymbol{\tau}}(s^{\prime}_n).d\vec{s}^{\prime}_n
	\right) \nonumber \\
	&=& \zeta\exp \left(\int_{s_0}^{s_n} \vec{\boldsymbol{\tau}}(s^{\prime}_n).d\vec{s}^{\prime}_n \right)\zeta\exp \left(-\int_{s_0}
	^{s_n} \vec{\boldsymbol{\tau}}(s^{\prime}_n).d\vec{s}^{\prime}_n\right) \nonumber \\
	&=& \textbf{I} 
\end{eqnarray}

\noindent
On the other hand, in Method II, we carry out a transformation of the nuclear wave function $(\boldsymbol{\psi^{ad}})$ 
in the adiabatic representation to another wave function $(\boldsymbol{\phi}^d)$ for 
the diabatic representation and those can be related through an orthogonal transformation matrix, $\boldsymbol{A}$ in the following way:
\begin{eqnarray}
	\boldsymbol{\psi}^{ad}(s_n) = \textbf{A}(s_n)\boldsymbol{\phi}^d(s_n)
	\label{nuwvfntrsf}
\end{eqnarray}
Using the above expression of $\boldsymbol{\psi^{ad}}$ in Eq. (\ref{adiase}), the SE in Eq. (\ref{adiase}) becomes:
\begin{eqnarray}
	-\dfrac{1}{2}(\vec{\boldsymbol{\nabla}}_n + \vec{\boldsymbol{\tau}})^2 \textbf{A}\boldsymbol{\phi}^d 
	+ (\textbf{u}-E)\textbf{A}\boldsymbol{\phi}^d = 0
	\label{diabsepre} 
\end{eqnarray}
\begin{eqnarray}
	(\vec{\boldsymbol{\nabla}}_n + \vec{\boldsymbol{\tau}})^2 \textbf{A}\boldsymbol{\phi}^d &=& (\vec{\boldsymbol{\nabla}}_n + 
	\vec{\boldsymbol{\tau}}).(\vec{\boldsymbol{\nabla}}_n + \vec{\boldsymbol{\tau}})\textbf{A}\boldsymbol{\phi}^d \nonumber \\
	&=& (\vec{\boldsymbol{\nabla}}_n + \vec{\boldsymbol{\tau}}).\left\lbrace\textbf{A}\vec{\boldsymbol{\nabla}}_n 
	\boldsymbol{\phi}^d + (\vec{\boldsymbol{\nabla}}_n\textbf{A})\boldsymbol{\phi}^d + \vec{\boldsymbol{\tau}}\textbf{A}
	\boldsymbol{\phi}^d \right\rbrace  \nonumber \\
	&=& \textbf{A}\boldsymbol{\nabla}_n^2
	\boldsymbol{\phi}^d + 2(\vec{\boldsymbol{\nabla}}_n\textbf{A}).\vec{\boldsymbol{\nabla}}_n\boldsymbol{\phi}^d +
	2\vec{\boldsymbol{\tau}}.\textbf{A}(\vec{\boldsymbol{\nabla}}_n\boldsymbol{\phi}^d) \nonumber \\
	&\quad& + (\boldsymbol{\nabla}_n^2 \textbf{A})\boldsymbol{\phi}^d +(\vec{\boldsymbol{\nabla}}_n.\vec{\boldsymbol{\tau}})\textbf{A}\boldsymbol{\phi}^d + 2\vec{\boldsymbol{\tau}}.
	(\vec{\boldsymbol{\nabla}}_n\textbf{A})\boldsymbol{\phi}^d  + \boldsymbol{\tau}^2\textbf{A}\boldsymbol{\phi}^d \nonumber 
\end{eqnarray}
which can be simplified to:
\begin{eqnarray}
	(\vec{\boldsymbol{\nabla}}_n + \vec{\boldsymbol{\tau}})^2 \textbf{A}\boldsymbol{\phi}^d &=& 
	\textbf{A}\boldsymbol{\nabla}_n^2\boldsymbol{\phi}^d + 2(\vec{\boldsymbol{\nabla}}_n\textbf{A} + 
	\vec{\boldsymbol{\tau}}\textbf{A}).\vec{\boldsymbol{\nabla}}_n\boldsymbol{\phi}^d  \nonumber \\ 
	&\quad &+\left\lbrace (\vec{\boldsymbol{\tau}} + \vec{\boldsymbol{\nabla}}_n).(\vec{\boldsymbol{\nabla}}_n\textbf{A} + \vec{\boldsymbol{\tau}}\textbf{A})\right\rbrace
	\boldsymbol{\phi}^d  
\end{eqnarray}
\noindent
One can choose the matrix $\textbf{A}$ such that it satisfies the following equation:
\begin{eqnarray}
	\vec{\boldsymbol{\nabla}}_n\textbf{A} + \vec{\boldsymbol{\tau}}\textbf{A} = 0,
	\label{adteq_p}
\end{eqnarray}
known as the ADT condition. Therefore Eq. (\ref{diabsepre}) reduces to:
\begin{eqnarray}
	-\dfrac{1}{2}\textbf{A}\boldsymbol{\nabla}_n^2\boldsymbol{\phi}^d + (\textbf{u}-E)\textbf{A}\boldsymbol{\phi}^d = 0. 
	\label{diabsepre2}
\end{eqnarray}
\noindent
The nature of ADT matrix, $\textbf{A}$ can be found out by multiplying Eq. (\ref{adteq_p}) by $\textbf{A}^{\dagger}$ from the left:
\begin{eqnarray}
	\textbf{A}^{\dagger}\vec{\boldsymbol{\nabla}}_n\textbf{A} + \textbf{A}^{\dagger}\vec{\boldsymbol{\tau}}\textbf{A} = 0,
	\label{amatorth1}
\end{eqnarray}
\noindent
and similarly, by taking dagger of Eq. (\ref{adteq_p}) and then, multiplying $\textbf{A}$ from the right, we get:
\begin{eqnarray}
	(\vec{\boldsymbol{\nabla}}_n\textbf{A}^{\dagger})\textbf{A} - \textbf{A}^{\dagger}\vec{\boldsymbol{\tau}}\textbf{A} = 0.
	\label{amatorth2}
\end{eqnarray}
Adding Eq. (\ref{amatorth2}) with (\ref{amatorth1}), we obtain:
\begin{eqnarray}
	\textbf{A}^{\dagger}\vec{\boldsymbol{\nabla}}_n\textbf{A} +
	(\vec{\boldsymbol{\nabla}}_n\textbf{A}^{\dagger})\textbf{A} =0 
	\nonumber  \\
	\Rightarrow \vec{\boldsymbol{\nabla}}_n(\textbf{A}^{\dagger}\textbf{A}) = 0 \nonumber \\
	\Rightarrow \textbf{A}^{\dagger}\textbf{A} = \text{const,}  \nonumber
\end{eqnarray}
which can be normalized to unity ($\textbf{I}$) and therefore, $\textbf{A}$ is an orthogonal matrix.\\

\noindent
Being $\textbf{A}$ is an orthogonal matrix, Eq. (\ref{diabsepre2}) is left multiplied by $\textbf{A}^{\dagger}$ and the 
SE is transformed into diabatic representation:
\begin{eqnarray}
	-\dfrac{1}{2}\boldsymbol{\nabla}_n^2\boldsymbol{\phi}^d + (\textbf{W}-E)\boldsymbol{\phi}^d = 0 
	\label{diabse2}
\end{eqnarray}
\begin{eqnarray}
	\textbf{W} = \textbf{A}^{\dagger}\textbf{u}\textbf{A}
\end{eqnarray}
The $\textbf{W}$ matrix is, in principle, full matrix with non-zero off-diagonal terms in the diabatic representation. Thus, the kinetic couplings in Eq. \ref{adiase} in adiabatic representation turn into potential couplings in Eq. (\ref{diabse2}) in 
diabatic representation through various nuclear DOFs. \\

\noindent
In summary, both the methods (I) and (II) provide the routes for adiabatic-to-diabatic transformation (ADT). Indeed, one can show that the two methods are equivalent resulting into the same diabatic representation as below:\\

\noindent
The condition in Eq. (\ref{adteq_p}) is known as ADT equation, which can be written by exploiting the form of the solution given in Eq. (\ref{vecdiffeqn}):
\begin{eqnarray}
	\textbf{A}(s_n) = \zeta \exp \left(-\int_{s_0}^{s_n} \vec{\boldsymbol{\tau}} (s^{\prime}_n) 
	.d\vec{s^{\prime}_n}\right)\textbf{A}(s_0).  
	\label{Amatsol2}
\end{eqnarray}
While choosing appropriate boundary conditions, $\textbf{A}(s_0) = \textbf{I}$, Eq. \ref{Amatsol2} is simplified as:
\begin{eqnarray}
	\textbf{A}(s_n) = \zeta \exp \left(-\int_{s_0}^{s_n} \vec{\boldsymbol{\tau}}(s^{\prime}_n)
	.d\vec{s^{\prime}_n} \right).
	\label{Amatsol3}
\end{eqnarray}
Equating Eqs. (\ref{Amatsol1}) and (\ref{Amatsol3}), we get:
\begin{eqnarray}
	\tilde{\textbf{A}}(s_n|s_0) = \textbf{A}(s_n). 
	\label{Amatquiv}
\end{eqnarray}
On recalling the expression for the total molecular wave function (see Eq. \ref{boexp}), we can write in the following matrix form:
\begin{eqnarray}
	\Psi(s_e,s_n) = \boldsymbol{\xi}^T(s_e|s_n) \boldsymbol{\psi^{ad}}(s_n). 
	\label{eq:comp_1}
\end{eqnarray}

\vspace{0.2 in}
\noindent
Using Eqs. (\ref{elecbasischng}), (\ref{Amatquiv}) and (\ref{nuwvfntrsf}) consecutively, Eq. \ref{eq:comp_1} can be rewritten as:
\begin{eqnarray}
	\Psi(s_e,s_n) &=& \boldsymbol{\xi}^T(s_e|s_0) \tilde{\textbf{A}}^{\dagger}\boldsymbol{\psi^{ad}}(s_n) \nonumber \\
	&=& \boldsymbol{\xi}^T(s_e|s_0)\textbf{A}^{\dagger}\boldsymbol{\psi^{ad}}(s_n)  \nonumber \\
	&=& \boldsymbol{\xi}^T(s_e|s_0)\boldsymbol{\phi}^d(s_n)    
	\label{eq:comp_2}
\end{eqnarray}
Comparing the above expression (Eq. \ref{eq:comp_2}) with the matrix form of Eq. (\ref{totpsidiab}), we affirm that:
\begin{eqnarray}
	\boldsymbol{\tilde{\phi}^d}(s_n) = \boldsymbol{\phi^d}(s_n). \nonumber
\end{eqnarray}
As a result, the diabatic wavefunctions in the two different methods becomes actually same which would lead to the same diabatic representation of the SE. Therefore, it is needless to mention that the potentials in this particular representation from the two methods are also same:
\begin{eqnarray}
	\textbf{V}=\textbf{W}. \nonumber
\end{eqnarray}

\subsection{Properties of Path Dependent Diabatic Hamiltonian Matrices}
\noindent
\underline{\bf Method I:}\\

\noindent
While depicting the path-dependence of diabatic Hamiltonian matrices, we consider the scalar form of ADT equations characterized by nuclear coordinates $p$ and $q$:
\begin{subequations}
	\begin{eqnarray}
		\dfrac{\partial}{\partial p}\textbf{A}(p,q) + \boldsymbol{\tau}_p\textbf{A}(p,q) &=& 0 \\
		\dfrac{\partial}{\partial q}\textbf{A}(p,q) + \boldsymbol{\tau}_q\textbf{A}(p,q) &=& 0
	\end{eqnarray}
\end{subequations}
\noindent
The integral form of each of the above equations is given by Eq. (\ref{Amatsol2}):
\begin{eqnarray}
	\textbf{A}(s|\Gamma) = \zeta \exp \left(-\int_{s_0}^s \vec{\boldsymbol{\tau}} (s^{\prime}_n) 
	.d\vec{s^{\prime}_n} \right)\textbf{A}(s_0),  
	\label{Amatsol4}
\end{eqnarray}
where $\Gamma$ is the contour of integration from $s_0$ to $s$ and $\zeta$ is the ordering operator for integration. We can carry out two consecutive steps of integration in 
two different paths $\Gamma_1$ and $\Gamma_2$ to reach a specific point. Subsequently, we will find the corresponding ADT matrices, $\textbf{A}_1(p,q)$ and
$\textbf{A}_2(p,q)$ by employing Eq. (\ref{Amatsol4}).\\

\noindent
The two different paths on the $p-q$ plane are chosen as follows:
\begin{eqnarray}
	\vspace{-0.8cm}
	\textbf{S.I:} \qquad (p_0,q_0) \rightarrow (p_0 +\Delta p,q_0) \rightarrow (p_0+\Delta p,q_0+\Delta q)
	\nonumber 
\end{eqnarray}
\begin{eqnarray}
	\textbf{S.II:} \qquad (p_0,q_0) \rightarrow (p_0,q_0 +\Delta q) \rightarrow (p_0+\Delta p,q_0+\Delta q)
	\nonumber
\end{eqnarray}
which is presented in the following figure (Figure \ref{fig:path}):		

\begin{center}
	\begin{figure}[!htp]
		\centering
		\includegraphics[width=0.5\textwidth,angle=0]{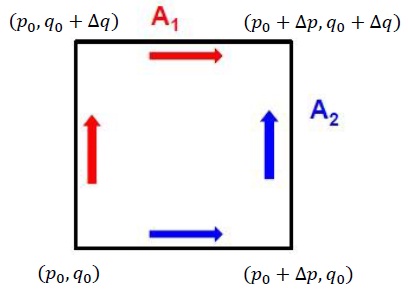}
		\caption{Two different paths of calculating ADT matrix. The figure has been adapted from Ref. [\protect\citen{joy_pccp}] and permission has been granted from The Royal Society of Chemistry (RSC).}
		\label{fig:path}
	\end{figure}
\end{center}
The corresponding ADT matrices for both the paths are given by:
\begin{subequations}
	\begin{eqnarray}
		\textbf{S.I:} \qquad \textbf{A}(p_0 +\Delta p,q_0) = \exp \left(-\int_{p_0}^{p_0 +\Delta p} 
		\boldsymbol{\tau}_p (p,q_0)dp\right)\textbf{A}(p_0,q_0)
		\label{pathdep1a}
	\end{eqnarray}
	\begin{eqnarray}
		\textbf{A}(p_0 +\Delta p,q_0 + \Delta q) = \exp \left(-\int_{q_0}^{q_0 +\Delta q} 
		\boldsymbol{\tau}_q (p_0 + \Delta p,q)dq\right)\textbf{A}(p_0 + \Delta p,q_0) \nonumber \\
		\label{pathdep1b}
	\end{eqnarray}
\end{subequations}

\begin{subequations}
	\begin{eqnarray}
		\textbf{S.II:} \qquad \textbf{A}(p_0,q_0 +\Delta q) = \exp \left(-\int_{q_0}^{q_0 +\Delta q} 
		\boldsymbol{\tau}_q (p_0,q)dq\right)\textbf{A}(p_0,q_0)
		\label{pathdep2a}
	\end{eqnarray}
	\begin{eqnarray}
		\textbf{A}(p_0 +\Delta p,q_0 + \Delta q) = \exp \left(-\int_{p_0}^{p_0 +\Delta p} 
		\boldsymbol{\tau}_p (p,q_0 + \Delta q)dp\right)\textbf{A}(p_0, q_0 + \Delta q) \nonumber \\
		\label{pathdep2b}
	\end{eqnarray}
\end{subequations}

\noindent
When Eqs. (\ref{pathdep1a}) and (\ref{pathdep2a}) are substituted in Eqs. (\ref{pathdep1b}) and
(\ref{pathdep2b}), respectively, the corresponding ADT matrices for two different chosen paths take the following form:
\begin{subequations}
	\begin{eqnarray}
		&\textbf{S.I:} \qquad \textbf{A}_1 \equiv \textbf{A}(p_0 +\Delta p,q_0 + \Delta q)& \nonumber \\
		&= \exp \left(-\int_{q_0}^{q_0 +\Delta q}\boldsymbol{\tau}_q (p_0 + \Delta p,q)dq
		-\int_{p_0}^{p_0 +\Delta p}\boldsymbol{\tau}_p (p,q_0)dp\right)\textbf{A}(p_0,q_0)& \quad
		\label{pathdep3a}
	\end{eqnarray}
	\begin{eqnarray}
		&\textbf{S.II:} \qquad \textbf{A}_2 \equiv \textbf{A}(p_0 +\Delta p,q_0 + \Delta q)& \nonumber \\
		&= \exp \left(-\int_{p_0}^{p_0 +\Delta p}\boldsymbol{\tau}_p (p,q_0 + \Delta q)dp
		-\int_{q_0}^{q_0 +\Delta q}\boldsymbol{\tau}_q (p_0,q)dq\right)\textbf{A}(p_0,q_0)& \quad
		\label{pathdep3b}
	\end{eqnarray}
\end{subequations}

\noindent
While taking dagger on Eq. (\ref{pathdep3a}), we can obtain:
\begin{eqnarray}
	&\,& \textbf{A}_1^{\dagger} =\textbf{A}^{\dagger}(p_0,q_0) \times  \nonumber \\
	&\,& \exp \left(\int_{q_0}^{q_0 +\Delta q}
	\boldsymbol{\tau}_q (p_0 + \Delta p,q)dq + \int_{p_0}^{p_0 +\Delta p}\boldsymbol{\tau}_p (p,q_0)dp\right)
\end{eqnarray}
and then, it is multiplied with Eq. (\ref{pathdep3b}) to achieve: 
\begin{subequations}
	\begin{eqnarray}
		\textbf{A}_1^{\dagger}\textbf{A}_2 &=& \textbf{A}^{\dagger}(p_0,q_0)  \exp \Big[\int_{p_0}^{p_0 +\Delta p}
		\Big(\boldsymbol{\tau}_p (p,q_0 + \Delta q)-\boldsymbol{\tau}_p (p,q_0)\Big)dp \nonumber \\
		&+& \int_{q_0}^{q_0 +\Delta q} \Big(\boldsymbol{\tau}_q (p_0 + \Delta p,q) - \boldsymbol{\tau}_q (p_0,q)
		\Big)dq\Big] \textbf{A}(p_0,q_0)
		\label{pathdep4a}
	\end{eqnarray}
\end{subequations}

\noindent	
Considering a midpoint $(\bar{p},\bar{q})$ in the planer interval as $p_0 \leq \bar{p} \leq p_0 + \Delta p$
and $q_0 \leq \bar{q} \leq q_0 + \Delta q$ and assuming that $\Delta p, \Delta q$ are small enough, we employ the 
definition of numerical differentiation as well as integration around the midpoint $(\bar{p},\bar{q})$ 
and write Eq. (\ref{pathdep4a}) as:
\begin{subequations}
	\begin{eqnarray}
		&\,& \textbf{A}_1^{\dagger}\textbf{A}_2 = \textbf{A}^{\dagger}(p_0,q_0) \times \nonumber \\
		&\,& \exp \left( -\Delta q \int_{p_0}^{p_0 +\Delta p} \dfrac{\partial \boldsymbol{\tau}_p (p,\bar{q})}{\partial q}dp 
		+ \Delta p \int_{q_0}^{q_0 +\Delta q} \dfrac{\partial \boldsymbol{\tau}_q (\bar{p},q)}{\partial p}dq
		\right)\textbf{A}(p_0,q_0) \nonumber 
	\end{eqnarray}
	\begin{eqnarray}
		= \textbf{A}^{\dagger}(p_0,q_0) \exp \left[ -\Delta p\Delta q \left( 
		\dfrac{\partial \boldsymbol{\tau}_p (\bar{p},\bar{q})}{\partial q} - 
		\dfrac{\partial \boldsymbol{\tau}_q (\bar{p},\bar{q})}{\partial p} \right)\right]\textbf{A}(p_0,q_0)
		\label{pathdep4b}
	\end{eqnarray}
\end{subequations}
which in turn takes the following form:
\begin{eqnarray}
	\textbf{A}_1^{\dagger}\textbf{A}_2 = \textbf{A}^{\dagger}(p_0,q_0) \exp \left(-\Delta p\Delta q
	(\text{curl} \boldsymbol{\tau})_{pq}\right)\textbf{A}(p_0,q_0)
	\label{pathdep4c}
\end{eqnarray}
\noindent
The argument of the exponential in Eq. (\ref{pathdep4c}) is not a null matrix and therefore, the product matrix $(\textbf{B} = \textbf{A}_1^{\dagger}\textbf{A}_2)$ does not appear as unit matrix. The ADT matrices, $\textbf{A}_1(p,q)$ and $\textbf{A}_2(p,q)$, are not the same at the point
$(p_0 +\Delta p, q_0 +\Delta q)$ for two different chosen paths of integration, {\bf S.I} and {\bf S.II}. 

\begin{itemize}
	\item If the curl of the NAC matrix is zero (i.e. $\text{curl} \boldsymbol{\tau}$), both the chosen path (even any path) will produce the same ADT matrices.
	\item On the contrary, the product matrix
	$(\textbf{B} = \textbf{A}_1^{\dagger}\textbf{A}_2)$ is an orthogonal one as shown below:
	\begin{eqnarray}
		\textbf{B}^{\dagger}\textbf{B} &=& \textbf{A}^{\dagger}(p_0,q_0) \exp \left(\Delta p\Delta q
		(\text{curl} \boldsymbol{\tau})_{pq}\right)\textbf{A}(p_0,q_0) \times \nonumber \\
		&\qquad& \textbf{A}^{\dagger}(p_0,q_0) 
		\exp \left(-\Delta p\Delta q(\text{curl} \boldsymbol{\tau})_{pq}\right)\textbf{A}(p_0,q_0) \nonumber \\
		&=& \textbf{I}
	\end{eqnarray}
	\noindent
	where $\text{curl} \vec{\boldsymbol{\tau}}$ is a skew - symmetric matrix, i.e.
	$(\text{curl} \vec{\boldsymbol{\tau}})^{\dagger} = -\text{curl} \vec{\boldsymbol{\tau}}$. 
	Therefore, the ADT matrices, $\textbf{A}_1(p,q)$ and $\textbf{A}_2(p,q)$, are related through an orthogonal transformation matrix, $\textbf{B}$.~\citep{mukherjee_jpca117}
	
\end{itemize}

\noindent
Therefore, there are infinite number of possible different diabatic potential matrices depending upon the chosen paths, but as these are related by orthogonal transformation matrix, observables will be identical on dynamical calculations. \\

\noindent
\underline{\bf Method II:}\\

\noindent
Since BBO theory based diabatization involves skew-symmetric non-Abelian non-adiabatic coupling matrices (NACM) of three or higher dimensional sub-Hilbert spaces,
the solutions of the coupled differential (ADT) equations appear as path-dependent over two-dimensional cross-sections of any multi-dimensional space of nuclear degrees of freedom. As for example, if one carries out even multi-step integration in two different paths (let's say paths I and II, see Figure \ref{fig:path}) towards a specific final ending point and the resulting ADT matrices are assigned as $A_1$ and $A_2$\citep{mukherjee_jpca117} at the same point, the associated diabatic PESs matrices at the end point take the following forms:

\begin{eqnarray}
	W_1 = A_{1}{^\dagger} U A_1 \quad \text{and} \quad W_2 = A_{2}{^\dagger} U A_2, 
\end{eqnarray}

\noindent
leading to 
\begin{eqnarray}
	A_1 W_1 A_{1}{^\dagger} = A_2 W_2 A_{2}{^\dagger} = U,
	\label{eq:pdO}
\end{eqnarray}
\noindent
which results into
\begin{eqnarray}
	W_1=BW_2{B}^{\dagger}.
	\label{eq:pdW}
\end{eqnarray}
The ADT matrices $A_1$($p$,$q$) and $A_2$($p$,$q$) are orthogonal transformation matrices and thereby, the product $B\,(=A_1^{\dagger}A_2)$ in Eq. \ref{eq:pdW} also appears to be an orthogonal matrix:~\citep{soumya_bz,mantu_jcp159_path}
\begin{eqnarray}
	B^{\dagger}B=A_2^{\dagger}A_1A_1^{\dagger}A_2=I.
\end{eqnarray}
Consequently, infinite number of different diabatic potential matrices ($W$) is possible depending upon the chosen paths. On the contrary, those matrices are
related by orthogonal transformation matrix ($B$) and therefore, one can get same observables on dynamical calculations (see Eq. \ref{eq:pdW}). In other words,
the above discussion affirms path independence of the calculated observables originating from different diabatic PESs matrices ($W_1$ and $W_2$).~\citep{joy_pccp,soumya_bz,mantu_jcp159_path} 

\section{Curl Condition and Quantization of NACTs}\label{ssec:removable}
\noindent
While performing BBO based diabatization of adiabatic PESs and NACTs, it is indeed necessary to verify whether the involved number of electronic states
($N$) over the interested domain of nuclear space constitutes a ``true" SHS within the numerical precision. We can examine the existence of such SHS 
by calculating matrix elements of derived Curl Condition of the associated NACTs as depicted below. Over $ p $ - $ q $ ($ p $ and $ q $ are nuclear degrees of freedom) nuclear plane, Eq. \ref{adteq_p} can be written as,
\begin{equation}
	\nabla_pA + \tau_pA = 0 
	\label{adt_cond1}
\end{equation}
\begin{equation}
	\nabla_qA + \tau_qA = 0 
	\label{adt_cond2}
\end{equation}
\noindent
Since the elements of ADT matrix ($ A $) are analytical functions with respect to nuclear coordinates ($ p $ and $ q $), the cross-derivatives on Eqs. \ref{adt_cond1} and \ref{adt_cond2} can lead to the following equation known as Curl Condition: \\

\noindent
(a) Curl
condition:
\begin{equation}
	\dfrac{\partial}{\partial p}\tau^{ij}_q-\dfrac{\partial}{\partial q}\tau^{ij}_p=(\tau_q\tau_p)_{ij}-
	(\tau_p\tau_q)_{ij},
	\label{eq:curl}
\end{equation}
\noindent
which is known as molecular field equation (similar to Yang-Mills field equation) over $p-q$ plane that can be represented as,
\begin{equation}
	F_{pq}^{ij}=\Big[\dfrac{\partial}{\partial p}\tau^{ij}_q-\dfrac{\partial}{\partial q}\tau^{ij}_p\Big]-\Big[(\tau_q\tau_p)_{ij}-
	(\tau_p\tau_q)_{ij}\Big],
	\label{eq:curl1}
\end{equation}
\noindent
and such components should acquire the magnitude of zero (0) as the condition for the existence of SHS.\\

\noindent
(b) Quantization of non-adiabatic coupling:\\

\noindent
In case of symmetry driven or accidental degeneracy between the electronic states at a point or along a seam, Hellmann-Feynmann~\citep{hellmann,feynmann} theorem (see Eq. \ref{Eq:helfen}) depicts the singularity of NACT. Such singular function obeys the Cauchy's residue theorem along a closed contour of the singular point or seam:
\begin{equation}
	\int_0^{2\pi} \tau^{ij}_{\phi}(\rho,\phi)d\phi = n\pi\,\, \text{or}\,\, n.(2\pi),
	\label{eq:quant}
\end{equation}

\noindent
which is known as quantization of NACT. In the above expression, $n$ represents the number of CIs, $i,j$ specifies two electronic states for a specific sub-Hilbert space [$\lbrace i,j \rbrace
\in N$], $p$ and $q$ designate Cartesian coordinates, and $\rho$ and $\phi$ define their polar counterparts. On the other hand, 
the scalar components of NACTs along Cartesian coordinates, $p$ and $q$, and the polar transformed circular coordinate, $\phi$ are denoted
as $\tau^{ij}_p$, $\tau^{ij}_q$ and $\tau^{ij}_\phi(\rho,\phi)$, respectively. \\

\noindent
If the conditions (Eqs. \ref{eq:curl} and \ref{eq:quant})
are validated by the NACT components, one can consider those states form a SHS as well as depict the number of CI(s) in the sub-space.
Since the entire magnitude of {\it ab initio} NACTs is incorporated in first principle based diabatization using ADT, the validity of the above conditions
predicts absence of any non-removable component for the chosen sub-Hilbert space. It affirms the calculated ADT matrix $A$ as well as the diabatic PES matrix $W$ as theoretically ``exact'' and numerically ``accurate'' ones. On the contrary, when those conditions (Eqs. \ref{eq:curl} and \ref{eq:quant}) are not satisfied
over the chosen domain of nuclear CS, expansion of the SHS is indeed necessary to achieve desired level of accuracy.

\subsection{Sub-Hilbert Space: Theoretical and Algorithmic Details on the Calculation of Curl Condition}

\noindent
While computing NACTs and adiabatic PESs, the accuracy of electronic structure calculation plays a major role as implemented in the quantum chemistry packages like Molpro~\citep{MOLPRO_brief} or Columbus~\citep{columbus}. In order to calculate the adiabatic PESs, complete active space self-consistent field (CASSCF) followed by multi-reference configuration interaction (MRCI) approaches can be opted to solve electronic SE at each nuclear geometry. In CASSCF (coupled-perturbed multi-configuration self-consistent field, CP-MCSCF) and MRCI (numerical finite difference, DDR) approaches, NACTs are evaluated by taking derivatives of the energies and derivatives of wavefunction, respectively. Since we are interested to calculate multiple adiabatic PESs and couplings between those states, the use of multi-reference electronic structure calculations (CASSCF/CP-MCSCF or MRCI/DDR) is a necessity. \\

\noindent
The accuracy of NACTs can be explored by performing closed contour integration of those {\it ab initio} quantities over the nuclear CS, where such integration leads to $n\pi$ (for symmetry driven or accidental JT CI/seam) or $n.2\pi$ (for RT CI/seam) or 0 [for Pseudo Jahn-Teller (PJT)] validating the quantization properties of NACTs. In other words, the following integral needs to be evaluated:
\begin{equation}
	\int_0^{2\pi} \tau^{ij}_{\phi}(\rho_0,\phi)d\phi = 0\,\,\text{or}\,\,n\pi\,\, \text{or}\,\, n.2\pi,
\end{equation}
where $\tau^{ij}_{\phi}$ symbolizes the NACT between $\lbrace i,j\rbrace$ electronic states along the circular nuclear coordinate $\phi$ for a fixed value of $\rho\,(=\rho_0)$. For any molecular process or chemical reaction, the above integral needs to be satisfied by the {\it ab initio} calculated NACTs if the 
electronic structure calculations are executed using complete active space (CAS) with appropriate choice of basis set for a specific system. In other words, the quantization rule validates the accuracy of {\it ab initio} calculated NACTs.\\

\noindent	
On the other hand, the accuracy of diabatic Hamiltonian matrix (obtained by ADT calculations) is dependent on the ``numerical" truncation of Hilbert space for a specific system. In principle, the diabatic Hamiltonian is theoretically exact for the $M$-dimensional complete Hilbert space, but for practical purposes,
a $N$-dimensional truncated Hilbert space can be used to generate finite [$N\,(<< M)$] dimensional ``accurate" diabatic Hamiltonian matrix. Even though the 
curl condition (difference between Mathematical and Commutator Curls) appears as zero (0) by construction for the complete matrix (Hilbert) space, one needs to 
verify the validity of such condition for a truncated Hilbert space prior to the construction of BBO based diabatic Hamiltonian (BBO-DH).\\

\noindent
While computing the curl condition for a truncated Hilbert space, direct use of {\it ab initio} calculated NACTs may often leads to erroneous results due to
numerical error at few nuclear geometries, more specifically, at the close vicinity of CI domain. As a matter of fact, taking derivatives of singular or very small values of {\it ab initio} NACTs for calculating Mathematical Curl may introduce inaccuracies and practical difficulty. On the contrary, while solving the 
ADT equations using backward differentiation formula [integration of differential 
equations for (a) two state, (b) three state, (c) four state and (d) five state (see the following section for details)], the numerical inaccuracies of NACTs over nuclear geometries could be wiped out to obtain the ADT angles fulfilling the quantization rule as discussed below. \\

\noindent
(a) Once the {\it ab initio} NACTs are calculated between the multiple electronic states as functions of vibrational modes, those quantities are substituted in the ADT equations (see Eqs. \ref{eq:adt-two}, \ref{eq:adt-three} and Appendix~\ref{ssec:adt-eq}) and solved to obtain the ADT angles.\\

\noindent
(b) Those ADT angles are plugged into the
explicit expressions of NACTs (see Eqs. \ref{eq:nact-two}, \ref{eq:nact-three} and Appendix~\ref{ssec:adt-eq}) and Mathematical Curls (see Eqs. \ref{eq:curl-two}, \ref{eq:curl-three} and Appendix~\ref{ssec:adt-eq}) to evaluate those quantities with greater accuracy. \\

\noindent
(c) The recalculated NACTs are substituted
in the commutator relation (RHS of Eq. \ref{eq:curl}) to evaluate the matrix elements of ADT Curl. \\

\noindent
(d) The matrix elements of Curl Condition are computed by subtracting Mathematical and ADT Curls (Eq. \ref{eq:curl1}). \\

\noindent
Such approach is a unique tool of BBO based diabatization for predicting the size of the truncated space known as sub-Hilbert space ensuring the ``accuracy" of diabatic Hamiltonian for a molecular process or chemical reaction.

\noindent
\section{Explicit Expressions of ADT Equations, NACTs and Curl Equations}\label{ssec:exp}

\noindent
We present the explicit expressions of ADT Equations, NACTs and Mathematical Curls for two to three state SHS,~\citep{mukherjee_jcp143,sarkar_jcp124} where the expressions for four to six state SHS are illustrated in the Appendix~\ref{ssec:adt-eq}~\citep{mukherjee_jpca117,sarkar_jpca112,soumya_jpca121,soumya_jcp150}. \\

\noindent
The ADT equations for two ($N=2$) and three ($N=3$) electronic state sub-Hilbert space take the following form:~\citep{mukherjee_jcp143,sarkar_jcp124}

\begin{equation}
	\boldsymbol{\nabla}_R\Theta^{12}= -\boldsymbol{\tau}^{12}
	\label{eq:adt-two}
\end{equation}

\begin{eqnarray}
	\boldsymbol{\nabla}_R\Theta^{12} &=& -{\boldsymbol{\tau}}^{12} - \tan\Theta^{13}
	({\boldsymbol{\tau}}^{13}\sin\Theta^{12}
	+{\boldsymbol{\tau}}^{23}\cos\Theta^{12}),\nonumber\\
	\boldsymbol{\nabla}_R\Theta^{13} &=& -({\boldsymbol{\tau}}^{13}\cos\Theta^{12}- {\boldsymbol{\tau}}^{23}\sin\Theta^{12}),\nonumber\\
	\boldsymbol{\nabla}_R\Theta^{23} &=& -\frac{1}{\cos\Theta^{13}}({\boldsymbol{\tau}}^{13}\sin\Theta^{12}				+{\boldsymbol{\tau}}^{23}\cos\Theta^{12}), 
	\label{eq:adt-three}
\end{eqnarray}

\noindent
On the other hand, NACT expressions can be written as~\citep{mukherjee_jcp143,sarkar_jcp124}:
\begin{eqnarray}
	\boldsymbol{\tau^{12}}=-\boldsymbol{\nabla}_R\Theta^{12}
	\label{eq:nact-two}
\end{eqnarray} 
\begin{eqnarray}
	\boldsymbol{\tau^{12}}&=&-\boldsymbol{\nabla}_R\Theta^{12}+\sin\Theta^{13}\boldsymbol{\nabla}_R\Theta^{23}\nonumber \\
	\boldsymbol{\tau^{13}}&=&-\cos\Theta^{13}\sin\Theta^{12}\boldsymbol{\nabla}_R\Theta^{23}-\cos\Theta^{12}\boldsymbol{\nabla}_R\Theta^{13}\nonumber \\
	\boldsymbol{\tau^{23}}&=&-\cos\Theta^{12}\cos\Theta^{13}\boldsymbol{\nabla}_R\Theta^{23}+\sin\Theta^{12}\boldsymbol{\nabla}_R\Theta^{13}
	\label{eq:nact-three}
\end{eqnarray}
\noindent
Finally, the Mathematical Curl equations take the following forms:
\begin{eqnarray}
	Z_{\rho\phi}^{12}&=&\nabla_\phi\nabla_\rho\Theta^{12}-
	\nabla_\rho\nabla_\phi\Theta^{12}
	\label{eq:curl-two}
\end{eqnarray}
\begin{small}
	\begin{eqnarray}
		Z_{\rho\phi}^{12}&=&-\cos\Theta^{23}(\nabla_\phi\Theta^{23}\nabla_\rho\Theta^{13}-
		\nabla_\rho\Theta^{23}\nabla_\phi\Theta^{13})\nonumber\\
		Z_{\rho\phi}^{13}&=&\sin\Theta^{12}\cos\Theta^{23}(\nabla_\phi\Theta^{12}\nabla_\rho\Theta^{13}-
		\nabla_\rho\Theta^{12}\nabla_\phi\Theta^{13})+\cos\Theta^{12}\sin\Theta^{23}(\nabla_\phi\Theta^{23}\nabla_\rho\Theta^{13}\nonumber\\
		&&-
		\nabla_\rho\Theta^{23}\nabla_\phi\Theta^{13})-\cos\Theta^{12}(\nabla_\phi\Theta^{12}\nabla_\rho\Theta^{23}-\nabla_\rho\Theta^{12}\nabla_\phi\Theta^{23})\nonumber\\
		Z_{\rho\phi}^{23}&=&\cos\Theta^{12}\cos\Theta^{23}(\nabla_\phi\Theta^{12}\nabla_\rho\Theta^{13}-
		\nabla_\rho\Theta^{12}\nabla_\phi\Theta^{13})-\sin\Theta^{12}\sin\Theta^{23}(\nabla_\phi\Theta^{23}\nabla_\rho\Theta^{13}\nonumber\\
		&&-
		\nabla_\rho\Theta^{23}\nabla_\phi\Theta^{13})+\sin\Theta^{12}(\nabla_\phi\Theta^{12}\nabla_\rho\Theta^{23}-\nabla_\rho\Theta^{12}\nabla_\phi\Theta^{23})
		\label{eq:curl-three}
	\end{eqnarray}
\end{small}

\noindent
\section{Molecular Symmetry Adaptation of Nonadiabatic Coupling Terms}\label{ssec:msgroup}
\vspace{0.2 cm}
\noindent 
While performing the electronic structure calculations for a molecular process or chemical reaction, the IREPs of the adiabatic states can be assigned as per the point group of the system at the equilibrium geometry and therefore, it is possible to define the IREP of the adiabatic potential energy at that point. When one moves from the equilibrium geometry to any other geometry of the nuclear CS, one can assign the IREPs of the adiabatic PESs, which would also be found out from the correlation of the associated (point) groups. It is to note that the IREPs of the adiabatic PESs change from equilibrium geometry to other geometry, in particular, because the symmetry of the molecular process changes from one nuclear configuration to another. On the contrary, in case of NACTs, since those terms are evaluated by taking the derivatives of the electronic eigenfunctions with respect to nuclear coordinates, it is not straightforward to define the IREPs of those coupling terms. While defining the IREPs of the NACTs, we need to consider higher symmetry groups known as molecular symmetry (MS),~\citep{longuet-higgins_mp6} where the symmetry elements of the point group are modified by multiplying Rigid Rotation and Nuclear Permutations of the molecular/chemical processes. For high amplitude nuclear (vibrational) motions of a species, the point group concept has limited applicability, where 
a larger set of symmetry operations is required to examine the floppy molecular processes. If 
the time scale of a given laboratory experiment appears shorter to allow certain nuclear permutations,\citep{longuet-higgins_mp6} it is necessary to construct the MS group using the feasible ones among the elements of complete group of Hamiltonian. \\

\noindent
Generally, a molecular symmetry (MS) group consists of $h$ symmetry operations, $\hat{O}_i$, 
$i=1,2, \ldots ,h$, namely, the identity operation $E$, all feasible permutations $P$ 
of the spatial and spin coordinates of the equivalent nuclei, the inversion $E^{\star}$ 
of all nuclear ($\textbf{R}$) and electronic ($\textbf{r}$) coordinates, and the 
inversion-permutation $P\boldsymbol\cdot E^{\star}$ operation. Any such operation, $\hat{O}$ can be expressed as a product of 
three different operations,\citep{bunker_book} i.e., $\hat{O} = \hat{O}_a \hat{O}_b \hat{O}_c$, where $\hat{O}_a$ is associated with 
vibronic DOFs, $\hat{O}_b$ involves rotation of the Euler angles and $\hat{O}_c$ leads to permutation of the nuclear spin. On the contrary, 
in the point group, symmetry operations transform only the vibronic DOFs of a molecule. \\

\noindent
As a representative example of symmetric triatomic molecule, we consider Na$_3$ cluster possessing D$_{3h}$
symmetric equilibrium structure. In this molecular system, the body fixed $x-y-z$ axes are situated in the cluster 
in such a way so that plane of symmetry, $\sigma_h$ passes through the molecular plane ($xy$ plane) 
and $z$ axis coincides with the threefold symmetry axis, $C_3$, which are demonstrated in Figure \ref{fig:pointgrpd3h}
using coordinate axes, symmetry elements and stereographic projection for $D_{3h}$ point
group. For such case, the Cartesian coordinates $x-y-z$ symbolize the normal modes, $Q_x$ (bending), $Q_y$ (antisymmetric 
stretching) and $Q_s$ (symmetric stretching), respectively. On the other hand, Figure \ref{fig:na3vib} depicts 
atomic displacements for those vibrational modes. In case of Na$_3$ cluster, polar counterparts of $x\,(=Q_x)$ and $y\,(=Q_y)$ coordinates ($\rho$ and $\phi$)
are chosen keeping the magnitude of $z\,(=Q_s)$ fixed at 3.7 \r{A}. The MS group [D$_{3h}$(M)] and the corresponding point group (D$_{3h}$) 
can be related through the following relations\citep{bunker_book}:

\begin{small}
	\begin{subequations} 
		\begin{eqnarray}                                                                                
			E &= &E\cdot R^0 \cdot p_0,\\
			(123) &=&C_3^1 \cdot R^{\frac{2\pi}{3}} \cdot p_{123}, \label{eq:msop}\\
			(23) &=&C_{2x} \cdot R^{\pi_z} \cdot p_{23},\\
			E^{\star} &=&\sigma_{xy} \cdot R^{\pi_z}\cdot p_0,\\
			(132)^{\star} &=&S_3^{-1} \cdot R_z^{\frac{\pi}{3}} \cdot p_{132},\\
			(23)^{\star} &=&\sigma_{xz} \cdot R_y^{\pi} \cdot p_{23},
		\end{eqnarray}
	\end{subequations}
\end{small}

\noindent
where ``$R^{\phi}$''s represent molecular rotations and ``$p$''s 
signify nuclear permutations. While defining 1D IREPs of NACTs for C$_{2v}$(M) group,
Al-Jabour and his coworkers\citep{al-jabour_jpca114} introduced two important theorems, 
which were extended for D$_{3h}$(M) group
by Adhikari and coworkers for 2D IREPs.\citep{mukherjee_jpca117}\\

\begin{figure}[!htp]
	\centering
	\includegraphics[width=0.5\textwidth,angle=0]{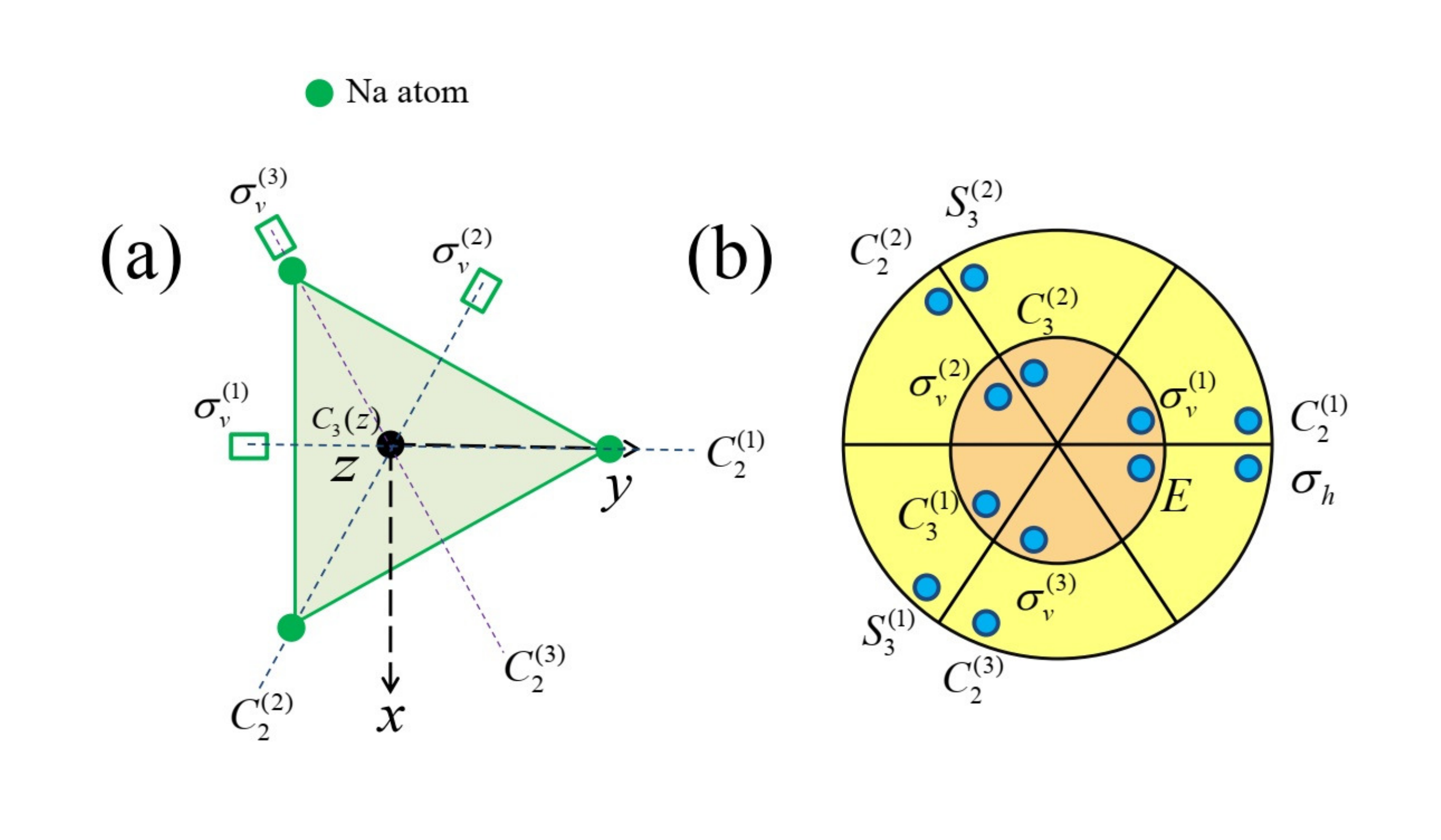}
	\caption{The left panel (a) shows symmetry elements of D$_{3h}$ point group and the coordinate axes 
		in Na$_3$ system. The right one (b) presents the effects of those symmetry operations in Na$_3$ cluster.
		The figure has been adapted from Ref. [\protect\citen{joy_pccp}] and permission has been granted from The Royal Society of Chemistry (RSC).}
	\label{fig:pointgrpd3h}
\end{figure}

\begin{figure}[!htp]
	\centering
	\includegraphics[width=0.4\textwidth,angle=0]{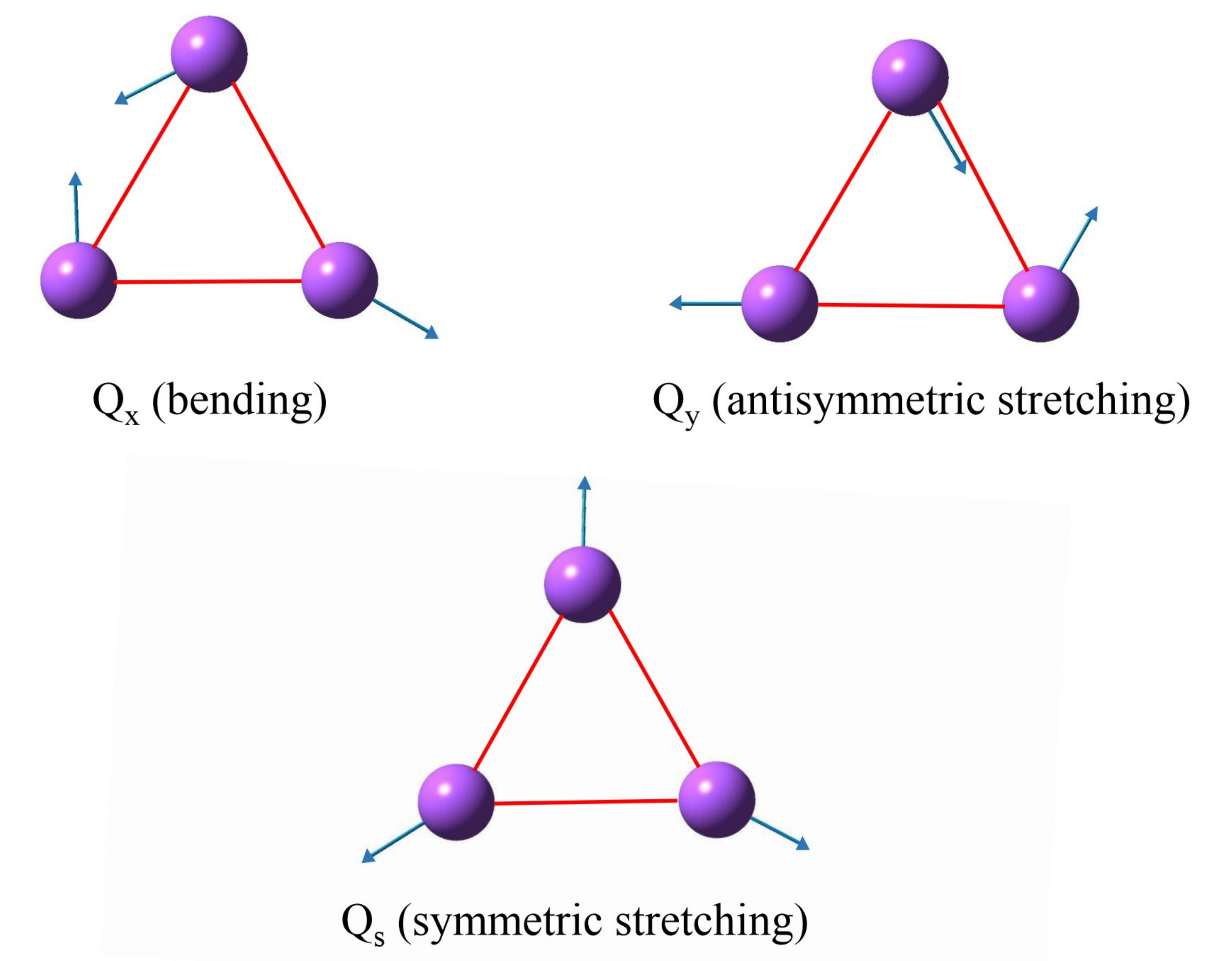}
	\caption{Vectorial displacements of the atoms in Na$_3$ cluster are depicted for three 
		vibrational modes, namely $Q_x$ (bending), $Q_y$ (antisymmetric stretching) and $Q_s$ (symmetric stretching). The figure has been taken from Ref. [\protect\citen{joy_pccp}] and permission has been granted from The Royal Society of Chemistry (RSC).}
	\label{fig:na3vib}
\end{figure}

\noindent
{\bf Theorem 1:} If the IREP of a NACT, 
$\tau_l^{ij}$ is known along a specific symmetry adapted nuclear degree of freedom (DOF), $R_l$, the IREP of the another one, $\tau_k^{ij}$ along a different symmetry adapted nuclear 
coordinate, $R_k$ can be found with the aid of following relationship:\citep{al-jabour_jpca114}

\begin{small}
	\begin{eqnarray}
		\Gamma(\tau_k^{ij}) &=&\Gamma\left(\frac{\partial}{\partial R_k}\right)\times
		\Gamma\left(\frac{\partial}{\partial R_l}\right)\times\Gamma(\tau_l^{ij}),
		\label{eq:theo1}
	\end{eqnarray}
\end{small}

\noindent
where
\begin{small}
	\begin{eqnarray}
		\tau_k^{ij}&=&\Big\langle\xi_i \Big|\frac{\partial\xi_j}{\partial R_k}\Big\rangle.
		\label{eq:taudef}
	\end{eqnarray}
\end{small}

\noindent
Using the above criterion, we can determine tentative combinations of the IREPs by 
assigning each of the NACTs [$\tau_k$s ($k=\rho$ and $\phi$ or $x$ and $y$)] to the totally
symmetric one ($A_1^{\prime}$), successively. In case of D$_{3h}$(M) MS group,
if $\tau_{\rho}$ is considered as $A_1^{\prime}$, Table \ref{tb:charac} and Eq. \ref{eq:theo1} show
that $\tau_{\phi} \in A_1^{\prime\prime}$ and $\tau_x,\,\tau_y \in E^{\prime\prime}$. In a similar way,
if $\tau_{\phi} \in A_1^{\prime}$, one can obtain $\tau_{\rho} \in A_1^{\prime\prime}$ and 
$\tau_x,\,\tau_y \in E^{\prime}$. Nevertheless, for other two possibilities 
($\tau_x$ $\in$ $A_1^{\prime}$ or $\tau_y$ $\in$ $A_1^{\prime}$), $\tau_y$ or $\tau_x$ belongs to {\it reducible representation} contradicting the construction of character table\citep{kettle_book} as presented in Table \ref{tb:charac}.\\


\begin{table}[!htp]
	\caption{\label{tb:charac}Extended character table of D$_{3h}(M)$ molecular symmetry group with
		different components of NACT. This table is adapted from Ref. [\protect\citen{mukherjee_jpca117}].}
	\begin{scriptsize}
		\begin{center}
			\begin{tabular} {|c|c|c|c|c|c|c|c|c|c|c|}
				\hline
				\multirow{3}{*} {D$_{3h}$(M)} & {        $E$         } & (123) & (12) & {     ${E^{\star}}$     } & 
				${(123)^{\star}}$ & ${(12)^{\star}}$ & coord. & deriv. & 1st& 2nd\\
				& & (132) & (13) & &  ${(132)^{\star}}$ &  ${(13)^{\star}}$ & & & comb. & comb.\\
				& & & (23) & & &  ${(23)^{\star}}$ & & & &\\ \hline
				$A_1^{\prime}$ & 1 &1 &1 &1 &1 &1 & & & $\tau_{\rho}$ & $\tau_{\phi}$\\ \hline
				$A_2^{\prime}$ &  1 &  1 & -1 &  1 &  1 & -1 & $\phi$ & ${\frac{\partial}{\partial \phi}}$&  &\\ 
				\hline
				$E^{\prime}$ & 2 & -1 & 0 & 2 & -1 & 0 & $x,y$ & ${\frac{ \partial}{\partial x}}$,
				${\frac{\partial}{\partial y}}$ & & $\tau_x$, $\tau_y$ \\ \hline
				$A_1^{\prime\prime}$ & 1 &  1 &  1 & -1 & -1 & -1 & & &  $\tau_{\phi}$ &  $\tau_{\rho}$\\ \hline
				$A_2^{\prime\prime}$ & 1 &  1 & -1 & -1 & -1 &  1 &$\rho$ & ${\frac{ \partial}{ \partial \rho}}$& &\\ 
				\hline
				$E^{\prime\prime}$ & 2 & -1 & 0 & -2 & 1 & 0 & & & $\tau_x$,$\tau_y$ &\\ \hline
			\end{tabular}
		\end{center}
	\end{scriptsize}
\end{table}

\noindent
Among the two possible combinations as presented above, the second one ($\tau_{\phi} \in A_1^{\prime}$,
$\tau_{\rho} \in A_1^{\prime\prime}$ and $\tau_x,\,\tau_y \in E^{\prime}$) fulfills the quantization rule, which is illustrated below:\\

\noindent
The nodal patterns of NACTs\citep{kettle_book,bunker_book} for all possible IREPs are presented in Figure \ref{fig:nodal}, which show functional dependence
on the normal mode coordinates, $Q_x$ (bending) and $Q_y$ (asymmetric stretching) of Na$_3$ cluster. The inner and outer concentric circles of Figure \ref{fig:nodal}
depict $Q_x-Q_y$ plane for positive and negative values of $Q_s$ (symmetric stretching), respectively. While obtaining the correct combination of IREPs from the two possible ones (see Table \ref{tb:charac}), the smaller circles of Figure \ref{fig:nodal} are considered and thereby, the following quantization rule\citep{baer_jcp120} is applied along a closed path $L_g$ of nuclear coordinate $\textbf{R}$ encapsulating a single CI:
\begin{eqnarray}
	\oint{d\textbf{R}\cdot\boldsymbol{\tau}(\textbf{R};L_g)} &=&\pm\pi  
	\label{eq:contour}
\end{eqnarray}
\noindent

\begin{figure}[!htp]
	\centering
	
	
	\includegraphics[width=0.65\textwidth,angle=0]{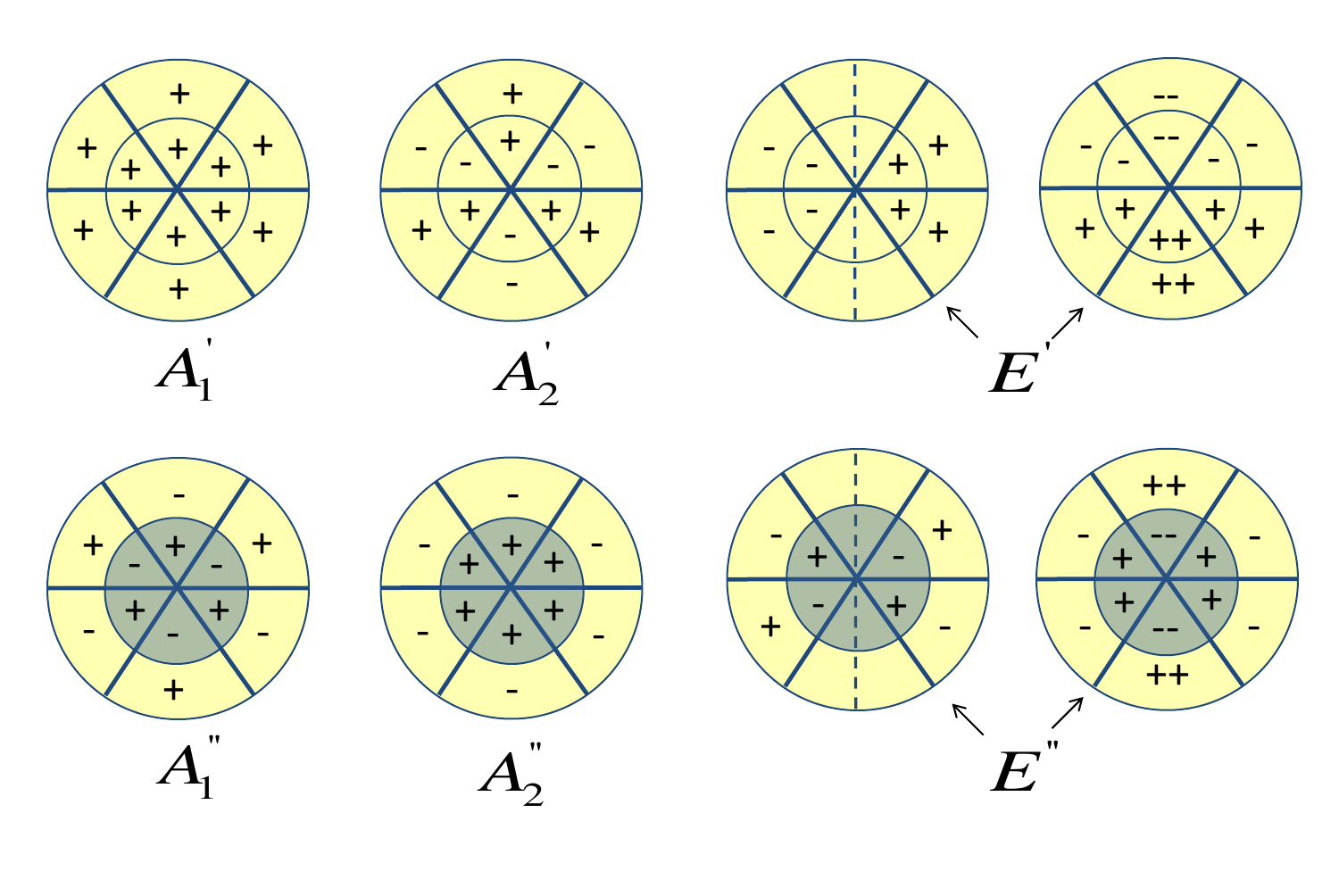}
	
	\vspace{-0.15 in}
	
	\caption{Nodal patterns of IREPs of $D_{3h}$ point group are presented. The figure has been granted from Ref. [\protect\citen{joy_pccp}] and permission has been taken from The Royal Society of Chemistry (RSC).}
	\label{fig:nodal}
\end{figure}

\noindent
On the other hand, Figure \ref{fig:loop} depicts that the closed contour can be constructed by 
three components,\citep{al-jabour_jpca114} namely, one torsional (tor) and two radial (rad) lines and then, the contour integral can be written as:
\begin{small}
	\begin{eqnarray}
		\oint{d\textbf{R}\cdot\boldsymbol{\tau}(\textbf{R};L_g)} &=& I_{\rm{rad},1}+I_{\rm{tor}}+I_{\rm{rad},2} = \pm\pi, \nonumber
	\end{eqnarray}
	\noindent
	where 
	\begin{eqnarray}
		&~& I_{\rm{rad},1}=\int\limits_0^{\rho_1}d\rho \ \tau_{\rho}(\rho,\phi_1;L_g), \;
		I_{\rm{tor}}=\int\limits_{\phi_1}^{\phi_2}{d\phi \ \tau_{\phi}(\rho_1,\phi;L_g)}
		\nonumber\\
		&~& \text{and} \qquad
		I_{\rm{rad},2}=\int\limits_{\rho_1}^0 d\rho \ \tau_{\rho}(\rho,\phi_2;L_g). \nonumber
	\end{eqnarray} 
\end{small}

\noindent
Figures \ref{fig:loop} (a-b) show that for the first combination ($\tau_{\rho} \in A_1^{\prime}$, $\tau_{\phi} \in A_1^{\prime\prime}$, $\tau_x,\,\tau_y \in E^{\prime\prime}$), $I_{rad,1}+I_{rad,2}=0$ and $I_{tor}=0$. On the contrary, for the second combination ($\tau_{\phi} \in A_1^{\prime}$,
$\tau_{\rho} \in A_1^{\prime\prime}$ and $\tau_x,\,\tau_y \in E^{\prime}$), the integral produces non-zero residue [see Figures \ref{fig:loop} (c-d)]. Therefore, it is worth mentioning that only the second set of IREPs provides physically meaningful data. \\

\begin{figure}
	\centering
	\includegraphics[width=0.6\textwidth,angle=0]{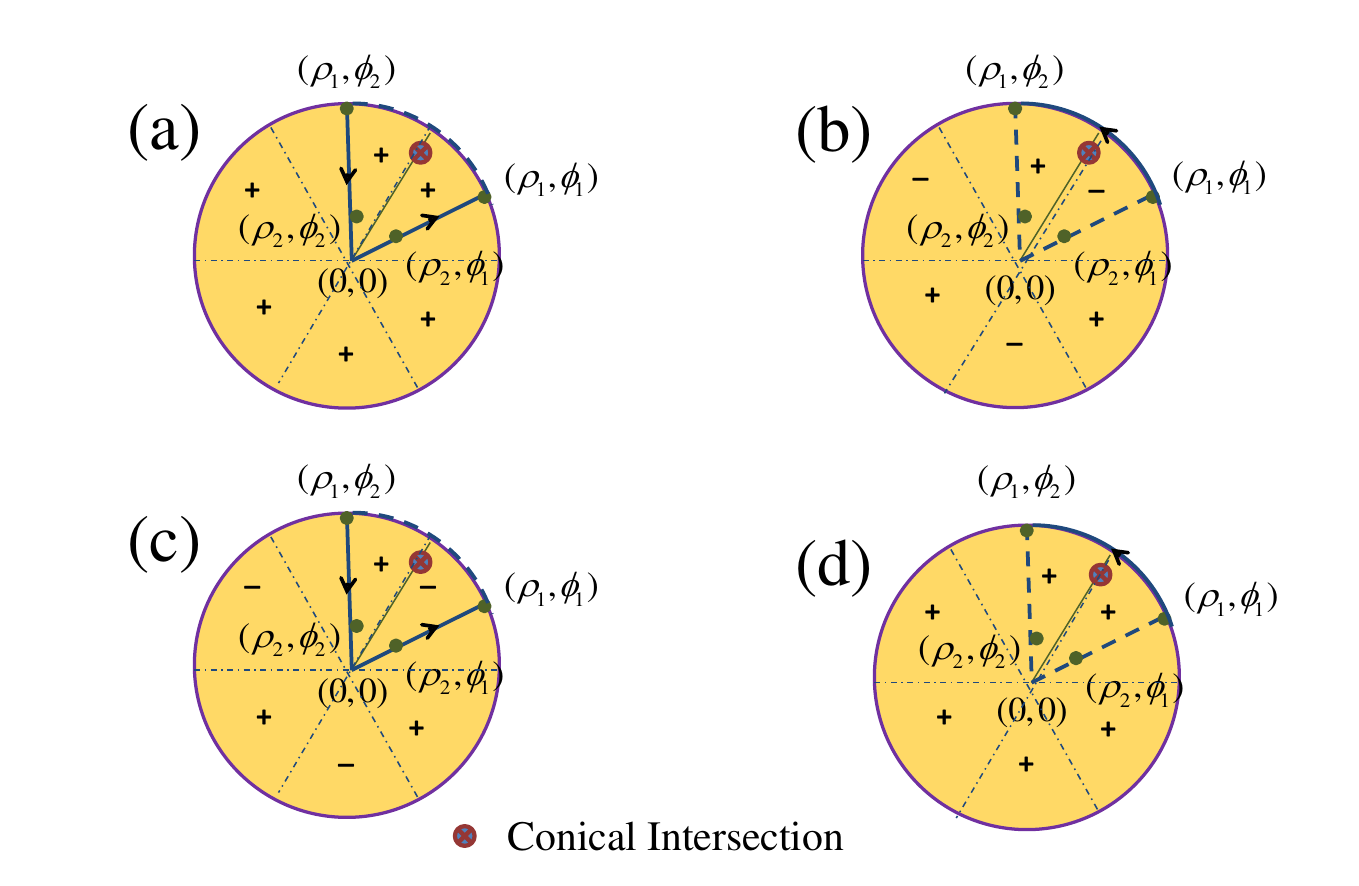}
	
	
	\caption{For 1st combination of IREPs of the NACTs, the radial and torsional parts of a closed loop $L_g$ 
		enclosing a CI point in ($\rho$, $\phi$) plane are presented in upper panel ((a) and (b)). On the other hand,
		the same quantities for the 2nd combination are illustrated in lower panel ((c) and (d)). Figures are taken from Ref. [\protect\citen{mukherjee_jpca117}].}
	\label{fig:loop}
\end{figure}

\noindent
{\bf Theorem 2:} For a specific nuclear DOF ($l$), if we are able to known or assume the IREP of a particular NACT within two electronic states ($p$ and $q$), $\tau_l^{pq}$, it is possible to speculate the IREP of another one ($\tau_l^{qr}$) within two electronic states,
$q$ and $r$.\\

\noindent
If we consider a loop-type sequence of $N$ molecular states with same spin multiplicity, namely, $N$ doublet states
of Na$_3$ cluster, $D_a$, $D_b$, $D_c$,....., $D_y$, $D_{z=a}$, multiplication of the NACTs ($\tau_k^{a,b}$, $\tau_k^{b,c}$,.....,
$\tau_k^{y,z=a}$) can be expressed as:
\begin{eqnarray}
	\tau_k^{a,b,c,{\text...},y,z=a} &=&\tau_k^{a,b} \tau_k^{b,c}{\text...}\tau_k^{y,z=a},  
	\label{eq:theo2a}
\end{eqnarray}
\noindent
where the associated IREPs are related as:
\begin{small}
	\begin{eqnarray}
		\Gamma(\tau_k^{a,b,c,{\text...},y,z=a}) &=& \Gamma(\tau_k^{a,b})\times\Gamma(\tau_k^{b,c})\times{\text...}\times\Gamma(\tau_k^{y,z=a})\nonumber\\
		&=&\Gamma(\xi_a)^2\times\Gamma(\xi_b)^2\nonumber\\
		&\times&{\text...}\times\Gamma(\xi_y)^2\times\Gamma\left(\frac{\partial}{\partial R_k}\right)^N
		\label{eq:theo2b}
	\end{eqnarray}
\end{small}
\noindent
In case of 1D IREPs\citep{al-jabour_jpca114}, the above relation takes the following form:
\begin{eqnarray}
	\Gamma(\tau_k^{a,b,c,{\text...},y,z=a}) &=& \left\{\begin{array}{c c}{\Gamma\left(\frac{ \partial}{\partial R_k}\right)} 
		& \quad \text{when $N$ is odd}\\
		{A_1^{\prime}}  & \quad \text{when $N$ is even}\end{array}\right.
	\label{eq:theo2c}
\end{eqnarray}

\noindent
On the other hand, if a conical intersection (CI) point is present between two electronic states $i$ and $i+1$, the corresponding electronic wavefunctions
change their signs while encapsulating that point. It is worth mentioning that those adiabatic wavefunctions and the NACTs, $\tau_k^{i,j}$ and $\tau_k^{(i+1),j}$ belong to same IREP (crossing-rule), which is given as, 
\begin{eqnarray}
	\Gamma(\tau_k^{i,j})&=&\Gamma(\tau_k^{(i+1),j}). 
	\label{eq:cross}
\end{eqnarray}

\noindent
For Na$_3$ cluster, the three electronic states, namely, $2^{2}E^{\prime}$ and $1^{2}A_{1}^{\prime}$ are non-adiabatically coupled with each other, where multiple CIs can be observed within $2^{2}E^{\prime}$ state. Hence, for such cluster, the second theorem appears as : 
\begin{eqnarray}
	\Gamma(\tau_k^{1,2,3,1}) &=&\Gamma(\tau_k^{12})\times\Gamma(\tau_k^{23})\times\Gamma(\tau_k^{31}) 
	\nonumber\\
	&=&\Gamma(\tau_k^{12})\times\Gamma(\tau_k^{23})^2
	\label{eq:loop}
\end{eqnarray}
\noindent
with the aid of the conditions, $\Gamma(\tau_k^{31})=\Gamma(\tau_k^{13})$
and $\Gamma(\tau_k^{13})=\Gamma(\tau_k^{23})$ as per the crossing rule (Eq. 
\ref{eq:cross}). As per the second combination of Table \ref{tb:charac},
it can be observed that
$\Gamma(\tau_{\rho}^{23})$ = $A_1^{\prime\prime}$ and $\Gamma(\tau_{\phi}^{23})$ = 
$A_1^{\prime}$. Hence, by employing Eqs. \ref{eq:theo2c} and \ref{eq:loop}, the IREPs of 
$\tau_{\rho}^{12}$ and $\tau_{\phi}^{12}$ are given by,
\begin{eqnarray}
	&&\Gamma(\tau_{\rho}^{12}) \times (A_1^{\prime\prime})^2 = \Gamma\left(\dfrac{\partial}{\partial\rho}\right)
	=A_2^{\prime\prime}\nonumber\\
	&&\Rightarrow\,\, \Gamma(\tau_{\rho}^{12}) = A_2^{\prime\prime}
\end{eqnarray} 
\vspace{-0.3 cm}
and
\vspace{-0.3 cm}
\begin{eqnarray}
	&&\Gamma(\tau_{\phi}^{12}) \times (A_1^{\prime})^2 = \Gamma\left(\dfrac{\partial}{\partial\phi}\right)
	=A_2^{\prime}\nonumber\\
	&&\Rightarrow\,\, \Gamma(\tau_{\phi}^{12}) = A_2^{\prime}
\end{eqnarray} 

\noindent
On the other hand, for 2D IREPs, the following relation can be obtained for $\Gamma(\tau_x^{23})$:\citep{mukherjee_jpca117} 
\begin{subequations}
	\begin{eqnarray}
		\Gamma(\tau_x^{23}) &=& \Gamma\left(\Big\langle\xi_2 \Big| \frac{\partial\xi_3}{\partial x}
		\Big\rangle\right) \nonumber \\
		\Rightarrow \Gamma(\tau_x^{23}) &=& \Gamma(\xi_2)\times\Gamma\left(\frac{\partial}
		{ \partial x}\right)\times\Gamma(\xi_3) \nonumber \\ 
		\Rightarrow E^{\prime} &=& \Gamma(\xi_2)\times E^{\prime}\times\Gamma(\xi_3).
		\label{eq:taux1}
	\end{eqnarray}
	\begin{eqnarray}
		\text{Therefore,} \quad \Gamma(\xi_2)\times\Gamma(\xi_3) &=& A_1^{\prime}
		\label{eq:taux2}
	\end{eqnarray}
	
	\noindent
	In a similar way, one can write for $\Gamma(\tau_x^{13})$,
	\begin{eqnarray}
		\Gamma(\xi_1)\times\Gamma(\xi_3) &=& A_1^{\prime}.
		\label{eq:taux3}
	\end{eqnarray}
\end{subequations}
\noindent
The above expressions (Eqs \ref{eq:taux2} and \ref{eq:taux3}) imply that,
\begin{eqnarray}
	\Gamma(\xi_1)\times\Gamma(\xi_2)\times\Gamma(\xi_3) &=& A_1^{\prime}.
	\label{eq:taux}
\end{eqnarray}
\noindent
Finally, with the help of Eqs \ref{eq:theo2b} and \ref{eq:taux}, we arrive at the following relation for the Na$_3$ case: 
\begin{small}
	\begin{eqnarray}
		\Gamma(\tau_x^{1,2,3,1}) &=& 
		\Gamma(\xi_1)^2\times\Gamma(\xi_2)^2\times\Gamma(\xi_3)^2\times
		\Gamma\left(\frac{\displaystyle \partial}{\displaystyle \partial x}\right)^3 \nonumber\\
		&=& A_1^{\prime}\times \Gamma\left(\frac{\displaystyle \partial}{\displaystyle \partial x}\right)^3 
		= \Gamma\left(\frac{\displaystyle \partial}{\displaystyle \partial x}\right)^3 
		=({E^\prime})^3 
		\label{eq:loop2da}
	\end{eqnarray}
\end{small}
\noindent
On the other hand, we can write from Table \ref{tb:charac} and Eq. \ref{eq:loop},
\begin{eqnarray}
	\Gamma(\tau_x^{1,2,3,1}) &=& \Gamma(\tau_x^{12})\times\Gamma(\tau_x^{23})^2\nonumber \\
	&=& \Gamma(\tau_x^{12})\times({E^\prime})^2
	\label{eq:loop2db}
\end{eqnarray}
\noindent
and then, by comparing Eqs. \ref{eq:loop2da} and \ref{eq:loop2db}, we come across $\Gamma(\tau_x^{12})$ =
$E^\prime$ as well as $\Gamma(\tau_y^{12})$ = $E^\prime$. The IREPs for 
all kind of NACTs for Na$_3$ cluster\citep{mukherjee_jpca117,mukherjee_cp440} are presented in Table \ref{tb:taucomb}. 

\begin{scriptsize}
	\begin{table}[!htp]
		\footnotesize
		\caption{\label{tb:taucomb}Extended character table of molecular symmetry group D$_{3h}$(M) is presented
			with different NAC elements.~\citep{mukherjee_jpca117}}
		\begin{center}
			\begin{tabular} {|c|c|c|c|c|c|c|c|c|c|c|}
				\hline
				\multirow{3}{*} {D$_{3h}$(M)} & {$E$} & (123) & (12) & {$E^{\star}$} & $(123)^{\star}$ & $(12)^{\star}$ & coord. & deriv. &$\tau_k^{23}$ & $\tau_k^{12}$\\
				& & (132) & (13) & &  $(132)^{\star}$ &  $(13)^{\star}$ & & & $\tau_k^{13}$ &\\
				& & & (23) & & &  $(23)^{\star}$ & & & &\\ \hline
				$A_1^{\prime}$ & 1 &1 &1 &1 &1 &1 & & &$\tau_{\phi}$ & \\ \hline
				$A_2^{\prime}$ &  1 &  1 & -1 &  1 &  1 & -1& $\phi$ & $\frac{\partial}{ \partial \phi}$ & &$\tau_{\phi}$ \\ \hline
				$E^{\prime}$ & 2 & -1 & 0 & 2 & -1 & 0 & $x,y$ & $\frac{\partial}{ \partial x}$, $\frac{\partial}
				{ \partial y}$ & $\tau_x$,$\tau_y$ & $\tau_x$,$\tau_y$\\ \hline
				$A_1^{\prime\prime}$ & 1 &  1 &  1 & -1 & -1 & -1 & & & $\tau_{\rho}$ & \\ \hline
				$A_2^{\prime\prime}$ & 1 &  1 & -1 & -1 & -1 &  1 &$\rho$ & $\frac{\partial}{ \partial \rho}$ & &$\tau_{\rho}$\\ \hline
				$E^{\prime\prime}$ & 2 & -1 & 0 & -2 & 1 & 0 & & & &\\ \hline
			\end{tabular}
		\end{center}
	\end{table} 
\end{scriptsize}

\noindent
\section{Extended Born-Oppenheimer Theory for Abelian and Non-abelian Cases}\label{ssec:exp}
\noindent 
The single-surface extended Born-Oppenheimer (EBO) equation is formulated
to capture the effects of other electronic states on a specific one through nonadiabatic couplings. Such formulation is useful to explore the nature of nonadiabatic coupling contributing on a specific state at sufficiently low energy process.  \\

\noindent
EBO formalism presents an intermediate representation between adiabatic and diabatic ones, which includes the contribution of 
nonadiabatic effects originated from other electronic states on a specific single-surface equation. While formulating such equation, one needs to
incorporate the effect of off-diagonal elements of $\vec{\boldsymbol{\tau}}$ matrix on the diagonal and then, classical inaccessibility condition for a specific state from others can be invoked considering the molecular process occurring at and around a particular energy. It is worth mentioning that 
such a concept is applicable only when $\vec{\boldsymbol{\tau}}$ can be expressed as the product of a vector function with an antisymmetric scalar matrix. In other words, two different components of NACM commute with each other, curl $\vec{\boldsymbol{\tau}}=\vec{\boldsymbol{0}}$, which is true for two coupled electronic states (Abelian case). On the other hand, for non-Abelian situation (three or higher dimensional sub-Hilbert spaces), curl $\vec{\boldsymbol{\tau}}$ acquires non-zero values, which demands the necessity of 
an approximation to express the NACM in product form. The parametric representation of a vector equation of a conical surface at the close vicinity of CI point(s) is employed and hence, the determinant (Jacobian) for coordinate transformation (let say, between Cartesian and polar) appears as zero (0) at the close vicinity of those point(s). As a matter of fact, curl $\vec{\boldsymbol{\tau}}\simeq\vec{\boldsymbol{0}}$ turns into zero (0) around a degenerate point for non-Abelian cases.\cite{sarkar_jcp124,sarkar_jpca112} A transformation matrix, $\textbf{G}\,\,(\boldsymbol{\psi}=\textbf{G}\boldsymbol{\chi})$ should exist, which diagonalizes the product form of antisymmetric vector matrix ($\vec{\boldsymbol{\tau}}$) with eigenvalues i$\vec{\boldsymbol{\omega}}$ and accordingly, the nuclear SE in Eq. \ref{adiase} can be written as,
\begin{eqnarray}
	\left[-\dfrac{1}{2}(\vec{\boldsymbol{\nabla}}_{n}+\vec{\boldsymbol{\Omega}})^2+\textbf{V}-\textbf{E}\right]
	\boldsymbol{\chi}=0,
	\label{Eq:ImNACT}
\end{eqnarray}
where 
$\vec{\boldsymbol{\Omega}}=i\vec{\boldsymbol{\omega}}=\textbf{G}^{*}\vec{\boldsymbol{\tau}}\textbf{G}$ and $\textbf{V}=\textbf{G}^{*}\textbf{U}\textbf{G}$. It is important to note that the derivation of $\vec{\boldsymbol{\omega}}$ can be achieved only when the calculated angular component of 
$\vec{\boldsymbol{\omega}}$ obeys the relationship of gauge invariance (GI) as follows,
\begin{eqnarray}
	\int_0^{2\pi} \omega_{\phi}d\phi = n\pi.
	\label{eq:gauge}
\end{eqnarray}

\vspace{0.3 cm}
\noindent
If ${\textbf{G}}$ slowly varies with respect to nuclear coordinates, it commutes with $\vec{\boldsymbol{\nabla}}_{n}$.
On the other hand, $\textbf{V}$ appears as a full matrix possessing both diagonal and off-diagonal terms. 
As per BO approximation, other electronic states can be considered as classically forbidden 
due to the width of a specific energy and hence, their contribution in the total wavefunction can be taken as negligibly small i.e.,  
$\vert\psi_{i}\vert\gg\vert\psi_{j}\vert$, where $j\neq i$, but $i,j\in N$.
While invoking this approximation, $\textbf{V}\boldsymbol{\chi}$ in Eq. \ref{Eq:ImNACT} can be written as,
\begin{eqnarray}
	(\textbf{V}\boldsymbol{\chi})_{i}&=&\lbrace(\textbf{G}^{*}\textbf{U}\textbf{G})(\textbf{G}^{*}\Psi)\rbrace_{i}\nonumber\\
	&=&\sum_{j=1}^{N}\text{G}_{ij}^{*}u_{j}\psi_{j}\nonumber\\
	&=&u_{i}\chi_{i}-u_{i}\sum_{j\neq i}^{N}\text{G}_{ij}^{*}\psi_{j}+\sum_{j\neq i}^{N}\text{G}_{ij}^{*}
	u_{j}\psi_{j}\nonumber\\
	&=&u_{i}\chi_{i}+\sum_{j\neq i}^{N}\text{G}_{ij}^{*}(u_{j}-u_{i})\psi_{j}.
	\label{eq:ebopot}
\end{eqnarray} 

\noindent
In Eq. \ref{eq:ebopot}, the first term symbolizes product of the $i^{\text{th}}$ state adiabatic PES with the
wavefunction, ${\chi_{i}}$, while the rest of the summation provides a negligible contribution appearing from other states
at a specific energy. For a two-dimensional sub-Hilbert space, this summation is exactly zero (0)
[$u_{1}=u_{2}$] at a degenerate point and the EBO equations are completely decoupled. Even for three state
degeneracy at a single point [$u_{1}=u_{2}=u_{3}$], the above summation is identically zero. In any case, one can approximate the 
second term of Eq. \ref{eq:ebopot} considering the molecular process at a specific energy and the EBO equation can be written as: \\
\begin{equation}
	-\frac{1}{2}\left(\vec{\nabla}_n + i\vec{\omega}_{i}\right)^2\chi_i
	+ \left(u_i - E\right)\chi_i = 0, \quad i =1,2,....,N
	\label{Eq:EBO}
\end{equation}  

\noindent
EBO treatment decouples SEs of `$N$' number of nuclear wavefunctions, $\chi$s to formulate `$N$' 
independent nuclear SEs. 

\vspace{0.3 cm}

\noindent
For two electronic state sub-Hilbert space, components of $\vec{\boldsymbol{\tau}}$ commute
with each other (Abelian case) and hence, $\vec{\boldsymbol{\tau}}$
can be written as a product of a vector function and an antisymmetric scalar matrix. Therefore, one can
formulate the EBO equations in terms of the eigenvalues of NACM ($\pm i\vec{\tau}_{12}$) by imposing the 
classical inaccessibility of the other electronic state beyond the width of a specific energy.

\vspace{0.3 cm}

\noindent
For non-Abelian cases, the determinant for the transformation between coordinates (as for example, Cartesian and polar), namely, the Jacobian turns into zero (0) at the CI point(s) considering parametric representation of a vector equation for a conical surface. Thereby, one may approximate curl
$\vec{\boldsymbol{\tau}}\simeq\vec{\boldsymbol{0}}$ at and around those point(s) for sub-Hilbert
spaces, N$\geq$3\cite{sarkar_jcp124,sarkar_jpca112} and it is possible to derive the corresponding EBO equations. \\

\noindent
For multiple electronic states ($N\geq 3$),~\cite{sarkar_jcp124,sarkar_jpca112} the following conditions need to be checked: (a) when CIs are at the same point of nuclear CS and
intensely coupled through singular functions of NACTs, the GI condition should be perfectly
fulfilled; (b) if CIs are at different nuclear CS, substantially apart from each other and
almost decoupled due to specific NACM element(s) with small magnitudes, the corresponding eigenvalues of NACM
show gauge invariance; (c) when the CIs
are not at the same position, but close to each other and substantially coupled through singular functions
of NACTs, the required GI condition is difficult to attain for the formulation of EBO equations.\\

\noindent
While exploring the validity of EBO equation, it is necessary to study the fulfillment of three conditions, namely, curl $\vec{\boldsymbol{\tau}}\simeq\vec{\boldsymbol{0}}$ at and around CI(s),
the GI of the eigenvalues of NACM ($\vec{\Omega}_{i}=i\vec{\omega}_{i}$) and the classical inaccessibility of a state from other states for a process occurring at specific energy, among which the first two appear as complementary with each other. As representative examples, we demonstrated the applicability of such EBO formalism (for non-Abelian systems) for two realistic molecular Hamiltonian namely, NO$_{3}$ radical and 1,3,5-trifluorobenzene radical cation (TFBz$^{+}$) by computing 
curl $\vec{\boldsymbol{\tau}}$ as functions of pairwise normal modes
and the GI for the eigenvalues ($\vec{\Omega}_{i}=i\vec{\omega}_{i}$) of the NACM over the chosen domain of nuclear plane.

\subsection{Two Electronic State Sub-Hilbert Space: Abelian Case} 
\noindent
For a two coupled electronic manifold, the 2 $\times$ 2 antisymmetric square matrix, $\vec{\boldsymbol{\tau}}$ takes the following form,
\begin{eqnarray}
	\vec{\boldsymbol{\tau}} = 
	\begin{pmatrix}
		0 & \vec{\tau}_{12} \\
		-\vec{\tau}_{12} & 0
	\end{pmatrix}=
	\vec{\tau}_{12}
	\begin{pmatrix}
		0 & 1 \\
		-1 & 0
	\end{pmatrix}.
	\label{Eq:taum}
\end{eqnarray}

\noindent
with curl $\vec{\boldsymbol{\tau}}=\vec{\boldsymbol{0}}$ for any arbitrary pair of nuclear geometry (Abelian case). In other words, 
$\vec{\boldsymbol{\tau}}$ can be expressed as a product of a vector function and an antisymmetric scalar matrix. As a consequence, the EBO equations
(Eq. \ref{Eq:EBO}) can be formulated by computing the eigenvalues of NACM ($\pm i\vec{\omega}_{i}$) followed by imposition of the classical inaccessibility of the other electronic states beyond the width of a specific energy.

\subsection{Three Electronic State Sub-Hilbert Space: Non-Abelian Case\cite{sarkar_jcp124,sarkar_jpca112}} 
\noindent
For three state sub-Hilbert space, the expressions of elements of NACM 
($\vec{\tau}_{ij}=\langle\xi_i\vert\vec{\nabla}_n\vert\xi_j\rangle$) takes the following forms in terms of ADT angles (see Eq. \ref{eq:nact-three}),
\begin{subequations}
	\begin{eqnarray}
		\vec{\tau}_{12}&=&-\cos\Theta_{13}\cos\Theta_{23}\vec{\nabla}_n\Theta_{12}-\sin\Theta_{23}\vec{\nabla}_n\Theta_{13}\\
		\vec{\tau}_{13}&=&\cos\Theta_{13}\sin\Theta_{23}\vec{\nabla}_n\Theta_{12}-\cos\Theta_{23}\vec{\nabla}_n\Theta_{13}\\
		\vec{\tau}_{23}&=&-\sin\Theta_{13}\vec{\nabla}_n\Theta_{12}-\vec{\nabla}_n\Theta_{23}
	\end{eqnarray}
	\label{Eq:NACMebo}
\end{subequations}

\noindent
where the explicit forms of curl $\vec{\boldsymbol{\tau}}$ appear as (see Eq. \ref{eq:curl-three}):
\begin{subequations}
	\begin{eqnarray}
		\text{curl} \, \tau_{12}^{pq} = (\vec{\nabla}_n \times \vec{\tau}_{12})_{\perp}
		&=& \sin\Theta_{13}\cos\Theta_{23}(\nabla_p\Theta_{12}\nabla_q\Theta_{13} - \nabla_q\Theta_{12}
		\nabla_p\Theta_{13}) \nonumber \\
		&+& \sin\Theta_{23}\cos\Theta_{13}(\nabla_p\Theta_{12}\nabla_q\Theta_{23} - \nabla_q\Theta_{12}
		\nabla_p\Theta_{23}) \nonumber \\
		&-& \cos\Theta_{23}(\nabla_p\Theta_{13}\nabla_q\Theta_{23} - \nabla_q\Theta_{13}\nabla_p\Theta_{23})
	\end{eqnarray}
	\begin{eqnarray}
		\text{curl} \, \tau_{13}^{pq} = (\vec{\nabla}_n \times \vec{\tau}_{13})_{\perp} 
		&=& -\sin\Theta_{13}\sin\Theta_{23}(\nabla_p\Theta_{12}\nabla_q\Theta_{13} - \nabla_q\Theta_{12}
		\nabla_p\Theta_{13}) \nonumber \\
		&+& \cos\Theta_{23}\cos\Theta_{13}(\nabla_p\Theta_{12}\nabla_q\Theta_{23} - \nabla_q\Theta_{12}
		\nabla_p\Theta_{23}) \nonumber \\
		&+& \sin\Theta_{23}(\nabla_p\Theta_{13}\nabla_q\Theta_{23} - \nabla_q\Theta_{13}\nabla_p\Theta_{23})
	\end{eqnarray}
	\begin{eqnarray}
		\text{curl} \, \tau_{23}^{pq} = (\vec{\nabla}_n \times \vec{\tau}_{23})_{\perp}
		&=& -\cos\Theta_{13}(\nabla_p\Theta_{12}\nabla_q\Theta_{13} - \nabla_q\Theta_{12}
		\nabla_p\Theta_{13}), \qquad 
	\end{eqnarray}
	\label{Eq:curltau}
\end{subequations}
\noindent
considering $p$ and $q$ as two Cartesian coordinates.\\

\noindent
Using the parametric representation of a vector equation of a surface containing CI point(s), the transformation determinant
between coordinates (as for example, Cartesian and polar), namely, the Jacobian attains the zero (0) value at those point(s)
and hence, the equations for curl $\vec{\boldsymbol{\tau}}$ become null vector ($\vec{\boldsymbol{0}}$) at the close vicinity of those point(s) for non-Abelian sub-Hilbert
spaces, N$\geq$3.\cite{sarkar_jcp124,sarkar_jpca112} 
If curl $\tau_{ij}^{pq}$ can be approximately equated to zero at the neighborhood of CI point(s), one can write the following relations 
of the ADT angles using Eq. \ref{Eq:curltau}:

\begin{small}
	\begin{eqnarray}
		\left(\frac{\nabla_p\Theta_{13}}{\nabla_p\Theta_{12}}\right) = \left(\frac{\nabla_q\Theta_{13}}{\nabla_q\Theta_{12}}\right) 
		\quad
		\text{and} \quad \left(\frac{\nabla_p\Theta_{23}}{\nabla_p\Theta_{12}}\right) = \left(\frac{\nabla_q\Theta_{23}}
		{\nabla_q\Theta_{12}}\right),
		\label{Eq:eboassum}
	\end{eqnarray}
\end{small}
\noindent
which lead to the following product form of NACM, i.e., Eq. \ref{Eq:NACMebo} can be expressed as,
\begin{scriptsize}
	\begin{equation}
		\vec{\tau} = \vec{\nabla}_n\Theta_{12} 
		\begin{pmatrix}
			0 & -\sin\Theta_{23}\left(\frac{\nabla_p\Theta_{13}}{\nabla_p\Theta_{12}}\right) 
			& -\cos\Theta_{23}\left(\frac{\nabla_p\Theta_{13}}{\nabla_p\Theta_{12}}\right) \\ 
			& -\cos\Theta_{13}\cos\Theta_{23} & + \cos\Theta_{13}\sin\Theta_{23} \\ \\ \\ 
			\sin\Theta_{23}\left(\frac{\nabla_p\Theta_{13}}{\nabla_p\Theta_{12}}\right) 
			& 0 & -\sin\Theta_{13}-\left(\frac{\nabla_p\Theta_{23}}{\nabla_p\Theta_{12}}\right) \\ 
			+ \cos\Theta_{13}\cos\Theta_{23} & & \\ \\ \\ 
			\cos\Theta_{23}\left(\frac{\nabla_p\Theta_{13}}{\nabla_p\Theta_{12}}\right) 
			& \sin\Theta_{13}+\left(\frac{\nabla_p\Theta_{23}}{\nabla_p\Theta_{12}}\right) & 0 \\ 
			- \cos\Theta_{13}\sin\Theta_{23} & & \\ 
		\end{pmatrix}
		\label{Eq:NACMebo1}
	\end{equation}
\end{scriptsize}

\noindent
It is important to note that the eigenvalues are $-i\vec{\omega}$, $\vec{0}$ and $i\vec{\omega}$, where 
\begin{small}
	\begin{equation}
		\vec{\omega} = \vec{\nabla}_n\Theta_{12}\left[1+\left(\frac{\nabla_p\Theta_{13}}{\nabla_p\Theta_{12}}
		\right)^2 
		+\left(\frac{\nabla_p\Theta_{23}}{\nabla_p\Theta_{12}}\right)^2
		+2\sin\Theta_{13}\left(\frac{\nabla_p\Theta_{23}}{\nabla_p\Theta_{12}}\right)\right]^{1/2}.
	\end{equation} 
\end{small}

\subsection{Four\cite{sarkar_jpca112} Electronic State Sub-Hilbert Space}

\noindent
In case of four coupled electronic manifold, one representative element of NACM 
($\vec{\tau}_{ij}=\langle\xi_i\vert\vec{\nabla}_n\vert\xi_j\rangle$) can be written as functions of ADT angles as follows,
\begin{eqnarray}
	\vec{\tau}_{12}=-\cos\Theta_{13}\cos\Theta_{23}\cos\Theta_{14}\cos\Theta_{24}\vec{\nabla}\Theta_{12}-\sin\Theta_{23}
	\cos\Theta_{14}\cos\Theta_{24}\vec{\nabla}\Theta_{13}-\sin\Theta_{24}\vec{\nabla}\Theta_{14}, \qquad
	\label{Eq:fst_tau12}
\end{eqnarray}

\noindent
and the associated ``curl'' equation takes the following forms,
\begin{scriptsize}
	\begin{eqnarray}
		\text{curl}\,\,\tau_{12}^{pq}&=&\sin\Theta_{13}\cos\Theta_{23}\cos\Theta_{14}\cos\Theta_{24}
		(\nabla_{p}\Theta_{12}\nabla_{q}\Theta_{13}-\nabla_{q}\Theta_{12}\nabla_{p}\Theta_{13})+\cos\Theta_{13}\sin\Theta_{23}
		\cos\Theta_{14}\cos\Theta_{24}\nonumber\\
		&&(\nabla_{p}\Theta_{12}\nabla_{q}\Theta_{23}-\nabla_{q}\Theta_{12}\nabla_{p}\Theta_{23})+\cos\Theta_{13}\cos\Theta_{23}
		\sin\Theta_{14}\cos\Theta_{24}(\nabla_{p}\Theta_{12}\nabla_{q}\Theta_{14}-\nabla_{q}\Theta_{12}\nabla_{p}\Theta_{14})
		\nonumber\\
		&&+\cos\Theta_{13}\cos\Theta_{23}\cos\Theta_{14}\sin\Theta_{24}(\nabla_{p}\Theta_{12}\nabla_{q}\Theta_{24}
		-\nabla_{q}\Theta_{12}\nabla_{p}\Theta_{24})-\cos\Theta_{23}\cos\Theta_{14}\cos\Theta_{24}
		(\nabla_{p}\Theta_{13}\nabla_{q}\Theta_{23}\nonumber\\
		&&-\nabla_{q}\Theta_{13}\nabla_{p}\Theta_{23})+\sin\Theta_{23}\sin\Theta_{14}\cos\Theta_{24}(\nabla_{p}\Theta_{13}
		\nabla_{q}\Theta_{14}-\nabla_{q}\Theta_{13}\nabla_{p}\Theta_{14})+\sin\Theta_{23}\cos\Theta_{14}\sin\Theta_{24}\nonumber\\
		&&(\nabla_{p}\Theta_{13}\nabla_{q}\Theta_{24}-\nabla_{q}\Theta_{13}\nabla_{p}\Theta_{24})-\cos\Theta_{24}
		(\nabla_{p}\Theta_{14}\nabla_{q}\Theta_{24}-\nabla_{q}\Theta_{14}\nabla_{p}\Theta_{24}).
		\label{Eq:fst_curl12}
	\end{eqnarray}
\end{scriptsize}

\noindent
On the other hand, based on the parametric representation of vector function for a conical surface and zero (0) value of the Jacobian, the following identities are valid at the CI point(s) as per the Eq. \ref{Eq:fst_curl12}:  

\begin{eqnarray}
	\left(\frac{\nabla_p\Theta_{13}}{\nabla_p\Theta_{12}}\right) &=& \left(\frac{\nabla_q\Theta_{13}}{\nabla_q\Theta_{12}}\right) 
	\text{,} \left(\frac{\nabla_p\Theta_{14}}{\nabla_p\Theta_{12}}\right) = \left(\frac{\nabla_q\Theta_{14}}{\nabla_q\Theta_{12}}\right) 
	\text{,} \left(\frac{\nabla_p\Theta_{23}}{\nabla_p\Theta_{12}}\right) = \left(\frac{\nabla_q\Theta_{23}}{\nabla_q\Theta_{12}}\right) 
	\text{,} \nonumber\\
	\left(\frac{\nabla_p\Theta_{24}}{\nabla_p\Theta_{12}}\right) &=& \left(\frac{\nabla_q\Theta_{24}}
	{\nabla_q\Theta_{12}}\right) \text{,} \left(\frac{\nabla_p\Theta_{34}}{\nabla_p\Theta_{12}}\right) = \left(\frac{\nabla_q\Theta_{34}}{\nabla_q\Theta_{12}}\right)
	\label{Eq:fst_equal}
\end{eqnarray} 

\noindent
It is interesting to note that the vector matrix, $\vec{\boldsymbol{\tau}}$ can be partitioned into a product form with a vector
function and a scalar matrix, which provides eigenvalues for a four dimensional SHS to
formulate the EBO equation:
\begin{equation}
	\omega_{i}=\pm\sqrt{\dfrac{K}{2}\pm\dfrac{1}{2}\sqrt{K^{2}-4L^{2}}},
\end{equation}

\noindent
\begin{footnotesize}
	\begin{eqnarray}
		K&=&\vec{\tau}_{12}.\vec{\tau}_{12}+\vec{\tau}_{13}.\vec{\tau}_{13}+\vec{\tau}_{23}.\vec{\tau}_{23}+\vec{\tau}_{14}.
		\vec{\tau}_{14}+\vec{\tau}_{24}.\vec{\tau}_{24}+\vec{\tau}_{34}.\vec{\tau}_{34}\nonumber\\
		&=&\vec{\nabla}\Theta_{12}.\vec{\nabla}\Theta_{12}\Bigg\lbrace 1+\left(\frac{\nabla_p\Theta_{13}}{\nabla_p\Theta_{12}}\right)
		^{2}+\left(\frac{\nabla_p\Theta_{23}}{\nabla_p\Theta_{12}}\right)^{2} +\left(\frac{\nabla_p\Theta_{14}}{\nabla_p\Theta_{12}}\right)^{2}+\left(\frac{\nabla_p\Theta_{24}}{\nabla_p\Theta_{12}}\right)^{2}+
		\left(\frac{\nabla_p\Theta_{34}}{\nabla_p\Theta_{12}}\right)^{2}\nonumber\\
		&&+2\cos\Theta_{23}\sin\Theta_{14}\cos\Theta_{24}\left(\frac{\nabla_p\Theta_{13}}{\nabla_p\Theta_{12}}\right)
		\left(\frac{\nabla_p\Theta_{34}}{\nabla_p\Theta_{12}}\right)+2\sin\Theta_{23}\sin\Theta_{14}
		\left(\frac{\nabla_p\Theta_{13}}{\nabla_p\Theta_{12}}\right)\left(\frac{\nabla_p\Theta_{24}}
		{\nabla_p\Theta_{12}}\right)\nonumber\\
		&&+2\sin\Theta_{24}\left(\frac{\nabla_p\Theta_{23}}{\nabla_p\Theta_{12}}\right)\left(\frac{\nabla_p\Theta_{34}}{\nabla_p\Theta_{12}}\right)+2\sin\Theta_{13}\left(\frac{\nabla_p\Theta_{23}}{\nabla_p\Theta_{12}}\right)+
		2\cos\Theta_{13}\cos\Theta_{23}\sin\Theta_{14}\left(\frac{\nabla_p\Theta_{24}}{\nabla_p\Theta_{12}}\right)\nonumber\\
		&&+2\left(\sin\Theta_{13}\sin\Theta_{24}-\cos\Theta_{13}\sin\Theta_{23}\sin\Theta_{14}\cos\Theta_{24}\right)
		\left(\frac{\nabla_p\Theta_{34}}{\nabla_p\Theta_{12}}\right)\Bigg\rbrace
	\end{eqnarray}
\end{footnotesize}

\noindent
and
\begin{footnotesize}
	\begin{eqnarray}
		L&=&\vec{\tau}_{14}.\vec{\tau}_{23}-\vec{\tau}_{13}.\vec{\tau}_{24}+\vec{\tau}_{12}.\vec{\tau}_{34}\nonumber\\
		&=&\vec{\nabla}\Theta_{12}.\vec{\nabla}\Theta_{12}\Bigg\lbrace\left(\frac{\nabla_p\Theta_{23}}
		{\nabla_p\Theta_{12}}\right)\left(\frac{\nabla_p\Theta_{24}}{\nabla_p\Theta_{12}}\right)+\sin\Theta_{13}
		\left(\frac{\nabla_p\Theta_{14}}{\nabla_p\Theta_{12}}\right)+\cos\Theta_{13}\sin\Theta_{23}\cos\Theta_{14}
		\left(\frac{\nabla_p\Theta_{24}}{\nabla_p\Theta_{12}}\right)\nonumber\\
		&&-\cos\Theta_{23}\cos\Theta_{14}
		\left(\frac{\nabla_p\Theta_{13}}{\nabla_p\Theta_{12}}\right)\left(\frac{\nabla_p\Theta_{24}}
		{\nabla_p\Theta_{12}}\right)+\cos\Theta_{13}\cos\Theta_{23}\cos\Theta_{14}\cos\Theta_{24}\left(\frac{\nabla_p\Theta_{34}}{\nabla_p\Theta_{12}}\right)\nonumber\\
		&&+\sin\Theta_{23}\cos\Theta_{14}\cos\Theta_{24}\left(\frac{\nabla_p\Theta_{13}}{\nabla_p\Theta_{12}}\right)
		\left(\frac{\nabla_p\Theta_{34}}{\nabla_p\Theta_{12}}\right)+\sin\Theta_{24}\left(\frac{\nabla_p\Theta_{14}}{\nabla_p\Theta_{12}}\right)\left(\frac{\nabla_p\Theta_{34}}{\nabla_p\Theta_{12}}\right)\Bigg\rbrace
	\end{eqnarray}
\end{footnotesize}

\noindent
\section{EBO Formalism for Model Calculation of Na$_3$ Cluster}\label{sec:spec}
\vspace{0.2 cm}

\noindent 
The workability of EBO formalism for a tri-state system has been explored using the model diabatic Hamiltonian for Na$_3$ cluster
proposed by Cocchini {\it et al.}\citep{cocchini_jcp88} Such system exhibits profound nonadiabatic couplings within 
$2\,^2E^{\prime}$ and $1\,^2A_1^{\prime}$ states over the $\rho-\phi$ nuclear plane constituted with bending ($Q_x$) and asymmetric 
stretching ($Q_y$) modes. The model diabatic Hamiltonian is represented as:\citep{cocchini_jcp88} 

\begin{footnotesize}
	\begin{eqnarray}
		V^{\rm{dia}}(\rho,\phi)=
		\begin{pmatrix}
			\dfrac{\rho^2}{2} + U_2 & U_1 & W_1-W_2 \\
			U_1 & \dfrac{\rho^2}{2} - U_2 & W_1+W_2 \\
			W_1-W_2 & W_1+W_2 & \epsilon_0 + \dfrac{\rho^2}{2}
		\end{pmatrix}, 
		\label{eq:diaham_mod}
	\end{eqnarray}
\end{footnotesize}

\noindent
where

\begin{footnotesize}
	\begin{minipage}{0.45\textwidth}
		\begin{flushleft}
			\begin{eqnarray}
				U_1(\rho,\phi) &=& K\rho \cos\phi + \dfrac{1}{2}g \rho^2\cos(2\phi) \nonumber \\
				U_2(\rho,\phi) &=& K\rho \sin\phi - \dfrac{1}{2}g \rho^2\sin(2\phi) \nonumber
			\end{eqnarray}
		\end{flushleft}
	\end{minipage}
	\hspace{0.2 in}
	\begin{minipage}{0.45\textwidth}
		\begin{flushright}
			\begin{eqnarray}
				W_1(\rho,\phi) &=& P\rho \cos\phi + \dfrac{1}{2}f \rho^2\cos(2\phi) \nonumber \\
				W_2(\rho,\phi) &=& P\rho \sin\phi - \dfrac{1}{2}f \rho^2\sin(2\phi) \nonumber
			\end{eqnarray}
		\end{flushright}
	\end{minipage}
\end{footnotesize}

\vspace{0.5 cm}

\noindent
It is important to note that JT coupling constants [$K\,\,(=4.9)$ and 
$g\,\,(=0.035)$]\citep{cocchini_jcp88} and PJT coupling parameters
[$P\,\,(=3.46)$ and $f\,\,(=0.025)$]\citep{cocchini_jcp88} of the above expressions appear as dimensionless, since the nuclear coordinates are dimensionless. On the other hand, the energy separation ($\Delta$) between second and third states modulates the adjustable parameter, $\epsilon_0$ ($=2\Delta$). \\

\noindent
In presence of only JT interaction ($P=f=0$, no PJT couplings), the 3 $\times$ 3 model Hamiltonian reduces to a 2 $\times$ 2 block matrix as follows:
\begin{eqnarray}
	V^{\rm{dia}}(\rho,\phi)=
	\begin{pmatrix}
		\dfrac{\rho^2}{2} + U_2 & U_1 & 0 \\
		U_1 & \dfrac{\rho^2}{2} - U_2 & 0 \\
		0 & 0 & \epsilon_0 + \dfrac{\rho^2}{2}
	\end{pmatrix}, 
	\label{eq:diaham_mod_two}
\end{eqnarray}
and accordingly, the adiabatic PESs take the following form:
\begin{eqnarray}
	u_1(\rho,\phi) &=& \frac{\rho^2}{2}-\sqrt{K^2\rho^2+\frac{g^2\rho^4}{4}+Kg\rho^3\cos(3\phi)}, \\ 
	u_2(\rho,\phi) &=& \frac{\rho^2}{2}+\sqrt{K^2\rho^2+\frac{g^2\rho^4}{4}+Kg\rho^3\cos(3\phi)},
\end{eqnarray}
The 2 $\times$ 2 diabatic Hamiltonian leads to the following explicit expressions of $\rho$ and $\phi$ components of NACTs:

\begin{subequations}
	\begin{eqnarray}
		\tau^{12}_{\rho} &=& -\frac{Kg\rho^2}{4\nu^2}\sin(3\phi) \\
		\tau^{12}_{\phi} &=& -\frac{-2K^2\rho^2+g^2\rho^4+Kg\rho^3\cos(3\phi)}{4\nu^2}
	\end{eqnarray}
\end{subequations}

\noindent
With the above form of the NAC matrices, the curl appears identically zero, i.e. curl $\vec{\boldsymbol{\tau}}$ = 0  as $[(\boldsymbol{\tau^{\rho}}\boldsymbol{\tau^{\phi}})_{12}-(\boldsymbol{\tau^{\phi}}\boldsymbol{\tau^{\rho}})_{12}]=\boldsymbol{0}$, where
the corresponding ADT matrix takes the following form:
\begin{equation}
	\boldsymbol{A} = 
	\begin{pmatrix}
		\sqrt{\frac{\nu+U_2}{2\nu}} & -\sqrt{\frac{\nu-U_2}{2\nu}} \\  
		\sqrt{\frac{\nu-U_2}{2\nu}} & \sqrt{\frac{\nu+U_2}{2\nu}}
	\end{pmatrix}, 
\end{equation}
\noindent
with $\nu = (K^2\rho^2+\frac{1}{4}g^2\rho^4+Kg\rho^3\cos(3\phi))^{1/2}$ and $U_2 = K\rho \sin\phi - \dfrac{1}{2}g \rho^2\sin(2\phi)$.\\

\noindent
On the other hand, for PJT situation ($K=g=0$, no JT couplings), the diabatic Hamiltonian (Eq. \ref{eq:diaham_mod}) leads to the following 
analytic expressions of $\rho$ and $\phi$ components of NACTs:

\begin{footnotesize}
	\begin{subequations}
		\begin{eqnarray}
			\tau^{12}_{\rho} = -\frac{A_0}{\sqrt{2wB_-}}, \quad
			\tau^{13}_{\rho} = -\frac{A_0}{\sqrt{2wB_+}}, \quad
			\tau^{23}_{\rho} = \frac{A_1\Delta}{2\sqrt{2w}(\Delta^2+2w)},  \\
			\tau^{12}_{\phi} = \frac{A_2}{\sqrt{2wB_-}}, \quad
			\tau^{13}_{\phi} = \frac{A_2}{\sqrt{2wB_+}}, \quad
			\tau^{23}_{\phi} = -\frac{3\rho A_0\Delta}{2\sqrt{2w}(\Delta^2+2w)},  
		\end{eqnarray}
	\end{subequations}
\end{footnotesize}
\noindent
where
\begin{footnotesize}
	\begin{eqnarray}
		&\,& A_0(\rho,\phi)=Pf\rho^2\sin(3\phi) \nonumber \\
		&\,& A_1(\rho,\phi)=2P^2\rho+f^2\rho^3+3Pf\rho^2\cos(3\phi)  \nonumber \\
		&\,& A_2(\rho,\phi)=2P^2\rho^2-f^2\rho^4-Pf\rho^3\cos(3\phi) \nonumber \\
		&\,& B_{\pm}(\rho,\phi)=2w+(\Delta \pm \sqrt{\Delta^2+2w})^2 \nonumber \\
		&\,& w(\rho,\phi)= W_1^2 + W_2^2 = P^2\rho^2 + \frac{f^2\rho^4}{4} + Pf\rho^3\cos(3\phi) \nonumber 
	\end{eqnarray}
\end{footnotesize}

\noindent
On the other hand, analytical forms of curls can be represented as:

\begin{footnotesize}
	\begin{eqnarray}
		\text{curl} \, \tau^{12}_{\rho\phi} = -\frac{(3\rho A_0^2 - A_1A_2)\Delta}{4w(\Delta^2+2w)
			\sqrt{B_+}}; \quad
		\text{curl} \, \tau^{13}_{\rho\phi} = \frac{(3\rho A_0^2 - A_1A_2)\Delta}{4w(\Delta^2+2w)\sqrt{B_-}};
		\quad \text{curl} \, \tau^{23}_{\rho\phi} = 0. 
		\label{eq:curltau}
	\end{eqnarray}
\end{footnotesize}

\noindent
The above functional forms show that curl $\tau^{ij}_{\rho\phi}$s become identically zero (0) at $\Delta=0$, i.e., at the point of three-state degeneracy between $A$ and $E$ states. Furthermore, numerically calculated curls using the gradient of ADT angles take negligibly small magnitude (curl $\tau^{ij}_{\rho\phi} \approx 0$) up to certain nonzero $\Delta$ ($\leq$ 172 cm$^{-1}$). In other words, a common diagonalizing matrix, $G$ is expected to exist at $\Delta=0$
for both the components of NACT matrices ($\tau_{\rho}$ and $\tau_{\phi}$), and therefore, the adiabatic SE can be transformed 
to the following form:

\begin{footnotesize}
	
	\vspace{-0.1 in}
	
	\begin{eqnarray}
		&\,& -\frac{\hbar^2}{2}\left(\boldsymbol{\nabla}_R + \boldsymbol{\Omega}\right)^2\Phi
		+ \left(V - E\right)\Phi = 0, \label{eq:modelebo}\\
		&\,& \text{where} \quad \psi^{\rm{ad}}=G\Phi, \quad \boldsymbol{\Omega}=G^{\dagger}\boldsymbol{\tau} G
		\quad \text{and} \quad V=G^{\dagger}UG. \nonumber
	\end{eqnarray}
	
	\vspace{-0.05 in}
	
\end{footnotesize}

\noindent
In case three state degenerate point ($\Delta=0$), one can write the diagonalizing matrix ($G$) as follows:
\begin{eqnarray}
	G = 
	\begin{pmatrix}
		0 & \frac{-i}{\sqrt{2}} & \frac{i}{\sqrt{2}} \\
		\frac{-1}{\sqrt{2}} & \frac{1}{2} & \frac{1}{2} \\
		\frac{1}{\sqrt{2}} & \frac{1}{2} & \frac{1}{2}
	\end{pmatrix} \nonumber
\end{eqnarray}

\noindent
For low value of $\Delta$ ($\leq$172 cm$^{-1}$), the magnitude of curl $\boldsymbol{\tau}$ 
also appears as negligible (curl $\boldsymbol{\tau} \approx 0$), which affirms the validity of the above formulation. While imposing classical inaccessibility condition for each electronic state from others at the close neighborhood of a 
specific energy ($\vert\psi_i\vert \gg \vert\psi_j\vert$ ($i\neq j$)), the single-surface 
EBO equations turn into,

\begin{scriptsize}
	\begin{subequations}
		\begin{eqnarray}
			&-&\frac{\hbar^2}{2m}\left\lbrace\left(\frac{\partial}{\partial\rho}+i\omega_{\rho}\right)^2 
			+\frac{1}{\rho}\left(\frac{\partial}{\partial\rho}+i\omega_{\rho}\right) 
			+ \frac{1}{\rho^2}\left(\frac{\partial}{\partial\phi}+i\omega_{\phi}\right)^2\right\rbrace\Phi_1 
			+ (u_1-E)\Phi_1 = 0, \\ 
			&-&\frac{\hbar^2}{2m}\left\lbrace\frac{\partial^2}{\partial\rho^2}
			+ \frac{1}{\rho}\frac{\partial}{\partial\rho} 
			+ \frac{1}{\rho^2}\frac{\partial^2}{\partial\phi^2}\right\rbrace\Phi_2 + (u_2-E)\Phi_2 = 0, \\
			&-&\frac{\hbar^2}{2m}\left\lbrace\left(\frac{\partial}{\partial\rho}-i\omega_{\rho}\right)^2 
			+\frac{1}{\rho}\left(\frac{\partial}{\partial\rho}-i\omega_{\rho}\right) 
			+ \frac{1}{\rho^2}\left(\frac{\partial}{\partial\phi}+i\omega_{\phi}\right)^2\right\rbrace\Phi_3 
			+ (u_3-E)\Phi_3 = 0,
		\end{eqnarray}
		\label{eq:ebo_eqs}
	\end{subequations}
\end{scriptsize}

\noindent
where adiabatic PESs are symbolized by $u_1$, $u_2$ and $u_3$, and components of NACM eigenvalues are represented as $\omega_{\rho}$ and $\omega_{\phi}$.
At this junction, it is worth mentioning that the angular component, $\omega_{\phi}$ ($=\frac{A_2}{2w}$) appears as gauge invariant ($\int_0^{2\pi} \omega_{\phi} d\phi=2\pi$) for three state degeneracy, affirming the workability of the EBO equations (Eq. \ref{eq:ebo_eqs}) for the 
specific domain of nuclear CS. \\

\noindent
Once the SEs for ordinary BO, EBO (Eq. \ref{eq:ebo_eqs}) and diabatic Hamiltonian (Eq. \ref{eq:diaham_mod},  
with $\Delta= 0$ and $172$ cm$^{-1}$) for the above mentioned model systems are in hand, photoabsorption spectra have been computed
for the lowest state, which are depicted in Figure \ref{fig:na3modelspec}. The spectral envelops obtained from EBO and diabatic
surfaces at $\Delta=0$\citep{akpaul_jcp131} exhibit good agreement, but approximate accord can be observed between the profiles
for $\Delta=172$ cm$^{-1}$ case. On the other hand, ordinary BO equation generated spectral bands exhibit 
large deviations from the diabatic ones for both the cases, namely, $\Delta= 0$ and $172$ cm$^{-1}$. \\

\begin{figure}[!htp]
	\centering
	\includegraphics[width=0.55\textwidth,angle=0]{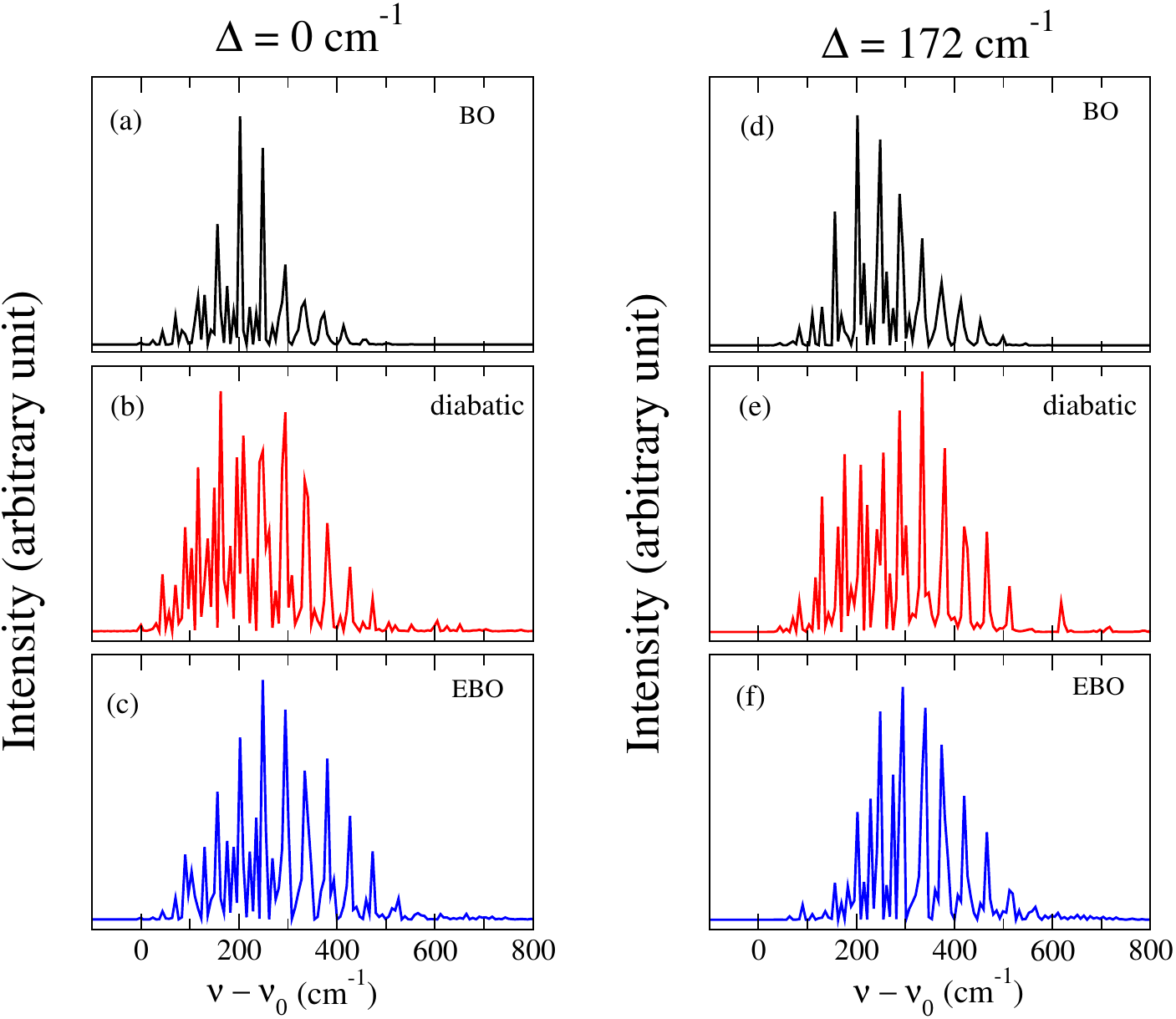}
	\caption{The left panel [(a), (b) and (c)] depicts the photoabsorption spectra of Na$_3$ cluster employing
		ordinary BO equations, diabatic SEs and single-surface EBO equations, respectively for $\Delta=0$ between $E^{\prime}$ and $A_1^{\prime}$ states. On the other hand, the right panel [(d), (e) and (f)] presents
		the same quantities for  $\Delta=172$ cm$^{-1}$. In case of $\Delta=0$, good agreement between the EBO and diabatic spectral profiles can be observed, whereas for $\Delta=172$ cm$^{-1}$, those bands show approximate matching. Nevertheless, BO profiles encounter substantial deviation from the diabatic ones.
		The figures have been taken from Ref. [\protect\citenum{joy_pccp}] with permission from American Institute of Physics (AIP).}
	\label{fig:na3modelspec}
\end{figure}

\noindent
\section{EBO Formalism for Model Calculations on Quasi-scattering Processes}\label{sec:spec}
\vspace{0.2 cm}

\noindent
In order to explore the usability of three-state EBO formalism, quasi-Jahn-Teller scattering models have been considered, where 
two CIs are located among three adiabatic PESs over two-dimensional nuclear CSs~\cite{sarkar_jcp124}. Among the two models, model A shows
two well-separated CIs between two different pairs of states and model B demonstrates non-adiabatic couplings between three states at a single point of the CS.

\begin{figure}[!htp]
	\centering
	
	\includegraphics[width=0.7\textwidth,angle=0]{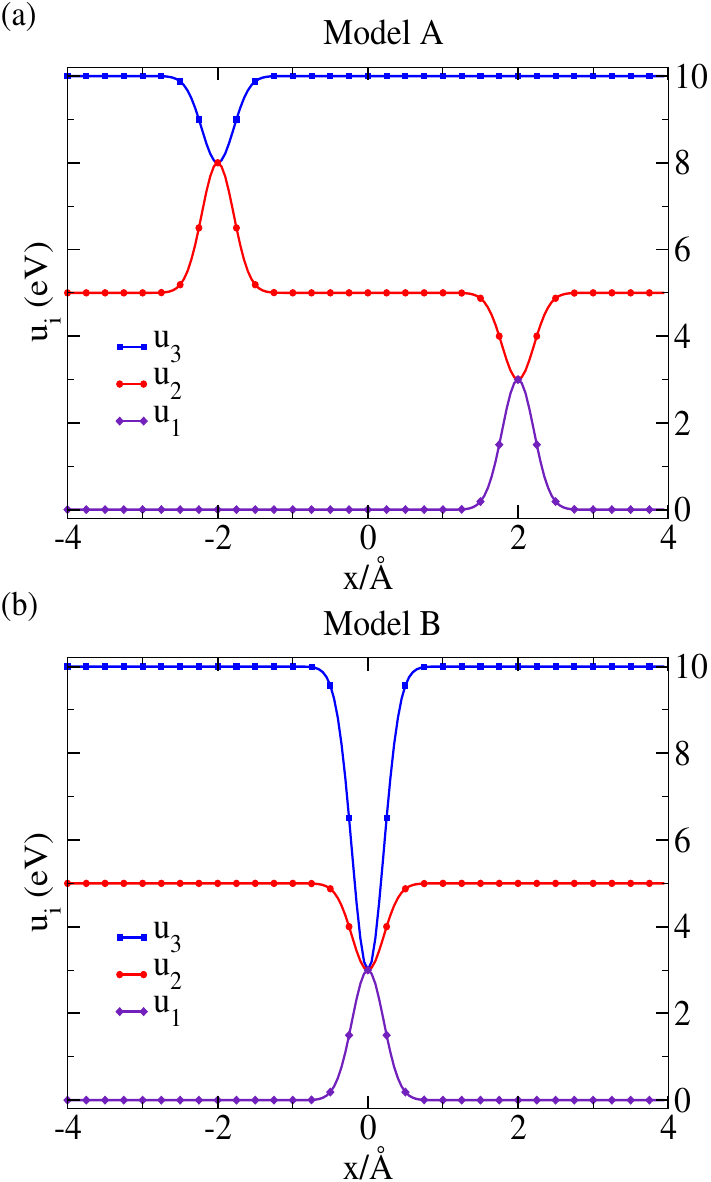}
	
	\caption{The upper panel (a) shows adiabatic PESs for model A with two well-separated CIs by a finite distance, whereas the lower panel (b) depicts two CIs at the same nuclear configuration. The figures have been taken from Ref. [\protect\citen{joy_pccp}] with permission from AIP.}
	\label{fig:ebomodels}
\end{figure}

\subsubsection{Model A}\label{sssec:modA}

\noindent
In the first model of non-adiabatically coupled three electronic states, a CI is assumed between first and second states, whereas another CI 
is present between second and third ones at different nuclear geometries. It is important to note that the separation between the two CIs
has been tuned to investigate two extreme situations, namely, two $2\times2$ and one $3\times3$ BBO systems. Such model PESs can be observed
for C$_2$H molecular system with respect to the NACT profiles, where the couplings have been computed as functions of C-C-H bond angle and C-H bond 
distance~\cite{baer_jcp117}. The model
is manifested in Figure \ref{fig:ebomodels}(a) with the three adiabatic PESs, where those take the following forms:

\begin{footnotesize}
	\begin{eqnarray}
		u_1(x,y) &=& \frac{1}{2}\mu\left[\omega_0-\tilde{\omega}_1(x)\right]^2y^2 + A_1\times f_1(x,y), 
		\nonumber\\
		u_2(x,y) &=& \frac{1}{2}\mu\omega_0^2y^2 - (D_1-A_1) \times f_1(x,y) + A_1\times f_2(x,y) + D_1,
		\nonumber\\
		u_3(x,y) &=& \frac{1}{2}\mu\omega_0^2y^2 - (D_1-A_1) \times f_2(x,y) + D_2, \nonumber\\
		\tilde{\omega}_1(x) &=& \omega_1\exp\left[-\frac{(x-x_0)^2}{\sigma_1^2}\right] \nonumber\\
		f_1(x,y) &=& \exp \left[-\frac{(x-x_0)^2+y^2}{\sigma^2}\right] \nonumber\\
		f_2(x,y) &=& \exp \left[-\frac{(x+x_0)^2+y^2}{\sigma^2}\right] \nonumber
	\end{eqnarray}
\end{footnotesize}

\noindent
In the above expression, the parameters are $\mu=0.58$ amu, $A_1=3.0$ eV, $D_1=5.0$ eV, $D_2=10.0$ eV, 
$\omega_0=39.14\times10^{13}$ s$^{-1}$, $\omega_1=7.83\times10^{13}$ s$^{-1}$, $\sigma=0.3$ \r{A}, and
$\sigma_1=0.75$ \r{A}. As a matter of fact, the above model demonstrates a two-arrangement channel system showing 
$x\rightarrow \infty$ and $-\infty$ as the reactant and product asymptotes,
respectively. The separation between the two CIs is $2x_0$ \r{A}$\,$, which is variable to 
explore the flexibility of the model indicating a $3\times3$ BBO system turning into two $2\times2$ BBO systems.\\

\noindent
While formulating the EBO equations for the above three electronic manifold, linearly independent functional forms of ADT angles
have been chosen to introduce the couplings among them, which fulfill the following properties:

\begin{itemize}
	\item The functional forms should be simple enough to efficiently capture the
	essential physical process of the system, more specifically, around the CIs between the three states.
	
	\item Those functions should depend on the separation between the CIs ($2x_0$) in such a way so that 
	the given $3\times3$ BBO system can be transformed into two simpler $2\times2$ BBO systems.
	
	\item The associated NACM eigenvalues should obey the correct gauge-invariance condition.
\end{itemize}

\begin{table}[t]
	\caption{Reactive state-to-state transition probabilities with various total energies ($E$) are compared in tabulated form 
		while executing the calculations on diabatic surfaces as well as EBO based surfaces keeping the distance between (1,2) and (2,3) CIs 
		by 4.0 \r{A}. The upper row for each total energy $E$ depicts the diabatic results
		whereas the lower one indicates the corresponding EBO results. The data are taken from Ref. 
		[\protect\citen{joy_pccp}].}
	\begin{center}
		{\begin{tabular}{@{}lccccc}\hline
				$E$(eV) & 0$\rightarrow$0 & 0$\rightarrow$1 & 0$\rightarrow$2 & 0$\rightarrow$3 & 0$\rightarrow$4 \\
				\hline
				1.25 & 0.0005$^a$ & 0.0274 & 0.0022 & 0.0145 & 0.0004 \\
				& 0.0008$^b$ & 0.0217 & 0.0020 & 0.0147 & 0.0007 \\ \\
				1.50 & 0.0008 & 0.0300 & 0.0084 & 0.0414 & 0.0021 \\
				& 0.0040 & 0.0300 & 0.0032 & 0.0412 & 0.0123 \\ \\
				1.75 & 0.0002 & 0.0357 & 0.0021 & 0.0761 & 0.0016 \\
				& 0.0006 & 0.0357 & 0.0075 & 0.0763 & 0.0049 \\
				\hline
		\end{tabular}}
	\end{center}
	{$^a$Diabatic}
	{$^b$EBO}
	\label{Tab:ebomodelA}
\end{table}

\noindent
Based on the above criteria, the functional forms of ADT angles have been taken as: 
\begin{eqnarray}
	\Theta_{12}(x,y)=\frac{1}{2}\tan^{-1}\left(\frac{y}{x-x_0}\right) \quad 
	\Theta_{23}(x,y)=\frac{1}{2}\tan^{-1}\left(\frac{y}{x+x_0}\right) \nonumber
\end{eqnarray}

\noindent
where $\Theta_{13}(x,y)$ is taken as zero (0) value for both the cases. In other words, such functional forms result into
a breakable $3\times3$ to two $2\times2$ BBO models. In order to generate the diabatic Hamiltonian, $2\times2$ BBO model necessitates 
$\boldsymbol{\nabla}\Theta_{12}\neq0$ and $\boldsymbol{\nabla}\Theta_{23}\approx0$ around the first 
CI as well as $\boldsymbol{\nabla}\Theta_{23}\neq0$ and $\boldsymbol{\nabla}\Theta_{12}\approx0$ 
around the other one, where $\boldsymbol{\nabla}\Theta_{13}$ can be taken as negligible for the entire
range, $0\leq\theta\leq2\pi$. If the separation of CIs becomes infinity ($x_0\rightarrow \infty$) or the CIs coalesce ($x_0\rightarrow 0$), 
the eigenvalues ($\boldsymbol{\omega}$) of NACM are expected to fulfill the gauge invariance
condition $\oint\boldsymbol{\omega}.d\textbf{R}=\pi$ and $2\pi$ bearing the signature of two-state and
three-state degeneracy, respectively. For $x_0\rightarrow \infty$, we have performed dynamical studies on 
diabatic as well as EBO based PESs to compute state-to-state reactive transition
probabilities at various total energies. It is important to note that the computed transition
probabilities perfectly obey the correct symmetry (even $\rightarrow$ odd or odd $\rightarrow$ even) as well as
good agreement can be observed between diabatic and EBO results (see Table \ref{Tab:ebomodelA}). The calculated results indicate that the EBO equations formulated by introducing zero curl
are able to produce correct symmetry-allowed transition probabilities, provided the eigenvalues of NACM are gauge invariant.\\

\subsubsection{Model B}\label{sssec:modB}

\noindent
In Figure \ref{fig:ebomodels}b, three BO states are coupled at a single nuclear configuration point, which can be considered as the 
second model (model B). As a representative example, one can choose Na$_3$ cluster possessing CIs between three coupled PESs
(lower and upper $2\,^2E^{\prime}$ states and $1\,^2A_1^{\prime}$ state) at close vicinity~\cite{cocchini_jcp88,mukherjee_jpca117}. 
For such model system, the adiabatic PESs can be written as: 

\begin{footnotesize}
	\begin{eqnarray}
		u_1(x,y) &=& \frac{1}{2}\mu\left[\omega_0-\tilde{\omega}_1(x)\right]^2y^2 + A_1\times f(x,y), 
		\nonumber\\
		u_2(x,y) &=& \frac{1}{2}\mu\omega_0^2y^2 - (D_1-A_1) \times f(x,y) + D_1,
		\nonumber\\
		u_3(x,y) &=& \frac{1}{2}\mu\omega_0^2y^2 - (D_2-A_1) \times f(x,y) + D_2, \nonumber
	\end{eqnarray}
\end{footnotesize}
with
\begin{footnotesize}
	\begin{eqnarray}
		\tilde{\omega}_1(x) = \omega_1\exp\left[-\left(\frac{x}{\sigma_1}\right)^2\right]; \quad
		f(x,y) = \exp \left[-\frac{x^2+y^2}{\sigma^2}\right] \nonumber
	\end{eqnarray}
\end{footnotesize}

\noindent
Such model deals with the couplings between three BO states at a point and therefore, one needs to consider all the three 
ADT angles around that region. Similar to model A, model B requires careful choice on the functional form of the ADT angles 
leading to meaningful EBO equation and accurate diabatic PESs. If the ADT angles are considered as 
$\Theta_{12}=\Theta_{13}=\Theta_{23}=\frac{1}{2}\tan^{-1}\left(\frac{y}{x}\right)$, the eigenvalue of the NACM becomes
$\boldsymbol{\omega}=\boldsymbol{\nabla}\Theta_{12}\sqrt{3+2\sin\Theta_{12}}$. It is interesting to note that gauge
invariance condition for the three-state CI, i.e.
$\oint \boldsymbol{\omega}.d\textbf{R}\approx 2\pi$ is approximately fulfilled for the computed NACM eigenvalue. 
Once the EBO and diabatic surfaces were in hand, those have been employed to carry out dynamical studies
and the calculated state-to-state vibrational transition probabilities for various total energies are presented in Table 
\ref{Tab:ebomodelB} As per the tabulated data, EBO generated probabilities obey the
correct symmetry as well as exhibit quantitative agreement with the diabatic ones.

\begin{table}[t]
	\caption{Reactive state-to-state transition probabilities with various total energies ($E$) are compared in tabulated form
		while performing the dynamical calculations on both diabatic as well as EBO based surfaces where (1,2) and (2,3) CIs 
		are located in same point leading to three-state degeneracy. The upper row for each total energy $E$ designates the diabatic
		results and the lower one specifies the corresponding EBO results. The results are taken from Ref. 
		[\protect\citen{joy_pccp}].}
	\begin{center}
		{\begin{tabular}{@{}lcccccc}\hline
				$E$(eV) & 0$\rightarrow$0 & 0$\rightarrow$1 & 0$\rightarrow$2 & 0$\rightarrow$3 & 0$\rightarrow$4 & 0$
				\rightarrow$5 
				\\\hline
				1.25 & 0.0227$^a$ & 0.0000 & 0.0677 & 0.0000 & 0.0061 & \\
				& 0.0228$^b$ & 0.0035 & 0.0727 & 0.0131 & 0.0090 & \\ \\
				1.50 & 0.0314 & 0.0000 & 0.0391 & 0.0000 & 0.0905 & 0.0000 \\
				& 0.0384 & 0.0030 & 0.0704 & 0.0017 & 0.0898 & 0.0060 \\ \\
				1.75 & 0.0250 & 0.0000 & 0.0562 & 0.0000 & 0.1280 & 0.0000 \\
				& 0.0251 & 0.0006 & 0.0590 & 0.0012 & 0.1388 & 0.0179 \\
				\hline
		\end{tabular}}
	\end{center}
	{$^a$Diabatic}
	{$^b$EBO}
	\label{Tab:ebomodelB}
\end{table}

\noindent
\section{Contemporary Diabatization Methodologies}

\noindent
Though we have discussed the detailed formulation of BBO based diabatization theory including non-adiabatic coupling terms (NACTs) explicitly, it would be appropriate to outline other contemporary diabatization schemes to incorporate the couplings among the electronic states on various molecular and chemical processes. 

\vspace{0.2 cm}
\noindent 
\begin{enumerate} 
	\item Vibronic Coupling Model (VCM): As per VCM\citep{kdc_acp57} methodology,
	diagonal and off-diagonal elements of diabatic Hamiltonian are expanded 
	with respect to the ground adiabatic PES ($V_0(\textbf{R})$) minima or more preferably, the CI point 
	($\textbf{R}_0$) as functions of linear, bilinear or even higher order polynomials of nuclear DOFs.\citep{koppel_jcp110,koppel_jcp150,koppel_mp104,viel_jcp120,bhatt_jpca118} It is worth mentioning that 
	Zeng {\it et al.}\citep{hickman_pccp20,issaka_pccp21} expanded the VCM based diabatic Hamiltonian 
	upto arbitrary order and applied the formalism for trigonal and tetragonal systems. In case of 
	linear VCM, the Taylor series expanded diabatic matrix elements at the close vicinity of a reference nuclear 
	geometry, $\textbf{R}_0$ can be expanded as: 	
	
	\begin{footnotesize}
		\begin{eqnarray}
			W_{ii}(\textbf{R}) &=& \hat{T}_{\rm nuc} + V_0(\textbf{R}) + W_{ii}(\textbf{R}_0) + \sum_s \kappa_s^{(i)}R_s \nonumber\\
			W_{ij}(\textbf{R}) &=& \sum_s\lambda_s^{(i,j)}R_s \quad i\neq j,
			\label{eq:lvc}
		\end{eqnarray}
	\end{footnotesize}
	
	\noindent
	where diagonal and off-diagonal diabatic Hamiltonian elements are represented as $W_{ii}(\textbf{R})$ and $W_{ij}(\textbf{R})$, respectively.
	On the other hand, the combining coefficients ($\kappa_s^{(i)}$ and $\lambda_s^{(i,j)}$) indicate linear vibronic coupling constants.
	It needs to be noted that being a Taylor series expansion, the VCM based diabatization approach is applicable only in a limited domain of Frank-Condon
	(FC) region close to the reference geometry $\textbf{R}_0$.
	
	\item Exact Factorisation (EF) Method: In the EF approach proposed and developed by Gross {\it et al.}\citep{gross_prl105,gross_jcp137,maitra_jcp150},
	simultaneous solution of the electronic and nuclear SEs takes place using time-dependent PESs, vector potential and a coupling operator.
	It is necessary to note that the wavefunctions in EF methodology cannot represent a specific quantum state and therefore,  
	the physical significance of such functions is ambiguous with respect to the adiabatic states and the
	correlations within them. Agostini and coworkers\citep{agostini_jpcl8,agostini_epjb91} have applied such approach to investigate the contribution of CI(s)
	on dynamical calculations for 2D model systems as well as 
	realistic species. For the later cases (realistic species), the electronic structure 
	calculations have been carried out with the aid of Time-Dependent Density Functional Theory (TDDFT).\citep{gross_jpcl8}

	\item Direct Dynamics (DD): Apart from the VCM or EF based diabatization followed by nuclear dynamics, direct dynamical methods, like 
	on-the-fly diabatization version of MCTDH,\citep{worth_irpc27,richings_jcp148} DD variational multi-configurational 
	Gaussian (DD-vMCG),\citep{worth_irpc34,worth_jpca119,worth_cpl683} etc. are also used now-a-days, where the
	following expression is numerically solved to compute the ADT matrix, $A(\textbf{R})$ at every step during time propagation:\citep{worth_jpca119}
	\begin{eqnarray}
		\boldsymbol{\nabla}_R A(\textbf{R}) = -\boldsymbol{\tau} (\textbf{R})A(\textbf{R}). 
		\label{eq:ddadt}
	\end{eqnarray}
	
	\noindent
	The solution of the above equation can be written as:\citep{baer_mp40,baer_phys_rep358}
	
	\begin{eqnarray}
		A(\textbf{R}+\Delta \textbf{R}) = \exp	\Big(-\int_{\textbf{R}}^{\textbf{R}+\Delta \textbf{R}} \boldsymbol{\tau}.
		d\textbf{R} \Big) A (\textbf{R})  \nonumber
	\end{eqnarray}
	
	
	\noindent
	In order to obtain a more formal solution of Eq. \ref{eq:ddadt}, ADT equations (see Eq. \ref{adteq_p}) of BBO formalism~\citep{bijit_irpc38} can be employed
	engaging all the electronic states forming the sub-space instead of only two states possessing CI point/seam.

	\item Simultaneous Electron-Nuclear Dynamics: Apart from the above mentioned diabatization approaches, Takatsuka and coworkers~\citep{takatsuka_book,takatsuka_rev,takatsuka_jcp160} developed a unified formalism to decompose the components of nonadiabatic interactions along fast and slow modes in multi-dimensional multi-state dynamical studies. Such mathematical framework provides (local) collective coordinates needed for the interconnection between slow and fast modes and therefore, it can be involved to investigate multi-dimensional nonadiabatic transition, resonance, chaos and many other intrinsic phenomenon. The derivatives of nuclear coordinate vectors, $\dot{\mu}$ can naturally or
	artificially govern the nonadiabatic dynamics at a given point $x(t)\mu(t)$. The nuclear coordinates are transformed into a new set of variables, $s_I$ and 
	the relative symmetry between the vectors, $s_I$ and $\dot{\mu}$ dictates the excited state chemical reactions in many-dimensional and (quasi-)degenerate nonadiabatic dynamics. 
\end{enumerate}

\noindent
\section{Features of BBO Theory: Advantages and Challenges}
\noindent
The theoretically ``exact" and numerically ``accurate" first principle based BBO theory is equally applicable for the generation of diabatic Hamiltonian
for spectroscopic systems as well as scattering processes using {\it ab initio} computed adiabatic surfaces and NACTs.\citep{bijit_irpc38,joy_pccp}
In other words, one can employ the BBO formalism at and around the FC domain as well as at the asymptotic limit of nuclear CS, where 
bond breaking and formation occur. Therefore, it is indeed necessary to thoroughly compare the BBO theory with other contemporary methodologies, which are presented in the earlier subsection.

\vspace{0.2 cm}
\noindent 
\begin{enumerate}
	\item In BBO formalism, the BO expanded total molecular wavefunction takes the following form, 
	\begin{eqnarray}
		\Psi(\textbf{r},\textbf{R}) = \sum_i \psi_i^{\rm ad}(\textbf{R}) \xi_i(\textbf{r};\textbf{R}) &=& \sum_i \psi_i^{\rm ad}(\textbf{R})\, A(\textbf{R}) \tilde{\xi}_i(\textbf{r})\nonumber\\
		&=& \sum_i \psi_i^{\rm dia}(\textbf{R}) \tilde{\xi}_i(\textbf{r}), \nonumber
		\label{eq:lvc1}
	\end{eqnarray}
	\noindent
	where $\tilde{\xi}_i(\textbf{r})$s represent functions of electronic coordinates ($\textbf{r}$). In other words,
	the derivatives of $\tilde{\xi}_i(\textbf{r})$s with respect to nuclear coordinates 
	($\textbf{R}$) do not survive. The exact form of $A(\textbf{R})$ matrix can be evaluated using Eq. \ref{adteq_p}
	and then, the ADT matrix needs to be incorporated into Eq. \ref{diabse2} to calculate the diabatic PES matrix at each nuclear configuration, $\textbf{R}$.
	On the contrary, in VCM approach, diabatic potential matrix elements ($W_{ij}(\textbf{R})$) are expressed in Taylor series expansion containing
	linear, bilinear or even higher order terms around the ground state minima or a CI point ($\textbf{R}_0$). Nevertheless, 
	the BBO-DH matrix elements (diabatic PESs and coupling terms) are highly accurate upto all order
	even around the accidental CIs for spectroscopic systems. Furthermore, such a complete theory can also be used to construct
	global diabatic PESs for scattering processes, which involves even the asymptotic domain of nuclear CS.

	\item The adiabatic PESs, NACTs as well as BBO-DH matrix elements (diabatic PESs and couplings) for realistic molecules often exhibit even-parity functional form  
	along one of the components in most of the degenerate pair of asymmetric stretching/bending 
	modes or even some specific non-degenerate asymmetric stretching modes.~\citep{mukherjee_jcp143,bijit_cp515,soumya_acs,soumya_jcp150,mukherjee_jpca117,bijit_irpc38,
		naskar_jctc,joy_pccp,soumya_bz,soumya_tfb} In other words, only even power polynomials can be used to fit 
	the potential couplings for those modes, where such unique property cannot be explored for the 
	well-known traditional JT model~\citep{JT_prs161,van-vleck,lh-opik,opik,
		kdc_acp57,mahapatra_jcp110,Koppel_jcp117_1,mahapatra_pccp11,viel_jcp120,mahapatra_cp347,hickman_pccp20} 
	(linear, bilinear or higher order). For such specific normal mode (say $Q_{l}$) of $Q_{k}-Q_{l}$ nuclear plane,
	one of the off-diagonal elements take the following form:
	\begin{eqnarray}
		W_{ij}(Q_k,Q_l)=\sum_{n=0}^N\sum_{m=0}^Mc_{nm}Q_k^nQ_l^{2m}, \qquad  n,m \equiv \{0,1,2,...\} 
		\label{eq:even-dia}
	\end{eqnarray}
	\noindent
	where $i$, $j$ represent two electronic states and $Q_k$, $Q_l$ signify two normal modes. The above fitted polynomial clearly shows that
	the contributing coefficients ($c_{01}$, $c_{11}$, $c_{21}$, ...), associated with $Q_l$, $Q_kQ_l$, $Q_k^2Q_l$, ... appear as zero within the numerical 
	precision. As a matter of fact, according to BBO formalism, the first non-zero term of some specific modes in the diabatic coupling can be quadratic (not linear), which can be observed for a wide range of molecules.~\citep{mukherjee_jcp143,bijit_cp515,soumya_acs,soumya_jcp150,mukherjee_jpca117,bijit_irpc38,
		naskar_jctc,joy_pccp,soumya_bz,soumya_tfb}

	\item In BBO formalism, the quantized nature of NACTs indicates the number of enclosed JT CI(s)/seam(s) (symmetry allowed or accidental) 
	or RT interaction(s) over the nuclear plane inside a closed contour. The resulting ADT angles at the end of the closed loop 
	attain the magnitude of zero (0) or $n\pi$
	or $n.2\pi$ ($n$ is an integer) bearing the signature of the encapsulated degenerate point(s). Such unique feature of BBO theory is helpful
	to explore the nature and accuracy of the diabatic Hamiltonian. On the other hand, other diabatization methods do not have similar procedure 
	to unveil the accuracy of diabatic PESs.
	
	
	\item The BBO theory has been truly generalized by developing an algorithm and user-friendly OpenMP parallelized program, ``ADT" (see \href{https://pubs.acs.org/doi/10.1021/acs.jctc.9b00948}{https://
		pubs.acs.org/doi/10.1021/acs.jctc.9b00948} or 
	\href{https://github.com/AdhikariLAB/ADT-Program}{https://github.com/ AdhikariLAB/ADT-Program})~\cite{naskar_jctc}. User can install and execute such software
	package to formulate analytic expressions of various ADT quantities (like ADT equations, ADT matrix elements, diabatic 
	potential energy matrix elements, etc.) as well as to compute the numerical values of those parameters 
	for `N' coupled electronic states with `M' number of nuclear DOFs. During diabatization, user can provide the {\it ab initio} 
	inputs, namely, adiabatic PESs and NACTs to the program or such input files can be generated with the aid of in-built python codes interfacing standard quantum chemistry packages, like Molpro,\citep{MOLPRO_brief} Columbus,\citep{columbus} Gamess\citep{gamess} and Gaussian\citep{g16}. At this junction, it needs to mention
	that calculation of highly accurate {\it ab initio} inputs, namely, adiabatic PESs and NACTs needs sophisticated electronic structure approaches, which are further used in BBO based diabatization. Till now, the ``ADT" algorithm~\citep{naskar_jctc} is not interfaced with DD approaches~\citep{worth_irpc27,richings_jcp148,worth_irpc34,worth_jpca119,worth_cpl683}, but such interfacing is expected to be done in near future to reduce the computation time and cost. 
	
\end{enumerate}

\bibliographystyle{aipnum4-1}
\bibliography{ref_book}

\chapter{Generalization of Beyond Born Oppenheimer Theory in External Fields}\label{ssec:mol}

\noindent
The field-dressed potential energy surfaces (PESs) raise immense interest among the scientific community with the idea of carrying out chemical 
processes in presence of laser fields. Even though the first experiment was attempted in 1980s~\citep{hering_prl44} using low intensity and continuous (as opposed to pulsed) lasers,
George and coworkers~\citep{zimmerman_jcp66,george_book} carried out the first theoretical/numerical investigations to study atom-diatom inelasic collisions governed by
continuous lasers and such formulations were later extended for other types of collisions~\citep{light_jcp70,chu_jcp75,zimmerman_jcp87,last_cpl101,last_jcp82}. In case of F+H$_2$ reaction, it was observed that 
fields with different frequencies alters the quantum yields of the reactive (electronic non-adiabatic) process F($^2P_{1/2}$) + H$_2$ $\rightarrow$ HF($v_f$) + H$_2$ significantly.~\citep{last_jcp82} Later on, various theoretical frameworks were introduced, which mainly deal with short-pulsed high-intensity fields applied on various systems in different situations.~\citep{manz_jcp113,barth_pra75,yonehara_jcp128} \\

\noindent
Adhikari {\it et al.}~\citep{sarkar_jcp126,sarkar_jcp127_1,sarkar_jcp127_2,paul_jcp132,paul_jpca113,paul_pra81,PAUL_PR496} have developed a rigorous first principle based theory to apply for general field dressed poly-atomic system. In order to carry out such development, the following two aspects are taken care: (a) Extension of the Born-Oppenheimer (BO) treatment has been carried out to incorporate quantum states of light beyond the usual coherent-state distribution.~\citep{paul_jcp132,paul_jpca113,paul_pra81,gavrila_book,corkum_fd113,cohen_rmp70} Such states are widely known as non-coherent Fock (or, also, as the non-classical photon) states. It is important to explore the effects of such states
on measured magnitudes such as kinetic energy release (KER) distributions and time-dependent dissociation probabilities (DP). This type of photon states mainly arise
in presence of short-pulsed, high intensity laser systems and thereby, significantly affect the KER distribution of the products. (b) It has been explored whether external fields can affect conical intersections (CIs),~\citep{sarkar_jcp126,sarkar_jcp127_1,sarkar_jcp127_2} where such studies have been executed comprehensively within the new gauge. \\

\noindent
The present chapter deals with Beyond BO (BBO) approach for a molecular system, which interacts with an external field. This treatment is carried out in two different ways: (a) using a time-independent (or field-free) electronic basis set, termed as the time-dependent ordinary treatment; (b) employing a time-dependent (or field-dressed) electronic basis set, known as  time-dependent field-dressed treatment.

\section{Time-Dependent Ordinary BBO Theory~\citep{PAUL_PR496}}

\noindent
For both the time-dependent BBO treatments (ordinary and field-dressed), the time-dependent Schr\"odinger Equation (SE) for the total wave-function ($\boldsymbol{\Psi}(\boldsymbol{r},\boldsymbol{R},t)$) takes the following form:\\
\begin{equation}
	i\hbar\frac{\partial \boldsymbol{\Psi}(\boldsymbol{r},\boldsymbol{R},t)}{\partial t}=\Big(-\frac{\hbar^2}{2m}\nabla_\textbf{R}^2+\boldsymbol{H_{e}}(\boldsymbol{r}|\boldsymbol{R},t)\Big)\boldsymbol{\Psi}(\boldsymbol{r},\boldsymbol{R},t),
	\label{eq:mf1_1}
\end{equation}
where the electronic Hamiltonian $\boldsymbol{H_{e}}(\boldsymbol{r}|\boldsymbol{R})$ appears as time dependent. On the other hand, one can assume that the
effect of the external field can be represented in terms of the interaction of the electric field with the electronic Hamiltonian. As a result, the associated time-dependent electronic Hamiltonian takes the form:~\citep{baer_cpl35,baer_jcp119}
\begin{equation}
	\boldsymbol{H_{e}}(\boldsymbol{r}|\boldsymbol{R},t) = \boldsymbol{H_{e0}}(\boldsymbol{r}|\boldsymbol{R}) + \boldsymbol{\mu}_e (\boldsymbol{r},\boldsymbol{R}).\boldsymbol{E}(t)
	\label{eq:mf1_2}
\end{equation}

\noindent
It is important to note that $\boldsymbol{\Psi}(\boldsymbol{r},\boldsymbol{R},t)$ can be expanded in terms of a time-independent electronic basis set originated from the adiabatic eigenfunctions of $\boldsymbol{H_{e0}}(\boldsymbol{r}|\boldsymbol{R})$:
\begin{equation}
	\boldsymbol{\Psi}(\boldsymbol{r},\boldsymbol{R},t) = \boldsymbol{\xi}^T(\boldsymbol{r}|\boldsymbol{R})\boldsymbol{\psi}(\boldsymbol{R},t)
	\label{eq:mf1_3}
\end{equation}
\noindent
where $\boldsymbol{\xi}^T(\boldsymbol{r}|\boldsymbol{R})$
and $\boldsymbol{\psi}(\boldsymbol{R},t)$ appear as
row and column vectors, respectively, and $\boldsymbol{\psi}(\boldsymbol{R},t)$ signifies field-dressed nuclear wave-functions. While substituting Eq. \ref{eq:mf1_3} in Eq. \ref{eq:mf1_1}, it produces time-dependent SE for the nuclear wave-functions as follows:
\begin{eqnarray}
	i\hbar\boldsymbol{\xi}^T(\boldsymbol{r}|\boldsymbol{R})\dfrac{\partial \boldsymbol{\psi}(\boldsymbol{R},t)}{\partial t}&=&-\dfrac{\hbar^2}{2m}\nabla^2(\boldsymbol{\xi}^T(\boldsymbol{r}|\boldsymbol{R},t)\boldsymbol{\psi}(\boldsymbol{R},t))\nonumber\\
	&&+\boldsymbol{H_{e}}(\boldsymbol{r}|\boldsymbol{R},t)\boldsymbol{\xi}^T(\boldsymbol{r}|\boldsymbol{R},t)\boldsymbol{\psi}(\boldsymbol{R},t)
	\label{eq:mf1_4}
\end{eqnarray}
or
\begin{equation}
	i\hbar\boldsymbol{\xi}^T\dfrac{\partial \boldsymbol{\psi}}{\partial t}= -\dfrac{\hbar^2}{2m}(\boldsymbol{\xi}^T\nabla^2\boldsymbol{\psi}+\nabla^2\boldsymbol{\xi}^T\boldsymbol{\psi}+2\nabla\boldsymbol{\xi}^T\nabla\boldsymbol{\psi})+
	\boldsymbol{H_{e}}\boldsymbol{\xi}^T\boldsymbol{\psi}
	\label{eq:mf1_5}
\end{equation}
\noindent
While projecting Eq. \ref{eq:mf1_5} by $\boldsymbol{\xi}^\dagger(\boldsymbol{r}|\boldsymbol{R})$ (column vector containing the complex conjugate eigenfunctions) and integrating over the electronic coordinates $\boldsymbol{r}$, one can arrive the following time-dependent adiabatic SE for the nuclear wavefunctions:
\begin{equation}
	i\hbar\dfrac{\partial \boldsymbol{\psi}}{\partial t}=-\dfrac{\hbar^2}{2m}(\nabla+\boldsymbol{\tau(R)})^2\boldsymbol{\psi}+\Big(\boldsymbol{u(R)}+\boldsymbol{\tilde{\mu}(R)}.\boldsymbol{\tilde{E}}(t)\Big)\boldsymbol{\psi}
	\label{eq:mf1_6}
\end{equation}
\noindent
where $\boldsymbol{\tau(R)}$ and $\boldsymbol{u(R)}$ correspond to the time-independent non-adiabatic coupling matrix (NACM) and diagonal adiabatic PESs matrix. On the other hand, the associated 
dipole moment matrix takes the following form:
\begin{equation}
	\boldsymbol{\tilde{\mu}}_{jk}\boldsymbol{(R)}=e\langle\xi_j\boldsymbol{(r|R)}\vert\boldsymbol{r}\vert\xi_k\boldsymbol{(r|R)}\rangle
	\label{eq:mf1_6}
\end{equation}
\noindent
Similar to the time-independent approach, the adiabatic-to-diabatic transformation (ADT)  can be carried out to eliminate $\boldsymbol{\tau}$ matrix and $\boldsymbol{\psi}(\boldsymbol{R},t)$ is replaced by 
$\boldsymbol{A(s)}\boldsymbol{\Phi}(\boldsymbol{R},t)$, which leads to the following time-dependent nuclear SE:~\citep{baer_cpl35}
\begin{equation}
	i\hbar\dfrac{\partial \boldsymbol{\Phi}}{\partial t}=-\dfrac{\hbar^2}{2m}\nabla^2\boldsymbol{\Phi}+\boldsymbol{\tilde{W}}_e(\boldsymbol{R},t)\boldsymbol{\Phi},
	\label{eq:mf1_7}
\end{equation}
where $\boldsymbol{\tilde{W}}_e=\boldsymbol{A}^\dagger\boldsymbol{\tilde{H}}_e\boldsymbol{A}$ and $\boldsymbol{\tilde{H}}_e(\boldsymbol{R},t)=\boldsymbol{u(R)}+\boldsymbol{\tilde{\mu}(R)}.\boldsymbol{\tilde{E}}(t)$. It is important to note that $\boldsymbol{A}$ is time-independent in the field-free case.

\section{Time-Dependent Field-Dressed BBO Theory~\citep{PAUL_PR496}}

\noindent
In case of field-dressed condition, the time-dependent electronic basis set, $\boldsymbol{\xi}(\boldsymbol{r}|\boldsymbol{R},t)$ 
can be used to expand the total wavefunction as follows:~\citep{baer_cpl35,baer_jcp119}

\begin{equation}
	\boldsymbol{\Psi}(\boldsymbol{r},\boldsymbol{R},t) = \boldsymbol{\xi}^T(\boldsymbol{r}|\boldsymbol{R},t)\boldsymbol{\psi}(\boldsymbol{R},t)
	\label{eq:mf2_1}
\end{equation}

\noindent
On the other hand, the electronic eignefunctions obey the following time-dependent SE:
\begin{equation}
	i\hbar\frac{\partial}{\partial t}|\boldsymbol{\xi}_j(\boldsymbol{r}|\boldsymbol{R},t)\rangle = \boldsymbol{H_e}(\boldsymbol{r}|\boldsymbol{R},t)|\boldsymbol{\xi}_j(\boldsymbol{r}|\boldsymbol{R},t)\rangle
	\label{eq:mf2_2}
\end{equation}

\noindent
At $t=0$, the electric field is turned on and therefore,
$\boldsymbol{H_e}$ is independent of time (and diagonal) at $t \leq 0$. As a consequence, for $t \leq 0$, the $j^{th}$ eigenfunction, $\xi_j(\boldsymbol{r}|\boldsymbol{R},t)$ takes the following form:

\begin{equation}
	|\xi_j(\boldsymbol{r}|\boldsymbol{R},t)\rangle = exp\Big(-\frac{i\,\boldsymbol{u}_j(\boldsymbol{R})t}{\hbar}\Big)|\xi_{j0}(\boldsymbol{r}|\boldsymbol{R})\rangle,
	\label{eq:mf2_3}
\end{equation}

\noindent
where $\xi_{j0}(\boldsymbol{r}|\boldsymbol{R})$s are the eigenvalues of the following $j^{th}$ eigenvalue equation:
\begin{equation}
	(\boldsymbol{H_{e0}}(\boldsymbol{r}|\boldsymbol{R})-\boldsymbol{u}_j(\boldsymbol{R}))|\xi_{j0}(\boldsymbol{r}|\boldsymbol{R})\rangle=0.
\end{equation}
\noindent
Here $\boldsymbol{u}_j(\boldsymbol{R})$s are time-independent potential energy surfaces. \\

\noindent
While solving Eq. \ref{eq:mf2_2} for $t > 0$ (considering
Eq. \ref{eq:mf2_3} as initial condition), $\xi_{j}(\boldsymbol{r}|\boldsymbol{R},t)$ can be expressed in terms of $L$ eigenfunctions of $\boldsymbol{H_{e}}$ as calculated at $t=0$. Therefore,

\begin{equation}
	|\xi_{j}(\boldsymbol{r}|\boldsymbol{R},t)\rangle = \sum_{k=1}^{L} |\xi_{k0}(\boldsymbol{r}|\boldsymbol{R})\rangle\boldsymbol{\omega}_{kj}(\boldsymbol{R},t),
	\label{eq:mf2_4}
\end{equation}

\noindent
where the square matrix, $\boldsymbol{\omega}_{kj}(\boldsymbol{R},t)$ possesses the expansion coefficients. If the expression of $|\xi_{j}(\boldsymbol{r}|\boldsymbol{R},t)\rangle$ of Eq. \ref{eq:mf2_4} is substituted in Eq. \ref{eq:mf2_2}, it results into the differential equation for $\boldsymbol{\omega}(\boldsymbol{R},t)$,
\begin{equation}
	i\hbar\frac{\partial}{\partial t}\boldsymbol{\omega}(\boldsymbol{R},t) = \tilde{\boldsymbol{H_e}}(\boldsymbol{R},t)\boldsymbol{\omega}(\boldsymbol{R},t),
	\label{eq:mf2_5}
\end{equation} 
\noindent
where $\tilde{\boldsymbol{H_e}}$ is represented as:
\begin{equation}
	\tilde{\boldsymbol{H_e}}(\boldsymbol{R},t)_{kj}=\langle\xi_{k0}(\boldsymbol{r}|\boldsymbol{R})|\boldsymbol{H_{e}}(\boldsymbol{r}|\boldsymbol{R},t)|\xi_{j0}(\boldsymbol{r}|\boldsymbol{R})\rangle
	\label{eq:mf2_6}
\end{equation}

\noindent
The solution of Eq. \ref{eq:mf2_5} is as follows,
\begin{eqnarray}
	\boldsymbol{\omega}(\boldsymbol{R},t) = \varrho\, exp\lbrace-\frac{i}{\hbar}\int_{-\infty}^t dt^{\prime}.\boldsymbol{H_e}(\boldsymbol{R},t)\rbrace\boldsymbol{I}
	\label{eq:mf2_7}
\end{eqnarray}
\noindent
where $\varrho$ represents time ordering operator and $\boldsymbol{I}$ is the unit matrix. On the other hand, for $t \leq 0$, the matrix $\boldsymbol{\omega}(\boldsymbol{R},t)$ becomes:
\begin{eqnarray}
	\boldsymbol{\omega}_{kj} (\boldsymbol{R},t \leq 0) = exp\Big(\frac{i\boldsymbol{u}_j(\boldsymbol{R})t}{\hbar}\Big)\delta_{jk}.
	\label{eq:mf2_8}
\end{eqnarray}

\noindent
If Eqs. \ref{eq:mf2_1} and \ref{eq:mf2_2} are substituted in Eq. \ref{eq:mf1_1}, the resulting equation is integrating over the electronic coordinates, which produces the following field-dressed nuclear SE:
\begin{equation}
	i\hbar\boldsymbol{\xi}^T(\boldsymbol{r}|\boldsymbol{R},t)\dfrac{\partial \boldsymbol{\psi}(\boldsymbol{R},t)}{\partial t}=-\dfrac{\hbar^2}{2m}\nabla^2(\boldsymbol{\xi}^T(\boldsymbol{r}|\boldsymbol{R},t)\boldsymbol{\psi}(\boldsymbol{R},t))
	\label{eq:mf3_1}
\end{equation}
\noindent
It is interesting to note that the above equation is entirely dictated by the nuclear kinetic energy operator. Eq. \ref{eq:mf3_1} turns into the following form:
\begin{equation}
	i\hbar\boldsymbol{\xi}^T\dfrac{\partial \boldsymbol{\psi}}{\partial t}= -\dfrac{\hbar^2}{2m}(\boldsymbol{\xi}^T\nabla^2\boldsymbol{\psi}+(\nabla^2\boldsymbol{\xi}^T)\boldsymbol{\psi}+2\nabla\boldsymbol{\xi}^T\nabla\boldsymbol{\psi})
	\label{eq:mf3_2}
\end{equation}
\noindent
Projecting the above equation by $\boldsymbol{\xi}^\dagger(\boldsymbol{r}|\boldsymbol{R},t)$ leads to the following expression:
\begin{equation}
	i\hbar\dfrac{\partial \boldsymbol{\psi}}{\partial t}=-\dfrac{\hbar^2}{2m}(\nabla+\boldsymbol{\tilde{\tau}})^2\boldsymbol{\psi}.
	\label{eq:mf3_3}
\end{equation}
\noindent
In the above equation, the field-dressed NACM ($\boldsymbol{\tilde{\tau}}(\boldsymbol{R},t)$) can be written as:
\begin{equation}
	\boldsymbol{\tilde{\tau}} = \boldsymbol{\omega}^\dagger\boldsymbol{\tau}\boldsymbol{\omega}+
	\boldsymbol{\omega}^\dagger\nabla\boldsymbol{\omega},
	\label{eq:mf3_4}
\end{equation}
where $\boldsymbol{\tau}$ is the field-free NACM. As a matter of fact, Eq. \ref{eq:mf3_3} presents the SE in absence of any potential energy matrix and the nuclear dynamics is solely governed by a vector potential. \\

\noindent
As per the structures of Eqs. \ref{eq:mf3_3} and \ref{eq:mf3_4}, it appears easier to remove the (full) field-dressed NACM, $\boldsymbol{\tilde{\tau}}$ compared to its field-free analogue, $\boldsymbol{\tau}$. In order to perform the ADT for Eq. 
\ref{eq:mf3_3}, we follow the similar procedure as the time-independent case, but for the present case (time-dependent field-dressed), ADT matrix ($\boldsymbol{\tilde{A}}(\boldsymbol{R},t)$) is considered as time-dependent. Considering the orthogonal transformation,  $\boldsymbol{\Psi}=\boldsymbol{\tilde{A}}\boldsymbol{\Phi}$, Eq. \ref{eq:mf3_3} produces the following expression:
\begin{equation}
	i\hbar\boldsymbol{\tilde{A}}\dfrac{\partial \boldsymbol{\Phi}}{\partial t}=-\dfrac{\hbar^2}{2m}\Big[\boldsymbol{\tilde{A}}\nabla^2+2(\boldsymbol{G\tilde{A}})\nabla+(\boldsymbol{G^2\tilde{A}})\Big]\boldsymbol{\Phi}-i\hbar\dfrac{\partial \boldsymbol{\tilde{A}}}{\partial t}\boldsymbol{\Phi}.
	\label{eq:mf4_1}
\end{equation}
\noindent
In the above expression, $\boldsymbol{G}$ represents the space covariant derivative operator acting only on $\boldsymbol{\tilde{A}}$, which is given in the form,
\begin{equation}
	\boldsymbol{G}=\nabla+\boldsymbol{\tilde{\tau}}.
	\label{eq:mf4_2}
\end{equation}
\noindent
Then, we add as well as subtract a product form of undetermined matrix $\boldsymbol{\tilde{\tau}}_t$ and $\boldsymbol{\tilde{A}}\boldsymbol{\psi}$ simultaneously so that Eq. \ref{eq:mf4_1} turns into:
\begin{equation}
	i\hbar\boldsymbol{\tilde{A}}\dfrac{\partial \boldsymbol{\Phi}}{\partial t}=-\dfrac{\hbar^2}{2m}\Big[\boldsymbol{\tilde{A}}\nabla^2+2(\boldsymbol{G\tilde{A}})\nabla+(\boldsymbol{G^2\tilde{A}})\Big]\boldsymbol{\Phi}-(\boldsymbol{G_t\tilde{A}}-\boldsymbol{\tilde{\tau}_t\tilde{A}})\boldsymbol{\Phi},
	\label{eq:mf4_3}
\end{equation}
where $\boldsymbol{G_t}$ signifies the corresponding time covariant derivative operator as follows,
\begin{equation}
	\boldsymbol{G_t}=i\hbar\dfrac{\partial}{\partial t}+\boldsymbol{\tilde{\tau}}_t
	\label{eq:mf4_4}
\end{equation}
and $\boldsymbol{\tilde{\tau}}_t$ is considered as a Hermitian operator. \\

\noindent
At this junction, the $\boldsymbol{\tilde{A}}$ matrix can be chosen to be a solution of the space-time first-order differential equations, where the space and time dependent equations are as follows
\begin{equation}
	\boldsymbol{G\tilde{A}}=\boldsymbol{0}	\Rightarrow \nabla\boldsymbol{\tilde{A}}+\boldsymbol{\tilde{\tau}\tilde{A}}=\boldsymbol{0}
	\label{eq:mf4_5}
\end{equation}
and
\begin{equation}
	\boldsymbol{G_t\tilde{A}}=\boldsymbol{0}	\Rightarrow i\hbar\dfrac{\partial\boldsymbol{\tilde{A}} }{\partial t}+\boldsymbol{\tilde{\tau}_t\tilde{A}}=\boldsymbol{0}
	\label{eq:mf4_6}
\end{equation}
\noindent
On the other hand, if Eqs. \ref{eq:mf4_5} and \ref{eq:mf4_6} are plugged into Eq. \ref{eq:mf4_3}, the corresponding diabatic equation appears as:
\begin{equation}
	i\hbar\dfrac{\partial \boldsymbol{\Phi}}{\partial t} = \Big(-\dfrac{\hbar^2}{2m}\nabla^2+\boldsymbol{\tilde{W}}_e\Big)\boldsymbol{\Phi},
	\label{eq:mf4_7}
\end{equation}
where $\boldsymbol{\tilde{W}}_e(\boldsymbol{R},t)$ symbolizes the recovered field-dressed diabatic Hamiltonian:
\begin{equation}
	\boldsymbol{\tilde{W}}_e=\boldsymbol{\tilde{A}^\dagger\tilde{\tau}_t\tilde{A}}.
	\label{eq:mf4_8}
\end{equation}

\noindent
It is necessary to mention that the two approaches, namely, the time-dependent ordinary and field-dressed approaches produce
exactly the same diabatic potential indicating that the field-dressed approach have limited
advantages over the first straightforward simple field-free approach. The second approach becomes important in cases where
the approximations are used to cut down the number of SE equations (Eq. \ref{eq:mf4_8}), which reduces the computational time. Such approximations are thoroughly discussed in the next subsection.

\section{Approximations for Solution of ADT Matrix~\citep{PAUL_PR496}}

\noindent
Eq. \ref{eq:mf4_8} contains $L$ equations with the following $L$ unknown functions:

\begin{equation}
	\tilde{\Phi}^T=(\phi_1, \phi_2, \cdots, \phi_j,\cdots,\phi_L)^T
	\label{eq:mf5_1}
\end{equation}

\noindent
Eq. \ref{eq:mf4_8} is valid for the first $N$ equations ($N < L$) till the RHS is considered as full vector, $\boldsymbol{\tilde{\Phi}}$. While converting the reduced set of $N$ equalities into an ordinary system of $N$ equations, the lower tail of $\boldsymbol{\tilde{\Phi}}$ (last $L-N$ elements) can be considered as negligible and therefore, can be ignored. On the other hand, the squared system of $L$ equations (see Eq. \ref{eq:mf4_8}) appears as reduced squared system of $N$ equation as follows:

\begin{equation}
	i\hbar\dfrac{\partial \boldsymbol{\tilde{\Phi}}_r}{\partial t} = \Big(-\dfrac{\hbar^2}{2m}\nabla^2+\boldsymbol{\tilde{W}}_r\Big)\boldsymbol{\tilde{\Phi}}_r,
	\label{eq:mf5_2}
\end{equation} 

\noindent
Even though there are several ways to generate the reduced $N\times N$ diabatic PESs matrix, the present approach is more advantageous due to the formulation of $\boldsymbol{\tilde{W}}_r(\boldsymbol{R},t)$ from original unaltered $L \times L$ PESs matrix,   
\begin{equation}
	\boldsymbol{\tilde{W}}_r=\boldsymbol{\tilde{A}}_r^\dagger\boldsymbol{\tilde{\tilde{H}}}_e\boldsymbol{\tilde{A}}_r
	\label{eq:mf5_3}
\end{equation}

\noindent
where $\boldsymbol{\tilde{\tilde{H}}}_e=\boldsymbol{\omega}^\dagger\boldsymbol{\tilde{H}}_e\boldsymbol{\omega}$. In other words, the diabatic PESs matrix of reduced dimension, $\boldsymbol{\tilde{W}}_r(\boldsymbol{R},t)$ can be generated with the aid of a rectangular $N\times L$ ADT matrix, $\boldsymbol{\tilde{A}}_r(\boldsymbol{R},t)$. Since the original full matrix, $\boldsymbol{\tilde{H}}_e(\boldsymbol{R},t)$ is still employed, the errors appearing due to reduced final diabatic PESs matrix is negligible. In order to derive $\boldsymbol{\tilde{A}}_r$, one can assume a space-time contour, $\Gamma_{st}$ and the solution is propagated as per the following equation:

\begin{eqnarray}
	\boldsymbol{\tilde{A}}_r(\boldsymbol{R},t\vert \Gamma_{st})  &=& \varrho\, \Pi_{n=1}^{M}\Big[exp\Big\lbrace-\int_{{\boldsymbol{R}}_{n-1}}^{{\boldsymbol{R}}_{n}}\boldsymbol{\tilde{\tau}}(\boldsymbol{R},t_n)\cdot d\boldsymbol{R}\Big\rbrace \nonumber\\
	&&exp\Big\lbrace\dfrac{i}{\hbar}\int_{t_{n-1}}^{t_n}\boldsymbol{\tilde{\tilde{H}}}_e(\boldsymbol{R}_{n-1},t)dt\Big\rbrace\Big] \boldsymbol{\tilde{A}}_r(\boldsymbol{R}_0,t_0)
	\label{eq:mf5_4}
\end{eqnarray}

\noindent
where $M$ represents the number of space-time segments along the contour, $\Gamma_{st}$:
\begin{equation}
	\Gamma_{st} \equiv (\boldsymbol{R}_0,t_0) \rightarrow (\boldsymbol{R}_0 + \delta\boldsymbol{R},t_0 + \delta t) \rightarrow (\boldsymbol{R}_M, t_M-\delta t) \rightarrow  (\boldsymbol{R}_M, t_M).
	\label{eq:mf5_5}
\end{equation}

\noindent
Eq. \ref{eq:mf5_4} applies for one single contour, but the ADT matrix needs to be calculated for the entire space-time region. Even though there are numerous ways to construct such networks, some of them appear as more efficient than others, but all paths are expected to produce same results. In other words, all the paths are initiated from one single point ($\boldsymbol{R}_0, t_0$) with the same initial ADT matrix, $\boldsymbol{\tilde{A}}_r(\boldsymbol{R}_0,t_0)$. 

\section{Are External Fields Capable of Creating or Annihilating CIs in a Given Region of Space–Time Nuclear Configuration Space?}

\noindent
External fields can shift the potential energy curves or surfaces as well as can change the spatial distributions of NACTs over nuclear space. In other words, it has been explored that whether the strong external field is capable of removing or creating singularities. \\

\noindent
Lemma 1.1: A space–time region defined by a closed contour made-up of one Cartesian coordinate ($X$) and the time $t$ is not hosting any CI over $R-t$ plane.\\

\noindent
Proof: Generally, the Curl equation arises from the necessity that the ADT matrix, $\tilde{\boldsymbol{A}}(X,t)$ will be an analytic function with respect to
the space coordinates and time. It is interesting to note that the Curl equation:
\begin{equation}
	\boldsymbol{F}_{Xt}(X,t)=i\hbar\dfrac{\partial\tilde{\boldsymbol{\tau}}_X}{\partial t} - \dfrac{\partial}{\partial R}\boldsymbol{\tilde{\tilde{H}}}_e+[\boldsymbol{\tilde{\tilde{H}}}_e,\tilde{\boldsymbol{\tau}}_X]=\boldsymbol{0}
	\label{eq:mf6_1}
\end{equation}
is fulfilled at any point in the $X-t$ plane for any number of states. \\

\noindent
It is evident from the fulfillment of the $X-t$ Curl equation at each and every point over the chosen domain of $X-t$ plane that 
the pathological points (Curl expressions are invalid) are absent over the $X-t$ region. If the functions involved in the Curl 
equation ($\boldsymbol{\omega}(X,t)$, $\boldsymbol{\tilde{\tilde{H}}}_e(X,t)$ and $\boldsymbol{\tau}(X)$) are examined, those quantities
as well as their derivatives (see Eq. \ref{eq:mf6_1}) are well-defined at every point. On the contrary, CI point is a pathological point and therefore, CIs
are not included on the $X-t$ plane.  \\

\noindent
Lemma 1.2: If a planar, spacial region, $\Lambda_{XY}$ is defined by a closed contour constructed by two Cartesian coordinates ($X$, $Y$) and if one CI is present at time, $t=0$, one CI will present also at any later time, $t > 0$.\\

\noindent
Proof: As representative case, we assume that one CI is present at the origin in $\Lambda_{XY}$ so that the $t$ axis hosts such CI at
$\lbrace$$X=0$, $Y=0$, $t=0$$\rbrace$. The intention is to prove the appearance of similar CI at any later time $t>0$ (after activation of the field) along the $t$
axis. In order to validate such statement, a planar (spatial) region is constructed at time $t=t_s$ parallel to the $\lbrace$$X$, $Y$, $t=0$$\rbrace$ plane.
After that, the $t$-axis is surrounded with a small contour, $\Gamma_{XY}(t=t_s)$ over that plane and then, the spatial ($X$, $Y$) component of
the tensorial Curl equation is analyzed for any point enclosed by $\Gamma_{XY}(t=t_s)$. As per the analysis:
\begin{equation}
	\dfrac{\partial}{\partial Y}\tilde{\boldsymbol{\tau}}_X(\tilde{X},\tilde{Y}|t_s) - \dfrac{\partial}{\partial X}\tilde{\boldsymbol{\tau}}_Y(\tilde{X},\tilde{Y}|t_s)-[\tilde{\boldsymbol{\tau}}_X, \tilde{\boldsymbol{\tau}}_Y](\tilde{X},\tilde{Y}|t_s)=\boldsymbol{0}
	\label{eq:mf6_2}
\end{equation} 

\noindent
According to Lemma 1.1, one needs to look for the pathological points, where Eq. \ref{eq:mf6_2} is not valid. Such equation encounters three matrices, namely,
$\boldsymbol{\omega}(\tilde{X},\tilde{Y}|t_s)$, $\tilde{\boldsymbol{\tau}}_X(\tilde{X},\tilde{Y})$ and $\tilde{\boldsymbol{\tau}}_Y(\tilde{X},\tilde{Y})$
forming $\tilde{\boldsymbol{\tau}}_u(\tilde{X},\tilde{Y}|t_s)$; $(u = X,Y)$. It is interesting to note that $\boldsymbol{\omega}(\tilde{X},\tilde{Y}|t_s)$ 
represents a regular matrix, but the other two are singular at the origin ($(\tilde{X},\tilde{Y})=(0,0)$) at initial time ($t=0$) as well as later time ($t=t_s$), since
those are time-independent. As a consequence, Eq. \ref{eq:mf6_2} is well-defined at every point except for the point of singularity, $(\tilde{X},\tilde{Y},t_s)=(0,0,t_s)$, indicating the existence of a CI (single CI) at this point on the above mentioned plane at $t=t_s$. Such finding unveils that, for the case, the t-axis generates a (virtual) seam with infinite number of CI points. \\

\noindent
Lemma 1.3: If a planar, spatial region $\Lambda_{XY}$, defined by a closed contour constituted with two Cartesian coordinates ($X$, $Y$) and  if numerous CIs are present at $n$ points ($\tilde{X}_j$, $Y_j$, $j$ = $\lbrace 1, n\rbrace$) at time, $t=0$, same number of CIs can be observed at any later time $t=t_s$.\\

\noindent
Proof: A situation is considered at $t=t_s$, where Eq. \ref{eq:mf6_2} is still valid except at the $n$ pathological points ($\tilde{\boldsymbol{\tau}}_u(\tilde{X},\tilde{Y}|t_s)$; $(u = X,Y)$; $j$ = $\lbrace 1, n\rbrace$ are not well-defined). While applying Eq. \ref{eq:mf3_4} for the present $n$ points, we obtain:

\begin{equation}
	\boldsymbol{\tilde{\tau}}(\tilde{X},\tilde{Y},t_s) = \boldsymbol{\omega}^\dagger\boldsymbol{\tau}(\tilde{X},\tilde{Y})\boldsymbol{\omega}+
	\boldsymbol{\omega}^\dagger\nabla\boldsymbol{\omega},
\end{equation}

\noindent
As per the above equation, all pathological points are encountered at any time, which already exist at $t=0$. In summary, pathological points can neither be formed nor annihilated by 
applying an external field.

\section{Numerical Verification of CIs in Presence of External Field}

\noindent
The potential term due to the external field ($\boldsymbol{E}(\boldsymbol{r}\vert\boldsymbol{R},t)$) for a single electron system can be expressed as a line integral in the electronic space:
\begin{equation}
	U_E(\boldsymbol{r}\vert\boldsymbol{R},t)=-\dfrac{e}{\hbar}\int_{\Gamma_e}d\boldsymbol{r}\cdot\boldsymbol{E}(\boldsymbol{r}\vert\boldsymbol{R},t)
	\label{eq:mf5e_1}
\end{equation}

\noindent
Here, $e$ represents the electronic charge and $\Gamma_e$ signifies the corresponding electronic contour. On the other hand, the electronic Hamiltonian in presence of external field interacting with electrons takes the following form:
\begin{equation}
	\boldsymbol{H}_e(\boldsymbol{r}\vert\boldsymbol{R},t)=\boldsymbol{H}_{e0}(\boldsymbol{r}\vert\boldsymbol{R})+U_E(\boldsymbol{r}\vert\boldsymbol{R},t)
	\label{eq:mf5e_2}
\end{equation}

\noindent
Prior to discuss any numerical example, one needs to specify the
intense/strong field, which indicates the population of
several electronic time independent states. For the present case, the field is capable for populating four states, whereas incorporation of even ten states reaches to the same outcome.  \\

\noindent
In order to carry out the numerical calculations, a model is considered
based on the Mathieu equation:
\begin{equation}
	\Big(-\dfrac{1}{2}\dfrac{d^2}{d \theta^2}-\rho\cos(2\theta-\phi)-u_j(\phi,\rho)\Big)\xi_j(\theta\vert\phi,\rho)=0
	\label{eq:mf5e_3}
\end{equation}
\noindent
where $\theta$ represents an electronic coordinate, and $\phi$ and $\rho$ signify the nuclear coordinates. It is important to note that
the variable $\rho$ shows linear dependence on the
nuclear radial coordinate. Generally $\rho$ is kept as constant so that
the space segments of the contour appear only as a function
of $\phi$. The detailed solution of Eq. \ref{eq:mf5e_3} are
present elsewhere~\citep{baer_book,englman_ijqc90}.\\

\noindent
Once the Mathieu equation is introduced, the next task is to propose the electric field and the associated electric potential, which are employed in this model. In order to present those equations, the electronic contour, $\Gamma_e$ is considered to be a circle so that the associated electronic coordinate becomes same as previously mentioned angular vector, $\theta$. As a consequence, Eq. \ref{eq:mf5e_1} is integrated along the angular (tangential) component, whereas the planar components of the field appears as:

\begin{subequations}
	\begin{eqnarray}
		E_{x_e}(x_e,y_e\vert t) &=& -x_e\tilde{E}(t)\\
		E_{y_e}(x_e,y_e\vert t) &=& y_e\tilde{E}(t).
	\end{eqnarray}
	\label{eq:mf5e_4}
\end{subequations}

\noindent
Such formulations guarantee that curl $\boldsymbol{E}$ = 0 and
div $\boldsymbol{E}$ = 0. Employing such choice, the tangential
component of the field appears as:
\begin{equation}
	E_\theta(\theta,t)=q_e^2\tilde{E}(t)\sin 2\theta.
	\label{eq:mf5e_5}
\end{equation} 

\noindent
Substituting this expression in Eq. \ref{eq:mf5e_1}, one can obtain the following functional form of the electric potential:
\begin{equation}
	U_E(\theta\vert\phi,t)(\equiv U_E(\theta\vert t))=-\dfrac{e}{2\hbar}q_e^2\tilde{E}(t)\cos 2\theta
	\label{eq:mf5e_6}
\end{equation}

\noindent
where $q_e$ represents the electronic radial coordinate. On the other hand, $\tilde{E}(t)$ takes the following form:
\begin{equation}
	\tilde{E}(t)=E_0\sin^2(\pi t/T)\sin(2\pi t/T),
	\label{eq:mf5e_7}
\end{equation}

\noindent
where $T$ represents the cycle of the field and $E_0$ contains the
other constants in Eq. \ref{eq:mf5e_6}. In all the calculations,
the field intensity is taken as $E_0=20$, which is sufficient to populate several electronic states. \\

\noindent
In the space-time framework (space and time segments are
mixed), the $\boldsymbol{D}$-matrix (or $\boldsymbol{A}$ matrix) needs be calculated using Eq. \ref{eq:mf5_4}, where a series of products of exponentiated line integrals
is taken ($M$ in Eq. \ref{eq:mf5_4} is considered as 4) over the space coordinate, $\phi$ and time coordinate, $t$. In order to compute the 
$\boldsymbol{D}$ matrix elements, one needs to divide $\phi$ and $t$ into small 
sub-segments, and diagonalize $\tilde{\boldsymbol{\tau}}(\phi,t)$ and
$\tilde{\boldsymbol{H}}_e(\phi,t)$ along a grids of angle and time, respectively. Figure \ref{fig:mf_8} presents the fourth eigenvalues, $\tilde{\tilde{\boldsymbol{\tau}}}(\phi,t)$ and $\tilde{\tilde{\boldsymbol{H}}}_e(\phi,t)$ originated from $\tilde{\boldsymbol{\tau}}(\phi,t)$ and
$\tilde{\boldsymbol{H}}_e(\phi,t)$, respectively along $\tilde{\Gamma}$. The red-coloured and grey components represent $\tilde{\tilde{\boldsymbol{\tau}}}(\phi,t)$ and $\tilde{\tilde{\boldsymbol{H}}}_e(\phi,t)$, respectively. On the other hand, the contour, $\tilde{\Gamma}$ is provided in the inset.  \\

\noindent
If one assumes the spatial segments of $\tilde{\Gamma}$, (namely, 
those associated with the fixed-time segments of the contour) to be identical to the field-free, time-independent one, ${\Gamma}$, the 
two $\boldsymbol{D}$-matrices appearing from field-free and field-dressed cases 
should be identical if and only if the number
and/or the positions of the CIs is not affected by the field. Such outcome is reflected in Figure \ref{fig:mf_8}. While repeating such calculations for any
space-time contour in the assigned domain of nuclear CS, one obtains the same $\boldsymbol{D}$ matrices for both cases (field-dressed and field-free), which validates the outcome presented in Figure \ref{fig:mf_8}. Along all the contours,
even a large number of space/time (i.e., $\phi$/$t$) flippings take place, the final $\boldsymbol{D}$-matrix appears as the unit matrix $-\boldsymbol{I}$. It is important to note that an identical $\boldsymbol{D}$ matrix is computed for the field-free Mathieu equation.

\begin{figure}[!htp]
	\centering
	\includegraphics[width=1.0\textwidth,angle=0]{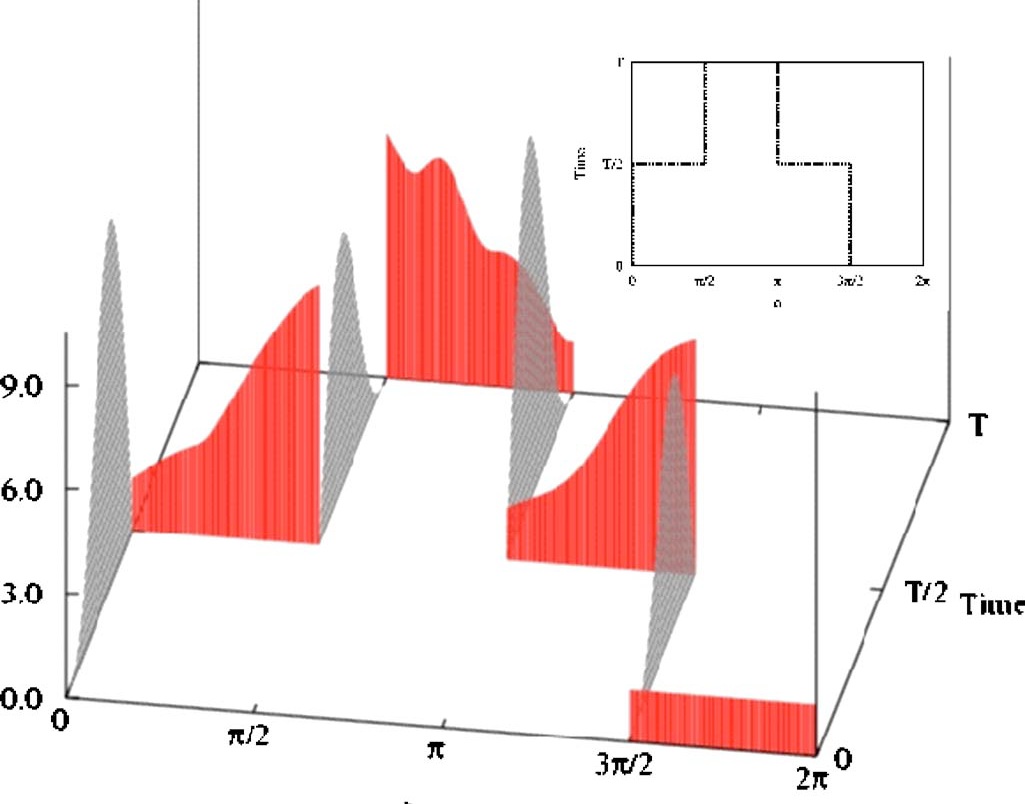}
	\caption{The fourth eigenvalues $\tilde{\tilde{\boldsymbol{\tau}}}(\phi,t)$ and $\tilde{\tilde{\boldsymbol{H}}}_e(\phi,t)$ associated with two matrices, $\tilde{\boldsymbol{\tau}}(\phi,t)$ and
		$\tilde{\boldsymbol{H}}_e(\phi,t)$ are presented along a closed contour, $\tilde{\Gamma}$. The red and grey area show $\tilde{\tilde{\boldsymbol{\tau}}}(\phi,t)$ and $\tilde{\tilde{\boldsymbol{H}}}_e(\phi,t)$, respectively. The contour,
		$\tilde{\Gamma}$ is given in inset. This figure has been taken from Ref. [\protect\citen{sarkar_jcp126}] and permission has been taken from AIP.}
	\label{fig:mf_8}
\end{figure}

\bibliographystyle{aipnum4-1}
\bibliography{ref_book}

\chapter{Spectroscopic Systems: BBO Based Diabatic Hamiltonian and Quantum Dynamics}\label{sec:spec}

\noindent
Vibronic Coupling Model (VCM)\citep{kdc_acp57} had been developed to derive
diagonal and off-diagonal elements of diabatic Hamiltonian to generate photodetachment (PD)/photoelectron (PE)/photoabsorption (PA)/Mass-Analyzed Threshold Ionization (MATI) spectra originating from multi-mode multi-electronic systems. While developing such model,\citep{koppel_jcp110,koppel_jcp150,koppel_mp104,viel_jcp120,bhatt_jpca118} adiabatic PESs are 
expanded in Taylor series with respect to the ground adiabatic PES ($V_0(\textbf{R}_0)$) minima or CI ($\textbf{R}_0$) and then, the coefficients
of linear and bilinear coupling terms are obtained by fitting with the functional forms of couplings calculated in diabatic basis. Zeng {\it et al.}\citep{hickman_pccp20,issaka_pccp21} generalized the VCM based quasi-diabatic Hamiltonian 
upto arbitrary order and implemented the formalism for trigonal and tetragonal systems. Since VCM based quasi-diabatization approach is based on 
Taylor series expansion, such model, in principle, is applicable only in a limited domain of Frank-Condon
(FC) region close to the reference geometry $\textbf{R}_0$. On the contrary, the construction of diabatic Hamiltonian employing Beyond Born-Oppenheimer (BBO) theory is based on {\it ab initio} calculated adiabatic potential energy surfaces (PESs) and non-adiabatic coupling terms (NACTs) among all the coupled electronic states and vibrational degrees of freedom (DOFs) over the entire/interested domain of nuclear configuration space (CS), which can not only include the effect of symmetry-driven as well as accidental Jahn-Teller (JT) conical intersections (CIs) and Renner-Teller (RT) interactions, but also locate their positions in the CS within highest level of accuracy.\\

\noindent
The first principle based BBO formalism has been employed to construct diabatic Hamiltonian  
for various spectroscopically interesting molecules~\citep{mukherjee_jpca117,mukherjee_cp440,sardar_cp416,mukherjee_jcp143,
	soumya_jpca121,bijit_mp115,soumya_acs,bijit_cp515,
	soumya_jcp150,soumya_comptc,bijit_jpcs1148,bijit_irpc38,joy_pccp,soumya_bz,soumya_tfb,saikat_pyra} exhibiting a wide variety of nonadiabatic interactions, namely, JT\citep{JT_prs161,bersuker_book,Pablo_prb71,Pablo_prl96,Hizhnyakov_jcp147,Hizhnyakov_cpl525} CIs/seams (symmetry driven~\citep{bijit_mp115} degeneracies), JT type accidental CIs/seams (symmetry allowed~\citep{bijit_mp115} or no symmetry~\citep{bijit_mp115} degeneracies), RT\citep{renner_zp92,Hizhnyakov_jcp147} 
interactions and PJT\citep{bersuker_book} interactions. In the last two decades, the BBO based diabatic Hamiltonians (BBO-DH) of 
alkali trimers (Na$_3$, K$_3$),\citep{mukherjee_jpca117,mukherjee_cp440} nitrogen oxides (NO$_{2}$, 
NO$_{3}$),\citep{mukherjee_jcp143,bijit_cp515} benzene radical cation 
(C$_6$H$_6^+$),~\citep{soumya_jpca121,soumya_bz} 1,3,5-trifluorobenzene radical cation (1,3,5-C$_6$H$_3$F$_3^+$),\citep{soumya_jcp150,soumya_tfb} pyrazine 
molecule (C$_4$N$_2$H$_4$)~\citep{saikat_pyra} and ortho-difluorobenzene (o-C$_6$H$_4$F$_2^+$)~\citep{saikat_cpc26} were constructed to calculate the PD/PE/PA/MATI spectra of their neutral/anionic counterpart using Time-Dependent Discrete Variable Representation (TDDVR)~\citep{SSMP107,SSJPCA118,SSPCCP13,SSJTCC12,SSPCCP10,subhankar_ijqc111,subhankar_jcp130,souvik_irpc37,soumya_bz,soumya_tfb,saikat_pyra} approaches.   \\

\noindent
In order to carry out quantum dynamical calculations for predicting photoelectron/photoabsorption/photo-\\
emission/MATI spectra of realistic molecular species, TDDVR methodology appears as the most suitable approach due to involvement of variable number of time-dependent basis functions along different nuclear coordinates depending upon their role on simultaneous ongoing classical and quantum dynamics. In the recent years, quantum dynamics of
nonadiabatic processes~\citep{GHH,PPP,JC1,WHM,YAR,MBB,BS1} has been receiving attention of the allied scientific community, specifically for relatively large molecular systems. Though theoretical investigation of nonadiabatic dynamics involving CI(s) requires rigorous quantum mechanical (QM) treatment
\cite{GSM}, traditional full QM treatment appears as impractical even for three - or four 
- particle systems, especially when the atoms are heavy, the energy is high and multiple PESs are involved. In order to get rid of such shortcomings, 
a quantum-classical molecular dynamics approach~\cite{SACPL309,SAJCP11148,SACP300} were implemented to include quantum correction to the classical mechanics leading to a 
first principle based theory of dynamics. Such a theory can efficiently distinguish dynamical strong quantum regimes of a system from quasi-classical one, where the later incorporates mild quantum corrections as per the requirement. The quantum-classical approaches with time-dependent basis functions/grid-points effectively attain convergence very fast saving a huge computation time. \\

\noindent
At the end of the nineties of the last century, Adhikari and Billing~\citep{SAJCP113,SACPL321} introduced a unique approach, namely, TDDVR methodology to perform multi-state (electronic) multi-dimensional (vibrational) quantum dynamics considering all the parameters of
Gauss-Hermite (GH) basis functions as time dependent. Later on, such approach was reformulated~\citep{SAJCP118,SACP300,souvik_irpc37} using time-independent width parameters of GH basis functions, where rest of the parameters, viz., center of the wavepacket and its momentum were introduced as time-dependent. In this formalism, blending of classical and quantum dynamics effectively becomes beneficial to reduce the necessity of higher number 
of basis functions (grid points) for each mode in contrary to Fast-Fourier Transformation (FFT) based algorithms. Over the decades, TDDVR dynamics were being carried out over VCM based quasi-diabatic PESs and couplings to generate high resolution spectral envelops of a variety of prototypical multi-mode multi-surface molecular systems, viz., pyrazine,~\citep{SAJCP123,SAJCP125} butatriene,~\citep{SSPCCP10} 
benzene,~\citep{subhankar_ijqc111,subhankar_jcp130,subhankar_jcs124,soumya_bz} allene,~\citep{SSMP107} monofluoro 
benzene,~\citep{SSCPL496} isomers of difluoro benzene,~\citep{SSPCCP13} trifluoro 
acetonitrile~\citep{SSJTCC12} and many others. In addition, the better accuracy and computational advantage of the TDDVR approach over certain versions of the Multi-Configuration Time-Dependent Hartree (MCTDH) method were depicted for two well-known aromatic molecules, napthalene and anthracene.~\citep{SSJPCA118} Adhikari and coworkers have executed TDDVR dynamics using BBO generated diabatic Hamiltonians of benzene radical cation (C$_6$H$_6^+$),~\citep{soumya_bz}  trifluorobenzene radical cation (1,3,5-C$_6$H$_3$F$_3^+$),~\citep{soumya_tfb} pyrazine (C$_4$H$_4$N$_2$),~\citep{saikat_pyra} and ortho-difluorobenzene (o-C$_6$H$_4$F$_2^+$)~\citep{saikat_cpc26}
compute the PE / PA / MATI spectra of their neutral or higher negative analogue. Once those quantities (adiabatic PESs and NACTs) are in hand, those are employed in BBO based diabatization and the newly generated first principle based diabatic Hamiltonian is used for the execution of dynamical calculations with the aim of generating PE/PA/MATI spectra. 

\noindent
\section{Outline of BBO Theory for Spectroscopic Processes}	

\noindent
Since the molecular properties originate from the total electron-nuclear 
Hamiltonian, one can start the formulation of BBO theory from the time independent molecular SE. Even though an extensive discussion
on the theoretical background of such theory is done in Chapter 4, we present a brief outline of such formalism here for the sake of completeness as well as to bring its compatibility for constructing diabatic Hamiltonian to calculate molecular spectra. Such spectroscopically interesting Hamiltonian
involves multiple electronic states as well as multiple nuclear degrees of freedom (normal modes).
The formalism starts from the following time independent molecular SE:
\begin{eqnarray}
	\hat{H}(\textbf{q}_e,\textbf{Q}_n)\Psi(\textbf{q}_e,\textbf{Q}_n) = E\Psi(\textbf{q}_e,\textbf{Q}_n), 
	\label{eq:mse_1}
\end{eqnarray}
where $\textbf{q}_e$ and $\textbf{Q}_n$ collectively represent the electronic and nuclear coordinate vectors, 
respectively. In the above equation, the non-relativistic molecular Hamiltonian [$\hat{H}(\textbf{q}_e,\textbf{Q}_n)$] can be partitioned into two parts, namely, nuclear kinetic energy operator [$\hat{T}_{\rm nuc}(\textbf{Q}_n)$] and electronic 
Hamiltonian [$\hat{H}_{\rm{el}}(\textbf{q}_e;\textbf{Q}_n)$] as follows,
\begin{eqnarray}
	\hat{H}(\textbf{q}_e,\textbf{Q}_n) = \hat{T}_{\rm nuc}(\textbf{Q}_n) + \hat{H}_{\rm{el}}(\textbf{q}_e;\textbf{Q}_n).
	\label{eq:hampart}
\end{eqnarray} 

\noindent 
On the other hand, well known Born-Oppenheimer-Huang (BOH) expansion\citep{BO,BH} tells that total molecular wavefunction [$\Psi(\textbf{q}_e,\textbf{Q}_n)$] for any $n$-state SHS achieves the following form:

\begin{equation}
	\Psi(\textbf{q}_e,\textbf{Q}_n) = \sum_{i=1}^n \psi_i^{\rm{ad}}(\textbf{Q}_n) \xi_i(\textbf{q}_e;\textbf{Q}_n), 
	\label{eq:bo-expansion_1} 
\end{equation}

\noindent
where the combining coefficients \{$\psi_i^{\rm{ad}}(\textbf{Q}_n)$\} represent nuclear wavefunctions. In the above BOH expanded molecular 
wavefunction, the eigenfunctions of the electronic SE can be presented as
\{$\xi_i(\textbf{q}_e;\textbf{Q}_n)$\}:

\begin{equation}
	\hat{H}_{\rm{el}}(\textbf{q}_e;\textbf{Q}_n)\xi_i(\textbf{q}_e;\textbf{Q}_n) = u_i(\textbf{Q}_n)
	\xi_i(\textbf{q}_e;\textbf{Q}_n), \qquad  
	\langle \xi_i | \xi_j \rangle_{\textbf{q}_e} = \delta_{ij},
	\label{eq:elese_1}
\end{equation}

\noindent
The adiabatic PESs ($u_i(\textbf{Q}_n)$) originate from the eigenvalues of electronic SE (\{$u_i(\textbf{Q}_n)$\}), whereas the integral, $ \langle \xi_i | 
\xi_j \rangle_{\textbf{q}_e}$ depicts the inner product between two electronic states over electronic 
coordinates only.\\

\noindent
If Eqs. \ref{eq:hampart}, \ref{eq:bo-expansion_1} and \ref{eq:elese_1} are plugged in the time 
independent molecular SE (Eq. \ref{eq:mse_1}) followed by projecting with various electronic eigenfunctions, 
the kinetically coupled (adiabatic) nuclear SE takes the following matrix form:

\begin{equation}
	-\frac{\hbar^2}{2}\left(\boldsymbol{\nabla}_{\textbf{Q}_n}+\boldsymbol{\tau}\right)^2\psi^{\rm{ad}} + \left(U-E\right)
	\psi^{\rm{ad}} = 0,
	\label{eq:adsem}
\end{equation}
where $U_{ij}(\textbf{Q}_n)\,(=u_j(\textbf{Q}_n)\delta_{ij})$ symbolizes the adiabatic
PESs matrix elements. On the other hand, for $n$ electronic state sub-Hilbert space, the skew-symmetric NACM can be written as:
\begin{equation}\label{eq:taumatrix}
	\boldsymbol{\tau} = 
	\begin{pmatrix}
		0 & \boldsymbol{\tau}_{12} & \boldsymbol{\tau}_{13} & \cdots\\
		-\boldsymbol{\tau}_{12} & 0.& \boldsymbol{\tau}_{23} & \cdots \\
		- \boldsymbol{\tau}_{13} & -\boldsymbol{\tau}_{23} & 0 & \cdots \\
		\cdots & \cdots & \cdots & \cdots
	\end{pmatrix}
\end{equation}
\noindent
It is interesting to note that the NACM elements exhibits singular behavior at the close neighborhood of CI point(s) or along seam(s) of degeneracy 
between two electronic states. Such singular characteristics can be unveiled from the following expression as per Hellmann-Feynman theorem:\citep{hellmann,feynman_pr56}
\begin{eqnarray}
	\boldsymbol{\tau}_{ij} = \dfrac{\langle\xi_i\vert\boldsymbol{\nabla}_{\textbf{Q}_n} 
		\hat{H}_{\rm{el}} \vert \xi_j\rangle_{\textbf{q}_e}}{u_j-u_i}\,\, .
	\label{eq:hf}
\end{eqnarray}

\noindent
In order to overcome the bottleneck of such singularity(ies), a paradigm shift from adiabatic to diabatic SE can be done using the transformation $\psi^{\rm ad}=A\psi^{\rm dia}$ leading to the orthogonal features of $A$ for the chosen SHS. Using such transformation, one can arrive the following matrix form of diabatic nuclear SE:
\begin{eqnarray}
	\left(-\dfrac{\hbar^2}{2}\nabla_Q^2 + W-E\right)\psi^{\rm{dia}} = 0, \quad W= A^{\dagger}U A
	\label{eq:diabse}
\end{eqnarray}
invoking the ADT condition:
\begin{eqnarray}
	\boldsymbol{\nabla}_{\textbf{Q}_n}A + \boldsymbol{\tau}A = 0.
	\label{eq:adteq}
\end{eqnarray}

\noindent
The explicit expression of ADT equations for two and three electronic states are presented in Eqs. \ref{eq:adt-two} and \ref{eq:adt-three}, whereas the ADT equations for four to six states are presented in Appendix~\ref{ssec:adt-eq}. \\

\noindent
In order to carry out the BBO based diabatization, {\it ab initio} computed NACTs are substituted into ADT equations and then, those equations are numerically solved to obtain 
the ADT angles. Once the mixing angles are in hand, those are used to form the ADT matrices and thereby, similarity transformation (Eq. \ref{eq:diabse}) is executed for the construction of smooth, single-valued and continuous diabatic PESs and couplings.

\noindent
\section{Theoretical Background of TDDVR Methodology}\label{ssec:tddvr}

\noindent
Once diabatic Hamiltonian is constructed, those matrices (PESs and couplings) can be used in nuclear dynamical calculations. In this context, TDDVR approach appears as the most suitable one. Such formalism can be applied for low dimensional model problems~\citep{SSJCS122,
	SSMP107} as well as multi-dimensional multi-surface realistic systems,~\citep{subhankar_ijqc111,subhankar_jcp130,subhankar_jcs124,SSJPCA118} equally, which endorses
that the complexity of the Hamiltonian is not a matter of concern. In case of a multi-dimensional multi-surface molecular system,
the time dependent Schr\"odinger Equation (TDSE) in the diabatic representation takes the following form: 
\begin{eqnarray}
	i\hbar \frac{\partial}{\partial t}\Xi (\{Q_k\},t) = [{\bf \widehat{T}}_{nuc} {\{Q_k\}} 
	+ {\bf \hat{W}}(\{Q_k\})] \Xi (\{Q_k\},t), \label{eq:tdse2}
\end{eqnarray}
where ${\bf \hat{T}}_{nuc} {\{Q_k\}}$ (= $\hat{T}_{nuc} {\{Q_k\}} \cdot {\bf I}$) and 
${\bf \hat{W}}(\{Q_k\})$ represent kinetic energy operator and diabatic PESs matrices, respectively, which 
are functions of the normal mode coordinates, ${\{Q_k\}}$. On the other hand, the nuclear wavefunction for $l$ 
electronic states appears as: 
\begin{eqnarray}
	\Xi(\{Q_k\},t)\equiv \left( \begin{array}{c}
		\psi^d_1 (\{Q_k\},t)\\
		\psi^d_2 (\{Q_k\},t)\\
		\ \ \vdots \\
		\psi^d_l (\{Q_k\},t)\\
		\ \ \vdots \\
	\end{array}\right) \label{eq:2dwf}
\end{eqnarray}
where $\int\Xi^{\dagger}(\{Q_k\},t) \Xi (\{Q_k\},t)\prod_{k=1}^p dQ_k$= 1 holds true at any time $t$. While expanding the 
$l^{th}$ element of the above column matrix (see Eq. (\ref{eq:2dwf})), i.e., the 
wavefunction for the $l^{th}$ surface ($\psi^d_l(\{Q_k\},t)$), one can obtain the following form,
\begin{equation}
	\psi^d_l(\{Q_k\},t)=\sum_{i_1i_2...i_p} c^d_{i_1i_2....i_p,l}(t) \prod_{k=1}^p \chi_{i_k}(Q_k,t),
	\label{eq:psil2}
\end{equation}
where $p$ represents the number of normal modes and $\chi_{i_k}(Q_k,t)$ symbolizes the TDDVR basis function associated with $k^{th}$ normal mode. On the 
other hand, that basis function is originated from multiplication of Discrete Variable Representation (DVR) basis and time-dependent plane wave as follows,
\begin{eqnarray}
	\chi_{i_k}(Q_k,t)&=& \phi(Q_k,t) \sum_{n=0}^{N_k} \zeta_n^{\ast}(x_{i_k}) \zeta_n(x_k) \label{eq:psiik2d}\\
	&=& \sum_{n=0}^{N_k} \zeta_n^{\ast}(x_{i_k}) \Phi_n(Q_k,t), \label{eq:xiphi}
\end{eqnarray}
where	
\begin{eqnarray}
	\phi(Q_k,t) &=& \pi^{1/4} \exp \Big( \frac{i}{\hbar}\big\{ p_{Q^c_k}(t) [Q_k - Q^c_k(t)]\big\}\Big).
\end{eqnarray}
In Eq. \ref{eq:psiik2d}, the following harmonic oscillator eigenfunctions can be used as the primitive bases to construct DVR functions,
\begin{eqnarray}
	\zeta_n (x_k) &=& \Big(\frac{2ImA_k}{\pi\hbar}\Big)^{1/4} \frac{1}{\sqrt{n!2^n \sqrt\pi}} 
	\exp \{-(x_k)^2/2\} H_n(x_k), \label{eq:xin}
\end{eqnarray}
where  
\begin{eqnarray}
	x_k &=& \sqrt{\frac {2ImA_k}{\hbar}}(Q_k - Q_k^c(t)). \label{eq:xk}
\end{eqnarray}
\noindent	
In a similar way, $x_{i_k}$	takes the following form,
\begin{eqnarray}
	x_{i_k}&=& \sqrt{\frac {2ImA_k}{\hbar}}(Q_{i_k}(t) - Q^c_k(t)).  \label{eq:xiik}
\end{eqnarray}


\noindent
It is interesting to note that throughout the above expressions (Eqs. (\ref{eq:psil2}) - (\ref{eq:xiik})), the center of the wavepacket 
($\{Q^c_k\}$) and its momentum ($\{p_{Q^c_k}\}$) are considered as time-dependent, whereas imaginary part of 
the width ($\{ImA_k\}$) is taken as time-independent.~\citep{SAJCP118} In Eq. \ref{eq:xiphi}, the Gauss-Hermite (G-H) basis
for the $k^{th}$ mode is denoted as $\Phi_n(Q_k,t)$ satisfying the orthonormal condition,
\begin{equation}
	\int dQ_k \Phi_m^{\ast} (Q_k,t) \Phi_n(Q_k,t) = \delta_{mn}^k, \label{eq:dskp_ortho}
\end{equation}
\noindent	
where GWP is the ground state of G-H basis. On the other hand, the TDDVR 
basis functions, $\chi_{i_k}$s, in Eq. (\ref{eq:psiik2d}) for the $k^{th}$ mode constitute an orthogonal set, but 
not normalized,
\begin{equation}
	\int dQ_k \chi_{i_k}^{\ast}(Q_k,t) \chi_{i_k^{\prime}}(Q_k,t)= \delta_{i_ki_k^{\prime}}
	A_{i_ki_k^{\prime}}^k, \label{eq:dskp}
\end{equation}
where prefactor $A^k$ indicates the normalization factor. If Eq. (\ref{eq:xiik}) is rearranged, one can obtain
the TDDVR grid-point, $Q_{i_k}$ using the root ($x_{i_k}^k$) of $N_k^{th}$ Hermite polynomial,
$H_{N_k}(x_k)$~\citep{SACPL321}: 
\begin{eqnarray}
	Q_{i_k}(t) = Q^c_k(t) + \sqrt{\frac{\hbar}{2 Im A_k}} x_{i_k}.  \label{eq:qik2d}
\end{eqnarray}
\noindent	
Even though roots $(x_{i_k} s)$ are fixed values, $Q_{i_k}$s appear as
time-dependent due to involvement of time in the variables, $Q^c_k(t)$.\\

\noindent
If the TDDVR 
representation of wavefunctions (Eqs.  (\ref{eq:2dwf}) - (\ref{eq:xiik})) are plugged in the TDSE (Eq. 
(\ref{eq:tdse2})), the classical path picture as well as quantum equation of motion appear simultaneously.
For the quantum equation of motion on the $l^{th}$ PES, the following matrix form of TDDVR equation can be obtained,
\begin{eqnarray}
	i\hbar {\bf A \dot{C_l^d}} &=&{\bf H^t_{ll} C_l^d+ A \sum_{l^{\prime}\neq l} W_{ll^{\prime}} C^d_{l^{\prime}}} \label{eq:com1}
\end{eqnarray}
which can be transformed into a more compact form with the help of a similarity transformation,
\begin{eqnarray}
	i\hbar {\bf \dot{D_l^d}(t) = A^{-1/2} H^t_{ll} A^{-1/2} D_l^d+\sum_{l^{\prime}\neq l} 
		W_{ll^{\prime}} D^d_{l^{\prime}}}
	\label{eq:ihbard_p}
\end{eqnarray}
considering ${\bf D_l^d = A^{1/2}C_l^d}$. The superscript `d' specifies the diabatic representation and the elements of ${\bf D_l^d}$ are denoted as $d^d_{i_1i_2....i_p,l}$. The expressions of the TDDVR coefficients, 
$d^d_{i_1i_2....i_p,l}$ take the following form,
\begin{eqnarray}
	i\hbar \dot{d}^d_{i_1i_2....i_p,l} &=& \frac{1}{2}\Big\{\sum_k\dot{p}_{Q^c_k}
	\sqrt{\frac{\hbar}{ImA^k}} \bar{X}_{i_ki_k}^k\Big\}d^d_{i_1i_2....i_p,l}
	+ \Big\{\sum_k \frac{\mu (\dot{s}_c^k)^2}{2}\Big\} d^d_{i_1i_2....i_p,l}\nonumber\\
	&+& W_{ll}(i_1i_2....i_p) d^d_{i_1i_2....i_p,l}
	+ \sum_{l^{\prime} \neq l} W_{ll^{\prime}}(i_1i_1....i_p) d^d_{i_1i_2....i_p,l^{\prime}}\nonumber\\
	&+& \sum_k\Big\{\frac{\hbar Im A^k}{2\mu} \sum_{i_1^{\prime}i_2^{\prime}....i_p^{\prime}}
	\big\{\bar{Y}_{i_ki_k^{\prime}}^k\big\}
	d^d_{i_1^{\prime}i_2^{\prime}...i_p^\prime,l} \prod_{k^{\prime}\neq k}^p 
	\delta_{i_{k^{\prime}}i_{k^{\prime}}^{\prime}}\Big\}, \label{eq:ihbard}
\end{eqnarray}
where
\begin{eqnarray}
	\bar{X}_{i_k,{i_k^{\prime}}}^k=\frac{X_{i_k,{i_k^{\prime}}}^k}{\sqrt{A_{i_k,i_k}^k
			A_{i_k^{\prime},i_k^{\prime}}^k}}, \hspace{0.2cm} 
	\bar{Y}_{i_k,i_k^{\prime}}^k=\frac{Y_{i_k,i_k^{\prime}}^k}
	{\sqrt{A_{i_k,i_k}^k A_{i_k^{\prime},i_k^{\prime}}^k}}, \hspace{0.2cm} =\frac{2E_{i_k,i_k^{\prime}}^k-F_{i_k,i_k^{\prime}}^k}
	{\sqrt{A_{i_k,i_k}^k A_{i_k^{\prime},i_k^{\prime}}^k}} \nonumber
\end{eqnarray}

\begin{eqnarray}
	d^d_{i_1i_2....i_p,l} &=& c^d_{i_1i_2....i_p,l}\prod_{k=1}^p (A_{i_k,i_k}^k)^{\frac{1}{2}}, \nonumber\\
	A_{i_k,i_k^{\prime}}^k &=& \sum_{n=0}^{N^k} \xi_n^{\ast}(x_{i_k}^k) \xi_n(x_{i_k^{\prime}}^k), \nonumber\\
	X_{i_k,i_k^{\prime}}^k&=&\sum_{n=0}^{N^k-1} 
	\xi_{n+1}^{\ast}(x_{i_k}^k) \sqrt{n+1} \xi_n(x_{i_k^{\prime}}^k)
	+\sum_{n=1}^{N^k} \xi_{n-1}^{\ast}(x_{i_k}^k) \sqrt{n} \xi_n(x_{i_k^{\prime}}^k), \nonumber\\
	E_{i_k,i_k^{\prime}}^k&=&\sum_{n=0}^{N^k} \xi_n^{\ast}(x_{i_k}^k) 2 n \xi_n(x_{i_k^{\prime}}^k), \nonumber\\
	F_{i_k,i_k^{\prime}}^k&=&\sum_{n=0}^{N^k-2} \xi_{n+2}^{\ast}(x_{i_k}^k)
	\sqrt{(n+1)(n+2)}\xi_n(x_{i_k^{\prime}}^k)
	+\sum_{n=2}^{N^k} \xi_{n-2}^{\ast}(x_{i_k}^k) \sqrt{n(n-1)}\xi_n(x_{i_k^{\prime}}^k)\nonumber \nonumber\\
	&+&\sum_{n=0}^{N^k} \xi_n^{\ast}(x_{i_k}^k) (2n+1) \xi_n(x_{i_k^{\prime}}^k) \nonumber 
\end{eqnarray}

\noindent
The time-independent components (${\bf \{A^k\}}$, ${\bf \{X^k\}}$, ${\bf \{E^k\}}$, 
${\bf \{F^k\}}$) of the TDDVR Hamiltonian matrix [Eq. \ref{eq:ihbard}] need to be calculated once for all, which are highly significant due to the quantum-classical nature of this approach. 
The diagonal ${\bf \{X^k\}}$ matrix is multiplied with the ``classical" variables, 
$\{\dot{p}_{Q^c_k}(t)\}$ and hence, no non-linearity appears in the quantum equations of motion
due to ``classical" motion. In other words, ``classical" mechanics can only govern
the convergence of the quantum dynamics. On the other hand, the full matrices, 
${\bf \{E^k\}}$ and ${\bf \{F^k\}}$ posses the couplings among the grid-points, which are free from
any non-linearity due to ``classical" propagation due to 
time-independent variable, $\{Im A^k\}$. Eq. (\ref{eq:ihbard}) clearly shows that 
${\bf \{E^k\}}$ and ${\bf \{F^k\}}$ matrices couple grid-points or basis function for the {\it k} th mode only indicating that one can employ parallel computation while solving the differential equation 
for each mode numerically. This can become helpful for reducing the computational cost even for large 
dimensional calculations.\\

\noindent 
On the other hand, the classical equations of motion (EOMs) for the $k^{th}$ mode can be written as,
\begin{eqnarray}
	\dot{Q^c_k}(t) &=&\frac{p_{Q^c_k}(t)}{\mu}, \label{eq:Qceq1}\\
	\dot{p}_{Q^c_k}(t) &=& -\frac{dW(\{Q_k\})}{dQ_k}\Big|_{Q_k(t)=Q_k^c(t)} \,.
	\label{eq:Pceq2}
\end{eqnarray}

\vspace{0.5cm}

\noindent	
The first principle based rigorous expression of $\dot{p}_{Q^c_k}$ for multi-dimensional multi-
surface systems can be derived employing Dirac-Frenkel variational principle,~\citep{PAM} in the following way:
\begin{eqnarray}
	I &=& \int \Big( -i \hbar \frac{\partial \Xi^{\ast}(\{Q_k\},t)}{\partial t} - H(\{p_{Q_k}\},\{Q_k\}) 
	\Xi^{\ast}(\{Q_k\},t)\Big)\nonumber\\
	&\times& \Big( i \hbar \frac{\partial \Xi(\{Q_k\},t)}{\partial t} - H(\{p_{Q_k}\},\{Q_k\}) 
	\Xi(\{Q_k\},t)\Big) \prod_{k=1}^p dQ_k,
\end{eqnarray}	

\vspace{0.5cm}

\noindent
with respect to $\{\dot{p}_{Q^c_k}\}$. The explicit form of $\{\dot{p}_{Q^c_k}\}$ appears as:

\clearpage 
\begin{eqnarray}
	\dot{p}_{Q^c_k}(t) &=& \bigg[\frac{2 (Im A^k)^2}{\mu}\Big\{ \sum_l \sum_{i_ki_k^{\prime}}
	{c^d}_{i_1i_2..i_k..i_p,l}^{\ast}(t) \frac{\sum_{i_k^{\prime\prime}} c^d_{i_1i_2..i_k^{\prime\prime}..i_p,l}(t)
		S_{i_k^{\prime\prime}i_k^{\prime}}^{(2)}}{A_{i_k^{\prime}i_k^{\prime}}}
	{S^{\ast}}^{(1)}_{i_ki_k^{\prime}}\nonumber\\
	&+& \sum_l \sum_{i_ki_k^{\prime}} \frac{\sum_{i^{\prime\prime}} 
		{c^d}_{i_1i_2..i_k^{\prime\prime}..i_p,l}^{\ast}(t)
		{S^{\ast}}_{i_k^{\prime\prime}i_k}^{(2)}}{A_{i_ki_k}} c^d_{i_1i_2..i_k^{\prime}..i_p,l}(t)
	{S^{\ast}}^{(1)}_{i_ki_k^{\prime}}\nonumber\\
	&-& 2 \sum_l \sum_{i_ki_k^{\prime}}{c^d}_{i_1i_2..i_k..i_p,l}^{\ast}(t) c^d_{i_1i_2..i_k^{\prime}..i_p,l}(t)
	{S^{\ast}}_{i_ki_k^{\prime}}^{(3)}\Big\}\nonumber\\
	&-& \frac{\hbar Im A^k}{\mu}\Big\{ \sum_l \sum_{i_ki_k^{\prime}} {c^d}_{i_1i_2..i_k..i_p,l}^{\ast}(t)
	\frac{\sum_{i_k^{\prime\prime}} c^d_{i_1i_2..i_k^{\prime\prime}..i_p,l}(t) R_{i_k^{\prime\prime}i_k^{\prime}}}
	{A_{i_k^{\prime}i_k^{\prime}}} {S^{\ast}}^{(1)}_{i_ki_k^{\prime}} \nonumber\\
	&+& \sum_l\sum_{i_ki_k^{\prime}} \frac{\sum_{i_k^{\prime\prime}}{c^d}_{i_1i_2..i_k^{\prime\prime}..i_p,l}^{\ast}(t)
		R^{\ast}_{i_k^{\prime\prime}i_k}} {A_{i_ki_k}} {c^d}_{i_1i_2..i_k^{\prime}..i_p,l}(t)
	{S^{\ast}}^{(1)}_{i_ki_k^{\prime}}\nonumber\\
	&-& 2 \sum_l \sum_{i_ki_{k^{\prime}}}{c^d}_{i_1i_2..i_k..i_p,l}^{\ast}(t) {c^d}_{i_1i_2..i_k^{\prime}..i_p,l}(t)
	T^{\ast}_{i_ki_k^{\prime}}\Big\}\bigg]\nonumber\\
	&\bigg/&\bigg[ \sum_l\sum_{i_ki_k^{\prime}} {c^d}_{i_1i_2..i_k..i_p,l}^{\ast}(t)
	\frac{\sum_{i_k^{\prime\prime}} {c^d}_{i_1i_2..i_k^{\prime\prime}..i_p}(t) S_{i_k^{\prime\prime}i_k^{\prime}}^{(1)}}
	{A_{i_k^{\prime}i_k^{\prime}}} {S^{\ast}}^{(1)}_{i_ki_k^{\prime}}\nonumber\\
	&+& \sum_l\sum_{i_ki_k^{\prime}} \frac{\sum_{i_k^{\prime\prime}} {c^d}_{i_1i_2..i_k^{\prime\prime}..i_p,l}^{\ast}(t)
		{S^{\ast}}^{(1)}_{i_k^{\prime\prime}i_k}}{A_{i_ki_k}} {c^d}_{i_1i_2..i_k^{\prime}..i_p,l}
	{S^{\ast}}^{(1)}_{i_ki_k^{\prime}}\nonumber\\
	&-&2 \sum_l \sum_{i_ki_k^{\prime}}{c^d}_{i_1i_2..i_k..i_p,l}^{\ast}(t) {c^d}_{i_1i_2..i_k^{\prime}..i_p,l}(t)
	{S^{\ast}}_{i_ki_k^{\prime}}^{(2)}\bigg] \label{eq:pdotsck}
\end{eqnarray}
where ${c^d}^{\ast}$ is the complex conjugate of ${c^d}$ and
\begin{eqnarray}
	R_{i_ki_k^{\prime}}&=&\sum_p {\zeta^{\ast}}_p(x_{i_k}^k) \zeta_p(x_{i_k^{\prime}}^k) 2p\nonumber\\
	S_{i_ki_k^{\prime}}^{(n)} &=& \sum_{pq} {\zeta^{\ast}}_p(x_{i_k}^k) \zeta_q(x_{i_k^{\prime}}^k)
	\int {\Phi^{\ast}}_q(Q_k,t)(Q_k-Q^c_k(t))^n\Phi_p(Q_k,t)dQ_k\nonumber\\
	T_{i_ki_k^{\prime}} &=& \sum_{pq} \zeta_p(x_{i_k}^k) {\zeta^{\ast}}_q(x_{i_k^{\prime}}^k) 2p
	\int {\Phi^{\ast}}_p(Q_k,t)(Q_k-Q^c_k(t))\Phi_q(Q_k,t)dQ_k \qquad 
\end{eqnarray}
with
\begin{eqnarray}
	\int {\Phi^{\ast}}_p(Q_k,t)(Q_k-Q^c_k(t))\Phi_q(Q_k,t)dQ_k &=&
	\frac{1}{2} \sqrt{\frac{\hbar}{Im A^k}} \Big\{ \sqrt{p+1} \delta_{p+1,q} 
	+ \sqrt{p} \delta_{p-1,q}\Big\}\nonumber\\
	\int {\Phi^{\ast}}_p(Q_k,t)(Q_k-Q^c_k(t))^2\Phi_q(Q_k,t)dQ_k 
	&=& \frac{\hbar}{4 Im A^k} \Big\{\sqrt{(p+1)(p+2)} \delta_{p+2,q} \nonumber\\
	&+& (2p+1) \delta_{p,q} + \sqrt{p(p-1)}\delta_{p-2,q}\Big\}\nonumber\\
	\int {\Phi^{\ast}}_p(Q_k,t)(Q_k-Q^c_k(t))^3\Phi_q(Q_k,t)dQ_k 
	&=& \frac{1}{8} \Big(\frac{\hbar}{Im A^k}\Big)^{3/2}            
	\Big\{\sqrt{(p+1)(p+2)(p+3)} \delta_{p+3,q} \nonumber\\
	&+& 3(p+1)\sqrt{p+1} \delta_{p+1,q} +3p\sqrt{p} \delta_{p-1,q}\nonumber\\
	&+& \sqrt{p(p-1)(p-2)}\delta_{p-3,q}\Big\}
\end{eqnarray}
The time-independent matrices, $R$, $S^{(n)}$, $T$ and $A$ in Eq. (\ref{eq:pdotsck}) need to be 
calculated only once. On the other hand, time-dependence of $\dot{p}_{Q^c_k}$ arises due to the 
time-dependent coefficient $\{ c^d_{i_1i_2..i_k..i_p,l}(t)\}$. While substituting 
$S^{(1)}_{i_ki_k^{\prime}}$ by $S^{(1)}_{i_ki_k^{\prime}} \delta_{i_ki_k^{\prime}}$ [as the 
first order term $(Q_k - Q^c_k(t))$ is diagonal in DVR~\citep{SAIJQC434}], $\dot{p}_{Q^c_k}$ in 
Eq. (\ref{eq:pdotsck}) takes the following simplified form:

\begin{eqnarray}
	\dot{p}_{Q^c_k}(t) &=& \sum_l \sum_{i_ki_k^{\prime}} {c^d}_{i_ii_2..i_k..i_p,l}^{\ast}(t)
	{c^d}_{i_1i_2..i_k^{\prime}..i_p,l}(t)\nonumber\\ &\times& \Big\{\frac{2 (Im A^k)^2}{\mu}
	\Big[S_{i_ki_k^{\prime}}^{(2)} \frac{{S^{(1)}}^{\ast}_{i_ki_k}}{A_{i_ki_k}}- S^{(3)}_{i_ki_k^{\prime}}\Big]
	- \frac{\hbar Im A^k}{\mu} \Big[R_{i_ki_k^{\prime}} \frac{{S^{(1)}}^{\ast}_{i_ki_k}}
	{A_{i_ki_k}}-T^{\ast}_{i_ki_k^{\prime}}\Big]\Big\}\nonumber\\
	&/&\Big[\sum_l \sum_{i_k} {c^d}_{i_1i_2..i_k..i_p,l}^{\ast}(t) {c^d}_{i_1i_2..i_k..i_p,l}(t)
	\frac{{S^{(1)}}^{\ast}_{i_ki_k} S^{(1)}_{i_ki_k}}{A_{i_ki_k}}\nonumber\\
	&-&\sum_i \sum_{i_ki_k^{\prime}} {c^d}_{i_1i_2..i_k..i_p,l}^{\ast}(t) {c^d}_{i_1i_2..i_k^{\prime}..i_p,l}(t)
	{S^{\ast}}_{i_ki_k^{\prime}}^{(2)}\Big] \, .
	\label{eq:pdotq}
\end{eqnarray}

\noindent
Initially ($t=0$), since the denominator of Eq. (\ref{eq:pdotq}) is very small, the large 
``quantum force" results into numerical inaccuracies in ``classical" variables (see Eq. 
(\ref{eq:Qceq1}) and (\ref{eq:pdotq})) and quantum equation of motion (see Eq. (\ref{eq:ihbard})).
In order to overcome such bottleneck, it is necessary to propagate the ``classical" equation of 
motion in Newtonian sense (see Eq. (\ref{eq:Qceq1}) and Eq. (\ref{eq:Pceq2})) until the amplitudes of 
the wavefunction are spread over the expansion coefficient ($\{ {c^d}_{i_1i_2..i_k..i_p,l}(t)\}$). Once 
the amplitudes are sufficiently distributed, one can carry out dynamics employing the quantum force 
as calculated in Eq. (\ref{eq:pdotq}).

\subsection{Initialization and Propagation}\label{ssec:ini}


\vspace{0.2cm}

\noindent
On the process of initialization of TDDVR dynamics, the starting wavefunction is defined in the adiabatic representation for a specific electronic state, which is represented as the product of Gaussian wavepackets ($\{\Phi_{GWP}(Q_k)\}$) and their center is located around the 
equilibrium geometry ($\bf \lbrace Q_0=0\rbrace$): 
\begin{eqnarray}
	\prod_{k=1}^p\Phi_{GWP}(Q_k) &=& \sum_{i_1i_2....i_p} c^{ad}_{i_1i_2....i_p,l}
	\prod_{k=1}^p\chi_{i_k}(Q_k,t)\nonumber\\
	&=& \sum_{i_1i_2....i_p} c^{ad}_{i_1i_2....i_p,l}
	\prod_{k=1}^p\sum_{n=0}^{N_k} \zeta_n(x_{i_k})\Phi_n(Q_k,t).  
	\label{eq:GWP}
\end{eqnarray}
While using the following property of TDDVR basis,
\begin{eqnarray}
	\chi_{i_k}(Q_{i_k},t)=\sum_{n=0}^{N_k} \zeta_n^{\ast}(x_{i_k})
	\Phi_n(Q_k,t)\cdot\delta(Q_k-Q_{i_k}), 
	\label{eq:proba}
\end{eqnarray}
Eq. (\ref{eq:GWP}) turns into the following form,
\begin{eqnarray}
	\prod_{k=1}^p\Phi_{GWP}(Q_{i_k}) &=& c^{ad}_{i_1i_2....i_p,l} 
	\prod_{k=1}^p\sum_{n=0}^{N_k}\zeta_n(x_{i_k})\Phi_n(Q_{i_k})\nonumber\\
	&=& c^{ad}_{i_1i_2....i_p,l} \prod_{k=1}^pA_{i_ki_k}.  
	\label{eq:GWPf}
\end{eqnarray}
Hence, using Eq. (\ref{eq:GWPf}), one can arrive the following expression of the vector ($d^{ad}_n$) on one of the surfaces (let say $n$):
\begin{eqnarray}
	d^{ad}_{i_1i_2....i_p,n}= c^{ad}_{i_1i_2....i_p,n}\sqrt{\prod_{k=1}^pA_{i_ki_k}}=\frac{\prod_{k=1}^p\Phi_{GWP}
		(Q_{i_k})} {{\sqrt{\prod_{k=1}^pA_{i_ki_k}}}}, 
	\label{eq:drearr}
\end{eqnarray}
where the initial expansion coefficient vectors ({\bf $d^{ad}_{i_1i_2....i_p,l}$}, where $l\ne n$) on other surfaces (except state $n$) take the following form,
\begin{eqnarray}
	\cdots\,, d^{ad}_{i_1i_2....i_p,n-1}=0\,,\quad d^{ad}_{i_1i_2....i_p,n+1}=0 \cdots.
	\label{eq:drearr1}
\end{eqnarray}
\noindent	
In order to perform dynamical evolution of any molecular system, the set of differential equations (see Eqs. (\ref{eq:ihbard_p}), 
(\ref{eq:Qceq1}), (\ref{eq:Pceq2})) are simultaneously solved with initial coefficient vectors, $\boldsymbol{D^d}$.\\

\noindent
Since the initial amplitudes of the wavepacket are in the adiabatic representation (see Eqs. (\ref{eq:drearr}) and (\ref{eq:drearr1})), those quantities need to be transformed to the diabatic ones for propagation by employing the transformation matrix, $\boldsymbol{X}$. In other words, $\boldsymbol{D^d}=\boldsymbol{X}\boldsymbol{D^{ad}}$, where 
$\boldsymbol{V}^{ad}=\boldsymbol{X}^\dagger\boldsymbol{W}\boldsymbol{X}$ with $\boldsymbol{X}$ being the diagonalizing matrix for the diabatic PESs matrix.
\\

\noindent
While calculating the diabatic population on a particular surface ($l$), one needs to perform the following grid summation over the amplitudes,
\begin{eqnarray}
	P^{d}_l(t) = \sum_{i_1i_2....i_p} |d^{d}_{i_1i_2....i_p,l}|^2, l = 1, 2, ... l, .  
	\label{eq:Gdsum_d}
\end{eqnarray}
whereas in adiabatic representation, the population is defined as,
\begin{eqnarray}
	P^{ad}_l(t) = \sum_{i_1i_2....i_p} |d^{ad}_{i_1i_2....i_p,l}|^2, l = 1, 2, ... l, .  
	\label{eq:Gdsum_ad}
\end{eqnarray}
at any time, $t$.

\subsection{Calculation of Theoretical PE, PD, PA and MATI Spectra}\label{ssec:spectra}

\vspace{0.2cm}

\noindent
While performing nuclear dynamics by TDDVR methodology, TDSE is numerically integrated to calculate the nuclear wavefunctions (Eq. (\ref{eq:2dwf})) at different
time, which are further employed to compute the autocorrelation function ($C(t)$). Once $C(t)$ s are in hand, the following Fourier 
transformation is executed,
\begin{eqnarray}
	I(\omega) \propto \omega \int_{-\infty}^{\infty} C(t) \exp(i\omega t) dt, \label{eq:abs}
\end{eqnarray}
where
\begin{subequations}
	\begin{eqnarray}
		C(t) &=& \langle \Xi(t) | \Xi(0)\rangle \label{eq:autocor1_6}\\
		&=& \Big\langle \Xi^{\star}\Big(\frac{t}{2}\Big) 
		\Big| \Xi\Big(\frac{t}{2}\Big)\Big\rangle \label{eq:autocor2_6}.
	\end{eqnarray}
\end{subequations}
It is interesting to note that Eq. (\ref{eq:autocor2_6}) is more accurate and computationally cost-effective with respect to the other one (Eq. (\ref{eq:autocor1_6})), though the former one 
(Eq. (\ref{eq:autocor2_6})) can be applied only for real initial wavefunction and totally symmetric Hamiltonian. In order to encapsulate the effects of phenomenological broadening of 
the experimental spectral lines arising due to the resolution of the spectrometer in the time-dependent picture, the autocorrelation function is multiplied by a time-dependent damping function as follows:
\begin{eqnarray}
	h(t)=\exp\Big[-\Big(\frac{|t|}{\tau}\Big)^{2}\Big] \label{eq:tdf}
\end{eqnarray}
where the unit of $\tau$ is 10$^{-14}$ sec. The value of $\tau$ should be properly tuned to correctly describe 
the experimental broadening. The details of the calculation of theoretical spectra is extensively illustrated 
elsewhere.~\citep{SAIJQC434,souvik_irpc37}

\noindent
\section{Two State : NO$_2$}	

\vspace{0.2 cm}
\noindent 
The two low-lying electronic states ($X^{2}A_{1}$ and $A^{2}B_{2}$) of nitrogen dioxide (NO$_2$) radical exhibits JT and RT interactions, which dictate the details of its photodetachment spectrum.\citep{mahapatra_jcp110,jackels_jcp63,kaldor_cpl185,haller_jms111,schinke_cp346,joyeux_jcp119,sanrey_jcp125,mukherjee_jcp143,
	sardar_cp416,takatsuka_jcp132,takatsuka_pccp13} Adhikari and coworkers had generated highly precise diabatic surfaces of NO$_2$\citep{mukherjee_jcp143} using MRCI based adiabatic PESs and NACTs obtained from MOLPRO\citep{MOLPRO_brief} quantum chemistry package followed by ADT calculation. Finally, nuclear dynamics had been performed to compute the theoretical photodetachment spectrum~\citep{mukherjee_jcp143} of  NO$_2$ and such spectrum is compared with the experimental one\citep{weaver_jcp90}.\\

\noindent
In NO$_2$ molecule, one can observe JT type accidental CIs and RT interactions over three 2D nuclear planes ($Q_i-Q_j$) consisting of 
symmetric bending ($Q_1$), symmetric ($Q_2$) and asymmetric ($Q_3$) stretching normal modes belonging to $a_{1}$, $a_{1}$ and $b_{2}$ irreducible representations (IREPs). While exploring JT interactions in NO$_2$, it should be noted that it exhibits
$C_{2v}$/$C_s$ symmetry and therefore, all the electronic states and normal modes can be assigned under 1D IREPs. In other words, no symmetry-driven JT CI can be observed in this radical, whereas JT type accidental CIs and RT interactions are located between the electronic states. In order to construct JT/RT active diabatic Hamiltonian, the direct product of irreducible representations for the electronic state and the vibrational mode should contain the totally symmetric representation ($A_{1}$). Therefore, the couplings between $X^{2}A_{1}$ and $A^{2}B_{2}$ states via $b_{2}$ symmetric $Q_3$ normal mode are allowed through accidental CIs between those states:

$$A_{1} \in A_{1} \otimes b_{2} \otimes  B_{2}.$$

\noindent
In {\it ab initio} calculation, JT type accidental CIs and 
RT interactions are observed over three 2D nuclear configuration space ($Q_i-Q_j$) consisting of 
symmetric bending ($Q_1$), symmetric ($Q_2$) and asymmetric ($Q_3$) stretching normal modes. For each pair of normal modes, 
we transform the {\it ab initio} quantities (adiabatic PESs and NACTs) in terms of the corresponding polar coordinates
($\rho$ and $\phi$) to explore the nature of the CIs. The $\rho$ component of NACT ($\tau_{\rho}^{12}$) over the polar plane ($\rho-\phi$) associated with
$Q_1$ and $Q_3$ normal modes is depicted in Figure \ref{fig:no2nact}a showing the existence of one JT type seam passing through the $C_{2v}$ point with three peaks [one at $C_{2v}$ 
(peak 1) and the other two at $C_{s}$ (peaks 2 and 3) nuclear geometries]. On the other hand, the RT 
intersections are prominent at the linear geometries (peaks 4 and 5).\citep{sardar_cp416,mukherjee_jcp143} \\

\noindent
Once the {\it ab initio} quantities (adiabatic PESs and NACTs) are in hand, those are plugged into the ADT equations (see Eq. \ref{eq:adt-two}) and solved those equations. While performing the first principle based ADT calculation along a closed contour at $\rho = 0.45$ \AA, 
the resulting ADT angle ($\Theta_{12}$) is displayed in Figure \ref{fig:no2nact}b, which attends the magnitude 
of $3\pi$ at the end of the closed contour ($\phi=2\pi$) bearing the signature of JT type CI seam ($\pi$) 
and RT interaction ($2\pi$).\citep{mukherjee_jcp143,naskar_jctc}
On the other hand, for other two pairs, namely, $Q_1-Q_2$ and $Q_2-Q_3$, two JT type accidental CIs and one RT intersection
are located for each case over the Franck-Condon (FC) domain of nuclear configuration space, which can be found out from the singularities of 
the NACT landscapes.\citep{mukherjee_jcp143} 
Finally, the ADT matrix has been used to generate a two-state three-mode diabatic Hamiltonian, where PESs and coupling elements are smooth, single-valued and continuous 
over the chosen portion of nuclear CS. Moreover, it is worth mentioning that along $Q_3$ normal mode, the diabatic coupling element exhibits symmetric form (see Figure \ref{fig:no2even}), which can only be fitted with even parity polynomial\citep{mukherjee_jcp143} unlike the traditional JT model~\citep{mahapatra_jcp110}. \\

\begin{figure}[!htp]
	\centering
	\includegraphics[ height=0.7\textheight, angle=0]{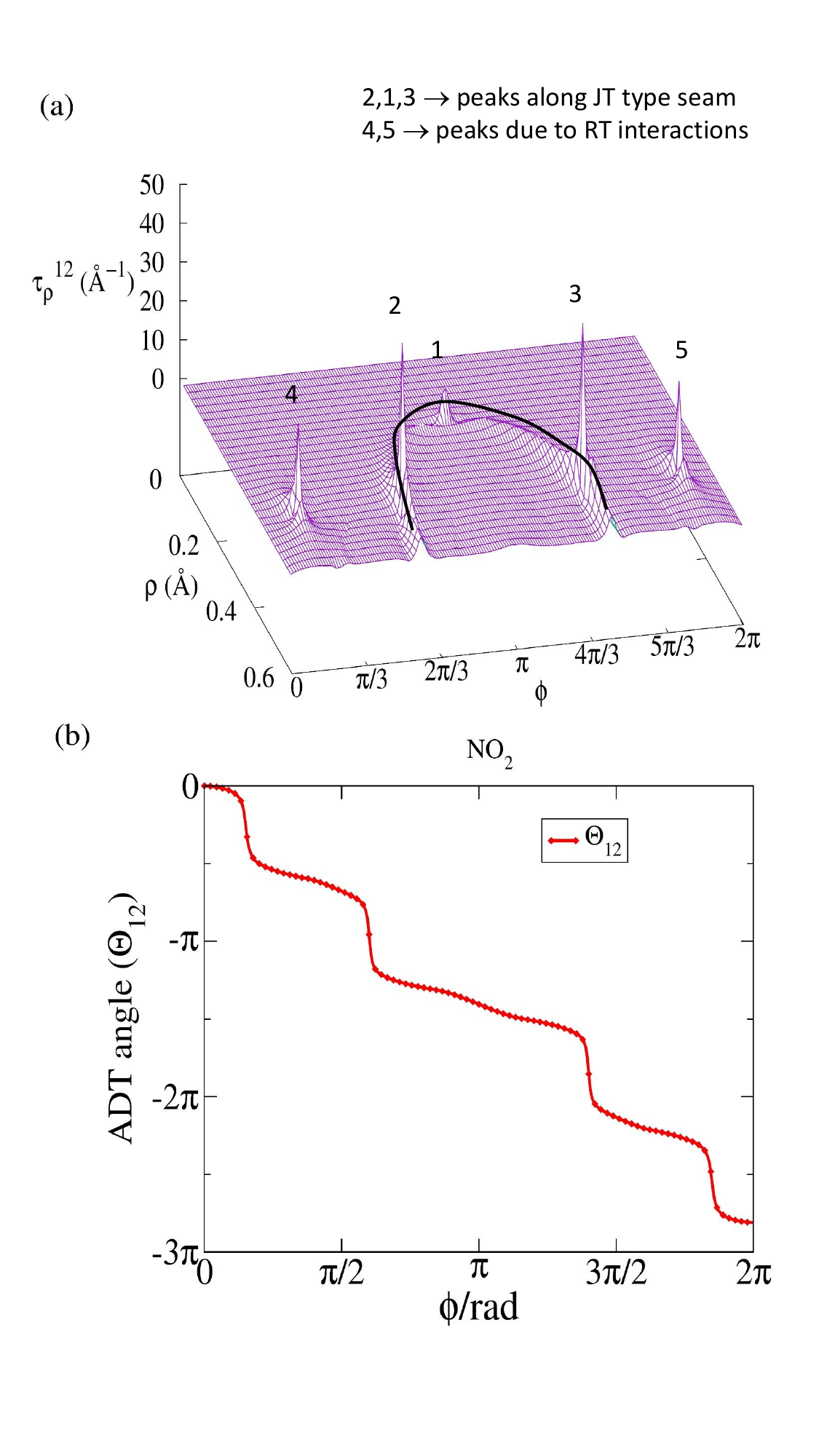}
	\caption{For $Q_{1}-Q_{3}$ pair of NO$_2$ radical, (a) $\tau_{\rho}^{12}$ shows singularities
		at the close vicinity of the semicircular JT type seam (peaks 1 to 3) as well as in the RT regions 
		(peaks 4 and 5) over $\rho-\phi$ configuration space. 
		Panel (b) depicts 1D variation of ADT angle, $\Theta_{12}$ along $\phi$ coordinate at a fixed 
		$\rho$ value 0.45 \AA. Such angle acquires the numerical value of $3\pi$ due to the JT and RT couplings. The results are adopted from Ref. [\protect\citen{joy_pccp}] and permission has been taken from Royal Society of Chemistry (RSC).}
	\label{fig:no2nact}
\end{figure}

\begin{figure}[!htp]
	\centering
	\includegraphics[ height=0.5\textheight, angle=0]{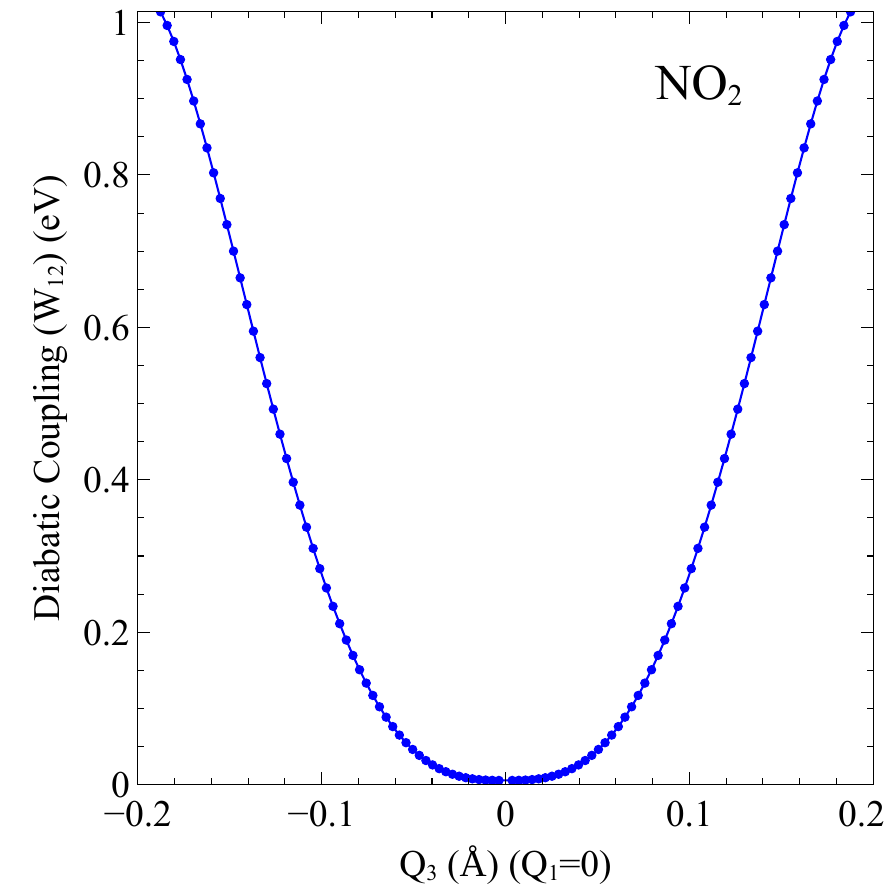}
	\caption{For NO$_2$ radical, diabatic coupling ($W_{12}$) shows even parity functional forms along $Q_{3}$ normal mode. The results are adopted from Ref. [\protect\citen{joy_pccp}] and permission has been taken from RSC.	}
	\label{fig:no2even}
\end{figure}

\noindent
The accuracy of the diabatic Hamiltonian and the existence of two state SHS can be validated by computing Curl condition.~\citep{mantu_ijqc_curl}
For two state SHS, ADT Curl (Commutator, $C_{\rho\phi}^{12}$ = $(\tau_\phi\tau_\rho)_{12}-
(\tau_\rho\tau_\phi)_{12}$) is ``trivially" zero (0) and the calculated Mathematical Curl ($Z_{\rho\phi}^{12}$ = $\dfrac{\partial}{\partial \phi}\tau^{12}_\rho-\dfrac{\partial}{\partial \rho}\tau^{12}_\phi$) has been shown to be numerically zero (0)~\citep{mantu_ijqc_curl}  using the cross-derivatives on the recalculated NACTs from ADT angles. Since both ADT Curl ($C_{\rho\phi}^{12}$) and numerically calculated Mathematical Curl ($Z_{\rho\phi}^{12}$) become zero (0), Curl Condition ($F_{\rho\phi}^{12}$) is obeyed (0) for two state SHS of NO$_2$. \\

\noindent
Once the first principle based accurate diabatic PESs for NO$_2$ radical are obtained, nuclear dynamics has been carried out
to generate bands of $\tilde{X}^2A_1$ and $\tilde{A}^2B_2$ states in the photodetachment spectrum.
Dynamical calculations have been executed at the close neighborhood of equilibrium geometry (FC region), where Gaussian wave 
packets (GWPs) can be used to represent molecular vibrations. Therefore, 
a product wavefunction with GWPs for each normal mode of NO$_2^-$ has been taken as the 
initial wavefunction for solving the time dependent SE. During dynamical calculations,\citep{sardar_pccp13,sardar_jcs124,sardar_jpca118,jpca-119-12392-SA,jcp-147-074105-SA,mandal_irpc37} time dependent nuclear wavefunctions vis-\'a-vis autocorrelation function have been calculated
at different time and then, Fourier transformation has been carried out to generate the photodetachment spectra of NO$_2$ radical. The overall
spectral profile\citep{mukherjee_jcp143} exhibits good agreement with the experimental one recorded by Weaver et al.\citep{weaver_jcp90} (see Figure \ref{fig:no2spectra}). The asymmetric stretching mode ($Q_3$) 
is mainly responsible for vibronic transitions in the spectral profile, whereas symmetric bending ($Q_1$) and stretching ($Q_2$) vibrations dictate the peak progressions. It is necessary to mention that absence of $Q_2$ normal mode in the diabatic Hamiltonian broadens the spectral band of $\tilde{X}^2A_1$ state.\citep{mukherjee_jcp143} In summary, our theoretical results specify the details of the effects due to the JT and RT couplings to the overall nonadiabaticity of NO$_2$, which modulates its photodetachment spectrum.\citep{mukherjee_jcp143} \\

\begin{figure}[!htp]
	\centering
	\includegraphics[width=0.7\textwidth,angle=0]{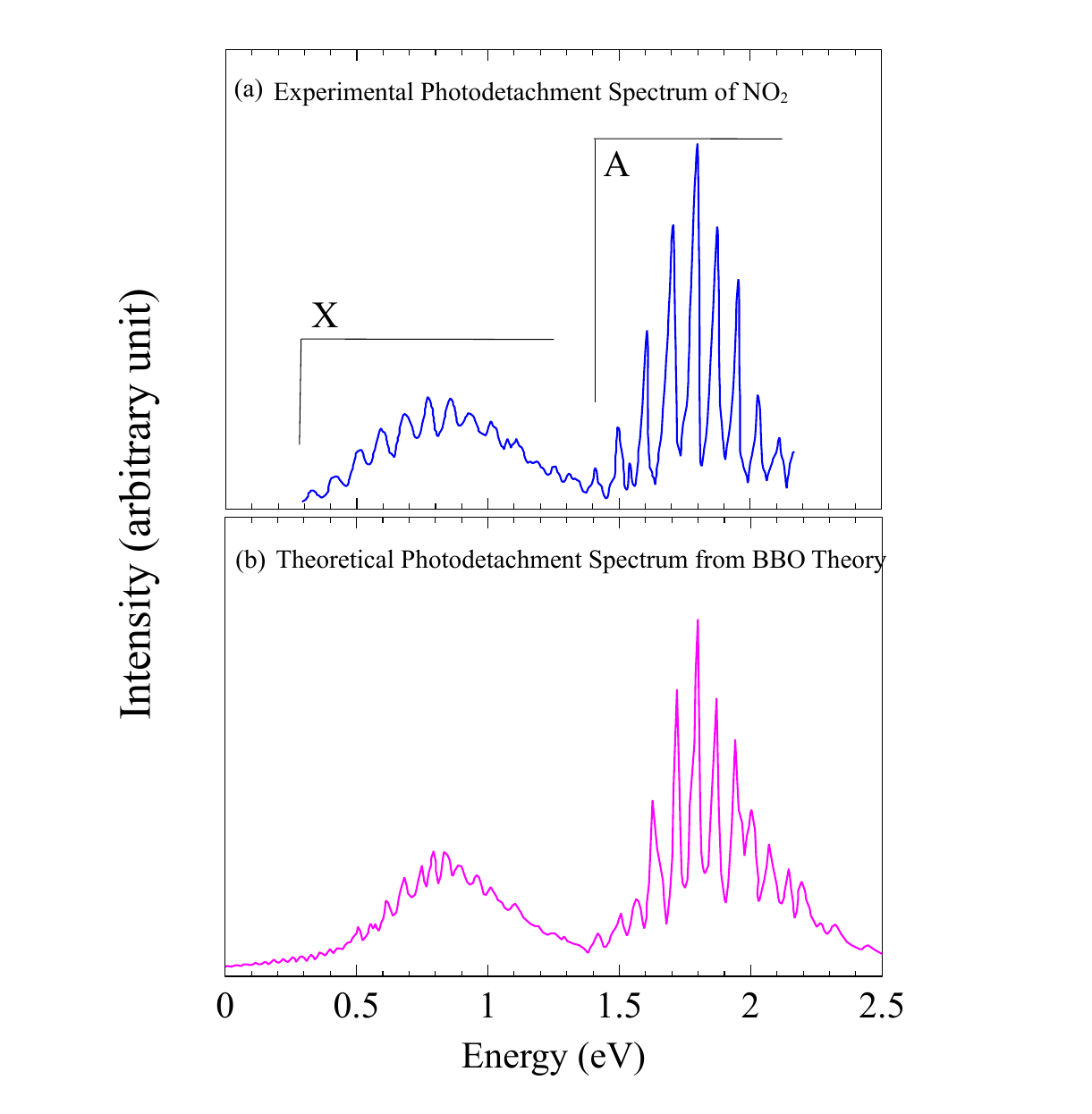}
	\caption{The BBO calculated spectrum (panel (b)) incorporating all nonadiabatic interactions (JT and RT couplings) is compared with the experimental finding (panel a)\citep{weaver_jcp90}. The results are taken from Ref. [\protect\citen{joy_pccp}] and permission has been taken from RSC.}
	\label{fig:no2spectra}
\end{figure}


\noindent
\section{Three State : Na$_3$}	

\vspace{0.2 cm}
\noindent 
While unveiling the applicability of BBO-DH for a three coupled electronic manifold, prototypical alkali trimer, like Na$_3$ has been chosen to investigate its inherent JT and PJT couplings and their effects on the spectral profiles. Over the last three decades,
several theoretical\citep{sadygov_jcp110,vivie-rieldle_cp223,meiswinkel_cp144,koizumi_prl83,cocchini_jcp88,hauser_jcp129,hauser_cp375} as well as experimental\citep{delacretaz_prl56,ernst_pst112} research groups are involved in the study of Na$_3$ cluster due to its simple electronic structure and presence of only three vibrational modes. 
Adhikari and coworkers~\citep{akpaul_jcp131} had performed dynamical calculations employing model diabatic Hamiltonian constructed by Cocchini {\it et al.}\citep{cocchini_jcp88} and then, theoretical 
spectra were calculated using the associated BO, EBO and diabatic surfaces. Later on, the same group had generated first principle based adiabatic PESs and NACTs of three excited electronic states ($2\,^2E^{\prime}$ and $1\,^2A_1^{\prime}$) 
using MRCI methodology as implemented in MOLPRO\citep{MOLPRO_brief}. Such electronic structure study has been followed by ADT calculations over 2D
nuclear plane ($\rho-\phi$) arising from   
bending ($Q_x$) - asymmetric stretching ($Q_y$) pair keeping the value of symmetric stretching ($Q_s$) mode fixed at 3.7 \AA. The Na$_3$ cluster exhibits accidental CIs within two 
branches of 2$^2E^{\prime}$ state as well as profound PJT
interaction among 2$^2E^{\prime}$ and 1$^2A_1^{\prime}$ states. 
Among the accidental CIs, three appear at C$_{2v}$ geometry, whereas the other three symmetry-driven CIs merge at D$_{3h}$ 
point. Therefore, one can consider total four CIs over the nuclear CS.
\citep{mukherjee_jcp143}\\

\noindent
On the process of exploring the conditions to understand the properties related to the distortions of Na$_3$ molecule, it appears that the vibrational modes ($e^\prime$ symmetric $Q_x-Q_y$) are responsible for JT interactions. The JT effects originate due to the coupling of electronic states with vibrational modes of appropriate symmetry. In other words, the direct product of irreducible representations for the electronic states and the vibrational modes contains the totally symmetric representation ($A_{1}^{\prime}$). The JT distortion  takes place between $^2E^{\prime}$ states via $e^{\prime}$ normal modes:

$$A_{1}^{\prime} \in e^\prime \otimes [(E^{\prime})^2] \equiv e^\prime \otimes (A_1^{\prime} + A_2^{\prime} + E^{\prime})$$

\noindent 
While performing BBO theory based diabatization of three excited states 
($2\,^2E^{\prime}$ and $1\,^2A_1^{\prime}$) of Na$_3$ cluster, the adiabatic PESs and MS adapted NACTs 
(see Section 4.6 for details) were substituted in
the ADT equations\citep{baer_jcp66,sarkar_jcp124} for three state electronic manifold (see Eq. \ref{eq:adt-three}). Furthermore, those equations had been numerically solved over the 2D nuclear planes to compute the ADT angles. Once the mixing angles were obtained, those ADT angles
had been plugged into the ADT matrix, which had been further used to calculate diabatic Hamiltonian for Na$_3$ cluster. It has been observed that
the diabatic coupling elements show even parity functional forms along 
$Q_y$ normal mode, which can be fitted only with even power polynomials of
$Q_y$ coordinate.\citep{mukherjee_jpca117,bijit_irpc38}\\

\begin{figure}[!htp]
	\centering
	\includegraphics[ height=0.5\textheight, angle=0]{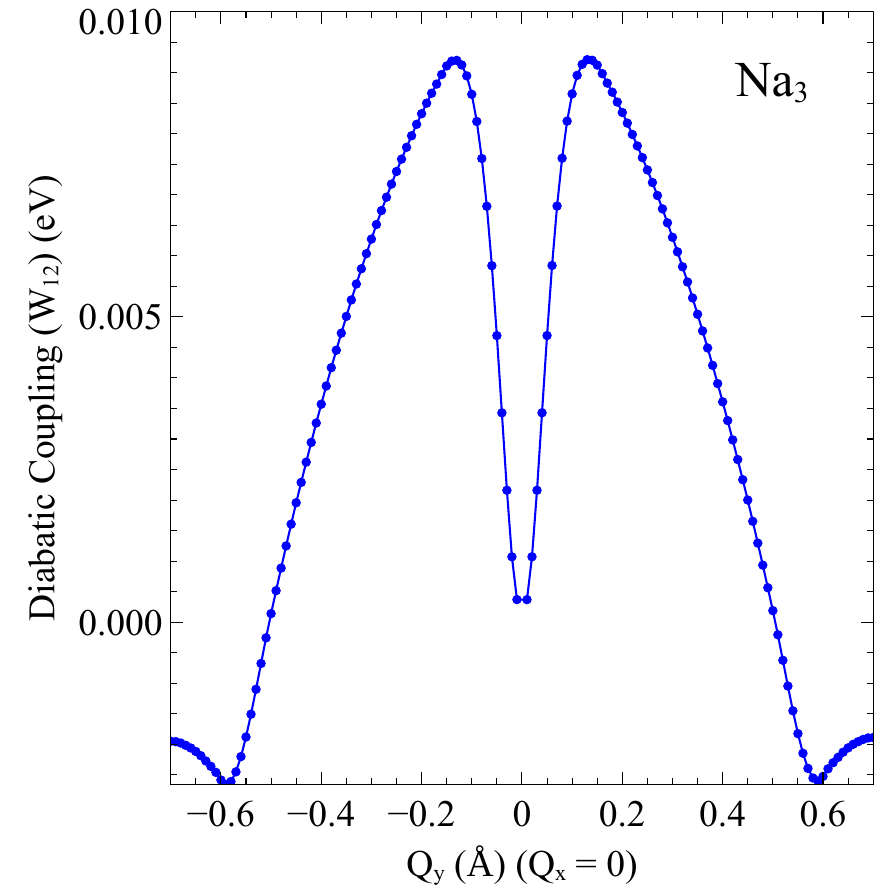}
	\caption{For Na$_3$ cluster, diabatic coupling ($W_{12}$) shows symmetric functional forms along $Q_{y}$ normal mode, which can only be fitted with even power polynomials. The results are adopted from Ref. [\protect\citen{joy_pccp}] and permission has been taken from RSC.}
	\label{fig:na3even}
\end{figure}

\noindent
After carrying out BBO theory based diabatization, nuclear wave packet dynamics had been performed over $Q_x-Q_y$ plane with the aid of GWP as the time-independent solution of the unperturbed Hamiltonian 
[$\hat{H}_0(Q_x,Q_y)=\hat{T}_{\rm nuc} + (1/2)(Q_x^2 + Q_y^2)$]. While executing the dynamical calculations, FFT-Lanczos method had been employed to compute the autocorrelation function, $C(t)$s and such function was further Fourier transformed to compute
the theoretical spectral envelop. In order to obtain individual ($2^2E^{\prime}$ and $1^2A^{\prime}_1$) state spectra, three different calculations had been carried out and then, the resulting bands were convoluted to obtain the overall spectra.\citep{mukherjee_jpca117} The experimental two-photon ionization (TPI) spectrum for system B of Na$_3$ cluster\citep{ernst_pst112} was well reproduced by the calculation of Adhikari {\it et al}~\citep{mukherjee_jpca117} as displayed in Figure \ref{fig:Na3spec}. 
In addition, it had been explored that the diabatic population 
profiles show a crossing with each other as function of time affirming the existence of CIs between those states.\citep{mukherjee_jpca117}

\begin{figure}[!htp]
	\centering
	\includegraphics[width=0.5\textwidth,angle=0]{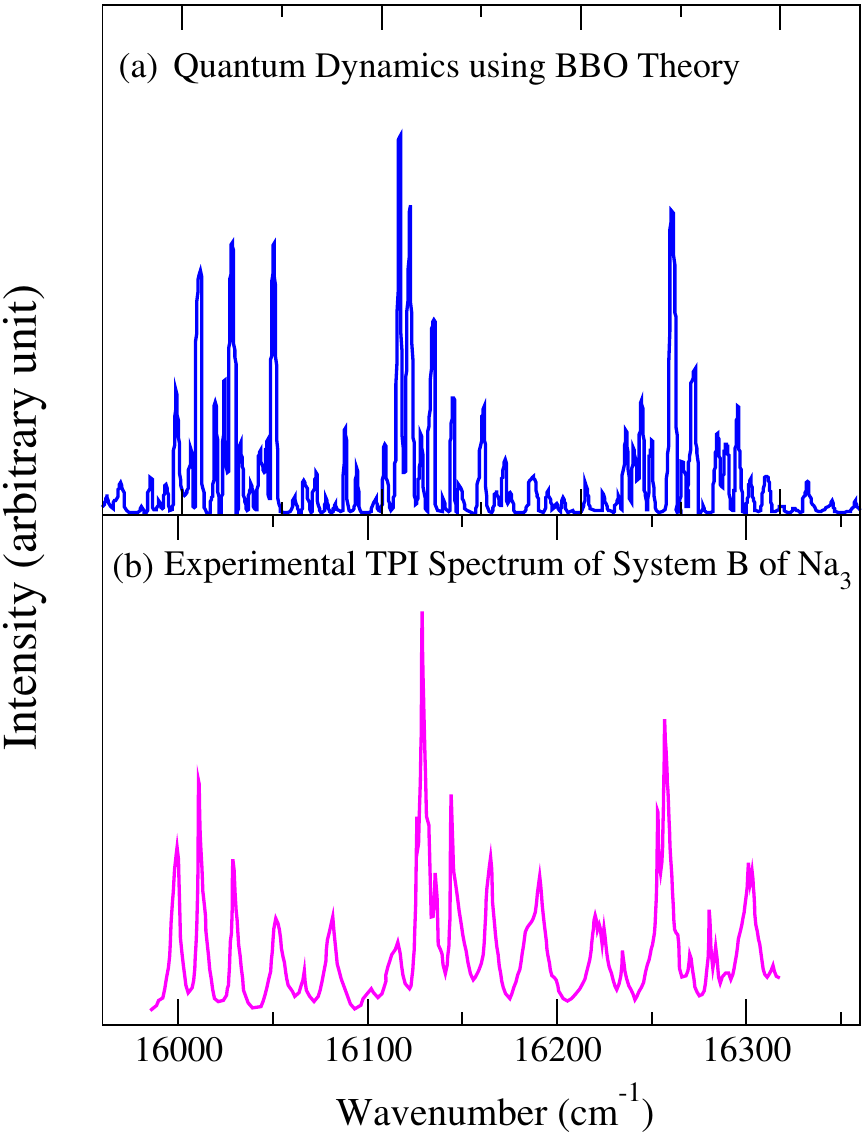}
	\caption{The diagram presents a comparison of absorption spectrum for system B of Na$_3$ cluster originated from
		from BBO based diabatic surfaces [panel (a)] with the experimental two-photon ionization spectrum
		\citep{ernst_pst112} [panel (b)]. This figure has been taken from Ref. [\protect\citen{joy_pccp}] and permission has been taken from RSC.}
	\label{fig:Na3spec}
\end{figure}

\noindent
\section{Four State : C$_4$N$_2$H$_4$}	

\noindent
Over the last few decades, immense interest has been grown to investigate the photophysics of pyrazine molecule, which is enriched with 
vibronic phenomena, namely, JT type CI on a point(s) or along a seam(s)
and PJT couplings. Several experimental techniques, namely, Resonance Raman (RR),~\citep{stock_jcp103} Fluorescence~\citep{domcke_jcp93} and PA~\citep{yamazaki_fdcs75} spectroscopy were employed to investigate the underneath electronic structure and nuclear dynamics of such molecule. In other words,
the electronic structure and PA spectra of pyrazine appear as a long standing point of interest for theoretical chemists also over several decades.~\citep{seidner_jcp96,domcke_jcp98,guo_jrs4074,kanno_4,woywood_jcp100,worth_jcp109,Burghardt_jcp129,vendrell_jcp134,richings_15} Adhikari and coworkers~\citep{saikat_pyra} have employed first principle based BBO theory to explore the wide variety of non-adiabatic interactions present within the 
low-lying four excited electronic states [$1^1B_{1u}$ ($S_1$), $1^1B_{2u}$ ($S_2$), $1^1A_{u}$ ($S_3$) and $1^1B_{2g}$ ($S_4$)] and to construct theoretically ``exact" and numerically ``accurate" 4$\times$4 diabatic PESs matrices of pyrazine molecule. It is important to note that ``1-2", ``2-3" and ``3-4" 
CI(s)/seam(s) can be observed over six 2D nuclear planes constituted with four normal modes ($Q_1$, $Q_{6a}$, $Q_{9a}$ and $Q_{10a}$). The direct product of irreducible representations for the electronic state and the vibrational mode must belong to the totally symmetric representation ($A_{g}$ for D$_{2h}$ point group). Therefore, $b_{1g}$ symmetric $Q_{10a}$ normal modes can couple $1^1B_{1u}$ and $1^1A_{u}$ electronic states due to the following relation: 

$$A_{g} \in B_{1u} \otimes b_{1g} \otimes  A_{u}$$
\noindent
Among all the normal mode pairs of C$_4$N$_2$H$_4$ molecule, $Q_{9a}-Q_{1}$ pair is constituted with non-degenerate normal modes, namely, in-plane C-H bending and out-of-plane C-H bending. Over the polar counterpart, $\rho-\phi$ ($Q_{9a}$ and $Q_{1}$) one can observe accidental CI(s)/seam(s), but 
no symmetry driven JT interaction is present at the equilibrium nuclear configuration ($\rho =0.0$). Figure \ref{fig:1-9a_1-2_adia}(a-c) presents 1D curves of adiabatic PESs for $1^1B_{1u}$ and $1^1B_{2u}$ electronic states ($u_1$ and $u_2$) and the corresponding $\phi$ component of NACTs ($\tau_{\phi}^{12}$) along $\phi$ 
coordinate keeping the value of $\rho$ fixed at 0.31 \AA, 0.41 \AA and 0.51 \AA.~\citep{saikat_pyra} It is interesting to note that two seams between $\phi=0$ and $\phi=\pi$ are gradually approaching 
to each other at lower $\rho$ value leading to only one seam passing through $\phi=\pi/2$. Similarly, other two 
seams between $\phi=\pi$ and $\phi=2\pi$ join at $\phi=3\pi/2$ to generate only one seam.~\citep{saikat_pyra} The locations of the seams are clearly identified in
the 2D functional form of $\tau_{\phi}^{12}$ in Figure \ref{fig:1-9a_1-2_adia}d exhibiting singular NACTs along two semi-circular seams. This nuclear plane also exhibits two
``3-4" ($1^1A_{u}$ - $1^1B_{2g}$) accidental CIs at $\rho$ = 0.15 \AA\, and $\phi=\pi/2$ and $3\pi/2$. On the other hand, another nuclear plane, $Q_{10a}-Q_{6a}$ is extremely rich in showing ``1-2" ($1^1B_{1u}$ - 
$1^1B_{2u}$) accidental CI, ``2-3" accidental seams ($1^1B_{2u}$-$1^1A_{u}$) and ``3-4" ($Q_{10a}-Q_{6a}$) accidental seams.~\citep{saikat_pyra}\\

\begin{figure}[!htp]
	\centering
	\includegraphics[height=0.55\textheight, angle=0]{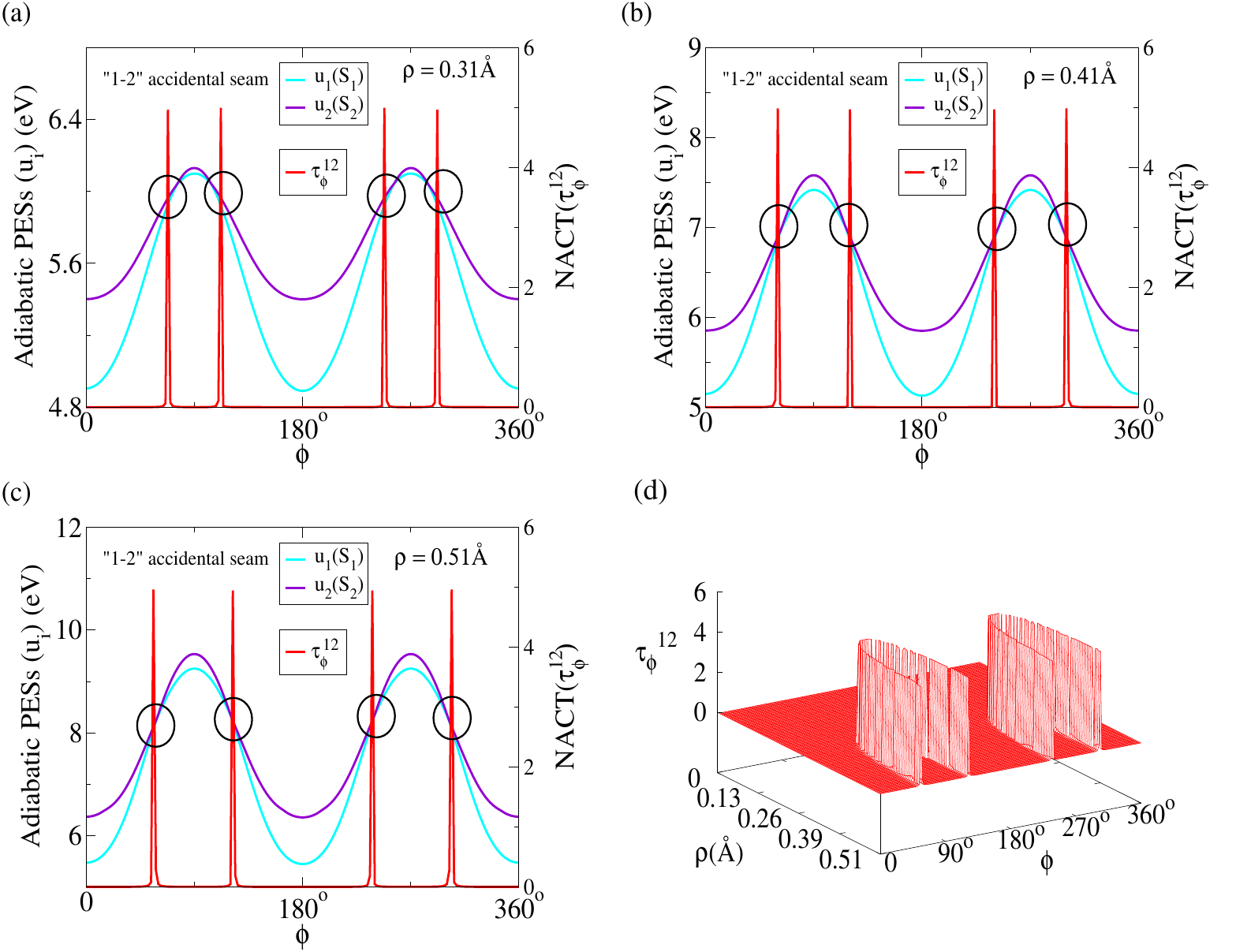}
	\caption{The diagram exhibits the location of ``1-2" ($1^1B_{1u}$ - $1^1B_{2u}$) accidental seams over the polar plane ($\rho$-$\phi$) of $Q_{9a}-Q_{1}$ pair. 1D curves of adiabatic PESs ($u_1$ and $u_2$) and the corresponding $\tau_{\phi}^{12}$ are presented along $\phi$ coordinate at three values of $\rho$ = (a) 0.31 \AA, (b) 0.41 \AA\, and (c) 0.51 \AA. $\tau_{\phi}^{12}$. where the later ones (NACTs) shows singular feature at the points of degeneracies. Diagram (d) shows the associated 2D functional forms of $\tau_{\phi}^{12}$. This figure has been taken from Ref. [\protect\citen{saikat_pyra}] and permission has been taken from Wiley.}
	\label{fig:1-9a_1-2_adia}
\end{figure}

\noindent
The {\it ab initio} calculated functions (adiabatic PESs and NACTs) had been plugged into the ADT equations for four electronic state SHS (see Appendix~\ref{ssec:adt-eq}) and then, those equations were solved to obtain the ADT angles, which lead to ADT and diabatic PESs matrices. 
Since $Q_{9a}-Q_{1}$ exhibits two ``1-2" accidental seams and two ``3-4" accidental CIs, the functional forms
of both the ADT angles, $\Theta_{12}$ and $\Theta_{34}$ attend the magnitude of $2\pi$ (see Figure \ref{fig:adt}a) at the end of the closed contour ($\phi=2\pi$). In case of $Q_{10a}-Q_{6a}$, the
magnitudes of ADT angles, $\Theta_{12}$, $\Theta_{23}$ and $\Theta_{34}$ reach the value as $\pi$, $2\pi$ and $2\pi$, respectively at $\phi=2\pi$ which is clear from Figure \ref{fig:adt}b. For both cases, the 1D curves (Figures \ref{fig:adt}) of the ADT angles appear as topological ($e^{\pm i\omega_{\phi}(\phi)}$) instead of geometric\citep{herzberg_dfs35,pancharatnam_pias44,berry_prca392} [$e^{\pm in(\phi/2)}$, 
$n$ is the number of enclosed CI(s)/seam(s)]. The diabatic PESs and couplings have been obtained from CASSCF as well as MRCI computed adiabatic PESs and CP-MCSCF calculated NACTs, which appear as smooth, continuous and single-valued.\\

\begin{figure}[!htp]
	\centering
	\includegraphics[ height=0.3\textheight, width=0.9\textwidth, angle=0]{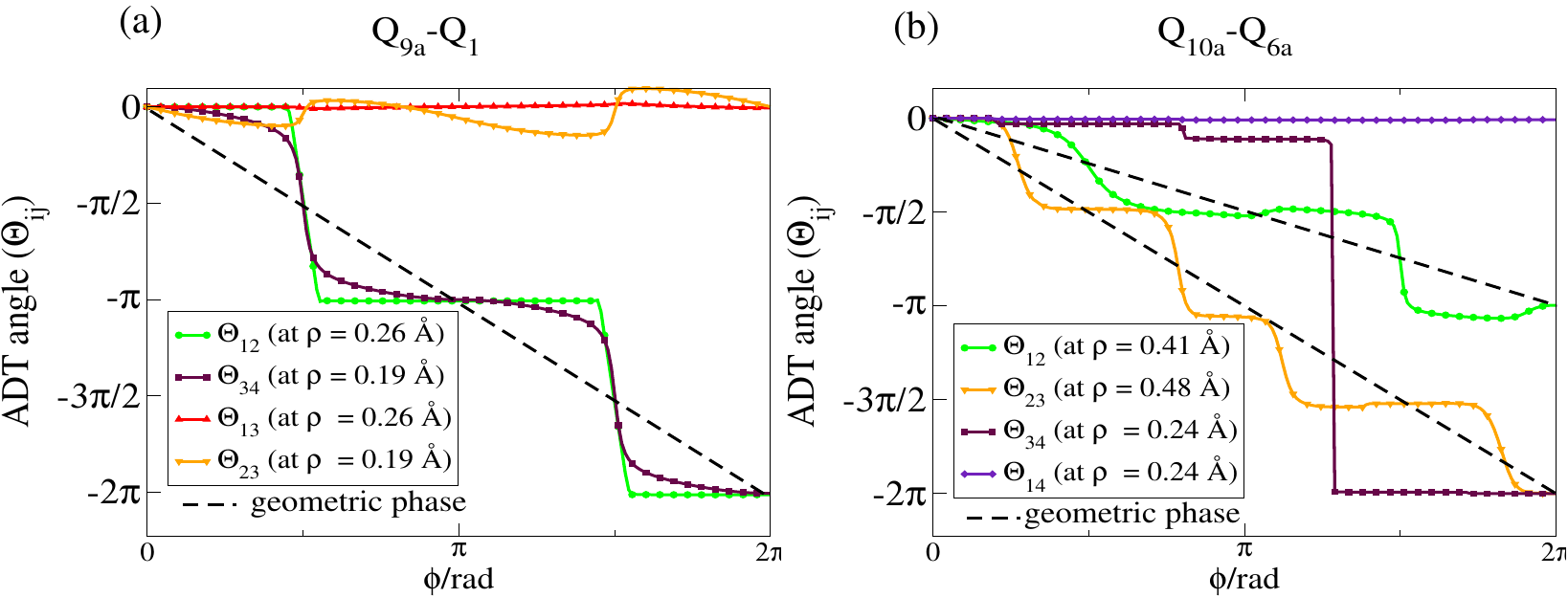}
	\caption{The left panel (a) depicts 1D functional forms of ADT angles, $\Theta_{12}$, $\Theta_{34}$, $\Theta_{13}$ and $\Theta_{23}$ for $Q_{9a}-Q_{1}$ pair, whereas the right one (b) shows $\Theta_{12}$, $\Theta_{23}$, $\Theta_{34}$ and $\Theta_{14}$ for $Q_{10a}-Q_{6a}$ pair along $\phi$ at various fixed values of $\rho$. This figure has been adapted from Ref. [\protect\citen{saikat_pyra}] and permission has been taken from Wiley.}    
	\label{fig:adt}
\end{figure}

\noindent
While depicting the size of the sub-Hilbert space for C$_4$N$_2$H$_4$ molecule, Mathematical Curls ($Z_{\rho\phi}^{ij}$) and ADT Curls ($C_{\rho\phi}^{ij}$) were computed involving the ADT angles~\citep{saikat_pyra} into their explicit expressions (see Appendix~\ref{ssec:adt-eq}). 
Figures \ref{fig:pyr}a and \ref{fig:pyr}c show Mathematical Curls associated with the ``1-2" and ``3-4" interactions ($Z_{\rho\phi}^{12}$ and $Z_{\rho\phi}^{34}$), respectively, whereas the 
corresponding ADT Curls ($C_{\rho\phi}^{12}$ and $C_{\rho\phi}^{34}$) are depicted in Figures \ref{fig:pyr}b and \ref{fig:pyr}d. Since the functional forms of Mathematical and ADT Curls are appeared superimposable with each other for both the cases, Curl Conditions ($F_{\rho\phi}^{12}$ and $F_{\rho\phi}^{34}$) turn into the values close to zero ($<$ 10$^{-12}$ $\AA^{-2}$) validating the existence of four-state SHS. \\

\begin{figure}[htp]
	\centering
	\includegraphics[width=0.9\textwidth]{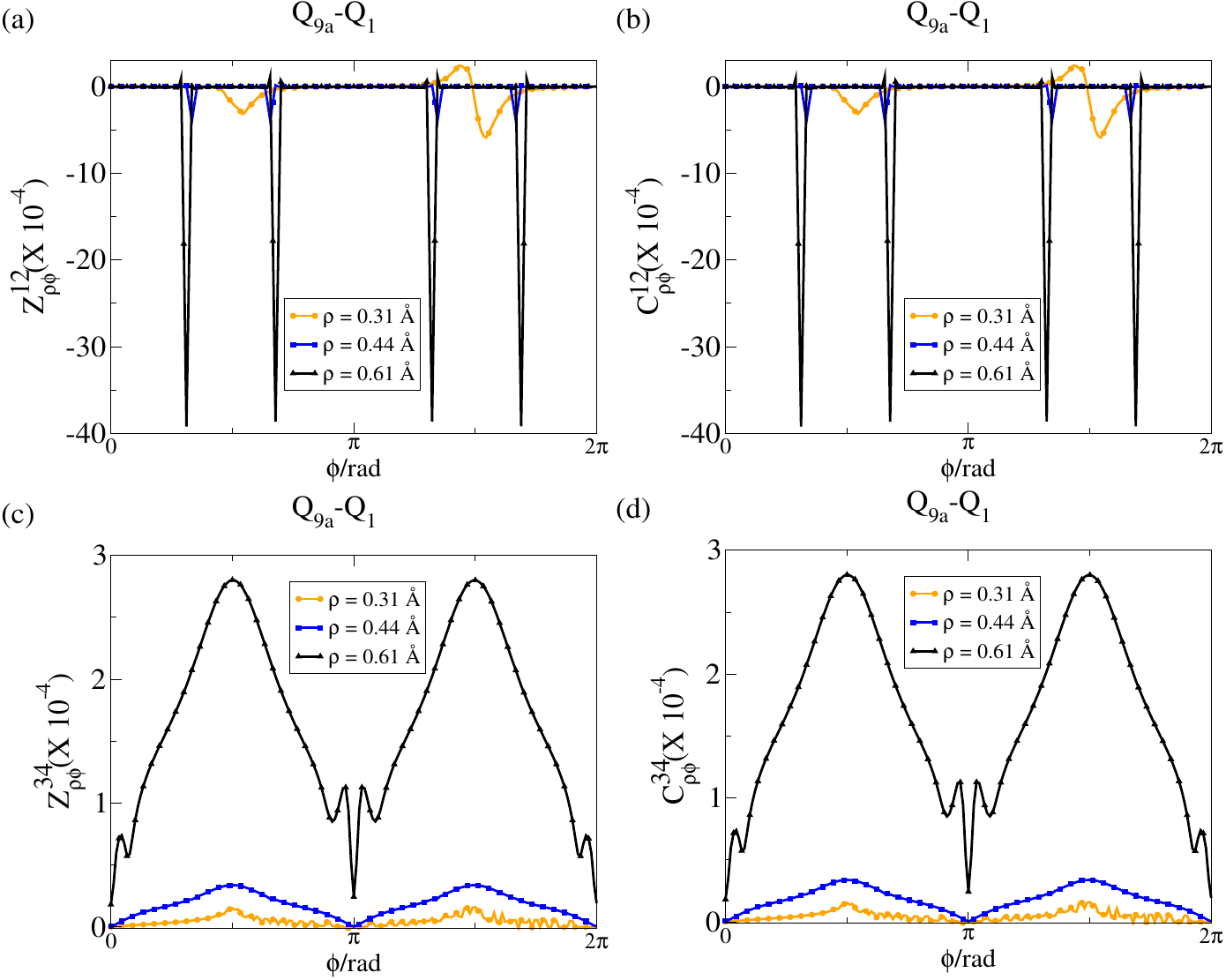}
	\caption{The left column shows Mathematical Curls, (a) $Z_{\rho\phi}^{12}$
		and (c) $Z_{\rho\phi}^{34}$, whereas the right one depicts the corresponding ADT Curls, 
		(b) $C_{\rho\phi}^{12}$ and (d) $C_{\rho\phi}^{34}$ along the $\phi$ coordinate at three different $\rho$ values for $Q_{9a}-Q_{1}$ normal mode pair. This figure has been adapted from Ref. [\protect\citen{mantu_curl}] and permission has been taken from Wiley.}
	\label{fig:pyr}
\end{figure}

\noindent
Once diabatic PESs and couplings were obtained, nuclear dynamics were performed over the diabatic Hamiltonian by placing the initial wavepacket on each of the electronic ($1^1B_{1u}$, $1^1B_{2u}$, $1^1A_{u}$ and $1^1B_{2g}$) states to calculate the autocorrelation functions for each case. The time-dependent autocorrelation functions were transformed into energy space (Fourier Transformation) and then, the spectra for each state were convoluted to obtain the overall spectra. TDDVR calculated envelops arising from BBO theory based diabatic Hamiltonian 
exhibit better agreement with the experimental ones~\citep{yamazaki_fdcs75} in comparison with the MCTDH results obtained from 2$\times$2 VCM Hamiltonian~\citep{woywood_jcp100,raab_jcp110} as presented in Figure \ref{fig:tddvr-2}. It is worth mentioning that the PA spectra for
$1^1B_{2u}$ state are broadened mainly due to strong PJT couplings between second and
the remaining three electronic states ($1^1B_{1u}$, $1^1A_{u}$ and $1^1B_{2g}$). The first principle based study unveils the ``accuracy" and ``exactness" of BBO formalism vis-\'a-vis TDDVR dynamics for ``correct" assignments of peaks in PA spectra for spectroscopically important system, like pyrazine.

\begin{figure}[!htp]
	\centering
	\includegraphics[height=0.5\textheight, angle=0]{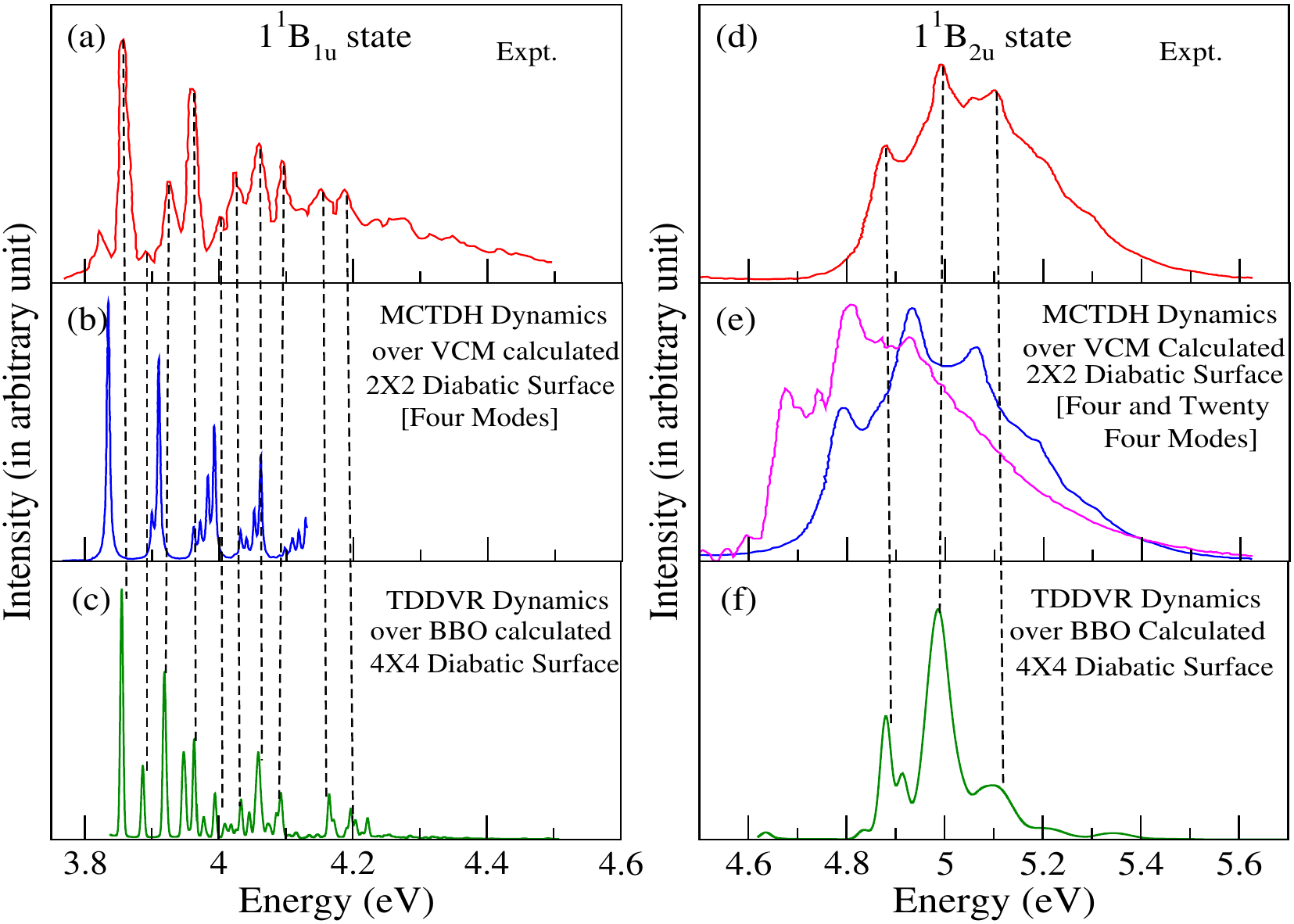}
	\caption{The diagram depicts (a) experimental PA spectrum~\citep{yamazaki_fdcs75} of pyrazine molecule for
		$1^1B_{1u}$ ($S_1$) state in comparison with (b) MCTDH spectrum calculated by VCM based 
		2$\times$2 quasi-diabatic surfaces~\citep{woywood_jcp100} as well as (c) TDDVR spectrum evaluated from
		BBO based 4$\times$4 diabatic surfaces. In a similar way, the right panel (d-e) demonstrates spectral bands for $1^1B_{2u}$ ($S_2$) state, where MCTDH calculated spectra have been obtained from both four~\citep{woywood_jcp100} and twenty-four~\citep{raab_jcp110} mode Hamiltonian. This figure has been taken from Ref. [\protect\citen{saikat_pyra}] and permission has been taken from Wiley.}
	\label{fig:tddvr-2}
\end{figure}

\noindent
\section{Five State : NO$_3$, o-C$_6$H$_4$F$_2^+$}	

\vspace{0.2 cm}
\noindent 
Nitrate radical (NO$_3$) and ortho-difluorobenzene radical cation (o-C$_6$H$_4$F$_2^+$) are two representative spectroscopically interesting systems, 
which were extensively investigated over the past few decades~\citep{mahapatra_cp347,faraji_jcp_struct,sardar_pccp,bijit_mp115,saikat_cpc26} due to the highly complex spectral bands of NO$_3$ and o-C$_6$H$_4$F$_2$ molecules. In the following section, BBO based diabatization and TDDVR dynamics are explored for NO$_3$~\citep{bijit_mp115} and o-C$_6$H$_4$F$_2^+$~\citep{saikat_cpc26} systems.

\subsection{NO$_3$ Radical}
\vspace{0.2 cm}
\noindent
Due to the presence of profound vibronic coupling 
between the low-lying (within 2 eV) five electronic states, $\tilde{X}^2A_2^{\prime}$ ($1^2B_2$), 
$\tilde{A}^2E^{\prime\prime}$ ($1^2A_2$ and $1^2B_1$)
and $\tilde{B}^2E^{\prime}$ ($1^2A_1$ and $2^2B_2$) of NO$_{3}$ radical, it shows highly complex spectral features. Several theoretical groups ~\citep{eisfeld_jcp113,mayer_jcp100,mahapatra_cp347,okumura_ps73,stanton_jcp126,soumya_jpca121,bijit_mp115,
	bijit_cp515} attempted to explore and understand experimentally recorded complex spectral envelop obtained from various techniques, like
dispersed fluorescence spectroscopy,\citep{nelson_jpc87,ishiwata_jpc87} Fourier-transform infrared spectroscopy,\citep{ishiwata_jcp82,hirota_jcp107}
photoelectron spectroscopy\cite{weaver_jcp94} and also from cavity ringdown spectroscopy (CRDS)\citep{deev_jcp122,codd_jcp142}. Recently, Adhikari and coworkers~\citep{soumya_jpca121,bijit_mp115} 
unveiled that JT interactions exist within the $\tilde{A}^2E^{\prime\prime}$ and $\tilde{B}^2E^{\prime}$ states at equilibrium D$_{3h}$ point, whereas accidental CIs can be observed within `3-4' and `4-5'  electronic states at $C_{s}$ and $C_{2v}$ geometries. Apart from the symmetry-allowed and accidental JT type CIs,
NO$_{3}$ shows substantial PJT coupling between $\tilde{X}^2A_2^{\prime}$ and $\tilde{B}^2E^{\prime}$ states.\citep{soumya_jpca121,bijit_mp115}
Such highly nonadiabatically coupled electronic manifold appears to be challenging even to predict the correct ground state equilibrium geometry of this radical, but finally, Eisfeld and Morokuma concluded to be of D$_{3h}$ symmetry.\citep{eisfeld_jcp113}\\

\noindent
Adhikari and coworkers carried out an extensive study to investigate 
all the contributing nonadiabatic interactions present over the nuclear planes constituted with symmetric stretching ($Q_1$), umbrella ($Q_2$) and degenerate asymmetric stretching ($Q_{3x}$ and $Q_{3y}$) modes.\citep{soumya_jpca121,bijit_mp115} It had been explored from the JT condition that $e^\prime$ symmetric ($Q_{3x}$, $Q_{3y}$) vibrational modes are mainly responsible for the JT distortion, which originates from the coupling of electronic states with vibrational modes of appropriate symmetry. In other words, the direct product of irreducible representations for the electronic state and the vibrational mode contains the totally symmetric representation ($ A_{1}^{\prime} $). The JT distortion  takes place between $^2E^{\prime}$ states via $e^{\prime}$ normal modes:

$$A_{1}^{\prime} \in e^\prime \otimes [(E^{\prime})^2] \equiv e^\prime \otimes (A_1^{\prime} + A_2^{\prime} + E^{\prime})$$

\noindent
The adiabatic PESs and NACTs were computed as functions of polar counterparts ($\rho$ and $\phi$)
of each normal mode pair ($Q_i$ and $Q_j$) with the use of MRCI approach in MOLPRO quantum chemistry package\citep{MOLPRO_brief}. Those {\it ab initio} calculated NACTs were substituted into the ADT equations for five-state SHS (see Appendix~\ref{ssec:adt-eq}) resulting into mixing angles at each nuclear configuration points over those 2D nuclear planes.\citep{bijit_cp515} NO$_3$ radical possesses one ``2-3" JT CI as well as one ``4-5" JT and three accidental CIs over $Q_{3x}-Q_{3y}$ plane, which are evident from the 2D plots of $\phi$ components of NACTs as depicted in Figure \ref{fig:no3nact}. Figure \ref{fig:no3adt} shows that angular variation ($\phi$) of ADT angles [$\Theta_{23}$ ($Q_{3x}$, $Q_{3y}$) and $\Theta_{45}$ ($Q_{3x}$, $Q_{3y}$)] at fixed $\rho\,\,$(= 0.21 \AA) fulfills the quantization property of NACTs, where such angles attend the magnitude of $\pi$ and $4\pi$, respectively. This calculation validates the presence of enclosed `2-3' CI (JT CI at D$_{3h}$ point) and `4-5' CIs (one JT CI at D$_{3h}$ geometry and three 
accidental CIs at equivalent C$_{2v}$ points) inside the circular loop. 
On the other hand, the diagonal elements of ADT matrices, namely, $A_{22}\,[=A_{33}]$ and $A_{44}\,[=A_{55}]$ exhibit one and four sign inversions, respectively, indicating the presence of aforementioned CIs.
As per Figure \ref{fig:no3adt}, the functional behaviour of ADT angles 
shows molecule-specific nonadiabatic features [$\omega_{\phi}(\phi)$, see Eq. \ref{eq:gauge} of Subchapter 4.7] that could be termed as topological 
($e^{\pm i\omega_{\phi}(\phi)}$) phase rather than geometric ($e^{\pm in\phi}$)\citep{herzberg_dfs35,pancharatnam_pias44,berry_prca392} (indicated by the dashed lines). Apart from the `2-3' and `4-5' CIs,
two ``3-4" C$_s$ CIs and one ``3-4" C$_s$ CI can be observed over
$Q_2-Q_{3x}$ and $Q_2-Q_{3y}$ planes, respectively.\citep{soumya_jpca121,bijit_cp515} The diabatic Hamiltonian matrix elements appear as smooth, single-valued and continuous
over the nuclear CS. It is worthwhile to note that diabatic PESs as well as the couplings show symmetric functional forms along $Q_{3x}$ normal mode, which can be fitted only with even power polynomial terms,\citep{soumya_acs} in contrast to the traditional JT model (see Figure \ref{fig:no3even} for $Q_{3x}$ mode). \\

\begin{figure}[!htp]
	\centering
	\includegraphics[ width=1.1\textwidth, angle=0]{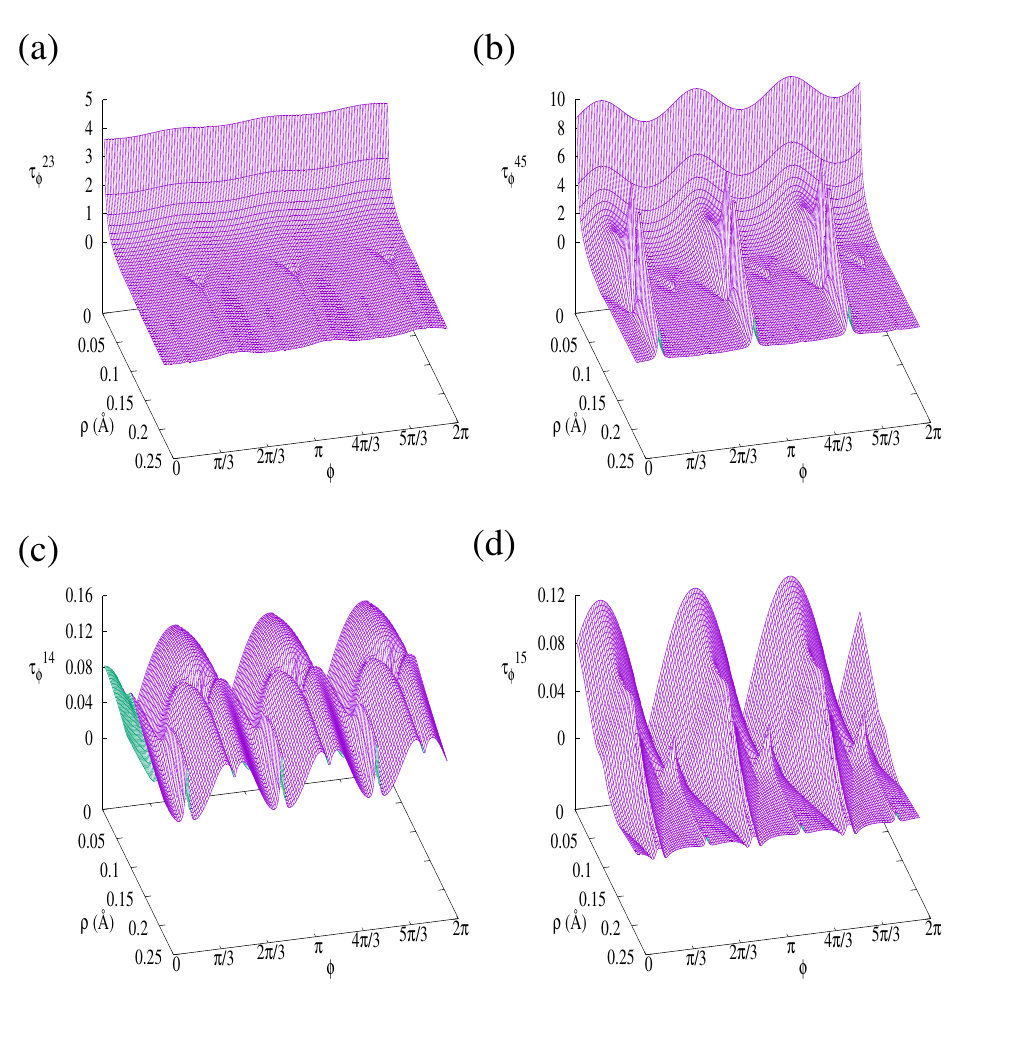}
	\caption{2D functional forms of (a) $\tau_{\phi}^{23}$, (b) $\tau_{\phi}^{45}$, (c) $\tau_{\phi}^{14}$ and (b) $\tau_{\phi}^{15}$ are presented over $\rho-\phi$ plane of $Q_{3x}-Q_{3y}$ pair. This data has been taken from Ref. [\protect\citen{bijit_mp115}].}
	\label{fig:no3nact}
\end{figure}

\begin{figure}[!htp]
	\centering
	\includegraphics[ width=0.7\textwidth, angle=0]{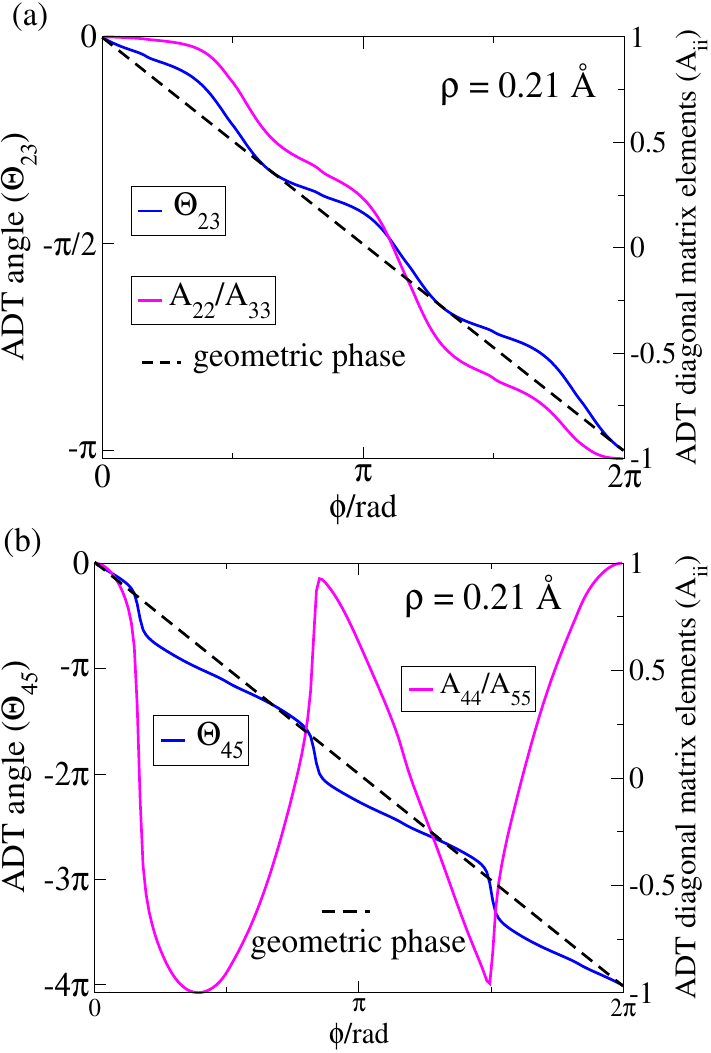}
	\caption{1D curves of (a) $\Theta_{23}$ and (b) 
		$\Theta_{45}$ for $Q_{3x}-Q_{3y}$ normal mode pair are plotted along $\phi$ coordinate at definite $\rho$ (= 0.21 \AA) attaining the value of $\pi$ and $4\pi$, respectively due to the presence of encapsulated one `2-3' JT CI and four `4-5' CIs (one JT and three accidental ones). On the other hand, (a) $A_{22}\,[=A_{33}]$ and (b) $A_{44}\,[=A_{55}]$ encounter one and four sign inversions, respectively, according to the number of encapsulated CI. This data has been taken from Ref. [\protect\citen{joy_pccp}] and permission has been taken from RSC}
	\label{fig:no3adt}
\end{figure}

\begin{figure}[!htp]
	\centering
	\includegraphics[ width=0.7\textwidth, angle=0]{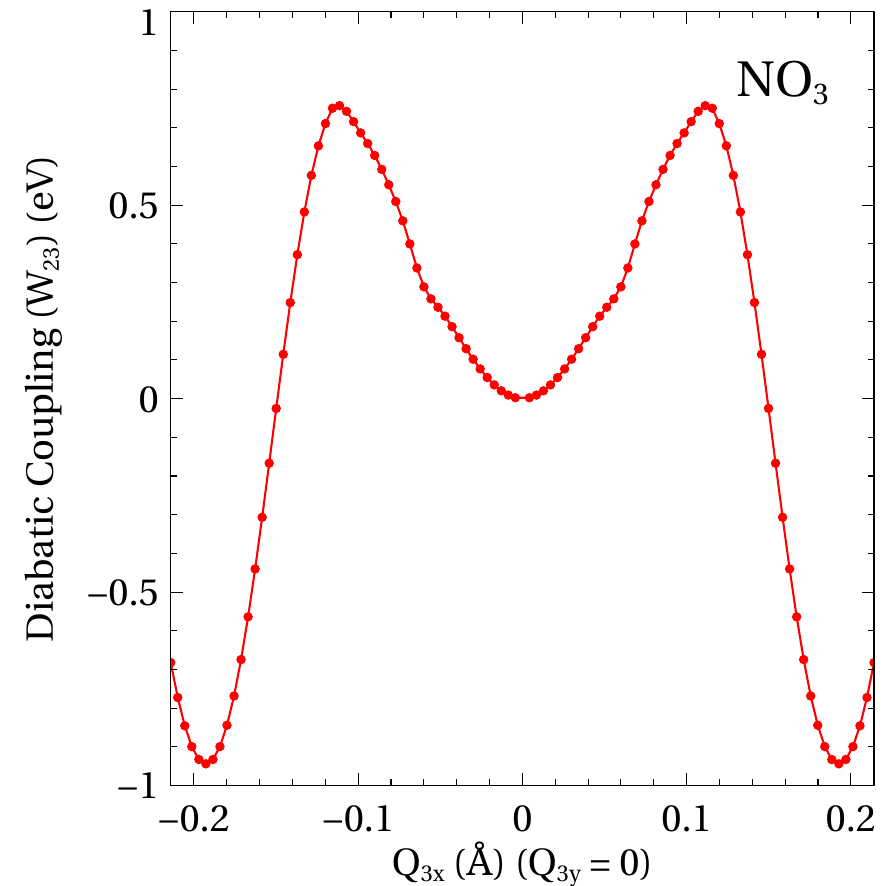}
	\caption{1D curve of diabatic coupling element, namely, $W_{23}$ among $\tilde{A}^2E^{\prime\prime}$ states along $Q_{3x}$ mode of NO$_3$ radical is presented. The coupling term exhibits symmetric functional form, which can be fitted with even parity polynomials. This data has been taken from Ref. [\protect\citen{joy_pccp}] and permission has been taken from RSC.}
	\label{fig:no3even}
\end{figure}

\begin{figure}[!htp]
	\centering
	\includegraphics[width=0.6\textwidth,angle=0]{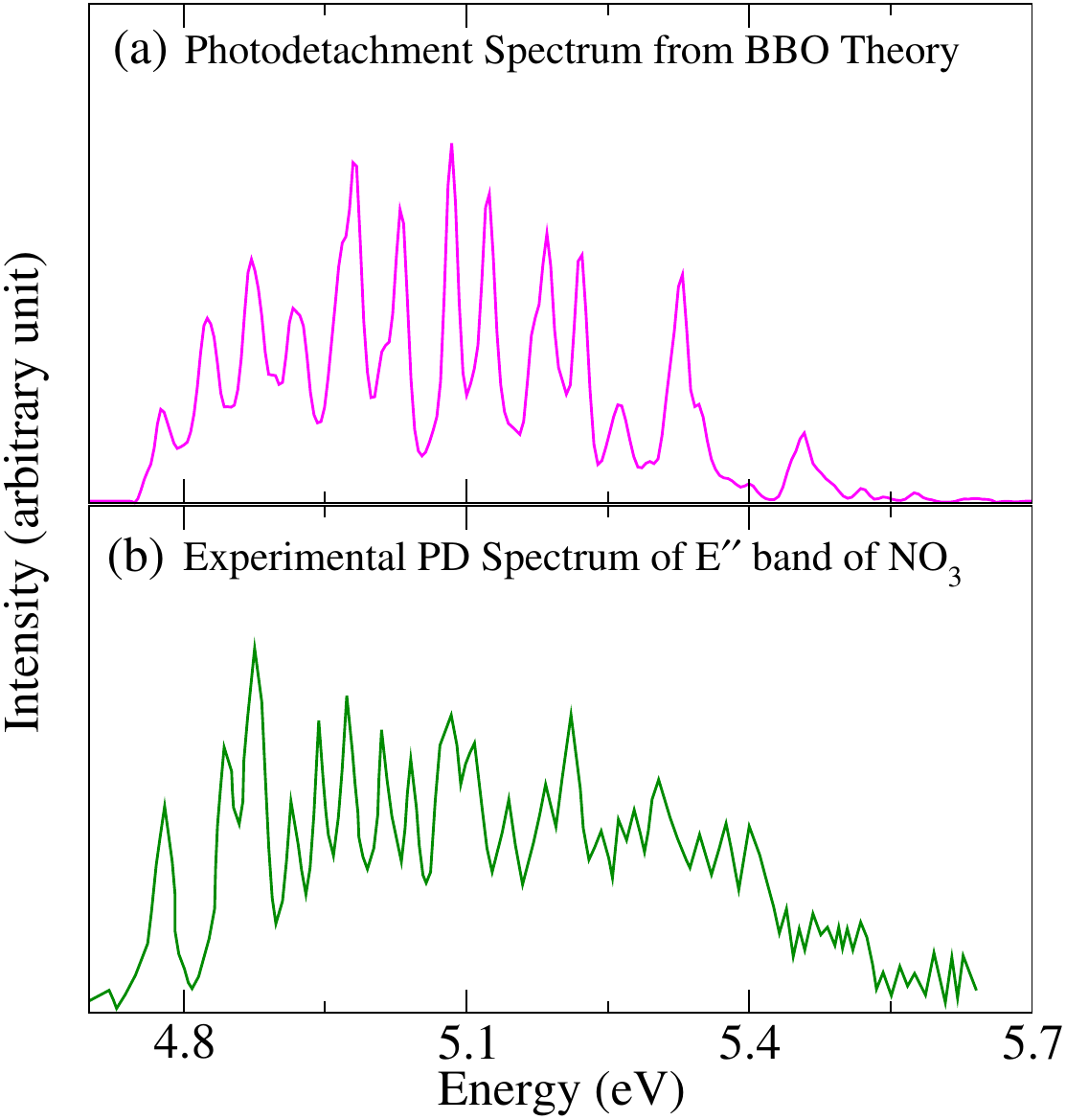}
	\caption{The upper column presents (a) theoretically calculated photodetachment spectra of NO$_3$ molecule 
		originating from BBO theory based diabatic PESs for $E^{\prime\prime}$ states, which is compared with the experimental one\citep{weaver_jcp94} in panel b. The results are taken from Ref. [\protect\citen{joy_pccp}] and permission has been taken from RSC.}
	\label{fig:no3spec}
\end{figure}

\noindent
In order to generate the photodetachment spectra of NO$_3$ molecule,
dynamical calculations were performed over the BBO theory based diabatic PESs of five coupled electronic manifold. The initial wavepacket was constituted with the product of GWP for each normal mode of NO$_3^-$, which was placed over each of the five low-lying electronic states ($\tilde{X}^2A_2^{\prime}$, $\tilde{A}^2E^{\prime \prime}$ and $\tilde{B}^2E^{\prime}$) one by one, produced the time-dependent wavefunction vis-\'a-vis autocorrelation functions ($C(t)$s) at various time. Finally, Fourier transformation of $C(t)$s were carried out to generate the spectral envelop for each wavepacket initiated from various electronic states. Then, the individual state spectra were convoluted to get the entire spectral profile. Figure \ref{fig:no3spec} depicts theoretical spectrum for $E^{\prime \prime}$ state\citep{bijit_cp515} obtained from BBO-DH in comparison with the experimental outcome\citep{weaver_jcp94}. For most of the peaks, both the spectra exhibit satisfactory agreement with each other, but some deviations appear due to the hot bands. It is interesting to note that the highly structured profile of 
$E^{\prime \prime}$ state is enriched with $A_2^\prime \rightarrow E^{\prime\prime}$ dipole forbidden transitions, which are mainly governed by umbrella mode, $Q_2$. On the other hand, $Q_3$ vibration mediated
JT distortion, more specifically along $Q_{3y}$ normal mode as well as other symmetry allowed accidental CIs produce some anharmonicity in the PESs and thereby, result into many dipole forbidden transitions. 

\subsection{o-C$_6$H$_4$F$_2^+$ Radical Cation}
\vspace{0.2 cm}
Over the last five decades, a wide variety of experimental techniques, namely, calorimetry, NMR, X-ray photoelectron and photoabsorption spectroscopy, and many others were employed to explore the geometrical parameters, normal mode frequencies, vibrational levels of various electronic states as well as spectral properties of o-C$_6$H$_4$F$_2$ (o-DFBz) and o-DFBz$^+$ species.~\citep{scott_jcp,abraham_jcsb,gerritsen_sapms,clark_jesc,shirley_jacs,gilbert_cpl,bieri_jesc,tsuchiya_cpl,tsuchiya_jcp,philis_jpb,goly_cp,kwon_jcp_2003}
Clark and co-workers\cite{clark_jesc} as well as Shirley group\cite{shirley_jacs} recorded the X-ray photoelectron spectra (XPS) to unveil the binding energies of core molecular orbitals of the titled system. On the other hand, Gilbert and Sandorfy\cite{gilbert_cpl} measured photoabsorption spectra, where
those bands originates from the lowest ionization potentials of the $\pi$ electrons. In the early eighties, 
Bieri {\it et al.}\cite{bieri_jesc} carried out the photoelectron spectra of low-lying eighteen (18) electronic states of the titled system. Furthermore, several other experimental techniques, namely, photoabsorption spectroscopy\cite{tsuchiya_cpl,tsuchiya_jcp,philis_jpb} and Mass-Analysed Threshold Ionization (MATI) spectroscopy\cite{kwon_jcp_2003} were utilized to reveal the complex electronic structure and energy spectrum of the neutral analogue. On the other hand, K\"oppel and coworkers\cite{faraji_jcp_struct} derived VCM based quasi-diabatic Hamiltonian involving five low-lying doublet electronic states ($\tilde{X}^2B_1$, $\tilde{A}^2A_2$, $\tilde{B}^2B_1$, $\tilde{C}^2A_1$ and $\tilde{D}^2B_2$) of the cationic species, o-DFBz$^+$ as functions of twelve coupling modes ($Q_6$, $Q_7$, $Q_8$, $Q_{12+15}$, $Q_{13}$, $Q_{14}$, $Q_{20}$, $Q_{22}$, $Q_{24}$, $Q_{25}$ and $Q_{26}$). The MCTDH dynamics\cite{faraji_jcp_dyna} on this Hamiltonian had been performed to calculate the photoelectron and MATI spectra of the neutral molecule, 
o-DFBz, and then, the spectral envelops were compared with the available experimental spectral profiles\cite{bieri_jesc,kwon_jcp_2003}. Later on, Adhikari and co-workers\cite{sardar_pccp} executed TDDVR dynamics over the same VCM based model Hamiltonian\cite{faraji_jcp_struct} for the cationic species. The simulated spectra were compared together with experimental\cite{bieri_jesc,kwon_jcp_2003} as well as theoretical\cite{faraji_jcp_dyna} results to depict the accuracy of TDDVR dynamics. \\

\noindent
Adhikari {\it et al.}~\citep{saikat_cpc26} carried out \textit{ab initio} calculation followed by BBO based diabatization over eight (8) 2D nuclear planes constituted with eight (8) non-adiabatically active normal modes (i.e., $Q_6$, $Q_7$, $Q_8$, $Q_{13}$ , $Q_{14}$, $Q_{20}$, $Q_{24}$ and $Q_{26}$), where those normal modes are defined as in plane C-C-C bending ($Q_6$ and $Q_8$), out of plane C-C-C bending ($Q_7$), out of plane C-H bending ($Q_{13}$ and $Q_{14}$), in plane C-F stretching ($Q_{20}$) and in plane C-C stretching ($Q_{24}$ and $Q_{26}$). Among these normal modes, the totally symmetric modes ($Q_8$, $Q_{20}$, $Q_{24}$ and $Q_{26}$) behave as tuning ones, where the non-totally symmetric ones ($Q_6$, $Q_7$, $Q_{13}$ and $Q_{14}$) serve as coupling modes. 
The adiabatic PESs were calculated using CASSCF methodology, followed by MRCI with the aid of correlation-consistent polarized valence double-$\xi$ (cc-pvDZ) basis set in no symmetry situation ($C_1$ condition). The state-average calculation incorporates the low-lying five electronic states ($\tilde{X}^2B_1$, $\tilde{A}^2A_2$, $\tilde{B}^2B_1$, $\tilde{C}^2A_1$ and $\tilde{D}^2B_2$), with an active space of eleven electrons in six orbitals (11e, 6o), which
involes lone pairs of fluorine and the $\pi$ orbitals of the o-DFBz$^+$ moiety.\\

\noindent
While scanning adiabatic PESs and NACTs, the pair of normal modes ($Q_k$, $Q_l$) for a nuclear plane were converted into their mass weighted dimensionless polar counterparts ($\rho$ and 
$\phi$), where $\rho$ and $\phi$ coordinates span from 0.2 to 10.0 and 0 to $2\pi$, respectively, possessing 50 × 181 grid points at and around the FC region of 
the nuclear CS. In reality, the normal mode coordinates range from -0.8 $\AA$ to +0.8 $\AA$. It is interesting to note that various non-adiabatic 
interactions, such as accidental conical intersections (CIs), accidental seams, and pseudo-Jahn-Teller (PJT) couplings can be observed over the  eight chosen 
normal mode ($Q_6$, $Q_7$, $Q_8$, $Q_{13}$ , $Q_{14}$, $Q_{20}$, $Q_{24}$ and $Q_{26}$), whereas the other four appear to be less contributing ones ($Q_{12}$, $Q_{15}$, $Q_{22}$ and $Q_{25}$)~\citep{saikat_cpc26}. Figure \ref{fig:q6-q7_12} presents 1D curves of adiabatic PESs of $\tilde{X}^2B_1$ and $\tilde{A}^2A_2$ state and the associated $\phi$ 
component of NACTs ($\tau^{\phi}_{12}$) as functions of $\rho$ coordinate at fixed value of $\phi$ = 8$^\circ$. The $\tau^{\phi}_{12}$ exhibits a sharp rise
at the point of CI ($\rho$ = 0.67 \AA) for $Q_6-Q_7$ plane. For this nuclear plane, the figure clearly depicts the presence of accidental CI between electronic 
states (i.e., $\tilde{X}^2B_1$ and $\tilde{A}^2A_2$) along a particular value of $\phi$ coordinate. On the other hand, Figure \ref{fig:q6-q7_34}(a–c) manifests 1D
variation of adiabatic PESs for $\tilde{B}^2B_1$ - $\tilde{C}^2A_1$ electronic states (u$_3$ and u$_4$) and the associated $\phi$ component of NACTs 
($\tau^{\phi}_{34}$) along $\phi$ coordinate keeping the values of $\rho$ fixed at three different values (0.24 \AA, 0.382 \AA\, and 0.54 \AA). As per these three figures 
(\ref{fig:q6-q7_34}(a–c)), the semicircular ``3-4" accidental seam is located over $\rho$-space from 3.6 to 0.67 \AA\, and $\phi$-space from 24$^{\circ}$ to 
160$^{\circ}$, respectively. Furthermore, in Figure \ref{fig:q6-q7_34}(d), the overall characteristic of the seam and the CI are manifested in the 2D functional 
form of $\tau^{\phi}_{34}$, exhibiting singular NACTs along the semi-circular seam as well as at the CI point. For the sake of better clarification, 
Figure \ref{fig:q6-q7_34}(e) are presented to show 1D variation of adiabatic PESs and the corresponding NACTs ($\tau^{\phi}_{34}$) along the $\rho$ coordinate at 
a fixed magnitude of $\phi$ (= 270$^{\circ}$). Finally, two ``4-5" ($\tilde{C}^2A_1$ - $\tilde{D}^2B_2$) accidental CIs are portrayed in Figure \ref{fig:q6-q7_45} 
at $\rho$ = 0.67 \AA\, with $\phi = 90^{\circ}$ and $262^{\circ}$. Not only JT interactions, but also profound PJT couplings are reflected in the functional forms of NACTs, specifically, $\tau^{\phi}_{13}$, 
$\tau^{\phi}_{14}$, $\tau^{\phi}_{15}$, $\tau^{\phi}_{23}$, $\tau^{\phi}_{24}$, $\tau^{\phi}_{25}$ and $\tau^{\phi}_{35}$. \\

\begin{figure}[htp]
	\centering
	\includegraphics[ width=8 cm, height=0.35\linewidth,angle=0]{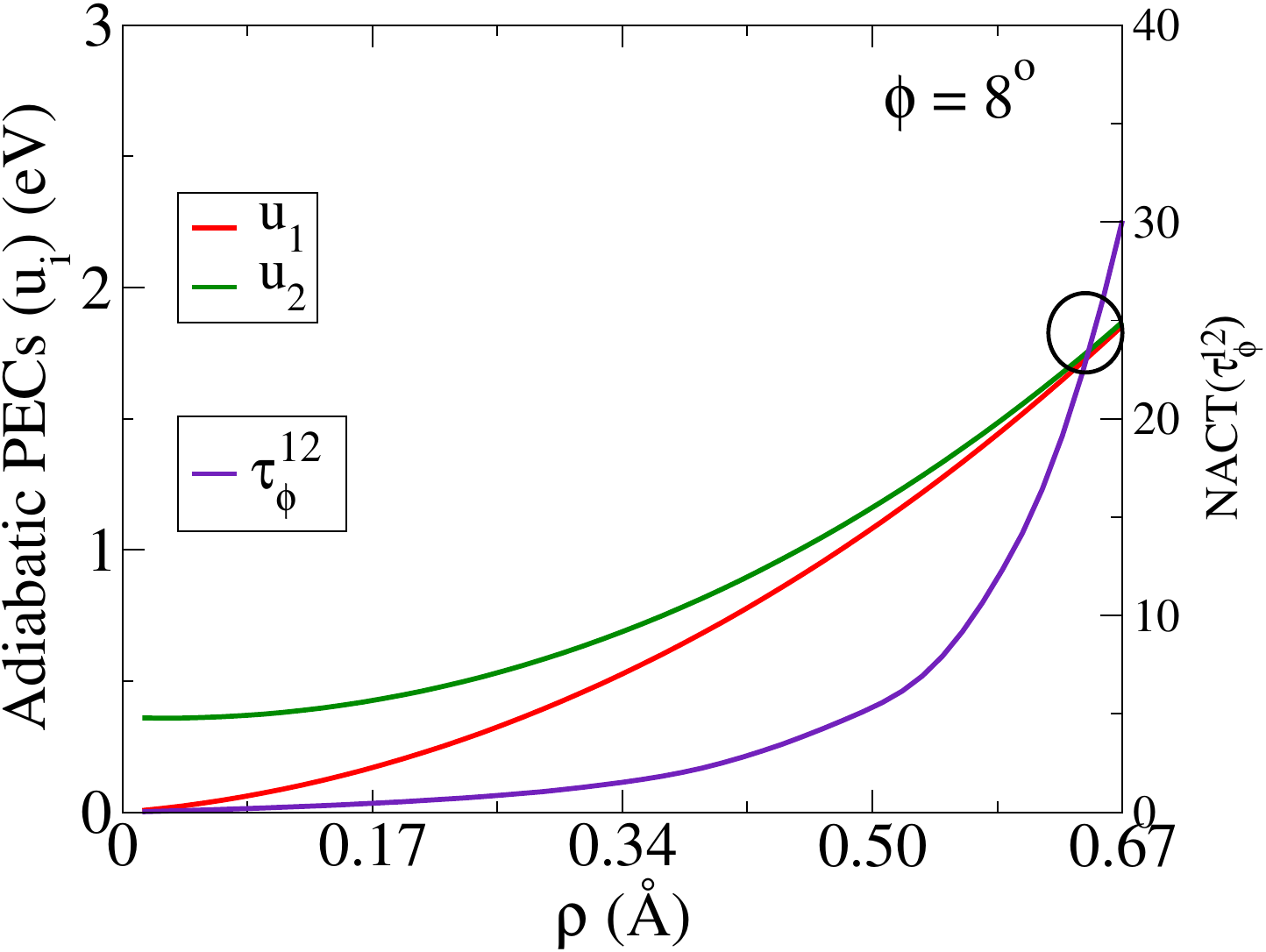}
	\caption{The 1D curves of adiabatic PESs (u$_1$ and u$_2$) and the
		corresponding NACTs ($\tau^{\phi}_{12}$) between $\tilde{X}^2B_1$ and $\tilde{A}^2A_2$ electronic states are
		depicted along $\rho$ coordinate at $\phi = 8^{\circ}$. The $\tau^{\phi}_{12}$ rises sharply
		at the point of CI ($\rho$ = 0.67 \AA). The figure is adapted from Ref. [\protect\citen{saikat_cpc26}] and permission has been taken from Wiley.}
	\label{fig:q6-q7_12}
\end{figure}

\begin{figure}[htp]
	\centering
	\includegraphics[ width=16 cm, height=1.0\linewidth,angle=0]{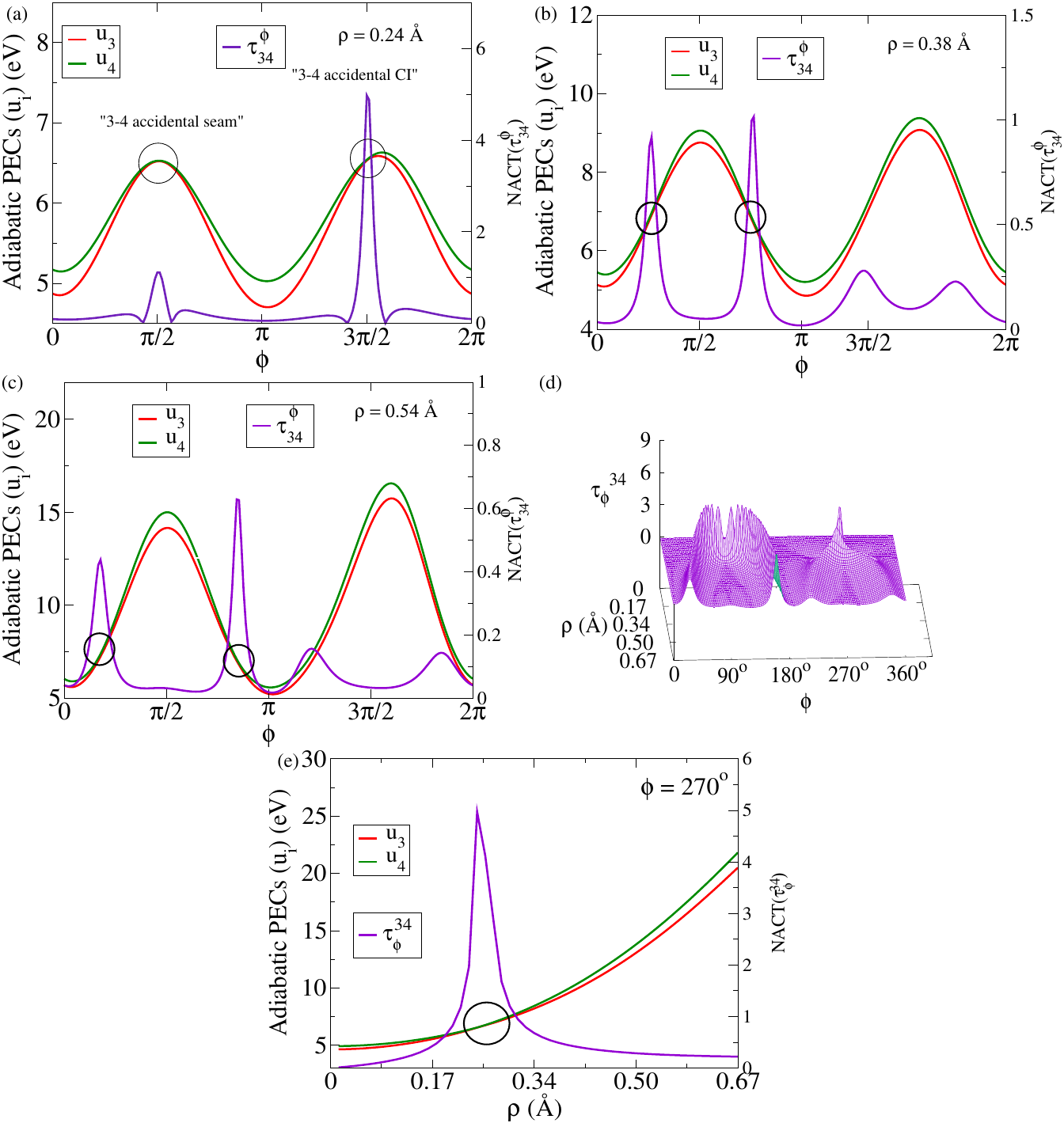}
	\caption{ For the ``3-4" ($\tilde{B}^2B_1$ - $\tilde{C}^2A_1$) electronic states, 1D curves of adiabatic PESs (u$_3$ and u$_4$) and the corresponding $\phi$ component of NACTs ($\tau^{\phi}_{34}$) are presented as function of $\phi$ coordinate at three different values of $\rho$ = (a) 0.24 \AA\, (b) 0.382 \AA\, and (c) 0.54 \AA, where $\tau^{\phi}_{34}$ exhibits singular feature at the points of seam. Diagram (d) shows the location of ``3-4" seam over 2D functional form of $\tau^{\phi}_{34}$ and (e) manifests the ``3-4" CI along $\rho$ coordinate at $\phi = 270^{\circ}$ ($\tau^{\phi}_{34}$, which shows a steep rise at the point of CI i.e., $\rho$ = 3.6). The figure is adapted from Ref. [\protect\citen{saikat_cpc26}] and permission has been taken from Wiley.}
	\label{fig:q6-q7_34}
\end{figure}

\begin{figure}[htp]
	\centering
	\includegraphics[ width=15 cm, height=0.35\linewidth,angle=0]{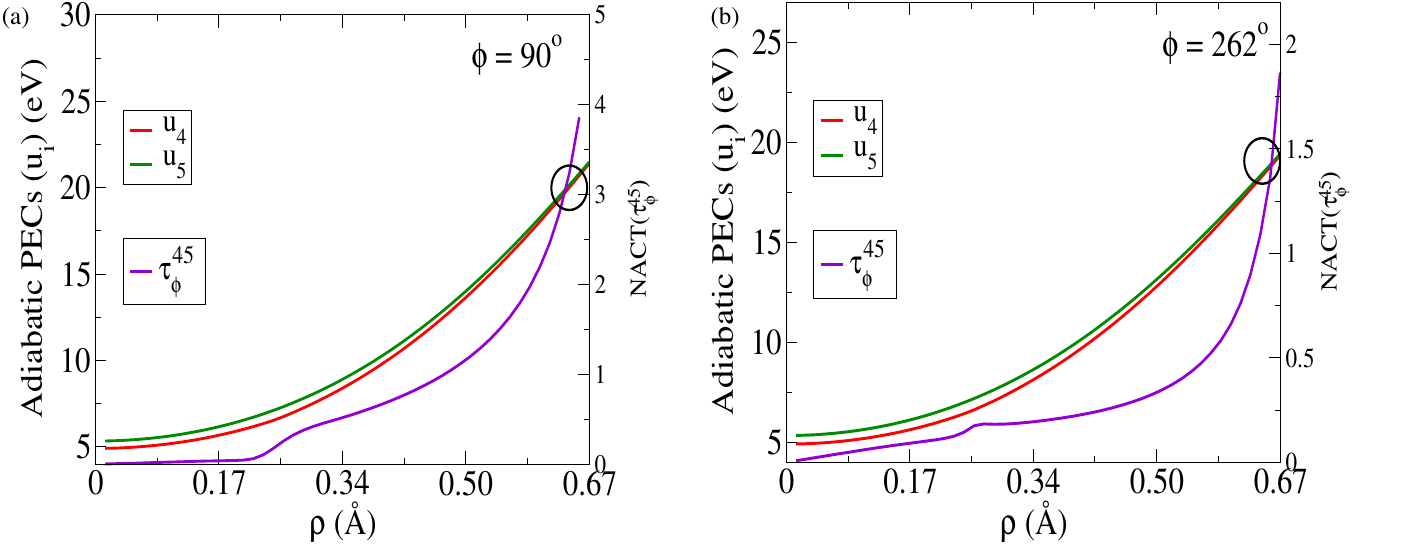}
	\caption{The 1D curves of adiabatic PESs (u$_4$ and u$_5$) and the
		corresponding NACTs ($\tau^{\phi}_{45}$ ) between $\tilde{C}^2A_1$ and $\tilde{D}^2B_2$ electronic states are
		plotted along $\rho$ coordinate at $\phi = 90^{\circ}$ and $262^{\circ}$, where $\tau^{\phi}_{45}$ rises sharply
		at the close vicinity of the CI ($\rho$ = 0.67 \AA). The figure is adapted from Ref. [\protect\citen{saikat_cpc26}] and permission has been taken from Wiley.}
	\label{fig:q6-q7_45}
\end{figure}

\noindent
Since $Q_6$-$Q_7$ normal mode pair shows one ``1-2" CI, one ``3-4" CI, one ``3-4" seam as well as two ``4-5" CIs, the closed contour integration along the $\phi$ coordinate at a fixed values of $\rho$ leads to $\pi$, 2$\pi$ and 2$\pi$, respectively for the associated ADT angles, namely $\Theta_{12}$, $\Theta_{34}$ and $\Theta_{45}$ (see Figure \ref{fig:adt_ang}(a)). On the other hand, $\Theta_{23}$ representing ``2-3" PJT is also presented, which acquires zero (0) value at $\phi$ = 2$\pi$ in spite of non-zero 
intermediate values at other $\phi$-coordinates. It is worth mentioning that the ADT angles appear topological in nature 
(depicted by the solid line) rather than geometric one (represented by the dashed line) in Figure \ref{fig:adt_ang}. 
On the other hand, diagonal elements of ADT matrices, A$_{11}$[=A$_{22}$], A$_{33}$ and
A$_{55}$ are presented in Figure \ref{fig:adt_ang}(b), where the associated 1D curves undergo one (1), two (2) and two (2) sign inversions, respectively encapsulating one ``1-2" accidental CI, one ``3-4" accidental CI along with one seam and two ``4-5" accidental CIs, respectively. It needs to mention that the numerical values of Curl condition becomes zero (0) (see Figure \ref{fig:curl_con}) validating the existence of five state SHS for o-C$_6$H$_4$F$_2^+$ radical cation. \\

\begin{figure}[h!]
	\centering
	\includegraphics[  height=0.25\textheight, angle=0]{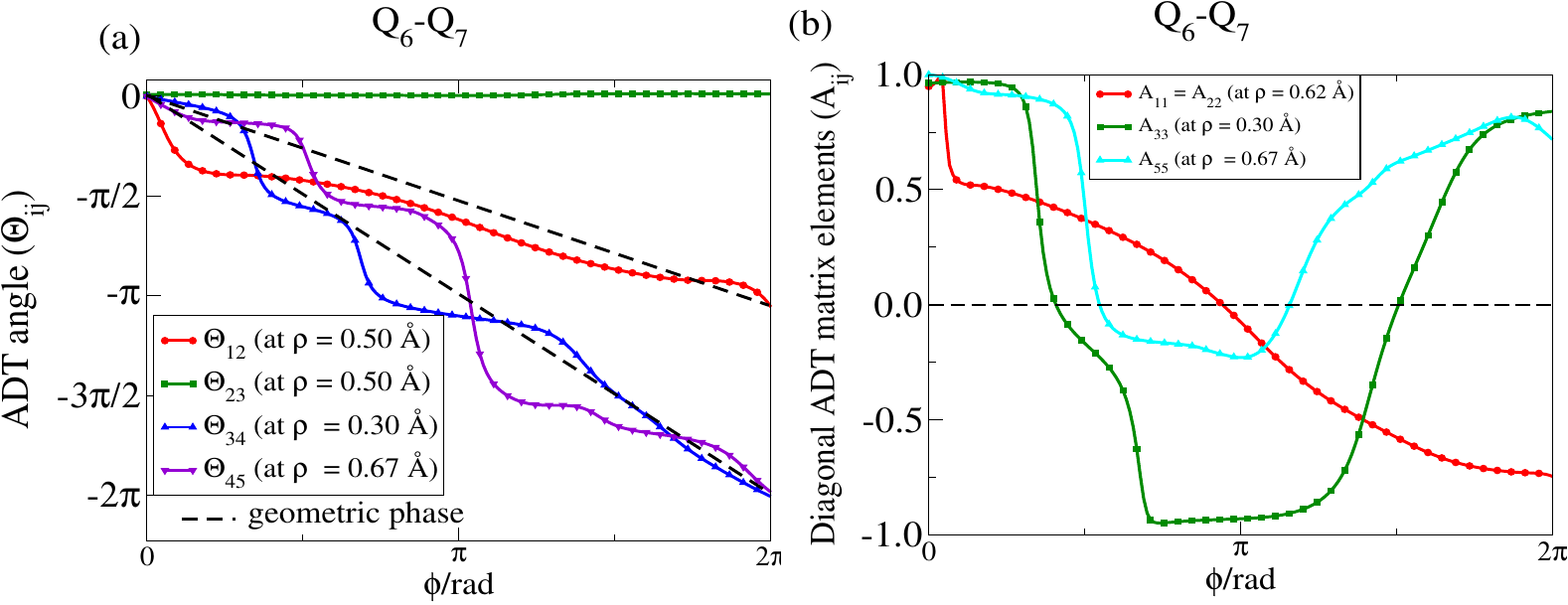}
	\caption{1D functional forms of ADT angles, $\Theta_{12}$, $\Theta_{23}$, $\Theta_{34}$ and $\Theta_{45}$, and diagonal elements of ADT matrices, namely, A$_{11}$ [= A$_{22}$],  A$_{33}$ and A$_{55}$ for $Q_{6}-Q_{7}$ normal mode pairs are plotted as function of the angular coordinate, $\phi$ at specific $\rho$-values, which affirms the presence or absence of the CIs/seams encapsulated by the closed loop of integration. The figure is adapted from Ref. [\protect\citen{saikat_cpc26}] and permission has been taken from Wiley.}
	\label{fig:adt_ang}
\end{figure}
\clearpage

\begin{figure}[h!]
	\centering
	
	\begin{subfloat} 
		\centering
		\includegraphics[width=0.3\linewidth]{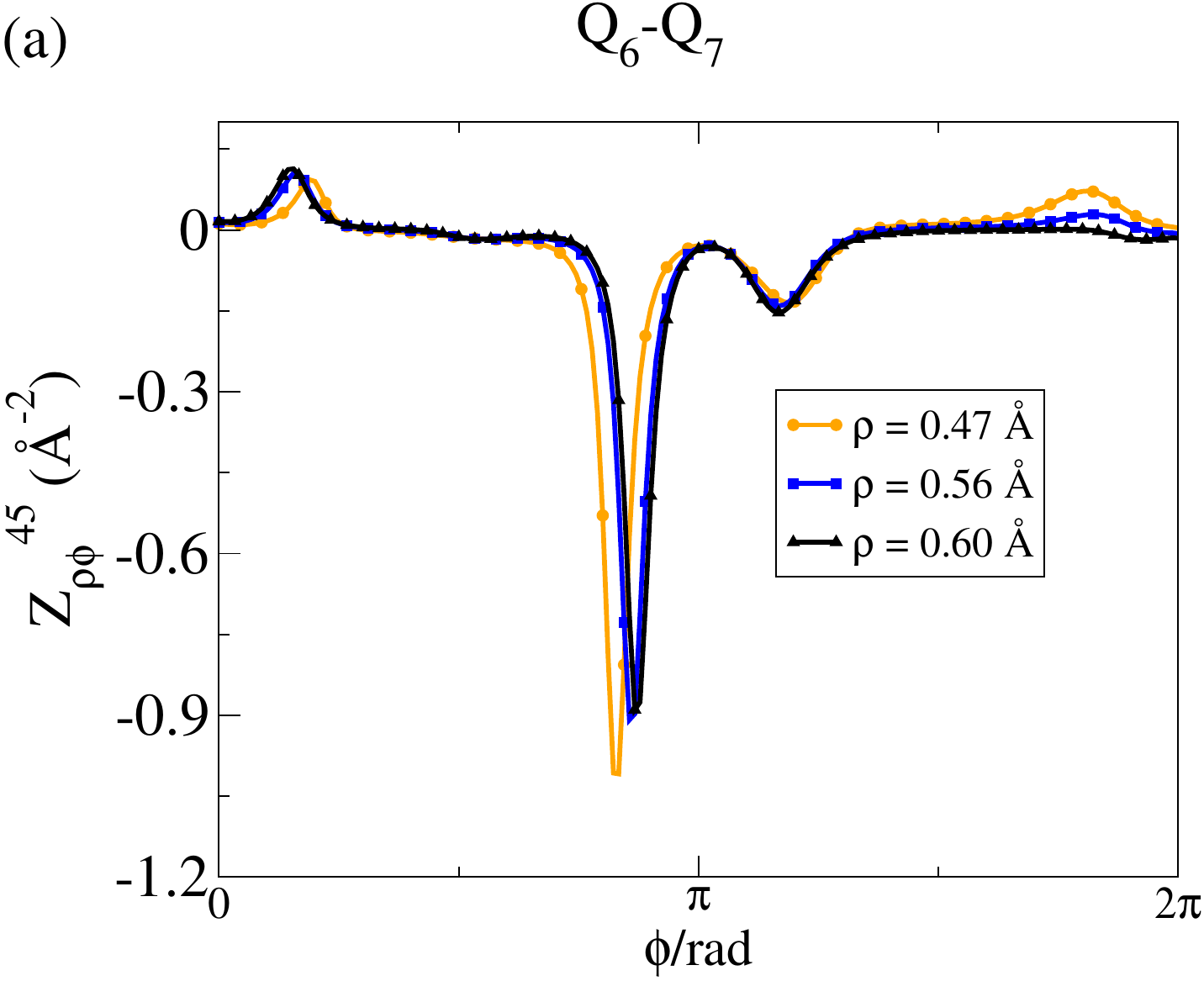} 
	\end{subfloat}
	\hfill
	\begin{subfloat}
		\centering
		\includegraphics[width=0.3\linewidth]{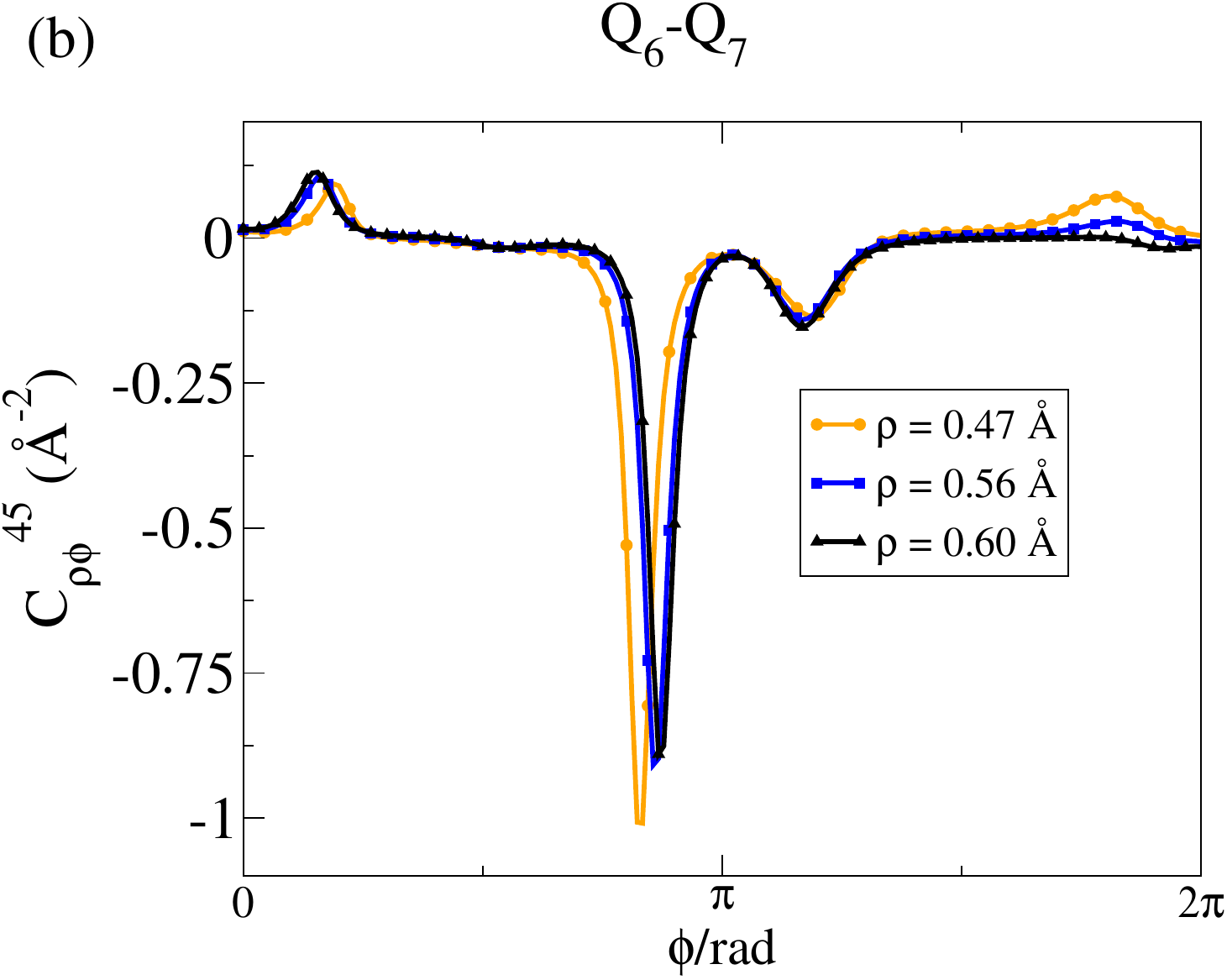}
	\end{subfloat}
	\hfill
	\begin{subfloat}
		\centering
		\includegraphics[width=0.3\linewidth]{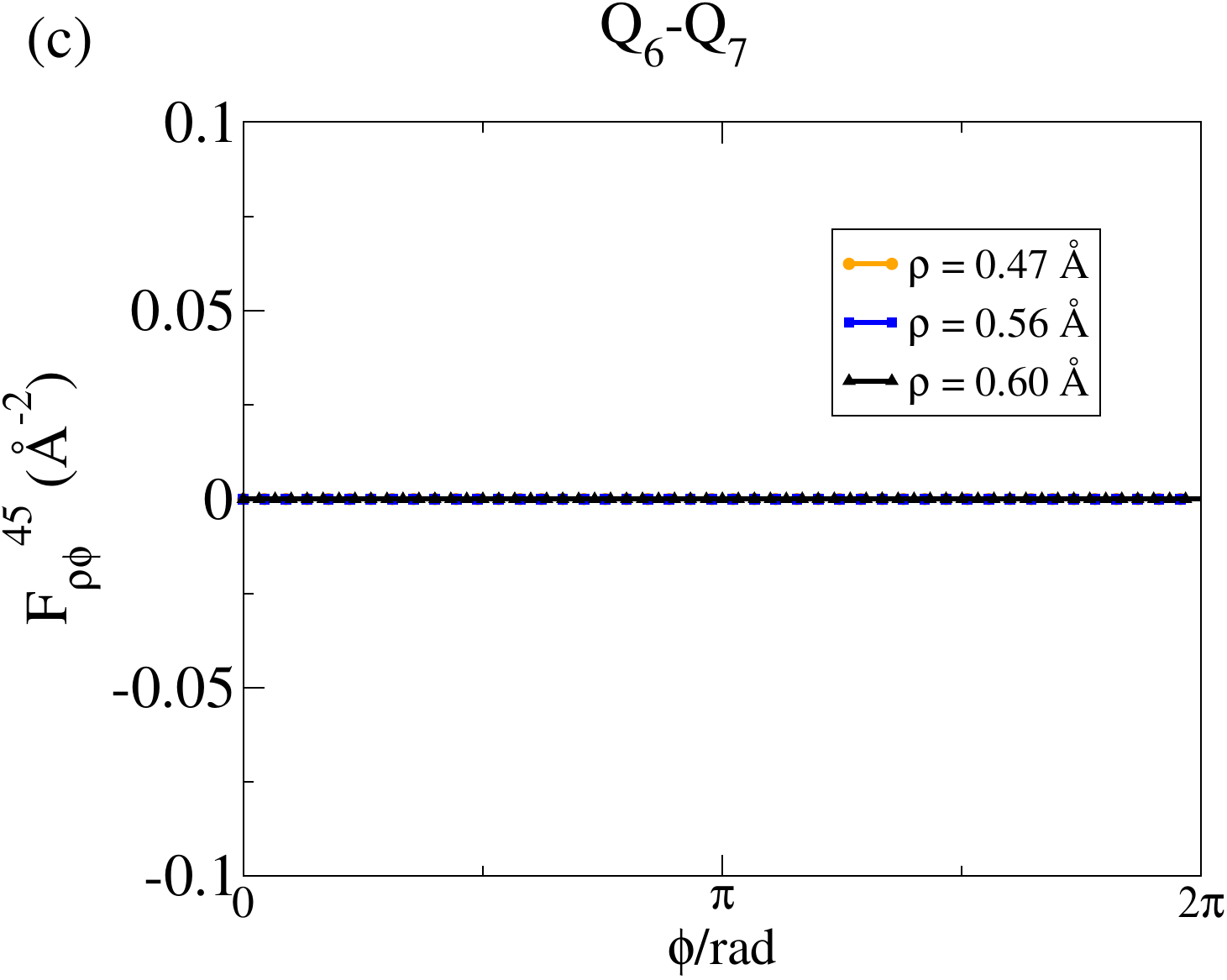}
	\end{subfloat}
	
	\vspace{2em} 
	
	\begin{subfloat}
		\centering
		\includegraphics[width=0.3\linewidth]{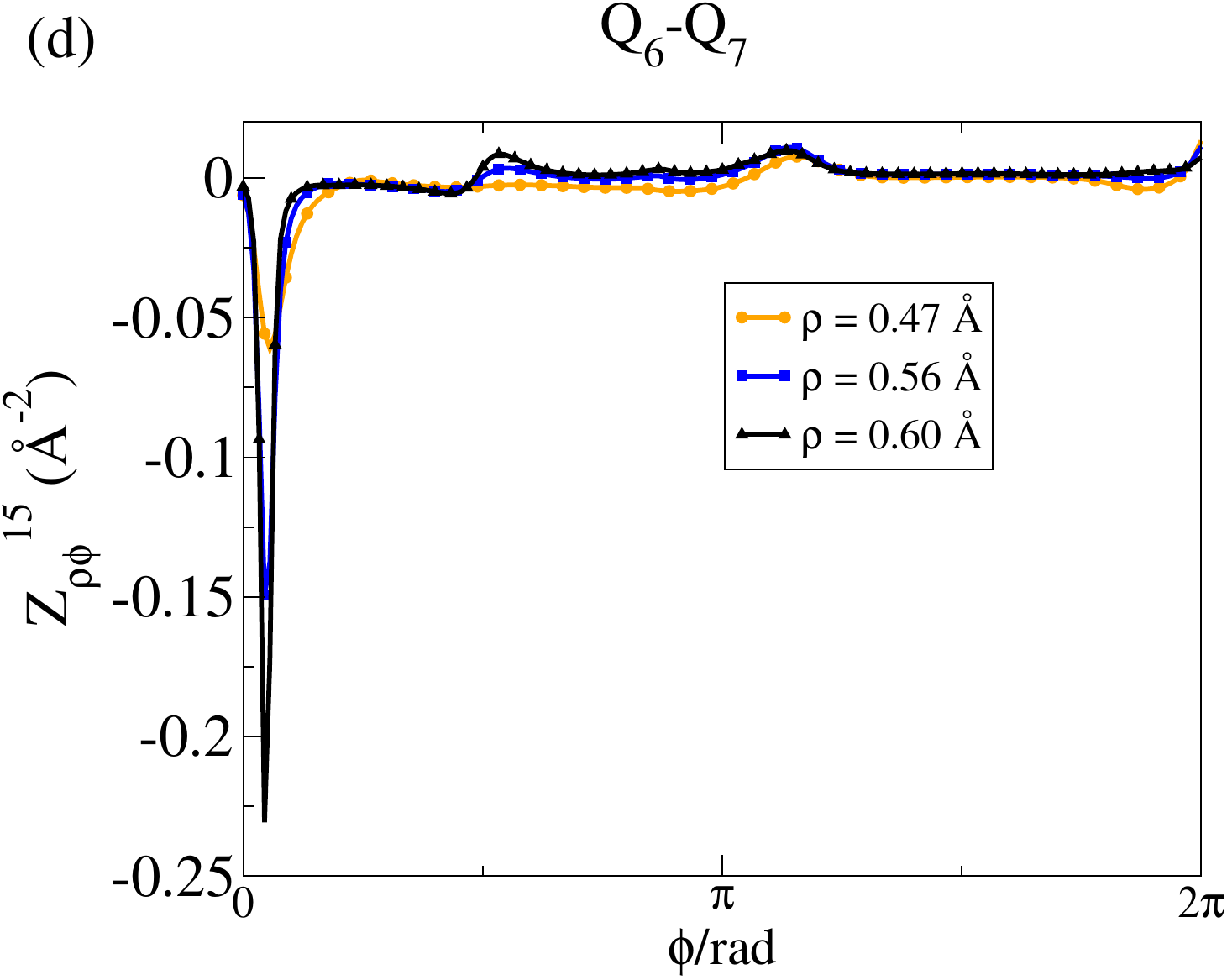}
	\end{subfloat}
	\hfill
	\begin{subfloat}
		\centering
		\includegraphics[width=0.3\linewidth]{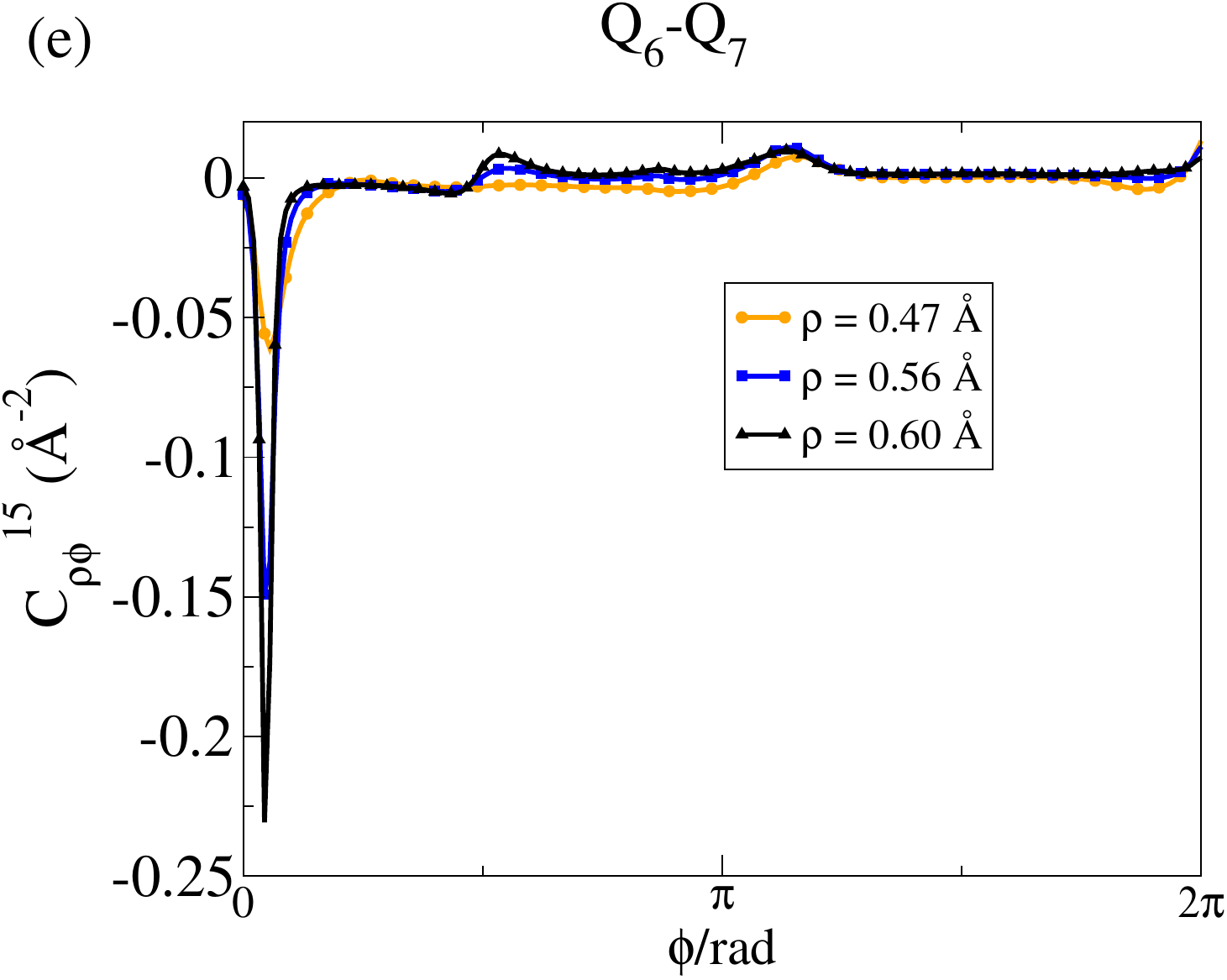}
	\end{subfloat}
	\hfill
	\begin{subfloat}
		\centering
		\includegraphics[width=0.3\linewidth]{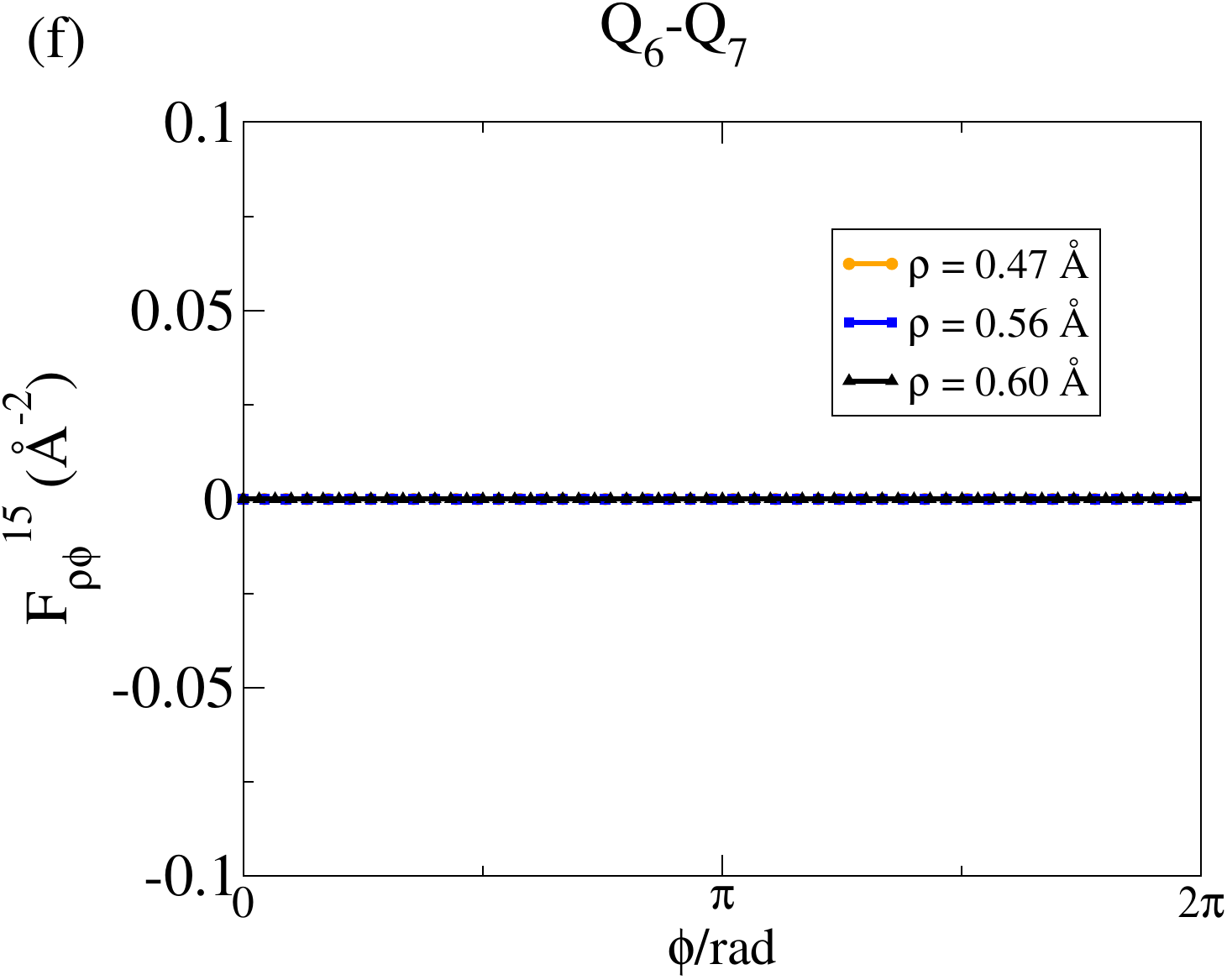}
	\end{subfloat}
	
	\caption{The Mathematical Curls [(a) Z$_{\rho\phi}^{45}$ and (d) Z$_{\rho\phi}^{15}$], ADT Curls [(b) C$_{\rho\phi}^{45}$ and (e) C$_{\rho\phi}^{15}$] and the Curl Condition [(c) F$_{\rho\phi}^{45}$ and (f) F$_{\rho\phi}^{15}$] corresponding to ``4-5" accidental CIs and ``1-5" PJT interactions are shown as functions of $\phi$-coordinate for three different values of $\rho$-coordinate of Q$_6$-Q$_7$ normal mode pair. The figure is adapted from Ref. [\protect\citen{saikat_cpc26}] and permission has been taken from Wiley.}
	\label{fig:curl_con}
\end{figure}

\noindent
Once BBO based diabatization was completed, the initial product wavepacket was placed on each of the five electronic states, $\tilde{X}^2B_1$, $\tilde{A}^2A_2$, $\tilde{B}^2B_1$, $\tilde{C}^2A_1$ and $\tilde{D}^2B_2$ to perform five separate dynamical calculations yielding individual state spectra, which were then convoluted to produce the overall spectral profile for the five lowest electronic states of o-C$_6$H$_4$F$_2$ molecule as depicted in left panel of Figure \ref{fig:spectra}. The individual state spectra for
$\tilde{X}^2B_1$ and $\tilde{A}^2A_2$ states (see Figure \ref{fig:spectra}(c)) are merged into Figure \ref{fig:spectra}(b) showing the convoluted spectra for  $\tilde{B}^2B_1$, $\tilde{C}^2A_1$ and $\tilde{D}^2B_2$ states. On the other hand, the entire spectral profile is presented in Figure \ref{fig:spectra}(a). Furthermore, in the right panel, TDDVR calculated photoelectron spectra of o-C$_6$H$_4$F$_2$ obtained from BBO-DH are compared with the existing theoretical\cite{faraji_jcp_struct,faraji_jcp_dyna} and experimental\cite{bieri_jesc} spectra. It is interesting to note that the TDDVR calculated profile exhibits good agreement with the experimental data reported by Beiri et al.\cite{bieri_jesc}. Most peaks in BBO based theoretical calculated spectra exhibit better accord with the experimental findings in comparison with the spectra obtained from MCTDH dynamics over VCM-based quasi-diabatic Hamiltonian\cite{faraji_jcp_struct,faraji_jcp_dyna} [see Figure \ref{fig:spectra}(d-f)].
The non-adiabatic interactions (JT CI/seam, accidental CI/seam and PJT interactions) present in this molecule are mainly responsible for the structured spectral bands of the first two (i.e., $\tilde{X}^2B_1$-$\tilde{A}^2A_2$) and the other three (i.e.,  $\tilde{B}^2B_1$-$\tilde{C}^2A_1$-$\tilde{D}^2B_2$) states. In addition, the vibrational fine structure in the spectrum originates from the electronic transitions between different vibrational levels of the neutral and cationic forms of the o-DFBz molecule.

\begin{figure}[h!]
	\centering
	
	\begin{subfloat} 
		\centering
		\includegraphics[width=0.45\linewidth]{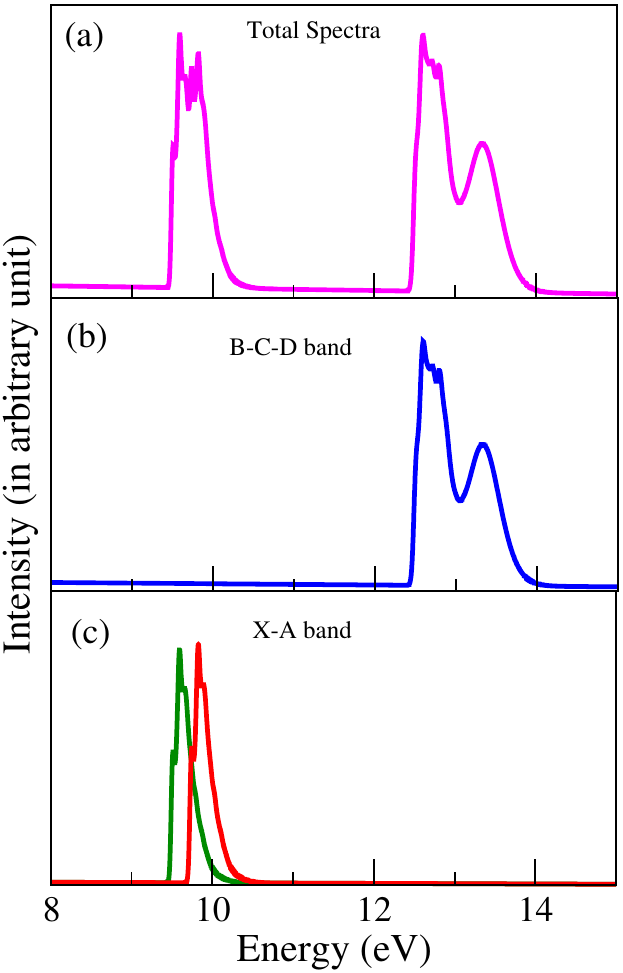} 
	\end{subfloat}
	\hspace*{-0.5 cm}
	\begin{subfloat} 
		\centering
		\includegraphics[width=0.45\linewidth]{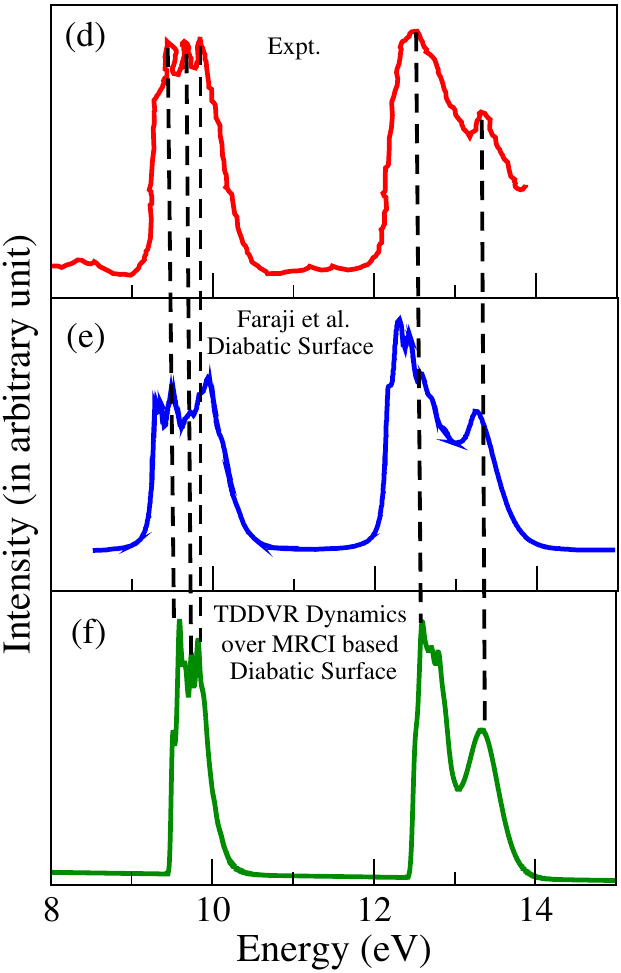} 
	\end{subfloat}
	
	\caption{The left column (a-c) demonstrates TDDVR computed spectral envelops calculated from BBO constructed 5$\times$5 diabatic Hamiltonian for $\tilde{X}^2B_1$ - $\tilde{A}^2A_2$ (X-A spectra) and $\tilde{B}^2B_1$ - $\tilde{C}^2A_1$ - $\tilde{D}^2B_2$ (B-C-D spectra), whereas the right column (d-f) presents the comparison of calculated spectra obtained from (f) TDDVR dynamics over BBO diabatic Hamiltonian with (d) the experimental one\cite{bieri_jesc} as well as with (e) other theoretical spectra\cite{faraji_jcp_dyna} originated from MCTDH dynamics over VCM based quasi-diabatic surfaces~\cite{faraji_jcp_struct}. The figure is taken from Ref. [\protect\citen{saikat_cpc26}] and permission has been taken from Wiley.}
	\label{fig:spectra}
\end{figure}

\noindent
Mass-analyzed threshold ionization (MATI) technique can be used to identify zero kinetic energy ions produced by pulsed field ionization of neutral molecules, which are
excited to high Rydberg states via an intermediate electronic state.\cite{kwon_jcp} MATI spectra provide detailed vibrational information of this molecular cation, since PE spectra alone is not sufficient to provide adequate insight. In the recent work,~\citep{saikat_cpc26} MATI spectra were calculated by initiating the product-type wave-function on the ground state  ($\tilde{X}^2B_1$) to carry out quantum dynamics by considering 2$\times$2  diabatic Hamiltonian for the lowest two electronic states,  $\tilde{X}^2B_1$-$\tilde{A}^2A_2$ of o-DFBz$^+$ cation. Figure \ref{fig:mati} exhibits the calculated MATI spectra depicting good agreement with both the experimentally measured\cite{kwon_jcp_2003} and the reported theoretical\cite{faraji_jcp_dyna} spectra. The experimental ``0-0" peak at 74995 cm$^{-1}$, along with few other peaks at 75287 cm$^{-1}$, 75547 cm$^{-1}$ and 75746 cm$^{-1}$ etc. are well captured in the theoretical spectral profile, appearing at 75027 cm$^{-1}$ (``0-0" peak), 75264 cm$^{-1}$, 75552 cm$^{-1}$ and 75780 cm$^{-1}$, respectively. Few other small peaks are also observed in the calculated MATI spectra~\citep{saikat_cpc26} in comparison with the experimental one~\citep{kwon_jcp_2003}, which are mainly originated from  fundamental transitions as well as numerous overtones and combination bands. In comparison to the PE spectra, the MATI spectra exhibit more prominent vibronic fine structure, where the intense peaks arise from electronic transitions appearing due to more populated vibrational states corresponding to low-frequency modes.

\begin{figure*}[!htp]
	\centering
	\includegraphics[ height=0.50\textheight, angle=0]{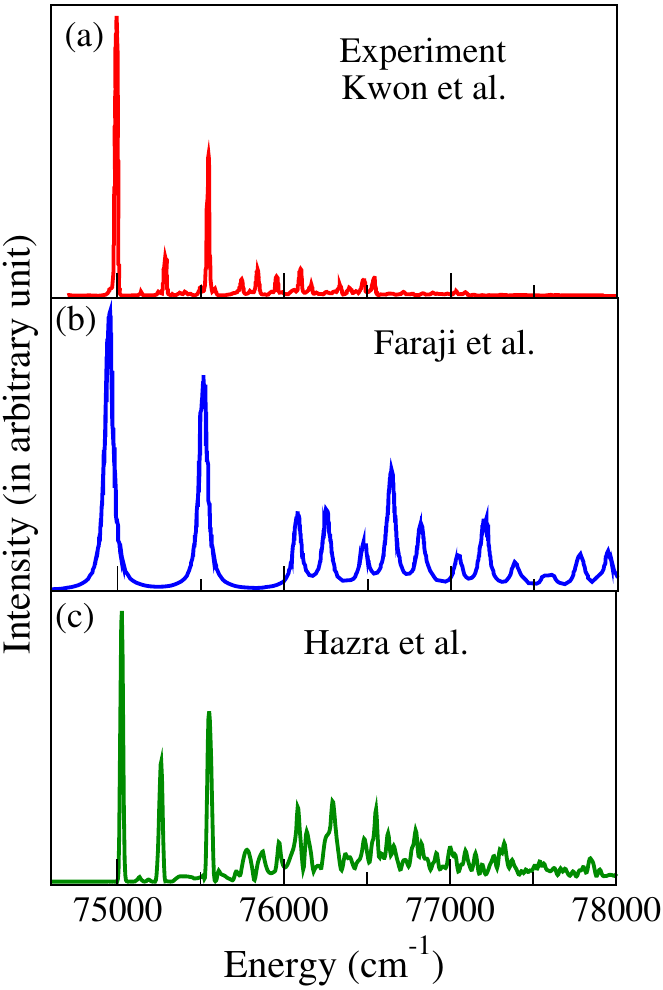}
	\caption{The MATI spectra calculated from the ground state ($\tilde{X}^2B_1$) of the [2$\times$2] diabatic Hamiltonian for the o-DFBz$^+$ molecule are shown in comparison with other theoretical result\cite{faraji_jcp_dyna} and experimentally measured one\cite{kwon_jcp_2003}. The figure is adapted from Ref. [\protect\citen{saikat_cpc26}] and permission has been taken from Wiley.}
	\label{fig:mati}
\end{figure*}

\noindent
\section{Six State : 1,3,5-C$_6$H$_3$F$_3^+$}
\label{ssec:sixst}
\vspace{0.2 cm}
\noindent 
The intense vibronic couplings prevailing within the low-lying six electronic states
of 1,3,5-C$_6$H$_3$F$_3^+$ (TFBz$^{+}$) radical cation, namely, $\tilde{X}^{2}E^{\prime\prime}$, 
$\tilde{A}^{2}A_{2}^{\prime\prime}$, 
$\tilde{B}^{2}E^{\prime}$ and $\tilde{C}^{2}A_{2}^{\prime}$ in the FC region are the key mechanistic 
elements for dictating the position and width of bands in the photoelectron spectrum.\citep{gilbert_cpl17,bieri_jesrp23} Several groups\citep{dujardin_jcp79,braden_jpca104,palmer_jms49} selectively chose the neutral analogue of this radical cation to explore the impact of fluorine substitution during the emission process from the excited states as well as to study the JT and PJT effects, which have substantial implications on the dynamical process. 
Recently, it has been unveiled that among the lowest six electronic states, $\tilde{X}^{2}E^{\prime\prime}$ and $\tilde{B}^{2}E^{\prime}$ encounter JT splitting resulting into lowering 
of symmetry followed by stabilization in energy due to the $e^{\prime}$ vibrational modes.\citep{soumya_tfb,mahapatra_pccp11} On the other hand, $e^{\prime\prime}$ modes are responsible for PJT effects, which are more predominant 
within $\tilde{A}^{2}A_{2}^{\prime\prime}$ and $\tilde{B}^{2}E^{\prime}$ 
states.\citep{soumya_jcp150} Similarly, $e^{\prime\prime}$, $e^{\prime}$ modes are also responsible for 
$\tilde{X}^{2}E^{\prime\prime}-\tilde{A}^{2}A_{2}^{\prime\prime}$ and $\tilde{B}^{2}E^{\prime}-\tilde{C}^{2}A_{2}^{\prime}$ PJT interactions.\citep{mahapatra_pccp11} \\

\noindent
As per the recent findings,~\citep{soumya_tfb} the $e^{\prime}$ normal modes, namely,
asymmetric C-C stretching ($Q_{9x}-Q_{9y}$) and C-C-C in-plane bending ($Q_{13x}-Q_{13y}$) are responsible
for the JT interactions with doubly degenerate electronic states, $\tilde{X}^{2}E^{\prime\prime}$ and 
$\tilde{B}^{2}E^{\prime}$, which can be validated using symmetrized direct product analysis
for the electronic states ($\tilde{X}^{2}E^{\prime\prime}$ and $\tilde{B}^{2}E^{\prime}$)
and the $e^{\prime}$ normal modes ($Q_{9x}$, $Q_{9y}$, $Q_{13x}$ and $Q_{13y}$):

$$A_{1}^{\prime} \in e^{\prime} \otimes
[E^{\prime\prime} \otimes E^{\prime\prime}]
$$
and
$$A_{1}^{\prime} \in e^{\prime} \otimes
[E^{\prime} \otimes E^{\prime}]
$$\\


\noindent
In case of $Q_{13x}-Q_{13y}$ nuclear plane, JT distortions are prominent within $\tilde{X}^{2}E^{\prime\prime}$ 
(``1-2'') and $\tilde{B}^{2}E^{\prime}$ (``4-5'') electronic states near equilibrium nuclear configuration 
($\rho \rightarrow 0.0$) leading to the stabilization of energy for the respective lower sheets (ground and third excited states) by lowering the symmetry of the nuclear geometry. Figure \ref{fig:13x-13y-adia-nact} depicts 1D functional forms of adiabatic PESs and the corresponding $\phi$ component of NACTs as functions of
the $\rho$ coordinate at a specific value of $\phi\,(=60^{\circ})$. The presence of ``1-2'' and ``4-5'' JT CIs can be observed in Figures \ref{fig:13x-13y-adia-nact}a and \ref{fig:13x-13y-adia-nact}b, respectively, where the NACTs 
($\tau_{\phi}^{12}$ and $\tau_{\phi}^{45}$) attain singular characteristics at the close vicinity of those CIs. On the other hand, 
1D functional forms of the corresponding ADT angles, $\Theta_{12}$ and 
$\Theta_{45}$ along $\phi$ coordinate (at $\rho\,=\,0.08$ \AA) attain the magnitude of $\pi$ at $\phi=2\pi$ (see 
Figure \ref{fig:adt-combined}a). In addition, ADT angles associated with the PJT interactions ($\Theta_{13}$ and 
$\Theta_{56}$) become zero (0) at $\phi=2\pi$, even though such angles take non-zero magnitude at intermediate values of $\phi$. Furthermore, Figure \ref{fig:adt-combined}b exhibits relevant diagonal elements of ADT matrices, namely, A$_{11}$ 
[= A$_{22}$] and A$_{44}$[= A$_{55}$] leading to one sign inversion along the $\phi$ coordinate 
affirming the presence of ``1-2'' and ``4-5'' JT CIs inside the closed contour. In case of TFBz$^+$ cation, the adiabatic PESs, NACTs, diabatic PESs and couplings exhibit even parity functional variation along various asymmetric normal modes, like $Q_{16x}$ (see Figure \ref{fig:tfbeven})\\

\begin{figure*}[!htp]
	\centering
	\includegraphics[ height=0.25\textheight, angle=0]{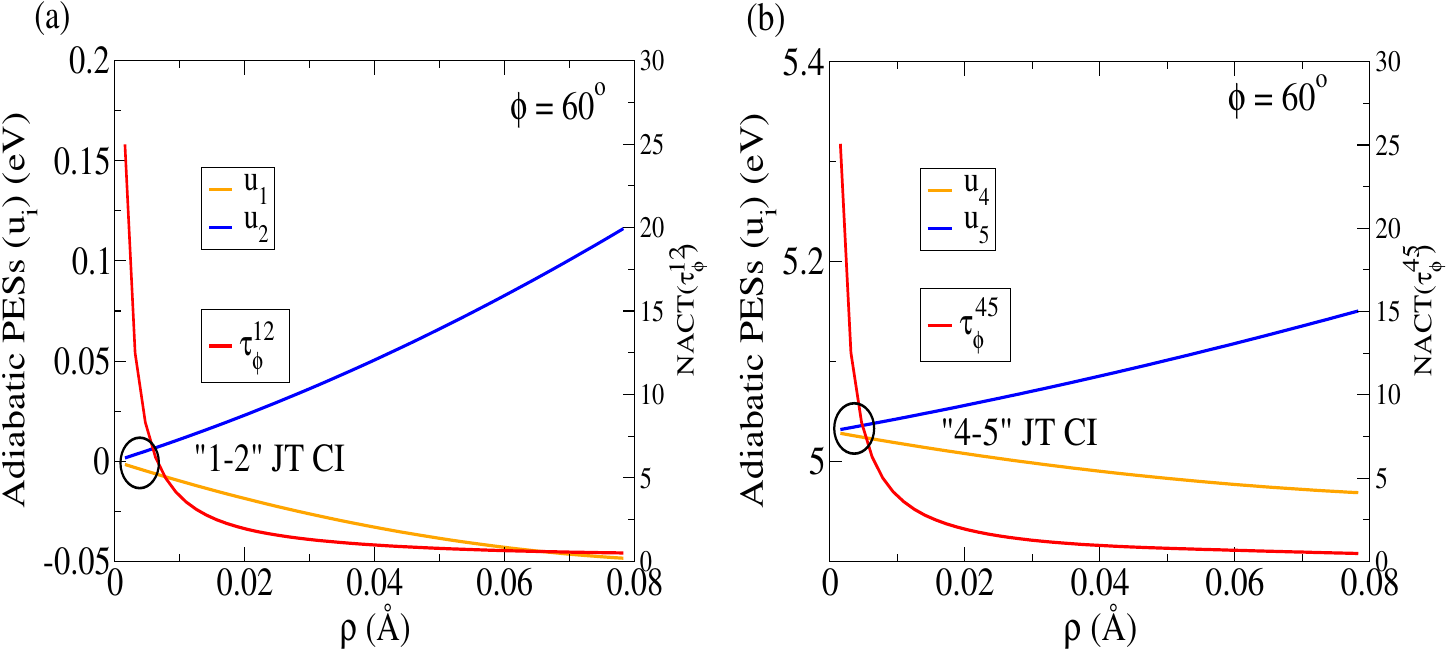}
	\caption{In case of $Q_{13x}-Q_{13y}$ pair of TFBz$^{+}$, (a) left and (b) right panels display 1D cuts of adiabatic PESs of $\tilde{X}^{2}E^{\prime\prime}$ ($u_1$ and $u_2$)
		and $\tilde{B}^{2}E^{\prime}$ ($u_4$ and $u_5$) electronic states and the associated NACTs 
		($\tau_{\phi}^{12}$ and $\tau_{\phi}^{45}$) along $\rho$ coordinate at $\phi = 60^{\circ}$. The ``1-2''
		and ``4-5'' JT CIs are located at the equilibrium geometry ($\rho \rightarrow 0.0$). The figure is adapted from Ref. [\protect\citen{soumya_tfb}] and permission has been taken from RSC.}
	\label{fig:13x-13y-adia-nact}
\end{figure*}

\begin{figure}[!htp]
	\centering
	\includegraphics[ height=0.25\textheight, angle=0]{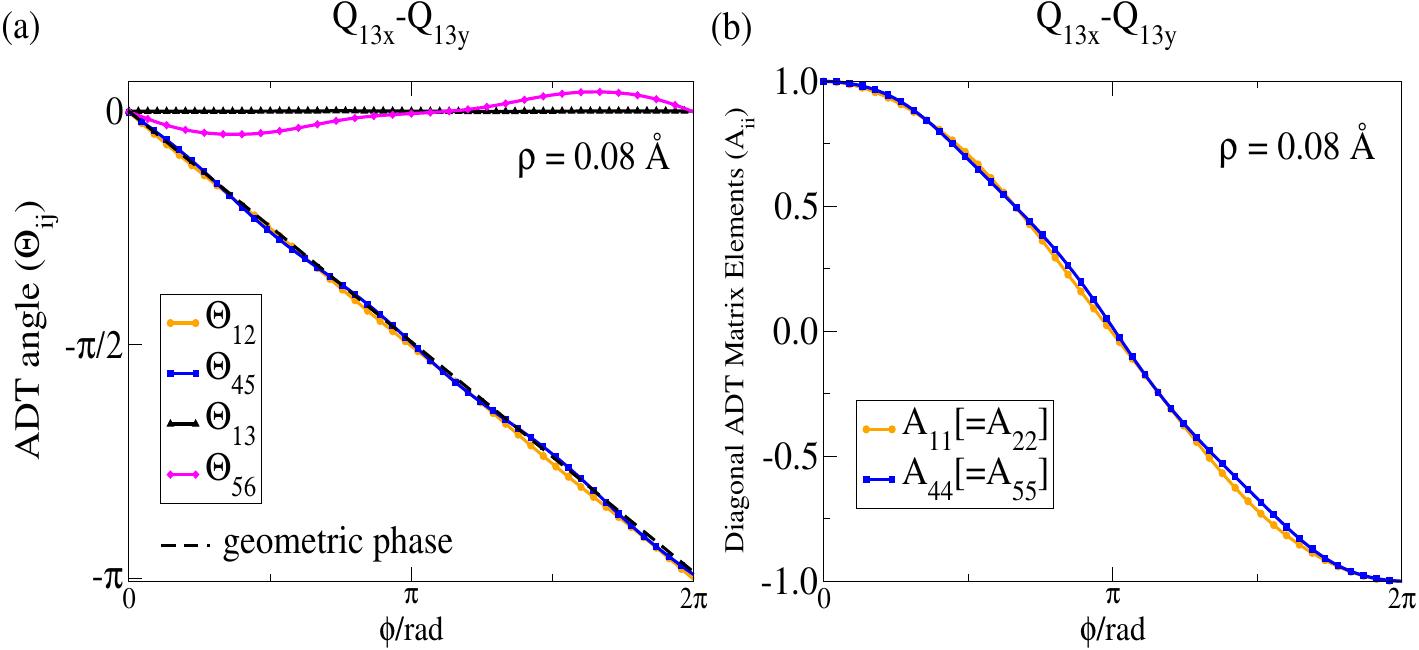}
	\caption{The left diagram (a) displays 1D functional form of ADT angles, $\Theta_{12}$, $\Theta_{45}$, $\Theta_{13}$ and $\Theta_{56}$ for $Q_{13x}-Q_{13y}$ pair along
		$\phi$ coordinate at a fixed value of $\rho\,$(= 0.08 \AA). The right diagram (b) demonstrates 1D cuts of diagonal elements of ADT matrices, viz., (b) A$_{11}$[= A$_{22}$] and A$_{44}$[= A$_{55}$] for the same pair. In each case, the ADT angles and diagonal elements of the ADT matrices unveil the presence or absence of the encapsulated CIs/seams inside the closed loop of integration. The figure has been adapted from Ref. [\protect\citen{soumya_tfb}] and permission has been taken from RSC.}    
	\label{fig:adt-combined}
\end{figure}

\begin{figure}[!htp]
	\centering
	\includegraphics[ width=0.7\textwidth, angle=0]{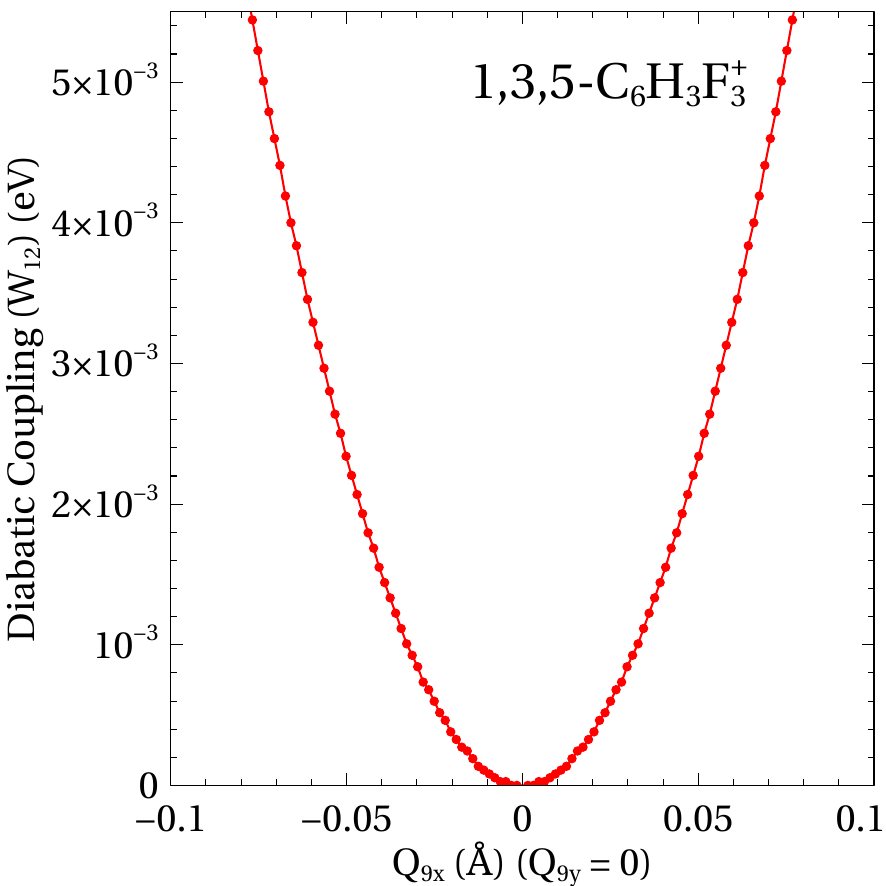}
	\caption{1D functional form of diabatic coupling element, namely, $W_{12}$ among $\tilde{X}^{2}E^{\prime\prime}$ states along $Q_{9x}$ mode of TFBz$^+$ radical cation is displayed. The coupling term possesses symmetric functional form, which can be fitted only with even parity polynomials. The figure has been adapted from Ref. [\protect\citen{joy_pccp}] and permission has been taken from RSC.}
	\label{fig:tfbeven}
\end{figure}

\noindent
The calculated ADT angles~\citep{soumya_tfb}
were further employed to calculate Mathematical Curls ($Z_{\rho\phi}^{ij}$) and ADT Curls ($C_{\rho\phi}^{ij}$) using their explicit expressions 
in terms of those mixing angles (see Appendix~\ref{ssec:adt-eq}). Figure \ref{fig:tfb6} presents 1D variation of those quantities
($Z_{\rho\phi}^{56}$ and $C_{\rho\phi}^{56}$) along $\phi$ coordinate for three different values of $\rho$ acquiring large non-zero values,
whereas their difference ($F_{\rho\phi}^{56}$) appears as negligibly zero ($<$ 10$^{-12}$ $\AA^{-2}$) validating the six state SHS as a ``true" subspace.\\

\begin{figure}[htp]
	\centering
	\includegraphics[width=1.0\textwidth]{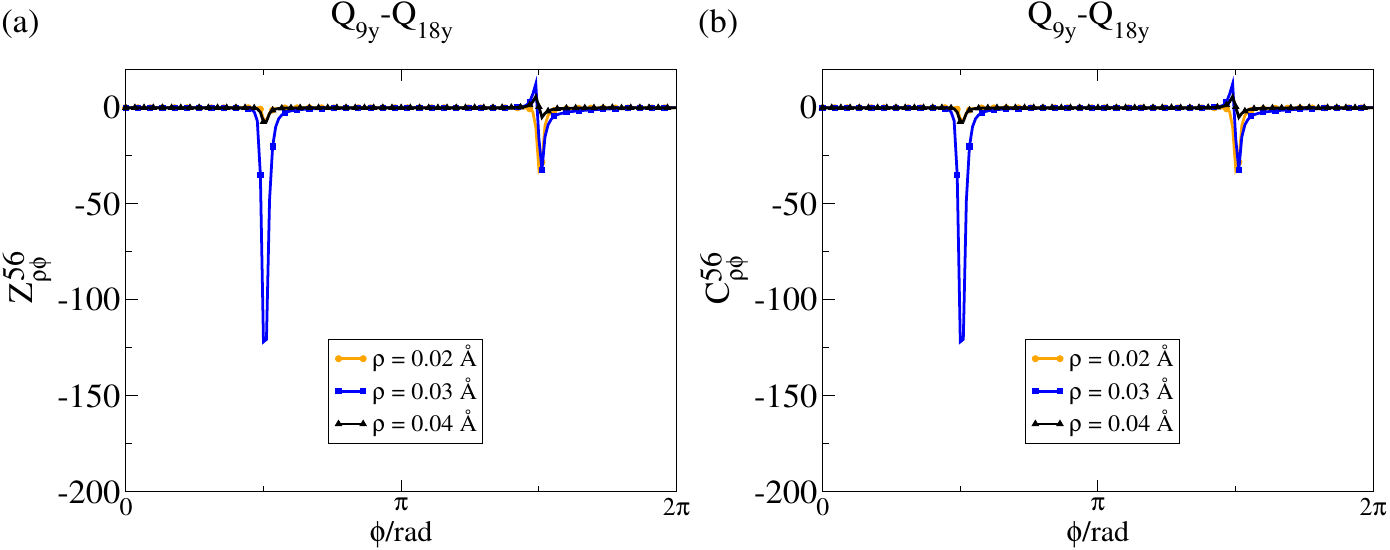}
	\caption{1D cuts of Mathematical Curl ($Z_{\rho\phi}^{56}$) and ADT Curl ($C_{\rho\phi}^{56}$) associated with ``5-6" accidental seam are plotted
		along $\phi$ coordinate at three different $\rho$ values for $Q_{9y}-Q_{18y}$ plane of 1,3,5-C$_6$H$_3$F$_3^+$ radical cation. The figure has been adapted from Ref. [\protect\citen{mantu_curl}] and permission has been taken from Wiley.}
	\label{fig:tfb6}
\end{figure}

\noindent
TDDVR dynamics over BBO-DH provides theoretically calculated PE spectra as presented in Figure \ref{fig:spec}, which are found to be in good accord with the available experimental spectra recorded by 
Gilbert {\it et al.}~\citep{gilbert_cpl17} as well as the outcome of MCTDH dynamics using VCM based quasi-diabatic Hamiltonian.~\citep{mahapatra_pccp11}
As a matter of fact, most of the peaks in our theoretical spectra exhibit better agreement 
with experiment,~\citep{gilbert_cpl17} which justifies the ``exactness" and ``accuracy" of the BBO formalism in conjunction with TDDVR approach over the VCM model~\citep{mahapatra_pccp11} followed by MCTDH dynamics. It is interesting to observe that the experimental spectrum for $\tilde{X}^{2}E^{\prime\prime}$ state contains three peaks at
9.7086 eV, 9.7696 eV and 9.8281 eV (Figure \ref{fig:spec}a) among which the third one has the 
highest intensity. Similarly, our theoretically calculated spectra (Figure \ref{fig:spec}c) also show
three peaks at 9.7223 eV, 9.7670 eV and 9.8273 eV, where the third peak exhibits the highest intensity. 
On the other hand, Figure \ref{fig:spec}f depicts our theoretically calculated PE spectra for $\tilde{A}^{2}A_{2}^{\prime\prime}$ state in comparison with the 
experimental ones (Figure \ref{fig:spec}d) and other theoretical findings (Figure \ref{fig:spec}e). 
In both cases ($\tilde{X}^{2}E^{\prime\prime}$ and $\tilde{A}^{2}A_{2}^{\prime\prime}$), 
one can observe a broad spectral envelop at relatively higher energy ($>$ 10.0 eV and $>$ 12.7 eV) without any well-defined peak positions, which are mainly
arising from the overtones of various vibrational frequencies. Finally, the PE spectra for $\tilde{B}^{2}E^{\prime}-\tilde{C}^{2}A_{2}^{\prime}$ states (Figure \ref{fig:spec}(g-i)) appear as a broad band with diffused structure, which is originating from profound electron-nuclear (nonadiabatic)
couplings. \\

\begin{figure*}[!htp]
	\centering
	\includegraphics[height=0.4\textheight, angle=0]{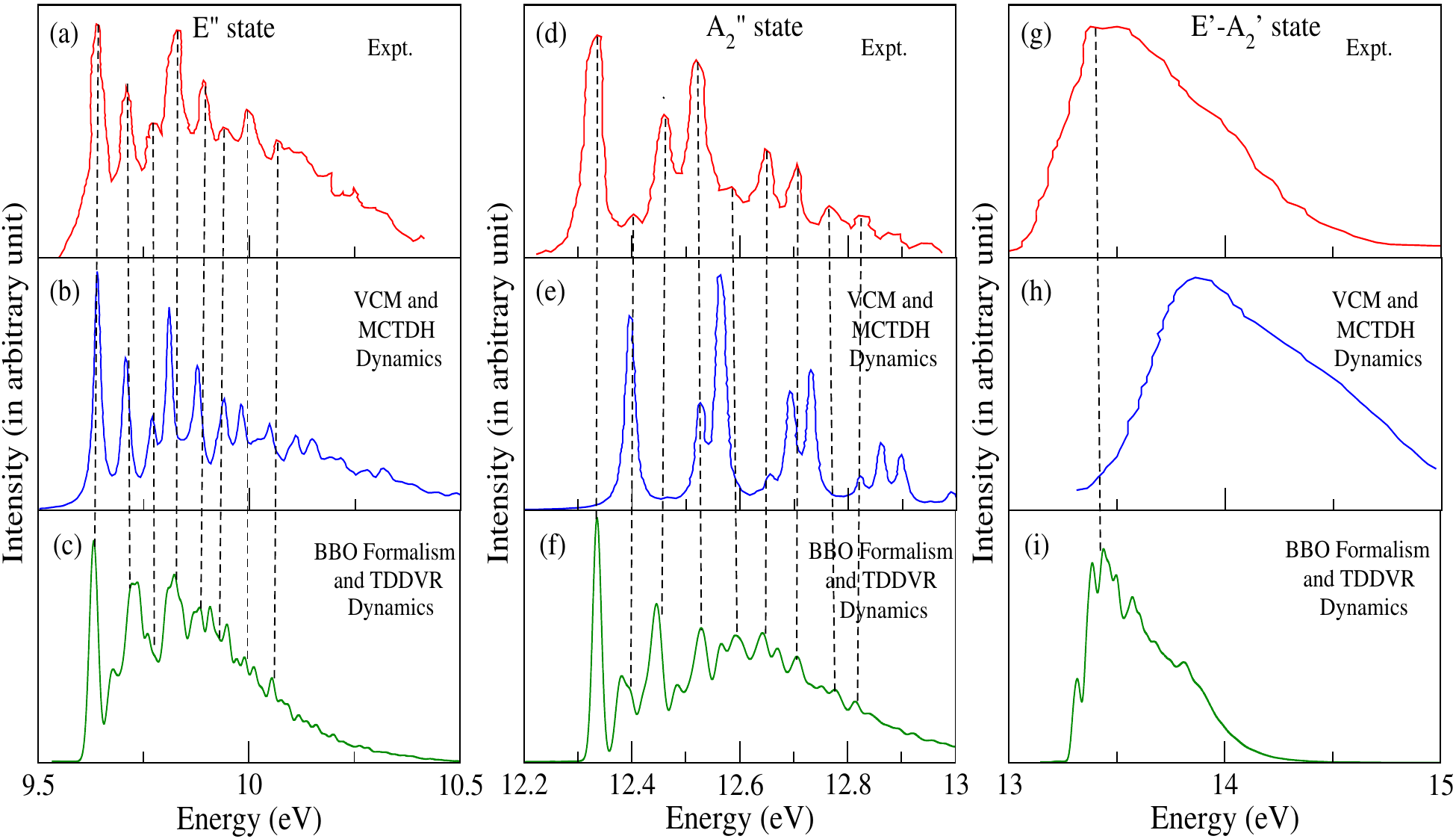}
	\caption{The left column displays (c) theoretical spectrum for the ground degenerate state ($\tilde{X}^{2}E^{\prime\prime}$) obtained from TDDVR dynamics over BBO-DH in comparison with (a)  experimental PE spectrum of TFBz~\citep{gilbert_cpl17}  
		and other (b) theoretical spectrum obtained from MCTDH approach on VCM based quasi-diabatic surfaces~\citep{mahapatra_pccp11}. The middle column (d-f) and the right 
		one (g-i) depict the same quantities for $\tilde{A}^{2}A_{2}^{\prime\prime}$ and  $\tilde{B}^{2}E^{\prime}-\tilde{C}^{2}A_{2}^{\prime}$ states. For all the
		three cases, TDDVR calculated spectra exhibit better peak by peak correspondence with the experimental ones over MCTDH~\citep{mahapatra_pccp11} dynamics. The figure has been taken from Ref. [\protect\citen{soumya_tfb}] and permission has been taken from RSC.}
	\label{fig:spec}
\end{figure*}

\section{Path Dependence of BBO Based Diabatic Hamiltonians and Associated Observables: C$_6$H$_6^+$ Radical Cation}
\vspace{0.2 cm}

\subsection{Brief Overview}

\noindent
C$_6$H$_6^+$ (Bz$^+$) radical cation drew attention of the allied scientific communities due to its highly complex spectral bands originating from intense vibronic couplings between its low-lying five electronic states ($\tilde{X}^2E_{1g}$, $\tilde{B}^2E_{2g}$ and $\tilde{C}^2A_{2u}$) through its vibrational modes. In order to explore the dynamical features, several theoretical calculations\citep{koppel_jcp89,Koppel_jcp117_1,Koppel_jcp117_2,subhankar_ijqc111,subhankar_jcp130,subhankar_jcs124,johnson_jcp117} were carried out considering nonadiabatic interactions prevalent within those states. It is interesting to note that the low lying doublet
states of Bz$^{+}$, namely, $\tilde{X}^2E_{1g}$ and $\tilde{B}^2E_{2g}$ exhibit JT stabilization along
the $e^{\prime}$ modes.\citep{soumya_jpca121,Koppel_jcp117_1} On the other hand, profound PJT couplings
and accidental CIs can be observed within $\tilde{B}^2E_{2g}$ and $\tilde{C}^2A_{2u}$ electronic 
states over the FC region of nuclear CS.\citep{soumya_jpca121,koppel_jcp89}

\subsection{{\it Ab Initio} Calculations and BBO Theory Based Diabatization}

\noindent
In order to perform BBO based diabatization followed by nuclear dynamics, seven normal modes were considered including two  $e_{2g}$ (in $D_{6h}$) 
symmetric degenerate pairs, namely, in-plane asymmetric stretching ($Q_{16x}$ and $Q_{16y}$) and asymmetric ring deformation ($Q_{18x}$ and $Q_{18y}$).
The symmetrized direct products for the $\tilde{X}^2E_{1g}$ and $\tilde{B}^2E_{2g}$ electronic states with $e_{2g}$ normal modes depict the JT active interactions, 
$$A_{1g} \in e_{2g} \otimes
[E_{1g} \otimes E_{1g}]
$$
and
$$A_{1g} \in e_{2g} \otimes
[E_{2g} \otimes E_{2g}]
$$

\noindent
Adhikari and coworkers\citep{subhankar_ijqc111,subhankar_jcp130,subhankar_jcs124} performed TDDVR dynamical calculations to predict the theoretical 
photoelectron spectra of benzene molecule using VCM based model diabatic Hamiltonian\citep{koppel_jcp89,Koppel_jcp117_1,Koppel_jcp117_2}.  Later on, the same group~\citep{soumya_bz} carried out extensive {\it ab initio} electronic structure 
calculations to compute {\it ab initio} adiabatic PESs and NACTs followed by BBO-DH for lowest five electronic states ($\tilde{X}^2E_{1g}$, $\tilde{B}^2E_{2g}$ and $\tilde{C}^2A_{2u}$) over various 2D nuclear planes constituted with seven normal modes, namely, breathing ($Q_2$), ring deformation ($Q_8$) degenerate in-plane asymmetric stretching ($Q_{16x}$ and $Q_{16y}$), degenerate in-plane bending ($Q_{18x}$ and $Q_{18y}$) and one
component of degenerate out-of-plane bending ($Q_{19x}$) modes.
The 2D functional features of $\phi$ components of NACTs over the $\rho-\phi$ plane of $Q_{16x}-Q_{16y}$ pair are depicted in
Figure \ref{fig:bz-nact}. While carrying out BBO based diabatization, five state ADT equations 
(see Appendix~\ref{ssec:adt-eq}) are solved numerically to calculate mixing angles followed by construction of ADT matrices and
diabatic potential energy matrices.\citep{soumya_jpca121,soumya_bz} The 1D functional forms of ADT angles are demonstrated in Figures \ref{fig:bz-angle}a and \ref{fig:bz-angle}c for $Q_{16x}-Q_{16y}$ and $Q_{18x}-Q_{19x}$ pairs, where the ADT angles ($\Theta_{12}$ and $\Theta_{34}$) attain the magnitude of $\pi$ due to ``1-2" and ``3-4" CIs/seams. On the other hand, the ADT angles appearing due to PJT couplings ($\Theta_{15}$ and $\Theta_{25}$ for $Q_{16x}-Q_{16y}$, and $\Theta_{13}$ and $\Theta_{23}$ for $Q_{18x}-Q_{19x}$) become zero (0) at $\phi=2\pi$, even though such angles attain non-zero magnitudes at other values of $\phi$ (see Figures. \ref{fig:bz-angle}a and \ref{fig:bz-angle}c). On the other hand, it is 
worth mentioning that A$_{11}$[= A$_{22}$] and A$_{33}$[= A$_{44}$] exhibit one sign flipping 
due to the presence of ``1-2" and ``3-4" JT CIs/seams, respectively 
for those nuclear planes (see Figures. \ref{fig:bz-angle}b and \ref{fig:bz-angle}d for $Q_{16x}-Q_{16y}$ and $Q_{18x}-Q_{19x}$ pairs). The diabatic PESs and couplings computed from those ADT angles display symmetric functional behaviour along $Q_{16x}$ and $Q_{18y}$ normal modes, which can only be fitted with even power 
polynomials\citep{soumya_acs} in contrast to traditional JT model (see Figure \ref{fig:bzeven} for $Q_{16x}$ mode). \\

\begin{figure}[!htp]
	\centering
	\includegraphics[ height=0.75\textheight, angle=0]{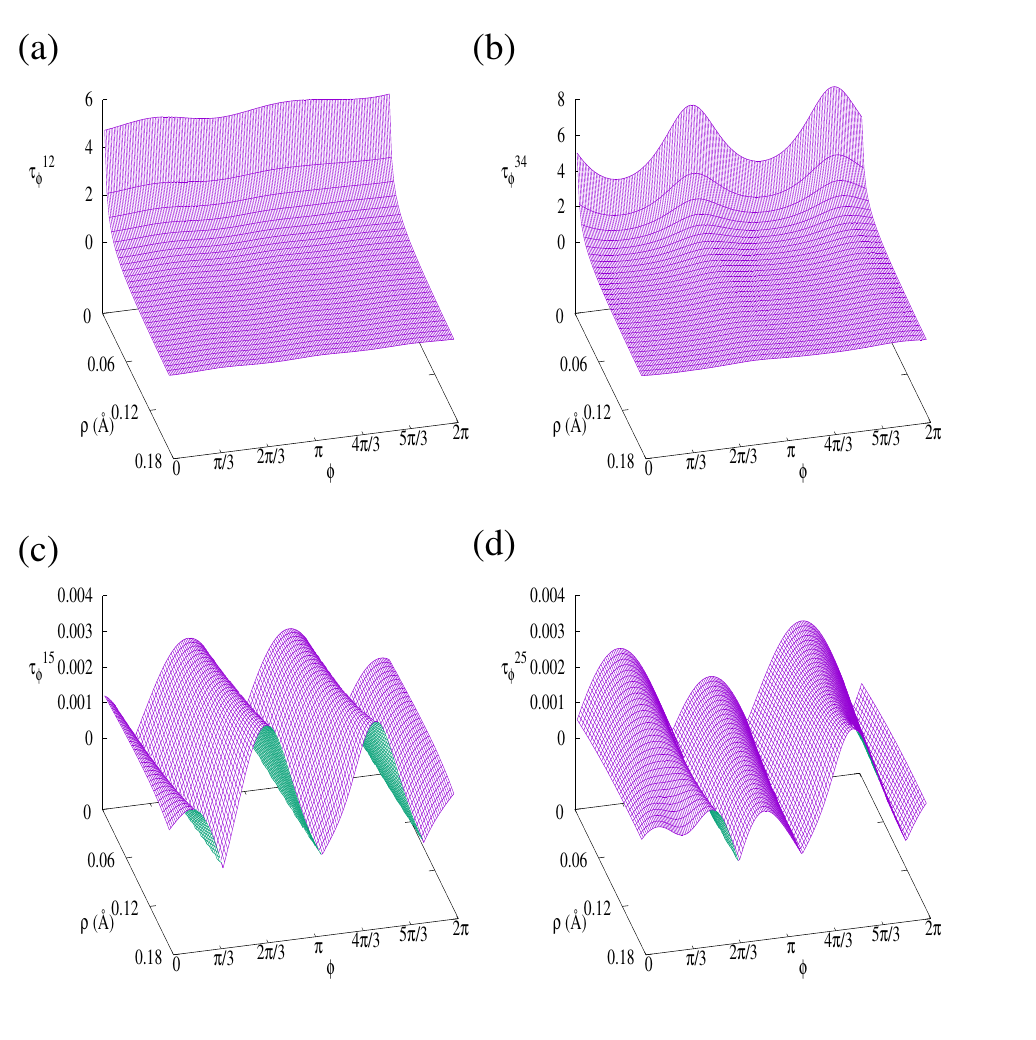}
	\caption{2D functional features of (a) $\tau_{\phi}^{12}$, (b) $\tau_{\phi}^{34}$, (c) $\tau_{\phi}^{15}$ and (b) $\tau_{\phi}^{25}$ are plotted over $\rho-\phi$ plane of $Q_{16x}-Q_{16y}$ pair. The permission for the results of Ref. [\protect\citen{soumya_bz}] has been taken from AIP.}
	\label{fig:bz-nact}
\end{figure}

\begin{figure}[!htp]
	\centering
	\includegraphics[ height=0.5\textheight, angle=0]{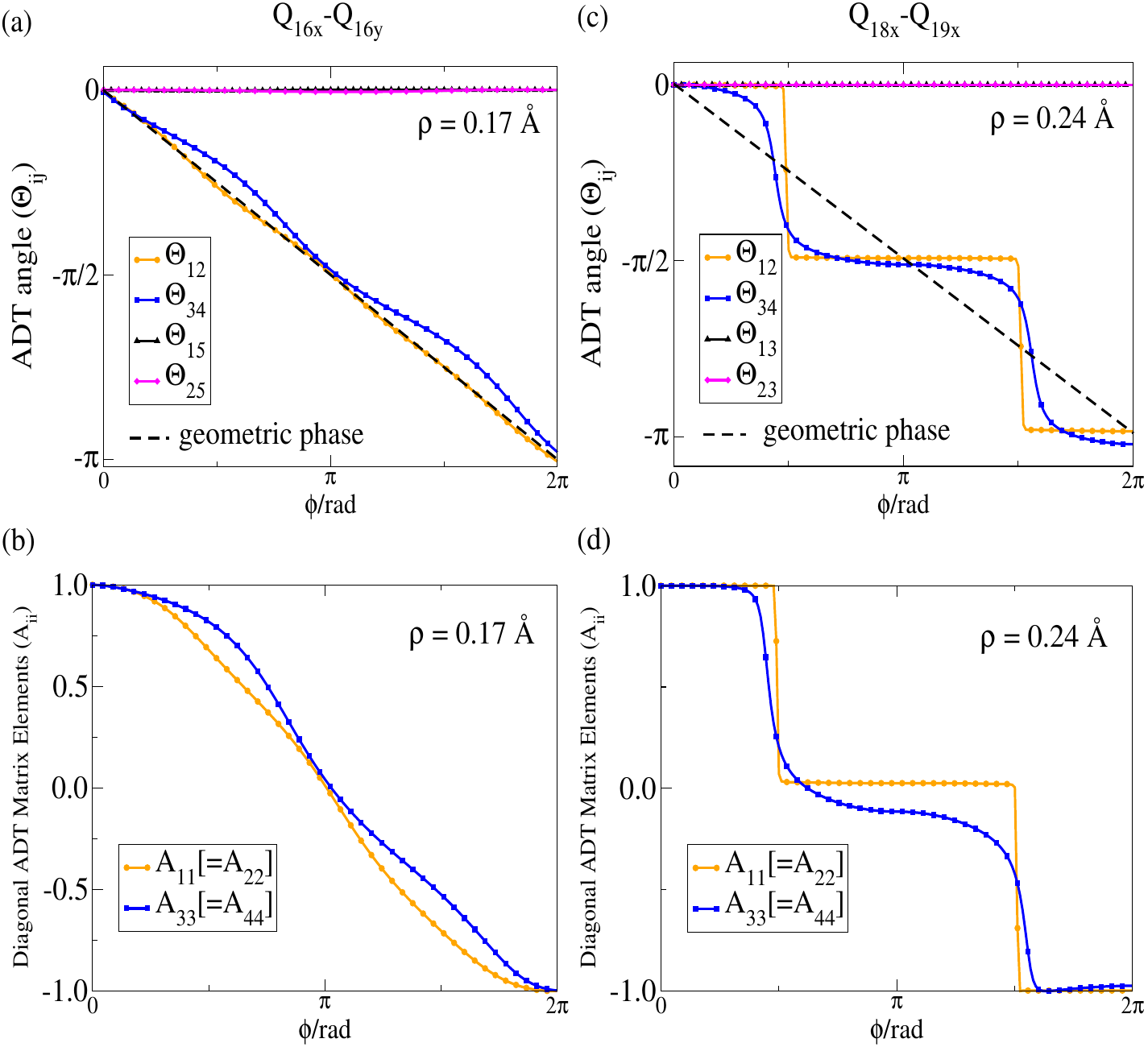}
	\caption{The left panel depicts 1D variation of (a) ADT angles ($\Theta_{12}$, $\Theta_{34}$, 
		$\Theta_{15}$ and $\Theta_{25}$) and (b) diagonal elements of ADT matrix, 
		A$_{11}$[= A$_{22}$] and A$_{33}$[= A$_{44}$] for $Q_{16x}-Q_{16y}$ pair 
		along $\phi$ axis keeping the value of $\rho$ fixed at 0.17 \AA.
		On the other hand, the right panel (c-d) shows the same quantities 
		($\Theta_{12}$, $\Theta_{34}$, $\Theta_{13}$, $\Theta_{23}$ and 
		A$_{11}$[= A$_{22}$], A$_{33}$[= A$_{44}$]) for $Q_{18x}-Q_{19x}$ pair at $\rho$ = 0.24 \AA. For all cases, ADT angles and diagonal ADT matrix elements reflect the quantization property of NACTs. The results have been taken from Ref. [\protect\citen{soumya_bz}] with permission from AIP.}
	\label{fig:bz-angle}
\end{figure}

\begin{figure}[!htp]
	\centering
	\includegraphics[ width=0.7\textwidth, angle=0]{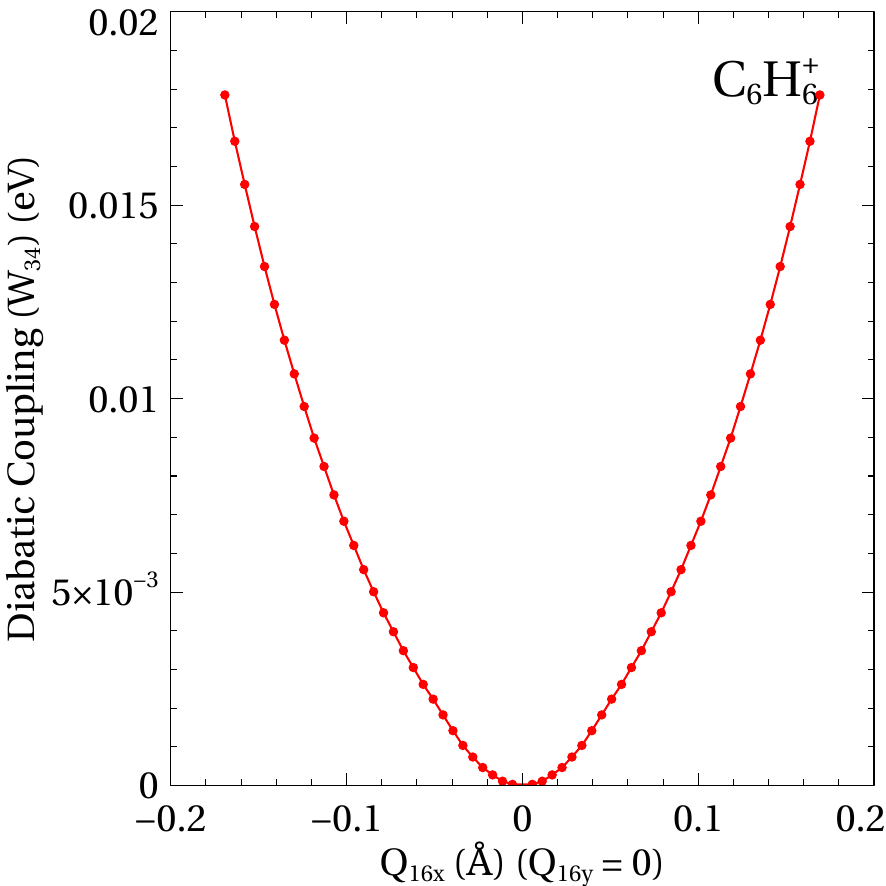}
	\caption{1D plot of diabatic coupling element, namely, $W_{34}$ among $\tilde{B}^2E_{2g}$ states along $Q_{16x}$ mode of Bz$^+$ radical cation is depicted. The coupling term shows symmetric functional form, which can be fitted with even parity polynomials. The results have been taken from Ref. [\protect\citen{joy_pccp}] with permission from RSC.}
	\label{fig:bzeven}
\end{figure}

\noindent
Once the ADT angles
are in hand,~\citep{soumya_jpca121} those are plugged into the explicit expressions of Curl Conditions to evaluate the Mathematical Curls ($Z_{\rho\phi}^{ij}$) (see Appendix~\ref{ssec:adt-eq}) and ADT Curls ($C_{\rho\phi}^{ij}$) to validate the Curl Condition ($F_{\rho\phi}^{ij} = 0$) for the five state SHS. Figure \ref{fig:bz_curl} depicts 1D curves of $Z_{\rho\phi}^{45}$ and $C_{\rho\phi}^{45}$ (exhibiting ``4-5" accidental CI) as well as $Z_{\rho\phi}^{15}$ and $C_{\rho\phi}^{15}$ (representing ``1-5" PJT interaction) along $\phi$ coordinates at three different values of $\rho$. The differences ($F_{\rho\phi}^{ij}$) between those Curls ($Z_{\rho\phi}^{ij}$ and $C_{\rho\phi}^{ij}$) appear almost zero (0) value for both cases ($<$ 10$^{-15}$ $\AA^{-2}$ for ``4-5" and ``1-5" interactions).~\citep{mantu_ijqc_curl} \\

\begin{figure}[htp]
	\centering
	\includegraphics[width=0.9\textwidth]{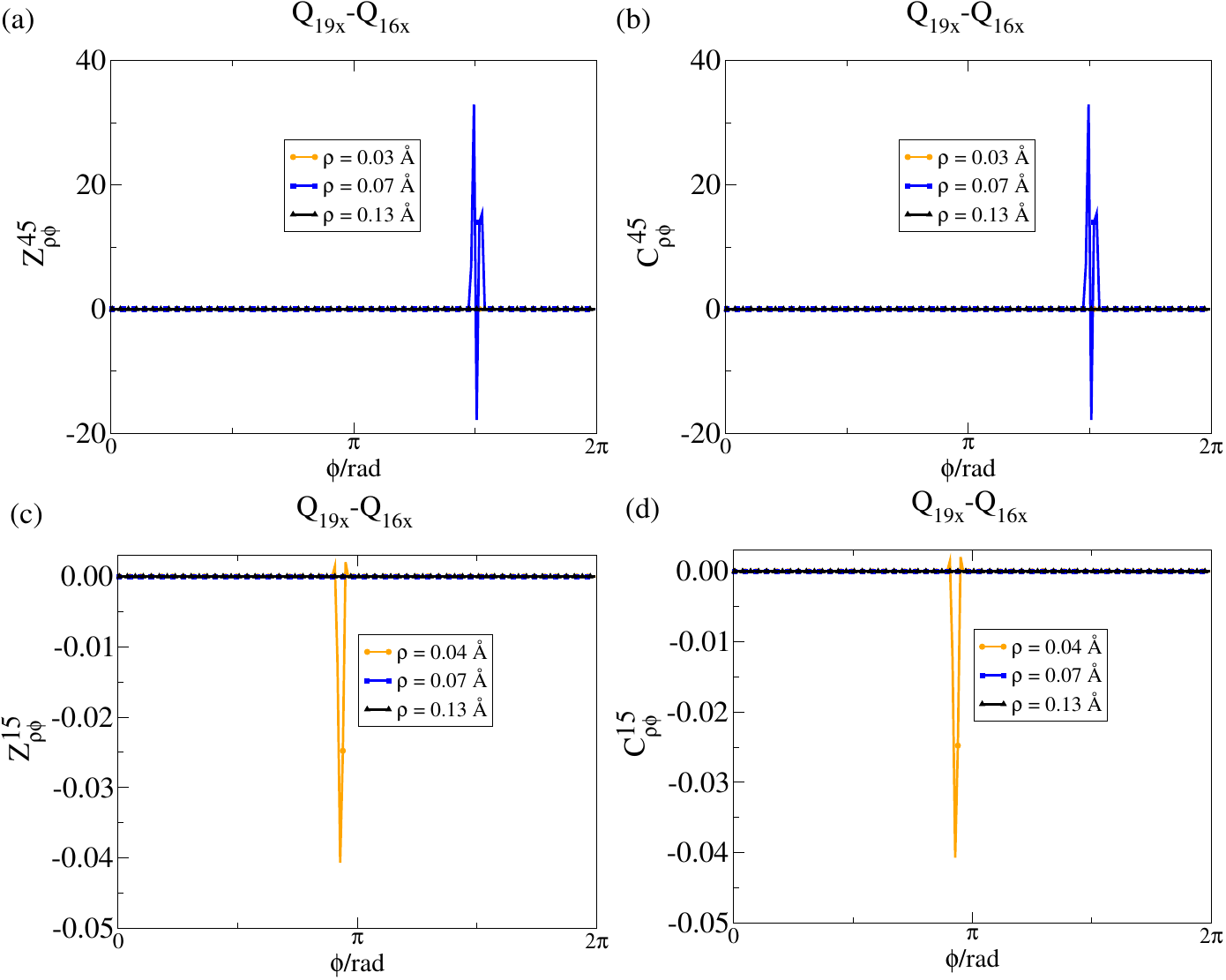}
	\caption{The Mathematical Curls [(a) $Z_{\rho\phi}^{45}$ and (c) $Z_{\rho\phi}^{15}$] and ADT Curls
		[(b) $C_{\rho\phi}^{45}$ and (d) $C_{\rho\phi}^{15}$] indicating``4-5" CI and ``1-5" PJT interactions are demonstrated
		along the angular coordinate ($\phi$) at three different values of $\rho$ for $Q_{19x}$-$Q_{16x}$ plane of C$_6$H$_6^+$. The figures have been adapted from Ref. [\protect\citen{mantu_curl}] with permission from Wiley.}
	\label{fig:bz_curl}
\end{figure}

\subsection{Quantum Dynamics over Path-Dependent Diabatic Hamiltonians with Unique Observables}

\noindent
The path independence of computed observables obtained from diabatic Hamiltonians constructed along different paths was explored. TDDVR based quantum dynamics~\citep{SSJPCA118,souvik_irpc37} 
was carried out over highly ``accurate" BBO-DH of 
low-lying five electronic states of C$_6$H$_6^+$ radical cation ($\tilde{X}^2E_{1g}$, $\tilde{B}^2E_{2g}$ and $\tilde{C}^2A_{2u}$) originating from 
three approaches of contour integration along paths I and II ($\lbrace W_{I}^{ij}\rbrace$, $\lbrace W_{IIa}^{ij}\rbrace$ and $\lbrace W_{IIb}^{ij}\rbrace$). In path I,
ADT equations were solved along $\phi$ coordinate from $\phi_{min}\,(=0)$ to $\phi_{max}\,(=\phi_{min}+2\pi)$ at the lowest value of $\rho$ ($=\rho_{min}$) and then, along the $\rho$ coordinate for each value of $\phi$ (see Figure \ref{fig:contour}a). On the other hand, in path II, ADT equations were solved first along $\rho$ coordinate from higher ($\rho_{max}$) to lower ($\rho_{min}$) at $\phi =\phi_{min}\,(=0)$ and then, along the $\phi$ grid from $\phi_{min}\,(=0)$ to $\phi_{max}\,(=\phi_{min}+2\pi)$ for each $\rho$ value (see Figure \ref{fig:contour}b). Finally the computed ADT angles ($\lbrace\Theta^{I}_{ij}\rbrace$) obtained from path I were plugged into the explicit expressions of NACTs for five state SHS (see Appendix~\ref{ssec:adt-eq}) to regenerate the components of NACTs ($\tau^\rho$ and $\tau^\phi$) and then, the ADT equations were again solved along path II with the help of recalculated NACTs to compute a new set of ADT angles ($\lbrace\Theta^{IIb}_{ij}\rbrace$), ADT matrix ($\lbrace A^{IIb}_{ij}\rbrace$) and diabatic PESs matrix ($\lbrace W^{IIb}_{ij}\rbrace$) elements. \\

\begin{figure}[!htp]
	\centering
	\includegraphics[ height=0.25\textheight, angle=0]{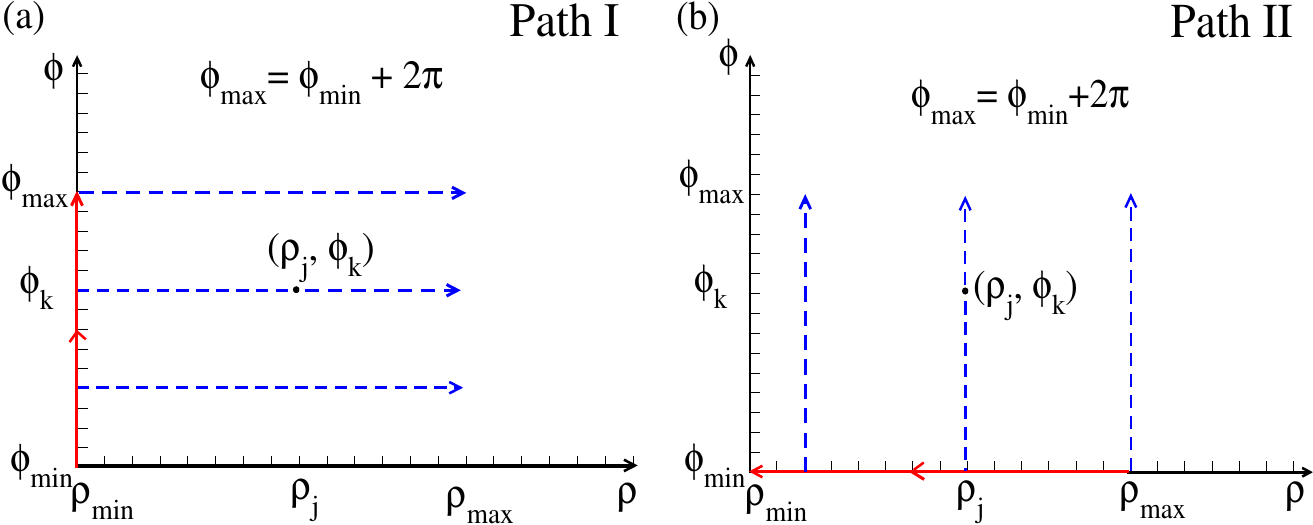}
	\caption{The two different paths over the $\rho-\phi$ space are depicted in diagrams (a) and (b) along which 
		we solve the ADT equations for the calculation of ADT angles, ADT matrices and diabatic PESs matrices. For each case, those stiff linear coupled differential
		(ADT) equations are first integrated along the bold red line and then, along the dotted blue lines. The results have been taken from Ref. [\protect\citen{mantu_jcp159_path}] with permission from AIP.}    
	\label{fig:contour}
\end{figure}

\noindent
Once the diabatic Hamiltonian along path I ($\lbrace W_{I}^{ij}\rbrace$) are in hand, the initial wavepacket was placed on each of the electronic states, $\tilde{X}^2E_{1g}$, $\tilde{B}^2E_{2g}$ and $\tilde{C}^2A_{2u}$
to carry out five different calculations for the generation of the individual state spectra and then, those were convoluted to obtain the overall photoelectron spectral profile for five low-lying electronic states. The TDDVR calculated spectral profiles of C$_{6}$H$_{6}$ obtained from BBO-DH are demonstrated in Figure \ref{fig:tddvr},  
showing very good peak by peak correspondence with the experimental one observed by Baltzer {\it et al.}~\citep{Baltzer_cp224}. Furthermore, peak positions of 
our theoretical spectra (see Figure \ref{fig:tddvr}c and \ref{fig:tddvr}f) show better accord with the experimental results~\citep{Baltzer_cp224} compared to the
MCTDH dynamics results obtained from VCM based quasi-diabatic PESs and couplings~\citep{Koppel_jcp117_1,Koppel_jcp117_2,Koppel_aqc44} (see Figure 
\ref{fig:tddvr}b and \ref{fig:tddvr}e). Figure \ref{fig:tddvr}c shows that the spectral envelop for $\tilde{X}^2E_{1g}$ states possesses the most intense 
0-0 peak at 9.2424 eV, which is in very close proximity of the experimental one (9.2453 eV). On the other hand, a broad vibrational structure with few low intense peaks 
can be observed on the high binding energy side ($>$ 9.4 eV), which can be considered as the hot bands. In addition, for $\tilde{B}^2E_{2g}$ and $\tilde{C}^2A_{2u}$ states, the peak maxima of the 0-0 band (11.4886 eV) as well as the other vibrational excitations (11.5746 eV, 11.6446 eV, 11.6856 eV, etc.) match well with the experimental results. 
A pronounced shoulder is present on the low binding energy compared to 0-0 peak for $\tilde{B}^2E_{2g}$ state, which is also observed in the experimentally recorded spectral band. It is worth mentioning that the broad features at high energy binding side (around 12 eV) of $\tilde{B}^2E_{2g}-\tilde{C}^2A_{2u}$ band (see Figure 
\ref{fig:tddvr}f) do not contain any well defined peak maxima, which bear the signature of $\tilde{C}^2A_{2u}$ electronic states. \\

\begin{figure}[!htp]
	\centering
	\includegraphics[height=0.45\textheight, angle=0]{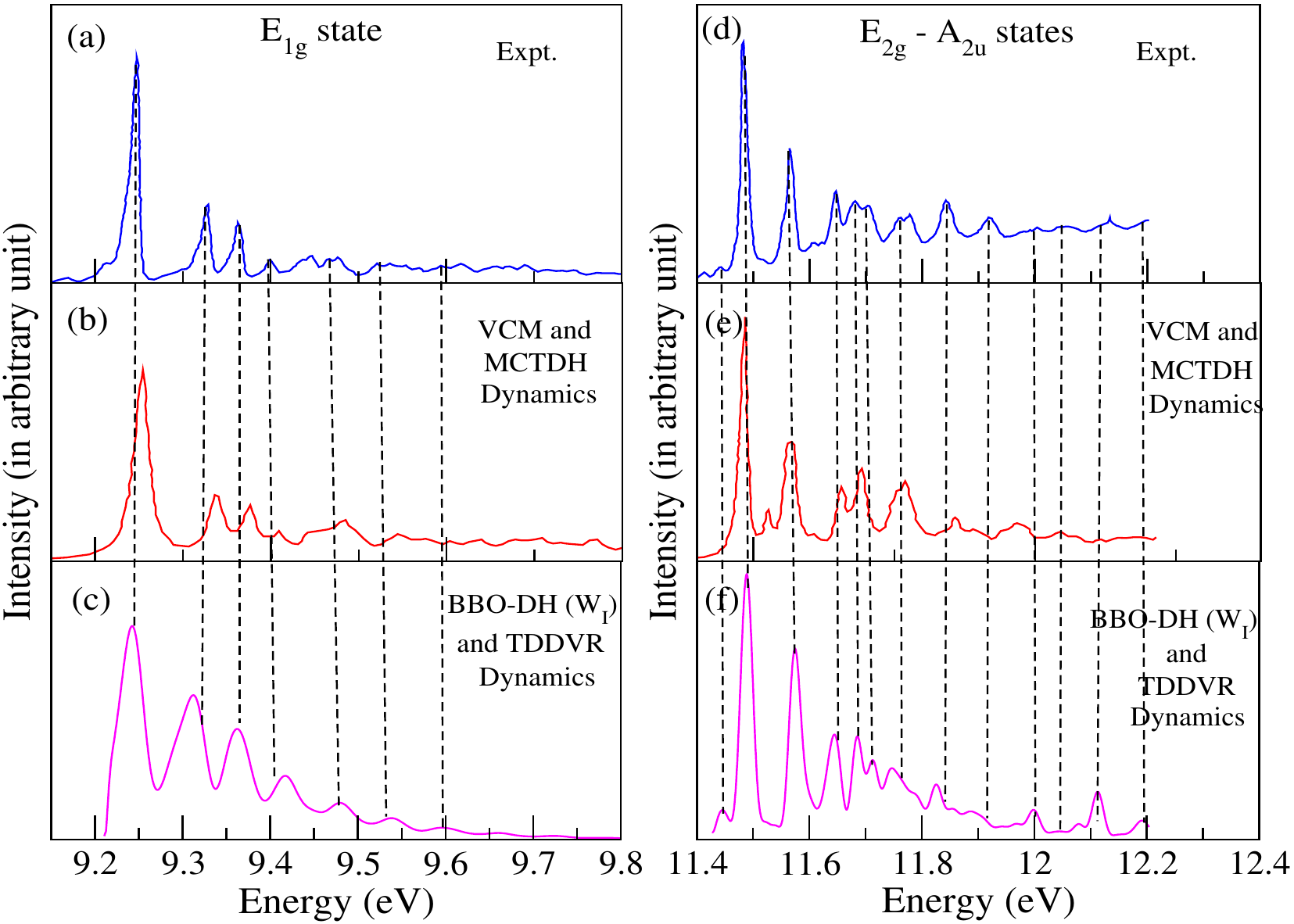}
	\caption{The left column presents (a) the experimental PE spectrum of benzene~\citep{Baltzer_cp224} for the ground degenerate state ($\tilde{X}^2E_{1g}$), which is compared with the theoretical spectra by (b) MCTDH approach on VCM based quasi-diabatic surfaces~\citep{Koppel_jcp117_2,
			Koppel_aqc44}
		and (c) TDDVR methodology on BBO-DH ($\lbrace W_{I}\rbrace$). The right panel (d-f) portrays the same quantities for $\tilde{B}^2E_{2g}$ and $\tilde{C}^2A_{2u}$ states. For all cases, TDDVR based theoretical spectra computed from BBO-DH exhibit better agreement with the experimental one. The results have been taken from Ref. [\protect\citen{soumya_bz}] with permission from AIP.}
	\label{fig:tddvr}
\end{figure}

\noindent
On the other hand, in order to explore the path independence property of the calculated observables using $\lbrace W_{I}^{ij}\rbrace$, $\lbrace W_{IIa}^{ij}\rbrace$ and $\lbrace W_{IIb}^{ij}\rbrace$, three different sets of PE
spectra of C$_{6}$H$_{6}$ molecule were generated.~\citep{mantu_jcp159_path}} 
The left column of Figure \ref{fig:tddvr_path} presents a comparative analysis between experimental spectra of $\tilde{X}^2E_{1g}$ state of C$_{6}$H$_{6}$ molecule recorded by 
Baltzer {\it et al.}~\citep{Baltzer_cp224} and the TDDVR based spectra obtained from three different diabatic PESs matrices ($W_{I}$, $W_{IIa}$ and $W_{IIb}$). It is worth mentioning that all the three theoretical spectra show good agreement within themselves as well as with the experimental one~\citep{Baltzer_cp224}. On the other hand, the right column of Figure \ref{fig:tddvr_path} depicts similar satisfactory matching of
TDDVR calculated spectra for $\tilde{B}^2E_{2g}$-$\tilde{C}^2A_{2u}$ electronic states and experimental one~\citep{Baltzer_cp224}. It is interesting to observe that for the $\tilde{B}^2E_{2g}$-$\tilde{C}^2A_{2u}$ electronic states,
such agreement is prominent even at the region of hot bands ($>$ 11.9 eV). Like the peak positions, intensity patterns of the spectral profiles 
for the five electronic states, $\tilde{X}^2E_{1g}$, $\tilde{B}^2E_{2g}$ and $\tilde{C}^2A_{2u}$ show similarities for all the three
sets of calculations on $W_{I}$, $W_{IIa}$ and $W_{IIb}$.

\begin{figure}[!htp]
\centering
\includegraphics[height=0.5\textheight, angle=0]{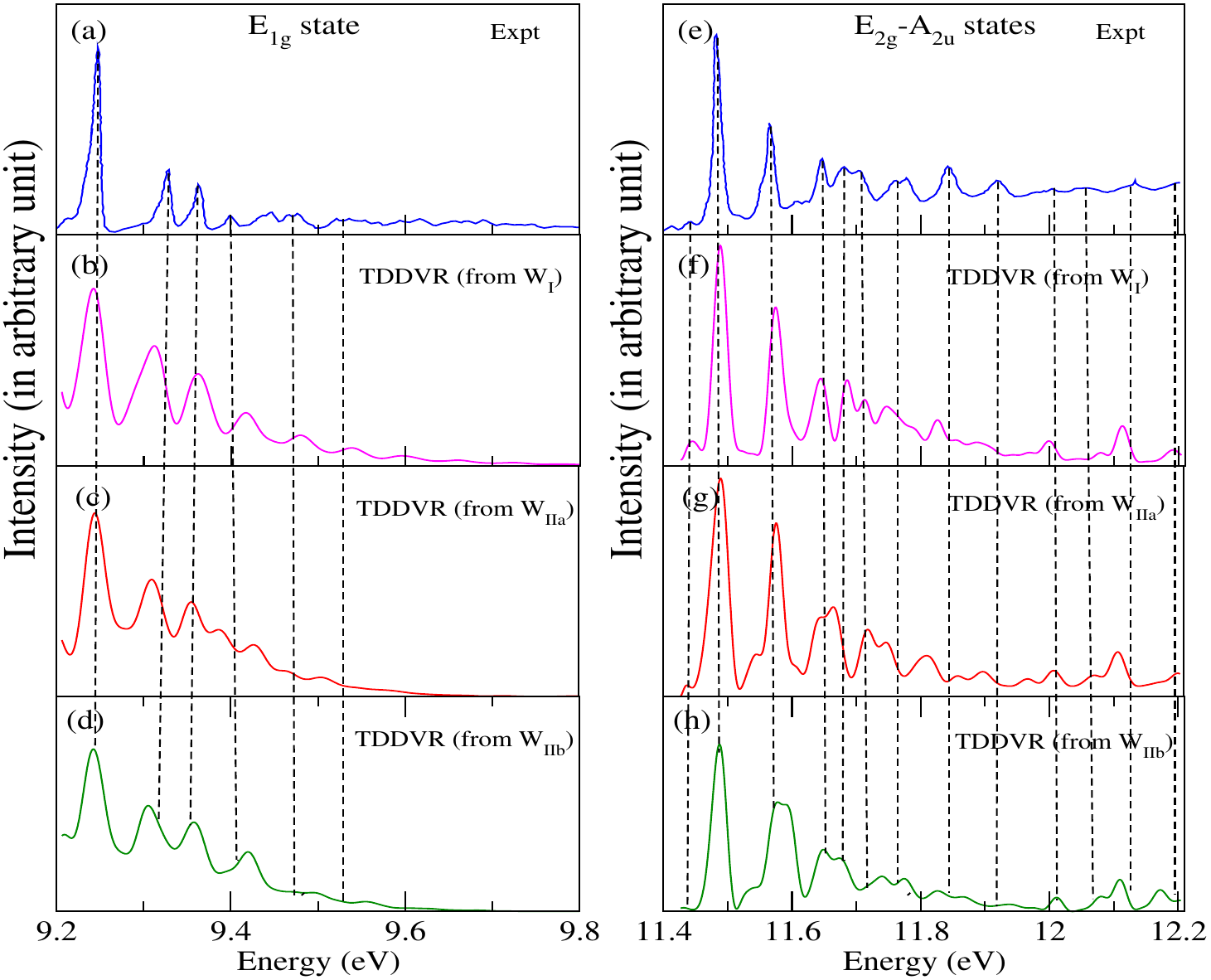}
\caption{The left column compares (b-d) TDDVR calculated PE spectra of benzene obtained from three different diabatic PESs matrices ($W_{I}$, $W_{IIa}$ and $W_{IIb}$) with the (a) experimental one~\citep{Baltzer_cp224} for $\tilde{X}^2E_{1g}$ state. The right panel presents the same 
	observables [(f-h) TDDVR calculated and (e) experimental spectral profiles~\citep{Baltzer_cp224}] for $\tilde{B}^2E_{2g}$ - $\tilde{C}^2A_{2u}$ electronic states. The figures have been taken from Ref. [\protect\citen{mantu_jcp159_path}] with permission from AIP.}
\label{fig:tddvr_path}
\end{figure}

\section{Conclusion}

\noindent
Over the last two decades, the role of electron-nuclear couplings among the low-lying electronic states on the PA/PD/PE/MATI spectra were explored with the 
aid of BBO treatment. After locating the CIs between the electronic states and constructing multi-state multi-mode BBO-DH, TDDVR based quantum dynamics were employed for a series of molecular systems, namely,
NO$_2$, Na$_3$, C$_4$N$_2$H$_4$, NO$_3$, C$_6$H$_6^+$, o-C$_6$H$_4$F$_2^+$ and 1,3,5-C$_6$H$_3$F$_3^+$ molecules/cations to evaluate the spectral features.\\

\noindent
In case of
NO$_2$ radical, two lowest electronic states ($X^{2}A_{1}$ and $A^{2}B_{2}$) possess JT and RT interactions governing its photodetachment spectrum.\citep{mahapatra_jcp110,jackels_jcp63,kaldor_cpl185,haller_jms111,schinke_cp346,joyeux_jcp119,sanrey_jcp125,mukherjee_jcp143,
sardar_cp416,takatsuka_jcp132,takatsuka_pccp13} Adhikari and coworkers constructed highly precise diabatic Hamiltonian for NO$_2$\citep{mukherjee_jcp143} employing MRCI based adiabatic PESs and NACTs obtained from MOLPRO\citep{MOLPRO_brief} quantum chemistry package followed by ADT calculation. On the other hand, first principle based adiabatic PESs and NACTs were generated for three excited electronic states ($2\,^2E^{\prime}$ and $1\,^2A_1^{\prime}$) of Na$_3$ system
using MRCI methodology. Electronic structure as well as ADT calculations were carried out over 2D
nuclear plane ($\rho-\phi$) originating from bending ($Q_x$) - asymmetric stretching ($Q_y$) pair at fixed value of symmetric stretching ($Q_s$) mode (3.7 \AA). 
The Na$_3$ cluster shows accidental CIs within two branches of 2$^2E^{\prime}$ state as well as profound PJT interaction among 2$^2E^{\prime}$ and 1$^2A_1^{\prime}$ states.\citep{mukherjee_jcp143} In order to explore the workability of four electronic state BBO formalism, pyrazine molecule were taken as a benchmark system.~\citep{saikat_pyra} The first principle based BBO theory was employed to unveil its wide variety of non-adiabatic interactions present within the low-lying four excited electronic states [$1^1B_{1u}$ ($S_1$), $1^1B_{2u}$ ($S_2$), $1^1A_{u}$ ($S_3$) and $1^1B_{2g}$ ($S_4$)] and to construct 4$\times$4 diabatic PES matrices, where the ``1-2", ``2-3" and ``3-4" CI(s)/seam(s) can be located over six 2D nuclear CSs constituted with four normal modes ($Q_1$, $Q_{6a}$, $Q_{9a}$ and $Q_{10a}$). While examining the validity of five state BBO formalism, three species, NO$_3$, C$_6$H$_6^+$ and o-C$_6$H$_4$F$_2^+$ were selectively chosen to study their rich nonadiabatic features. For NO$_3$ radical,~\citep{soumya_jpca121,bijit_mp115} JT interactions were unveiled within $\tilde{A}^2E^{\prime\prime}$ and $\tilde{B}^2E^{\prime}$ states at equilibrium D$_{3h}$ point and accidental CIs can be observed within `3-4' and `4-5'  electronic states at $C_{s}$ and $C_{2v}$ geometries. Apart from the JT and accidental CIs,
NO$_{3}$ is enriched with substantial PJT coupling between $\tilde{X}^2A_2^{\prime}$ and $\tilde{B}^2E^{\prime}$ states.\citep{soumya_jpca121,bijit_mp115} 
While exploring Bz$^+$ radical cation, the low lying doublet
states, namely, $\tilde{X}^2E_{1g}$ and $\tilde{B}^2E_{2g}$ were observed to show profound ``1-2" and ``3-4" JT distortion along
the $e^{\prime}$ modes.\citep{soumya_jpca121,Koppel_jcp117_1,soumya_bz} In addition, strong PJT couplings
and accidental CIs are also present within $\tilde{B}^2E_{2g}$ and $\tilde{C}^2A_{2u}$ electronic 
states (mainly ``4-5" PJT) over the FC region of nuclear CS.\citep{soumya_jpca121,koppel_jcp89,soumya_bz} Moreover, o-dfbz$^+$ was considered as a prototypical system exhibiting rich variety of nonadiabatic interactions, including JT-type accidental CIs/seams (``1-2", ``3-4" and ``4-5") and PJT interactions among these $\tilde{X}^2B_1$, $\tilde{A}^2A_2$, $\tilde{B}^2B_1$, $\tilde{C}^2A_1$ and $\tilde{D}^2B_2$ states over eight (8) non-adiabatically active nuclear planes ($Q_6$-$Q_7$, $Q_{24}$-$Q_{26}$, $Q_6$-$Q_{13}$, $Q_7$-$Q_8$, $Q_{8}$-$Q_{13}$, $Q_{14}$-$Q_{26}$, $Q_{20}$-$Q_{24}$ and $Q_{20}$-$Q_{26}$).~\citep{saikat_cpc26} Finally, intense vibronic couplings (``1-2" JT/accidental CI(s)/seam(s), ``4-5" JT/accidental CI(s)/seam(s), ``1-3", ``2-3", ``4-6" and ``5-6" PJT couplings and many others) were explored within the low-lying six electronic states of TFBz$^{+}$ radical cation, namely, $\tilde{X}^{2}E^{\prime\prime}$, $\tilde{A}^{2}A_{2}^{\prime\prime}$, 
$\tilde{B}^{2}E^{\prime}$ and $\tilde{C}^{2}A_{2}^{\prime}$.~\citep{soumya_tfb}\\

\noindent
For each system, once the {\it ab initio} quantities (adiabatic PESs and NACTs) were in hand, those had been plugged into the ADT equations (see Eqs. \ref{eq:adt-two}, \ref{eq:adt-three} and Appendix~\ref{ssec:adt-eq}) to solve those equations for the generation of ADT angles vis-\'a-vis ADT matrices and diabatic PESs matrices. 
For all cases, the ADT angles attend the magnitude of 0 or $n\pi$ or $n.2\pi$ at the end of the contours representing the PJT, JT and RT situations, respectively, with $n$ being the number of CIs. It is important to mention that the functional forms of ADT angles appear topological (depends on the NACT profiles) in nature.~\citep{bijit_mp115,soumya_bz,soumya_tfb} On the other hand, those ADT angles were used to regenerate the NACTs and then, Mathematical as well as ADT curls were calculated validating the Curl condition and affirming the existence of sub-Hilbert spaces for each specific molecular system.~\citep{mantu_ijqc_curl,saikat_cpc26} Furthermore, BBO based diabatization procedure depicts an unique finding, i.e., even parity functional forms of diabatic PESs and couplings along some components of asymmetric normal modes of systems, namely, $Q_3$ mode of NO$_2$, $Q_y$ mode of Na$_3$, $Q_{3x}$ mode of NO$_3$, $Q_{16x}$ mode of C$_6$H$_6^+$ and $Q_{9x}$ mode of 1,3,5-C$_6$H$_3$F$_3^+$ show exclusive symmetric feature in contrary to the ``traditional" JT model.~\citep{joy_pccp}\\

\noindent
After constructing diabatic Hamiltonians of the molecular species/radical cation, quantum dynamics need to be carried out to compute the PE/PA/PD/MATI spectra of the corresponding neutral analogue. Dynamical calculations are performed at the close vicinity of equilibrium geometry (FC region), where Gaussian wave packets (GWPs) are utilized to represent molecular vibrations. In other words, a product wavefunction with GWPs for each normal mode of the neutral/anionic species is considered as the initial wavefunction for solving the time dependent SE. While performing dynamical calculations, time dependent nuclear wavefunctions vis-\'a-vis autocorrelation function are computed
at different time followed by Fourier transformation to generate the PE/PD/PA/MATI spectra for individual state. Then, the individual state spectra are convoluted to get the entire spectral profile. It is worth mentioning that for each system, theoretically calculated spectral profiles~\citep{mukherjee_jcp141,mukherjee_jpca117,saikat_pyra,bijit_cp515,saikat_cpc26,soumya_tfb,soumya_bz} exhibit better agreement with the experimental ones~\citep{weaver_jcp90,ernst_pst112,woywood_jcp100,weaver_jcp94,bieri_jesc,Baltzer_cp224,gilbert_cpl17} in comparison with MCTDH based spectra~\citep{raab_jcp110,faraji_jcp_dyna,Koppel_jcp117_2,
Koppel_aqc44,mahapatra_pccp11} obtained from VCM calculated diabatic Hamiltonian. Finally, as per the theoretical prediction, the path-dependent diabatic Hamiltonians are expected to show up unique observables (spectra), which has been explored to confirm the prediction.~\citep{mantu_jcp159_path}  \\

\noindent
In summary, theoretically ``exact" and numerically ``accurate" BBO constructed diabatic PES matrix followed by quantum dynamics successfully depicts the contribution of nonadiabatic coupling in
various prototype systems. Such theoretical approach consistently depict the importance of nonadiabatic effects to reshape spectral profiles exhibiting good accord with experimental data.

\bibliographystyle{iopart-num}
\bibliography{ref_book}

\chapter{Reactive Scattering Processes: BBO Based Diabatic Hamiltonian and Nuclear Dynamics}\label{sec:scat}

\noindent
The profiles of cross-sections and rate coefficients for several prototype chemical reactions of atom-diatom (A+BC) collision exhibit substantial deviations between theoretically computed and experimentally measured ones in many cases even at reasonably lower collision energy. Such deviations could be originated due to the neglect of electron-nuclear coupling in the theoretical framework as well as numerical calculations on a specific surface. Such approaches are designated as single surface adiabatic scattering dynamics based on Quasi-Classical Trajectory (QCT) and/or Quantum Mechanical (QM) approaches. In this chapter, a series of reaction attributes is presented for various reactive processes, where calculated cross-sections/rate coefficients exhibit satisfactory agreement with sophisticated experimentally measured data while carrying out the scattering dynamics involving multiple electronic states. Furthermore, during such theoretical calculations on multiple electronic states,
it is indeed necessary to include electron-nuclear couplings from the first principle in the respective Hamiltonian. In subsequent paragraphs, recently performed theoretically calculated adiabatic and diabatic results are manifested for few specific reactive processes in comparison with the corresponding experimental data. Such analysis unveils the importance of incorporation of electron-nuclear coupling in the Hamiltonian. \\


\noindent
In the 1980s, the landmark molecular beam experiment by Lee \textit{et al.}~\cite{jcp-53-3385-LEE,jcp-82-3045-LEE,jcp-82-3067-LEE} and subsequently, Toennies \textit{et al.}~\cite{jcp-94-4676-TOENN,jcp-101-2106-TOENN,jcp-104-2743-TOENN, jpcA-101-6415-TOENN,jcp-108-9694-TOENN} recorded velocity contour maps of HF and also DF product angular distributions. On the other hand,
the temperature dependent rate coefficients for the F + H$_2$/D$_2$ reaction were measured by the multiphoton dissociation techniques \cite{jcp-72-4811-HOUSTON,jcp-72-4815-HEIDNER} in the temperature range 190-370 K, whereas the rate coefficient over the  temperature range 240-373 K was speculated by discharge flow/resonance fluorescence~\cite{jpc-93-4068-ANDERSON}. At the same time, Sims \textit{et al}~\cite{nchem-6-141-SIMS}
reported the rate coefficient for this reaction at the low temperature (11-295 K). Over the extended temperature range (220-960 K), the kinetic isotopic effect for the F + H$_2$/D$_2$ reaction was analyzed by Bedjanian~\cite{bedjaninan}. It is worth mentioning that 
theoretical calculations on single adiabatic surface for rate coefficients\cite{aql2,jcp-142-024303-ZHANG,fxz_pes,Naskar2022_JPCA_126} show deviation far more from the experimental results~\cite{jpc-93-4068-ANDERSON,nchem-6-141-SIMS,bedjaninan} compared to the recent diabatic one\cite{FH2_dynamics}, where both electron-nuclear as well as spin-orbit (SO) couplings are incorporated.
In addition, the diabatic Hamiltonians constructed only with SO coupling\cite{werner_jcp104,werner_jcp109} versus the one\cite{FH2_dynamics} both with nonadiabatic as well as SO couplings display that the latter approach is able to better reproduce the experimental rate constant. The presence of Jahn-Teller (JT) type conical intersection (CI) between the ground (1$^2A^\prime$) and the first excited electronic state (2$^2A^\prime$) as well the existence of additional Renner-Teller (RT) CI(s) between the first (2$^2A^\prime$) and second excited electronic states (1$^2A^{\prime\prime}$) affirm the necessity for the calculation of 
multi-state (diabatic) Hamiltonian.\cite{FH2_Surf,cpl-517-92-ANITA}\\


\noindent
The initial experimental detection of H$_3^+$ by Oka~\cite{OKA:PRL} 
drew the attention of allied scientific community. Even though 
the species shows molecular simplicity, the quantum mechanical calculations appear as ``difficult" due to the presence of an extraordinary deep well ($\sim$ - 4.0 eV) along the reactant channel on the ground electronic state of the H$_3^+$ potential energy surface (PES).\cite{H3+_Surf} The guided molecular beam technique yielded
integral cross-sections (ICSs) for the reaction H$^+$ + D$_2$, where
Fehsenfeld \textit{et al.}~\cite{jcp-61-2150-FOUR} and Villinger \textit{et al.}~\cite{jcp-76-1590-HEN2} successfully measured the rate constants for H$^+$ + D$_2$ as well as D$^+$ + H$_2$  reaction at lower collision energy range (< 0.3 eV). On the other hand, Gerlich~\cite{advcp-82-1-GER} used merged beam technique to record 
the rate constants for the isotope exchange reaction, D$^+$ + H$_2$ at the low temperature regime. In this perspective, theoretically computed\cite{Gerlich_1990,HONV_2011_Er,ortho-para_2024} rate constants at low collisional energies on the ground adiabatic PES 
demonstrate the appropriate prediction of ortho-para transition on the collision process, H$^+$+H$_2$. The charge transfer processes involved in D$^+$+H$_2$ reaction were studied experimentally to obtain ICSs, where for such reaction (D$^+$+H$_2$), the effect of multiple electronic state on reactive and non-reactive charge transfer processes were illustrated\cite{jcp-147-074105-SA} by initiating the collision process from the ground electronic state. On the other hand, for H$^+$+H$_2$ reaction, such effects were explored\cite{H3+_dynamics_2021} by allowing the collision from the excited electronic state. When the calculations of the ICSs for various processes, namely, charge transfer and non charge transfer were performed, the role of electron-nuclear coupling among the electronic states demands the formulation of diabatic Hamiltonian considering first three low-lying electronic states (1$^1A^\prime$, 2$^1A^\prime$ and 3$^1A^\prime$).\cite{H3+_Surf} Dynamical calculations using such diabatic Hamiltonian reveal that 
in order to reproduce the experimentally measured ICS data, the nonadiabatic couplings play a crucial role in monitoring the reaction mechanism.~\cite{jcp-147-074105-SA} Since the ground and the first excited states are coupled through the symmetry driven CI as well as the accidental CI along a seam, the construction of multi-state diabatic Hamiltonian is indeed necessary for dynamical calculation.\cite{H3+_Surf} \\
%



\noindent
Chemical reaction involving HeH$_2^+$ species acts as a prototype system both for experimental measurements\cite{Klein1964_JCP_41,chupka1968_1,chupka1968_2,Tang2005_JCP_122,Turner2022_JCP_81} and theoretical calculations\cite{Maiti2002_JCP_117,Kraemer2002_CP_276,Mrugala2003_JCP_118,Tiwari2005_CPL_414,Panda2005_JCP_122,Panda2006_CPL_419,Tang2007_JCP_127}, which are necessary to elucidate the underlying reaction mechanism. Initially, Chupka \textit{et al.}\cite{chupka1968_1,chupka1968_2,chupka1972} depicted vibrational state assisted cross-sections of the proton transfer reaction, He + H$_2^+ (v;j) \rightarrow$ HeH$^+ (v^\prime;j^\prime)$ + H, where the endothermicity ($\sim$ 0.79 eV) in the product channel plays a crucial role. Moreover, various experiments were performed employing electron impact ionization\cite{CPL_37_1976_Schneider, CPL_49_1977_Pacak}, molecular beam\cite{JCP_94_1991_Pollard, JCP_95_1991_Pollard} and other techniques\cite{CPL_96_1983_Pijkeren, JCSFT2_83_1987_Herman, Tang2005_JCP_122} to analyze the vibrational enhancement of the cross-sections and its decaying trend with the increase of collision energy for higher initial vibrational states. Over the past few decades, various theoretical studies based on QCT and QM calculations were carried out to explore the reaction dynamics in HeH$_2^+$ system on its ground electronic state.\cite{Baer1987_JPC_91, Aquilanti2000_CPL_318, Panda2005_JCP_122, Tang2005_JCP_122, Xu2008_CPL_464, JCP_137_2012_Fazio,JCP_98_1993_Kumar} On the contrary, calculated ICSs from QCT and QM approaches for the reaction, He + H$_2^+ (v=0;j=0) \rightarrow$ HeH$^+ (v^\prime;j^\prime)$ + H 
show consistently lower values compared to the result obtained from experimental observations \cite{chupka1968_1,chupka1968_2,JCP_81_1984_Turner,Achtenhagen,Tang2005_JCP_122}. Such differences display the necessity to construct the multi-state diabatic Hamiltonian\cite{HeH2_Surf}, where the nonadiabatic coupling terms (NACTs) among the first four electronic states were included to study the proton-exchange reaction. While carrying out quantum reactive scattering dynamics employing such diabatic Hamiltonian\cite{HeH2_Surf}, the computed reactive ICSs show satisfactory quantitative agreement with experimental ones in comparison with the calculations performed on other available adiabatic/diabatic Hamiltonian for this system.\cite{HeH2+_dynamics_2023} This indicates the crucial contribution of electron-nuclear coupling in the reaction mechanism of He + H$_2^+$ collision. It is worthwhile to mention that the lowest four electronic states (1$^2A^\prime$, 2$^2A^\prime$, 3$^2A^\prime$ and 4$^2A^\prime$) of HeH$_2^+$ are strongly coupled, wherein CIs are prevalent between the first two electronic states at C$_{2v}$ geometries\cite{HeH2_Surf,IJQC_121_2021_Mondal,MolPhys_118_2020_Gupta} and additional CI(s) are found between the second and third as well as the third and fourth electronic states at collinear geometries\cite{MolPhys_118_2020_Gupta,MolPhys_119_2021_ravi,HeH2_Surf}.


\vspace{0.3cm}

\noindent
The hydrogen exchange reaction in H$_3$ system appears as a prototype system for collision-induced reaction dynamics, which was thoroughly studied as a benchmark one in theoretical~\cite{Kuppermann_1975,Walker_1978,Charutz_Baer_1997,Kendrick_2001,Aoiz_2005} and experimental investigations~\cite{Zare_92,Nieh_Valentini,Y.T_Lee_90,Banares_Exp_Theo_1998,Bean_Zare_DCS_2002,Mielke_Sutherland_2003,Dai_Wang_Science_2003,Althorpe_Zare_ICS_2004,Pomerantz_2004_ref_20,zare_2005}. 
Adiabatic calculations for the H + D$_2$ reaction on the ground state PES at relatively higher (total) energies (1.5~eV~$\le$~E$_{\mathrm{tot}}$~$\le$~4.5~eV) unveiled substantial deviations from experimentally recorded cross-section~\cite{Pomerantz_2004_ref_20,zare_2005}. Due to the strong coupling among the low lying doublet electronic states (1$^2A^\prime$, 2$^2A^\prime$ and 3$^2A^\prime$) of H$_3$ through CI or seam~\cite{Halasz_Baer2003,Baer_2004,Ghosh_H3_2025}, the applicability of adiabatic calculations~\cite{Kuppermann_1975,Walker_1978,Charutz_Baer_1997,Kendrick_2001,Aoiz_2005} on the ground state is limited even at lower total energy (0.5~eV~$\le$~E$_{\mathrm{tot}}$~$\le$~1.5~eV) compared to the experimental result. Deviations can be observed between computed reaction attributes (such as integral and differential cross-sections) 
and experimental data~\cite{Pomerantz_2004_ref_20,Althorpe_2015_DCS,Jankunas_Althrope_2013} over the energy range (1.5~eV~$\le$~E$_{\mathrm{tot}}$~$\le$~3.5~eV). As a consequence, the inclusion of coupling among the adiabatic states is indeed necessary to explore mechanism of the reaction.\cite{DMBE_1987,Ghosh_H3_2025} Calculations of \textit{ab initio} adiabatic PESs and NACTs were executed for the three electronic states followed by construction of the diabatic Hamiltonian\cite{Ghosh_H3_2025} to find out the mechanism of this prototype reaction of H$_3$ system. The experimental cross-section/rate constants were compared with those quantities calculated from the diabatic Hamiltonian\cite{Ghosh_H3_2025} to understand the contribution of electron-nuclear coupling.

\vspace{0.3 cm}

\noindent
In order to construct the diabatic Hamiltonians for performing scattering calculations on various reactive processes, Beyond Born-Oppenheimer (BBO) theory was employed including electron-nuclear couplings among the electronic states. Once the diabatic Hamiltonians are derived for several systems, it is necessary to use sophisticated approach for scattering calculation for obtaining various quantities, cross-section and rate constants.~\cite{Scat_Review} Over the last few decades, the formulation of methodologies for quantum reactive scattering calculations on triatomic systems were applied employing both time-independent \cite{jcp-92-3419-BAER} and time-dependent \cite{jcp-90-4351-BAER} approaches, namely, the S-matrix formalism \cite{jcp-91-1528-MILL}, R-matrix propagation method\cite{jcp-85-4744-LIT2,jcp-90-265-LIT2}, Newton variational principle employing T operator\cite{prA-4-1924-MBK2} etc. In case of Jacobi coordinate\cite{jcp-132-084112-GUO,jcp-125-054102-GOM,jcp-88-6233-MILL,jcp-91-1528-MILL} based methodologies, it is necessary to convert the 
wavefunction from reactant to product coordinates\cite{jcp-106-1742-ZHA1,cpl-262-519-ZHA1,jcp-105-6072-ZHA1,jcp-106-7629-ALT1,jcp-107-7816-ALT1}, whereas the hyperspherical coordinates\cite{jcp-79-1916-JOH,Xu2000_JCP_112,cpc-184-270-SA,jcp-99-2674-BILL,jcp-100-1085-BILL,cp-191-247-MAR,science-286-1713-HYP2,jcp-111-6665-HYP2,cpl-324-466-HYP2,jpcA-118-4837-SA,jcp-84-5962-HYP1} treat all the channels even handedly. Furthermore, while extending the methodology from single state adiabatic to multi-state diabatic scattering cases, such extension using hyperspherical coordinate based wavepacket approach appears as quite straightforward. On the other hand, all the channels of different electronic states can be denoted without ambiguity. With the aid of Johnson's hyperspherical coordinate\cite{jcp-79-1916-JOH}, Billing and co-workers\cite{jcp-99-2674-BILL,jcp-100-1085-BILL} developed the time-dependent wave packet (TDWP) formalism for the first time depicting its workability on single adiabatic PES of D + H$_2$ system for total angular momentum quantum numbers $J = 0$ and $J = 1$. In the subsequent research works, Adhikari and co-workers\cite{cpc-184-270-SA} generalized this framework for any non-zero total angular momentum ($J \neq 0$ cases)  \cite{jpcA-118-4837-SA,jcp-142-024304-SA,cpl-675-85-SA,pccp-20-478-SA,pccp-21-20166-SA} by including all helicity quantum numbers ($K$)\cite{heh2+_dyna} through OpenMP-parallelization algorithm. In addition, such approach was further extended for multi electronic state diabatic Hamiltonian, where the propagation of partial waves are performed by OpenMP scheme and the solution of differential equations for different electronic states are obtained by employing MPI technique.\cite{jpca-119-12392-SA}

\noindent
\section{Outline of BBO Theory for Chemical Reactions}

\subsection{Adiabatic-to-Diabatic Transformation Formalism}

\noindent
In the adiabatic representation of BO treatment~\cite{BO,BH}, the total molecular wave function for a M-dimensional Hilbert Space can be written in terms of the product of nuclear wave functions (\{$\psi_i^{\rm ad}$\}) and electronic eigenfunctions (\{$\xi_i$\}), which is defined as the Born-Oppenheimer-Huang (BOH) expansion: 

\vspace{-0.8cm}

\begin{eqnarray}
	\Psi(\textbf{r},\textbf{R}) = \sum_{i=1}^M \psi_i^{\rm{ad}}(\textbf{R}) \xi_i(\textbf{r};\textbf{R}),
	\label{eq:bo-expansion} 
\end{eqnarray}

\noindent
The total electron-nuclear Hamiltonian, $ \hat{H}(\mathbf{r}, \mathbf{R}) $ can be segregated into two operators, nuclear kinetic energy operator $ \hat{T}_{\text{nuc}}(\mathbf{R}) $ and electronic Hamiltonian, $ \hat{H}_{\text{el}}(\mathbf{r}; \mathbf{R}) $ as follows:

\begin{eqnarray}
	\hat{H}(\textbf{r},\textbf{R}) = \hat{T}_{\rm nuc}(\textbf{R}) + \hat{H}_{\rm{el}}(\textbf{r};\textbf{R}),
	\label{tothamil}
\end{eqnarray} 
where $\mathbf{r}$ and $\mathbf{R}$ symbolize the electronic and nuclear coordinates, respectively. 

\vspace{0.3 cm}

\noindent
On the other hand, electronic Schr\"{o}dinger equation (SE) with  orthonormal set of electronic eigenfunctions, $\{\xi_i\}$ can be written as:
\begin{eqnarray}
	\hat{H}_{\rm{el}}(\textbf{r};\textbf{R})\xi_i(\textbf{r};\textbf{R}) = u_i(\textbf{R})
	\xi_i(\textbf{r};\textbf{R}), \qquad  
	\langle \xi_i | \xi_j \rangle_{\textbf{r}} = \delta_{ij},
	\label{eq:elese}
\end{eqnarray}

\noindent
where the eigenvalues at any fixed nuclear geometry, $u_i(\mathbf{R})$ are termed as adiabatic potential energy surface(s) (PESs). 

\vspace{0.3 cm}

\noindent
While substituting the molecular wave function [Eq.~(\ref{eq:bo-expansion})], the total electron-nuclear Hamiltonian [Eq.~(\ref{tothamil})] and the electronic eigenvalue equation [Eq.~(\ref{eq:elese})] into the molecular SE [$\hat{H}(\textbf{r},\textbf{R})	\Psi(\textbf{r},\textbf{R}) = E \hat{H}(\textbf{r},\textbf{R})$] followed by projection with the
electronic eigenfunctions $\{\xi_i\}$ leads to the matrix form of the kinetically coupled adiabatic nuclear Schr\"{o}dinger equation \cite{Baer2006_bbo_book_sect211} as follows:
\begin{eqnarray}
	-\frac{\hbar^2}{2m}\left(\boldsymbol{\nabla}_R+\boldsymbol{\tau}\right)^2\psi^{\rm{ad}} + \left(U-E\right)\psi^{\rm{ad}} 
	= 0,
	\label{eq:adsem}
\end{eqnarray}

\noindent
where $E$ signifies the total molecular energy, $U$ indicates the adiabatic PESs matrix ($U_{ij} = u_i \delta_{ij}$) and $\boldsymbol{\tau}$ designates the skew-symmetric nonadiabatic coupling matrix (NACM) showing following anti-symmetric property:

\begin{eqnarray}
	\boldsymbol{\tau}_{ij} &=& \langle\xi_i\vert\boldsymbol{\nabla}_R\xi_j\rangle_\textbf{r} = - \langle\xi_j\vert\boldsymbol{\nabla}_R\xi_i\rangle_\textbf{r}.
	\label{eq:nacm}
\end{eqnarray}

\noindent
Furthermore, when electronic SE [Eq.~(\ref{eq:elese})] is differentiated with respect to the nuclear coordinates ($\boldsymbol{\nabla}_R$) and then, projected by the electronic eigenfunctions ($\langle \xi_i |$), the following form of NACTs can be achieved:

\begin{eqnarray}
	\boldsymbol{\tau}_{ij} = \frac{\langle\xi_i\vert\boldsymbol{\nabla}_R \hat{H}_{\rm{el}}\vert\xi_j\rangle_\textbf{r}}{u_j-u_i},
	\label{eq:HF}
\end{eqnarray}

\vspace{0.3cm}

\noindent
which indicates singular behavior at and around the degenerate points ($u_i$ = $u_j$), where such relation is a reminiscent of well-known Hellmann-Feynman theorem~\cite{clusius1941einfuhrung,feynman1939forces,epstein1954note}. Due to appearance of such singular behavior, adiabatic representation of the SE produces numerical inaccuracies and non-conservation of energy~\cite{mukherjee2024quasi}. Hence elimination of these 
singular terms is a necessity by converting the adiabatic SE [Eq.~(\ref{eq:adsem})] into diabatic framework by employing 
an orthogonal transformation, $\psi^{\rm{ad}}=A\psi^{\rm{dia}}$ as demonstrated below:

\vspace{-0.8cm}

\begin{eqnarray}
	\left(-\dfrac{\hbar^2}{2m}\nabla_R^2 + W-E\right)\psi^{\rm{dia}} = 0.
	\label{eq:diasem}
\end{eqnarray}
In the above expression, $\psi^{\rm{dia}}$ defines the diabatic nuclear wavefunction and $ W $ indicates the diabatic PESs matrix,
\begin{eqnarray}
	W = A^{\dagger}U A.
	\label{eq:diapot1}
\end{eqnarray}

\noindent
In other words, such transformation leads to a paradigm shift from the kinetically coupled adiabatic nuclear SE to potentially coupled diabatic one. In order to carry out such process, the matrix $ A $ needs to fulfill the following criterion:
\begin{eqnarray}
	\boldsymbol{\nabla}_RA + \boldsymbol{\tau}A = 0,
	\label{eq:adteq}
\end{eqnarray}
which is commonly known as the ADT condition.~\cite{Baer1975_CPL_35,TB,Sarkar2006_JCP_124_EBO} Since \textbf{$\tau$} signifies
a skew-symmetric matrix, Eq.~\ref{eq:adteq} shows that $A$ is an orthogonal matrix.\\

\noindent
The explicit expression of ADT equations for two and three electronic states are given in Eqs. \ref{eq:adt-two} and \ref{eq:adt-three}, whereas the ADT equations for four to six states are demonstrated in Appendix \ref{ssec:adt-eq}. \\

\noindent
While carrying out BBO based diabatization, {\it ab initio} calculated NACTs are substituted into ADT equations and then, those equations are numerically integrated to compute
the ADT angles. Once the mixing angles are obtained, those are used to construct the ADT matrices followed by similarity transformation (Eq. \ref{eq:diabse}) for the generation of smooth, single-valued and continuous diabatic PESs and couplings.

\subsection{Diabatic Hamiltonian: Nonadiabatic and Spin-Orbit Couplings}

\noindent
Since the spin-orbit (SO) coupling has crucial role in a chemical reaction, such coupling should be incorporated as perturbation in total molecular Hamiltonian. At first, the molecular wavefunction is written in terms of Born-Oppenheimer Huang expansion (Eq. \ref{eq:bo-expansion}) for non-relativistic Hamiltonian and then, the kinetically coupled nuclear SE is formulated to generate the diabatic Hamiltonian through ADT formalism. Once such forms of the equations are available, the SO couplings originated from the adiabatic basis are incorporated as perturbation prior to transformation to the diabatic basis.

\vspace{0.3 cm}

\noindent
The total molecular Hamiltonian, $ \hat{H}(\mathbf{r}, \mathbf{R}) $ can be partitioned into nuclear kinetic energy operator $ \hat{T}_{\text{nuc}}(\mathbf{R}) $, the electronic Hamiltonian $ \hat{H}_{\text{el}}(\mathbf{r}; \mathbf{R}) $ and the spin-orbit Hamiltonian $ \hat{H}_{\text{SO}}(\mathbf{r}; \mathbf{R}) $, as follows:

\begin{eqnarray}
	\hat{H}(\textbf{r},\textbf{R}) = \hat{H}_{\rm{0}} + \hat{H}_{\rm{SO}}(\textbf{r};\textbf{R})
	\label{eq:molham},
\end{eqnarray} 
where
\begin{eqnarray}
	\hat{H}_{\rm{0}} &=& \hat{T}_{\rm nuc}(\textbf{R}) + \hat{H}_{\rm{el}}(\textbf{r};\textbf{R})
	\label{eq:molham_1}.
\end{eqnarray} 

\vspace{0.3cm}

\begin{eqnarray}
	-\frac{\hbar^2}{2}\left(\boldsymbol{\nabla}_R+\boldsymbol{\tau}\right)^2\psi^{\rm{ad}} + \left(U + H^{ad}_{\rm{SO}}  -E\right)\psi^{\rm{ad}} 
	= 0,
	\label{eq:adsem1}
\end{eqnarray}
where $H^{ad}_{\rm{SO}} $ indicates the SO coupling matrix appearing from the  adiabatic basis (\{$\xi_i$\}) and the rest of the terms have been defined in Subsection \ref{ssec:adia_th}.

\vspace{0.3cm}

\noindent
Since the NACM elements in Eq. \ref{eq:adsem1} is expected to exhibit singularity around the close vicinity of degenerate point(s) in the nuclear configuration space (CS), ADT is necessary to be performed to calculate the diabatic Hamiltonian in presence of SO coupling so that the potentially coupled SE converts into smooth and continuous functions of nuclear coordinates. While carrying out such transformation of nuclear SE, the example of F + H$_2$ reactive system showing both the effects (nonadiabatic and SO couplings) is briefly demonstrated, where three electronic states (1$^2$A$^\prime$, 2$^2$A$^\prime$ and 1$^2$A$^{\prime\prime}$) form a sub-Hilbert space (SHS) at low collision energies.\cite{FH2_Surf} For the two possible spin projections ($M_s = \pm \tfrac{1}{2}$) of the F ($^2P_{1/2}$ and $^2P_{3/2}$) atom, the corresponding adiabatic states and their SO coupled counterparts are defined as $|\xi_1\rangle$, $|\xi_2\rangle$, $|\xi_3\rangle$, $|\overline{\xi_1}\rangle$, $|\overline{\xi_2}\rangle$ and $|\overline{\xi_3}\rangle$. The $6 \times 6$ matrix representations of the adiabatic potential energy, $U_{ij} = u_i \delta_{ij}$ and the skew-symmetric nonadiabatic coupling, $\boldsymbol{\tau}_{ij} = \langle \xi_i | \boldsymbol{\nabla}_R \xi_j \rangle_{\mathbf{r}} = -\boldsymbol{\tau}_{ji}$, are thoroughly discussed in Appendix \ref{ssec:so_coup}.

\vspace{0.3cm}

\noindent
On the other hand, the general form of the adiabatic SO (coupling) Hamiltonian matrix ($H^{ad}_{SO}$) can be expressed as:

\begin{eqnarray}
	H_{\rm so}^{\rm ad} = 
	\begin{array}{|c|cccccc|}
		• & |\xi_1\rangle & |\xi_2\rangle & |\xi_3\rangle & |\overline{\xi_1}\rangle & |\overline{\xi_2}\rangle & |\overline{\xi_3}\rangle \\ 
		\hline 
		|\xi_1\rangle & 0 & D & -iB & 0 & -iF_1 & F \\ 
		|\xi_2\rangle & D & 0 & -iG & iF_1 & 0 & -C \\ 
		|\xi_3\rangle & iB & iG & 0 & -F & C & 0 \\ 
		|\overline{\xi_1}\rangle & 0 & -iF_1 & -F & 0 & D & iB \\ 
		|\overline{\xi_2}\rangle & iF_1 & 0 & C & D & 0 & iG \\ 
		|\overline{\xi_3}\rangle & F & -C & 0 & -iB & -iG & 0 
		\label{so_full}
	\end{array},
	\label{eq:so_1}
\end{eqnarray}

\vspace{0.3cm}

\noindent
where $D$, $B$, $G$, $F_1$, $F$ and $C$ specify the real valued functions of nuclear coordinates. The elements of the adiabatic PESs, nonadiabatic coupling and SO coupling matrices are originated\cite{jcp-153-174301-SA} from \textit{ab initio} calculations at the MRCI level, as implemented in the \textsc{MOLPRO} quantum chemistry package\cite{MOL}.\\


\noindent
The $6\times6$ SO coupling matrix (see Eq. \ref{eq:so_1}) can be written in terms of two non-interacting $3\times3$ sub-blocks leading to reduction of the effective dimensionality of the system to $3\times3$. In order to achieve such a block-diagonal structure, the time-reversal-adapted Kramers basis needs to be employed and the irreducible representations (IREPs) of each Kramers pair are determined under the $C_s$ double group. The basis elements associated with the same fermionic IREP are then assembled together~\cite{Yarkony_jpc100}. One can introduce the so-called $\lambda$ basis~\cite{Yarkony_jpc100,FH2_dynamics} by considering appropriate linear combinations of the original adiabatic electronic basis functions for two set of spin states:

\begin{subequations}
	\begin{eqnarray}
		|\xi_{1\lambda}\rangle = \frac{1}{\sqrt{2}} (|\xi_1 \rangle -\lambda | \overline{\xi_1} \rangle )\\
		|\xi_{2\lambda}\rangle = \frac{1}{\sqrt{2}} (|\xi_2 \rangle -\lambda | \overline{\xi_2} \rangle )\\
		|\xi_{3\lambda}\rangle = \frac{i}{\sqrt{2}} (|\xi_3 \rangle +\lambda | \overline{\xi_3} \rangle )
	\end{eqnarray}
\end{subequations}
where $\lambda=\pm1$. The above transformation can take the following matrix form: 
\begin{eqnarray}
	|\xi_{\lambda}\rangle = K_\lambda|\xi\rangle,
\end{eqnarray}
where the 6$\times$6  K$_\lambda$ matrix is depicted in Eq. \ref{eq:kmat} in Appendix \ref{ssec:so_coup}.

\vspace{0.3cm}

\noindent
The SO coupling  matrix (Eq. \ref{so_full}) under such transformation is as follows:

\begin{small}
	\begin{eqnarray}
		(K_\lambda^T)^{\dagger} H_{\rm so}^{\rm ad} K_\lambda^T = 
		\begin{pmatrix}
			0 & D+iF_1 & B+iF & 0 & 0 & 0 \\
			D-iF_1 & 0 & -iC+G & 0 & 0 & 0 \\
			B-iF & iC+G & 0 & 0 & 0 & 0 \\
			0 & 0 & 0 & 0 & D-iF_1 & B-iF \\
			0 & 0 & 0 & D+iF_1 & 0 & iC+G \\
			0 & 0 & 0 & B+iF & -iC+G & 0
		\end{pmatrix}.
		\label{eq:soblockdiag}
	\end{eqnarray}
\end{small}

\vspace{0.3cm}

\noindent
In order to reduce the computational complexity, the approach proposed by Werner and co-workers\cite{jcp-113-11084-WERNER} is followed, where the spin-orbit (SO) matrix elements are calculated exclusively for linear configurations. In these particular geometries, the parameters $B$, $C$ and $D$ become virtually zero (0), and $F_1$ reduced to $F$. \cite{FH2_dynamics} The resulting form of the $H_{\rm so}^{\rm ad}$ matrix and the explicit expression of the reduced block diagonal potential matrix is demonstrated in Appendix \ref{ssec:so_com}.

\vspace{0.3cm}

\noindent
It needs to be noted that the $K_\lambda$ matrix commutes with the kinetic part of the Hamiltonian (see Appendix \ref{ssec:so_coup}) and under such conditions, the diabatic representation takes the following form with the help of orthogonal transformation of the adiabatic SE $\psi^{\rm{ad}}=A\psi^{\rm{dia}}$ :\cite{BaerBook2006,Baer2002_PhysRep,baer_cpl35,irpc-38-287-SA,pccp-22-27496-SA,jctc-16-1666-SA}

\begin{eqnarray}
	\left(-\dfrac{\hbar^2}{2}\nabla_R^2 + \tilde{W}-E\right)\psi^{\rm{dia}} = 0,
	\label{eq:diasem1}
\end{eqnarray}
where $\psi^{\rm dia}$ signifies the diabatic nuclear wavefunction (as defined earlier) and $\tilde{W}$ represents the diabatic potential energy matrix\cite{BaerBook2006,Baer2002_PhysRep,baer_cpl35,irpc-38-287-SA,pccp-22-27496-SA,jctc-16-1666-SA} including SO coupling,

\begin{eqnarray}
	\tilde{W} = A^{\dagger}\tilde{U} A =  
	A^{\dagger}UA + 
	A^{\dagger}
	\begin{pmatrix}
		0 & iF & iF \\
		-iF & 0 & G \\
		-iF & G & 0
	\end{pmatrix} A.
	\label{eq:diapot}
\end{eqnarray}
\noindent
It is worth mentioning that validity of Eq. \ref{eq:diasem1} is guaranteed subject to the fulfillment of the ADT condition shown in Eq.~\ref{eq:adteq}.

\noindent
\section{Theoretical Background of Fully Close Coupled (FCC) Approach}	
\noindent
The atom-diatom (A + BC) reaction dynamics are carried out with the aid of 3D time-dependent wavepacket approach (TDWP)\cite{jcp-100-1085-BILL,cpc-184-270-SA} using the framework of Johnson's hyperspherical coordinates,~\cite{Johnson1980_JCP_73}
namely, hyperradius, $\rho$ and hyperangles, $\theta$ and $\phi$. Since 
the channels in hyperspherical coordinate system is even-handed, it is 
advantageous for analyzing the reactive and non-reactive processes occurring either on the ground or on the excited electronic states at the asymptotic hyperradius ($\rho \ge 15$~Bohr) ranging over the hyperangles, $0 \le \theta \le \frac{\pi}{2}$ and $ \frac{2\pi \cdot (I-1)}{3} \le \phi \le \frac{2\pi\cdot I}{3}$, $I = 1, 2$ and $3$, where $I$ symbolizes different channels. BBO based diabatic Hamiltonian (BBO-DH) vis-\'a-vis the quantum dynamics for reactive and non-reactive pathways of selected triatomic scattering systems are discussed in Subsections \ref{sec:Application-Surf} and \ref{sec:Application-Dyna}. \\

\noindent
If the wavepacket associated with the reactants is initialized (see Appendix \ref{ssec:init} and \ref{ssec:orth} for details) in channel 2 (reactive channel) of the ground electronic state, channels 1 and 3 represent the product ones. In order to maintain the clarity, the channel index is removed in all subsequent discussions.
The wavepacket can be expressed in hyperspherical coordinates using the 3\textit{j}-symbol and the modified spherical harmonics ($C_{j\mu}$) and such wavepacket can be written as:

\begin{eqnarray}
	\Phi^a_K&=&\sqrt{2\pi \ \sin\eta \ J(Rr\eta|\rho \theta \phi)(2l+1)}\phi_{vj}(r) \ \chi(R)(-1)^{j-l}\nonumber\\
	&\times& \sum_{\mu}
	\bigg(\begin{matrix}
		j&l&J\\
		\mu&0&-\mu
	\end{matrix} \bigg)
	C_{j\mu}(\eta)A_{K\mu},
	\label{eq:wp_hsp}
\end{eqnarray}
where, $\chi(R)$ signifies the Gaussian wave packet and all the remaining symbols are depicted elsewhere~\cite{jcp-147-074105-SA,cpc-184-270-SA}. \\


\noindent
The adiabatic Hamiltonian, $\hat{H}_a$ attains the following form in terms of Johnson's hyperspherical coordinates~\cite{jcp-79-1916-JOH} as follows:

\begin{eqnarray}
	\hat{H}_a&=&\Big[-\frac{\hbar^2}{2\mu_R}\frac{\partial^2}{\partial \rho^2}+\frac{2}{\mu_R \rho^2}\hat{L}^2(\theta, \phi)+
	\frac{\hat{J}^2-\hat{J}^2_z}{\mu_R \rho^2 \cos^2\theta}+\frac{\hat{J}^2_z-4\cos\theta \ \hat{J}_z\hat{P}_{\phi}}{2\mu_R 
		\rho^2 \sin^2\theta}\nonumber\\&+&\frac{\sin\theta}{\mu_R \rho^2\cos^2\theta}\frac{1}{2}[\hat{J}^2_++\hat{J}^2_-]
	+\Delta V(\rho, \theta)\Big]
	+\hat{V}_a(\rho, \theta, \phi),
	\label{eq:ham}
\end{eqnarray}
where $\hat{V}_a(\rho,\theta,\phi)$ represents the single surface adiabatic potential for the triatomic reactive systems, whereas the other operators are extensively demonstrated elsewhere.~\cite{cpc-184-270-SA,jcp-147-074105-SA}\\

\noindent
The wavepacket $\Phi^a_K$ [Eq.~(\ref{eq:wp_hsp})] and the Hamiltonian [Eq.~(\ref{eq:ham})] are substituted into the time-dependent Schr\"odinger equation (TDSE) to achieve the following set of coupled differential equations:

\begin{eqnarray}
	i\hbar\frac{\partial \Phi^a_K}{\partial t}&=&\bigg\{-\frac{\hbar^2}{2\mu_R}\frac{\partial^2}{\partial \rho^2}
	+\frac{2}{\mu_R \rho^2}\hat{L}^2(\theta,\phi)
	+\frac{\hbar K (\hbar K-4\cos\theta \ \hat{P}_{\phi})}{2\mu_R \rho^2 \ \sin^2\theta}\nonumber\\
	&+&\frac{\hbar^2[J(J+1)-K^2]} {\mu_R\rho^2\cos^2\theta}+V_{a}(\rho,\theta,\phi)+\Delta V(\rho,\theta)\bigg\}\Phi^a_K\nonumber\\
	&+&\frac{\sin\theta}{\mu_R \rho^2 \cos^2 \theta}[M_{K,K+2} \Phi^a_{K+2}+M_{K,K-2}\Phi^a_{K-2}],
	\label{eq:ham2_1}
\end{eqnarray}

\noindent
where the last term of Eq.~\ref{eq:ham2_1} couples $\Phi^a_K$ component and $\Phi^a_{K\pm2}$ through the coupling element $M_{K,K\pm2}$ as follows:
\begin{eqnarray}
	M_{K,K\pm 2}=\frac{\hbar^2}{2}\sqrt{(J \mp K)(J\pm K+1)(J\mp K-1)(J\pm K+2)}
	\label{eq:ham7}
\end{eqnarray}
and $\Delta V(\rho,\theta)$ signifies an additional potential term appearing from the kinetic energy part of the Hamiltonian while performing transformation from Jacobi to hyperspherical coordinates (see Appendix \ref{ssec:trans-JH} for details).\\

\noindent
Furthermore, the multi-state diabatic Hamiltonian ($\hat{H}_d$) for a triatomic reactive system takes the following form (see Appendix \ref{ssec:haml-coup} for details):
\begin{eqnarray}
	\hat{H}_d&=&\Big[-\frac{\hbar^2}{2\mu_R}\frac{\partial^2}{\partial \rho^2}+\frac{2}{\mu_R \rho^2}\hat{L}^2(\theta, \phi)+
	\frac{\hat{J}^2-\hat{J}^2_z}{\mu_R \rho^2 \cos^2\theta}+\frac{\hat{J}^2_z-4\cos\theta \ \hat{J}_z\hat{P}_{\phi}}{2\mu_R 
		\rho^2 \sin^2\theta}\nonumber\\&+&\frac{\sin\theta}{\mu_R \rho^2\cos^2\theta}\frac{1}{2}[\hat{J}^2_++\hat{J}^2_-]
	+\Delta V(\rho, \theta)\Big]\bf{\hat{I}}
	+\bf{\hat{V}_d}(\rho, \theta, \phi),
	\label{eq:ham3}
\end{eqnarray}
\noindent
where $\bf{\hat{I}}$ and $\bf{\hat{V}_d}(\rho, \theta, \phi)$ defines ($n\times n$) unit and diabatic potential energy matrices, respectively.\\ 

\noindent
The $K$-component adiabatic wave functions ($\Phi^a_{K}$) are transformed into the diabatic ones ($\mathbf{\Phi}^d_{K} = \mathbf{X}\mathbf{\Phi}^a_{K}$) for the different surfaces employing the transformation matrix $\mathbf{X}$, which is used to diagonalize the BBO-DH according to $\mathbf{\hat{V}_{a}} = \mathbf{X}^\dagger \mathbf{\hat{V}_{d}} \mathbf{X}$. As a result, the solutions of the TDSE in the diabatic representation, namely, $\Phi_{K,i}^d$ ($i = 1, 2, 3, \ldots, n$) can be written from the following set of coupled differential equations:

\begin{eqnarray}
	i\hbar\frac{\partial}{\partial t}
	\begin{bmatrix}
		\Phi_{K,1}^d \\ 
		\Phi_{K,2}^d \\ 
		\Phi_{K,3}^d  \\
		.\\
	\end{bmatrix} 
	&=&\bigg\{-\frac{\hbar^2}{2\mu_R}\frac{\partial^2}{\partial \rho^2}
	+\frac{2}{\mu_R \rho^2}\hat{L}^2(\theta,\phi)
	+\frac{\hbar K (\hbar K-4\cos\theta \ \hat{P}_{\phi})}{2\mu_R \rho^2 \ \sin^2\theta}\nonumber\\
	&+&\frac{\hbar^2[J(J+1)-K^2]} {\mu_R\rho^2\cos^2\theta}+\Delta V(\rho,\theta)\bigg\}
	\begin{bmatrix}
		\Phi_{K,1}^d \\ 
		\Phi_{K,2}^d \\ 
		\Phi_{K,3}^d \\
		.\\
	\end{bmatrix}\nonumber\\
	&+&
	\begin{bmatrix}
		V_{11}&V_{12}&V_{13}&.\\
		V_{21}&V_{22}&V_{23}&.\\
		V_{31}&V_{32}&V_{33}&.\\
		.&.&.&.\\
	\end{bmatrix}\begin{bmatrix}
		\Phi_{K,1}^d \\ 
		\Phi_{K,2}^d \\ 
		\Phi_{K,3}^d \\
		.\\
	\end{bmatrix}\nonumber\\
	&+&\frac{\sin\theta}{\mu_R \rho^2 \cos^2 \theta}\left( M_{K,K+2} 
	\begin{bmatrix}
		\Phi_{K+2,1}^d \\ 
		\Phi_{K+2,2}^d \\ 
		\Phi_{K+2,3}^d \\
		.\\
	\end{bmatrix}
	+ M_{K,K-2}
	\begin{bmatrix}
		\Phi_{K-2,1}^d \\ 
		\Phi_{K-2,2}^d \\ 
		\Phi_{K-2,3}^d \\
		.\\
		
	\end{bmatrix}\nonumber\right) .\\
	\label{eq:ham6_m}
\end{eqnarray}

\noindent
The other operators manifested in Eq.~\ref{eq:ham6_m} are demonstrated elsewhere~\cite{FH2_dynamics,HeH2+_dynamics_2023}. 

\vspace{0.2cm}

\subsection{Propagation of Wavefunction, Computation details and Absorbing potential}
\noindent
The Fast Fourier Transform (FFT) technique~\cite{jcomphy-52-35-DKRR} is widely used to formulate kinetic energy operator, which scales as $cN\log N$. It is to be noted that $N$ number of grid points are spanned over the $\rho$, $\theta$ and $\phi$ coordinates and such large number of grid points appears as the principal computational bottleneck in the propagation process. Such high computational cost can be significantly decreased by implementing parallel version of the FFT algorithm. On the other hand, while performing time propagation of the wavefunction, the iterative Lanczos reduction scheme~\cite{jcp-85-5870-TJPR} is adapted, 
where the recursive process is controlled by governing the magnitude of the wavefunction vectors. Furthermore, generation of the Lanczos vectors is also parallelized to improve efficiency. In order to minimize the numerical instabilities originating due to the singularities in the kinetic term of the Hamiltonian, the wavefunction is propagated using a sine transformation. While carrying out such process, the 
$\theta$ grid is extended from $\pi/2$ to $\pi$, where the amplitudes define as inverse mirror images of those in the interval $0$ to $\pi/2$ due to the antisymmetric nature about $\theta = \pi/2$. As a result, 
the wavefunction appears as zero (0) at $\theta = 0$, $\pi/2$, and $\pi$, which is consistent with the sine transformation technique~\cite{cpc-184-270-SA}. On the other hand, the steep increase in computational time with higher total angular momentum values ($J$) is substantially resolved by employing Open Multi-Processing (OpenMP) parallelization over the set of $K$-components corresponding to each specific $J$ during the propagation. 

\vspace{0.3 cm}

\noindent
In order to perform scattering calculations, the grid cannot be extended to infinity and in order to prevent unphysical reflections, the outgoing portion of the wavepacket must be removed from the grid immediately after projection before arriving at the boundary. Such goal can be achieved by using a negative imaginary potential, namely, $-i\hat{V}_{\mathrm{Im}}(\rho)$, which is placed at few points towards the end of the $\rho$ grid but far from the interaction region. Accordingly, the modified total potential takes the following form:

\begin{eqnarray}
	\hat{V}(\rho, \theta, \phi)=\hat{V}_0(\rho, \theta, \phi)-i\hat{V}_{Im}(\rho).
	\label{eq:impot1}
\end{eqnarray}

\noindent
It is worth mentioning that the chosen form of the imaginary potential is not unique, which largely depends on the condition that the outcomes of scattering calculation should become independent of the particular choice of the absorbing potential. One such form of the absorbing potential~\cite{jcp-90-4351-BAER,cp-191-247-MAR,FH2_dynamics} is expressed as:

\[
V_{Im}(\rho)= 
\begin{cases}
	V_{op}\cdot(\rho - \rho_{st}),& \rho_{st} \leq \rho\\
	0,              & \text{otherwise}.
\end{cases}
\]

\noindent
Here, $\rho_{st}$ symbolizes the starting point for such a potential. A thorough discussion on propagation of wavepackets are demonstrated in Appendix \ref{ssec:prop-co} and \ref{ssec:prop-ki}.

\subsection{Analysis of the wavepacket}
\noindent
When the atom and the centre of mass of the diatom is largely separated, 
$R = R^{\ast}$,~\cite{jcp-100-1085-BILL,cpc-184-270-SA} and the amplitudes of the $K$-component diabatic wavefunction, $\Phi^d_{K,i}(\rho, \theta, \phi)$, are converted back into their adiabatic counterparts, $\Phi^a_{K,i}(\rho, \theta, \phi)$ [$\mathbf{\Phi}^a_{K}$ = $\mathbf{X^\dagger}\mathbf{\Phi}^d_{K}$]. Then, it is projected onto the asymptotic eigenstates at each time step. It is to be noted that the scattering amplitude corresponding to a particular channel, defined by the vibrational, rotational, and orbital quantum numbers $v^{\prime}$, $j^{\prime}$ and $l^{\prime}$,respectively, as follows:
%
\begin{eqnarray}
	u^J_{v^{\prime}j^{\prime}l^{\prime}}(R=R^\ast;t)&=&4R\int \ dr \ \int \ d\eta \ r \sin\eta 
	\ \rho^{-5/2} (\sin2\theta)^{-1/2} \phi_{v^{\prime}j^{\prime}}(r)\nonumber\\
	&\times&\sum_{K\mu^{\prime}}g_{j^{\prime}l^{\prime}\mu^{\prime}}
	A^{\ast}_{K\mu^{\prime}} C_{j^{\prime}\mu^{\prime}}(\eta)\Phi^a_K(\rho, \theta, \phi).
	\label{eq:proj1}
\end{eqnarray}

\noindent
Since the outgoing wavepacket is projected onto the asymptotic eigenstates at a fixed value of R (=R$^\ast$) in Jacobi coordinates, the amplitude of the K component waves ($ \Phi^a_{K,i}(\rho, \theta, \phi) $) are interpolated on the grid to calculate the value of the ($ \Phi^a_{K,i}(\rho, \theta, \phi) $) on the ($\rho$, $\theta$, $\phi$)-space at R=R$^\ast$.

\vspace{0.3cm}

\noindent
Once the wavepacket completely traverses the projection region and then, it is absorbed at the boundary, the scattering amplitudes are Fourier transformed from the time domain to the energy domain and we obtain:

\begin{eqnarray}
	b^J_{v^{\prime}j^{\prime}l^{\prime}}(E;R)=\frac{1}{\sqrt{2\pi}}\int 
	\ dt \ u^J_{v^{\prime}j^{\prime}l^{\prime}}(R;t)
	\exp(iEt/\hbar)\,,
	\label{eq:proj5}
\end{eqnarray}
which are expanded in terms of the incoming and outgoing waves as:
\begin{eqnarray}
	b^J_{v^{\prime}j^{\prime}l^{\prime}}=A^{in}_{v^{\prime}j^{\prime}l^{\prime}}k_{v^{\prime}j^{\prime}}R 
	\ h^-(k_{v^{\prime}j^{\prime}}R)+A^{out,J}_{v^{\prime}j^{\prime}l^{\prime}}k_{v^{\prime}j^{\prime}}R 
	\ h^+ (k_{v^{\prime}j^{\prime}}R),
	\label{eq:proj6}
\end{eqnarray}
where
\begin{eqnarray}
	h^{\pm} (k_{v^{\prime}j^{\prime}}R)=-n_l(k_{v^{\prime}j^{\prime}}R)\pm ij_l(k_{v^{\prime}j^{\prime}}R).
	\label{eq:proj7}
\end{eqnarray}
In the above expression, $j_l$ and $n_l$ represent spherical Bessel and Neumann functions, respectively. Since the projection onto the incoming component of the scattering amplitude must be zero at asymptotic time, the reaction probability corresponding to a transition from the initial state ($vjl$) to the product state ($v^{\prime}j^{\prime}l^{\prime}$) can be obtained as the ratio of the outgoing to the incoming flux for a given total angular momentum, $J$ as follows:

\begin{eqnarray}
	P^J_{v^{\prime}j^{\prime}l^{\prime}\leftarrow vjl}=\frac{F^J_{v^{\prime}j^{\prime}l^{\prime}}}{F_{vjl}},
	\label{eq:proj8_1}
\end{eqnarray}
where
\begin{eqnarray}
	F^J_{v^{\prime}j^{\prime}l^{\prime}}&=
	&\frac{1}{\mu_{out}}k_{v^{\prime}j^{\prime}}|A^{out,J}_{v^{\prime}j^{\prime}l^{\prime}}|^2,\\
	F_{vjl}&=&\frac{1}{\mu_{in}}k_{vj}|c^l_E|^2.
	\label{eq:proj9}
\end{eqnarray}
Here, $c^l_E$ defines the scattering amplitude of the initial wavepacket at energy $E$, while $k_{vj}$ and $k_{v^{\prime} j^{\prime}}$ demonstrate the wavevectors corresponding to the initial ($vj$) and final ($v^{\prime} j^{\prime}$) ro-vibrational states for each channel, respectively. On the other hand, reduced masses of the reactants and products are defined by $\mu_{\text{in}}$ and $\mu_{\text{out}}$, respectively. The weighting in the energy domain is then associated with the $k$ domain through the following relation:

\begin{eqnarray}
	|c^l_E|^2=\bigg(\frac{\mu_{in}}{\hbar k}\bigg)^2|c^l_k|^2,
	\label{eq:proj10}
\end{eqnarray}
where the expression for $|c^l_k|^2$ for a Gaussian wavepacket takes the following form:
\begin{eqnarray}
	|c^l_k|^2=\sqrt{2/\pi}\ \sigma \ \exp[-2\sigma^2(k-k_0)^2]
	\label{eq:proj11}
\end{eqnarray}
with $\sigma$ being the width of the Gaussian wavepacket.

\vspace{0.3 cm}

\noindent
Once one obtains the reaction probabilities (see Eq.~\ref{eq:proj8_1}), the ro-vibrationally resolved cross-sections are computed as functions of the collision energy by taking the summation of the state-to-state reaction probabilities, $P^J_{v^{\prime}j^{\prime}l^{\prime}\leftarrow vjl}$, over all values of $J$ up to the convergence limit, $J_{\text{max}}$.

\begin{eqnarray}
	\sigma^{v^{\prime}j^{\prime}}_{vj} (E_{col}) = \frac{\pi}{\kappa_{v j}^{2}(2j+1)}{\sum_{J = 0}^{J_{max}}{(2 J + 1)
			P_{v^{\prime} j^{\prime} \leftarrow v j}^{J}(E_{col})}}.
	\label{eq:state_resol_ics}
\end{eqnarray}
In a similar fashion, the total integral cross-sections (ICSs) for the various rearrangement channels are computed as functions of the collision energy by summing over the ro-vibrationally resolved reaction probabilities.

\begin{eqnarray}
	\sigma_{vj} (E_{col}) = \frac{\pi}{\kappa_{v j}^{2}(2j+1)} \sum_{j^\prime =
		0}^{j^{\prime}_{max}}\sum_{v^\prime =
		0}^{v^{\prime}_{max}}{\sum_{J = 0}^{J_{max}}{(2 J + 1)
			P_{v^{\prime} j^{\prime} \leftarrow v j}^{J}(E_{col})}}.
	\label{eq:ics}
\end{eqnarray}

\noindent
On the other hand, one can evaluate the state-to-state rate coefficients~\cite{cpl-254-341-AOIZ,jcp-91-1528-MILL} ($k_{v,j\rightarrow v^{\prime},j^{\prime}}(T)$) as follows:

\begin{eqnarray}
	k_{v,j\rightarrow v^{\prime},j^{\prime}}(T) = \left( \frac{8k_{B}T}{\pi \mu_{R}}
	\right)^{\frac{1}{2}} \frac{1}{(k_{B}T)^{2}} \times
	\int_{E_{0}}^{E_{max}} [\sigma_{vj}^{v^{\prime}j^{\prime}} (E_{col}) \exp\left[{-E_{col}/k_{B}T}\right] \, \, E_{col} \, {dE_{col}}. \, 
	\label{eq:rate_threshold}\\
	\nonumber
\end{eqnarray}

\noindent
A detailed discussion on the analysis procedure is presented in Appendix \ref{ssec:ana1} to \ref{ssec:ana3}.

\section{\textit{Ab Initio} Adiabatic and BBO based Diabatic Hamiltonian} \label{sec:Application-Surf}
\begin{figure}[!htp]
	\centering
	\includegraphics[width=0.8\linewidth, height=0.4\textheight]{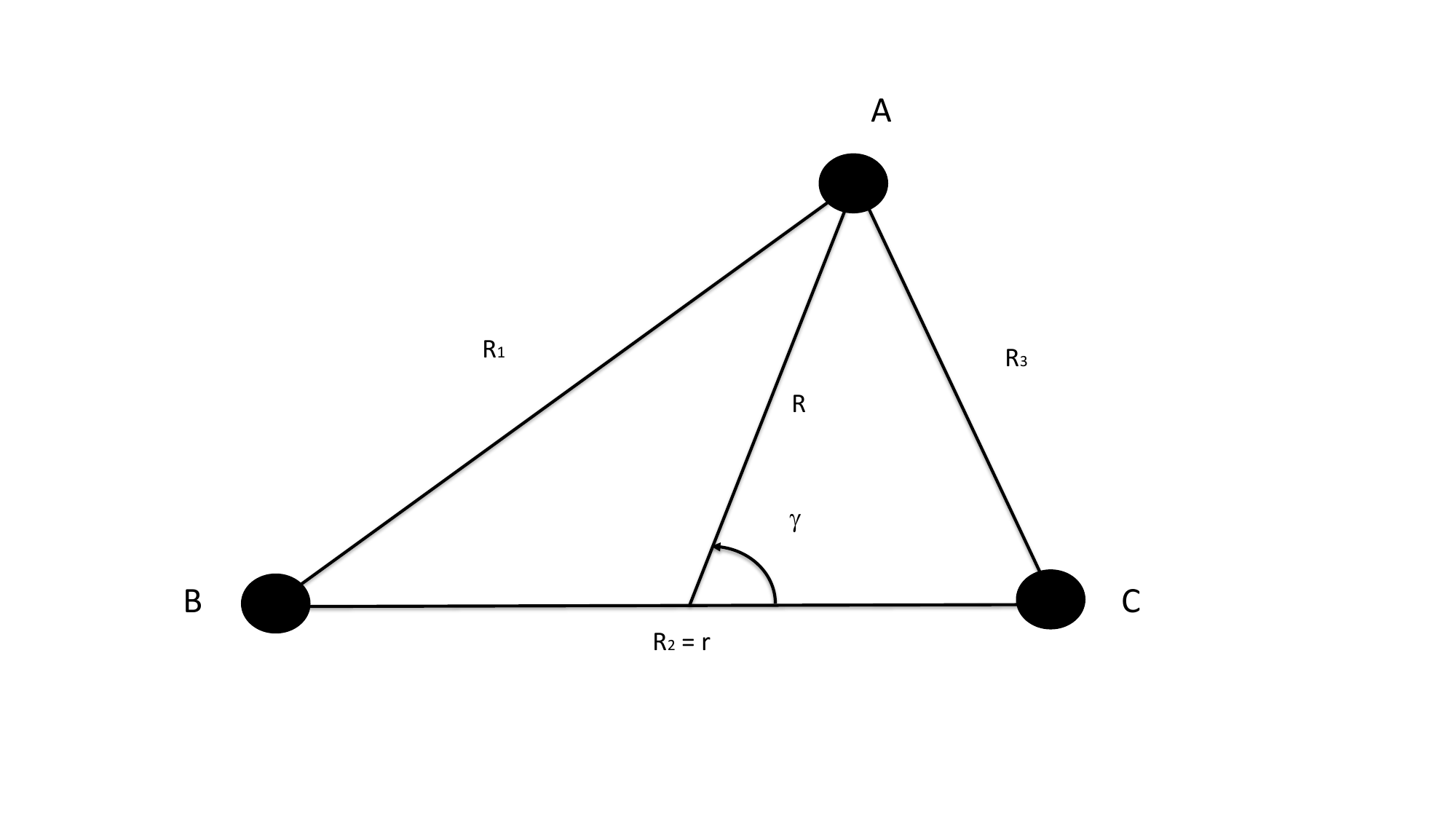}
	\caption{Schematic diagram of the Jacobi coordinates for the  triatomic system is presented, where the side lengths, R$_1$, R$_2$ and R$_3$ are defined in terms of hyperspherical coordinates. This figure is taken from Ref.~\citenum{Ghosh_H3_2025} and permission has been taken from American Chemical Society (ACS).}
	\label{fig:jacobi}
\end{figure}

\noindent
\textit{Ab initio} calculation in Jacobi coordinates is carried out to locate the conical intersection (symmetry-driven as well as accidental) among the electronic states for triatomic reactive species. On the contrary, the hyperspherical coordinates are used to generate \textit{ab initio} calculated adiabatic PESs and the NACTs for those systems (see Figure \ref{fig:jacobi}). Hyperradius $ \rho $ (0$ \le \rho \le $ 25 Bohr)
dictates the overall size of the triangle constituted by three atoms,  whereas the shape and orientation of the triangle are governed by the two hyperangles, $ \theta $ (0$ \le \theta \le$ 90$ ^\circ $) and $\phi $ (0$ \le \phi \le$ 360$ ^\circ $). Such coordinate system can evenhandedly describes all the reaction channels.

\noindent
While carrying out the \textit{ab initio} calculations, 
the inter-atomic distances ($R_1$, $R_2$ and $R_3$) 
of triatom can be expressed in terms of the three hyperspherical coordinates as follows,

\begin{subequations}
	\begin{eqnarray}
		R_1 &=& \dfrac{\rho}{\sqrt{2}}d_3\left[1+\sin\theta\cos(\phi+\epsilon_3)\right]^{1/2}, \\
		R_2 &=& \dfrac{\rho}{\sqrt{2}}d_1\left(1+\sin\theta\cos\phi\right)^{1/2}, \\
		R_3 &=& \dfrac{\rho}{\sqrt{2}}d_2\left[1+\sin\theta\cos(\phi-\epsilon_2)\right]^{1/2},
	\end{eqnarray}
	\label{eq:hyper}
\end{subequations}

\noindent
where $d_i = \sqrt{m_i(m_j+m_k)/\mu M}, \quad M = \sum_i^3 m_i, \quad \mu = \sqrt{(m_1m_2m_3/M)},   
\quad\epsilon_2 = 2\tan^{-1}(m_3/\mu)$ and $\epsilon_3 = 2\tan^{-1}(m_2/\mu)$. 

\subsection{FH$_2$}

\noindent
Stark and Werner (SW)~\citep{cpl-223-215-AOIZ,SW_jcp104,jcp-104-6531-CAST} first computed global \textit{ab initio} adiabatic and diabatic PESs of FH$_2$ system using multi-reference configuration interaction (MRCI) method with Davidson correction. Such PESs were later refined by incorporating spin-orbit (SO) interactions for describing asymptotic channel more accurately. In the subsequent works, SO coupling of the F atom was corrected in Hartke-Stark-Werner (HSW)\cite{HSW_cpl280}, Alexander-Stark-Werner (ASW)\cite{ASW_jcp113} as well as Li-Werner-Alexander (LWA)\cite{Werner_jcp127} PESs, where the ASW is a set of diabatic PESs originated from  6$\times$6 SO electronic basis set.\cite{ASW_jcp113} Indeed, accurate ground adiabatic PES was generated by Xu-Xie-Zhang (XXZ)\cite{cjcp-19-96-XU} using large scale MRCI calculation with the SO coupling effect. On the other hand, Chen-Sun-Zhang (CSZ)\cite{jcp-142-024303-ZHANG} also computed adiabatic PESs with the aid of spin-unrestricted coupled-cluster method [UCCSDT(2)$_\text{Q}$], whereas Li \textit{et al.}\cite{epjd-69-22-AJCV} introduced a double many-body expansion (DMBE) PES.

\vspace{0.3cm}

\begin{figure*}
	\centering
	\subfloat[\label{fig:tauth_prime_FH2}]{\includegraphics[width=0.5\textwidth]{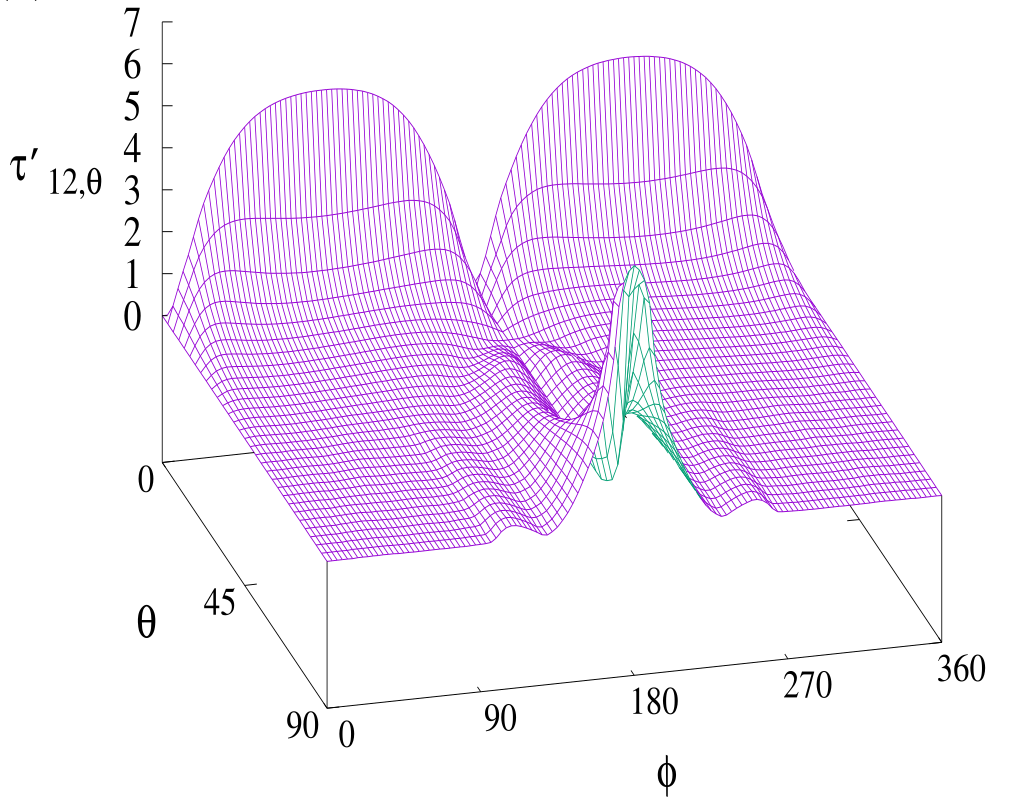}}
	\subfloat[\label{fig:tauph_prime_FH2}]{\includegraphics[width=0.5\textwidth]{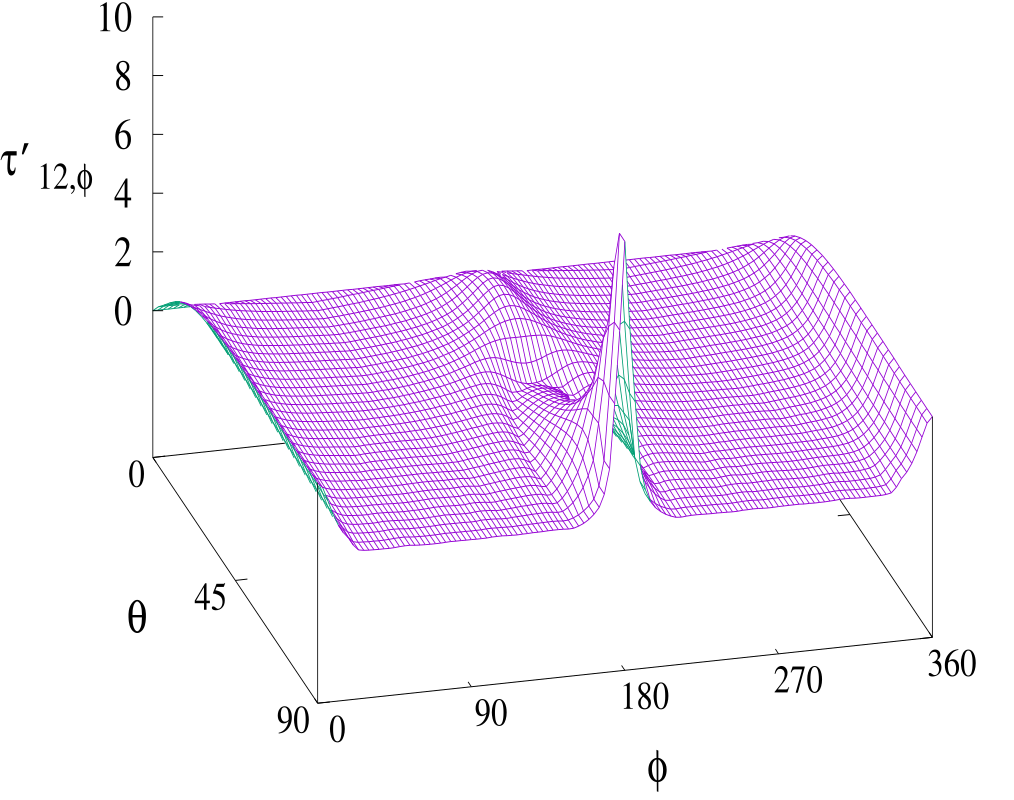}} 
	\caption{Functional forms of the swapped NACTs, (a) $\tau^{\prime}_{12,\theta}$ and (b) $\tau^{\prime}_{12,\phi}$ originated from the DDR methodology are presented as functions of the hyerangles $\theta$ and $\phi$ at $\rho=4.5$ Bohr. The figures are taken from Ref.~\citenum{FH2_Surf} with permission from American Institute of Physics (AIP).}
	\label{fig:tau_fh2}
\end{figure*}

\begin{figure}[!htp]
	\centering
	\includegraphics[width=0.6\textwidth , angle=0]{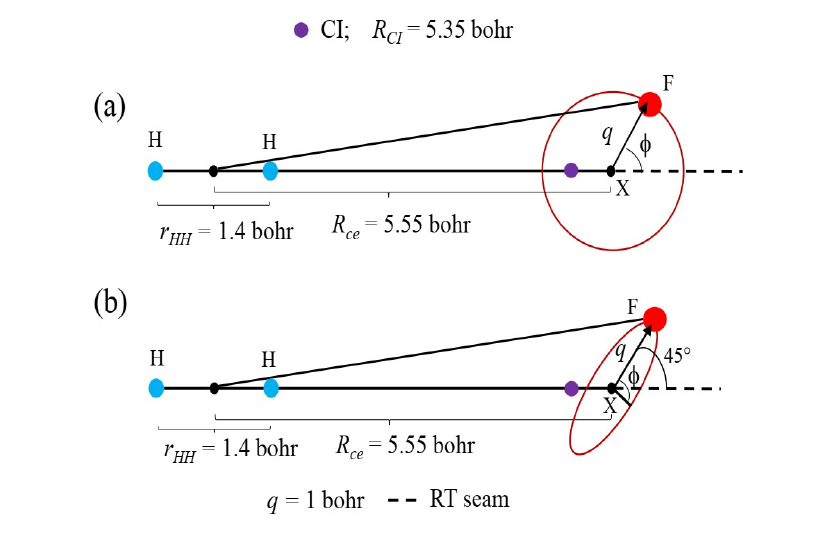}
	\caption{The diagrams display contour of radius, $q = 1.0$ Bohr
		traversed by the F atom taking X as center, which lies on the collinear 
		H$_2$ axis. In this arrangement, the magnitude of $r_{HH}$ (H-H distance), $R_{ce}$ (distance from the center of closed contour, X to the center of mass of H-H)
		and $R_{CI}$ (distance from the CI to the center of mass of H-H) appears as 1.401 Bohr, 5.55 Bohr and 
		5.35 Bohr, respectively. In upper diagram (a), the contour remains in the plane with the H-H-X axis, whereas in lower one (b), the loop is tilted by $45^{\circ}$ with respect to the same axis. The figures are taken from Ref.~\citenum{joy_pccp} with permission from the Royal Society of Chemistry (RSC).}
	\label{fig:fh2loop2}
\end{figure}

\noindent
On the other hand, Adhikari and co-workers carried out extensive \textit{ab initio} calculations to generate adiabatic PESs, nonadiabatic and SO couplings for the lowest three electronic states ($1^{2}A'$, $1^{2}A''$ and $2^{2}A'$) followed by construction of the diabatic Hamiltonian using BBO theory.~\cite{FH2_Surf} While performing such \textit{ab initio} electronic structure calculations, multi-configuration self-consistent field (MCSCF) approach was employed with an active space of (9e,10o) and Dunning's correlation-consistent quadruple-$\zeta$ (aug-cc-pVQZ or AVQZ) basis set. It is worth mentioning that the active space is constituted with 6$\sigma$ and 2$\pi$ orbitals, which represents 2$s$, 3$s$, 2$p$ and 3$p$ orbitals of F atom, and 1$\sigma_g$ and 1$\sigma_u$ orbitals of H$_2$. The adiabatic PESs were further calculated at the MRCI+Q level to incorporate dynamic correlation and reproduce the ground-state collinear barrier height~\cite{SW_jcp104}. In addition, SO  coupling matrix elements for the six spin components associated with the three electronic states were computed employing Breit-Pauli operator as implemented in \textsc{MOLPRO}. On the other hand, numerical finite-difference (DDR) approach was used to compute the NACTs among those lowest three electronic states with the same (9e,10o) active space and basis set. For
comparison, few \textit{ab initio} points were evaluated using analytic gradient method, like state-averaged coupled-perturbed MCSCF (CP-MCSCF) employing the 6-311++G** basis set with the same active space. Figure ~\ref{fig:tau_fh2} depicts $\theta$ and $\phi$ components of NACTs as functions of hyperangles ($\theta$,$\phi$) for a fixed $\rho$ = 4.5 Bohr. JT CI between the ground state (1$^2A'$) and the first excited state (2$^2A'$) as well as additional RT CIs/seams between the first (2$^2A'$) and the second excited state (1$^2A''$) are explored.\cite{ijqc-112-2561-ANITA} Figure \ref{fig:fh2loop2} depicts
location of JT and RT CIs through closed in-plane and out-of-plane contour integrations, respectively, around the linear F + H$_2$ configuration in Jacobi coordinate. In the planar loop configuration (see Figure ~\ref{fig:fh2loop2} (a)), the minimum-energy conical intersection (MECI) between the $1^{2}A'$ and $2^{2}A'$ states is present
 5.35 Bohr apart from the H$_2$ center of mass, whereas the RT CI seam between the $2^{2}A'$ and $1^{2}A''$ states is explored with the tilted loop as shown in Figure \ref{fig:fh2loop2} (b).

\vspace{0.3 cm}

\begin{figure}[!htp]
	\centering
	\subfloat[\label{fig:diab11_FH2}]{\includegraphics[width=0.5\textwidth]{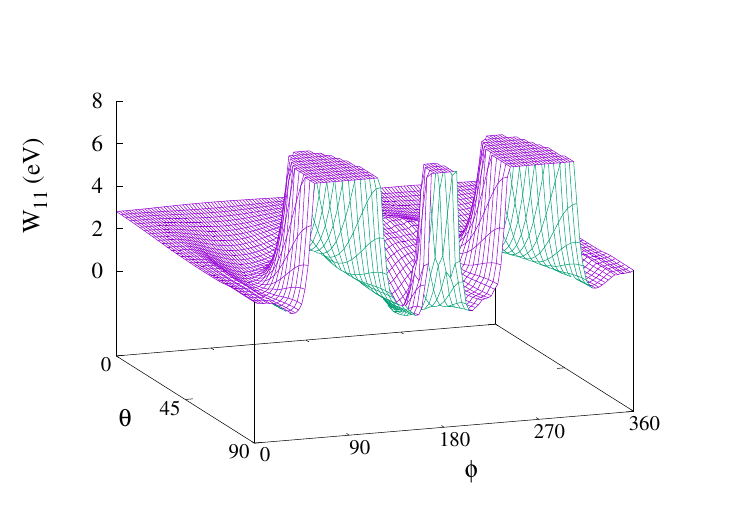}}
	\subfloat[\label{fig:diab22_FH2}]{\includegraphics[width=0.5\textwidth]{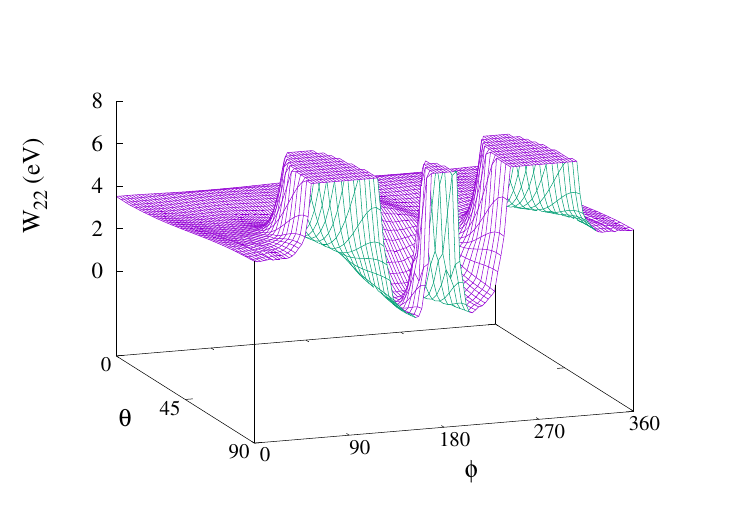}}\\
	\subfloat[\label{fig:diab12_FH2}]{\includegraphics[width=0.5\textwidth]{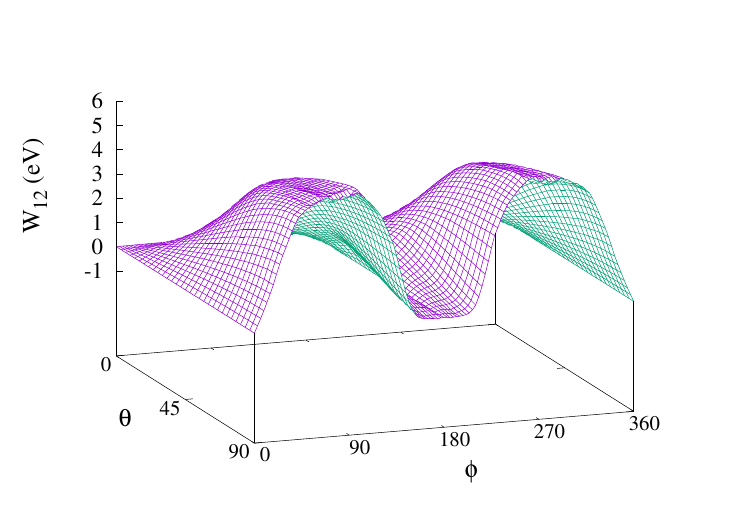}}	
	\caption{The diabatic PES matrix elements (a): $W_{11}$, (b): $W_{22}$ and (c) $W_{12}$ are plotted as functions 
		of $\theta$ and $\phi$ for $\rho=4.5$ Bohr. The figures are adapted from Ref. \citenum{FH2_Surf} with the permission of AIP.}
	\label{fig:diabsurf_fh2}
\end{figure}

\begin{figure*}
	\centering
	\includegraphics[width=1.0\textwidth]{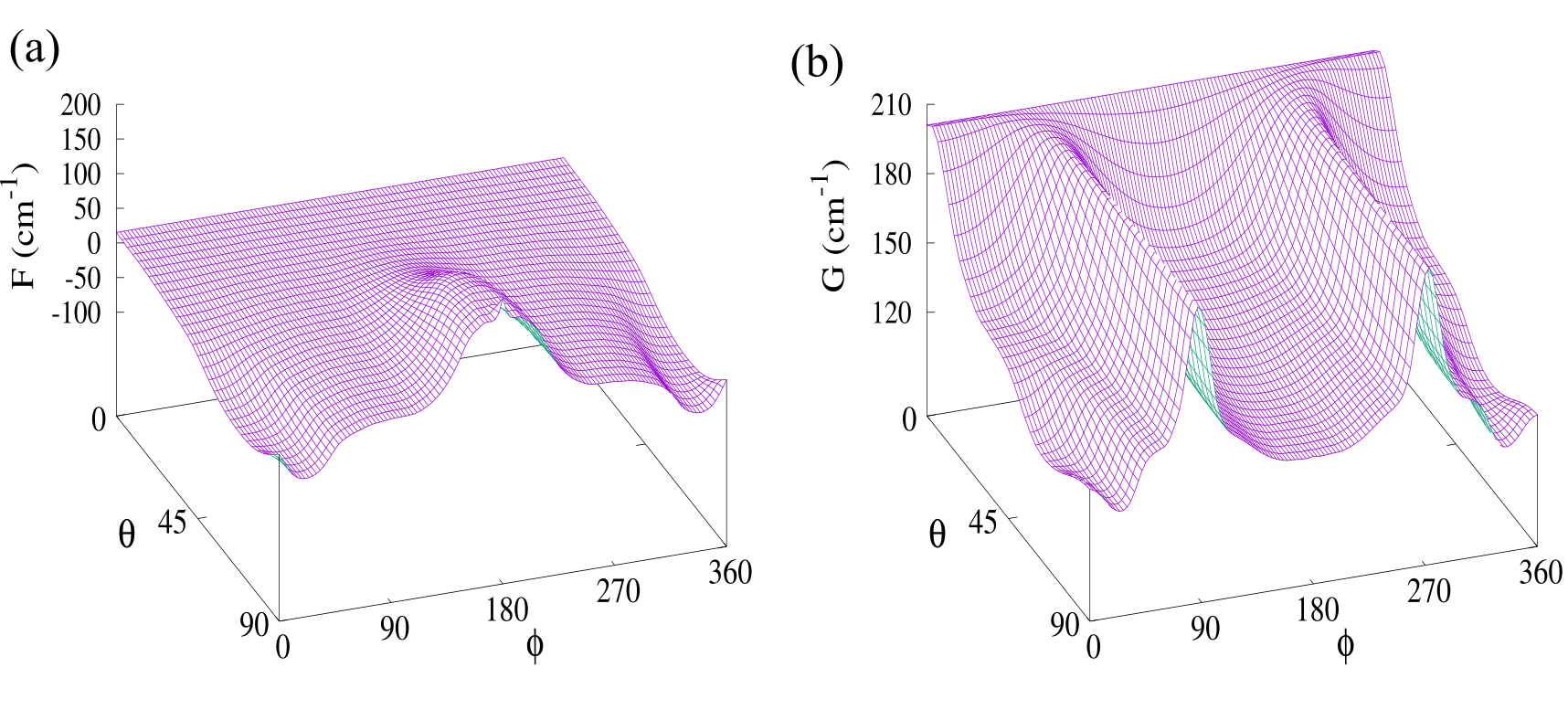} 
	\caption{SO matrix elements, (a) F and (b) G are presented as functions of hyperangle for $\rho$ = 4.5 Bohr. The figures are taken from Ref.~\citenum{FH2_Surf} with the permission of AIP.}
	\label{fig:fg_fh2}
\end{figure*}

\noindent
Once the \textit{ab initio} NACTs are in hand, the topologically correct mixing angles among the electronic states were calculated using
the ADT equations for three coupled electronic manifold (see Eq. \ref{eq:adt-three}). Then, adiabatic PESs and NACTs were diabatized to generate the global diabatic Hamiltonian matrix in hyperspherical coordinates. The diabatic PES matrices without SO coupling are depicted
in Figure \ref{fig:diabsurf_fh2}, where the matrix elements are smooth, single-valued and continuous functions of the nuclear coordinates. In addition, in order to get more complete and accurate description of the F + H$_2$ system, the SO coupling are also included in diabatic PES matrices.\cite{FH2_Surf} The elements of SO coupling matrix, F and G are originated from the same adiabatic basis along the linear geometry in Jacobi coordinates, which are interpolated in hyperspherical (coordinates) as depicted in Figure \ref{fig:fg_fh2}.\cite{FH2_Surf}

\subsection{H$_3^+$}

\begin{figure}[!htp]
	\centering
	\subfloat[\label{fig:nact_hyp_phi12_H3+}]{\includegraphics[width=.4\textwidth,angle=270]{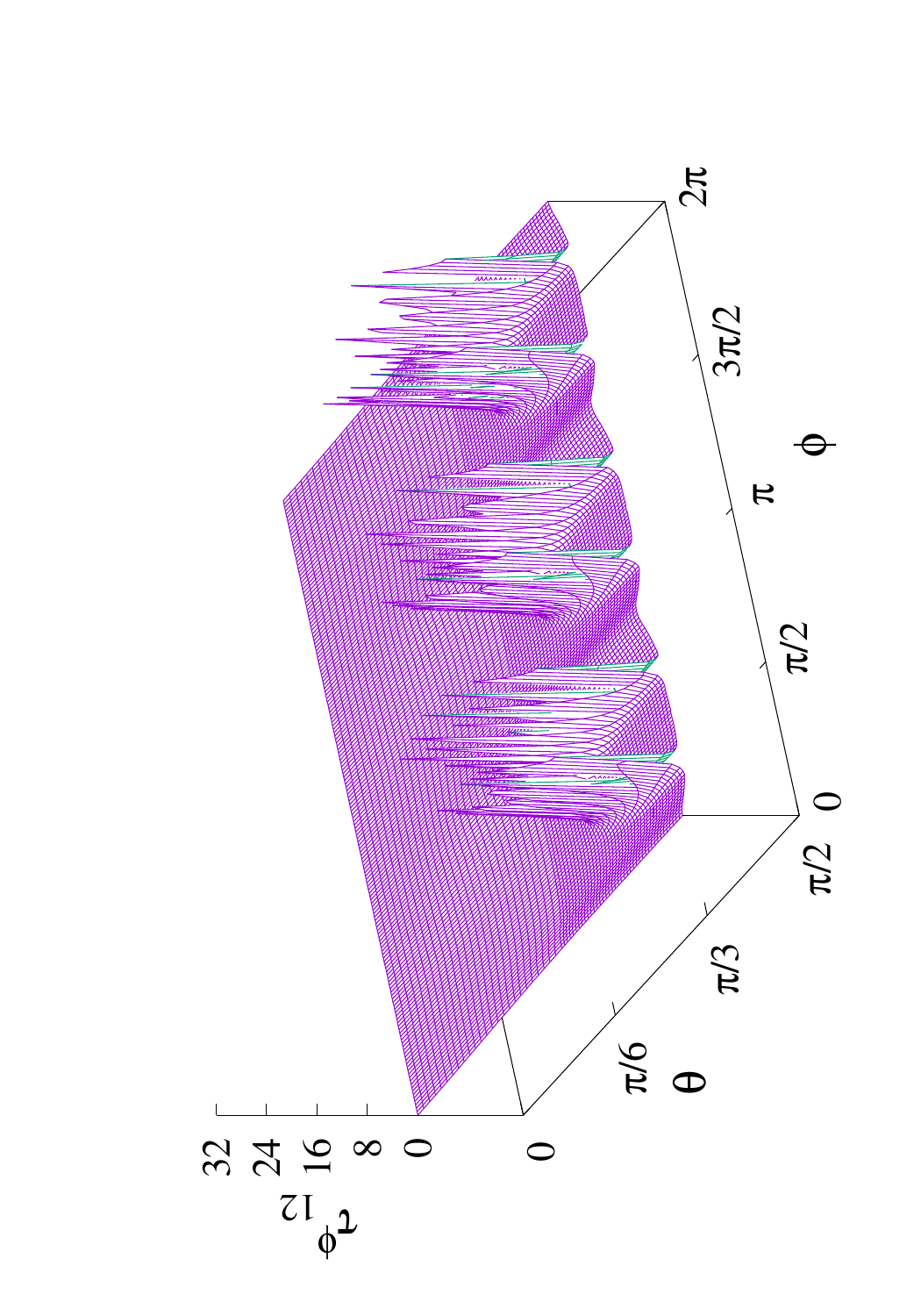}}\hspace*{-2 cm}
	\subfloat[\label{fig:nact_hyp_theta12_H3+}]{\includegraphics[width=.4\textwidth,angle=270]{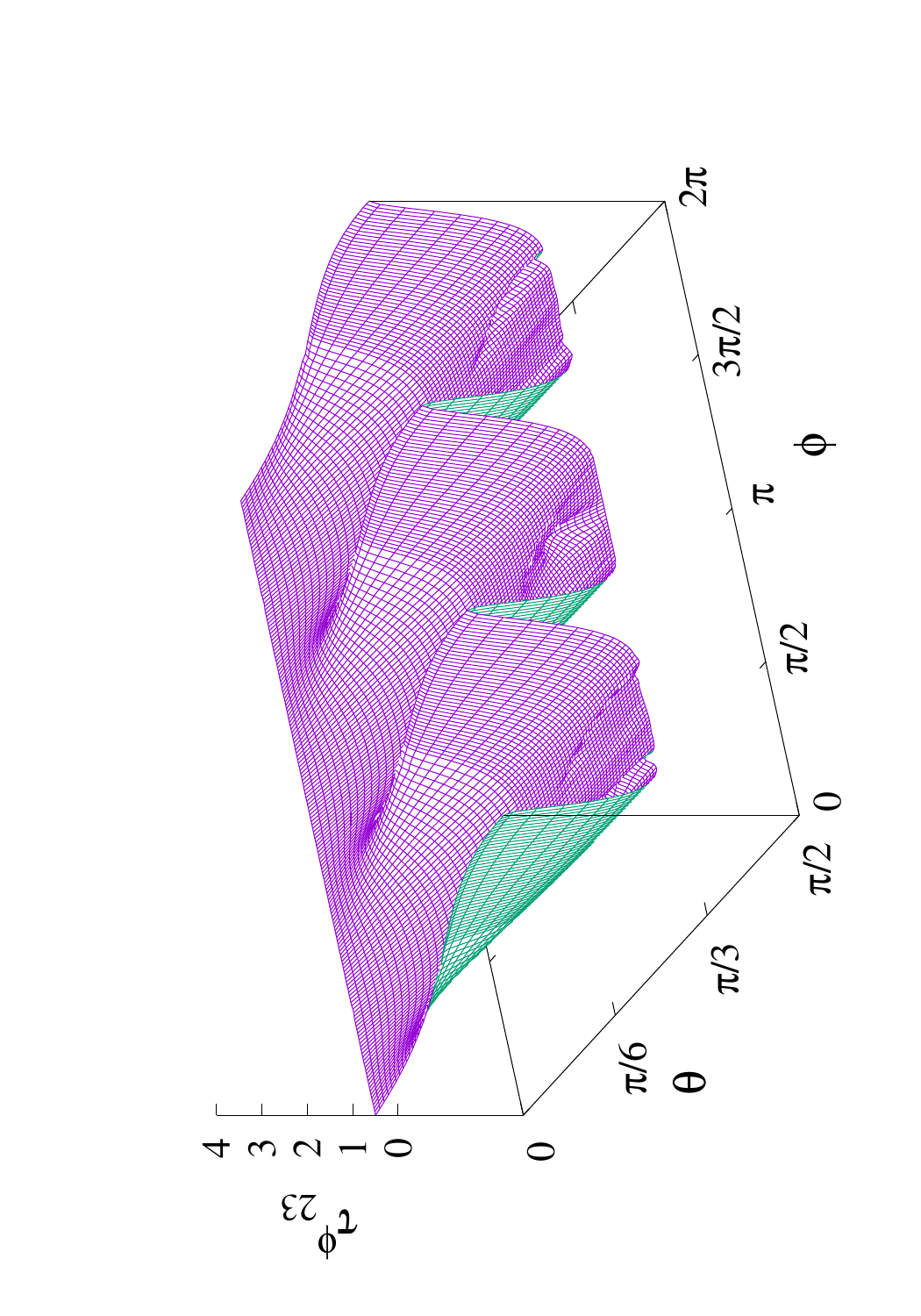}}\\
	\caption{$\phi$ components of NACTs for  H$ _3^+ $ obtained from the DDR methodology are presented in hyperspherical coordinates, namely, 
		(a) $ \tau^{\phi}_{12}$, (b) $ \tau^{\phi}_{23}$ over $ \theta $-$ \phi $ configuration space for a fixed value of $ \rho $ = 10.0 Bohr. The figures are taken from Ref.~\citenum{Scat_Review} with the permission of Wiley.}
	\label{fig:nact_hyp_h3+}
\end{figure}

\begin{figure}[!htp]
	\centering
	\includegraphics[width=.45\textwidth,angle=270]{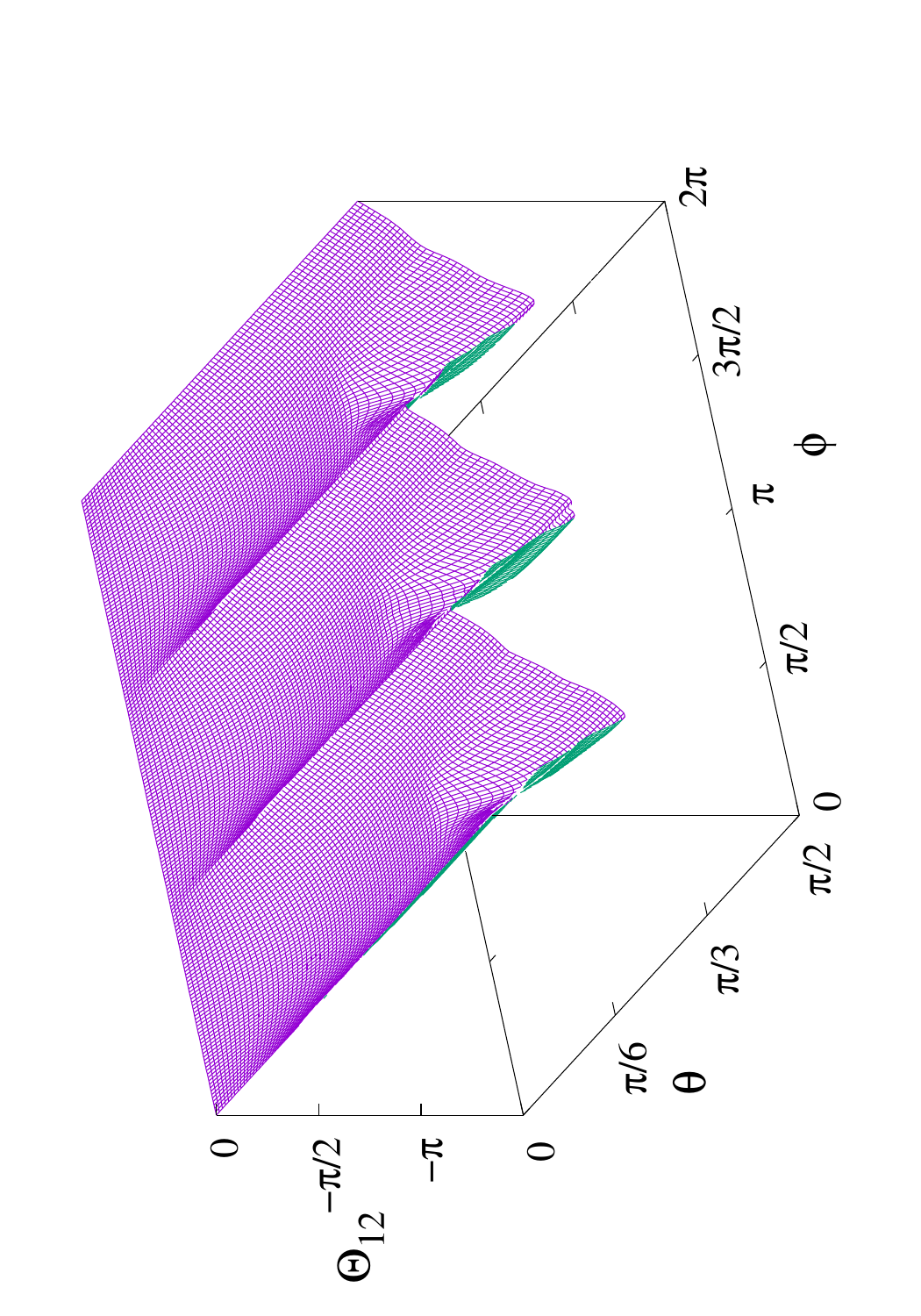}\hfill
	\caption{ Variation of ($\Theta_{12}$) over the $\theta-\phi$ CS is depicted as functions of hyperspherical coordinates at a fixed hyperradius $ \rho $ = 10.0 Bohr for the  H$ _3^+ $ system. The data are adapted from Ref.~\citenum{mukherjee_jcp141}.}
	\label{fig:adt_angle_h3+}
\end{figure}

\begin{figure}[!htp]
	\centering
	\includegraphics[width=0.7\textwidth]{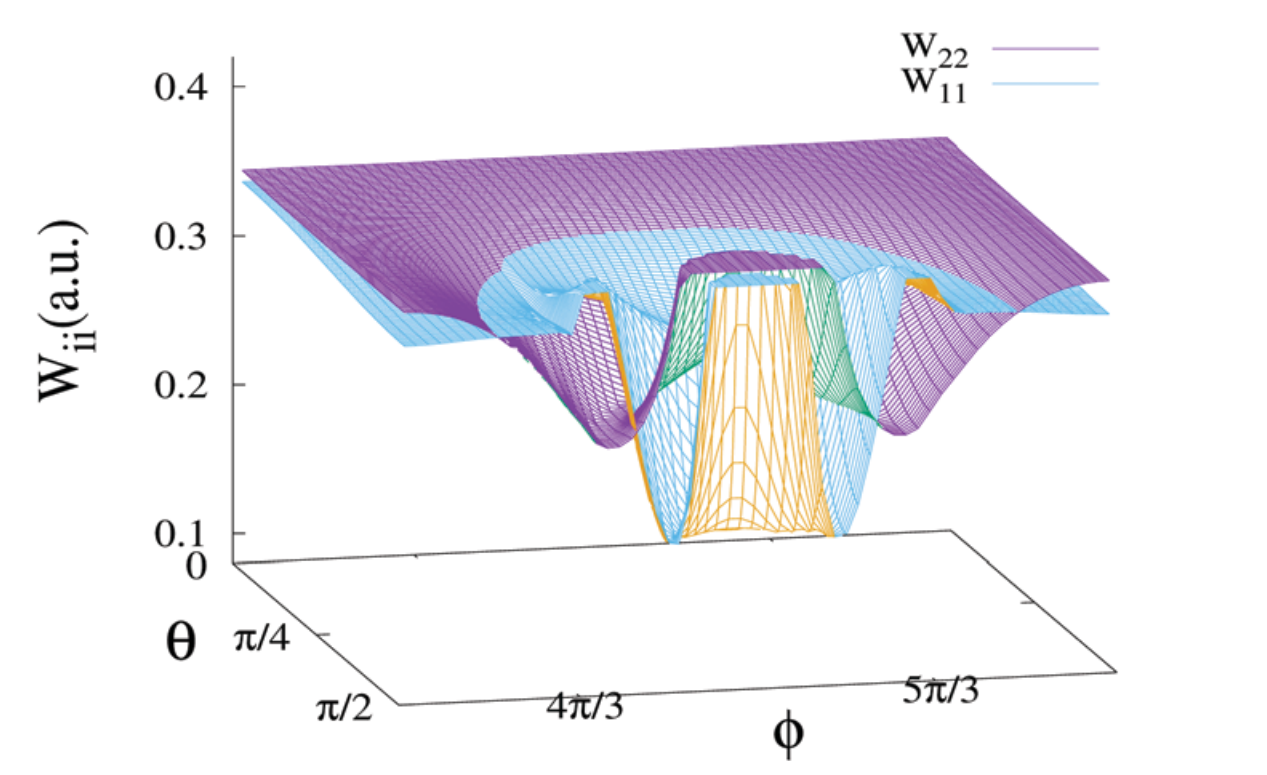}
	\caption{The diabatic PES matrix elements  $W_{11}$ and  $W_{22}$ are plotted in hyperspherical coordinates as functions of $\theta$ and $\phi$ at $\rho=10.0$ Bohr. The figures are reproduced from Ref.~\citenum{pccp-22-27496-SA} with permission from RSC.}
	\label{fig:diabsurf_h3}
\end{figure}

\noindent
In the realm of diabatization, several research groups~\cite{jcp-54-4297-TUL,Kamisaka_JCP_2002,jcp-126-074309-AJCV} took attempts to incorporate the nonadiabatic effects among the coupled (low lying singlet) electronic states semi-empirically. While constructing the BBO-theory-based diabatic Hamiltonian for this system, \textit{ab initio} calculations are performed for the lowest three singlet electronic states ($1^{1}A'$, $2^{1}A'$ and $3^{1}A'$) along with the NACTs among those states.~\cite{jcp-141-204306-SA} Such calculations for the three adiabatic PESs are executed in $C_{s}$ symmetry with the use of Dunning’s correlation-consistent quintuple-$\zeta$ (cc-pV5Z) basis set~\cite{dunning_basis}. First, MCSCF calculations are opted over the lowest six electronic states, considering two electrons distributed among ten active orbitals, followed by more accurate three state MRCI calculation. These surfaces are generated as functions of the hyperangles $\theta$ ($0 \le \theta \le 90^{\circ}$) and $\phi$ ($0 \le \phi \le 360^{\circ}$) over a grid of fixed hyperradii ($1.5 \le \rho \le 20$ Bohr). One can observe a semi-circular accidental CI seam of $C_{s}$ symmetry passing through a $C_{2v}$ geometry, which is located between the ground ($1^{1}A'$) and first excited ($2^{1}A'$) states of the H$_3^{+}$ system in the $\theta$-$\phi$ space across the hyperradius range of $5 \le \rho \le 12$ Bohr.~\cite{jcp-141-204306-SA} Figure 
~\ref{fig:nact_hyp_h3+} presents the $\phi$ components of the NACT between $1^{1}A'$-$2^{1}A'$ and $2^{1}A'$-$3^{1}A'$ states as functions of hyperangles ($\theta$ and $\phi$) at $\rho$ = 10 Bohr. Furthermore, 
symmetry driven $2^{1}A'$-$3^{1}A'$ D$_{3h}$ CI is observed in Jacobi coordinates by integrating the ADT equations along a circular path of radius $q = 0.8$ Bohr.\cite{joy_pccp} During the closed contour integration, $H-H$ bond length ($r_{HH}$) is taken as 2 Bohr, separation vector ($R$) from the contour center to the $H-H$ center of mass is fixed at 1.732 Bohr, and the angle $\gamma$ between $r_{HH}$ and $R$ appears as $90^\circ$. The \textit{ab initio} calculated NACT(s) are plugged into the ADT equations (see Eq. \ref{eq:adt-three})\cite{ADT_JCTC_16} and then those are integrated leading to the mixing angles between the electronic states. On the other hand, two-dimensional (2D) variation of ADT angle ($\Theta_{12}$) is plotted as function of $\theta-\phi$ in Figure \ref{fig:adt_angle_h3+}, where such angle attains the magnitude of $\pi$ bearing the signature of $1^{1}A'$-$2^{1}A'$ CI/seam. Finally, diabatic PES matrices are computed employing \textit{ab initio} calculated adiabatic PESs and ADT angles over the $\theta-\phi$ plane for each fixed $\rho$ values. Figure ~\ref{fig:diabsurf_h3} manifests diabatic PES(s) as functions of hyperangles ($\theta$, $\phi$) at $\rho$ = 10 Bohr. It is important to note that the global diabatic PES matrix for H$_3^+$ system is available in terms of hyperspherical coordinates elsewhere\cite{H3+_Surf,jcp-147-074105-SA}.

\subsection{HeH$_2^+$}

\begin{figure}[!htp]
	\centering
	\subfloat[\label{fig:nact_hyp_phi12}]{\includegraphics[width=.5\textwidth]{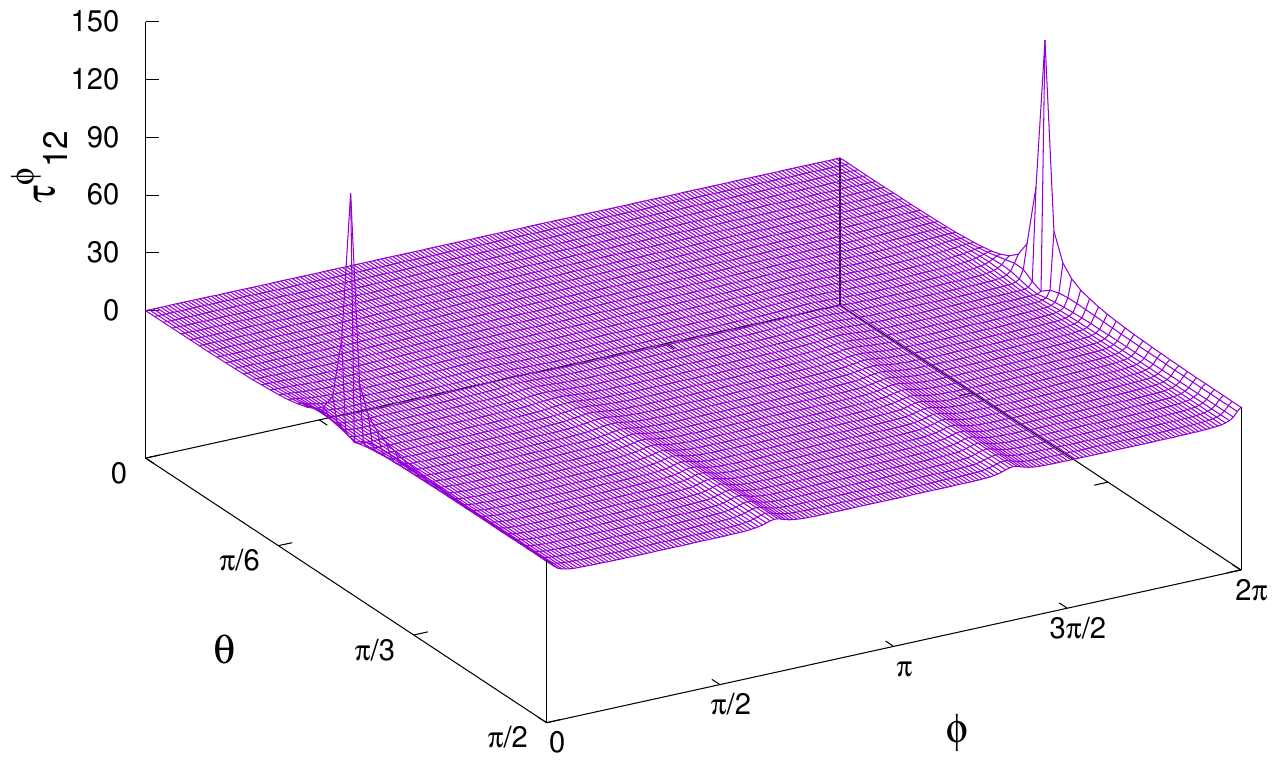}}\hfill
	\subfloat[\label{fig:nact_hyp_theta12}]{\includegraphics[width=.5\textwidth]{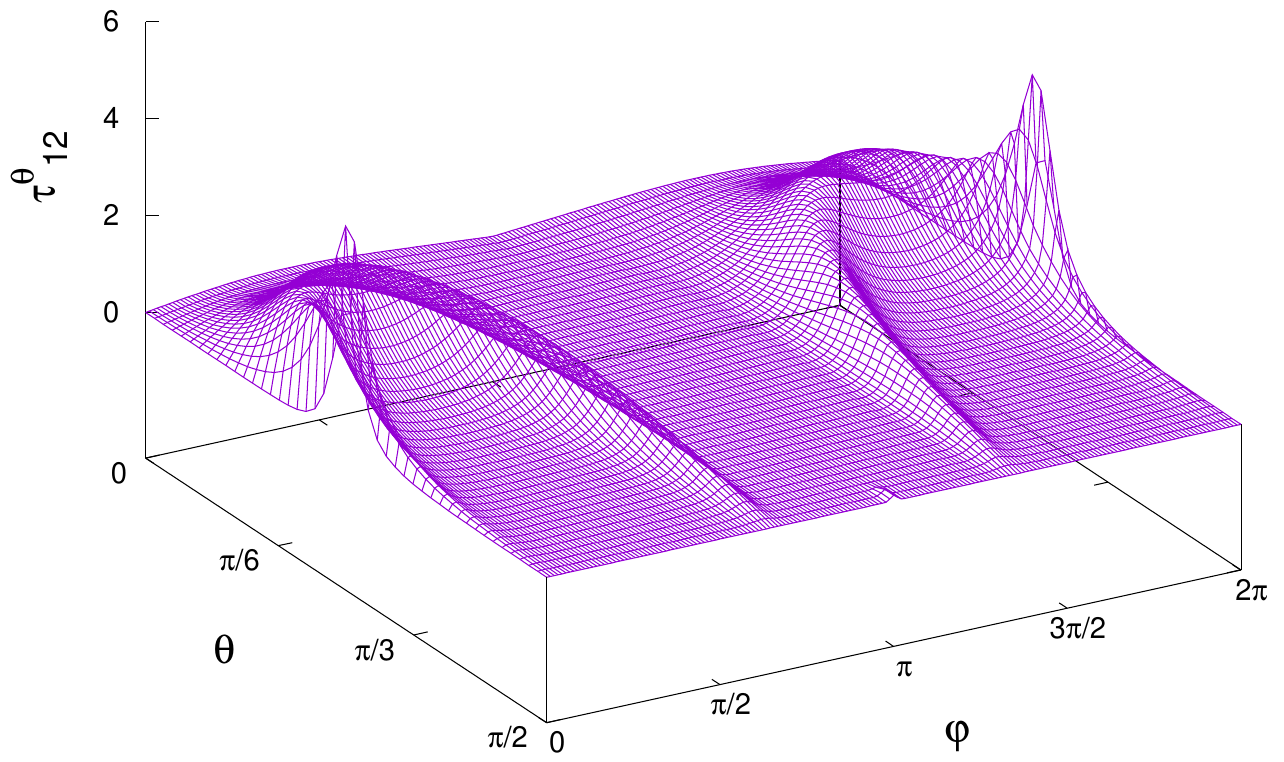}}\\
	\caption{$ \phi$  and $\theta$ components of NACTs for  He-H$ _2^+ $ system are presented as functions of hyperspherical coordinates:  
		(a) $ \tau^{\phi}_{12}$ and (b) $ \tau^{\theta}_{12}$ over $ \theta $-$ \phi $ configuration space for a fixed value of $ \rho $ = 10.0 Bohr. Such figures are taken from Ref.~\citenum{HeH2_Surf} with the permission from ACS.}
	\label{fig:nact_hyp_heh2+}
\end{figure}

\begin{figure}[!htp]
	\centering
	\hspace{0.7cm}\subfloat[\label{fig:12CI_fig_heh2+}]{\includegraphics[width=.43\textwidth,height=0.31\textwidth]{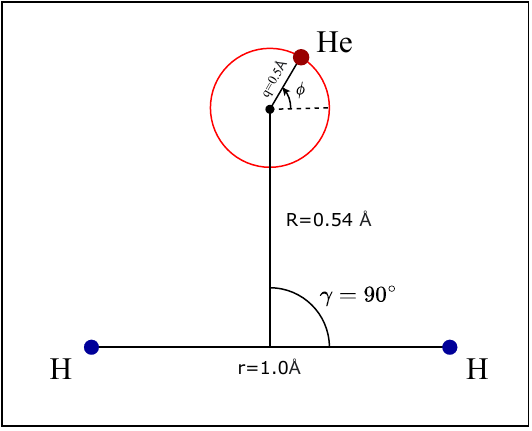}}\hfill
	\subfloat[\label{fig:12CI_angle_heh2+}]{\includegraphics[width=.47\textwidth]{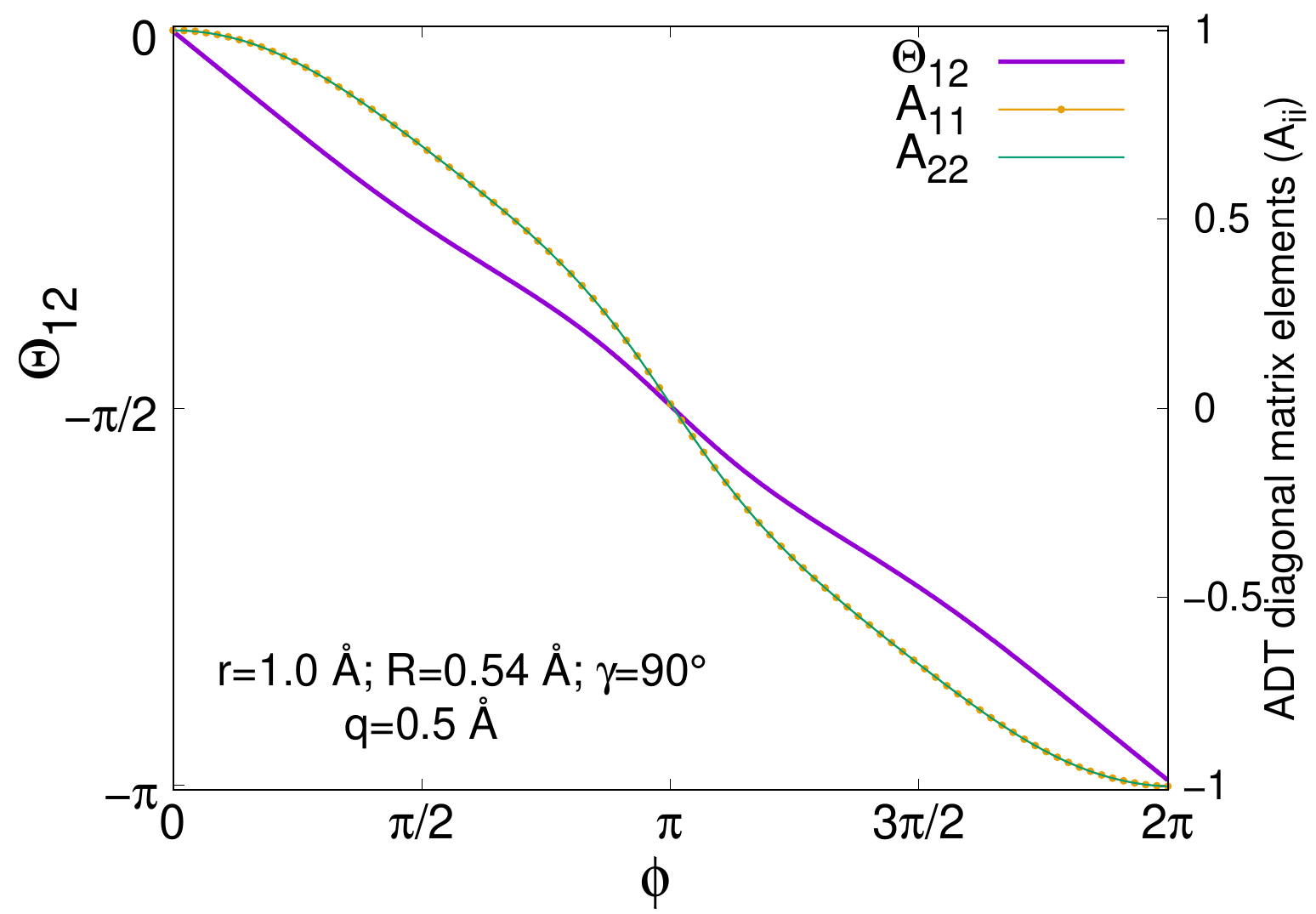}}
	\caption{(a) A schematic representation in Jacobi coordinates is portrayed for identifying the location of 1-2 CI in C$_{2v}$ configuration; (b) the ADT angle and the diagonal elements of the ADT matrix (A$_{11}$ and A$_{22}$) achieve values of $\pi$ and $-1$, respectively, upon close contour integration, thereby affirming the presence of 1-2 CI inside the contour. Such figures are adapted from Ref.~\citenum{HeH2_Surf} with the permission from ACS.} 		
	\label{fig:12CI_heh2+}
\end{figure}

\begin{figure}[!htp]
	\centering
	\subfloat[\label{fig:W11}]{\includegraphics[width=.5\textwidth]{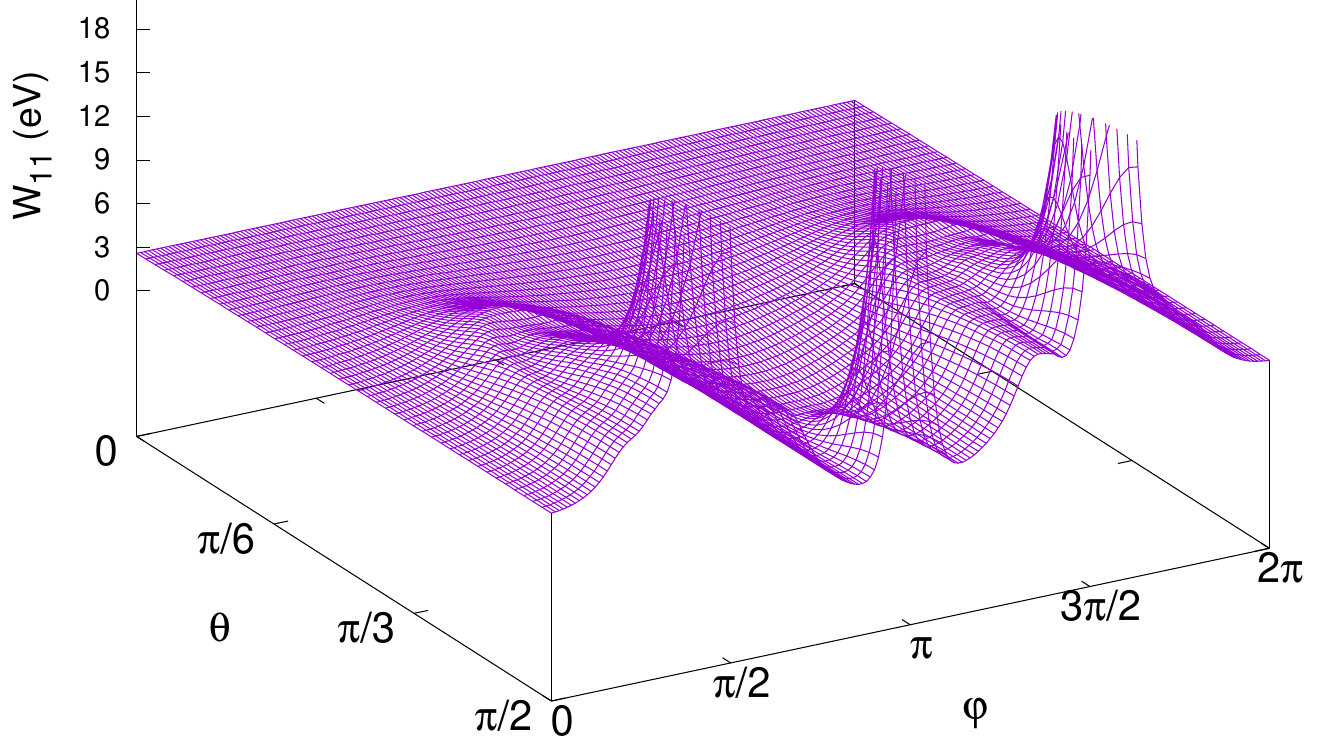}}\hfill
	\subfloat[\label{fig:W12}]{\includegraphics[width=.45\textwidth]{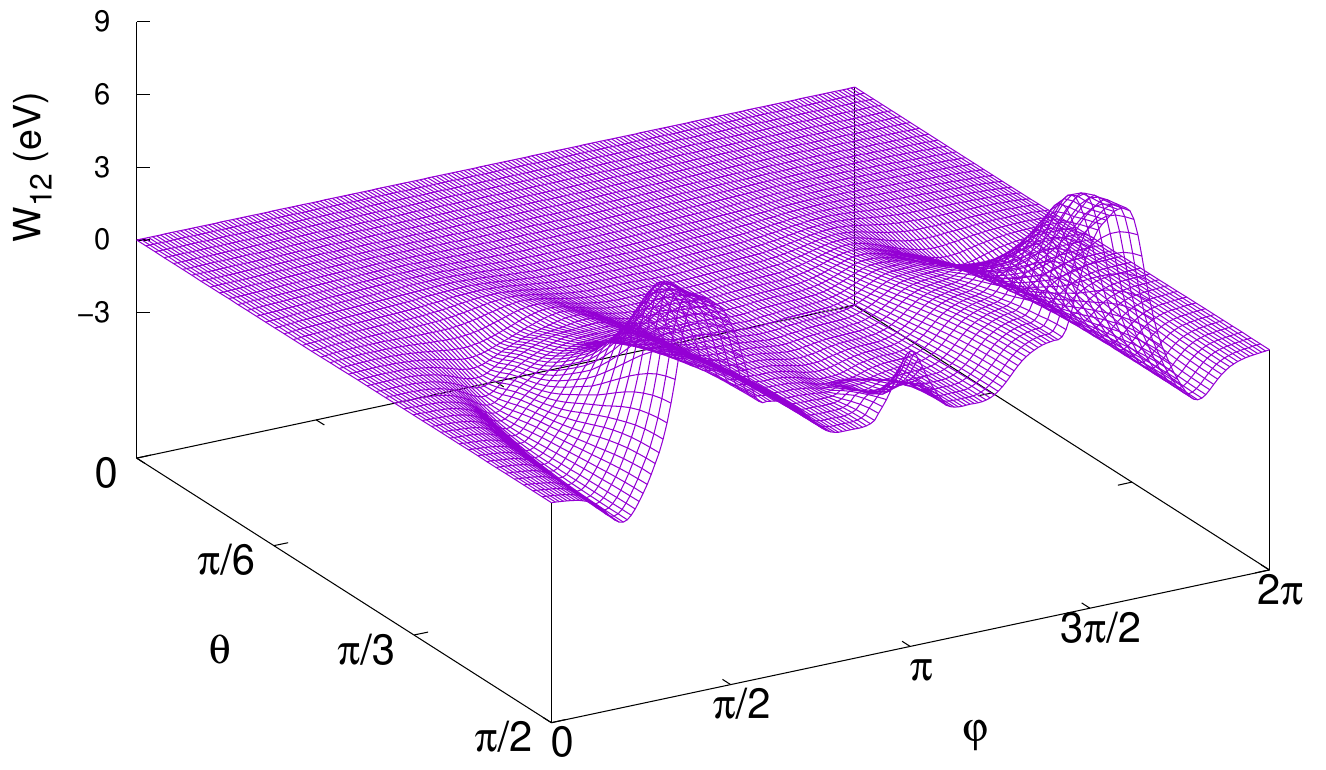}}\hfill
	\subfloat[\label{fig:W33}]{\includegraphics[width=.5\textwidth]{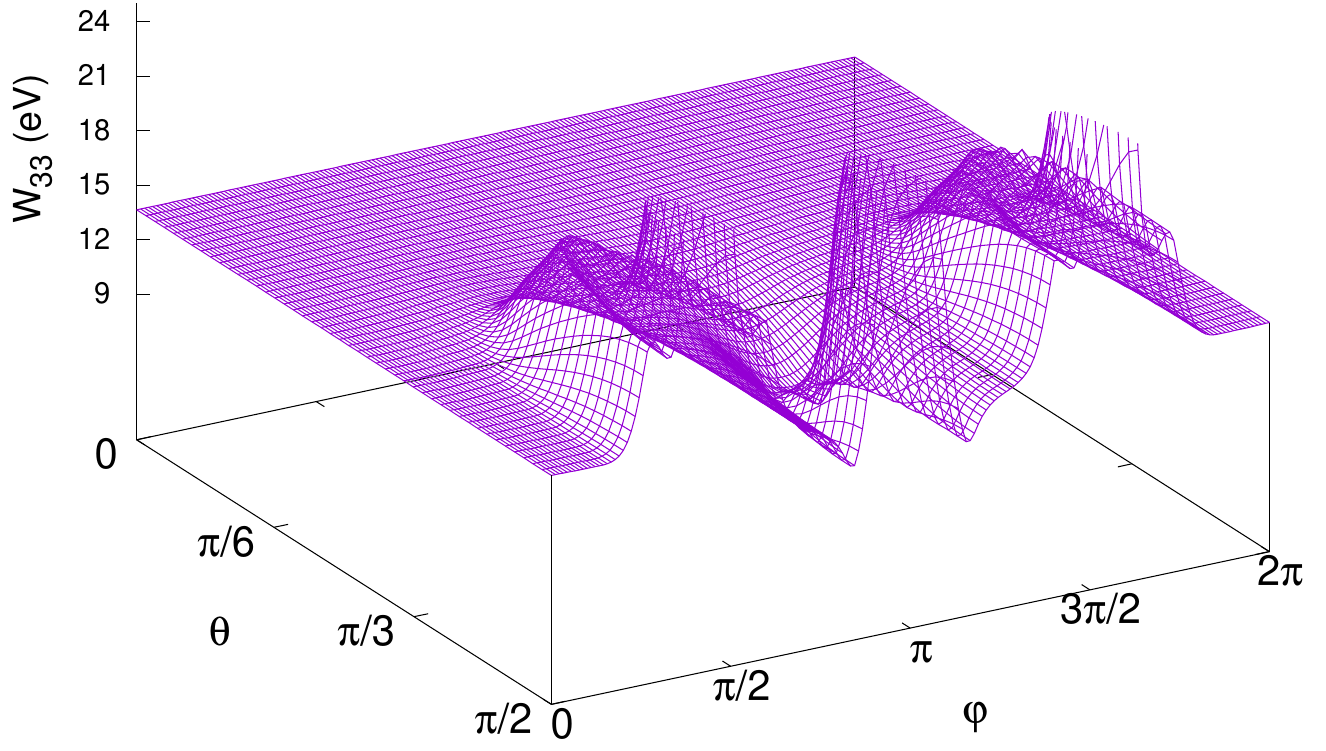}} \hfill
	\subfloat[\label{fig:W34}]{\includegraphics[width=.45\textwidth]{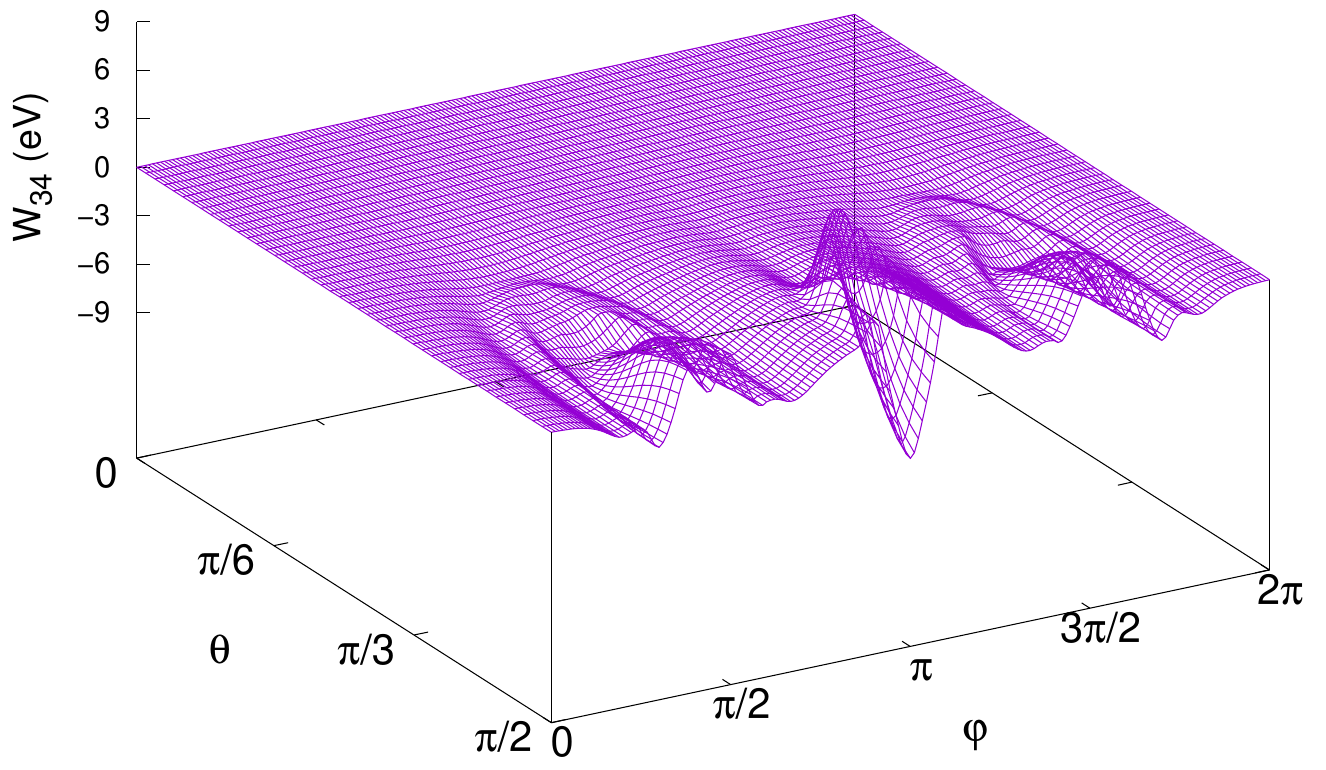}}\\
	\caption{Diabatic PES matrix elements are plotted as functions of hyperangles $\theta$ and $ \phi $ for $ \rho $ = 7 Bohr: (a) W$_{11}$, (b): W$_{12}$, (c) W$_{33}$ and  (d) W$_{34}$. The figures are taken from Ref.~\citenum{HeH2_Surf} with the permission from ACS.}
	\label{fig:dia_coup_HeH2+}
\end{figure}

\noindent 
In the last few decades, several calculations were carried out on the HeH$_2^+$ system~\cite{Russek1989,Furlan1990}, which explored the crossings between the ground and excited states as well as computed the nonadiabatic couplings. Such calculations was followed by the development of diabatic representations for the lowest excited states using the diatomics-in-molecules (DIM) approach. ~\cite{Schneider1979,Schneider1983} Double Many-Body Expansion (DMBE) method were employed to generate analytically fitted quasi-diabatic PESs for the two lowest excited states.~\cite{Aguado1993} In the subsequent works, diabatic PES matrices were constructed incorporating multiple electronic states for analyzing the excited states proton-transfer reaction, He$^+$ + H$_2$ $\rightarrow$ HeH$^+$ + H.\cite{Sidis1996}  
For such system, topology of the CIs and the associated ADT angles are examined to locate the seams of intersection between the lowest excited states.~\cite{MolPhys_118_2020_Gupta,IJQC_121_2021_Mondal} With the help of MOLPRO quantum chemistry package,~\citep{MOL}, MRCI level of calculations are performed by Ravi \textit{et al.}~\cite{Ravi2021_molphys_119} exposing multiple (ten) CIs among the electronic states and validating the contribution of NACTs in the accurate (qualitative and quantitative) description of the species. The NACTs among are computed as functions of hyperangles ($\theta$, $\phi$) for a set of fixed hyperradii ($\rho$). Once the adiabatic PESs and NACTs are obtained, global diabatic Hamiltonian\cite{HeH2_Surf} is constructed by involving the NACTs between the lowest four (doublet) adiabatic PESs (1$^2$A$^\prime$, 2$^2$A$^\prime$, 3$^2$A$^\prime$ and 4 $^2$A$^\prime$). While performing the \textit{ab initio} calculations, adiabatic PESs are computed employing aug-cc-pVQZ basis set by state-averaged SA-6-MCSCF methodology with the CAS (3e, 8o), whereas such study is followed by internally contracted MRCI calculations for those four low-lying electronic states. NACTs are evaluated at the same MRCI level using the DDR method as implemented in MOLPRO. Figure 
~\ref{fig:nact_hyp_heh2+} shows few representative $\phi$- and $\theta$-components of NACTs at the fixed hyperradius, $\rho = 10.0$~Bohr. On the other hand, Figures \ref{fig:nact_hyp_phi12} and \ref{fig:nact_hyp_theta12} show $\phi$ and $\theta$ components of $\tau_{12}$ exhibiting peaks at the $C_{2v}$ configuration ($\phi = 0^\circ$ or $360^\circ$, $\theta \approx 45^\circ$). Such steep rise of NACTs reflects an accidental CI between the ground and first excited states, which is further affirmed by carrying out loop-integral quantization leading to ADT angles (see Figures \ref{fig:12CI_fig_heh2+} and ~\ref{fig:12CI_angle_heh2+}). Figure \ref{fig:12CI_angle_heh2+} explores that the ADT angle attains the value of 
$\pi$ upon closed-contour integration, whereas the diagonal elements of the ADT matrix ($A_{11}$ and $A_{22}$) exhibit sign inversion from $+1$ to $-1$, confirming the presence of a conical intersection between the ground and first excited states. In a similar way, presence of additional CIs between the second (2$^2$A$^\prime$) and third (3$^2$A$^\prime$) as well as the third (3$^2$A$^\prime$) and fourth (4$^2$A$^\prime$) electronic state at linear geometries can also be validated.\cite{HeH2_Surf} On the other hand, 2D functional forms of mixing angles between the electronic states $(\Theta_{ij})$ over $\theta$-$\phi$ plane for fixed $\rho$ are evaluated by integrating the 
ADT equations for four coupled electronic states using the \textit{ab initio} calculated NACTs (see Appendix \ref{ssec:adt-eq}). Such mixing angles vis-\'a-vis \textit{ab initio} calculated adiabatic PESs are utilized to generate the diabatic PES matrix as functions of $\theta$ and $\phi$ for a series of $\rho$ values as demonstrated in Figure \ref{fig:dia_coup_HeH2+} for a specific case.

\subsection{H$_3$}

\begin{figure}[!htp]
	\centering
	\includegraphics[width=0.6\linewidth]{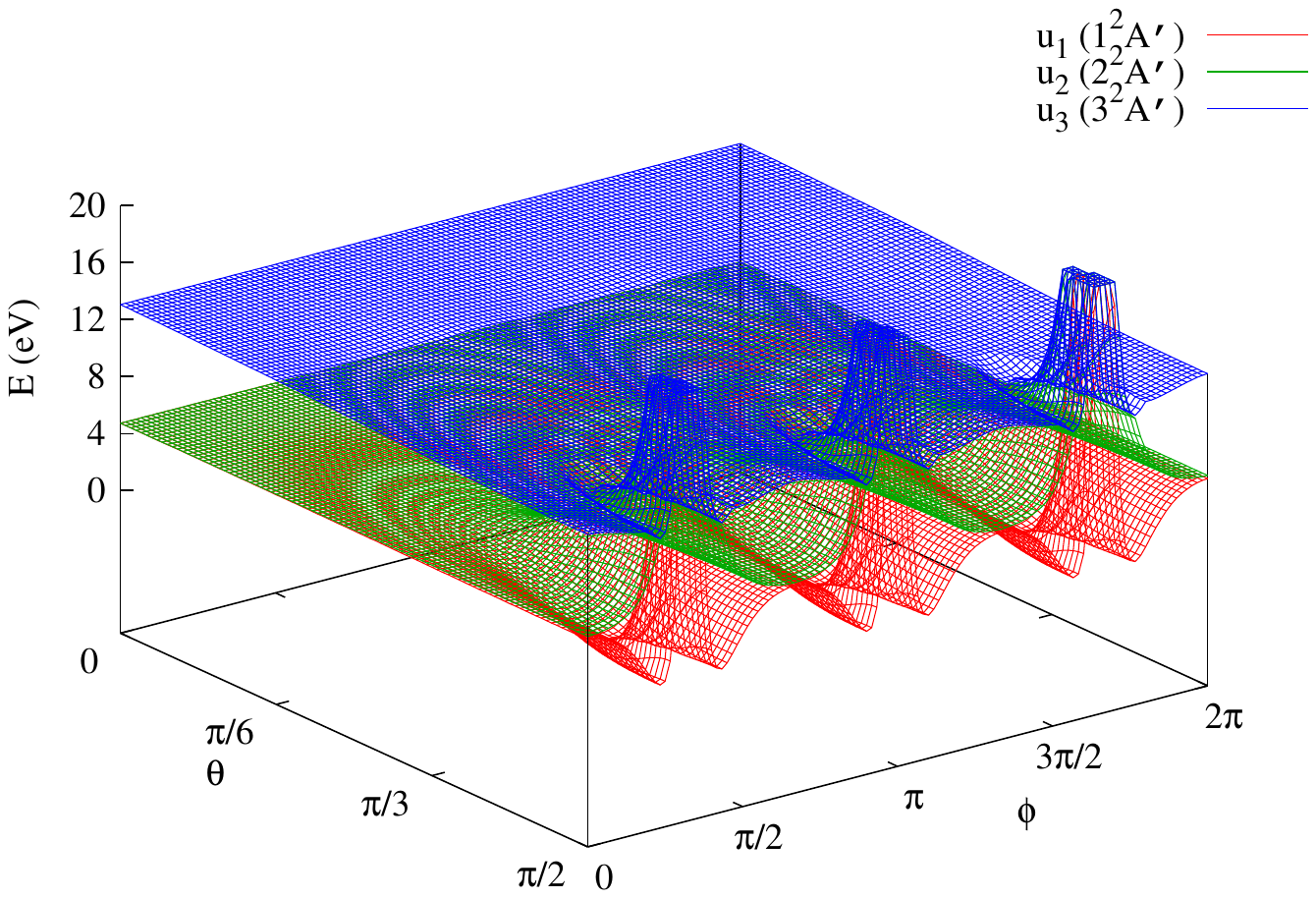}
	\caption{The lowest three adiabatic PESs (\( u_1 \),
		\( u_2 \) and \( u_3 \)) for the H$ _3$ system are plotted over the $ \theta $-$ \phi $ 
		plane at fixed $ \rho $ = 9 Bohr. The figure is adapted from
		Ref.~\citenum{Ghosh_H3_2025} with permission from ACS.}
	\label{fig:pesadia4final}
\end{figure}

\begin{figure}[!htp]
	\centering
	\includegraphics[width=.5\textwidth]{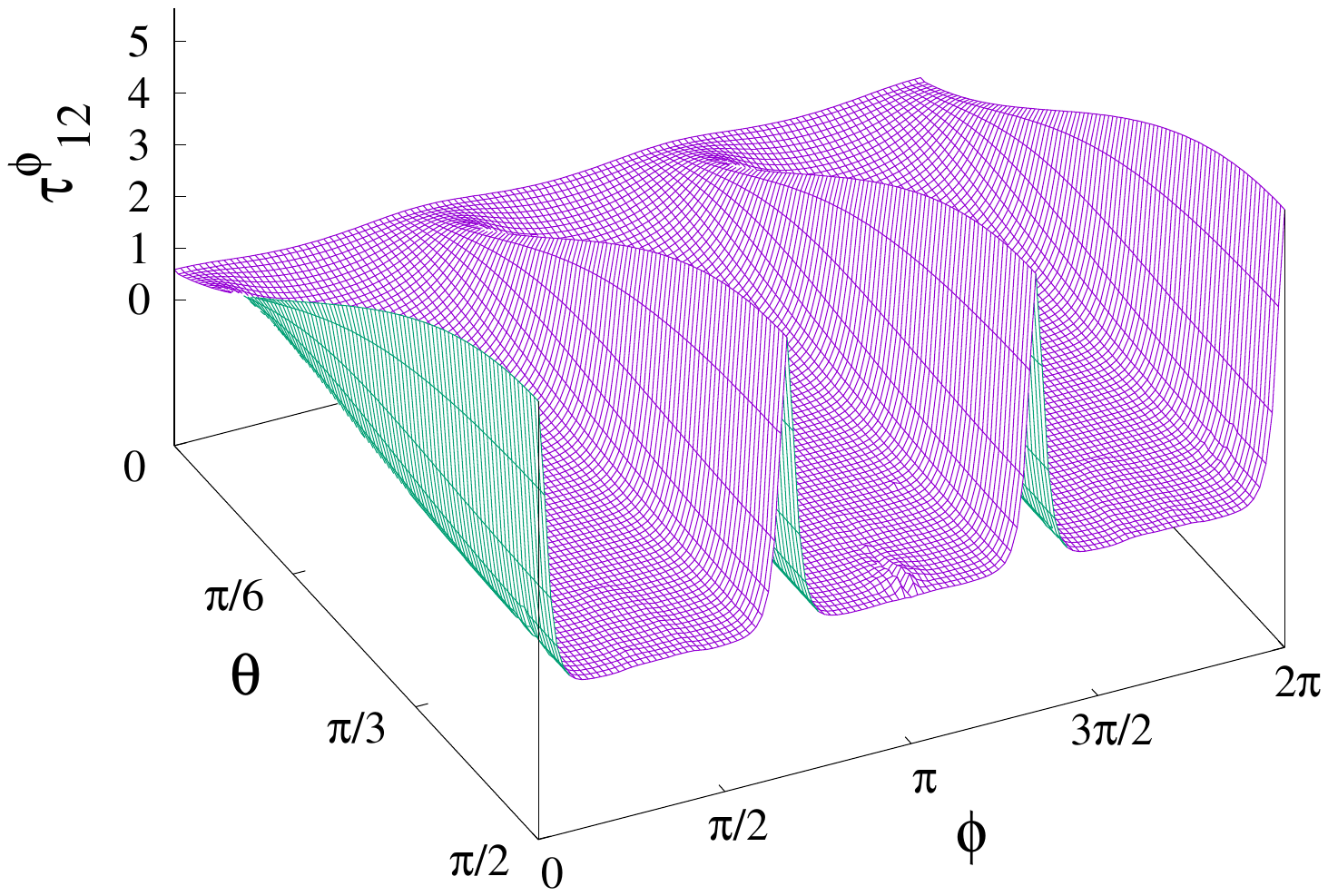}\hfill
	\caption{Hyperspherical $\phi$-component of NACT ($ \tau^{\phi}_{12}$) for the  H$ _3$ system is displayed as function of $ \theta $ and $ \phi $ coordinates, where $ \rho $ is fixed at 9 Bohr. The figure is adapted from Ref.~\citenum{Ghosh_H3_2025} with permission from ACS.}
	\label{fig:nact_hyp}
\end{figure}

\begin{figure}[!htp]
	\centering
	\subfloat[\label{fig:12CI_fig}]{\includegraphics[width=.45\textwidth]{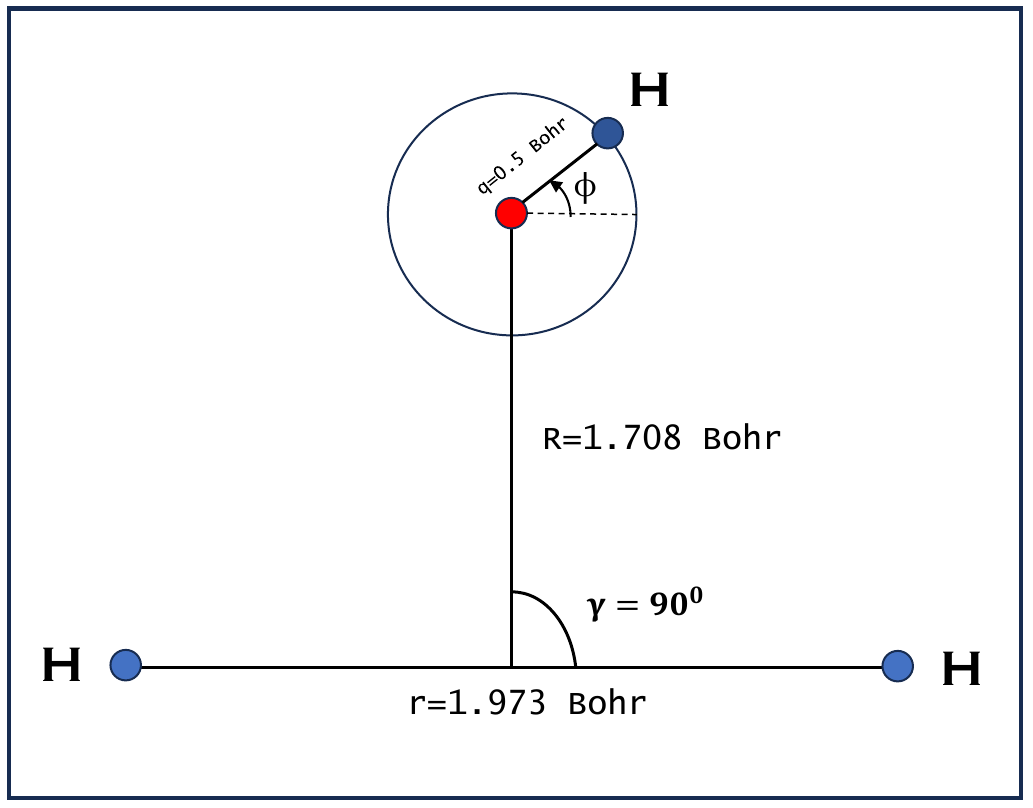}}\hfill
	\subfloat[\label{fig:12CI_angle}]{\includegraphics[width=.5\textwidth]{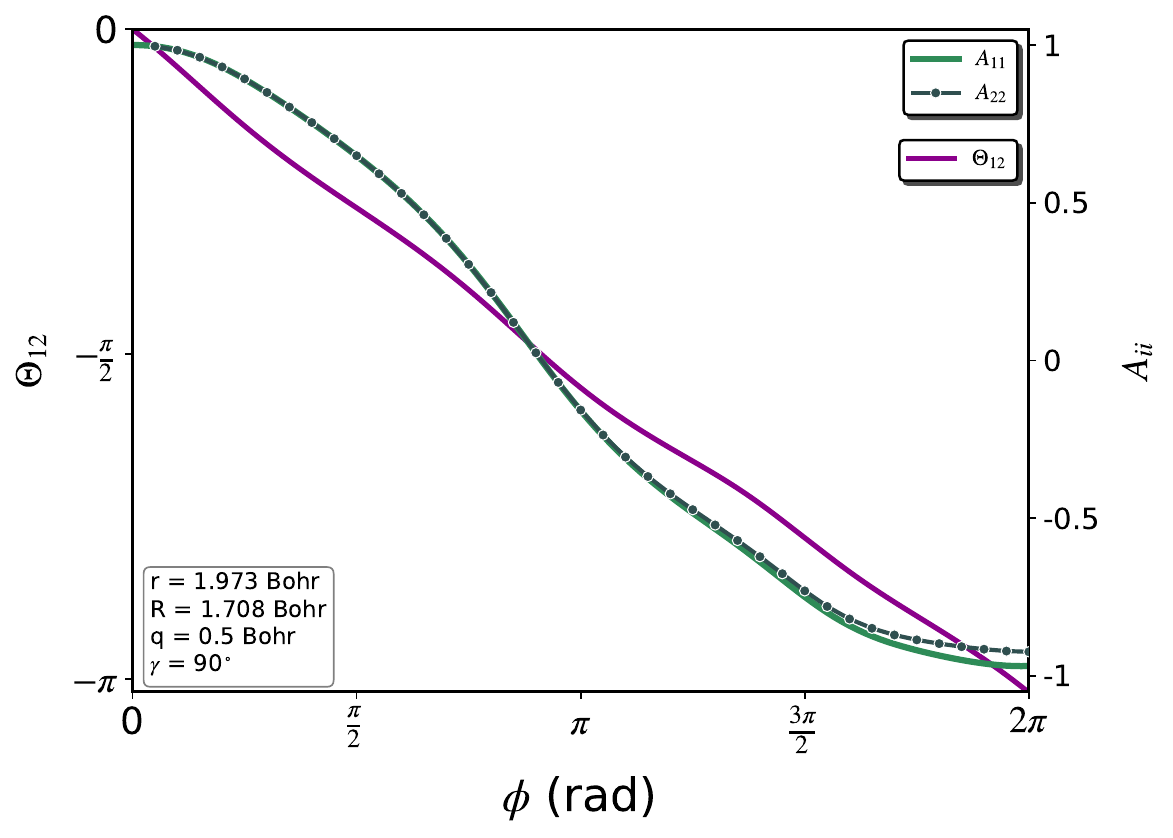}}
	\caption{(a) The closed contour with radius $q = 0.5$ Bohr is illustrated enclosing the D$_{3h}$ 1-2 CI in Jacobi coordinates.
		(b) The ADT angle ($\Theta_{12}$) and the diagonal ADT matrix elements (A$_{11}$ and A$_{22}$) attain the values of $-\pi$ and $-1$, respectively, at the end of closing contour, thereby depicting the topological signature of the 1-2 CI. The figures are adapted from Ref.~\citenum{Ghosh_H3_2025} with permission from ACS.}
	\label{fig:12CI}
\end{figure}

\noindent 
In the past few decades, extensive investigations were carried out on the exchange reaction between H/D and H$_2$/D$_2$/HD over the ground adiabatic PES \cite{LSTH_1978,DMBE_1987,BKMP2_1996,CCI_2002} of H$_3$ system. Even though such calculations appear as quite accurate at lower collision energies, construction of multi-state diabatic Hamiltonian is indeed necessary for dynamical calculation around the energy range where the low lying electronic states become close to each other energetically through multiple CIs. Several attempts\cite{DMBE_1987,Abrol_Kuppermann_2002} were taken to construct diabatic PES matrix, but the accuracy needs to be improved. In order to increase the accuracy, global diabatic PES matrix is constructed, which is highly precise not only at interaction region, but also at the domain of asymptote. For the H$_3$ system~\cite{Ghosh_H3_2025}, the lowest three (doublet) adiabatic PESs (1$^2  $A$ ^\prime $, 2$^2  $A$ ^\prime $ and 3$^2  $A$ ^\prime $) are calculated at the CASSCF followed by MRCI level of approaches, where the NACTs between those electronic states are evaluated with the aid of MRCI wavefunctions via the finite-difference (DDR) procedure as implemented in the MOLPRO quantum chemistry package~\cite{MOLPRO_brief,MOLPRO_2020}. The state-averaged multi-configurational self-consistent field (SA-MCSCF) level of calculation is performed with the use of converged CAS constituted with three electrons distributed among sixteen orbitals, namely, twelve a${'}$ and four a${''}$ orbitals. On the other hand, six electronic states (6) 
are averaged and thereby, denoted as SA-6-CAS(16o,3e) using aug-cc-pVQZ basis set. At $ \rho$ = 9.0 Bohr, the 2D PESs for three electronic states and the $\phi$-component of NACTs are plotted in Figures \ref{fig:pesadia4final} and \ref{fig:nact_hyp}, respectively, as function of $ \theta $-$ \phi $. While locating CIs between the electronic states, the $\phi$-component of the NACTs ($\tau_{12}^{\phi}$) are calculated in Jacobi coordinates ($r$, $R$, $\gamma$) along a circular contour of radius $q$ encapsulating the CI.\cite{Ghosh_H3_2025} After solving ADT equations for three coupled electronic states (see Eq. \ref{eq:adt-three}), the mixing angle reaches $\pi$ and diagonal elements of the ADT matrix attains the value of minus one (-1) validating the resence of D$_{3h}$ CI between the ground and first excited states (see Figure \ref{fig:12CI}). In addition, the accidental CIs between the ground and first excited states as well as the first and second excited states at C$_{2v}$ configurations are also explored.\cite{Ghosh_H3_2025,Baer_2004} 

\begin{figure}[!htp]
	\centering
	\subfloat[\label{fig:w11}]{\includegraphics[width=.5\textwidth]{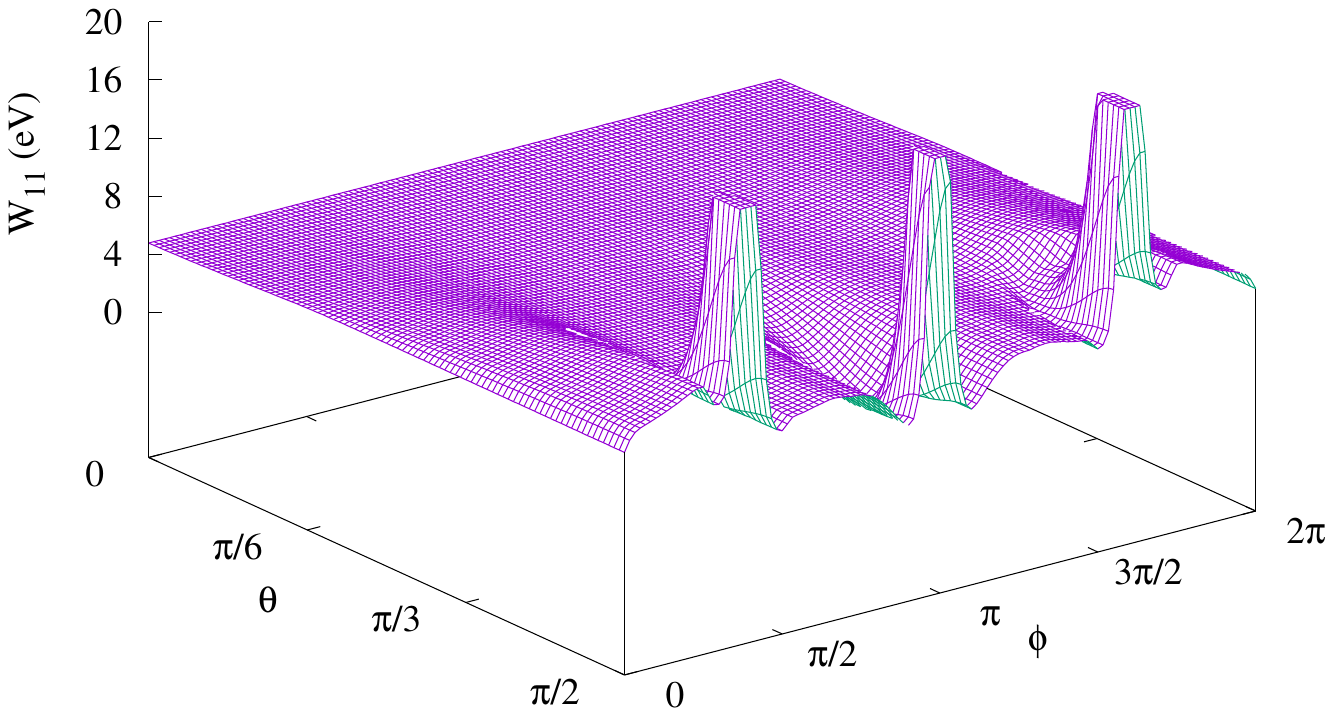}}\hfill
	\subfloat[\label{fig:w22}]{\includegraphics[width=.45\textwidth]{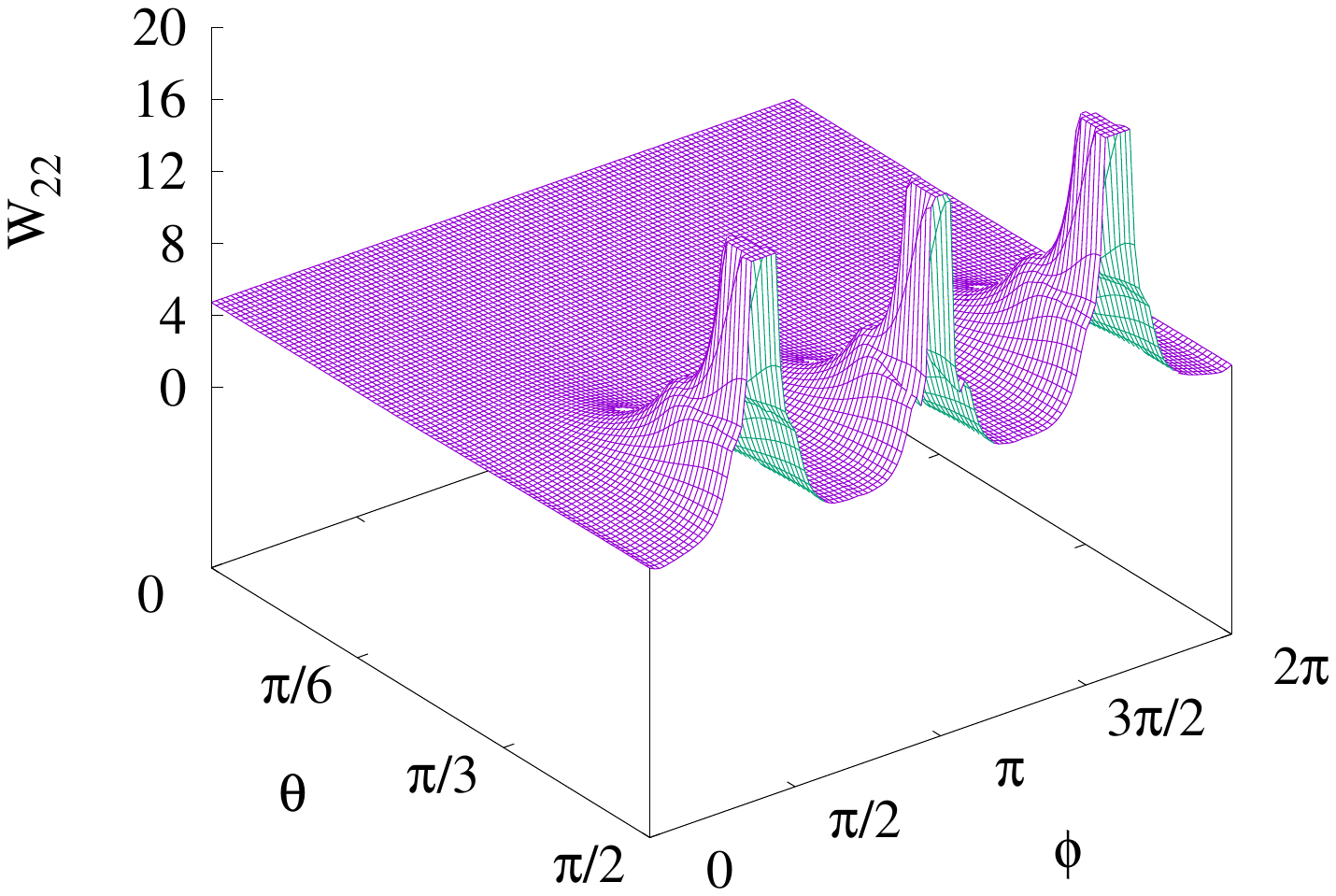}}\\
	\subfloat[\label{fig:w12}]{\includegraphics[width=.45\textwidth]{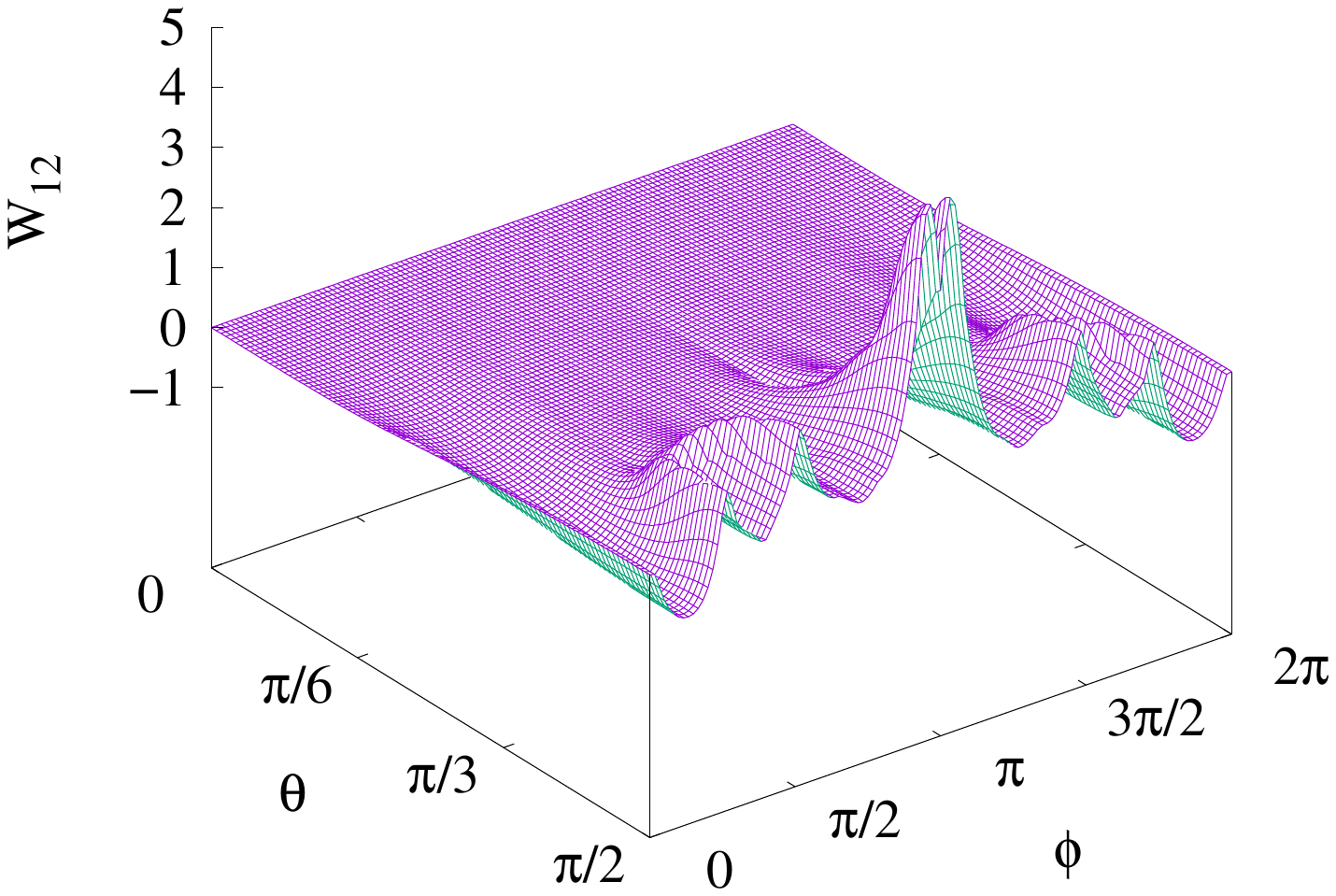}}\hfill
	\caption{Diabatic PES matrix elements are plotted as functions of hyperangles $\theta$ and $ \phi $ for fixed $ \rho $ = 9 Bohr: (a) W$_{11}$ and (b): W$_{22}$ and (c): W$_{12}$. The figures are adapted from Ref.~\citenum{Ghosh_H3_2025} with permission from ACS.}
	\label{fig:dia_mat}
\end{figure}

\noindent
The \textit{ab initio} calculated NACTs are substituted into the differential equations (Eq.~\ref{eq:adt-three}) and then, those are  numerically integrated\cite{ADT_JCTC_16} to compute the mixing angles over over $\theta$-$\phi$ plane for each fixed value of $\rho$. Such angles and \textit{ab initio} adiabatic PESs are employed for the construction of diabatic PESs matrix through an orthogonal transformation [see Eq.~\ref{eq:diapot1}]. In Figure \ref{fig:dia_mat}, the diabatic PESs matrix elements (W$_{11}$, W$_{12}$ and W$_{22}$) are demonstrated as functions of $\theta$ and $\phi$ for $\rho$ = 9 Bohr, where the diabatic PESs and couplings are single valued, smooth and continuous functions of the hyperangles.

\section{Reactive Scattering Calculations} \label{sec:Application-Dyna}

\subsection{F + H$_2$/D$_2$ reaction}

\begin{figure}[!htp]
	\begin{center}
		\includegraphics[width=1.00\linewidth,angle=0]{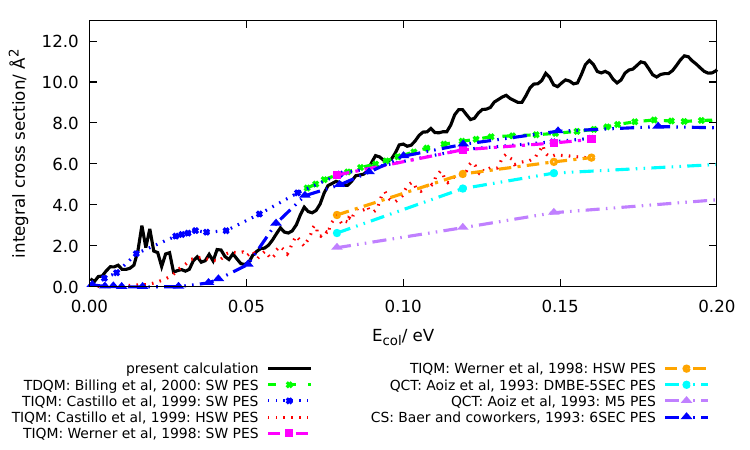}
	\end{center}
	\noindent
	\caption{Total integral cross-sections for F+H$_2$ ($v=0; j=0$) reaction are plotted as functions of collision energy in comparison with previous theoretical calculations  obtained from different 
		dynamical approaches and various potential energy surfaces.
The figures are adapted from Ref.~\citenum{FH2_adia_dynamics} with permission of ACS.}
	\label{fig:ICS_comp}
\end{figure}

\begin{figure}[!htp]
	\begin{center}
		\includegraphics[width=1.00\linewidth,angle=0]{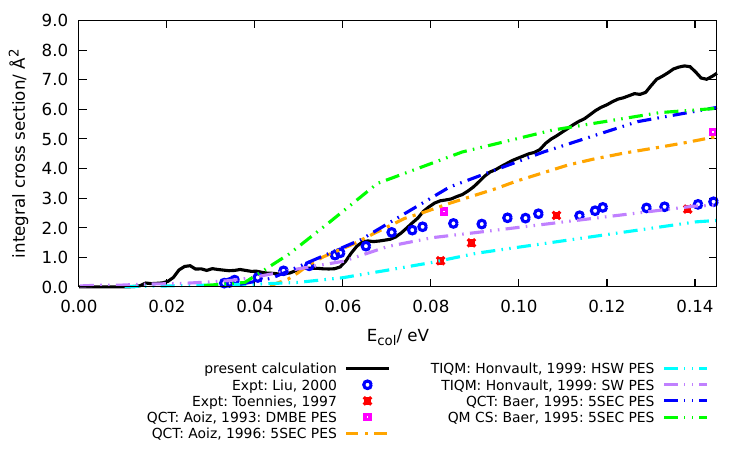}
	\end{center}
	\noindent
	\caption{Total integral cross-sections for F + D$_2$ ($v=0; j=0$) reaction are plotted as function of 
		collision energy in comparison with previous theoretical calculations and experimental 
		measurements. The figures are taken from Ref.~\citenum{FH2_adia_dynamics} with permission of ACS.}
	\label{fig:ICS_comp_fd2}
\end{figure}

\noindent
Adhikari and coworkers performed FCC-3D-TDWP quantum scattering calculation on the ground adiabatic PES involving all the helicity quantum numbers ($K$) for the F + H$_2$ ($v$ = 0; $j$ = 0) and F + D$_2$ ($v$ = 0; $j$ = 0) reactions.~\cite{FH2_adia_dynamics} While doing such study, reaction probabilities upto the converged total angular momentum ($J$) are used to determine the ICSs. Figures \ref{fig:ICS_comp} and \ref{fig:ICS_comp_fd2} depict product cross-section for FH and FD species, respectively. It is interesting to note that the ICS profiles 
exhibit good accord with earlier theoretical studies based on QM centrifugal sudden (CS),~\cite{jpc-99-16523-BAER} time independent quantum mechanical (TIQM),~\cite{jcp-109-7224-WERNER,jcp-111-4013-AOIZ} and time dependent quantum mechanical (TDQM)~\cite{mp-98-1771-BILLING} approaches. Such calculations involve several variants of FH$_2$ PES, including the M5,~\cite{tcap-6A-1-MUCKER} 6SEC,~\cite{jcp-94-7136-TRU} SW,~\cite{jcp-104-6515-WERNER} and HSW~\cite{cpl-280-430-HARTKE} PESs. 
In case of F + D$_2$ reaction, the ICS profiles\cite{cpl-254-341-AOIZ,jpc-99-16523-BAER,cpl-204-359-AOIZ,cpl-303-657-HON} originated from various (ground) adiabatic PESs\cite{tcap-6A-1-MUCKER,jcp-94-7136-TRU,jcp-104-6515-WERNER,cpl-280-430-HARTKE} exhibit some agreement with crossed-beam experimental measurements~\cite{jcp-113-3633-LIU} over the limited range of collision energy (0.03-0.08~eV). On the contrary, experimentally measured rate constants\cite{jpc-93-4068-ANDERSON,bredjaninan,jcp-72-4811-HOUSTON} over the temperature range (200-400 K) as depicted in Figure \ref{fig:rateground_low} show substantial difference from the calculated rate constants originated from adiabatic PESs. Furthermore, rate constants involving SO coupling (excluding the effect of non-adiabaticity) also show substantial deviation from the experimentally measured ones (see Figure \ref{fig:rateground_low}). Such differences mainly arise from absence of nonadiabatic couplings in the underlying Hamiltonian, which motivates for the calculation of  
reaction attributes using multi-sheeted diabatic Hamiltonian\cite{FH2_Surf}.

\vspace{0.3 cm}

\noindent
The wave packet is initialized on the ground electronic state [$1^2A^\prime$] and then, quantum dynamics are carried out 
on global diabatic PES matrix, which includes the contributions of 
\textit{ab initio} calculated adiabatic PESs, nonadiabatic and SO couplings for F+H$_2$ system. Such dynamical calculations lead to 
various kinds of reaction attributes leading to cross-sections and rate constants of the product (HF) formation as well as the non-reactive excitation process as defined below\cite{FH2_dynamics}:

\begin{itemize}
	\item Reactive  process [Ground state ($1^2A^\prime$)]:
	\begin{center}
		Without SO coupling: F($^2P$)+H$ _2\, (v=0; j=0) \  [1^2A^\prime]$ $ \rightarrow $ HF $(v^\prime; j^\prime)$+H $ \  [1^2A^\prime]$ \\
		With SO coupling: F($^2P_{3/2}$)+H$ _2\, (v=0; j=0)\   [1^2A^\prime]$ $ \rightarrow $ HF $(v^\prime; j^\prime) $+H $\ [1^2A^\prime]$.
	\end{center}
	\item Non-reactive excitation process [Excited state ($1^2A^{\prime\prime}$)]:
	\begin{center}
		Without SO coupling: F($^2P$)+H$ _2 \, (v=0; j=0)  \  [1^2A^\prime]$ $ \rightarrow $ F$ $($^2P$)+H$ _2 $ $(v^\prime; j^\prime)  \  [1^2A^{\prime\prime}]$\\   
		With SO coupling: F($^2P_{3/2}$)+H$ _2 \, (v=0; j=0)  \  [1^2A^\prime]$ $ \rightarrow $ F$ ^* $($^2P_{1/2}$)+H$ _2 $ $(v^\prime; j^\prime) \  [1^2A^{\prime\prime}]$	
	\end{center}
\end{itemize}

\begin{figure}[!htp]	
	\centering
	\includegraphics[width=1.0\linewidth]{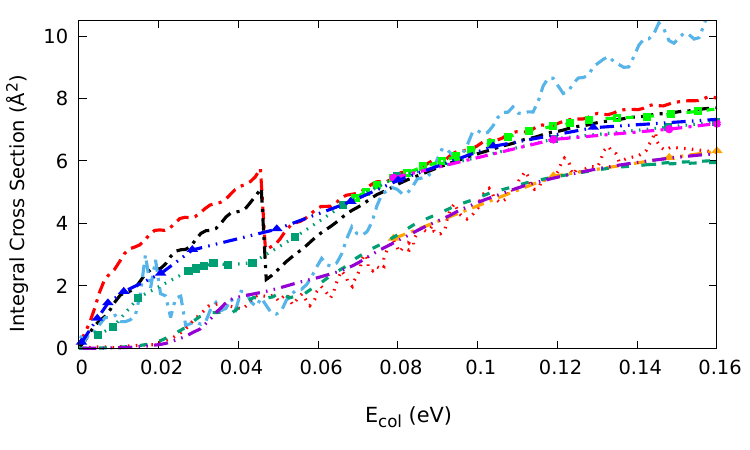}
	\caption{Total integral cross-sections for the product formation (F + H$ _2 $ ($v=0$; $j=0$) $ \rightarrow $
		HF+H) reaction are presented on the ground state (1$^2A^\prime$) of BBO based diabatic calculation as function of collision
		energy for without (\protect\includegraphics{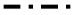}) and with (\protect\includegraphics{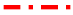})	the
		incorporation of SO coupling in comparison with our previous adiabatic
		calculation\cite{Naskar2022_JPCA_126} (\protect\includegraphics{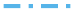}) as well as other theoretical
		results: TIQM by Aoiz \textit{et. al.}\cite{jcp-111-4013-AOIZ} on SW PES\cite{jcp-104-6515-WERNER}
		(\protect\includegraphics{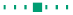}) and HSW PES\cite{cpl-280-430-HARTKE} (\protect\includegraphics{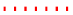});
		TIQM by Castillo \textit{et. al.}\cite{jcp-109-7224-WERNER} on SW PES (\protect\includegraphics{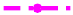})
		and HSW PES (\protect\includegraphics{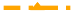}); TDQM by Billing \textit{et al.}\cite{mp-98-1771-BILLING}
		(\protect\includegraphics{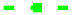}); QM CS calculation  by Baer and co-workers\cite{Rosenman1996_CPC_257}   on SW PES\cite{jcp-104-6515-WERNER} (\protect\includegraphics{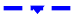});TIQM by Alexander
		\textit{et. al.}\cite{jcp-113-11084-WERNER}(\protect\includegraphics{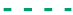});
		TDWP by Zhang \textit{et al.}\cite{jcp-119-12921-HAN}(\protect\includegraphics{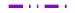}). The figure is taken from Ref.~\citenum{FH2_dynamics} with permission of ACS.}
	\label{fig:cross}
\end{figure}

\begin{figure}[!htp]
	\centering
	\includegraphics[width=1.0\linewidth]{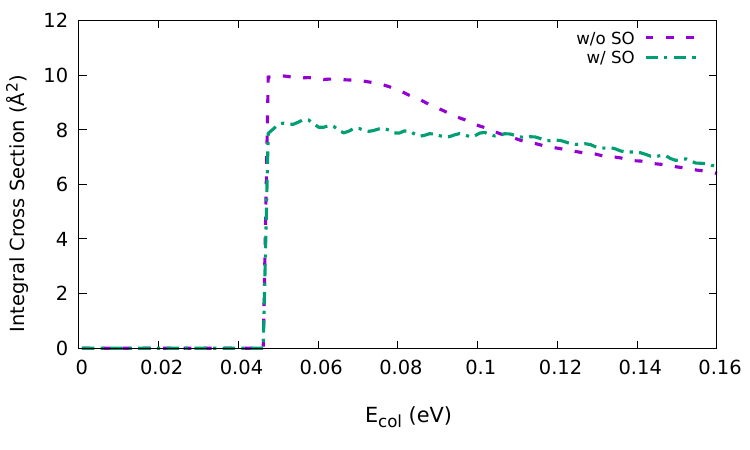}
	\caption{Total integral cross sectiosn for the excited state  non-reactive species [F$  $+H$
		_2 $ [$1^2A^{\prime\prime}$] (without SO) and F$^*  $+H$ _2 $ [$1^2A^{\prime\prime} $] (with SO)] are presented as function of collision energy with and without the involvement of SO coupling. The figure is adapted from Ref.~\citenum{FH2_dynamics} with permission of ACS.}
	\label{fig:cross_excited}
\end{figure}

\begin{figure}[tph!]
	\centering
	\includegraphics[width=1.0\linewidth]{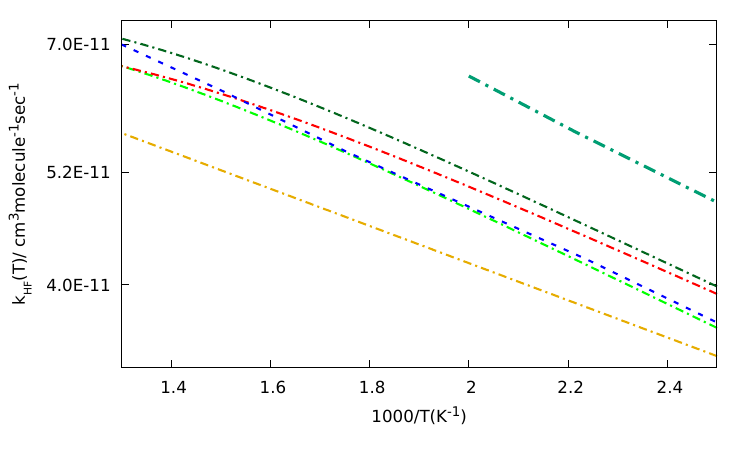}  
	\caption{Temperature dependent rate constant for the product formation 
		(F + H$ _2 $ ($v=0; j=0$) $ \rightarrow $ HF+H) reaction on the ground state (1$^2A^\prime$) are portrayed, where diabatic calculation is carried out
		without (\protect\includegraphics{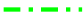}) and with (\protect\includegraphics{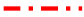}) 
		the inclusion of SO coupling. The profiles are compared with our previous 
		adiabatic calculation\cite{Naskar2022_JPCA_126} (\protect\includegraphics{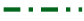}) along 
		with existing experimental findings: 
		Wurzberg \textit{et al.}~\cite{jcp-72-4811-HOUSTON}(\protect\includegraphics{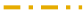}),
		Heidner \textit{et al.}~\cite{jcp-72-4815-HEIDNER}(\protect\includegraphics{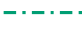}),
		Bedjanian\cite{bredjaninan} (\protect\includegraphics{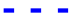}). The figure is taken from Ref.~\citenum{FH2_dynamics} with permission of ACS.}
	\label{fig:rateground_high}
\end{figure}

\begin{figure}[tph!]
	\centering
	\includegraphics[width=1.0\linewidth]{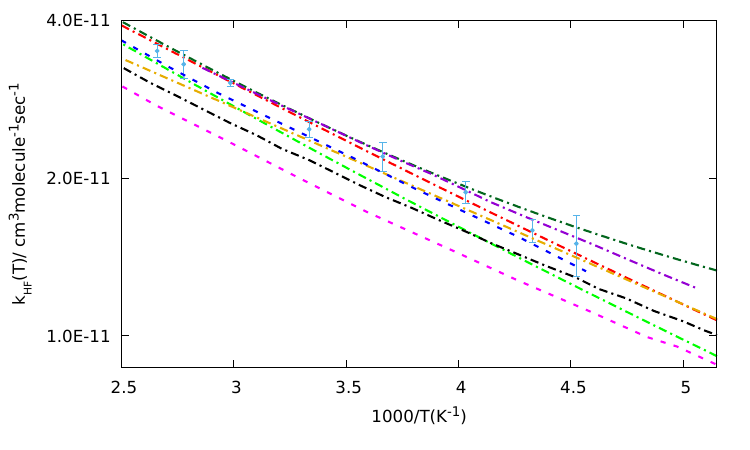}  
	\caption{Temperature dependent rate constant for the product formation (F + H$ _2 $ ($v=0; j=0$) $ \rightarrow $ HF+H) reaction on the ground state (1$^2A^\prime$) of diabatic calculation without (\protect\includegraphics{Chapters/Chapter7/Figures/1.pdf}) and with (\protect\includegraphics{Chapters/Chapter7/Figures/2.pdf}) the inclusion of SO coupling in comparison with our previous adiabatic calculation\cite{Naskar2022_JPCA_126} (\protect\includegraphics{Chapters/Chapter7/Figures/3.pdf}) along with existing experimental findings: Wurzberg \textit{et al.}~\cite{jcp-72-4811-HOUSTON}(\protect\includegraphics{Chapters/Chapter7/Figures/9.pdf}), Anderson \textit{et al.}~\cite{jpc-93-4068-ANDERSON}(\protect\includegraphics{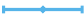}), Bedjanian\cite{bredjaninan} (\protect\includegraphics{Chapters/Chapter7/Figures/10.pdf}); and theoretical results: Aquilanti \textit{et al.}\cite{aql2} (PES III) (\protect\includegraphics{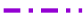}), TIQM by Fu \textit{et al.}\cite{fxz_pes}(\protect\includegraphics{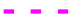}) and Chen \textit{et al.}\cite{jcp-142-024303-ZHANG} (\protect\includegraphics{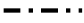}) Adapted with permission of ACS from Ref.~\citenum{FH2_dynamics}.}
	\label{fig:rateground_low}
\end{figure}

\noindent
Figure \ref{fig:cross} depicts calculated ICSs as function of collision energy for the product (HF) formation of the reactive process with and without the inclusion of SO couplings. On the other hand, ICSs for the
non-reactive excitation process are shown in Figure \ref{fig:cross_excited}. For all cases, employed diabatic surfaces matrices include nonadiabatic couplings.\cite{FH2_dynamics} The ICS profiles show good agreement with other theoretical results at intermediate as well as higher ranges of collision energies.\cite{jcp-111-4013-AOIZ,jcp-109-7224-WERNER} Once the calculated ICSs with and without the inclusion of SO coupling are in hand, those are employed to calculate temperature dependent rate constants. At higher domain of temperature ($\sim$400K - 600K),
calculated rate constants without SO coupling exhibit close agreement with the experimentally recorded values of Bedjanian\textit{ et al.}\cite{bedjaninan} (see Figure \ref{fig:rateground_high}). On the other hand, while including SO coupling in diabatic Hamiltonian, the ICS profiles demonstrate better agreement at the lower temperaure regime ($\sim$200K - 300K), in comparison with recent experimental result\cite{bedjaninan} as presented in Figure \ref{fig:rateground_low}.\cite{FH2_dynamics} Table.~\ref{tab:ratemeandiff} presents the mean and RMS deviation of theoretically calculated rate constants in comparison with the experimental data.

\vspace{0.3cm}

\begin{small}
	\begin{table}[!tpb]
		\centering
			\caption{
			The table summarizes mean (root mean square)  relative deviation of theoretically computed rate constants from
			experimentally recorded ones (Bedjanian\cite{bredjaninan} and Wurzberg \textit{et
				al.}~\cite{jcp-72-4811-HOUSTON}). While calculating  such deviation from experiment, the
			exact temperature range of the available theoretical rate constants is only taken into account (see
			Figure \ref{fig:rateground_low}). The table is taken from Ref.~\citenum{FH2_dynamics} with permission of ACS.}
		\begin{tabular}{|c|c|c|}
			\hline
			
			Reference & Mean (Root mean square)    &  Mean (Root mean square)\\ 
			& relative deviation &  relative deviation \\
			& from data of  Bedjanian\cite{bredjaninan} & from data by  Wurzberg \textit{et al.}~\cite{jcp-72-4811-HOUSTON}   \\	
			\hline
			Present diabatic calculation (w/o SO)    & 0.051 (0.056)   & 0.049 (0.061) \\
			Present diabatic calculation (w/ SO)        &  0.066 (0.066) & 0.075 (0.082)  \\
			Previous adiabatic calculation\cite{Naskar2023_JPCA_127}  & 0.103 (0.105)  & 0.112 (0.113) \\
			Aquilanti \textit{et al.} (PES III)\cite{aql2}    & 0.093 (0.093) & 0.097 (0.098)  \\
			Chen \textit{et al.}\cite{jcp-142-024303-ZHANG}     & 0.091 (0.092) & 0.087 (0.088)  \\
			Fu \textit{et al.}\cite{fxz_pes}         &  0.179 (0.179)  &  0.176 (0.177) \\		
			\hline
		\end{tabular}
		\label{tab:ratemeandiff}
	\end{table}
\end{small}

\subsection{H$^+$/D$^+$/H + H$_2$/H$_2^+$ reaction}
\noindent
Adhikari and coworkers performed multi-state quantum reactive scattering calculations over \textit{ab initio} BBO-theory-based diabatic PES matrix\cite{H3+_Surf} for the collision process, D$^+$ + H$_2$ ($v=0$; $j=0$) system with the aid of fully close coupled (FCC) TDWP approach.\cite{jcp-147-074105-SA} While initiating dynamical calculation from the ground electronic state ($1^{1}A'$) with roto-vibrational state, \{$v = 0; j = 0$\}, the following competing reaction processes, namely, reactive non-charge transfer (RNCT), non-reactive charge transfer (NRCT) and reactive charge transfer (RCT) are studied, which  are as follows:\cite{jcp-147-074105-SA}

\begin{subequations}
	\begin{eqnarray}
		D^+ \ + \ H_2 \ (v=0; j=0) & \rightarrow & DH (v^\prime; j^\prime) \ + \ H^+ \hspace{0.4 cm}(\text{RNCT})\\
		& \rightarrow & H_2^+ (v^\prime; j^\prime) \ + \ D \hspace{0.4 cm} (\text{NRCT}) \\ 
		& \rightarrow & DH^+ (v^\prime; j^\prime) \ + \ H \hspace{0.4 cm} (\text{RCT}) .
	\end{eqnarray}
\end{subequations}

\begin{figure}[!htp]
	\centering
	\includegraphics[width=0.45\textwidth,angle=0]{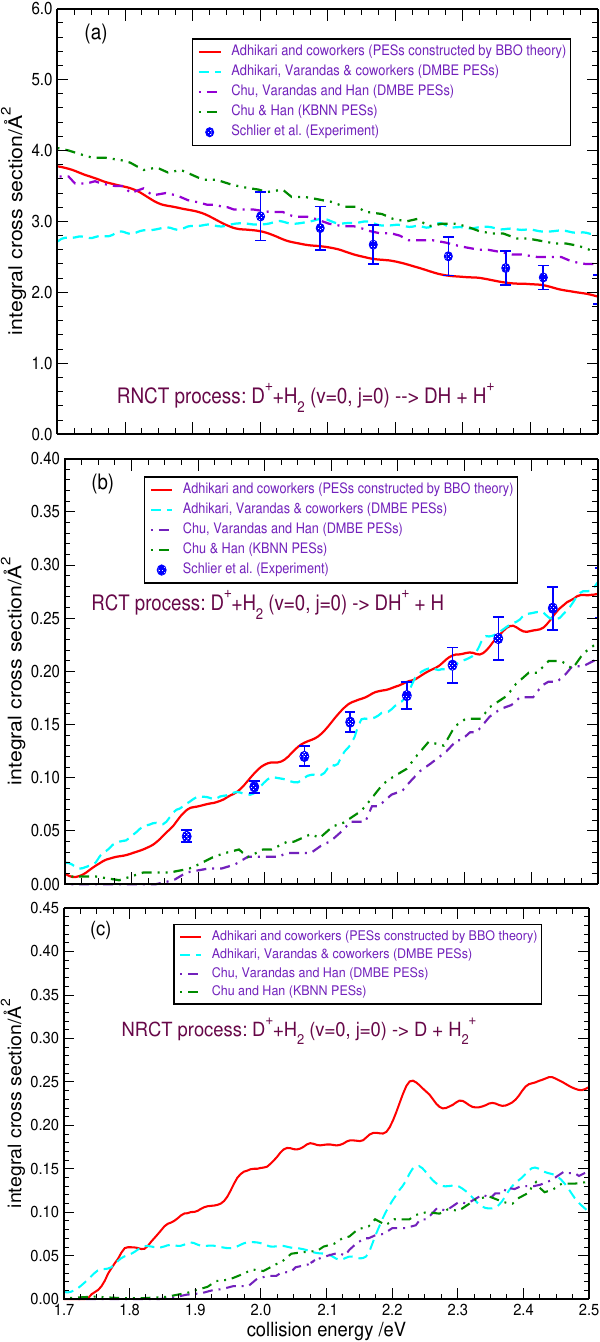}
	\caption{Total integral cross-section for (a) RNCT; (b) RCT and (c) NRCT processes of D$^+$ + H$_2$ reaction is displayed as a function of collision energy.\citep{jcp-147-074105-SA} The figure is taken from Ref.~\citenum{pccp-22-27496-SA} with the permission of RSC.}
	\label{fig:all_cross}
\end{figure}

\noindent
Figure \ref{fig:all_cross} depicts theoretically calculated ICS profiles for the RNCT, RCT and NRCT processes over the collision energy range, 1.7-2.5 eV and the outcomes are compared with earlier theoretical results\cite{jpca-119-12392-SA,viegas_jcp126,cpl-471-222-HAN3,jpcA-109-2050-HAN1,jcp-116-654-KAM} as well as with available (RNCT and RCT) experimentally measured data\citep{cp-111-401-SCH}. Over the entire collision energy range (1.7-2.5 eV), the ICS profiles exhibit decreasing and increasing trend for RNCT (see Figure \ref{fig:all_cross}a) and RCT (see Figure \ref{fig:all_cross}b) processes, respectively, which are computed from the dynamical calculation on the BBO based diabatic Hamiltonian matrices\cite{H3+_Surf}. Those profiles show very good agreement with the experimental curves\citep{cp-111-401-SCH} in comparison with other theoretically calculated ones\cite{jcp-147-074105-SA,jpca-119-12392-SA,cpl-471-222-HAN3,jpcA-109-2050-HAN1}. On the other hand, ICS profile for NRCT process shows similarity with that of RCT, which shows increasing trend with the collisional energy. In both cases (RCT and NRCT), the profiles possess  threshold collision energies ($E_{\mathrm{col}} \approx 1.8$ eV) affirming the accessibility of the first excited state through the CI (and the corresponding NACT) with the ground electronic state. The distinct differences between the ICS profiles for the RCT process originated from the BBO constructed diabatic Hamiltonian\cite{H3+_Surf} and other quasi-diabatic Hamiltonians\cite{KBNN,jcp-126-074309-AJCV} are governed by semi-empirical or \textit{ab initio} way of incorporation of nonadiabatic effects while generating multi-state Hamiltonians.\\

\noindent
On the contrary, while initiating the collision process from the first excited electronic state, $2^{1}A'$ of H$_3^+$ species, with asymptotic roto-vibrational level of H$_2^+$ ($v = 0; j = 0$), reaction attributes for the charge transfer processes from the first excited state [H$_2^+$ ($^2\Sigma_g^+$)] to the ground [H$_2$ ($^1\Sigma_g^+$)] are calculated involving the nonadiabatic couplings among the electronic states ($1^{1}A'$, $2^{1}A'$ and $3^{1}A'$), where the competing charge transfer processes are as follows:

\vspace{0.3 cm}

\noindent
(i) Reactive charge transfer (RCT): \hspace{0.1 in} $\rm H^{(1)} + \left(H^{(2)}H^{(3)}\right)^+ \rightarrow \left(H^{(2)/(3)}\right)^+ 
+ \left(H^{(1)}H^{(3)/(2)}\right)$; 

\noindent
(ii) Non-reactive charge transfer (NRCT): \hspace{0.05 in} $\rm H^{(1)} + \left(H^{(2)}H^{(3)}\right)^+ \rightarrow \left(H^{(1)}\right)^+ 
+ \left(H^{(2)}H^{(3)}\right)$.

\vspace{0.3cm}

\begin{figure}[!htp]
	\centering
	\includegraphics[height=0.55\textheight,width=1.0\linewidth,angle=0]{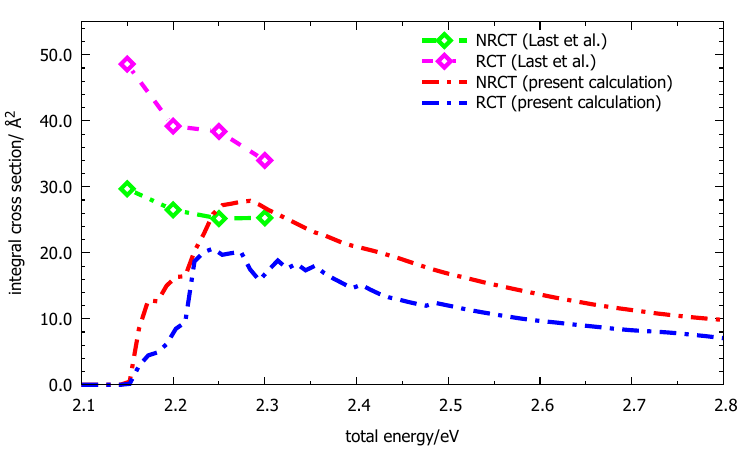}
	\caption{Total integral cross-section of the RCT and NRCT process associated with H + H$_2^+$ ($v=0$; $j=0$) reaction are depicted. The figure is adapted from Ref.~\citenum{H3+_dynamics_2021} with the permission of ACS.}
	\label{fig:ICS}
\end{figure}

\begin{figure}[!htp]
	\centering
	\subfloat[\label{fig:vprime0}]{\includegraphics[width=0.5\linewidth,angle=0]{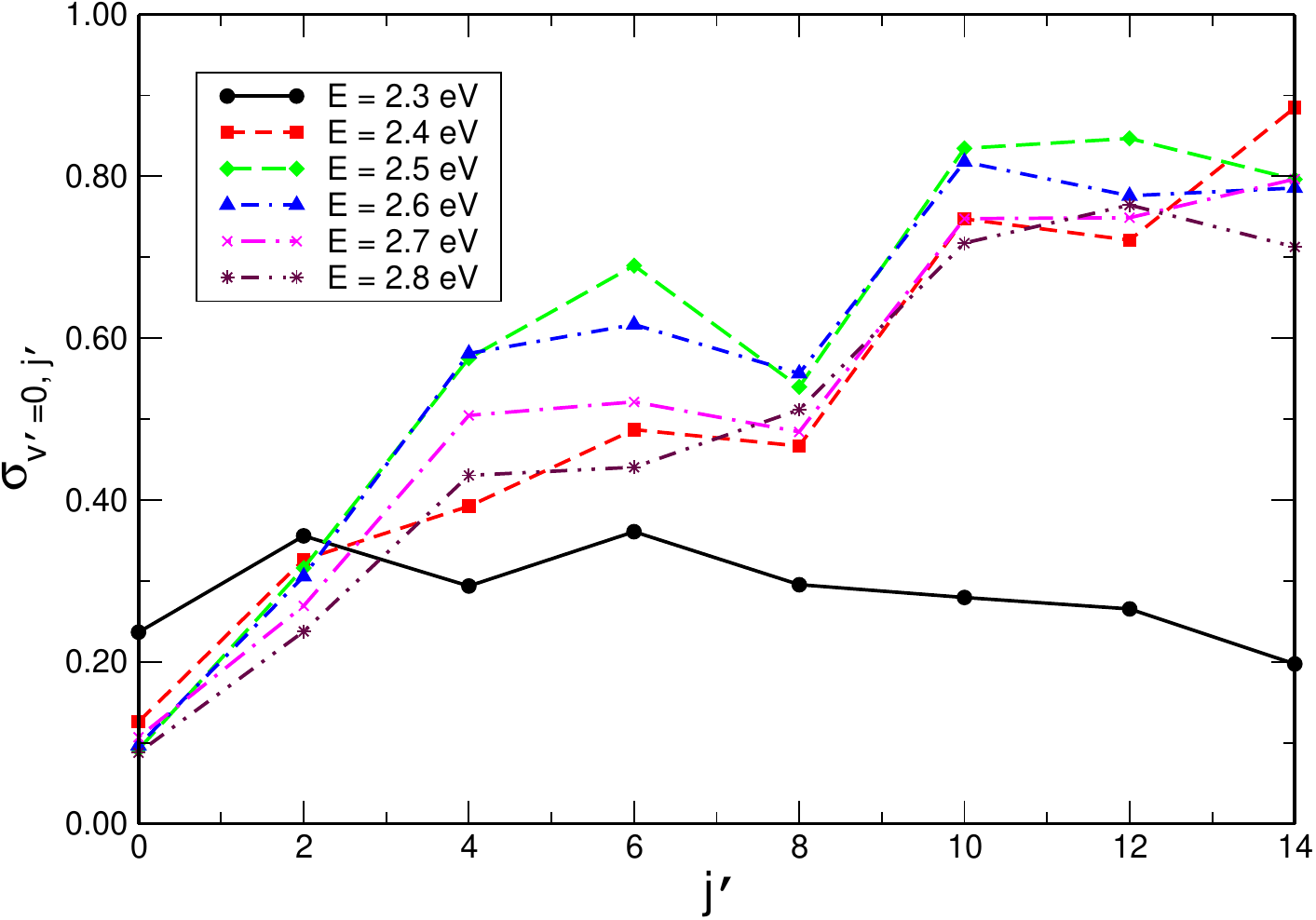}}\hfill
	\subfloat[\label{fig:vprime3}]{\includegraphics[width=0.5\linewidth,angle=0]{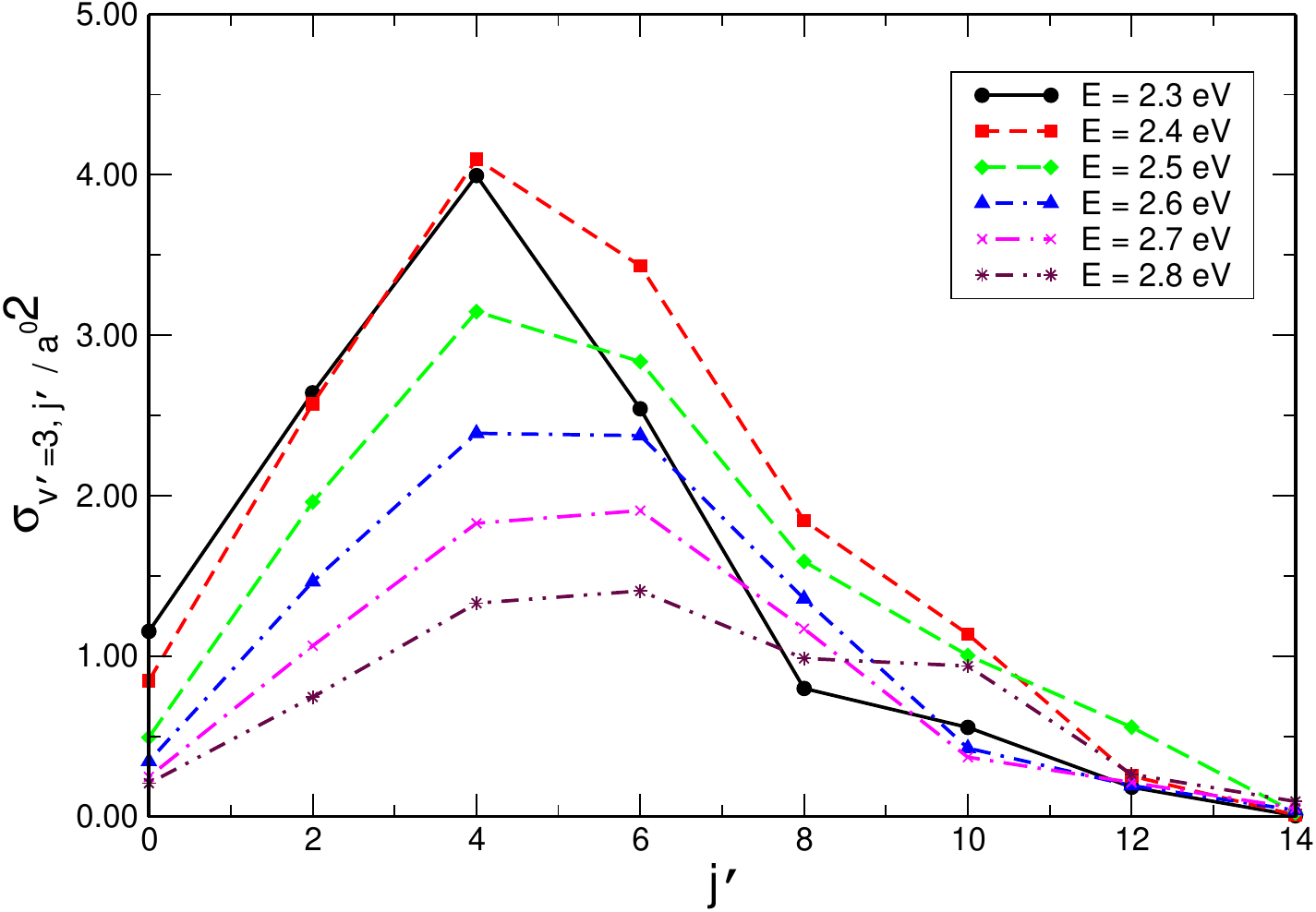}}\\
	\subfloat[\label{fig:vprime4}]{\includegraphics[,width=0.5\linewidth,angle=0]{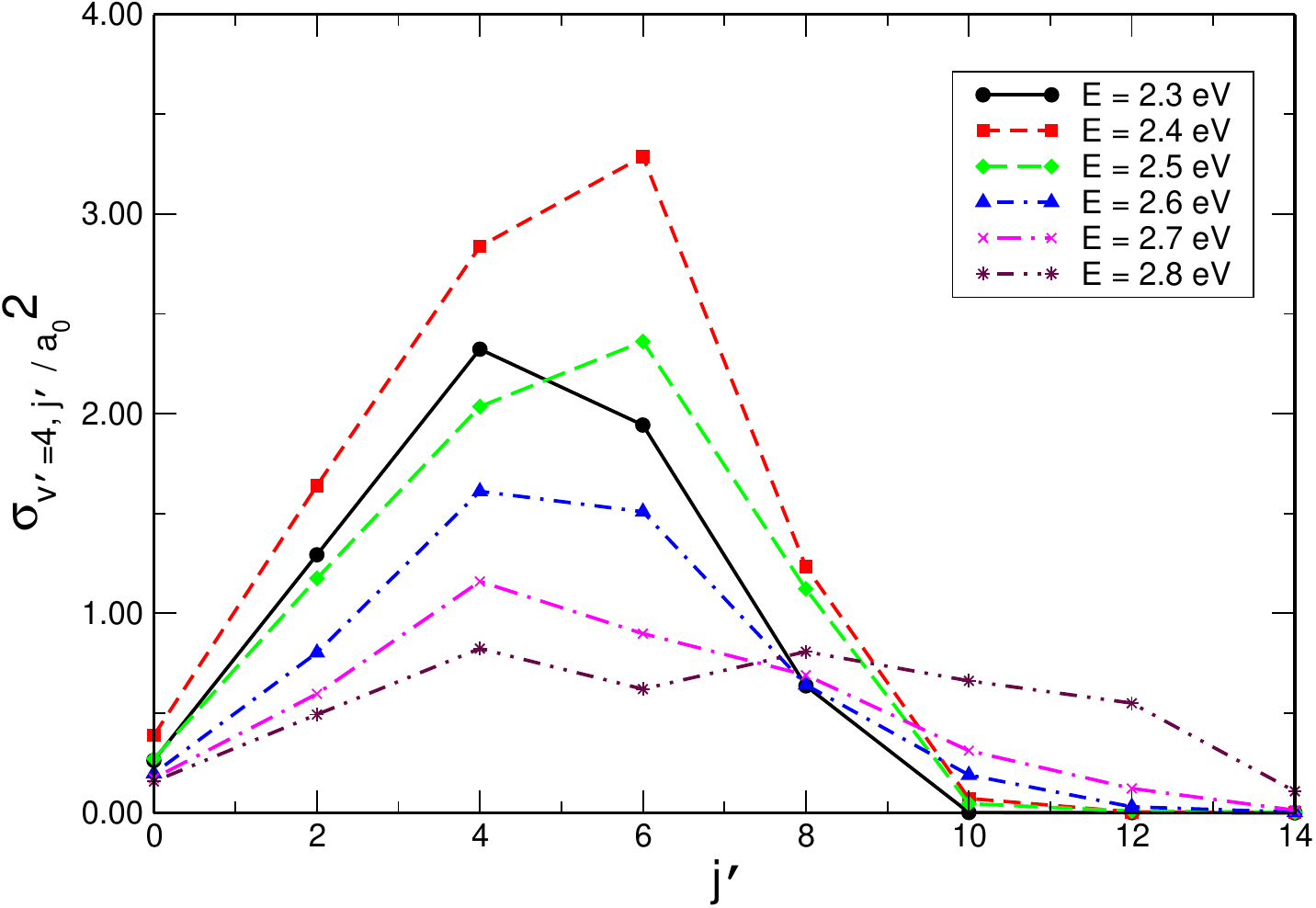}}\hfill
	\subfloat[\label{fig:vprime5}]{\includegraphics[width=0.5\linewidth,angle=0]{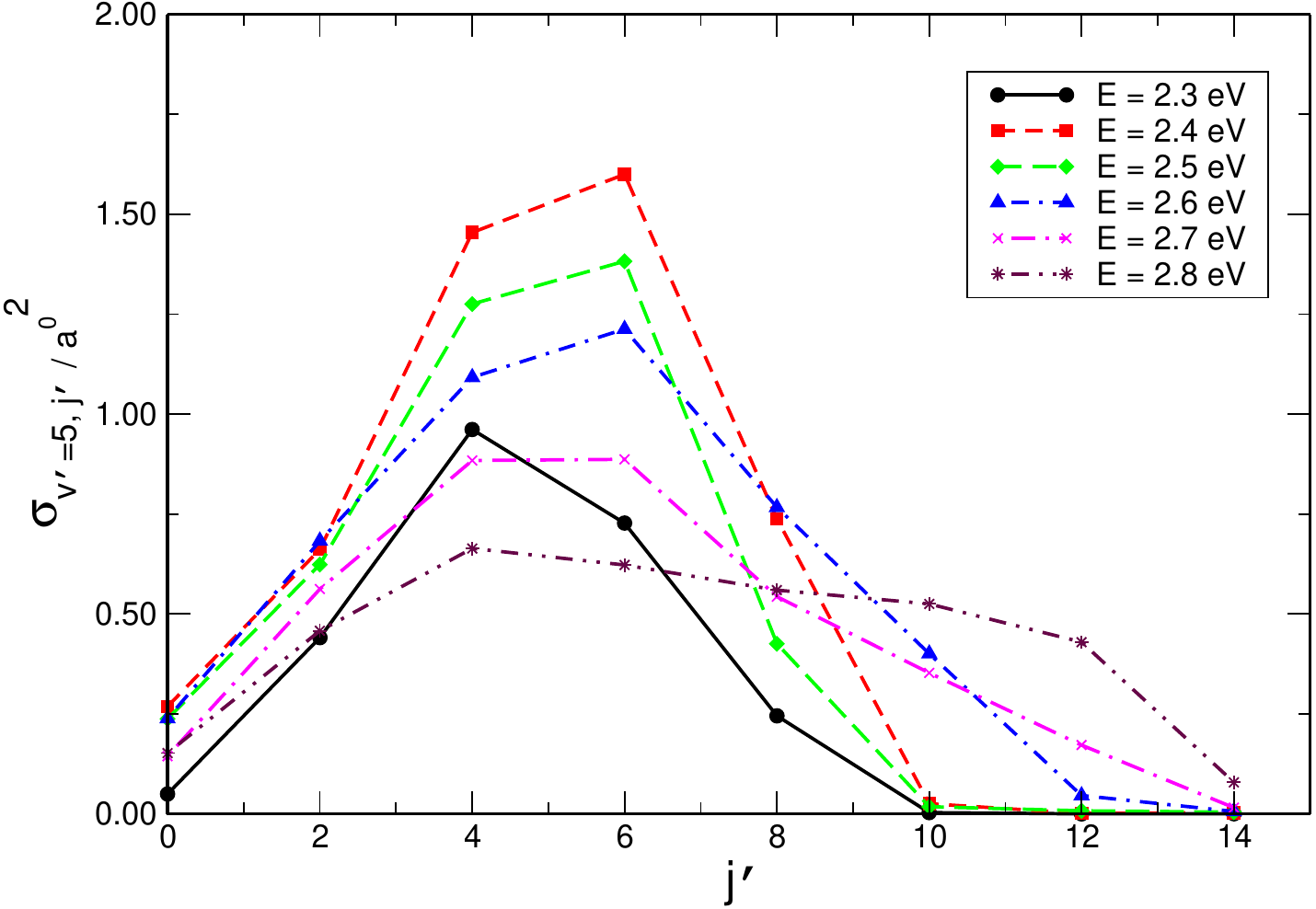}}\\
	\caption{State-to-state integral cross-section for \textit{v$^\prime$} = 0, 3, 4 and 5 (a)-(d) are plotted as function of final rotational states (\textit{j$^\prime$}) for the NRCT process of H + H$_2^+$ ($v=0$; $j=0$) reaction. The figure clearly demonstrates the enhancement in cross-section for  v$^\prime$ =3 and 4. The figures are taken from Ref.~\citenum{H3+_dynamics_2021} with the permission of ACS.}
	\label{fig:vib_enh_H3+}
\end{figure}

\begin{figure}[!htp]
	\centering
	\includegraphics[height=0.65\textheight,width=0.5\linewidth,angle=0]{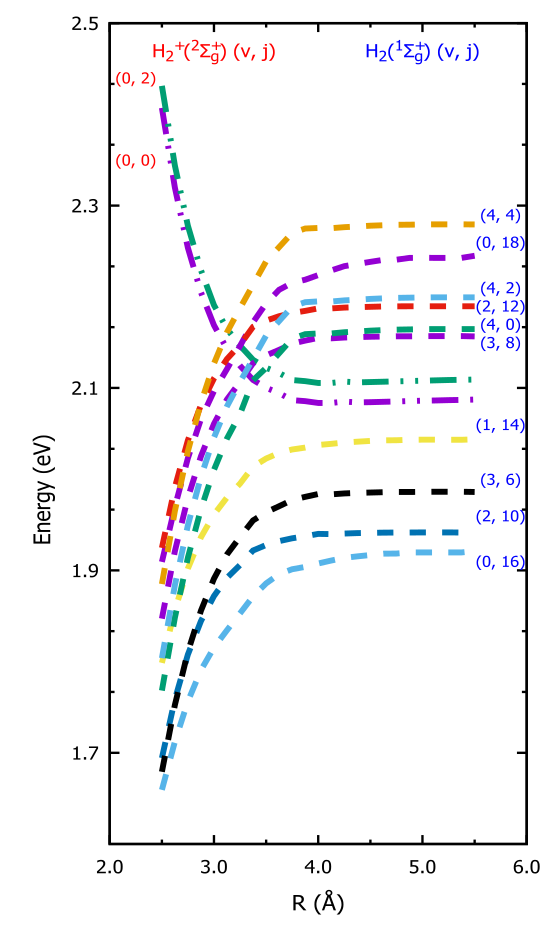}
	\caption{The diagram shows curves depicting the electronic-vibrational-rotational eigenvalue as function of the distance (R) of the third atom from the centre of mass of the diatom, computed using adiabatic PESs. The dashed-dotted lines represent eigenstates of H$_2^+$ ($^2\Sigma_g^+$) while the dashed lines signify H$_2$ ($^1\Sigma_g^+$) eigenstates. The figures are adapted from Ref.~\citenum{H3+_dynamics_2021} with the permission of ACS.}
	\label{fig:eigen_H3+}
\end{figure}

\noindent
Figure \ref{fig:ICS} depicts the variation of theoretically calculated ICSs for the charge transfer processes in comparison with other theoretical outcomes\cite{jcp-107-1451-LAST} as function of total energy. Once the ICSs are in hand, those are employed to compute the temperature dependent rate constant, where the values are presented in Table \ref{tab:mult_temp_rate}. The average rate constant over the range of temperature (250-1000 K) exhibits reasonable agreement with the available experimental data\cite{jcp-70-2877-KARPAS}. When the excited state species [H$_2^+$ ($^2\Sigma_g^+$)] relaxes and come down to ground state [H$_2$ ($^1\Sigma_g^+$)], calculated cross-section for NRCT process shows substantial incorporation of nonadiabatic effects.\cite{H3+_dynamics_2021} On the other hand, 
Figure \ref{fig:vib_enh_H3+} shows enhanced cross-section for the intermediate vibrational states ($v^\prime = 3$-4). The effective overlap of electronic-vibrational-rotational eigenvalues between $v^\prime = 3$, 4 levels of the ground electronic state of H$_2$ and  those of the H$_2^+$ species ($v^\prime = 0$) is mainly responsible for such enhancement, which is demonstrated as function of $R$ in Figure \ref{fig:eigen_H3+}.

\begin{table}[!htp]
	\Large
	\centering
		\caption{Rate constants calculated from theoretical study\cite{H3+_dynamics_2021} are summarized in the temperature range 250K - 1000K. Data are taken from Ref.~\citenum{H3+_dynamics_2021} with the permission of ACS.}
	\begin{tabular}{||c|c|c|c||}
		\hline \hline \hline
		\shortstack{Temperature \\ (K)} & \shortstack{$k(T)$ / \\ cm$^3$molecule$^{-1}$sec$^{-1}$ \\ (Present calculation)} & \shortstack{Temperature \\ (K)} & \shortstack{$k(T)$ / \\ cm$^3$molecule$^{-1}$sec$^{-1}$ \\ (Present calculation)} \\
		\hline \hline
		250 & 1.19 $\times$ 10$^{-10}$ & 650 & 9.81 $\times$ 10$^{-10}$ \\ 
		\hline \hline
		300 & 2.04 $\times$ 10$^{-10}$ & 700 & 10.85 $\times$ 10$^{-10}$ \\ 
		\hline \hline
		350 & 3.05 $\times$ 10$^{-10}$ & 750 & 11.82 $\times$ 10$^{-10}$ \\ 
		\hline \hline
		400 & 4.15 $\times$ 10$^{-10}$ & 800 & 12.75 $\times$ 10$^{-10}$ \\ 
		\hline \hline          
		450 & 5.30 $\times$ 10$^{-10}$ & 850 & 13.62 $\times$ 10$^{-10}$ \\ 
		\hline \hline
		500 & 6.46 $\times$ 10$^{-10}$ & 900 & 14.43 $\times$ 10$^{-10}$ \\ 
		\hline \hline
		550 & 7.61 $\times$ 10$^{-10}$ & 950 & 15.19 $\times$ 10$^{-10}$ \\ 
		\hline \hline
		600 & 8.73 $\times$ 10$^{-10}$ & 1000 & 15.90 $\times$ 10$^{-10}$ \\ 
		\hline \hline \hline
		\multicolumn{2}{||c|}{\shortstack{Average \\ (Present Calculation)}} & \multicolumn{2}{c||}{8.93 $\times$ 10$^{-10}$}  \\ 
		\hline \hline   
		\multicolumn{2}{||c|}{Experiment} &  \multicolumn{2}{c||}{(6.4 $\pm$ 1.2) $\times$ 10$^{-10}$} \\ 
		\hline     
	\end{tabular}
	\label{tab:mult_temp_rate}
\end{table}


\subsection{He + H$_2^+$ reaction}
\begin{figure}[!htp]
	\centering
ch 	\includegraphics[width=1.0\linewidth]{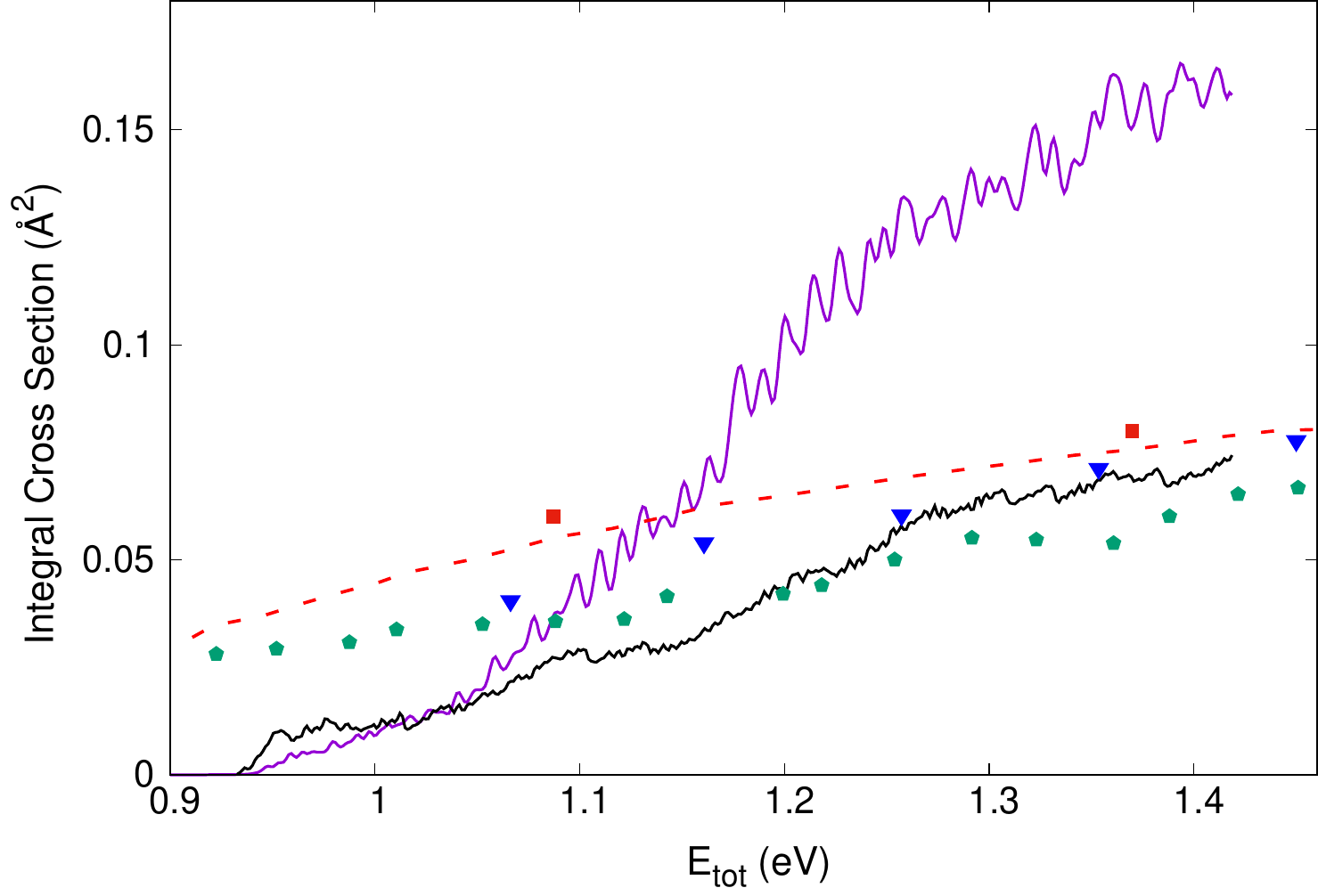}
	\caption{Total integral cross-section for the He+H$ _2^+(v=0,j=0) \rightarrow $ HeH$ ^+ $ +H reaction is plotted as
		function of total energy obtained from the ground electronic adiabatic PES (\protect\includegraphics[scale=1]{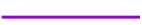})
		as well as the 2-state diabatic PESs (\protect\includegraphics[scale=1]{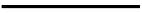}) compared with the available experimental results:
		(\protect\includegraphics{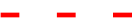}) Chupka and Russell (Expt.), \cite{chupka1968_1, chupka1968_2} 
		(\protect\includegraphics{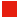}) Turner et al. (Expt.) \cite{Turner1984_JCP_81}, 
		(\protect\includegraphics{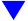}) Achtenhagen et al. (Expt.) \cite{Achtenhagen}, 
		(\protect\includegraphics{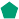}) Tang et al. (Expt.) \cite{Tang2005_JCP_122} The figure is adapted from Ref.~\citenum{HeH2+_dynamics_2023} and permission is taken from AIP.
	}
	\label{fig:cross_HeH2+}
\end{figure}

\noindent
The FCC-3D-TDQM wavepacket calculations\cite{cpc-184-270-SA} are performed to explore the proton-transfer reaction, He + H$_2^+$ ($v=0$; $j=0$) $\rightarrow$ HeH$^+$ + H using {\it ab initio} calculated ground adiabatic and two-state BBO based diabatic PES matrix for HeH$_2^+$ system~\cite{heh2+_surface_article} over the total energy range, 0.94~$\le$~E$_{\rm tot}$~$\le$~1.4~eV. When the total reaction probabilities are computed for various total angular momentum ($J$) on the adiabatic and diabatic PES(s), the cross-sections are evaluated for both the cases (adiabatic and diabatic) over the converged range of total energy. Figure \ref{fig:cross_HeH2+} demonstrates that the calculated ICSs on the adiabatic PES is slightly lower than that obtained from the diabatic one around the threshold energies (0.94 $\le$ E$_{tot}$ $\le$ 1.05 eV), but for $E_{\text{tot}} \ge 1.1$ eV onwards, adiabatic cross-section acquires higher values compared to the diabatic case as well as the experimental ones\cite{chupka1968_1, chupka1968_2, chupka1972, Achtenhagen, Turner1984_JCP_81, Tang2005_JCP_122}. In the total energy range, 1.1-1.4 eV, diabatic calculation based cross-section profile aligns more closely with experimental observations\cite{chupka1968_1, chupka1968_2, chupka1972, Achtenhagen, Turner1984_JCP_81, Tang2005_JCP_122} in comparison with the corresponding adiabatic results. In other words, contribution of nonadiabatic couplings among the electronic states is significant for 
construction of diabatic Hamiltonian vis-\'a-vis reaction dynamics.


\subsection{H + D$_2$/H$_2$ reaction}

\begin{figure}[!htp]
	\centering
	\includegraphics[height=0.32\textheight,width=.8\textwidth]{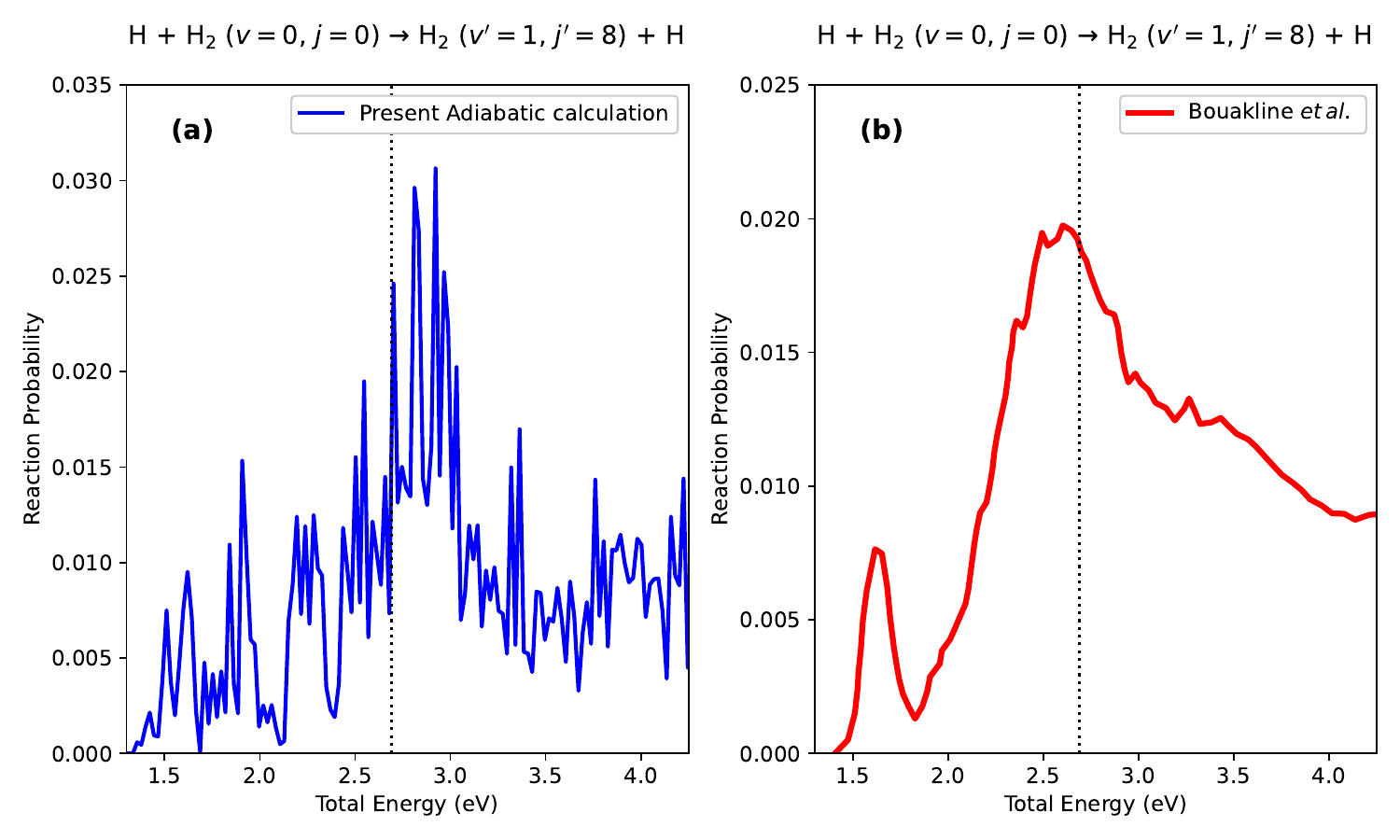}\hfill\\
	\caption{Ro-vibrationally resolved reaction probabilities at \textit{J} = 0 are plotted, which are obtained through adiabatic PES: (a) present; (b) Bouakline \textit{et al.} (see bottom right panel of FIG.~2 in Ref.\citenum{Althorpe_2008}). 
		The vertical dotted line signifies the minimum energy of the 12-D$_{3h}$ CI. The figuers are adapted from Ref.\citenum{Ghosh_H3_2025} with the permission of ACS. }
	\label{fig:reaction_adia_v1_j8}
\end{figure}

\begin{figure}[!htp]
	\centering
	\includegraphics[height=0.32\textheight,width=.8\textwidth]{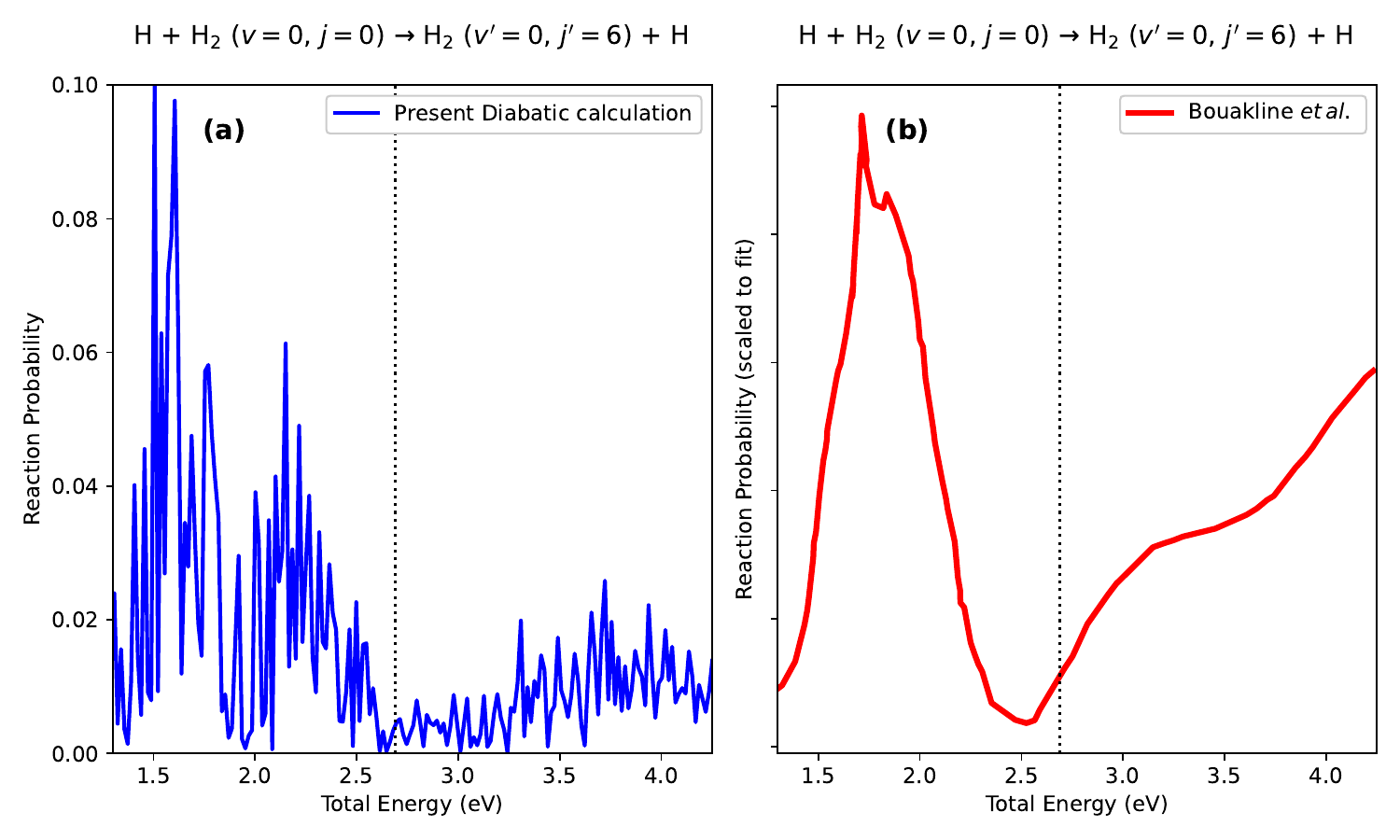}\hfill
	\caption{Ro-vibrationally resolved reaction probabilities at \textit{J} = 0 are presented, which are originated from diabatic PESs: 
		(a) present; (b) Bouakline \textit{et al.} (see top left panel of FIG.~4 in Ref.\citenum{Althorpe_2008}). 
		The vertical dotted line symbolizes the minimum energy of the 12-D$_{3h}$ CI. The figures are taken from Ref.\citenum{Ghosh_H3_2025} with the permission of ACS.}
	\label{fig:reaction_dia_v0_j6}
\end{figure}

\begin{figure}[!htp]
	\centering
	\includegraphics[width=.5\textwidth]{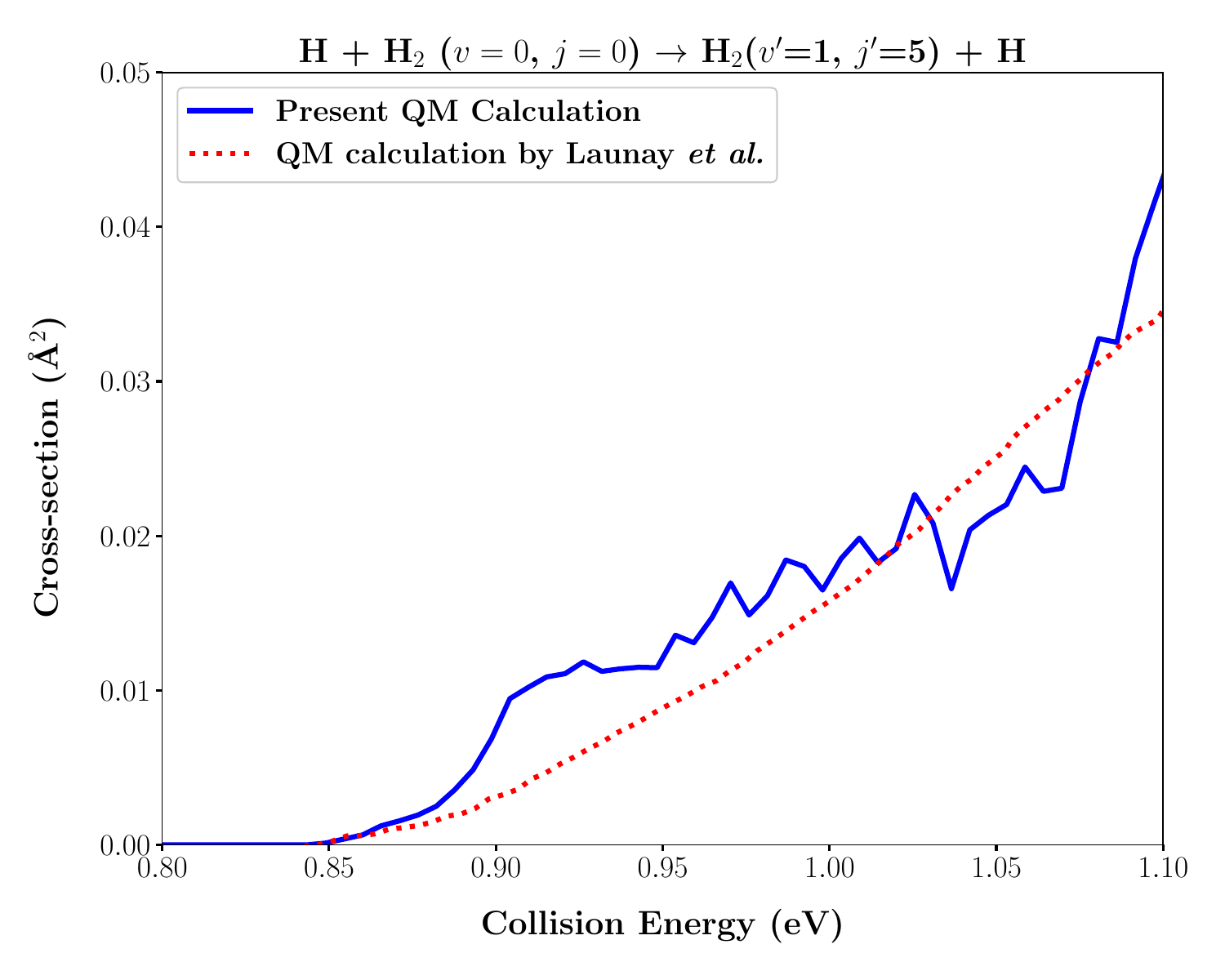}\hfill \\
	\caption{Ro-vibrationally resolved integral cross-sections obtained by adiabatic calculations are compared with other theoretical result of Launay \textit{et al.}\cite{Launay_1989}. Figure is adapted from Ref.\citenum{Ghosh_H3_2025} with permission from ACS.}
	\label{fig:reaction_cs_v1_j5}
\end{figure}

\noindent
3D-TDWP quantum dynamics is carried out on the \textit{ab initio} calculated adiabatic PESs as well as the constructed multi-state diabatic PES matrix to compute the state-resolved reaction probabilities at $J$ = 0 for the H + H$_2$ ($v=0; j=0$) $\rightarrow$ H$_2$ ($v^\prime$; $j^\prime$) + H reaction over the total energy range, 0.6 $\le$ E$_{tot}$ $\le$ 4.5 eV. Those probabilities originated from adiabatic and diabatic PESs matrices are compared with earlier theoretical results\cite{Althorpe_2008} in Figures  \ref{fig:reaction_adia_v1_j8} and \ref{fig:reaction_dia_v0_j6}, respectively. The observed differences between the two sets of calculations (adiabatic and diabatic) are mainly governed by the explicit incorporation of nonadiabatic effects in the diabatic Hamiltonian over adiabatic one apart from accuracy of the newly calculated adiabatic and constructed diabatic PESs. Once the reaction probabilities are calculated from adiabatic calculations, state-resolved integral cross-sections are evaluated for H + H$_2$ ($v = 0; j = 0$) $\rightarrow$ H$_2$ ($v^\prime$ = 1; $j^\prime$ = 5) + H reaction with the aid of FCC-TDWP approach, where the probabilities are fully converged at $J$ = 50 over the energy range, 0.34 $\le$ E$_{col}$ $\le$ 1.1 eV. Figure \ref{fig:reaction_cs_v1_j5} shows the presently calculated ICS profiles in comparison with other theoretical result\cite{Launay_1989} for different collision energy upto 1.1 eV exhibiting good accord with each other. On the other hand, the role of nonadiabatic coupling in the reaction, H + D$_2 $($v=0; j=0$) $\rightarrow$ HD ($v^\prime$; $j^\prime$) + D at relatively higher collision energies (0.4~eV $\leq E_{\mathrm{col}} \leq$ 3.26~eV) is ongoing on the BBO based diabatic Hamilonian\cite{Ghosh_H3_2025}. The evaluated cross-section and rate constants will be compared with the existing theoretical predictions~\cite{Althorpe_2015_DCS,zare_2005} and available experimental findings~\cite{zare_2005,Pomerantz_2004_ref_20}.

\section{Conclusion}

\noindent
In the last few decades, reaction attributes (cross-sections and rate constants) of triatomic chemical reactions are computed including 
electron-nuclear couplings among the low-lying electronic states, where the CIs are analyzed and diabatic Hamiltonian is constructed by BBO treatment. Once the BBO based diabatic PESs and couplings are in hand, 
reactive scattering theory is implemented for quantum dynamics calculations. 

\vspace*{0.3 cm}

\noindent
The presence of CIs between two electronic states can be explored 
by closed contour integration for different triatomic reactive species.
For FH$_2$ system, one can locate the presence of a JT CI between the ground state (1$^{2}A'$) and the first excited state (2$^{2}A'$), together with additional RT CIs/seams between 2$^{2}A'$ and 1$^{2}A''$ states using in-plane and out-of-plane contours.\cite{FH2_Surf} On the other hand, H$_3^{+}$ system exhibits a semi-circular accidental CI seam
propagating through the $C_{2v}$ geometry between the $1^{1}A'$ and $2^{1}A'$ states, whereas the symmetry driven CI can be observed between the $2^{1}A'$-$3^{1}A'$ states at D$_{3h}$ geometries.\cite{H3+_Surf,joy_pccp} In a similar way, HeH$_2^+$ system shows CIs between the ground (1$^2$A$^\prime$) and first excited (2$^2$A$^\prime$) electronic states at C$_{2v}$ geometries and also between the first (2$^2$A$^\prime$) and second (3$^2$A$^\prime$) as well as second (3$^2$A$^\prime$) and third (4$^2$A$^\prime$) excited electronic states at linear geometries.\cite{HeH2_Surf} On the other hand, H$_3$ exhibits both the symmetry driven and accidental CIs/seam between the 1$^2$A$^\prime$ and 2$^2$A$^\prime$ states as well as the accidental C$_{2v}$ CI seam between the $2^{2}A'$-$3^{2}A'$ states.\cite{Ghosh_H3_2025}

\vspace{0.3 cm}

\noindent
Once the \textit{ab initio} calculated adiabatic PESs and NACTs are in hand, ADT equations are solved and the resulting (ADT) angles are employed in the orthogonal transformation to construct the diabatic Hamiltonian. In other words, employing Eqs.~\ref{eq:diapot1}, \ref{eq:adteq}, \ref{eq:adt-two}, \ref{eq:adt-three} as well as Appendix \ref{ssec:adt-eq}, ADT angles are computed and thereafter, multi-state diabatic Hamiltonians are generated for the representative triatomic systems such as FH$_2$\cite{FH2_Surf}, H$_3^+$~\cite{H3+_Surf}, HeH$_2^+$~\cite{HeH2_Surf} and H$_3$~\cite{Ghosh_H3_2025}. It is worth mentioning that the construction of diabatic PESs~\cite{FH2_Surf,H3+_Surf,HeH2_Surf,Ghosh_H3_2025} as well as reactive scattering calculations~\cite{cpc-184-270-SA,FH2_dynamics,HeH2+_dynamics_2023,jcp-147-074105-SA,H3+_dynamics_2021} are performed in hyperspherical coordinates~\cite{Johnson1983_JCP_79}, where all the channels (reactive and non-reactive) are being taken even-handedly leading to the reduction of the computational cost for electronic structure as well as scattering calculation.  

\vspace{0.2 cm}

\noindent
Once the multi-state diabatic Hamiltonians are obtained, fully close coupled 3D-TDWP quantum-dynamics calculations are carried out to investigate the triatomic reactions, F + H$_2$/D$_2$, D$^{+}$ + H$_2$, H + H$_2^+$, He + H$_2^+$, and H + H$_2$/D$_2$, which capture the influence of NACTs on reaction probabilities, cross-sections and rate constants. For the F + H$_2$ ($v = 0; j = 0$) reaction, the computed rate constants on diabatic PESs including both nonadiabatic and SO couplings~\cite{FH2_dynamics,FH2_Surf} show good accord with experiment~\cite{bedjaninan} across the temperature range (200K - 600K).~\cite{FH2_dynamics} On the other hand, in case of D$^{+}$ + H$_2$ ($v = 0; j = 0$) collision, ICSs (RNCT and RCT processes) originated from the multi-state diabatic Hamiltonian exhibit excellent agreement with experimental findings\citep{cp-111-401-SCH} over the entire collision energy range (1.7-2.5 eV) in comparison to other theoretical ones\cite{jpca-119-12392-SA,viegas_jcp126,cpl-471-222-HAN3,jpcA-109-2050-HAN1,jcp-116-654-KAM}. Such findings emphasizes the crucial role of accurately including NACTs in the diabatic Hamiltonian.\cite{jcp-147-074105-SA} While initiating the charge transfer reaction for the H + H$_2^+$ ($v = 0; j = 0$) system from the first excited electronic state ($2^{1}A'$), the calculated ICSs ~\cite{H3+_dynamics_2021} for the competing RCT and NRCT processes match well with other theoretical results\cite{jcp-107-1451-LAST}. Moreover, the rate constant averaged over the temperature range (250 K-1000 K) shows excellent correspondence with the experimental outcomes~\cite{jcp-70-2877-KARPAS}. For the He + H$_2^+$ ($v$ = 0; $j$ = 0) reaction, calculated cross-section obtained from two-state diabatic Hamiltonian~\cite{HeH2_Surf} follow the experimental trend~\citep{chupka1968_1, chupka1968_2} much better over the energy range, 0.9 - 1.4~eV compared to other theoretical calculations as well as with the results appeared from the same ground adiabatic surface. Since the nonadiabaticity increase the endothermicity of the product channel, the agreement with the experiment appear very close with each other. For H + H$_2$ ($v=0$; $j=0$) $\rightarrow$ H$_2$ ($v'$; $j'$) + H reaction, the recent \textit{ab initio} constructed single-state adiabatic and multi-state diabatic PES matrix are used to perform scattering calculations at $J$ = 0 for evaluating state-resolved reaction probabilities over the total energy range, 0.6 - 4.5~eV. Such findings show substantial difference from the earlier work~\cite{Althorpe_2008} due to the explicit involvement of nonadiabatic couplings. The state-resolved cross-sections obtained from single state adiabatic PES are presented with other existing theoretical calculation\cite{Launay_1989}, where the ICS profiles are in good agreement with each other. At present, the investigation of H + D$_2$ ($v=0$; $j=0$) $\rightarrow$ HD ($v'$; $j'$) + D reaction is ongoing over the collision energy range, 0.4 - 3.0~eV with the same diabatic Hamiltonian~\cite{Ghosh_H3_2025} to compute ICSs as well as rate constant. In future, such results will be used for comparison with the available theoretical and experimental data~\cite{Pomerantz_2004_ref_20,zare_2005,Mielke_Sutherland_2003}.

\vspace{0.3 cm}

\noindent
In summary, workability of BBO based \textit{ab initio} constructed diabatic PES matrix followed by quantum dynamics and the underlying role of nonadiabatic coupling are successfully explored for triatomic reactive systems such as F + H$_2$, D$^+$ + H$_2$, He + H$_2^+$ and H + D$_2$. Such theoretical unveils how nonadiabatic effects (for some cases, SO couplings) reshape the ICS profiles exhibiting much closer agreement with experimental data.

\bibliographystyle{aipnum4-1}
\bibliography{ref_book}

\chapter{Classical Trajectory Theory for Multi-Surface Systems}\label{sec:traj}

\noindent
Along with the quantum dynamics, semi-classical trajectory surface hopping (TSH) dynamics were introduced and developed parallely, specifically to monitor the 
triatomic, tetratomic and gas-surface chemical reactions. The TSH approach was first introduced by Tully and Preston~\citep{TULLY:JCP55}, which is based on the semiclassical dynamics of Bjerre and Nikitin on Na + N$_2$~\citep{nikitin_cpl1} reaction involving transition probabilities computed by Landau-Zener treatment. On the other hand, a wide variety of theoretical developments were carried out in the arena of TSH approach.~\cite{muckerman_jcp65,tully_jcp93,mclafferty_jcp79,ali_jcp78,kuntz_jcp95,tully_jcp101,rothlisberger_prl98,werner_cp349,martinez_jcp154,klos_jcp142} It is worth mentioning that the TSH formalism was primarily developed for the nonadiabatic dynamics of coupled states within the same spin multiplicity as well as
different multiplicities exhibiting prominent spin-orbit effects.~\citep{karplus_cpl32,yoshida_jpca101,schatz_jpca112,gross_prb81} In order to analyze laser-catalyzed reactions and laser-induced dissociations, the TSH approach was extended for the dressed electronic states by an electric field, whereas such field-induced TSH approach was implemented to a variety of systems.~\citep{morokuma_cpl169,voth_jap71,zerbetto_jpca112,mitric_pra79,herman_cp183} Adhikari and coworkers~\citep{soumya_hopp} 
implemented Tully's fewest switch algorithm to perform TSH dynamics on D$^{+}$ + H$_{2}$ and H + H$_{2}^{+}$ reactions
using the {\it ab initio} (MRCI) calculated adiabatic PESs and NACTs~\citep{mukherjee_jcp141} of H$_3^+$ for the first time. It is necessary to mention that 
the energy conservation of such formalism was guaranteed with the aid of kinetic energy adjustment algorithm of Thiel {\it et al.}~\citep{thiel_cp349} Furthermore,
a new set of equations were introduced to analyze the momentum conservation during the hopping process.~\citep{soumya_hopp} The TSH based cross-sections for reactive non-charge transfer (RNCT) agree quite well with quantum mechanical results~\citep{jcp-147-074105-SA,ghosh_jpca125} and experimental measurements~\citep{cp-111-401-SCH,karpas_jcp70}, but substantial deviation can be observed for the excited state cross-sections, namely, reactive charge transfer (RCT) and non-reactive charge transfer (NRCT) processes. Later on, a different approach of semi-classical dynamics~\citep{friction_jpca128} were proposed, where quantum mechanical NACTs were explicitly employed as equivalent to friction in classical mechanics. In order to explore the validity of such newly developed theory, the (D$^+$, H$_2$) collision dynamics were investigated over the ground and first excited PESs of H$_3^+$ system,~\citep{mukherjee_jcp141} which demonstrated
slowing down of the particles with the formation of the triatomic species, DH$_2^+$ as well as other products (RNCT, RCT and NRCT).~\citep{friction_jpca128}     
Even though three different forms of friction terms (classical, semi-classical and quantum frictions) were explored in these calculations, the results of quantum friction show satisfactory accord with the outcome of full quantum dynamics~\citep{jcp-147-074105-SA} and experimental findings~\citep{cp-111-401-SCH}. 

\section{Trajectory Surface Hopping Approach}	
\noindent
In this section, theoretical background of TSH approach~\citep{TULLY:JCP55,klos_jcp142} as well as energy~\citep{thiel_cp349} and momentum~\citep{soumya_hopp} conservation of the trajectories along with their implementation~\citep{soumya_hopp} on D$^{+}$ + H$_{2}$ and H + H$_{2}^{+}$ reactions are presented. Furthermore, the outcome of TSH calculations are compared with the results of
Fully Closed Coupled (FCC) 3D wavepacket based quantum results~\citep{jcp-147-074105-SA,ghosh_jpca125} and experimental findings~\citep{cp-111-401-SCH}.

\subsection{Initialization of Positions and Momenta}\label{ssec:ini}
\noindent
In case of the triatomic reactive system, D$^{+}$ + H$_{2}$ or H + H$_{2}^{+}$, the initial positions and momenta are determined as prescribed by Porter {\it et al.}~\citep{miller_jcp63}
for a rotating Morse oscillator. While initializing the trajectories, two sets of coordinates and their associated momenta are required for the diatom (H$_{2}$ or 
H$_{2}^{+}$) and the triatomic species (D$^{+}$ + H$_{2}$ or H + H$_{2}^{+}$). The diatom (H$_{2}$ or H$_{2}^{+}$) of length $r$ can be treated as a rotating Morse oscillator leading to the first set of initial parameters ($r_x$, $r_y$, $r_z$, $p_x$, $p_y$ and $p_z$). In a similar way, the third atom (D$^{+}$ or H) and the 
center of mass of diatom form another rotating Morse oscillator of length $R$ with another set of coordinates and momenta ($R_x$, $R_y$,
$R_z$, $P_x$, $P_y$, $P_z$). The brief outline of the initialization is depicted in the following paragraphs, where the formalism is thoroughly discussed elsewhere.~\citep{miller_jcp63,soumya_hopp}\\

\noindent
While initializing the position and momentum of the diatom with specific roto-vibrational states and the incoming atom with particular kinetic energy, angular representation of 
the coordinates and momenta of the triatomic species can be conveniently used as random distribution in the configuration space (CS) with a constant total energy of the triatomic species. Therefore, a canonical transformation is needed to obtain another set of positions and momenta ($Q_{k}^r$, $P_{k}^r$, $k=1,2,3$) [$P_k^r$ s are constants of motion], which can be used to redefine the polar coordinates of the diatom ($r$, $\Theta$, $\Phi$) and their associated momenta ($p_r$, $p_{\Theta}$, $p_{\Phi}$). $P_k^r$ s are invariant with time and therefore, time-dependence in the old coordinates ($r$, $\Theta$, $\Phi$, $p_r$, $p_{\Theta}$, $p_{\Phi}$) appears
only through the canonically transformed $Q_{k}^r(t)$ ($Q_{k}^r$, $P_{k}^r$ $\rightarrow$ $r$, $\Theta$, $\Phi$, $p_r$, $p_{\Theta}$, $p_{\Phi}$). The polar coordinates ($r$, $\Theta$ and $\Phi$) and their momenta ($p_r$, $p_\Theta$ and $p_\Phi$) are transformed into the corresponding Cartesian analogues [$r_x$, $r_y$, $r_z$, $p_x$, $p_y$ and $p_z$] through the following matrix equations:~\citep{miller_jcp63,soumya_hopp}
\begin{eqnarray}
	\begin{pmatrix}
		r_x\\
		r_y\\
		r_z\\
	\end{pmatrix} =r
	\begin{pmatrix}
		-\sin q_2^r \sin Q_3^r + \lambda\cos q_2^r \cos Q_3^r\\
		\sin q_2^r \cos Q_3^r + \lambda \cos q_2^r \sin Q_3^r\\
		\sqrt{1-\lambda^2}\cos q_2^r\\
	\end{pmatrix}
	\label{eq:cartin}
\end{eqnarray}

\begin{eqnarray}
	\begin{pmatrix}
		p_x\\
		p_y\\
		p_z\\
	\end{pmatrix} = p_r 
	\begin{pmatrix}
		-\sin q_2^r\sin Q_3^r + \lambda \cos q_2^r \cos Q_3^r \\
		\sin q_2^r\cos Q_3^r + \lambda \cos q_2^r \sin Q_3^r \\
		\sqrt{1-\lambda^2}\cos q_2^r \\
	\end{pmatrix}
	+\dfrac{L}{r}
	\begin{pmatrix}
		-\cos q_2^r\sin Q_3^r - \lambda\sin q_2^r\cos Q_3^r \\
		\cos q_2^r\cos Q_3^r - \lambda\sin q_2^r\sin Q_3^r \\
		-\sqrt{1-\lambda^2}\sin q_2^r
	\end{pmatrix}, \qquad
	\label{eq:momin}
\end{eqnarray}
where $q_2^r=Q_2^r+L\Delta_{L}$ ($L$ signifies the angular momentum of the diatom) and $\lambda = P_{3}^r/P_{2}^r$. Similar formalism is equally applicable 
for the other set of coordinates ($R_x$, $R_y$, $R_z$, $P_x$, $P_y$, $P_z$). In order to assign the initial values
of the twelve quantities ($\lbrace$ $r_x$, $r_y$, $r_z$, $p_x$, $p_y$, $p_z$ $\rbrace$ and $\lbrace$ $R_x$, $R_y$, $R_z$, $P_x$, $P_y$, $P_z$ $\rbrace$), random numbers are used. \\

\noindent
The Miller's approach as discussed above is utilized to compute the initial values of coordinates ($r_x$, $r_y$, $r_z$, $R_x$, $R_y$ and $R_z$) and momenta ($p_x$, $p_y$, $p_z$, $P_x$, $P_y$ and $P_z$) while executing classical trajectory based calculations. It is evident from Eqs. \ref{eq:cartin} and \ref{eq:momin} that the initial parameters ($r_x$, $r_y$, $r_z$, $p_x$, $p_y$ and $p_z$) depend on one radial coordinate ($r$) and two angular
coordinates ($q_2^r$ and $Q_3^r$). The radial coordinate $r$ and the associated momentum ($p_r$) 
are functions of another angular coordinate ($Q_1^r$).~\citep{miller_jcp63,soumya_hopp} Furthermore, 
$\lambda$ and $\sqrt{1-\lambda^2}$ can be expressed as $\cos\theta_{arb}^r$ and $\sin\theta_{arb}^r$, where
$\theta_{arb}^r$ symbolizes an arbitrary angle. Overall, one has to choose four random numbers ($O_1$, $O_2$, $O_3$
and $O_4$) between zero (0) and $2\pi$ for those angles ($Q_1^r$, $q_2^r$, $Q_3^r$ and $\theta_{arb}^r$) to propose a set of initial values for the diatomic variables ($r_x$, $r_y$, $r_z$, $p_x$, $p_y$ and $p_z$). While initializing the second group of coordinates and momenta ($R_x$, $R_y$, $R_z$, $P_x$, $P_y$, 
$P_z$), another set of three random numbers ($O_5$, $O_6$ and $O_7$) are employed for the corresponding angular coordinates, $q_2^R$, $Q_3^R$ and 
$\theta_{arb}^R$. Since the trajectory needs to be started from the asymptote, a large value ($\sim$5 \AA) can be assigned for the $R$ coordinate instead of using 
any random number ($Q_1^R$). The expression of the associated momenta, $P_R$ takes the following form:  
\begin{equation}
	P_R = \sqrt{2E_{KE}\mu_{tri}},
\end{equation}
where $\mu_{tri}$ represents triatomic center of mass and $E_{KE}$ signifies the kinetic energy of the incoming atom.

\subsection{Trajectory Surface Hopping between Electronic States}\label{ssec:prop}

\noindent
Once the value of initial coordinates ($r_x$, $r_y$, $r_z$, $R_x$, $R_y$, $R_z$) and the associated momenta ($p_x$, $p_y$, $p_z$, $P_x$, $P_y$, $P_z$) are 
in hand, those are converted to 
the corresponding Johnson's hyperspherical coordinates~\citep{johnson_jcp79} to perform TSH dynamics. The classical Hamilton's equations $\Big[\dot{q}(t) =\frac{p(t)}{\mu}$ and $\dot{p}(t) = -\frac{du}{dq(t)}\Big]$ for triatomic Hamiltonian in terms of those coordinates ($\rho$, $\theta$, $\phi$, $\alpha$, $\beta$ and $\gamma$) and momenta ($P_\rho$, $P_\theta$, $P_\phi$, $P_\alpha$, $P_\beta$ and $P_\gamma$) can be written as:~\citep{johnson_jcp79}
\begin{subequations}
	\begin{eqnarray}
		\dot{\rho}&=&\dfrac{P_\rho}{\mu_t}\\
		\dot{\theta}&=&\dfrac{4P_\theta}{\mu_t\rho^2}\\
		\dot{\phi}&=&\dfrac{4[P_\phi-\dfrac{1}{2}P_\gamma\cos\theta]}{\mu_t\rho^2\sin^2\theta}\\
		\dot{\alpha}&=&\dfrac{2P_\alpha(1+\sin\theta\cos(2\gamma))}{\mu_t\rho^2\cos^2\theta}\\
		\dot{\beta}&=&\cos^{-1}(\dfrac{P_\gamma}{J})\\
		\dot{\gamma}&=&\dfrac{P_\gamma-2P_\phi\cos\theta}{\mu_t\rho^2\sin^2\theta}-\Big(\dfrac{2P_\gamma(1+\sin\theta\cos(2\gamma))}{\mu_t\rho^2\cos^2\theta}\Big)
	\end{eqnarray} 
	\label{eq:eom1}
\end{subequations}
\begin{subequations}
	\begin{eqnarray}
		\dot{P}_\rho &=&\dfrac{2}{\rho}\Big(H-\dfrac{P^2_\rho}{2\mu_t}-u_I\Big)-\dfrac{\partial u_I}{\partial\rho}\quad \quad I =1,2,3 \\
		\dot{P}_\theta &=& \dfrac{\cos\theta}{\mu_t\rho^2\sin^3\theta}(4P^2_\phi+ P^2_\gamma - 4P_\phi P_\gamma\cos\theta)-\dfrac{2P_\gamma P_\phi}{\mu_t\rho^2\sin\theta}\nonumber \\
		&&-\Big[\dfrac{2(P^2_\alpha - P^2_\gamma)\sin\theta}{\mu_t\rho^2\cos^3\theta}[1+\sin\theta\cos(2\gamma)]+\dfrac{(P^2_\alpha-P^2_\gamma)}{\mu_t\rho^2\cos\theta}\cos(2\gamma)\Big]-\dfrac{\partial u_I}
		{\partial\theta}\qquad \quad\\
		\dot{P}_\phi &=&-\dfrac{\partial u_I}{\partial\phi}\\
		\dot{P}_\alpha &=& 0\\
		\dot{P}_\beta &=& 0\\
		\dot{P}_\gamma &=& \dfrac{2(P^2_\alpha-P^2_\gamma)\sin\theta\sin(2\gamma)}{\mu_t\rho^2\cos^2\theta}
	\end{eqnarray}
	\label{eq:eom2} 
\end{subequations}

\noindent
where $\mu_t$ represents the triatomic reduced mass.\\

\noindent
Furthermore, while computing adiabatic population on various electronic states
($m$, $n$ $\in$ $o$), the following time-dependent Liouville equation~\citep{klos_jcp142} of density matrix ($s$) is used: 
\begin{equation}
	i\hbar\dot{s}_{mn}=\sum_o [s_{on}(U_{mo}-i\hbar\dot{\boldsymbol{C}}\boldsymbol{\tau}_{mo})-
	s_{mo}(U_{on}-i\hbar\dot{\boldsymbol{C}}\boldsymbol{\tau}_{on})],
	\label{eq:lio}
\end{equation} 
where $\boldsymbol{C}$ collectively signifies the hyperspherical coordinates ($\rho$, $\theta$ and $\phi$), 
$\boldsymbol{\tau}_{mo}$ or $\boldsymbol{\tau}_{on}$ present the NACM elements and 
$U_{mo}\,(=\langle m\vert\hat{H}_e\vert o\rangle)$ or $U_{on}\,(=\langle o\vert\hat{H}_e\vert n\rangle)$ signify the adiabatic PESs matrix elements, which are non-zero only 
for diagonal cases ($m=o$ or $o=n$). In order to perform TSH dynamics, {\it ab initio} calculated adiabatic PESs and NACTs are plugged into the above
equation (Eq. \ref{eq:lio}) and then, it is numerically integrated to calculate the density matrix ($s_{mn}$) as function of time. The resulting density matrix elements are 
further employed to evaluate the switching probability
($S_{mn}$) for transitions from initial ($n$) to final ($m$) state with the aid of the following empirical equation:~\citep{klos_jcp142}
\begin{equation}
	S_{mn} = -\dfrac{2\text{Re}(s_{mn}^*\dot{\boldsymbol{C}}\boldsymbol{\tau}_{mn})\Delta t}{s_{nn}}
	\label{eq:prob}
\end{equation}   
\noindent
If the calculated probability, $S_{mn}$ at any time step appears as negative, it is considered to be zero (0). On the other hand, since the maximum probability 
cannot exceed one (1), a random number $\zeta$ between zero (0) and one (1) is necessary to be invoked to dictate the hopping between any two electronic states. According to Eq. \ref{eq:prob}, the magnitude of switching probability becomes high at the close neighbourhood of degenerate point(s)/seam(s) due to singular features of
NACTs indicating highly probable transition between any two electronic states. If the condition $S_{mn}\,>\,\zeta$ is satisfied for two adjacent surfaces (as for example, $1\rightarrow 2$, $2\rightarrow 1$, $2\rightarrow 3$, ...), hopping takes place
from the initial ($n$) to final ($m$) state.~\citep{tully_jcp93} On the other hand, transition 
between non-adjacent surfaces ($1\rightarrow 3$ or $3\rightarrow 1$) occurs if the relation $S_{12}\,<\,\zeta\,<\,S_{12}+S_{13}$ or $S_{32}\,<\,\zeta\,<\,S_{32}+S_{31}$ fulfills for those states. In this perspective, it is necessary to mention that direct switching between ground and second excited states 
($1\rightarrow 3$ or $3\rightarrow 1$) is highly improbable for the titled reactions (D$^+$ + H$_2$ and H + H$_2^+$) due to large energy 
spacing between those surfaces.

\subsection{Conservation of Energy by Momentum Adjustment in Trajectory Surface Hopping}\label{ssec:enr}

\noindent
In the TSH calculation, once switching occurs between two electronic states ($n$
to $m$), readjusted diatomic ($\boldsymbol{p}$) and triatomic ($\boldsymbol{P}$) momenta for the $m^{th}$ electronic state can be written in the following form:~\citep{thiel_cp349,klos_jcp142}
\begin{eqnarray}
	\boldsymbol{p_{m}} &=& \boldsymbol{p_{n}}-\omega_{nm}\boldsymbol{\tau_{nm}^{r}} \quad \quad \boldsymbol{P_{m}} = \boldsymbol{P_{n}}-\omega_{nm}\boldsymbol{\tau_{nm}^{R}}, \quad \text{$\omega_{nm}$ is an adjustment factor} \quad \qquad
	\label{eq:momscal}
\end{eqnarray}
\noindent
where the above equations appear to be true for every components of $\boldsymbol{p}$ ($p_x$, $p_y$, $p_z$) or $\boldsymbol{P}$ 
($P_x$, $P_y$, $P_z$). Since the gain/loss in potential energy ($\Delta U = u_m-u_n$) is balanced by the loss/gain in kinetic energy ($\Delta E_{KE} = 
\omega_{nm}^2a_{nm}-\omega_{nm}b_{nm}$) preserving the conservation of total energy, the following equation can be written:
\begin{equation}
	\Delta E_{KE}+\Delta U=\omega_{nm}^2a_{nm}-\omega_{nm}b_{nm}+(u_m-u_n)=0,
	\label{eq:scal}
\end{equation} 
\noindent
where $a_{nm}=\dfrac{1}{2}\Big[\dfrac{|\boldsymbol{\tau_{nm}^{r}}|^2}{\mu_{di}}+\dfrac{|\boldsymbol{\tau_{nm}^{R}}|^2}{\mu_{tri}}\Big]$ ($\mu_{di}$ and 
$\mu_{tri}$ represent diatomic reduced mass and triatomic center of mass, respectively) and $b_{nm}=
\boldsymbol{v_{di}}.\boldsymbol{\tau_{nm}^{r}}+\boldsymbol{v_{tri}}.\boldsymbol{\tau_{nm}^{R}}$ ($\boldsymbol{v_{di}}$ and $\boldsymbol{v_{tri}}$ signify diatomic and 
triatomic velocities, respectively for the initial state). The solution of Eq. \ref{eq:scal} takes the following form:
\begin{equation}
	\omega_{nm}=\dfrac{1}{2a_{nm}}\Big[b_{nm}\pm \sqrt{b_{nm}^2+4a_{nm}(u_n-u_m)}\Big],
	\label{eq:sol}
\end{equation}
\noindent
where positive and negative signs are applicable for $b_{nm}<0$ and $b_{nm}\geq 0$, respectively. For negative values of $b_{nm}^2+4a_{nm}(u_n-u_m)$,
no real solution exists due to insufficient kinetic energy of the system (before hopping) to compensate the potential 
energy change due to the transition and therefore, the hopping is rejected. Once switching of the nuclei occurs from initial to final electronic state, the trajectory starts to evolve on the new surface and this procedure is repeated until the hyperradius, $\rho$ reaches a predefined threshold value when the triatomic system 
leading to non-reactive or reactive processes including charge-transfer.

\subsection{Conservation of Linear Momenta on the Process of Hopping~\citep{soumya_hopp}}\label{ssec:momcon}

\noindent
In order to ensure the conservation of total energy during hopping, the momenta readjustment formalism by Thiel {\it et al.}~\citep{thiel_cp349} is employed as per the previous subsection, where the components of diatomic and triatomic momenta ($\boldsymbol{p}$ and $\boldsymbol{P}$) are redefined from initial ($n$) to final ($m$) electronic state. Since the components of $\boldsymbol{p}$ and $\boldsymbol{P}$ are different in different electronic states to 
preserve the energy conservation for $n^{th}$ and $m^{th}$ electronic states, the conservation of linear momenta between the two sets of quantities, $\lbrace p_{x,n}$, $p_{y,n}$, $p_{z,n}$, $P_{x,n}$, $P_{y,n}$ and $P_{z,n}\rbrace$ and $\lbrace p_{x,m}$, $p_{y,m}$, $p_{z,m}$, $P_{x,m}$, $P_{y,m}$ and $P_{z,m} \rbrace$ can be achieved by employing a scaling factor. On the process of hopping, the diatomic and triatomic kinetic energies are
scaled as $E_{di}\pm|\Delta E_{di}|$ and $E_{tri}\pm|\Delta E_{tri}|$, respectively, where
\begin{subequations}
	\begin{eqnarray}
		\Delta E_{di} &=& \dfrac{|\boldsymbol{\tau_{nm}^{r}}|^2\omega_{nm}^2}{2\mu_{di}}-\omega_{nm}\boldsymbol{v_{di}}.\boldsymbol{\tau_{nm}^{r}}, \qquad \text{(from Eq. \ref{eq:momscal})}\\
		\Delta E_{tri} &=& \dfrac{|\boldsymbol{\tau_{nm}^{R}}|^2\omega_{nm}^2}{2\mu_{tri}}-\omega_{nm}\boldsymbol{v_{tri}}.\boldsymbol{\tau_{nm}^{R}} \qquad \text{(from Eq. \ref{eq:momscal})}
	\end{eqnarray}
\end{subequations}
and total change in kinetic energy during hopping appears as $\Delta E_{KE}\,(=
\omega_{nm}^2a_{nm}-\omega_{nm}b_{nm},\,\,\text{see}$ $\text{Eq. \ref{eq:scal}})$. 
On the other hand, the associated magnitude of momenta become $\sqrt{\dfrac{2(E_{di}\pm|\Delta E_{di}|)}{\mu_{di}}}$ and $\sqrt{\dfrac{2(E_{tri}\pm|\Delta E_{tri}|)}{\mu_{tri}}}$ and prior to hopping, 
components of scaled initial momenta attend the following 
expressions:
\begin{subequations}
	\begin{eqnarray}
		p_{i,n}^s &=& \dfrac{p_{i,n}}{|\boldsymbol{p_{n}}|}\sqrt{\dfrac{2(E_{di}\pm|\Delta E_{di}|)}{\mu_{di}}} = p_{i,n}\sqrt{1\pm\Big(\dfrac{\Delta p}{|\boldsymbol{p_{n}}|}\Big)^2}\qquad i=x,y,z \nonumber \\ \\
		P_{i,n}^s &=& \dfrac{P_{i,n}}{|\boldsymbol{P_{n}}|}\sqrt{\dfrac{2(E_{tri}\pm|\Delta E_{tri}|)}{\mu_{tri}}} = P_{i,n}\sqrt{1\pm\Big(\dfrac{\Delta P}{|\boldsymbol{P_{n}}|}\Big)^2} \qquad i=x,y,z, \nonumber \\
	\end{eqnarray}
\end{subequations}
where positive and negative signs are employed for transition to lower and upper states, respectively. The above scaling factors result into new 
magnitude of the initial momenta ($p_{x,n}^s$, $p_{y,n}^s$, $p_{z,n}^s$, $P_{x,n}^s$, $P_{y,n}^s$ and $P_{z,n}^s$), which can be compared with the 
final ones ($p_{x,m}$, $p_{y,m}$, $p_{z,m}$, $P_{x,m}$, $P_{y,m}$ and $P_{z,m}$) to explore the conservation of linear momenta. In order to show the deviation between scaled initial momenta and final momenta, the root mean square deviation of the initial (scaled) and final momenta is 
computed at the time of hopping for each collision/total energy:
\begin{eqnarray}
	\chi_{p,i} &=& \sqrt{\dfrac{\sum_{ii=1}^{N_{hop}}(p_{i,m,ii}-p_{i,n,ii}^s)^2}{N_{hop}}} \qquad i=x,y,z \\
	\chi_{P,i} &=& \sqrt{\dfrac{\sum_{ii=1}^{N_{hop}}(P_{i,m,ii}-P_{i,n,ii}^s)^2}{N_{hop}}} \qquad i=x,y,z,
\end{eqnarray}
where $\chi_{p,i}$ and $\chi_{P,i}$ ($i=x,y,z$) are root mean square deviations between before (scaled) and after hopp momenta, and $N_{hop}$ signifies total number of hops for a set of trajectories. The above quantities
($\chi_{p,i}$ and $\chi_{P,i}$, $i=x,y,z$) are calculated for the D$^+$ + H$_2$ and H + H$_2^+$ reactions~\citep{soumya_hopp} and the outcomes
are displayed for nine collision energies (1.7 eV - 2.5 eV) of D$^+$ + H$_2$
and thirteen total energies (2.2 eV - 2.8 eV) of H + H$_2^+$ reactions in Tables \ref{Tab:momscale1} and \ref{Tab:momscale2}, respectively. The magnitudes of all the six quantities ($\chi_{p,x}$, $\chi_{p,y}$, $\chi_{p,z}$, $\chi_{P,x}$, $\chi_{P,y}$ and $\chi_{P,z}$) depict the deviation of the order, 
10$^{-6}$ or 10$^{-7}$ validating the conservation of linear momentum.

\begin{table}[ht!]
	\caption{Root mean square deviations of linear momenta in the collision energy range 1.7 eV to 2.5 eV for D$^+$ + H$_2$ reaction are presented.~\citep{soumya_hopp} The results are adopted from Ref. [\protect\citen{soumya_hopp}].}
	\begin{center}
		\scalebox{0.8}{\begin{tabular}{|c|c|c|c|c|c|c|}\hline
				$E_{col}$ (eV) & $\chi_{p,x}$ & $\chi_{p,y}$ & $\chi_{p,z}$ & $\chi_{P,x}$ & $\chi_{P,y}$ & $\chi_{P,z}$ \\ 
				\hline
				1.7 &  5.2677 $\times$ 10$^{-7}$ & 5.1672 $\times$ 10$^{-7}$ & 5.3294 $\times$ 10$^{-7}$ & 7.1530
				$\times$ 10$^{-7}$  & 8.5243 $\times$ 10$^{-7}$ & 8.6290 $\times$ 10$^{-7}$ \\
				1.8 & 5.1637 $\times$ 10$^{-7}$ & 4.8391 $\times$ 10$^{-7}$ & 5.1547 $\times$ 10$^{-7}$ & 6.8163
				$\times$ 10$^{-7}$  & 8.6979 $\times$ 10$^{-7}$ & 8.7511 $\times$ 10$^{-7}$ \\
				1.9 & 4.5859 $\times$ 10$^{-7}$ & 4.1539 $\times$ 10$^{-7}$ & 4.9886 $\times$ 10$^{-7}$ & 6.3213
				$\times$ 10$^{-7}$  & 8.3776 $\times$ 10$^{-7}$ & 8.3682 $\times$ 10$^{-7}$ \\
				2.0 & 4.7916 $\times$ 10$^{-7}$ & 4.3061 $\times$ 10$^{-7}$ & 4.8584 $\times$ 10$^{-7}$ & 6.1972
				$\times$ 10$^{-7}$  & 8.4977 $\times$ 10$^{-7}$ & 8.6352 $\times$ 10$^{-7}$ \\
				2.1 & 4.9708 $\times$ 10$^{-7}$ & 4.4200 $\times$ 10$^{-7}$ & 5.0326 $\times$ 10$^{-7}$ & 7.3403
				$\times$ 10$^{-7}$  & 8.5488 $\times$ 10$^{-7}$ & 8.6890 $\times$ 10$^{-7}$ \\
				2.2 & 4.7639 $\times$ 10$^{-7}$ & 4.4295 $\times$ 10$^{-7}$ & 4.9074 $\times$ 10$^{-7}$ & 6.6735
				$\times$ 10$^{-7}$  & 8.7300 $\times$ 10$^{-7}$ & 8.8896 $\times$ 10$^{-7}$ \\
				2.3 & 4.8477 $\times$ 10$^{-7}$ & 4.5083 $\times$ 10$^{-7}$ & 5.0323 $\times$ 10$^{-7}$ & 6.9058
				$\times$ 10$^{-7}$  & 8.7031 $\times$ 10$^{-7}$ & 9.0892 $\times$ 10$^{-7}$ \\
				2.4 & 4.9343 $\times$ 10$^{-7}$ & 4.3554 $\times$ 10$^{-7}$ & 4.8766 $\times$ 10$^{-7}$ & 6.5504
				$\times$ 10$^{-7}$  & 9.0925 $\times$ 10$^{-7}$ & 9.1940 $\times$ 10$^{-7}$  \\ 
				2.5 & 4.8393 $\times$ 10$^{-7}$ & 4.3836 $\times$ 10$^{-7}$ & 4.9650 $\times$ 10$^{-7}$ & 6.7775
				$\times$ 10$^{-7}$  & 9.1886 $\times$ 10$^{-7}$ & 9.3631 $\times$ 10$^{-7}$\\ \hline
		\end{tabular}}
	\end{center}
	\label{Tab:momscale1}
\end{table}

\begin{table}[ht!]
	\caption{Root mean square deviations of linear momenta in the total energy range 2.2 eV to 2.8 eV for H + H$_2^+$ reaction are tabulated.~\citep{soumya_hopp} The results are adopted from Ref. [\protect\citen{soumya_hopp}].}
	\begin{center}
		\scalebox{0.8}{\begin{tabular}{|c|c|c|c|c|c|c|}\hline
				$E_{tot}$ (eV) & $\chi_{p,x}$ & $\chi_{p,y}$ & $\chi_{p,z}$ & $\chi_{P,x}$ & $\chi_{P,y}$ & $\chi_{P,z}$ \\ 
				\hline
				2.2 & 4.3166 $\times$ 10$^{-7}$ & 4.1324 $\times$ 10$^{-7}$ & 4.3806 $\times$ 10$^{-7}$ & 6.6505
				$\times$ 10$^{-7}$  & 1.1354 $\times$ 10$^{-6}$ & 1.0926 $\times$ 10$^{-6}$ \\
				2.25 & 4.3164 $\times$ 10$^{-7}$ & 4.1373 $\times$ 10$^{-7}$ & 4.3808 $\times$ 10$^{-7}$ & 6.6930
				$\times$ 10$^{-7}$  & 1.1387 $\times$ 10$^{-6}$ & 1.1039 $\times$ 10$^{-6}$ \\
				2.3 & 4.2948 $\times$ 10$^{-7}$ & 4.1834 $\times$ 10$^{-7}$ & 4.4131 $\times$ 10$^{-7}$ & 6.7468
				$\times$ 10$^{-7}$  & 1.1396 $\times$ 10$^{-6}$ & 1.1054 $\times$ 10$^{-6}$ \\
				2.35 & 4.2913 $\times$ 10$^{-7}$ & 4.1907 $\times$ 10$^{-7}$ & 4.4433 $\times$ 10$^{-7}$ & 6.8420
				$\times$ 10$^{-7}$  & 1.1463 $\times$ 10$^{-6}$ & 1.1084 $\times$ 10$^{-6}$ \\
				2.4 & 4.3679 $\times$ 10$^{-7}$ & 4.2241 $\times$ 10$^{-7}$ & 4.4730 $\times$ 10$^{-7}$ & 6.8822
				$\times$ 10$^{-7}$  & 1.1475 $\times$ 10$^{-6}$ & 1.1135 $\times$ 10$^{-6}$  \\ 
				2.45 & 4.3755 $\times$ 10$^{-7}$ & 4.2357 $\times$ 10$^{-7}$ & 4.4473 $\times$ 10$^{-7}$ & 6.9435
				$\times$ 10$^{-7}$  & 1.1538 $\times$ 10$^{-6}$ & 1.1272 $\times$ 10$^{-6}$  \\ 
				2.5 & 4.3835 $\times$ 10$^{-7}$ & 4.2897 $\times$ 10$^{-7}$ & 4.4659 $\times$ 10$^{-7}$ & 6.9863
				$\times$ 10$^{-7}$  & 1.1699 $\times$ 10$^{-6}$ & 1.1283 $\times$ 10$^{-6}$\\ 
				2.55 & 4.4242 $\times$ 10$^{-7}$ & 4.3148 $\times$ 10$^{-7}$ & 4.5130 $\times$ 10$^{-7}$ & 7.0952
				$\times$ 10$^{-7}$  & 1.1721 $\times$ 10$^{-6}$ & 1.1342 $\times$ 10$^{-6}$\\
				2.6 & 4.4267 $\times$ 10$^{-7}$ & 4.3659 $\times$ 10$^{-7}$ & 4.5667 $\times$ 10$^{-7}$ & 7.1289
				$\times$ 10$^{-7}$  & 1.1775 $\times$ 10$^{-6}$ & 1.1416 $\times$ 10$^{-6}$\\ 
				2.65 & 4.4574 $\times$ 10$^{-7}$ & 4.4304 $\times$ 10$^{-7}$ & 4.6353 $\times$ 10$^{-7}$ & 7.1963
				$\times$ 10$^{-7}$  & 1.1881 $\times$ 10$^{-6}$ & 1.1462 $\times$ 10$^{-6}$\\ 
				2.7 & 4.4776 $\times$ 10$^{-7}$ & 4.4319 $\times$ 10$^{-7}$ & 4.6310 $\times$ 10$^{-7}$ & 7.2640
				$\times$ 10$^{-7}$  & 1.1866 $\times$ 10$^{-6}$ & 1.1545 $\times$ 10$^{-6}$\\
				2.75 & 4.4768 $\times$ 10$^{-7}$ & 4.4791 $\times$ 10$^{-7}$ & 4.6345 $\times$ 10$^{-7}$ & 7.3740
				$\times$ 10$^{-7}$  & 1.2028 $\times$ 10$^{-6}$ & 1.1623 $\times$ 10$^{-6}$\\ 
				2.8 & 4.5294 $\times$ 10$^{-7}$ & 4.4965 $\times$ 10$^{-7}$ & 4.6589 $\times$ 10$^{-7}$ & 7.4314
				$\times$ 10$^{-7}$  & 1.2101 $\times$ 10$^{-6}$ & 1.1692 $\times$ 10$^{-6}$\\
				\hline
		\end{tabular}}
	\end{center}
	\label{Tab:momscale2}
\end{table}

\subsection{TSH Calculation of Integral Cross-Sections and Rate Constant}\label{ssec:ics}
\noindent
Integral Cross-Sections (ICSs) at a particular collision energy along a specific channel ($I_C$) of a reaction process can be computed by adding the reaction probabilities [$P_{v^{\prime} j^{\prime} \leftarrow v j}^{J,I}(E_{col})$] of all final rovibrational states for all values of 
total angular momentum, $J$ as follows:
\begin{eqnarray}
	\sigma^I(E_{col}) = \frac{\pi}{\kappa_{v j}^{2}} \sum_{j^\prime =
		0}^{j^{\prime}_{max}}\sum_{v^\prime =
		0}^{v^{\prime}_{max}}{\sum_{J = 0}^{J_{max}}{(2 J + 1)
			P_{v^{\prime} j^{\prime} \leftarrow v j}^{J,I}(E_{col})}},
	\label{eq:ics}
\end{eqnarray} 
where $E_{col}$ indicates the collision energy and 
$\kappa_{v j}$ represents the wave-vector, $\sqrt{\dfrac{2\mu_{tri} E_{col}}{\hbar}}$ ($\mu_{tri}$ is triatomic 
center of mass). In order to achieve convergence in the ICS profiles of TSH methodology, it is necessary to perform 
dynamical calculations over a large number of trajectories on the adiabatic PESs. In ``semi-classical" limit, total angular momentum quantum number, $J$ 
can be approximated to the orbital angular 
momentum quantum number, $l_{v^\prime,j^\prime}$ ($J \sim l_{v^\prime,j^\prime}$).~\citep{Billing_book} The following expression can be employed for orbital angular
momentum quantum number:~\citep{Billing_book}
\begin{equation}
	l_{v^\prime,j^\prime} = l_{max}O_{v^\prime,j^\prime} = \Big(\dfrac{b_{max}\mu_{tri}v_0}{\hbar} - \dfrac{1}{2}\Big)O_{v^\prime,j^\prime}
\end{equation}
where $b_{max}$, $\mu_{tri}$ and $v_0$ represent maximum impact parameter, triatomic center of mass and initial relative velocity, respectively. A specific random number, $O_{v^\prime,j^\prime}$ (ranges from 0 to 1) is invoked for a particular trajectory leading to a product or reactant at 
ro-vibrational state ($v^\prime,j^\prime$) indicating the dependence of $l_{v^\prime,j^\prime}$ on the final ro-vibrational state.\\

\noindent
On the other hand, reactant/product formation probabilities can be written as $(l_{v^\prime,j^\prime}+1)^21/N_T$ ($N_T$ represents total number of trajectories leading to
charge or non-charge transfer reactants/products) in ``semi-classical" limit, where the rovibrationally resolved [$\sigma^I_{v^\prime,j^\prime}(E_{col})$], rotationally summed vibrationally resolved [$\sigma^I_{v^\prime}(E_{col})$] and total integral [$\sigma^I(E_{col})$] cross-sections achieve the following form:
\begin{eqnarray}
	\sigma^I_{v^\prime,j^\prime}(E_{col}) &=& \frac{\pi}{\kappa_{v j}^{2}} \sum_{N_{v^\prime, j^\prime}}
	(l_{v^\prime,j^\prime}+1)^2(1/N_T)
	\label{eq:ics_tsh1}
\end{eqnarray}
\begin{eqnarray}
	\sigma^I_{v^\prime}(E_{col}) &=& \frac{\pi}{\kappa_{v j}^{2}} \sum_{j^\prime =
		0}^{j^{\prime}_{max}}\sum_{N_{v^\prime, j^\prime}}(l_{v^\prime,j^\prime}+1)^2(1/N_T)
	\label{eq:ics_tsh2}
\end{eqnarray}
\begin{eqnarray}
	\sigma^I(E_{col}) &=& \frac{\pi}{\kappa_{v j}^{2}}\sum_{v^\prime =
		0}^{v^{\prime}_{max}}\sum_{j^\prime =
		0}^{j^{\prime}_{max}}\sum_{N_{v^\prime, j^\prime}}(l_{v^\prime,j^\prime}+1)^2(1/N_T).
	\label{eq:ics_tsh3}
\end{eqnarray}
In the above expression, $N_{v^\prime, j^\prime}$ symbolizes the number of trajectories leading to the formation of product at a specific rovibrational state ($v^\prime,j^\prime$) and therefore, one can write:
\begin{equation}
	\sum_{v^\prime = 0}^{v^{\prime}_{max}}\sum_{j^\prime =  0}^{j^{\prime}_{max}}N_{v^\prime, j^\prime} = N_T.
	\label{eq:traj}
\end{equation}

\noindent
Once the ICSs are obtained, the temperature (T) dependent rate constant can be computed with the use of the following expression:
\begin{equation}
	k^I(T) = \Big(\dfrac{8k_BT}{\pi\mu_{tri}}\Big)^{1/2}\dfrac{1}{(k_BT)^2}\int_{E_0}^{E_{max}}E_{col}\,\sigma^I(E_{col})\,e^{-E_{col}/k_BT}dE_{col},
	\label{eq:rate}
\end{equation}
\noindent
where $E_{col}(D^{+}+H_{2})$ $ = (E_{tot}-E_{vj})$ eV 
and $E_{col}(H+H_{2}^{+}) = (E_{tot}-E_{vj}-1.95)$ eV. $E_{tot}$ and $E_{vj}$ denote total and initial rovibrational energies, respectively. In case of H + H$_{2}^{+}$ process, the reaction initiates
from the electronically first excited state of H$_3^+$ PESs. Asymptotically, H$_2^+ (^2\Sigma_g^+)$+H is $\sim$ 1.95 eV higher compared to the ground electronic state, H$_2 (^1\Sigma_g^+)$+H$^+$. Therefore, such amount of energy 
needs to be subtracted from initial kinetic energy to obtain TSH based cross-sections and rate constants for charge-transfer and non charge-transfer processes of the said reaction.

\subsection{Results and Discussion}\label{ssec:res}

\noindent
\subsubsection{TSH Dynamics of D$^+$ + H$_2$ Reaction}

\noindent
TSH based total ICS~\citep{soumya_hopp} is computed for three competing processes of D$^+$ + H$_2$ reaction with the help of Eq. \ref{eq:ics_tsh3}
and then, the numerical results are compared with quantum calculated ICS profiles by Adhikari {\it et al.}~\citep{jcp-147-074105-SA}. 
It is worth mentioning that quantum scattering calculations for D$^+$ + H$_2$ reaction~\citep{jcp-147-074105-SA} 
are performed over Beyond Born-Oppenheimer based Diabatic Hamiltonian (BBO-DH) of H$_3^+$~\citep{mukherjee_jcp141}. In order to perform such quantum dynamical calculations, Fully Closed Coupled (FCC) equations of motion are employed incorporating all the
helicity quantum number ($K$ = $-J$ to $J$ for total angular momentum, $J$).
It should be noted that adiabatic and diabatic representations 
of same {\it ab initio} H$_3^+$ Hamiltonian~\citep{mukherjee_jcp141} are used for TSH and quantum dynamics,~\citep{jcp-147-074105-SA} respectively. \\

\begin{figure}[!htp]
	\centering
	\includegraphics[height=0.75\textheight, angle=0]{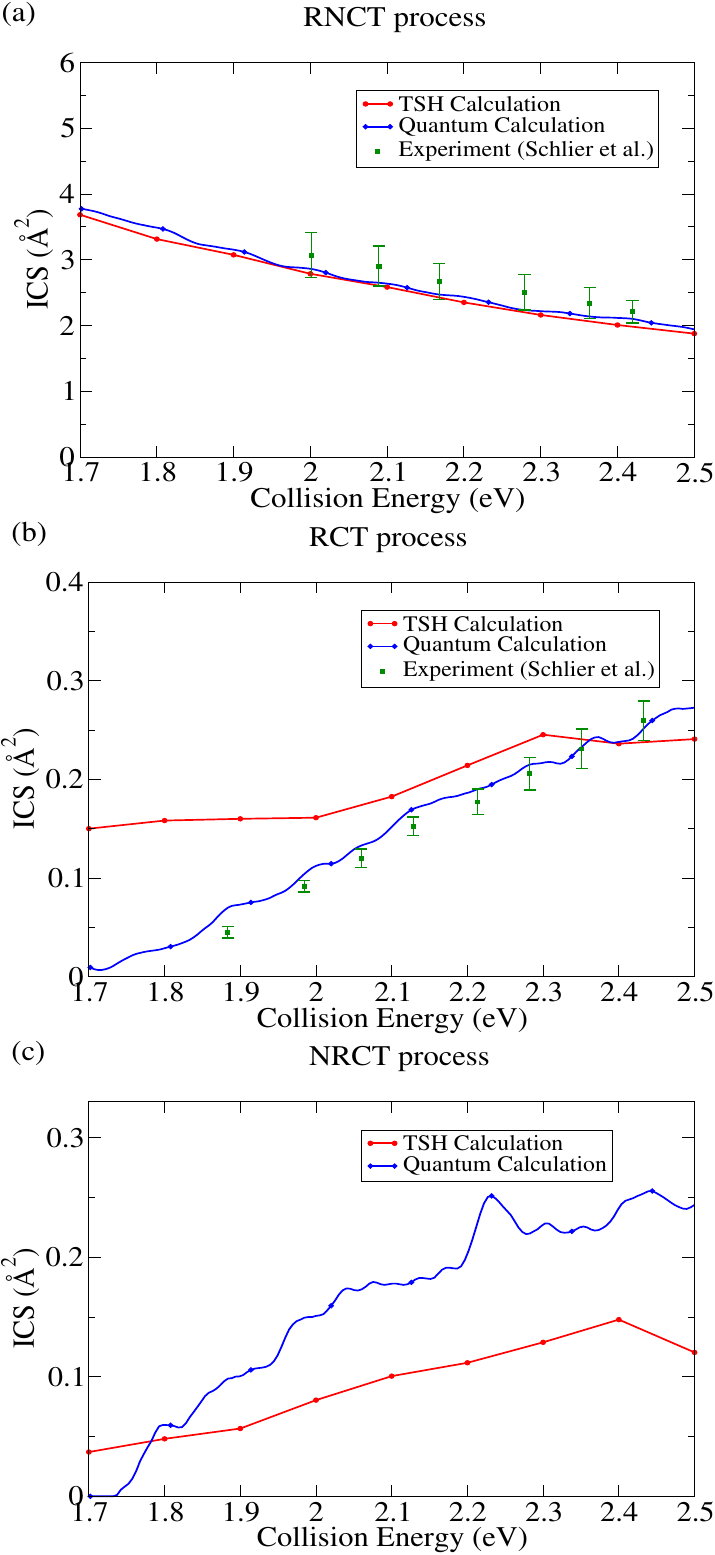}
	\caption{TSH calculated total ICSs~\citep{soumya_hopp} are plotted as functions of collision energies (1.7 eV to 2.5 eV) for the (a) RNCT, (b) RCT and (c) NRCT processes. In case of RNCT, TSH,~\citep{soumya_hopp} Quantum~\citep{jcp-147-074105-SA}
		and experimental~\citep{cp-111-401-SCH} profiles exhibit a good agreement with each other. 
		For RCT process, TSH calculated ICSs~\citep{soumya_hopp} exhibit deviations
		from quantum results~\citep{jcp-147-074105-SA} as well as experimental outcomes~\citep{cp-111-401-SCH}. Such deviation is high at the collision energy domain of hopping (region of prominent quantum effects), specifically, below 2.1 eV. Moreover, ICS profile of NRCT process~\citep{soumya_hopp} originated from TSH calculation also display dissimilarities with respect to quantum 
		results~\citep{jcp-147-074105-SA}. The figures are adopted from Ref. [\protect\citen{soumya_hopp}] and permission has been taken from Elsevier.}
	\label{fig:tot-ics}
\end{figure}

\noindent
The variation of TSH calculated ICS~\citep{soumya_hopp} for RNCT process is depicted in panel (a) of Fig. \ref{fig:tot-ics} in comparison with the quantum results~\citep{jcp-147-074105-SA} and experimental data~\citep{cp-111-401-SCH} as function of collision energy over the range 1.7 
to 2.5 eV. The figure depicts that TSH based ICS profile shows good agreement with the quantum as well as experimental cross-sections over the 
entire collision energy range. On the other hand, TSH computed ICS~\citep{soumya_hopp} for RCT acquires non-zero magnitude around 1.7 eV, gradually increases upto 2.2 eV and finally, reached a plateau at higher collision energies. Nevertheless, in the collision energy $\sim$ 1.7 eV, quantum calculated profile for RCT process~\citep{jcp-147-074105-SA} 
(see Fig. \ref{fig:tot-ics}b) starts with very small magnitudes followed by steep increase with the increase in collision energy, which indicates
the presence of a threshold value for initiating such process. In case of quantum calculated results, the ICS profile
perfectly follow the experimental results~\citep{cp-111-401-SCH} over the domain of collision energy range. 
In summary, semi-classical dynamics, like TSH is unable to capture the presence of threshold energy for RCT process. In a similar way, TSH and quantum~\citep{jcp-147-074105-SA} based ICSs for NRCT process are plotted in Fig. \ref{fig:tot-ics}c, which also display deviations over the range of collision energy. Moreover, ICS profile 
for NRCT process~\citep{soumya_hopp} obtained from TSH approach fails to display any threshold energy.

\noindent
\subsubsection{TSH Dynamics of H + H$_2^+$ Reaction}

\begin{figure}[!htp]
	\centering
	\includegraphics[height=0.23\textheight, angle=0]{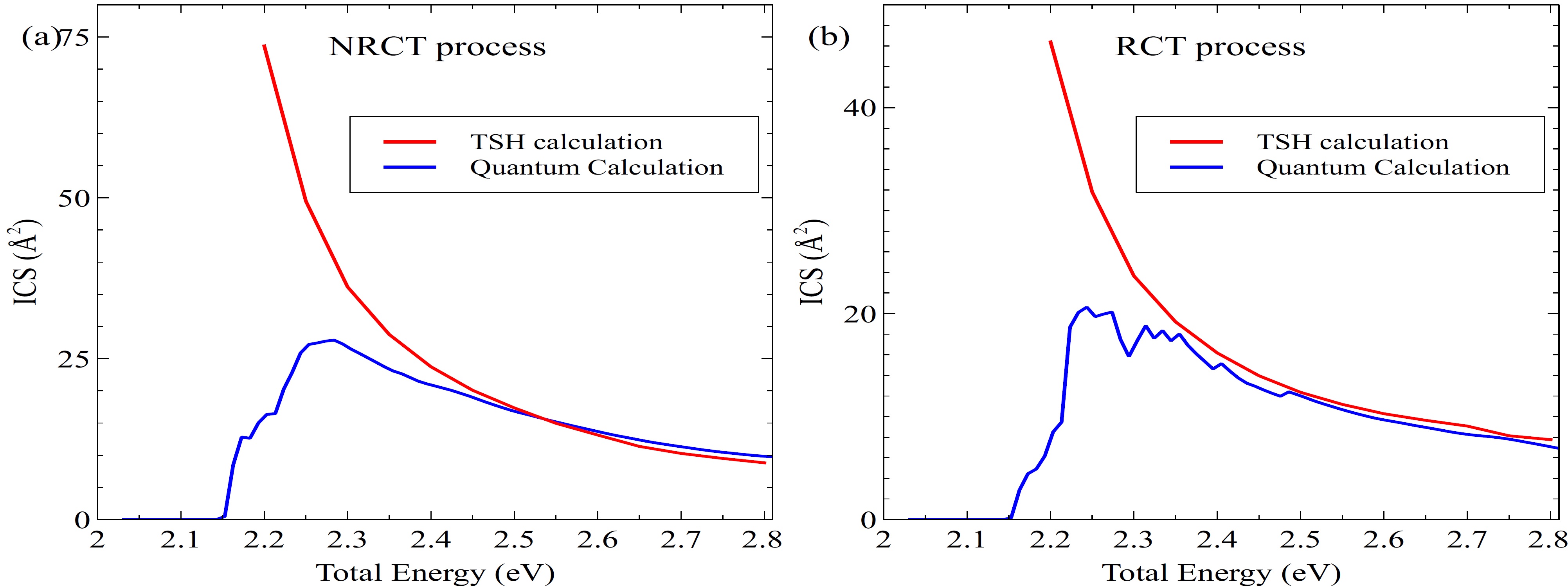}
	\caption{TSH~\citep{soumya_hopp} and quantum~\citep{ghosh_jpca125} calculated ICSs are displayed as functions of total energies for the (a) NRCT and (b) RCT processes. For both cases, TSH~\citep{soumya_hopp} and quantum results~\citep{ghosh_jpca125} show close agreement at higher energies ($E_{tot} \ge 2.3$ eV), but show reverse trend at lower energy domain. The figures are adopted from Ref. [\protect\citen{soumya_hopp}] and permission has been taken from Elsevier.}
	\label{fig:tot-ics-h+h2+}
\end{figure}

\noindent
TSH based cross-sections (see Eq. \ref{eq:ics_tsh3}) and rate constants are calculated by Adhikari and coworkers~\citep{soumya_hopp} for 
both the charge-transfer processes (NRCT and RCT) of H + H$_2^+$ reaction within total energy range 2.2 eV to 2.8 eV initiating the process from first excited state [H$_2^+ (^2\Sigma_g^+)$+H]. Once the ICSs and rate constants are
in hand, the outcomes are compared with the quantum results obtained by FCC 3D wave-packet dynamics~\citep{ghosh_jpca125} (see Figs \ref{fig:tot-ics-h+h2+}a 
and \ref{fig:tot-ics-h+h2+}b). It is worth mentioning that for both the processes (NRCT and RCT), 
quantum ICS profiles~\citep{ghosh_jpca125} initiate from zero (0) at threshold energy ($\sim$ 2.15 eV), pass through the maxima and then, decrease gradually at higher energies leading to the formation of a plateau. The minimum energy conical intersection (MECI) between the ground and first excited electronic
states appears about the same energy, 2.15 eV.~\citep{mukherjee_jcp141} On the other hand, TSH based ICS curves~\citep{soumya_hopp} do not exhibit any threshold value, start with very high magnitude followed by a decreasing trend. In other words, TSH based ICS profiles for NRCT 
(Fig. \ref{fig:tot-ics-h+h2+}a) and RCT (Fig. \ref{fig:tot-ics-h+h2+}b) processes
exhibit good accord with the quantum calculated results~\citep{ghosh_jpca125} at higher energy domain  
(2.3 eV $\le$ $E_{tot}$ $\le$ 2.8 eV), but TSH computed ICSs exhibit sharp increase at lower energies
($<$ 2.3 eV) indicating the absence of any threshold value. 

\begin{figure}[!htp]
	\centering
	\includegraphics[height=0.3\textheight, angle=0]{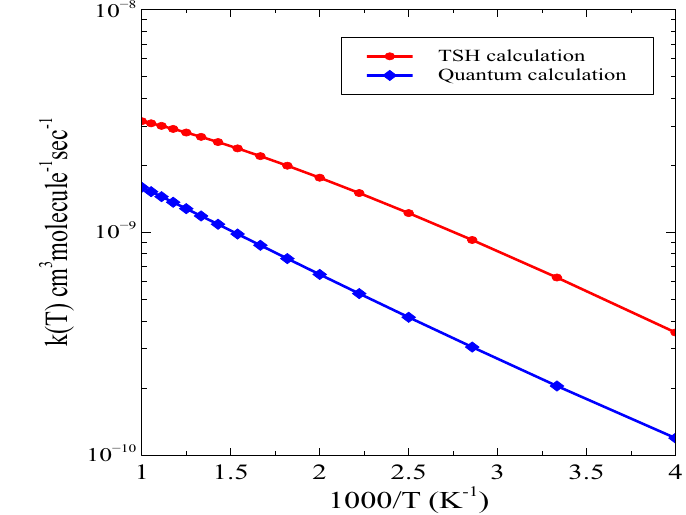}
	\caption{TSH~\citep{soumya_hopp} and quantum calculated~\citep{ghosh_jpca125} rate constants ($k$) are presented as functions of inverse temperature. The magnitudes of TSH based rate constant are relatively higher than the quantum ones with almost same slopes. The figure is adopted from Ref. [\protect\citen{soumya_hopp}] and permission has been taken from Elsevier.}
	\label{fig:tot-rate}
\end{figure}

\noindent
On the other hand, temperature ($T$) dependent rate constants of the titled reaction (H + H$_2^+$)~\citep{soumya_hopp} are computed 
over the range of 250 K to 1000 K employing TSH based ICSs of charge-transfer processes (NRCT and RCT). Fig. \ref{fig:tot-rate} shows
total rate constants obtained from TSH and FCC 3D wave-packet calculation as function of inverse temperature. 
The rate constants for different temperatures 
(250 K - 1000 K) and the average ones are also summarized in Table \ref{Tab:rateconst} in comparison with experimental findings~\citep{karplus_cpl32}. It is 
interesting to note that quasi-classical rate constants at individual temperature as well as the average quantity ($\sim$ 10$^{-9}$)~\citep{soumya_hopp} attend relatively higher magnitude with respect to the quantum calculated ones~\citep{ghosh_jpca125} ($\sim$ 10$^{-10}$), but
slopes of both the profiles representing the activation energy of the charge transfer reactions due to non-adiabatic processes are almost same. On the other hand, the quantum calculated average rate constant shows good agreement with the experimental outcome by Karplus {\it et al.},~\citep{karplus_cpl32} but the magnitude of TSH~\citep{soumya_hopp} one is relatively higher. 

\begin{table}[ht!]
	\caption{TSH~\citep{soumya_hopp} and quantum~\citep{ghosh_jpca125} calculated rate constants (in cm$^{3}$ molecule$^{-1}$ s$^{-1}$ unit) are presented in the temperature range 250 K to 1000 K. The results are adopted from Ref. [\protect\citen{soumya_hopp}].}
	\begin{center}
		\scalebox{0.8}{\begin{tabular}{|c|c|c|c|c|c|}\hline
				temp & k(T) (TSH)~\citep{soumya_hopp} &  k(T) (Quantum)~\citep{ghosh_jpca125} & temp & k(T) (TSH)~\citep{soumya_hopp}  & k(T)(Quantum)~\citep{ghosh_jpca125} \\ 
				\hline
				250 & 3.54 $\times$ 10$^{-10}$ & 1.19 $\times$ 10$^{-10}$ & 650 & 2.38 $\times$ 10$^{-9}$ &  9.81 $\times$ 10$^{-10}$\\
				300 & 6.26 $\times$ 10$^{-10}$ &  2.04 $\times$ 10$^{-10}$ & 700 & 2.54 $\times$ 10$^{-9}$ &  10.85 $\times$ 10$^{-10}$\\
				350 & 9.22 $\times$ 10$^{-10}$ &  3.05 $\times$ 10$^{-10}$ & 750 & 2.68 $\times$ 10$^{-9}$ &  11.82 $\times$ 10$^{-10}$\\
				400 & 1.22 $\times$ 10$^{-9}$ & 4.15 $\times$ 10$^{-10}$ & 800 & 2.80 $\times$ 10$^{-9}$ &  12.75 $\times$ 10$^{-10}$\\
				450 & 1.50 $\times$ 10$^{-9}$ &  5.30 $\times$ 10$^{-10}$ & 850 & 2.91 $\times$ 10$^{-9}$ &  13.62 $\times$ 10$^{-10}$\\
				500 & 1.76 $\times$ 10$^{-9}$ &  6.46 $\times$ 10$^{-10}$ & 900 & 3.00 $\times$ 10$^{-9}$ & 14.43 $\times$ 10$^{-10}$\\
				550 & 1.99 $\times$ 10$^{-9}$ &  7.61 $\times$ 10$^{-10}$ & 950 & 3.08 $\times$ 10$^{-9}$ &  15.19 $\times$ 10$^{-10}$\\
				600 & 2.19 $\times$ 10$^{-9}$ &  8.73 $\times$ 10$^{-10}$ & 1000 & 3.15 $\times$ 10$^{-9}$ &  15.90 $\times$ 10$^{-10}$\\ \hline
				\multicolumn{3}{|c}{average (TSH)~\citep{soumya_hopp}: \quad 2.07 $\times$ 10$^{-9}$} & \multicolumn{3}{c|}{average (quantum)~\citep{ghosh_jpca125}: \quad 8.93 $\times$ 10$^{-10}$}\\ 
				\multicolumn{6}{|c|}{experiment~\citep{karpas_jcp70}: \quad (6.4 $\pm$ 1.2) $\times$ 10$^{-10}$}\\ \hline
		\end{tabular}}
	\end{center}
	\label{Tab:rateconst}
\end{table}
\clearpage

\section{Non-adiabatic Couplings as Friction in Quasi-Classical Trajectory Calculations}\label{ssec:quas}	
\noindent
\subsection{NACTs As Friction}	
\noindent
In the Beyond Born-Oppenheimer (BBO) regime, the adiabatic SE possesses two kinds of forces, namely, (a) one originated from the PESs by electrostatic interactions and (b) the another appeared 
from the NACTs. The latter ones are mainly associated with velocities and hence, can be denoted as Forces of Friction. In order to explore such possibility, classical mechanics~\citep{ravi_epjd,ravi_mp119} can be applied to justify the idea of considering NACTs as frictional forces. As a matter of fact, application of classical mechanics is more advantageous, since the solutions
of classical equations of motion (EOMs) can be achieved also for singular NACTs. Classical EOMs containing forces of friction are thoroughly discussed 
elsewhere~\citep{goldstein_classical_book,landau_mechanics_book}, where one benchmark example can be considered as the damped vibrations affected by force of friction proportional to the velocity. \\

\noindent
The one-dimensional classical equation of motion containing the friction term ($\beta\dot{x}$) can be written as:
\begin{equation}
	m\ddot{x} + \beta\dot{x} + \dfrac{dV(x)}{dx} = 0.
	\label{eq:qc1}
\end{equation} 
In the above expression, the terms $\dot{x}\,(=dx/dt)$ and $\ddot{x}\,(=d^2x/dt^2)$ represent velocity and acceleration, respectively, where $V(x)$ symbolizes the potential energy and $\beta$ denotes the corresponding (x-component) friction coefficient. So far, the friction term is assumed to be positive in any numerical treatment.\\

\noindent
Eq. \ref{eq:qc1} can also be expressed as an energy equation:
\begin{equation}
	[\frac{1}{2m}(m\dot{x} + \beta x)^2 + V(x)] = E + G(x,\dot{x}),
	\label{eq:qc2}
\end{equation}      
\noindent
where for $\beta = 0$, $E$ is a constant and $G(x,\dot{x})$ can be written as:
\begin{equation}
	G(x,\dot{x}) = \int x [-\frac{1}{m}\beta\frac{dV}{dx} + \frac{d\beta}{dt}(\frac{1}{m}\beta x + \dot{x})]
	\label{eq:qc3}
\end{equation}

\noindent
The similarity between total Schr\"odinger Equation (SE) and the classical EOM is not obvious since the former matrix-equation (molecular SE) needs to be applied for a group of coupled equations. On the contrary, classical EOM is applicable for a single isolated state. Still similarity between these two frameworks can be envisaged by introducing extended BO (EBO) approximation as developed by Baer and Englman~\citep{baer_cpl265} as well as Sarkar and Adhikari~\citep{sarkar_jcp124} for two/three coupled equations. At low
enough energies, such approximation yields the following single surface equation:

\begin{equation}
	\frac{1}{2m}\Big(\frac{\hbar}{i}\frac{\partial}{\partial x}+i\hbar\tau_x^{12}(x)\Big)^2\Psi(x) + (u_1(x)-E)\Psi(x) = 0
	\label{eq:qc4}
\end{equation}

\noindent
It is worth mentioning that the solution of Eq. \ref{eq:qc4} can be obtained for all the conditions/situations envisaged in the present study. On the other hand, 
Schr\"odinger correspondence principle can be applied for Eqs. \ref{eq:qc2} and \ref{eq:qc4} to unveil the connection between quantum mechanical x-component of $\tau_x^{12}(x)$ (see Eq. \ref{eq:qc4}) and the corresponding classical friction coefficient, $\beta x$ (which is responsible for the classical dissipative force or friction force). Hence, one can write the following equation, 

\begin{equation}
	\hbar\tau_x^{12}(x|s) = \beta(x)x \Rightarrow \beta(x) = \dfrac{\hbar \tau_x^{12}(x|s)}{x}
	\label{eq:qc5}
\end{equation}

\noindent
Similar relation can also be written for other Cartesian coordinates, like y and z for three-dimensional coordinate system. At this junction, it needs to emphasize that the NACTs are pure quantum mechanical entities, which can be used as classical friction components during solution of classical EOMs for the molecular systems. In other words, the whole treatment turns into a semi-classical one.\\

\noindent
As per the basic principles of classical mechanics~\citep{landau_mechanics_book}, the friction coefficient must be positive (or zero) in all situations and such condition should be applicable for Eqs. \ref{eq:qc1} - \ref{eq:qc3}. On the other hand, according to quantum mechanics, NACTs (numerator of Eq. \ref{eq:qc5}) vis-\'a-vis the Cartesian coordinates (denominator of 
Eq. \ref{eq:qc5}) might be either negative or positive and therefore, both negative and positive values of $\beta(x)$ or 
$\beta(y)$ or $\beta(z)$ are possible. Since the friction coefficient must be positive (or zero) in Classical Mechanics, only the positive ratios, $\beta(x)$ or 
$\beta(y)$ or $\beta(z)$ are physically realizable. On the contrary, in the semiclassical treatment of classical EOMs, the NACTs as friction could be included as negative or positive quantity. In other words, the first possibility is that the friction terms are assumed to be positive “all the way”, namely:
\[\begin{array}{*{20}{c}}
	{}&{}&{{\beta _{qj}}(q|s)}
	\label{eq:qc6}
\end{array} = \hbar \left| {{\tau _{qjj + 1}}(q|s)/q} \right|\quad q = x, y, z\],

\noindent
which will be termed as the classical friction. \\

\noindent
The second possibility, which is introduced under the influence of quantum mechanics can be expressed as:

\[{\beta _{qj}}(q|s) = \left\{ {\begin{array}{*{20}{c}}
		{\hbar {\tau _{qjj + 1}}(q|s)/q\begin{array}{*{20}{c}}
				{}&{when\begin{array}{*{20}{c}}
						{}&{{\tau _{qjj + 1}}(q|s)/q \ge 0}
				\end{array}}
		\end{array}}\\
		{\begin{array}{*{20}{c}}
				{}&{}
			\end{array}0\begin{array}{*{20}{c}}
				{\begin{array}{*{20}{c}}
						{}&{}&{}
				\end{array}}&{when\begin{array}{*{20}{c}}
						{}&{{\tau _{qjj + 1}}(q|s)/q \le 0}
				\end{array}}
		\end{array}}
\end{array}} \right.\]	

\noindent
Such possibility is reminiscent of Schr\"odinger’s cat and this type of friction will be designated as semi-classical friction.\\

\noindent
Finally, the third possibility is to assume the friction term to be as it is: 
\[\begin{array}{*{20}{c}}
	{}&{}&{{\beta _{qj}}(q|s)}
\end{array} = \hbar {\tau _{qjj + 1}}(q|s)/q\]
\noindent
This possibility can be justified by the fact that the often singular NACTs are capable not only to slow down the interacting atomic/ionic particles but also to accelerate the system. Such type of friction will be denoted as quantum friction.     \\

\noindent
The quantum version of friction is more suitable for application due to the artificial singularities encountered by the particle while crossing a
q-axis (q = x, y, z). If  $\beta(q)$ becomes positive, the large acceleration is soon abolished once the particle crosses the q-axis and then, $\beta(q)$ becomes negative leading to the slowing down of the particle at similar rate. In other words, the quantum friction annihilates the artificial singularity. \\

\noindent
It is necessary to emphasize that similar to the EBO equations (eq. \ref{eq:qc4}), the semi-classical friction approach as depicted
in Eq. \ref{eq:qc2} does not conserve energy.  During the slowing down process of an damped system, it releases energy as heat, whereas in the present quasi-classical approach, the energy would be released as light (or, better, as photons), which was discussed in the production of HeH$^+$ in (He, H$_2^+$) collisions~\citep{ravi_mp119,hopper_jcp_73}. The dissipation process becomes fasten once three or more quasi-ions are in relative proximity due to appearance of singular NACTs
at certain regions in configuration space. As a matter of fact, the fast-moving quasi-ions are continuously slowed down and gradually looses their excess
energy at various rates. On the other hand, influence of molecular friction on reaction dynamics in condensed phases is being investigated for several decades~\citep{kramers_physica_7} and recently, a lot of interest has been grown in electronic friction originating from non-adiabatic coupling particularly in gas-metal interactions~\citep{head_jcp103,spiering_jpcl10,smits_jcp161,sah_jcp161,saalfrank_cr106}. 
The approach of considering a single potential energy surface (PES) appears useful for metallic systems due to presence of continuum of electronic states acting as a bath. While the electronic friction in gas-metal system appears to be applicable in the weak non-adiabatic limit, the present approach is valid in two or more electronic states over the entire configuration space for a molecular system.  \\

\noindent
Adhikari and coworkers~\citep{friction_jpca128} solve the classical EOMs by incorporating friction terms equivalent to the NACTs, which 
exhibits that the (D$^+$, H$_2$) dynamics slows down enough to arrive at the bottom of the potential well. In other words, a classical mechanical study including the NACTs would lead the formation of the triatomic species, (DH$_2$)$^+$ at the minima of the potential.  
In case of (DH$_2$)$^+$ system, a wide variety of NACTs are observed over the entire nuclear configuration space, where one can find accidental seams between the ground (1$^1$A$^\prime$) and the first excited (2$^1$A$^\prime$) states at the C$_{2v}$ (isosceles) as well as C$_s$ (acute) geometries over the global PESs. On the other hand, the CIs between the first (2$^1$A$^\prime$) and the second (3$^1$A$^\prime$) excited states are located at D$_{3h}$ (equilateral) nuclear configurations as illustrated elsewhere.~\citep{mukherjee_jcp141} Finally, the classical dynamical calculations are carried out over the global three-dimensional (3D) PES of H$_3^+$ to calculate the trapping percentage in the global minimum (-9.31 eV) as well as other reaction attributes (reaction probabilities and cross sections) of various processes resulting from (D$^+$, H$_2$) interactions.

\noindent
\subsection{Hamilton's Equation of Motion Including Friction and Analysis}	
\noindent
For D$^{+}$ + H$_{2}$~\citep{mukherjee_jcp141,jcp-147-074105-SA}, the various reaction processes can be expressed as:
\begin{itemize}
	
	\item  Creation of Triatomic Species (trapping in the minima of the ground state):\\
	D$^+$ + H$_2$ $\rightarrow$ DH$_2^+$,
	
	\item  Non-Reactive Non-Charge Transfer (NRNCT) (reactant itself):\\
	D$^+$ + H$_2$ $\rightarrow$ D$^+$ + H$_2$,

	\item  Reactive Non-Charge Transfer (RNCT):\\
	D$^+$ + H$_2$ $\rightarrow$ H$^+$ + DH,
	
	\item Reactive Charge Transfer (RCT) (first excited state):\\
	D$^+$ + H$_2$ $\rightarrow$ DH$^+$ + H, 
	
	\item Non-Reactive Charge Transfer (NRCT) (first excited state):\\
	D$^+$ + H$_2$ $\rightarrow$ D + H$_2^+$
	
\end{itemize}

\noindent
The classical trajectory calculations are executed over the single electronic state (ground) PES as well as the two state (ground and first excited) PESs at E$_{coll}$ = 1.7, 2.5 and 5.0 eV taking the maximum impact parameter ($b_{max}$) of 3 \AA. While performing such study, NACTs are incorporated as friction coefficients. It is to note that a cut-off value of  0.05 \AA\, is employed at the origin of the Cartesian coordinates to avoid any artifact originating at x or y or z  = 0. On the other hand, the trajectories are initialized on the ground adiabatic surface of DH$_2^+$, where such electronic state asymptotically represents D$^+$ + H$_2$ $(v = 0, j = 0)$. $v$ and $j$ symbolize the vibrational and rotational quantum numbers, respectively.  \\

\noindent
Since eighteen (18) equations of motion (EOMs) are possible for positions and momenta on a single PES~\citep{baer_aqc89}, one needs to consider thirty-six (36) EOMs for two electronic states~\citep{friction_jpca128} as depicted below:

\begin{eqnarray*}
	&&\dot{p}_{1x_{k}} + \frac{\hbar}{m_{k}x_{k}}p_{1x_{k}}\tau_{12}^{1x_{k}} + \frac{dV_{1}(x_{k},y_{k},z_{k})}{dx_{k}} = 0 \\
	&&\dot{p}_{2x_{k}} - \frac{\hbar}{m_{k}x_{k}}p_{2x_{k}}\tau_{12}^{2x_{k}} + \frac{dV_{2}(x_{k},y_{k},z_{k})}{dx_{k}} = 0 \\
	&&\dot{p}_{1y_{k}} + \frac{\hbar}{m_{k}y_{k}}p_{1y_{k}}\tau_{12}^{1y_{k}} + \frac{dV_{1}(x_{k},y_{k},z_{k})}{dy_{k}} = 0 \\
	&&\dot{p}_{2y_{k}} - \frac{\hbar}{m_{k}y_{k}}p_{2y_{k}}\tau_{12}^{2y_{k}} + \frac{dV_{2}(x_{k},y_{k},z_{k})}{dy_{k}} = 0 \\
	&&\dot{p}_{1z_{k}} + \frac{\hbar}{m_{k}z_{k}}p_{1z_{k}}\tau_{12}^{1z_{k}} + \frac{dV_{1}(x_{k},y_{k},z_{k})}{dz_{k}} = 0 \\
	&&\dot{p}_{2z_{k}} - \frac{\hbar}{m_{k}z_{k}}p_{2z_{k}}\tau_{12}^{2z_{k}} + \frac{dV_{2}(x_{k},y_{k},z_{k})}{dz_{k}} = 0 \\
	&&\dot{x}_{1k} = \frac{p_{1x_{k}}}{m}\\
	&&\dot{y}_{1k} = \frac{p_{1y_{k}}}{m}\\
	&&\dot{z}_{1k} = \frac{p_{1z_{k}}}{m}\\
	&&\dot{x}_{2k} = \frac{p_{2x_{k}}}{m}\\
	&&\dot{y}_{2k} = \frac{p_{2y_{k}}}{m}\\
	&&\dot{z}_{2k} = \frac{p_{2z_{k}}}{m}
\end{eqnarray*}

\noindent
In the two-electronic state classical trajectory calculations, three different friction conditions are imposed to explore the effects of NACTs as frictional force.
In the first case (I), the modulus value of $\left|\tau_{q}/q\right|$ is considered throughout the trajectory in accordance with Landau-Lifshitz (LL) condition (classical friction). On the other hand, for the second case (II), $\tau_{q}/q$ is incorporated in the classical EOMs, only when the 
product of $\tau_{q}$ and $1/q$ is naturally positive (semiclassical friction). In other words, if the quantity $\tau_{q}/q$ is 
positive, it is included in the classical EOMs, otherwise set to zero. Finally, 
the calculations are also carried out without including any sign correction in the friction term, i.e., choice III (quantum friction), where no constraint is imposed on the sign of $\tau_{q}$ and $q$.\\

\noindent
The cross-section ($\sigma$) values for any dynamical process are computed employing the following expression: 

\begin{equation}
	\sigma = \frac{\pi}{\kappa_{vj}^{2}}\sum_{v^{'} = 0}^{v_{\max}^{'}}{\sum_{j^{'} = 0}^{j_{\max}^{'}}{\sum_{N_{v^{'}j^{'}}}^{}\left( l_{v^{'}j^{'}} + 1 \right)^{2}}}\left( \frac{1}{N_{T}} \right)
\end{equation}

\noindent
where  $N_{v^\prime, j^\prime}$ signifies the number of trajectories leading to the reactant/product at a specific rovibrational state ($v^\prime$, $j^\prime$). $N_T$ 
symbolizes total number of trajectories. The orbital angular momentum quantum number ($l$) can be written as:

\begin{equation}
	l_{v^\prime, j^\prime} = l_{max}O_{v^\prime, j^\prime} = (\frac{b_{max}\mu_{tri}v_{0}}{\hbar} - \frac{1}{2})O_{v^\prime, j^\prime}
\end{equation}

\noindent
where $b_{max}$, $\mu_{tri}$, $v_{0}$ and $O_{v^\prime, j^\prime}$ define maximum impact parameter, triatomic center of mass, initial relative velocity and pseudo random number, respectively. 

\subsection{Results and Discussion}	

\noindent
While performing the classical trajectory calculations, 6000 trajectories are executed with different initial conditions for a specific collision energy ($E_{coll}$) after affirming the convergence with respect to the incoming atom distance, $R$ to collide with the diatom. The probability followed by cross-sections are computed for five different processes, namely, trapping (formation of DH$_2^+$), NRNCT, RNCT, RCT and NRCT, where RCT and NRCT processes occur on the first excited electronic state. In the earlier quantum dynamical
calculations,~\citep{jcp-147-074105-SA} the maximum value of the total angular momentum quantum number ($J$) were considered until the reaction probabilities are fully converged upto a particular value of the collision energy ($E_{coll}$). On the other hand, in the trajectory surface hopping (TSH) calculations,~\citep{soumya_hopp} $b_{max}$ is considered as 3 \AA\, and in a similar way, the present QCT calculations with friction are also executed with $b_{max}$ = 3 \AA. It is to note that the value of $b_{max}$ used in QCT calculation and $J_{max}$ involved in quantum calculation are of the same order to reach required convergence upto the specific collisional energies. \\

\begin{figure}[!htp]
	\centering
	\includegraphics[height=0.6\textheight, angle=0]{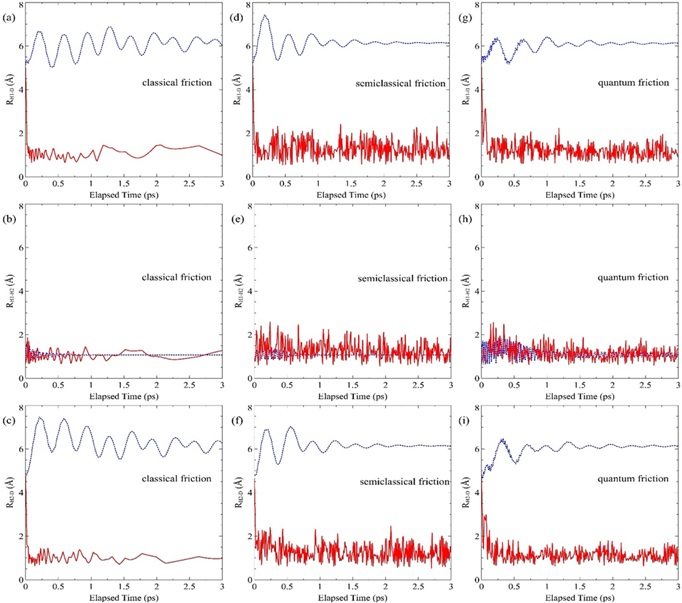}
	\caption{Variation of interatomic distances, (a,d,g) $R_{H1-D}$, (b,e,h) $R_{H1-H2}$ and (c,f,i) $R_{H2-D}$ is presented with respect to elapsed time for a representative trajectory at $E_{coll}$ = 1.7 eV and $b_{max}$  = 3 \AA. The results on the ground and first excited states (using three types of friction) are indicated by red solid and blue dotted lines, respectively. For trapped trajectory, the ground state bond lengths appear as $R_{H1-D}$ = 0.88 \AA, $R_{H1-H2}$ = 0.96 \AA\, and $R_{H2-D}$ = 0.72 \AA\, for classical friction; $R_{H1-D}$ = 1.31 \AA, $R_{H1-H2}$ = 1.77 \AA\, and $R_{H2-D}$ = 0.92 \AA\, for semiclassical friction and $R_{H1-D}$ = 1.19 \AA, $R_{H1-H2}$ = 1.20 \AA\, and $R_{H2-D}$ = 0.77 \AA\, for quantum friction. The bond lengths in the excited state  correspond to H$_2^+$ + D as the asymptote. The figures are adopted from Ref. [\protect\citen{friction_jpca128}] and permission has been taken from ACS.}
	\label{fig:qct_1}
\end{figure}

\noindent
In order to illustrate the damping effects of friction for slowing down the trajectories leading to trapping of the triatomic species, the various side lengths of the triatomic species~\citep{friction_jpca128} in the ground as well as the excited electronic states are plotted in Fig. \ref{fig:qct_1}, for a representative trajectory with $E_{coll}$ = 1.7 eV.
It is worth mentioning that the side lengths, $R_{H1-D}$ and $R_{H2-D}$ appear larger for the excited state due to different asymptotes for the ground and excited states.\\

\begin{figure}[!htp]
	\centering
	\includegraphics[height=0.6\textheight, angle=0]{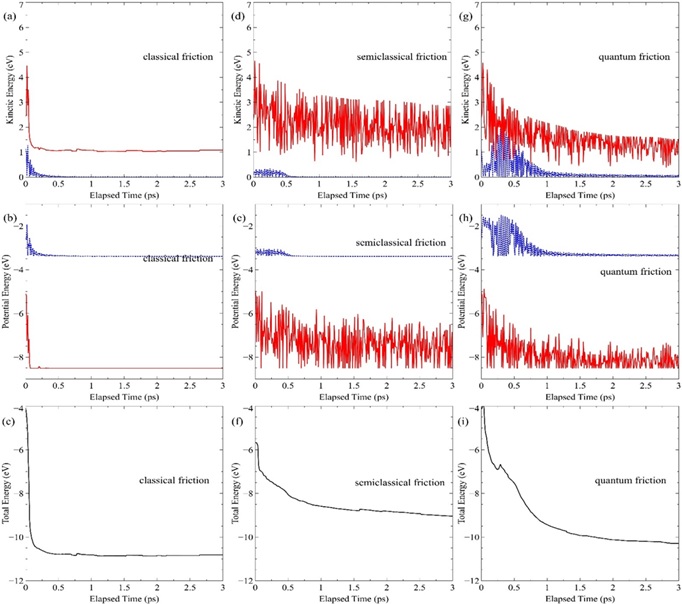}
	\caption{Variation of (a,d,g) kinetic energy, (b,e,h) potential energy and (c,f,i) total energy is plotted with respect to time for a representative trajectory at $E_{coll}$ = 1.7 eV and $b_{max}$ = 3 \AA. The results on the ground and first excited states (using three types of friction) are denoted by red solid and blue dotted lines, respectively. In the trapped condition, the ground state potential energies ($V_{min}^T$) appear around -8.51 eV, -7.44 eV and -8.51 eV for classical, semiclassical and quantum friction, respectively. On the other hand, total energies become -10.45 eV, -9.71 eV and -10.61 eV, respectively. The figures are adopted from Ref. [\protect\citen{friction_jpca128}] and permission has been taken from ACS.}
	\label{fig:qct_2}
\end{figure}

\begin{figure}[!htp]
	\centering
	\includegraphics[height=0.6\textheight, angle=0]{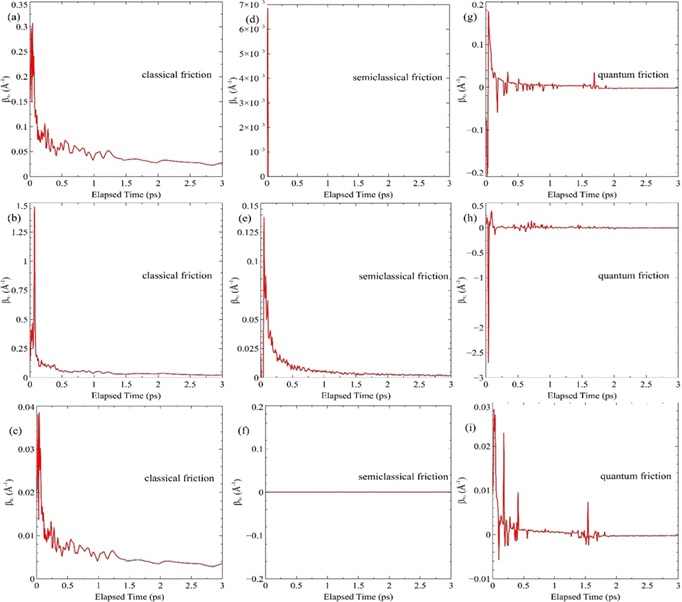}
	\caption{Variation of the x-component of friction coefficient is demonstrated with time for a representative trajectory at $E_{coll}$ = 1.7 eV and $b_{max}$ = 3 \AA\, for the ground electronic state of DH$_2^+$ (using three types of friction).  The figures are adopted from Ref. [\protect\citen{friction_jpca128}] and permission has been taken from ACS.}
	\label{fig:qct_3}
\end{figure}

\begin{figure}[!htp]
	\centering
	\includegraphics[height=0.6\textheight, angle=0]{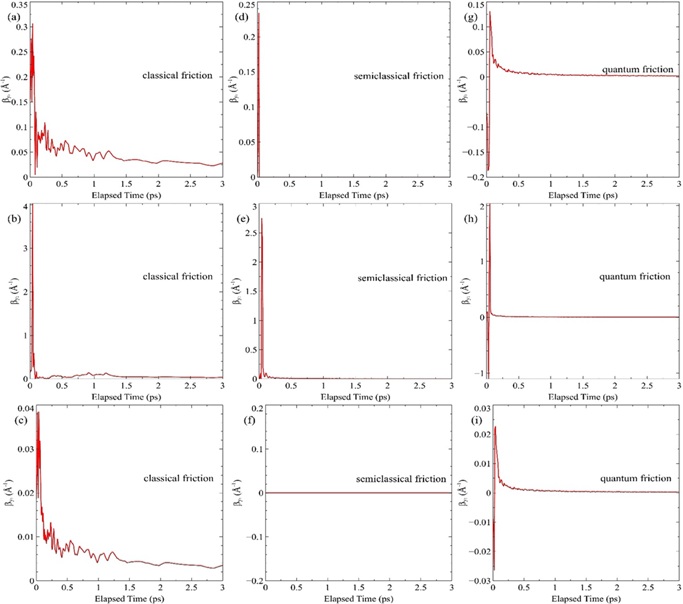}
	\caption{Variation of the y-component of friction coefficient is depicted with time for a representative trajectory at $E_{coll}$ =1.7 eV ($b_{max}$ = 3 \AA) for the ground electronic state of DH$_2^+$ (employing three types of friction). The figures are adopted from Ref. [\protect\citen{friction_jpca128}] and permission has been taken from ACS.}
	\label{fig:qct_4}
\end{figure}

\noindent
On the other hand, Fig. \ref{fig:qct_2} depicts the variation of kinetic energy, potential energy and total energy with respect to time. It is interesting to observe that the kinetic and potential energy oscillate as function of time, but the total energy decreases steadily over a period of time reaching the lowest possible energy around 3 ps. In other words, this particular trajectory is “trapped”, resulting in the formation of DH$_2^+$. In order to unveil the role of friction in determining the collisional outcome in a 2-state calculation, the x- and y-components of the friction coefficient in the ground and the excited electronic states are plotted in Figs. \ref{fig:qct_3} and \ref{fig:qct_4}, respectively, which shows that the dynamics is largely performed over the (x, y) plane and z-component of friction remains zero. \\

\noindent
For untrapped trajectories, the ground state of a specific trajectory leads to either NRNCT or RNCT (non charge-transferred reactants/products), whereas the excited state component of the same trajectory results into either NRCT or RCT (charge transferred products). For $E_{coll}$ = 1.7 eV and $b_{max}$ = 3 \AA, out of 6000 trajectories with classical friction, 5236 are trapped and 764 are untrapped.~\citep{friction_jpca128} Among the ground state components of untrapped trajectories, 699 and 65 produce NRNCT and RNCT species. On the other hand, among the excited state components, 657 and 107 lead to NRCT and RCT reactants/products. In case of semiclassical friction, the numbers appear as 2949 (trapping), 2257 (NRNCT), 794 (RNCT), 2594 (NRCT) and 457 (RCT), respectively. In absence of any sign correction (quantum friction), the numbers are drastically changed to 167 (trapping), 3741 (NRNCT), 2092 (RNCT), 3193 (NRCT) and 2640 (RCT), respectively. \\

\noindent
In case of $E_{coll}$ = 2.5 eV and $b_{max}$ = 3 \AA, classical friction is responsible for generation of 4234 trapped and 1766 untrapped species.~\citep{friction_jpca128}
Among the ground state components of untrapped paths, NRNCT and RNCT processes encounter 1653 and 113 trajectories, whereas NRCT and RCT processes exhibit 
1493 and 273 trajectories. For semiclassical friction, the numbers are 1301 (trapping), 3505 (NRNCT), 1194 (RNCT), 3994 (NRCT) and 705 (RCT), respectively. In case of quantum friction, the numbers appear as 70 (trapping), 4157 (NRNCT), 1773 (RNCT), 3577 (NRCT) and 2353 (RCT), respectively.\\

\noindent
For $E_{coll}$ = 5.0 eV and $b_{max}$ = 3 \AA, out of 6000 trajectories using classical friction, only 578 trajectories are trapped.~\citep{friction_jpca128} Among the ground state components of untrapped trajectories, 5059 and 363 lead to NRNCT and RNCT reactants/products. On the other hand, among the excited state components, 4390 and 1032 paths result into NRCT and RCT species. In case of semiclassical friction, the numbers are 16 (trapping), 4687 (NRNCT), 1297 (RNCT), 4775 (NRCT) and 1209 (RCT), respectively. Finally, while considering quantum friction, the numbers are 1 (trapping), 4615 (NRNCT), 1384 (RNCT), 4111 (NRCT) and 1888 (RCT), respectively.\\

\begin{figure}[!htp]
	\centering
	\includegraphics[height=0.5\textheight, angle=0]{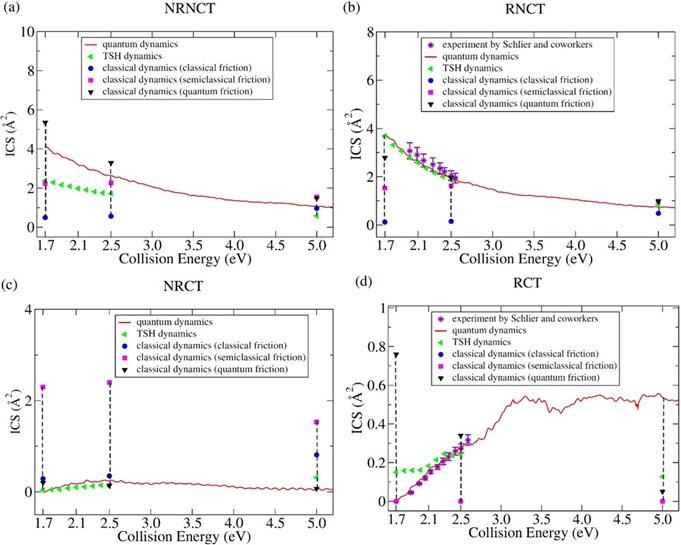}
	\caption{The calculated ICSs (using classical, semiclassical and quantum friction) at $b_{max}$ = 3 \AA\, are presented along with the earlier reported quantum dynamical results~\citep{jcp-147-074105-SA}, TSH findings~\citep{soumya_hopp} and the available experimental data~\citep{cp-111-401-SCH} for NRNCT, RNCT, NRCT and RCT processes. The figures are adopted from Ref. [\protect\citen{friction_jpca128}] and permission has been taken from ACS.}
	\label{fig:qct_5}
\end{figure}

\noindent
Once the classical trajectory calculated cross-sections for two electronic states are in hand, the computed results are presented in Table \ref{Tab:qct_ics} and Fig. \ref{fig:qct_5} in comparison with the earlier reported quantum dynamical results,~\citep{jcp-147-074105-SA} trajectory surface hopping findings~\citep{soumya_hopp} and 
available experimental outcomes~\citep{cp-111-401-SCH}. While carrying out the classical trajectory calculations on a single adiabatic PES with friction,~\citep{baer_aqc89} 
trapping cross section decreases from 1.2 \AA$^2$ at $E_{coll}$ = 1.7 eV to 0.4 \AA$^2$ at $E_{coll}$ = 2.5 eV. On the other hand, RNCT cross-section appears around
3.0 \AA$^2$ at $E_{coll}$ = 2.5 eV,~\citep{jcp-147-074105-SA} which shows good agreement with quantum dynamical (1.94 \AA$^2$), TSH (1.88 \AA$^2$)  and experimental (2.04 \AA$^2$) results. In case of two electronic state study,~\citep{friction_jpca128} the cross section for the RNCT process at $E_{coll}$ = 2.5 eV appears to be 0.15 \AA$^2$ for 
classical friction, 1.62 \AA$^2$ for semiclassical friction and 1.95 \AA$^2$ for quantum friction. Using the same conditions, ICS for RCT is found to be 0.0 \AA$^2$ (classical friction), 0.0 \AA$^2$  (semiclassical friction) and 0.34 \AA$^2$  (quantum friction) in comparison with 0.27 \AA$^2$ from  quantum dynamics~\citep{jcp-147-074105-SA}, 0.24 \AA$^2$  from TSH calculations~\citep{soumya_hopp}, and 0.27 \AA$^2$  from experiment~\citep{cp-111-401-SCH}. Finally, the trapping cross-sections are found to be  4.37 \AA$^2$, 0.92 \AA$^2$ and 0.13 \AA$^2$ at $E_{coll}$ = 2.5 eV for classical, semiclassical and quantum friction, respectively. As a matter of fact, the trapping cross-section exhibits a decaying trend from (a) 6.96 \AA$^2$ at $E_{coll}$ = 1.7 eV to 4.37 \AA$^2$ at $E_{coll}$ = 2.5 eV to 1.29 \AA$^2$ at 5.0 eV for classical friction; (b) from 3.53 \AA$^2$ to 0.92 \AA$^2$ to 0.10 \AA$^2$ for semiclassical friction, and (c) from 0.53 \AA$^2$ to 0.13 \AA$^2$ to 0.36 \AA$^2$ for quantum friction. It is important to note that quantum friction calculated results~\citep{friction_jpca128} agree well with the quantum dynamical outcome~\citep{jcp-147-074105-SA} at higher energies ($E_{coll}$ = 5.0 eV) also. The cross section for NRCT process~\citep{friction_jpca128} is overestimated by the quantum friction model, but for the RCT process, the ICS is underestimated with respect to the exact quantum results~\citep{jcp-147-074105-SA} over the entire range. \\

\begin{table}[!htp]
	\vspace{-0.5 cm}
	\caption{ICS values originated from two electronic state classical trajectory calculations for $b_{max}$ = 3 \AA\, are compared
		with the earlier quantum dynamics results,~\citep{jcp-147-074105-SA} TSH outcomes~\citep{soumya_hopp} and experimental findings~\citep{cp-111-401-SCH}.
		The results are taken from Ref. [\protect\citen{friction_jpca128}].}
	\scalebox{0.6}{\begin{tabular}{|c|c|c|c|c|c|c|c|c|c|}\hline
			& type          & NRNCT~\citep{friction_jpca128} & RNCT~\citep{friction_jpca128} & NRCT~\citep{friction_jpca128} & RCT~\citep{friction_jpca128}  & Trap~\citep{friction_jpca128} & \begin{tabular}[c]{@{}l@{}}Quantum\\  Dynamics~\citep{jcp-147-074105-SA}\end{tabular}                                                & TSH~\citep{soumya_hopp}                                            & Expt.~\citep{cp-111-401-SCH}                                                                         \\\hline
			\multirow{3}{*}{1.7 eV} & classical     & 0.49  & 0.12 & 0.29 & 0.0  & 6.96 & \multirow{3}{*}{\begin{tabular}[c]{@{}l@{}}NRNCT: 4.13, RNCT: 3.77,\\ NRCT: 0.0, RCT: 0.0\end{tabular}}   & \multirow{3}{*}{\begin{tabular}[c]{@{}l@{}}NRNCT: 2.38, RNCT: 3.69,\\  NRCT: 0.04, RCT: 0.15\end{tabular}} & \multirow{3}{*}{---}                                                                   \\
			& semiclassical & 2.23  & 1.53 & 2.30 & 0.0  & 3.53 &                                                                                                                                  &                                                                                                                                  &                                                                                        \\
			& quantum       & 5.35  & 2.79 & 0.20 & 0.76 & 0.53 &                                                                                                                                  &                                                                                                                                  &                                                                                        \\ \hline
			\multirow{3}{*}{2.5 eV} & classical     & 0.56  & 0.15 & 0.35 & 0.0  & 4.37 & \multirow{3}{*}{\begin{tabular}[c]{@{}l@{}}NRNCT: 2.62, RNCT: 1.94,\\    NRCT: 0.24, RCT: 0.27\end{tabular}} & \multirow{3}{*}{\begin{tabular}[c]{@{}l@{}}NRNCT: 1.72, RNCT: 1.88,\\ NRCT: 0.12, RCT: 0.24\end{tabular}} & \multirow{3}{*}{\begin{tabular}[c]{@{}l@{}}RNCT: 2.04,\\    \\ RCT: 0.27\end{tabular}} \\
			& semiclassical & 2.26  & 1.62 & 2.40 & 0.0  & 0.92 &                                                                                                                                  &                                                                                                                                  &                                                                                        \\
			& quantum       & 3.29  & 1.95 & 0.14 & 0.34 & 0.13 &                                                                                                                                  &                                                                                                                                  &                                                                                        \\ \hline
			\multirow{3}{*}{5.0 eV} & classical     & 0.96  & 0.48 & 0.81 & 0.0  & 1.29 & \multirow{3}{*}{\begin{tabular}[c]{@{}l@{}}NRNCT: 1.06, RNCT: 0.74,\\ NRCT: 0.07, RCT: 0.53\end{tabular}} & \multirow{3}{*}{\begin{tabular}[c]{@{}l@{}}NRNCT: 0.57, RNCT: 0.85,\\  NRCT: 0.32, RCT: 0.13\end{tabular}} & \multirow{3}{*}{}                                                                      \\
			& semiclassical & 1.54  & 0.99 & 1.53 & 0.0  & 0.10 &                                                                                                                                  &                                                                                                                                  &                                                                                        \\
			& quantum       & 1.48  & 0.98 & 0.07 & 0.05 & 0.36 &                                                                                                                                  &                                                                                                                            &                \\ \hline                                                                       
	\end{tabular}}
	\label{Tab:qct_ics}
\end{table}

\section{Conclusion}\label{ssec:qct_conc}	
\noindent
In order to explore the workability of TSH formalism, two prototypical triatomic reactions, namely, 
D$^{+}$ + H$_{2}$ ($v=0$, $j=0$)~\citep{jcp-147-074105-SA} and H + H$_{2}^{+}$ ($v=0$, $j=0$)~\citep{ghosh_jpca125} [initiating from the ground (H$_2 (^1\Sigma_g^+)$ + D$^+$) and first excited 
(H$_2^+ (^2\Sigma_g^+)$ + H) electronic states of H$_3^+$] are chosen to compute integral
cross-sections and/or rate constants of various charge transfer and non-charge transfer processes (RNCT, RCT and NRCT).~\citep{soumya_hopp} 
While performing such calculations, adiabatic PESs and NACTs of the lowest three singlet states (1$^1A^\prime$, 2$^1A^\prime$ and 
3$^1A^\prime$) of H$_3^+$ are employed obtained from accurate {\it ab initio} calculation (MRCI).~\citep{mukherjee_jcp143} The energy conservation is imposed on the process of hopping according to Thiel's algorithm~\citep{thiel_cp349} along with setting of new equations~\citep{soumya_hopp} for exploring momentum conservation. It is worth mentioning that 
for both the reaction processes (D$^{+}$ + H$_{2}$ and H + H$_2^+$), the root mean square (rms) deviation of momentum conservation appears in the order of 10$^{-6}$ or 10$^{-7}$ affirming the scaling algorithm of momentum conservation. For D$^{+}$ + H$_{2}$ ($v=0$, $j=0$) reaction, TSH based ICS for RNCT process~\citep{soumya_hopp}  
matches well with the quantum results~\citep{jcp-147-074105-SA}, but the same quantities (TSH based ICSs)~\citep{soumya_hopp} for the other two cases, RCT and NRCT show deviation from quantum ones~\citep{jcp-147-074105-SA}. On the other hand, in case of H + H$_{2}^{+}$ ($v=0$, $j=0$) reaction, TSH calculated ICS for NRCT and RCT processes~\citep{soumya_hopp} exhibit steep rise at the lower
energy domain ($E_{tot}$ $<$ 2.3 eV) in contrary to the quantum calculated ones~\citep{ghosh_jpca125}. Furthermore, TSH calculated temperature dependent rate constants~\citep{soumya_hopp} appear higher than the quantum counterparts~\citep{ghosh_jpca125} with almost same slope. As a matter of fact, TSH algorithm is unable to capture the accurate chemistry of multi-state processes at the domain of strong non-adiabatic couplings. \\

\noindent
On the other hand, semi-classical dynamics are carried out considering  quantum mechanical NACTs as equivalent to friction in classical mechanics.~\citep{friction_jpca128} The (D$^+$, H$_2$) collision dynamics is studied over a two-state PES, where the nuclei are substantially slowed down due to nonadiabatic coupling leading to 
formation of the triatomic species DH$_2^+$ as well as NRNCT, RNCT, RCT and NRCT reactants/products. The two-state classical trajectory calculations with quantum friction results are quite close while comparing with quantum~\citep{jcp-147-074105-SA} and experimental~\citep{cp-111-401-SCH} ones at $E_{coll}$ = 2.5 eV. In addition, the use of semiclassical friction also produces comparable results, but classical friction seems to overestimate the trapping cross section and underestimate the RNCT cross section. In other words, the concept of quantum friction satisfactorily reproduces the quantum~\citep{jcp-147-074105-SA} and experimental~\citep{cp-111-401-SCH} results in comparison to the other ones.

\bibliographystyle{aipnum4-1}
\bibliography{ref_book}

\chapter{Phase Transition in Solid State Compounds: Diabatic Hamiltonian and Dynamics}\label{sec:phase}
\noindent
Perovskites with general chemical formula ABO$_3$ show an extensive range of functional properties in materials science and condensed matter physics. Among these, the benchmark systems, namely, lanthanum manganite (LaMnO$_3$, LMO) and barium titanate (BaTiO$_3$, BTO) possesses distinct structural and electronic characteristics. LMO acts as precursor for colossal magnetoresistance (CMR) materials and acquires a distorted orthorhombic perovskite structure due to the cooperative Jahn-Teller effect associated with the Mn$^{3+}$ ions. Such structural distortion originates from the degeneracy of \(e_g\) orbitals of Mn$^{3+}$ resulting into strong electron-lattice interactions, which are responsible for 
significantly influence the material's transport and magnetic properties.~\cite{Tokura_1999} It is important to note that the 
interplay of orbital ordering, lattice distortions and electron correlations in LMO have crucial contribution in its magnetoelectronic behavior, which makes it a cornerstone in the study of correlated electron systems.~\cite{Dagotto_2001} On the contrary, the well-known ferroelectric perovskite, BTO shows temperature-driven phase transition from a cubic to a tetragonal structure including spontaneous electric polarization.~\cite{Lines_2001} Such phase transition is dictated by the 
off-center displacement of Ti$^{4+}$ ions within the oxygen octahedral unit leading to a strong dipolar interaction. The ferroelectricity in BTO is responsible for its widespread utilization in a broad spectrum of applications, including capacitors, piezoelectric transducers and non-volatile memory devices.~\cite{Jaffe_1971} The complex interplay between lattice distortions, electron correlations and orbital interactions leads to the strikingly different physical properties observed in these perovskites. This intricate coupling plays a crucial role for exploring emergent quantum phenomena as well as drives technological advancements in electronic and spintronic devices.~\cite{Ramesh_2007}\\

\noindent
It is important to note that transition metal oxides and halides with Jahn-Teller (JT) active ions, such as Mn$^{3+}$ (3$d^4$) and Cu$^{2+}$ (3$d^9$), exhibit a wide variety of magnetic phenomena, like colossal magnetoresistance and high-temperature superconductivity, with the cooperative electron-lattice JT effect.\cite{JT,JT1,VanVleck1,VanVleck,Opik,Longuet-Higgins,O'Brien,Bersuker1,Sovphys,Cooperative-JT-Englman-I,Cooperative-JT-Englman-II,JCS,REnglman,IBersuker-VPolinger,IBersukerbook,JT-review-Bersuker,IBB} Such 
properties drive structural phase transitions and superexchange interactions\cite{Kugel-Khomskii,Oles-Feiner,Natalia_JETP1,Natalia_JETP2} through the stabilization of magnetic order. Several research groups\cite{domcke_cp387,mukhanov_jsc53,pitzer_jcp70,mukhanov_rjpca88,domcke_jcp152} investigated the contribution of JT couplings on the vibronic spectra of transition metal halides, including MnF$_3$, CoF$_3$ and VF$_{3,4}$. In LMO, JT effect governs the transition from a pseudo-cubic perovskite ($Pm \bar 3 m$) to an orthorhombic ($Pbnm$) structure at $\sim$780 K,\cite{Phase} leading to the stabilization of an $A$-type antiferromagnetic phase below the N\'eel temperature ($T_{\rm N} \sim$ 140 K).\cite{Hirota,Natalia_JETP,joy_jcp} Further optical studies explore resonant excitations associated with intersite electronic transitions and their temperature-dependent shifts, indicating a strong vibronic contribution.\cite{Natalia_JETP,MultiRaman,Natalia_JETP1, Natalia_JETP2, Natalia_JETP3} The JT theorem was initially formulated in the 1930s,\cite{JT,JT1} which depicts geometric distortions in electronically degenerate polyatomic systems. Such formalism was first 
applied to transition metal complexes by Van Vleck.\cite{VanVleck1,VanVleck} The `Mexican hat' adiabatic potential energy surface (PES) is generated by linear vibronic interactions,\cite{Longuet-Higgins,Opik} where strong coupling (with and without anharmonic terms) is responsible for pseudo-rotations or static JT distortions.\cite{O'Brien} On the other hand, the tunneling effects in the EPR spectra of Cu$^{2+}$ complexes was first revealed by Bersuker.\cite{Bersuker1} Quadratic JT coupling also plays crucial role in the warping of APES, forming additional conical intersections (CIs),\cite{quadratic-coupling1,quadratic-coupling-strong1,quadratic-coupling-Li3,quadratic-coupling-Na3,mukherjee_jpca117,quadratic-coupling-Exe,quadratic-coupling-Txt2,quadratic-coupling-Exe-nonadiabatic} where \textit{ab initio} studies explore the dominant role of cubic anharmonicity in these systems.\cite{warping-Exe} In addition, excited-state JT effects create centrifugally stabilized states,\cite{JCS} where theoretical models are used to explore the optical spectra vis-\'a-vis non-adiabatic interactions.\cite{Hizhnyakov1,Hizhnyakov2,Hizhnyakov3,Hizhnyakov4}
In presence of strong coupling distorting molecular structures, JT effects arise often in conjunction with pseudo JT (PJT) interactions.\cite{Bersuker-Yarkony-book,Hidden-JT1,Hidden-JT-PJT,IBersuker-VPolinger, IBB, PJT0,PJT1,PJT2,PJT3} In recent works, JT model has been extended incorporating cubic anharmonicity, which provides 
in-depth understanding about the high-energy excitations in LMO.\cite{joy_jcp} Such work affirms the necessity of calculation of \textit{ab initio} adiabatic PES and nonadiabatic coupling to illustrate these transitions.\cite{Natalia_JETP, Natalia_JETP1, Natalia_JETP2, Natalia_JETP3} It is worth mentioning that first-principle based 
diabatization of  adiabatic PESs and nonadiabatic coupling terms (NACTs) is indeed necessary to unveil the role of electron-nuclear interactions predicting the photoelectron spectra of MnO$_6^{10-}$ and related systems.\\

\noindent
On the other hand, TiO$_6^{8-}$ unit in BTO \cite{Bersuker18112018,Bersuker_review_2021} exhibits strong Pseudo Jahn-Teller (PJT) interactions between the ground ($^1A_{1g}$) and first excited ($^1T_{1u}$) states resulting into temperature-dependent structural distortions and spontaneous polarization. Such phenomena are 
central to its ferroelectric properties, as depicted by I. B. Bersuker.~\cite{Bersuker_1966} Electronic structure and ferroelectric behavior of BaTiO$_3$ were investigated theoretically, where the studies reveal the influence of PJT effects on the material's polarization, flexoelectricity, and dipolar distortions. On the other hand, structural and electronic properties were examined through various computational approaches such as Density Functional Theory (DFT) and Linearized Augmented-Plane-Wave (LAPW) methods\cite{Kolezynski_2005, Qi-Jun_Liu_2013, bersuker_pbcm457, Bersuker_2015}. Further studies were carried out to examine the role of spin-orbit (SO) coupling in doped BTO crystals exploring significant Rashba spin-orbit interactions in Bi adlayers and mixed perovskites. On the contrary, the effect of SO coupling in pure BTO remain is still an unexplored area\cite{Mirhosseini_phys_rev_2010, Zhong_Adv_Mat, Ali_BFO_BTO}. In addition, experimental techniques, such as Raman spectroscopy, dielectric measurements and terahertz time-domain spectroscopy were performed to unveil the relationship between optical and low-frequency dielectric properties of BTO \cite{Pinczuki_1967, Turik_Shevchenko_1979, R_Cord_1985, Roberts_1987, Nasser_Expt_2000, Wan_2008}. The {\it ab initio} calculated
 adiabatic PESs and NACTs followed by Beyond Born-Oppenheimer (BBO) theory based diabatization\cite{Mantu_BTO} affirm the origin of ferroelectricity through TiO$_6^{8-}$ unit of the material, BTO \cite{Kolezynski_2005, Qi-Jun_Liu_2013, Bersuker_2015}. Adhikari and coworkers\cite{Mantu_BTO} investigated the ferroelectric properties of BTO by studying the TiO$_6^{8-}$ unit theoretically, where the good correspondence are observed with the experimental photoemission spectra and order parameter of ferroelectricity.  \\

\section{Theoretical Background of Dynamic JT Theory}\label{ssec:dyJT}	

\noindent
Even though a detailed discussion of the theoretical background of dynamic JT theory is presented in Chapter \ref{sec:symm}, a brief overview of the methodology is discussed in the following sections for the sake of completeness.

\subsection{Jahn Teller Theory} \label{sec:jt}
\vspace{0.2 cm}

\noindent   
As per the JT theorem,\citep{JT} any nonlinear molecule with degenerate electronic states exhibits geometrical distortion leading to lifting of degeneracy and lowering of symmetry, which reduce the total energy. The resulting electronic Hamiltonian can be expressed as a Taylor series in terms of the nuclear coordinates ($\{Q_k\}$) as follows:
\begin{eqnarray}
	\hat{H} = \hat{H}^0 + \sum_k \hat{V}_kQ_k + 
	\dfrac{1}{2}\sum_{kl}\hat{V}_{kl}Q_kQ_l + \cdots,
	\label{Eq:ham_taylor}
\end{eqnarray}

\noindent
where $\hat{H}^0$ represents the electronic Hamiltonian at the equilibrium geometry $(\{Q_k\}=0)$. The terms $\hat{V}_k \left(=\frac{\partial H}{\partial Q_k}\right)$ and $\hat{V}_{kl} \left(=\frac{\partial^2 H}{\partial Q_k \partial Q_l}\right)$ symbolize first- and second-order coupling coefficients. The first-order JT coupling [Eq. \ref{Eq:ham_taylor}] can be written as:
\begin{eqnarray}
	V^{ij}=\sum_{k} Q_{k} \langle \phi^{i,(0)} \vert \hat{V}_{k}\vert \phi^{j,(0)}\rangle,
	\label{eq:jtpot}
\end{eqnarray}

\noindent
where the set $\{\phi^{n,(0)}\}$ represents degenerate electronic eigenfunctions. A molecule is JT active along a mode $Q_k$ for non-zero coupling integral $\langle \phi^{i,(0)} | \hat{V}_k | \phi^{j,(0)} \rangle$. Such integral survives only when the product representation $\Gamma(\phi^{i,(0)}) \otimes \Gamma(\hat{V}_k) \otimes \Gamma(\phi^{j,(0)})$ contains the totally symmetric irreducible representation (IREP) $\Gamma^0$. If $\Gamma^\alpha$ and $\Gamma^\beta$ denote the IREPs of the electronic state and displacement $Q_k$, respectively, the JT activity condition can be written as:  
\begin{eqnarray}
	\Gamma^0 \in \Gamma^{\beta} \otimes \left[ (\Gamma^{\alpha})^2 \right] 
	\label{Eq:JTcond}
\end{eqnarray}

\noindent 
Hence, $Q_k$ is JT active only if its IREP $\Gamma^\beta$ is present in the symmetrized product $\Gamma^\alpha \times \Gamma^\alpha$, where $\Gamma^\alpha$ represents a non-totally symmetric one. 

\subsection{Outline of the Linear Jahn-Teller $\boldsymbol{E \otimes e}$ Model}\label{ssec: longuet-higgins Exe}
\vspace{0.2 cm}

\noindent
According to the $E \otimes e$ JT model \citep{Longuet-Higgins}, doubly degenerate $e_g$ modes [$Q_x = \rho \cos\phi$, $Q_y = \rho \sin\phi$] exhibit JT splitting of singly (e.g., high-spin d$^4$, Mn$^{3+}$) or triply (d$^9$, Cu$^{2+}$) occupied $E_g$ states ($|x^2 - y^2\rangle$, $|3z^2 - r^2\rangle$). The splitting is mediated by a perturbation of the form~\citep{longuet-higgins_prsl344}:

\begin{equation}
	H^{\prime} = 2A\rho\cos\phi(\vert X\rangle\langle X\vert-\vert Y\rangle\langle Y\vert)+2A\rho\sin\phi(\vert X\rangle\langle Y\vert+\vert Y\rangle\langle X\vert).
	\label{Eq:pert}
\end{equation}

\noindent
Such perturbation splits the unperturbed adiabatic PES (APES) into a doubly branched ‘Mexican hat’ form, with $|X\rangle \sim \cos\theta$ and $|Y\rangle \sim \sin\theta$ designating the $E_g$ states. Since the electronic motion is assumed to be circular, Eq.~\ref{Eq:pert} can be simplified as:  
\begin{equation}
	H^{\prime} = 2A \rho \cos (2 \theta -\phi).
\end{equation}

\begin{figure}
	\centering
	\includegraphics[width=0.9 \textwidth , width=1 \textwidth, angle=0]{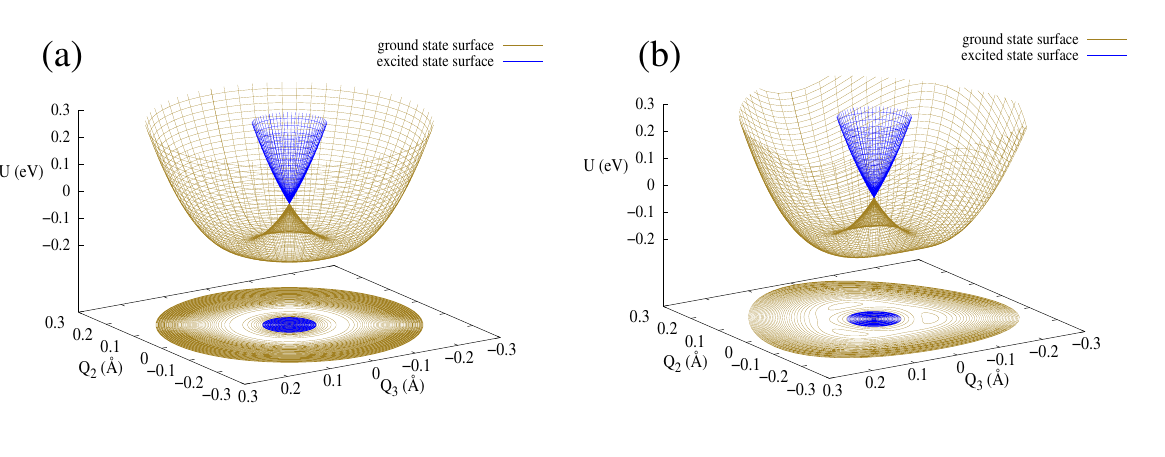}
	\caption{Diagrams display doubly branched `Mexican Hat' APESs for LaMnO$_3$ (a) without and (b) with cubic perturbation, employing experimental JT parameters\citep{Natalia_JETP,MultiRaman} ($\omega_{\text{b\_vib}} = 500$ cm$^{-1}$, $A=$ 2.5 $\times$ 10$^{4}$ cm$^{-1}$ \AA$^{-1}$). The figure is adapted from Ref.~[\protect\citen{Mantu_Solid-state_Review}] and permission has been taken from Wiley.}
	\label{fig:jt}
\end{figure}

\noindent
The degeneracy of the doubly degenerate electronic states is lifted by
JT distortion yielding eigenvalues $\pm A\rho$. The generated vibronic APESs form a doubly branched ‘Mexican hat’ potential, which is represented as:  
\begin{eqnarray}
	U^{i=1,2}= \pm A \rho + \frac{1}{2} M \omega_{\text{b\_vib}}^2 \rho^2, 
	\label{Eq: two valued surface}
\end{eqnarray}

\noindent
where $A$, $M$, and $\omega_{\text{b\_vib}}$ designate the linear JT coupling constant, mass of the vibrating atom, and bond vibration frequency, respectively. Figure~\ref{fig:jt}a shows $-A\rho$ and $A\rho$ JT energy splittings of the lower and upper APES branches ($U^1$ and $U^2$). The minimum of $U^1$ appears at $\left\vert \rho^g_0 \right\vert= \frac{A}{M \omega_{\text{b\_vib}}^2}$ resulting into a circular trough of degenerate distorted geometries. Such trough is stabilized by the JT energy $E_{\rm JT}$ $=-\frac{A^2}{2 M \omega_{\text{b\_vib}}^2}$. Since the minima spanning all $\phi$ at fixed $\rho$ are continuous, zero barriers are responsible for nuclear tunneling between configurations known as the dynamic JT or Longuet-Higgins-\"Opik-Pryce-Sack (LHOPS) effect \citep{Longuet-Higgins}.\\

\noindent
Linear $E \otimes e$ interaction couples electronic and nuclear motions leading to the formation of pseudo-rotational states labeled by $l = m + \frac{1}{2}j$, with $m$ and $j$ as vibrational and electronic angular quantum numbers. The corresponding radial Schrödinger equations (SE) for ground and excited APES branches are expressed as:\\

\begin{subequations}
	\begin{align}
		-\frac{\hbar^2}{2M} \left[\frac{d^2 \psi_g(\rho)}{d \rho^2} - \frac{l^2}{\rho^2} \psi_g(\rho)\right] - E^{g} \psi_g(\rho) + \left( -A \rho + \frac{1}{2} M \omega_{\text{b\_vib}}^2 \rho^2 \right) \psi_g(\rho) = 0, \: \: \: \text{for ground state}
	\end{align}
	\text{and}
	\begin{align}
		-\frac{\hbar^2}{2M} \left[\frac{d^2 \psi_e(\rho)}{d \rho^2} - \frac{l^2}{\rho^2} \psi_e(\rho)\right] - E^{e} \psi_e(\rho) + \left( A \rho + \frac{1}{2} M \omega_{\text{b\_vib}}^2 \rho^2 \right) \psi_e(\rho) = 0, \: \: \: \text{for excited state} \nonumber \\
	\end{align}
	\label{Eq:radial_lower_excited_surface_equ}
\end{subequations}

\noindent
where $\psi_{g/e}(\rho)$ symbolize radial vibronic wavefunctions and $\frac{\hbar^2 l^2}{2M \rho^2}$ denotes the centrifugal term for half-integer $l$.

\subsection{Static JT Effect at Lower Adiabatic PES Minima}
\vspace{0.2 cm}

\noindent
The lower branch of the `Mexican hat' adiabatic PES creates a minima at $\left\vert \rho^g_0 \right\vert = \frac{A}{M \omega_{\text{b\_vib}}^2}$
by balancing the linear JT force ($F_{JT} = -A$) and the elastic force of harmonic vibration ($F_{\text{b\_vib}} = M \omega_{\text{b\_vib}}^2 \rho$). In presence of strong JT coupling, such minima move to higher $\rho$ values (e.g., 0.3 \AA \citep{Opik} for [Cu(H$_2$O)$_6$]$^{2+}$ and 0.21 \AA\citep{joy_jcp} for LaMnO$_3$). At and around these minima, centrifugal energy ($\frac{\hbar^2 l^2}{2M \rho^2}$) being as an approximately constant quantity can be treated as a parameter in the ground state SE (see Eq. \ref{Eq:radial_lower_excited_surface_equ}a). In other words, at the close vicinity of the minima, the nuclear SE eigenvalues can be expressed in harmonic approximation~\citep{Longuet-Higgins, joy_jcp}:

\begin{eqnarray}
	E_{p l}^{g} &=& \hbar \omega_{\text{b\_vib}} \left( p + \frac{1}{2} \right) - \frac{A^2}{2 M \omega_{\text{b\_vib}}^2} + \frac{\hbar^2 l^2 M \omega_{\text{b\_vib}}^4}{2 A^{2}},  \quad p = 0, 1, 2,... , \nonumber \\
	\label{Eq: Eigenvalue of lower surface}
\end{eqnarray}


\noindent
where $p$ designates the vibrational quantum number of the harmonic oscillator displaced due to JT interaction and the term $\frac{\hbar^2 l^2 M \omega_{\text{b\_vib}}^4}{2 A^{2}}$ symbolizes the centrifugal energy for a pseudo-rotational level ($l$) at the minima of the ground state APES.\\

\noindent
On the other hand, incorporation of an anharmonic term ($A_3 \rho^3 \cos3\phi$) in the SE leads to three minima and associated barriers (for $\phi = \frac{\pi}{3}, \pi, \frac{5\pi}{3}$ when $A_3 > 0$, and $\phi = 0, \frac{2\pi}{3}, \frac{4\pi}{3}$ when $A_3 < 0$) resulting into the warping of the `Mexican hat' APES (see Figure \ref{fig:jt}b). Such warping
breaks the infinite degeneracy of the ground vibronic state and hence, roto-vibrational motion becomes hindered, described by the SE~\citep{O'Brien}:

\vspace*{-0.3 cm}

\begin{eqnarray}
	&&-\frac{\hbar^2}{2M} \left[ \frac{\partial^2}{\partial \rho^2} +  \frac{1}{\rho^2} \frac{\partial^2}{\partial \phi^2} \right] \xi_g (\rho, \phi) - E \xi_g(\rho, \phi) \nonumber \\
	&& + \left( - A \rho + \frac{1}{2} M \omega_{\text{b\_vib}}^2 \rho^2 + A_3 \rho^3 \cos(3 \phi) \right) \xi_g(\rho, \phi) = 0.
	\label{Anharmonic-lower-schroedinger-eqn}
\end{eqnarray}

\noindent
As a result, the eigenvalue of the SE (Eq. \ref{Anharmonic-lower-schroedinger-eqn}), $E$, gets split into $E = \epsilon_0^g + \epsilon$. $\epsilon_0^g$ represents the displaced harmonic oscillator energy originating from $E \otimes e$ vibronic coupling, and $\epsilon$ symbolizes the modified eigenvalue of the cubic perturbed pseudo-rotational level with a barrier. The effect of the barrier on the eigenvalue spectrum can be illustrated by the following equation:  

\begin{eqnarray}
	&&\alpha \frac{\partial^2}{\partial \phi^2} \gamma (\phi) + \beta \cos (3 \phi) \gamma(\phi) + \epsilon (\rho)\gamma(\phi) = 0, 
	\label{Eq: ground state Mathieu Eq}
\end{eqnarray}
\text{where} 
\begin{eqnarray}
	&&\alpha =\langle w_{g}^0(\rho) \vert \frac{\hbar^2}{2 M \rho^2} \vert w_{g}^0(\rho)\rangle,  \nonumber \\
	&&\beta = \langle w_{g}^0(\rho) \vert - A_3 \rho^3 \vert w_{g}^0(\rho) \rangle.
	\label{Eq: ground state Mathieu alpha}
\end{eqnarray}

\noindent
In case of elastically stabilized ground vibronic adiabatic PES, the splitting ($\alpha$) and barrier ($\beta$) parameters are approximately expressed as $\alpha = \frac{\hbar^2}{2 M \rho_0^2}$ and $\beta = - A_3 \rho_0^3$. While solving the secular equation (Eq. \ref{Eq: ground state Mathieu Eq}) over a range of $\beta$ values, cubic perturbed rotational eigenvalue profiles are obtained for various pseudo-rotational levels ($\lbrace l \rbrace$). If the pseudo-rotation becomes very slow (low $l$, $\epsilon < \beta$), the species is trapped in one of the minima, widely known as static JT effect~\citep{O'Brien}.

\subsection{Centrifugally Stabilized Excited Vibronic State and Anharmonic Effects}
\vspace{0.2 cm}


\noindent
The radial vibronic SE for the upper branch of the adiabatic PES (see Eq. \ref{Eq:radial_lower_excited_surface_equ}b) takes the following form \citep{JCS}:

\vspace*{-0.6 cm}

\begin{eqnarray}
	H \psi_e(\rho)= \left(- \frac{\hbar^2}{2M} \frac{\partial^2}{\partial \rho^2} + \frac{\hbar^2 l^2}{2M\rho^2} + A \rho + \frac{1}{2} M \omega^2_{\text{b\_vib}} \rho^2\right)\psi_e(\rho)=E\psi_e(\rho),
	\label{Eq: Radial Hamiltonian}
\end{eqnarray}

\noindent
where $A\rho$ represents the JT destabilization energy for the excited state and $\frac{\hbar^2 l^2}{2M\rho^2}$ corresponds to the centrifugal energy for the pseudo-rotational level $l$.\\

\noindent
In case of ground and excited state `Mexican hat' adiabatic PESs, minima appear far and near from the center, respectively. It is interesting to note that the JT interaction energy ($A \rho$) exhibits a linear increasing trend with $\rho$, while the centrifugal term ($\frac{\hbar^2 l^2}{2M\rho^2}$) diminishes sharply away from the center. In presence of two opposing forces, excited state minima ($\{\rho_0^l\}$) are formed.  
The elastic potential energy of bond vibration ($\frac{1}{2} M \omega^2_{\text{b\_vib}} \rho^2$) has less contribution near $\rho \sim 0$ leading to centrifugally stabilized bound vibronic states \cite{JCS}, which show dependence on pseudo-rotational quantum numbers ($l$). While incorporating anharmonicity, the cubic perturbation ($A_3 \rho^3 \cos 3\phi$) modifies the SE taking the following form:

\begin{eqnarray}
	&&-\frac{\hbar^2}{2M} \left[ \frac{\partial^2}{\partial \rho^2} + \frac{1}{\rho^2} \frac{\partial^2}{\partial \phi^2} \right]\xi_e(\rho, \phi)  - E \xi_e(\rho, \phi) \nonumber \\
	&& + \left(A \rho + \frac{1}{2} M \omega_{\text{b\_vib}}^2 \rho^2 + A_3 \rho^3 \cos(3 \phi) \right) \xi_e(\rho, \phi) = 0.
	\label{Eq: Hamiltonian of cubic perturbed excited state}
\end{eqnarray}

\noindent
In order to determine the positions of minima for cubic perturbed centrifugally stabilized bound states\cite{joy_jcp}, it is necessary to calculate the full cubic perturbed pseudo-rotational energy spectrum across the radial coordinate $\rho$. If Eq. \ref{Eq: Hamiltonian of cubic perturbed excited state} is averaged radially, the angular SE (Mathieu equation under C$_{3v}$ symmetry) can be solved as:

\begin{eqnarray}
	\alpha_{i} \frac{\partial^2}{\partial \phi^2} \gamma (\phi) + \beta_{i} \cos (3 \phi) \gamma(\phi) + \epsilon (\rho_i)\gamma(\phi) = 0, 
	\label{Eq:Excited state Mathieu Equation}
\end{eqnarray}
resulting into the energy spectrum of the excited state APES. \\

\noindent
In the excited state angular SE, $\alpha_i = \frac{\hbar^2}{2 M \rho_i^2}$ and $\beta_i = -A_3 \rho_i^3$ appear as variable terms contrary to the ground state. The eigenvalues of elastically coupled excited states in presence of harmonic bond vibration ($\omega_{\text{b\_vib}}$) are written as\cite{joy_jcp}:

\begin{eqnarray}
	E_{s, l}^{\text{cp + b\_vib}} = \epsilon_{l} (\rho_{0}^{l}) + A \rho_{0}^{l} + \hbar \left[ 
	\frac{1}{M} \left( \frac{\partial^2 U_{+}}{\partial \rho^2}  \right)_{\rho=\rho_{0}^{l}} + \omega^2_{\text{b\_vib}}  
	\right]^{\frac{1}{2}} \left( s + \frac{1}{2} \right),
	\label{Eq: cp_bond_eigenvalue}
\end{eqnarray}   

\noindent
where $s$ and $l$ denote the cubic perturbed vibronic and roto-vibronic quantum numbers, and $U_{+}$ (=$\epsilon_{l} (\rho) + A \rho$) corresponds to the effective potential originating from the combination of cubic perturbed centrifugal and linear JT energies.

\section{Outline of BBO Theory for Perovskites}\label{ssec:bboJT}	


\noindent
The underlying physics of molecular processes and chemical reactions are dictated by the total electron-nuclear Hamiltonian ($\hat{H}$). The associated molecular Schr\"odinger Equation (SE) in terms of the total Hamiltonian ($\hat{H}$) and wavefunction ($\Psi$) acquires the following form:
\begin{eqnarray}
	\hat{H}(\textbf{r$_e$},\textbf{R}_n)\Psi(\textbf{r$_e$},\textbf{R}_n) = E\Psi(\textbf{r$_e$},\textbf{R}_n),
	\label{mse}
\end{eqnarray}

\noindent
where the electronic and nuclear coordinate vectors are collectively defined as \textbf{r$_e$} and \textbf{R$_n$}, respectively. In the non-relativistic framework, the molecular Hamiltonian $\hat{H}(\textbf{r}_e,\textbf{R}_n)$ can be partitioned into the nuclear kinetic energy operator $\hat{T}_{\rm nuc}(\textbf{R}_n)$ and the electronic Hamiltonian $\hat{H}_{\rm{el}}(\textbf{r}_e;\textbf{R}_n)$ as follows:

\begin{eqnarray}
	\hat{H}(\textbf{r$_e$},\textbf{R}_n) = \hat{T}_{\rm nuc}(\textbf{R}_n) + \hat{H}_{\rm{el}}(\textbf{r$_e$};\textbf{R}_n)
	; \quad \hat{T}_{\rm nuc} = -\frac{\hbar^2}{2}\nabla^2_{R},
	\label{eq:hpart}
\end{eqnarray} 

\noindent
where the derivatives with respect to the mass-weighted nuclear coordinates (\textbf{R$_n$}) are shown as $\boldsymbol{\nabla_{R}}$. \\

\noindent
For a complete matrix (Hilbert) space of dimension \( M \), the total molecular wavefunction \( \Psi(\mathbf{r}_e, \mathbf{R}_n) \) can be expanded as linear combination of electronic eigenfunctions \( \{ \xi_i(\mathbf{r}_e; \mathbf{R}_n) \} \) according to the Born-Oppenheimer-Huang~\citep{BO,BH} expansion:

\begin{equation}
	\Psi(\textbf{r}_e,\textbf{R}_n) = \sum_{i=1}^M \psi_i^{\rm{ad}}(\textbf{R}_n) \xi_i(\textbf{r}_e;\textbf{R}_n), 
	\label{eq:bo-expansion} 
\end{equation}

\noindent
where the coefficients \( \psi_i^{\text{ad}}(\mathbf{R}_n) \) signify the nuclear wavefunctions. On the other hand, the solutions to the electronic SE, namely, the electronic eigenfunctions \( \{ \xi_i(\mathbf{r}_e; \mathbf{R}_n) \} \) can be obtained from:

\begin{equation}
	\hat{H}_{\rm{el}}(\textbf{r}_e;\textbf{R}_n)\xi_i(\textbf{r}_e;\textbf{R}_n) = u_i(\textbf{R}_n)
	\xi_i(\textbf{r}_e;\textbf{R}_n), \qquad  
	\langle \xi_i | \xi_j \rangle_{\textbf{r}_e} = \delta_{ij},
	\label{eq:elese}
\end{equation}

\noindent
with the eigenvalues \( u_i(\mathbf{R}_n) \). Such eigenvalues are widely known as adiabatic PESs, which shows dependence on the nuclear coordinates. In 
Eq. \ref{eq:elese}, the product of two electronic eigenfunctions over the electronic coordinates represents the inner product \( \langle \cdots \rangle_{\mathbf{r}_e} \).\\

\noindent
While plugging the BO expanded molecular wavefunction (Eq. \ref{eq:bo-expansion}), total Hamiltonian (Eq. \ref{eq:hpart}) 
and electronic SE (Eq. \ref{eq:elese}) in the molecular SE (Eq. \ref{mse}) followed by projection with various electronic eigenfunctions,
the kinetically coupled nuclear SE (known as adiabatic nuclear SE) is expressed in the following matrix form:
\begin{equation}
	-\frac{\hbar^2}{2}\left(\boldsymbol{\nabla}_{R}+\boldsymbol{\tau}\right)^2\psi^{\rm{ad}} + \left(U-E\right)
	\psi^{\rm{ad}} = 0,
	\label{eq:adsem}
\end{equation}
where the elements of diagonal adiabatic PESs matrix and skew-symmetric non-adiabatic coupling matrix [$\boldsymbol{\tau}_{ij}(\textbf{R}_n) (= \langle\xi_i\vert\boldsymbol{\nabla}_{R}\xi_j\rangle_{\textbf{r}_e})$] appear as:
\begin{equation}
	U_{ij}(\textbf{R}_n)\,=u_j(\textbf{R}_n)\delta_{ij},
	\label{eq:adiael}
\end{equation} 
\begin{equation}
	\langle 
	\xi_j | \boldsymbol{\nabla}_{R}\xi_i\rangle_{\textbf{r}_e} = -\langle 
	\xi_i | \boldsymbol{\nabla}_{R}\xi_j\rangle_{\textbf{r}_e}.
	\label{eq:skew}
\end{equation} 

\vspace{0.5 cm}

\noindent
In absence of non-adiabatic coupling matrix ($\tau$) in Eq.~\ref{eq:adsem}, one can come across the BO approximation. On the other hand, according to Hellmann-Feynman theorem,~\cite{hellmann,feynmann} the NACTs (see Eq. \ref{eq:skew}) exhibit singular behavior at the close vicinity of degenerate point(s) \{$u_i \simeq u_j$\} over the nuclear configuration space (CS):     
\begin{eqnarray}
	\boldsymbol{\tau}_{ij} = \dfrac{\langle\xi_i\vert\boldsymbol{\nabla}_{R} 
		\hat{H}_{\rm{el}} \vert \xi_j\rangle_{\textbf{r}_e}}{u_j-u_i}\,\, .
	\label{eq:hf}
\end{eqnarray}

\noindent
As a matter of fact, a paradigm shift from adiabatic to diabatic representation of nuclear SE is indeed necessary to transform the often singular kinetic (NACTs) couplings to smooth, single-valued and continuous potential (diabatic) couplings. Such operation needs to be carried out to avoid the 
numerical inaccuracies originating in dynamical calculations with adiabatic SE. The changeover (adiabatic to diabatic) is executed employing the transformation $\psi^{\rm ad}=A\psi^{\rm dia}$, where $\psi^{\rm ad}$ and $\psi^{\rm dia}$ corresponds to the adiabatic and diabatic basis, respectively. While using such relationship ($\psi^{\rm ad}=A\psi^{\rm dia}$) in Eq. \ref{eq:adsem}, the diabatic representation of nuclear SE takes the following form:
\begin{eqnarray}
	\left(-\dfrac{\hbar^2}{2}\nabla_{R}^2 + W-E\right)\psi^{\rm{dia}} = 0, \quad W = A^{\dagger}U A,
	\label{eq:diaben}
\end{eqnarray}

\noindent
under the Adiabatic-to-Diabatic Transformation (ADT) condition:~\citep{baer_phys_rep358} 
\begin{eqnarray}
	\boldsymbol{\nabla}_{R}A + \boldsymbol{\tau}A = 0.
	\label{eq:adteq}
\end{eqnarray}
\noindent
The above-mentioned ADT condition ensures  
the orthogonal property of $A$ matrix [$SO(M)$] due to the skew-symmetric nature of $\boldsymbol{\tau}$ matrix for $M$ dimensional Hilbert space. The generalized form of $A$ matrix is presented and discussed in Chapter \ref{sec:bbo}~\citep{bijit_irpc38,saikat_pyra}.

\vspace{0.2 cm}

\noindent
While plugging the generalized form of ADT ($A$) and the skew-symmetric $\boldsymbol{\tau}$ matrix in ADT condition (Eq. \ref{eq:adteq}), $^NC_2\,(=P)$ number of first order differential equations are obtained as follows:  
\begin{eqnarray}
	\boldsymbol{\nabla}_R\Theta_{ij} = \sum_{m=1}^{P} c^{(m)}\boldsymbol{\tau}^{(m)}.
	\label{eq:ADTgen}
\end{eqnarray}

\noindent
In the above linear combination, $\boldsymbol{\tau}^{(m)}$ represents NACT between different electronic states and the associated coefficients, $\{c^{(m)}\}$s are functions of ADT angles $\{\Theta_{kl}\}$. In order to perform symbolic manipulation as well as numerical calculations of the ADT quantities from \textit{ab initio} calculated adiabatic PESs and NACTs, one can employ ADT program package \cite{naskar_jctc} for any chemical or physical process involving $N$ electronic states and $M$ vibrational modes.

\section{Calculation of Spectral profile}\label{ssec:spec}

\noindent
Once diabatic PESs and couplings for MO$_6^{z-}$ are obtained with the aid of the BBO theory, multi-state multi-mode Time-Dependent Discrete Variable Representation (TDDVR) dynamics~\citep{SAJCP113, SAJCP121, SAIJQC105, SSJPCA118, souvik_irpc37, joy_jpca} are performed to compute the photoelectron/photoemission (PE) spectra of the MO$_6^{(z+1)-}$ moiety. While calculating the spectral features, the autocorrelation function ($C(t)$) is obtained employing nuclear wavefunctions at different times:

\begin{subequations}
	\begin{eqnarray}
		C(t) &=& \langle \Xi(t) | \Xi(0)\rangle \label{eq:autocor1}\\
		&=& \Big\langle \Xi^{\star}\Big(\frac{t}{2}\Big) 
		\Big| \Xi\Big(\frac{t}{2}\Big)\Big\rangle \label{eq:autocor2}.
		\label{Eq:Auto-Correlation}
	\end{eqnarray}
\end{subequations} 

\noindent 
Then, such autocorrelation function is Fourier transformed to evaluate the spectral intensity as per the following expression:
\begin{eqnarray}
	I(\omega) \propto \omega \int_{-\infty}^{\infty} C(t) \exp(i\omega t) dt. \label{eq:abs}
	\label{Eq:Fourier}
\end{eqnarray}
where the integration time limit ranges from 0 to 100 fs for dynamical calculation. \\

\noindent 
In order to achieve the desired level of computational accuracy, one needs to take the product of autocorrelation function with a damping function to encapsulate the effects of spectrometer resolution and to reduce Gibbs phenomena artifacts:
\begin{eqnarray}
	h(t)=\exp\Big[-\Big(\frac{|t|}{\tau}\Big)^{2}\Big]. \label{eq:tdf}
	\label{Eq:damping}
\end{eqnarray}
These autocorrelation function is further multiplied by a cosine function to get the spectral intensity:
\begin{eqnarray}
	g(t)= \cos^{n}\Big(\frac{\pi t}{2T}\Big),
	\label{Eq:cos-function}
\end{eqnarray}
\noindent
where $n=2$ and \textit{T} represents propagation length. In order to obtain further insight on TDDVR spectral calculations, one can go through Section \ref{ssec:tddvr} in detail~\citep{SSJPCA118, souvik_irpc37, soumya_tfb}

\section{Calculation of Ferroelectricity}\label{ssec:Ferr}

\noindent
For TiO$_6^{8-}$ unit of BaTiO$_3$ crystal, the \textit{ab initio} computed adiabatic PESs~\cite{Mantu_BTO} can be used to analyze the structural distortion. While carrying out such study, an equation derived from the mean-field Hamiltonian, $H = H_{0} - \lambda Q \langle Q \rangle$ is employed, where the potential energy takes the form, 
$U = U_{0} - \lambda Q \langle Q \rangle$. In such form of potential energy, $\langle Q \rangle$ acts as a parameter resulting into a system of coupled transcendental equations\cite{bersuker_cm5}:
\begin{eqnarray}
	\langle Q_{j}\rangle &=& \frac{\int Q_{j} \exp^{-U(Q)/{kT}} d^{3}Q}{Z}, \qquad Z= \int \exp^{-U(Q)/{kT}} d^{3}Q,
	\label{Eq:Partition_function}
\end{eqnarray}

\noindent
where $j$ corresponds to one of the axes ($x$ or $y$ or $z$) of distortion. In TiO$_6^{8-}$, the phase transition occurs due to the change of geometry from tetragonal to cubic or the reverse and therefore, the titanium ion ($Ti^{4+}$) encounters displacement along a specific direction. If distortion of the system occurs along z-axis, displacements along the x and y axes appear as zero ($Q_{x}=Q_{y}= 0$).
The partition function for the displacement along the z-axis take the following form:

\begin{eqnarray}
	\langle Q_{z}\rangle &=& \frac{\int Q_{z} \exp^{-U(Q_{z})/{kT}} dQ_{z}}{Z}, \qquad Z= \int \exp^{-U(Q_{z})/{kT}} dQ_{z}.
	\label{Eq:Partition_function1}
\end{eqnarray}

\noindent
Such theoretical formalism is implemented to compute the ferroelectric properties originating from the structural distortions.

\section{Two and Four Coupled Electronic States: LaMnO$_3$ and BaTiO$_3$} \label{sec:perovskite}

\noindent
\subsection{Roto-Vibrational Spectra of LaMnO$_3$} \label{ssec:Roto-Vibrational}

\noindent
The experimental dielectric function spectra of LaMnO$_3$~\citep{Natalia_JETP} show
an anomalous temperature dependence near the N\'{e}el temperature ($T_{\rm N} \approx 140$ K). On the other hand, optical ellipsometry measurements along the room-temperature $a$-axis in the orthorhombic $Pbnm$ symmetry~\citep{Natalia_JETP, Natalia_JETP1, Natalia_JETP2, Natalia_JETP3, Berhard-TbMnO} explore strong intersite optical electronic transition (d$^4$d$^4$ $\rightarrow$ d$^3$d$^5$) around 1.96 eV, along with several higher-energy satellite peaks appeared at 2.29 $\pm$ 0.02 eV, 2.41 $\pm$ 0.02 eV, 2.66 $\pm$ 0.02 eV and 2.84 $\pm$ 0.02 eV. It is worth mentioning that such spectral features show distinct temperature-dependent behavior below $T_{\rm N}$ with a downward shift of peak energies by 80, 20 and 110 meV, respectively, vis-\'a-vis a concurrent change in linewidths~\citep{Natalia_JETP, Natalia_JETP1, Natalia_JETP2, Natalia_JETP3}. This behavior is mainly originated from transitions between pseudo-rotational vibronic levels of both the lower and upper branches of a characteristic `Mexican hat' APESs.

\noindent
\subsubsection{9.5.1.1 JT Effect on Ground Electronic State}

\noindent
Since the ground-state minimum is generated by the resultant between bond vibrational forces and the linear JT effect, the parameters, $\alpha$ (tunneling splitting) and $\beta$ (barrier height) turns into  
$\alpha$  = $\frac{\hbar^2}{2M\rho^2}$ and $\beta$ = $-A_3 \rho_0^3$. 
It is evident from polarized Raman scattering measurements on untwinned LMO crystals that phonon density of states (PDOS) features are present 
up to the fourth harmonic, showing sidebands near a low-energy vibrational mode at 25 cm$^{-1}$. The tunnel splitting $\alpha$ linked to the MnO$_6^{9-}$ complex is mainly responsible for such shift. In addition, the value of the barrier parameter $\beta \approx 375$ cm$^{-1}$ is evaluated as function of the temperature ($\sim$780 K) of the structural transition from the high-temperature pseudo-cubic phase ($Pm\bar{3}m$) to the orthorhombic phase ($Pbnm$).\\

\vspace{0.2 cm}

\noindent
In order to solve Eq. \ref{Eq: ground state Mathieu Eq}, the angular wavefunction $\gamma(\phi)$ can be expanded in terms of the following Fourier series:

\begin{equation}
	\gamma(\phi)=\sum_l a_l e^{i l \phi}
\end{equation}

\vspace{0.2 cm}

\noindent
According to the boundary condition $\gamma(\phi + 2\pi) = -\gamma(\phi)$, $l + \frac{1}{2}$ must be an integer, which indicates that the angular momentum values $l$ are half-integers offset by $\frac{1}{2}$. The coefficients $a_l$ follows the recurrence relation:

\begin{equation}
	a_l(\epsilon-\alpha l^2)  + \frac{1}{2}\beta(a_{l-3}+a_{l+3})=0. 
	\label{eq:eigen}
\end{equation}

\begin{figure}[htp]
	\centering
	\includegraphics[width=0.9\textwidth, angle=0]{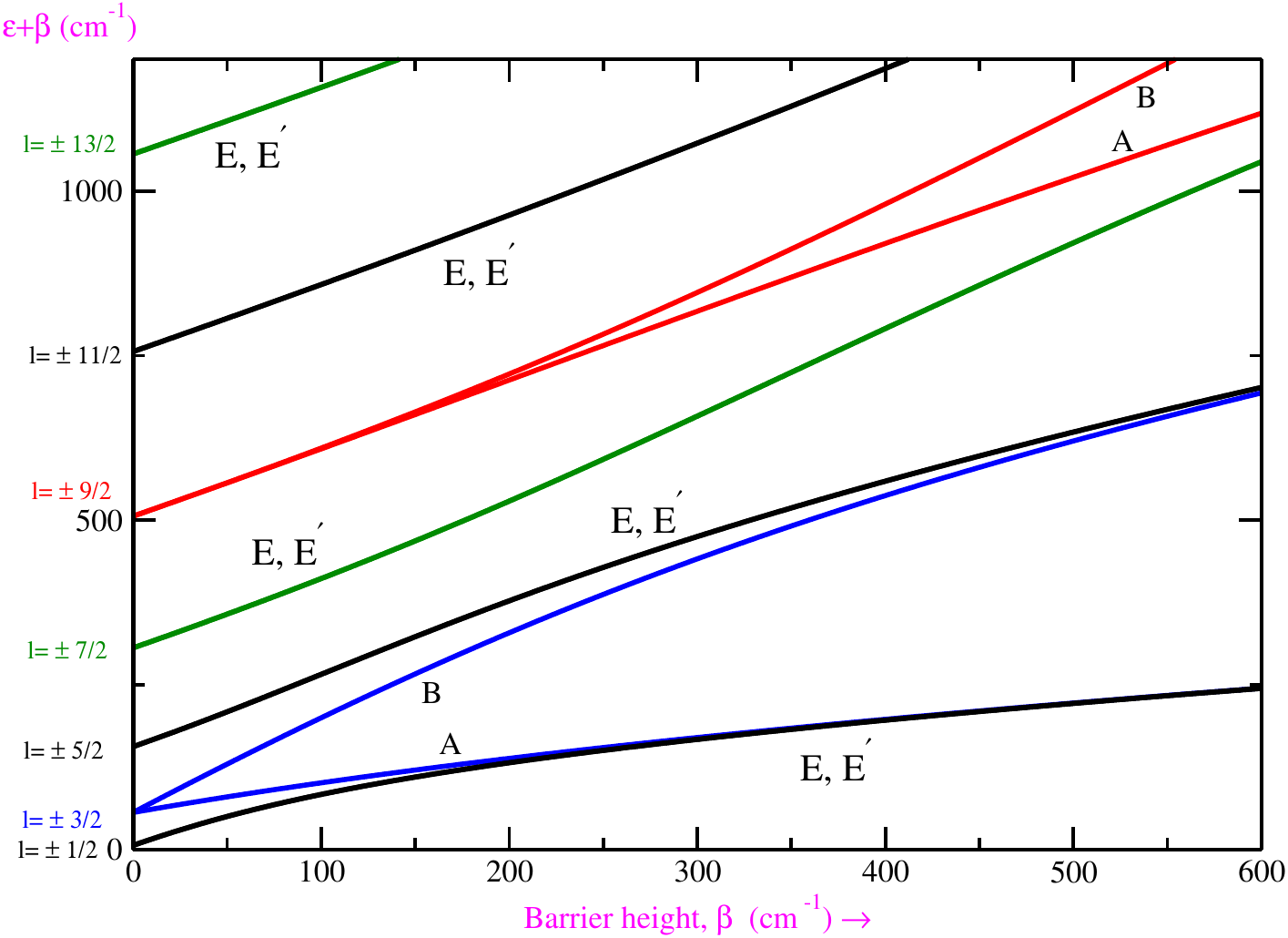}
	\caption{ The pseudo-rotational eigenvalue profile of JT-split ground state are presented. The results are collected from Ref.~[\protect\citen{Mantu_Solid-state_Review}] and permission has been taken from Wiley.}
	\label{fig:Barrier}
\end{figure}

\noindent
Figure \ref{fig:Barrier} depicts eigenvalues $\epsilon$ as functions of the potential barrier parameter $\beta$ for LMO crystal, which is obtained from Eq. \ref{eq:eigen}. The symmetry of the solutions correspond to $C_{3v}$ point group, assuming $\phi$ as the rotation angle around a symmetry axis. More particularly, when $l = \pm \frac{3}{2}, \pm \frac{9}{2}, \ldots$ holds true, the solutions appear as the totally symmetric representation $A$ for $a_l = +a_{-l}$, and as representation $B$ for $a_l = -a_{-l}$. In addition, sets such as $l = +\frac{1}{2}, -\frac{5}{2}, +\frac{7}{2}, \ldots$, are converted according to one $E$-type representation (designated as $E$). On the other hand, set like $l = -\frac{1}{2}, +\frac{5}{2}, -\frac{7}{2}, \ldots$, transforms as a second E-type representation (denoted as $\acute{E}$).

\noindent

\begin{table}[htp]
	\begin{small}
		
		\noindent
		\caption{Comparison of theoretical and experimental peak positions ($\nu_{\text{Exp}}$) for LaMnO$_3$ is tabulated based on eigenvalues of the lower ``Mexican Hat" PES sheet with $\alpha = 25\ \text{cm}^{-1}$, $\beta = 375\ \text{cm}^{-1}$ and an energy shift of 1.96 eV. The data are taken from Ref. [\protect\citen{joy_jcp}].}
		\begin{center}
			\begin{tabular}{|c|c|c|c|c|}
				\hline 
				\multirow{2}{*}{\makecell{Roto-vibrational \\ level ($\pm l$)}} & \multirow{2}{*}{\makecell{$\epsilon + \beta$ [cm$^{-1}$] \\ ($\beta=375$ cm$^{-1}$)}} & \multirow{2}{*}{\makecell{$\epsilon_{\text{average}}$ (eV)}} & \multirow{2}{*}{\makecell{ $ \text{shift}+ \epsilon_{\text{average}}$ \\ (eV)}} & \multirow{2}{*}{\makecell{$\nu_{Exp}$ \\ (eV)}} \\
				& & &  & \\
				\hline 
				$\pm 17/2$ &  2194.34 & \multirow{2}{*}{\makecell{0.253}} & \multirow{2}{*}{\makecell{2.21}} & \multirow{2}{*}{\makecell{2.29 $\pm$ 0.02}}\\ 
				
				$\pm 19/2$ &  2639.26 &  &  &\\ 
				\hline 
				$\pm 23/2$ &  3686.66 & \multirow{2}{*}{\makecell{0.448}} & \multirow{2}{*}{\makecell{2.41}} & \multirow{2}{*}{\makecell{2.41 $\pm$ 0.02}}\\ 
				
				$\pm 25/2$ &  4285.82 &  & &\\ 
				\hline 
				$\pm 29/2$ &  5634.67 &\multirow{2}{*}{\makecell{0.699}} & \multirow{2}{*}{\makecell{2.66}} & \multirow{2}{*}{\makecell{2.66 $\pm$ 0.02}}\\ 
				
				$\pm 31/2$ &  6384.24 &  & &\\ 
				\hline 
				$\pm 35/2$ &  8045.86 &\multirow{2}{*}{\makecell{1.006}} & \multirow{2}{*}{\makecell{2.97}} & \multirow{2}{*}{\makecell{2.84 $\pm$ 0.02}}\\ 
				
				$\pm 37/2$ &  8945.00 & & & \\
				\hline
			\end{tabular} 
		\end{center}
		\label{Table: comparison between theory and experiment LaMnO3 ground state}
	\end{small}
\end{table}
\noindent
Table \ref{Table: comparison between theory and experiment LaMnO3 ground state} indicates that the theoretical predictions for roto-vibrational excitations linked to the lower, cubic-perturbed sheet of the ``Mexican hat" PES appear at energies of 0.253, 0.448, 0.699, and 1.006 eV in presence of the estimated barrier height $\beta = 375\ \text{cm}^{-1}$. Such values show good agreement with the experimental outcomes. As a matter of fact, Table \ref{Table: comparison between theory and experiment LaMnO3 ground state} exhibits satisfactory accord (within experimental uncertainty) between theory and the experimentally observed peaks at 2.41 $\pm$ 0.02 eV and 2.66 $\pm$ 0.02 eV. It is necessary to mention that the theoretically predicted peak position at 2.21 eV slightly underestimates the measured value of 2.29 $\pm$ 0.02 eV.  
Moreover, calculations were also carried out with a higher barrier parameter, $\beta = 650\ \text{cm}^{-1}$. The adjustment imparts negligible effect on the previously identified theoretical excitation energies for $\beta = 375\ \text{cm}^{-1}$, indicating their robustness to moderate changes in barrier height.

\noindent
\subsubsection{9.5.1.2 Centrifugally Stabilized Excited Vibronic State}
\begin{figure}[htp!]
	\centering
	\includegraphics[height=0.5 \textwidth , width=0.6 \textwidth, angle=0]{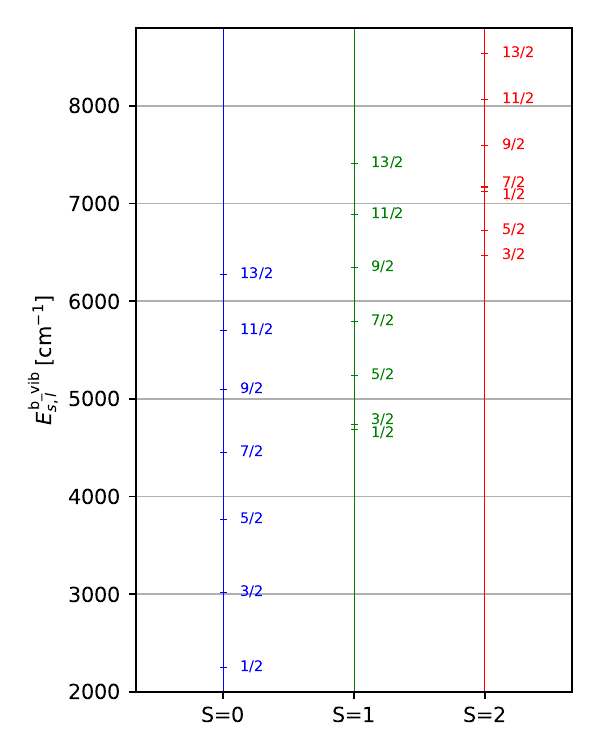}
	\caption{ Roto-vibronic eigenvalues of the JT-split excited state: energy levels exhibits centrifugally stabilized motion coupled with harmonic bond vibrations in manganites. The results are adapted from Ref.~[\protect\citen{Mantu_Solid-state_Review}] and permission has been taken from Wiley.}
	\label{fig:b_vib}
\end{figure}

\vspace{0.2 cm}
\noindent
The eigenvalues for centrifugally stabilized bound states show substantial deviations from those of conventional harmonic bond vibrations. Such divergence mainly originates from the fact that frequency of oscillation in the centrifugally stabilized case explicitly depends on the pseudo-rotational quantum number $l$ (unlike in harmonic bond vibrations). Such study explores the relevant quantum numbers as $l = \pm \frac{1}{2}, \pm \frac{3}{2}, \pm \frac{5}{2}, \ldots$ for pseudo-rotation and $s = 0, 1, 2, \ldots$ for vibrational excitation (see Eqs. \ref{Eq:Excited state Mathieu Equation} and \ref{Eq: cp_bond_eigenvalue}). While performing such study, the radial component of the nuclear wavefunction, $\omega_e(\rho)$ obeys the standard form of a harmonic oscillator with a modified frequency influenced by pseudo-rotational effects. One of the major findings of such analysis is the entanglement of pseudo-rotational energy levels upon incorporation of harmonic bond vibrations resulting into unexpected shifts in energy level ordering. It is important to note that in the second vibrationally excited state ($s = 2$), the energy level associated with $l = \frac{1}{2}$ lies below that of $l = \frac{7}{2}$. If the bond vibrations are excluded, such ordering becomes reversed, where $l = \frac{1}{2}$ typically lies above $l = \frac{7}{2}$. Such entanglement effect is widely observed in the second excited vibrational state unveiling the significant interplay between centrifugal stabilization and harmonic vibrational coupling (see Figure \ref{fig:b_vib}).\\

\begin{figure}[htp]
	\centering
	\includegraphics[width=0.9\textwidth, angle=0]{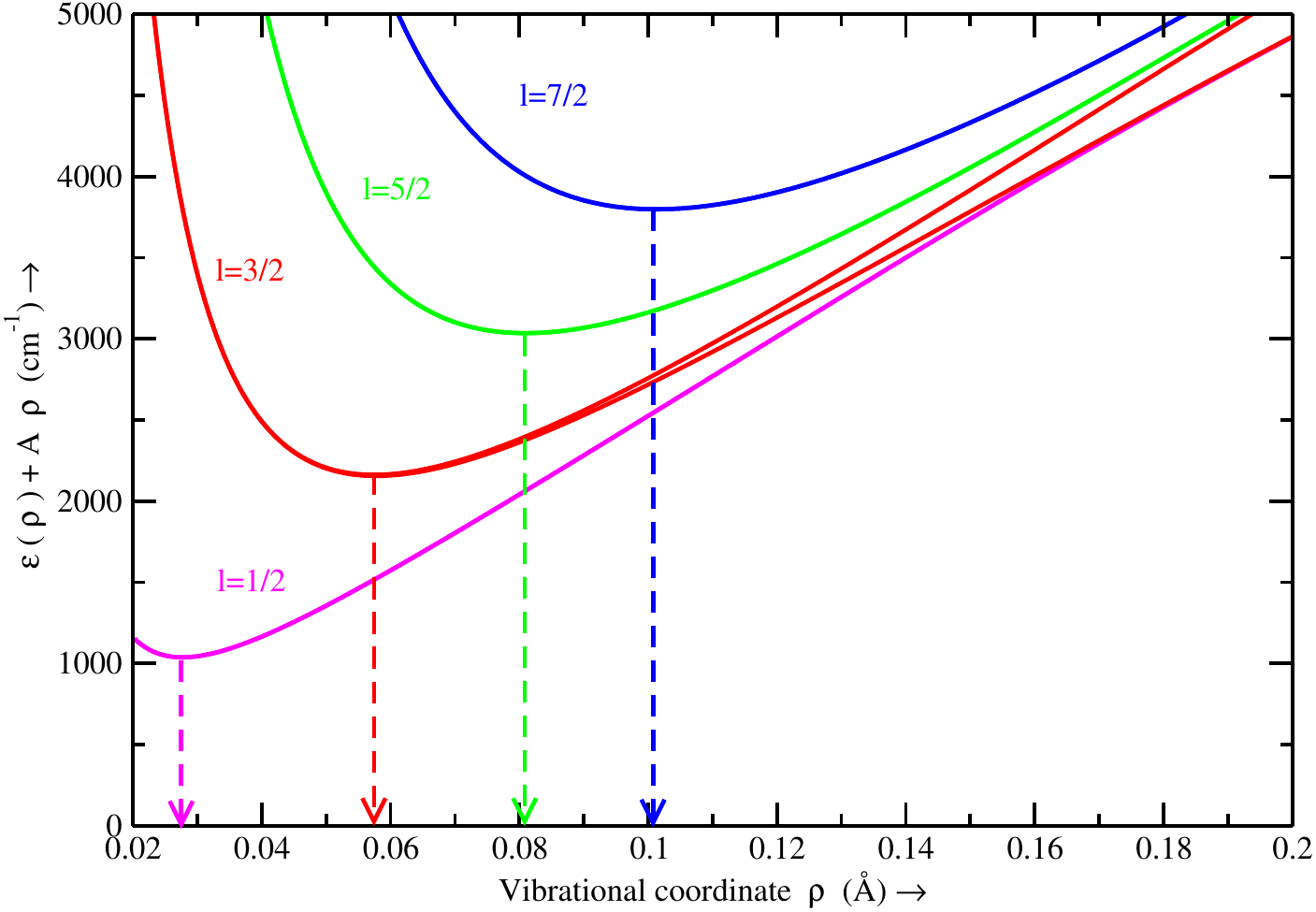}
	\caption{ Effective potential $\epsilon_l (\rho)+ A\rho $ is plotted as functions of $\rho$ coordinates. The results are taken from Ref.~[\protect\citen{Mantu_Solid-state_Review}] with permission from Wiley.}
	\label{fig:potential}
\end{figure}

\noindent
Figure \ref{fig:potential} exhibits the the combined effect of cubic perturbations on the centrifugal potential energies ($\epsilon_l (\rho)$) for different pseudo-rotational quantum numbers ($l$) and the linear JT interaction term A$\rho$. Such opposing forces are responsible for the formation of a bound state, namely, the cubic-perturbed centrifugal potential typically acting as a repulsive barrier and the JT linear term introducing an attractive component. The resultant of these two forces generates a potential well leading to the formation of localized vibronic bound states. Table \ref{Table: comparison between theory and experiment} displays calculated average roto-vibrational energies for pseudo-rotational levels $\pm\frac{1}{2}$ and $\pm\frac{3}{2}$ at 0.327 eV, and $\pm\frac{9}{2}$ and $\pm\frac{11}{2}$ at 0.672 eV, which are in very good agreement with the corresponding experimental energy shifts of 
0.33 eV (2.29-1.96 eV) and 0.70 eV (2.66-1.96 eV), respectively. In case of pseudo-rotational levels $\pm\frac{5}{2}$ and $\pm\frac{7}{2}$, the theoretical value of 0.511 eV shows slight deviation from the observed shift of 0.45 eV. On the contrary, for $\pm\frac{13}{2}$ and $\pm\frac{15}{2}$, the predicted energy of 0.817 eV differs modestly from the experimental value of 0.88 eV. While repeating the calculations with the modified $A_3$ parameter derived from a higher barrier value $\beta = 650\ \text{cm}^{-1}$, the theoretical peak positions remain essentially unchanged compared to those evaluated using $\beta = 375\ \text{cm}^{-1}$.

\begin{table}[htp]
	\caption{Comparison of theoretical and experimental peak positions ($\nu_{\text{Exp}}$) for LaMnO$_3$ is depicted, based on eigenvalues of the cubic-perturbed upper sheet of the ``Mexican Hat" potential with an energy shift of 1.96 eV. The data are taken from Ref.~[\protect\citen{joy_jcp}].}
	\begin{center}
		\begin{tabular}{|c|c|c|c|c|}
			\hline 
			\multirow{2}{*}{\makecell{Roto-vibrational \\ level ($\pm l$)}} & \multirow{2}{*}{\makecell{$\overline{E}^{\text{cp + b$\_$vib}}_{s=0, \pm l}$ [cm$^{-1}$]}} & \multirow{2}{*}{\makecell{$\langle \overline{E}^{\text{cp + b$\_$vib}}_{s=0, \pm l; \: \pm l+1}  \rangle$ \\ (eV)}} & \multirow{2}{*}{\makecell{$ \text{shift} + \langle \overline{E}^{\text{cp + b$\_$vib}}_{s=0, \pm l; \: \pm l+1}  \rangle$ \\ (eV)}} & \multirow{2}{*}{\makecell{$\nu_{Exp}$ \\ (eV)}}\\
			& & & &\\
			\hline 
			$\pm 1/2$ & 2254.09& \multirow{2}{*}{\makecell{0.327}} & \multirow{2}{*}{\makecell{2.29}} & \multirow{2}{*}{\makecell{2.29 $\pm$ 0.02}}\\ 
			
			$\pm 3/2$ & 3021.01 &  &  &\\ 
			\hline 
			$\pm 5/2$ & 3774.73 & \multirow{2}{*}{\makecell{0.511}} & \multirow{2}{*}{\makecell{2.47}} & \multirow{2}{*}{\makecell{2.41 $\pm$ 0.02}}\\ 
			
			$\pm 7/2$ & 4469.21 &  &  &\\ 
			\hline 
			$\pm 9/2$ & 5116.46 & \multirow{2}{*}{\makecell{0.672}} & \multirow{2}{*}{\makecell{2.63}} & \multirow{2}{*}{\makecell{2.66 $\pm$ 0.02}} \\ 
			
			$\pm 11/2$ & 5729.75 &  &  &\\ 
			\hline 
			$\pm 13/2$ & 6312.56 & \multirow{2}{*}{\makecell{0.817}} & \multirow{2}{*}{\makecell{2.78}} & \multirow{2}{*}{\makecell{2.84 $\pm$ 0.02}}\\ 
			
			$\pm 15/2$ & 6871.97 & & &\\
			\hline
		\end{tabular} 
	\end{center}
	\label{Table: comparison between theory and experiment}
\end{table}

\noindent
\subsubsection{9.5.1.3 \textit{Ab Initio} Hamiltonian: Ground and Excited Roto-Vibrational Spectra of LMO}

\vspace{0.2 cm}

\noindent
In order to achieve in-depth insight on the origin of high-energy satellite transitions, \textit{ab initio}~\citep{joy_jpca} adiabatic PES for the MnO$_6^{9-}$ unit were used to compute \text{E}$\otimes$\text{e} roto-vibrational eigenvalues. Such calculation unveils the underlying roto-vibronic structure responsible for the observed spectral complexity.\\

\noindent
Over the last twenty years, plane-wave-based density functional theory (DFT) and hybrid DFT approaches,~\citep{maier_pccp7, maier_prb72} typically local density approximation (LDA) or generalized gradient approximation (GGA) in the form of Perdew-Wang (PW91) and Perdew-Burke-Ernzerhof (PBE) functionals were employed to compute the structural and electronic properties of perovskite oxides. On the contrary, Adhikari and coworkers adopted multireference electronic structure methods including complete active space self-consistent field (CASSCF), second-order Rayleigh--Schr\"odinger perturbation theory (RS2c) and multireference configuration interaction with single and double excitations [MRCI(SD)] to generate accurate adiabatic PESs for the two lowest-lying electronic states of the MnO$_6^{9-}$ unit in LaMnO$_3$~\citep{joy_jpca}. \\

\noindent
While investigating the microscopic origin of high-energy spectral features in LaMnO$_3$,  the vibronic structure of the  MnO$_6^{9-}$ unit was explored using state-of-the-art quantum chemical methods. Optimization of the equilibrium geometry and vibrational frequencies of the JT active doubly degenerate $e_g$ modes ($Q_x$, $Q_y$) were performed at the CCSD/ANO-R1 level as implemented in the MOLPRO package~\citep{MOLPRO_brief}. Then the APESs and NACTs of the electronically degenerate $^5E_g$ states were scanned as functions of those modes. It is important to mention that the APESs and the corresponding NACTs were calculated employing CASSCF, RS2c, and MRCI(SD) methods with a def2-TZVP basis set with (5o,4e) as complete active space (CAS). Such CAS comprises of five Mn$^{3+}$ 3d orbitals and four electrons, which was used with state averaging over the two lowest $^5E_g$ states [SA2-CAS(5o,4e)]~\citep{joy_jpca}. Finally, the adiabatic PESs were fitted by an analytic functional form with polar coordinate (Eq.~\ref{Eq:fitted_APES}) to find out the vibronic parameters, namely, linear coupling constant $A$ and cubic anharmonicity $A_3$:

\begin{equation}
	\begin{aligned}
		U_{i} (\rho, \, \phi) = 
		& F \{ \frac{1}{2} \, M \, \omega_{b\_{vib}}^{2}\, \rho^{2} +  A \, \rho  + A_{3} \, \rho^{3} \,cos(3\, \phi)  + \\
		& B_1 \, \rho \,cos(\phi) 
		+ B_2 \, \rho \,sin(\phi)  +  C \, \rho^{3} \,sin(3\, \phi) \}, \quad i=1,2
	\end{aligned}
	\label{Eq:fitted_APES}
\end{equation}

\noindent
where, $M = 15.9999$ amu and $\omega_{b\_{vib}} = 405.24$ cm$^{-1} $, were obtained from CCSD/ANO-R1 calculations. The coefficients~\citep{joy_jpca} $A,\ A_3,\ B_1,\ B_2$ and $C$ were
evaluated by fitting the surfaces with a negative value of $A$ for ground state ($U_1$) and positive for excited state ($U_2$), indicating the JT distortion. Such parameters were further utilized to generate
the full roto-vibrational eigenvalue spectrum for both surfaces. On the other hand, in order to incorporate the environmental effects of La atom into the Hamiltonian, the mass ratio of a La-centered coordination environment to a cluster of eight MnO$_6^{9-}$ units without La was considered as an empirical scaling factor. Even though there are negligible differences in the local geometry with and without the La atom, the mass correction factor, $F$, can be expressed as~\citep{joy_jpca}: 

\begin{equation*}
	F = \dfrac{M_{\lbrace La+8Mn+12O\rbrace}}{M_{\lbrace 8Mn+12O\rbrace}} = 1.2199.
\end{equation*}

\noindent
The above empirical factor is able to appropriately rescale the vibrational normal mode frequencies and to adjust the parameters of the \textit{ab initio} roto-vibrational adiabatic and diabatic Hamiltonians leading to the eigenvalue spectra. Such involvement of correction terms affirm a more realistic modeling of the MnO$_6^{9-}$ unit within the solid-state environment of LaMnO$_3$. It is interesting to note that theoretically calculated results exhibit excellent agreement with previous vibronic model studies~\citep{joy_jcp} and experimental optical ellipsometry data~\citep{Natalia_JETP}, as presented in Tables~\ref{Table:ground} and~\ref{Table:excited}. The JT-split excited vibronic states are mainly responsible for the fine structure in LaMnO$_3$'s high-energy optical spectra, validating the theoretical framework employing various electronic structure approaches as consistent and reliable one.

\begin{table}
	\caption{Comparison of theoretical peak positions for MnO$_6^{9-}$ obtained from model-based~\citep{joy_jcp} and \textit{ab initio} ground PES calculations~\citep{joy_jpca} with experimental~\citep{Natalia_JETP} satellite transitions in LaMnO$_3$ is presented. A shift of 1.96 eV~\citep{Natalia_JETP, Natalia_JETP1, Natalia_JETP2, Natalia_JETP3, Berhard-TbMnO} is applied to account for intersite charge transfer. The theoretical data are taken from Ref.~[\protect\citen{joy_jpca}].}
	\begin{small}

		\begin{center}
			\begin{tabular}{|c|c|c|c|c|c|}
				\hline
				\multicolumn{2}{|c|}{\multirow{4}{*}{\makecell{Modelled, \textit{ab-initio}\\  and experimental \\ peaks}}} & \multicolumn{4}{c|}{\multirow{2}{*}{ \makecell{pseudo-rotational levels $ \lbrace \pm l; \pm l \pm 1 \rbrace$ } }} \\
				\multicolumn{2}{|c|}{} &  \multicolumn{4}{c|}{}\\
				\cline{3-6}
				\multicolumn{2}{|c|}{\multirow{4}{*}{\makecell{ }}} &\multirow{2}{*}{\makecell{$\pm \frac{17}{2}$ ; $\pm \frac{19}{2}$}} & \multirow{2}{*}{\makecell{$\pm \frac{23}{2}$ ; $\pm \frac{25}{2}$}} & \multirow{2}{*}{\makecell{$\pm \frac{29}{2}$ ; $\pm \frac{31}{2}$}} & \multirow{2}{*}{\makecell{$\pm \frac{35}{2}$ ; $\pm \frac{37}{2}$}} \\
				\multicolumn{2}{|c|}{} & & & & \\
				\hline
				\multicolumn{2}{|c|}{$\nu_{\text{LaMnO}_3}^{\text{Model}}$\cite{joy_jcp} (eV)} & 2.21 & 2.41 & 2.66 &  2.97\\
				\hline
				\multirow{3}{*}{\makecell{ $\nu_{\textit{ab-initio}}^{[\text{MnO}_6]^{9-}}$\cite{joy_jpca} \\ (eV)}} & CASSCF & 2.06 & 2.11  & 2.18 & 2.29 \\
				\cline{2-6}
				& RS2c & 2.07 & 2.13 & 2.21 & 2.33 \\
				\cline{2-6}
				& MRCI(SD) & 2.07 & 2.14 & 2.24 & 2.37\\
				\hline
				\multirow{3}{*}{\makecell{ $\nu_{\text{Scaled} \: \textit{ab-initio}}^{[\text{MnO}_6]^{9-}}$\cite{joy_jpca} \\ (eV)}} & CASSCF & 2.06 & 2.13 & 2.19 & 2.31 \\
				\cline{2-6}
				& RS2c & 2.08 & 2.13 & 2.22 & 2.35 \\
				\cline{2-6}
				& MRCI(SD) & 2.08 & 2.15 & 2.24 & 2.38\\
				\hline
				\multicolumn{2}{|c|}{$\nu_{Exp}$\cite{Natalia_JETP} (eV)} & 2.29 $\pm$ 0.02 & 2.41 $\pm$ 0.02 & 2.66 $\pm$ 0.02 & 2.84 $\pm$ 0.02\\
				\hline
			\end{tabular}
		\end{center}
	\end{small}
	\label{Table:ground}
\end{table}

\begin{table}
	\caption{Comparison of theoretical peak positions for MnO$_6^{9-}$ obtained from model-based~\citep{joy_jcp} and \textit{ab initio} excited PES calculations~\citep{joy_jpca} with experimental~\citep{Natalia_JETP} satellite transitions in LaMnO$_3$ is depicted. A uniform 1.96 eV~\citep{Natalia_JETP, Natalia_JETP1, Natalia_JETP2, Natalia_JETP3, Berhard-TbMnO} energy shift is used to account for intersite charge transfer. The calculated values are adapted from Ref.~[\protect\citen{joy_jpca}].}
	\begin{small}

		\begin{center}
			\begin{tabular}{|c|c|c|c|c|c|}
				\hline
				\multicolumn{2}{|c|}{\multirow{4}{*}{\makecell{Modelled, \textit{ab-initio}\\  and experimental \\ peaks }}} & \multicolumn{4}{c|}{\multirow{2}{*}{ \makecell{pseudo-rotational levels $ \lbrace \pm l; \pm l \pm 1 \rbrace$ } }} \\
				\multicolumn{2}{|c|}{} &  \multicolumn{4}{c|}{}\\
				\cline{3-6}
				\multicolumn{2}{|c|}{\multirow{4}{*}{\makecell{ }}} &\multirow{2}{*}{\makecell{$\pm \frac{1}{2}$ ; $\pm \frac{3}{2}$}} & \multirow{2}{*}{\makecell{$\pm \frac{5}{2}$ ; $\pm \frac{7}{2}$}} & \multirow{2}{*}{\makecell{$\pm \frac{9}{2}$ ; $\pm \frac{11}{2}$}} & \multirow{2}{*}{\makecell{$\pm \frac{13}{2}$ ; $\pm \frac{15}{2}$}} \\
				\multicolumn{2}{|c|}{} & & & & \\
				\hline
				\multicolumn{2}{|c|}{$\nu_{\text{LaMnO}_3}^{\text{Model}}$\cite{joy_jcp} (eV)} & 2.29 & 2.47 & 2.63  & 2.78 \\
				\hline
				\multirow{3}{*}{\makecell{ $\nu_{\textit{ab-initio}}^{[\text{MnO}_6]^{9-}}$\cite{joy_jpca} \\ (eV)}} & CASSCF & 2.25 & 2.41  & 2.55 & 2.67 \\
				\cline{2-6}
				& RS2c & 2.27 & 2.44 & 2.59 & 2.72 \\
				\cline{2-6}
				& MRCI(SD) & 2.25 & 2.41 & 2.55 & 2.68\\
				\hline
				\multirow{3}{*}{\makecell{ $\nu_{\text{Scaled} \: \textit{ab-initio}}^{[\text{MnO}_6]^{9-}}$\cite{joy_jpca} \\ (eV)}} & CASSCF & 2.29 & 2.47 & 2.63 & 2.78 \\
				\cline{2-6}
				& RS2c & 2.31 & 2.51 & 2.68 & 2.83 \\
				\cline{2-6}
				& MRCI(SD) & 2.29 & 2.47 & 2.63 & 2.78\\
				\hline
				\multicolumn{2}{|c|}{$\nu_{Exp}$\cite{Natalia_JETP} (eV)} & 2.29 $\pm$ 0.02 & 2.41 $\pm$ 0.02 & 2.66 $\pm$ 0.02 & 2.84 $\pm$ 0.02\\
				\hline
			\end{tabular}
		\end{center}
	\end{small}
	\label{Table:excited}
\end{table}

\noindent
\subsection{Adiabatic Hamiltonian for TiO$_6^{8-}$ unit of BTO}\label{ssec:tio68-}
\vspace{0.2 cm}

\noindent
Over the past few decades, perovskite materials, in particular, BTO appear as a challenging system within the condensed matter and materials science communities, mainly due to their prominent ferroelectric behavior. Such ferroelectricity mainly originates from  the presence of strong PJT interactions involving low-lying electronic states of the constituent, TiO$_6^{8-}$ unit. \citep{bersuker_cm5,bersuker_pbcm457,Mantu_BTO} Since the titanate complex features a $d^0$ electronic configuration, the ground state becomes totally symmetric, $^1A_{1g}$, where the triply degenerate excited states belong to $^1T_{1u}$ IREP under $O_h$ symmetry at the configuration ($\{Q_k\} = 0$).\\

\begin{figure} [!htp]
	\centering
	\includegraphics[ width=0.9\textwidth, angle=0]{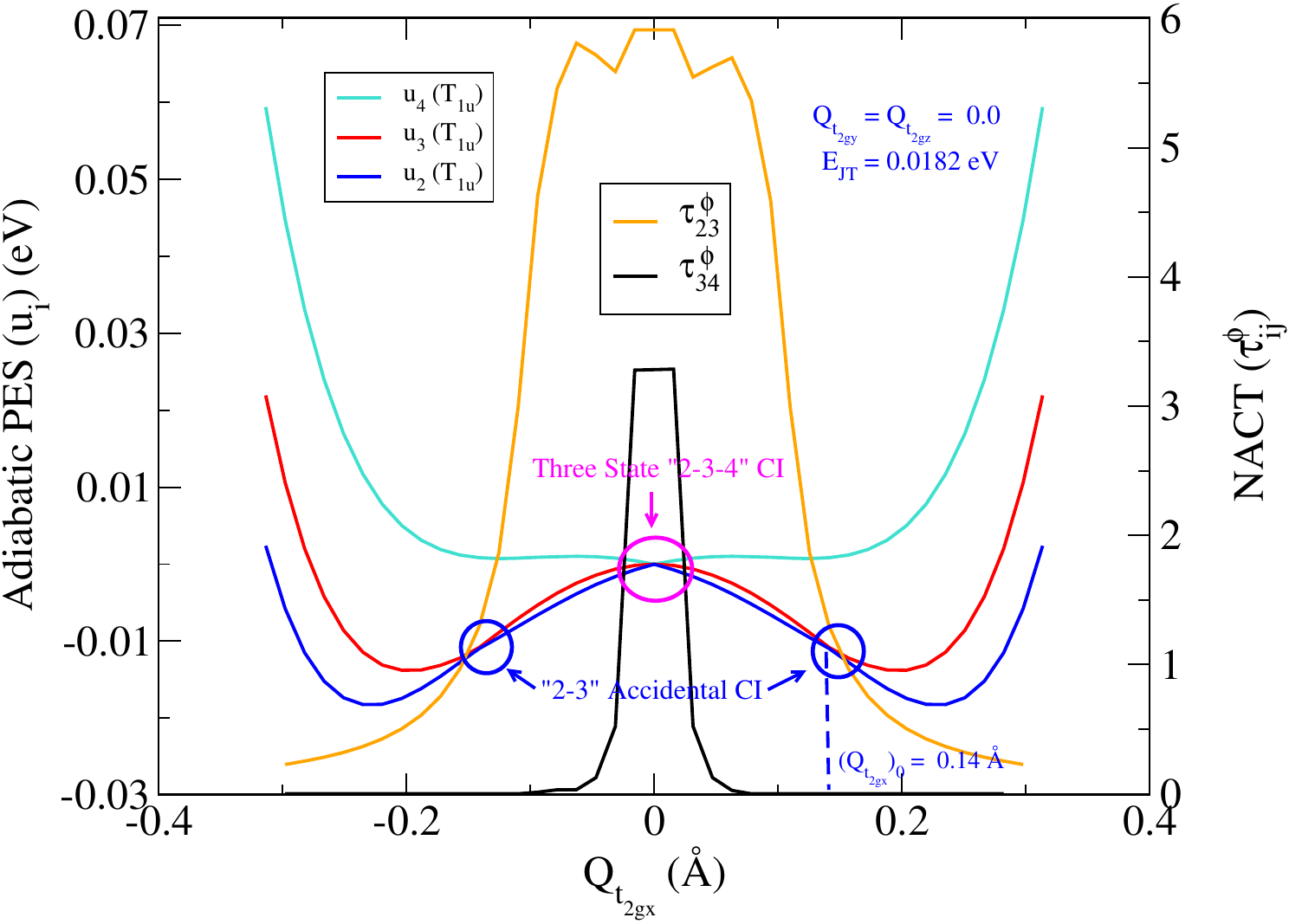}
	\caption{1D cuts of adiabatic PESs ($u_2$, $u_3$, $u_4$) and NACTs ($\tau_{23}^{\phi}$, $\tau_{34}^{\phi}$) for $T_{1u}$ states of TiO$_6^{8-}$ are presented along $Q_{t_{2gx}}$ keeping the other $t_{2g}$ modes at zero; curves along $Q_{t_{2gy}}$ and $Q_{t_{2gz}}$ are identical. The results are taken from Ref.~[\protect\citen{Mantu_Solid-state_Review}] with permission from Wiley.}
	\label{fig:tio6adt}
\end{figure}

\begin{figure}[!htp]
	\centering
	\includegraphics[ width=0.9\textwidth, angle=0]{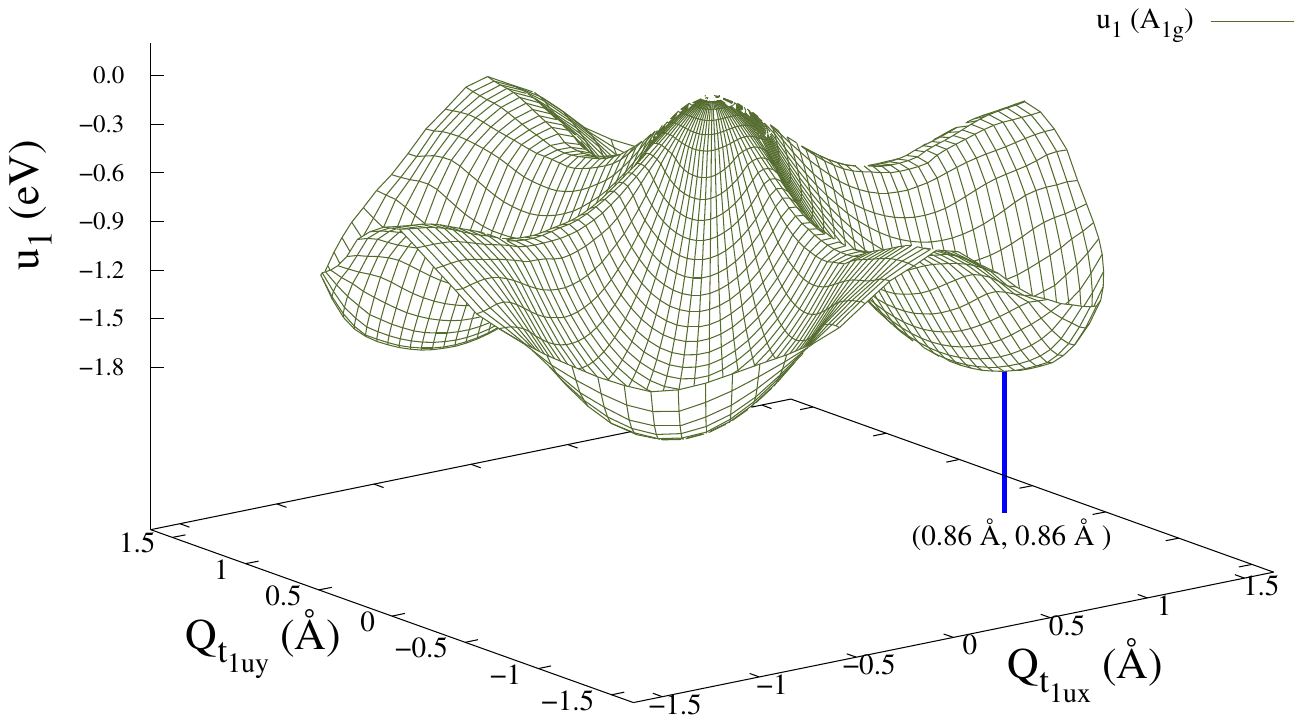}
	\caption{2D plot of the ground adiabatic surface $u_1$ are presented over the $Q_{t_{1ux}}$-$Q_{t_{1uy}}$ plane, exhibiting PJT-distorted minima at $Q_{t_{1ux}} = Q_{t_{1uy}} = 0.86$ \AA. The results are adapted from Ref.~[\protect\citen{Mantu_Solid-state_Review}], where permission has been taken from Wiley.}
	\label{fig:2d_u_surface}
\end{figure}

\noindent
 The $g$-symmetric vibrational modes, like $t_{2g}$ introduce JT distortions within the triply degenerate $^1T_{1u}$ excited states, as predicted by JT condition (see Eq.~\ref{Eq:JTcond}). On the contrary, 
 $u$-symmetric modes, namely, $t_{1u}$ symmetric ones facilitate PJT interactions between the non-degenerate $^1A_{1g}$ ground state and the $^1T_{1u}$ excited manifold. In order to achieve further insight about the underlying vibronic interactions, Sah \textit{et al}~\cite{Mantu_BTO} executed an exhaustive computational analysis of the TiO$_6^{8-}$ cluster with the aid of multireference quantum chemical framework. Frequencies of the vibrational normal modes were computed at the coupled-cluster level with singles and doubles (CCSD) with ANO-R1 basis set, which were validated against earlier theoretical benchmarks (see Table~\ref{tab:thresh}). On the other hand, the scanning of adiabatic PESs and NACTs was performed along the specified normal mode coordinates through a combination of CASSCF and MRCI methodologies using (6e,6o) as the active space. The active space comprises with six electrons distributed in six orbitals involving 
the highest occupied $t_{1u}$ and lowest unoccupied $t_{2g}^*$ orbitals. 
At the undistorted situation, the high-symmetry reference structure with equivalent Ti-O bond lengths of $2.0$ $\AA$ matches with both experimental~\cite{Kwei_Bond_order_exp} findings and previous theoretical~\cite{Heliyon_bond_order_theor} models. Scanning of {\it ab initio} quantities were systematically carried out across six independent pair of degenerate normal modes ($Q_{t_{1ux}}$--$Q_{t_{1uy}}$, $Q_{t_{1uy}}$--$Q_{t_{1uz}}$, $Q_{t_{1uz}}$--$Q_{t_{1ux}}$, $Q_{t_{2gx}}$--$Q_{t_{2gy}}$, $Q_{t_{2gy}}$--$Q_{t_{2gz}}$, $Q_{t_{2gz}}$--$Q_{t_{2gx}}$).  
It is evident from the adiabatic PESs that both JT and PJT effects are prominent involving multiple vibrational modes. It is interesting to note that $^1T_{1u}$ electronic state shows significant JT coupling 
along the $t_{2g}$ modes, with a characteristic vibrational frequency of 266.81 cm$^{-1}$. Additionally, strong PJT couplings are prominent 
between the $^1A_{1g}$ and $^1T_{1u}$ states via $t_{1u}$ modes, occurring at 123.87 cm$^{-1}$. Figure~\ref{fig:tio6adt} displays 
these interactions, where one-dimensional (1D) cuts of the adiabatic PESs along with the NACTs, $\tau_{ij}^{\phi}$ are plotted across the $Q_{t_{2gx}}$--$Q_{t_{2gy}}$ coordinate space. From this analysis,
the JT stabilization energy appears as relatively small, $\sim$0.0182 eV, whereas PJT interactions control the vibronic behavior in the ground electronic state. Figure~\ref{fig:2d_u_surface} shows the PES projected onto the $Q_{t_{1ux}}$--$Q_{t_{1uy}}$ plane, which reveals the presence of four symmetry-equivalent minima, corresponding to elongated octahedral distortions at $Q_{t_{1ux}} = Q_{t_{1uy}} = \pm 0.86 \AA$, with a stabilization energy of $\simeq$ 1.578 eV. In presence of simultaneous displacements along all three $t_{1u}$ modes, the system undergoes eight global minima positioned at $Q_{t_{1ux}} = Q_{t_{1uy}} = Q_{t_{1uz}} = \pm 0.51 \AA$, resulting into a total stabilization energy of approximately 2.582 eV.~\cite{Mantu_BTO} Such energy value differs
substantially from the earlier predicted 4 eV by Green's function-based DFT calculations~\cite{Bersuker_2015}.

\begin{table}[!htp]
	\caption{A comparison of molecular parameters is presented, where $Q_0$ represents the saddle point, $K_0$ shows the force constant for $Q$ displacement, and $\{Q_{t_{1ux}}^{gm}, Q_{t_{1uy}}^{gm}, Q_{t_{1uz}}^{gm}\}$ designates the global minimum. The data are taken from Ref. [\protect\citen{Mantu_BTO}].
	}
	\label{tab:thresh}
	\begin{center}
		\begin{small}
			\begin{tabular}{|c|c|c|}
				\hline
				Parameters& CCSD & Green's Function DFT~\cite{Bersuker_2015}  \\
				\hline
				$\hbar w_{e}$ ($t_{1u}$) ($cm^{-1}$) & 123.87  & 193 (also Ref.~[\citen{A_Mahmoud}]) \\
				$\hbar w_{e}$ ($t_{2g}$) ($cm^{-1}$) & 266.81 &---  \multirow{3}{*}{}  \\
				\cline{1-2}
				& MRCI(SD) & \multirow{4}{*}{} \\
				\cline{1-2}
				$Q_{t_{1ux}}^{gm}$= $Q_{t_{1uy}}^{gm}$= $Q_{t_{1uz}}^{gm}$ (\AA) & 0.51 & 0.14  \\
				
				$Q_o$ (\AA) & 0.64 &0.16 \\
				
				$K_o$ (eV/\AA) & 44.18 &55 \\

				$E_{PJT}$ ($eV$) &  2.582& 4.0 \\
				$E_{JT}$ ($eV$) &  0.0182& --- \\
				\hline
				
			\end{tabular}
		\end{small}
	\end{center}
\end{table}

\noindent
\subsection{Diabatic Hamiltonian and Calculated Observables} \label{Sec:Diabatic}

\vspace{0.2 cm}

\noindent 
In order to compute the photoelectron spectra of MnO$_6^{9-}$ unit of LMO crystal as well as photoemission spectra of TiO$_6^{8-}$ unit of BTO crystal, multi-state multi-mode diabatic Hamiltonians were calculated incorporating the electron-nuclear coupling. The BBO formalism 
can be suitably applied to the construction of diabatic Hamiltonians for
the  fundamental structural units in perovskite materials like LMO and BTO, namely, MnO$_6^{9-}$ and TiO$_6^{8-}$, respectively. In other words, such approach can successfully encapsulate the non-adiabatic interactions between electronic and nuclear motions in the accurate computation of photoelectron or photoemission (PE) spectral envelopes of these systems. The detailed implementation of the BBO theory to the MnO$_6^{9-}$ unit\citep{joy_jpca} and the TiO$_6^{8-}$ unit~\cite{Mantu_BTO} is thoroughly discussed in the following subsections. Furthermore, spectra bands were simulated and critically compared with theoretical models as well as experimental results to examine the validity of the BBO approach in depicting complex vibronic interactions in transition metal oxides.

\noindent
\subsubsection{9.5.3.1 MnO$_6^{9-}$: Photoelectron Spectra}\label{ssec:mno69-}

\begin{figure}[!htp]
	\centering
	\includegraphics[ width=0.9\textwidth, angle=0]{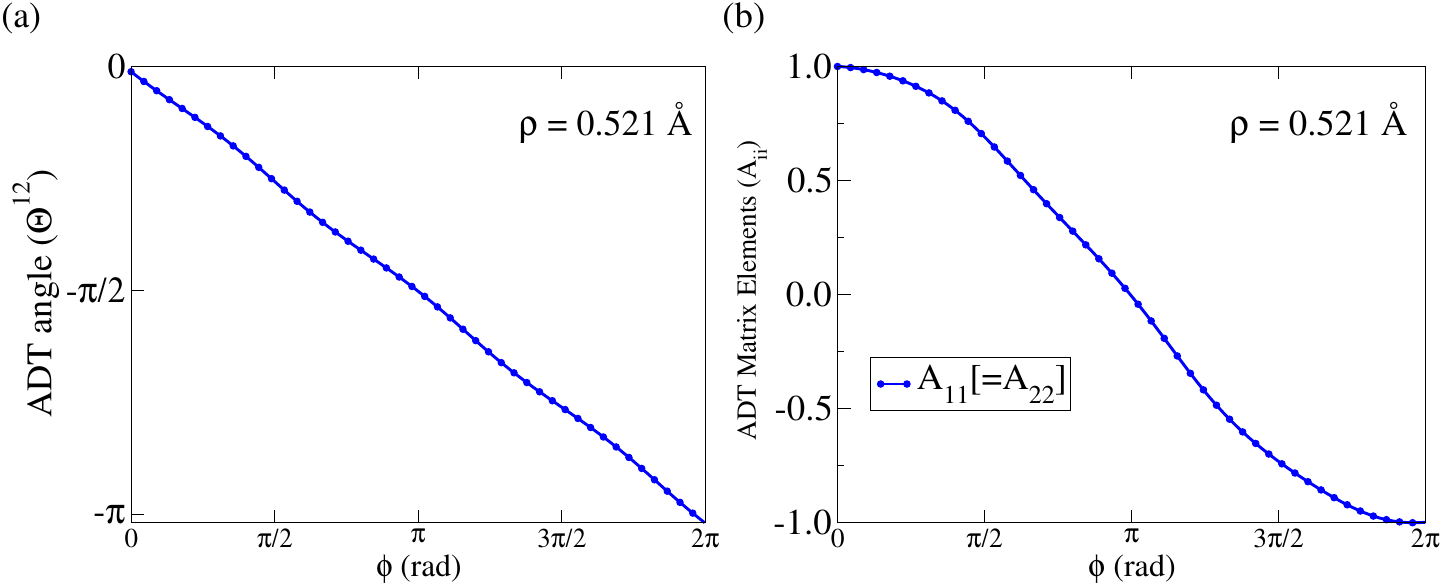}
	\caption{For the MnO$_6^{9-}$ unit at $\rho$ = 0.521 $\AA$   in the $Q_x$--$Q_y$ plane, panels (a) and (b) depict the 1D variation of the ADT angle ($\Theta^{12}$) and the diagonal elements of the ADT matrix, respectively, with respect to the $\phi $ coordinate.
		The results are taken from Ref.~[\protect\citen{Mantu_Solid-state_Review}] with permission from Wiley.} 
	\label{fig:mno6adt}
\end{figure}

\begin{figure}[!htp]
	\centering
	\includegraphics[ width=0.9\textwidth, angle=0]{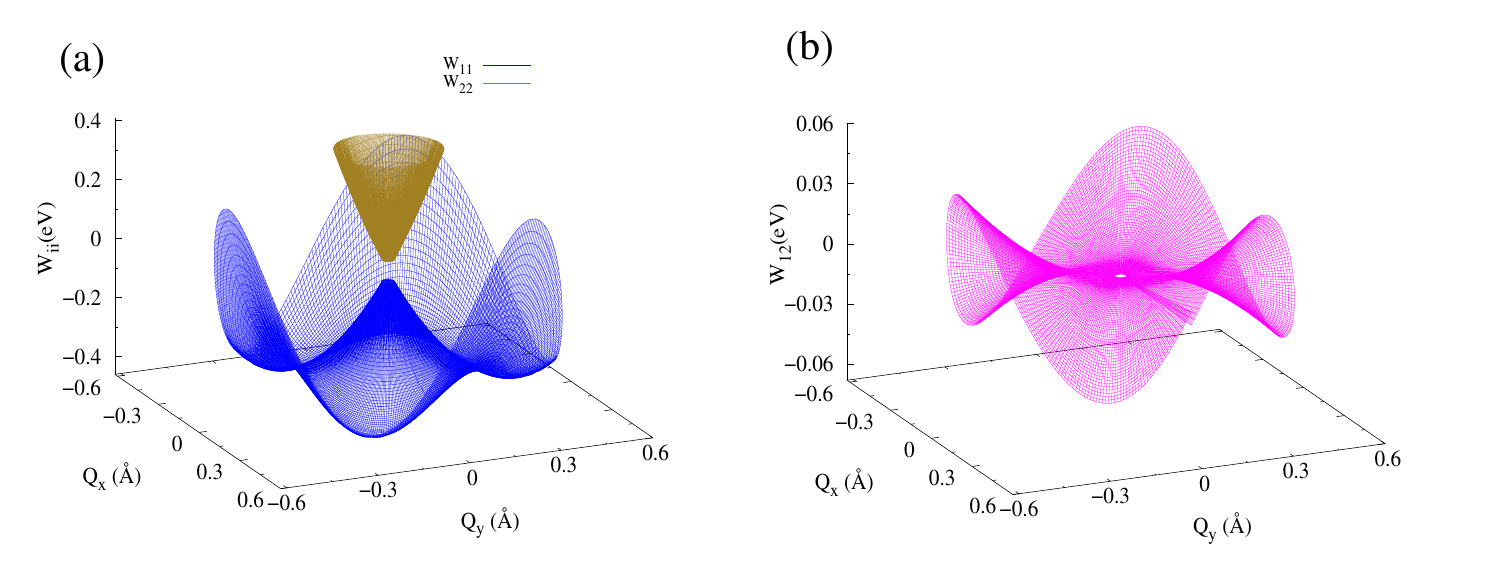}
	\caption{ The Diabatic PES matrix elements originated from MRCI(SD)-based adiabatic PESs and CP-MCSCF NACTs: (a) diagonal elements W$_{11}$ and W$_{22}$; (b) off-diagonal coupling W$_{12}$, are shown as functions of the nuclear  DOFs ($Q_x$ and $Q_y$). 
			The results are adapted from Ref.~[\protect\citen{Mantu_Solid-state_Review}], where the permission has been taken from Wiley.}
	\label{fig:2d_diab_11_22_LMO}
\end{figure}

\begin{figure}[!htp]
	\centering
	\includegraphics[ width=0.5\textwidth, angle=0]{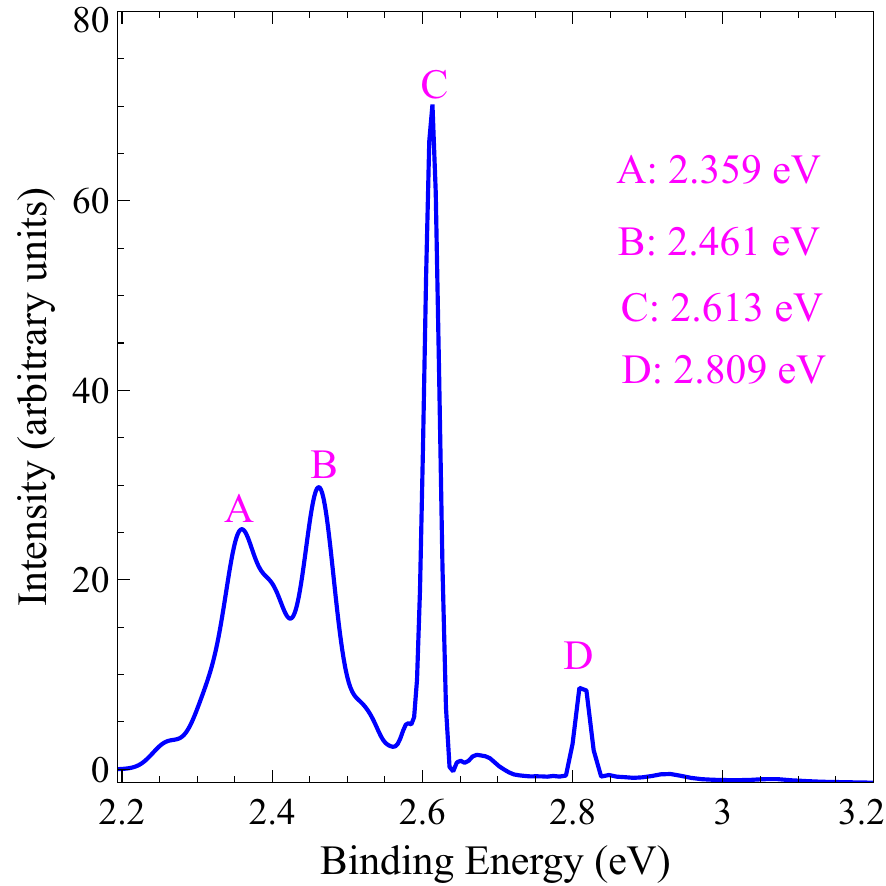}
	\caption{The PE spectra of MnO$_6^{10-}$, calculated using the scaled MRCI-based diabatic Hamiltonian of MnO$_6^{9-}$ are depicted with the peak positions labeled in the figure. 	The results are adapted from Ref.~[\protect\citen{Mantu_Solid-state_Review}], where the permission has been taken from Wiley.}
	\label{fig:spectra_mno6}
\end{figure}

\noindent
In order to unveil the nature of electron-nuclear coupling from an alternative perspective, a BBO theory based diabatic Hamiltonian for the MnO$_6^{9-}$ unit was computed. The \textit{ab initio} APESs and NACTs were calculated by MRCI(SD) and CP-MCSCF, respectively. The evaluated ADT angle accumulate a phase of $\pi$ upon enclosing a closed loop in nuclear coordinate space (specifically along the azimuthal coordinate $\phi = 2\pi$) surrounding the JT conical intersection (CI), as displayed in Figure~\ref{fig:mno6adt}a.~\citep{joy_jpca} Such quantization property endorses the presence of a JT CI, which is further validated by the sign change of the diagonal elements of the ADT matrix, namely $A_{11}$ and $A_{22}$ in Figure~\ref{fig:mno6adt}b. The diagonal elements (W$_{11}$ and W$_{22}$) and the off-diagonal coupling (W$_{12}$) of diabatic Hamiltonian appear as smoothly varying functions with respect to the nuclear DOFs ($Q_x$, $Q_y$), which guarantees     
the numerical accuracy for solving SE in the diabatic representation, as portrayed in Figure \ref{fig:2d_diab_11_22_LMO}. It is necessary to mention that the resulting diabatic PESs and coupling terms appear as smooth, continuous, and single-valued functions of the nuclear coordinates, ensuring their suitability for subsequent dynamical simulations. Once the diabatic Hamiltonian was in hand, high-resolution photoelectron (PE) spectra were computed with the aid of accurate multi-state (electronic), multi-dimensional (vibrational) quantum dynamics based on the Time-Dependent Discrete Variable Representation (TDDVR) formalism.~\citep{SSJPCA118,souvik_irpc37} The resulting BBO based photoelecton spectra (see Figure~\ref{fig:spectra_mno6}) show a peak-by-peak correspondence with the higher-energy satellite features in the experimental PE spectra of the LMO complex, justifying the fidelity of the theoretical model.~\citep{joy_jpca}

\noindent
\subsubsection{9.5.3.2 TiO$_6^{8-}$ : Photoemission Spectra} \label{ssec:photoemission}

\begin{figure}[!htp]
	\centering
	\includegraphics[ width=1.1\textwidth,  height=0.5\textheight, angle=0]{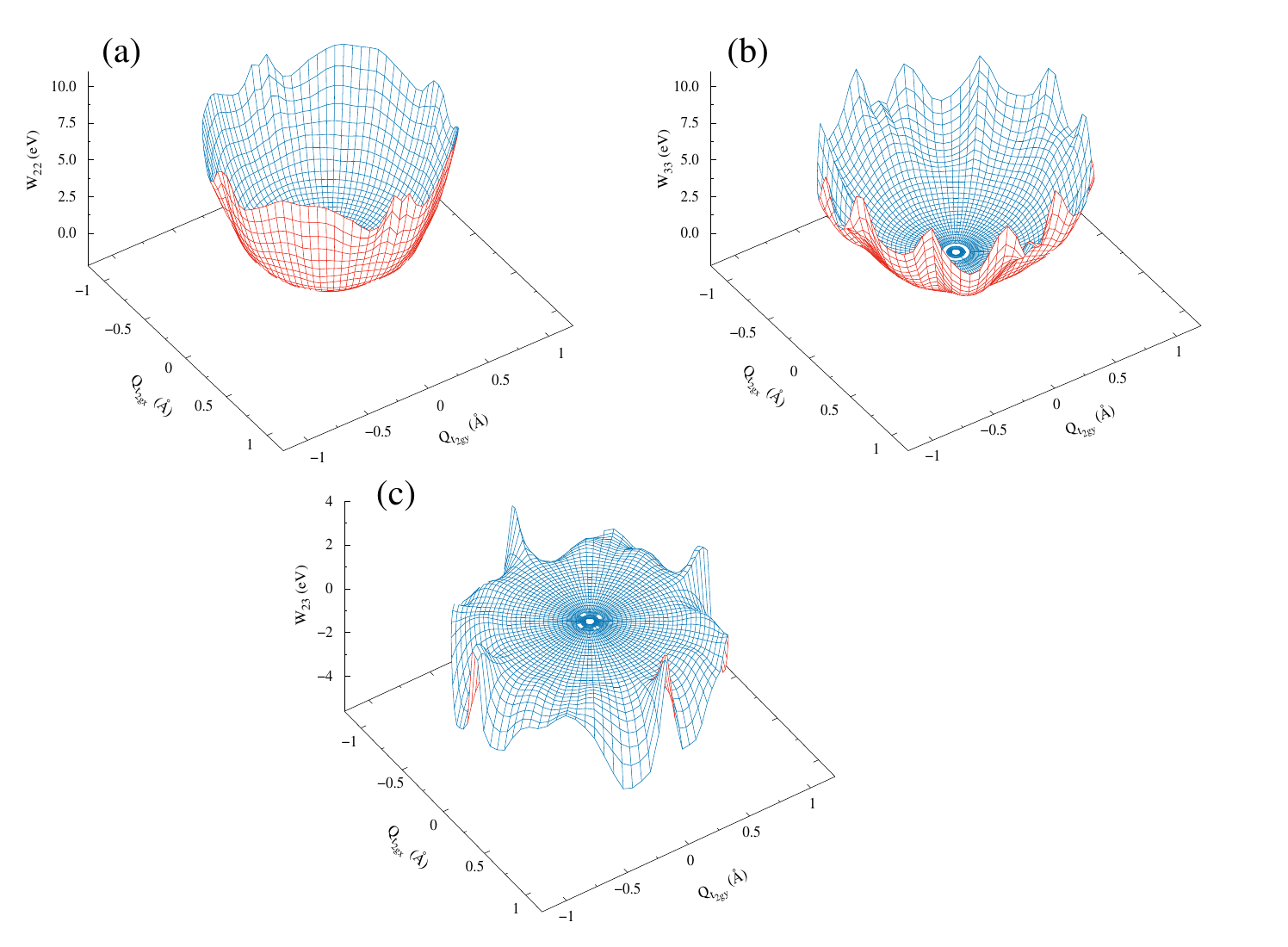}
	\caption{The 2D diabatic surfaces in the $Q_{t_{2gx}}-Q_{t_{2gy}}$ plane of the TiO$_6^{8-}$ unit in BTO crystal, namely  (a) W$_{22}$ and (b) W$_{33}$ together with the coupling element (c) W$_{23}$, which appear as smooth functions of the nuclear coordinates. The results are taken from Ref.~[\protect\citen{Mantu_Solid-state_Review}] with permission from Wiley. }
	\label{fig:t2g_BTO}
\end{figure}

\begin{figure}[!htp]
	\centering
	\includegraphics[ width=1.1\textwidth, height=0.5\textheight, angle=0]{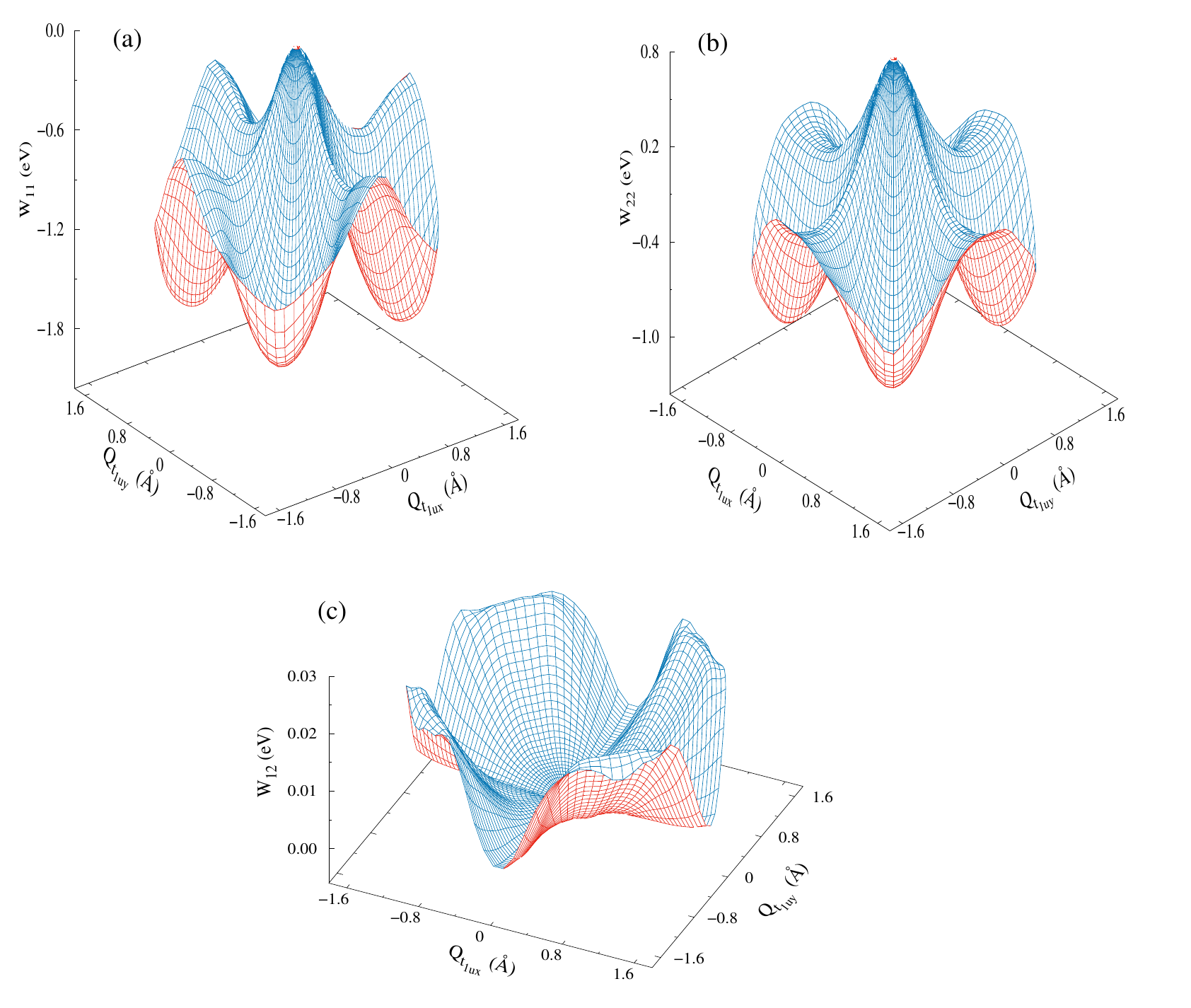}
	\caption{ The $Q_{t_{1ux}}-Q_{t_{1uy}}$ plane of the TiO$_6^{8-}$ unit in BTO crystal depicts the diabatic surfaces (a) W$_{11}$ and (b) W$_{22}$ along with their coupling (c) W$_{12}$, where all appear as smooth 2D functions of the nuclear coordinates. The results are taken from Ref.~[\protect\citen{Mantu_Solid-state_Review}] with permission from Wiley. }
	\label{fig:t1u_BTO}
\end{figure}

\begin{figure}[!htp]
	\centering
	\includegraphics[ width=0.8\textwidth, angle=0]{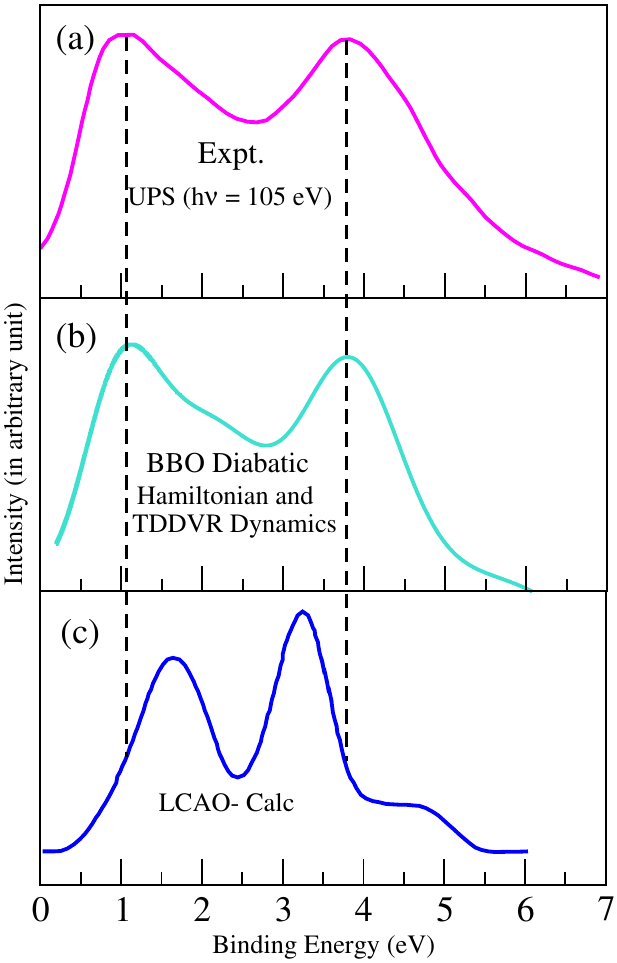}
	\caption{Comparison of valence electronic spectra: (a) experimental UPS of BaTiO$_3$~\cite{R_Cord_1985}, (b) theoretical result by Sah \textit{et al}~\cite{Mantu_BTO}, and (c) bulk BaTiO$_3$ calculation by Pertosa \textit{et al.}~\cite{Pertosa_Theory_Spectra} is demonstrated. The results are taken from Ref.~[\protect\citen{Mantu_Solid-state_Review}] with permission from Wiley.}
	\label{fig:photoemission}
\end{figure}

\noindent
A rigorous quantum dynamical calculations were carried out to explore the JT and PJT effects in the TiO$_6^{8-}$ moiety, which is a fundamental unit in perovskite-type transition metal oxides. The diabatic Hamiltonian encapsulates electron-nuclear couplings within $t_{1u}$ and $t_{2g}$ vibrational modes among the low lying electronic states, $A_{1g}$ and $T_{1u}$. Such Hamiltonian serves as input to simulate wavepacket evolution on both ground ($A_{1g}$) and excited ($T_{1u}$) electronic states.~\cite{Mantu_BTO} Figure \ref{fig:t2g_BTO} depicts W$_{22}$, W$_{33}$ and the coupling W$_{23}$ over the $Q_{t_{2gx}}-Q_{t_{2gy}}$ plane, which explores the characteristic four-fold symmetry associated with $t_{2g}$ distortions. On the other hand, Figure \ref{fig:t1u_BTO} displays the diabatic PESs W$_{11}$ and W$_{22}$ and their coupling W$_{12}$ for the $Q_{t_{1ux}}-Q_{t_{1uy}}$ plane associated with the $t_{1u}$ modes. While executing TDDVR calculation, a comprehensive convergence analysis of the TDDVR grid were carried out to find out the spectral resolution and then, the final optimized basis set is chosen to obtain numerically accurate results. The autocorrelation functions were Fourier transformed to get the PE spectra, which show good agreement with the experimental ones (see Figure \ref{fig:photoemission}) including a notable energy separation ($\sim$2.75 eV) consistent with the calculated PJT stabilization energy. Such study also involves phenomenological damping via a $\tau$ parameter (see Eq. \ref{Eq:damping}) to incorporate approximate lattice-induced energy dissipation effects, increasing the realistic features of the modeled spectral profiles. Even though the present approach provides valuable mechanistic insight into local vibronic interactions within the TiO$_6^{8-}$ unit, limitations arise from cooperative interactions intrinsic to the extended BTO crystal lattice. As a matter of fact, the work represents a significant advancement in understanding the role of JT and PJT effects in modulating the electronic structure and PE spectra of transition metal oxides, indicating a future platform for further theoretical and spectroscopic investigations.

\subsection{Spontaneous Polarisation in BTO due to PJT Interaction}\label{Spontaneous}

\begin{figure}[!htp]
	\centering
	\includegraphics[ width=0.9\textwidth, angle=0]{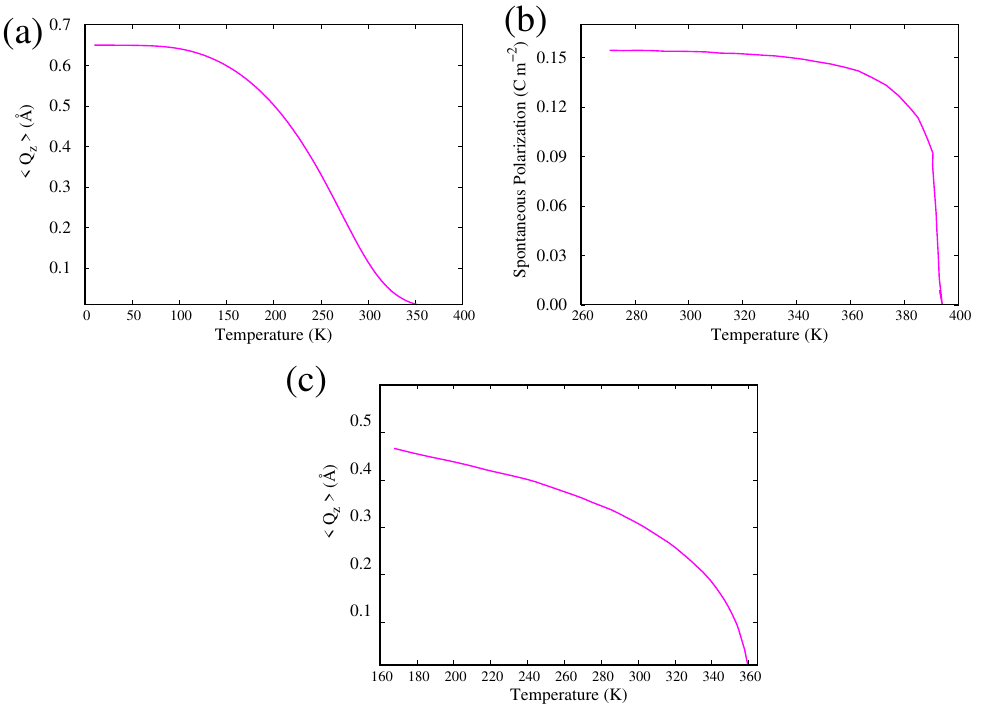}
	\caption{Temperature dependence of order parameter $\langle Q_z \rangle$ shows a second-order phase transition, which is compared with polarization data from (a) Sah \textit{et al}\cite{Mantu_BTO}, (b) Merz~\cite{Walter_spont_1949} and (c) Polinger~\cite{Polinger_2013}.  The results are taken from Ref.~[\protect\citen{Mantu_Solid-state_Review}] with permission from Wiley.}
	\label{fig:ferroelectric}
\end{figure}

\noindent
Eq. \ref{Eq:Partition_function1} was used to explore the PJT effect on the microscopic origin of ferroelectricity in the TiO$_6^{8-}$ units of BTO. When the temperature is lower than the Curie temperature ($T_C \approx 350$ K), one can observe structural distortion to a non-centrosymmetric phase leading to spontaneous polarization via the off-center displacement of Ti$^{4+}$ ions and counter-motion of oxygen atoms. While performing quantum dynamical simulations using adiabatic PESs, the phase transition~\cite{Mantu_BTO} and Curie temperature were reproduced, which show good accord with prior experimental~\cite{Walter_spont_1949} and theoretical study~\cite{Polinger_2013} (see Figure \ref{fig:ferroelectric}). It is important to note that the PJT-induced vibronic coupling between electronic states of opposite parity increases Ti--O covalency and stabilizes the polar phase, providing a quantum-level illustration for ferroelectric behavior in perovskite oxides.

\section{Conclusion}\label{sec:jt-conc}

\noindent
In order to investigate the origin of roto-vibrational features in the dielectric function spectra of LaMnO$_3$, quantum mechanical analysis was carried out for both the ground and the less-studied first excited states of JT-active ions, such as Mn$^{3+}$ with the aid of model and \textit{ab initio} Hamiltonian. In case of TiO$_6^{8-}$ unit of the BaTiO$_3$ crystal, the PJT interaction between the non-degenerate ground and the lower sheet of degenerate excited symmetry state is governed by 
$t_{1u}$ modes. In addition, the order parameter were also explored, associated with the ferroelectric properties of this crystal. \\

\noindent
In Sections \ref{ssec:dyJT} and \ref{ssec:bboJT}, an overview of the theoretical frameworks underlying the JT and BBO methodologies were presented. While doing the analysis, the linear $E \otimes e$ JT model was involved around the minima of the lower adiabatic PESs, whereas excited states were found to be stabilized by centrifugal effects anharmonicities. The computation of adiabatic PESs and NACTs, as well as their transformation into smooth diabatic surfaces are also discussed in detail. Sections \ref{ssec:spec} and \ref{ssec:Ferr} are mainly focused on the theoretical algorithm for electronic spectra and unveil ferroelectric behavior, representing the methodologies used to compute spectral profiles and polarization characteristics, respectively.\\

\noindent
Section \ref{ssec:Roto-Vibrational} deals with JT effect in a perovskite system employing two complementary methodologies: (a) The first approach is dedicated to analysis of roto-vibrational spectra using a model Hamiltonian,~\citep{joy_jcp} where the parameters were obtained by fitting to experimental ellipsometric dielectric function data.~\citep{Natalia_JETP} Such model is widely known as centrifugally stabilized cubic-perturbed excited-state Hamiltonian, effectively capturing key spectral features; (b) In the second approach, {\it ab initio} APESs were calculated for the distorted octahedral MnO$_6^{9-}$  embedded in LaMnO$_3$. On the other hand, while parameterizing the same model Hamiltonian using the \textit{ab initio} APESs, it reproduced the roto-vibrational spectral features with notable accuracy.~\citep{joy_jpca} It is worth mentioning that even though influence of the La atom was approximately encapsulated via an empirical mass scaling factor, collective lattice effects such as cooperative interactions were not been incorporated, which could further improve the predictive capability of the model.\\

\noindent
Section \ref{ssec:tio68-} deals with the calculation of diabatic PESs employing \textit{ab initio} APESs and NACTs of TiO$_6^{8-}$ unit of BaTiO$_3$, where dominant role of PJT interactions is observed, highlighting their contribution to large structural distortions and the origin of ferroelectric behavior. In comparison with the relatively weak JT effect, PJT coupling between the $A_{1g}$ ground state and the $^1T_{1u}$ excited state via $t_{1u}$ modes appears as the vibronic mechanism. 
In Section \ref{Sec:Diabatic}, the peak positions of the photoelectron spectra for the MnO$_6^{10-}$ species exhibit reasonable agreement with the roto-vibrational peak positions as obtained from the ellipsometric spectra of the MnO$_6^{9-}$ ion.~\cite{joy_jpca} In addition, the calculated photoemission spectra originated from combination of four electronic states for the TiO$_6^{8-}$ cluster in the BaTiO$_3$ crystal, show close agreement with the experimental spectra for a BaTiO$_3$ single crystal.~\cite{Mantu_BTO} Finally Section \ref{Spontaneous} depicts the spontaneous polarization leading to the ferroelectric properties of BaTiO$_3$ below its Curie temperature, with the aid of \textit{ab initio} APESs. Such findings indicate the polar instability and the off-center displacement of transition metal ions to the PJT effect.

\bibliographystyle{aipnum4-1}
\bibliography{ref_book}

\chapter{Role of Phonons and Electron-Hole Pair Excitations in Gas--Metal Surface Interactions: An Effective Hamiltonian Approach}\label{sec:surface}

\section{Introduction}\label{sec:intro}

\noindent
Heterogeneous catalysis plays a dictating role in a vast majority of surface reaction. Those macroscopic processes are constituted with 
elementary steps of reactions on the catalytic surface, mostly dissociative chemisorption. In these phenomena, molecular bonds are cleaved upon surface impact, leading to adsorbed fragments. In other words, interactions between gas-phase molecules and surface atoms form
a fundamental class of phenomena controlling a 
 wide variety of physical and chemical processes. Moreover, a precise  investigation of the prototypical reactions, namely, dissociation of \ce{H2} and \ce{D2} on metal surfaces becomes a challenging problem in spite of the simplicity of incoming diatomic molecules. These systems can be considered as essential benchmarks for validating theoretical frameworks before extended to more complex gas-surface scattering phenomena, motivating extensive experimental~\cite{HA2,CTD,CTH,CTR2,CTR,HAM,EXPNEW2,EXPNEW3,Wodtke,Chadwick} and theoretical~\cite{HO1,CDGJK,HA1,SAL,SAG,TSS,TSS1,TSS2,TSSM,FNGJK1,SPO,MWMF,QD-AIMD-H2-D2-Cu-review-2016,H2-Cu-multiple,PSMFS,VV-Hartree-2021,EAM-SCM-QC-Mark-2021,SRP-DF-new-H2-Cu-Kroes-2021,Kroes-Review-2021,SAMARK,Joy_D2_jcp_2022,Mantu_jcp_2024,Bauke_jcp_2024,Mantu_JPCC_2026} investigations.\\

\noindent	
In recent days, the advancement of this field has been developed by a synergistic relationship between experiment and theory. Several experimental techniques, like high-resolution techniques, including molecular beam scattering~\cite{HA2,CTR,CTH,CTR2,CTD} and associative desorption measurements~\cite{HAM,EXPNEW2,EXPNEW3,Wodtke} generated the critical datasets necessary for the construction of chemically accurate potential energy surfaces (PESs)~\cite{SC1,CRP,CRP3,CRP4,MWMF,PSMFS,EAM-SCM-QC-Mark-2021,NN-PES-perspective-2021,Kroes-Review-2021}. In order to interpret the microscopic mechanisms underlying these experimental observables, significant evolution~\cite{SMIRPC,FNGJK1,MWMF,H2-Cu-multiple,PSMFS,VV-Hartree-2021,EAM-SCM-QC-Mark-2021,Kroes-Review-2021,Mantu_jcp_2024,Bauke_jcp_2024,Mantu_JPCC_2026} was introduced in the dynamical methodologies. On the other hand, contemporary theoretical approaches employ high-dimensional quantum dynamics (QD)~\cite{QD-method-1993,SPO,SRP-DF-24-CH4-step-Ni-2018,SRP-DF-23-CH4-step-Ni-2019,SMIRPC,QD-AIMD-H2-D2-Cu-review-2016,Kroes-Review-2021} and quasi-classical trajectory (QCT) simulations~\cite{SC1,MWMF,SRP-DF-8-H2-D2-Cu-LDFA-ODF-elhp-2018,H2-Cu-multiple,HD-NNP-perspective-2020,EAM-SCM-QC-Mark-2021}, often complemented by direct \textit{ab initio} molecular dynamics (AIMD) or density functional theory molecular dynamics (DFMD)~\cite{FNGJK,FNGJK1,PBE-19-CH4-on-Ni-AIMD-2014,QD-AIMD-H2-D2-Cu-review-2016,SRP-DF-17-H2-Cu-GLO-AIMDEF-elhp-2017} to provide inner physical insight on the scattering events. The accuracy of these simulations depends entirely on the quality of the PES~\cite{Kroes_2025}. Even though the generalized gradient approximation (GGA) within density functional theory (DFT) are widely used in surface scattering phenomena~\cite{Zhang-QD-H2-Cu111,H2-Cu111-CRP,6D-QD-HD-NNP-B-Jiang-H-Guo,Kroes-Review-2021}, its limitations necessitate the formulation of more advanced techniques, such as GGA-exchange plus correlation functionals~\cite{GGA-exchange+correlation-8-H2-Metal-2015,GGA-exchange+correlation-2-CH3OH-Cu-2019} and meta-GGA (mGGA) methods~\cite{mGGA-1-H2-Cu-Ag-2019}.
A major breakthrough was introduced by specific reaction parameter (SRP) DFT approach, which effectively tunes the functional to specific systems, significantly enriching PES reliability~\cite{SRP-origin,SRP-DFT-Truhlar,SC1,CDGJK,FNGJK1,SRP-DF-17-H2-Cu-GLO-AIMDEF-elhp-2017,SRP-DF-8-H2-D2-Cu-LDFA-ODF-elhp-2018,PSMFS,SBH10-SRP-DF-database,SRP-DF-new-H2-Cu-Kroes-2021,SRP-DF-38-limitation-2020}. Moreover, alternative high-level methods are also emerging, such as embedded correlation wavefunction (ECW) theory~\cite{ECW-3-O2-Al-2018} and diffusion Monte Carlo (DMC)~\cite{DMC-1-H2-Cu-2017,DMC-3-2020} to explore the complex landscapes of PES. In order to make the PESs computationally tractable for dynamics, fitting procedures are essential. Techniques like the corrugation reducing procedure (CRP)~\cite{CRP,SRP-DF-new-H2-Cu-Kroes-2021,H2-Cu111-CRP,Kroes-7D-QD,MWMF,PSMFS,EAM-SCM-QC-Mark-2021} are well-known with a surge in machine learning applications in recent years. High dimensional neural network PESs (HD-NNPs)~\cite{HD-NN-tutorial-review,HD-NNP-perspective-2020,NN-PES-perspective-2021,H2-Cu-multiple,AtNNmethod} and embedded atom neural networks (EANNs)~\cite{H2-Cu-multiple} PESs can also be used for efficient 6D QD simulations across a variety of systems~\cite{6D-QD-HD-NNP-B-Jiang-H-Guo,6D-QD-HD-NNP-D-Zhang-2}.\\

\noindent	
A critical aspect of accurate modeling is consideration of non-zero surface temperatures and exploration of role of surface lattice vibrations or electron-hole pair (elhp) excitation. In the early days, these thermal effects were captured by utilizing model Hamiltonians such as the single oscillator (SO)~\cite{HA1,HO1}, modified surface oscillator (MSO)~\cite{SAL,UZER1} and generalized Langevin oscillator (GLO)~\cite{GLO-1,GLO-2,CRP-H2-Pd-thermally-distorted-2005,SRP-DF-9-gas-surface-GLO-new-2019}. These models addressed energy dissipation, whereas recent frameworks are introduced to integrate specific lattice dynamics. Jackson and co-workers developed the lattice reconstruction sudden (LRS) model~\cite{AKT1,AKT} and trapping-mediated dissociation frameworks~\cite{CH4-trapping-1-Ni-Ir-Pt-Jackson-2020,CH4-trapping-2-Ir-Jackson-2021} for polyatomic systems like \ce{CH4}. Later on, Somers and co-workers proposed a thermally expanded static corrugation model (SCM) based on the embedded atom method (EAM)~\cite{MWMF,PSMFS,EAM-SCM-QC-Mark-2021,Bauke_jcp_2022,Bauke_jcp_2023,Bauke_jcp_2024}. As per the dynamical simulations, it is indeed necessary to include
elastic and inelastic scattering probabilities to correctly describe temperature effects in \ce{H2}/\ce{D2} chemisorption on Cu(111). Furthermore,
 Adhikari and colleagues~\cite{SAG,TSS,TSS1,TSS2,TSSM,SAMARK,SM-JTCC,SM-MOLPHYS,VV-Hartree-2021,Joy_D2_jcp_2022,Mantu_jcp_2024,Mantu_JPCC_2026} developed the field with a time- and temperature-dependent effective Hartree potential (EHP) model.
 Such many-oscillator framework, involving linear and quadratic couplings exhibits a non-negligible dependence on surface vibrational state populations governed by the Bose-Einstein (BE) distribution. In case of \ce{H2}/\ce{D2} interacting with Cu(111) or Ni (1nn) at elevated temperatures, the anharmonicity of the interaction potential appears as one of the dictating factor for dynamics~\cite{VV-Hartree-2021}. The time propagation of this model Hamiltonian within a mean-field scheme is carried out employing 6D QD with the help of time-dependent discrete variable representation (TDDVR)~\cite{SMIRPC} or split-operator (SPO) DVR approach~\cite{SPO,SPO-DVR-code-2}. Such calculations predicted that the quantum nature of surface lattice vibrations sufficiently alters reaction probabilities, particularly, in the threshold region as well as illustrates the diffraction probability.\\
 
\noindent
Beyond the domain of Born–Oppenheimer static surface (BOSS) approximation, nonadiabatic effects originating due to electron–hole pair (elhp)~\cite{JCT,JCTullyrev,Saalfrankrev,Wodtkeelhp,JCTully-elhp1,JCTully-elhp2,O2-Pd100-JM1,Tremblay6,Kroes-Review-2021} excitations are widely encountered in chemisorption dynamics and inelastic gas–surface scattering. The non-Born–Oppenheimer static surface (NBOSS) framework involves such interactions with the help of two established approaches: electronic friction (EF)~\cite{JIJ1,SRP-DF-8-H2-D2-Cu-LDFA-ODF-elhp-2018} and independent electron surface hopping (IESH)~\cite{IESH-2009,JCTully-elhp1}. In the realm of EF formalism, incorporation of the local density friction approximation (LDFA)~\cite{Kroes-Review-2021} and orbital-dependent friction (ODF)~\cite{RPBE-1-N2-Ru-ODF-elhp-2019,ODF-5,Kroes-Review-2021} leads to the evolution of the nuclei on the ground-state Born–Oppenheimer PES under the influence of dissipative friction and temperature-dependent stochastic forces successfully encapsulating the elhp excitations. Due to the 
computational efficiency and use of a single ground-state PES, 
the EF approach appears as one of the most widely adopted methods for treating electronic nonadiabaticity. On the other hand, the IESH approach~\cite{Dynamical-models-review-2018,IESH-2009,JCTully-elhp1} explicitly deals with elhp excitations through stochastic transitions between electronic states leading to an alternative framework for simulating electronically nonadiabatic gas–surface dynamics. \\

\noindent	
In order to explain the relative contributions of surface phonons and elhp excitations in energy dissipation, numerous studies were carried out in the non-Born–Oppenheimer moving surface (NBOMS) framework involving both the mechanisms simultaneously. MDEF (GLO+ODF) calculations for CO/Cu(100)~\cite{JCT} GLO+LDFA studies of H$_2$/Cu(111)~\cite{SRP-DF-17-H2-Cu-GLO-AIMDEF-elhp-2017}, N$_2$/W(110)~\cite{GGA-exchange+correlation-1-N2-W-elhp-2019} and CO$_2$/Ni(100)~\cite{PBE-6-CO2-surface-mode-elhp-2018} consistently depict the dominating role of the surface phonons in the energy dissipation process. In case of H$_2$/Cu(111),
the combined involvement of phonons and elhp excitations decreases the vibrational transition probabilities relative to the BOSS model~\cite{RPBE-1-N2-Ru-ODF-elhp-2019}, whereas 
elhp excitations are indeed necessary for reproducing the experimentally observed increasing trend of the v= 0$\rightarrow$ 1 transition probability with surface temperature. In a similar way, the AIMD EF approach is able to 
validate the combined effects of phonons and elhp excitations for H$_2$/Cu(111)~\cite{SRP-DF-17-H2-Cu-GLO-AIMDEF-elhp-2017}, dissociative chemisorption of N$_2$ on Fe(110)~\cite{RPBE-2-H2-Pd-N-Ag-N2-Fe-AIMDEF-elhp-2016}, and the dissociation and scattering of HCl on and from Au(111)~\cite{GGA-exchange+correlation-5-HCl-Au-AIMD-2019}. Nevertheless, HD-NNP-based MDEF (GLO+ODF) calculations for N$_2$ /Ru(0001) explored the significant role of elhp excitations in dissociative chemisorption resulting into reasonably better agreement with experiment in comparison with BOMS model at elevated surface temperatures.~\cite{RPBE-1-N2-Ru-ODF-elhp-2019} In a nutshell, such investigations affirms that surface phonons often provide the dominant channel for energy dissipation and the contribution of elhp excitations is crucial for precise determination of the specific reaction systems and experimental observables. \\

\noindent
In the effective Hamiltonian, linear and quadratic couplings~\cite{VV-Hartree-2021} between surface modes and molecular DOFs were included through the embedded diatomics-in-molecules (EDIM) interaction potential put forward by Truong et al~\cite{TNT} and the states associated with the surface modes with the help of BE statistics at finite surface temperature. TDDVR~\cite{SMIRPC} dynamics were used to carry out QD on the resulting Hamiltonian. Furthermore, elhp couplings were also taken into account within a Fermi-Dirac-weighted mean-field effective Hamiltonian using Fermi-Thomas model potential.~\cite{SM-MOLPHYS} While performing QD followed by the analysis of outgoing wavepacket, such study shows the extent of elhp contributions effects in the dissociation and transition probabilities through their coupling with molecular DOFs over the surface mode-mediated processes. During such formulation of EHP, the ``accuracy" of  linear and quadratic couplings between surface modes and molecular DOFs were computed by using SCM potential. The quantum states of surface modes show the distribution, governed by the BE probability factors at finite surface temperature. Finally, 6D QD were performed with SPO-DVR methodology to evaluate reaction attributes for H$_2$/D$_2$ colliding with Cu surface.

\section{Theoretical Formulation}\label{sec:theory}

\vspace{0.2 cm}

\noindent
Since our aim is to formulate a Hamiltonian for the collision process between a diatomic species and the atoms of a metal surface, the wavefunction for the entire system takes the following form:

\vspace{-0.8 cm}

\begin{eqnarray}
	\zeta (\lbrace \chi_m \rbrace,\{Q_k\}, \lbrace r_e \rbrace,t) = \sum_{ijk} C_{ijk} (t) \cdot \Psi^i_m(\lbrace \chi_m \rbrace, t)\cdot \Phi^j_{vib}(\{Q_k\},t)\cdot \Phi^k_{elhp}\left( \lbrace r_e \rbrace, t \right),
	\label{eq:Ansatz}
\end{eqnarray}

\noindent
 where $\Psi^i_m(\lbrace \chi_m \rbrace, t)$ represents the diatomic wavefunction with the molecular degrees of freedom (DOFs), $\lbrace \chi_m \rbrace$, $\Phi^j_{vib}(\{Q_k\},t)$ symbolizes the wavefunction for the surface modes, ($\lbrace Q_k \rbrace$) and  $\Phi^k_{elhp}\left( \lbrace r_e \rbrace, t \right)$ denotes for elhp excitations expressed by the electronic coordinates $(\lbrace r_e \rbrace$).\\

\noindent
When there is enough small recoil effect from the surface to the molecule vis-\'a-vis the contributions from the higher configuration space are negligible, the Equation~(\ref{eq:Ansatz}) could be simplified as per the mean-field approximation. In practical scenario, such situation typically arises during the collision of light diatomic molecule with a surface composed of heavy metal atoms. Under these conditions, the mean-field approximation can be used for a theoretically sound and computationally tractable framework to construct Hamiltonian and to perform quantum dynamical (QD) calculations. Its validity can then be explored through comparison with experimental observations. 

\subsection{Mean-field approximation} \label{ssec:Mean-field}

\noindent
Within the mean-field framework, an effective Hamiltonian is constructed to incorporate the influence of surface modes and elhp excitations at finite surface temperature. Thermal effects are involved through Bose-Einstein (BE) and Fermi-Dirac (FD) probability factors defining the initial distributions of phonons and elhp states, respectively. Such formulation is helpful to generate a unified time-dependent treatment of the molecular nuclear coordinates ({$\lbrace \chi_m\rbrace$}), lattice vibrational modes ($\lbrace Q_k\rbrace$) and elhp excitations ($\lbrace r_e\rbrace$), which assists for the dynamical exploration of the full manifold of configurations corresponding to their respective quantum states.  \\

\noindent
 While using this approach, each subsystem (molecular DOFs, phonons and electronic exciton) is only internally linked across its accessible configurations based on the assumption of negligible small recoil effect from the surface atoms. As a matter of fact, thermal effects are taken into account in the Hamiltonian through the coupling of the molecular DOFs to surface phonons and elhp excitations due to the negligible recoil of the surface. In other words, the ansatz for the total wavefunction takes the following form:
 
\begin{eqnarray}
	\zeta (\lbrace \chi_m \rbrace,\{Q_k\}, \lbrace r_e \rbrace,t) \simeq	\Psi_m(\lbrace \chi_m \rbrace, t)\cdot \Phi_{vib}(\{Q_k\},t)\cdot \Phi_{elhp}\left( \lbrace r_e \rbrace, t \right),
		\label{eq:meanfield}
\end{eqnarray}

\noindent
where the molecular coordinates are expressed as {$\chi_m$} (= {$x, y, z, X, Y, Z$}) constituted with both internal and external degrees of freedom relative to the surface (see Figure~\ref{fig:Geometry}). To be specific, ($x, y, z$) symbolizes the diatomic bond vector, while ($X, Y, Z$) designates the center-of-mass (COM) position of the molecule fixed and measured by taking the topmost layer of the metal surface taken as reference at $Z$ = 0. On the other hand, the set {$\lbrace Q_k \rbrace$} corresponds to the normal or phonon modes corresponding to surface lattice vibrations and {$\lbrace r_e\rbrace$} represents the electronic coordinates related to elhp excitations.

\subsubsection{10.2.1.1. Time-dependent Schr\"odinger equation and effective Hamiltonian} \label{ssec:Effective_Hamiltonian}

\noindent
The total Hamiltonian for the molecule-surface interactions takes the following form:
\begin{eqnarray}
	\widehat H = \widehat T_m + \widehat {V}_0 + \widehat H_0^s +  \widehat H^{el}_0 + \widehat {V}^{vib}_I + \widehat {V}^{elhp}_I.
	\label{eq:hamtotsys}
\end{eqnarray}

\noindent
Here $\hat{T}_m$ represents the kinetic-energy operator of the incident gas-phase diatomic molecule relative to the surface. On the other hand, $\hat{V}_0$ symbolizes the interaction potential between the diatomic molecule and the rigid surface. The Hamiltonian of the bare solid surface is manifested by $\widehat H_0^{s}$, whereas the free-electron Hamiltonian of the metal surface are depicted by $\widehat H_0^{el}$. The term $\hat{V}^{vib}_{I}$ is associated with the coupling between the molecular DOFs and the surface vibrational modes, while $\hat{V}^{elhp}_{I}$ depicts the interaction originating from elhp excitations. \\

\noindent
While projecting the time-dependent Schr\"odinger equation (TDSE) onto the molecular subspace and integrating out the surface vibrational and elhp DOF, one can obtain the equation of motion for the molecular DOF. Employing Equations~(\ref{eq:meanfield}) and ~(\ref{eq:hamtotsys}), one can achieve:

\begin{eqnarray}
	i\hbar|\dot{\Psi}_m \rangle  = (\widehat {T}_m + \widehat {V}_0 + \epsilon^0_s + \epsilon^0_{el} + \langle \Phi_{vib} | \widehat {V}^{vib}_I | \Phi_{vib} \rangle  + \langle\Phi_{elhp} | \widehat {V}^{elhp}_I | \Phi_{elhp}\rangle) |\Psi_m \rangle.
	\label{eq:TDSE1}
\end{eqnarray}

\noindent
It is important to mention that $\epsilon^0_s$ and $\epsilon^0_{el}$ represents normal or phonon mode frequencies and energy of free electrons on the metal surface, respectively. \\ 

\noindent
Since the interaction picture can be considered as a convenient framework for describing the molecule–surface coupling, one can transform Equation~(\ref{eq:TDSE1}) into the same representation leading to the following form of effective Hamiltonian:

\begin{eqnarray}
	\widehat{H}_{eff} (\lbrace \chi_m \rbrace, t) =&-&  \frac{\hbar^2}{2 \mu} \left( \frac{\partial^2}{\partial x^2} + \frac{\partial^2}{\partial y^2} +\frac{\partial^2}{\partial z^2} \right) - \frac{\hbar^2}{2 M} \left( \frac{\partial^2}{\partial X^2} + \frac{\partial^2}{\partial Y^2} +\frac{\partial^2}{\partial Z^2} \right) \nonumber \\
	&+& \widehat {V}_0 (\lbrace \chi_m \rbrace)+ \widehat {V}_{eff}^{vib}(\lbrace \chi_m \rbrace, t) + \widehat {V}_{eff}^{elhp}(\lbrace \chi_m \rbrace, t).
	\label{Eq: initial_effective_Hamiltonian}
\end{eqnarray}

\noindent
The kinetic energy operators are functions of the reduced mass ($\mu$) for internal coordinates of diatom and the total mass ($M$) of diatomic molecule for its COM motion. The potential, $\widehat {V}_0(\lbrace \chi_m \rbrace$) describes the Born-Oppenheimer static surface (BOSS) interaction with a rigid--non-vibrating surface atoms (either at 0 K or  linearly expanded at 925 K) and two additional time-dependent effective Hartree potentials: $\widehat {V}_{I}^{vib}(\lbrace \chi_m \rbrace, t)$ arising from coupling between molecular motion and surface modes, and $\widehat {V}_{I}^{elhp}(\lbrace \chi_m \rbrace, t)$ originating from elhp excitations. Thus, the projected TDSE yields a mean-field molecular Hamiltonian that incorporates both static potential energy contributions and dynamical dissipative effects arising from the coupling of the molecular DOFs to surface vibrational modes and elhp excitations. On the process of formulating effective Hamiltonian,  we treat the coupling of the molecular DOF to surface vibrational modes in Sec.~\ref{ssec:Interaction-potential} to \ref{ssub:VV-EHP} and elhp excitations in Sec.~\ref{ssec:elp}, respectively.

\subsubsection{10.2.1.2. Mean field vs sudden approximation}

\noindent
Employing the sudden approximation~\cite{OLKR}, the coupling between molecular DOFs and surface mode vibrations is included by assuming the collision occurring on a timescale much shorter than that of lattice motion. Consequently, the surface atoms are considered as effectively frozen while performing a single scattering event and the lattice vibrational coordinates ($\{Q\}$) appear only as static parameters. It is important to note that the effect of surface temperature is introduced by sampling these configurations with the help of Maxwell-Boltzmann (MB) distributions, where dynamical calculations are executed independently on each sampled surface. Then, the observables are computed by averaging over the results originating due to the large ensemble of such configurations. Even though this treatment overlooks explicit time-dependent couplings and higher-order correlations between molecular DOFs and surface modes, it remains effective in encapsulating thermal effects in gas--surface scattering dynamics due to consideration of ensemble average over temperature dependent configurations. Nevertheless, in the mean-field approach, the time- and temperature- dependence of the Hamiltonian originate from the quantum mechanical treatment of surface atom vibrations and their distributions over quantum states through BE probability factors.

\subsubsection{10.2.1.3. Interaction potential between molecular DOFs and surface modes} \label{ssec:Interaction-potential}

\noindent
Since this formulation is performed to explore the interaction potential between molecular DOF - surface mode couplings and its effect on the dissociation and elastic/inelastic scattering of diatom on and from metal surface, it is indeed necessary to put forward a self-consistent treatment by including first and second derivatives of the interaction potential in terms of surface vibrational modes. In the subsequent work, the interaction potential is expressed as functions of surface mode coordinates ($Q_k$) as follows:~\cite{TSS,TSS1,TSS2,TSSM,SAMARK,VV-Hartree-2021}

\begin{eqnarray}
	V_{I}^{vib}(\lbrace \chi _m\rbrace) = \sum_k \lambda_k V_{k,1}(\lbrace \chi _m\rbrace)\: Q_k + \frac{1}{2} \sum_{k l} \gamma_{k l} V_{k l,2}(\lbrace \chi _m\rbrace) \: Q_k Q_l +  \cdot \cdot \cdot,
	\label{Eq: Interaction_potential_in_normal_mode}
\end{eqnarray}

\noindent
where $V_{k,1} (\lbrace \chi _m\rbrace)$ and $V_{k l,2}(\lbrace \chi _m\rbrace)$ represents the first and second derivative of the interaction potential. On the other hand, the switching parameters, $\lambda_k$ and $\gamma_{kl}$, consider the sign of the first and second derivatives, respectively. Such derivatives represents the coupling elements between molecular DOFs-surface modes appearing during diatom-metal collision process. \\

\noindent
The coupling elements, $V_{k,1}(\lbrace \chi _m\rbrace)$ and $V_{kl,2}(\lbrace \chi _m\rbrace)$, are devised employing the chain rule of differentiation with respect to surface modes ($\lbrace Q_{k/l}\rbrace$) as follows: 

\begin{eqnarray}
	V_{k,1} (\lbrace \chi _m\rbrace)=\frac{\partial V^{vib}_I (\lbrace \chi _m\rbrace)}{\partial Q_k} 
	&=&\sum_{a\alpha}\dfrac{\partial(V_{a\alpha}^\text{Cu-H}(r_{a\alpha})-V_{a\alpha}^\text{Cu-H}(r_{a\alpha}^\text{eq}))}{\partial 
		Q_{k}} \nonumber\\
	&=&\sum_{a\alpha i}\left[\dfrac{\partial V_{a\alpha}^\text{Cu-H}(r_{a\alpha})}{\partial X_{\alpha i}}-\dfrac{\partial 
		V_{a\alpha}^\text{Cu-H}(r_{a\alpha}^\text{eq})}{\partial X_{\alpha i}}\right]\cdot
	\dfrac{\partial X_{\alpha i}}{\partial Q_{k}}
	\label{Eq:potfirst}
\end{eqnarray}

\begin{eqnarray}
	V_{k l, 2}(\lbrace \chi _m\rbrace) = \frac{\partial^2 V^{vib}_I(\lbrace \chi _m\rbrace)}{\partial Q_k \partial Q_l} = && \sum_{a\alpha i}\left[\dfrac{\partial V_{a\alpha}^\text{Cu-H}(r_{a\alpha})}{\partial X_{\alpha i}}-\dfrac{\partial 
		V_{a\alpha}^\text{Cu-H}(r_{a\alpha}^\text{eq})}{\partial X_{\alpha i}}\right]\cdot
	\dfrac{\partial^2 X_{\alpha i}}{\partial Q_{k}\partial Q_{l}} \nonumber \\
	&+& 
	\sum_{a\alpha \beta i j}\left[\dfrac{\partial^2 V_{a\alpha}^\text{Cu-H}(r_{a\alpha})}{\partial X_{\alpha i}\partial X_{\beta j}}-\dfrac{\partial^2 
		V_{a\alpha}^\text{Cu-H}(r_{a\alpha}^\text{eq})}{\partial X_{\alpha i}\partial X_{\beta j}}\right]\cdot
	\dfrac{\partial X_{\alpha i}}{\partial Q_{k}}\cdot \dfrac{\partial X_{\beta j}}{\partial Q_{l}},
	\label{Eq:second_derivative}
\end{eqnarray}

\noindent
where $r_{a\alpha}$ indicates the distance between the gas atom ($a$) and the metal atom ($\alpha$). In the ideal static lattice case, such distance is expressed by $r^{\mathrm{eq}}_{a\alpha} \left (= \sqrt{\sum_i (X_{ai} - X^{\mathrm{eq}}_{\alpha i})^2} \right)$, whereas at non-zero temperature, the same is written as $r_{a\alpha} \left (= \sqrt{\sum_i (X_{ai} - X_{\alpha i})^2}\right)$. In addition, $V^\text{Cu-H}_{a \alpha}$ shows the sum of two body interaction potential between each gas ($a$) - metal ($\alpha$) atom pair, which depends on the gas-metal atom distance ($r_{a \alpha}$).\\

\noindent
 The derivative of the potential, $V_{a\alpha}^\text{Cu-H}(r_{a\alpha}^\text{eq}$), with respect to the metal atom position ($X_{\alpha i}$ or $X_{\beta j}$) at the distance ($r^\text{eq}_{a\alpha}$) is zero and therefore, Equations~(\ref{Eq:potfirst}) and (\ref{Eq:second_derivative}) take the following form: 
\begin{eqnarray}
	V_{k,1} (\lbrace \chi _m\rbrace)=\sum_{a\alpha i}\left[\dfrac{\partial V_{a\alpha}^\text{Cu-H}(r_{a\alpha})}{\partial X_{\alpha i}}\right]\cdot
	\dfrac{\partial X_{\alpha i}}{\partial Q_{k}}
	\label{Eq:potfirst1}
\end{eqnarray}
\begin{eqnarray}
	V_{k l, 2}(\lbrace \chi _m\rbrace) = && \sum_{a\alpha i}\left[\dfrac{\partial V_{a\alpha}^\text{Cu-H}(r_{a\alpha})}{\partial X_{\alpha i}}\right]\cdot\dfrac{\partial^2 X_{\alpha i}}{\partial Q_{k}\partial Q_{l}} + \sum_{a\alpha \beta i j}\left[\dfrac{\partial^2 V_{a\alpha}^\text{Cu-H}(r_{a\alpha})}{\partial X_{\alpha i}\partial X_{\beta j}}\right]\cdot
	\dfrac{\partial X_{\alpha i}}{\partial Q_{k}}\cdot \dfrac{\partial X_{\beta j}}{\partial Q_{l}}
	\label{Eq: second_derivative1}
\end{eqnarray}
 
\noindent
While deriving $\frac{\partial X_{\alpha i}}{\partial Q_k}$  $  \left( \text{or} \ \frac{\partial X_{\beta j}}{\partial Q_l}\right)$  and $\frac{\partial^2 X_{\alpha i}}{\partial Q_k \partial Q_l}$, the following relationship between local and normal/phonon mode is used:
\begin{eqnarray}
	X_{\alpha i} - X_{\alpha i}^{eq} = \frac{1}{\sqrt {m_{\alpha}}} \sum_{k} T_{\alpha i; k} Q_k.
	\label{Eq: transformation_matrix_equation}
\end{eqnarray}

\noindent 
As a consequence, the first and cross derivative of the metal atom position ($X_{\alpha i}$) with respect to normal/phonon mode ($\lbrace Q_{k/l} \rbrace$) can be expressed as:
\begin{eqnarray}
	\frac{\partial X_{\alpha i}}{\partial Q_{k}} = m_{\alpha}^{-1/2} T_{\alpha i; k}, \quad \frac{\partial}{\partial Q_l} \left(\frac{\partial X_{\alpha i}}{\partial Q_k}\right) = 0 
	\label{Eq: dx_alpha_i/dQ_k/l11}
\end{eqnarray}

\noindent
which in turn, the first (Equation~(\ref{Eq:potfirst1})) and second (Equation~(\ref{Eq: second_derivative1})) derivatives of the interaction potential are arrived at: 
\begin{eqnarray}
	V_{k,1} (\lbrace \chi _m\rbrace)=\sum_{a\alpha i}\left[\dfrac{\partial V_{a\alpha}^\text{Cu-H}(r_{a\alpha})}{\partial X_{\alpha i}}\right]\cdot
	m_{\alpha}^{-1/2} T_{\alpha i; k},
	\label{Eq:potfirst12}
\end{eqnarray}
\begin{eqnarray}
V_{k l, 2}(\lbrace \chi _m\rbrace) = \sum_{a\alpha \beta i j}\left[\dfrac{\partial^2 V_{a\alpha}^\text{Cu-H}(r_{a\alpha})}{\partial X_{\alpha i}\partial X_{\beta j}}\right] \cdot {(m_{\alpha}m_{\beta})}^{-1/2} T_{\alpha i; k} \cdot T_{\beta j; l}.
\label{Eq: second_derivative5}
\end{eqnarray}

\noindent
The interaction potential (SCM)~\cite{MWMF} is used to compute Cu--H coupling at non-zero surface temperature. The indices $\alpha/ \beta (= 1, 2, \cdot \cdot \cdot, N_{\alpha/\beta})$, $a (= 1, 2)$ and $k/l (= 7, 8, \cdot \cdot \cdot, 3N)$ designate surface atom, gas atom and surface mode, respectively. $X_{ai}$ and $X_{\alpha i}$ signify the gas and metal atom coordinates for a particular direction, $i$, where $X_{\alpha i}^{eq}$ is associated with equilibrium position of surface atom. $m_{\alpha/\beta}$ represents mass of metal atom; $Q_{k/l}$ designates the normal/phonon mode ($k/l$); $T_{\alpha i/\beta j; k/l}$ is the transformation matrix for local ($\alpha i/\beta j$) to normal/phonon ($k/l$) mode of the surface.  \\

\noindent
While finding the expression of $\widehat {V}_{eff}^{vib}(\lbrace \chi_m \rbrace, t)$ (=$\langle \Phi_{vib} | \widehat {V}^{vib}_I(\lbrace \chi_m \rbrace) | \Phi_{vib} \rangle$) (see Equation~(\ref{eq:TDSE1})), the surface modes ($Q_{k/l}$s) are written in terms of bosonic annihilation ($b_{k/l}$)/creation ($b_{k/l}^{\dagger}$) operators such as $Q_{k/l} = A_{k/l} (b_{k/l} + b_{k/l}^{\dagger})$, where $A_{k/l} = \sqrt{\frac{\hbar}{2 \omega_{k/l}}}$ and [$b_k, b_{k}^{\dagger}$]=1. As a consequence, the interaction potential in Equation (\ref{Eq: Interaction_potential_in_normal_mode}) takes the following form:~\cite{VV-Hartree-2021}
\begin{eqnarray}
	\widehat V_{I}^{vib}(\lbrace \chi _m\rbrace) = \sum_k \lambda_k V_{k,1}(\lbrace \chi _m\rbrace) A_k (b_k  + b_k^{\dagger}) + \frac{1}{2} \sum_{k l} \gamma_{k l} A_k A_l V_{k l, 2}(\lbrace \chi _m\rbrace) \left( b_k^{\dagger} b_{l}^{\dagger} +  b_k^{\dagger} b_{l} + b_k b_{l}^{\dagger} + b_k b_{l}\right).
	\label{Eq: Interaction_potential_second_quantization}
\end{eqnarray}

\noindent
While applying $b_k$ and $b_k^{\dagger}$ on surface mode eigenfunction, one vibrational quantum can be annihilated or created and these operators are coined as vibration-translation (VT) operators, Such operators show the forcing generated by the translation motion of diatom. During the application of $b_k^{\dagger} b_{l}$ and $b_k b_{l}^{\dagger}$ on eigenstates, total number of vibrational quanta remains conserved and thereby, such operators are termed as vibration-vibration (VV) operators. Since simultaneous two quantum excitation or de-excitation processes are less probable except at the domain of high energy collision and very high surface temperature situations, two quanta operators ($b_k^{\dagger} b_{l}^{\dagger}$ and $b_k b_{l}$) have been dropped from Equation (\ref{Eq: Interaction_potential_second_quantization}) as follows:~\cite{VV-Hartree-2021}
\begin{eqnarray}
	\widehat V_{I}^{vib}(\lbrace \chi _m\rbrace) = \sum_k \lambda_k V_{k,1}(\lbrace \chi _m\rbrace) A_k (b_k  + b_k^{\dagger}) + \frac{1}{2} \sum_{k l} \gamma_{k l} A_k 
	A_l V_{k l, 2}(\lbrace \chi _m\rbrace) \left( b_k^{\dagger} b_{l} + b_k b_{l}^{\dagger}\right).
	\label{Eq: Interaction_potential_final}
\end{eqnarray}

\noindent
In presence of the linear and quadratic perturbation of molecular-surface interaction, evolution of surface modes eigenstates from an initial state ($\Phi_{vib}(t_0)$) at $t_0$ to arbitrary one ($\Phi_{vib}(t)$) at $t$ is governed by the evolution operator, $U(t, t_0)$ as:   

\begin{eqnarray}
	\Phi_{vib}(t)= U(t, t_0) \Phi_{vib}(t_0).
	\label{Eq: Initial_wavefunction_phonon}
\end{eqnarray}

\noindent
On the other hand, in the interaction picture, the evolution operator takes the following form:
\begin{eqnarray}
	U(t, t_0) = \exp \left( -i \int \frac{H^s_0 dt}{\hbar}  \right) U_{I} (t, t_0),
	\label{Eq: evolution_operator_interaction_picture}
\end{eqnarray}

\noindent
where $\widehat H_0^{s}$ is the Hamiltonian of the bare solid surface. \\

\noindent
Employing Equations (\ref{Eq: Initial_wavefunction_phonon}) and (\ref{Eq: evolution_operator_interaction_picture}), the corresponding TDSE for the surface modes is expressed as: 

\begin{eqnarray}
	i \hbar \dot{U_{I}} &=&  \Bigg[\exp \left( i \int \frac{H^s_0 dt}{\hbar}  \right) \widehat V_{I}^{vib}(\lbrace \chi _m\rbrace) \exp \left( -i \int \frac{H^s_0 dt}{\hbar}  \right)\Bigg] U_{I},\nonumber \vspace{0.6 cm} \\ 
	&=& V_{I}^{\prime}(\lbrace \chi _m\rbrace, t) U_{I},
	\label{Eq: TDSE_interaction}
\end{eqnarray}

\noindent
where $ \int H_0^{s} dt = \sum_{k} \hbar \theta_{k} \left(  b_k^{\dagger} b_k + \frac{1}{2}\right)$, $\theta_k = \int {\omega}_{k} dt$ and $\omega_{k}$s represent frequencies of surface modes.\\

\noindent
In other words, the molecule-surface interaction potential in the above Equation (\ref{Eq: TDSE_interaction}) is converted using Baker-Hausdorff formula under the interaction framework as depicted below:

\begin{eqnarray}
	V_{I}^{\prime}(\lbrace \chi _m\rbrace, t) &=& \sum_k \lambda_k V_{k,1}(\lbrace \chi _m\rbrace) A_k (b_k F_k^{-} + b_k^{\dagger} F_k^{+})  
	  +\frac{1}{2} \sum_{k\neq l} \gamma_{k l} A_k A_l V_{k l, 2}(\lbrace \chi _m\rbrace) \left( b_k^{\dagger} b_{l} F^{+}_{k l} + b_k b_{l}^{\dagger} F^{-}_{k l}\right) \nonumber \\
	&=& V^{VT}_I(\lbrace \chi _m\rbrace, t) +V^{VV}_I(\lbrace \chi _m\rbrace, t)
	\label{Eq: Interaction_potential_interaction_picture_I}
\end{eqnarray}

\noindent
where $F^{+}_{k} = \exp \left( i \theta_k \right)$, $F_{k}^{-} = \left( F^{+}_{k} \right)^{*}$ and $F^{+}_{k l} =  \exp \left[ i (\theta_k - \theta_l) \right] $, $F^{-}_{k l} = \left( F^{+}_{k l} \right)^{*}$ 
represent the time dependent modulatory terms appeared in interaction picture accompanied with VT opertors 
($b_k / b_k^{\dagger}$) and VV opertors ($b_k^{\dagger} b_{l} / b_k b_{l}^{\dagger}$ ), respectively.\\

\noindent
In a similar way, in the interaction picture (Equation (\ref{Eq: Interaction_potential_interaction_picture_I})), the total evolution operator for the surface mode vibrations is denoted as the product of VT and VV components:

\begin{eqnarray}
	U_{I} = U_{VV} U_{VT}.
	\label{Eq: Evolution_VT_VV}
\end{eqnarray}

\noindent
As a consequence, the TDSEs for the collection of surface modes vibrations (i.e., bath) can be segregated into:~\cite{VV-Hartree-2021}

\vspace{-0.6 cm}

\begin{eqnarray}
	i \hbar \frac{d U_{VT}}{dt} = \left[ U_{VV}^{\dagger} V^{VT}_{I}(\lbrace \chi _m\rbrace, t) U_{VV}\right] U_{VT},  \: \: \text{for VT process}
	\label{Eq: VT_TDSE_m}
\end{eqnarray}

\noindent
and

\vspace{-0.6 cm}

\begin{eqnarray}
	i \hbar \frac{d U_{VV}}{dt} = V^{VV}_{I}(\lbrace \chi _m\rbrace, t) U_{VV}, \: \: \: \text{for VV process}.
	\label{Eq: VV_TDSE_m}
\end{eqnarray}

\noindent
In absence of VV process, Equation (\ref{Eq: VT_TDSE_m}) takes the following modified form:

\begin{eqnarray}
	i \hbar \frac{d U_{VT}}{dt} =  V^{VT}_{I}(\lbrace \chi _m\rbrace, t) U_{VT},  \: \: \text{for VT process}
	\label{Eq: VT_TDSE_m11}
\end{eqnarray}

\subsubsection{10.2.1.4. Time- and temperature-dependent effective Hartree potential} \label{ssec:EHP} 

\noindent
For surface modes, the time- and temperature-dependent effective Hartree potential (EHP), averaged over the initial phonon-state distribution $\lbrace n_0 \rbrace$, is defined as:~\cite{TSS,TSS1,TSS2,TSSM,SM-JTCC,SM-MOLPHYS,SAMARK,VV-Hartree-2021}

\vspace{-0.6 cm}

\begin{eqnarray}
	V^{vib}_{eff} (\lbrace \chi _m\rbrace,t,T_s) = \sum_{\lbrace n_0 \rbrace} p_{\lbrace n_0 \rbrace} (T_s) \ \langle V^{\prime}_I(\lbrace \chi _m\rbrace,t) \rangle_{\lbrace n_0 \rbrace}, 
	\label{Eq: Effective_Hartree_potential_I+II}
\end{eqnarray}

\noindent
where $p_{\lbrace n_0 \rbrace}(T_s)$  is the probability factor for a set of surface mode vibrations ($\lbrace k \rbrace$): 
\begin{eqnarray}
	p_{\lbrace n_0 \rbrace} = \prod_{k=1}^{M} p_{n_k^0}^{(k)}.
	\label{Eq: probability_initial_state}
\end{eqnarray}

\noindent
The normalized probability, $p_{n_k^0}^{(k)}$ for a specific occupation number ($n_k^0$) of a particular surface mode ($\omega_k$) can be defined either in terms of BE or MB probability distribution as:

\vspace{-0.6 cm}

\begin{eqnarray}
	p_{n_k^0}^{(k)} = \frac{\tilde{p}_{n_k^0}^{(k)}}{N_{\text{BE/MB}}}, \: \: \text{and}  \: \: \: \quad
	\sum_k \sum_{n_k^0 =0}^{\infty} p_{n_k^0}^{(k)} =1,
	\label{Eq:norm_prob}
\end{eqnarray}

\noindent
where the normalization factor for a specific distribution (BE/MB) has been denoted by $N_{\text{BE/MB}}$.

\vspace{0.2 cm}

\noindent
The BE probability factor, $\tilde{p}_{n_k^0}^{(k)}$ of a particular surface mode ($k$) with  quantum state, $n_k^0$ is described as:

\vspace{-0.4 cm}

\begin{eqnarray}
	\tilde{p}_{n_k^0}^{(k)} &\propto& \frac{1}{\exp \left[ \hbar \omega_k \left( n_k^0 + {\color {black} \frac{1}{2}} \right) \beta \right] - 1 } \propto \sum_{q=1}^{\infty} \left( z_k^{n_k^0} \right)^q {\color {black} \left( z_k \right)^{\frac{q}{2}}} 
	\label{Eq: BEP_new_factor_app}
\end{eqnarray}

\noindent
and the MB probability factor can be written as:
\begin{eqnarray}
	\tilde{p}_{n_k^0}^{(k)} &\propto &   \exp \left[ - \hbar \omega_k \left( n_k^0 + {\color {black} \frac{1}{2}} \right) \beta\right] \propto z_k^{n_k^0} {\color {black} \cdot z_k^{\frac{1}{2}}},
	\label{Eq: MBP_new_factor_app}
\end{eqnarray}

\noindent	
where $z_k = \exp \left(- \beta \hbar \omega_k \right)$ and $\beta = \frac{1}{k_B T_s}$. The set of surface modes frequencies ($\lbrace \omega_k \rbrace$) can be computed by diagonalizing the force constant (Hessian) matrix, which is obtained from metal-metal interaction potential.

\vspace{0.2 cm}

\noindent
 The normalization factors for BE and MB probability distribution are:

\begin{footnotesize}
\begin{eqnarray}
	N_{\text{BE}} = \sum_k \sum_{n_k^0 =0}^{\infty} \frac{1}{\exp \left[ \hbar \omega_k \left( n_k^0 + \frac{1}{2} \right) \beta \right] 
		- 1 } &=& \sum_k \sum_{n_k^0 =0}^{\infty}  \sum_{q=1}^{\infty} \left( z_k^{n_k^0} \right)^q \left( z_k \right)^{\frac{q}{2}} = \sum_k \sum_{q=1}^{\infty} \frac{\left( z_k \right)^{\frac{q}{2}}}{\left( 1 - z_k^q \right)}\nonumber \\
     \label{Eq: final_form_of_norm_BE}
\end{eqnarray}
\end{footnotesize}

\vspace*{-1 cm}

\noindent
and

\vspace*{-1 cm}

\begin{small}
\begin{eqnarray}
N_{\text{MB}} = \sum_k \sum_{n_k^0 =0}^{\infty}  \exp \left[ - \hbar \omega_k \left( n_k^0 +  \frac{1}{2} \right) \beta\right] &=& \sum_k \sum_{n_k^0 =0}^{\infty} z_k^{n_k^0}  \cdot z_k^{\frac{1}{2}} = \sum_k \frac{z_k^{\frac{1}{2}}}{\left( 1 - z_k\right)}.
\label{Eq: final_form_of_norm_MB}
\end{eqnarray}
\end{small}

\noindent
The explicit initial ($\lbrace n_0 \rbrace$) state-dependent Hartee potential emerging from interactions ($V_{I}^{vib}(\lbrace \chi _m\rbrace$) among molecular DOFs and surface mode vibrations is described as (see Equations~ (\ref{eq:TDSE1}) and (\ref{Eq: TDSE_interaction})):

\vspace{-0.6 cm}

\begin{eqnarray}
\langle V_{I}^{\prime}(\lbrace \chi _m\rbrace,t)\rangle_{\lbrace n_0 \rbrace} = \langle \Phi_{vib} \vert V_{I}^{\prime}(\lbrace \chi _m\rbrace,t) \vert \Phi_{vib} \rangle.
\label{Eq: state_dependent_Hartree_potential}
\end{eqnarray}

\noindent
Again, the time- and temperature- dependent EHP can be decomposed into linear VT ($V_{eff}^{VT} (\lbrace \chi _m\rbrace, t, T_s)$)~\cite{TSS,TSS1,TSS2,TSSM,SM-JTCC,SM-MOLPHYS,SAMARK} and quadratic VV ($V_{eff}^{VV} (\lbrace \chi _m\rbrace, t, T_s)$)~\cite{VV-Hartree-2021}  parts as by using Equations (\ref{Eq: Interaction_potential_interaction_picture_I}), (\ref{Eq: Effective_Hartree_potential_I+II}) and (\ref{Eq: state_dependent_Hartree_potential}):

\vspace{-0.6 cm}

\begin{eqnarray}
V_{eff} (\lbrace \chi _m\rbrace, t, T_s)   
&=& \sum_{\lbrace n_0 \rbrace} p_{\lbrace n_0 \rbrace}(T_s) \langle V^{VT}_I (\lbrace \chi _m\rbrace,t)\rangle_{\lbrace n_0 \rbrace} + 
\sum_{\lbrace n_0 \rbrace} p_{\lbrace n_0 \rbrace} (T_s) \langle V^{VV}_I(\lbrace \chi _m\rbrace,t) \rangle_{\lbrace n_0 \rbrace} \nonumber 
\\
&=&V_{eff}^{VT} (\lbrace \chi _m\rbrace, t, T_s)+V_{eff}^{VV} (\lbrace \chi _m\rbrace, t, T_s). 
\label{Eq: Effective_Hartree_potential_VT_VV} 
\end{eqnarray}

\subsubsection{10.2.1.5. EHP for linear (VT) molecular DOF–surface vibrational mode coupling } \label{sssec:VT_EHP} 

\noindent
The initial state $\lbrace n_0 \rbrace$  dependent Hartree potential appearing due to linear (VT) coupling between diatomic DOFs and surface modes can be represented as \cite{TSS,TSS1,TSSM,SM-MOLPHYS,SMIRPC,SAMARK}:

\begin{eqnarray}
	 \langle V^{VT}_I (\lbrace \chi _m\rbrace,t)\rangle_{\lbrace n_0 \rbrace}&=& \langle \Phi_{vib} (t)\vert V_{I}^{VT}(\lbrace \chi _m\rbrace,t) \vert 
	\Phi_{vib} (t)\rangle ,  \nonumber \\
	&=& \sum_{\lbrace n^{\prime} \rbrace} \sum_{\lbrace n \rbrace} \alpha_{\lbrace n^{\prime} \rbrace 
		\leftarrow  \lbrace n_0 \rbrace }^{*} (t) \alpha_{\lbrace n \rbrace \leftarrow  \lbrace n_0 \rbrace } 
	(t) \langle  \lbrace n^{\prime} \rbrace \vert 
	V_{I}^{VT} \vert \lbrace n \rbrace \rangle,
	\label{Eq: VT_state_dependent_effective_potential2}
\end{eqnarray}  

\noindent
where the amplitude $\alpha_{\lbrace n \rbrace} (t)$ of a particular state ($\lbrace n \rbrace$) generating from the initial one ($\lbrace n_0 \rbrace$) under VT evolution operator ($U_{VT}$) takes the following form:

\begin{eqnarray}
	\alpha_{\lbrace n \rbrace \leftarrow  \lbrace n_0 \rbrace } (t) = \langle \lbrace n \rbrace \vert U_{VT} \vert \lbrace n_0 \rbrace \rangle .
	\label{Eq: amplites_VT} 
\end{eqnarray}

\noindent
The interaction potential associated with VT molecular DOFs-surface modes coupling can be expressed as \cite{SAMARK}:

\vspace{-0.6 cm}

\begin{eqnarray}
	V_{I}^{VT}(\lbrace \chi _m\rbrace,t) &=&  \sum_k \lambda_k  A_k (b_k F_k^{-}  + b_k^{\dagger} F_k^{+}) V_{k,1}(\lbrace \chi _m \rbrace), 
	\label{Eq: Interaction_potential_VT_I}
\end{eqnarray}

\noindent
where $F_k^{+} = \exp (i \omega_k t ) $,  $F_k^{-} = \left( F_k^{+} \right)^{*} $
and $V_{k,1}(\lbrace \chi _m \rbrace) = \frac{\partial V_I^{vib}(\lbrace \chi _m \rbrace)}{\partial Q_k}  \vert_{eq}$. The first derivative of interaction potential $V_{k,1}(\lbrace \chi _m \rbrace)$ (see Equation~(\ref{Eq: Interaction_potential_VT_I})) with respect to surface modes can be computed with the help of the SCM potential~\cite{MWMF}.\\

\noindent
The state-averaged EHP arising from such VT processes can be expressed as:

\begin{eqnarray}
	V_{eff}^{VT} (\lbrace \chi _m\rbrace, t, T_s) = \sum_{\lbrace n_0 \rbrace} p_{\lbrace n_0 \rbrace}(T_s) \langle V^{VT}_I (\lbrace \chi _m\rbrace,t)\rangle_{\lbrace n_0 \rbrace} ,
	\label{Eq: state_averaged_Hartree_potential_VT}
\end{eqnarray}

\noindent
where the distribution ($p_{\lbrace n_0 \rbrace}$) will be of either BE or MB type (see Equations (\ref{Eq:norm_prob}), (\ref{Eq: BEP_new_factor_app}) and (\ref{Eq: MBP_new_factor_app}) in Sec. \ref{ssec:EHP}). While plugging Equation (\ref{Eq: Interaction_potential_VT_I}) in Equation (\ref{Eq: VT_state_dependent_effective_potential2}), and further in Equation (\ref{Eq: state_averaged_Hartree_potential_VT}) for BE and MB situation, the EHP appears as~\cite{SAMARK}:

\begin{footnotesize}
	\begin{eqnarray}
		V_{eff}^{VT} (\lbrace \chi _m\rbrace, t, T_s)  &=& \sum_k   \lambda_k   \sum_{n^0_k} \sum_{n_k}\ \ A_{k} \: p_{n_k^0}  V_{k,1} (\lbrace \chi _m\rbrace) 
		[n_k^{1/2}\alpha^{\ast(k)}_{n_{k-1} \leftarrow n_k^0  }(t) F_k^- + (n_k + 1)^{1/2} F_k^+ \alpha^{\ast(k)}_{n_{k+1} \leftarrow 
			n_k^0 }(t)] \alpha_{n_k \leftarrow n_k^0 }^{(k)}(t)\nonumber \\
		& =& \frac{1}{N_{\text{BE/MB}}}\sum_k   \lambda_k   \sum_{n^0_k} \sum_{n_k}\   \tilde{p}_{n_k^0} \ A_k\ V_{k,1}(\lbrace \chi _m\rbrace) 
		[n_k^{1/2}\alpha^{\ast(k)}_{n_{k-1} \leftarrow n_k^0  }(t) F_k^- + (n_k + 1)^{1/2} F_k^+ \alpha^{\ast(k)}_{n_{k+1} \leftarrow 
			n_k^0 }(t)] \alpha_{n_k \leftarrow n_k^0 }^{(k)}(t). \nonumber \\ 
		\label{Eq: VT_effective_I1}
	\end{eqnarray}
\end{footnotesize}

\noindent
While achieving Equation (\ref{Eq: VT_effective_I1}), one can use the following expression:
\begin{eqnarray}
	\alpha_{\{n\}}(t)=\prod_{k=1}^M \alpha_{n_k}^{(k)}(t) \: \text{and} \: \sum_{n_k}|\alpha_{n_k}^{(k)}(t)|^2=1,
	\label{Eq: alpha_prod1}
\end{eqnarray}

\noindent
where $\alpha_{n_k}^{(k)}(t)$ represents the exact solution of the Linearly Forced Harmonic Oscillator (LFHO) \cite{PPJS} indicating the amplitude for the $n_k$th quantum state of the VT coupled mode, $k$. \\

\noindent
Finally, while replacing the analytic expressions of $\alpha_{n_k}^{(k)}(t)$ in Equation (\ref{Eq: VT_effective_I1}), and using several algebric operations, we obtain the final expression of 
time- and temperature- dependent 6D EHP [$V_{eff}^{VT} (\lbrace \chi_m \rbrace, t, T_s)$] originating from linear VT molecular DOFs-surface modes coupling (For details, see Appendix \ref{app:BB}):

\begin{eqnarray}
	V_{eff, \: \text{BE}}^{VT} (\lbrace \chi_m \rbrace, t, T_s) = \frac{1}
	{N_{\text{BE}}} \sum_k \omega_k^{-1} \epsilon_k (\lbrace \chi_m \rbrace, t)  \sum^{\infty}_{q=1}\frac{z_k^{q/2}}{(1-z^q_k)}, 
	\hspace{1cm} \text{for BE case} 
	\label{Eq: V_eff_VT_BEP}
\end{eqnarray}

\noindent
and
\begin{eqnarray}
	V_{eff, \: \text{MB}}^{VT} (\lbrace \chi_m \rbrace, t, T_s) =  \frac{1}
	{N_{\text{MB}}} \sum_k \omega_k^{-1} \epsilon_k (\lbrace  \chi_m \rbrace, t)  \frac{z_k^{1/2}}{(1- z_k)}, \hspace{1cm} \text{for MB 
		case}
	\label{Eq: V_eff_VT_MBP}
\end{eqnarray}

\noindent
where $N_{\text{BE}}$ $\left(= \sum_k \sum_{q=1}^{\infty} \frac{\left( z_k \right)^{\frac{q}{2}}}{\left( 1 - z_k^q \right)}\right)$, $N_{\text{MB}}$ $\left(= \sum_k \frac{z_k^{\frac{1}{2}}}{\left( 1 - z_k\right)}\right)$ denote the normalization factors of BE and MB probability distribution (see Equations (\ref{Eq: final_form_of_norm_BE}) and (\ref{Eq: final_form_of_norm_MB}) in Sec.~\ref{ssec:EHP}), respectively. On the other hand, the explicit time dependent quantity, $\epsilon_k (\lbrace \chi_m \rbrace, t)$ can be written as\cite{SAMARK}:
\begin{eqnarray}
	\epsilon_k(\lbrace \chi_m \rbrace, t) = \lambda_k V_{k,1} (\lbrace \chi_m \rbrace) \int_{t_0}^{t} dt^{\prime} V_{k,1} (\lbrace \chi_m \rbrace) \sin \left[ \omega_k \left( t^{\prime} 
	- t \right) \right].
	\label{Eq: epsilon1}
\end{eqnarray} 

\noindent
The essential features of the EHP ($V_{eff, \: \text{BE/MB}}^{VT}$) for linear VT coupling model can be summarized as follows:

\begin{itemize}
	\item [(i)]  The temperature dependent term in EHP takes different forms for BE $\left(\sum_{q=1}^{\infty} \frac{\left( z_k \right)^{\frac{q}{2}}}{\left( 1 - z_k^q \right)}\right)$ and MB $\left( \frac{z_k^{\frac{1}{2}}}{\left( 1 - z_k\right)}\right)$ cases. As a consequence, these two different factors would modulate the scattering outcomes at a specific surface temperature in a different manner. In addition, the frequency spectrum ($\lbrace\omega_k\rbrace$) of the surface mode vibrations could also play a dictating role on the nature of reaction probability.
	
	\item [(ii)] The explicit time-dependent quantity $\epsilon_k (\lbrace \chi_m \rbrace, t) $, corresponding to the $k^{\text{th}}$ surface vibrational mode of frequency $\omega_k$ governs the importance of that mode to the EHP throughout the collision process. 
\end{itemize}

\subsubsection{10.2.1.6. EHP for bi-linear (VV) molecular DOF–surface vibrational mode coupling} \label{ssub:VV-EHP} 

\noindent
The EHP is originating from higher-order anharmonic (known as VV) coupling among molecular DOFs and surface modes, which can be averaged over initial surface vibrational states ($\lbrace n_0 \rbrace$) as follows:

\begin{eqnarray}
	V_{eff}^{VV} (\lbrace \chi_m \rbrace, t, T_s) =  \sum_{\lbrace n_0 \rbrace} p_{\lbrace n_0 \rbrace}(T_s) \langle V^{VV}_I(\lbrace \chi _m\rbrace,t) \rangle_{\lbrace n_0 \rbrace}, 
	\label{Eq: state_averaged_effective_Hartree_potential_VV}
\end{eqnarray}

\noindent
where the averaging coefficient, $p_{\lbrace n_0 \rbrace}(T_s)$ represents the product form of BE or MB probability factors (see Equations (\ref{Eq: probability_initial_state}) and (\ref{Eq:norm_prob}) of Sec.~\ref{ssec:EHP}) over various quantum states ($\lbrace n_0 \rbrace$) of surface modes ($k$). The state-dependent Hartree potential $\langle V^{VV}_I(\lbrace \chi _m\rbrace,t) \rangle_{\lbrace n_0 \rbrace}$ is depicted as functions of the evolution operator ($ U_{I} = U_{VV} U_{VT}$) propagating the initial surface-mode eigenfunction $\Phi_{vib} (t_0)$ as follows:   

\begin{eqnarray}
	\langle V^{VV}_I (\lbrace \chi _m\rbrace,t)\rangle_{\lbrace n_0 \rbrace} &=& \langle \Phi_{vib} (t)\vert V_I^{VV}(\lbrace \chi _m \rbrace,t) \vert 
	\Phi_{vib} (t) \rangle \nonumber \\
	&=&  \langle \Phi_{vib} (t_0) \vert U_{VT}^{\dagger}  U^{\dagger}_{VV} V_I^{VV}(\lbrace \chi _m \rbrace,t) \ U_{VV}  
	U_{VT} \vert \Phi_{vib} (t_0) \rangle.
	\label{Eq: state_dependent_Hartree_potential_VV}
\end{eqnarray}

\noindent
Within the framework of second-quantization, the bi-linear (anharmonic) potential $V_I^{VV} (\lbrace \chi _m \rbrace,t)$ appears as the following expression in the interaction picture (see Equation (\ref{Eq: Interaction_potential_interaction_picture_I})):

\begin{eqnarray}
	V_I^{VV}(\lbrace \chi _m \rbrace,t) &=& \frac{1}{2}\sum_{k \neq l}^{M} \gamma_{k l} A_k A_l V_{k l, 2} (\lbrace \chi _m \rbrace) \left( b_k^{\dagger} b_{l} 
	F^{+}_{k l} + b_k b_{l}^{\dagger} F^{-}_{k  l}\right), \nonumber \\  
   \label{Eq: Interaction_potential_for_VV_process1}
   \end{eqnarray}
   
   \noindent
   where $F^{+}_{k l} = \exp [i (\theta_k - \theta_l) ]$, $F_{k l}^{-} = \left( F^{+}_{k l} \right)^{*} $ and $\theta_{k/l} (= \int \omega_{k/l} dt)$ symbolizes the phase originating from the modulation of surface mode frequency spectrum. \\

\noindent
$M(M-1)$ number of VV operators, $b_k b_{l}^{\dagger}/b_k^{\dagger} b_{l}$ in Equation (\ref{Eq: Interaction_potential_for_VV_process1}) constitute closed set of $M^2$ number of such operators with respect to their commutation (see Appendix \ref{app:CC} and \ref{app:CC_1}), and therefore, an operator algebra method is introduced to derive the expression of state-dependent Hartree potential. In such formalism, VV coupling is considered through the time evolution of bi-linear operators, where the action of $b_k$ and $b_k^{\dagger}$ on the evolution ($U_{VT}$ and $U_{VV}$) operators are explored by employing closed commutation relationships within the set of VV operators, $\lbrace b_k b_{l}^{\dagger}/b_k^{\dagger} b_{l} \rbrace$.~\cite{VV-Hartree-2021} \\   

\noindent
The TDSE of surface vibrational state ($\Phi_{vib} (t) $) for VV process can be written as follows (see Equation~\ref{Eq: VV_TDSE_m} in Sec.~\ref{ssec:Interaction-potential}):

\begin{eqnarray}
	i \hbar \frac{d U_{VV}}{dt} = V^{VV}_{I}(\lbrace \chi _m \rbrace,t)\: U_{VV} 
	= \frac{1}{2} \sum_{k l} \gamma_{k l} A_k A_l V_{k l, 2}(\lbrace \chi _m \rbrace) \left( b_k^{\dagger} b_{l} F^{+}_{k l} + b_k b_{l}^{\dagger} F^{-}_{k l}\right) U_{VV}.
	\label{Eq: TDSE_VV_process}
\end{eqnarray}  

\noindent
The solution, $U_{VV}$ of Equation (\ref{Eq: TDSE_VV_process}) can be expanded in terms of the matrix notation of the evolution operators (see Appendix \ref{app:CC} and \ref{app:CC_1}) as below:

\begin{eqnarray}
	R_{U_{VV}} = \prod_{k l}^{M} \exp \left[ \alpha_{k l} (t, t_0)\: \delta_{k l}^{M \times M} \right],
	\label{Eq: R_UVV}
\end{eqnarray}

\noindent
where $\alpha_{k l}$ and $\delta_{k l}^{M \times M}$ represents the amplitudes of bi-linearly perturbed harmonic oscillators and matrix ($M \times M$) representation of $b_k^{\dagger} b_{l}/b_k b_{l}^{\dagger}$ operators, repectively (See Appendix \ref{app:CC_1}). Furthermore, the matrix representation of evolution operator ($R_{U_{VV}}$) can be computed by solving the TDSE (see Equation (\ref{Eq: TDSE_VV_process})) for VV process,

\begin{eqnarray}
	i \hbar \frac{d }{dt} R_{U_{VV}}(t, t_0) = V_{I}^{VV}(\lbrace \chi _m \rbrace,t)\: R_{U_{VV}} (t_0, t_0),
	\label{Eq: dRdt}
\end{eqnarray}

\noindent
where $R_{U_{VV}}(t_0, t_0) = I$, represents the unit matrix.\\

\noindent
In order to compute EHP, the following operator algebric relationships, namely, the transformation of bosonic operators under the evolution operator ($U_I = U_{VV} U_{VT}$) involving linear and bi-linear couplings among molecular DOFs and surface modes, are formulated as (see Appendix \ref{app:CC}):

\vspace{-1.0 cm}

\begin{subequations}
	\begin{align}
		U_{I}^{\dagger} b_k U_{I} = (b_k + i \xi_{k}^{+}) + \sum_{j=1}^{M} Q_{kj}^{VV} (b_j + i \xi_{j}^{+})
	\end{align}
	\begin{align}
		U_{I}^{\dagger} b_k^{\dagger} U_{I} = (b_k^{\dagger} - i \xi_{k}^{-}) + \sum_{j=1}^{M} Q_{kj}^{VV^{*}} 
		(b_j^{\dagger} - i \xi_{j}^{-}),
	\end{align}
	\label{Eq: evolution of bosonic operators}
\end{subequations}

\noindent
where $\xi_{k/j}^{\pm}$ and $Q_{kj}^{VV}$ signify the oscillatory terms associated with modified VT coupled vibrational state ($k/j$) possessing VV couplings. In addition, the matrix element, $T_{Q^{VV}}$ ($\equiv\lbrace Q_{kj}^{VV} \rbrace$) takes the following form (see Equation~\ref{Eq: Q_VV_coupling matrix1} in Appendix \ref{app:CC}):

\begin{eqnarray}
	T_{Q^{VV}} = R_{U_{VV}} - I.
	\label{Eq: Q_VV_coupling matrix}
\end{eqnarray}   

\noindent
Once the Equation~(\ref{Eq: dRdt}) is numerically solved as function of time, time-dependent elements ($\lbrace Q_{kj}^{VV} \rbrace$) are calculated with the help of Equation~(\ref{Eq: Q_VV_coupling matrix}). Then, those elements are further employed in EHP (see Equation~(\ref{Eq: state_averaged_effective_Hartree_potential_VV_BE}) or Equation~(\ref{Eq: state_averaged_effective_Hartree_potential_VV_MB})).\\

\noindent
While plugging Equations (\ref{Eq: evolution of bosonic operators}a) and (\ref{Eq: evolution of bosonic operators}b), one can compute the state-dependent EHP for bi-linear couplings between molecular DOFs and surface modes (see Appendix \ref{app:CC}). The effects of such state dependency is averaged out by taking the BE and MB probability distribution in account (Equations (\ref{Eq: probability_initial_state}) - (\ref{Eq: MBP_new_factor_app}) in Sec.~\ref{ssec:EHP}). As a result, the expression for time- and temperature- dependent 6D EHP [$V_{eff}^{VV} (\lbrace \chi_m \rbrace, t, T_s)$] is converted into the following form under VV coupling scheme:~\cite{VV-Hartree-2021}


\begin{eqnarray}
	V_{eff, BE}^{VV} (\lbrace \chi_m \rbrace, t, T_s) &=&   \sum_{\lbrace n_0 \rbrace} p_{\lbrace n_0 \rbrace}(T_s) \langle V^{VV}_I(\lbrace \chi _m\rbrace,t) \rangle_{\lbrace n_0 \rbrace} \nonumber \\
	&=& \frac{1}{2} \frac{1}{N_{\text{BE}}} \sum_{k \neq l} \gamma_{k l} V_{kl, \:2} (\lbrace \chi _m \rbrace) A_k A_l \Bigg\lbrack F_{kl}^+ \Bigg(Q_{l k}^{VV} \sum_{q} \frac{z_k^{\frac{3}{2}q}}{(1-z_k^q)^2}  + Q_{k l}^{VV^{*}} \sum_{q} \frac{z_{l}^{\frac{3}{2}q}}{(1-z_{l}^q)^2}  \nonumber \\
	&+& \sum_{j} Q^{VV^{*}}_{kj} Q_{l j}^{VV} \sum_{q} \frac{z_j^{\frac{3}{2}q}}{(1-z_j^q)^2} + \xi_{k}^-  \: \xi_{l}^+ 
	+ \xi_{k}^- \sum_j Q_{l j}^{VV} \xi^+_{j} + \xi_{l}^+ \sum_j Q_{k j}^{VV^{*}} \xi^-_{j}  \nonumber \\
	&+& \sum_{jj^{\prime}} Q_{kj}^{VV^{*}} Q_{l j^{\prime}}^{VV} \xi_{j}^- \: \xi_{j^{\prime}}^+\Bigg)
	+ F_{k l}^- \Bigg( Q^{VV^{*}}_{lk} \Bigg\lbrace \sum_{q} \frac{z_{k}^{\frac{3}{2}q}}{(1-z_{k}^q)^2} +1  \Bigg\rbrace  +   Q^{VV}_{kl} \Bigg\lbrace \sum_{q} \frac{z_{l}^{\frac{3}{2}q}}{(1-z_{l}^q)^2} +1 \Bigg\rbrace \nonumber \\
	&+& \sum_j Q_{kj}^{VV} Q^{VV^{*}}_{l j} \sum_{q} \Bigg\lbrace \frac{z_j^{\frac{3}{2}q}}{(1-z_j^q)^2}+1 \Bigg\rbrace + \xi_{k}^+ \: \xi_{l}^- +  \xi_{k}^+ \sum_j Q^{VV^{*}}_{l j} \xi_{j}^-   \nonumber \\
	&+& \xi_{l}^- \sum_j Q^{VV}_{k j} \xi_{j}^+ + \sum_{j j^{\prime}} 
	Q_{kj}^{VV} Q^{VV^{*}}_{l j^{\prime}} \xi^+_{j} \: \xi^-_{j^{\prime}} \Bigg)  \Bigg\rbrack, \qquad \: \text{for} \: \: \text{BE case} 
	\label{Eq: state_averaged_effective_Hartree_potential_VV_BE}
\end{eqnarray}

\noindent
and

\begin{eqnarray}
	V_{eff, MB}^{VV} (\lbrace \chi_m \rbrace, t, T_s) &=&  \sum_{\lbrace n_0 \rbrace} p_{\lbrace n_0 \rbrace}(T_s) \langle V^{VV}_I(\lbrace \chi _m\rbrace,t) \rangle_{\lbrace n_0 \rbrace} \nonumber \\
	&=& \frac{1}{2} \frac{1}{N_{\text{MB}}} \sum_{k \neq l} \gamma_{k l} V_{kl, \: 2}(\lbrace \chi _m \rbrace) A_k A_{l}
	\Bigg[ F_{kl}^+ \Bigg(Q_{l k}^{VV}  \frac{z_k^{\frac{3}{2}}}{(1-z_k)^2}  + Q_{k l}^{VV^{*}}  \frac{z_{l}^{\frac{3}{2}}}{(1-z_{l})^2}  \nonumber \\
	&+& \sum_{j} Q^{VV^{*}}_{kj} Q_{l j}^{VV}  \frac{z_j^{\frac{3}{2}}}{(1-z_j)^2} + \xi_{k}^- \: \xi_{l}^+ 
	+ \xi_{k}^- \sum_j Q_{l j}^{VV} \xi^+_{j} + \xi_{l}^+ \sum_j Q_{k j}^{VV^{*}} \xi^-_{j}  \nonumber \\
	&+& \sum_{jj^{\prime}} Q_{kj}^{VV^{*}} Q_{l j^{\prime}}^{VV} \xi_{j}^- \: \xi_{j^{\prime}}^+\Bigg) 
	+ F_{k l}^- \Bigg(Q^{VV^{*}}_{lk} \Bigg\lbrace \frac{z_{k}^{\frac{3}{2}}}{(1-z_{k})^2} +1  \Bigg\rbrace  +  Q^{VV}_{kl} \Bigg\lbrace  \frac{z_{l}^{\frac{3}{2}}}{(1-z_{l})^2} +1 \Bigg\rbrace \nonumber \\
	&+& \sum_j Q_{kj}^{VV} Q^{VV^{*}}_{l j} \Bigg\lbrace\frac{z_j^{\frac{3}{2}}}{(1-z_j)^2} + 1 \Bigg\rbrace 
	+ \xi_{k}^+ \: \xi_{l}^- +  \xi_{k}^+ \sum_j Q^{VV^{*}}_{l j} \xi_{j}^-   \nonumber \\
	&+& \xi_{l}^- \sum_j Q^{VV}_{k j} \xi_{j}^+ + \sum_{j j^{\prime}} 
	Q_{kj}^{VV} Q^{VV^{*}}_{l j^{\prime}} \xi^+_{j} \: \xi^-_{j^{\prime}} \Bigg)   \Bigg], \: \: \qquad \text{for} \: \: \text{MB case} 
	\label{Eq: state_averaged_effective_Hartree_potential_VV_MB}
\end{eqnarray}

\noindent
where the normalization factors for BE and MB probability distribution are represented as $N_{\text{BE}}$ $(= \sum_k \sum_{q=1}^{\infty} \frac{\left( z_k \right)^{\frac{q}{2}}}{\left( 1 - z_k^q \right)})$ and $N_{\text{MB}}$ $(= \sum_k \frac{z_k^{\frac{1}{2}}}{\left( 1 - z_k\right)})$, respectively (see Equations (\ref{Eq: final_form_of_norm_BE}) and (\ref{Eq: final_form_of_norm_MB}) in Sec.~\ref{ssec:EHP}). \\

\noindent
It is important to mention that bi-linearly (VV) coupled EHP obeys the following characteristics:

\begin{itemize}
	\item [(i)] Non-negligible effect on the reaction probability can be arisen from the second derivative  ($V_{kl, \:2}(\lbrace \chi _m \rbrace)$) of the interaction potential. Moreover, the bosonic phase factors, $F_{kl}^{\pm}$ and bi-linear evolution matrix elements, $Q_{kl}^{VV}$ associated with the VV interaction between different vibrational states ($k$ and $l$) changes the EHP as a function of time during the collision process. 

	\item [(ii)] Incorporation of BE and MB distribution functions produces the EHP as a temperature dependent quantity with the functional form, $\sum_{q=1}^{\infty} \frac{\left( z_k \right)^{\frac{3}{2}q}}{\left( 1 - z_k^q \right)^2}$ and $\frac{z_k^{\frac{3}{2}}}{\left( 1 - z_k\right)^2}$, respectively. Such expressions modify the scattering outcomes at a particular surface temperature ($T_s$).
	
	\item [(iii)] The surface mode frequencies ($\omega_k$s) in the time- and temperature-dependent EHP play a crucial role in predicting the transition and reaction probabilities.
	
\end{itemize}

\subsubsection{10.2.1.7. Electron-hole pair coupling between molecular DOFs and electrons on surface atoms} \label{ssec:elp}

\noindent
While investigating the contribution of electron-hole pair (elhp) couplings between ``free" electrons on metal surface and  diatomic (in particular, hetronuclear) molecule, the electronic Hamiltonian for the molecule-surface interacting system takes the following form:

\begin{eqnarray}
	H_e=H_0 + F,
	\label{eq:aa}
\end{eqnarray}

\noindent
where $H_0$ represents the Hamiltonian for non-interacting $N$ electrons depicted by one electron eigenfunctions ($\Phi_{k,n}(r_e)$) with $r_e$ being electronic coordinate and $F$ signifies the interaction potentials between  electrons on the surface and charge - separated  on the molecule. \\

\noindent
Since the electrons are considered to move on a 2D surface (X, Y), the one-electron wavefunctions 
\begin{eqnarray}
	\Phi_{k,n}(r_e)=\sum_{l=1}^m C_{k-G_l,n} \: e^{i(k-G_l)\cdot r_e},
	\label{eq:expansion_eq}
\end{eqnarray}
are originated from simple 2D Bloch model periodic potential: 

\vspace{-0.6 cm}

\begin{eqnarray}
	U(r_e)=\sum_l U_{G_l}e^{iG_l\cdot r_e},
	\label{eq:ad}
\end{eqnarray}
\noindent
where summation of the wavefunctions is taken over all reciprocal lattice vectors $G_l $ within the first Brillouin zone of a unit cell. \\

\begin{figure}
  	\centering
  	\includegraphics[width=0.6\textwidth, height= 0.3\textwidth]{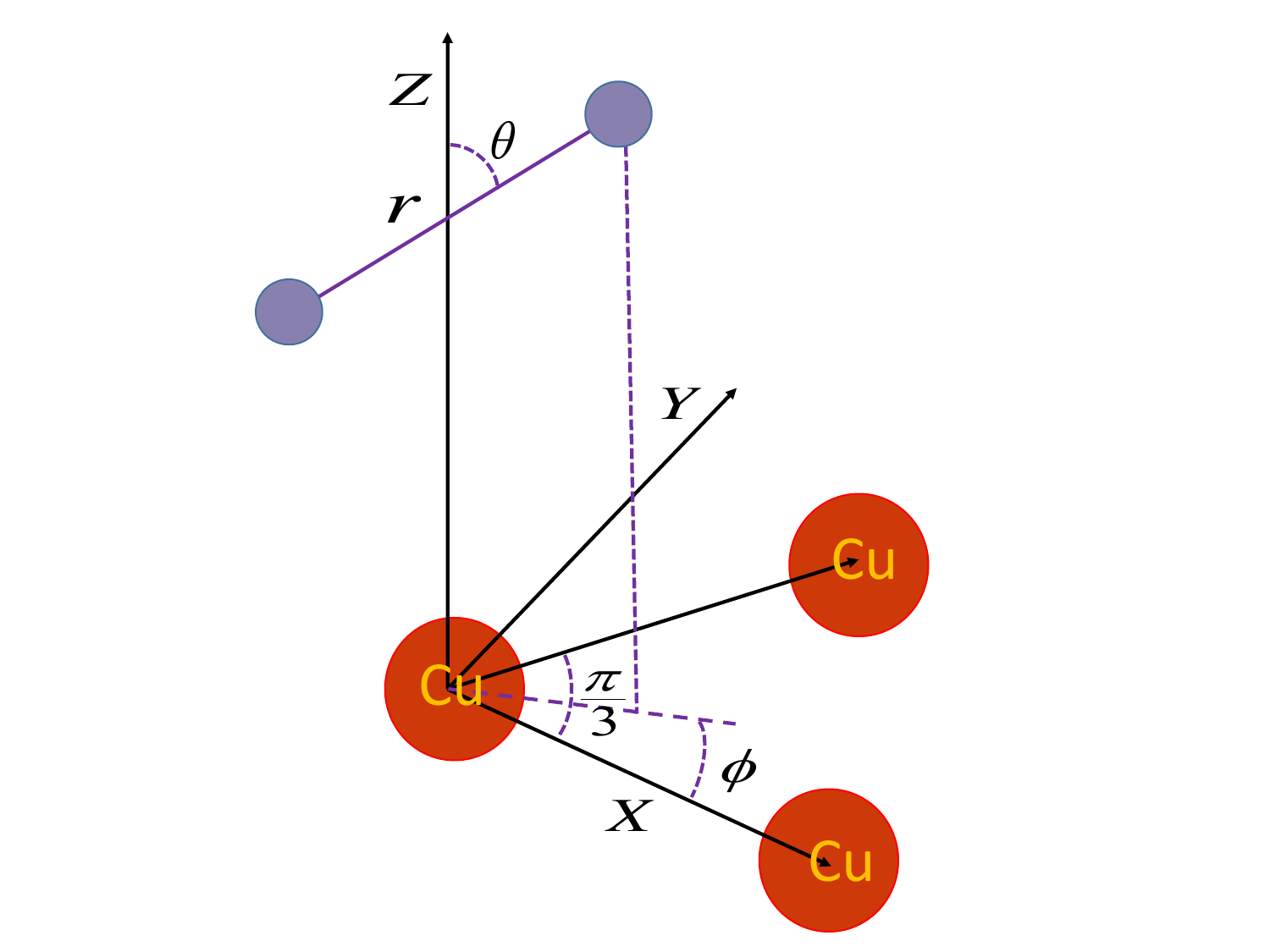}
  	\caption{Schematic representation of the six coordinates used in the quantum dynamical model of a diatomic molecule interacting with a metal surface is presented, exhibiting both the molecular and lattice reference frames.}
  	\label{fig:Geometry}
  \end{figure}

\noindent
On the other hand, the interaction potential, $F$ can be written in terms of the Fermi-Thomas screened coulomb potential as follows:

\begin{eqnarray}
	F=-\frac{e^2\delta(Z,r)}{R}e^{-k_cR},
	\label{eq:ab}
\end{eqnarray}

\noindent
where $R$ ($=\sqrt{X^2+Y^2+Z^2}$) denotes the distance between the electron and the excess charge $e\delta(Z,r)$ on the molecule, $Z$ indicates the separation between the center of the mass of the molecule and the surface, where $r$ presents the diatomic bond length (see Figure \ref{fig:Geometry}). The screening wavenumber $k_c$ is linked with Fermi vector $k_F$ through the screening constant $s_p$:

  \begin{eqnarray}
  	k_c =s_pk_F.
  \end{eqnarray}

\noindent
When the expanded form of one electron wavefunction as depicted in Equation (\ref{eq:expansion_eq}) and the periodic 2D Bloch potential as depicted in Equation (\ref{eq:ad}) are plugged into the time independent Schr\"{o}dinger equation (TISE), the expansion coefficients $C_{k-G_l,n}$ and eigenenergies $\epsilon_{k,n}$ are computed by solving TISE with the aid of a five-band model with free electron energies, $\epsilon_k^0,\epsilon_{k-b_1}^0,\epsilon_{k+b_1}^0,\epsilon_{k-b_2}^0,\epsilon_{k+b_2}^0$:

\begin{eqnarray}
	(\epsilon_{k-G_j}^0- \epsilon_{k,n}) C_{k-G_j,n} + \sum_{l=1}^m U_{G_l-G_j} C_{k-G_l,n} = 0	
\end{eqnarray}

\begin{small}
	\begin{eqnarray}
		\left(
		\begin{array}{ccccc}
			\epsilon_{k-b_1}^0 -\epsilon_{k,n}& 0 & U_0 & 0 & 0 \\ 
			0 & \epsilon_{k-b_2}^0 -\epsilon_{k,n} & U_0 & 0 & 0 \\ 
			U_0 & U_0 &\epsilon_{k}^0 -\epsilon_{k,n}  & U_0 & U_0 \\ 
			0 & 0 & U_0 & \epsilon_{k+b_2}^0 -\epsilon_{k,n} & 0 \\ 
			0 & 0 & U_0 & 0 &\epsilon_{k+b_1}^0 -\epsilon_{k,n}  
		\end{array} 
		\right) 
		\left(
		\begin{array}{c}
			C_{k-b_1,n} \\ 
			C_{k-b_2,n} \\ 
			C_{k,n} \\ 
			C_{k+b_2,n} \\ 
			C_{k+b_1,n}
		\end{array} 
		\right) 
		&=&
		\left(
		\begin{array}{c}
			0 \\ 
			0 \\ 
			0 \\ 
			0 \\ 
			0
		\end{array} 
		\right),\nonumber \\
	\end{eqnarray}
\end{small}

\noindent
where the kinetic energies ($\epsilon_k^0,\epsilon_{k-b_1}^0,\epsilon_{k+b_1}^0,\epsilon_{k-b_2}^0,\epsilon_{k+b_2}^0$) are expressed in terms of the wavevectors, $k_{X/Y}= -(\frac{\pi}{d_{2D}})i+ (\frac{\pi}{d_{2D}})j$. Moreover, the periodic potential~\cite{BILEH2} takes the following form:

\begin{eqnarray}
	U(r_e)=2U_0 \: cos\: (r_e \cdot b_1) + 2U_0 \: cos \: (r_e \cdot b_2),
	\label{eq:ad1}
\end{eqnarray}

\noindent
with 2$U_0$ as the band gap. The reciprocal lattice vectors $G_l$ and periodic potential are expressed in terms of the basis vectors,
$b_1 = (\frac{2\pi}{d_{2D}})i+ (\frac{2\pi}{d_{2D}})j $ and $b_2= (\frac{2\pi}{d_{2D}})i - (\frac{2\pi}{d_{2D}})j$ with the lattice constant, $d_{2D}$. \\

\noindent
In the interaction picture, for the transition from state $(k,n)$ to state $(k^\prime, n^\prime)$, the matrix element can be written as,
\begin{eqnarray}
	H_{(k,n),(k^\prime, n^\prime)}=\langle \Phi_{k,n}\mid F\mid\Phi_{k^\prime, n^\prime}\rangle,
	\label{eq:ac}
\end{eqnarray}
where $\Phi_{k,n}$ signifies the one electron wavefunction of the lattice with eigen energy, $\epsilon_{k,n}$.
If the motion of the electrons are considered to be restricted in two dimensions ($X, Y$) and the interaction potential is expanded in Fourier series (see Equation~\ref{eq:ab4} of Appendix~\ref{app-G}), the matrix element~\cite{BILEH2} is obtained as:
 
\begin{eqnarray}
	H_{(k,n),(k^\prime, n^\prime)} (Z,r) = \sum_{j\in {1,2}}\frac{e^2\delta(Z_j,r)}{2\pi 
		d_{2D}}\sum_{m=-\infty}^{m=+\infty}\sum_{h=-\infty}^{h=+\infty}\frac{1}{\sqrt{h^2+m^2+(d_{2D}k_c)^2}}\times \nonumber \\ 
	e^{-Z_j\lbrace k_c^2+(h/d_{2D})^2+(m/d_{2D})^2\rbrace^{1/2}}\times \langle \Phi_{k,n}\mid e^{im(X/d_{2D})+ih(Y/d_{2D})}\mid\Phi_{k^\prime, 
		n^\prime}\rangle ,
	\label{eq:af}
\end{eqnarray}

\noindent
where $Z_j$s show the distances between the surface and the atoms of the diatom. Using 2D Bloch model, the selection rule for the evaluation of the matrix element 
for electronic transition can be expressed as (see Appendix \ref{app-G}). \\

\noindent
Once the matrix element of the interaction Hamiltonian is in hand (see Equation (\ref{eq:af})), total Hamiltonian (see Equation~(\ref{eq:aa})) can be expressed in terms of second quantization~\cite{DPI,GDB2}:
\begin{eqnarray}
	H_e=\sum_i \epsilon_i c_i^\dagger c_i + \sum_{ij}H_{ij}c_i^\dagger c_j,
	\label{eq:ag}
\end{eqnarray}
where $i=(k,n)$ and $j=(k^\prime, n^\prime)$. It is important to mention that the Fermi operators ($c_i$, $c_j^\dagger$) fulfill the following commutation relation:
\begin{eqnarray}
	\lbrace c_i,c_j^\dagger \rbrace=\delta_{ij}.
	\label{eq:ah}
\end{eqnarray}
On the other hand, the time evolution operator $U_e(t, t_0)$ of the electronic system in presence of the interaction Hamiltonian is controlled by:
\begin{eqnarray}
	i\hbar \frac{\partial }{\partial t} U_e(t, t_0) =H_eU_e(t, t_0),
	\label{eq:ai}
\end{eqnarray}
where the time-dependent wavefunction can be represented as:
\begin{eqnarray}
	\Phi_{elhp}(t)=U_e(t, t_0)\Phi_{elhp}(t_0),
	\label{eq:aj}
\end{eqnarray}
with $\Phi_{elhp}(t_0)$ as the initial function. Hence, the effective Hamiltonian to encapsulate the contribution of elhp excitation on the 
molecule can be computed within mean-field approach:
\begin{eqnarray}
	V_{eff}^{elhp}(r,Z,t)=\langle \Phi_{elhp}(t)\mid H_e\mid \Phi_{elhp}(t)\rangle .
	\label{eq:ak}
\end{eqnarray}
While substituting Equation (\ref{eq:ag}) in Equation (\ref{eq:ak}), one can obtain (see Appendix \ref{app-DD}):
\begin{eqnarray}
	V_{eff}^{elhp}(r,Z,t)&=&\sum_i \epsilon_i n_i+\sum_i H_{ii}n_i +\sum_{ij}H_{ij}\left(Q_{ij}(t)^\star n_j+Q_{ji}(t)\right) \nonumber \\
	&+&\sum_{ij}\sum_k Q_{ik}(t)^\star Q_{jk}(t)n_kH_{ij},
	\label{eq:al}
\end{eqnarray}
where $Q_{ij}(t)$ is expressed as
\begin{eqnarray}
	Q_{ij}(t)=R_{ij}(t)-\delta_{ij},
	\label{eq:am}
\end{eqnarray}
and $R_{ij}(t)$ is originated from the solution of coupled differential equations:
\begin{eqnarray}
	i\hbar \frac {dR(t)}{dt}  = B(t)R(t),
	\label{eq:an}
\end{eqnarray}
with $R(t_0)=I$ (unit matrix) and
\begin{eqnarray}
	B_{ij}(t)=H_{ij}\exp\left(\frac{i}{\hbar}\left(\epsilon_i-\epsilon_j\right)t\right).
	\label{eq:ao}
\end{eqnarray}
The occupation number ($n_i$) of the exciton can be calculated from Fermi - Dirac (FD) Probability  factor at surface temperature, T$_s$:

\begin{eqnarray}
	n_i (T_s)=\frac {1}{1+ \exp\left[ \left( \epsilon_i-\mu\right) /kT_s\right]} \:,
	\label{eq:ap}
\end{eqnarray}
where $\mu$ represents the Fermi energy. 
Since the first term in the effective potential [Equation (\ref{eq:al})] represents the initial electron energy and the second term 
$H_{ii}=0$ (the transition between same state is forbidden), the effective potential for the elhp coupling is expressed as:
\begin{eqnarray}
	V_{eff}^{elhp}(r,Z,t,T_s)&=&\sum_{ij}H_{ij}(r,Z)\left(Q_{ij}^\star (t) n_j(T_s)+Q_{ji} (t)n_i(T_s)\right) \nonumber \\
	&+&\sum_{ij}\sum_k Q_{ik}^\star (t) Q_{jk} (t) n_k(T_s) H_{ij}(r,Z).
	\label{eq:aq}
\end{eqnarray}

\subsection{Initialization, propagation and projection}\label{ssec:propagation}

\noindent
Once the time- and temperature- dependent EHP is in hand involving linear (see Equations~(\ref{Eq: V_eff_VT_BEP}) and (\ref{Eq: V_eff_VT_MBP})) and/or bi-linear (see Equations~(\ref{Eq: state_averaged_effective_Hartree_potential_VV_BE}) and (\ref{Eq: state_averaged_effective_Hartree_potential_VV_MB})) as well as elhp (see Equation~(\ref{eq:aq})) couplings. Such couplings are present between the molecular DOFs and surface vibrational modes as well as free electrons on surface and excess charge on the molecule, respectively. The associated initial wavefunction employed in the 6D QD~\cite{TSS1,TSSM,Mantu_jcp_2024,Bauke_jcp_2024} can be expressed as:

\begin{eqnarray}
	\Psi_{m}(r, \theta, \phi, X, Y, Z, t_0) = \frac{1}{r} \, g_\nu(r) \, Y_{j, m_j}(\theta, \phi) \, \Phi_{\text{GWP}}(X, t_0) \, \Phi_{\text{GWP}}(Y, t_0) \, \Phi_{\text{GWP}}(Z, t_0),
	\label{eq:initial_wavefunction}
\end{eqnarray}

\noindent
where internal DOFs of the diatomic molecule are depicted by the internuclear separation $r$ and its orientation as dictated by the polar and azimuthal angles, $\theta$ and $\phi$, respectively. The Cartesian coordinates ($x$, $y$ and $z$) are related with polar ones (r, $\theta$ and $\phi$) through standard transformation.  The radial function $ g_\nu(r) $ signifies the vibrational eigenstate of the diatom, while the angular part $ Y_{j, m_j}(\theta, \phi) $ represents the spherical harmonics appearing due to the rotational state. The function $ \Phi_{\text{GWP}}(Z, t_0) $ demonstrates a Gaussian wavepacket along the translational coordinate ($ Z $-axis) indicating an incident diatom initially propagating with momentum, $ p_0 = \hbar k_0 $.

\subsubsection{10.2.2.1. TDDVR propagation and  analysis}

\noindent 
While employing the time-dependent discrete variable representation (TDDVR)~\cite{SMIRPC} method, the initial wavefunction is expressed in term of TDDVR basis functions with converged number of grid points. The time propagation is performed with the aid of the Lanczos propagation technique. In the asymptotic region, the incoming  and scattered wavefunctions are far from the interaction zone, where the scattering outcomes can be computed in terms of flux components. The energy-resolved probabilities for elastic and inelastic transitions between specific quantum states are calculated by considering the ratio of the outgoing to incoming flux coefficients, $C^+_{k_{out},v^{\prime}j^{\prime} m_{j^{\prime}}}$ and $C^-_{k_{in},vjm_j}$, respectively (see Appendix~\ref{app:JJ}):

\begin{eqnarray}
	P_{vjm_j}^{v^{\prime}j^{\prime}m_{j^{\prime}}}(E) = \frac{ \frac{k_{out}}{M}|C^+_{k_{out}, v^{\prime}j^{\prime}
			m_{j^{\prime}}}|^{2}}{\frac{k_{in}}{M}|C^-_{k_{in},vjm_j}|^{2}}. 
	\label{eq:transprob11}
\end{eqnarray}

\noindent
The sticking probability ($P_{\text{stick}}(E)$) on the surface takes the following form:

\begin{eqnarray}
	P_{\text{stick}}(E)= 1- \sum_{v\prime j\prime m_{j\prime}} P_{v j m_j}^{v\prime j\prime m_{j\prime}} (E)
	\label{eq:pstick1}
\end{eqnarray}

\noindent
The asymptotic wave function at a particular total energy ($E$) is studied by using Equations (\ref{eq:transprob11}) and (\ref{eq:pstick1}) to calculate $P_{v j m_j}^{v\prime j\prime m_{j\prime}} (E)$ and $P_{\text{stick}}(E)$.

\subsubsection{10.2.2.2. SPO-DVR propagation and analysis}

\noindent
The wavepacket is propagated employing the split-operator discrete variable representation (SPO-DVR)~\cite{SPO,SPO-DVR-code-2} method. In order to eliminate artificial reflections from the grid boundaries, absorbing optical potentials are used in both the reactive and asymptotic scattering regions. State-to-state scattering probabilities for specific rovibrational and diffraction channels are analyzed from the scattering matrix (S-matrix) located in the asymptotic region. On the other hand, the sticking probability is computed as the complement of the total scattering probability. In addition, dissociation probabilities can be evaluated through flux analysis along a dividing surface, which is placed at large internuclear separations.

\section{ Model interaction potentials: TDDVR quantum dynamics}\label{sec:TDDVR_Probability}

\noindent
 Quantum dynamics (QD) calculations were carried out with the help of effective Hamiltonian, in which the initial state distributions of the surface modes were included through Maxwell-Boltzmann (MB) or Bose--Einstein (BE) as well as the electron-hole pair (elhp) states through Fermi--Dirac (FD) probability factors to successfully incorporate the effect of surface temperature. Once such effective Hamiltonian is in hand, embedded diatomic in molecule (EDIM) potentials for metal-metal as well as gas-metal interactions were used to capture the contribution of surface modes. On the other hand, Fermi-Thomas model potential is incorporated in the surface-electron excitation to molecule at a specific surface temperature ($T_s$). When the construction of the effective Hamiltonians was over, TDDVR QD had been carried out to find out sticking and survival/excitation probability on such 6D Hamiltonians. 
 
\subsection{Linear coupling between molecular DOFs and surface modes}

\subsubsection{10.3.1.1. Reaction probability for H$_2$($v=0, j=0$) on Cu(100) surface} \label{ssec:v0j0_reaction}

\begin{figure}[htp!]
	\centering
	\includegraphics[width=0.5\textwidth,angle=0]{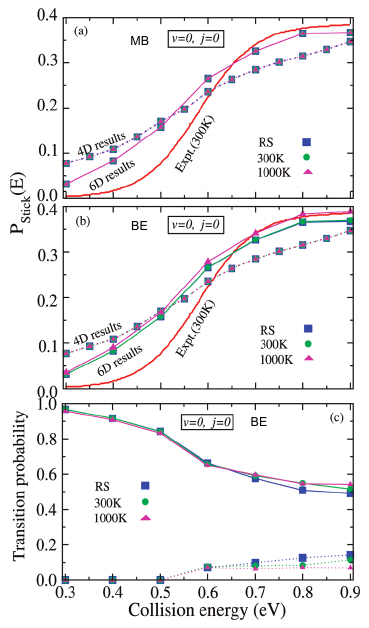}
	\caption{Figure presents the sticking probability (P$_{\text{stick}}$(E)) for the H$_2$(v = 0, j = 0)--Cu(100) system calculated using 4D $\otimes$ 2D and 6D quantum dynamics, with (a) MB and (b) BE probability factors, as functions of incident energy, which is compared with experimental~\cite{HAM1} data. Panel (c) depicts the energy-resolved state-to-state transition probabilities for vibrationally elastic (v = 0 $\rightarrow$ $v^\prime$ = 0, solid line) and inelastic (v = 0 $\rightarrow$ $v^\prime$ = 1, dashed line) processes. This figure has been taken from Ref. [\protect\citen{TSS1}] and permission has been taken from American Chemical Society (ACS).}
	\label{fig:JPCA_2011_2}
\end{figure}

\noindent
While performing QD calculations on an EHP based on the MB probability factor for the initial surface modes distribution (see Equation~(\ref{Eq: V_eff_VT_MBP})), the sticking probabilities for H$_2$ \((v = 0, j = 0)\) on Cu(100) (Figure~\ref{fig:JPCA_2011_2}a) were obtained at various values of incident kinetic energy either with rigid surface (RS) situation or in the presence of surface modes at different surface temperatures ($T_s$= 300  and 1000 K).~\cite{TSS1} The calculated 6D results exhibited close agreement with the experimentally~\cite{HAM1} fitted reaction probabilities at higher collisional energies for $T_s = 300$ K, whereas the 4D $\otimes$ 2D calculations were unable to reproduce the experimental trend over the entire energy range. It is important to mention that for both approaches, significant temperature dependence was not observed. \\

\noindent
On the contrary, while constructing the EHP using the BE probability factor (see Equation~(\ref{Eq: V_eff_VT_BEP})), the calculated 6D sticking probabilities encountered (Figure~\ref{fig:JPCA_2011_2}b) a temperature dependence, more specifically, at higher energies. Furthermore, the corresponding 4D $\otimes$ 2D results appeared almost insensitive to temperature. Hence, in order to get accurate description of temperature effects, it necessitates a fully coupled 6D treatment of all molecular DOFs and quantum states of surface modes distributed through BE probability factor. Figure~\ref{fig:JPCA_2011_2}c depicts the 6D state-to-state transition probabilities, calculated for both RS at 0 K and finite-temperature surfaces (300 and 1000 K for BE case), which possessed negligible temperature dependence up to incident energies of $\sim 0.6$ eV. On the other hand, 
over the energy range (0.6--0.9 eV), survival and  excitation probabilities exhibited increasing and decreasing trend, respectively, with increasing surface temperature. 

\subsubsection{10.3.1.2. Survival probability for H$_2$($v=1, j=1$) on Cu(100) surface} \label{ssec:reaction}

\begin{figure}[!htp]
	\centering
	\includegraphics[width=0.6\textwidth,angle=0]{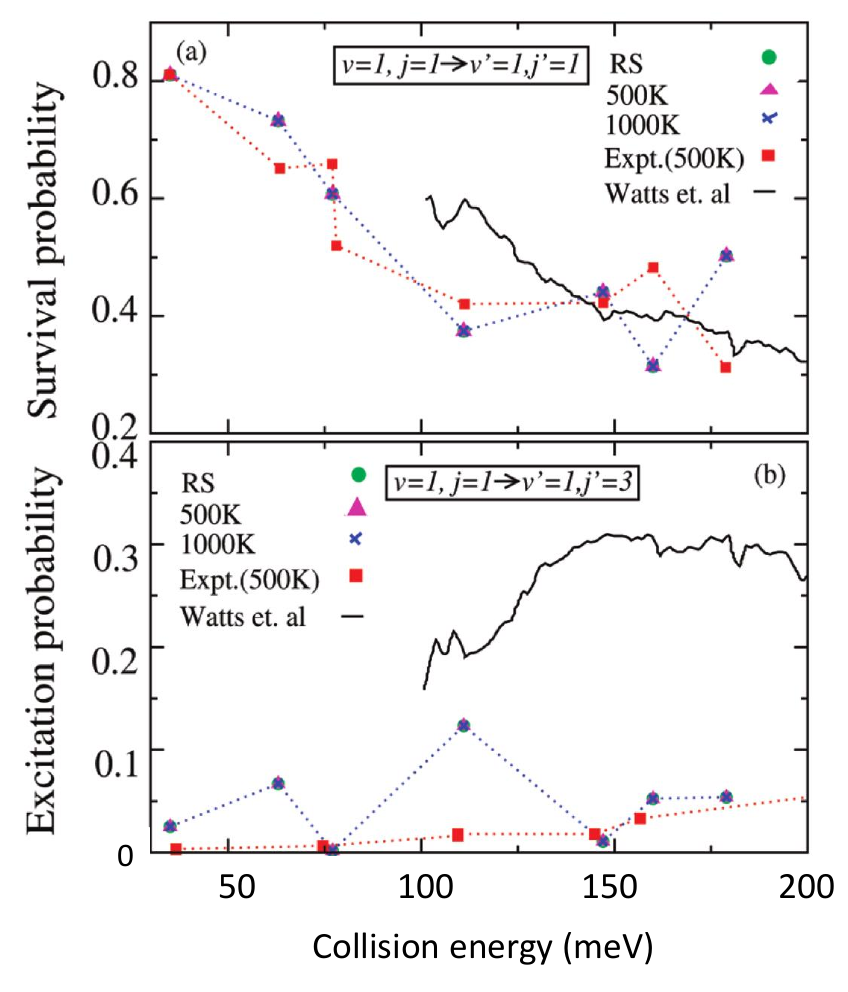}
	\caption{Figure depicts the calculated (a) survival (v = 1, j = 1 $\rightarrow$ $v^\prime$ = 1, $j^\prime$ = 1) and (b) excitation (v = 1, j = 1 $\rightarrow$ $v^\prime$ = 1, $j^\prime$ = 3) probabilities of H$_2$ scattering from the Cu(100) surface as functions of incident kinetic energy with experimental~\cite{CYL2} and previous theoretical results~\cite{GJ3} . The calculations are originated from an effective Hamiltonian incorporating surface temperature through the BE probability factor. This figure has been adapted from Ref. [\protect\citen{TSS1}] with permission from American Chemical Society (ACS).}
	\label{fig:JPCA_2011}
\end{figure}

\noindent
The incoming diatom H$_2$($v=1, j=1$) colliding with Cu(100) surface was initialized employing BE based EHP (see Equation~(\ref{Eq: V_eff_VT_BEP})) and hence, Figure~\ref{fig:JPCA_2011} portrays the TDDVR-computed (a) survival ($v^\prime = 1, j^\prime = 1$) and (b) excitation ($v^\prime = 1, j^\prime = 3$) probabilities as functions of the initial kinetic energy of the incident molecule.~\cite{TSS1} Such outcomes were obtained
within the RS approximation as well as by considering surface modes at temperatures, 500 K and 1000 K. Those results were compared with experimental data at 500 K as well as previous theoretical results. It is interesting to note that the computed survival probabilities match well with experiment over the entire collisional energy range
vis-\'a-vis earlier theoretical studies (see Figure~\ref{fig:JPCA_2011}a). On the contrary, the excitation probabilities showed noticeable deviations from experiment. As a matter of fact, 
the calculated results exhibited improved agreement compared to previous calculations (see Figure~\ref{fig:JPCA_2011}b), but both survival and excitation probabilities were found to be almost invariant with respect to surface temperature.

\newpage

\subsubsection{10.3.1.3. Reaction probability for D$_2$ (v=0, j=0) on Cu(111) surface}

\begin{figure}[!htp]
	\centering
	\includegraphics[width=0.5\textwidth,angle=0]{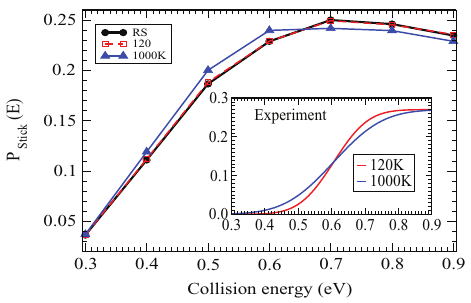}
	\caption{Computed sticking probability (P$_{\text {Stick}}$ (E)) using 6D quantum dynamics considering BE Probability factor in case of D$_2$ (v = 0, j = 0)-Cu(111) system is presented as a function of initial collision energy in comparison with experimental~\cite{CTH} results. This figure has been adapted from Ref. [\protect\citen{Tapas_JCP_2012}] with permission from American Institute of Physics (AIP).}
	\label{fig:JCP_2012_1}
\end{figure}

\noindent
While performing the 6D QD calculations with the EHP obtained from BE distributed quantum states of surface modes (see Equation~(\ref{Eq: V_eff_VT_BEP})), the calculated sticking probabilities are presented in Figure~\ref{fig:JCP_2012_1} for incoming D$_2$(v=0, j=0) scattered from the Cu(111) surface over the temperature range of 120 and 1000 K.~\cite{Tapas_JCP_2012}. Even though the 4D$\otimes$2D treatment had been found to show only minor broadening at both low and high incident kinetic energies, the full 6D results carried the signature of pronounced broadening. The computed probability shows consistency with experimental observations. The increased surface temperature has enhanced reactivity at low collision energies leading to the reduction of reactivity at higher energies.

\begin{figure}[!htp]
	\centering
	\includegraphics[width=0.5\textwidth,angle=0]{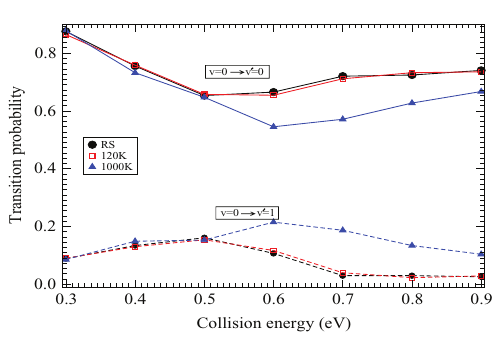}
	\caption{The energy-resolved state-to-state transition probabilities (calculated with the help of BE Probability factor) for vibrationally elastic, $v = 0 \rightarrow v' = 0$ (solid line), and inelastic, $v = 0 \rightarrow v' = 1$ (dashed line), collisions are depicted. This figure has been taken from Ref. [\protect\citen{Tapas_JCP_2012}] with permission from American Institute of Physics (AIP).}
	\label{fig:JCP_2012_2}
\end{figure}

\noindent
Figure~\ref{fig:JCP_2012_2} displays the state-resolved scattering probabilities for incident D$_2$$(v = 0, j = 0)$, where the probability of remaining in the initial vibrational state ($v' = 0$) exhibited a reduction with increasing surface temperature. Vibrational excitation ($v' = 1$) channels became more populated while scattering.

\subsection{ Linear coupling between molecular DOFs and surface vibrational modes, as well as coupling between molecular DOFs and elhp excitations}
		
\subsubsection{10.3.2.1. Reaction probability for H$_2$/D$_2$(v=0, j=0) on Ni(100) surface} \label{ssec:reaction1}

\noindent
In order to include surface temperature effects originating from the coupling of the molecular DOFs to surface vibrational modes and elhp excitations, the applicability of theoretical framework (see Equation~(\ref{Eq: V_eff_VT_BEP}) for BE and Equation~(\ref{eq:aq}) for FD) was explored by calculating dissociation and transition probabilities for the H$_2$/D$_2$--Ni(100) system.~\cite{SM-MOLPHYS}
While following this approach, effects from the surface modes were included through the BE probability factor dictating the initial occupation of surface states. Furthermore, the 
effects of elhp interactions were taken into account with the aid of FD probability factor to represent the distribution of free electrons on the metal surface. QD calculations were executed employing a model consisting of eighty-nine (89) Ni atoms arranged in three layers of the Ni(100) surface at RS as well as over thermally activated surface at 1000 K. Reaction probabilities were calculated at various values of initial kinetic energy of the incident diatom.\\

\begin{figure}[!htp]
 	\centering
 	\includegraphics[width=0.5\textwidth,angle=0]{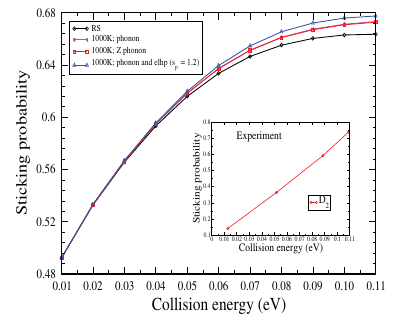}
 	\caption{Calculated sticking probability using 4D $\otimes$ 2D quantum dynamics for D$_2$ (v = 0, j = 0)--Ni(100) system is presented as function of initial collision energy for RS and surface at 1000 K. This figure has been taken from Ref. [\protect\citen{SM-MOLPHYS}] with permission from Taylor \& Francis.}
 	\label{fig:Mol_Phys_2015_1}
 \end{figure}

\noindent
 Figure~\ref{fig:Mol_Phys_2015_1} depicts dissociation probabilities of D$_2$ (v = 0, j = 0) at 1000 K by including: (i) only perpendicular (Z) surface modes, (ii) all surface modes and (iii) both surface modes and elhp coupling.~\cite{SM-MOLPHYS} According to the outcome, inclusion of only Z modes had produced dissociation probabilities similar to those obtained with all surface modes exploring the dominant contribution of perpendicular vibrations. The results obtained from RS and finite-temperature surfaces are clearly distinct, more specifically at higher collision energies, where thermal effects accelerated the reactivity. It is worth mentioning that inclusion of elhp coupling (with screening constant $s_p = 1.2$) resulted into a slight increase in dissociation probability at higher energies. As a matter of fact, its overall effect remained secondary compared to surface mode mediated effects. The temperature dependence could not be validated directly 
 due to lack of experimental~\cite{AVH} data at elevated temperatures.

\begin{figure}[!htp]
	\centering
	\includegraphics[width=0.5\textwidth,angle=0]{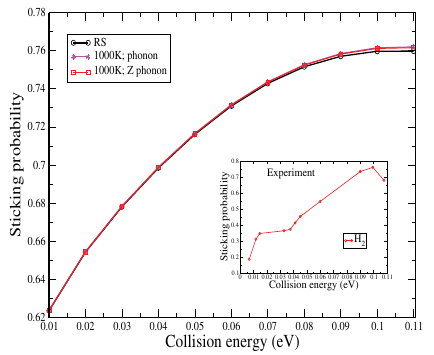}
	\caption{Calculated sticking probability using 4D $\otimes$2D quantum dynamics for H$_2$ (v = 0, j = 0)--Ni(100) system is displayed as a function of initial collision energy for RS and surface at 1000 K. This figure has been adapted from Ref. [\protect\citen{SM-MOLPHYS}] with permission from Taylor \& Francis.}
	\label{fig:Mol_Phys_2015_2}
\end{figure}

\noindent
On the contrary, in case of H$_2$ (v = 0, j = 0) scattering on Ni(100) surface (see Figure~\ref{fig:Mol_Phys_2015_2}), dissociation probabilities encountered weak sensitivity to surface temperature. Additionally, RS, Z-mode-only and all-mode results are nearly coincided over the entire collision energy range. Such behavior is observed to be consistent with experimental observations and originated from the lower mass of H$_2$ relative to D$_2$. Both theoretical and experimental~\cite{AVH} studies consistently validated that dissociation probabilities for H$_2$ are higher in comparison with D$_2$ across all collision energies. \\

\begin{figure}[!htp]
	\centering
	\includegraphics[width=0.5\textwidth,angle=0]{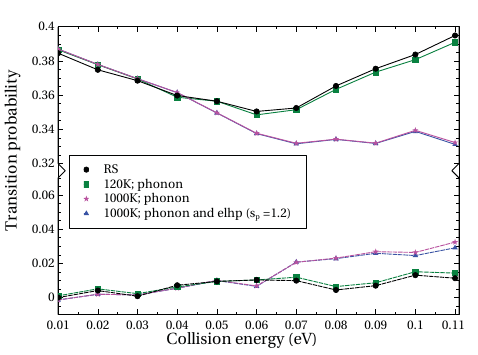}
	\caption{The energy-resolved state-to-state transition probabilities using 4D $\otimes$ 2D quantum dynamics for vibrationally elastic, v = 0 $\rightarrow$  $v^\prime$ = 0 (solid line), and inelastic, v = 0 $\rightarrow$ $v^\prime$ = 1 (dash line), collisions are presented. This figure has been adapted from Ref. [\protect\citen{SM-MOLPHYS}] with permission from Taylor \& Francis.}
	\label{fig:Mol_Phys_2015_3}
\end{figure}

\noindent
Figure~\ref{fig:Mol_Phys_2015_3} presents state-to-state vibrational transition probabilities for D$_2$ ($v^\prime = 0, 1$), which further highlights the effect of surface temperature. At 1000 K, excitation and survival probabilities exhibited increasing and decreasing trends across the studied energy range (0.01–0.11 eV). On the other hand, the results at RS and 120 K remained largely unchanged. At higher collision energies (0.07–0.11 eV), excitation probabilities achieved saturated value at approximately 0.03. In addition, results obtained with combined surface modes and elhp coupling show very close agreement with those including only surface modes and thereby, the results affirmed the minor role of elhp effects.

\subsubsection{10.3.2.2. Reaction probability for D$_2$(v=0, j=0) on Cu(111) surface}

\begin{figure}[!htp]
	\centering
	\includegraphics[width=0.5\textwidth,angle=0]{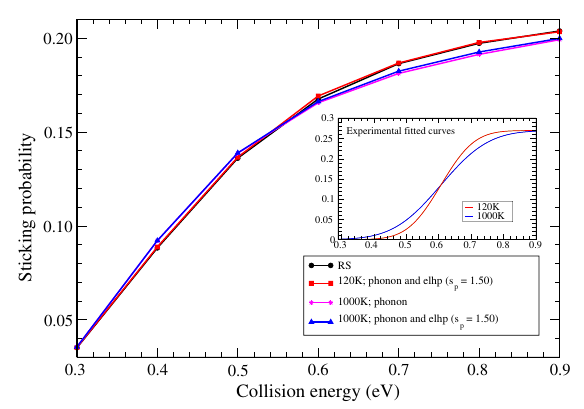}
	\caption{Calculated sticking probabilities for D$_2$ ($v = 0$, $j = 0$) on the Cu(111) surface are plotted, which are obtained using 4D$\otimes$2D quantum dynamics, as a function of collision energy for the rigid surface (RS) and for surfaces at 120~K and 1000~K. This figure has been taken from Ref. [\protect\citen{SM-JTCC}] with permission from World Scientific Publishing.}
	\label{fig:JTCC_2015_1}
\end{figure}

\noindent
While using the ground PES constructed by fitted EDIM analytic functions for  D$_2$(v=0, j=0)-Cu(111) system, the effects of surface temperature originating from surface modes (using Equation~(\ref{Eq: V_eff_VT_BEP}) for BE) and elhp couplings (employing Equation~(\ref{eq:aq}) for FD) are explored in Figure~\ref{fig:JTCC_2015_1} and Table~\ref{Tab:table1}.~\cite{SM-JTCC} In the selected domain of collisional energies, the sticking probabilities for D$_2$ molecule on the RS and 120~K surface, involving both surface modes and elhp couplings, was found to be nearly identical. Nevertheless, in case of 1000~K surface, the inclusion of either the effects of surface modes alone or combined surface modes and elhp contributions resulted into slight deviations from the RS and 120~K cases. On the other hand, the sticking probability at 1000~K was slightly higher ar lower collision energies, but reverse trend was observed at higher (collisional) energies. Overall, the elhp coupling is responsible for negligible influence on the sticking probability even at elevated temperature. If only surface modes were considered, the sticking probability profiles for the RS and the 1000~K surface undergo an intersection near a collision energy of 0.557~eV.
Table~\ref{Tab:table1} summarizes the influence of elhp coupling with increasing screening constant. At $s_p = 1.50$, the outcomes involving both surface modes and elhp effects superimpose with those obtained from surface modes contributions alone. In addition, 
the crossing points between the sticking profiles of the RS and the 1000~K surface exhibit dependence on the screening constant at various values of collision energies for different $s_p$ s. With the increase in surface temperature, the population of higher electronic states exhibited an increasing trend indicating a progressively more significant role of elhp excitation. Since the molecule--surface interaction was illustrated with the aid of a screened Coulomb (Fermi--Thomas) potential, the excess charge on the diatom interact with the metal electrons and it can be promoted to higher energy levels. Since the excess charge on D$_2$ appears as very small ($\sim 0.146\,e$), the resulting elhp coupling becomes weak. As a result, its contribution to the sticking and transition probabilities for elhp over linear coupling acquire negligible value(s) even at elevated surface temperatures.\\

\begin{table}[h!]
	\centering
	\caption{The crossing points of the profiles for sticking probabilities due to RS and surface at 1000~K are summarized.~\cite{SM-JTCC}}
	\begin{tabular}{lc}
		\\ \hline
		\textbf{Coupling} & \textbf{Crossing point} \\
		\hline
		Surface mode & 0.557 (eV) \\
		Surface mode + elhp ($s_p = 1.75$) & 0.557 (eV) \\
		Surface mode + elhp ($s_p = 1.60$) & 0.561 (eV) \\
		Surface mode + elhp ($s_p = 1.50$) & 0.568 (eV) \\
		Surface mode + elhp ($s_p = 1.40$) & 0.586 (eV) \\
		Surface mode + elhp ($s_p = 1.35$) & 0.617 (eV) \\
		Surface mode + elhp ($s_p = 1.30$) & 0.695 (eV) \\
		Expt.\ fitted curves~\cite{CTH} & 0.607 (eV) \\
		\hline
	\end{tabular}
	\label{Tab:table1}
\end{table}

\newpage

\subsection{Linear and bi-linear couplings between molecular DOFs and surface modes }

\subsubsection{10.3.3.1. Reaction probability of D$_2$(v=0, j=0, 2) on Cu(111) surface}

\begin{figure}[!htp]
	\centering
	\includegraphics[width=0.9\textwidth,angle=0]{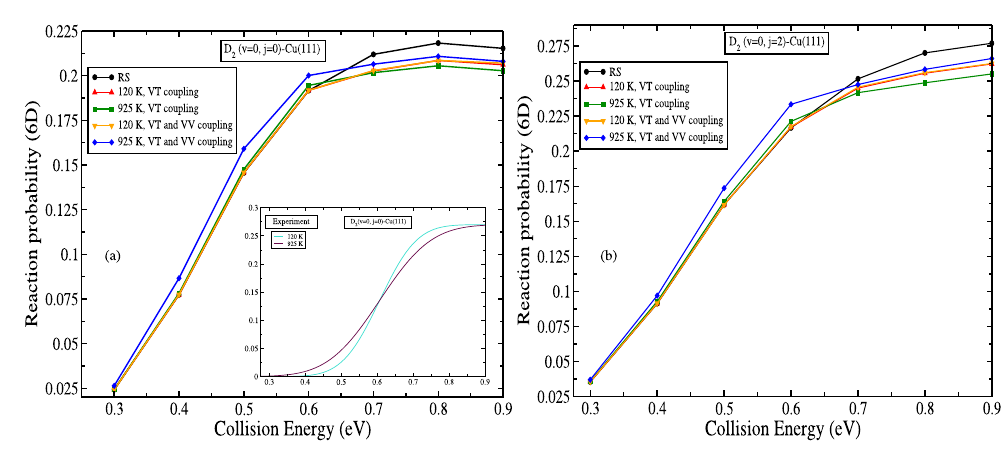}
	\caption{Comparison of combined VT--VV and VT-only coupling effects on 6D QD reaction probabilities for the rigid surface (RS) and at surface temperatures of 120~K and 925~K is depicted for the dissociation of (a) D$_2$ ($v=0$, $j=0$) and (b) D$_2$ ($v=0$, $j=2$) on Cu(111). Experimental curves~\cite{CTH} for 120~K and 925~K are also presented as insets for the collision of D$_2$ ($v=0$, $j=0$) on Cu(111). The initial vibrational state distribution is demonstrated using BE probability factors. This figure has been taken from Ref. [\protect\citen{VV-Hartree-2021}] with permission from Elsevier.}
	\label{fig:Chem_Phys_2022_1}
\end{figure}

\noindent
With the help of EHP defined in Equation~(\ref{Eq: V_eff_VT_BEP}) for linear coupling and Equation~(\ref{Eq: state_averaged_effective_Hartree_potential_VV_BE}) for bi-linear coupling, as well as with the EDIM interaction potential and BE probability factors for the surface-mode quantum states, 
6D QD calculations were executed to compute reaction probabilities [see Figures~\ref{fig:Chem_Phys_2022_1}(a) and \ref{fig:Chem_Phys_2022_1}(b)] for the rigid-surface (RS) limit and for surface temperatures of 120 and 925 K.~\cite{VV-Hartree-2021} The combined effect of linear vibration-translation (VT) and bi-linear vibration-vibration (VV) couplings was presented in comparison with the linear (VT) contribution alone for the dissociation process of D$_2$ \((v = 0, j = 0)\) and D$_2$ \((v = 0, j = 2)\) on the Cu(111) surface. According to the results, the incorporation of bi-linear VV coupling between the molecular DOFs and surface modes at 120 K resulted into no noticeable change in the reaction probabilities for either initial state, $(v=0, j=0)$ and $(v=0, j=2)$. On the contrary, inclusion of the bi-linear VV term in addition to the linear term at 925 K led to significant enhancement of the reaction probabilities across the entire collision-energy range. Such behavior mainly originates from the initial surface vibrational-state distribution by BE statistics as well as more accurate representation of the effective Hamiltonian. In such framework, higher-order anharmonic (bi-linear, VV) couplings between molecular DOFs and surface modes appear increasingly relevant at elevated temperatures. As a consequence, the modified many-oscillator model within the mean-field framework can successfully capture the role of higher-order anharmonic (bi-linear, VV) molecule-surface modes couplings at elevated surface temperatures.\\

\begin{figure}[!htp]
	\centering
	\includegraphics[width=0.9\textwidth,angle=0]{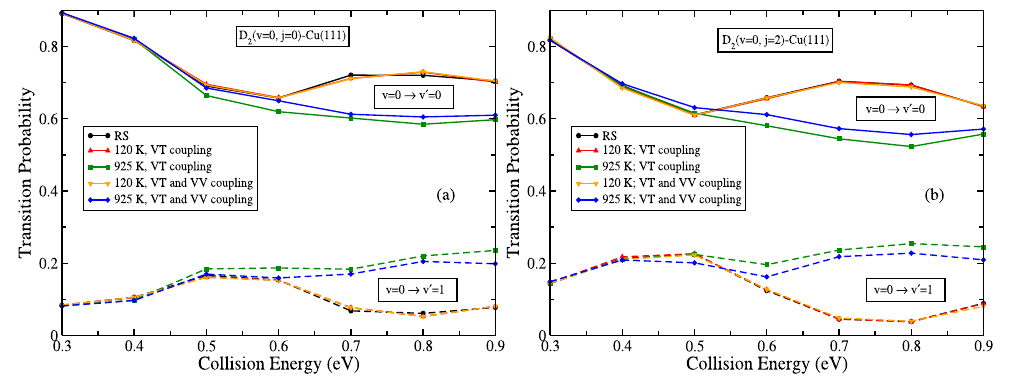}
	\caption{Comparison of VT--VV and VT-only coupling effects on 6D QD energy-resolved transition probabilities is presented for elastic, $v = 0 \rightarrow v^\prime = 0$ (solid lines) and inelastic, $v = 0 \rightarrow v^\prime = 1$ (dashed lines) channels in D$_2$ ($v = 0, j = 0, 2$) scattering on Cu(111) at RS, 120~K and 925~K. This figure has been adapted from Ref. [\protect\citen{VV-Hartree-2021}] with permission from Elsevier. }
	\label{fig:Chem_Phys_2022_2}
\end{figure}

\noindent
In case of D$_2$ ($v = 0, j = 0$) and D$_2$ ($v = 0, j = 2$), Figures~\ref{fig:Chem_Phys_2022_2}(a)--(b) depict the energy-resolved 6D QD state-to-state vibrationally elastic ($v = 0 \rightarrow v^\prime = 0$) and inelastic ($v = 0 \rightarrow v^\prime = 1$) transition probabilities for D$_2$ scattering on Cu(111) at various values of collision energy under RS conditions and at surface temperatures of 120~K and 925~K. Transition probabilities obtained from VT-only and combined VT--VV treatments depicted negligible influence of bi-linear (VV) couplings on both survival and excitation probabilities at 120 K. On the contrary, incorporation of VV coupling at 925 K significantly altered the transition probabilities for both initial states resulting into an overall enhancement of reactivity with increasing surface temperature.

\section{Realistic interaction potentials: SPO-DVR quantum dynamics}\label{sec:chemisorption}

\noindent
The 6D QD calculations were executed employing EHPs, where the initial state distributions of the surface modes were taken into account through BE or MB probability factors to consider the effect of surface temperature. In order to achieve realistic EHPs with higher accuracy, the contribution of surface modes is included, and the frequencies and displacement vectors of those modes were evaluated through Vienna \textit{Ab initio} Simulation Package (VASP). On the other hand, CRP and SCM potentials were included to depict the gas-metal interactions maintaining the surface atoms at equilibrium for rigid surface and vibrating around the equilibrium at non-zero surface temperature. While employing these surface modes and interaction potentials, 6D EHP-based Hamiltonians were formulated with BE and MB probability factors [see Equations~(\ref{Eq: V_eff_VT_BEP}) and (\ref{Eq: V_eff_VT_MBP}), respectively]. At the same time, the corresponding QD calculations were executed using the SPO-DVR~\cite{SPO,SPO-DVR-code-2} method.

\subsection{ Linear coupling between molecular DOFs and surface vibrational modes}

\subsubsection{10.4.1.1. Reaction probability for H$_2$(v=0, j=0) on Cu(111) surface} \label{ssec:scattering}

\begin{figure}[!htp]
	\centering
	\includegraphics[width=0.9\textwidth,angle=0]{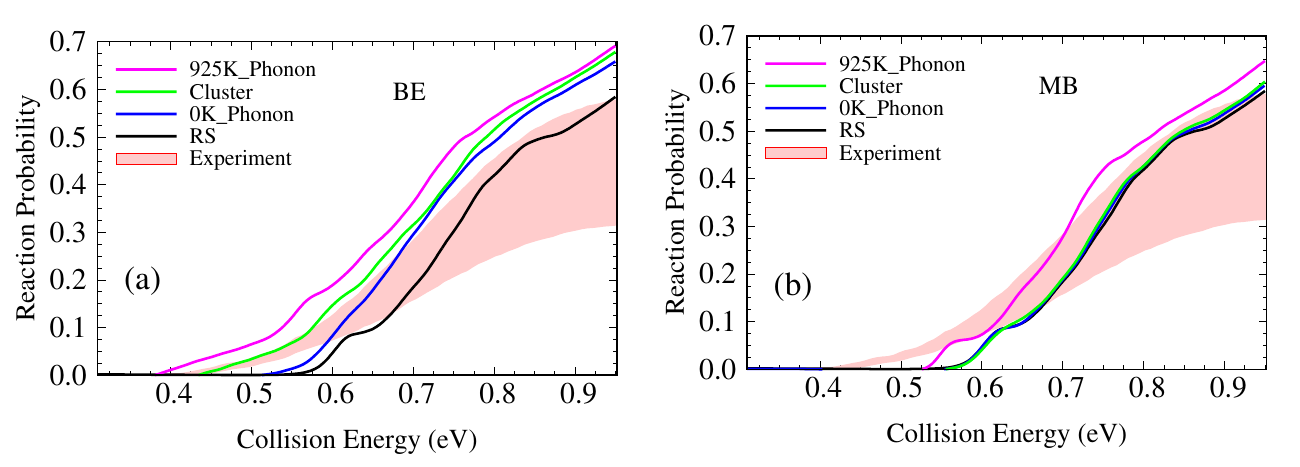}
	\caption{Reaction probabilities for H$_2$ on Cu(111) based on the rigid surface (RS) and EHP originated from VASP-SRP48 calculated normal-mode frequencies are presented at a surface temperature of 925~K. Panels (a) and (b) depict the BE and MB probability factors, respectively. This figure has been adapted from Ref. [\protect\citen{Mantu_jcp_2024}] with permission from AIP.}
	\label{fig:Mantu_JCP_2024_1}
\end{figure}

\noindent
While constructing various EHPs~\cite{Mantu_jcp_2024}, surface mode frequencies and displacement vectors were calculated from SRP48-based normal mode analyses of a 31-atom Cu cluster, where the required phonons were evaluated from periodic $3 \times 3$ Cu(111) slabs at 0 and 925~K as well as a 925~K EAM representation fitted to DFT data by Sheng \textit{et al.}~\cite{Sheng_2011}. All phonon characteristics were computed within the quasi-harmonic approximation employing Phonopy~\cite{togo_Firstprinciples_2023,togo_Implementation_2023} with 0.01~\AA\ displacements, interfaced with Large-scale Atomic/Molecular Massively Parallel Simulator (LAMMPS)~\cite{LAMMPS_2022} and VASP~\cite{kresse_ab_1993,kresse_efficient_1996,kresse_efficiency_1996} for EAM and SRP48 potentials, respectively. In case of such EHPs, gas--surface interactions were included through the CRP-fitted BOSS PES for 0~K and thermally expanded lattices at 925 K, whereas the gas--metal coupling through the SCM potential is originated from surface atom vibrations. In this theoretical framework, four distinct EHP formulations (cluster approximation, 0 K and 925 slab, and EAM) were used at various treatments of thermally expanded lattice and phonons. In the subsequent works, 6D QD investigations using the SPO-DVR method were performed on such EHPs for H$_2$ on Cu(111) initializing the the molecule in its rovibrational ground state ($v=0$, $j=0$). Figures ~\ref{fig:Mantu_JCP_2024_1}(a) and \ref{fig:Mantu_JCP_2024_1}(b) depict the reaction probability profiles at $T_s = 925$~K, which were analyzed in comparison with the RS (0~K crystal) and experimental~\cite{Wodtke} data. Table~\ref{Tab:Table_2} presents the corresponding threshold energies for reaction probability. At finite-temperature (925 K), BE-based profiles exhibit a noticeable shift of threshold for reaction probability toward lower energies vis-\'a-vis pronounced broadening across the entire collisional energy range. Nevertheless, MB-based outcomes shows comparatively less broadening even in the case of 925 K phonons and surface temperature. It is interesting to note that the EHP originated from slab-derived 925~K phonons together with thermal lattice expansion resulted into the highest chemisorption probabilities with the lowest threshold energy, which are in better agreement with experimental ones. Such results depict significant improvement relative to RS case. Such behavior mainly originates from the  combined effects of realistic finite-temperature phonons, lattice expansion through the 925~K CRP potential and quantum states of surface mode  distributed through BE probability factor. Such effects are able to reduce the dissociation barrier corresponding to the thermally expanded PES, improving upon earlier EHP constructions~\cite{SAMARK,Joy_D2_jcp_2022}.\\

\begin{table}[!htp]
	\caption{Threshold energy values for chemisorption probabilities obtained for the BE and MB cases are compared, for different approaches used for the H$_2$ on Cu(111) systems.~\cite{Mantu_jcp_2024}}
	\centering
	{\footnotesize
		\begin{tabular}{l c c c c}\\
		    \hline\hline
			Method & Surface temperature (K) & CRP potential & Threshold energy (eV) BE & Threshold energy (eV) MB \\
			\hline
			SRP48 925 K phonons & 925 & 925 K BOSS PES & 0.384 & 0.527 \\
			SRP48 cluster        & 925 & 0 K BOSS PES          & 0.436 & 0.557 \\
			SRP48 0 K phonons    & 925 & 0 K BOSS PES          & 0.514 & 0.556 \\
			QD-BOSS              & $\cdots$ & 0 K BOSS PES     & 0.496 &  \\
			Experiment~\cite{Wodtke}    & 925 &                       & 0.392 &  \\
			\hline\hline
		\end{tabular}
	}
	\label{Tab:Table_2}
\end{table}

\begin{figure}[!htp]
	\centering
	\includegraphics[width=0.9\textwidth,angle=0]{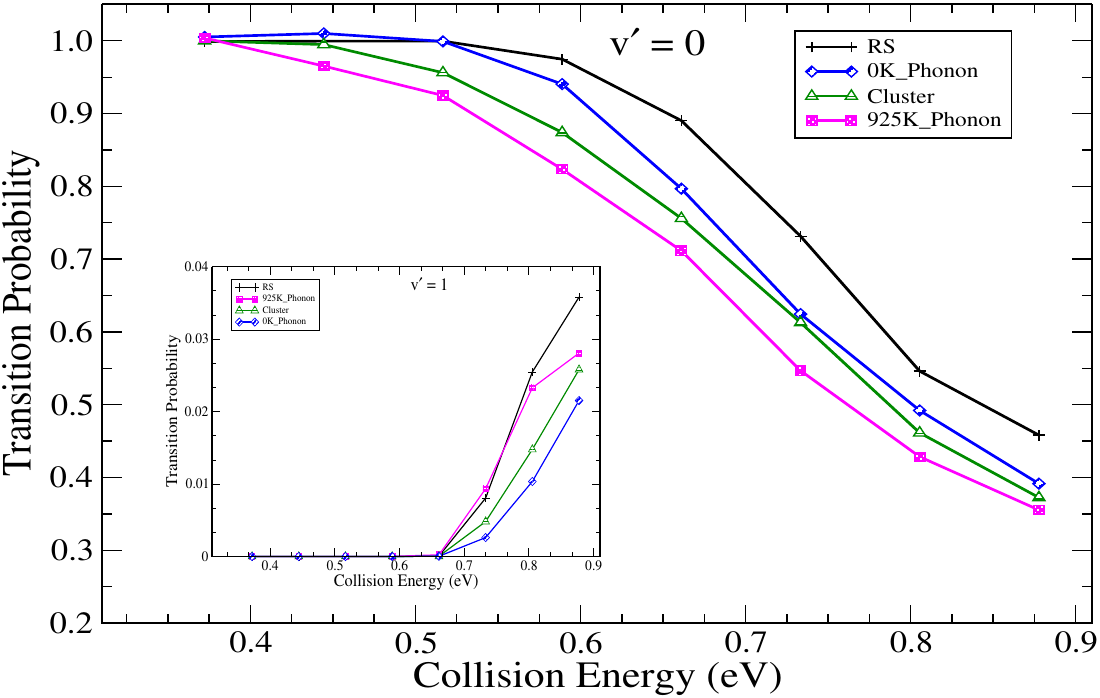}
	\caption{Energy-resolved state-to-state vibrationally elastic and inelastic transition
		probabilities are presented. This figure has been taken from Ref. [\protect\citen{Mantu_jcp_2024}] with permission from AIP.}
	\label{fig:Mantu_JCP_2024_2}
\end{figure}

\noindent
Figure~\ref{fig:Mantu_JCP_2024_2} displays the vibrational state-to-state scattering probabilities for H$_2$ $(v' = 0, 1)$ at $T_s = 925$~K, which depicts reduction of survival probabilities while including finite-temperature. On the other hand, 925~K phonon-based approach exhibits the lowest survival as various collisional energies due to enhanced reactivity. On the other hand, the RS produced the lowest chemisorption probability leading to the highest survival $(v' = 0)$ and excitation $(v' = 1)$ probabilities. Even though excitation probabilities $(v' = 1)$ remain small in all cases, relatively higher values are observed for the 925~K phonon and surface temperature.

\subsubsection{10.4.1.2. Diffraction probability for H$_2$(v=0, j=0) on Cu(111) surface} \label{ssec:diffraction_H2}

\begin{figure}[!htp]
	\centering
	\includegraphics[width=0.9\textwidth,angle=0]{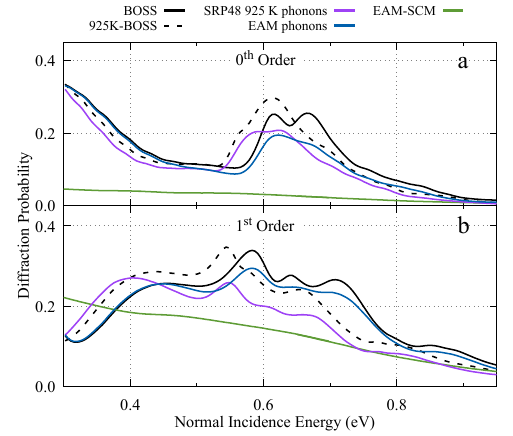}
	\caption{Diffraction probabilities for rovibrational ground-state H$_2$ scattering from Cu(111) are presented for (a) the specular and (b) first-order channels. Results are shown for BOSS at 0~K and 925~K-expanded lattice, EHP calculations with SRP48 phonons and EAM phonons, and EAM--SCM results. This figure has been taken from Ref. [\protect\citen{Bauke_jcp_2024}] with permission from AIP.}
	\label{fig:Bauke_JCP_2024_1}
\end{figure}

\noindent
Even though diffraction probability is considered as an inherently quantum mechanical observable, it is measured in molecular beam experiments~\cite{Chadwick} of H$_2$ scattering from Cu surfaces.
While employing QD methods, diffraction is naturally included without invoking additional approximations, unlike quasi-classical approaches. Hence, it enables an analysis of thermal effects originating from different treatments of surface corrugation. Figure~\ref{fig:Bauke_JCP_2024_1} depicts the zeroth-order specular diffraction peak as well as sum of the six first-order peaks as functions of the normal incidence energy, which can be associated with de Broglie wavelengths.~\cite{Bauke_jcp_2024}
At $T_s = 0$~K, both EHP and EAM based on SCM interaction potential can be treated as the BOSS case corresponding to scattering from metal atoms fixed at their ideal lattice positions.
 While moving the lattice constant from 0 to 925 K CRP potential, the BOSS curve is shifted toward lower energies (larger wavelengths) as per Bragg’s law. On the other hand, at $T_s = 925$~K, EHP-based results are similar to the BOSS trends, which revealed the dictating role of the static potential $V_0$ in the 
  diffraction peak positions and overall profiles, with only minor reductions in intensity due to increased reaction probabilities.
 In other words, the effective corrugation introduced by thermal motion is mostly confined within the EHP-based framework. Similar trend can be observed while incorporating the EAM phonons in the SCM interaction potential without considering lattice expansion.

\noindent
On the contrary, sudden approximation based EAM-SCM results exhibited pronounced thermal effects, most notably a strong attenuation of the specular diffraction peak. Such deviation mainly arises from the distinct treatment of surface motion. 
EHP takes average over surface atoms displacement around equilibrium positions, whereas EAM-SCM is based on significantly displaced surface configurations. As a result, such modification
enriches surface corrugation and reduces specular intensity, also pronounced for higher-order diffraction channels. On the other hand, the substantial reduction in specular intensity appeared from thermal attenuation effects as discussed in molecular beam experiments. It is worth mentioning that due to differences in temperature, incidence energy range and initial rovibrational states, a direct quantitative comparison between diffraction experiments on H$_2$ scattering from Cu(111) by Chadwick and Alexandrowicz~\cite{Chadwick}, and EHP based results is not feasible. Nevertheless, such validation could be examined in the upcoming calculations. It is indeed necessary to extend EAM-SCM calculations in lower temperatures, which requires careful treatment of surface configurations, particularly below the Debye temperature ($\sim$300~K). On the contrary, standard thermostat-based methods  introduce artifacts at this domain. In other words, EHP approach can be employed for accurate describing of the diffraction at lower surface temperatures. 

\subsubsection{10.4.1.3. Reaction probability for D$_2$ (v=0, j=0,2) on Cu(111) surface}\label{ssec:scattering_D2}

\begin{figure}[!htp]
	\centering
	\includegraphics[width=0.9\textwidth,angle=0]{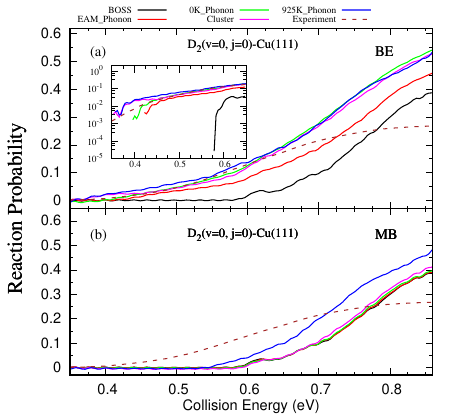}
	\caption{Reaction probabilities for D$_2$ ($v=0, j=0$) on Cu(111) at $T_s = 925$~K, computed using BE and MB probability factors within EHPs based on cluster normal modes, slab phonons (0 and 925~K), and EAM phonons are presented, which are are compared with the BOSS case (0~K ideal lattice) and experimental~\cite{CTH} data. BE results are shown on a logarithmic scale (inset). This figure has been taken from Ref. [\protect\citen{Mantu_JPCC_2026}] with permission from ACS.}
	\label{fig:Mantu_D2_JPCC_2026_1}
\end{figure}

\begin{figure}[!htp]
	\centering
	\includegraphics[width=0.9\textwidth,angle=0]{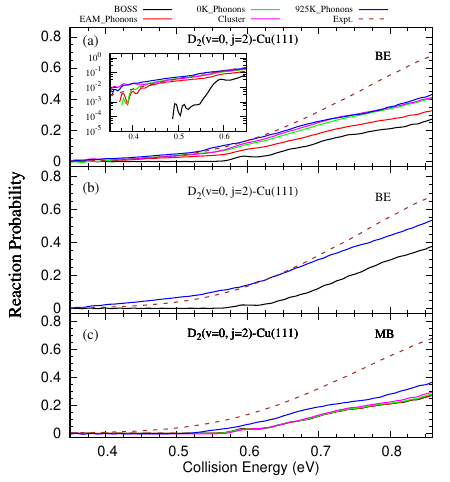}
	\caption{Reaction probabilities for D$_2$ ($v=0, j=2$) on Cu(111) at $T_s = 925$~K, calculated using BE and MB probability factors within EHPs based on cluster normal modes, slab phonons (0 and 925~K), and EAM phonons are presented and compared with the BOSS case and experimental~\cite{Wodtke} data. BE results are shown on a logarithmic scale (inset). This figure has been adapted from Ref. [\protect\citen{Mantu_JPCC_2026}] with permission from ACS.}
	\label{fig:Mantu_D2_JPCC_2026_2}
\end{figure}

\noindent
The time-dependent wavepacket calculations were executed with the help of SPO-DVR methodology to compute chemisorption probabilities for the \ce{D2}-Cu(111) system using  EHPs (see Equations~(\ref{Eq: V_eff_VT_BEP}) and (\ref{Eq: V_eff_VT_MBP})) based on cluster approximation, 0 K and 925 slab, and EAM.
Figures~\ref{fig:Mantu_D2_JPCC_2026_1} and \ref{fig:Mantu_D2_JPCC_2026_2} depict calculated dissociation probabilities in comparison with experimental data for the initial rovibrational states $(v=0,j=0)$~\cite{CTH} and $(v=0,j=2)$~\cite{Wodtke}. Keeping the surface temperature at 925~K, EHPs constructed with BE probability factor resulted into threshold energies (see Table~\ref{tab:thresh}) and reaction profiles (Figures~\ref{fig:Mantu_D2_JPCC_2026_1} and \ref{fig:Mantu_D2_JPCC_2026_2}), which show closer accord with experimental ones, particularly for the 925~K phonon and cluster-mode situations.~\cite{Mantu_JPCC_2026} Nevertheless, 
the 0~K phonon model exhibited a mismatch with respect to experimental results toward higher collision energies, whereas the EAM potential produced relatively broader distributions. In case of
$(v=0,j=2)$ initial state, consistent behavior is observed, but averaging over the $|m_j|$ components yielded chemisorption probabilities in similar trend with the $(v=0,j=0)$ case and improved correspondence with experimental threshold values. In the close neighborhood of threshold energies for reaction probability, 
quantum effects are found to be dominant one. On the other hand, at higher collision energies, the discrepancies among different models got reduced leading to converged profiles. On the contrary, 
MB statistics based EHPs showed weak dependence on the choice of phonon representation at 925~K, where the calculated chemisorption probabilities exhibited close resemblance with those obtained from the BOSS approximation. The probability profile encountered a slight broadening only for the MB-based EHP incorporating 925~K phonons. In summary, the BE-based formulation resulted into systematically higher chemisorption probabilities with lower threshold energies than the MB-based approach, which shows improved agreement with experimental observations. Such enhancement is mainly originated from the differences in statistical weight factors, which are responsible for the increase in the magnitude of the interaction potential in the BE framework. Hence, large chemisorption probabilities were observed compared to the MB description.\\

\begin{table}[h!]
	\centering
	\caption{Threshold energy values for BE and MB cases obtained using different EHPs for D$_2$ ($v=0$, $j=0,2$) on Cu(111) are summarized.~\cite{Mantu_JPCC_2026}}
	\begin{tabular}{lcccc}
		\\
		\hline \hline
		\textbf{Model} & \textbf{BE ($j=0$)} & \textbf{MB ($j=0$)} & \textbf{BE ($j=2$)} & \textbf{MB ($j=2$)} \\
		\hline
		925 K phonons (925 K BOSS) & 0.36 & 0.53 & 0.35 & 0.52 \\
		Cluster (0 K BOSS) & 0.36 & 0.59 & 0.36 & 0.57 \\
		0 K phonons (0 K BOSS) & 0.40 & 0.58 & 0.38 & 0.56 \\
		EAM phonons (925 K BOSS) & 0.42 & 0.58 & 0.39 & 0.53 \\
		QD--BOSS (0 K BOSS) & \multicolumn{2}{c}{0.58} & \multicolumn{2}{c}{0.49} \\
		Experiment & \multicolumn{2}{c}{0.37~\cite{CTH}} & \multicolumn{2}{c}{0.30~\cite{Wodtke}} \\
		\hline \hline
	\end{tabular}
	\label{tab:thresh}
\end{table}

\begin{figure}[!htp]
	\centering
	\includegraphics[width=0.9\textwidth,angle=0]{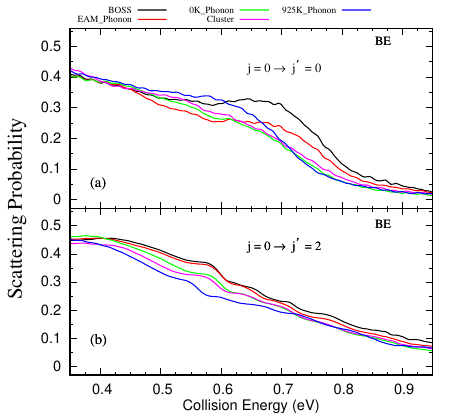}
	\caption{Scattering probabilities for rovibrational ground-state D$_2$ on Cu(111) are displayed for (a) rotationally elastic ($j=0 \rightarrow j'=0$) and (b) inelastic ($j=0 \rightarrow j'=2$) channels. Results incorporate ideal lattice BOSS and EHP calculations using cluster normal modes, 925~K phonons, 0~K phonons and EAM phonons. This figure has been adapted from Ref. [\protect\citen{Mantu_JPCC_2026}] with permission from ACS.}
	\label{fig:Mantu_D2_JPCC_2026_3}
\end{figure}

\begin{figure}[!htp]
	\centering
	\includegraphics[width=0.9\textwidth,angle=0]{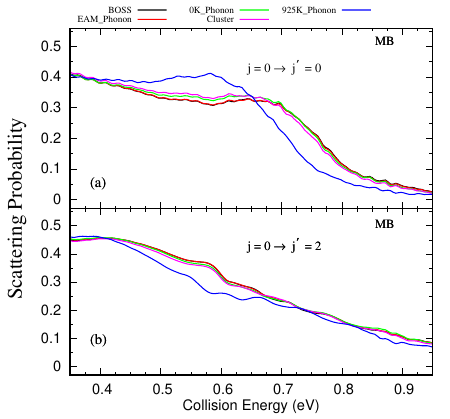}
	\caption{State-resolved scattering probabilities for rovibrational ground-state D$_2$ on Cu(111), computed for (a) rotationally elastic ($j=0 \rightarrow j'=0$) and (b) rotationally inelastic ($j=0 \rightarrow j'=2$) channels are depicted in comparison with the results obtained from ideal lattice BOSS reference and corresponding EHP-based predictions employing cluster modes, 925~K phonons, 0~K phonons and EAM phonons. This figure has been adapted from Ref. [\protect\citen{Mantu_JPCC_2026}] with permission from ACS.}
	\label{fig:Mantu_D2_JPCC_2026_MB}
\end{figure}

\noindent
On the other hand, BE probability factor based EHPs for differently calculated surface modes and lattice expansion parameter were also used to compute state-resolved scattering probabilities of the outgoing D$_2$ molecule from the Cu(111) surface, which are presented in Figure~\ref{fig:Mantu_D2_JPCC_2026_3} initiating the wavepacket at $\lbrace v=0, j=0\rbrace$.~\cite{Mantu_JPCC_2026} The survival $\lbrace j=0 \rightarrow j'=0 \rbrace$ and the excitation $\lbrace j=0 \rightarrow j'=2\rbrace$ probabilities explored a net transfer of energy from the molecule to the surface. Such characteristics are supported by the lower probabilities predicted by the EHP relative to the BOSS model over the entire collisional energy range. On the contrary, for the MB case, the survival as well as excitation probabilities calculated from different models show striking similarities with those obtained from the BOSS approximation, except few excitations for the 925~K phonon case (see Figure~\ref{fig:Mantu_D2_JPCC_2026_MB}).

\subsubsection{10.4.1.4. Diffraction probability for D$_2$ (v=0, j=0) on Cu(111) surface Surface}\label{ssec:Diffraction_D2}

\begin{figure}[htp!]
	\centering
	\includegraphics[width=0.9\textwidth,angle=0]{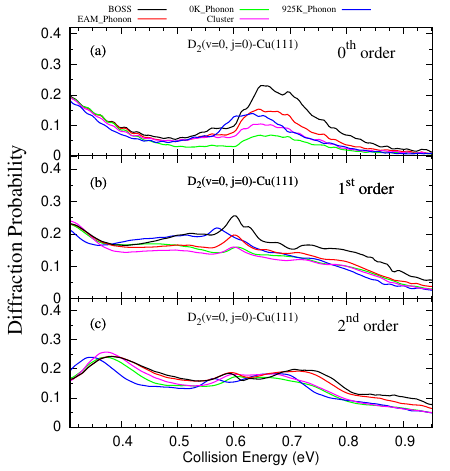}
	\caption{Diffraction probabilities for rovibrational ground-state D$_2$ scattering from Cu(111) are displayed for (a) specular, (b) first-order, and (c) second-order channels in comparison with  results obtained from ideal lattice BOSS and EHP-based calculations with cluster normal modes, 925~K phonons, 0~K phonons, and EAM phonons. This figure has been adapted from Ref. [\protect\citen{Mantu_JPCC_2026}] with permission from ACS.}
	\label{fig:Mantu_D2_JPCC_2026_5}
\end{figure}

\begin{figure}[htp!]
	\centering
	\includegraphics[width=0.9\textwidth,angle=0]{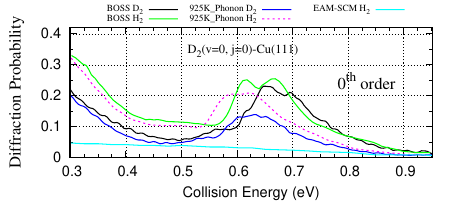}
     \caption{Comparison of specular (zeroth-order) diffraction probabilities for rovibrational ground-state H$_2$ and D$_2$ scattering from Cu(111) is presented including results of BOSS H$_2$ and BOSS D$_2$, EHP-based calculations with 925~K phonons for D$_2$ and H$_2$, and EAM--SCM~\cite{Bauke_jcp_2022} results for H$_2$. This figure has been taken from Ref. [\protect\citen{Mantu_JPCC_2026}] with permission from ACS.}
	\label{fig:Mantu_D2_JPCC_2026_6}
\end{figure}

\noindent
Figure~\ref{fig:Mantu_D2_JPCC_2026_5} depicts the zeroth (specular), first- and second-order diffraction probabilities at various values of collision energies for incident D$_2$ with initial states $(v=0, j=0)$.~\cite{Mantu_JPCC_2026} It is interesting to note that the BOSS model predicts higher diffraction probabilities than the BE-based EHP approaches throughout the entire range of collisional energy, whereas 
the corresponding reaction probabilities encountered the opposite trend. The diffraction profiles exhibit peak positions near the BOSS threshold energies. At elevated surface temperature (925~K), lattice vibrations enhance leading to increase of molecule--surface interactions. At the same time, zeroth-order diffraction intensity shows reduction, which is qualitatively consistent with thermal attenuation trends as predicted by earlier molecular beam studies. Previous theoretical~\cite{Bauke_jcp_2024} work on H$_2$ diffraction from Cu(111) indicates the 
possibility of quantitative comparison, but such study is beyond the present scope of investigation due to differences in temperature and collision energy regimes, as well as high computational cost of EHP-based quantum dynamics. On the contrary, the calculated attenuation behavior is compared for several models highlighting clearer signatures in the EAM-SCM framework (see Figure~\ref{fig:Mantu_D2_JPCC_2026_6}). In order to carry out further quantitative comparison, simulations at lower collision energies are indeed necessary with refined dynamical parameters and improved coupling descriptions, particularly incorporating  bi-linear over linear couplings. Furthermore, the limitations of surface modeling approaches need to be addressed below the Debye temperature.

\section{Conclusions and outlook} \label{sec:conclusion}

\noindent
The present section summarizes a comprehensive account of theoretical developments and numerical calculations to obtain understanding of chemisorption and scattering of diatomic molecules on/from metal surfaces, with extensive discussion on the roles of surface temperature, lattice vibrations (phonons) and elhp excitations. Over the past decades, the field was upgraded from the rigid-surface (RS) treatment toward increasingly complicated descriptions based on sudden approximation as well as time- and temperature-dependent EHPs. In other words, such calculations enable a more realistic treatment of gas--surface interactions in comparison with experimental results at various conditions.\\

\noindent
Within mean-field approximation, the formulation of EHP is highly relevant to include linear VT and bi-linear VV couplings between molecular DOFs and surface modes as well as to incorporate screened coulomb potential (Fermi-Thomas) for elhp coupling.
The Hartree product of the nuclear wavefunctions among molecular DOFs, surface modes and elhp excitation is considered as a 
physically insightful approximation due to the huge mass difference between incoming gas molecule (H$_2$/D$_2$) and the metal atoms (Cu/Ni) on the surface. It is essential to employ the Bose--Einstein (BE) and Fermi--Dirac (FD) probability factors to predict the distribution of quantum states of surface atoms as well as elhp excitations for treating the thermal effects of the surface on the broadening of reaction probabilities of gas molecule and the redistribution of ro-vibrational populations of the scattered diatom. The critical role of phonons in determining the overall reactivity and energy transfer mechanisms was explored by analyzing the linear (VT) and anharmonic (VV) couplings at elevated surface temperatures. In this theoretical framework, it is indeed necessary to use accurate surface modes, interaction potentials and QD methodologies for quantitative investigations, in particular, to explore the agreement with experimental data.\\

\noindent
Initially, EDIM potentials for metal-metal and gas-metal interactions were employed to compute the surface modes vis-\'a-vis to generate the EHP involving linear and bi-linear couplings at non-zero surface temperature, where TDDVR approach was used to carry out the QD calculations. Such treatment produced reaction and scattering probabilities to explore the agreement with experimental data. On the other hand, in order to model gas-surface interaction originating from the charge separation in the gas molecule interacting with surface electrons, Fermi-Thomas screened coulomb potential was used. Such EHP due to elhp excitations (over and above phonon coupling) was used to carry out TDDVR QD calculations. \\

\noindent
In order to carry out accurate dynamical calculations, the EHP is constructed by including SRP48-functional-based surface modes obtained from cluster approximation for 0 K normal mode and slab calculations for 0 and 925 K phonons, together with gas--surface interactions provided by the CRP-fitted BOSS PES for static lattice configurations and the SCM potential to take care the couplings between molecular DOFs and surface vibrational modes. In the subsequent work, the SPO-DVR formalism was used for propagating 6D wavepackets on time- and temperature- dependent Hamiltonians. In addition, the scattering matrices were evaluated within the 
SPO-DVR approach, which allows accurate determination of state-to-state transition probabilities, including diffraction and dissociation probability. Such calculations provide detailed mechanistic insight into gas--surface interactions. As a consequence, 
SPO-DVR-based dynamics on such accurate Hamiltonian exhibited good accord with the experimental findings, more specifically, in the threshold energy regime and in capturing quantum diffraction phenomena. The calculated results establishes the benchmark modeling and methodology for gas--surface scattering and chemisorption dynamics.\\

\noindent
In future, construction of accurate EHP will be carried out by including bi-linear (VV) over linear (VT) coupling between molecular DOFs and surface modes, which involves SRP48-functional-based phonon modes as well as SCM interaction potential. Such calculations would be followed by SPO-DVR QD to acquire even better agreement of reaction, scattering and diffraction probabilities with experimental data, both at lower and higher collisional energy domain.


\bibliographystyle{aipnum4-1}
\bibliography{myrefs_H2, refs_new,refs_article}


%
\begin{appendices}
	\chapter{Properties of NACTs: Correlation with Vector Calculus}\label{sec:app1}

\section{Divergence, Gradient and Curl}

\noindent
The gradient of a given scalar function $ f(x,y,z) $ is denoted by $ \vec{\nabla} f $ and it is defined as,
\begin{eqnarray}
	\text{grad}f = \vec{\nabla}f = [\frac{\partial f}{\partial x}, \frac{\partial f}{\partial y}, \frac{\partial f}{\partial z}] = \frac{\partial f}{\partial x}\hat{i} + \frac{\partial f}{\partial y}\hat{j}+\frac{\partial f}{\partial z}\hat{k},
\end{eqnarray}
where $ x,y,z $ are Cartesian coordinates and $ f $ is differentiable with respect to those coordinates. In other words, $\vec{\nabla}f$ is a vector field obtained from the scalar field, $f$.\\

\noindent
On the other hand, the divergence of a vector field, $ \vec{v} $, can be represented as,
\begin{eqnarray}
	\text{div}\,\,\vec{v} = \vec{\nabla}\cdot\vec{v} &=& [\frac{\partial }{\partial x}, \frac{\partial }{\partial y}, \frac{\partial }{\partial z}] \cdot[v_1, v_2, v_3]\\
	&=& (\frac{\partial }{\partial x}\hat{i} + \frac{\partial }{\partial y}\hat{j} + \frac{\partial }{\partial z}\hat{k})\cdot(v_1\hat{i}+v_2\hat{j}+v_3\hat{k})\\
	&=& \frac{\partial v_1}{\partial x} + \frac{\partial v_2}{\partial y} + \frac{\partial v_3}{\partial z}
\end{eqnarray}

\noindent
On the other hand, Curl of a vector field can be represented as below:\\

\noindent
Let $ \vec{f}(x,y,z)\,[= f_1\hat{i} + f_2\hat{j} + f_3\hat{k}]$ is a differentiable vector function of $ x,y,z $. Then, the Curl of the vector function, $\vec{f}$, is as:
\begin{eqnarray}
	\text{curl}\,\,\vec{f} &=& \vec{\nabla}\times\vec{f} = \begin{vmatrix}
		\hat{i} & \hat{j} & \hat{k}\nonumber\\
		\frac{\partial }{\partial x} & \frac{\partial }{\partial y}  & \frac{\partial }{\partial z} \\ f_1 & f_2 & f_3
	\end{vmatrix}\\
	&=& \Big(\frac{\partial f_3}{\partial y} - \frac{\partial f_2}{\partial z}\Big)\hat{i} + \Big(\frac{\partial f_1}{\partial z} - \frac{\partial f_3}{\partial x}\Big)\hat{j} + \Big(\frac{\partial f_2}{\partial x} - \frac{\partial f_1}{\partial y}\Big)\hat{k}
\end{eqnarray}

\begin{itemize}

	\item If $\vec{f}$ is represented as:
	\begin{equation}
		\vec{f} = \begin{pmatrix}
			f_1 (x,y,z)\\
			f_2 (x,y,z) \\
			f_3 (x,y,z)
		\end{pmatrix} = \begin{pmatrix}
			xy \\
			-\sin z \\
			1
		\end{pmatrix} = xy\hat{i} - \sin z\hat{j}+\hat{k}, 
	\end{equation}
	
	\noindent
	the calculated Curl of the vector field ($\vec{f}$) will be:
	\begin{eqnarray}
		\text{curl}\,\,\vec{f} &=& \vec{\nabla}\times\vec{f} = \begin{vmatrix}
			\hat{i} & \hat{j} & \hat{k}\nonumber\\
			\frac{\partial }{\partial x} & \frac{\partial }{\partial y}  & \frac{\partial }{\partial z} \\ xy & \sin z & 1
		\end{vmatrix}\\
		&=& \cos z \hat{i} - x\hat{k} =
		\begin{pmatrix}
			\cos z \\
			0 \\
			-x
		\end{pmatrix}
	\end{eqnarray}

	\item If $\vec{f}$ is taken as $\begin{pmatrix}
		x \\
		y
	\end{pmatrix}$, the associated Curl will be $\text{curl}\,\,\vec{f} = \vec{\nabla}\times\vec{f} = (\dfrac{\partial f_2}{\partial x}-\dfrac{\partial f_1}{\partial y})\hat{k}=0$., i.e. vector field becomes curl free, but non-zero divergence ($\vec{\nabla}\cdot\vec{f}=2$). Such vector field is known as irrotational vector field.\\
	
	\noindent
	\item On the other hand, if $\vec{f}$ is taken as $\begin{pmatrix}
		y \\
		x
	\end{pmatrix}$, the associated Curl will be $\text{curl}\,\,\vec{f} = \vec{\nabla}\times\vec{f} = (\dfrac{\partial f_2}{\partial x}-\dfrac{\partial f_1}{\partial y})\hat{k}=1+1=2$., i.e. vector field has non-zero curl, but zero divergence ($\vec{\nabla}\cdot\vec{f}=0$). Such vector field is known as incompressible vector field.\\

	\item If a particle is rotating with $\vec{\omega}$ angular velocity around a center in 3D coordinate space, the radical velocity can be represented as:
	\begin{equation}
		\vec{v} = \vec{\omega} \times \vec{r},
	\end{equation}
	
	\noindent
	where $\vec{\omega}$ (along the $z$ axis) and $\vec{r}$ (on the $x-y$ plane) can be represented as:
	\begin{equation}
		\vec{\omega} = 0\hat{i} + 0\hat{j} + \omega_z\hat{k}
	\end{equation}
	
	\begin{equation}
		\vec{r} = x\hat{i} + y\hat{j} + 0\hat{k}.
	\end{equation}
	
	\noindent
	Therefore, the velocity vector turns into,
	\begin{equation}
		\vec{v} = -\omega_z y \hat{i} + \omega_z x \hat{j}
	\end{equation}
	and the corresponding Curl becomes,
	\begin{eqnarray}
		\text{curl}\,\,\vec{v} &=& \vec{\nabla}\times\vec{v} = \begin{vmatrix}
			\hat{i} & \hat{j} & \hat{k}\nonumber\\
			\frac{\partial }{\partial x} & \frac{\partial }{\partial y}  & \frac{\partial }{\partial z} \\ -\omega_z y & \omega_z x & 0
		\end{vmatrix}\\
		&=& 0\hat{i} + 0\hat{j} + 2\omega_z\hat{k} 
	\end{eqnarray}
	\noindent
	Thus, Curl of $\vec{v}$ is along the $z$ axis.
\end{itemize}

\noindent
In case of a scalar field (potential), Curl of a gradient ($\vec{\nabla} \times \vec{\nabla}f$) is zero for all $f$, which is known as potential flow (like irrotational gravitational flow). On the other hand, divergence of a Curl of a vector field ($\vec{\nabla} \cdot(\vec{\nabla}\times \vec{f})$) is zero for all $\vec{f}$. In other words, curl of any vector field is always divergence free.

\section{Divergence, Stoke's and Green's Theorems}

\subsection{Divergence Theorem}
\begin{figure}
	\centering
	\includegraphics[width=0.4\textwidth]{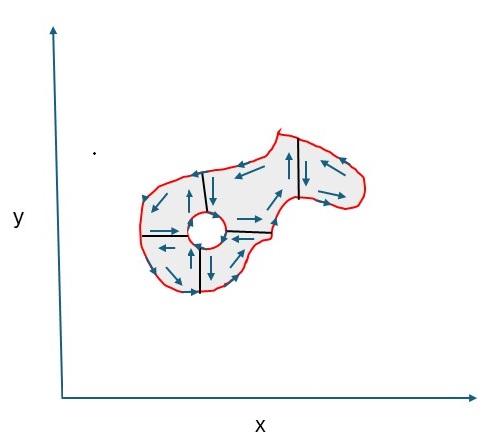}
	\caption{A three-dimensional space is partitioned into infinitesimal boxes. The figure has been adapted from the book of Kreszig~\citep{kreyszig}.} 
	\label{fig:fig24}
\end{figure}

\noindent
Let the volume, $ V $ is a bounded region in space enclosed  by a piecewise smooth surface $ S $. If $ \vec{F}(x,y,z) $ is a vector function, which is continuous and its first derivatives (partial) are defined in $ V $. For such a vector function, $ \vec{F}(x,y,z) $, we have the following equality, \begin{eqnarray}
	\iiint\limits_V \text{div}\vec{F}dV = \iint\limits_S\vec{F}\cdot\vec{n}dA,
	\label{eq:diven}
\end{eqnarray}
which is known as Gauss's divergence theorem. In other words, the flux of a vector field, $\vec{F}$ through a closed surface, $S$ is equal to the integral of the divergence, $\text{div}\vec{F}$ over the entire enclosed volume, $V$. In Eq. \ref{eq:diven}, $\vec{F}\cdot\vec{n}$ represents the flux of $\vec{F}$ through the surface, $S$. If a three-dimensional space is partitioned into infinitesimal boxes (see Figure \ref{fig:fig24}), the inner components cancel out each other. In other words, the unbalanced components of  $\vec{F}$ on the outer surface should be equal to the $\text{div}\vec{F}$ over the entire enclosed volume. \\

\newpage							

\begin{itemize}			
	
	\item If $ \vec{F} = [F_1, F_2, F_3] $ and outer unit normal vector of $ S $ is $ \vec{n} = [\cos\alpha, \cos\beta, \cos\gamma] $, then one can write using Eqs. \ref{eq:diven}: \begin{eqnarray}
		\iiint\limits_V\Big( \frac{\partial F_1}{\partial x} + \frac{\partial F_2}{\partial y } + \frac{\partial F_3}{\partial z}\Big)dxdydz &=& \iint\limits_S(F_1\cos\alpha + F_2\cos\beta + F_3\cos\gamma)dA \qquad \quad\\ &=& \iint\limits_S (F_1dydz + F_2dzdx + F_3dxdy) \qquad
	\end{eqnarray}
	
	\item	Let, $ \vec{F} = 7x\hat{i} -z\hat{k} $. Verify divergence theorem for a sphere of radius 2\\
	
	\noindent
	$\text{div} \vec{F} = \text{div}[7x\hat{i} - z\hat{k}]  = 7-1 = 6$.
	\[\iiint\limits_V \text{div} \vec{F}dV = \iiint\limits_V 6dV = 6\frac{4}{3}\pi2^3 = 64\pi \]
	
	On the other hand, equation of a sphere S: $ x^2 + y^2 + z^2 = 4 $\\
	
	The parametric equation of this sphere is given by, \begin{eqnarray}
		2\cos v \cos u \hat{i} + 2\cos v \sin u \hat{j} + 2\sin v \hat{k} = \vec{r} \nonumber
	\end{eqnarray}
	where, $ 0\le u\le 2\pi, -\frac{\pi}{2}\le v \le \frac{\pi}{2} $\\$ \therefore $
	\begin{eqnarray}
		\vec{r} &=& [2\cos v \cos u, 2\cos v \sin u , 2\sin v ]\nonumber\\
		\vec{r}_u &=& \dfrac{\partial\vec{r}}{\partial u} = [-2\cos v \sin u, 2\cos v \cos u , 0 ]\nonumber\\
		\vec{r}_v &=& \dfrac{\partial\vec{r}}{\partial v} = [-2\sin v \cos u, -2\sin v \sin u , 2\cos v ]\nonumber
	\end{eqnarray}
	\[\vec{N} = \vec{r_u}\times\vec{r_v} =[4\cos^2v\cos u , 4\cos^2v\sin u , 4\cos v \sin v] \]
	$ \therefore\quad \vec{F} = [14\cos v \cos u, 0, -2\sin v] $\\
	$ \therefore\quad \vec{F}\cdot\vec{N} = 56\cos^3v\cos^2u - 8\cos v \sin^2 v $
	
	\begin{eqnarray}
		\iint\limits_S \vec{F}\cdot\vec{n}dA &=& \int_{v=-\frac{\pi}{2}}^{\frac{\pi}{2}}\int_{u = 0}^{2\pi}\vec{F}(s)\cdot\vec{N}dudv\\
		&=&\int_{v=-\frac{\pi}{2}}^{\frac{\pi}{2}}\int_{u = 0}^{2\pi}(56\cos^3v\cos^2u - 8\cos v \sin^2 v)dudv \qquad\\
		&=& 64\pi
	\end{eqnarray}
	
	\noindent
	Here we verify the Divergence theorem with the above example.
	
	\item \underline{Mass Continuity Equation}:
	
	\noindent
	Total mass in an enclosed volume is represented as $\iiint_{V}\rho dV$, where $\rho$ indicates the density of the system. Since increase of mass inside the volume due to the flow can be equated to the outgoing mass from the outer surface, one can write
	
	\begin{eqnarray}
		\dfrac{d}{dt} \iiint_{V}\rho dV  &=& -\iint\limits_S\rho\vec{F}\cdot\vec{n}dA\nonumber \\
		&=& - \iiint\limits_V\vec{\nabla}\cdot(\rho\vec{F})dV\nonumber\\
		\Rightarrow \iiint\limits_V \Big(\dfrac{d\rho}{dt}+\vec{\nabla}\cdot(\rho\vec{F})dV\Big) &=&0 
	\end{eqnarray}
	
	\noindent
	Therefore, the mass continuity equation can be represented as:
	\begin{equation}
		\dfrac{d\rho}{dt}+\vec{\nabla}\cdot(\rho\vec{F})dV = 0,
	\end{equation}
	which indicates conservation of mass in a flowing fluid.				
	
\end{itemize}

\noindent
At this junction, it is important to note that Gauss's divergence theorem is not applicable for NACTs ($\vec{\boldsymbol{\tau}}$) in case of Jahn-Teller (JT) couplings, but could be useful for Pseudo Jahn-Teller (PJT) interactions. Since the NACTs could be singular for JT case, it is not possible to show the above equality.

\subsection{Stoke's Theorem}

\begin{figure}[htp]
	\centering
	\includegraphics[width=0.6\textwidth]{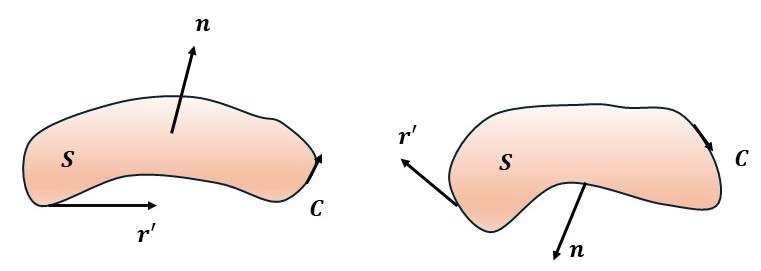} 
	\caption{Demonstration of Stokes's theorem. The figure has been adapted from the book of Kreszig.~\citep{kreyszig}}
	\label{fig:fig28}
\end{figure}

\noindent
Let $ S $ be a piecewise  smooth surface closed by a curve, $ C $. If $ \vec{F}(x,y,z) $ is a continuous vector function with continuous first order partial derivatives, the following equality can be shown: \[\iint\limits_S(\text{curl }\vec{F})\cdot\hat{n}dA = \oint_C \vec{F}\cdot\vec{r}\,\,^{\prime}(s)ds\]
where $ \hat{n} $ is unit normal vector of $ S,\ \vec{r}\,\,^{\prime}(=\frac{d\vec{r}}{ds}) $ is tangent vector along the curve $C$, where $ s $  is a parameter. Thus, the above equation
can be written as:
\[\iint\limits_R \big[\big(\frac{\partial F_3}{\partial y} - \frac{\partial F_2}{\partial z}\big)N_1 + \big(\frac{\partial F_1}{\partial z} - \frac{\partial F_3}{\partial x}\big)N_2  + \big(\frac{\partial F_2}{\partial x} - \frac{\partial F_1}{\partial y}\big)N_3 \big]dudv = \oint_C (F_1dx + F_2dy + F_3dz)\]

\noindent
where $ \vec{F}=[F_1,F_2,F_3]$ and therefore, $\text{curl }\vec{F}=\Big[\frac{\partial F_3}{\partial y} - \frac{\partial F_2}{\partial z},\frac{\partial F_1}{\partial z} - \frac{\partial F_3}{\partial x},\frac{\partial F_2}{\partial x} - \frac{\partial F_1}{\partial y}\Big]$. Similarly,
$  \vec{n}dA = \vec{N}dudv$ and $ \vec{N} = [N_1,N_2,N_3]$.\\

\noindent
On the other hand, $\oint_C \vec{F}\cdot\vec{r}\,\,^{\prime}(s)ds=\oint_C \vec{F}\cdot d\vec{r}=
\oint_C (F_1\hat{i}+F_2\hat{j}+F_3\hat{k})\cdot (dx\hat{i}+dy\hat{j}+dz\hat{k})=\oint_C (F_1dx + F_2dy + F_3dz)$\\

\noindent
As NACTs ($\vec{\boldsymbol{\tau}}$) are vector functions over the Hilbert space either for JT or PJT interactions, there is a possibility to apply Stoke's theorem on such (vector) functions to depict the role of electron-nuclear coupling. On the contrary, for JT coupling, since NACTs are singular at a specific point or along a seam in the Hilbert space, this theorem cannot be implemented as the function has to be defined and continuous at each point of the configuration space. On the other hand, for PJT interaction, the NACTs are smooth functions of the nuclear coordinates and therefore, the Stoke's theorem can be applied to find out its contribution over the configuration space for the molecular species or process. 

\subsubsection*{\large A2.3 Green's Theorem}

\begin{figure}[htp]
	\centering
	\includegraphics[width=0.4\textwidth]{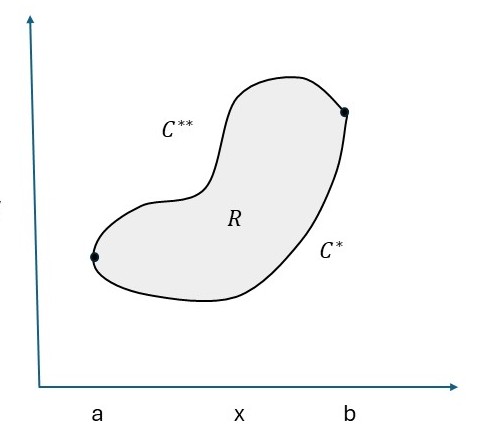}
	\caption{\small The closed path $C$ has two parts, $C^*$ and $C^{**}$. i.e. $C=C^*+C^{**}$. The figure has been adapted from the book of Kreszig.~\citep{kreyszig}}
	\label{fig:green}
\end{figure}

\noindent		
Let	$ R $ is a closed region in the x-y plane and the boundary $ C $ consists finite numbers of smooth curves (say $C^*$ and $C^{**}$, see Figure \ref{fig:green}). In that space, let $ F_1 (x,y)$ and $ F_2(x,y) $ are two continuous functions with partial derivatives $ \frac{\partial F_1}{\partial y} $ and $ \frac{\partial F_2}{\partial x} $ everywhere in the domain containing $ R $. Then, the following integrals show equality known as Green's Theorem:  \begin{eqnarray}\label{eq:green}
	\iint\limits_R \Big( \frac{\partial F_2}{\partial x} - \frac{\partial F_1}{\partial y}\Big)dxdy = \oint\limits_C (F_1dx + F_2dy) 
\end{eqnarray}

\noindent
The integration is carried out along the entire boundary $ C $ of $ R $ in such a way that $ R $ is on the left of the boundary as we advance in the direction of integration. Green's theorem can be used to get the surface integral using a closed line integral. \\

\noindent
If $ \vec{F} = [F_1, F_2] = F_1\hat{i} + F_2\hat{j}$, Eq \ref{eq:green} becomes,\\
\begin{eqnarray}
	\iint\limits_R(\text{curl}\vec{F})\cdot\hat{k}dxdy = \oint\limits_C\vec{F}\cdot\vec{dr},
\end{eqnarray}
where $\hat{k}dxdy$ is the surface vector ($k$ is unit vector along the $z$ axis) and $\vec{dr}$ is $dx\hat{i}+dy\hat{j}$.\\

\noindent
\underline{Green's Theorem excluding some portion (A) of the region}:\\

\noindent
This is important if a region contains singularity. The specific part of the region (A) needs to be eliminated for applying Green's Theorem.\\

\begin{figure}[htp]
	\centering
	\includegraphics[width=0.4\textwidth]{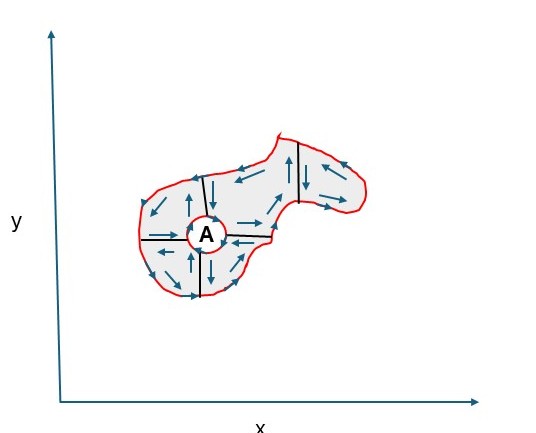}
	\caption{Demonstration of Green's theorem excluding some portion (A) of the region.  The figure has been adapted from the book of Kreszig.~\citep{kreyszig}}
	\label{fig:fig28}
\end{figure}

\noindent
For the above case, the theorem is applied in each sub region and the results are added. The integrals cancels each other at the common boundaries and the line integrals only for external and internal boundary survives. \\

\noindent
Green's theorem is also not applicable for NACTs, ($\vec{\boldsymbol{\tau}}$), since $\vec{F}$ (here, it is $\vec{\boldsymbol{\tau}}$) must be analytical function to fulfill such relation.

\section{Helmholz Decomposition}

\noindent
If $f(x,y,z)$ represents a scalar field, $\vec{\nabla}f$ symbolizes a vector field. On the contrary, if $\vec{F}$ indicates 
any vector field, it may or may not be equal only with $\vec{\nabla}f$. If it is possible to express $\vec{F}$ as $\vec{\nabla}f$, such vector field is known as gradient flow, which actually represents a conservative field with no gain or loss in energy. On the other hand, while taking curl of a vector field, $\vec{F}$, the magnitude [$=\vec{\nabla}\times(\vec{\nabla}f)$] becomes zero (0), since the mixed partial derivatives cancel out each other leading to no rotational component (characteristic of conservative fields). In other words, gradient flows are curl free. 
In case of gradient flow ($\vec{F}=\vec{\nabla}f$), $\vec{\nabla}\times\vec{F}$ is always zero and therefore, it is irrotational. On the other hand, if both 
$\vec{\nabla}\times\vec{F}$ as well as $\vec{\nabla}.\vec{F}$ take zero values, the force becomes irrotational as well as incompressible and such flow is known as potential flow. In other words, potential flow is a subset of gradient flow. Gravitational potential can be considered as one of the example of potential flow. \\

\noindent
On the contrary, as a matter of fact, the field, $\vec{F}$ can be decomposed into:
\begin{equation}
	\vec{F} = -\vec{\nabla}\phi + \vec{\nabla}\times\vec{A},
	\label{eq:helmlotz}
\end{equation}  
where $\vec{A}$ represents another vector field. Such decomposition is known as Helmholtz or Hedge decomposition. The second component is known as solenoidal. In case of a moving charged particle in presence of a magnetic field, the associated vector field ($\vec{F}$) has two components, one is related to scalar potential and other one is from vector potential. \\

\noindent
The force ($\vec{F}$) in a molecular process or chemical reaction originates from: (a) the Coloumbic interaction of the sub-atomic species (electron-electron, nucleus-nucleus and electron-nucleus) that can be equated to the derivative of a scalar function (see the first term of Eq. \ref{eq:helmlotz}); and (b) the Curl of a vector function known as nonadiabatic coupling, NACT ($\vec{\boldsymbol{\tau}}$). Considering the dimension of the Hilbert space, the vector function, $\vec{\boldsymbol{\tau}}$ can be decomposed into longitudinal or removable component and transverse or non-removable component. Adiabatic-to-Diabatic Transformation (ADT) can at best remove the longitudinal component, whereas the second part is generally negligible or can be made negligible by expanding the nuclear
configuration space (CS) and/or the sub-Hilbert space.~\citep{mead_jcp77} 

\section{Potential Flow and Laplace's Equation}	

\noindent
In case of steady, incompressible, irrotational flow, a two-dimensional (2D) vector field can be expressed as:

\begin{equation}
	\vec{F} = \begin{pmatrix}
		F_1 \\
		F_2
	\end{pmatrix}
\end{equation}

\noindent
For steady flow, the following relation holds true:
\begin{equation}
	\dfrac{\partial \vec{F}}{\partial t} = 0,
\end{equation}
since $\vec{F}$ does not change with time. On the other hand, incompressible and irrotational properties of the flow lead to the following expressions:
\begin{equation}
	\vec{\nabla}.\vec{F} = 0
	\label{eq:lap1}
\end{equation}
\begin{equation}
	\vec{\nabla}\times\vec{F} = 0
	\label{eq:lap2}
\end{equation}

\noindent
From Eqs. \ref{eq:lap1} and \ref{eq:lap2}, one can write,
\begin{equation}
	\dfrac{\partial F_1}{\partial x} + \dfrac{\partial F_2}{\partial y} = 0 \qquad \text{(incompressible)}
	\label{eq:lap3}
\end{equation}
\begin{equation}
	\dfrac{\partial F_2}{\partial x} - \dfrac{\partial F_1}{\partial y} = 0 \qquad \text{(irrotational)}
	\label{eq:lap4}
\end{equation}

\noindent
The above two equations are satisfied only when $\vec{F} = \vec{\nabla}.\phi$ is considered. For a real-valued potential, $\phi$ satisfies the following equation:
\begin{equation}
	\nabla^2\phi = 0,
	\label{eq:lap5}
\end{equation}
which is known as Laplace's equation. Such equation can be proven using Eq. \ref{eq:lap1} as follows:
\begin{eqnarray}
	\vec{\nabla}.\vec{F} &=& \begin{pmatrix}
		\dfrac{\partial}{\partial x} & \dfrac{\partial}{\partial y} 
	\end{pmatrix}.\begin{pmatrix}
		\dfrac{\partial \phi}{\partial x} \\
		\dfrac{\partial \phi}{\partial y}
	\end{pmatrix}\nonumber\\
	&=& \dfrac{\partial^2 \phi}{\partial x^2}+\dfrac{\partial^2 \phi}{\partial y^2} = \nabla^2\phi = 0.
\end{eqnarray} 

\noindent
On the other hand, if we consider a complex potential, $\Phi(z)$, it can be segregated into two parts as follows:
\begin{equation}
	\Phi(z) = \phi(x,y) + i\psi(x,y),
\end{equation}
where the real part [$\phi(x,y)$] is known as potential function and imaginary part [$\psi(x,y)$] is denoted as stream function. If $\Phi(z)$ is an analytic function, each of the real and imaginary parts satisfies the Laplace's equation:
\begin{equation}
	\nabla^2\phi = 0
\end{equation}
\begin{equation}
	\nabla^2\psi = 0
\end{equation}

\noindent
In other words, in case of a potential flow, the following three equations hold true:

\begin{equation}
	\dfrac{d\vec{F}}{dt} = 0,
\end{equation}
\begin{equation}
	\vec{\nabla}.\vec{F} = 0
\end{equation}
\begin{equation}
	\vec{\nabla}\times\vec{F} = 0
\end{equation}	

\noindent
for steady state, incompressible and irrotational nature. \\

\noindent
It is evident from the above formulation that Laplace equation is valid only for analytic functions. Since NACTs often exhibit singular nature at the close vicinity of degenerate point(s), the associated force field does not fulfill Laplace's equation.

\section{Application of Divergence and Stokes's Theorem for Potential Flow}

\begin{figure}[htp]
	\centering
	\includegraphics[width=0.6\textwidth]{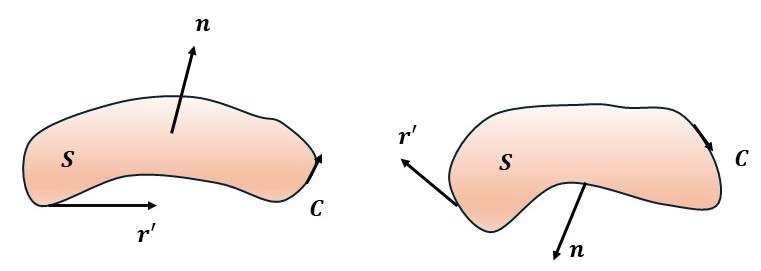} 
	\caption{Potential flow inside a closed loop}
\end{figure}		

\noindent
As per the above figure, a vector field (potential flow), $\vec{F}$ is passing through a surface, $S$. As per the Stokes's theorem, one can write using the properties of potential flow:

\begin{equation}
	\iint\limits_S(\text{curl}\vec{F})\cdot\hat{n}dA = \oint_C \vec{F}\cdot\vec{r}\,\,^{\prime}(s)ds = 0,
\end{equation}
which means zero net circulation. On the other hand, according to divergence theorem:

\begin{eqnarray}
	\iiint\limits_T \text{div}\vec{F}dV = \iint\limits_S\vec{F}\cdot\vec{n}dA = 0,
\end{eqnarray}
which means zero net flux over the surface $S$. \\

\noindent
If we consider, $\vec{F}=\vec{\nabla}\Phi$, one can introduce the complex potential as follows,
\begin{equation}
	\Phi(z) = \phi(x,y)+i\psi(x,y)
	\label{eq:complex}
\end{equation}
If $\Phi(z)=z = x+iy$ holds true, one can write $z^2 = x^2 - y^2 + i2xy$, where the real component, $x^2 - y^2$ and imaginary component, $2xy$ represent potential and stream flow, respectively. For a complex potential as presented in Eq. \ref{eq:complex}, one can write:
\begin{equation}
	\vec{\nabla}\Phi=\begin{pmatrix} 
		\dfrac{\partial\phi}{\partial x} \vspace{0.3 cm} \\ 
		\dfrac{\partial\phi}{\partial y}
	\end{pmatrix} =\begin{pmatrix}
		\dfrac{\partial\psi}{\partial y} \vspace{0.3 cm} \\ 
		-\dfrac{\partial\psi}{\partial x}
	\end{pmatrix} 
	\label{eq:complex1}
\end{equation}

\noindent
Since $\vec{F}=\vec{\nabla}\Phi$ and as per Eq. \ref{eq:complex1}, the following equations can be written:
\begin{equation}
	\vec{F}=\vec{\nabla}\Phi=\begin{pmatrix}
		2x \\
		-2y
	\end{pmatrix}
\end{equation}
which leads to the following relation:

\begin{equation}
	\vec{\nabla}.\vec{F} = 2-2 = 0
\end{equation}
\begin{equation}
	\vec{\nabla}\times\vec{F} = 0+0 = 0
\end{equation}
\begin{equation}
	\dfrac{\partial^2\phi}{\partial x^2}+\dfrac{\partial^2\phi}{\partial y^2}=2-2=0
\end{equation}
\begin{equation}
	\dfrac{\partial^2\psi}{\partial x^2}+\dfrac{\partial^2\psi}{\partial y^2}=0+0=0
\end{equation}

\noindent
In other words, Laplace equation is valid for real as well as conservative force fields.

\section{Cauchy's Residue Theorem}

\noindent
If $f(z)$ is an analytic function in the region $A$ except for a set of isolated points with singularities and $C$ represents a curve encapsulating the area, $A$, Cauchy's residue theorem predicts:

\begin{equation}
	\int_C f(z)dz = 2\pi i B,
\end{equation}

\noindent
where $B = \Sigma \,\text{residues of}\, f(z)\, \text{inside}\, C$. It is to note that the closed curve ($C$) does not intersect or encounter any singularities and is oriented counterclockwise. \\

\noindent
\underline{Proof}~\citep{Shankaraiah} :

\begin{figure}[htp]
	\centering
	\includegraphics[width=0.6\textwidth]{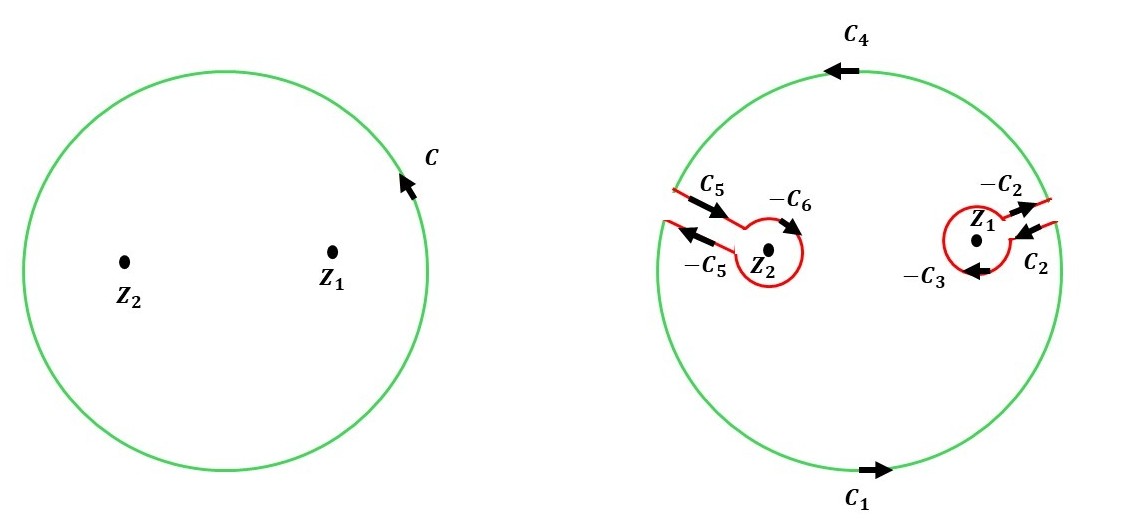} 
	\caption{Circular loops enclosing poles of singularity. The figure has been adapted Ref. [\protect\citen{Shankaraiah}].}
	\label{fig:cauchy}
\end{figure}

\noindent
Figure \ref{fig:cauchy}(i) depicts the curve $C$ surrounding two poles, $z_1$
and $z_2$. On the other hand, Figure \ref{fig:cauchy}(ii) exhibits a curve ($\tilde{C}$), which excludes the two poles, $z_1$ and $z_2$. \\

\noindent
As per Figure \ref{fig:cauchy}(ii), the closed curve, $\tilde{C}$ can be written as sum of the smaller parts of the curve:
\begin{equation}
	\tilde{C} = C_1+C_2-C_3-C_2+C_4+C_5-C_6-C_5
\end{equation}

\noindent
Since no poles are present inside $\tilde{C}$ in Figure \ref{fig:cauchy}(ii), Cauchy’s integral theorem depicts:

\begin{eqnarray}
	\int_{\tilde{C}} f(z)dz&=&\int_{C_1+C_2-C_3-C_2+C_4+C_5-C_6-C_5} f(z)dz  \vspace {0.5 cm}\nonumber\\
	&=&\int_{C_1} f(z)dz + \int_{C_2} f(z)dz - \int_{C_3} f(z)dz -\int_{C_2} f(z)dz\vspace {0.5 cm} \nonumber\\
	&& + \int_{C_4} f(z)dz + \int_{C_5} f(z)dz -\int_{C_6} f(z)dz-\int_{C_5} f(z)dz \vspace {0.5 cm} \nonumber\\
	&=& 0
\end{eqnarray}
which leads to,
\begin{equation}
	\int_{C_1+C_4} f(z)dz = \int_{C_3+C_6} f(z)dz = \int_{C} f(z)dz
	\label{eq:cont}
\end{equation}

\noindent
If we consider the function, $f(z)$ as the following Laurent expansion around $z_1$, one can obtain:
\begin{equation}
	f(z) = \cdots \cdots \dfrac{b_2}{(z-z_1)^2} + \dfrac{b_1}{(z-z_1)} + a_0 + a_1(z-z_1) + a_2(z-z_1)^2 \cdots \cdots  
\end{equation}
and the associated integrals take the following forms:
\begin{eqnarray}
	\int_{C_3} f(z)dz &=& \int_{c_3} \cdots \cdots \dfrac{b_2}{(z-z_1)^2} + \dfrac{b_1}{(z-z_1)} + a_0 + a_1(z-z_1) + a_2(z-z_1)^2 \cdots \cdots dz \nonumber\\
	&=&2\pi i b_1\nonumber\\
	&=&2\pi iRes(f,z_1) \qquad \text{see Figure \ref{fig:cauchy}(ii)} \\
	\int_{C_6} f(z)dz &=& 2\pi iRes(f,z_2) \qquad \text{see Figure \ref{fig:cauchy}(ii)}
\end{eqnarray}

\noindent
Using Eq. \ref{eq:cont}, one can obtain:
\begin{equation}
	\int_{C} f(z)dz = 2\pi i[Res(f,z_1)+Res(f,z_2)]	
\end{equation}

\noindent
The above formulation validates the residue theorem for the case of two poles, which can be generalized to any number of poles.\\

\noindent
Since NACTs ($\vec{\boldsymbol{\tau}}$) are often singular function at the close vicinity of degenerate electronic states, the associated contour integral along a closed loop follows Cauchy's residue theorem. In case of NACTs, one can write:
\begin{equation}
	\int_0^{2\pi} \tau^{ij}_{\phi}(\phi \vert \bar{\rho}, \bar{\theta})d\phi = 0\,\, \text{or}\,\, n\pi\,\, \text{or}\,\, n.(2\pi),
	\label{eq:quant_app}
\end{equation}
which reflects the presence of Pseudo Jahn-Teller (PJT), Jahn-Teller (JT), or Renner-Teller (RT) interactions, respectively.

\bibliographystyle{aipnum4-1}
\bibliography{ref_book}

	\chapter{Explicit Expressions in BBO Formulation}\label{sec:bbo}

\section{Detailed Expressions of ADT Equations, NACTs and Curl Equations}\label{ssec:adt-eq}

\noindent
ADT equations for four~\citep{sarkar_jpca112}, five~\citep{soumya_jpca121} and six~\citep{soumya_tfb} electronic states SHS are as follows:\\

\noindent
{\bf Four State ($N=4$)}:~\citep{sarkar_jpca112}
\begin{scriptsize}
	\begin{subequations}
		\begin{eqnarray*}
			\boldsymbol{\nabla}_R\Theta^{12}&=& -\boldsymbol{\tau}^{12}-\sin\Theta^{12}\tan\Theta^{13}\boldsymbol{\tau}^{13}-\cos\Theta^{12}\tan\Theta^{13}
			\boldsymbol{\tau}^{23}-\sin\Theta^{12}\sec\Theta^{13}\tan\Theta^{14}\boldsymbol{\tau}^{14}-\cos\Theta^{12}\sec\Theta^{13}\tan\Theta^{14}
			\boldsymbol{\tau}^{24} \\
			\boldsymbol{\nabla}_R\Theta^{13}&=& -\cos\Theta^{12}\boldsymbol{\tau}^{13}+\sin\Theta^{12}\boldsymbol{\tau}^{23}-\cos\Theta^{12}\sin\Theta^{13}
			\tan\Theta^{14}\boldsymbol{\tau}^{14}+\sin\Theta^{12}\sin\Theta^{13}\tan\Theta^{14}\boldsymbol{\tau}^{24}-\cos\Theta^{13}\tan\Theta^{14}\boldsymbol{\tau}^{34}\\
			\boldsymbol{\nabla}_R\Theta^{23}&=&-\cos\Theta^{13}[\boldsymbol{\tau}^{13}\sin\Theta^{12}\sec^2\Theta^{13}+\cos\Theta^{23}\sec\Theta^{14}(\boldsymbol{\tau}^{34}-\boldsymbol{\tau}^{24}\sin\Theta^{12}\tan\Theta^{13})\tan\Theta^{24}\nonumber\\
			&&+\sin\Theta^{12}\sec\Theta^{13}\tan\Theta^{13}\boldsymbol{\tau}^{14}\tan\Theta^{14}+\sin\Theta^{23}\boldsymbol{\tau}^{14}\sec\Theta^{14}\tan\Theta^{24}+\cos\Theta^{12}\lbrace\boldsymbol{\tau}^{23}\sec^2\Theta^{13}\nonumber \\
			&&+\tan\Theta^{13}\cos\Theta^{23}\boldsymbol{\tau}^{14}\sec\Theta^{14}
			\tan\Theta^{24}+\sec\Theta^{13}(\tan\Theta^{13}\boldsymbol{\tau}^{24}
			\tan\Theta^{14}\nonumber \\
			&&+\sin\Theta^{23}\boldsymbol{\tau}^{24}\sec\Theta^{14}
			\tan\Theta^{24})\rbrace]\\
			\boldsymbol{\nabla}_R\Theta^{14}&=&-\cos\Theta^{12}\cos\Theta^{13}\boldsymbol{\tau}^{14}+\sin\Theta^{12}\cos\Theta^{13}\boldsymbol{\tau}^{24}
			+\sin\Theta^{13}\boldsymbol{\tau}^{34}\\
			\boldsymbol{\nabla}_R\Theta^{24}&=&\sin\Theta^{23}(-\sin\Theta^{12}\sin\Theta^{13}\boldsymbol{\tau}^{24}\sec\Theta^{14}+
			\cos\Theta^{13}\boldsymbol{\tau}^{34}\sec\Theta^{14})-
			\sin\Theta^{12}\cos\Theta^{23}\boldsymbol{\tau}^{14}\sec\Theta^{14}\nonumber \\
			&&+\cos\Theta^{12}(\sin\Theta^{13}
			\sin\Theta^{23}\boldsymbol{\tau}^{14}\sec\Theta^{14}-\cos\Theta^{23}\boldsymbol{\tau}^{24}\sec\Theta^{14}) \\
			\boldsymbol{\nabla}_R\Theta^{34}&=&\sin\Theta^{12}\lbrace-\sin\Theta^{23}\boldsymbol{\tau}^{14}\sec\Theta^{14}\sec\Theta^{24}+\sin\Theta^{13}\cos\Theta^{23}\boldsymbol{\tau}^{24}\sec\Theta^{14}\sec\Theta^{24}\rbrace+\cos\Theta^{12}[-\sec\Theta^{24}\nonumber\\
			&&\lbrace\sin\Theta^{13}\cos\Theta^{23}\boldsymbol{\tau}^{14}\sec\Theta^{14}+\sin\Theta^{23}
			\boldsymbol{\tau}^{24}\sec\Theta^{14}\rbrace]-\cos\Theta^{13}\cos\Theta^{23}\boldsymbol{\tau}^{34}\sec\Theta^{14}\sec\Theta^{24}
		\end{eqnarray*}
		\label{eq:adteq1}
	\end{subequations}
\end{scriptsize}			

\newpage
\noindent
{\bf Five State ($N=5$)}:~\citep{soumya_jpca121}
\begin{scriptsize}
	\begin{subequations}
		\begin{eqnarray*}
			\boldsymbol{\nabla}_R\Theta^{12}&=&-{\boldsymbol{\tau}}^{12}-\sin\Theta^{12}({\boldsymbol{\tau}}^{13}\tan\Theta^{13}+{\boldsymbol{\tau}}^{14}\sec\Theta^{13}
			\tan\Theta^{14}+{\boldsymbol{\tau}}^{15}\sec\Theta^{13}\sec\Theta^{14}\tan\Theta^{15})-\cos\Theta^{12}\lbrace{\boldsymbol{\tau}}^{23}
			\tan\Theta^{13}\nonumber \\
			&&+\sec\Theta^{13}({\boldsymbol{\tau}}^{24}\tan\Theta^{14}+{\boldsymbol{\tau}}^{25}\sec\Theta^{14}\tan\Theta^{15})
			\rbrace\\
			\boldsymbol{\nabla}_R\Theta^{13}&=&-\cos\Theta^{13}({\boldsymbol{\tau}}^{34}\tan\Theta^{14}+{\boldsymbol{\tau}}^{35}\sec\Theta^{14}\tan
			\Theta^{15})+\sin\Theta^{12}({\boldsymbol{\tau}}^{23}+{\boldsymbol{\tau}}^{24}\sin\Theta^{13}\tan\Theta^{14}+{\boldsymbol{\tau}}^{25}
			\sin\Theta^{13}\sec\Theta^{14}\nonumber \\
			&&\tan\Theta^{15})-\cos\Theta^{12}\lbrace{\boldsymbol{\tau}}^{13}+\sin\Theta^{13}({\boldsymbol{\tau}}^{14}\tan\Theta^{14}+{\boldsymbol{\tau}}^{15}\sec\Theta^{14}
			\tan\Theta^{15})\rbrace\\
			\boldsymbol{\nabla}_R\Theta^{23}&=&-\cos\Theta^{13}[{\boldsymbol{\tau}}^{13}\sin\Theta^{12}\sec^2\Theta^{13}+\cos\Theta^{23}\lbrace\sec\Theta^{14}({\boldsymbol{\tau}}^{34}-{\boldsymbol{\tau}}^{24}\sin\Theta^{12}\tan\Theta^{13})\tan\Theta^{24}
			+({\boldsymbol{\tau}}^{35}\nonumber \\
			&&-{\boldsymbol{\tau}}^{25}\sin\Theta^{12}\tan\Theta^{13})(\tan\Theta^{14}\tan\Theta^{15}\tan\Theta^{24}+
			\sec\Theta^{15}\sec\Theta^{24}\tan\Theta^{25})\rbrace+\sin\Theta^{12}\sec\Theta^{13}\lbrace\tan\Theta^{13}
			\nonumber \\
			&&({\boldsymbol{\tau}}^{14}\tan\Theta^{14}+{\boldsymbol{\tau}}^{15}\sec\Theta^{14}\tan\Theta^{15})+\sin\Theta^{23}({\boldsymbol{\tau}}^{14}\sec\Theta^{14}\tan\Theta^{24}
			+{\boldsymbol{\tau}}^{15}\tan\Theta^{14}\tan\Theta^{15}\tan\Theta^{24}\nonumber \\
			&&+{\boldsymbol{\tau}}^{15}\sec\Theta^{15}\sec\Theta^{24}\tan\Theta^{25})\rbrace+\cos\Theta^{12}\lbrace{\boldsymbol{\tau}}^{23}\sec^2\Theta^{13}+\tan\Theta^{13}\cos\Theta^{23}({\boldsymbol{\tau}}^{14}\sec\Theta^{14}
			\tan\Theta^{24}\nonumber \\
			&&+{\boldsymbol{\tau}}^{15}\tan\Theta^{14}\tan\Theta^{15}\tan\Theta^{24}+{\boldsymbol{\tau}}^{15}\sec\Theta^{15}\sec\Theta^{24}\tan\Theta^{25})+\sec\Theta^{13}(\tan\Theta^{13}({\boldsymbol{\tau}}^{24}
			\tan\Theta^{14}\nonumber \\
			&&+{\boldsymbol{\tau}}^{25}\sec\Theta^{14}\tan\Theta^{15})+\sin\Theta^{23}({\boldsymbol{\tau}}^{24}\sec\Theta^{14}
			\tan\Theta^{24}+{\boldsymbol{\tau}}^{25}\tan\Theta^{14}\tan\Theta^{15}\tan\Theta^{24}\nonumber \\
			&&+{\boldsymbol{\tau}}^{25}\sec\Theta^{15}\sec\Theta^{24}\tan\Theta^{25}))\rbrace]\\
			\boldsymbol{\nabla}_R\Theta^{14}&=&-{\boldsymbol{\tau}}^{45}\cos\Theta^{14}\tan\Theta^{15}-\cos\Theta^{12}\cos\Theta^{13}
			({\boldsymbol{\tau}}^{14}+{\boldsymbol{\tau}}^{15}\sin\Theta^{14}\tan\Theta^{15})+\sin\Theta^{12}\cos\Theta^{13}({\boldsymbol{\tau}}^{24}+
			{\boldsymbol{\tau}}^{25}\sin\Theta^{14}\nonumber \\
			&&\tan\Theta^{15})+\sin\Theta^{13}({\boldsymbol{\tau}}^{34}+{\boldsymbol{\tau}}^{35}\sin\Theta^{14}\tan\Theta^{15})
			\nonumber\\	
			\boldsymbol{\nabla}_R\Theta^{24}&=&\sin\Theta^{23}\lbrace-\sin\Theta^{12}\sin\Theta^{13}({\boldsymbol{\tau}}^{24}\sec\Theta^{14}+
			{\boldsymbol{\tau}}^{25}\tan\Theta^{14}\tan\Theta^{15})+\cos\Theta^{13}({\boldsymbol{\tau}}^{34}\sec\Theta^{14}\nonumber \\
			&&+{\boldsymbol{\tau}}^{35}\tan\Theta^{14}\tan\Theta^{15})\rbrace+\sec\Theta^{15}\lbrace-{\boldsymbol{\tau}}^{45}\cos\Theta^{14}\cos\Theta^{24}+\cos\Theta^{24}({\boldsymbol{\tau}}^{25}\cos\Theta^{13}
			\sin\Theta^{14}\nonumber \\
			&&+{\boldsymbol{\tau}}^{35}\sin\Theta^{13})\sin\Theta^{14}+({\boldsymbol{\tau}}^{35}\cos\Theta^{13}-{\boldsymbol{\tau}}^{25}\sin\Theta^{12}\sin\Theta^{13})\sin\Theta^{23}\sin\Theta^{24}\rbrace\tan\Theta^{25}\nonumber \\
			&&-\sin\Theta^{12}\cos\Theta^{23}({\boldsymbol{\tau}}^{14}\sec\Theta^{14}+{\boldsymbol{\tau}}^{15}\tan\Theta^{14}\tan\Theta^{15}+{\boldsymbol{\tau}}^{15}\sec\Theta^{15}\sin\Theta^{24}\tan\Theta^{25})\nonumber \\	&&+\cos\Theta^{12}\lbrace\sin\Theta^{13}
			\sin\Theta^{23}({\boldsymbol{\tau}}^{14}\sec\Theta^{14}+{\boldsymbol{\tau}}^{15}\tan\Theta^{14}\tan\Theta^{15})\nonumber \\
			&&+{\boldsymbol{\tau}}^{15}\sec\Theta^{15}(-\cos\Theta^{13}\cos\Theta^{24}\sin\Theta^{14}+\sin\Theta^{13}
			\sin\Theta^{23}\sin\Theta^{24})\tan\Theta^{25}\nonumber \\
			&&-\cos\Theta^{23}({\boldsymbol{\tau}}^{24}\sec\Theta^{14}+{\boldsymbol{\tau}}^{25}\tan\Theta^{14}\tan\Theta^{15}+{\boldsymbol{\tau}}^{25}\sec\Theta^{15}\sin\Theta^{24}
			\tan\Theta^{25})\rbrace\\
			\boldsymbol{\nabla}_R\Theta^{34}&=&\sin\Theta^{12}\lbrace-\sin\Theta^{23}({\boldsymbol{\tau}}^{14}\sec\Theta^{14}\sec\Theta^{24}+{\boldsymbol{\tau}}^{15}\tan\Theta^{14}\tan\Theta^{15}\sec\Theta^{24}+{\boldsymbol{\tau}}^{15}\sec\Theta^{15}\tan\Theta^{24}\tan\Theta^{25})\nonumber\\
			&&+\sin\Theta^{13}\cos\Theta^{23}({\boldsymbol{\tau}}^{24}\sec\Theta^{14}\sec\Theta^{24}+{\boldsymbol{\tau}}^{25}\tan\Theta^{14}\tan\Theta^{15}\sec\Theta^{24}+{\boldsymbol{\tau}}^{25}\sec\Theta^{15}\tan\Theta^{24}\tan\Theta^{25})\rbrace\nonumber \\
			&&-\sec\Theta^{15}\sec\Theta^{25}[\cos\Theta^{34}\lbrace{\boldsymbol{\tau}}^{45}\cos\Theta^{14}\cos\Theta^{24}-{\boldsymbol{\tau}}^{35}\sin\Theta^{13}\sin\Theta^{14}\cos\Theta^{24}+\sin\Theta^{12}({\boldsymbol{\tau}}^{15}\cos\Theta^{23}\nonumber \\
			&&+{\boldsymbol{\tau}}^{25}\sin\Theta^{13}\sin\Theta^{23})\sin\Theta^{24}\rbrace+\sin\Theta^{12}(-{\boldsymbol{\tau}}^{25}\sin\Theta^{13}
			\cos\Theta^{23}+{\boldsymbol{\tau}}^{15}\sin\Theta^{23})\sin\Theta^{34}]\tan\Theta^{35}+\cos\Theta^{12}
			\nonumber \\
			&&[-\sec\Theta^{24}\lbrace\sin\Theta^{13}\cos\Theta^{23}({\boldsymbol{\tau}}^{14}\sec\Theta^{14}+{\boldsymbol{\tau}}^{15}\tan\Theta^{14}\tan\Theta^{15})+\sin\Theta^{23}
			({\boldsymbol{\tau}}^{24}\sec\Theta^{14}+{\boldsymbol{\tau}}^{25}\tan\Theta^{14}\nonumber \\
			&&\tan\Theta^{15})\rbrace-\sec\Theta^{15}({\boldsymbol{\tau}}^{15}\sin\Theta^{13}\cos\Theta^{23}+{\boldsymbol{\tau}}^{25}\sin\Theta^{23})\tan\Theta^{24}\tan\Theta^{25}-\sec\Theta^{15}\sec\Theta^{25}\lbrace\cos\Theta^{34}\nonumber \\
			&&({\boldsymbol{\tau}}^{25}\cos\Theta^{23}-{\boldsymbol{\tau}}^{15}\sin\Theta^{13}
			\sin\Theta^{23})\sin\Theta^{24}+({\boldsymbol{\tau}}^{15}\sin\Theta^{13}\cos\Theta^{23}+{\boldsymbol{\tau}}^{25}\sin\Theta^{23})\sin\Theta^{34}\rbrace\tan\Theta^{35}]\nonumber \\
			&&+\cos\Theta^{13}[\sec\Theta^{15}\sec\Theta^{25}\cos\Theta^{34}\lbrace\cos\Theta^{24}(-{\boldsymbol{\tau}}^{15}\cos\Theta^{12}+{\boldsymbol{\tau}}^{25}\sin\Theta^{12})\sin\Theta^{14}+{\boldsymbol{\tau}}^{35}\sin\Theta^{23}\sin\Theta^{24}\rbrace\nonumber \\
			&&\tan\Theta^{35}-\cos\Theta^{23}({\boldsymbol{\tau}}^{34}\sec\Theta^{14}\sec\Theta^{24}+{\boldsymbol{\tau}}^{35}\tan\Theta^{14}\tan\Theta^{15}\sec\Theta^{24}+{\boldsymbol{\tau}}^{35}\sec\Theta^{15}\tan\Theta^{24}\tan\Theta^{25}\nonumber \\
			&&+{\boldsymbol{\tau}}^{35}\sec\Theta^{15}\sec\Theta^{25}\sin\Theta^{34}\tan\Theta^{35})]\\
			\boldsymbol{\nabla}_R\Theta^{15}&=&-{\boldsymbol{\tau}}^{15}\cos\Theta^{12}\cos\Theta^{13}\cos\Theta^{14}+{\boldsymbol{\tau}}^{25}\sin
			\Theta^{12}\cos\Theta^{13}\cos\Theta^{14}+{\boldsymbol{\tau}}^{35}\sin\Theta^{13}\cos\Theta^{14}
			+{\boldsymbol{\tau}}^{45}\sin\Theta^{14}\\
			\boldsymbol{\nabla}_R\Theta^{25}&=&\sec\Theta^{15}[\cos\Theta^{24}\lbrace-\cos\Theta^{23}({\boldsymbol{\tau}}^{25}\cos\Theta^{12}
			+{\boldsymbol{\tau}}^{15}\sin\Theta^{12})+({\boldsymbol{\tau}}^{35}\cos\Theta^{13}\nonumber \\
			&&+({\boldsymbol{\tau}}^{15}\cos\Theta^{12}-{\boldsymbol{\tau}}^{25}
			\sin\Theta^{12})\sin\Theta^{13})\sin\Theta^{23}\rbrace+\lbrace{\boldsymbol{\tau}}^{45}\cos\Theta^{14}+({\boldsymbol{\tau}}^{15}\cos\Theta^{12}\cos\Theta^{13}\nonumber \\
			&&-{\boldsymbol{\tau}}^{25}\sin\Theta^{12}
			\cos\Theta^{13}-{\boldsymbol{\tau}}^{35}\sin\Theta^{13})\sin\Theta^{14}\rbrace\sin\Theta^{24}]\nonumber\\
			\boldsymbol{\nabla}_R\Theta^{35}&=&\sec\Theta^{15}\sec\Theta^{25}[-\cos\Theta^{34}\lbrace{\boldsymbol{\tau}}^{35}\cos\Theta^{13}
			\cos\Theta^{23}+\cos\Theta^{23}({\boldsymbol{\tau}}^{15}\cos\Theta^{12}-{\boldsymbol{\tau}}^{25}\sin\Theta^{12})\sin\Theta^{13}
			+({\boldsymbol{\tau}}^{25}\cos\Theta^{12}\nonumber \\
			&&+{\boldsymbol{\tau}}^{15}\sin\Theta^{12})\sin\Theta^{23}\rbrace+\lbrace{\boldsymbol{\tau}}^{45}\cos\Theta^{14}\cos\Theta^{24}+\cos\Theta^{24}
			({\boldsymbol{\tau}}^{15}\cos\Theta^{12}\cos\Theta^{13}-{\boldsymbol{\tau}}^{25}\sin\Theta^{12}\cos\Theta^{13}
			\nonumber \\
			&&-{\boldsymbol{\tau}}^{35}\sin\Theta^{13})\sin\Theta^{14}+(\cos\Theta^{23}({\boldsymbol{\tau}}^{25}\cos\Theta^{12}+{\boldsymbol{\tau}}^{15}\sin\Theta^{12})-({\boldsymbol{\tau}}^{35}\cos\Theta^{13}
			+({\boldsymbol{\tau}}^{15}\cos\Theta^{12}\nonumber \\
			&&-{\boldsymbol{\tau}}^{25}\sin\Theta^{12})\sin\Theta^{13})\sin\Theta^{23})\sin\Theta^{24}\rbrace\sin\Theta^{34}]\nonumber\\
			\boldsymbol{\nabla}_R\Theta^{45}&=&\sec\Theta^{15}\sec\Theta^{25}\sec\Theta^{35}[\cos\Theta^{34}\lbrace-{\boldsymbol{\tau}}^{45}
			\cos\Theta^{14}\cos\Theta^{24}+\cos\Theta^{24}(-{\boldsymbol{\tau}}^{15}\cos\Theta^{12}\cos\Theta^{13}\nonumber\\
			&&+{\boldsymbol{\tau}}^{25}\sin\Theta^{12}\cos\Theta^{13}+{\boldsymbol{\tau}}^{35}\sin\Theta^{13})\sin\Theta^{14}-(\cos\Theta^{23}({\boldsymbol{\tau}}^{25}\cos\Theta^{12}+
			{\boldsymbol{\tau}}^{15}\sin\Theta^{12})- \nonumber\\ &&({\boldsymbol{\tau}}^{35}\cos\Theta^{13}+({\boldsymbol{\tau}}^{15}\cos\Theta^{12}-{\boldsymbol{\tau}}^{25}\sin
			\Theta^{12})\sin\Theta^{13})\sin\Theta^{23})\sin\Theta^{24}\rbrace-\nonumber\\&&\lbrace{\boldsymbol{\tau}}^{35}\cos\Theta^{13}\cos\Theta^{23}+\cos\Theta^{23}({\boldsymbol{\tau}}^{15}\cos\Theta^{12}-{\boldsymbol{\tau}}^{25}\sin\Theta^{12})\sin\Theta^{13}\nonumber\\&&+({\boldsymbol{\tau}}^{25}\cos\Theta^{12} +{\boldsymbol{\tau}}^{15}\sin\Theta^{12})\sin\Theta^{23}\rbrace\sin\Theta^{34}]\nonumber
		\end{eqnarray*}
	\end{subequations}	
\end{scriptsize}

\noindent
{\bf Few Six State ($N=6$) Equations}:~\citep{soumya_tfb}
\begin{scriptsize}
	\begin{subequations}
		\begin{eqnarray}
			\boldsymbol{\nabla}_R\Theta^{15}&=&-\cos\Theta^{12}\cos\Theta^{13}\cos\Theta^{14}\boldsymbol{\tau}^{15}+\sin\Theta^{12}\cos\Theta^{13}
			\cos\Theta^{14}\boldsymbol{\tau}^{25}+\sin\Theta^{13}\cos\Theta^{14}\boldsymbol{\tau}^{35}+\sin\Theta^{14}\boldsymbol{\tau}^{45}\nonumber \\
			&&-\cos\Theta^{12}\cos\Theta^{13}\cos\Theta^{14}\sin\Theta^{15}\tan\Theta^{16}\boldsymbol{\tau}^{16}
			+\sin\Theta^{12}\cos\Theta^{13}\cos\Theta^{14}\sin\Theta^{15}\tan\Theta^{16}\boldsymbol{\tau}^{26}\nonumber \\
			&&+\sin\Theta^{13}\cos\Theta^{14}\sin\Theta^{15}\tan\Theta^{16}\boldsymbol{\tau}^{36}+\sin\Theta^{14}\sin\Theta^{15}
			\tan\Theta^{16}\boldsymbol{\tau}^{46}-\cos\Theta^{15}\tan\Theta^{16}\boldsymbol{\tau}^{56}\\
			\boldsymbol{\nabla}_R\Theta^{16}&=&-\cos\Theta^{12}\cos\Theta^{13}\cos\Theta^{14}\cos\Theta^{15}\boldsymbol{\tau}^{16}+
			\sin\Theta^{12}\cos\Theta^{13}\cos\Theta^{14}\cos\Theta^{15}\boldsymbol{\tau}^{26}+\sin\Theta^{13}\cos\Theta^{14}\cos\Theta^{15}
			\boldsymbol{\tau}^{36}\nonumber \\
			&&+\sin\Theta^{14}\cos\Theta^{15}\boldsymbol{\tau}^{46}+\sin\Theta^{15}\boldsymbol{\tau}^{56}
		\end{eqnarray}
	\end{subequations}
\end{scriptsize}

\newpage
\noindent
NACT expressions for four,~\citep{sarkar_jpca112} five~\citep{soumya_jpca121} and six ~\citep{soumya_tfb} electronic states SHS are as follows:\\

\noindent
{\bf Four State ($N=4$)}:~\citep{sarkar_jpca112}
\begin{scriptsize}
	\begin{subequations}
		\begin{eqnarray*}
			\boldsymbol{\tau^{12}}&=&-\cos\Theta^{13}\cos\Theta^{23}\cos\Theta^{14}\cos\Theta^{24}\boldsymbol{\nabla}_R\Theta^{12}-\sin\Theta^{23}\cos\Theta^{14}\cos\Theta^{24}\boldsymbol{\nabla}_R\Theta^{13}-\sin\Theta^{24}\boldsymbol{\nabla}_R\Theta^{14}\\
			\boldsymbol{\tau^{13}}&=&\cos\Theta^{13}\sin\Theta^{23}\cos\Theta^{14}\cos\Theta^{34}\boldsymbol{\nabla}_R\Theta^{12}+\cos\Theta^{13}\cos\Theta^{23}\cos\Theta^{14}\sin\Theta^{24}\sin\Theta^{34}\boldsymbol{\nabla}_R\Theta^{12}-\cos\Theta^{23}\cos\Theta^{14}\cos\Theta^{34}\boldsymbol{\nabla}_R\Theta^{13}\nonumber\\
			&&+\sin\Theta^{23}\cos\Theta^{14}\sin\Theta^{24}\sin\Theta^{34}\boldsymbol{\nabla}_R\Theta^{13}\\
			\boldsymbol{\tau^{23}}&=&-\sin\Theta^{13}\cos\Theta^{24}\cos\Theta^{34}\boldsymbol{\nabla}_R\Theta^{12}-\cos\Theta^{13}\sin\Theta^{23}\sin\Theta^{14}\sin\Theta^{24}\cos\Theta^{34}\boldsymbol{\nabla}_R\Theta^{12}-\cos\Theta^{13}\cos\Theta^{23}\sin\Theta^{14}\sin\Theta^{34}\boldsymbol{\nabla}_R\Theta^{12}\nonumber\\
			&&+\cos\Theta^{23}\sin\Theta^{14}\sin\Theta^{24}\cos\Theta^{34}\boldsymbol{\nabla}_R\Theta^{13}-\sin\Theta^{23}\sin\Theta^{14}\sin\Theta^{34}\boldsymbol{\nabla}_R\Theta^{13}-\cos\Theta^{24}\cos\Theta^{34}\boldsymbol{\nabla}_R\Theta^{23}-\sin\Theta^{34}\boldsymbol{\nabla}_R\Theta^{24}\\
			\boldsymbol{\tau^{14}}&=&-\cos\Theta^{13}\sin\Theta^{23}\cos\Theta^{14}\sin\Theta^{34}\boldsymbol{\nabla}_R\Theta^{12}+\cos\Theta^{13}\cos\Theta^{23}\cos\Theta^{14}\sin\Theta^{24}\cos\Theta^{34}\boldsymbol{\nabla}_R\Theta^{12}+\cos\Theta^{23}\cos\Theta^{14}\sin\Theta^{34}\boldsymbol{\nabla}_R\Theta^{13}\nonumber\\
			&&+\sin\Theta^{23}\cos\Theta^{14}\sin\Theta^{24}\cos\Theta^{34}\boldsymbol{\nabla}_R\Theta^{13}-\cos\Theta^{24}\cos\Theta^{34}\boldsymbol{\nabla}_R\Theta^{14}\\
			\boldsymbol{\tau^{24}}&=&\sin\Theta^{13}\cos\Theta^{24}\sin\Theta^{34}\boldsymbol{\nabla}_R\Theta^{12}+\cos\Theta^{13}\sin\Theta^{23}\sin\Theta^{14}\sin\Theta^{24}\sin\Theta^{34}\boldsymbol{\nabla}_R\Theta^{12}-\cos\Theta^{13}\cos\Theta^{23}\sin\Theta^{14}\cos\Theta^{34}\boldsymbol{\nabla}_R\Theta^{12}\nonumber\\
			&&-\cos\Theta^{23}\sin\Theta^{14}\sin\Theta^{24}\sin\Theta^{34}\boldsymbol{\nabla}_R\Theta^{13}-\sin\Theta^{23}\sin\Theta^{14}\cos\Theta^{34}\boldsymbol{\nabla}_R\Theta^{13}+\cos\Theta^{24}\sin\Theta^{34}\boldsymbol{\nabla}_R\Theta^{23}-\cos\Theta^{34}\boldsymbol{\nabla}_R\Theta^{24}\\
			\boldsymbol{\tau^{34}}&=&-\sin\Theta^{13}\sin\Theta^{24}\boldsymbol{\nabla}_R\Theta^{12}+\cos\Theta^{13}\sin\Theta^{23}\sin\Theta^{14}\cos\Theta^{24}
			\boldsymbol{\nabla}_R\Theta^{12}-\cos\Theta^{23}\sin\Theta^{14}\cos\Theta^{24}\boldsymbol{\nabla}_R\Theta^{13}\nonumber\\
			&&-\sin\Theta^{24}\boldsymbol{\nabla}_R\Theta^{23}-\boldsymbol{\nabla}_R\Theta^{34}
		\end{eqnarray*}   
	\end{subequations}
\end{scriptsize}

\noindent
{\bf Five State ($N=5$)}:~\citep{soumya_jpca121}
\begin{scriptsize}
	\begin{subequations}
		\begin{eqnarray}
			\boldsymbol{\tau^{12}}&=&-\boldsymbol{\nabla}_R\Theta^{12}+\boldsymbol{\nabla}_R\Theta^{23}\sin\Theta^{13}+\boldsymbol{\nabla}_R\Theta^{24}\cos\Theta^{13}
			\sin\Theta^{14}\cos\Theta^{23}-\boldsymbol{\nabla}_R\Theta^{34}\cos\Theta^{24}\big\lbrace-\sec\Theta^{13}\sin\Theta^{14}
			\sin\Theta^{23}+\sin\Theta^{13}(\tan\Theta^{13}\sin\Theta^{14}\sin\Theta^{23}\nonumber \\
			&&+\tan\Theta^{24})\big\rbrace+\boldsymbol{\nabla}_R\Theta^{25}
			\cos\Theta^{13}\cos\Theta^{14}\sin\Theta^{15}\cos\Theta^{23}\cos\Theta^{24}+\boldsymbol{\nabla}_R\Theta^{35}\cos\Theta^{24}\cos\Theta^{25}\big[\cos\Theta^{13}\cos\Theta^{23}\sin\Theta^{34}\big\lbrace-\sec\Theta^{14}\sin\Theta^{15}\tan\Theta^{24}\nonumber \\
			&&+\sin\Theta^{14}(\tan\Theta^{14}\sin\Theta^{15}
			\tan\Theta^{24}+\sec\Theta^{24}\tan\Theta^{25})\big\rbrace+\sec\Theta^{24}\cos\Theta^{34}\big\lbrace\sec\Theta^{13}\sin\Theta^{23}
			(\sec\Theta^{14}\sin\Theta^{15}-\sin\Theta^{14}(\tan\Theta^{14}\sin\Theta^{15}+\sin\Theta^{24}
			\nonumber \\
			&&\tan\Theta^{25}))+\sin\Theta^{13}(-\tan\Theta^{13}\sec\Theta^{14}\sin\Theta^{15}\sin\Theta^{23}-\cos\Theta^{24}
			\tan\Theta^{25}+\tan\Theta^{13}\sin\Theta^{14}\sin\Theta^{23}(\tan\Theta^{14}\sin\Theta^{15}+\sin\Theta^{24}\tan\Theta^{25}))\big\rbrace\big]\nonumber \\
			&&+\boldsymbol{\nabla}_R\Theta^{45}\cos\Theta^{13}\cos\Theta^{24}\cos\Theta^{35}\big\lbrace\cos\Theta^{14}\sin\Theta^{15}
			\sin\Theta^{23}\sec\Theta^{24}\cos\Theta^{25}\sin\Theta^{34}+\cos\Theta^{23}\cos\Theta^{34}(-\sin\Theta^{14}\sec\Theta^{24}\sin\Theta^{25}\nonumber \\
			&&+\cos\Theta^{14}
			\sin\Theta^{15}\tan\Theta^{24}\cos\Theta^{25})-\sin\Theta^{25}\sin\Theta^{34}(\tan\Theta^{13}+\sin\Theta^{14}\sin\Theta^{23}\tan\Theta^{24})
			+(-\sin\Theta^{14}\sin\Theta^{23}+\tan\Theta^{13}\tan\Theta^{24})\tan\Theta^{35}\big\rbrace\nonumber \\
			\boldsymbol{\tau^{13}}&=&-\boldsymbol{\nabla}_R\Theta^{13}\cos\Theta^{12}-\boldsymbol{\nabla}_R\Theta^{23}\sin\Theta^{12}\cos\Theta^{13}+\boldsymbol{\nabla}_R\Theta^{24}\sin\Theta^{14}(\sin\Theta^{12}\sin\Theta^{13}\cos\Theta^{23}-\cos\Theta^{12}\sin\Theta^{23})+\boldsymbol{\nabla}_R\Theta^{34}\big\lbrace\sin\Theta^{14}\cos\Theta^{24}\nonumber \\
			&&(\cos\Theta^{12}\cos\Theta^{23}+\sin\Theta^{12}\sin\Theta^{13}\sin\Theta^{23})+\sin\Theta^{12}\cos\Theta^{13}\sin\Theta^{24}\big\rbrace+\boldsymbol{\nabla}_R\Theta^{25}\cos\Theta^{14}\sin\Theta^{15}\cos\Theta^{24}(\sin\Theta^{12}\sin\Theta^{13}\cos\Theta^{23}-\cos\Theta^{12}\nonumber \\
			&&\sin\Theta^{23})+\boldsymbol{\nabla}_R\Theta^{35}\cos\Theta^{24}\cos\Theta^{25}\big[\cos\Theta^{14}\sin\Theta^{15}\big\lbrace\sec\Theta^{24}\cos\Theta^{34}(\cos\Theta^{12}\cos\Theta^{23}+\sin\Theta^{12}\sin\Theta^{13}\sin\Theta^{23})
			+(-\sin\Theta^{12}\sin\Theta^{13}\cos\Theta^{23}\nonumber \\
			&&+\cos\Theta^{12}\sin\Theta^{23})\tan\Theta^{24}
			\sin\Theta^{34}\big\rbrace+\big\lbrace\sin\Theta^{12}\cos\Theta^{13}\cos\Theta^{34}+\sin\Theta^{14}(\sec\Theta^{24}(\sin\Theta^{12}\sin\Theta^{13}\cos\Theta^{23}-\cos\Theta^{12}\sin\Theta^{23})
			\sin\Theta^{34}-\cos\Theta^{34}\nonumber \\
			&&(\cos\Theta^{12}\cos\Theta^{23}+\sin\Theta^{12}\sin\Theta^{13}\sin\Theta^{23})\tan\Theta^{24})\big\rbrace\tan\Theta^{25}\big]-\boldsymbol{\nabla}_R\Theta^{45}
			\cos\Theta^{24}\cos\Theta^{35}\big[\cos\Theta^{12}\sin\Theta^{23}\cos\Theta^{34}(-\sin\Theta^{14}
			\sec\Theta^{24}\sin\Theta^{25}\nonumber \\
			&&+\cos\Theta^{14}\sin\Theta^{15}\tan\Theta^{24}\cos\Theta^{25})+\sin\Theta^{12}\big\lbrace-\sin\Theta^{13}
			\cos\Theta^{14}\sin\Theta^{15}\sin\Theta^{23}\sec\Theta^{24}\cos\Theta^{25}\sin\Theta^{34}+\sin\Theta^{13}\sin\Theta^{14}\sin\Theta^{23}\nonumber \\
			&&(\tan\Theta^{24}\sin\Theta^{25}\sin\Theta^{34}
			+\tan\Theta^{35})+\cos\Theta^{13}(-\sin\Theta^{25}\sin\Theta^{34}+\tan\Theta^{24}\tan\Theta^{35})\big\rbrace
			+\cos\Theta^{23}\big\lbrace\sin\Theta^{12}\sin\Theta^{13}\cos\Theta^{34}(\sin\Theta^{14}\sec\Theta^{24}
			\sin\Theta^{25}\nonumber \\
			&&-\cos\Theta^{14}\sin\Theta^{15}\tan\Theta^{24}\cos\Theta^{25})+\cos\Theta^{12}(-\cos\Theta^{14}\sin\Theta^{15}\sec\Theta^{24}\cos\Theta^{25}\sin\Theta^{34}
			+\sin\Theta^{14}(\tan\Theta^{24}\sin\Theta^{25}\sin\Theta^{34}+\tan\Theta^{35}))\big\rbrace\big]\nonumber\\
			\boldsymbol{\tau^{23}}&=&\boldsymbol{\nabla}_R\Theta^{13}\sin\Theta^{12}-\boldsymbol{\nabla}_R\Theta^{23}\cos\Theta^{12}\cos\Theta^{13}+\boldsymbol{\nabla}_R\Theta^{24}\sin\Theta^{14}(\cos\Theta^{12}\sin\Theta^{13}\cos\Theta^{23}+\sin\Theta^{12}\sin\Theta^{23})-\boldsymbol{\nabla}_R\Theta^{34}\big\lbrace\sin\Theta^{12}\sin\Theta^{14}\cos\Theta^{23}\nonumber \\
			&&\cos\Theta^{24}-\cos\Theta^{12}(\sin\Theta^{13}\sin\Theta^{14}\sin\Theta^{23}\cos\Theta^{24}+\cos\Theta^{13}
			\sin\Theta^{24})\big\rbrace+\boldsymbol{\nabla}_R\Theta^{25}\cos\Theta^{14}\sin\Theta^{15}\cos\Theta^{24}(\cos\Theta^{12}\sin\Theta^{13}\cos\Theta^{23}+\sin\Theta^{12}\nonumber\\
			&&\sin\Theta^{23})+\boldsymbol{\nabla}_R\Theta^{35}\big[\cos\Theta^{34}\big\lbrace\cos\Theta^{14}\sin\Theta^{15}\cos\Theta^{25}(-\sin\Theta^{12}\cos\Theta^{23}+\cos\Theta^{12}\sin\Theta^{13}\sin\Theta^{23})+(\cos\Theta^{12}\cos\Theta^{13}\cos\Theta^{24}+
			\sin\Theta^{14}\nonumber \\
			&&(\sin\Theta^{12}\cos\Theta^{23}-\cos\Theta^{12}\sin\Theta^{13}\sin\Theta^{23})\sin
			\Theta^{24})\sin\Theta^{25}\big\rbrace+(\cos\Theta^{12}\sin\Theta^{13}\cos\Theta^{23}+\sin\Theta^{12}\sin\Theta^{23})(-\cos\Theta^{14}\sin\Theta^{15}\sin\Theta^{24}\cos\Theta^{25}\nonumber \\
			&&+\sin\Theta^{14}\sin\Theta^{25})\sin\Theta^{34}\big]-\boldsymbol{\nabla}_R\Theta^{45}\big[\cos\Theta^{12}\cos\Theta^{24}\cos\Theta^{35}\big\lbrace
			\sin\Theta^{13}\cos\Theta^{23}\cos\Theta^{34}(\sin\Theta^{14}\sec\Theta^{24}\sin\Theta^{25}-\cos\Theta^{14}\sin\Theta^{15}\tan\Theta^{24}\nonumber \\
			&&\cos\Theta^{25})-\sin\Theta^{34}(\cos\Theta^{13}\sin\Theta^{25}
			+\sin\Theta^{13}\sin\Theta^{23}(\cos\Theta^{14}\sin\Theta^{15}\sec\Theta^{24}\cos\Theta^{25}-\sin\Theta^{14}\tan\Theta^{24}\sin\Theta^{25}))+(\sin\Theta^{13}\sin\Theta^{14}\sin\Theta^{23}\nonumber \\
			&&
			+\cos\Theta^{13}\tan\Theta^{24})\tan\Theta^{35}\big\rbrace-\sin\Theta^{12}\cos\Theta^{24}\cos\Theta^{35}\big\lbrace\sin\Theta^{23}\cos\Theta^{34}(-\sin\Theta^{14}\sec\Theta^{24}\sin\Theta^{25}+\cos\Theta^{14}\sin\Theta^{15}\tan\Theta^{24}\cos\Theta^{25})\nonumber\\
			&&+\cos\Theta^{23}(-\cos\Theta^{14}\sin\Theta^{15}\sec\Theta^{24}\cos\Theta^{25}\sin\Theta^{34}+\sin\Theta^{14}(\tan\Theta^{24}\sin\Theta^{25}\sin\Theta^{34}+\tan\Theta^{35}))\big\rbrace\big]\nonumber \\
			\boldsymbol{\tau^{14}}&=&-\boldsymbol{\nabla}_R\Theta^{14}\cos\Theta^{12}\cos\Theta^{13}+\boldsymbol{\nabla}_R\Theta^{24}\cos\Theta^{14}(-\sin\Theta^{12}\cos\Theta^{23}+\cos\Theta^{12}\sin\Theta^{13}\sin\Theta^{23})-\boldsymbol{\nabla}_R\Theta^{34}\cos\Theta^{14}\cos\Theta^{24}(\cos\Theta^{12}\sin\Theta^{13}\cos\Theta^{23}\nonumber\\
			&&+\sin\Theta^{12}\sin\Theta^{23})+\boldsymbol{\nabla}_R\Theta^{25}\sin\Theta^{15}\cos\Theta^{24}\big\lbrace\sin\Theta^{12}\sin\Theta^{14}\cos\Theta^{23}-\cos\Theta^{12}(\sin\Theta^{13}\sin\Theta^{14}\sin\Theta^{23}
			+\cos\Theta^{13}\tan\Theta^{24})\big\rbrace+\boldsymbol{\nabla}_R\Theta^{35}\cos\Theta^{13}\nonumber\\
			&&\cos\Theta^{24}\cos
			\Theta^{25}\big[\cos\Theta^{34}(\sin\Theta^{12}\sec\Theta^{13}\sin\Theta^{23}+\cos\Theta^{12}\tan\Theta^{13}\cos\Theta^{23})(\sin\Theta^{14}\sin\Theta^{15}\sec\Theta^{24}+\cos\Theta^{14}\tan\Theta^{24}
			\tan\Theta^{25})+\sin\Theta^{34}\nonumber\\
			&&\big\lbrace-\sin\Theta^{12}\sec\Theta^{13}\cos\Theta^{23}(\sin\Theta^{14}\sin\Theta^{15}\tan\Theta^{24}+\cos\Theta^{14}\sec\Theta^{24}\tan\Theta^{25})+\cos\Theta^{12}(\sin\Theta^{15}(-1+\tan\Theta^{13}\sin\Theta^{14}\sin\Theta^{23}\tan\Theta^{24})\nonumber \\
			&&+\tan\Theta^{13}\cos\Theta^{14}\sin\Theta^{23}\sec\Theta^{24}\tan\Theta^{25})\big\rbrace\big]+\boldsymbol{\nabla}_R\Theta^{45}\cos\Theta^{13}\cos\Theta^{24}\cos\Theta^{35}\big[\cos\Theta^{12}\big\lbrace\sin\Theta^{15}\cos\Theta^{25}(\cos\Theta^{34}+\tan\Theta^{13}\sin\Theta^{14}\cos\Theta^{23}\nonumber \\
			&&\sec\Theta^{24}\sin\Theta^{34}
			-\tan\Theta^{13}\sin\Theta^{14}\sin\Theta^{23}\tan\Theta^{24}\cos\Theta^{34})+\tan\Theta^{13}\cos\Theta^{14}(-\sin\Theta^{23}\sec\Theta^{24}\sin\Theta^{25}\cos\Theta^{34}+\cos\Theta^{23}(\tan\Theta^{24}\sin\Theta^{25}\nonumber\\
			&&\sin\Theta^{34}+\tan\Theta^{35}))\big\rbrace+\sin\Theta^{12}\sec\Theta^{13}\big\lbrace\sin\Theta^{14}\sin\Theta^{15}\cos\Theta^{25}(\sin\Theta^{23}\sec\Theta^{24}\sin\Theta^{34}+\cos\Theta^{23}\tan\Theta^{24}\cos\Theta^{34})+\cos\Theta^{14}(\cos\Theta^{23}\sec\Theta^{24}\nonumber\\
			&&\sin\Theta^{25}\cos\Theta^{34}+\sin\Theta^{23}(\tan\Theta^{24}\sin\Theta^{25}\sin\Theta^{34}+\tan\Theta^{35}))\big\rbrace\big]\nonumber
		\end{eqnarray}
	\end{subequations}
\end{scriptsize}

\begin{scriptsize}
	\begin{subequations}
		\begin{eqnarray}			\boldsymbol{\tau^{24}}&=&\boldsymbol{\nabla}_R\Theta^{14}\sin\Theta^{12}\cos\Theta^{13}-\boldsymbol{\nabla}_R\Theta^{24}\cos\Theta^{14}(\cos\Theta^{12}\cos\Theta^{23}+\sin\Theta^{12}\sin\Theta^{13}\sin\Theta^{23})+\boldsymbol{\nabla}_R\Theta^{34}\cos\Theta^{14}\cos\Theta^{24}(\sin\Theta^{12}\sin\Theta^{13}\cos\Theta^{23}\nonumber \\
			&&-\cos\Theta^{12}\sin\Theta^{23})+\boldsymbol{\nabla}_R\Theta^{25}\sin\Theta^{15}\cos\Theta^{24}\big\lbrace\cos\Theta^{12}\sin\Theta^{14}\cos\Theta^{23}+\sin\Theta^{12}(\sin\Theta^{13}\sin\Theta^{14}\sin\Theta^{23}+\cos\Theta^{13}\tan\Theta^{24})\big\rbrace-\boldsymbol{\nabla}_R\Theta^{35}\big[\cos\Theta^{34}\nonumber\\
			&&(\sin\Theta^{12}\sin\Theta^{13}\cos\Theta^{23}-\cos\Theta^{12}\sin\Theta^{23})(\sin\Theta^{14}\sin\Theta^{15}\cos\Theta^{25}+\cos\Theta^{14}\sin\Theta^{24}\sin\Theta^{25})+\big\lbrace-\sin\Theta^{12}\cos\Theta^{13}\sin\Theta^{15}\cos\Theta^{24}\cos\Theta^{25}\nonumber\\
			&&+(\cos\Theta^{12}\cos\Theta^{23}+\sin\Theta^{12}\sin\Theta^{13}\sin\Theta^{23})(\sin\Theta^{14}\sin\Theta^{15}\sin\Theta^{24}\cos\Theta^{25}+\cos\Theta^{14}\sin\Theta^{25})\big\rbrace
			\sin\Theta^{34}\big]+\boldsymbol{\nabla}_R\Theta^{45}\cos\Theta^{13}\cos\Theta^{24}\cos\Theta^{35}\nonumber\\
			&&\big[\sin\Theta^{15}\cos\Theta^{25}\big\lbrace\sin\Theta^{14}\sec\Theta^{24}\sin\Theta^{34}(\cos\Theta^{12}\sec\Theta^{13}\sin\Theta^{23}-\sin\Theta^{12}\tan\Theta^{13}\cos\Theta^{23})+\cos\Theta^{34}(-\sin\Theta^{12}+\sin\Theta^{14}(\cos\Theta^{12}\sec\Theta^{13}\nonumber\\
			&&\cos\Theta^{23}+\sin\Theta^{12}\tan\Theta^{13}\sin\Theta^{23})\tan\Theta^{24})\big\rbrace+\cos\Theta^{14}\big\lbrace\sin\Theta^{12}\tan\Theta^{13}(\sin\Theta^{23}\sec\Theta^{24}\sin\Theta^{25}\cos\Theta^{34}-\cos\Theta^{23}(\tan\Theta^{24}\sin\Theta^{25}\sin\Theta^{34}
			\nonumber\\
			&&+\tan\Theta^{35}))+\cos\Theta^{12}\sec\Theta^{13}(\cos\Theta^{23}\sec\Theta^{24}\sin\Theta^{25}\cos\Theta^{34}+\sin\Theta^{23}(\tan\Theta^{24}\sin\Theta^{25}\sin\Theta^{34}+\tan\Theta^{35}))\big\rbrace\big]\nonumber \\
				\boldsymbol{\tau^{34}}&=&\boldsymbol{\nabla}_R\Theta^{14}\sin\Theta^{13}+\boldsymbol{\nabla}_R\Theta^{24}\cos\Theta^{13}\cos\Theta^{14}\sin\Theta^{23}-\boldsymbol{\nabla}_R\Theta^{34}\cos\Theta^{13}\cos\Theta^{14}\cos\Theta^{23}\cos\Theta^{24}+\boldsymbol{\nabla}_R\Theta^{25}\cos\Theta^{13}\sin\Theta^{15}\cos\Theta^{24}(-\sin\Theta^{14}\nonumber\\
			&&\sin\Theta^{23}+\tan\Theta^{13}\tan\Theta^{24})+\boldsymbol{\nabla}_R\Theta^{35}\cos\Theta^{13}\big[\cos\Theta^{14}\sin\Theta^{25}(\cos\Theta^{23}\sin\Theta^{24}\cos\Theta^{34}+\sin\Theta^{23}\sin\Theta^{34})+\sin\Theta^{15}\cos\Theta^{25}\big\lbrace\sin\Theta^{14}\cos\Theta^{23}\nonumber \\
			&&\cos\Theta^{34}+\sin\Theta^{34}(\sin\Theta^{14}\sin\Theta^{23}\sin\Theta^{24}+\tan\Theta^{13}\cos\Theta^{24})\big\rbrace\big]+\boldsymbol{\nabla}_R\Theta^{45}\cos\Theta^{13}\cos\Theta^{24}\cos\Theta^{35}\big[\sin\Theta^{15}\cos\Theta^{25}\big\lbrace\sin\Theta^{14}\cos\Theta^{23}\sec\Theta^{24}\nonumber\\
			&&\sin\Theta^{34}-\cos\Theta^{34}(\tan\Theta^{13}+\sin\Theta^{14}\sin\Theta^{23}\tan\Theta^{24})\big\rbrace+\cos\Theta^{14}\big\lbrace-\sin\Theta^{23}\sec\Theta^{24}\sin\Theta^{25}\cos\Theta^{34}+\cos\Theta^{23}(\tan\Theta^{24}\sin\Theta^{25}\sin\Theta^{34}\nonumber\\
			&&+\tan\Theta^{35})\big\rbrace\big]\nonumber\\
			\boldsymbol{\tau^{15}}&=&-\boldsymbol{\nabla}_R\Theta^{15}\cos\Theta^{12}\cos\Theta^{13}\cos\Theta^{14}+\boldsymbol{\nabla}_R\Theta^{25}
			\cos\Theta^{15}\cos\Theta^{24}\big\lbrace-\sin\Theta^{12}\cos\Theta^{23}+\cos\Theta^{12}(\sin\Theta^{13}
			\sin\Theta^{23}+\cos\Theta^{13}\sin\Theta^{14}\tan\Theta^{24})\big\rbrace\nonumber\\
			&&-\boldsymbol{\nabla}_R\Theta^{35}\cos\Theta^{15}\cos\Theta^{25}\big[\sin\Theta^{12}(\sin\Theta^{23}\cos\Theta^{34}-\cos\Theta^{23}\sin\Theta^{24}\sin\Theta^{34})+\cos\Theta^{12}\big\lbrace\sin\Theta^{13}\cos\Theta^{23}\cos\Theta^{34}+(-\cos\Theta^{13}\sin\Theta^{14}\cos\Theta^{24}\nonumber \\
			&&+\sin\Theta^{13}\sin\Theta^{23}
			\sin\Theta^{24})\sin\Theta^{34}\big\rbrace\big]-\boldsymbol{\nabla}_R\Theta^{45}\cos\Theta^{15}\cos\Theta^{25}\cos\Theta^{35}\big[\sin\Theta^{12}(\cos\Theta^{23}
			\sin\Theta^{24}\cos\Theta^{34}+\sin\Theta^{23}\sin\Theta^{34})+\cos\Theta^{12}\nonumber \\
			&&\lbrace\cos\Theta^{13}\sin\Theta^{14}\cos\Theta^{24}\cos\Theta^{34}+\sin\Theta^{13}(-\sin\Theta^{23}\sin\Theta^{24}\cos\Theta^{34}+\cos\Theta^{23}\sin\Theta^{34})\big\rbrace\big]\nonumber\\
			\boldsymbol{\tau^{25}}&=&\boldsymbol{\nabla}_R\Theta^{15}\sin\Theta^{12}\cos\Theta^{13}\cos\Theta^{14}-\boldsymbol{\nabla}_R\Theta^{25}
			\cos\Theta^{15}\cos\Theta^{24}\big\lbrace\cos\Theta^{12}\cos\Theta^{23}+\sin\Theta^{12}(\sin\Theta^{13}
			\sin\Theta^{23}+\cos\Theta^{13}\sin\Theta^{14}\tan\Theta^{24})\big\rbrace\nonumber\\
			&&+\boldsymbol{\nabla}_R\Theta^{35}\cos\Theta^{15}\cos\Theta^{25}\big\lbrace-\cos\Theta^{12}\sin\Theta^{23}\cos\Theta^{34}
			+\sin\Theta^{12}(-\cos\Theta^{13}\sin\Theta^{14}\cos\Theta^{24}+\sin\Theta^{13}\sin\Theta^{23}\sin\Theta^{24})\sin\Theta^{34}+\cos\Theta^{23}\nonumber\\
			&&(\sin\Theta^{12}\sin\Theta^{13}\cos\Theta^{34}
			+\cos\Theta^{12}\sin\Theta^{24}\sin\Theta^{34})\big\rbrace+\boldsymbol{\nabla}_R\Theta^{45}\cos\Theta^{15}\cos\Theta^{25}\cos\Theta^{35}\big\lbrace\sin\Theta^{12}
			\cos\Theta^{13}\sin\Theta^{14}\cos\Theta^{24}\cos\Theta^{34}\nonumber\\
			&&+\sin\Theta^{12}\sin\Theta^{13}(-\sin\Theta^{23}\sin\Theta^{24}\cos\Theta^{34}+\cos\Theta^{23}\sin\Theta^{34})-\cos\Theta^{12}(\cos\Theta^{23}\sin\Theta^{24}\cos\Theta^{34}+\sin\Theta^{23}
			\sin\Theta^{34})\big\rbrace\nonumber\\
			\boldsymbol{\tau^{35}}&=&\boldsymbol{\nabla}_R\Theta^{15}\sin\Theta^{13}\cos\Theta^{14}+\boldsymbol{\nabla}_R\Theta^{25}\cos\Theta^{13}
			\cos\Theta^{15}\cos\Theta^{24}(\sin\Theta^{23}-\tan\Theta^{13}\sin\Theta^{14}\tan\Theta^{24})-\boldsymbol{\nabla}_R\Theta^{35}\cos\Theta^{13}\cos\Theta^{15}\cos\Theta^{25}\nonumber\\
			&&\big\lbrace\cos\Theta^{23}\cos\Theta^{34}+
			\sin\Theta^{34}(\sin\Theta^{23}\sin\Theta^{24}+\tan\Theta^{13}\sin\Theta^{14}\cos\Theta^{24})\big\rbrace
			+\boldsymbol{\nabla}_R\Theta^{45}\cos\Theta^{13}\cos\Theta^{15}\cos\Theta^{25}\cos\Theta^{35}\big\lbrace-\cos\Theta^{23}
			\sin\Theta^{34}\nonumber \\
			&&+\cos\Theta^{34}(\sin\Theta^{23}\sin\Theta^{24}+\tan\Theta^{13}\sin\Theta^{14}\cos\Theta^{24})\big\rbrace\nonumber\\
			\boldsymbol{\tau^{45}}&=&\sin\Theta^{14}\boldsymbol{\nabla}_R\Theta^{15}+\cos\Theta^{14}
			\cos\Theta^{15}\sin\Theta^{24}\boldsymbol{\nabla}_R\Theta^{25}+\cos\Theta^{14}\cos\Theta^{15}
			\cos\Theta^{24}\cos\Theta^{25}\sin\Theta^{34}\boldsymbol{\nabla}_R\Theta^{35}\nonumber \\
			&&-\cos\Theta^{14}\cos\Theta^{15}
			\cos\Theta^{24}\cos\Theta^{25}\cos\Theta^{34}\cos\Theta^{35}\boldsymbol{\nabla}_R\Theta^{45}\nonumber
		\end{eqnarray}
	\end{subequations}
\end{scriptsize}

\noindent
{\bf Six State ($N=6$)}:~\citep{soumya_tfb}
\begin{scriptsize}
	\begin{subequations}
		\begin{eqnarray*}
			\boldsymbol{\tau^{56}}&=& \sin\Theta^{15}\boldsymbol{\nabla}_R\Theta^{16}+\cos\Theta^{15}\cos\Theta^{16}\sin\Theta^{25}\boldsymbol{\nabla}_R\Theta^{26}+\cos\Theta^{15}\cos\Theta^{16}\cos\Theta^{25}\cos\Theta^{26}\sin\Theta^{35}\boldsymbol{\nabla}_R\Theta^{36}+\cos\Theta^{15}\cos\Theta^{16}\nonumber\\
			&&\cos\Theta^{25}\cos\Theta^{26}\cos\Theta^{35}\cos\Theta^{36}\sin\Theta^{45}\boldsymbol{\nabla}_R\Theta^{46}-\cos\Theta^{15}\cos\Theta^{16}\cos\Theta^{25}\cos\Theta^{26}\cos\Theta^{35}\cos\Theta^{36}\cos\Theta^{45}\cos\Theta^{46}\boldsymbol{\nabla}_R\Theta^{56}
		\end{eqnarray*}   
		\label{eq:six}
	\end{subequations}
\end{scriptsize}

\newpage
\noindent
The expressions of Mathematical Curls for four,~\citep{sarkar_jpca112} five~\citep{mantu_ijqc_curl} and six~\citep{mantu_ijqc_curl} electronic states SHS are as follows:\\

\noindent
{\bf Four State ($N=4$)}:~\citep{sarkar_jpca112}
\begin{scriptsize}
	\begin{subequations}
		\begin{eqnarray*}
			Z_{\rho\phi}^{12}&=&\sin\Theta^{13}\cos\Theta^{23}\cos\Theta^{14}\cos\Theta^{24}(\nabla_\phi\Theta^{13}\nabla_\rho\Theta^{12}-
			\nabla_\rho\Theta^{13}\nabla_\phi\Theta^{12})+\cos\Theta^{13}\sin\Theta^{23}\cos\Theta^{14}\cos\Theta^{24}(\nabla_\phi\Theta^{23}\nabla_\rho\Theta^{12}-\nabla_\rho\Theta^{23}\nabla_\phi\Theta^{12})\nonumber\\
			&&+\cos\Theta^{13}\cos\Theta^{23}\sin\Theta^{14}\cos\Theta^{24}(\nabla_\phi\Theta^{14}\nabla_\rho\Theta^{12}-
			\nabla_\rho\Theta^{14}\nabla_\phi\Theta^{12})+\cos\Theta^{13}\cos\Theta^{23}\cos\Theta^{14}\sin\Theta^{24}(\nabla_\phi\Theta^{24}\nabla_\rho\Theta^{12}-\nabla_\rho\Theta^{24}\nabla_\phi\Theta^{12})\nonumber\\
			&&-\cos\Theta^{23}\cos\Theta^{14}\cos\Theta^{24}(\nabla_\phi\Theta^{23}\nabla_\rho\Theta^{13}-\nabla_\rho\Theta^{23}\nabla_\phi\Theta^{13})+
			\sin\Theta^{23}\sin\Theta^{14}\cos\Theta^{24}(\nabla_\phi\Theta^{14}\nabla_\rho\Theta^{13}-\nabla_\rho\Theta^{14}\nabla_\phi\Theta^{13})\nonumber\\
			&&+\sin\Theta^{23}\cos\Theta^{14}\sin\Theta^{24}(\nabla_\phi\Theta^{24}\nabla_\rho\Theta^{13}-\nabla_\rho\Theta^{24}\nabla_\phi\Theta^{13})-\cos\Theta^{24}
			(\nabla_\phi\Theta^{24}\nabla_\rho\Theta^{14}-\nabla_\rho\Theta^{24}\nabla_\phi\Theta^{14})\\
			Z_{\rho\phi}^{13}&=&(-\sin\Theta^{13}\sin\Theta^{23}\cos\Theta^{14}\cos\Theta^{34}-\sin\Theta^{13}\cos\Theta^{23}\cos\Theta^{14}\sin\Theta^{24}\sin\Theta^{34})
			(\nabla_\phi\Theta^{13}\nabla_\rho\Theta^{12}-\nabla_\rho\Theta^{13}\nabla_\phi\Theta^{12})+(\cos\Theta^{13}\cos\Theta^{23}\nonumber\\
			&&\cos\Theta^{14}\cos\Theta^{34}
			-\cos\Theta^{13}\sin\Theta^{23}\cos\Theta^{14}\sin\Theta^{24}\sin\Theta^{34}) (\nabla_\phi\Theta^{23}\nabla_\rho\Theta^{12}-\nabla_\rho\Theta^{23}\nabla_\phi\Theta^{12})+(-\cos\Theta^{13}\sin\Theta^{23}\sin\Theta^{14}\cos\Theta^{34}\nonumber\\
			&&-\cos\Theta^{13}\cos\Theta^{23}\sin\Theta^{14}\sin\Theta^{24}\sin\Theta^{34})(\nabla_\phi\Theta^{14}\nabla_\rho\Theta^{12}-\nabla_\rho\Theta^{14}\nabla_\phi\Theta^{12})+(-\cos\Theta^{13}\sin\Theta^{23}\cos\Theta^{14}\sin\Theta^{34}\nonumber\\
			&&+\cos\Theta^{13}\cos\Theta^{23}\cos\Theta^{14}\sin\Theta^{24}\cos\Theta^{34})(\nabla_\phi\Theta^{34}\nabla_\rho\Theta^{12}-\nabla_\rho\Theta^{34}\nabla_\phi\Theta^{12})+\cos\Theta^{13}\cos\Theta^{23}\cos\Theta^{14}
			\cos\Theta^{24}\sin\Theta^{34}\nonumber\\
			&&(\nabla_\phi\Theta^{24}\nabla_\rho\Theta^{12}-\nabla_\rho\Theta^{24}\nabla_\phi\Theta^{12})+(\sin\Theta^{23}\cos\Theta^{14}\cos\Theta^{34}+\cos\Theta^{23}\cos\Theta^{14}\sin\Theta^{24}\sin\Theta^{34})(\nabla_\phi\Theta^{23}\nabla_\rho\Theta^{13}-\nabla_\rho\Theta^{23}\nabla_\phi\Theta^{13})\nonumber\\
			&&+(\cos\Theta^{23}\sin\Theta^{14}\cos\Theta^{34}-\sin\Theta^{23}\sin\Theta^{14}\sin\Theta^{24}\sin\Theta^{34})(\nabla_\phi\Theta^{14}\nabla_\rho\Theta^{13}-\nabla_\rho\Theta^{14}\nabla_\phi\Theta^{13})+\cos\Theta^{23}\cos\Theta^{14}\sin\Theta^{34}\nonumber\\
			&&(\nabla_\phi\Theta^{34}\nabla_\rho\Theta^{13}-\nabla_\rho\Theta^{34}\nabla_\phi\Theta^{13})+
			\sin\Theta^{23}\cos\Theta^{14}\cos\Theta^{24}\sin\Theta^{34}(\nabla_\phi\Theta^{24}\nabla_\rho\Theta^{13}-\nabla_\rho\Theta^{24}\nabla_\phi\Theta^{13})\nonumber\\
			&&+\sin\Theta^{24}\sin\Theta^{34}(\nabla_\phi\Theta^{24}\nabla_\rho\Theta^{14}-\nabla_\rho\Theta^{24}\nabla_\phi\Theta^{14})-\cos\Theta^{24}\cos\Theta^{34}(\nabla_\phi\Theta^{34}\nabla_\rho\Theta^{14}-\nabla_\rho\Theta^{34}\nabla_\phi\Theta^{14})\\
			Z_{\rho\phi}^{23}&=&(-\cos\Theta^{13}\cos\Theta^{24}\cos\Theta^{34}+\sin\Theta^{13}\sin\Theta^{23}\sin\Theta^{14}\sin\Theta^{24}\cos\Theta^{34}+\sin\Theta^{13}\cos\Theta^{23}\sin\Theta^{14}\sin\Theta^{34})(\nabla_\phi\Theta^{13}\nabla_\rho\Theta^{12}\nonumber\\
			&&-\nabla_\rho\Theta^{13}\nabla_\phi\Theta^{12})+(\sin\Theta^{13}\sin\Theta^{24}\cos\Theta^{34}-\cos\Theta^{13}\sin\Theta^{23}\sin\Theta^{14}\cos\Theta^{24}\cos\Theta^{34})(\nabla_\phi\Theta^{24}\nabla_\rho\Theta^{12}-\nabla_\rho\Theta^{24}\nabla_\phi\Theta^{12})\nonumber\\
			&&+(\sin\Theta^{13}\cos\Theta^{24}\sin\Theta^{34}+\cos\Theta^{13}\sin\Theta^{23}\sin\Theta^{14}\sin\Theta^{24}\sin\Theta^{34}-\cos\Theta^{13}\cos\Theta^{23}\sin\Theta^{14}\cos\Theta^{34})(\nabla_\phi\Theta^{34}\nabla_\rho\Theta^{12}-\nabla_\rho\Theta^{34}\nabla_\phi\Theta^{12})\nonumber\\
			&&+(-\cos\Theta^{13}\cos\Theta^{23}\sin\Theta^{14}\sin\Theta^{24}\cos\Theta^{34}+\cos\Theta^{13}\sin\Theta^{23}\sin\Theta^{14}\sin\Theta^{34})(\nabla_\phi\Theta^{23}\nabla_\rho\Theta^{12}-\nabla_\rho\Theta^{23}\nabla_\phi\Theta^{12})\nonumber\\
			&&+(-\cos\Theta^{13}\sin\Theta^{23}\cos\Theta^{14}\sin\Theta^{24}\cos\Theta^{34}-\cos\Theta^{13}\cos\Theta^{23}\cos\Theta^{14}\sin\Theta^{34})(\nabla_\phi\Theta^{14}\nabla_\rho\Theta^{12}-\nabla_\rho\Theta^{14}\nabla_\phi\Theta^{12})\nonumber\\
			&&+(\sin\Theta^{23}\sin\Theta^{14}\sin\Theta^{24}\cos\Theta^{34}-\cos\Theta^{23}\sin\Theta^{14}\sin\Theta^{34})(\nabla_\phi\Theta^{23}\nabla_\rho\Theta^{13}-\nabla_\rho\Theta^{23}\nabla_\phi\Theta^{13})+(\cos\Theta^{23}\cos\Theta^{14}\sin\Theta^{24}\cos\Theta^{34}\nonumber\\
			&&-\sin\Theta^{23}\cos\Theta^{14}\sin\Theta^{34})
			(\nabla_\phi\Theta^{14}\nabla_\rho\Theta^{13}-\nabla_\rho\Theta^{14}\nabla_\phi\Theta^{13})\\
			Z_{\rho\phi}^{14}&=& (\sin\Theta^{13}\sin\Theta^{23}\cos\Theta^{14}\sin\Theta^{34}-\sin\Theta^{13}\cos\Theta^{23}\cos\Theta^{14}\sin\Theta^{24}\cos\Theta^{34})(\nabla_\phi\Theta^{13}\nabla_\rho\Theta^{12}-
			\nabla_\rho\Theta^{13}\nabla_\phi\Theta^{12})+(-\cos\Theta^{13}\cos\Theta^{23}\nonumber\\
			&&\cos\Theta^{14}\sin\Theta^{34}-\cos\Theta^{13}\sin\Theta^{23}\cos\Theta^{14}\sin\Theta^{24}\cos\Theta^{34})(\nabla_\phi\Theta^{23}\nabla_\rho\Theta^{12}-\nabla_\rho\Theta^{23}\nabla_\phi\Theta^{12})+(\cos\Theta^{13}\sin\Theta^{23}\sin\Theta^{14}\sin\Theta^{34}\nonumber\\
			&&-\cos\Theta^{13}\cos\Theta^{23}\sin\Theta^{14}\sin\Theta^{24}\cos\Theta^{34})(\nabla_\phi\Theta^{14}\nabla_\rho\Theta^{12}-
			\nabla_\rho\Theta^{14}\nabla_\phi\Theta^{12})+(-\cos\Theta^{13}\sin\Theta^{23}\cos\Theta^{14}\cos\Theta^{34}-\cos\Theta^{13}\nonumber\\
			&&\cos\Theta^{23}\cos\Theta^{14}\sin\Theta^{24}\sin\Theta^{34})(\nabla_\phi\Theta^{34}\nabla_\rho\Theta^{12}-\nabla_\rho\Theta^{34}\nabla_\phi\Theta^{12})+\cos\Theta^{13}\cos\Theta^{23}\cos\Theta^{14}\cos\Theta^{24}\cos\Theta^{34}(\nabla_\phi\Theta^{24}\nabla_\rho\Theta^{12}\nonumber\\
			&&-\nabla_\rho\Theta^{24}\nabla_\phi\Theta^{12})+(-\sin\Theta^{23}\cos\Theta^{14}\sin\Theta^{34}+\cos\Theta^{23}\cos\Theta^{14}\sin\Theta^{24}\cos\Theta^{34})
			(\nabla_\phi\Theta^{23}\nabla_\rho\Theta^{13}-\nabla_\rho\Theta^{23}\nabla_\phi\Theta^{13})\nonumber\\
			&&+(-\cos\Theta^{23}\sin\Theta^{14}\sin\Theta^{34}-\sin\Theta^{23}\sin\Theta^{14}\sin\Theta^{24}\cos\Theta^{34})(\nabla_\phi\Theta^{14}\nabla_\rho\Theta^{13}-\nabla_\rho\Theta^{14}\nabla_\phi\Theta^{13})+(\cos\Theta^{23}\cos\Theta^{14}\cos\Theta^{34}\nonumber\\
			&&-\sin\Theta^{23}\cos\Theta^{14}\sin\Theta^{24}\sin\Theta^{34})(\nabla_\phi\Theta^{34}\nabla_\rho\Theta^{13}-\nabla_\rho\Theta^{34}\nabla_\phi\Theta^{13})+\sin\Theta^{23}\cos\Theta^{14}\cos\Theta^{24}\cos\Theta^{34}(\nabla_\phi\Theta^{24}\nabla_\rho\Theta^{13}\nonumber\\
			&&-\nabla_\rho\Theta^{24}\nabla_\phi\Theta^{13})+\sin\Theta^{24}\cos\Theta^{34}(\nabla_\phi\Theta^{24}\nabla_\rho\Theta^{14}-\nabla_\rho\Theta^{24}\nabla_\phi\Theta^{14})+\cos\Theta^{24}\sin\Theta^{34}(\nabla_\phi\Theta^{34}\nabla_\rho\Theta^{14}-\nabla_\rho\Theta^{34}\nabla_\phi\Theta^{14})\\
			Z_{\rho\phi}^{24}&=& (\cos\Theta^{13}\cos\Theta^{24}\sin\Theta^{34}-\sin\Theta^{13}\sin\Theta^{23}\sin\Theta^{14}\sin\Theta^{24}\sin\Theta^{34}+\sin\Theta^{13}\cos\Theta^{23}\sin\Theta^{14}\cos\Theta^{34})(\nabla_\phi\Theta^{13}\nabla_\rho\Theta^{12}-
			\nabla_\rho\Theta^{13}\nabla_\phi\Theta^{12})\nonumber\\
			&&+(-\sin\Theta^{13}\sin\Theta^{24}\sin\Theta^{34}+\cos\Theta^{13}\sin\Theta^{23}\sin\Theta^{14}\cos\Theta^{24}\sin\Theta^{34})(\nabla_\phi\Theta^{24}\nabla_\rho\Theta^{12}-\nabla_\rho\Theta^{24}\nabla_\phi\Theta^{12})+(\sin\Theta^{13}\cos\Theta^{24}\cos\Theta^{34}+\nonumber\\
			&&\cos\Theta^{13}\sin\Theta^{23}\sin\Theta^{14}\sin\Theta^{24}\cos\Theta^{34}+\cos\Theta^{13}\cos\Theta^{23}\sin\Theta^{14}\sin\Theta^{34})(\nabla_\phi\Theta^{34}\nabla_\rho\Theta^{12}-\nabla_\rho\Theta^{34}\nabla_\phi\Theta^{12})+(\cos\Theta^{13}\cos\Theta^{23}\sin\Theta^{14}\nonumber\\
			&&\sin\Theta^{24}\sin\Theta^{34}+\cos\Theta^{13}\sin\Theta^{23}\sin\Theta^{14}\cos\Theta^{34})(\nabla_\phi\Theta^{23}\nabla_\rho\Theta^{12}-\nabla_\rho\Theta^{23}\nabla_\phi\Theta^{12})+(\cos\Theta^{13}\sin\Theta^{23}\cos\Theta^{14}\sin\Theta^{24}\sin\Theta^{34}-\nonumber\\
			&&\cos\Theta^{13}\cos\Theta^{23}\cos\Theta^{14}\cos\Theta^{34})(\nabla_\phi\Theta^{14}\nabla_\rho\Theta^{12}-\nabla_\rho\Theta^{14}\nabla_\phi\Theta^{12})+(\sin
			\Theta^{23}\sin\Theta^{14}\sin\Theta^{24}\sin\Theta^{34}-\cos\Theta^{23}\sin\Theta^{14}\cos\Theta^{34})\nonumber\\
			&&(\nabla_\phi\Theta^{23}\nabla_\rho\Theta^{13}-\nabla_\rho\Theta^{23}\nabla_\phi\Theta^{13})+(-\cos\Theta^{23}\cos\Theta^{14}\sin\Theta^{24}\sin\Theta^{34}-\sin\Theta^{23}\cos\Theta^{14}\cos\Theta^{34})(\nabla_\phi\Theta^{14}\nabla_\rho\Theta^{13}-\nabla_\rho\Theta^{14}\nabla_\phi\Theta^{13})\nonumber\\
			&&-\cos\Theta^{23}\sin\Theta^{14}\cos\Theta^{24}\sin\Theta^{34}(\nabla_\phi\Theta^{24}\nabla_\rho\Theta^{13}-\nabla_\rho\Theta^{24}\nabla_\phi\Theta^{13})+(-\cos\Theta^{23}\sin\Theta^{14}\sin\Theta^{24}\cos\Theta^{34}+\sin\Theta^{23}\sin\Theta^{14}\sin\Theta^{34})\nonumber\\
			&&(\nabla_\phi\Theta^{34}\nabla_\rho\Theta^{13}-\nabla_\rho\Theta^{34}\nabla_\phi\Theta^{13})-\sin\Theta^{24}\sin\Theta^{34}(\nabla_\phi\Theta^{24}\nabla_\rho\Theta^{23}-\nabla_\rho\Theta^{24}\nabla_\phi\Theta^{23})+cos\Theta^{24}\cos\Theta^{34}(\nabla_\phi\Theta^{34}\nabla_\rho\Theta^{23}-\nabla_\rho\Theta^{34}\nabla_\phi\Theta^{23})\nonumber\\
			&&+\sin\Theta^{34}(\nabla_\phi\Theta^{34}\nabla_\rho\Theta^{24}-\nabla_\rho\Theta^{34}\nabla_\phi\Theta^{24})
		\end{eqnarray*}   
	\end{subequations}
\end{scriptsize}

\begin{scriptsize}
	\begin{subequations}
		\begin{eqnarray*}
			Z_{\rho\phi}^{34}&=&-\cos\Theta^{13}\sin\Theta^{24}(\nabla_\phi\Theta^{13}\nabla_\rho\Theta^{12}-
			\nabla_\rho\Theta^{13}\nabla_\phi\Theta^{12})-\sin\Theta^{13}\cos\Theta^{24}(\nabla_\phi\Theta^{24}\nabla_\rho\Theta^{12}-
			\nabla_\rho\Theta^{24}\nabla_\phi\Theta^{12})\nonumber\\
			&&-\sin\Theta^{13}\sin\Theta^{23}\sin\Theta^{14}\cos\Theta^{24}(\nabla_\phi\Theta^{13}\nabla_\rho\Theta^{12}-
			\nabla_\rho\Theta^{13}\nabla_\phi\Theta^{12})+\cos\Theta^{13}\cos\Theta^{23}\sin\Theta^{14}\cos\Theta^{24}\nonumber\\
			&&(\nabla_\phi\Theta^{23}\nabla_\rho\Theta^{12}-\nabla_\rho\Theta^{23}\nabla_\phi\Theta^{12})+\cos\Theta^{13}\sin\Theta^{23}\cos\Theta^{14}\cos\Theta^{24}
			(\nabla_\phi\Theta^{14}\nabla_\rho\Theta^{12}-\nabla_\rho\Theta^{14}\nabla_\phi\Theta^{12})\nonumber\\
			&&-\cos\Theta^{13}\sin\Theta^{23}\sin\Theta^{14}\sin\Theta^{24}
			(\nabla_\phi\Theta^{24}\nabla_\rho\Theta^{12}-\nabla_\rho\Theta^{24}\nabla_\phi\Theta^{12})+\sin\Theta^{23}\sin\Theta^{14}\cos\Theta^{24}\nonumber\\
			&&(\nabla_\phi\Theta^{23}\nabla_\rho\Theta^{13}-\nabla_\rho\Theta^{23}\nabla_\phi\Theta^{13})-\cos\Theta^{23}\cos\Theta^{14}\cos\Theta^{24}
			(\nabla_\phi\Theta^{14}\nabla_\rho\Theta^{13}-\nabla_\rho\Theta^{14}\nabla_\phi\Theta^{13})+\cos\Theta^{23}\nonumber\\
			&&\sin\Theta^{14}\sin\Theta^{24}
			(\nabla_\phi\Theta^{24}\nabla_\rho\Theta^{13}-\nabla_\rho\Theta^{24}\nabla_\phi\Theta^{13})-\cos\Theta^{24}(\nabla_\phi\Theta^{24}\nabla_\rho\Theta^{23}-
			\nabla_\rho\Theta^{24}\nabla_\phi\Theta^{23})
		\end{eqnarray*}   
	\end{subequations}
\end{scriptsize}

\noindent
{\bf Five State ($N=5$)}:~\citep{mantu_ijqc_curl}
\begin{scriptsize}
	\begin{subequations}
		\begin{eqnarray*}
			Z_{\rho\phi}^{15}&=&\sin\Theta^{12}\cos\Theta^{13}\cos\Theta^{14}(\nabla_\phi\Theta^{12}\nabla_\rho\Theta^{15}-
			\nabla_\rho\Theta^{12}\nabla_\phi\Theta^{15})+\cos\Theta^{12}\sin\Theta^{13}\cos\Theta^{14}(\nabla_\phi\Theta^{13}\nabla_\rho\Theta^{15}-
			\nabla_\rho\Theta^{13}\nabla_\phi\Theta^{15})+\cos\Theta^{12}\cos\Theta^{13}\sin\Theta^{14}\nonumber\\
			&&(\nabla_\phi\Theta^{14}\nabla_\rho\Theta^{15}-\nabla_\rho\Theta^{14}\nabla_\phi\Theta^{15})+(\sin\Theta^{15}\cos\Theta^{23}\cos\Theta^{24}\sin\Theta^{12}-
			\cos\Theta^{12}\sin\Theta^{15}\cos\Theta^{24}\sin\Theta^{13}\sin\Theta^{23}-\cos\Theta^{12}\cos\Theta^{13}\sin\Theta^{15}\sin\Theta^{14}\nonumber\\
			&&\sin\Theta^{24})(\nabla_\phi\Theta^{15}\nabla_\rho\Theta^{25}-\nabla_\rho\Theta^{15}\nabla_\phi\Theta^{25})+(\cos\Theta^{15}\sin\Theta^{23}\cos\Theta^{24}\sin\Theta^{12}+\cos\Theta^{12}\cos\Theta^{15}\cos\Theta^{24}\sin\Theta^{13}\cos\Theta^{23})(\nabla_\phi\Theta^{23}\nabla_\rho\Theta^{25}-\nonumber\\
			&&\nabla_\rho\Theta^{23}\nabla_\phi\Theta^{25})+
			(\cos\Theta^{15}\cos\Theta^{23}\sin\Theta^{24}\sin\Theta^{12}-\cos\Theta^{12}\cos\Theta^{15}\sin\Theta^{24}\sin\Theta^{13}\sin\Theta^{23}+\cos\Theta^{12}\cos\Theta^{13}\cos\Theta^{15}\sin\Theta^{14}\cos\Theta^{24})\nonumber\\
			&&(\nabla_\phi\Theta^{24}\nabla_\rho\Theta^{25}-\nabla_\rho\Theta^{24}\nabla_\phi\Theta^{25})+(\cos\Theta^{12}\cos\Theta^{15}\cos\Theta^{24}\cos\Theta^{13}\sin\Theta^{23}-\cos\Theta^{12}\sin\Theta^{13}\cos\Theta^{15}\sin\Theta^{14}\sin\Theta^{24})(\nabla_\phi\Theta^{13}\nabla_\rho\Theta^{25}\nonumber\\
			&&-\nabla_\rho\Theta^{13}\nabla_\phi\Theta^{25})+(-\cos\Theta^{15}\cos\Theta^{23}\cos\Theta^{24}\cos\Theta^{12}-\sin\Theta^{12}\cos\Theta^{15}\cos\Theta^{24}\sin\Theta^{13}\sin\Theta^{23}-\sin\Theta^{12}\cos\Theta^{13}\cos\Theta^{15}\nonumber\\
			&&\sin\Theta^{14}\sin\Theta^{24})(\nabla_\phi\Theta^{12}\nabla_\rho\Theta^{25}-\nabla_\rho\Theta^{12}\nabla_\phi\Theta^{25})+(\sin\Theta^{12}\cos\Theta^{15}\cos\Theta^{23}\cos\Theta^{25}\cos\Theta^{34}\sin\Theta^{13}-\cos\Theta^{15}\cos\Theta^{25}\cos\Theta^{34}\cos\Theta^{12}\sin\Theta^{23}\nonumber\\
			&&-\sin\Theta^{12}\cos\Theta^{13}\cos\Theta^{15}\cos\Theta^{24}\cos\Theta^{25}\sin\Theta^{14}\sin\Theta^{34}+\cos\Theta^{15}\cos\Theta^{23}\cos\Theta^{25}\cos\Theta^{12}\sin\Theta^{24}\sin\Theta^{34}+\sin\Theta^{12}\cos\Theta^{15}\cos\Theta^{25}\sin\Theta^{13}\nonumber\\
			&&\sin\Theta^{23}\sin\Theta^{24}\sin\Theta^{34})(\nabla_\phi\Theta^{12}\nabla_\rho\Theta^{35}-\nabla_\rho\Theta^{12}\nabla_\phi\Theta^{35})+(\cos\Theta^{12}\sin\Theta^{15}\cos\Theta^{23}\cos\Theta^{25}\cos\Theta^{34}\sin\Theta^{13}+\sin\Theta^{15}\cos\Theta^{25}\cos\Theta^{34}\sin\Theta^{12}\nonumber\\
			&&\sin\Theta^{23}-\cos\Theta^{12}\cos\Theta^{13}\sin\Theta^{15}\cos\Theta^{24}\cos\Theta^{25}\sin\Theta^{14}\sin\Theta^{34}-\sin\Theta^{15}\cos\Theta^{23}\cos\Theta^{25}\sin\Theta^{12}\sin\Theta^{24}\sin\Theta^{34}+\cos\Theta^{12}\sin\Theta^{15}\cos\Theta^{25}\nonumber\\
			&&\sin\Theta^{13}\sin\Theta^{23}\sin\Theta^{24}\sin\Theta^{34})(\nabla_\phi\Theta^{15}\nabla_\rho\Theta^{35}-\nabla_\rho\Theta^{15}\nabla_\phi\Theta^{35})+(\cos\Theta^{12}\cos\Theta^{15}\sin\Theta^{23}\cos\Theta^{25}\cos\Theta^{34}\sin\Theta^{13}-\cos\Theta^{15}\cos\Theta^{25}\cos\Theta^{34}\nonumber\\
			&&\sin\Theta^{12}\cos\Theta^{23}-\cos\Theta^{15}\sin\Theta^{23}\cos\Theta^{25}\sin\Theta^{12}\sin\Theta^{24}\sin\Theta^{34}-\cos\Theta^{12}\cos\Theta^{15}\cos\Theta^{25}\sin\Theta^{13}\cos\Theta^{23}\sin\Theta^{24}\sin\Theta^{34})(\nabla_\phi\Theta^{23}\nabla_\rho\Theta^{35}-\nonumber\\
			&&\nabla_\rho\Theta^{23}\nabla_\phi\Theta^{35})+(\cos\Theta^{12}\cos\Theta^{15}\cos\Theta^{23}\sin\Theta^{25}\cos\Theta^{34}\sin\Theta^{13}+\cos\Theta^{15}\sin\Theta^{25}\cos\Theta^{34}\sin\Theta^{12}\sin\Theta^{23}+\cos\Theta^{12}\cos\Theta^{13}\cos\Theta^{15}\cos\Theta^{24}\nonumber\\
			&&\sin\Theta^{25}\sin\Theta^{14}\sin\Theta^{34}-\cos\Theta^{15}\cos\Theta^{23}\sin\Theta^{25}\sin\Theta^{12}\sin\Theta^{24}\sin\Theta^{34}+\cos\Theta^{12}\cos\Theta^{15}\sin\Theta^{25}\sin\Theta^{13}\sin\Theta^{23}\sin\Theta^{24}\sin\Theta^{34})\nonumber\\
			&&(\nabla_\phi\Theta^{25}\nabla_\rho\Theta^{35}-\nabla_\rho\Theta^{25}\nabla_\phi\Theta^{35})+(\cos\Theta^{12}\cos\Theta^{15}\cos\Theta^{23}\cos\Theta^{25}\sin\Theta^{34}\sin\Theta^{13}+\cos\Theta^{15}\cos\Theta^{25}\sin\Theta^{34}\sin\Theta^{12}\sin\Theta^{23}+\cos\Theta^{12}\nonumber\\
			&&\cos\Theta^{13}\cos\Theta^{15}\cos\Theta^{24}\cos\Theta^{25}\sin\Theta^{14}\cos\Theta^{34}+\cos\Theta^{15}\cos\Theta^{23}\cos\Theta^{25}\sin\Theta^{12}\sin\Theta^{24}\cos\Theta^{34}-\cos\Theta^{12}\cos\Theta^{15}\cos\Theta^{25}\sin\Theta^{13}\sin\Theta^{23}\nonumber\\
			&&\sin\Theta^{24}\cos\Theta^{34})(\nabla_\phi\Theta^{34}\nabla_\rho\Theta^{35}-\nabla_\rho\Theta^{34}\nabla_\phi\Theta^{35})+(-\cos\Theta^{12}\cos\Theta^{15}\cos\Theta^{23}\cos\Theta^{25}\cos\Theta^{34}\cos\Theta^{13}-\cos\Theta^{12}\sin\Theta^{13}\cos\Theta^{15}\cos\Theta^{24}\nonumber\\
			&&\cos\Theta^{25}\sin\Theta^{14}\sin\Theta^{34}-\cos\Theta^{12}\cos\Theta^{15}\cos\Theta^{25}\cos\Theta^{13}\sin\Theta^{23}\sin\Theta^{24}\sin\Theta^{34})
			(\nabla_\phi\Theta^{13}\nabla_\rho\Theta^{35}-\nabla_\rho\Theta^{13}\nabla_\phi\Theta^{35})+(-\cos\Theta^{12}\cos\Theta^{13}\cos\Theta^{15}\nonumber\\
			&&\sin\Theta^{24}\cos\Theta^{25}\sin\Theta^{14}\sin\Theta^{34}+\cos\Theta^{15}\cos\Theta^{23}\cos\Theta^{25}\sin\Theta^{12}\cos\Theta^{24}\sin\Theta^{34}-\cos\Theta^{12}\cos\Theta^{15}\cos\Theta^{25}\sin\Theta^{13}\sin\Theta^{23}\cos\Theta^{24}\sin\Theta^{34})\nonumber\\
			&&(\nabla_\phi\Theta^{24}\nabla_\rho\Theta^{35}-\nabla_\rho\Theta^{24}\nabla_\phi\Theta^{35})+\cos\Theta^{12}\cos\Theta^{13}\cos\Theta^{15}\cos\Theta^{24}\cos\Theta^{25}\cos\Theta^{14}\sin\Theta^{34}(\nabla_\phi\Theta^{14}\nabla_\rho\Theta^{35}-\nabla_\rho\Theta^{14}\nabla_\phi\Theta^{35})+(\sin\Theta^{12}\cos\Theta^{13}\nonumber\\
			&&\cos\Theta^{15}\cos\Theta^{24}\cos\Theta^{25}\cos\Theta^{34}\cos\Theta^{35}\sin\Theta^{14}-\cos\Theta^{15}\cos\Theta^{23}\cos\Theta^{25}\cos\Theta^{34}\cos\Theta^{35}\cos\Theta^{12}\sin\Theta^{24}-\sin\Theta^{12}\cos\Theta^{15}\cos\Theta^{25}\cos\Theta^{34}\nonumber\\
			&&\cos\Theta^{35}\sin\Theta^{13}\sin\Theta^{23}\sin\Theta^{24}+\sin\Theta^{12}\cos\Theta^{15}\cos\Theta^{23}\cos\Theta^{25}\cos\Theta^{35}\sin\Theta^{13}\sin\Theta^{34}-\cos\Theta^{15}\cos\Theta^{25}\cos\Theta^{35}\cos\Theta^{12}\sin\Theta^{23}\sin\Theta^{34})\nonumber\\
			&&(\nabla_\phi\Theta^{12}\nabla_\rho\Theta^{45}-\nabla_\rho\Theta^{12}\nabla_\phi\Theta^{45})+(\cos\Theta^{12}\sin\Theta^{13}\cos\Theta^{15}\cos\Theta^{24}\cos\Theta^{25}\cos\Theta^{34}\cos\Theta^{35}\sin\Theta^{14}+\cos\Theta^{12}\cos\Theta^{15}\cos\Theta^{25}\cos\Theta^{34}\cos\Theta^{35}\nonumber\\
			&&\cos\Theta^{13}\sin\Theta^{23}\sin\Theta^{24}-\cos\Theta^{12}\cos\Theta^{15}\cos\Theta^{23}\cos\Theta^{25}\cos\Theta^{35}\cos\Theta^{13}\sin\Theta^{34})
			(\nabla_\phi\Theta^{13}\nabla_\rho\Theta^{45}-\nabla_\rho\Theta^{13}\nabla_\phi\Theta^{45})+(\cos\Theta^{12}\cos\Theta^{13}\sin\Theta^{15}\nonumber\\
			&&\cos\Theta^{24}\cos\Theta^{25}\cos\Theta^{34}\cos\Theta^{35}\sin\Theta^{14}+\sin\Theta^{15}\cos\Theta^{23}\cos\Theta^{25}\cos\Theta^{34}\cos\Theta^{35}\sin\Theta^{12}\sin\Theta^{24}-\cos\Theta^{12}\sin\Theta^{15}\cos\Theta^{25}\cos\Theta^{34}\cos\Theta^{35}\nonumber\\
			&&\sin\Theta^{13}\sin\Theta^{23}\sin\Theta^{24}+\cos\Theta^{12}\sin\Theta^{15}\cos\Theta^{23}\cos\Theta^{25}\cos\Theta^{35}\sin\Theta^{13}\sin\Theta^{34}+\sin\Theta^{15}\cos\Theta^{25}\cos\Theta^{35}\sin\Theta^{12}\sin\Theta^{23}\sin\Theta^{34})\nonumber\\
			&&(\nabla_\phi\Theta^{15}\nabla_\rho\Theta^{45}-\nabla_\rho\Theta^{15}\nabla_\phi\Theta^{45})+(\cos\Theta^{12}\cos\Theta^{13}\cos\Theta^{15}\sin\Theta^{24}\cos\Theta^{25}\cos\Theta^{34}\cos\Theta^{35}\sin\Theta^{14}-\cos\Theta^{15}\cos\Theta^{23}\cos\Theta^{25}\cos\Theta^{34}\cos\Theta^{35}\nonumber\\
			&&\sin\Theta^{12}\cos\Theta^{24}+\cos\Theta^{12}\cos\Theta^{15}\cos\Theta^{25}\cos\Theta^{34}\cos\Theta^{35}\sin\Theta^{13}\sin\Theta^{23}\cos\Theta^{24})
			(\nabla_\phi\Theta^{24}\nabla_\rho\Theta^{45}-\nabla_\rho\Theta^{24}\nabla_\phi\Theta^{45})+(\cos\Theta^{12}\cos\Theta^{13}\cos\Theta^{15}\nonumber\\
			&&\cos\Theta^{24}\sin\Theta^{25}\cos\Theta^{34}\cos\Theta^{35}\sin\Theta^{14}+\cos\Theta^{15}\cos\Theta^{23}\sin\Theta^{25}\cos\Theta^{34}\cos\Theta^{35}\sin\Theta^{12}\sin\Theta^{24}-\cos\Theta^{12}\cos\Theta^{15}\sin\Theta^{25}\cos\Theta^{34}\cos\Theta^{35}\nonumber\\
			&&\sin\Theta^{13}\sin\Theta^{23}\sin\Theta^{24}+\cos\Theta^{12}\cos\Theta^{15}\cos\Theta^{23}\sin\Theta^{25}\cos\Theta^{35}\sin\Theta^{13}\sin\Theta^{34}+\cos\Theta^{15}\sin\Theta^{25}\cos\Theta^{35}\sin\Theta^{12}\sin\Theta^{23}\sin\Theta^{34})\nonumber\\
			&&(\nabla_\phi\Theta^{25}\nabla_\rho\Theta^{45}-\nabla_\rho\Theta^{25}\nabla_\phi\Theta^{45})+(\cos\Theta^{12}\cos\Theta^{13}\cos\Theta^{15}\cos\Theta^{24}\cos\Theta^{25}\sin\Theta^{34}\cos\Theta^{35}\sin\Theta^{14}+\cos\Theta^{15}\cos\Theta^{23}\cos\Theta^{25}\sin\Theta^{34}\cos\Theta^{35}\nonumber\\
			&&\sin\Theta^{12}\sin\Theta^{24}-\cos\Theta^{12}\cos\Theta^{15}\cos\Theta^{25}\sin\Theta^{34}\cos\Theta^{35}\sin\Theta^{13}\sin\Theta^{23}\sin\Theta^{24}-\cos\Theta^{12}\cos\Theta^{15}\cos\Theta^{23}\cos\Theta^{25}\cos\Theta^{35}\sin\Theta^{13}\cos\Theta^{34}\nonumber\\
			&&-\cos\Theta^{15}\cos\Theta^{25}\cos\Theta^{35}\sin\Theta^{12}\sin\Theta^{23}\cos\Theta^{34})(\nabla_\phi\Theta^{34}\nabla_\rho\Theta^{45}-\nabla_\rho\Theta^{34}\nabla_\phi\Theta^{45})+(\cos\Theta^{12}\cos\Theta^{13}\cos\Theta^{15}\cos\Theta^{24}\cos\Theta^{25}\cos\Theta^{34}\sin\Theta^{35}\nonumber\\
			&&\sin\Theta^{14}+\cos\Theta^{15}\cos\Theta^{23}\cos\Theta^{25}\cos\Theta^{34}\sin\Theta^{35}\sin\Theta^{12}\sin\Theta^{24}-\cos\Theta^{12}
			\cos\Theta^{15}\cos\Theta^{25}\cos\Theta^{34}\sin\Theta^{35}\sin\Theta^{13}\sin\Theta^{23}\sin\Theta^{24}+\cos\Theta^{12}\nonumber\\
			&&\cos\Theta^{15}\cos\Theta^{23}\cos\Theta^{25}\sin\Theta^{35}\sin\Theta^{13}\sin\Theta^{34}+\cos\Theta^{15}\cos\Theta^{25}\sin\Theta^{35}\sin\Theta^{12}\sin\Theta^{23}\sin\Theta^{34})(\nabla_\phi\Theta^{35}\nabla_\rho\Theta^{45}-\nabla_\rho\Theta^{35}\nabla_\phi\Theta^{45})\nonumber\\
			&&-\cos\Theta^{12}\cos\Theta^{13}\cos\Theta^{15}\cos\Theta^{24}\cos\Theta^{25}\cos\Theta^{34}\cos\Theta^{35}\cos\Theta^{14}(\nabla_\phi\Theta^{14}\nabla_\rho\Theta^{45}-\nabla_\rho\Theta^{14}\nabla_\phi\Theta^{45})+(\cos\Theta^{15}\sin\Theta^{23}\cos\Theta^{25}\cos\Theta^{34}\cos\Theta^{35}\nonumber\\
			&&\sin\Theta^{12}\sin\Theta^{24}+\cos\Theta^{12}\cos\Theta^{15}\sin\Theta^{23}\cos\Theta^{25}\cos\Theta^{35}\sin\Theta^{13}\sin\Theta^{34}-\cos\Theta^{15}\cos\Theta^{25}\cos\Theta^{35}\sin\Theta^{12}\cos\Theta^{23}\sin\Theta^{34}+\cos\Theta^{12}\cos\Theta^{15}\nonumber\\
			&&\cos\Theta^{25}\cos\Theta^{34}\cos\Theta^{35}\sin\Theta^{13}\cos\Theta^{23}\sin\Theta^{24})(\nabla_\phi\Theta^{23}\nabla_\rho\Theta^{45}-\nabla_\rho\Theta^{23}\nabla_\phi\Theta^{45})+\cos\Theta^{12}\cos\Theta^{13}\cos\Theta^{15}\cos\Theta^{14}\sin\Theta^{24}(\nabla_\phi\Theta^{14}\nabla_\rho\Theta^{25}\nonumber\\
			&&-\nabla_\rho\Theta^{14}\nabla_\phi\Theta^{25})
		\end{eqnarray*}
	\end{subequations}
\end{scriptsize}

\begin{scriptsize}
	\begin{subequations}
		\begin{eqnarray*}
			Z_{\rho\phi}^{45}&=&\cos\Theta^{14}(\nabla_\phi\Theta^{14}\nabla_\rho\Theta^{15}-\nabla_\rho\Theta^{14}\nabla_\phi\Theta^{15})-\sin\Theta^{14}\cos\Theta^{15}
			\sin\Theta^{24}(\nabla_\phi\Theta^{14}\nabla_\rho\Theta^{25}-\nabla_\rho\Theta^{14}\nabla_\phi\Theta^{25})-\cos\Theta^{14}\sin\Theta^{15}
			\sin\Theta^{24}(\nabla_\phi\Theta^{15}\nabla_\rho\Theta^{25}\nonumber\\
			&&-\nabla_\rho\Theta^{15}\nabla_\phi\Theta^{25})+\cos\Theta^{14}\cos\Theta^{15}
			\cos\Theta^{24}(\nabla_\phi\Theta^{24}\nabla_\rho\Theta^{25}-\nabla_\rho\Theta^{24}\nabla_\phi\Theta^{25})-\sin\Theta^{14}\cos\Theta^{15}\cos\Theta^{24}\cos\Theta^{25}
			\sin\Theta^{34}(\nabla_\phi\Theta^{14}\nabla_\rho\Theta^{35}\nonumber\\
			&&-\nabla_\rho\Theta^{14}\nabla_\phi\Theta^{35})-\cos\Theta^{14}\sin\Theta^{15}\cos\Theta^{24}\cos\Theta^{25}
			\sin\Theta^{34}(\nabla_\phi\Theta^{15}\nabla_\rho\Theta^{35}-\nabla_\rho\Theta^{15}\nabla_\phi\Theta^{35})-\cos\Theta^{14}\cos\Theta^{15}\sin\Theta^{24}\cos\Theta^{25}
			\sin\Theta^{34}\nonumber\\
			&&(\nabla_\phi\Theta^{24}\nabla_\rho\Theta^{35}-\nabla_\rho\Theta^{24}\nabla_\phi\Theta^{35})-\cos\Theta^{14}\cos\Theta^{15}\cos\Theta^{24}\sin\Theta^{25}
			\sin\Theta^{34}(\nabla_\phi\Theta^{25}\nabla_\rho\Theta^{35}-\nabla_\rho\Theta^{25}\nabla_\phi\Theta^{35})\nonumber\\
			&&+\cos\Theta^{14}\cos\Theta^{15}\cos\Theta^{24}\cos\Theta^{25}
			\cos\Theta^{34}(\nabla_\phi\Theta^{34}\nabla_\rho\Theta^{35}-\nabla_\rho\Theta^{34}\nabla_\phi\Theta^{35})+\sin\Theta^{14}\cos\Theta^{15}\cos\Theta^{24}
			\cos\Theta^{25}\cos\Theta^{34}\cos\Theta^{35}(\nabla_\phi\Theta^{14}\nabla_\rho\Theta^{45}\nonumber\\
			&&-\nabla_\rho\Theta^{14}\nabla_\phi\Theta^{45})+\cos\Theta^{14}\sin\Theta^{15}\cos\Theta^{24}
			\cos\Theta^{25}\cos\Theta^{34}\cos\Theta^{35}(\nabla_\phi\Theta^{15}\nabla_\rho\Theta^{45}-\nabla_\rho\Theta^{15}\nabla_\phi\Theta^{45})\nonumber\\
			&&+\cos\Theta^{14}\cos\Theta^{15}\sin\Theta^{24}
			\cos\Theta^{25}\cos\Theta^{34}\cos\Theta^{35}(\nabla_\phi\Theta^{24}\nabla_\rho\Theta^{45}-\nabla_\rho\Theta^{24}\nabla_\phi\Theta^{45})+\cos\Theta^{14}\cos\Theta^{15}\cos\Theta^{24}
			\sin\Theta^{25}\cos\Theta^{34}\cos\Theta^{35}\nonumber\\
			&&(\nabla_\phi\Theta^{25}\nabla_\rho\Theta^{45}-\nabla_\rho\Theta^{25}\nabla_\phi\Theta^{45})+\cos\Theta^{14}\cos\Theta^{15}\cos\Theta^{24}
			\cos\Theta^{25}\sin\Theta^{34}\cos\Theta^{35}(\nabla_\phi\Theta^{34}\nabla_\rho\Theta^{45}-\nabla_\rho\Theta^{34}\nabla_\phi\Theta^{45})\nonumber\\
			&&+\cos\Theta^{14}\cos\Theta^{15}\cos\Theta^{24}
			\cos\Theta^{25}\sin\Theta^{34}\sin\Theta^{35}(\nabla_\phi\Theta^{35}\nabla_\rho\Theta^{45}-\nabla_\rho\Theta^{35}\nabla_\phi\Theta^{45})
		\end{eqnarray*}
	\end{subequations}
\end{scriptsize}

\noindent
{\bf Six State ($N=6$)}:~\citep{mantu_ijqc_curl}
\begin{scriptsize}
	\begin{subequations}
		\begin{eqnarray*}
			Z_{\rho\phi}^{56}&=&\cos\Theta^{15}(\nabla_\phi\Theta^{15}\nabla_\rho\Theta^{16}-\nabla_\rho\Theta^{15}\nabla_\phi\Theta^{16})-\sin\Theta^{15}\cos\Theta^{16}
			\sin\Theta^{25}(\nabla_\phi\Theta^{15}\nabla_\rho\Theta^{26}-\nabla_\rho\Theta^{15}\nabla_\phi\Theta^{26})-\cos\Theta^{15}\sin\Theta^{16}\sin\Theta^{25}
			\nonumber\\
			&&(\nabla_\phi\Theta^{16}\nabla_\rho\Theta^{26}-\nabla_\rho\Theta^{16}\nabla_\phi\Theta^{26})+\cos\Theta^{15}\cos\Theta^{16}\cos\Theta^{25}
			(\nabla_\phi\Theta^{25}\nabla_\rho\Theta^{26}-\nabla_\rho\Theta^{25}\nabla_\phi\Theta^{26})-\sin\Theta^{15}\cos\Theta^{16}\cos\Theta^{25}\cos\Theta^{26}\sin\Theta^{35}\nonumber\\
			&&(\nabla_\phi\Theta^{15}\nabla_\rho\Theta^{36}-\nabla_\rho\Theta^{15}\nabla_\phi\Theta^{36})-\cos\Theta^{15}\sin\Theta^{16}\cos\Theta^{25}\cos\Theta^{26}\sin\Theta^{35}(\nabla_\phi\Theta^{16}\nabla_\rho\Theta^{36}-\nabla_\rho\Theta^{16}\nabla_\phi\Theta^{36})-\cos\Theta^{15}\cos\Theta^{16}\sin\Theta^{25}\nonumber\\
			&&\cos\Theta^{26}\sin\Theta^{35}(\nabla_\phi\Theta^{25}\nabla_\rho\Theta^{36}-\nabla_\rho\Theta^{25}\nabla_\phi\Theta^{36})-\cos\Theta^{15}\cos\Theta^{16}\cos\Theta^{25}\sin\Theta^{26}\sin\Theta^{35}(\nabla_\phi\Theta^{26}\nabla_\rho\Theta^{36}-\nabla_\rho\Theta^{26}\nabla_\phi\Theta^{36})\nonumber\\
			&&+\cos\Theta^{15}\cos\Theta^{16}\cos\Theta^{25}\cos\Theta^{26}\cos\Theta^{35}(\nabla_\phi\Theta^{35}\nabla_\rho\Theta^{36}-\nabla_\rho\Theta^{35}\nabla_\phi\Theta^{36})-\sin\Theta^{15}\cos\Theta^{16}\cos\Theta^{25}\cos\Theta^{26}\cos\Theta^{35}\cos\Theta^{36}\sin\Theta^{45}\nonumber\\
			&&(\nabla_\phi\Theta^{15}\nabla_\rho\Theta^{46}-\nabla_\rho\Theta^{15}\nabla_\phi\Theta^{46})-\cos\Theta^{15}\sin\Theta^{16}\cos\Theta^{25}\cos\Theta^{26}\cos\Theta^{35}\cos\Theta^{36}\sin\Theta^{45}(\nabla_\phi\Theta^{16}\nabla_\rho\Theta^{46}-\nabla_\rho\Theta^{16}\nabla_\phi\Theta^{46})-\nonumber\\
			&&\cos\Theta^{15}\cos\Theta^{16}\sin\Theta^{25}\cos\Theta^{26}\cos\Theta^{35}\cos\Theta^{36}\sin\Theta^{45}(\nabla_\phi\Theta^{25}\nabla_\rho\Theta^{46}-\nabla_\rho\Theta^{25}\nabla_\phi\Theta^{46})-\cos\Theta^{15}\cos\Theta^{16}\cos\Theta^{25}\sin\Theta^{26}\nonumber\\
			&&\cos\Theta^{35}\cos\Theta^{36}\sin\Theta^{45}(\nabla_\phi\Theta^{26}\nabla_\rho\Theta^{46}-\nabla_\rho\Theta^{26}\nabla_\phi\Theta^{46})-\cos\Theta^{15}\cos\Theta^{16}\cos\Theta^{25}\cos\Theta^{26}\sin\Theta^{35}\cos\Theta^{36}\sin\Theta^{45}\nonumber\\
			&&(\nabla_\phi\Theta^{35}\nabla_\rho\Theta^{46}-\nabla_\rho\Theta^{35}\nabla_\phi\Theta^{46})-\cos\Theta^{15}\cos\Theta^{16}\cos\Theta^{25}\cos\Theta^{26}
			\cos\Theta^{35}\sin\Theta^{36}\sin\Theta^{45}(\nabla_\phi\Theta^{36}\nabla_\rho\Theta^{46}-\nabla_\rho\Theta^{36}\nabla_\phi\Theta^{46})\nonumber\\
			&&+\cos\Theta^{15}\cos\Theta^{16}\cos\Theta^{25}\cos\Theta^{26}\cos\Theta^{35}\cos\Theta^{36}\cos\Theta^{45}(\nabla_\phi\Theta^{45}\nabla_\rho\Theta^{46}-\nabla_\rho\Theta^{45}\nabla_\phi\Theta^{46})+\sin\Theta^{15}\cos\Theta^{16}\cos\Theta^{25}\cos\Theta^{26}\cos\Theta^{35}\nonumber\\
			&&\cos\Theta^{36}\cos\Theta^{45}\cos\Theta^{46}(\nabla_\phi\Theta^{15}\nabla_\rho\Theta^{56}-\nabla_\rho\Theta^{15}\nabla_\phi\Theta^{56})+
			\cos\Theta^{15}\sin\Theta^{16}\cos\Theta^{25}\cos\Theta^{26}
			\cos\Theta^{35}\cos\Theta^{36}\cos\Theta^{45}\cos\Theta^{46}\nonumber\\
			&&(\nabla_\phi\Theta^{16}\nabla_\rho\Theta^{56}-\nabla_\rho\Theta^{16}\nabla_\phi\Theta^{56})+\cos\Theta^{15}\cos\Theta^{16}\sin\Theta^{25}\cos\Theta^{26}\cos\Theta^{35}\cos\Theta^{36}\cos\Theta^{45}\cos\Theta^{46}(\nabla_\phi\Theta^{25}\nabla_\rho\Theta^{56}-\nabla_\rho\Theta^{25}\nabla_\phi\Theta^{56})\nonumber\\
			&&+\cos\Theta^{15}\cos\Theta^{16}\cos\Theta^{25}\sin\Theta^{26}\cos\Theta^{35}\cos\Theta^{36}\cos\Theta^{45}\cos\Theta^{46}(\nabla_\phi\Theta^{26}\nabla_\rho\Theta^{56}-\nabla_\rho\Theta^{26}\nabla_\phi\Theta^{56})+\cos\Theta^{15}\cos\Theta^{16}\cos\Theta^{25}\cos\Theta^{26}\nonumber\\
			&&\sin\Theta^{35}\cos\Theta^{36}\cos\Theta^{45}\cos\Theta^{46}(\nabla_\phi\Theta^{35}\nabla_\rho\Theta^{56}-\nabla_\rho\Theta^{35}\nabla_\phi\Theta^{56})+\cos\Theta^{15}\cos\Theta^{16}\cos\Theta^{25}\cos\Theta^{26}\cos\Theta^{35}\sin\Theta^{36}\cos\Theta^{45}\cos\Theta^{46}\nonumber\\
			&&(\nabla_\phi\Theta^{36}\nabla_\rho\Theta^{56}-\nabla_\rho\Theta^{36}\nabla_\phi\Theta^{56})+\cos\Theta^{15}\cos\Theta^{16}\cos\Theta^{25}\cos\Theta^{26}\cos\Theta^{35}\cos\Theta^{36}\sin\Theta^{45}\cos\Theta^{46}(\nabla_\phi\Theta^{45}\nabla_\rho\Theta^{56}-\nabla_\rho\Theta^{45}\nabla_\phi\Theta^{56})\nonumber\\
			&&+\cos\Theta^{15}\cos\Theta^{16}\cos\Theta^{25}\cos\Theta^{26}\cos\Theta^{35}\cos\Theta^{36}\cos\Theta^{45}\sin\Theta^{46}(\nabla_\phi\Theta^{46}\nabla_\rho\Theta^{56}-\nabla_\rho\Theta^{46}\nabla_\phi\Theta^{56})
		\end{eqnarray*}
	\end{subequations}
\end{scriptsize} 

\section{Explicit forms of adiabatic potential energy matrix, nonadiabatic coupling matrix, K$_\lambda$-transformation matrix and spin-orbit Hamiltonian matrix}\label{ssec:so_coup}

\noindent
In case of two spin states of F atom ($^2P_{1/2}$ and $^2P_{3/2}$), the 6 x 6 adiabatic potential energy ($U_{ij} =u_i\delta_{ij}$) and skew-symmetric nonadiabatic coupling  ($\boldsymbol{\tau}_{ij} = \langle\xi_i\vert\boldsymbol{\nabla}_R\xi_j\rangle_\textbf{r} = -\boldsymbol{\tau}_{ji}$) matrices are written as follows :

\begin{eqnarray}
	U = 
	\begin{array}{|c|cccccc|}
		• & |\xi_1\rangle & |\xi_2\rangle & |\xi_3\rangle & |\overline{\xi_1}\rangle & |\overline{\xi_2}\rangle & |\overline{\xi_3}\rangle \\ 
		\hline 
		|\xi_1\rangle & u_1 & 0 & 0 & 0 & 0 & 0 \\ 
		|\xi_2\rangle & 0 & u_2 & 0 & 0 & 0 & 0 \\ 
		|\xi_3\rangle & 0 & 0 & u_3 & 0 & 0 & 0 \\ 
		|\overline{\xi_1}\rangle & 0 & 0 & 0 & u_1 & 0 & 0 \\ 
		|\overline{\xi_2}\rangle & 0 & 0 & 0 & 0 & u_2 & 0 \\ 
		|\overline{\xi_3}\rangle & 0 & 0 & 0 & 0 & 0 & u_3 
		\label{eq:umat}
	\end{array}
\end{eqnarray}

and
\begin{eqnarray}
	\boldsymbol{\tau} =
	\begin{pmatrix}
		0 & \boldsymbol{\tau}_{12} & \boldsymbol{\tau}_{13} & 0 & 0 & 0 \\
		-\boldsymbol{\tau}_{12} & 0 & \boldsymbol{\tau}_{23} & 0 & 0 & 0 \\
		-\boldsymbol{\tau}_{13} & -\boldsymbol{\tau}_{23} & 0 & 0 & 0 & 0 \\
		0 & 0 & 0 & 0 & \boldsymbol{\tau}_{12} & \boldsymbol{\tau}_{13} \\
		0 & 0 & 0 & -\boldsymbol{\tau}_{12} & 0 & \boldsymbol{\tau}_{23} \\
		0 & 0 & 0 & -\boldsymbol{\tau}_{13} & -\boldsymbol{\tau}_{23} & 0
	\end{pmatrix}.
	\label{eq:taumat}
\end{eqnarray}

\noindent
On the other hand, the $K_\lambda$ transformation matrix can be written as:

\begin{eqnarray}
	K_\lambda = \frac{1}{\sqrt{2}}
	\begin{pmatrix}
		1 & 0 & 0 & -1 & 0 & 0 \\
		0 & 1 & 0 & 0 & -1 & 0 \\
		0 & 0 & i & 0 & 0 & i \\
		1 & 0 & 0 & 1 & 0 & 0 \\
		0 & 1 & 0 & 0 & 1 & 0 \\
		0 & 0 & i & 0 & 0 & -i
	\end{pmatrix}.
	\label{eq:kmat}
\end{eqnarray}

\noindent
The form of the  spin-orbit (SO) Hamiltonian matrix $H^{ad}_{SO}$ in collinear geometries appears as: 
\begin{small}
	\begin{eqnarray}
		H_{\rm so}^{\rm ad} = 
		\begin{array}{|c|cccccc|}
			• & |\xi_1\rangle & |\xi_2\rangle & |\xi_3\rangle & |\overline{\xi_1}\rangle
			& |\overline{\xi_2}\rangle & |\overline{\xi_3}\rangle \\ 
			\hline 
			|\xi_1\rangle & 0 & 0 & 0 & 0 & -iF & F \\ 
			|\xi_2\rangle & 0 & 0 & -iG & iF & 0 & 0 \\ 
			|\xi_3\rangle & 0 & iG & 0 & -F & 0 & 0 \\ 
			|\overline{\xi_1}\rangle & 0 & -iF & -F & 0 & 0 & 0 \\ 
			|\overline{\xi_2}\rangle & iF & 0 & 0 & 0 & 0 & iG \\ 
			|\overline{\xi_3}\rangle & F & 0 & 0 & 0 & -iG & 0 \\ 
		\end{array},
		\label{eq:somat1}
	\end{eqnarray}
\end{small}
where $F=\langle \xi_1 | H_{\rm so} | \overline{\xi}_3 \rangle$ and $G=i\langle \xi_2 | H_{\rm so} | \xi_3 \rangle$. 

\vspace{0.3 cm}

\noindent
The diagonal form of the adiabatic potential as well as the SO coupling matrix are then transformed in terms of the $\lambda$-basis:\cite{FH2_dynamics} 

\begin{small}
	\begin{equation}
		V = (K_\lambda^T)^{\dagger}U K_\lambda^T + (K_\lambda^T)^{\dagger}H_{\rm so}^{\rm ad} K_\lambda^T = 
		\begin{pmatrix}
			u_1 & iF & iF & 0 & 0 & 0 \\
			-iF & u_2 & G & 0 & 0 & 0 \\
			-iF & G & u_3 & 0 & 0 & 0 \\
			0 & 0 & 0 & u_1 & -iF & -iF \\
			0 & 0 & 0 & iF & u_2 & G \\
			0 & 0 & 0 & iF & G & u_3
		\end{pmatrix}.
		\label{eq:vmat}
	\end{equation}
\end{small}

\vspace{0.3cm}

\noindent
Moreover, the kinetic energy operator in the $\lambda$-basis can be represented as:
\begin{eqnarray}
	\hat{T}_{\rm nuc} & =& -\frac{\hbar^2}{2} \left( (K_\lambda^T)^{\dagger} \left(\boldsymbol{\nabla}_R+ \boldsymbol{\tau}  \right)^2 K_\lambda^T \right)\nonumber\\
	& = & -\frac{\hbar^2}{2} \left( (K_\lambda^T)^{\dagger} \left(\boldsymbol{\nabla}_R+ \boldsymbol{\tau}  \right) K_\lambda^T (K_\lambda^T)^{\dagger} \left(\boldsymbol{\nabla}_R+ \boldsymbol{\tau}  \right) K_\lambda^T\right)\nonumber\\
	&\approx& -\frac{\hbar^2}{2}  \left(\boldsymbol{\nabla}_R+ \boldsymbol{\tau} \right)^2
	\label{eq:kinso}
\end{eqnarray}

\noindent
Since $K_\lambda$ represents a constant matrix, the commutator $[\boldsymbol{\nabla}_R, K_\lambda^T]$ vanishes. Moreover, the matrix elements of the commutator, $[\boldsymbol{\tau}, K_\lambda^T]$ can also be approximated to be zero, which allows the kinetic energy part of the Hamiltonian to remain unchanged in the $\lambda$ basis.

\vspace{0.3cm}

\noindent
In this context, for the F+H$_2$ system, the formulation involves  the SO coupling to the potential matrix in
the adiabatic representation, where the kinetic part of the Hamiltonian does not change under the time-reversal adapted Kramer's basis with the $C_s$ double group. In the adiabatic framework, reaction probabilities at total angular momentum without (J) and with spin orbit coupling (J$^\prime $=$ J+\frac{1}{2} $) can be compared for zero  as well as non-zero J(J$^\prime$) values [see Section III of Ref.~\citen{jcp-109-5710-WERNER}].

\vspace{0.3cm}

\noindent
While substituting Eq. \ref{eq:vmat} and \ref{eq:kinso} in Eq. \ref{eq:adsem1}, the final block diagonal form of the
SE in adiabatic representation takes the following form:

\begin{eqnarray}
	\left[ -\frac{\hbar^2}{2} \begin{pmatrix}
		\boldsymbol{\nabla}_R & \boldsymbol{\tau}_{12} & \boldsymbol{\tau}_{13}  \\
		-\boldsymbol{\tau}_{12} & \boldsymbol{\nabla}_R & \boldsymbol{\tau}_{23}  \\
		-\boldsymbol{\tau}_{13} & -\boldsymbol{\tau}_{23} & \boldsymbol{\nabla}_R 
	\end{pmatrix} ^2 + 
	\begin{pmatrix}
		u_1 -E & iF & iF   \\
		-iF & u_2 -E & G \\
		-iF & G & u_3 -E  
	\end{pmatrix}\right] \begin{pmatrix}
		\psi^{ad}_1  \\
		\psi^{ad}_2  \\
		\psi^{ad}_3  \\
	\end{pmatrix} = \boldsymbol{0}, \qquad
	\label{eq:semat}
\end{eqnarray}

\subsection{Commutator of non-adiabatic coupling matrix (NACM) and $K_\lambda$ matrix}\label{ssec:so_com}

\noindent
For two possible spin projections ($M_s = \pm\frac{1}{2}$) and three state sub-Hilbert space (1$^2$A$^\prime$, 2$^2$A$^\prime$ and 1$^2$A$^{\prime\prime}$ ) of F+H$_2$ system, the NACM ($\boldsymbol{\tau}$) can be written as:

\begin{eqnarray}
	\boldsymbol{\tau} =
	\begin{pmatrix}
		0 & \boldsymbol{\tau}_{12} & \boldsymbol{\tau}_{13} & 0 & 0 & 0 \\
		-\boldsymbol{\tau}_{12} & 0 & \boldsymbol{\tau}_{23} & 0 & 0 & 0 \\
		-\boldsymbol{\tau}_{13} & -\boldsymbol{\tau}_{23} & 0 & 0 & 0 & 0 \\
		0 & 0 & 0 & 0 & \boldsymbol{\tau}_{12} & \boldsymbol{\tau}_{13} \\
		0 & 0 & 0 & -\boldsymbol{\tau}_{12} & 0 & \boldsymbol{\tau}_{23} \\
		0 & 0 & 0 & -\boldsymbol{\tau}_{13} & -\boldsymbol{\tau}_{23} & 0
	\end{pmatrix}
	\label{eq:taumat6}
\end{eqnarray}

\noindent
Therefore, the commutator $[\boldsymbol{\tau},K_\lambda^T]$ can be represented as:

\begin{small}
	\begin{eqnarray}
		[\boldsymbol{\tau},K_\lambda^T] = \frac{1}{\sqrt{2}} 
		\begin{pmatrix}
			0 & 0 & (i-1)\boldsymbol{\tau}_{13} & 0 & 0 & (i-1)\boldsymbol{\tau}_{13} \\
			0 & 0 & (i-1)\boldsymbol{\tau}_{23} & 0 & 0 & (i-1)\boldsymbol{\tau}_{23} \\
			(i-1)\boldsymbol{\tau}_{13} & (i-1)\boldsymbol{\tau}_{23} & 0 & (i-1)\boldsymbol{\tau}_{13} & 
			(i-1)\boldsymbol{\tau}_{23} & 0\\
			0 & 0 & (i+1)\boldsymbol{\tau}_{13} & 0 & 0 & -(i+1)\boldsymbol{\tau}_{13} \\
			0 & 0 & (i+1)\boldsymbol{\tau}_{23} & 0 & 0 & -(i+1)\boldsymbol{\tau}_{23} \\
			(i+1)\boldsymbol{\tau}_{13} & (i+1)\boldsymbol{\tau}_{23} & 0 & -(i+1)\boldsymbol{\tau}_{13} 
			& -(i+1)\boldsymbol{\tau}_{23} & 0\\
		\end{pmatrix} 
	\end{eqnarray}
\end{small}

\noindent
In C$_s$ symmetry, since the $\Sigma$ state (1$^2$A$^\prime$) and $\Pi$ states (2$^2$A$^\prime$ and 1$^2$A$^{\prime\prime}$) encounter interchange over certain regions of the nuclear configuration space, the definitions of those states  get changed, namely, from \{1$^2$A$^\prime$, 2$^2$A$^\prime$, 1$^2$A$^{\prime\prime}$\} $\rightarrow$ \{1$^2$A$^\prime$, 1$^2$A$^{\prime\prime}$, 2$^2$A$^\prime$\}. As a consequence, \textit{ab initio} calculation demonstrates that a region of the profiles for $\boldsymbol{\tau}_{12}$ and $\boldsymbol{\tau}_{13}$ are also interchanged and hence, those needs to be swapped\cite{jcp-153-174301-SA}. The modified $\boldsymbol{\tau}_{12}$ ($\boldsymbol{\tau}^{\prime}_{12}$) incorporates the features of both the original $\boldsymbol{\tau}_{12}$ and $\boldsymbol{\tau}_{13}$, while the magnitude of $\boldsymbol{\tau}_{13}$ ($\boldsymbol{\tau}^{\prime}_{13}$) appears as zero. In addition,  $\boldsymbol{\tau}_{23}$ survives in the vicinity of linear geometries and higher energetic configurations ($> 3$ eV).
Therefore, in case of low-energy collision processes, we can approximately consider $[\boldsymbol{\tau},K_\lambda^T] \sim 0$.

\noindent
On the other hand, in 6x6 $\lambda$ basis, the kinetic energy operator, $\left(\boldsymbol{\nabla}_R+ \boldsymbol{\tau} \right)^2$ turns into:

\begin{center}	
	\begin{eqnarray}
		\left(\boldsymbol{\nabla}_R+ \boldsymbol{\tau} \right)_{6x6}^2=\left[\begin{matrix}
			\boldsymbol{\nabla}_R & \boldsymbol{\tau}_{12} & \boldsymbol{\tau}_{13} & 0 & 0 & 0 \\
			-\boldsymbol{\tau}_{12} & \boldsymbol{\nabla}_R & \boldsymbol{\tau}_{23} & 0 & 0 & 0 \\
			-\boldsymbol{\tau}_{13} & -\boldsymbol{\tau}_{23} & \boldsymbol{\nabla}_R & 0 & 0 & 0 \\
			0 & 0 & 0 & \boldsymbol{\nabla}_R & \boldsymbol{\tau}_{12} & \boldsymbol{\tau}_{13} \\
			0 & 0 & 0 & -\boldsymbol{\tau}_{12} & \boldsymbol{\nabla}_R & \boldsymbol{\tau}_{23} \\
			0 & 0 & 0 & -\boldsymbol{\tau}_{13} & -\boldsymbol{\tau}_{23} & \boldsymbol{\nabla}_R
		\end{matrix} \right]\left[\begin{matrix}
			\boldsymbol{\nabla}_R & \boldsymbol{\tau}_{12} & \boldsymbol{\tau}_{13} & 0 & 0 & 0 \\
			-\boldsymbol{\tau}_{12} & \boldsymbol{\nabla}_R & \boldsymbol{\tau}_{23} & 0 & 0 & 0 \\
			-\boldsymbol{\tau}_{13} & -\boldsymbol{\tau}_{23} & \boldsymbol{\nabla}_R & 0 & 0 & 0 \\
			0 & 0 & 0 & \boldsymbol{\nabla}_R & \boldsymbol{\tau}_{12} & \boldsymbol{\tau}_{13} \\
			0 & 0 & 0 & -\boldsymbol{\tau}_{12} & \boldsymbol{\nabla}_R & \boldsymbol{\tau}_{23} \\
			0 & 0 & 0 & -\boldsymbol{\tau}_{13} & -\boldsymbol{\tau}_{23} & \boldsymbol{\nabla}_R
		\end{matrix} \right]	\nonumber
	\end{eqnarray}
\end{center}
\begin{center}	
	\begin{eqnarray}
		=\left[\begin{matrix}
			\left(\boldsymbol{\nabla}_R+ \boldsymbol{\tau} \right)_{3x3}^2 & 0\\
			0 & \left(\boldsymbol{\nabla}_R+ \boldsymbol{\tau} \right)_{3x3}^2
		\end{matrix}\right],
	\end{eqnarray}
\end{center}
where

\begin{eqnarray}
	\left(\boldsymbol{\nabla}_R+ \boldsymbol{\tau} \right)_{3x3}^2 = \left[\begin{matrix}
		\boldsymbol{\nabla}_R & \boldsymbol{\tau}_{12} & \boldsymbol{\tau}_{13}  \\
		-\boldsymbol{\tau}_{12} & \boldsymbol{\nabla}_R & \boldsymbol{\tau}_{23}  \\
		-\boldsymbol{\tau}_{13} & -\boldsymbol{\tau}_{23} & \boldsymbol{\nabla}_R 
	\end{matrix} \right]^2
\end{eqnarray}

\noindent
The kinetic energy operator remains block diagonal and therefore, only one of the two blocks is taken for diabatization.

\bibliographystyle{aipnum4-1}
\bibliography{ref_book}

	\chapter{Additional Formulations of Triatomic Reactive Scattering}\label{sec:app3}

\section{Initial Wave Packet in Hyperspherical Coordinates}\label{ssec:init}
\begin{figure}[!htp]
	\centering
	\includegraphics[ width=14 cm, height=0.35 \textheight, angle=0]{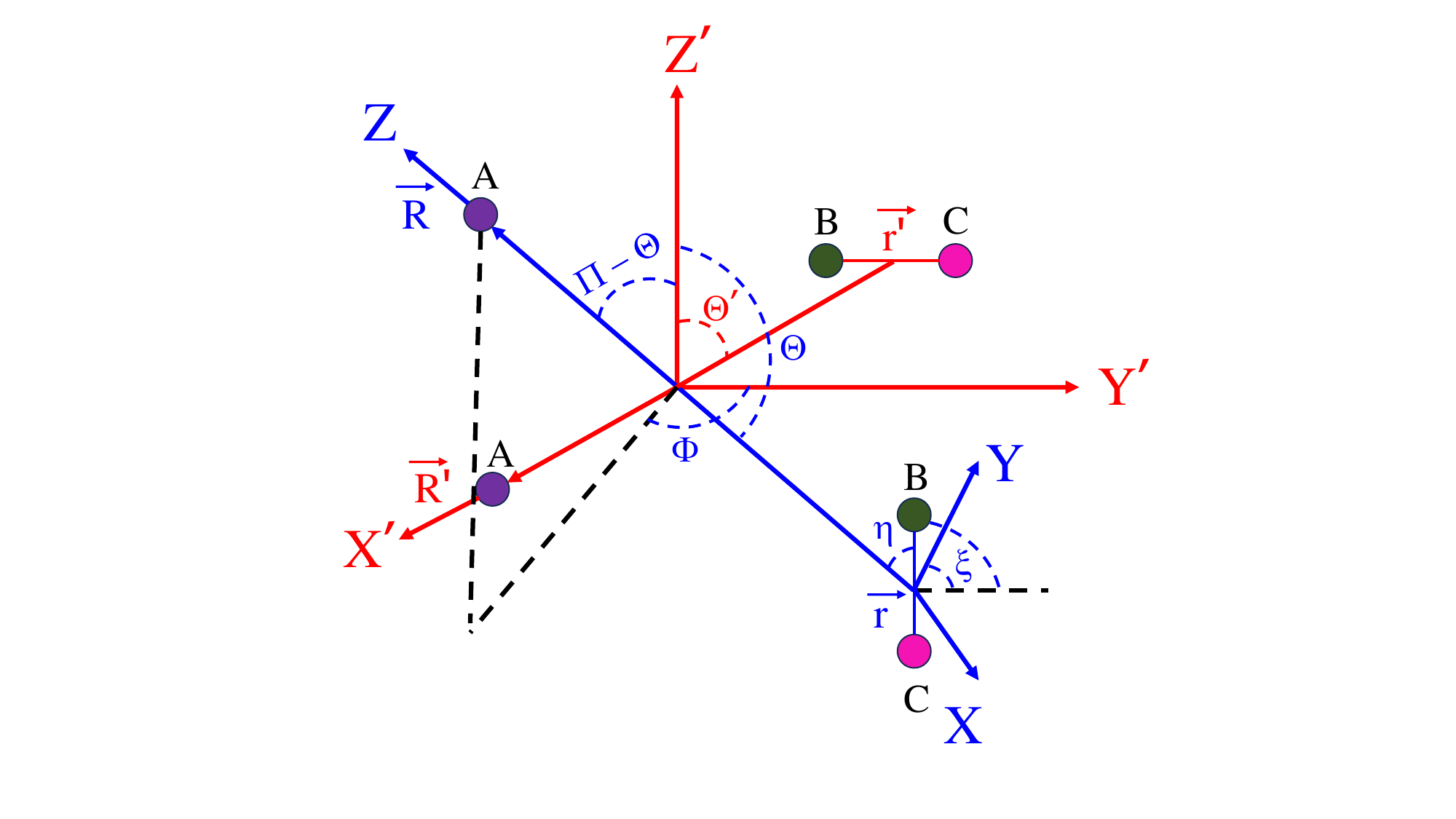}
	\caption{The position of the three atoms in the SFC system
		$\{X^\prime , Y^\prime , Z^\prime\}$  and the BFC system $\{X, Y, Z\}$, where the corresponding Jacobi vectors are $\{\vec{r^{\prime}},\vec{R^{\prime}}\}$ and $\{\vec{r},\vec{R}\}$, respectively. In SFC, the three atoms are on the X$^\prime$-Y$^\prime$ plane with $\Theta^\prime = \Pi/2$ and $\Phi^\prime$ = 0. The atom ``A'' of the BFC on projection over the X$^\prime$-Y$^\prime$ plane of the SFC makes the angle $\Phi$ with the Y$^\prime$-axis, where it's angle on the Z$^\prime$-axis is ($\Pi$ - $\Theta$). Detailed formulation is available in the  Ref.\citenum{cpc-184-270-SA}. The figure is adapted from Ref.\citenum{cpc-184-270-SA}.}  
	\label{fig:coordinate} 
\end{figure}

The channel (reactive/non-reactive) independent asymptotic solution to the reactive scattering of 
a triatomic system in Jacobi coordinates can take the following form:  
\begin{eqnarray}
	\Psi \rightarrow \frac{1}{R^{\prime}r^{\prime}}\sum_{\gamma^{\prime}vjl} 
	u^{Jl^{\prime}}_{\gamma \gamma^{\prime}}(R^{\prime}){\it y}^{JM}_{jl}(\hat{R^{\prime}},\hat{r^{\prime}}) \phi_{vj}(r^{\prime}),
	\label{eq:eq1}
\end{eqnarray}
\noindent
where ${\it y}^{JM}_{jl}$ and $\phi_{vj}$ represent Arthurs-Dalgarno (AD) states \cite{ADS}
and vibrational wave packets, respectively. The vibrational wave packet, $\phi_{vj}$ is dependent on the rotational quantum number $j$ through the centrifugal coupling term. The volume element appears as: 
\begin{eqnarray}
	R^{\prime 2}dR^{\prime} \ r^{\prime 2}dr^{\prime}  d\hat{R^{\prime}} d\hat{r^{\prime}},
	\label{eq:eq2}
\end{eqnarray}
\noindent
where
\begin{eqnarray}
	d\hat{R^{\prime}} \ d\hat{r^{\prime}}=\sin\Theta \ d\Theta \ d\Phi \sin\eta \ d\eta \ d\xi.
	\label{eq:eq3}
\end{eqnarray}
\noindent
The distance from the atom (A) to the center of mass of the diatomic molecule (BC) is $R^{\prime}$ and 
the orientation of $R^{\prime}$ in a space-fixed coordinate system is given by the angles $\Theta^{\prime}$ 
and $\Phi^{\prime}$ (see Figure.~\ref{fig:coordinate}). The bond distance of the BC molecule is $r$ and its orientation 
in a body-fixed coordinate system (with $R$ axis along $z$) is given by the angles $\eta$ 
and $\xi$. It is obvious that the function $(1/R)u^{Jl}_{\gamma \gamma^{\prime}}(R)$ is normalized 
on $R^2 \ dR$, $(1/r)\phi_{vj}(r)$ on $r^2 \ dr$ and the AD states ${\it y}^{Jl}_{jl}(\hat{R} 
\ \hat{r})$ on $d\hat{R} \ d\hat{r}$. \\

\noindent
In Jacobi coordinates, the wave packet can be initialized as: 
\begin{eqnarray}
	\Psi \rightarrow \frac{1}{r}\phi_{vj}(r){\it y}^{JM}_{jl}\frac{1}{R}\chi(R),
	\label{eq:eq4}
\end{eqnarray}
\noindent
where $\frac{1}{R} \ \chi(R)$ is a wave packet normalized on $R^2 \ dR$.\\

\noindent
In hyperspherical coordinates, the corresponding volume element is
\begin{eqnarray}
	\frac{1}{8}\rho^5 \sin\theta \cos\theta \ d\theta \ d\phi \ d\alpha \sin\beta \ d\beta \ d\gamma,
	\label{eq:eq5}
\end{eqnarray}
where $\rho$ is the hyperradius, and $\theta$ and $\phi$ are the two hyperangles. 
These three variables define the geometry of the $\rm ABC$ plane, while the orientation 
of this plane in space is defined by the three Euler angles $\alpha$, $\beta$ and $\gamma$.\\

\noindent
The wave function used in our approach is given by the following transformation:
\begin{eqnarray}
	\psi&=&(1/\sqrt{8})\rho^{5/2}\sqrt{\sin\theta \cos\theta} \ \Psi \nonumber\\
	&=&(1/\sqrt{8})\rho^{5/2}\sqrt{\sin\theta \cos\theta}\frac{1}{r}\phi_{vj}(r)
	{\it y}^{JM}_{jl}\frac{1}{R}\chi(R),
	\label{eq:eq6}
\end{eqnarray}
\noindent
where $\psi$ has the volume element: 
\begin{eqnarray}
	d\rho \ d\theta \ d\phi \ d\alpha \sin\beta \ d\beta \ d\gamma.
	\label{eq:eq7}
\end{eqnarray}

\noindent
Substituting the Jacobi variables in terms of hyperspherical coordinates for the reactant 
channel ($\rm A + BC$),
\begin{eqnarray}
	Rr=\frac{1}{2}\rho^2\sqrt{1-\sin^2\theta \cos^2\phi},
	\label{eq:eq8}
\end{eqnarray}
and the initial wavefuntion in Eq. (\ref{eq:eq6}) yields: 
\begin{eqnarray}
	\psi&=&\frac{1}{\sqrt{8}}\rho^{5/2}\sqrt{\sin\theta \cos\theta}\frac{2}{\rho^2\sqrt{1-\sin^2\theta \cos^2\phi}}
	\phi_{vj}(r){\it y}^{JM}_{jl}\chi(R) \nonumber\\
	&=&\sqrt{\frac{\rho}{2} \bigg(\frac{\sin\theta \cos\theta}{1-\sin^2\theta \cos^2\phi}\bigg)} 
	\phi_{vj}(r){\it y}^{JM}_{jl}\chi(R),
	\label{eq:eq9}
\end{eqnarray}
where the orientation angle of the diatom and the Jacobian are defined as:
\begin{eqnarray}
	\sin\eta=\frac{\cos\theta}{\sqrt{1-\sin^2\theta \cos^2\phi}}
	\label{eq:eq10}
\end{eqnarray}
and
\begin{eqnarray}
	J(Rr\eta|\rho \theta \phi)=\frac{\rho}{2}\frac{\sin\theta}{\sqrt{1-\sin^2\theta \cos^2\phi}},
	\label{eq:eq11}
\end{eqnarray}
respectively.\\

\noindent
The initial wavefuntion can therefore be written as:
\begin{eqnarray}
	\psi=\sqrt{\sin\eta \ J(Rr\eta|\rho \theta \phi)} \ \phi_{vj}(r){\it y}^{JM}_{jl}\chi(R).
	\label{eq:eq12}
\end{eqnarray}

\noindent
On the other hand, the wave packet $\psi$ in Eq. (\ref{eq:eq12}) can be expanded in 
terms of the Wigner $\bf D$ matrix as:
\begin{eqnarray}
	\psi=\sum^J_{K=-J} \Phi_K(\rho, \theta, \phi) \sqrt{\frac{2J+1}{8\pi^2}} D^{J\ast}_{MK} (\alpha, \beta, \gamma),
	\label{eq:eq13}
\end{eqnarray}
where $\sqrt{\frac{2J+1}{8\pi^2}} D^J_{MK}$ is normalized on $d\alpha \ \sin\beta \ d\beta \ d\gamma$. Thus, we have
\begin{eqnarray}
	\sum_K \int \ d\rho \ d\theta \ d\phi \ |\Phi_K|^2=1.
	\label{eq:eq14}
\end{eqnarray}
where the summation implicitly considers all allowed $K$ values.\\

\noindent
Equating Eq. (\ref{eq:eq13}) with Eq. (\ref{eq:eq12}), the functional form of the initial 
wave packet for the $K$ th component $\Phi_K(\rho, \theta, \phi)$ is given by: 
\begin{eqnarray}
	\Phi_K(\rho, \theta, \phi)&=&\sqrt{\frac{2J+1}{8\pi^2}} \ \int \ d\alpha \ \int \ d\beta \ \sin\beta \ \int \ d\gamma \ 
	D^J_{MK}(\alpha \beta \gamma)\nonumber\\
	&\times&\sqrt{\sin\eta \ J(Rr\eta|\rho \theta \phi)} \ \phi_{vj}(r){\it y}^{JM}_{jl}\chi(R).
	\label{eq:eq15}
\end{eqnarray}
This can be written in a more compact form by employing the AD states ${\it y}^{JM}_{jl}$ in terms of 
{\bf D} matrix and spherical harmonics ({\it Y}), namely,
\begin{eqnarray}
	{\it y}^{JM}_{jl}&=&(-1)^{j-l-M}\sqrt{(2J+1)(2l+1)/4\pi} \nonumber\\
	&\times&\sum_{\mu}\bigg(\begin{matrix}
		j&l&J\\
		\mu&0&\mu
	\end{matrix} \bigg)
	Y_{j\mu}(\eta, \xi) D^J_{-M,-\mu}(\Phi^\prime,\Theta^\prime,0),
	\label{eq:eq16}
\end{eqnarray}
\noindent
where $3-j$ symbols are the linear combining coefficients. 
Thus, by substituting 
\begin{eqnarray}
	Y_{j \mu}(\eta, \xi)=C_{j\mu}(\eta)\exp(i\mu\xi),
	\label{eq:eq17}
\end{eqnarray}
and
\begin{eqnarray}
	D^J_{M,-\mu}(\Phi^{\prime},\Theta^{\prime},\xi)&=&\exp(-iM\Phi^{\prime})d^J_{M,\mu}(\Theta^{\prime}) \exp(i\mu\xi),
	\label{eq:eq18}
\end{eqnarray}
Eq. (\ref{eq:eq16}) assumes the form: 
\begin{eqnarray}
	{\it y}^{JM}_{jl}&=&\sqrt{(2J+1)(2l+1)/4\pi} \ (-1)^{j-l-M}\nonumber\\
	&\times&\sum_{\mu}
	\bigg(\begin{matrix}
		j&l&J\\
		\mu&0&-\mu
	\end{matrix} \bigg)
	C_{j\mu}(\eta) D^J_{-M,-\mu}(\Phi^\prime,\Theta^\prime,\xi).
	\label{eq:eq19}
\end{eqnarray}
\noindent
Without loss of generality, it is possible to set the projection quantum number $M$ equal to zero and thereby,
\begin{eqnarray}
	D^J_{0, -\mu}(\Phi^{\prime}, \Theta^{\prime}, \xi)&=&\exp(i\mu\xi)d^J_{\mu0}(\Theta^{\prime})\nonumber\\
	&=&\sqrt{\frac{4\pi}{2J+1}} \ C_{J\mu}(\Theta^{\prime})\exp(i\mu\xi)
	\label{eq:eq20}
\end{eqnarray}
\noindent
and
\begin{eqnarray}
	D^J_{0,K}(\alpha, \beta, \gamma)=\exp(-iK\gamma)(-1)^K\sqrt{\frac{4\pi}{2J+1}}C_{JK}(\beta).
	\label{eq:eq21}
\end{eqnarray}
\noindent
If Eq. (\ref{eq:eq20}) is now substituted in Eq. (\ref{eq:eq19}), the AD functions, ${\it y}^{JM}_{jl}$, 
assume the form:
\begin{eqnarray}
	{\it y}^{JM}_{jl}=\sqrt{(2l+1)} \ (-1)^{j-l-M}\times\sum_{\mu}
	\bigg(\begin{matrix}
		j&l&J\\
		\mu&0&-\mu
	\end{matrix} \bigg)
	C_{j\mu}(\eta) C_{J\mu}(\Theta^{\prime}) \exp(i\mu\xi).
	\label{eq:eq22}
\end{eqnarray}
\noindent
By inserting next Eq. (\ref{eq:eq21}) and (\ref{eq:eq22}) into Eq. (\ref{eq:eq15}), 
the final form of the wave packet in hyperspherical coordinates becomes:
\begin{eqnarray}
	\Phi_K&=&\sqrt{2\pi \ \sin\eta \ J(Rr\eta|\rho \theta \phi)(2l+1)}\phi_{vj}(r) \ \chi(R)(-1)^{j-l}\nonumber\\
	&\times& \sum_{\mu}
	\bigg(\begin{matrix}
		j&l&J\\
		\mu&0&-\mu
	\end{matrix} \bigg)
	C_{j\mu}(\eta)A_{K\mu},
	\label{eq:eq23}
\end{eqnarray}
where
\begin{eqnarray}
	A_{K\mu}=(-1)^K \ \int^{2\pi}_0 \ d\gamma \int^{\pi}_0 \ d\beta \ \sin\beta \times \exp(-iK\gamma+i\mu\xi) C_{JK}(\beta)
	C_{J\mu}(\Theta^\prime),
	\label{eq:eq24}
\end{eqnarray}
with the orthonormality condition:
\begin{eqnarray}
	\sum_K \ A^{\ast}_{K\mu}A_{K\mu^{\prime}}=\delta_{\mu\mu^{\prime}} \ {\rm (See \ Appendix~\ref{ssec:orth})}.
	\label{eq:eq27}
\end{eqnarray}
\noindent
While constructing the $\bf{A}$ matrix, the following relationship between the 
orientation ($\Theta^{\prime}$ and $\xi$) and Euler ($\beta$ and $\gamma$) angles is employed:
\begin{eqnarray}
	\cos\Theta^{\prime}=-\sin\beta \ \cos\gamma,
	\label{eq:eq25}
\end{eqnarray}
\begin{eqnarray}
	\sin\xi=\cos\beta/\sin\Theta^{\prime}.
	\label{eq:eq26}
\end{eqnarray}
The part of the wave function corresponding to $\rm A-BC$ relative motion (translation) 
is described by a Gaussian wave packet,
\begin{eqnarray}
	\chi(R)=N\exp[-ik_0R-a(R-R_0)^2],
	\label{eq:eq28}
\end{eqnarray}
\noindent
where
\begin{eqnarray}
	a&=&1/(4\sigma^2-iq),\\
	q&=&2(R_0-R_f)/k_0,\\
	N&=&(A/\pi)^{1/4},\\
	A&=&8\sigma^2/(16\sigma^4+q^2).
	\label{eq:eq29}
\end{eqnarray}
The center of the wave packet is initially at $R=R_0$ and will, in the absence of forces, 
move in the negative $R$ direction with momentum $p_0$ = $\hbar$ $k_0$. The imaginary term 
in the denominator of the constant $a$ will keep the wave packet narrow until it reaches 
the ``focal point'', $R_f$. The initial collisional energy determines the wavevector ($k_0$) 
by employing the triatomic center of mass for the reactants, 
$\mu_R$ = $m_{\rm A}(m_{\rm B}+m_{\rm C})/(m_{\rm A}+m_{\rm B}+m_{\rm C})$, 
where the diatomic (BC) reduced mass is given by $\mu_r$ = $m_{\rm B}m_{\rm C}/(m_{\rm B}+m_{\rm C})$.

	\section{Orthonormality of the initial wave packet} \label{ssec:orth}

\noindent
The normalization of the wave packet (see Eq. (\ref{eq:eq14})) is defined as:
\begin{eqnarray}
	\sum \int d\rho \ d\theta \ d\phi |\Phi_k|^2=1.
	\label{eq:appendixB1}
\end{eqnarray}
If Eq. (\ref{eq:eq23}) is now considered, we may write:
\begin{eqnarray}
	\Phi_k^{\ast}\Phi_k&=&2\pi \ \sin\eta \ J(Rr\eta|\rho \theta \phi) (2l+1)\phi^{\ast}_{vj}(r)\phi_{vj}(r)\chi^{\ast}(R)\chi(R) 
	\nonumber\\
	&\times&\sum_{\mu}\sum_{{\mu}^{\prime}}
	\bigg(\begin{matrix}
		j&l&J\\
		\mu&0&-\mu
	\end{matrix} \bigg)
	\bigg(\begin{matrix}
		j&l&J\\
		\mu^{\prime}&0&-\mu^{\prime}
	\end{matrix} \bigg)
	C_{j\mu}(\eta)C_{j\mu^{\prime}}(\eta) A_{K\mu}A_{K\mu^{\prime}},
	\label{eq:appendixB2}
\end{eqnarray}
where $J(Rr\eta|\rho \theta \phi)$ is a Jacobi factor and
\begin{eqnarray}
	J(Rr\eta|\rho \theta \phi) \ d\rho \ d\theta \ d\phi&=&dR \ dr \ d\eta,\\
	\int dR \ \chi^{\ast}(R)\chi(R)&=&1,\\
	\int dr \phi^{\ast}_{vj}(r)\phi_{vj}(r)&=&1,
	\label{eq:appendixB3}
\end{eqnarray}
leading to
\begin{eqnarray}
	2\pi \sum_K \int &d \eta& \sin \eta \ (2l+1) \sum_{\mu} \sum_{\mu^{\prime}}
	\bigg(\begin{matrix}
		j&l&J\\
		\mu&0&-\mu
	\end{matrix} \bigg)\nonumber\\
	&\times&\bigg(\begin{matrix}
		j&l&J\\
		\mu^{\prime}&0&-\mu^{\prime}
	\end{matrix} \bigg)
	C_{j\mu}(\eta)C_{j\mu^{\prime}}(\eta) A_{K\mu}A_{K\mu^{\prime}} = 1.
	\label{eq:appendixB4}
\end{eqnarray}
Since
\begin{eqnarray}
	2\pi\int d \eta \ \sin \eta \ C_{j\mu} (\eta) C_{j\mu^{\prime}} (\eta) = 1 {\hspace {0.5cm}} \text{only for} 
	{\hspace {0.5cm}} \mu=\mu^{\prime}
	\label{eq:appendixB5}
\end{eqnarray}
and
\begin{eqnarray}
	(2l+1) \sum_{\mu} 
	\bigg(\begin{matrix}
		j&l&J\\
		\mu&0&-\mu
	\end{matrix} \bigg)
	\bigg(\begin{matrix}
		j&l&J\\
		\mu&0&-\mu
	\end{matrix} \bigg)=1,
	\label{eq:appendixB6}
\end{eqnarray}
the only possibility is
\begin{eqnarray}
	\sum_K A_{K \mu}A_{K \mu^{\prime}}=\delta_{\mu \mu^{\prime}}.
	\label{eq:appendixB7}
	\end{eqnarray}

\section{Transformations between Jacobi and Hyperspherical Coordinates}\label{ssec:trans-JH} 

\noindent
Since we need to use Jacobi coordinates asymptotically to define the initial wave function 
of the triatomic system and wish to perform the dynamics in hyperspherical coordinates, it 
is necessary to establish the relations between these two sets of coordinates. Consider a 
space-fixed coordinate system (X$^{\prime}$, Y$^{\prime}$, Z$^{\prime}$) such that the three 
atoms A, B, C initially lie on the (X$^{\prime}$, Y$^{\prime}$) plane. If the 
diatom (B-C) and the incoming atom A coincide with the Y$^{\prime}$ and X$^{\prime}$ axis, 
respectively, the orientation of the {\bf R} vector will be defined by the angles, 
$\Theta^{\prime}$=$\pi$/2 and $\Phi^{\prime}$=0 in the space-fixed coordinate 
system. On the other hand, the orientation of the {\bf r} vector for the diatomic molecule 
will be specified by the angles, $\eta$ and $\xi$ in the body-fixed coordinate system. \\

\noindent
In order to find the position vectors ${\bf r^{\prime}}$ and ${\bf R^{\prime}}$ at any point 
in the configuration space, we use the following relations:
\begin{eqnarray}
	{\bf r^{\prime}}&=&{\bf \tilde{O}}{\bf r},\\
	{\bf R^{\prime}}&=&{\bf \tilde{O}}{\bf R},
	\label{eq:appendixA1}
\end{eqnarray}
\noindent
where ${\bf \tilde{O}}$ specifies the rotation matrix defining the rotation of the triangle 
formed by the three atoms.\\

\noindent
The definition of the hyperspherical coordinates in terms of the above mass-weighted Cartesian 
coordinates is introduced as
\begin{eqnarray}
	r_x&=&-\frac{\rho}{\sqrt{2}}\bigg(\cos \frac{\theta}{2}+\sin \frac{\theta}{2} \bigg)\cos \frac{\phi}{2},\\
	R_x&=&\frac{\rho}{\sqrt{2}}\bigg(\cos \frac{\theta}{2}+\sin \frac{\theta}{2} \bigg)\sin \frac{\phi}{2},
\end{eqnarray}
\begin{eqnarray}
	r_y&=&\frac{\rho}{\sqrt{2}}\bigg(\cos \frac{\theta}{2}-\sin \frac{\theta}{2} \bigg)\sin \frac{\phi}{2},\\
	R_y&=&\frac{\rho}{\sqrt{2}}\bigg(\cos \frac{\theta}{2}-\sin \frac{\theta}{2} \bigg)\cos \frac{\phi}{2},\\
	r_z&=&R_z=0,
	\label{eq:appendixA2}
\end{eqnarray}
which in turn lead to 
\begin{eqnarray}
	r^2&=&r_x^2+r_y^2=\frac{1}{2}\rho^2(1+\sin \theta \cos \phi),\\
	R^2&=&R_x^2+R_y^2=\frac{1}{2}\rho^2(1-\sin \theta \cos \phi),\\
	r^2&+&R^2=\rho^2,
	\label{eq:appendixA3}
\end{eqnarray}
and
\begin{eqnarray}
	{\bf r \cdot R}=r_x R_x+r_y R_y=-\frac{1}{2}\rho^2 \sin \theta \sin \phi.
	\label{eq:appendixA4}
\end{eqnarray}
The angle between the {\bf r} and {\bf R} vectors is known as:
\begin{eqnarray}
	\cos \gamma^{\prime}=\frac{\bf r \cdot R}{rR}&=&\frac{-\sin \theta \sin \phi}{\sqrt{1-\sin^2\theta\cos^2\phi}},
	\label{eq:appendixA5}
\end{eqnarray}
which, at $\rho \rightarrow \infty$, becomes identical to the 
diatomic orientation angle ($\gamma^{\prime}$ = $\eta$).\\

\noindent
Since the AD states are expressed in term of spherical harmonics 
$Y_{j\mu}$($\gamma^{\prime},\xi$) with $\gamma^{\prime}$, $\xi$ 
defining the orientation of the diatomic molecule in a body-fixed 
coordinate system, the corresponding ${\bf R}$ vector along the 
Z-axis is given by 
\begin{eqnarray}
	{\bf R}=(0,0,R),
	\label{eq:appendixA6}
\end{eqnarray}
where the orientation of ${\bf R^{\prime}}$ in a space-fixed coordinate 
system $\Theta^{\prime}$ and $\Phi^{\prime}$ also define the AD states 
and the the angle $\Theta^{\prime}$ can be evaluated as:
\begin{eqnarray}
	{\bf R^{\prime}}&=&{\bf \tilde{O}}{\bf R}\\
	\cos\Theta^{\prime}&=&\frac{{\bf R \cdot}(0,0,1)}{|R^{\prime}|}=\frac{R_z^{\prime}}{R^{\prime}}.
	\label{eq:appendixA7}
\end{eqnarray}
\noindent
When we use the following form of Euler's rotation matrix 
\begin{eqnarray}
	{\bf \tilde{O}}\!\!=\!\!\!\left( \begin{array}{ccc}
		\cos \gamma \cos \alpha \cos \beta- \sin \alpha \sin \gamma& -\sin\gamma\cos\alpha\cos\beta-\sin\alpha\cos\gamma& 
		\sin\beta\cos\alpha \\
		\cos\gamma\cos\beta\sin\alpha+\sin\gamma\cos\alpha& \cos\alpha\cos\gamma-\sin\gamma\cos\beta\sin\alpha&
		\sin\beta\sin\alpha\\
		-\cos\gamma\sin\beta&\sin\gamma\sin\beta&\cos\beta\end{array} \right),
	\label{eq:appendixA8}
\end{eqnarray}
\noindent
in  Eq. (A14), we obtain
\begin{eqnarray}
	R_z^{\prime}=\sin\beta\frac{\rho}{\sqrt{2}}\bigg(\cos\frac{\theta}{2}\sin\bigg(\gamma-\frac{\phi}{2}\bigg)
	-\sin\frac{\theta}{2}\sin\bigg(\gamma+\frac{\phi}{2}\bigg)\bigg).
	\label{eq:appendixA9}
\end{eqnarray}
While employing $R^{\prime}=R$, the space-fixed orientation angle $\Theta^{\prime}$ 
obtained from Eq. (A9) and Eq. (A15) becomes 
\begin{eqnarray}
	\cos\Theta^{\prime}=\frac{\sin\beta\bigg(\cos\frac{\theta}{2}\sin\bigg(\gamma-\frac{\phi}{2}\bigg)
		-\sin\frac{\theta}{2}\sin\bigg(\gamma+\frac{\phi}{2}\bigg)\bigg)}{\sqrt{1-\sin\theta\cos\phi}}.
	\label{eq:appendixA10}
\end{eqnarray}
In the asymptotic limit, we substitute $\theta$=$\theta_0$=$\pi$/2 to obtain:
\begin{eqnarray}
	\cos\Theta^{\prime}=-\sin\beta\cos\gamma.
	\label{eq:appendixA11}
\end{eqnarray}
On the other hand, in order to determine $\xi$, we first rotate {\bf r} 
to {\bf r$^{\prime}$} (see Eq. A1) and define a rotation of R$^{\prime}$ 
such that:
\begin{eqnarray}
	{\bf TR^{\prime}}=\left( \begin{array}{c}
		0\\0\\R^{\prime}
	\end{array}  \right).
	\label{eq:appendixA13}
\end{eqnarray}
\noindent
With this rotated {\bf R$^{\prime}$} vector along the Z-axis in this 
body-fixed system, the orientation of {\bf r$^{\prime}$} in the same 
coordinate system becomes 
\begin{eqnarray}
	r^{\prime\prime}={\bf Tr^{\prime}}=\left( \begin{array}{c}
		r_x^{\prime\prime}\\
		r_y^{\prime\prime}\\
		r_z^{\prime\prime}
	\end{array} \right)=
	\left( \begin{array}{c}
		r\ \sin\eta \ \cos\xi\\
		r\ \sin\eta \ \sin\xi\\
		r\ \cos\eta 
	\end{array} \right),
	\label{eq:appendixA14}
\end{eqnarray}
\noindent
with angles $\eta$ and $\xi$. \\

\noindent
When we substitute Eq. (A1) in Eq. (\ref{eq:appendixA14}), we get  
\begin{eqnarray}
	\left( \begin{array}{c}
		r_x^{\prime\prime}\\
		r_y^{\prime\prime}\\
		r_z^{\prime\prime}
	\end{array} \right)
	={\bf T\tilde{O}}
	\left( \begin{array}{c}
		r_x\\
		r_y\\
		0
	\end{array} \right),
	\label{eq:appendixA15}
\end{eqnarray}
where the {\bf T} matrix is given by 
	\begin{eqnarray}
		{\bf T}=\left( \begin{array}{ccc}
			\cos\Theta^{\prime}\cos\Phi^{\prime}& \cos\Theta^{\prime}\sin\Phi^{\prime}& -\sin\Theta^{\prime}\\
			-\sin\Phi^{\prime}& \cos\Phi^{\prime}& 0\\
			\sin\Theta^{\prime}\cos\Phi^{\prime}& \sin\Theta^{\prime}\sin\Phi^{\prime}& \cos\Theta^{\prime}
		\end{array} \right),
		\label{eq:appendixA16}
	\end{eqnarray}
	\noindent
	with $\Theta^{\prime}$ and $\Phi^{\prime}$ specifying the orientation of {\bf R$^{\prime}$} 
	in the space-fixed coordinate system.
	\noindent
	Considering Eq. (\ref{eq:appendixA15}), we have 
	\begin{eqnarray}
		r_z^{\prime\prime}=\sum_{kl}T_{3k}\tilde{O}_{kl}r_l.
		\label{eq:appendixA17}
	\end{eqnarray}
	\noindent
	When we substitute Eq. (A14) in Eq. (\ref{eq:appendixA13}), we get the following equation by using 
	orthonormality condition of {\bf T} matrix
	\begin{eqnarray}
		T_{3j}=\frac{1}{R^{\prime}}\sum_l\tilde{O}_{jl}R_l,
		\label{eq:appendixA18}
	\end{eqnarray}
	\noindent
	Inserting Eq. (\ref{eq:appendixA18}) and employing the orthonormality condition of ${\bf \tilde{O}}$ matrix,
	$\sum_k\tilde{O}_{km}\tilde{O}_{kl}=\delta_{ml}$ in Eq. (\ref{eq:appendixA17}), we obtain
	\begin{eqnarray}
		r_z^{\prime\prime}&=&\frac{1}{R^{\prime}}\sum_{kl}\sum_m\tilde{O}_{km}\tilde{O}_{kl}R_m r_l\nonumber\\
		&=&\frac{1}{R^{\prime}}\sum_{lm}\bigg(\sum_k\tilde{O}_{km}\tilde{O}_{kl}\bigg)R_m r_l\nonumber\\
		&=&\frac{1}{R^{\prime}} (R_xr_x+R_yr_y).
		\label{eq:appendixA19}
	\end{eqnarray}
	\noindent
	When we substitute Eq. (\ref{eq:appendixA14}) and Eqs. (A3)-(A6) in Eq. (\ref{eq:appendixA19}), we obtain
	\begin{eqnarray}
		\cos\eta=\frac{-\sin\theta\sin\phi}{\sqrt{1-\sin^2\theta\cos^2\phi}}.
		\label{eq:appendixA21}
	\end{eqnarray}
	\noindent
	We can now use the same procedure to find $\sin\xi$ by considering $r_y^{\prime\prime}$,
	\begin{eqnarray}
		r_y^{\prime\prime}=\sum_k T_{2k} \sum_l \tilde{O}_{kl}r_l=\sum_kT_{2k}\tilde{O}_{k1}r_x+\sum_kT_{2k}\tilde{O}_{k2}r_y
		\label{eq:appendixA22}
	\end{eqnarray}
	\noindent
	with ($r_z=0$).\\
	
	\noindent
	Introducing the following relations
	\begin{eqnarray}
		T_{21}&=&-\sin\Phi^{\prime}=-\frac{R_y^{\prime}}{R^{\prime}\sin\Theta^{\prime}}
		=-\frac{1}{R^{\prime}\sin\Theta^{\prime}}(\tilde{O}_{21}R_x +\tilde{O}_{22}R_y),\nonumber\\
		T_{22}&=&\cos\Phi^{\prime}=\frac{R_x^{\prime}}{R^{\prime}\sin\Theta^{\prime}} 
		= \frac{1}{R^{\prime}\sin\Theta^{\prime}}(\tilde{O}_{11}R_x+\tilde{O}_{12}R_y),
		\label{eq:appendixA23}
	\end{eqnarray}
	\noindent
	leads to
	\begin{eqnarray}
		r_y^{\prime\prime}=r\sin\eta\sin\xi=\frac{\cos\beta}{R^{\prime}\sin\Theta^{\prime}}\frac{\rho^2}{2}\cos\theta,
		\label{eq:appendixA24}
	\end{eqnarray}
	\noindent
	where we have used
	\begin{eqnarray}
		r_yR_x-r_xR_y=(\rho^2/2)\cos\theta,\nonumber\\
		\tilde{O}_{11}\tilde{O}_{22}-\tilde{O}_{12}\tilde{O}_{21}=\cos\beta,
		\label{eq:appendixA25}
	\end{eqnarray}
	\noindent
	with
	\begin{eqnarray}
		rR^{\prime}=\frac{1}{2}\rho^2\sqrt{1-\sin^2\theta\cos^2\phi}
		\label{eq:appendixA26}
	\end{eqnarray}
	\noindent
	and
	\begin{eqnarray}
		\sin\eta=\frac{\cos\theta}{\sqrt{1-\sin^2\theta\cos^2\phi}},
		\label{eq:appendixA27}
	\end{eqnarray}
	\noindent
	to finally obtain
	\begin{eqnarray}
		\sin\xi=\cos\beta/\sin\Theta^{\prime}.
		\label{eq:appendixA28}
	\end{eqnarray}
	
\section{The Hamiltonian and the Coupled Equations}\label{ssec:haml-coup}

\noindent
The Hamiltonian operator describing a three-particle system expressed in 
terms of Johnson's hypersherical coordinates assumes the form:
\begin{eqnarray}
	\hat{H}&=&-\frac{\hbar^2}{2\mu_R}\frac{\partial^2}{\partial \rho^2}+\frac{2}{\mu_R \rho^2}\hat{L}^2(\theta, \phi)+
	\frac{1}{\mu_R \rho^2}\bigg[\frac{\hat{J}^2_x}{1-\sin\theta}+\frac{\hat{J}^2_y}{1+\sin\theta}+
	\frac{\hat{J}^2_z}{2\sin^2\theta}\bigg]\nonumber\\
	&-&\frac{2\cos\theta \ \hat{J}_z\hat{P}_{\phi}}{\mu_R \rho^2 \sin^2\theta}
	+V(\rho, \theta, \phi)+\Delta V(\rho, \theta),
	\label{eq:ham1}
\end{eqnarray}
where
\begin{eqnarray}
	\hat{J}_z&=&-i\hbar\bigg(\frac{\partial}{\partial \gamma}\bigg),\\
	\hat{P}_{\phi}&=&-i\hbar\bigg(\frac{\partial}{\partial \phi}\bigg),\\
	\hat{L}^2(\theta, \phi)&=&-\hbar^2\bigg[\frac{\partial^2}{\partial \theta^2}+\frac{1}{\sin^2 \theta}
	\frac{\partial^2}{\partial \phi^2}\bigg],\\
	\Delta V(\rho, \theta)&=&-\frac{\hbar^2}{2\mu_R \rho^2}\bigg[\frac{1}{4}+\frac{4}{\sin^2 2\theta}\bigg].
	\label{eq:ham2}
\end{eqnarray}
Although the angular momentum operators $\hat{J}_x$ and $\hat{J}_y$ are functions of 
the Euler angles $\alpha$, $\beta$ and $\gamma$, their forms are not needed because 
the Hamiltonian (Eq. (\ref{eq:ham1})) can be simplified using the following relation:
\begin{eqnarray}
	\hat{J}^2=\hat{J}^2_x+\hat{J}^2_y+\hat{J}^2_z,
	\label{eq:ham3}
\end{eqnarray}
leading to:
\begin{eqnarray}
	\hat{H}&=&-\frac{\hbar^2}{2\mu_R}\frac{\partial^2}{\partial \rho^2}+\frac{2}{\mu_R \rho^2}\hat{L}^2(\theta, \phi)+
	\frac{\hat{J}^2-\hat{J}^2_z}{\mu_R \rho^2 \cos^2\theta}+\frac{\hat{J}^2_z-4\cos\theta \ \hat{J}_z\hat{P}_{\phi}}{2\mu_R 
		\rho^2 \sin^2\theta}\nonumber\\&+&\frac{\sin\theta}{\mu_R \rho^2\cos^2\theta}\frac{1}{2}[\hat{J}^2_++\hat{J}^2_-]
	+V(\rho, \theta, \phi)+\Delta V(\rho, \theta),
	\label{eq:ham4}
\end{eqnarray}
where the raising and lowering operators have been introduced using the relation: 
\begin{eqnarray}
	\hat{J}^2_x-\hat{J}^2_y=\frac{1}{2}[\hat{J}^2_++\hat{J}^2_-].
	\label{eq:ham5}
\end{eqnarray}

\noindent
When we substitute the wave packet $\psi$ [Eq. (\ref{eq:eq13})] and the Hamiltonian 
[Eq. (\ref{eq:ham4})] into the time-dependent Schr\"{o}dinger equation (TDSE), the 
following set of coupled equations are obtained in terms of partial waves ($\Phi_K$):
\begin{eqnarray}
	i\hbar\frac{\partial \Phi_K}{\partial t}&=&\bigg\{-\frac{\hbar^2}{2\mu_R}\frac{\partial^2}{\partial \rho^2}
	+\frac{2}{\mu_R \rho^2}\hat{L}^2(\theta,\phi)
	+\frac{\hbar K (\hbar K-4\cos\theta \ \hat{P}_{\phi})}{2\mu_R \rho^2 \ \sin^2\theta}\nonumber\\
	&+&\frac{\hbar^2[J(J+1)-K^2]} {\mu_R\rho^2\cos^2\theta}+V(\rho,\theta,\phi)+\Delta V(\rho,\theta)\bigg\}\Phi_K\nonumber\\
	&+&\frac{\sin\theta}{\mu_R \rho^2 \cos^2 \theta}[M_{K,K+2} \Phi_{K+2}+M_{K,K-2}\Phi_{K-2}],
	\label{eq:ham6}
\end{eqnarray}
where the last term couples the component $\Phi_K$ to $\Phi_{K\pm2}$ via the coupling element $M_{K,K\pm2}$
\begin{eqnarray}
	M_{K,K\pm 2}=\frac{\hbar^2}{2}\sqrt{(J \mp K)(J\pm K+1)(J\mp K-1)(J\pm K+2)}.
	\label{eq:ham7}
\end{eqnarray}
A discrete representation of the partial waves $\Phi_K(\rho,\theta,\phi)$s on 
three-dimensional grids are introduced to propagate the coupled wave packets with time.

\section{Propagation and Computation Details} \label{ssec:prop-co}
\subsection{Propagation}
We propagate the wave packet(s) in time by employing the iterative Lanczos 
reduction or Chebysheb polynomial technique. Such an algorithm is known as a short time propagator,
capable of automatically controlling recursion by monitoring the magnitude
of the last few vectors, and accurate enough for the analysis of the outgoing
wave packet. In order to recover the wave packet after each propagation step,
we need to store the transformation matrix constituted with Lanczos recursion
vectors or the coefficients of Chebysheb polynomials. Since it is possible to store this matrix in the core memory, one
can avoid repeated calculation for half of such vectors are required for
the next time step. \\

\subsection{Lanczos Reduction Technique}

\noindent
Time dependent Schrodinger equation in matrix form can be written as:
\begin{equation}
	i\hbar \frac{{\partial \Phi }}{{\partial t}} = H\Phi
\end{equation}
where $\Phi$ and $H$ are vector of length N containing the value of the wave function at the N grid points, and $N \times N$ matrix, respectively. \\

\noindent
We wish to find a transformation matrix (Q) that transforms the Hamiltonian matrix ($H$) to a tridiagonal form ($T$):
\begin{equation}
	{Q^\dag }HQ = T\,\,\,\,\,\,\,\,\,\,\,\,\,\,\,\,\,\,\,{Q^\dag }Q = 1
\end{equation}
Assuming $Q$ in matrix form as
\begin{equation}
	Q = {\left[ {{q_1}\:{q_2} \ldots {q_j} \ldots {q_l}} \right]_{1 \times l}}
\end{equation}
where $q_i$  is vector of length $N$. The $T$-matrix becomes:
\begin{equation}
	T = {\left[ {\begin{array}{*{20}{c}}
				{{\alpha _1}}&{{\beta _1}}&0& \ldots &{\rm{ }}&{\rm{ }}& \ldots &{\rm{ }}&{\rm{ }}&0\\
				{{\beta _1}}&{{\alpha _2}}&{{\beta _2}}&{\rm{ }}&{\rm{ }}&{\rm{ }}&{\rm{ }}&{\rm{ }}&{\rm{ }}& \vdots \\
				0&{{\beta _2}}&{{\alpha _3}}&{{\beta _3}}&{\rm{ }}&{\rm{ }}&{\rm{ }}&{\rm{ }}&{\rm{ }}&{\rm{ }}\\
				\vdots &{\rm{ }}&{\rm{ }}& \ddots &{\rm{ }}&{\rm{ }}&{\rm{ }}&{\rm{ }}&{\rm{ }}&{\rm{ }}\\
				{\rm{ }}&{\rm{ }}&{\rm{ }}&{\rm{ }}&{\rm{ }}&{{\beta _{j - 1}}}&{}&{\rm{ }}&{\rm{ }}&{\rm{ }}\\
				{\rm{ }}&{\rm{ }}&{\rm{ }}&{\rm{ }}&{{\beta _{j - 1}}}&{{\alpha _j}}&{{\beta _j}}&{\rm{ }}&{\rm{ }}&{\rm{ }}\\
				\vdots &{\rm{ }}&{\rm{ }}&{\rm{ }}&{\rm{ }}&{{\beta _j}}&{{\alpha _{j + 1}}}&{\rm{ }}&{\rm{ }}& \vdots \\
				{\rm{ }}&{\rm{ }}&{\rm{ }}&{\rm{ }}&{\rm{ }}&{\rm{ }}&{\rm{ }}&{\rm{ }}&{\rm{ }}&{\rm{ }}\\
				{\rm{ }}&{\rm{ }}&{\rm{ }}&{\rm{ }}&{\rm{ }}&{\rm{ }}&{\rm{ }}&{\rm{ }}&{\rm{ }}&{{\beta _{l - 1}}}\\
				0& \ldots &{\rm{ }}&{\rm{ }}&{\rm{ }}&{\rm{ }}& \ldots &0&{{\beta _{l - 1}}}&{{\alpha _l}}
		\end{array}} \right]_{l \times l}}
\end{equation}
and
\begin{equation}
	\begin{array}{l}
		\,\,\,\,\,\,\,\,\,\,\,\,\,\,\,\,\,\,\,\,\,\,\,\,\,\,\,\,\,\,\,\,\,\,\,\,\,\,\,\,\,\,\,\,\,\,\,\,\,\,\,\,\,\,\,\,\,\,\,\,\,HQ\,{\rm{ = }}\,QT\\
		\Rightarrow H\left[ {\begin{array}{*{20}{c}}
				{{q_1}}&{{q_2}}& \ldots &{{q_j}}& \ldots &{{q_l}}
		\end{array}} \right] = \left[ {\begin{array}{*{20}{c}}
				{{q_1}}&{{q_2}}& \ldots &{{q_j}}& \ldots &{{q_l}}
		\end{array}} \right]T
	\end{array}
\end{equation}
The first column of the above equation appears as:
\begin{equation}
	H{q_1} = {\alpha _1}{q_1} + {\beta _1}{q_2}
\end{equation}
and the jth column of the matrix equation becomes:
\begin{equation}
	H{q_j} = {\beta _{j - 1}}{q_{j - 1}} + {\alpha _j}{q_j} + {\beta _j}{q_{j + 1}}
\end{equation}
Using the orthonormality condition of $Q$ matrix:
\begin{equation}
	\langle {q_i}|{q_j}\rangle  = {\delta _{ij}}
\end{equation}
we obtain the coefficients multiplying both sides of the above equation by,
\begin{equation}
	\begin{array}{l}
		q_j^\dag H{q_j} = {\beta _{j - 1}}\left( {q_j^\dag {q_{j - 1}}} \right) + {\alpha _j}\left( {q_j^\dag {q_j}} \right) + {\beta _j}\left( {q_j^\dag {q_{j + 1}}} \right)\\
		\Rightarrow {\alpha _j} = q_j^\dag H{q_j}
	\end{array}
\end{equation}
\noindent
Similarly, one can write
\begin{equation}
	\begin{array}{l}
		q_{j - 1}^\dag H{q_j} = {\beta _{j - 1}}\left( {q_{j - 1}^\dag {q_{j - 1}}} \right) + {\alpha _j}\left( {q_{j - 1}^\dag {q_j}} \right) + {\beta _j}\left( {q_{j - 1}^\dag {q_{j + 1}}} \right)\\
		\Rightarrow {\beta _{j - 1}} = q_{j - 1}^\dag H{q_j}
	\end{array}
\end{equation}
\noindent
We use the first column of the matrix equation:
\begin{equation}
	H{q_1} = {\alpha _1}{q_1} + {\beta _1}{q_2}
\end{equation}
\noindent
where $q_1$ is the initial wave function ($\phi_0$). $H$ operates on $q_1$ by FFT method:
\begin{equation}
	{\alpha _1} = q_1^\dag H{q_1}
\end{equation}
which leads to
\begin{equation}
	\Rightarrow {\beta _1}{q_2} = H{q_1} - {\alpha _1}{q_1}
	\label{eq:lc_beta}
\end{equation}
\noindent
Now one can obtain,
\begin{equation}
	\begin{array}{l}
		\langle {\beta _1}{q_2}|{\beta _1}{q_2}\rangle  = \beta _1^2\left( {q_2^\dag {q_2}} \right) = \beta _1^2 \Rightarrow {\beta _1} = \sqrt {\beta _1^2} 
	\end{array}
\end{equation}
\noindent
As $\alpha_1$ and $\beta_1$ can be evaluated in this way, from Eq. \ref{eq:lc_beta}, we obtain $q_2$:
\begin{equation}
	{q_2} = \frac{{\left( {H{q_1} - {\alpha _1}{q_1}} \right)}}{{{\beta _1}}}
\end{equation}
\noindent
Therefore, so on and so forth, we can calculate all the $\alpha$-terms upto $\alpha_j (j=l)$
and $\beta$-terms up to $\alpha_{(j-1)} (j=l)$, and $q$-vectors upto $q_j (j=l)$. Such a scheme can be used to evaluate the eigenvalue of a large system or its time propagation.
\begin{equation}
	i\hbar \dot \Phi  = H\Phi
\end{equation}             .
The transformation matrix ($Q$) is employed as:
\begin{equation}
	i\hbar {Q^\dag }\dot \Phi  = \left( {{Q^\dag }HQ} \right)\left( {{Q^\dag }\Phi } \right)
\end{equation} 
Substituting $\Psi  = {Q^\dag }\Phi \,(or\,\,\Phi  = Q\Psi )$, we obtain:
\begin{equation}
	i\hbar \dot \Psi  = T\Psi ,
\end{equation}
where 
\begin{equation}
	T = {Q^\dag }HQ
\end{equation}
\noindent
When $H$ is time independent, the TDSE can be written as:
\begin{equation}
	\begin{array}{l}
		\Phi (t + \Delta t) = \exp \left( { - \frac{{iHt}}{\hbar }} \right)\Phi (t)\\
		\Rightarrow \Phi (t + \Delta t) = Q\exp \left( { - \frac{{iTt}}{\hbar }} \right){Q^\dag }\Phi (t)
	\end{array}
\end{equation}
If we assume a matrix $S$ that diagonalizes the tridiagonal matrix $T$ to a diagonal matrix $D$ (the eigenvalue spectrum):
\begin{equation}
	D = {S^\dag }TS
\end{equation}
we get:
\begin{equation}
	\Phi (t + \Delta t) = QS\exp \left( { - \frac{{itD}}{\hbar }} \right){S^\dag }{Q^\dag }\Phi (t)
\end{equation}
where $\Phi (t + \Delta t)$ is the time propagated wave function from $\Phi (t)$.

\subsection{Chebysheb Polynomial Technique}

\noindent
Global propagators can be defined by their use of very long time steps, sometimes a single time step completes the calculation. A method that is capable of taking large time steps is the Chebyshev method.\\

\noindent
The main idea behind global propagators is to use a polynomial expansion of the exponential in the evolution operator:
\begin{equation}
	\hat U(t) = \exp \left( { - \frac{{i\hat Ht}}{\hbar }} \right)
\end{equation}
While expanding in terms of Chebyshev series, we obtain,
\begin{equation}
	\hat U(t) = \sum\limits_{n = 0}^N {{a_n}} {\phi _n}\left( { - \frac{{i\hat Ht}}{\hbar }} \right)
\end{equation}
\noindent
where $a_n$ and $\phi_n$ are expansion coefficient and complex Chebyshev polynomials, respectively:
\begin{equation}
	\begin{array}{l}
		{\phi _0}(x) = 1\\
		{\phi _1}(x) = x\\
		{\phi _2}(x) = 2{x^2} - 1\\
		{\phi _{n + 1}}(x) = 2x\,{\mkern 1mu} {\phi _n}(x) - {\phi _{n - 1}}(x){\mkern 1mu} ,\quad \,\, - 1 \le x \le 1
	\end{array}
\end{equation}
with the orthogonality relation:
\begin{equation}
	\int_{ - 1}^{ + 1} {\frac{{{\phi _m}(x){\mkern 1mu} {\phi _n}(x){\mkern 1mu} dx}}{{\sqrt {1 - {x^2}} }}}  = \left\{ {\begin{array}{*{20}{l}}
			{\frac{1}{2}{\mkern 1mu} \pi {\mkern 1mu} {\delta _{n,m}},}&{m \ne 0,n \ne 0}\\
			{\pi ,}&{m = n = 0.}
	\end{array}} \right.
\end{equation}
\noindent
In case of ${\phi _n}\left( { - \frac{{i\hat Ht}}{\hbar }} \right)\,$ the range becomes -i to i. Now the Hamiltonian is re-normalized by division with $\Delta {E_{grid}} = ({E_{max}} - {E_{min}})$ to preserve the range of the eigenvalues as -i to i. Therefore, $\hat H$ is shifted to ${\hat H_{norm}}$ as given below:
\begin{equation}
	\begin{array}{l}
		{{\hat H}_{norm}} = 2\left[ {\frac{{\hat H - \hat I\left( {\frac{{\Delta {E_{grid}}}}{2} + {E_{min}}} \right)}}{{\Delta {E_{grid}}}}} \right]\\
		\Rightarrow \hat H = \left( {\frac{{\Delta {E_{grid}}}}{2}} \right){{\hat H}_{norm}} + \left( {\frac{{\Delta {E_{grid}}}}{2} + {E_{min}}} \right)\hat I
	\end{array}
\end{equation}
The maximum momentum represented on the grid is ${P_{max}} = \frac{{\pi \hbar }}{{\Delta x}}$, where $\Delta x$ is the grid size. Now the maximum kinetic energy becomes
\begin{equation}
	\frac{{{\rm{P}}_{{\rm{max}}}^{\rm{2}}}}{{{\rm{2m}}}}\, = \frac{{{\pi ^2}{\hbar ^2}}}{{2m\Delta {x^2}}}
\end{equation}
\noindent
Let's assume ${V_{min}}$ and ${V_{max}}$ are minimum and maximum potential represented on the grid, respectively. Now, the range of eigenvalues of $\hat H$ becomes:
\begin{equation}
	\lambda  \in \left[ {{V_{min}},\;{V_{max}} + \frac{{{\pi ^2}{\hbar ^2}}}{{2m\Delta {x^2}}}} \right]
\end{equation}      
leading to 
\begin{equation}\begin{array}{l}
		{E_{max}} = {V_{max}} + \frac{{{\pi ^2}{\hbar ^2}}}{{2m\Delta {x^2}}}\\
		{E_{\min }} = {V_{\min }}
	\end{array}
\end{equation}
\noindent
The evolution of the wave function can be written as:
\begin{small}
	\begin{eqnarray}
		\Psi (t) &=& \hat U(t)\,{\mkern 1mu} \Psi (0)\, = \exp \left[ { - \frac{i}{\hbar }\hat Ht} \right]\Psi (0)\\
		&=& \exp \left[ { - \frac{i}{\hbar }\left( {\frac{{\Delta {E_{grid}}}}{2}} \right){\mkern 1mu} {{\hat H}_{norm}}{\mkern 1mu} t} \right] \times \exp \left[ { - \frac{i}{\hbar }\left( {\frac{{\Delta {E_{grid}}}}{2} + {E_{min}}} \right)\cdot\hat I\cdot} \right]\Psi (0)
	\end{eqnarray}
\end{small}
\noindent
While expanding in terms of polynomial, we obtain:
\begin{equation}
	\begin{array}{l}
		\Psi (t) \simeq \exp \left[ { - \frac{i}{\hbar }\left( {\frac{{\Delta {E_{grid}}}}{2} + {E_{min}}} \right)t} \right]\\
		\,\,\,\,\,\,\,\,\,\,\,\,\,\,\,\,\,\,\,\,\,\, \times {\rm{ }}\left[ {\sum\limits_{n = 0}^N {{a_n}} \left( {\frac{{\Delta {E_{grid}}\,t}}{{2\hbar }}} \right).{\Phi _n}\left\{ { - i\,{{\hat H}_{norm}}} \right\}} \right]\Psi (0)
	\end{array}
\end{equation}
where ${\Phi _n}$ represents complex Chebyshev polynomials and $\exp \left[ { - \frac{i}{\hbar }\left( {\frac{{\Delta {E_{grid}}}}{2} + {E_{min}}} \right)t} \right]$ corresponds to a phase shift compensating the shift in the energy scale. \\

\noindent
The expansion coefficients take the following form:
\begin{equation}
	{a_n}(\alpha ) = \int_{ - i}^{ + i} {\frac{{\exp (i\alpha x){\Phi _n}(x)}}{{{{\left( {1 - {x^2}} \right)}^{\frac{1}{2}}}}}} dx = {c_n}{J_n}(\alpha )
\end{equation}
where $c_n$ = 1, when n = 1 and $c_n$ = 2, when n > 1. On the other hand, ${J_n}(\alpha )$ represents Bessel function of the first kind of order n:
\begin{equation}
	{a_0}(\alpha ) = {J_0}(\alpha )
\end{equation}
\begin{equation}
	\alpha  = \frac{{\Delta {E_{grid}}{\mkern 1mu} t}}{{2\hbar }}
\end{equation}
\noindent
Examining the expansion coefficients as a function of $n$, one finds that when $n$ becomes larger than $\alpha$, ${J_n}(\alpha )$ decays exponentially and thus,  the maximum order ($N$) of the Bessel functions (determined by the size of the time-energy phase-space volume, $\alpha$ becomes:
\begin{equation}
	N \simeq \frac{{\Delta {E_{grid}}t}}{{2\hbar }}
\end{equation}
Employing the operation of ${\Phi _n}\left( { - i{{\hat H}_{norm}}} \right)$ on $\Psi (0)$, we obtain the recursion relation of the Chebyshev polynomials:
\begin{equation}
	{\phi _{n + 1}} =  - 2i{\hat H_{norm}}{\phi _n} + {\phi _{n - 1}}
\end{equation}
where
\begin{equation}
	{\phi _n} = {\Phi _n}\left( { - i{{\hat H}_{norm}}} \right)\Psi (0)
\end{equation}

\noindent
If ${\psi _0} = \Psi (0)$ is the initial wavefunction, the numerical iteration scheme is as follows:
\begin{eqnarray}
	{\phi _0} = {\psi _0} \rightarrow {{\hat H}_{norm}}{\phi _0} \rightarrow {\phi _1} =  - i\left[ {{{\hat H}_{norm}}{\phi _0}} \right]
	\rightarrow {{\hat H}_{norm}}{\phi _1} \nonumber \\ 
	\rightarrow {\phi _2} =  - 2i\left( {{{\hat H}_{norm}}{\phi _1}} \right) + {\phi _0} \rightarrow 
	{\phi _{n + 1}} =  - 2i\left( {{{\hat H}_{norm}}{\phi _n}} \right) + {\phi _{n - 1}}
\end{eqnarray}

\section{Evolution of Kinetic Energy Operator}\label{ssec:prop-ki}

\subsection{Fast Fourier Transformation}

\noindent
Although the actual bottleneck is the evaluation of the kinetic energy operator 
by using the Fast Fourier Transformation (FFT) method, it has the favorable 
property that the computational effort scales as $cN\log N$, where $N$ is the 
total number of grid points on the $\rho$, $\theta$ and $\phi$ coordinates. 
However, the CPU time is still very high for the recursive calculation of 
the kinetic energy operators for iterative Lanczos propagation on a number 
of grid points upto 2$\times$10$^6$ (256$\times$64$\times$128). In particular, 
because the domain of $\theta$ varies from 0 to $\pi$/2 in hyperspherical 
coordinate system, the grid size needs to be doubled in that coordinate so 
that the basis functions (plane waves) vanish at $\theta$ = 0 and $\pi$ 
leading to a sine transformation (see subsection~\ref{ssec:sine_trans}). Moreover, the 
computational cost increases tremendously as the total angular momentum 
$J$ increases, particularly because the projection of the time dependent 
wave packets on asymptotic states ($v^{\prime}, j^{\prime}$) arising due 
to various orbital angular momentum ($l^{\prime}$).\\

\noindent
Let us assume that the amplitudes of the wave packet on the $\theta$-grid ($0\leq \theta \leq 2\pi $)
are given by: 
\begin{equation}
	\chi ({\theta _1}),\chi ({\theta _2}), \ldots ,\chi ({\theta _{N - 1}}),\chi ({\theta _N})
\end{equation}
where $\theta_1$, $\theta_2$, .... $\theta_n$ are $N$ number of grid points. Those amplitudes can be written as linear combination of plane waves which form a complete basis set:
\begin{equation}
	\chi (\theta ) = \sum\limits_{n = 1}^{2M} {C_n^{}} \exp \left( { - i{\kappa _n}\theta } \right)
	\label{eq:a621_1}
\end{equation}
Such basis functions are continuous and their amplitudes match exactly at $\theta=0$ and $2\pi$. From Eq. \ref{eq:a621_1}, one can write:
\begin{equation}
	\begin{array}{l}
		\chi ({\theta _1}) = \sum\limits_{n = 1}^{2M} {C_n^{}} \exp \left( { - i{\kappa _n}{\theta _1}} \right),\\
		\chi ({\theta _2}) = \sum\limits_{n = 1}^{2M} {C_n^{}} \exp \left( { - i{\kappa _n}{\theta _2}} \right),\\
		\,\,\,\,\,\,\,\,\,\,\,\,\,\, \vdots \\
		\chi ({\theta _k}) = \sum\limits_{n = 1}^{2M} {C_n^{}} \exp \left( { - i{\kappa _n}{\theta _k}} \right),
	\end{array}
\end{equation}
where
\begin{equation}
	\begin{array}{l}
		{\kappa _n} = \frac{{2\pi }}{L}{\mkern 1mu} n,\qquad when\quad n \le M\\
		\,\,\,\,\,\, = \frac{{2\pi }}{L}(n - 1 - 2M),\qquad when\quad n > M
	\end{array}
\end{equation}
Thus, one can write
\begin{equation}
	\begin{array}{l}
		\int_\theta  \chi  (\theta )\exp (i{\kappa _m}\theta )\,d\theta  = \sum\limits_{n = 1}^{2M} {C_n^{}} \left[ {\int_\theta  {\exp } ( - i{\kappa _n}\theta )\exp (i{\kappa _m}\theta )\,d\theta } \right]\\
		\,\,\,\,\,\,\,\,\,\,\,\,\,\,\,\,\,\,\,\,\,\,\,\,\,\,\,\,\,\,\,\,\,\,\,\,\,\,\,\,\,\,\,\, = \sum\limits_{n = 1}^{2M} {C_n^{}} {\delta _{nm}} = C_m^{}
	\end{array}
\end{equation}
On the other hand, the Fourier coefficients can be generalized as:
\begin{equation}
	\begin{array}{l}
		C_m^{} = \int_\theta  \chi  (\theta )\exp (i{\kappa _m}\theta )\,d\theta \\
		\Rightarrow \,C_m^{}\,\, = \sum\limits_{l = 1}^{2N} \chi  ({\theta _l})\exp (i{\kappa _m}{\theta _l})\,\Delta \theta 
	\end{array}
\end{equation}
If we back substitute those $C_m$s into Eq. (3.1), we can regenerate the original wave function.\\

\noindent
In order to carry out FFT, one needs to calculate the following quantities:
\begin{equation}
	\begin{array}{l}
		\,\,\frac{d}{{d\theta }}\left[ {\exp ( - i{\kappa _n}\theta )} \right] = ( - i{\kappa _n})\exp ( - i{\kappa _n}\theta )\\
		\Rightarrow \frac{{{d^2}}}{{d{\theta ^2}}}\left[ {\exp ( - i{\kappa _n}\theta )} \right] = {(i{\kappa _n})^2}\exp ( - i{\kappa _n}\theta )\\
		\,\,\,\,\,\,\,\,\,\,\,\,\,\,\,\,\,\,\,\,\,\,\,\,\,\,\,\,\,\,\,\,\,\,\,\,\,\,\,\,\,\,\,\, = ( - \kappa _n^2)\exp ( - i{\kappa _n}\theta )\\
		\Rightarrow \frac{{{d^2}}}{{d{\theta ^2}}}\left[ {\chi (\theta )} \right] = \sum\limits_{n = 1}^{2M} {C_n^{}} ( - \kappa _n^2)\exp ( - i{\kappa _n}\theta )
	\end{array}
\end{equation}
\noindent
where the Fourier coefficients for second derivative of a function is represented as:

\begin{equation}
	C_m^{}( - \kappa _m^2) = ( - \kappa _m^2)\sum\limits_{l = 1}^{2N} \chi  ({\theta _l})\exp (i{\kappa _m}{\theta _l})\Delta \theta, \nonumber
\end{equation}
which is basically the Fourier transform of the function.\\

\noindent
Indeed, one can reduce the huge computational cost through parallelization. 
The computation for a set of partial waves ($K$s) is then carried out in 
different nodes by employing the {\bf MPI} thread, where the time propagation 
and its projection for each partial wave are performed through {\bf OpenMp} 
parallelization in a particular node. Fig. 2 displays the computational 
cost as a function of $J$ for the serial and parallel algorithms. The 
results speak for themselves by showing that the labor with parallelization 
is done much faster than with a serial implementation.

\subsection{Sine Transformation}\label{ssec:sine_trans}

\noindent
For a given $\rho$ and $\phi$, let us denote the amplitudes of the wave packet 
on the $\theta$ grid within the domain 0 to $\pi$/2 by
\begin{eqnarray}
	\chi(\theta_1), \chi(\theta_2), \cdot \cdot \cdot, \chi(\theta_{N-1}), \chi(\theta_N).
	\label{eq:appendixC1}
\end{eqnarray}

\noindent
It is now important to note that while performing Fast Fourier transformation (FFT) 
on those amplitudes to evaluate the kinetic energy operator(s), the basis function 
do not obey the boundary condition, {\em i.e.}, the corresponding Cosine and Sine 
functions do not vanish either at $\theta$ = 0 or $\pi$/2, respectively. In order 
to resolve this problem, we double the grid with appropriate amplitudes as given below:
\begin{eqnarray}
	\chi(\theta_1), \chi(\theta_2), \cdot \cdot \cdot, \chi(\theta_{N-1}), \chi(\theta_N), 
	-\chi(\theta_N), -\chi(\theta_{N-1}), \cdot \cdot \cdot, -\chi(\theta_2), -\chi(\theta_1).
	\label{eq:appendixC2}
\end{eqnarray}

\noindent
The following scheme explores the necessity of doubling the grid in a symmetric manner with negative sign. 
Any one of the above discrete amplitudes ($\chi(\theta_l)s$) can be written in terms of Fourier coefficient:
\begin{eqnarray}
	\chi(\theta_k)=\sum_{n=1}^{2N} C_n^k \exp(-i \kappa_n \theta_k), \hspace*{0.5cm} \kappa_n&=&\frac{2\pi}{L};  
	\hspace*{0.5cm} n\le N \nonumber\\ 
	&=&\frac{2\pi}{L}(n-1-2N);  \hspace*{0.5cm} n > N 
	\label{eq:appendixC3}
\end{eqnarray}
where
\begin{eqnarray}
	C_m^l = \sum_{l=1}^{2N} \chi(\theta_l)\exp(i\kappa_m\theta_l) \Delta \theta.
	\label{eq:appendixC4}
\end{eqnarray}
If we now substitute the anti-symmetric form of the amplitudes (see Eq. (\ref{eq:appendixC2})), Eq. (\ref{eq:appendixC4}) 
can be rewritten as:
\begin{eqnarray}
	C_m^l &=& \sum_{l=1}^{N} 2 i \sin (\kappa_m\theta_l) \Delta \theta; \hspace*{0.5cm} m \le N \nonumber\\
	C^l_{m^{\prime}}&=&-\sum_{l=1}^{N} 2 i \sin (\kappa_{2N-m+1}\theta_l) \Delta \theta; \hspace*{0.5cm} m^{\prime} > N 
	\label{eq:appendixC5}
\end{eqnarray}
Therefore, we have 
\begin{eqnarray}
	C_{2N}^l &=&-C_1^l,\nonumber\\
	C_{2N-1}^l &=&-C_2^l,\nonumber\\
	\cdot &\cdot& \cdot,\nonumber\\
	\cdot &\cdot& \cdot,\nonumber\\
	so \ on.
	\label{eq:appendixC6}
\end{eqnarray}
Accordingly, we have the following anti-symmetric string of Fourier 
coefficients for the $l$th amplitude, $\chi(\theta_l)$, of the wave packet:
\begin{eqnarray}
	C_1^l, C_2^l, \cdot \cdot \cdot, C_{N-1}^l, C_N^l, -C_N^l, -C_{N-1}^l, \cdot \cdot \cdot, -C_2^l, -C_1^l.
	\label{eq:appendixC7}
\end{eqnarray}
If we back substitute Eq. (\ref{eq:appendixC7}) in Eq. (\ref{eq:appendixC3}), 
we can rewrite the amplitudes, 
$\chi(\theta_1)$ and $\chi(\theta_{2N})$ as given below:
\begin{eqnarray}
	\chi(\theta_1)&=&-2 i \sum_{n=1}^N C_n^1 \sin(\kappa_n \theta_1), \nonumber\\
	\chi(\theta_{2N})&=&2 i \sum_{n=1}^N C_n^{2N} \sin(\kappa_n \theta_1). 
	\label{eq:appendixC8}
\end{eqnarray}
Thus, we find the following relationship:
\begin{eqnarray}
	\chi(\theta_{2N}) &=&-\chi(\theta_1),\nonumber\\
	\chi(\theta_{2N-1}) &=&-\chi(\theta_2),\nonumber\\
	\cdot &\cdot& \cdot,\nonumber\\
	\cdot &\cdot& \cdot,\nonumber\\
	so \ on.
	\label{eq:appendixC9}
\end{eqnarray}
\noindent
which has been assumed in Eq. (\ref{eq:appendixC2}) for the required Sine 
transformation. Moreover, it is important to note that in this scheme, either 
the amplitudes (Eq. (\ref{eq:appendixC8})) or the Fourier coefficients 
(Eq. (\ref{eq:appendixC5})) are expressed in terms of sine functions.

\section{Analysis of the outgoing wave packet}\label{ssec:ana1}
\noindent
It is possible to project the outgoing wave packet onto asymptotic eigenstates 
at a fixed value of the hyperradius $\rho$. Since, in hypersherical coordinates, 
all distances are represented by this single variable, the inelastic coupling 
between different vibrational states will vanish very slowly with $\rho$. 
An asymptotic propagation to large values of the hyperradius is therefore 
necessary before accurate reaction probabilities could be calculated. In the 
present work, we project the wave packet onto asymptotic eigenstates at a fixed 
value of $R$ which greatly simplifies or entirely removes the asymptotic propagation 
part of the problem\cite{jcp-99-2674-BILL,jcp-100-1085-BILL}. \\

\noindent
The scattering amplitude in the channel specified by vibrational, rotational 
and orbital quantum numbers $v^{\prime}$, $j^{\prime}$, $l^{\prime}$ is obtained as:
\begin{eqnarray}
	u^J_{v^{\prime}j^{\prime}l^{\prime}}(R;t)&=&4R\int \ dr \ \int \ d\eta \ r \sin\eta 
	\ \rho^{-5/2} (\sin2\theta)^{-1/2} \phi_{v^{\prime}j^{\prime}}(r)\nonumber\\
	&\times&\sum_{K\mu^{\prime}}g_{j^{\prime}l^{\prime}\mu^{\prime}}
	A^{\ast}_{K\mu^{\prime}} C_{j^{\prime}\mu^{\prime}}(\eta)\Phi_K(\rho, \theta, \phi),
	\label{eq:proj1}
\end{eqnarray}
where
\begin{eqnarray}
	g_{jl\mu}=\sqrt{2\pi(2l+1)} (-1)^{j-l} 
	\bigg(\begin{matrix}
		j&l&J\\
		\mu&0&-\mu
	\end{matrix} \bigg).
	\label{eq:proj2}
\end{eqnarray}
The integration is performed over the ($\theta$, $\phi$) grid using the relations:
\begin{eqnarray}
	\rho& = &\sqrt{2}d_i \ R/\sqrt{1-\sin\theta\cos(\phi-\epsilon_i)},\\
	r& = &\frac{d_i \ \rho}{\sqrt{2}}\sqrt{1 + \sin\theta\cos(\phi-\epsilon_i)},\\
	\sin \eta& = &\cos\theta/\sqrt{1-\sin^2\theta\cos^2(\phi-\epsilon_i)},
	\label{eq:proj3}
\end{eqnarray}
with the Jacobi factor (for fixed $R$) given by:
\begin{eqnarray}
	J(r, \eta|\theta, \phi) = \frac{d^2_iR\sin\theta}{[1-\sin\theta\cos(\phi-\epsilon_i)]
		\sqrt{1-\sin^2\theta\cos^2(\phi-\epsilon_i)}},
	\label{eq:proj4}
\end{eqnarray}
where $d_i$ and $\epsilon_i$ are channel dependent constants. Since the projection 
is performed for a fixed value of $R$ and the hyperradius is a function of $R$, 
$\theta$ and $\phi$, it is necessary to interpolate on the grid in order to find 
the corresponding value of $\Phi_K$.\\

\noindent
When the wave packet has passed the projection region the amplitudes are Fourier 
transformed from time to energy space:
\begin{eqnarray}
	b^J_{v^{\prime}j^{\prime}l^{\prime}}(E;R)=\frac{1}{\sqrt{2}\pi}\int 
	\ dt \ u^J_{v^{\prime}j^{\prime}l^{\prime}}(R;t)
	\exp(iEt/\hbar).
	\label{eq:proj5}
\end{eqnarray}
These amplitudes are now expanded on incoming and outgoing waves,
\begin{eqnarray}
	b^J_{v^{\prime}j^{\prime}l^{\prime}}=A^{in}_{v^{\prime}j^{\prime}l^{\prime}}k_{v^{\prime}j^{\prime}}R 
	\ h^-(k_{v^{\prime}j^{\prime}}R)+A^{out}_{v^{\prime}j^{\prime}l^{\prime}}k_{v^{\prime}j^{\prime}}R 
	\ h^+ (k_{v^{\prime}j^{\prime}}R),
	\label{eq:proj6}
\end{eqnarray}
where
\begin{eqnarray}
	h^{\pm} (k_{v^{\prime}j^{\prime}}R)=-n_l(k_{v^{\prime}j^{\prime}}R)\pm ij_l(k_{v^{\prime}j^{\prime}}R)
	\label{eq:proj7}
\end{eqnarray}
and $j_l$ and $n_l$ are spherical Bessel and Neumann functions, respectively. 
The projection on the incoming component of the scattered wave packet should, 
of course, be zero. The reaction probability from the initial state ($vjl$) 
to the product state ($v^{\prime}j^{\prime}l^{\prime}$) is obtained as the 
ratio of the outgoing and incoming fluxes,
\begin{eqnarray}
	P_{v^{\prime}j^{\prime}l^{\prime}\leftarrow vjl}=\frac{F_{v^{\prime}j^{\prime}l^{\prime}}}{F_{vjl}},
	\label{eq:proj8}
\end{eqnarray}
where
\begin{eqnarray}
	F_{v^{\prime}j^{\prime}l^{\prime}}&=
	&\frac{1}{\mu_{out}}k_{v^{\prime}j^{\prime}}|A^{out}_{v^{\prime}j^{\prime}l^{\prime}}|^2,\\
	F_{vjl}&=&\frac{1}{\mu_{in}}k_{vj}|c^l_E|^2,
	\label{eq:proj9}
\end{eqnarray}
$c^l_E$ is the component with energy $E$ contained in the initial wave packet, 
$k_{vj}$ the wave number of the initial channel $vj$, and $\mu_{in}$ and $\mu_{out}$ 
the appropriate center of masses for the reactants and products, respectively. The 
weight in energy space is related (see Appendix A8) to the weight in $k$ space 
through the relation: 
\begin{eqnarray}
	|c^l_E|^2=\bigg(\frac{\mu_{in}}{\hbar k}\bigg)^2|c^l_k|^2.
	\label{eq:proj10}
\end{eqnarray}
\noindent
$|c^l_k|^2$ can be obtained analytically for a Gaussian wave packet as:
\begin{eqnarray}
	|c^l_k|^2=\sqrt{2/\pi}\ \sigma \ \exp[-2\sigma^2(k-k_0)^2],
	\label{eq:proj11}
\end{eqnarray}
but this expression is only valid for a free particle, i.e., corresponding to a system 
initialized at very large separation. On the other hand, if the incoming or outgoing 
wave packet is initialized or analyzed at a moderate value of $R$, the effective potential 
at the same $R$ appears as $[V(R)+l(l+1)/2 \mu R^2]$, where the potential is averaged 
over the spherical DOFs, $\theta$ and $\phi$, to calculate the effective potential $V(R)$.  

\section{Absorbing Potential}\label{ssec:ana2}

\noindent
If absorbing boundaries are applied to absorb the wave-packet entering into different channels for a one-dimensional system defined in the range 
$- \infty  \le x \le  + \infty$, the scattering process takes place in the range $x_1 < x < x_r$, where the wave function is initially located at $x=x_m$. The scattering potential is defined around x = 0 and the negative imaginary potential is depicted along the interval $a < x < b$. $V_0 (x)$ depicts the scattering potential around $x = 0$. On the other hand, $\psi (x,t=0)$ represents initial wave-packet located at $x = x_m$, moving towards the potential. Due to the potential, one part of the wave-packet will be reflected and another part will be transmitted. \\

\noindent
The time-dependent Schrödinger equation can be written as:
\begin{equation}
	i\hbar \frac{{\partial \Psi (x,t)}}{{\partial t}} = \frac{{ - {\hbar ^2}}}{{2m}}\frac{{{\partial ^2}\Psi (x,t)}}{{\partial {x^2}}} + {V_0}(x)\Psi (x,t),
\end{equation}
where $m$ is mass of the particle. Since the particle is confined in a box ($x_1 < x < x_r$), one can write:
\begin{equation}
	\Psi ({x_l},t){\rm{ }} = \Psi ({x_r},t) = 0
\end{equation}
\noindent
Let us now consider the case, where in the vicinity of $x = x_f$, an imaginary negative short-range potential $[-iV_I(x)]$ is added. Thus, total potential of the system takes the following form:
\begin{equation}
	V\left( x \right) = {\rm{ }}{V_0}\left( x \right) - i{V_I}\left( x \right).
\end{equation}

\noindent
There are two conditions to be fulfilled by $V_I(x)$:
\begin{itemize}
	\item Absorption is large enough so that the reflected wave from the boundaries at $x = x_l$ and $x = x_r$ are negligibly small.
	\item The imaginary potential is smooth enough so that hardly any reflection takes place before the wave-packet penetrates into the potential.
\end{itemize}

If we consider $V_I(x) = 0$, when $x_1 \leq x \leq x_2$\\

\noindent
The outgoing wave packet must be removed from the grid just after the projection but before 
it reaches the grid boundary at $\rho_{max}$. This is accomplished by adding a negative 
imaginary potential to the last grid points for the hyperradius. This absorbing potential 
has no physical meaning, it is located far away from the interaction region but is a device 
used to avoid reflection. The form of the imaginary potential is arbitrary, the only 
requirement is that the scattering amplitudes should be independent of the type of 
function used. In the present work, we use an absorbing potential of the following form: 
\begin{eqnarray}
	V_{Im}(\rho) = -\frac{iV_{Im \ max}}{\cosh^2[(\rho_{max}-\rho)/\alpha]}, \ \ \ \ \rho \ge \rho_{Im},
	\label{eq:impot1}
\end{eqnarray}
where $\rho_{Im}$ is the starting point for the absorbing potential. The performance 
of the imaginary potential is not very sensitive to the choice of $V_{Im \ max}$, but 
with $\alpha$. The error at low energies indicates that these components are not absorbed 
efficiently. To improve the absorption, the magnitude of the absorbing potential should 
be decreased. This will, however, lead to reflection at higher energies. We therefore 
decided to decrease $\alpha$ linearly with time during the absorption of the wave packet. 
By doing so the slow moving energy components will experience an absorbing potential 
of a magnitude small enough to avoid reflection. This procedure led to a very uniform 
error distribution over the entire energy range.

\section{Time energy Fourier transform}\label{ssec:ana3}

\noindent
The relationship between the weight in $ k $ space of a given component 
of a Gaussian wave packet [see Eq. (\ref{eq:eq28})] and its weight in energy 
space is derived via the free particle propagator \cite{cpc-184-270-SA}
\begin{eqnarray}
	K(R,t=0;R^{\ast},t)=\bigg(\frac{\mu}{2 \pi i \hbar t}\bigg)^{\frac{1}{2}} \exp\bigg[\frac{i\mu(R^{\ast}-R)^2}{2\hbar t}\bigg].
	\label{eq:appendixD1}
\end{eqnarray}

\noindent
The amplitude of the wave packet at a specified time, $t$, and point in space, $R^{\ast}$, can be expressed as
\begin{eqnarray}
	\chi(t,R^{\ast})=\int^{\infty}_{-\infty}dR \ K(R,0;R^{\ast},t) \chi(t=0,R).
	\label{eq:appendixD2}
\end{eqnarray}
The Fourier transformation from time to energy space is now performed,
\begin{eqnarray}
	b(E;R^{\ast})&=&\frac{1}{\sqrt{2\pi}}\int^{\infty}_{-\infty}dt\chi(t,R^{\ast})\exp(iEt/\hbar)\nonumber\\
	&=&\frac{1}{\sqrt{2\pi}}Nc\int^{\infty}_{-\infty}dR\bigg[\int^{\infty}_{-\infty}dt \ t^{\gamma-1}\nonumber\\
	&\times&\exp\bigg(\frac{i\alpha}{2}\bigg(t-\frac{\beta^2}{t}\bigg)\bigg)\bigg]\exp[-ik_0R-a(R-R_0)^2],
	\label{eq:appendixD3}
\end{eqnarray}
where
\begin{eqnarray}
	c&=&\sqrt{\frac{\mu}{2\pi i \hbar}},\\
	\gamma&=&1/2,\\
	\alpha&=&\frac{2E}{\hbar}=\frac{\hbar k^2}{\mu},\\
	\beta&=&i\sqrt{\frac{\mu}{2E}}(R^{\ast}-R).
	\label{eq:appendixD4}
\end{eqnarray}
The integral in brackets can be solved analytically \cite{ISG} giving
\begin{eqnarray}
	I=\sqrt{\frac{2\pi i}{\alpha}}\exp[-i k (R^{\ast}-R)].
	\label{eq:appendixD5}
\end{eqnarray}
Substitution of this result together with the expressions for $c$ and $\alpha$ into Eq. (\ref{eq:appendixD3}) yields
\begin{eqnarray}
	b(E;R^{\ast})&=&\frac{\mu}{\hbar k}\exp(-i k R^{\ast})\frac{1}{\sqrt{2\pi}}N\int^{\infty}_{-\infty}dR \nonumber \\
	&\times&\ \exp[-ik_0R-a(R-R_0)^2]\exp(ikR)\nonumber\\
	&=&\frac{\mu}{\hbar k}\exp(-i k R^{\ast})\frac{1}{\sqrt{2\pi}}\int^{\infty}_{-\infty}dR \ \chi(t=0,R)\exp(ikR)\nonumber\\
	&=&\frac{\mu}{\hbar k}\exp(-i k R^{\ast})c_k.
	\label{eq:appendixD6}
\end{eqnarray}
Thus, we have shown that
\begin{eqnarray}
	|c_E|^2=\bigg(\frac{\mu}{\hbar k}\bigg)^2|c_k|^2.
	\label{eq:appendixD7}
\end{eqnarray}

\bibliographystyle{aipnum4-1}
\bibliography{ref_book}

\chapter{Time- and temperature- dependent EHP for linear coupling between molecular DOFs and surface modes} \label{app:BB}

\setcounter{equation}{0}
\renewcommand{\theequation}{D.\arabic{equation}}

\noindent
When bi-linear coupling (VV) between molecular DOFs and surface modes is absent, the TDSE for VT process takes the following form:

\vspace{-0.8 cm} 

\begin{eqnarray}
	i \hbar \frac{d U_{VT}}{dt} = V^{VT}_{I}(\lbrace \chi _m \rbrace,t) U_{VT}
	\label{Eq: VT_TDSE_m2}
\end{eqnarray}

\noindent
and VT evolution operator ($U_{VT}$) is written as\cite{PPJS,SAMARK}: 

\begin{eqnarray}
	U_{VT} = \prod_{k}^{M} U_{VT}^{(k)} = \prod_{k}^{M} \exp \left[ i \left( \beta_{k} + \alpha_{k}^+ \: b_k^{\dagger} +  \alpha_{k}^- \: b_k\right) \right], 
	\label{Eq: U_VT in general1}
\end{eqnarray} 

\noindent
where
\begin{eqnarray}
	\alpha_{k}^\pm &=& -\frac{A_k}{\hbar} \int_{t_0}^t dt^{\prime} V_{k,1} (\lbrace \chi_m \rbrace) \exp[\pm i\omega_k t^{\prime}],
	\label{Eq: alpha_plus_minus2}
\end{eqnarray}

\noindent
and
\begin{eqnarray}
	\beta_{k}=\frac{i}{\hbar}\int_{t_0}^t dt^{\prime} V_{k,1}(\lbrace \chi_m \rbrace) \{ \exp [i\omega_k t^{\prime}]\alpha_{k}^
	-(t^{\prime})-\exp[-i\omega_k t^{\prime}] \alpha_{k}^+(t^{\prime})\}.
	\label{Eq: beta2}
\end{eqnarray}

{\vspace{1 cm}

	\noindent
	The time- and temperature- dependent state averaged EHP emerging from linear (VT) molecular DOFs-surface modes coupling is denoted as:
	
	\begin{eqnarray}
		V_{eff}^{VT} (\lbrace \chi _m\rbrace, t, T_s)  = \sum_{\lbrace n_0 \rbrace} p_{\lbrace n_0 \rbrace}(T_s) \langle V^{VT}_I (\lbrace \chi _m\rbrace,t)\rangle_{\lbrace n_0 \rbrace},
		\label{Eq: state_averaged_Hartree_potential_VT1}
	\end{eqnarray}
	
	\noindent
	where $p_{\lbrace n_0 \rbrace}(T_s)$ represents the initial state distribution function for specific kind of statistics, i.e., either BE or MB type (see Equations (\ref{Eq:norm_prob}), (\ref{Eq: BEP_new_factor_app}) and (\ref{Eq: MBP_new_factor_app}) in Sec. \ref{ssec:EHP}). \\

	\noindent
	The initial state, $\lbrace n_0 \rbrace$  dependent Hartree potential originating from such VT coupled  surface modes is depicted in terms of initial wavefunction and evolution operator [ $\Phi_{vib} (t) = U_{VT} \: \Phi_{vib}(t_0)$] as~\cite{SAMARK}:
	
	\begin{eqnarray}
		\langle V^{VT}_I (\lbrace \chi _m\rbrace,t)\rangle_{\lbrace n_0 \rbrace} &=& \langle \Phi_{vib} (t) \vert V_{I}^{VT} (\lbrace \chi_m \rbrace, t) \vert 
		\Phi_{vib} (t) \rangle ,  \nonumber \\
		&=& \langle	\Phi_{vib}(t_0) \vert U_{VT}^{\dagger} V_{I}^{VT}(\lbrace \chi_m \rbrace, t) U_{VT} \vert  \Phi_{vib}(t_0) \rangle \nonumber \\
		&=& \langle \lbrace n_0 \rbrace \vert U_{VT}^{\dagger} V_{I}^{VT} (\lbrace \chi_m \rbrace, t) U_{VT} \vert \lbrace n_0 \rbrace \rangle \nonumber \\
		&=& \sum_{\lbrace n^{\prime} \rbrace} \sum_{\lbrace n \rbrace} \langle \lbrace n_0 \rbrace \vert U_{VT}^{\dagger} \vert \lbrace 
		n^{\prime} \rbrace \rangle \langle \lbrace n^{\prime} \rbrace \vert V_{I}^{VT} (\lbrace \chi_m \rbrace, t) \vert \lbrace n \rbrace \rangle \langle \lbrace n  
		\rbrace \vert U_{VT} \vert \lbrace n_0  \rbrace \rangle \nonumber \\
		&=& \sum_{\lbrace n^{\prime} \rbrace} \sum_{\lbrace n \rbrace} \alpha_{\lbrace n^{\prime} \rbrace 
			\leftarrow  \lbrace n_0 \rbrace }^{*} (t) \alpha_{\lbrace n \rbrace \leftarrow  \lbrace n_0 \rbrace } 
		(t) \langle  \lbrace n^{\prime} \rbrace \vert 
		V_{I}^{VT}(\lbrace \chi_m \rbrace, t) \vert \lbrace n \rbrace \rangle,
		\label{Eq: VT_state_dependent_effective_potential1}
	\end{eqnarray}  
	
	\noindent
	where $\alpha_{\lbrace n \rbrace} (t)$ denotes the amplitude of a particular surface vibrational eigenstate ($\lbrace n \rbrace$). Those amplitudes ($\alpha_{\lbrace n \rbrace \leftarrow  \lbrace n_0 \rbrace } (t)$) represent the time evolution of initial configuration ($\lbrace n_0 \rbrace$) under VT operator ($U_{VT}$) as presented below:

	\begin{eqnarray}
		\alpha_{\lbrace n \rbrace \leftarrow  \lbrace n_0 \rbrace } (t) = \langle \lbrace n \rbrace \vert U_{VT} \vert \lbrace n_0 \rbrace \rangle.
		\label{Eq: amplites_VT1} 
	\end{eqnarray}

	\noindent
	The interaction potential appearing from linear (VT) molecular DOFs and surface modes coupling is converted
	into the following form within second quantization treatment (see Equation (\ref{Eq: Interaction_potential_interaction_picture_I}) in Sec. \ref{ssec:Interaction-potential}):

	\begin{eqnarray}
		V_{I}^{VT}(\lbrace \chi_m \rbrace, t) &=& \sum_k \lambda_k  A_k (b_k F_k^{-}  + b_k^{\dagger} F_k^{+}) V_{k,1}(\lbrace \chi_m \rbrace), 
		\label{Eq: Interaction_potential_VT_I1}
	\end{eqnarray}
	
	\noindent
	where $F_k^{+} = \exp (i \omega_k t ) $,  $F_k^{-} = \left( F_k^{+} \right)^{*} $
	and $V_{k,1}(\lbrace \chi_m \rbrace) = \frac{\partial V_I(\lbrace \chi_m \rbrace)}{\partial Q_k}  \vert_{eq}$. \\

	\noindent
	Plugging Equation (\ref{Eq: Interaction_potential_VT_I1}) sequentially in Equation (\ref{Eq: VT_state_dependent_effective_potential1}) and Equation (\ref{Eq: state_averaged_Hartree_potential_VT1}) for BE and MB case, the EHP can be written as\cite{SAMARK}:

	\begin{footnotesize}
		\begin{eqnarray}
			V_{eff}^{VT} (\lbrace \chi _m\rbrace, t, T_s) &=& \sum_k   \lambda_k   \sum_{n^0_k} \sum_{n_k}\ \ A_{k} \: p_{n_k^0}  V_{k,1} (\lbrace \chi_m \rbrace) 
			[n_k^{1/2}\alpha^{\ast(k)}_{n_{k-1} \leftarrow n_k^0  }(t) F_k^- + (n_k + 1)^{1/2} F_k^+ \alpha^{\ast(k)}_{n_{k+1} \leftarrow 
				n_k^0 }(t)] \alpha_{n_k \leftarrow n_k^0 }^{(k)}(t)\nonumber \\
			& =& \frac{1}{N_{\text{BE/MB}}}\sum_k   \lambda_k   \sum_{n^0_k} \sum_{n_k}\   \tilde{p}_{n_k^0} \ A_k\ V_{k,1} (\lbrace \chi_m \rbrace) 
			[n_k^{1/2}\alpha^{\ast(k)}_{n_{k-1} \leftarrow n_k^0  }(t) F_k^- + (n_k + 1)^{1/2} F_k^+ \alpha^{\ast(k)}_{n_{k+1} \leftarrow 
				n_k^0 }(t)] \alpha_{n_k \leftarrow n_k^0 }^{(k)}(t). \nonumber \\ 
			\label{Eq: VT_effective_I}
		\end{eqnarray}
	\end{footnotesize}
	
	\noindent
	In order to arrive at Equation (\ref{Eq: VT_effective_I}), one can employ:
	\begin{eqnarray}
		\alpha_{\{n\}}(t)=\prod_{k=1}^M \alpha_{n_k}^{(k)}(t), \: \text{and} \: \sum_{n_k}|\alpha_{n_k}^{(k)}(t)|^2=1,
		\label{Eq: alpha_prod}
	\end{eqnarray}
	
	\noindent
	where $\alpha_{n_k}^{(k)}(t)$, the amplitude of the $n_k$th quantum state for a particular VT coupled mode ($k$), is represented in terms of the exact solution of the LFHO \citen{PPJS} as:

	\begin{eqnarray}
		\alpha_{n_k+1}^{\ast (k)}(t) = \exp(-i\beta_{k} - \frac{1}{2}\rho_k)[(n_k+1)! n_k^0!]^{1/2} (-i\alpha_{k}^-)^{n_k-n_k^0+1} 
		f(\rho_k,n_k+1),
		\label{Eq: alpha_n+1}
	\end{eqnarray}
	
	\noindent
	and
	\begin{eqnarray}
		\alpha_{n_k}^{(k)}(t) = \exp(i\beta_{k} - \frac{1}{2}\rho_k)[n_k! n_k^0!]^{1/2} (i \alpha_{k}^+)^{n_k-n_k^0} f(\rho_k,n_k).
		\label{Eq: alpha_n}
	\end{eqnarray}
	
	\noindent
	The terms $\alpha_{k}^\pm$, $\beta_{k}$ show in Equations (\ref{Eq: alpha_plus_minus2})-(\ref{Eq: beta2}) and $\rho_k = \alpha_{k}^+ \: \alpha_{k}^-$. \\

	\noindent
	The $f(\rho_k,n_k)$ is written as:
	\begin{eqnarray}
		f(\rho_k,n_k)=\frac{1}{n_k!} L_{n_k^0}^{n_k-n_k^0} (\rho_k), \hspace{0.5 cm} n_k \geq n_k^0,
		\label{Eq: laguerre_creation}
	\end{eqnarray}
	
	\noindent
	and
	\begin{eqnarray}
		f(\rho_k,n_k)=\frac{1}{n_k^0!}(-\rho_k)^{n_k^0-n_k} L_{n_k}^{n_k^0-n_k} (\rho_k), \hspace{0.5 cm} n_k < n_k^0,
		\label{Eq: laguerre_annhilation}
	\end{eqnarray}
	
	\noindent
	where $L_{n_k}^{n_k^0-n_k}$ represents the Laguerre-polynomial. \\
	
	\noindent
	After incorporating Equations (\ref{Eq: alpha_n+1})-(\ref{Eq: laguerre_annhilation}) in Equation (\ref{Eq: VT_effective_I}) and further, using the BE or the MB probability factor for the initial state distribution, one can arrive at \cite{SAMARK} 
	
	\begin{eqnarray}
		V_{eff}^{VT} (\lbrace \chi _m\rbrace, t, T_s) = \frac{1}{N_{\text{BE/MB}}}\sum_k \left( S_I^{(k)} (\lbrace \chi_m \rbrace,t, T_\text{s}) + S_{II}^{(k)} (\lbrace \chi_m \rbrace,t, T_\text{s}) \right),
		\label{Eq: V_eff_SI+SII}
	\end{eqnarray}
	
	\noindent
	where
	\begin{eqnarray}
		S_I^{(k)} (\lbrace \chi_m \rbrace, t, T_\text{s}) &=& - \omega_k^{-1} \epsilon_k (\lbrace \chi_m \rbrace, t) \exp(-\rho_k) \sum_{q=1}^{\infty} \left( z_k \right)^{q/2} 
		\sum_{n_k^0 = 1}^{\infty} \sum_{m=0}^{n_k^0 -1} \left( z_k^q \right)^{n_k^0} \frac{m!}{n_k^0!} \nonumber \\ &\times &
		\left( \rho_k \right)^{n_k^0 - m} 
		L_m^{n_k^0 - m} (\rho_k) L_{m-1}^{n_k^0 - m +1} (\rho_k) \hspace{2cm} \text{for BE case}
	\end{eqnarray}
	
	\noindent
	and
	\begin{eqnarray}
		S_I^{(k)} (\lbrace \chi_m \rbrace, t, T_\text{s}) &=& - \omega_k^{-1} \epsilon_k (\lbrace \chi_m \rbrace, t) \exp(-\rho_k) z_k^{1/2} 
		\sum_{n_k^0 = 1}^{\infty} \sum_{m=0}^{n_k^0 -1} z_k^{n_k^0} \frac{m!}{n_k^0!}\nonumber \\ &\times &
		\left( \rho_k \right)^{n_k^0 - m} L_m^{n_k^0 - m} (\rho_k) L_{m-1}^{n_k^0 - m +1} (\rho_k)  \hspace{2cm} \text{for MB case}.
	\end{eqnarray}
	
	\noindent
	The expression of $S_I^{(k)} (\lbrace \chi_m \rbrace, t, T_\text{s})$ is further rewritten as:
	\begin{eqnarray}
		S_I^{(k)} (\lbrace \chi_m \rbrace, t, T_\text{s}) &=& - \omega_k^{-1} \epsilon_k (\lbrace \chi_m \rbrace, t) \exp(-\rho_k) \sum_{q=1}^{\infty} \left( z_k \right)^{q/2}
		\sum_{n=0}^{\infty} \sum_{m=0}^{\infty} \left( z_k^q \right)^{m+n} \frac{n!}{\left( n+m \right)!} \nonumber \\ &\times &
		\left( \rho_k \right)^m L_n^m 
		(\rho_k) L_{n-1}^{m+1} (\rho_k) \hspace{4cm} \text{for BE case} 
		\label{Eq: S_I}
	\end{eqnarray}
	
	\noindent
	and
	\begin{eqnarray}
		S_I^{(k)} (\lbrace \chi_m \rbrace, t, T_\text{s}) &=& - \omega_k^{-1} \epsilon_k (\lbrace \chi_m \rbrace, t) \exp(-\rho_k) z_k^{1/2}
		\sum_{n=0}^{\infty} \sum_{m=0}^{\infty} z_k^{m+n} \frac{n!}{\left( n+m \right)!} \nonumber \\ &\times &
		\left( \rho_k \right)^m L_n^m (\rho_k) L_{n-1}^{m+1} (\rho_k) \hspace{4cm} \text{for MB case}.
		\label{Eq: S_I_MBP}
	\end{eqnarray}
	
	\noindent
	In a similar wat, the second term $S_{II}^{(k)} (\lbrace \chi_m \rbrace, t, T_\text{s})$ in Equation (\ref{Eq: V_eff_SI+SII}) can be written as:
	\begin{eqnarray}
		S_{II}^{(k)} (\lbrace \chi_m \rbrace, t, T_\text{s}) &=& \omega_k^{-1} \epsilon_k (\lbrace \chi_m \rbrace, t) \exp(-\rho_k) \sum_{q=1}^{\infty} \left( z_k \right)^{q/2} 
		\sum_{n_k^0 = 0}^{\infty} \sum_{m=n_k^0}^{\infty} \left( z_k^q \right)^{n_k^0} \frac{n_k^0!}{m!}\nonumber \\ &\times & \left( \rho_k \right)^{m-n_k^0} 
		L_m^{m-n_k^0} (\rho_k) L_m^{m-n_k^0+1} (\rho_k)\hspace{2cm} \text{for BE case} 
	\end{eqnarray}
	
	\noindent
	and
	\begin{eqnarray}
		S_{II}^{(k)} (\lbrace \chi_m \rbrace, t, T_\text{s}) &=& \omega_k^{-1} \epsilon_k (\lbrace \chi_m \rbrace, t) \exp(-\rho_k) z_k^{1/2} 
		\sum_{n_k^0 = 0}^{\infty} \sum_{m=n_k^0}^{\infty} z_k^{n_k^0} \frac{n_k^0!}{m!} \nonumber \\ &\times & \left( \rho_k \right)^{m-n_k^0} L_m^{m-n_k^0} (\rho_k) L_m^{m-n_k^0+1} (\rho_k), \hspace{2cm} \text{for MB case} 
	\end{eqnarray}
	
	\noindent
	which can be expressed in the following form:
	\begin{eqnarray}
		S_{II}^{(k)} (\lbrace \chi_m \rbrace, t, T_\text{s}) &=& \omega_k^{-1} \epsilon_k (\lbrace \chi_m \rbrace, t) \exp(-\rho_k)  \sum_{q=1}^{\infty} \left( z_k \right)^{q/2} 
		\sum_{n=0}^{\infty} \sum_{m=0}^{\infty} \left( z_k^q \right)^n \frac{n!}{(n+m)!} \nonumber \\ &\times & \left( \rho_k \right)^m L_n^{m} (\rho_k) 
		L_n^{m+1} (\rho_k)  \hspace{4cm} \text{for BE case}
		\label{Eq: S_II}
	\end{eqnarray}
	and
	\begin{eqnarray}
		S_{II}^{(k)} (\lbrace \chi_m \rbrace, t, T_\text{s}) &=& \omega_k^{-1} \epsilon_k (\lbrace \chi_m \rbrace, t) \exp(-\rho_k)  \sum_{q=1}^{\infty} z_k^{1/2} 
		\sum_{n=0}^{\infty} \sum_{m=0}^{\infty} z_k^n \frac{n!}{(n+m)!} \nonumber \\ &\times & 
		\left( \rho_k \right)^m L_n^{m} (\rho_k) 
		L_n^{m+1} (\rho_k)  \hspace{4cm} \text{for MB case}. 
		\label{Eq: S_II_MBP}
	\end{eqnarray}
	
	\noindent
	The explicit time dependent quantity, $\epsilon_k (t)$, arising in Equation (\ref{Eq: S_I}) - (\ref{Eq: S_II_MBP}) is given by:  
	
	\begin{eqnarray}
		\epsilon_k(\lbrace \chi_m \rbrace,t) = \lambda_k V_{k,1}(\lbrace \chi_m \rbrace) \int_{t_0}^{t} dt^{\prime} V_{k,1} (\lbrace \chi_m \rbrace) \sin \left[ \omega_k \left( t^{\prime} 
		- t \right) \right].
		\label{Eq: epsilon}
	\end{eqnarray}

	\vspace{0.3cm}
	
	\noindent
	Substitution of Equation (\ref{Eq: S_I}) and Equation (\ref{Eq: S_II}) in Equation (\ref{Eq: V_eff_SI+SII}) results into the expression of EHP into the following form for the BE distribution case:
	\begin{eqnarray}
		V_{eff, \: BE}^{VT} (\lbrace \chi _m\rbrace, t, T_s) & =&\frac{1}{N_\text{BE}}\sum_k \omega_k^{-1} \epsilon_k (\lbrace \chi_m \rbrace, t) \exp(- \rho_k) \nonumber \\
		& \times & \bigg\{ \sum_{q=1}^{\infty} \left( z_k 
		\right)^{q/2} \sum_{n=0}^{\infty} \sum_{m=0}^{\infty} \left( z_k^q  \right)^{n+m} \frac{n!}{(n+m)!}
		\left( \rho_k \right)^m L_n^{m} (\rho_k) L_n^{m} (\rho_k) \nonumber \\
		& + &\sum_{q=1}^{\infty} \left( z_k \right)^{q/2} \sum_{n=0}^{\infty} \sum_{m=0}^{\infty} \left( z_k^q  \right)^{n} \frac{n!}
		{(n+m)!}
		\left( \rho_k \right)^m L_n^{m} (\rho_k) L_n^{m} (\rho_k) \nonumber \\
		& +& \sum_{q=1}^{\infty} \left( z_k \right)^{q/2} \sum_{n=0}^{\infty} \sum_{m=0}^{\infty} \left( z_k^q  \right)^{n} \frac{n!}
		{(n+m)!}
		\left( \rho_k \right)^m L_n^{m} (\rho_k) L_{n-1}^{m} (\rho_k) \nonumber \\
		&-& \sum_{q=1}^{\infty} \left( z_k \right)^{q/2} \sum_{n=0}^{\infty} \sum_{m=0}^{\infty} \left( z_k^q  \right)^{n+m} \frac{n!}
		{(n+m)!}
		\left( \rho_k \right)^m L_n^{m} (\rho_k) L_n^{m+1} (\rho_k) \bigg\}.
		\label{Eq: V_eff_pre_reduced}
	\end{eqnarray}
	
	\vspace{0.3cm}
	
	\noindent
	In a similar way, the EHP for the MB distribution is turned into:
	\begin{eqnarray}
		V_{eff, \: MB}^{VT} (\lbrace \chi _m\rbrace, t, T_s) &=&\frac{1}{N_{\text{MB}}}\sum_k \omega_k^{-1} \epsilon_k (\lbrace \chi_m \rbrace, t) \exp(- \rho_k) \nonumber \\
		&  \times & \bigg\{ z_k^{1/2} \sum_{n=0}^{\infty} \sum_{m=0}^{\infty} z_k^{n+m} \frac{n!}{(n+m)!}
		\left( \rho_k \right)^m L_n^{m} (\rho_k) L_n^{m} (\rho_k) \nonumber \\
		& + & z_k^{1/2} \sum_{n=0}^{\infty} \sum_{m=0}^{\infty} z_k^{n} \frac{n!}
		{(n+m)!}
		\left( \rho_k \right)^m L_n^{m} (\rho_k) L_n^{m} (\rho_k) \nonumber \\
		& + &z_k^{1/2} \sum_{n=0}^{\infty} \sum_{m=0}^{\infty} z_k^{n} \frac{n!}
		{(n+m)!}
		\left( \rho_k \right)^m L_n^{m} (\rho_k) L_{n-1}^{m} (\rho_k) \nonumber \\
		&-& z_k^{1/2} \sum_{n=0}^{\infty} \sum_{m=0}^{\infty} z_k^{n+m} \frac{n!}
		{(n+m)!}
		\left( \rho_k \right)^m L_n^{m} (\rho_k) L_n^{m+1} (\rho_k) \bigg\}.
		\label{Eq: V_eff_pre_reduced_MBP}
	\end{eqnarray}
	
	\vspace{0.3cm}
	
	\noindent
	While formulating a more simplified expression of EHP for both BE and MB cases, the following identity :

	\begin{eqnarray}
		\sum_{p=0}^\infty p! \ L_p^\alpha(x) L_p^\alpha(y) \frac{z^p}{(p+\alpha)!}
		=\bigg[\frac{(xyz)^{-\frac{\alpha}{2}}}{(1-z)}\bigg] \exp \bigg[-\frac{z(x+y)}{(1-z)}\bigg]\times I_\alpha [2(xyz)^
		\frac{1}{2}/(1-z)],\nonumber
	\end{eqnarray}

	\noindent
	has been included in Equation (\ref{Eq: V_eff_pre_reduced}) and (\ref{Eq: V_eff_pre_reduced_MBP}), and we achieve:
	
	\begin{eqnarray}
		V_{eff, \: \text{BE}}^{VT} (\lbrace \chi_m \rbrace, t, T_s) &=& \frac{1}{N_{\text{BE}}} \sum_k \omega_k^{-1} \epsilon_k (\lbrace \chi_m \rbrace, t) \sum_{q=1}^\infty \frac{z_k^{q/2}}{(1-
			z_k^q)} 
		\exp{[\rho_k (1+z_k^q)/(z_k^q-1)]} \nonumber\\
		&\times& \{S^{q+}_k(t, T_\text{s}) + S^{q-}_k(t, T_\text{s}) -I_0(t^q_k)\}, \hspace{2cm} \text{for BE case}
		\label{Eq: V_eff_reduced}
	\end{eqnarray}
	
	\noindent
	and
	\begin{eqnarray}
		V_{eff, \: \text{MB}}^{VT} (\lbrace \chi_m \rbrace, t, T_s)&=& \frac{1}{N_{\text{MB}}} \sum_k \omega_k^{-1} \epsilon_k (\lbrace \chi_m \rbrace, t) \frac{z_k^{1/2}}{(1-z_k)} 
		\exp{[\rho_k (1+z_k)/(z_k-1)]} \nonumber\\
		&\times& \{S^{+}_k(t, T_\text{s}) + S^{-}_k(t, T_\text{s}) -I_0(t_k)\}, \hspace{2cm} \text{for MB case}
		\label{Eq: V_eff_reduced_MBP}
	\end{eqnarray}
	
	\vspace{0.5cm}
	
	\noindent
	where the terms $S_k^{q \pm}(t, T_\text{s})$ can be represented in terms of a polynomial of the modified Bessel-function of the first kind ($I_m(t_k^q)$) as $S_k^{q \pm}(t, T_\text{s}) = \displaystyle{\sum_{m=0}^\infty} (z_k^q)^{\pm m/2} I_m(t_k^q)$; $t_k^q = 2 \rho_k (z_k^q)^{1/2}/(1-z_k^q)$. \\
	
	\noindent
	Employing $S_k^{q+}(t, T_\text{s}) + S_k^{q-}(t, T_\text{s}) - I_0(t_k^q) = \exp[\rho_k(1+z_k^q)/(1-z_k^q)]$, ($q$ = 1, 2, 3, ....), the
	expression in Equation (\ref{Eq: V_eff_reduced}) and (\ref{Eq: V_eff_reduced_MBP}) are further simplified. The final form of time- and temperature- dependent 6D EHP $V_{eff}^{VT} ( \lbrace \chi_m \rbrace, t, T_s)$ within linear VT coupling model becomes \protect\citen{SAMARK}:
	\begin{eqnarray}
		V_{eff, \: \text{BE}}^{VT} (\lbrace \chi_m \rbrace, t, T_s) = \frac{1}
		{N_{\text{BE}}} \sum_k \omega_k^{-1} \epsilon_k (\lbrace \chi_m \rbrace, t)  \sum^{\infty}_{q=1}\frac{z_k^{q/2}}{(1-z^q_k)}, \: \text{for BE case} 
		\label{Eq: V_eff_final}
	\end{eqnarray}
	
	\noindent
	and
	\begin{eqnarray}
		V_{eff, \: \text{MB}}^{VT} (\lbrace \chi_m \rbrace, t, T_s) = \frac{1}
		{N_{\text{MB}}} \sum_k \omega_k^{-1} \epsilon_k (\lbrace \chi_m \rbrace, t)  \frac{z_k^{1/2}}{(1- z_k)}, \: \text{for MB case} 
		\label{Eq: V_eff_final_MBP}
	\end{eqnarray}
	
	\noindent
	where the normalization factors of BE and MB probability distribution are expressed as $N_{\text{BE}}$ $= \sum_k \sum_{q=1}^{\infty} \frac{\left( z_k \right)^{\frac{q}{2}}}{\left( 1 - z_k^q \right)}$ and  $N_{\text{MB}}$ $= \sum_k \frac{z_k^{\frac{1}{2}}}{\left( 1 - z_k\right)}$, respectively.

\newpage

\bibliographystyle{aipnum4-1}
\bibliography{myrefs_H2, refs_new,refs_article}


\chapter{Operator algebra techniques involved in the formulation of EHP originating from bi-linear molecular DOFs-surface modes couplings} \label{app:CC}
	
	\setcounter{equation}{0}
	\renewcommand{\theequation}{E.\arabic{equation}}
	
	\vspace{0.2 cm} 
	
	\noindent
	While deriving the initial state dependence of surface vibrational mode in EHP emerging from bi-linear couplings (VV) between molecular DOFs and surface modes, one can employ:~\cite{VV-Hartree-2021} (i) matrix representation of VV evolution [$U_{VV} (\equiv R_{U_{VV}})$] using coupling [$b_k^{\dagger} b_{l} /b_k b_{l}^{\dagger}  (\equiv \delta_{k l})$] operators; (ii) transformations of bosonic creation ($b_k^{\dagger}$) and annhilation ($b_k$) operators under $U_{I} (=U_{VV} U_{VT})$ (i.e.,  $ U_{I}^{\dagger} b_k U_{I} \:  U_{I}^{\dagger} b_k^{\dagger} U_{I} $).  \\

	\noindent
	The TDSE corresponding to VV process can be represented in the following matrix representation:\\

	\begin{eqnarray}
		i \hbar \frac{d }{dt} R_{U_{VV}}(t, t_0) = V^{VV}_{I}(\lbrace \chi_m \rbrace, t) \: R_{U_{VV}} (t_0, t_0),
		\label{Eq: dRdt_Appendix_B}
	\end{eqnarray}
	
	\noindent
	where $R_{U_{VV}}(t, t_0)$ and $R_{U_{VV}}(t_0, t_0)$ represent the matrix representation of the VV evolution operator ($U_{VV}$) at time, $t$ and $t_0$, respectively. Since the interaction potential ($V^{VV}_{I}(\lbrace \chi_m \rbrace, t)$) does not depend on time explicitly, the integral form of $U_{VV}$ can be written as:  \\
	
	\begin{eqnarray}
		U_{VV} = \exp \left[ \Delta^{VV}(t) \right],
		\label{Eq: A2}
	\end{eqnarray}
	
	\noindent
	where $\Delta^{VV}(t) = \sum_{i} \Delta^{VV}_{i} (t)$ (see Equation (\ref{Eq: Delta(t)1})). \\

	\noindent
	On the other hand, in the matrix representation, the interaction potential ($V_I^{VV}(\lbrace \chi _m \rbrace,t)$) involving bi-linear (VV) coupling between  molecular DOFs and surface modes can be depicted as (see Equation (\ref{Eq: Interaction_potential_for_VV_process1})):
	
	\begin{eqnarray}
		V_I^{VV}(\lbrace \chi _m \rbrace,t) &=& \frac{1}{2}\sum_{k \neq l}^{M} \gamma_{k l} A_k A_l V_{k l, 2}(\lbrace \chi _m \rbrace) \left( b_k^{\dagger} b_{l} 
		F^{+}_{k l} + b_k b_{l}^{\dagger} F^{-}_{k l}\right), \nonumber \\
		&=& \sum_{k \neq l}  \alpha_{k l} O_{k l}  
		\label{Eq: Interaction_potential_for_VV_process}
	\end{eqnarray}

	\noindent
	where the elements $\lbrace \alpha_{k l}\rbrace$ are designated as:\\

	\begin{eqnarray}
		\alpha_{k l} &=& \frac{1}{2} \gamma_{k l} A_k A_l V_{kl, \:2}(\lbrace \chi _m \rbrace) F_{k l}^{+},  \: \: \text{for} \: \: (k > l) \nonumber\\
		\alpha_{k l} &=& \frac{1}{2} \gamma_{k l} A_k A_l V_{kl, \:2}(\lbrace \chi _m \rbrace) F_{k l}^{-}, \: \: \text{for} \: \: (k < l) \nonumber
	\end{eqnarray}
	
	\noindent
	where $F^{+}_{k l} = \exp [i (\theta_k - \theta_l) ]$ and $F_{k l}^{-} = \left( F^{+}_{k l} \right)^{*} $. The angle, 
	$\theta_{k/l} (= \int \omega_{k/l} dt)$, represents the phase originating from the modulated surface mode frequency. The VV operator is denoted as $O_{k l}$ $\left( \equiv b^{\dagger}_{k} b_{l}/ b_{k} b_{l}^{\dagger}\right)$, which turns out as $R_{O_{k l}} = \delta_{k l}^{M \times M}$ in the matrix representation. \\
	
	\noindent
	Consequently, $\Delta^{VV}(t)$ (see Equation (\ref{Eq: A2})) can be represented as the linear combination of VV operators: 
	
	\begin{eqnarray}
		\Delta^{VV}(t) = \sum_{k l}^{M} \alpha_{k l} (t) b_k^{\dagger} b_l.
		\label{Eq: Delta(t)1}
	\end{eqnarray}
	
	\noindent
	Therefore, the general expression of VV evolution operator ($U_{VV}$) in the matrix representation takes the following form:
	
	\begin{eqnarray}
		R_{U_{VV}} &=& \prod_{k l}^{M} \exp \left[ \alpha_{k l} (t, t_0) \: b_k^{\dagger} b_l \right],\nonumber \\
		&=& \prod_{k l}^{M} \exp \left[ \alpha_{k l} (t, t_0) \: \delta_{k l}^{M \times M} \right],
		\label{Eq: R_UVV_Appendix_B1}
	\end{eqnarray}
	
	\noindent
		where the ($M \times M$) matrix representation of bi-linear VV coupling operator ($b^{\dagger}_{k} b_{l}/b_{k} b_{l}^{\dagger}$) is designated as $\delta_{k l}^{M \times M}$ and its $\lbrace k,l\rbrace^\text{th}$ element will be one (1). In order to depict $R_{U_{VV}}$ with a finite series representation,  each member of $R_{U_{VV}}$ (see Equation (\ref{Eq: R_UVV_Appendix_B1})) may be represented as:~\cite{VV-Hartree-2021}\\

\begin{eqnarray}
	\exp(\alpha_{k l} \delta_{k l}^{M \times M} ) = I + g_{k l} \cdot \delta_{k l}^{M \times M},
	\label{Eq: member of U_VV Appendix B1}
\end{eqnarray}

\noindent
where the relationship between $\alpha_{k l}$ and $g_{k l}$ is presented in the Appendix \ref{app:CC_1} (see Equation~(\ref{Eq: member of U_VV Appendix B})).\\

\noindent
Including Equation (\ref{Eq: member of U_VV Appendix B1}) in Equation (\ref{Eq: R_UVV_Appendix_B1}), we achieve
the finite series representation of $R_{U_{VV}}$ as follows: \\

\begin{eqnarray}
R_{U_{VV}} &=& \prod_{k l}^{M} \left( I + g_{k l} \delta_{k l}^{M \times M} \right)  \nonumber \\
&=& \prod_{k}^{M} \left( I + g_{k k} \delta_{k k}^{M \times M} \right) \times \prod_{k}^{M} \prod_{l \neq k}^{M} \left( I + g_{l k} \delta_{l k}^{M \times M} \right).
\label{Eq: U_VV new representation in expansion Appendix B}
\end{eqnarray}

\noindent
Since the matrix representations of $\delta_{k l}^{M \times M}$s show the following properties of orthogonality,

\begin{subequations}
	\begin{align}
	&\delta_{k k}^{M \times M} \cdot \delta_{k^{\prime} k^{\prime}}^{M \times M} = 0,\\ 
	&\delta_{k k^{\prime}}^{M \times M} \cdot \delta_{j k^{\prime}}^{M \times M} = 0, \\
	&\delta_{k k^{\prime}}^{M \times M} \cdot \delta_{k^{\prime} l}^{M \times M} = \delta_{k l}^{M \times M},
\end{align}
	\label{Eq: Properties_of_delta_matrix}
\end{subequations}

\noindent
the $R_{U_{VV}}$ matrix (as shown in Equation~(\ref{Eq: U_VV new representation in expansion Appendix B})) acquires the following form:

\begin{eqnarray}
R_{U_{VV}} &=& \left( I + \sum_{k}^{M} g_{k k} \delta_{k k}^{M \times M}  \right) \times \prod_{k}^{M} \left( I + \sum_{l \neq k} g_{l k} \delta_{l k}^{M \times M}  \right).
\label{Eq: R_U_VV_final_matrix_form_Appendix_B}
\end{eqnarray} 

\noindent
Alternately, Equation~(\ref{Eq: R_U_VV_final_matrix_form_Appendix_B})  can be expressed as:

\begin{eqnarray}
	R_{U_{VV}} = I +T_{Q^{VV}},
	\label{Eq: Q_VV_coupling matrix1}
\end{eqnarray}  

\noindent
where the matrix elements of $T_{Q^{VV}}$ can be written in terms of $g_{lk}$. As per example, for $M=3$ (no. of surface vibrational modes), the $T_{Q^{VV}}$ matrix is expressed as (see Equation~(\ref{Eq: U_VV b_1 U_VV}) and (\ref{Eq: U_VV b_2 U_VV}) in Appendix \ref{app:CC_bi}):

\begin{eqnarray}
	T_{Q^{VV}} &=& \Big((I +  g_{1 1} \delta_{11}^{3 \times 3} +g_{2 2} \delta_{22}^{3 \times 3}+g_{33} \delta_{33}^{3 \times 3}) \times (I +  g_{2 1} \delta_{21}^{3 \times 3} +g_{31} \delta_{31}^{3 \times 3}) \nonumber \\
	&\times& (I +  g_{1 2} \delta_{12}^{3 \times 3} +g_{32} \delta_{32}^{3 \times 3}) \times (I +  g_{13} \delta_{13}^{3 \times 3} +g_{23} \delta_{23}^{3 \times 3}) \Big) - I, \nonumber
	\label{eq:general_m=3+matrix1}
\end{eqnarray}

\noindent
or 
\begin{eqnarray}
T_{Q^{VV}}=\begin{pmatrix}
	g_{11} & (1+g_{11}) g_{12}  & (1+g_{11}) (g_{13}+ g_{12}g_{23}) \\
	(1+g_{22}) g_{21} & g_{2 2} + \left( 1 +  g_{2 2} \right) g_{2 1} g_{1 2} & \left( 1 + g_{2 2}  \right) ( g_{2 3} + g_{2 1} g_{1 3} + g_{2 1} g_{1 2} g_{2 3} ) \\
	(1+g_{33}) g_{31}& (1+g_{33}) (g_{32} +g_{31}g_{12})& g_{33}+(1+g_{33})(g_{31}g_{13}+g_{32}g_{23}+g_{31}g_{12}g_{23})
\end{pmatrix}.
\label{eq:general_m=3+matrix}
\end{eqnarray}

\noindent
On the other hand, while performing the  unitary transformation of $U_{VV}$, the bosonic operators ($b_k$ and $b_k^{\dagger}$) acquire the following general forms (see Appendix \ref{app_CC_2}):~\cite{VV-Hartree-2021}

\begin{subequations}
	\begin{align}
		U_{VV}^{\dagger} b_k U_{VV} = b_k + \sum_{j=1}^{M} Q_{k j}^{VV} b_j,
	\end{align}
	\begin{align}
		U_{VV}^{\dagger} b_{k}^{\dagger} U_{VV} = b_{k}^{\dagger} + \sum_{j=1}^{M} Q_{k j}^{VV^{*}} b_j^{\dagger}.
	\end{align}
	\label{Eq: Evolution of b_k and b_k_dagger under VV unitary transformation1}
\end{subequations}

	\noindent
The transformations of bosonic operators ($b_k$ and $b_k^{\dagger}$) under $U_{VT}$ operator can be represented as (see Appendix \ref{app:CC_VT}):~\cite{VV-Hartree-2021}

\begin{subequations}
	\begin{align}
		U_{VT}^{\dagger} b_k U_{VT} = \left( b_k + i \xi_{k}^{+} \right),
	\end{align}
	\begin{align}
		U_{VT}^{\dagger} b_k^{\dagger} U_{VT} = \left( b_k^{\dagger} - i \xi_{k}^{-} \right).
	\end{align}
	\label{Eq: UVT_b_k_U_VT1}
\end{subequations}

\noindent
In other words, the bosonic operators ($b_k$ and $b_k^{\dagger}$) under the unitary transformation with $U_I = U_{VV} U_{VT}$ acquire the following form (using Equations (\ref{Eq: Evolution of b_k and b_k_dagger under VV unitary transformation1}a)-(\ref{Eq: Evolution of b_k and b_k_dagger under VV unitary transformation1}b) and Equations (\ref{Eq: UVT_b_k_U_VT1}a)-(\ref{Eq: UVT_b_k_U_VT1}b)):~\cite{VV-Hartree-2021}

\begin{eqnarray}
	U_{I}^{\dagger} b_k U_I &=& U_{VT}^{\dagger} \left( U_{VV}^{\dagger} b_k U_{VV} \right) U_{VT}, \nonumber \\
	&=& U_{VT}^{\dagger} \left( b_k + \sum_{j=1}^{M} Q_{kj}^{VV} b_j  \right) U_{VT}, \nonumber \\
	&=& \left( b_k + i \xi_{k}^{+} \right) + \sum_{j=1}^{M} Q_{kj}^{VV} \left( b_j + i \xi_{j}^{+} \right).
	\label{Eq: Evolution of boson operator I Appendix C}
\end{eqnarray}

\noindent
In a similar way, one can write,

\begin{eqnarray}
	U_{I}^{\dagger} b_k^{\dagger} U_I = \left( b_k^{\dagger} - i \xi_{k}^{-} \right) + \sum_{j=1}^{M} Q_{kj}^{VV^{*}} \left( b_j^{\dagger} - i \xi_{j}^{-} \right). \nonumber \\
	\label{Eq: Evolution of boson operator II Appendix C}
\end{eqnarray}

\noindent
In this perspective, the initial state dependence of surface vibrational mode in EHP ($\langle V^{VV}_I (\lbrace \chi _m\rbrace,t)\rangle_{\lbrace n_0 \rbrace}$) originating from the bi-linear (VV) coupling  between molecular DOFs and surface modes can be expressed in terms of initial surface mode eigenfunction ($\Phi_{vib} (t_0)$) with the aid of corresponding evolution operator $U_{I} (= U_{VV} U_{VT})$ as:

\begin{eqnarray}
	\langle V^{VV}_I (\lbrace \chi _m\rbrace,t)\rangle_{\lbrace n_0 \rbrace} &=& \langle \Phi_{vib} (t)\vert V_I^{VV}(\lbrace \chi _m\rbrace, t) \vert 
	\Phi_{vib} (t) \rangle \nonumber \\
	&=&  \langle \Phi_{vib} (t_0) \vert U_{VT}^{\dagger} \: U^{\dagger}_{VV} \: V_I^{VV}(\lbrace \chi _m\rbrace, t) \: U_{VV} \: U_{VT} \vert \Phi_{vib} (t_0) \rangle.
	\label{Eq: state_dependent_Hartree_potential_VV1}
\end{eqnarray}

\noindent
Hence, using $\vert \Phi_{vib} (t_0) \rangle = \prod_{k=1}^{M} \vert n_k^0 \rangle$ and the unitary transformation of bosonic operators, i.e., $U_{I}^{\dagger} b_k U_{I}$ and $U_{I}^{\dagger} b_k^{\dagger} U_{I}$ [Equations (\ref{Eq: Evolution of boson operator I Appendix C}) and (\ref{Eq: Evolution of boson operator II Appendix C})], the expression of initial state dependent Hartree potential for bi-linear (VV) coupling between molecular DOFs and surface mode interactions takes the following form:~\cite{VV-Hartree-2021}\\

\begin{footnotesize}
\begin{eqnarray}
	\langle V^{VV}_I (\lbrace \chi _m\rbrace,t)\rangle_{\lbrace n_0 \rbrace} &=& \langle \Phi_{vib} (t_0) \vert U_{VT}^{\dagger}  U^{\dagger}_{VV} V_I^{VV}(\lbrace \chi _m\rbrace, t) U_{VV} U_{VT} \vert \Phi_{vib} (t_0) \rangle \nonumber \\
	&=& \frac{1}{2} \sum_{k \neq l} \gamma_{k l} V_{k l, 2}(\lbrace \chi _m \rbrace)  A_k A_l \bigg[ F_{k l}^{+}  \prod_{k=1}^{M} \langle n_k^0  \vert \left( U_I^{\dagger} b_k^{\dagger} U_I \right) \left( U_I^{\dagger} b_l U_I \right) \prod_{k=1}^{M} \vert n_k^0 \rangle 
	+ F_{k l}^{-} \prod_{k=1}^{M} \langle n_k^0  \vert \left( U_I^{\dagger} b_k U_I \right) \left( U_I^{\dagger} b_l^{\dagger} U_I \right) \prod_{k=1}^{M} \vert n_k^0 \rangle \bigg] \nonumber \\
	&=& \frac{1}{2} \sum_{k \neq l} \gamma_{k l} V_{k l, 2}(\lbrace \chi _m \rbrace) A_k A_l \bigg[ F_{k l}^{+} \bigg( \delta_{l k} n^0_{l} + Q_{l k}^{VV} n^0_{k} +  Q_{k l}^{VV^{*}} n^0_{l} + \sum_j Q_{kj}^{VV^{*}} Q_{l j}^{VV} n_j^{0} 
	+ \xi_{k}^{-} \: \xi_{l}^{+} + \xi_{k}^{-} \sum_{j} Q_{l j}^{VV} \xi_{j}^{+} \nonumber\\ 
	&+&  \xi_{l}^{+} \sum_{j} Q_{kj}^{VV^{*}} \xi_{j}^{-} + \sum_j \sum_{j^{\prime}} Q_{kj}^{VV^{*}} Q_{lj^{\prime}}^{VV} \xi_{j}^{-} \: \xi_{j^{\prime}}^{+} \bigg)
	+ F_{k l}^{-} \bigg( (n^0_{l} + 1) \delta_{l k} + (n^0_{k} + 1) Q^{VV^{*}}_{l k}  + (n^0_{l} + 1)Q_{kl}^{VV} \nonumber \\ &+&  \sum_j Q_{kj}^{VV} Q^{VV^{*}}_{l j} (n_j^0+1)  
	+ \xi_{k}^{+} \: \xi_{l}^{-} + \xi_{k}^+ \sum_j Q^{VV^{*}}_{l j} \xi_{j}^{-}
	+ \xi_{l}^{-} \sum_j Q_{kj}^{VV} \xi_{j}^+ 
	+ \sum_j \sum_{j^{\prime}} Q_{kj}^{VV} Q_{lj^{\prime}}^{VV^{*}} \xi_{j}^{+} \: \xi_{j^{\prime}}^{-} \bigg) \bigg], 
	\label{Eq: state_dependent_Hartree_potential_VV_final_expression}
\end{eqnarray}
\end{footnotesize}

\noindent
where the following operations of transformed VV operators (e.g., $b_k^{\dagger} \sum_j Q_{lj}^{VV} b_j$) on the vibrational eigenstates $\prod_{k=1}^{M} \vert n_k^0 \rangle$ are used to derive above Equation (\ref{Eq: state_dependent_Hartree_potential_VV_final_expression}):

\begin{eqnarray}
	\prod_{k=1}^{M} \langle n_k^0 \vert b_k^{\dagger} b_l \prod_{k=1}^{M} \vert n_k^0 \rangle &=& n_l^0 \delta_{lk}, \nonumber \\
	\prod_{k=1}^{M} \langle n_k^0 \vert b_k^{\dagger} \sum_j Q_{lj}^{VV} b_j \prod_{k=1}^{M} \vert n_k^0 \rangle  &=& n_k^0 Q_{lk}^{VV}, \nonumber \\
	\prod_{k=1}^{M} \langle n_k^0 \vert \sum_j Q_{kj}^{VV^{*}} b_j^{\dagger} b_l \prod_{k=1}^{M} \vert n_k^0 \rangle &=& Q_{kl}^{VV^{*}} n_l^0. 
\end{eqnarray}

\noindent
The state dependent EHP constitutes with various time dependent quantities (i.e., $\xi_{k/l}^{\pm}$, $F_{k l}^{\pm}$ and $Q_{l k}^{VV}/Q_{k l}^{VV^{*}}$). Incorporation of the ensemble average by taking all possible initial state configuration $(\lbrace n_0 \rbrace)$ with the aid of BE or MB probability factor, the Equation~(\ref{Eq: state_dependent_Hartree_potential_VV_final_expression}) lead to the EHP, where the surface temperature is introduced parametrically (see Sec.~\ref{ssub:VV-EHP}).

\section{Expansion of VV interaction to demonstrate the closed operator relationship for M=3 vibrational modes.} \label{app:CC_1}

	\noindent
	One of the key advantages of VV coupling operators ($\lbrace b^{\dagger}_{k} b_{l}/ b_{k} b_{l}^{\dagger} \rbrace$) indicates that $M(M-1)$ number of such operators form a closed set with $M^2$ ones with respect to their commutation relationships, which can be depicted by considering a few mode expansion. The VV interaction potential is expressed for $M=3$ surface mode case,
	
	\begin{eqnarray}
		V^{VV}_{I} (\lbrace \chi _m\rbrace,t) &=& \sum_{k \neq l}^{M=3} \alpha_{k l} b^{\dagger}_{k} b_l \nonumber \\
		&=& \alpha_{12} b^{\dagger}_{1} b_{2} + \alpha_{13} b^{\dagger}_{1} b_{3} + \alpha_{23} b^{\dagger}_{2} 
		b_{3} + \alpha_{21} b_{1} b^{\dagger}_{2} + \alpha_{31} b_{1} b^{\dagger}_{3} + \alpha_{32} b_{2} b^{\dagger}_{3},
		\label{Eq: low mode expansion H_VV}
	\end{eqnarray}
	
	\noindent
	where the inter-commutation relationships between various VV coupling operators are written as:

	\begin{eqnarray}
		[b_{1}^{\dagger} b_2, b_1 b_{2}^{\dagger}] &=& b_{1}^{\dagger} b_2 b_1 b_{2}^{\dagger} - b_1 b_{2}
		^{\dagger} b_{1}^{\dagger} b_2  \nonumber \\
		&=& b_2 b_{2}^{\dagger} b_{1}^{\dagger} b_1 - b_{2}^{\dagger} b_2 b_1 b_{1}^{\dagger}  \nonumber \\
		&=& \left( b_2 b_2^{\dagger} - b_2^{\dagger} b_2 \right) b_{1}^{\dagger} b_1 - b_{2}^{\dagger} b_2 \left( 
		b_1 b_1^{\dagger} - b_1^{\dagger} b_1 \right) \nonumber \\
		&=& b_{1}^{\dagger} b_1 - b_{2}^{\dagger} b_2 
		\label{Eq: closed set 1}
	\end{eqnarray}
	
	\begin{eqnarray}
		[b_{1}^{\dagger} b_2, \left( b_{1}^{\dagger} b_1 - b_{2}^{\dagger} b_2 \right) ] &=& -b_1^{\dagger}b_2 
		\left( b_1 b_1^{\dagger} - b_1^{\dagger} b_1 \right) - b_1^{\dagger} \left( b_2 b_2^{\dagger} - 
		b_2^{\dagger} b_2 \right) b_2 \nonumber \\
		&=& -2 b_1^{\dagger} b_2
		\label{Eq: closed set 2}
	\end{eqnarray}
	
	\begin{eqnarray}
		[b_1^{\dagger} b_3, b_1 b_3^{\dagger} ] &=& b_1^{\dagger} b_1 [b_3, b_3^{\dagger}] - [b_1, 
		b_1^{\dagger}]b_3^{\dagger}b_3 \nonumber \\
		&=& b_1^{\dagger}b_1 - b_3^{\dagger} b_3
		\label{Eq: closed set 3}
	\end{eqnarray}
	
	\begin{eqnarray}
		[b_2^{\dagger} b_3, b_2 b_3^{\dagger}] &=& b_2^{\dagger} b_2 [b_3, b_3^{\dagger}] - [b_2, 
		b_2^{\dagger}]b_3^{\dagger} b_3 \nonumber \\
		&=& b_2^{\dagger} b_2 - b_3^{\dagger} b_3
		\label{Eq: closed set 4}
	\end{eqnarray}
	
	\noindent
	The above Equations (\ref{Eq: closed set 1})-(\ref{Eq: closed set 4}) indicate that six (6) operators 
	form a closed set with nine (9) relationship with respect to their commutations. Hence, the  
	$U_{VV}$ acquires the following form for $M=3$ number of surface mode case:\\

	\begin{eqnarray}
		U_{VV} &=& \exp(\alpha_{11} b_1^{\dagger} b_1) \exp(\alpha_{22} b_2^{\dagger} b_2) \exp(\alpha_{33} b_3^{\dagger} 
		b_3) \nonumber \\
		&\times& \exp(\alpha_{12} b_1^{\dagger} b_2 ) \exp(\alpha_{13} b_1^{\dagger} b_3 ) \exp(\alpha_{23} b_2^{\dagger} b_3 ) \nonumber \\
		&\times& \exp(\alpha_{21} b_1 b_2^{\dagger}) \exp(\alpha_{31} b_1 b_3^{\dagger}) \exp(\alpha_{32} b_2 b_3^{\dagger}) 
	\end{eqnarray}

	\noindent
	Therefore, the general expression of VV evolution operator ($U_{VV}$) in the matrix representation (see Equation (\ref{Eq: R_UVV_Appendix_B1}) in Appendix~\ref{app:CC}) becomes:~\cite{VV-Hartree-2021}
	
	\begin{eqnarray}
		R_{U_{VV}} = \prod_{k l}^{M} \exp \left[ \alpha_{k l} (t, t_0) \delta_{k l}^{M \times M} \right],
		\label{Eq: R_UVV_Appendix_B}
	\end{eqnarray}
	
	\noindent
	where the representation of bi-linear VV coupling operator ($b^{\dagger}_{k} b_{l}/b_{k} b_{l}^{\dagger}$) is defined by ($M \times M$) matrix as $\delta_{k l}^{M \times M}$. For $M=3$ number of surface mode, nine (9) possible combination of $\delta_{k l}^{M \times M}$ is written as:
	
	\begin{eqnarray}
	{b_1^{\dagger} b_2}\equiv\delta_{12}^{3 \times 3} = 
		\begin{pmatrix}
			0 & 1 & 0 \\
			0 & 0 & 0 \\
			0 & 0 & 0
		\end{pmatrix}; \: \:
		{b_1 b_2^{\dagger}}  \equiv \delta_{12}^{3 \times 3} = 
		\begin{pmatrix}
			0 & 0 & 0 \\
			1 & 0 & 0 \\
			0 & 0 & 0
		\end{pmatrix}; \: \:
		{b_1^{\dagger} b_1} \equiv\delta_{11}^{3 \times 3} = 
		\begin{pmatrix}
			1 & 0 & 0 \\
			0 & 0 & 0 \\
			0 & 0 & 0
		\end{pmatrix} \nonumber \\ \nonumber \\ 
		{b_1^{\dagger} b_3} \equiv \delta_{13}^{3 \times 3} = 
		\begin{pmatrix}
			0 & 0 & 1 \\
			0 & 0 & 0 \\
			0 & 0 & 0
		\end{pmatrix}; \: \:
		{b_1 b_3^{\dagger}} \equiv \delta_{13}^{3 \times 3} = 
		\begin{pmatrix}
			0 & 0 & 0 \\
			0 & 0 & 0 \\
			1 & 0 & 0
		\end{pmatrix}; \: \:
		{b_2^{\dagger} b_2} \equiv \delta_{13}^{3 \times 3} = 
		\begin{pmatrix}
			0 & 0 & 0 \\
			0 & 1 & 0 \\
			0 & 0 & 0
		\end{pmatrix} \nonumber \\ \nonumber \\ 
		{b_2	^{\dagger} b_3} \equiv \delta_{23}^{3 \times 3} = 
		\begin{pmatrix}
			0 & 0 & 0 \\
			0 & 0 & 1 \\
			0 & 0 & 0
		\end{pmatrix}; \: \:
		{b_2b_3^{\dagger}} \equiv \delta_{b_2	 b_3^{\dagger}}^{3 \times 3} = 
		\begin{pmatrix}
			0 & 0 & 0 \\
			0 & 0 & 0 \\
			0 & 1 & 0
		\end{pmatrix}; \: \:
		{b_3^{\dagger} b_3} \equiv \delta_{33}^{3 \times 3} = 
		\begin{pmatrix}
			0 & 0 & 0 \\
			0 & 0 & 0 \\
			0 & 0 & 1
		\end{pmatrix}. \nonumber \\
	\end{eqnarray}
	
	\noindent
	For $M=3$ number of surface modes, Equation (\ref{Eq: R_UVV_Appendix_B}) can be written as: 
	
	\begin{eqnarray}
		R_{U_{VV}} = \exp \left[ \alpha_{12} \begin{pmatrix}
			0 & 1 & 0 \\
			0 & 0 & 0 \\
			0 & 0 & 0
		\end{pmatrix}   \right] \exp \left[ \alpha_{21} \begin{pmatrix}
			0 & 0 & 0 \\
			1 & 0 & 0 \\
			0 & 0 & 0
		\end{pmatrix}  \right] ..... \exp \left[ \alpha_{33}  \begin{pmatrix}
			0 & 0 & 0 \\
			0 & 0 & 0 \\
			0 & 0 & 1
		\end{pmatrix}   \right].
	\end{eqnarray}

	\noindent
	For $(k \neq l)$, considering the matrix form of $b_1 b_2^{\dagger}$, one can write: 
	
	\begin{eqnarray}
		\exp \left[ \alpha_{21} \begin{pmatrix}
			0 & 0 & 0 \\
			1 & 0 & 0 \\
			0 & 0 & 0
		\end{pmatrix}   \right] &=&  \begin{pmatrix}
			1 & 0 & 0 \\
			0 & 1 & 0 \\
			0 & 0 & 1
		\end{pmatrix} + \alpha_{21} \begin{pmatrix}
			0 & 0 & 0 \\
			1 & 0 & 0 \\
			0 & 0 & 0
		\end{pmatrix} + \frac{\alpha_{21}^2}{2!} \begin{pmatrix}
			0 & 0 & 0 \\
			1 & 0 & 0 \\
			0 & 0 & 0
		\end{pmatrix} \begin{pmatrix}
			0 & 0 & 0 \\
			1 & 0 & 0 \\
			0 & 0 & 0
		\end{pmatrix} + .... \nonumber \\ \nonumber \\
		&=& \begin{pmatrix}
			1 & 0 & 0 \\
			0 & 1 & 0 \\
			0 & 0 & 1
		\end{pmatrix} + \alpha_{21} \begin{pmatrix}
			0 & 0 & 0 \\
			1 & 0 & 0 \\
			0 & 0 & 0
		\end{pmatrix}. \nonumber \\ \nonumber \\
	\text{or} \qquad	\exp \left[ \alpha_{21} \delta_{21}^{3 \times 3} \right] &=& I + \alpha_{21} \delta_{21}^{3 \times 3} 
		\label{Eq:Low_mode_Non-diagonal_coefficient_of_U_VV}
	\end{eqnarray}

	\noindent
	For diagonal quadratic operator ($b_3^{\dagger} b_3$), i.e., $(k = l)$, the following expression is achieved: 
	
	\begin{eqnarray}
		\exp \left[ \alpha_{33}  \begin{pmatrix}
			0 & 0 & 0 \\
			0 & 0 & 0 \\
			0 & 0 & 1
		\end{pmatrix}   \right] &=& \begin{pmatrix}
			1 & 0 & 0 \\
			0 & 1 & 0 \\
			0 & 0 & e^{\alpha_{33}}
		\end{pmatrix} =  \begin{pmatrix}
			1 & 0 & 0 \\
			0 & 1 & 0 \\
			0 & 0 & 1
		\end{pmatrix} + \begin{pmatrix}
			0 & 0 & 0 \\
			0 & 0 & 0 \\
			0 & 0 & e^{\alpha_{33}}-1
		\end{pmatrix} \nonumber\\  \nonumber\\
		\text{or} \qquad \exp \left[ \alpha_{33} \delta_{33}^{3 \times 3} \right] &=& I + (\alpha_{33}-1) \delta_{33}^{3 \times 3}
	\label{Eq:Low_mode_diagonal_coefficient_of_U_VV}
	\end{eqnarray}

    \noindent
	While blending Equation (\ref{Eq:Low_mode_Non-diagonal_coefficient_of_U_VV}) and Equation~(\ref{Eq:Low_mode_diagonal_coefficient_of_U_VV}), each exponential term in $R_{U_{VV}}$ (Equation (\ref{Eq: R_UVV_Appendix_B})) may be expressed as the following generalized form: \\

	\begin{eqnarray}
		\exp(\alpha_{k l} \delta_{k l}^{M \times M}) = I + g_{k l} \cdot \delta_{k l}^{M \times M},
		\label{Eq: member of U_VV Appendix B}
	\end{eqnarray}
	
	\noindent
	where the relationship between $\alpha_{k l}$ and $g_{k l}$ is depicted as:
	\vspace{-0.4 cm}
	\begin{eqnarray}
		g_{k l} &=& \alpha_{k l},  \: \: (k \neq l) \nonumber\\
		g_{k k} &=& \exp (\alpha_{k k}) - 1. \: \: (l = k)  \nonumber
		\label{Eq:conditions}
	\end{eqnarray}

\section{Bi-linear coupling (VV) unitary transformation of bosonic operators ($\boldsymbol{U_{VV}^{\dagger} b_k U_{VV}}$ or $\boldsymbol{U_{VV}^{\dagger} b_k^{\dagger} U_{VV}}$)} \label{app_CC_2}

		\noindent
		While computing such transformations, operations, $b_k$ and $b_k^{\dagger}$, on $U_{VV}$ are explored by using commutator relationships among various possible combinations of bosonic operators ($b_{k/l}$ and $b_{k/l}^{\dagger}$):~\cite{VV-Hartree-2021}

		\begin{eqnarray}
			b_i R_{U_{VV}} &=& b_i \prod_{k}^{M} \left( I + \sum_{l}^{M} g_{k l} \delta_{k l}^{M \times M}  \right)
			\label{Eq: b_i_on_R_U_VV_Appendix_C} 
		\end{eqnarray}
		
			\noindent
		where $\lbrace g_{l k} \rbrace$ designates the series expansion coefficients for VV operator $(b_k^{\dagger} b_l)$ basis and $\delta_{k l}^{M \times M}$ represents the matrix representation of such bi-linear operator (see Equations~(\ref{Eq: R_UVV_Appendix_B}) and (\ref{Eq: member of U_VV Appendix B})). Using Equations~(\ref{Eq: Act of b_i on diagonal part of U_VV Appendix C}) and (\ref{Eq: Act of b_i on off-diagonal part of U_VV Appendix C}) of Appendix \ref{app:CC_bi}, one can achieve the following expression:\\

		\begin{eqnarray}
			b_i R_{U_{VV}} &=& b_i \left( I + \sum_{k=i}^{M} g_{k k} \delta_{k k}^{M \times M}  \right) \times \prod_{k}^{M} \left( I + \sum_{l \neq k} g_{l k} \delta_{l k}^{M \times M}  \right) \nonumber \\
			&=& \left( I + \sum_{k=i}^{M} g_{k k} \delta_{k k}^{M \times M}  \right) \left( 1 + g_{i i} \right) b_i  \prod_{k}^{M} \left( I + \sum_{l \neq k} g_{l k} \delta_{l k}^{M \times M}  \right)  \nonumber \\
			&=& \left( I + \sum_{k=i}^{M} g_{k k} \delta_{k k}^{M \times M}  \right) \times \prod_{k}^{M} \left( I + \sum_{l \neq k} g_{l k} \delta_{l k}^{M \times M}  \right) \left( 1 + g_{i i} \right) 	\left( b_i + \delta_{il} g_{ik} b_k \left( 1 - \delta_{ik}  \right) \right) \nonumber \\
			&=& R_{U_{VV}} \left( 1 + g_{i i} \right) 	\left( b_i + \delta_{il} g_{ik} b_k \left( 1 - \delta_{ik}  \right) \right) \nonumber \\
			&=& R_{U_{VV}} \left(b_i + \sum_{k=1}^{M} Q_{i k}^{VV} b_k \right)
			\label{Eq: b_i_on_R_U_VV_Appendix_Cx} 
		\end{eqnarray}

		\noindent
		Hence, Equation (\ref{Eq: b_i_on_R_U_VV_Appendix_Cx}) clearly shows that $b_i$ operator acquires the following form under VV unitary transformation ($U_{VV}^{\dagger} b_i U_{VV}$).~\cite{VV-Hartree-2021}

		{\setlength{\jot}{1pt}
			\begin{subequations}
				\begin{align}
					U_{VV}^{\dagger} b_i U_{VV} = b_i + \sum_{j=1}^{M} Q_{i j}^{VV} b_j \\  
					U_{VV}^{\dagger} b_{i}^{\dagger} U_{VV} = b_{i}^{\dagger} + \sum_{j=1}^{M} Q_{i j}^{VV^{*}} b_j^{\dagger}.
				\end{align}
				\label{Eq: U_VV b_1 U_VV1}
			\end{subequations}
		}

\section{The  $b_i$ operator on the diagonal and off-diagonal elements of $R_{U_{VV}}$} \label{app:CC_bi}

\noindent
\textbf{(i) Operation of $\boldsymbol{b_i}$ on the diagonal elements of $\boldsymbol{R_{U_{VV}}}$}, \textbf{i.e.,} $\boldsymbol{b_i \prod_{k}^{M} \exp \left( \alpha_{k k} O_{k k} \right)}$\\

\noindent
The commutation relationship between $b_i$ and the diagonal elements of ${R_{U_{VV}}}$ takes the following form:~\cite{VV-Hartree-2021}

\begin{eqnarray}
	\left[ b_i, \exp \left({\alpha_{k k} O_{k k}}\right)\right] &=&  b_i \left[\sum_{n=0}^{\infty} \frac{\left( \alpha_{k k} \right)^{n} \left( b_k^{\dagger} b_k \right)^n}{n!}\right] -\left[ \sum_{n=0}^{\infty} \frac{\left( \alpha_{k k} \right)^{n} \left( b_k^{\dagger} b_k \right)^n}{n!}\right] b_i \nonumber \\ \nonumber \\
	&=& 0, \quad \text{for} \quad k \neq i \nonumber \\ \nonumber \\
	&=& \sum_{n=0}^{\infty} \frac{\left( \alpha_{i i} \right)^{n}}{n!} b_i \left( b_i^{\dagger} b_i \right)^n - \sum_{n=0}^{\infty} \frac{\left( \alpha_{i i} \right)^{n}}{n!} \left( b_i^{\dagger} b_i \right)^n b_i, \quad \text{for} \quad k = i \nonumber \\ \nonumber \\ 
	&=& \sum_{n=0}^{\infty} \frac{\left( \alpha_{i i} \right)^{n}}{n!} \left( 1 + b_i^{\dagger} b_i \right)^n b_i - \sum_{n=0}^{\infty} \frac{\left( \alpha_{i i} \right)^{n}}{n!} \left( b_i^{\dagger} b_i \right)^n b_i, \quad \text{since} \: \left[b_i, b^{\dagger}_i \right] = 1 \nonumber \\
	&=& \left[ \exp \left(\alpha_{i i} \left( 1 + b_i^{\dagger} b_i \right)\right) - \exp \left(\alpha_{i i} \left(b_i^{\dagger} b_i \right)\right)  \right] b_i \nonumber \\
	&=& \left[ \exp (\alpha_{i i}) - 1   \right] \exp \left( \alpha_{i i} \: b_i^{\dagger} b_i  \right) b_i ,\nonumber \\ 
	\left[ b_i, \exp \left({\alpha_{k k} O_{k k}}\right)\right]&=& g_{i i} \exp \left( \alpha_{i i} O_{i i}  \right) b_i, \quad \text{where} \quad g_{i i}= \exp (\alpha_{i i}) - 1 \nonumber \\ 
	\text{or} \quad b_i \exp \left( {\alpha_{ii} O_{ii}}\right) &=& \left( 1 + g_{i i} \right) \exp \left( {\alpha_{ii} O_{ii}}\right) b_i.  
	\label{Eq: Act of b_i on diagonal member of U_VV Appendix C}
\end{eqnarray}

Since $\exp(\alpha_{k l} \delta_{k l}^{M \times M}) = I + g_{k l} \cdot \delta_{k l}^{M \times M}$ (see Equation~(\ref{Eq: member of U_VV Appendix B})), Equation~(\ref{Eq: Act of b_i on diagonal member of U_VV Appendix C}) is converted into:

\begin{eqnarray}
	\quad b_i \left( I + g_{i i} \delta_{i i}^{M \times M} \right) &= & \left( I + g_{i i} \delta_{i i}^{M \times M} \right)\left( 1 + g_{i i} \right) b_i. 
	\label{Eq: Act of b_i on diagonal member of U_VV Appendix Cx}
\end{eqnarray}

\noindent
Thus, the action of $b_i$ on the diagonal elements of $R_{U_{VV}}$ can be written as:~\cite{VV-Hartree-2021} \\

\begin{eqnarray}
	b_i \prod_{k}^{M} \exp \left( \alpha_{k k} O_{k k} \right) &=& b_i 
	\prod_{k}^{M} \left( I + g_{k k} \delta_{k k}^{M \times M} \right) \nonumber \\
	&=& b_i \left( I + g_{i i} \delta_{i i}^{M \times M} \right) \prod_{k \neq i}^{M}   \left( I + g_{k k} \delta_{k k}^{M \times M} \right) \nonumber \\
	&=&   \left( I + g_{i i} \delta_{i i}^{M \times M} \right) \left( 1 + g_{i i} \right) b_i \prod_{k \neq i}^{M}   \left( I + g_{k k} \delta_{k k}^{M \times M} \right), \quad [\text{see Equations (\ref{Eq: Act of b_i on diagonal member of U_VV Appendix C}) and (\ref{Eq: Act of b_i on diagonal member of U_VV Appendix Cx}})] \nonumber \\
	&=& \bigg[ \prod_{k}^{M} \left( I + g_{k k} \delta_{k k}^{M \times M} \right)\bigg] \left( 1 + g_{i i} \right) b_i,\nonumber \\ 
	&=&  \left( I + \sum_{k} g_{k k} \delta_{k k}^{M \times M} \right)  \left( 1 + g_{i i} \right) b_i,
	\label{Eq: Act of b_i on diagonal part of U_VV Appendix C}
\end{eqnarray}

\noindent
where the orthogonality properties of $\delta_{k k}^{M \times M}$ is employed (see Equation (\ref{Eq: Properties_of_delta_matrix}a)). \\

\noindent
\textbf{(ii) Operation of $\boldsymbol{b_i}$ on the off-diagonal elements of $\boldsymbol{R_{U_{VV}}}$}, \textbf{i.e.,} $\boldsymbol{b_i \prod_{k \neq l}^{M} \exp \left( \alpha_{l k} O_{l k} \right)}$ \\

\noindent
The commutation relationship between $b_i$ operator and the off-diagonal elements of $R_{U_{VV}}$ is presented as:~\cite{VV-Hartree-2021}

\begin{subequations}
\begin{eqnarray}
	\left[ b_i, \exp \left ({\alpha_{lk} O_{l k}}\right )\right] &=& \sum_{n=0}^{\infty} \frac{(\alpha_{lk})^{n}}{n!} b_i \left( b_k b_l^{\dagger} \right)^n - \sum_{n=0}^{\infty} \frac{(\alpha_{lk})^{n}}{n!} \left( b_k b_l^{\dagger} \right)^n b_i \nonumber \\ \nonumber \\
	&=& 0, \quad \text{if} \quad i=k  \label{Eq: Act of b_i on off-diagonal member of U_VV Appendix Ca}\\ \nonumber \\
	&=& \sum_{n=0}^{\infty} \frac{(\alpha_{lk})^{n}}{n!} \left[ n \left( b_l^{\dagger} \right)^{n-1} \delta_{il} +  \left( b_l^{\dagger} \right)^n b_i \right] \left( b_k \right)^{n} 
	- \sum_{n=0}^{\infty} \frac{(\alpha_{lk})^{n}}{n!} \left( b_l^{\dagger} \right)^n b_i  \left( b_k \right)^{n}, \quad \text{if} \quad i\neq k \nonumber \\ \nonumber \\
	&=& \sum_{n=1}^{\infty} \left[\frac{(\alpha_{lk})^{n-1}}{(n-1)!}  \left( b_l^{\dagger} b_k \right)^{n-1} \right] \alpha_{lk} \delta_{il} b_k +  \sum_{n=0}^{\infty} \frac{(\alpha_{lk})^{n}}{n!}\left( b_l^{\dagger} \right)^n b_i \left( b_k \right)^{n} 
	- \sum_{n=0}^{\infty} \frac{(\alpha_{lk})^{n}}{n!} \left( b_l^{\dagger} \right)^n b_i  \left( b_k \right)^{n}  \nonumber \\ \nonumber \\
		&=&\sum_{m=0}^{\infty} \left[\frac{(\alpha_{lk})^{m}}{(m)!}  \left( b_l^{\dagger} b_k \right)^{m} \right] \alpha_{lk} \delta_{il} b_k ,\quad \text{where} \quad m=n-1   \nonumber \\ \nonumber \\
		&=& \exp \left( \alpha_{lk} \: b_l^{\dagger} b_k \right) \ g_{lk} \: \delta_{il} \: b_k     
	 \label{Eq: Act of b_i on off-diagonal member of U_VV Appendix Cb}
\end{eqnarray}
\end{subequations}
\noindent
where the general relationship of bosonic operators, $b_i \left( b_l^{\dagger} \right)^{n}=n \left( b_l^{\dagger} \right)^{n - 1 } \delta_{il}+\left( b_l^{\dagger} \right)^{n} b_i$ is using with Kronecker delta, $\delta_{il}$. \\

\noindent
Equations~(\ref{Eq: Act of b_i on off-diagonal member of U_VV Appendix Ca}) and (\ref{Eq: Act of b_i on off-diagonal member of U_VV Appendix Cb}) are combined leading to:\\

\begin{eqnarray}
\left[ b_i, \exp \left ({\alpha_{lk} O_{l k}}\right )\right] &=& \exp \left( \alpha_{lk} \: b_l^{\dagger} b_k \right) \ g_{lk} \: \delta_{il} \: b_k \left( 1 - \delta_{ik}  \right),    \nonumber \\ \nonumber \\ 
\text{or} \quad b_i \exp \left ({\alpha_{lk} O_{l k}}\right ) &=&  \exp \left ({\alpha_{lk} O_{l k}}\right ) b_i + \exp \left( \alpha_{lk} \: b_l^{\dagger} b_k \right) \ g_{lk} \: \delta_{il} \: b_k \left( 1 - \delta_{ik}  \right).  
\label{Eq: Act of b_i on off-diagonal member of U_VV Appendix C}
\end{eqnarray}

\noindent
Employing Equation~(\ref{Eq: member of U_VV Appendix B}) on Equation~(\ref{Eq: Act of b_i on off-diagonal member of U_VV Appendix C}), one can achieve the following generalized form:\\

\begin{align}
 \quad b_i \left[ I + g_{lk} \delta_{lk} \right] = \left[ I + g_{lk} \delta_{lk} \right]  \left( b_i + \delta_{il} g_{lk} b_k \left( 1 - \delta_{ik}  \right) \right). 
\label{Eq: Act of b_i on off-diagonal member of U_VV Appendix Cx}
\end{align}

\noindent
Therefore, the off-diagonal part of $R_{U_{VV}}$ $ \left(\prod_{k \neq l}^{M} \exp \left( \alpha_{lk} O_{lk} \right) \right)$ takes the following form while operating, $b_i$:

\begin{eqnarray}
	b_i \prod_{k \neq l}^{M} \exp \left( \alpha_{lk} O_{lk} \right) &=& b_i 
	\prod_{k \neq l}^{M}  \left( I + g_{lk} \delta_{lk}^{M \times M} \right) \nonumber \\
	&=& b_i \left( I + g_{ik} \delta_{ik} \right) \prod_{i \neq k \neq l}^{M}
	\left( I + g_{lk} \delta_{lk}^{M \times M} \right) \nonumber \\
	&=& \left( I + \sum_{l \neq k} g_{lk} \delta_{lk}^{M \times M} \right)
	\left( b_i + \delta_{il} g_{ik} b_k \left( 1 - \delta_{ik}  \right) \right).  \nonumber \\
	\label{Eq: Act of b_i on off-diagonal part of U_VV Appendix C}
\end{eqnarray}

\noindent
The new term ($g_{i k} b_k$) appears while applying bosonic operator on the $\prod_{k \neq l}^{M}  \left( I + g_{lk} \delta_{lk}^{M \times M} \right)$ (a member of the product in $R_{U_{VV}}$) and consequently, would further operate on the remaining terms in the product. \\

\noindent
In case of $M=3$, operation of $b_i$ on diagonal and non-diagonal elements of $R_{U_{VV}}$ is studied employing Equations (\ref{Eq: Act of b_i on diagonal part of U_VV Appendix C}) and (\ref{Eq: Act of b_i on off-diagonal part of U_VV Appendix C}), respectively as follows:

\begin{eqnarray}
	b_1 R_{U_{VV}} &=& \left( I + \sum_{k}^{M=3} g_{k k} \delta_{k k}^{M \times M} \right) (1 + g_{1 1} ) b_1 \prod_{k=1}^{M=3} \left(  I + \sum_{l \neq k}^{M=3} g_{l k} \delta_{l k}^{M \times M} \right) \nonumber \\
	&=& \left( I + \sum_{k}^{M=3} g_{k k} \delta_{k k}^{M \times M} \right)  \prod_{k=1}^{M=3} \left(  I + \sum_{l \neq k}^{M=3} g_{l k} \delta_{l k}^{M \times M} \right) (1 + g_{1 1} ) \left( b_1 + g_{1 2} b_2 + g_{1 3} b_3 + g_{1 2} g_{2 3} b_3 \right) \nonumber \\
	&=& R_{U_{VV}} (1 + g_{1 1} ) \left( b_1 + g_{1 2} b_2 + g_{1 3} b_3 + g_{1 2} g_{2 3} b_3 \right).
	\label{Eq: b_k U_VV for M=3}
\end{eqnarray}

\noindent
Therefore, it is evident from Equation (\ref{Eq: b_k U_VV for M=3}) that $b_1$ operator acquires the following form under VV unitary transformation ($U_{VV}^{\dagger} b_1 U_{VV}$):

\begin{eqnarray}
	U_{VV}^{\dagger} b_1 U_{VV} &=& (1 + g_{1 1} ) \left( b_1 + g_{1 2} b_2 + g_{1 3} b_3 + g_{1 2} g_{2 3} b_3 \right) \nonumber \\
	&=& b_1 + \sum_{j=1}^{M=3} Q_{1 j}^{VV} b_j,
	\label{Eq: U_VV b_1 U_VV}
\end{eqnarray} 

\noindent
where $Q_{1 1}^{VV} = g_{1 1}$, $Q_{1 2}^{VV} = \left( 1 + g_{1 1}   \right) g_{1 2}$ and $Q_{1 3}^{VV} = \left( 1 + g_{1 1}  \right) \left( g_{1 3} + g_{1 2} g_{2 3} \right)$.\\

\noindent
In a similar way, one can write,

\begin{eqnarray}
	U_{VV}^{\dagger} b_2 U_{VV} &=& \left( 1 + g_{2 2}  \right) \left( b_2 + g_{2 1} b_1 + g_{2 1} g_{1 2} b_2 + g_{2 1} g_{1 3} b_3 + g_{2 1} g_{1 2} g_{2 3} b_3 + g_{2 3} b_3  \right) \nonumber \\
	&=& b_2 + \sum_{j=1}^{M=3} Q_{2 j}^{VV} b_j, 
	\label{Eq: U_VV b_2 U_VV}
\end{eqnarray}

\noindent
where $Q_{2 1}^{VV} = \left( 1 + g_{2 2}  \right) g_{2 1}$, $Q_{2 2}^{VV} = g_{2 2} + \left( 1 +  g_{2 2} \right) g_{2 1} g_{1 2}$  and  $Q_{2 3}^{VV} = \left( 1 + g_{2 2}  \right) \left( g_{2 3} + g_{2 1} g_{1 3} + g_{2 1} g_{1 2} g_{2 3}  \right)$. \\

\noindent
Thus, the bosonic operators ($b_k$ and $b_k^{\dagger}$) are transformed into the following general forms (for $M$ no. of surface vibrational mode) while using the VV unitary transformation ($U_{VV}^{\dagger}b_kU_{VV}$ and $U_{VV}^{\dagger}b_k^{\dagger}U_{VV}$):~\cite{VV-Hartree-2021}

{\setlength{\jot}{1pt}
\begin{subequations}
	\begin{align}
		U_{VV}^{\dagger} b_k U_{VV} = b_k + \sum_{j=1}^{M} Q_{k j}^{VV} b_j \\  
        U_{VV}^{\dagger} b_{k}^{\dagger} U_{VV} = b_{k}^{\dagger} + \sum_{j=1}^{M} Q_{k j}^{VV^{*}} b_j^{\dagger}.
	\end{align}
	\label{Eq: Evolution of b_k and b_k_dagger under VV unitary transformation}
\end{subequations}
}
			
	\section{VT unitary transformation of bosonic operators ($\boldsymbol{U_{VT}^{\dagger} b_k U_{VT}}$ and $\boldsymbol{U_{VT}^{\dagger} b_k^{\dagger} U_{VT}}$)} \label{app:CC_VT}

	\noindent
In order to investigate the actions of bosonic annihilation ($b_k$) and creation ($b_k^{\dagger}$) operators on $U_{VT}$, the expression of VT evolution operator ($U_{VT}$) can be formulated from Magnus expansion (see Equations (\ref{Eq: U_VT in general1}), (\ref{Eq: alpha_plus_minus2}) and (\ref{Eq: beta2}) in Appendix \ref{app:BB}):
	
	\begin{eqnarray}
		U_{VT} = \prod_{k}^{M} U_{VT}^{(k)} = \prod_{k}^{M} \exp \left[ i \left( \zeta_{k} + \xi_{k}^+ \: b_k^{\dagger} +  \xi_{k}^- \: b_k\right) \right],
		\label{Eq: U_VT in general4} 
	\end{eqnarray}
	
	where
	\begin{eqnarray}
		\xi_{k}^\pm &=& -\frac{A_k}{\hbar} \int_{t_0}^t dt^{\prime} V_{k,1}(\lbrace \chi_m \rbrace) \exp[\pm i\omega_k t^{\prime}],
		\label{Eq: xi_plus_minus2}
	\end{eqnarray}
	
	\noindent
	and
	\begin{eqnarray}
		\zeta_{k}=\frac{i}{\hbar}\int_{t_0}^t dt^{\prime} V_{k,1}(\lbrace \chi_m \rbrace) \{ \exp [i\omega_k t^{\prime}]\xi_{k}^
		-(t^{\prime})-\exp[-i\omega_k t^{\prime}] \xi_{k}^+(t^{\prime})\}
		\label{Eq: zeta2}
	\end{eqnarray}
	
	\noindent
	represent the phase factors of LFHO amplitude of modified VT coupled vibrational mode.  \\
	
	\noindent
	In order to compute $U_{VT}^{\dagger} b_k U_{VT}$ or $U_{VT}^{\dagger} b_k^{\dagger} U_{VT}$, the commutation relationships among $b_k/b_k^{\dagger}$ and $\exp \left(  i \xi_{k}^+  \: b_k^{\dagger}\right)$ or $\exp \left(  i \xi_{k}^- \: b_k\right)$ are employed to validate the following relationship:
	
	\begin{eqnarray}
			b_k \exp \left( i \xi_{k}^{+} \: b_k^{\dagger} \right) = \exp \left( i \xi_{k}^{+} \: b_k^{\dagger} \right) \left( \exp \left(- i \xi_{k}^{+} \: b_k^{\dagger} \right)b_k \exp \left( i \xi_{k}^{+} \: b_k^{\dagger} \right) \right)
		\label{Eq: b_k exp(i alpha b_k11)}
	\end{eqnarray}

\noindent
While applying Baker-Hausdorff formula, one can express:

\begin{eqnarray}
&b_k& \exp \left( i \xi_{k}^{+} \: b_k^{\dagger} \right) = \exp \left( i \xi_{k}^{+} \: b_k^{\dagger} \right) \left( b_k+ [b_k, i \xi_{k}^{+} b_k^{\dagger}] \right) \nonumber\\
\text{or} \quad &b_k& \exp \left( i \xi_{k}^{+} \: b_k^{\dagger} \right) = \exp \left( i \xi_{k}^{+} \: b_k^{\dagger} \right) \left( b_k + i \xi_{k}^{+}\right). \quad \text{since} \:\:  [b_k, i \xi_{k}^{+} b_k^{\dagger}] = i \xi_{k}^{+}
\label{Eq: b_k exp(i alpha b_k12)}
\end{eqnarray}

\noindent
In a similar way, the following expression can be written,	

	\begin{eqnarray}
		b_k^{\dagger} \exp \left( i \xi_{k}^{-} \: b_k \right) = \exp \left( i \xi_{k}^{-} \: b_k \right) \left( b_k^{\dagger} - i \xi_{k}^{-}\right).
		\label{Eq: b_k exp(i alpha b_k)}
	\end{eqnarray}

	\noindent
	Employing Equations (\ref{Eq: b_k exp(i alpha b_k12)}) and (\ref{Eq: b_k exp(i alpha b_k)}), the bosonic operators ($b_k$ and $b_k^{\dagger}$) under $U_{VT}$ operator are converted into:~\cite{VV-Hartree-2021}
		
	\begin{subequations}
		\begin{align}
			U_{VT}^{\dagger} b_k U_{VT} = \left( b_k + i \xi_{k}^{+} \right)
		\end{align}
		\begin{align}
			U_{VT}^{\dagger} b_k^{\dagger} U_{VT} = \left( b_k^{\dagger} - i \xi_{k}^{-} \right).
		\end{align}
		\label{Eq: UVT_b_k_U_VT}
	\end{subequations}

\newpage

\bibliographystyle{aipnum4-1}
\bibliography{myrefs_H2, refs_new,refs_article}


\chapter{Matrix element for electronic transitions using the 2D Bloch model}\label{app-G}

\setcounter{equation}{0}
\renewcommand{\theequation}{F.\arabic{equation}}

\noindent
The interaction potential ($F$) may be expressed by a Fermi--Thomas screened Coulomb form (see Equation~(\ref{eq:ab})) as follows:

\begin{eqnarray}
	F=-\frac{e^2\delta(Z,r)}{R}\exp(-k_cR),
	\label{eq:ab1}
\end{eqnarray}

\noindent
where ($R=\sqrt{\tilde{X}^2+\tilde{Y}^2+Z^2}$) designates the distance between electron and the effective excess charge ($e\delta(Z,r)$) corresponding to the gas molecule. Here, ($Z$) represents the distance between the molecular center of mass and the surface, whereas ($r$) designates the diatomic bond length. On the other hand, the quantity, $k_c$, indicates the screening wave vector.\\

\noindent
While involving the in-plane coordinates of the electron in the metal surface as ($X,Y$), the interaction potential can be expressed in terms of a two-dimensional Fourier series~\cite{BILEH2,GD1} as follows:

\begin{eqnarray}
	F=-\frac{e^2\delta(Z,r)}{\sqrt{\tilde{X}^2+\tilde{Y}^2+Z^2}}
	\exp\left(-k_c\sqrt{\tilde{X}^2+\tilde{Y}^2+Z^2}\right)
	=\sum_m\sum_h A_{hm}\exp(i  \tilde{X}+i  \tilde{Y}),
	\label{eq:ab2}
\end{eqnarray}

\noindent
where the dimensionless coordinates are denoted as ($\tilde{X}=m/d_{2D}$) and ($\tilde{Y}=h/d_{2D}$), with $d_{2D}$ being the lattice constant.\\

\noindent
The Fourier coefficients ($A_{hm}$) are computed by using the integral representation of the Kronecker delta function,\\

\begin{eqnarray}
	\delta_{hm}=\frac{1}{2\pi}\int_{-\infty}^{\infty}dX \exp\left(i(m-h)X\right),
	\label{eq:ab3}
\end{eqnarray}

\noindent
which results into the following expression of Fourier amplitudes~\cite{BILEH2},

\begin{eqnarray}
	A_{hm}=\frac{e^2\delta(Z,r)} {2\pi d_{2D}} \sum_{m=-\infty}^{+\infty} \sum_{h=-\infty}^{+\infty}
	\frac{\exp\left(-Z\sqrt{k_c^2+(h/d_{2D})^2+(m/d_{2D})^2}\right)}{\sqrt{h^2+m^2+(d_{2D}k_c)^2}}.
	\label{eq:ab4}
\end{eqnarray}

\noindent
Such coefficients can be subsequently employed to compute the matrix elements over the 2D Bloch states used as single-particle electronic wavefunctions.

\begin{table}[!htp]
	\begin{scriptsize}
		\begin{center}
			\caption{\label{tb:matrix}Matrix element for electronic transitions~\cite{BILEH2} is presented.}
			\vspace{0.2 cm}
			\begin{tabular}{lr}
				\hline
				Matrix Element  & Selection Rule \\
				\hline 
				$C_{k-b_1}C_{k^{\prime} -b_1} + C_{k-b_2} C_{k^{\prime} -b_2} + C_k C_{k^{\prime}} 
				+ C_{k+b_2}C_{k^{\prime} + b_2} + C_{k+b_1}c_{k^{\prime} +b_1}$ & $k^{\prime} = k-(m/d_{2D},n/d_{2D})$ \\ 
				$C_{k-b_2}C_{k^{\prime}} + C_k C_{k^{\prime} + b_2}$            & $k^{\prime} = k-b_2-(m/d_{2D},n/d_{2D})$ \\ 
				$C_{k+b_2}C_{k^{\prime}} + C_k C_{k^{\prime} - b_2}$            & $k^{\prime} = k+b_2-(m/d_{2D},n/d_{2D})$ \\ 
				$C_{k-b_1}C_{k^{\prime}} + C_k C_{k^{\prime} + b_1}$            & $k^{\prime} = k-b_1-(m/d_{2D},n/d_{2D})$ \\ 
				$C_{k+b_1}C_{k^{\prime}} + C_k C_{k^{\prime} - b_1}$            & $k^{\prime} = k+b_1-(m/d_{2D},n/d_{2D})$ \\ 
				$C_{k-b_2}C_{k^{\prime} + b_2}$                                 & $k^{\prime} = k-2b_2-(m/d_{2D},n/d_{2D})$ \\ 
				$C_{k+b_2}C_{k^{\prime} - b_2}$       & $k^{\prime} = k+2b_2-(m/d_{2D},n/d_{2D})$ \\ 
				$C_{k-b_2}C_{k^\prime -b_1} +C_{k+b_1}C_{k^\prime + b_2}$       & $k^{\prime} = k+b_1-b_2-(m/d_{2D},n/d_{2D})$ \\ 
				$C_{k-b_2}C_{k^\prime +b_1} +C_{k-b_1}C_{k^\prime + b_2}$       & $k^{\prime} = k-b_1-b_2-(m/d_{2D},n/d_{2D})$ \\ 
				$C_{k+b_2}C_{k^\prime -b_1} +C_{k+b_1}C_{k^\prime - b_2}$       & $k^{\prime} = k+b_1+b_2-(m/d_{2D},n/d_{2D})$ \\ 
				$C_{k+b_2}C_{k^\prime +b_1} +C_{k-b_1}C_{k^\prime - b_2}$       & $k^{\prime} = k+b_2-b_1-(m/d_{2D},n/d_{2D})$ \\ 
				$C_{k-b_1}C_{k^\prime + b_1}$                                   & $k^{\prime} = k-2b_1-(m/d_{2D},n/d_{2D})$ \\ 
				$C_{k+b_1}C_{k^\prime - b_1}$                                   & $k^{\prime} = k+2b_1-(m/d_{2D},n/d_{2D})$ \\ 
				\hline 
			\end{tabular} 
		\end{center}
		\label{tb:selec}
	\end{scriptsize}
\end{table}

}	
	
\newpage

\section{The effective potential for elhp coupling}\label{app-DD}

\noindent
In order to account the contribution of elhp excitation on the molecule, the effective potential can be formulated 
within mean-field approach:~\cite{SM-MOLPHYS,SM-JTCC}
\begin{eqnarray}
	V_{eff}^{elhp}(r,Z,t)=\langle \Phi_{elhp}(t)\mid H_e\mid \Phi_{elhp}\rangle .
	\label{eq:C1}
\end{eqnarray}
While including Equation (\ref{eq:ag}) in $V_{eff}^{elhp}$, one can obtain,
\begin{eqnarray}
	V_{eff}^{elhp}(r,Z,t)=\sum_i \epsilon_i\langle \Phi_{elhp}(t)\mid c_i^\dagger c_i\mid \Phi_{elhp}(t)\rangle + \sum_{ij}H_{ij}\langle \Phi_{elhp}(t)\mid c_i^\dagger c_j\mid \Phi_{elhp}(t)\rangle .
	\label{eq:C2}
\end{eqnarray}
On the other hand, the second term of the Equation (\ref{eq:C2}) can be written as:
\begin{eqnarray}
	\sum_{ij}H_{ij}\langle \Phi_{elhp}\mid c_i^\dagger c_j\mid \Phi_{elhp}(t)\rangle  = \sum_{ij}H_{ij}\langle U_e(t,t_0)\Phi_{elhp}(t_0)\mid c_i^\dagger c_j\mid U_e(t,t_0)\Phi_{elhp}(t_0)\rangle,  \nonumber \\
	=\sum_{ij}H_{ij}\langle \Phi_{elhp}(t_0)\mid U_e^\dagger(t,t_0)c_i^\dagger c_jU_e(t,t_0)\mid\Phi_{elhp}(t_0)\rangle,  \nonumber \\
	=\sum_{ij}H_{ij}\langle \Phi_{elhp}(t_0)\mid U_e^\dagger(t,t_0)c_i^\dagger U_e(t,t_0)U_e^\dagger(t,t_0)  c_jU_e(t,t_0)\mid\Phi_{elhp}(t_0)\rangle .
	\label{eq:C3}
\end{eqnarray}
Similar to the bosonic operators, fermionic operators also fulfill the following identity (see Equation~(\ref{Eq: Evolution of b_k and b_k_dagger under VV unitary transformation}a) and (\ref{Eq: Evolution of b_k and b_k_dagger under VV unitary transformation}b) in Appendix \ref{app:CC_bi})~\cite{SM-MOLPHYS,SM-JTCC}
\begin{eqnarray}
	U_e^\dagger(t,t_0)c_i^\dagger U_e(t,t_0)=c_i^\dagger + \sum_k Q_{ik}^\star c_k^\dagger,
	\label{eq:C4}
\end{eqnarray}
and
\begin{eqnarray}
	U_e^\dagger(t,t_0)c_j U_e(t,t_0)=c_j+\sum_{k^\prime} Q_{jk^\prime}^\star c_{k^\prime},
	\label{eq:C5}
\end{eqnarray}

\noindent
where Equations (\ref{eq:am}) and (\ref{eq:an}) denote $Q$ matrix elements.\\

\noindent
 While substituting Equations (\ref{eq:C4}) and (\ref{eq:C5}) in (\ref{eq:C3}), it appears:\\

\begin{small}
	\begin{eqnarray}
		&&\sum_{ij}H_{ij}\langle \Phi_{elhp}(t)\mid c_i^\dagger c_j\mid \Phi_{elhp}(t)\rangle =\sum_{ij}H_{ij}\langle \Phi_{elhp}(t_0)\mid \left(c_i^\dagger + \sum_k Q_{ik}^\star c_k^\dagger\right) \left(c_j+\sum_{k^\prime} Q_{jk^\prime}^\star c_{k^\prime}\right)\mid \Phi_{elhp}(t_0)\rangle \nonumber \\
		&&=\sum_{ij}H_{ij}\langle \Phi_{elhp}(t_0)\mid \left(c_i^\dagger c_j+\sum_{k^\prime}Q_{jk^\prime}c_i^\dagger c_{k^\prime}+\sum_k Q_{ik}^\star c_k^\dagger c_j + \sum_{kk^\prime}Q_{ik}^\star Q_{jk^\prime}c_k^\dagger c_{k^\prime} \right)\mid \Phi_{elhp}(t_0)\rangle \nonumber \\
		&&= \sum_{ij}H_{ij}\langle \Phi_{elhp}(t_0)\mid c_i^\dagger c_j\mid \Phi_{elhp}(t_0)\rangle +\sum_{ij}H_{ij}\sum_{k^\prime}Q_{jk^\prime}\langle \Phi_{elhp}(t_0)\mid c_i^\dagger c_{k^\prime}\mid \Phi_{elhp}(t_0)\rangle  \nonumber \\
		&&+\sum_{ij}H_{ij}\sum_k Q_{ik}^\star \langle \Phi_{elhp}(t_0)\mid c_k^\dagger c_j\mid \Phi_{elhp}(t_0) +\sum_{ij}H_{ij}\sum_{kk^\prime}Q_{ik}^\star Q_{jk^\prime}\langle \Phi_{elhp}(t_0)\mid c_k^\dagger c_{k^\prime}\mid \Phi_{elhp}(t_0).
		\label{eq:C6}
	\end{eqnarray}
\end{small}
Employing
\begin{eqnarray}
	\langle \Phi_{elhp}(t_0)\mid c_i^\dagger c_j\mid \Phi_{elhp}(t_0)\rangle=0, & i\neq j
	\label{eq:C7}
\end{eqnarray}
and
\begin{eqnarray}
	\langle \Phi_{elhp}(t_0)\mid c_i^\dagger c_i\mid \Phi_{elhp}(t_0)\rangle=n_i,
	\label{eq:C8}
\end{eqnarray}
one can obtain,
\begin{eqnarray}
	&\sum_{ij}H_{ij}\langle \Phi_{elhp}(t)\mid c_i^\dagger c_j\mid \Phi_{elhp}(t)\rangle = &\sum_i H_{ii}n_i+\sum_{ij}H_{ij}Q_{ji}n_i  
	+\sum_{ij}H_{ij}Q_{ij}^\star n_j \nonumber \\
	&&+\sum_{ij}H_{ij}\sum_k Q_{ik}^\star Q_{jk}n_k.
	\label{eq:C9}
\end{eqnarray}
The first term of the $V_{eff}^{elhp}(r,Z,t)$ is written as,
\begin{eqnarray}
	\sum_i\epsilon_i\langle \Phi_{elhp}(t)\mid c_i^\dagger c_i\mid \Phi_{elhp}(t)\rangle=\sum_i \epsilon_i n_i .
	\label{eq:C10}
\end{eqnarray}
Thus, the effective Hamiltonian for elhp coupling takes the following form~\cite{SM-MOLPHYS,SM-JTCC}:
\begin{eqnarray}
	V_{eff}^{elhp}(r,Z,t)&=&\sum_i \epsilon_i n_i+\sum_i H_{ii}n_i+\sum_{ij}H_{ij}\left( Q_{ij}^\star n_j + Q_{ji}n_i\right)\nonumber \\
	&+& \sum_{ij}\sum_k Q_{ik}^\star Q_{jk}n_kH_{ij}. 
	\label{eq:C11}
\end{eqnarray}

\newpage

\bibliographystyle{aipnum4-1}
\bibliography{myrefs_H2, refs_new,refs_article}


\chapter{Initialization and projection} \label{app:JJ}

\setcounter{equation}{0}
\renewcommand{\theequation}{G.\arabic{equation}}

\noindent
 The translational motion of the COM of the incoming gas molecule can be expressed by a Gaussian wavepacket ($\Phi_{\text{GWP}} \left( Z, t_0 \right)$) along $Z$ coordinate with an initial momentum, $p_0 \: (=\hbar k_0; k_0 $ is the wavevector) as presented below (see Ref. \citenum{SMIRPC}):

\begin{eqnarray}
	\Phi_{\text{GWP}} \left( Z, t_0 \right) &=& \left[ 2 \pi \Delta^2_Z \left( 1 + \frac{\hbar^2 t^2_0}{4 M^2 \Delta^4_Z} \right)\right] 
	\times \exp \left( - \frac{(Z-Z_{\text{foc}} -\hbar k_0 t_0/M)^2}{4 \Delta^2_Z - 2 i \hbar t_0/M} -i k_0 Z \right).
	\label{GWP_Z}
\end{eqnarray}

\noindent
The position of the center of the wavepacket is symbolized as $Z_0$, where $t_0$ represents the time at which focal point of the GWP remains around $Z_{\text{foc}}$ with a width of $\Delta_Z (=\frac{1}{2} \sqrt{\hbar/\text{Im} A_Z})$. The motion of molecule on the $X-Y$ plane is depicted by $\Phi_{\text{GWP}} \left( X, t_0 \right)$ and $\Phi_{\text{GWP}} \left( Y, t_0 \right)$ along $X$ and $Y$ coordinates without any momentum component. \\

\noindent
The 6D initial wavefunction (Equation (\ref{eq:initial_wavefunction})) can be written as functions of TDDVR basis function as follows~ \cite{SMIRPC}:

\begin{eqnarray}
	\Psi_{m}\left(x, y, z, X, Y, Z, t_0  \right) &=& \sum_{IJKLMN} C_{IJKLMN} (t_0)\: \psi_{I} (x)\: \psi_{J} (y)\: \psi_{K} (z)\: \psi_{L} (X)\: \psi_{M} (Y)\: \psi_{N} (Z) \nonumber \\
	&=&  \frac{1}{r} g_v \left(r \right) Y_{j, m_j} \left( \theta,\phi \right) \Phi_{\text{GWP}} \left( X, t_0 \right) \Phi_{\text{GWP}} \left( Y, t_0 \right) 
	\Phi_{\text{GWP}} \left( Z, t_0 \right), \nonumber \\
	\label{TDDVR basis function}
\end{eqnarray}

\noindent
where Cartesian coordinates ($x$, $y$ and $z$) are connected with polar ones (r, $\theta$ and $\phi$) with the help of standard transformation. While depicting the amplitude of the initial wavefunction only at TDDVR grid points, the function takes the following form due to the orthogonality relationship of the TDDVR basis functions as computed at the grid points: 

\begin{eqnarray}
	\Psi_{m}\left(x_I, y_J, z_K, X_L, Y_M, Z_N, t_0  \right) &=& \frac{C_{IJKLMN} (t_0)}{\text{Im}A}  A_{II}^x   A_{JJ}^y A_{KK}^z A_{LL}^X A_{MM}^Y A_{NN}^Z \nonumber \\
	&=&  \frac{g_v \left(r_{IJK} \right)}{r_{IJK}}  Y_{j, m_j} \left( \theta_{IJK},\phi_{IJK} \right) \Phi_{\text{GWP}} \left( X_L, t_0 \right) \nonumber \\
	&\times& \Phi_{\text{GWP}} \left( Y_M, t_0 \right) \Phi_{\text{GWP}} \left( Z_N, t_0 \right),  \nonumber \\
\end{eqnarray}

\noindent
from which the initial amplitude of the wavefunction can be expressed with the help of DVR amplitudes as follows:

\begin{eqnarray}
	d_{IJKLMN} (t_0) &=& \text{Im} A \: \frac{\frac{1}{r_{IJK}} g_v \left(r_{IJK} \right) Y_{j, m_j} \left( \theta_{IJK},\phi_{IJK} \right)}{\sqrt{A_{II}^x   A_{JJ}^y A_{KK}^z A_{LL}^X A_{MM}^Y A_{NN}^Z}} \nonumber \\ 
	&\times& \Phi_{\text{GWP}} \left( X_L, t_0 \right) \Phi_{\text{GWP}} \left( Y_M, t_0 \right) \Phi_{\text{GWP}} \left( Z_N, t_0 \right),
\end{eqnarray}

\noindent
where Im$A$ $=$ $(\hbar/2)^{3/2} (\text{Im} A_x \: \text{Im} A_y \: \text{Im} A_z \: \text{Im} A_X \: \text{Im} A_Y \: \text{Im} A_Z)^{-1/4}$. \\

\noindent
In the asymptotic limit, the scattered wave function remains far away from the interaction region, and therefore, the energy resolved elastic/inelastic state to state transition probabilities can be calculated from the ratio of the outgoing ($C^+_{k_{out},v^{\prime}j^{\prime} m_{j^{\prime}}}$) and the incoming ($C^-_{k_{in},vjm_j}$) fluxes (see Ref. \cite{SMIRPC}):

\begin{eqnarray}
	P_{vjm_j}^{v^{\prime}j^{\prime}m_{j^{\prime}}}(E) = \frac{ \frac{k_{out}}{M}|C^+_{k_{out}, v^{\prime}j^{\prime}
			m_{j^{\prime}}}|^{2}}{\frac{k_{in}}{M}|C^-_{k_{in},vjm_j}|^{2}}, 
	\label{eq:transprob1}
\end{eqnarray}

\noindent
where

\begin{subequations}
	\begin{eqnarray}
		C^+_{{k_{out}}_{,v^{\prime}j^{\prime}m_{j^{\prime}}}} &=& \frac{1}{({2\pi})^{3/2}} ImA \sum_{IJKLMN} 
		\left(\frac{d_{IJKLMN}} {\sqrt{A^x_{II}A^y_{JJ}A^z_{KK}A^X_{LL}A^Y_{MM}A^Z_{NN}}} \right)\nonumber\\ 
		&\times& \frac{1}{r_{IJK}} g_{v^{\prime}}^{\ast}(\displaystyle {r}_{IJK}) Y^{\ast}_{j^{\prime}m_{j^{\prime}}}
		(\displaystyle {\theta}_{IJK}, \displaystyle {\phi}_{IJK}) \exp(-ik_{out}Z_N) \exp\bigg\{\frac{i}{\hbar} 
		[P_{Z_c} (Z_N-Z_c)]\bigg\},\nonumber\\ \ \\
		C^-_{{k_{in},}_{vjm_j}} &=& \bigg(\frac{2}{\pi}\bigg)^{1/4}\bigg(\frac{1}{2\pi}\bigg) ImA \sqrt{\Delta_Z}
		\exp[i(k_{in}-k_{0})Z_c-\Delta_Z^2(k_{in}-k_0)^2]\nonumber\\
		&\times& \frac{ \frac{1}{r_{IJK}} g_{v}(\displaystyle {r}_{IJK}) Y_{jm_j} (\displaystyle {\theta}_{IJK}, 
			\displaystyle {\phi}_{IJK})}
		{\sqrt{A^x_{II}A^y_{JJ}A^z_{KK}A^X_{LL}A^Y_{MM}}} \Phi_{GWP}(X,t_{0})\Phi_{GWP}(Y,t_{0}).
	\end{eqnarray}
\end{subequations}

\newpage

\bibliographystyle{aipnum4-1}
\bibliography{myrefs_H2, refs_new,refs_article}

\begin{thebibliography}{3}%
\makeatletter
\providecommand \@ifxundefined [1]{%
 \@ifx{#1\undefined}
}%
\providecommand \@ifnum [1]{%
 \ifnum #1\expandafter \@firstoftwo
 \else \expandafter \@secondoftwo
 \fi
}%
\providecommand \@ifx [1]{%
 \ifx #1\expandafter \@firstoftwo
 \else \expandafter \@secondoftwo
 \fi
}%
\providecommand \natexlab [1]{#1}%
\providecommand \enquote  [1]{``#1''}%
\providecommand \bibnamefont  [1]{#1}%
\providecommand \bibfnamefont [1]{#1}%
\providecommand \citenamefont [1]{#1}%
\providecommand \href@noop [0]{\@secondoftwo}%
\providecommand \href [0]{\begingroup \@sanitize@url \@href}%
\providecommand \@href[1]{\@@startlink{#1}\@@href}%
\providecommand \@@href[1]{\endgroup#1\@@endlink}%
\providecommand \@sanitize@url [0]{\catcode `\\12\catcode `\$12\catcode
  `\&12\catcode `\#12\catcode `\^12\catcode `\_12\catcode `\%12\relax}%
\providecommand \@@startlink[1]{}%
\providecommand \@@endlink[0]{}%
\providecommand \url  [0]{\begingroup\@sanitize@url \@url }%
\providecommand \@url [1]{\endgroup\@href {#1}{\urlprefix }}%
\providecommand \urlprefix  [0]{URL }%
\providecommand \Eprint [0]{\href }%
\providecommand \doibase [0]{http://dx.doi.org/}%
\providecommand \selectlanguage [0]{\@gobble}%
\providecommand \bibinfo  [0]{\@secondoftwo}%
\providecommand \bibfield  [0]{\@secondoftwo}%
\providecommand \translation [1]{[#1]}%
\providecommand \BibitemOpen [0]{}%
\providecommand \bibitemStop [0]{}%
\providecommand \bibitemNoStop [0]{.\EOS\space}%
\providecommand \EOS [0]{\spacefactor3000\relax}%
\providecommand \BibitemShut  [1]{\csname bibitem#1\endcsname}%
\let\auto@bib@innerbib\@empty
\bibitem [{\citenamefont {Kreyszig}(2025)}]{kreyszig}%
  \BibitemOpen
  \bibfield  {author} {\bibinfo {author} {\bibfnamefont {E.}~\bibnamefont
  {Kreyszig}},\ }\href@noop {} {\emph {\bibinfo {title} {Advanced Engineering
  Mathematics, International Adaptation}}}\ (\bibinfo  {publisher} {john wiley
  \& sons},\ \bibinfo {year} {2025})\BibitemShut {NoStop}%
\bibitem [{\citenamefont {Mead}\ and\ \citenamefont
  {Truhlar}(1982)}]{mead_jcp77}%
  \BibitemOpen
  \bibfield  {author} {\bibinfo {author} {\bibfnamefont {C.~A.}\ \bibnamefont
  {Mead}}\ and\ \bibinfo {author} {\bibfnamefont {D.~G.}\ \bibnamefont
  {Truhlar}},\ }\href@noop {} {\bibfield  {journal} {\bibinfo  {journal} {J.
  Chem. Phys.}\ }\textbf {\bibinfo {volume} {77}},\ \bibinfo {pages} {6090}
  (\bibinfo {year} {1982})}\BibitemShut {NoStop}%
\bibitem [{\citenamefont {Shankaraiah}(2023)}]{Shankaraiah}%
  \BibitemOpen
  \bibfield  {author} {\bibinfo {author} {\bibfnamefont {G.}~\bibnamefont
  {Shankaraiah}},\ }\href@noop {} {\bibfield  {journal} {\bibinfo  {journal}
  {International Journal of Research Publication and Reviews}\ }\textbf
  {\bibinfo {volume} {4}},\ \bibinfo {pages} {167} (\bibinfo {year}
  {2023})}\BibitemShut {NoStop}%
\end{thebibliography}%


\begin{thebibliography}{7}%
\makeatletter
\providecommand \@ifxundefined [1]{%
 \@ifx{#1\undefined}
}%
\providecommand \@ifnum [1]{%
 \ifnum #1\expandafter \@firstoftwo
 \else \expandafter \@secondoftwo
 \fi
}%
\providecommand \@ifx [1]{%
 \ifx #1\expandafter \@firstoftwo
 \else \expandafter \@secondoftwo
 \fi
}%
\providecommand \natexlab [1]{#1}%
\providecommand \enquote  [1]{``#1''}%
\providecommand \bibnamefont  [1]{#1}%
\providecommand \bibfnamefont [1]{#1}%
\providecommand \citenamefont [1]{#1}%
\providecommand \href@noop [0]{\@secondoftwo}%
\providecommand \href [0]{\begingroup \@sanitize@url \@href}%
\providecommand \@href[1]{\@@startlink{#1}\@@href}%
\providecommand \@@href[1]{\endgroup#1\@@endlink}%
\providecommand \@sanitize@url [0]{\catcode `\\12\catcode `\$12\catcode
  `\&12\catcode `\#12\catcode `\^12\catcode `\_12\catcode `\%12\relax}%
\providecommand \@@startlink[1]{}%
\providecommand \@@endlink[0]{}%
\providecommand \url  [0]{\begingroup\@sanitize@url \@url }%
\providecommand \@url [1]{\endgroup\@href {#1}{\urlprefix }}%
\providecommand \urlprefix  [0]{URL }%
\providecommand \Eprint [0]{\href }%
\providecommand \doibase [0]{http://dx.doi.org/}%
\providecommand \selectlanguage [0]{\@gobble}%
\providecommand \bibinfo  [0]{\@secondoftwo}%
\providecommand \bibfield  [0]{\@secondoftwo}%
\providecommand \translation [1]{[#1]}%
\providecommand \BibitemOpen [0]{}%
\providecommand \bibitemStop [0]{}%
\providecommand \bibitemNoStop [0]{.\EOS\space}%
\providecommand \EOS [0]{\spacefactor3000\relax}%
\providecommand \BibitemShut  [1]{\csname bibitem#1\endcsname}%
\let\auto@bib@innerbib\@empty
\bibitem [{\citenamefont {Sarkar}\ and\ \citenamefont
  {Adhikari}(2008)}]{sarkar_jpca112}%
  \BibitemOpen
  \bibfield  {author} {\bibinfo {author} {\bibfnamefont {B.}~\bibnamefont
  {Sarkar}}\ and\ \bibinfo {author} {\bibfnamefont {S.}~\bibnamefont
  {Adhikari}},\ }\href@noop {} {\bibfield  {journal} {\bibinfo  {journal} {J.
  Phys. Chem. A}\ }\textbf {\bibinfo {volume} {112}},\ \bibinfo {pages} {9868}
  (\bibinfo {year} {2008})}\BibitemShut {NoStop}%
\bibitem [{\citenamefont {Mukherjee}, \citenamefont {Mukherjee},\ and\
  \citenamefont {Adhikari}(2017)}]{soumya_jpca121}%
  \BibitemOpen
  \bibfield  {author} {\bibinfo {author} {\bibfnamefont {S.}~\bibnamefont
  {Mukherjee}}, \bibinfo {author} {\bibfnamefont {B.}~\bibnamefont
  {Mukherjee}}, \ and\ \bibinfo {author} {\bibfnamefont {S.}~\bibnamefont
  {Adhikari}},\ }\href@noop {} {\bibfield  {journal} {\bibinfo  {journal} {J.
  Phys. Chem. A}\ }\textbf {\bibinfo {volume} {121}},\ \bibinfo {pages} {6314}
  (\bibinfo {year} {2017})}\BibitemShut {NoStop}%
\bibitem [{\citenamefont {Mukherjee}\ \emph {et~al.}(2022)\citenamefont
  {Mukherjee}, \citenamefont {Ravi}, \citenamefont {Dutta}, \citenamefont
  {Sardar},\ and\ \citenamefont {Adhikari}}]{soumya_tfb}%
  \BibitemOpen
  \bibfield  {author} {\bibinfo {author} {\bibfnamefont {S.}~\bibnamefont
  {Mukherjee}}, \bibinfo {author} {\bibfnamefont {S.}~\bibnamefont {Ravi}},
  \bibinfo {author} {\bibfnamefont {J.}~\bibnamefont {Dutta}}, \bibinfo
  {author} {\bibfnamefont {S.}~\bibnamefont {Sardar}}, \ and\ \bibinfo {author}
  {\bibfnamefont {S.}~\bibnamefont {Adhikari}},\ }\href@noop {} {\bibfield
  {journal} {\bibinfo  {journal} {Phys. Chem. Chem. Phys.}\ }\textbf {\bibinfo
  {volume} {24}},\ \bibinfo {pages} {2185} (\bibinfo {year}
  {2022})}\BibitemShut {NoStop}%
\bibitem [{\citenamefont {Sah}\ \emph {et~al.}(2023)\citenamefont {Sah},
  \citenamefont {Mukherjee}, \citenamefont {Naskar}, \citenamefont {Hazra},\
  and\ \citenamefont {Adhikari}}]{mantu_ijqc_curl}%
  \BibitemOpen
  \bibfield  {author} {\bibinfo {author} {\bibfnamefont {M.~K.}\ \bibnamefont
  {Sah}}, \bibinfo {author} {\bibfnamefont {S.}~\bibnamefont {Mukherjee}},
  \bibinfo {author} {\bibfnamefont {K.}~\bibnamefont {Naskar}}, \bibinfo
  {author} {\bibfnamefont {S.}~\bibnamefont {Hazra}}, \ and\ \bibinfo {author}
  {\bibfnamefont {S.}~\bibnamefont {Adhikari}},\ }\href@noop {} {\bibfield
  {journal} {\bibinfo  {journal} {International Journal of Quantum Chemistry}\
  }\textbf {\bibinfo {volume} {123}},\ \bibinfo {pages} {e27212} (\bibinfo
  {year} {2023})}\BibitemShut {NoStop}%
\bibitem [{\citenamefont {Naskar}\ \emph {et~al.}(2024)\citenamefont {Naskar},
  \citenamefont {Mukherjee}, \citenamefont {Ghosh},\ and\ \citenamefont
  {Adhikari}}]{FH2_dynamics}%
  \BibitemOpen
  \bibfield  {author} {\bibinfo {author} {\bibfnamefont {K.}~\bibnamefont
  {Naskar}}, \bibinfo {author} {\bibfnamefont {S.}~\bibnamefont {Mukherjee}},
  \bibinfo {author} {\bibfnamefont {S.}~\bibnamefont {Ghosh}}, \ and\ \bibinfo
  {author} {\bibfnamefont {S.}~\bibnamefont {Adhikari}},\ }\href@noop {}
  {\bibfield  {journal} {\bibinfo  {journal} {J. Phys. Chem. A}\ }\textbf
  {\bibinfo {volume} {128}},\ \bibinfo {pages} {1438} (\bibinfo {year}
  {2024})}\BibitemShut {NoStop}%
\bibitem [{\citenamefont {Alexander}, \citenamefont {Werner},\ and\
  \citenamefont {Manolopoulos}(1998)}]{jcp-109-5710-WERNER}%
  \BibitemOpen
  \bibfield  {author} {\bibinfo {author} {\bibfnamefont {M.~H.}\ \bibnamefont
  {Alexander}}, \bibinfo {author} {\bibfnamefont {H.}~\bibnamefont {Werner}}, \
  and\ \bibinfo {author} {\bibfnamefont {D.~E.}\ \bibnamefont {Manolopoulos}},\
  }\href@noop {} {\bibfield  {journal} {\bibinfo  {journal} {J. Chem. Phys.}\
  }\textbf {\bibinfo {volume} {109}},\ \bibinfo {pages} {5710} (\bibinfo {year}
  {1998})}\BibitemShut {NoStop}%
\bibitem [{\citenamefont {Mukherjee}\ \emph {et~al.}(2020)\citenamefont
  {Mukherjee}, \citenamefont {Naskar}, \citenamefont {Mukherjee}, \citenamefont
  {Ravi}, \citenamefont {Shamasundar}, \citenamefont {Mukhopadhyay},\ and\
  \citenamefont {Adhikari}}]{jcp-153-174301-SA}%
  \BibitemOpen
  \bibfield  {author} {\bibinfo {author} {\bibfnamefont {B.}~\bibnamefont
  {Mukherjee}}, \bibinfo {author} {\bibfnamefont {K.}~\bibnamefont {Naskar}},
  \bibinfo {author} {\bibfnamefont {S.}~\bibnamefont {Mukherjee}}, \bibinfo
  {author} {\bibfnamefont {S.}~\bibnamefont {Ravi}}, \bibinfo {author}
  {\bibfnamefont {K.~R.}\ \bibnamefont {Shamasundar}}, \bibinfo {author}
  {\bibfnamefont {D.}~\bibnamefont {Mukhopadhyay}}, \ and\ \bibinfo {author}
  {\bibfnamefont {S.}~\bibnamefont {Adhikari}},\ }\href@noop {} {\bibfield
  {journal} {\bibinfo  {journal} {J. Chem. Phys.}\ }\textbf {\bibinfo {volume}
  {153}},\ \bibinfo {pages} {174301} (\bibinfo {year} {2020})}\BibitemShut
  {NoStop}%
\end{thebibliography}%


\begin{thebibliography}{5}%
\makeatletter
\providecommand \@ifxundefined [1]{%
 \@ifx{#1\undefined}
}%
\providecommand \@ifnum [1]{%
 \ifnum #1\expandafter \@firstoftwo
 \else \expandafter \@secondoftwo
 \fi
}%
\providecommand \@ifx [1]{%
 \ifx #1\expandafter \@firstoftwo
 \else \expandafter \@secondoftwo
 \fi
}%
\providecommand \natexlab [1]{#1}%
\providecommand \enquote  [1]{``#1''}%
\providecommand \bibnamefont  [1]{#1}%
\providecommand \bibfnamefont [1]{#1}%
\providecommand \citenamefont [1]{#1}%
\providecommand \href@noop [0]{\@secondoftwo}%
\providecommand \href [0]{\begingroup \@sanitize@url \@href}%
\providecommand \@href[1]{\@@startlink{#1}\@@href}%
\providecommand \@@href[1]{\endgroup#1\@@endlink}%
\providecommand \@sanitize@url [0]{\catcode `\\12\catcode `\$12\catcode
  `\&12\catcode `\#12\catcode `\^12\catcode `\_12\catcode `\%12\relax}%
\providecommand \@@startlink[1]{}%
\providecommand \@@endlink[0]{}%
\providecommand \url  [0]{\begingroup\@sanitize@url \@url }%
\providecommand \@url [1]{\endgroup\@href {#1}{\urlprefix }}%
\providecommand \urlprefix  [0]{URL }%
\providecommand \Eprint [0]{\href }%
\providecommand \doibase [0]{http://dx.doi.org/}%
\providecommand \selectlanguage [0]{\@gobble}%
\providecommand \bibinfo  [0]{\@secondoftwo}%
\providecommand \bibfield  [0]{\@secondoftwo}%
\providecommand \translation [1]{[#1]}%
\providecommand \BibitemOpen [0]{}%
\providecommand \bibitemStop [0]{}%
\providecommand \bibitemNoStop [0]{.\EOS\space}%
\providecommand \EOS [0]{\spacefactor3000\relax}%
\providecommand \BibitemShut  [1]{\csname bibitem#1\endcsname}%
\let\auto@bib@innerbib\@empty
\bibitem [{\citenamefont {Arthurs}\ and\ \citenamefont {Dalgarno}(1960)}]{ADS}%
  \BibitemOpen
  \bibfield  {author} {\bibinfo {author} {\bibfnamefont {A.}~\bibnamefont
  {Arthurs}}\ and\ \bibinfo {author} {\bibfnamefont {A.}~\bibnamefont
  {Dalgarno}},\ }\href@noop {} {\bibfield  {journal} {\bibinfo  {journal}
  {Proc. R. Soc. Lond. Ser. A}\ }\textbf {\bibinfo {volume} {256}},\ \bibinfo
  {pages} {540} (\bibinfo {year} {1960})}\BibitemShut {NoStop}%
\bibitem [{\citenamefont {Adhikari}\ and\ \citenamefont
  {Varandas}(2013)}]{cpc-184-270-SA}%
  \BibitemOpen
  \bibfield  {author} {\bibinfo {author} {\bibfnamefont {S.}~\bibnamefont
  {Adhikari}}\ and\ \bibinfo {author} {\bibfnamefont {A.~J.~C.}\ \bibnamefont
  {Varandas}},\ }\href@noop {} {\bibfield  {journal} {\bibinfo  {journal}
  {Comput. Phys. Comm.}\ }\textbf {\bibinfo {volume} {184}},\ \bibinfo {pages}
  {270} (\bibinfo {year} {2013})}\BibitemShut {NoStop}%
\bibitem [{\citenamefont {Billing}\ and\ \citenamefont
  {Markovic}(1993)}]{jcp-99-2674-BILL}%
  \BibitemOpen
  \bibfield  {author} {\bibinfo {author} {\bibfnamefont {G.~D.}\ \bibnamefont
  {Billing}}\ and\ \bibinfo {author} {\bibfnamefont {N.}~\bibnamefont
  {Markovic}},\ }\href@noop {} {\bibfield  {journal} {\bibinfo  {journal} {J.
  Chem. Phys.}\ }\textbf {\bibinfo {volume} {99}},\ \bibinfo {pages} {2674}
  (\bibinfo {year} {1993})}\BibitemShut {NoStop}%
\bibitem [{\citenamefont {Billing}\ and\ \citenamefont
  {Markovic}(1994)}]{jcp-100-1085-BILL}%
  \BibitemOpen
  \bibfield  {author} {\bibinfo {author} {\bibfnamefont {G.~D.}\ \bibnamefont
  {Billing}}\ and\ \bibinfo {author} {\bibfnamefont {N.}~\bibnamefont
  {Markovic}},\ }\href@noop {} {\bibfield  {journal} {\bibinfo  {journal} {J.
  Chem. Phys.}\ }\textbf {\bibinfo {volume} {100}},\ \bibinfo {pages} {1085}
  (\bibinfo {year} {1994})}\BibitemShut {NoStop}%
\bibitem [{\citenamefont {Gradshteyn}\ and\ \citenamefont {Ryzhik}()}]{ISG}%
  \BibitemOpen
  \bibfield  {author} {\bibinfo {author} {\bibfnamefont {I.~S.}\ \bibnamefont
  {Gradshteyn}}\ and\ \bibinfo {author} {\bibfnamefont {I.~M.}\ \bibnamefont
  {Ryzhik}},\ }\href@noop {} {\emph {\bibinfo {title} {Table of Integrals
  Series and Products; Academic Press, New York (1965)}}}\BibitemShut {NoStop}%
\end{thebibliography}%


\begin{thebibliography}{2}%
\makeatletter
\providecommand \@ifxundefined [1]{%
 \@ifx{#1\undefined}
}%
\providecommand \@ifnum [1]{%
 \ifnum #1\expandafter \@firstoftwo
 \else \expandafter \@secondoftwo
 \fi
}%
\providecommand \@ifx [1]{%
 \ifx #1\expandafter \@firstoftwo
 \else \expandafter \@secondoftwo
 \fi
}%
\providecommand \natexlab [1]{#1}%
\providecommand \enquote  [1]{``#1''}%
\providecommand \bibnamefont  [1]{#1}%
\providecommand \bibfnamefont [1]{#1}%
\providecommand \citenamefont [1]{#1}%
\providecommand \href@noop [0]{\@secondoftwo}%
\providecommand \href [0]{\begingroup \@sanitize@url \@href}%
\providecommand \@href[1]{\@@startlink{#1}\@@href}%
\providecommand \@@href[1]{\endgroup#1\@@endlink}%
\providecommand \@sanitize@url [0]{\catcode `\\12\catcode `\$12\catcode
  `\&12\catcode `\#12\catcode `\^12\catcode `\_12\catcode `\%12\relax}%
\providecommand \@@startlink[1]{}%
\providecommand \@@endlink[0]{}%
\providecommand \url  [0]{\begingroup\@sanitize@url \@url }%
\providecommand \@url [1]{\endgroup\@href {#1}{\urlprefix }}%
\providecommand \urlprefix  [0]{URL }%
\providecommand \Eprint [0]{\href }%
\providecommand \doibase [0]{http://dx.doi.org/}%
\providecommand \selectlanguage [0]{\@gobble}%
\providecommand \bibinfo  [0]{\@secondoftwo}%
\providecommand \bibfield  [0]{\@secondoftwo}%
\providecommand \translation [1]{[#1]}%
\providecommand \BibitemOpen [0]{}%
\providecommand \bibitemStop [0]{}%
\providecommand \bibitemNoStop [0]{.\EOS\space}%
\providecommand \EOS [0]{\spacefactor3000\relax}%
\providecommand \BibitemShut  [1]{\csname bibitem#1\endcsname}%
\let\auto@bib@innerbib\@empty
\bibitem [{\citenamefont {Pechukas}\ and\ \citenamefont {Light}(1966)}]{PPJS}%
  \BibitemOpen
  \bibfield  {author} {\bibinfo {author} {\bibfnamefont {P.}~\bibnamefont
  {Pechukas}}\ and\ \bibinfo {author} {\bibfnamefont {J.~C.}\ \bibnamefont
  {Light}},\ }\href@noop {} {\bibfield  {journal} {\bibinfo  {journal} {J.\
  Chem.\ Phys.}\ }\textbf {\bibinfo {volume} {44}},\ \bibinfo {pages} {3897}
  (\bibinfo {year} {1966})}\BibitemShut {NoStop}%
\bibitem [{\citenamefont {Dutta}\ \emph {et~al.}(2021)\citenamefont {Dutta},
  \citenamefont {Mandal}, \citenamefont {Adhikari}, \citenamefont {Spiering},
  \citenamefont {Meyer},\ and\ \citenamefont {Somers}}]{SAMARK}%
  \BibitemOpen
  \bibfield  {author} {\bibinfo {author} {\bibfnamefont {J.}~\bibnamefont
  {Dutta}}, \bibinfo {author} {\bibfnamefont {S.}~\bibnamefont {Mandal}},
  \bibinfo {author} {\bibfnamefont {S.}~\bibnamefont {Adhikari}}, \bibinfo
  {author} {\bibfnamefont {P.}~\bibnamefont {Spiering}}, \bibinfo {author}
  {\bibfnamefont {J.}~\bibnamefont {Meyer}}, \ and\ \bibinfo {author}
  {\bibfnamefont {M.~F.}\ \bibnamefont {Somers}},\ }\href@noop {} {\bibfield
  {journal} {\bibinfo  {journal} {J. Chem. Phys.}\ }\textbf {\bibinfo {volume}
  {154}},\ \bibinfo {pages} {104103} (\bibinfo {year} {2021})}\BibitemShut
  {NoStop}%
\end{thebibliography}%


\begin{thebibliography}{1}%
\makeatletter
\providecommand \@ifxundefined [1]{%
 \@ifx{#1\undefined}
}%
\providecommand \@ifnum [1]{%
 \ifnum #1\expandafter \@firstoftwo
 \else \expandafter \@secondoftwo
 \fi
}%
\providecommand \@ifx [1]{%
 \ifx #1\expandafter \@firstoftwo
 \else \expandafter \@secondoftwo
 \fi
}%
\providecommand \natexlab [1]{#1}%
\providecommand \enquote  [1]{``#1''}%
\providecommand \bibnamefont  [1]{#1}%
\providecommand \bibfnamefont [1]{#1}%
\providecommand \citenamefont [1]{#1}%
\providecommand \href@noop [0]{\@secondoftwo}%
\providecommand \href [0]{\begingroup \@sanitize@url \@href}%
\providecommand \@href[1]{\@@startlink{#1}\@@href}%
\providecommand \@@href[1]{\endgroup#1\@@endlink}%
\providecommand \@sanitize@url [0]{\catcode `\\12\catcode `\$12\catcode
  `\&12\catcode `\#12\catcode `\^12\catcode `\_12\catcode `\%12\relax}%
\providecommand \@@startlink[1]{}%
\providecommand \@@endlink[0]{}%
\providecommand \url  [0]{\begingroup\@sanitize@url \@url }%
\providecommand \@url [1]{\endgroup\@href {#1}{\urlprefix }}%
\providecommand \urlprefix  [0]{URL }%
\providecommand \Eprint [0]{\href }%
\providecommand \doibase [0]{http://dx.doi.org/}%
\providecommand \selectlanguage [0]{\@gobble}%
\providecommand \bibinfo  [0]{\@secondoftwo}%
\providecommand \bibfield  [0]{\@secondoftwo}%
\providecommand \translation [1]{[#1]}%
\providecommand \BibitemOpen [0]{}%
\providecommand \bibitemStop [0]{}%
\providecommand \bibitemNoStop [0]{.\EOS\space}%
\providecommand \EOS [0]{\spacefactor3000\relax}%
\providecommand \BibitemShut  [1]{\csname bibitem#1\endcsname}%
\let\auto@bib@innerbib\@empty
\bibitem [{\citenamefont {Dutta}, \citenamefont {Mandal},\ and\ \citenamefont
  {Adhikari}(2022)}]{VV-Hartree-2021}%
  \BibitemOpen
  \bibfield  {author} {\bibinfo {author} {\bibfnamefont {J.}~\bibnamefont
  {Dutta}}, \bibinfo {author} {\bibfnamefont {S.}~\bibnamefont {Mandal}}, \
  and\ \bibinfo {author} {\bibfnamefont {S.}~\bibnamefont {Adhikari}},\
  }\href@noop {} {\bibfield  {journal} {\bibinfo  {journal} {Chem. Phys.}\
  }\textbf {\bibinfo {volume} {552}},\ \bibinfo {pages} {111371} (\bibinfo
  {year} {2022})}\BibitemShut {NoStop}%
\end{thebibliography}%


\begin{thebibliography}{4}%
\makeatletter
\providecommand \@ifxundefined [1]{%
 \@ifx{#1\undefined}
}%
\providecommand \@ifnum [1]{%
 \ifnum #1\expandafter \@firstoftwo
 \else \expandafter \@secondoftwo
 \fi
}%
\providecommand \@ifx [1]{%
 \ifx #1\expandafter \@firstoftwo
 \else \expandafter \@secondoftwo
 \fi
}%
\providecommand \natexlab [1]{#1}%
\providecommand \enquote  [1]{``#1''}%
\providecommand \bibnamefont  [1]{#1}%
\providecommand \bibfnamefont [1]{#1}%
\providecommand \citenamefont [1]{#1}%
\providecommand \href@noop [0]{\@secondoftwo}%
\providecommand \href [0]{\begingroup \@sanitize@url \@href}%
\providecommand \@href[1]{\@@startlink{#1}\@@href}%
\providecommand \@@href[1]{\endgroup#1\@@endlink}%
\providecommand \@sanitize@url [0]{\catcode `\\12\catcode `\$12\catcode
  `\&12\catcode `\#12\catcode `\^12\catcode `\_12\catcode `\%12\relax}%
\providecommand \@@startlink[1]{}%
\providecommand \@@endlink[0]{}%
\providecommand \url  [0]{\begingroup\@sanitize@url \@url }%
\providecommand \@url [1]{\endgroup\@href {#1}{\urlprefix }}%
\providecommand \urlprefix  [0]{URL }%
\providecommand \Eprint [0]{\href }%
\providecommand \doibase [0]{http://dx.doi.org/}%
\providecommand \selectlanguage [0]{\@gobble}%
\providecommand \bibinfo  [0]{\@secondoftwo}%
\providecommand \bibfield  [0]{\@secondoftwo}%
\providecommand \translation [1]{[#1]}%
\providecommand \BibitemOpen [0]{}%
\providecommand \bibitemStop [0]{}%
\providecommand \bibitemNoStop [0]{.\EOS\space}%
\providecommand \EOS [0]{\spacefactor3000\relax}%
\providecommand \BibitemShut  [1]{\csname bibitem#1\endcsname}%
\let\auto@bib@innerbib\@empty
\bibitem [{\citenamefont {Mengel}\ and\ \citenamefont
  {Billing}(1997)}]{BILEH2}%
  \BibitemOpen
  \bibfield  {author} {\bibinfo {author} {\bibfnamefont {S.~K.}\ \bibnamefont
  {Mengel}}\ and\ \bibinfo {author} {\bibfnamefont {G.~D.}\ \bibnamefont
  {Billing}},\ }\href@noop {} {\bibfield  {journal} {\bibinfo  {journal} {J.
  Phys. Chem. B}\ }\textbf {\bibinfo {volume} {101}},\ \bibinfo {pages} {10781}
  (\bibinfo {year} {1997})}\BibitemShut {NoStop}%
\bibitem [{\citenamefont {Billing}(1995)}]{GD1}%
  \BibitemOpen
  \bibfield  {author} {\bibinfo {author} {\bibfnamefont {G.~D.}\ \bibnamefont
  {Billing}},\ }\href@noop {} {\bibfield  {journal} {\bibinfo  {journal} {J.\
  Phys.\ Chem.}\ }\textbf {\bibinfo {volume} {99}},\ \bibinfo {pages} {15378}
  (\bibinfo {year} {1995})}\BibitemShut {NoStop}%
\bibitem [{\citenamefont {Mandal}\ \emph
  {et~al.}(2015{\natexlab{a}})\citenamefont {Mandal}, \citenamefont {Sahoo},
  \citenamefont {Ghosh},\ and\ \citenamefont {Adhikari}}]{SM-MOLPHYS}%
  \BibitemOpen
  \bibfield  {author} {\bibinfo {author} {\bibfnamefont {S.}~\bibnamefont
  {Mandal}}, \bibinfo {author} {\bibfnamefont {T.}~\bibnamefont {Sahoo}},
  \bibinfo {author} {\bibfnamefont {S.}~\bibnamefont {Ghosh}}, \ and\ \bibinfo
  {author} {\bibfnamefont {S.}~\bibnamefont {Adhikari}},\ }\href@noop {}
  {\bibfield  {journal} {\bibinfo  {journal} {Mol. Phys.}\ }\textbf {\bibinfo
  {volume} {113}},\ \bibinfo {pages} {3042} (\bibinfo {year}
  {2015}{\natexlab{a}})}\BibitemShut {NoStop}%
\bibitem [{\citenamefont {Mandal}\ \emph
  {et~al.}(2015{\natexlab{b}})\citenamefont {Mandal}, \citenamefont {Sahoo},
  \citenamefont {Ghosh},\ and\ \citenamefont {Adhikari}}]{SM-JTCC}%
  \BibitemOpen
  \bibfield  {author} {\bibinfo {author} {\bibfnamefont {S.}~\bibnamefont
  {Mandal}}, \bibinfo {author} {\bibfnamefont {T.}~\bibnamefont {Sahoo}},
  \bibinfo {author} {\bibfnamefont {S.}~\bibnamefont {Ghosh}}, \ and\ \bibinfo
  {author} {\bibfnamefont {S.}~\bibnamefont {Adhikari}},\ }\href@noop {}
  {\bibfield  {journal} {\bibinfo  {journal} {J. Theor. Comput. Chem.}\
  }\textbf {\bibinfo {volume} {14}},\ \bibinfo {pages} {1550028} (\bibinfo
  {year} {2015}{\natexlab{b}})}\BibitemShut {NoStop}%
\end{thebibliography}%


\begin{thebibliography}{1}%
\makeatletter
\providecommand \@ifxundefined [1]{%
 \@ifx{#1\undefined}
}%
\providecommand \@ifnum [1]{%
 \ifnum #1\expandafter \@firstoftwo
 \else \expandafter \@secondoftwo
 \fi
}%
\providecommand \@ifx [1]{%
 \ifx #1\expandafter \@firstoftwo
 \else \expandafter \@secondoftwo
 \fi
}%
\providecommand \natexlab [1]{#1}%
\providecommand \enquote  [1]{``#1''}%
\providecommand \bibnamefont  [1]{#1}%
\providecommand \bibfnamefont [1]{#1}%
\providecommand \citenamefont [1]{#1}%
\providecommand \href@noop [0]{\@secondoftwo}%
\providecommand \href [0]{\begingroup \@sanitize@url \@href}%
\providecommand \@href[1]{\@@startlink{#1}\@@href}%
\providecommand \@@href[1]{\endgroup#1\@@endlink}%
\providecommand \@sanitize@url [0]{\catcode `\\12\catcode `\$12\catcode
  `\&12\catcode `\#12\catcode `\^12\catcode `\_12\catcode `\%12\relax}%
\providecommand \@@startlink[1]{}%
\providecommand \@@endlink[0]{}%
\providecommand \url  [0]{\begingroup\@sanitize@url \@url }%
\providecommand \@url [1]{\endgroup\@href {#1}{\urlprefix }}%
\providecommand \urlprefix  [0]{URL }%
\providecommand \Eprint [0]{\href }%
\providecommand \doibase [0]{http://dx.doi.org/}%
\providecommand \selectlanguage [0]{\@gobble}%
\providecommand \bibinfo  [0]{\@secondoftwo}%
\providecommand \bibfield  [0]{\@secondoftwo}%
\providecommand \translation [1]{[#1]}%
\providecommand \BibitemOpen [0]{}%
\providecommand \bibitemStop [0]{}%
\providecommand \bibitemNoStop [0]{.\EOS\space}%
\providecommand \EOS [0]{\spacefactor3000\relax}%
\providecommand \BibitemShut  [1]{\csname bibitem#1\endcsname}%
\let\auto@bib@innerbib\@empty
\bibitem [{\citenamefont {Mandal}\ \emph {et~al.}(2019)\citenamefont {Mandal},
  \citenamefont {Ghosh}, \citenamefont {Sardar},\ and\ \citenamefont
  {Adhikari}}]{SMIRPC}%
  \BibitemOpen
  \bibfield  {author} {\bibinfo {author} {\bibfnamefont {S.}~\bibnamefont
  {Mandal}}, \bibinfo {author} {\bibfnamefont {S.}~\bibnamefont {Ghosh}},
  \bibinfo {author} {\bibfnamefont {S.~S.}\ \bibnamefont {Sardar}}, \ and\
  \bibinfo {author} {\bibfnamefont {S.}~\bibnamefont {Adhikari}},\ }\href@noop
  {} {\bibfield  {journal} {\bibinfo  {journal} {Int. Rev. Phys. Chem.}\
  }\textbf {\bibinfo {volume} {37}},\ \bibinfo {pages} {607} (\bibinfo {year}
  {2019})}\BibitemShut {NoStop}%
\end{thebibliography}%


\begin{thebibliography}{137}%
\makeatletter
\providecommand \@ifxundefined [1]{%
 \@ifx{#1\undefined}
}%
\providecommand \@ifnum [1]{%
 \ifnum #1\expandafter \@firstoftwo
 \else \expandafter \@secondoftwo
 \fi
}%
\providecommand \@ifx [1]{%
 \ifx #1\expandafter \@firstoftwo
 \else \expandafter \@secondoftwo
 \fi
}%
\providecommand \natexlab [1]{#1}%
\providecommand \enquote  [1]{``#1''}%
\providecommand \bibnamefont  [1]{#1}%
\providecommand \bibfnamefont [1]{#1}%
\providecommand \citenamefont [1]{#1}%
\providecommand \href@noop [0]{\@secondoftwo}%
\providecommand \href [0]{\begingroup \@sanitize@url \@href}%
\providecommand \@href[1]{\@@startlink{#1}\@@href}%
\providecommand \@@href[1]{\endgroup#1\@@endlink}%
\providecommand \@sanitize@url [0]{\catcode `\\12\catcode `\$12\catcode
  `\&12\catcode `\#12\catcode `\^12\catcode `\_12\catcode `\%12\relax}%
\providecommand \@@startlink[1]{}%
\providecommand \@@endlink[0]{}%
\providecommand \url  [0]{\begingroup\@sanitize@url \@url }%
\providecommand \@url [1]{\endgroup\@href {#1}{\urlprefix }}%
\providecommand \urlprefix  [0]{URL }%
\providecommand \Eprint [0]{\href }%
\providecommand \doibase [0]{http://dx.doi.org/}%
\providecommand \selectlanguage [0]{\@gobble}%
\providecommand \bibinfo  [0]{\@secondoftwo}%
\providecommand \bibfield  [0]{\@secondoftwo}%
\providecommand \translation [1]{[#1]}%
\providecommand \BibitemOpen [0]{}%
\providecommand \bibitemStop [0]{}%
\providecommand \bibitemNoStop [0]{.\EOS\space}%
\providecommand \EOS [0]{\spacefactor3000\relax}%
\providecommand \BibitemShut  [1]{\csname bibitem#1\endcsname}%
\let\auto@bib@innerbib\@empty
\bibitem [{\citenamefont {Born}\ and\ \citenamefont {Oppenheimer}(1927)}]{BO}%
  \BibitemOpen
  \bibfield  {author} {\bibinfo {author} {\bibfnamefont {M.}~\bibnamefont
  {Born}}\ and\ \bibinfo {author} {\bibfnamefont {R.}~\bibnamefont
  {Oppenheimer}},\ }\href@noop {} {\bibfield  {journal} {\bibinfo  {journal}
  {Ann. Phys.}\ }\textbf {\bibinfo {volume} {389}},\ \bibinfo {pages} {457}
  (\bibinfo {year} {1927})}\BibitemShut {NoStop}%
\bibitem [{\citenamefont {Born}\ and\ \citenamefont {Huang}(1954)}]{BH}%
  \BibitemOpen
  \bibfield  {author} {\bibinfo {author} {\bibfnamefont {M.}~\bibnamefont
  {Born}}\ and\ \bibinfo {author} {\bibfnamefont {K.}~\bibnamefont {Huang}},\
  }\href@noop {} {\emph {\bibinfo {title} {Dynamical Theory of Crystal
  Lattices}}}\ (\bibinfo  {publisher} {Oxford University Press, Oxford},\
  \bibinfo {year} {1954})\BibitemShut {NoStop}%
\bibitem [{\citenamefont {Top}\ and\ \citenamefont {Baer}(1977)}]{baer_jcp66}%
  \BibitemOpen
  \bibfield  {author} {\bibinfo {author} {\bibfnamefont {Z.~H.}\ \bibnamefont
  {Top}}\ and\ \bibinfo {author} {\bibfnamefont {M.}~\bibnamefont {Baer}},\
  }\href@noop {} {\bibfield  {journal} {\bibinfo  {journal} {J. Chem. Phys.}\
  }\textbf {\bibinfo {volume} {66}},\ \bibinfo {pages} {1363} (\bibinfo {year}
  {1977})}\BibitemShut {NoStop}%
\bibitem [{\citenamefont {Adhikari}\ and\ \citenamefont
  {Billing}(1999)}]{adhikari_jcp111}%
  \BibitemOpen
  \bibfield  {author} {\bibinfo {author} {\bibfnamefont {S.}~\bibnamefont
  {Adhikari}}\ and\ \bibinfo {author} {\bibfnamefont {G.~D.}\ \bibnamefont
  {Billing}},\ }\href@noop {} {\bibfield  {journal} {\bibinfo  {journal} {J.
  Chem. Phys.}\ }\textbf {\bibinfo {volume} {111}},\ \bibinfo {pages} {40}
  (\bibinfo {year} {1999})}\BibitemShut {NoStop}%
\bibitem [{\citenamefont {Varandas}\ and\ \citenamefont
  {Xu}(2000)}]{varandas_jcp112}%
  \BibitemOpen
  \bibfield  {author} {\bibinfo {author} {\bibfnamefont {A.~J.~C.}\
  \bibnamefont {Varandas}}\ and\ \bibinfo {author} {\bibfnamefont {Z.~R.}\
  \bibnamefont {Xu}},\ }\href@noop {} {\bibfield  {journal} {\bibinfo
  {journal} {J. Chem. Phys.}\ }\textbf {\bibinfo {volume} {112}},\ \bibinfo
  {pages} {2121} (\bibinfo {year} {2000})}\BibitemShut {NoStop}%
\bibitem [{\citenamefont {Curchod}\ and\ \citenamefont
  {Mart\'inez}(2018)}]{martinez_chemrev}%
  \BibitemOpen
  \bibfield  {author} {\bibinfo {author} {\bibfnamefont {B.~F.~E.}\
  \bibnamefont {Curchod}}\ and\ \bibinfo {author} {\bibfnamefont {T.~J.}\
  \bibnamefont {Mart\'inez}},\ }\href@noop {} {\bibfield  {journal} {\bibinfo
  {journal} {Chem. Rev.}\ }\textbf {\bibinfo {volume} {118}},\ \bibinfo {pages}
  {3305} (\bibinfo {year} {2018})}\BibitemShut {NoStop}%
\bibitem [{\citenamefont {Mukherjee}\ \emph
  {et~al.}(2018{\natexlab{a}})\citenamefont {Mukherjee}, \citenamefont
  {Mukherjee}, \citenamefont {Sardar}, \citenamefont {Shamasundar},\ and\
  \citenamefont {Adhikari}}]{bijit_cp515}%
  \BibitemOpen
  \bibfield  {author} {\bibinfo {author} {\bibfnamefont {B.}~\bibnamefont
  {Mukherjee}}, \bibinfo {author} {\bibfnamefont {S.}~\bibnamefont
  {Mukherjee}}, \bibinfo {author} {\bibfnamefont {S.}~\bibnamefont {Sardar}},
  \bibinfo {author} {\bibfnamefont {K.~R.}\ \bibnamefont {Shamasundar}}, \ and\
  \bibinfo {author} {\bibfnamefont {S.}~\bibnamefont {Adhikari}},\ }\href@noop
  {} {\bibfield  {journal} {\bibinfo  {journal} {Chem. Phys.}\ }\textbf
  {\bibinfo {volume} {515}},\ \bibinfo {pages} {350} (\bibinfo {year}
  {2018}{\natexlab{a}})}\BibitemShut {NoStop}%
\bibitem [{\citenamefont {Mukherjee}\ \emph {et~al.}(2021)\citenamefont
  {Mukherjee}, \citenamefont {Ravi}, \citenamefont {Naskar}, \citenamefont
  {Sardar},\ and\ \citenamefont {Adhikari}}]{soumya_bz}%
  \BibitemOpen
  \bibfield  {author} {\bibinfo {author} {\bibfnamefont {S.}~\bibnamefont
  {Mukherjee}}, \bibinfo {author} {\bibfnamefont {S.}~\bibnamefont {Ravi}},
  \bibinfo {author} {\bibfnamefont {K.}~\bibnamefont {Naskar}}, \bibinfo
  {author} {\bibfnamefont {S.}~\bibnamefont {Sardar}}, \ and\ \bibinfo {author}
  {\bibfnamefont {S.}~\bibnamefont {Adhikari}},\ }\href@noop {} {\bibfield
  {journal} {\bibinfo  {journal} {J. Chem. Phys.}\ }\textbf {\bibinfo {volume}
  {154}},\ \bibinfo {pages} {094306} (\bibinfo {year} {2021})}\BibitemShut
  {NoStop}%
\bibitem [{\citenamefont {Mukherjee}\ \emph {et~al.}(2022)\citenamefont
  {Mukherjee}, \citenamefont {Ravi}, \citenamefont {Dutta}, \citenamefont
  {Sardar},\ and\ \citenamefont {Adhikari}}]{soumya_tfb}%
  \BibitemOpen
  \bibfield  {author} {\bibinfo {author} {\bibfnamefont {S.}~\bibnamefont
  {Mukherjee}}, \bibinfo {author} {\bibfnamefont {S.}~\bibnamefont {Ravi}},
  \bibinfo {author} {\bibfnamefont {J.}~\bibnamefont {Dutta}}, \bibinfo
  {author} {\bibfnamefont {S.}~\bibnamefont {Sardar}}, \ and\ \bibinfo {author}
  {\bibfnamefont {S.}~\bibnamefont {Adhikari}},\ }\href@noop {} {\bibfield
  {journal} {\bibinfo  {journal} {Phys. Chem. Chem. Phys.}\ }\textbf {\bibinfo
  {volume} {24}},\ \bibinfo {pages} {2185} (\bibinfo {year}
  {2022})}\BibitemShut {NoStop}%
\bibitem [{\citenamefont {Baer}, \citenamefont {Niedner-Schatterburg},\ and\
  \citenamefont {Toennies}(1989)}]{baer_jcp91}%
  \BibitemOpen
  \bibfield  {author} {\bibinfo {author} {\bibfnamefont {M.}~\bibnamefont
  {Baer}}, \bibinfo {author} {\bibfnamefont {G.}~\bibnamefont
  {Niedner-Schatterburg}}, \ and\ \bibinfo {author} {\bibfnamefont {J.~P.}\
  \bibnamefont {Toennies}},\ }\href@noop {} {\bibfield  {journal} {\bibinfo
  {journal} {J. Chem. Phys.}\ }\textbf {\bibinfo {volume} {91}},\ \bibinfo
  {pages} {4169} (\bibinfo {year} {1989})}\BibitemShut {NoStop}%
\bibitem [{\citenamefont {Grimbert}, \citenamefont {Sidis},\ and\ \citenamefont
  {Cobut}(1998)}]{grimbert_jcp108}%
  \BibitemOpen
  \bibfield  {author} {\bibinfo {author} {\bibfnamefont {D.}~\bibnamefont
  {Grimbert}}, \bibinfo {author} {\bibfnamefont {V.}~\bibnamefont {Sidis}}, \
  and\ \bibinfo {author} {\bibfnamefont {V.}~\bibnamefont {Cobut}},\
  }\href@noop {} {\bibfield  {journal} {\bibinfo  {journal} {J. Chem. Phys.}\
  }\textbf {\bibinfo {volume} {108}},\ \bibinfo {pages} {6331} (\bibinfo {year}
  {1998})}\BibitemShut {NoStop}%
\bibitem [{\citenamefont {Baer}\ \emph {et~al.}(1996)\citenamefont {Baer},
  \citenamefont {Charutz}, \citenamefont {Kosloff},\ and\ \citenamefont
  {Baer}}]{baer_jcp105}%
  \BibitemOpen
  \bibfield  {author} {\bibinfo {author} {\bibfnamefont {R.}~\bibnamefont
  {Baer}}, \bibinfo {author} {\bibfnamefont {D.~M.}\ \bibnamefont {Charutz}},
  \bibinfo {author} {\bibfnamefont {R.}~\bibnamefont {Kosloff}}, \ and\
  \bibinfo {author} {\bibfnamefont {M.}~\bibnamefont {Baer}},\ }\href@noop {}
  {\bibfield  {journal} {\bibinfo  {journal} {J. Chem. Phys.}\ }\textbf
  {\bibinfo {volume} {105}},\ \bibinfo {pages} {9141} (\bibinfo {year}
  {1996})}\BibitemShut {NoStop}%
\bibitem [{\citenamefont {Baer}\ \emph {et~al.}(2000)\citenamefont {Baer},
  \citenamefont {Lin}, \citenamefont {Alijah}, \citenamefont {Adhikari},\ and\
  \citenamefont {Billing}}]{baer_pra62}%
  \BibitemOpen
  \bibfield  {author} {\bibinfo {author} {\bibfnamefont {M.}~\bibnamefont
  {Baer}}, \bibinfo {author} {\bibfnamefont {S.~H.}\ \bibnamefont {Lin}},
  \bibinfo {author} {\bibfnamefont {A.}~\bibnamefont {Alijah}}, \bibinfo
  {author} {\bibfnamefont {S.}~\bibnamefont {Adhikari}}, \ and\ \bibinfo
  {author} {\bibfnamefont {G.~D.}\ \bibnamefont {Billing}},\ }\href@noop {}
  {\bibfield  {journal} {\bibinfo  {journal} {Phys. Rev. A}\ }\textbf {\bibinfo
  {volume} {62}},\ \bibinfo {pages} {32506} (\bibinfo {year}
  {2000})}\BibitemShut {NoStop}%
\bibitem [{\citenamefont {Adhikari}\ \emph {et~al.}(2000)\citenamefont
  {Adhikari}, \citenamefont {Billing}, \citenamefont {Alijah}, \citenamefont
  {Lin},\ and\ \citenamefont {Baer}}]{adhikari_pra62}%
  \BibitemOpen
  \bibfield  {author} {\bibinfo {author} {\bibfnamefont {S.}~\bibnamefont
  {Adhikari}}, \bibinfo {author} {\bibfnamefont {G.~D.}\ \bibnamefont
  {Billing}}, \bibinfo {author} {\bibfnamefont {A.}~\bibnamefont {Alijah}},
  \bibinfo {author} {\bibfnamefont {S.~H.}\ \bibnamefont {Lin}}, \ and\
  \bibinfo {author} {\bibfnamefont {M.}~\bibnamefont {Baer}},\ }\href@noop {}
  {\bibfield  {journal} {\bibinfo  {journal} {Phys. Rev. A}\ }\textbf {\bibinfo
  {volume} {62}},\ \bibinfo {pages} {32507} (\bibinfo {year}
  {2000})}\BibitemShut {NoStop}%
\bibitem [{\citenamefont {Sarkar}\ and\ \citenamefont
  {Adhikari}(2006)}]{sarkar_jcp124}%
  \BibitemOpen
  \bibfield  {author} {\bibinfo {author} {\bibfnamefont {B.}~\bibnamefont
  {Sarkar}}\ and\ \bibinfo {author} {\bibfnamefont {S.}~\bibnamefont
  {Adhikari}},\ }\href@noop {} {\bibfield  {journal} {\bibinfo  {journal} {J.
  Chem. Phys.}\ }\textbf {\bibinfo {volume} {124}},\ \bibinfo {pages} {074101}
  (\bibinfo {year} {2006})}\BibitemShut {NoStop}%
\bibitem [{\citenamefont {Sarkar}\ and\ \citenamefont
  {Adhikari}(2008)}]{sarkar_jpca112}%
  \BibitemOpen
  \bibfield  {author} {\bibinfo {author} {\bibfnamefont {B.}~\bibnamefont
  {Sarkar}}\ and\ \bibinfo {author} {\bibfnamefont {S.}~\bibnamefont
  {Adhikari}},\ }\href@noop {} {\bibfield  {journal} {\bibinfo  {journal} {J.
  Phys. Chem. A}\ }\textbf {\bibinfo {volume} {112}},\ \bibinfo {pages} {9868}
  (\bibinfo {year} {2008})}\BibitemShut {NoStop}%
\bibitem [{\citenamefont {Sarkar}\ and\ \citenamefont
  {Adhikari}(2009)}]{sarkar_ijqc109}%
  \BibitemOpen
  \bibfield  {author} {\bibinfo {author} {\bibfnamefont {B.}~\bibnamefont
  {Sarkar}}\ and\ \bibinfo {author} {\bibfnamefont {S.}~\bibnamefont
  {Adhikari}},\ }\href@noop {} {\bibfield  {journal} {\bibinfo  {journal} {Int.
  J. Quantum Chem.}\ }\textbf {\bibinfo {volume} {109}},\ \bibinfo {pages}
  {650} (\bibinfo {year} {2009})}\BibitemShut {NoStop}%
\bibitem [{\citenamefont {Paul}\ \emph {et~al.}(2009)\citenamefont {Paul},
  \citenamefont {Sardar}, \citenamefont {Sarkar},\ and\ \citenamefont
  {Adhikari}}]{akpaul_jcp131}%
  \BibitemOpen
  \bibfield  {author} {\bibinfo {author} {\bibfnamefont {A.~K.}\ \bibnamefont
  {Paul}}, \bibinfo {author} {\bibfnamefont {S.}~\bibnamefont {Sardar}},
  \bibinfo {author} {\bibfnamefont {B.}~\bibnamefont {Sarkar}}, \ and\ \bibinfo
  {author} {\bibfnamefont {S.}~\bibnamefont {Adhikari}},\ }\href@noop {}
  {\bibfield  {journal} {\bibinfo  {journal} {J. Chem. Phys.}\ }\textbf
  {\bibinfo {volume} {131}},\ \bibinfo {pages} {124312} (\bibinfo {year}
  {2009})}\BibitemShut {NoStop}%
\bibitem [{\citenamefont {Dutta}\ \emph
  {et~al.}(2022{\natexlab{a}})\citenamefont {Dutta}, \citenamefont {Ravi},
  \citenamefont {Mukherjee}, \citenamefont {Ojha},\ and\ \citenamefont
  {Adhikari}}]{joy_jpca}%
  \BibitemOpen
  \bibfield  {author} {\bibinfo {author} {\bibfnamefont {J.}~\bibnamefont
  {Dutta}}, \bibinfo {author} {\bibfnamefont {S.}~\bibnamefont {Ravi}},
  \bibinfo {author} {\bibfnamefont {S.}~\bibnamefont {Mukherjee}}, \bibinfo
  {author} {\bibfnamefont {A.~K.}\ \bibnamefont {Ojha}}, \ and\ \bibinfo
  {author} {\bibfnamefont {S.}~\bibnamefont {Adhikari}},\ }\href@noop {}
  {\bibfield  {journal} {\bibinfo  {journal} {J. Phys. Chem. A}\ }\textbf
  {\bibinfo {volume} {126}},\ \bibinfo {pages} {691} (\bibinfo {year}
  {2022}{\natexlab{a}})}\BibitemShut {NoStop}%
\bibitem [{\citenamefont {Hellmann}(1937)}]{hellmann}%
  \BibitemOpen
  \bibfield  {author} {\bibinfo {author} {\bibfnamefont {H.}~\bibnamefont
  {Hellmann}},\ }\href@noop {} {\emph {\bibinfo {title} {{\it Einfuring in die
  Quantenchemie}}}}\ (\bibinfo  {publisher} {Franz Duetiche:Leipzig},\ \bibinfo
  {address} {Germany},\ \bibinfo {year} {1937})\BibitemShut {NoStop}%
\bibitem [{\citenamefont {Feynman}(1939)}]{feynman_pr56}%
  \BibitemOpen
  \bibfield  {author} {\bibinfo {author} {\bibfnamefont {R.}~\bibnamefont
  {Feynman}},\ }\href@noop {} {\bibfield  {journal} {\bibinfo  {journal} {Phys.
  Rev.}\ }\textbf {\bibinfo {volume} {56}},\ \bibinfo {pages} {340} (\bibinfo
  {year} {1939})}\BibitemShut {NoStop}%
\bibitem [{\citenamefont {Mukherjee}\ \emph {et~al.}(2024)\citenamefont
  {Mukherjee}, \citenamefont {Saha}, \citenamefont {Ghosh}, \citenamefont
  {Adhikari}, \citenamefont {Sathyamurthy},\ and\ \citenamefont
  {Baer}}]{friction_jpca128}%
  \BibitemOpen
  \bibfield  {author} {\bibinfo {author} {\bibfnamefont {S.}~\bibnamefont
  {Mukherjee}}, \bibinfo {author} {\bibfnamefont {S.}~\bibnamefont {Saha}},
  \bibinfo {author} {\bibfnamefont {S.}~\bibnamefont {Ghosh}}, \bibinfo
  {author} {\bibfnamefont {S.}~\bibnamefont {Adhikari}}, \bibinfo {author}
  {\bibfnamefont {N.}~\bibnamefont {Sathyamurthy}}, \ and\ \bibinfo {author}
  {\bibfnamefont {M.}~\bibnamefont {Baer}},\ }\href@noop {} {\bibfield
  {journal} {\bibinfo  {journal} {J. Phys. Chem. A}\ }\textbf {\bibinfo
  {volume} {128}},\ \bibinfo {pages} {7691} (\bibinfo {year}
  {2024})}\BibitemShut {NoStop}%
\bibitem [{\citenamefont {Longuet-Higgins}(1961)}]{longuet-higgins}%
  \BibitemOpen
  \bibfield  {author} {\bibinfo {author} {\bibfnamefont {H.~C.}\ \bibnamefont
  {Longuet-Higgins}},\ }\href@noop {} {\emph {\bibinfo {title} {{\it Advances
  in Spectroscopy}}}},\ Vol.~\bibinfo {volume} {2}\ (\bibinfo  {publisher}
  {Interscience Publishers, Ltd.},\ \bibinfo {address} {New York and London},\
  \bibinfo {year} {1961})\ Chap.\ \bibinfo {chapter} {Some Recent Developments
  in the Theory of Molecular Energy Levels}\BibitemShut {NoStop}%
\bibitem [{\citenamefont {Longuet-Higgins}(1975)}]{longuet-higgins_prsl344}%
  \BibitemOpen
  \bibfield  {author} {\bibinfo {author} {\bibfnamefont {H.~C.}\ \bibnamefont
  {Longuet-Higgins}},\ }\href@noop {} {\bibfield  {journal} {\bibinfo
  {journal} {Proc. R. Soc. London, Ser. A}\ }\textbf {\bibinfo {volume}
  {344}},\ \bibinfo {pages} {147} (\bibinfo {year} {1975})}\BibitemShut
  {NoStop}%
\bibitem [{\citenamefont {Herzberg}\ and\ \citenamefont
  {Longuet-Higgins}(1963)}]{herzberg_dfs35}%
  \BibitemOpen
  \bibfield  {author} {\bibinfo {author} {\bibfnamefont {G.}~\bibnamefont
  {Herzberg}}\ and\ \bibinfo {author} {\bibfnamefont {H.~C.}\ \bibnamefont
  {Longuet-Higgins}},\ }\href@noop {} {\bibfield  {journal} {\bibinfo
  {journal} {Discuss Faraday Soc.}\ }\textbf {\bibinfo {volume} {35}},\
  \bibinfo {pages} {77} (\bibinfo {year} {1963})}\BibitemShut {NoStop}%
\bibitem [{\citenamefont {Baer}(2006)}]{baer_book}%
  \BibitemOpen
  \bibfield  {author} {\bibinfo {author} {\bibfnamefont {M.}~\bibnamefont
  {Baer}},\ }\href@noop {} {\emph {\bibinfo {title} {{\it Beyond Born -
  Oppenheimer: Conical intersections and Electronic nonadiabatic coupling
  terms}}}}\ (\bibinfo  {publisher} {Wiley Interscience},\ \bibinfo {address}
  {New Jersey},\ \bibinfo {year} {2006})\BibitemShut {NoStop}%
\bibitem [{\citenamefont {Mead}\ and\ \citenamefont
  {Truhlar}(1982)}]{mead_jcp77}%
  \BibitemOpen
  \bibfield  {author} {\bibinfo {author} {\bibfnamefont {C.~A.}\ \bibnamefont
  {Mead}}\ and\ \bibinfo {author} {\bibfnamefont {D.~G.}\ \bibnamefont
  {Truhlar}},\ }\href@noop {} {\bibfield  {journal} {\bibinfo  {journal} {J.
  Chem. Phys.}\ }\textbf {\bibinfo {volume} {77}},\ \bibinfo {pages} {6090}
  (\bibinfo {year} {1982})}\BibitemShut {NoStop}%
\bibitem [{\citenamefont {Varandas}, \citenamefont {Tennyson},\ and\
  \citenamefont {Murrell}(1979)}]{varandas_cpl61}%
  \BibitemOpen
  \bibfield  {author} {\bibinfo {author} {\bibfnamefont {A.~J.~C.}\
  \bibnamefont {Varandas}}, \bibinfo {author} {\bibfnamefont {J.}~\bibnamefont
  {Tennyson}}, \ and\ \bibinfo {author} {\bibfnamefont {J.~N.}\ \bibnamefont
  {Murrell}},\ }\href@noop {} {\bibfield  {journal} {\bibinfo  {journal} {Chem.
  Phys. Lett.}\ }\textbf {\bibinfo {volume} {61}},\ \bibinfo {pages} {431}
  (\bibinfo {year} {1979})}\BibitemShut {NoStop}%
\bibitem [{\citenamefont {Mukherjee}\ \emph
  {et~al.}(2013{\natexlab{a}})\citenamefont {Mukherjee}, \citenamefont
  {Bandyopadhyay}, \citenamefont {Paul},\ and\ \citenamefont
  {Adhikari}}]{saikat_jpca117}%
  \BibitemOpen
  \bibfield  {author} {\bibinfo {author} {\bibfnamefont {S.}~\bibnamefont
  {Mukherjee}}, \bibinfo {author} {\bibfnamefont {S.}~\bibnamefont
  {Bandyopadhyay}}, \bibinfo {author} {\bibfnamefont {A.~K.}\ \bibnamefont
  {Paul}}, \ and\ \bibinfo {author} {\bibfnamefont {S.}~\bibnamefont
  {Adhikari}},\ }\href@noop {} {\bibfield  {journal} {\bibinfo  {journal} {J.
  Phys. Chem. A}\ }\textbf {\bibinfo {volume} {117}},\ \bibinfo {pages} {3475}
  (\bibinfo {year} {2013}{\natexlab{a}})}\BibitemShut {NoStop}%
\bibitem [{\citenamefont {Neumark}\ \emph {et~al.}(1985)\citenamefont
  {Neumark}, \citenamefont {Wodtke}, \citenamefont {Robinson}, \citenamefont
  {Hayden},\ and\ \citenamefont {Lee}}]{ytlee_jcp82_3045}%
  \BibitemOpen
  \bibfield  {author} {\bibinfo {author} {\bibfnamefont {D.~M.}\ \bibnamefont
  {Neumark}}, \bibinfo {author} {\bibfnamefont {A.~M.}\ \bibnamefont {Wodtke}},
  \bibinfo {author} {\bibfnamefont {G.~N.}\ \bibnamefont {Robinson}}, \bibinfo
  {author} {\bibfnamefont {C.~C.}\ \bibnamefont {Hayden}}, \ and\ \bibinfo
  {author} {\bibfnamefont {Y.~T.}\ \bibnamefont {Lee}},\ }\href@noop {}
  {\bibfield  {journal} {\bibinfo  {journal} {J. Chem. Phys.}\ }\textbf
  {\bibinfo {volume} {82}},\ \bibinfo {pages} {3045} (\bibinfo {year}
  {1985})}\BibitemShut {NoStop}%
\bibitem [{\citenamefont {Castillo}\ \emph {et~al.}(1998)\citenamefont
  {Castillo}, \citenamefont {Hartke}, \citenamefont {Werner}, \citenamefont
  {Aoiz}, \citenamefont {nares},\ and\ \citenamefont
  {Mart\'inez-Haya}}]{werner_jcp109}%
  \BibitemOpen
  \bibfield  {author} {\bibinfo {author} {\bibfnamefont {J.~F.}\ \bibnamefont
  {Castillo}}, \bibinfo {author} {\bibfnamefont {B.}~\bibnamefont {Hartke}},
  \bibinfo {author} {\bibfnamefont {H.-J.}\ \bibnamefont {Werner}}, \bibinfo
  {author} {\bibfnamefont {F.~J.}\ \bibnamefont {Aoiz}}, \bibinfo {author}
  {\bibfnamefont {L.~B.}\ \bibnamefont {nares}}, \ and\ \bibinfo {author}
  {\bibfnamefont {B.}~\bibnamefont {Mart\'inez-Haya}},\ }\href@noop {}
  {\bibfield  {journal} {\bibinfo  {journal} {J. Chem. Phys.}\ }\textbf
  {\bibinfo {volume} {109}},\ \bibinfo {pages} {7224} (\bibinfo {year}
  {1998})}\BibitemShut {NoStop}%
\bibitem [{\citenamefont {Hobey}\ and\ \citenamefont
  {Mclachlan}(1960)}]{hobey_jcp33}%
  \BibitemOpen
  \bibfield  {author} {\bibinfo {author} {\bibfnamefont {W.~D.}\ \bibnamefont
  {Hobey}}\ and\ \bibinfo {author} {\bibfnamefont {A.~D.}\ \bibnamefont
  {Mclachlan}},\ }\href@noop {} {\bibfield  {journal} {\bibinfo  {journal} {J.
  Chem. Phys.}\ }\textbf {\bibinfo {volume} {33}},\ \bibinfo {pages} {1695}
  (\bibinfo {year} {1960})}\BibitemShut {NoStop}%
\bibitem [{\citenamefont {Lichten}(1967)}]{litchen_pr164}%
  \BibitemOpen
  \bibfield  {author} {\bibinfo {author} {\bibfnamefont {W.}~\bibnamefont
  {Lichten}},\ }\href@noop {} {\bibfield  {journal} {\bibinfo  {journal} {Phys.
  Rev.}\ }\textbf {\bibinfo {volume} {164}},\ \bibinfo {pages} {131} (\bibinfo
  {year} {1967})}\BibitemShut {NoStop}%
\bibitem [{\citenamefont {Smith}(1969)}]{smith_pr179}%
  \BibitemOpen
  \bibfield  {author} {\bibinfo {author} {\bibfnamefont {F.~T.}\ \bibnamefont
  {Smith}},\ }\href@noop {} {\bibfield  {journal} {\bibinfo  {journal} {Phys.
  Rev.}\ }\textbf {\bibinfo {volume} {179}},\ \bibinfo {pages} {111} (\bibinfo
  {year} {1969})}\BibitemShut {NoStop}%
\bibitem [{\citenamefont {Thiel}\ and\ \citenamefont
  {K\"oppel}(1999)}]{koppel_jcp110}%
  \BibitemOpen
  \bibfield  {author} {\bibinfo {author} {\bibfnamefont {A.}~\bibnamefont
  {Thiel}}\ and\ \bibinfo {author} {\bibfnamefont {H.}~\bibnamefont
  {K\"oppel}},\ }\href@noop {} {\bibfield  {journal} {\bibinfo  {journal} {J.
  Chem. Phys.}\ }\textbf {\bibinfo {volume} {110}},\ \bibinfo {pages} {9371}
  (\bibinfo {year} {1999})}\BibitemShut {NoStop}%
\bibitem [{\citenamefont {K\"oppel}, \citenamefont {Gronki},\ and\
  \citenamefont {Mahapatra}(2001)}]{koppel_jcp150}%
  \BibitemOpen
  \bibfield  {author} {\bibinfo {author} {\bibfnamefont {H.}~\bibnamefont
  {K\"oppel}}, \bibinfo {author} {\bibfnamefont {J.}~\bibnamefont {Gronki}}, \
  and\ \bibinfo {author} {\bibfnamefont {S.}~\bibnamefont {Mahapatra}},\
  }\href@noop {} {\bibfield  {journal} {\bibinfo  {journal} {J. Chem. Phys.}\
  }\textbf {\bibinfo {volume} {150}},\ \bibinfo {pages} {2377} (\bibinfo {year}
  {2001})}\BibitemShut {NoStop}%
\bibitem [{\citenamefont {Viel}\ and\ \citenamefont
  {Eisfeld}(2004)}]{viel_jcp120}%
  \BibitemOpen
  \bibfield  {author} {\bibinfo {author} {\bibfnamefont {A.}~\bibnamefont
  {Viel}}\ and\ \bibinfo {author} {\bibfnamefont {W.}~\bibnamefont {Eisfeld}},\
  }\href@noop {} {\bibfield  {journal} {\bibinfo  {journal} {J. Chem. Phys.}\
  }\textbf {\bibinfo {volume} {120}},\ \bibinfo {pages} {4603} (\bibinfo {year}
  {2004})}\BibitemShut {NoStop}%
\bibitem [{\citenamefont {Abedi}, \citenamefont {Maitra},\ and\ \citenamefont
  {Gross}(2010)}]{gross_prl105}%
  \BibitemOpen
  \bibfield  {author} {\bibinfo {author} {\bibfnamefont {A.}~\bibnamefont
  {Abedi}}, \bibinfo {author} {\bibfnamefont {N.~T.}\ \bibnamefont {Maitra}}, \
  and\ \bibinfo {author} {\bibfnamefont {E.~K.~U.}\ \bibnamefont {Gross}},\
  }\href@noop {} {\bibfield  {journal} {\bibinfo  {journal} {Phys. Rev. Lett.}\
  }\textbf {\bibinfo {volume} {105}},\ \bibinfo {pages} {123002} (\bibinfo
  {year} {2010})}\BibitemShut {NoStop}%
\bibitem [{\citenamefont {Abedi}, \citenamefont {Maitra},\ and\ \citenamefont
  {Gross}(2012)}]{gross_jcp137}%
  \BibitemOpen
  \bibfield  {author} {\bibinfo {author} {\bibfnamefont {A.}~\bibnamefont
  {Abedi}}, \bibinfo {author} {\bibfnamefont {N.~T.}\ \bibnamefont {Maitra}}, \
  and\ \bibinfo {author} {\bibfnamefont {E.~K.~U.}\ \bibnamefont {Gross}},\
  }\href@noop {} {\bibfield  {journal} {\bibinfo  {journal} {J. Chem. Phys.}\
  }\textbf {\bibinfo {volume} {137}},\ \bibinfo {pages} {22A530} (\bibinfo
  {year} {2012})}\BibitemShut {NoStop}%
\bibitem [{\citenamefont {Worth}\ \emph {et~al.}(2008)\citenamefont {Worth},
  \citenamefont {Meyer}, \citenamefont {K\"oppel}, \citenamefont {Cederbaum},\
  and\ \citenamefont {Burghardt}}]{worth_irpc27}%
  \BibitemOpen
  \bibfield  {author} {\bibinfo {author} {\bibfnamefont {G.~A.}\ \bibnamefont
  {Worth}}, \bibinfo {author} {\bibfnamefont {H.-D.}\ \bibnamefont {Meyer}},
  \bibinfo {author} {\bibfnamefont {H.}~\bibnamefont {K\"oppel}}, \bibinfo
  {author} {\bibfnamefont {L.~S.}\ \bibnamefont {Cederbaum}}, \ and\ \bibinfo
  {author} {\bibfnamefont {I.}~\bibnamefont {Burghardt}},\ }\href@noop {}
  {\bibfield  {journal} {\bibinfo  {journal} {Int. Rev. Phys. Chem.}\ }\textbf
  {\bibinfo {volume} {27}},\ \bibinfo {pages} {569} (\bibinfo {year}
  {2008})}\BibitemShut {NoStop}%
\bibitem [{\citenamefont {Richings}\ and\ \citenamefont
  {Habershon}(2018)}]{richings_jcp148}%
  \BibitemOpen
  \bibfield  {author} {\bibinfo {author} {\bibfnamefont {G.~W.}\ \bibnamefont
  {Richings}}\ and\ \bibinfo {author} {\bibfnamefont {S.}~\bibnamefont
  {Habershon}},\ }\href@noop {} {\bibfield  {journal} {\bibinfo  {journal} {J.
  Chem. Phys.}\ }\textbf {\bibinfo {volume} {148}},\ \bibinfo {pages} {134116}
  (\bibinfo {year} {2018})}\BibitemShut {NoStop}%
\bibitem [{\citenamefont {Richings}\ \emph {et~al.}(2015)\citenamefont
  {Richings}, \citenamefont {Polyak}, \citenamefont {Spinlove}, \citenamefont
  {Worth}, , \citenamefont {Burghardt},\ and\ \citenamefont
  {Lasorne}}]{worth_irpc34}%
  \BibitemOpen
  \bibfield  {author} {\bibinfo {author} {\bibfnamefont {G.~W.}\ \bibnamefont
  {Richings}}, \bibinfo {author} {\bibfnamefont {I.}~\bibnamefont {Polyak}},
  \bibinfo {author} {\bibfnamefont {K.}~\bibnamefont {Spinlove}}, \bibinfo
  {author} {\bibfnamefont {G.~A.}\ \bibnamefont {Worth}}, , \bibinfo {author}
  {\bibfnamefont {I.}~\bibnamefont {Burghardt}}, \ and\ \bibinfo {author}
  {\bibfnamefont {B.}~\bibnamefont {Lasorne}},\ }\href@noop {} {\bibfield
  {journal} {\bibinfo  {journal} {Int. Rev. Phys. Chem.}\ }\textbf {\bibinfo
  {volume} {34}},\ \bibinfo {pages} {269} (\bibinfo {year} {2015})}\BibitemShut
  {NoStop}%
\bibitem [{\citenamefont {Richings}\ and\ \citenamefont
  {Worth}(2015)}]{worth_jpca119}%
  \BibitemOpen
  \bibfield  {author} {\bibinfo {author} {\bibfnamefont {G.~W.}\ \bibnamefont
  {Richings}}\ and\ \bibinfo {author} {\bibfnamefont {G.~A.}\ \bibnamefont
  {Worth}},\ }\href@noop {} {\bibfield  {journal} {\bibinfo  {journal} {J.
  Phys. Chem. A}\ }\textbf {\bibinfo {volume} {119}},\ \bibinfo {pages} {12457}
  (\bibinfo {year} {2015})}\BibitemShut {NoStop}%
\bibitem [{\citenamefont {Mart\'{i}nez}, \citenamefont {Ben-Nun},\ and\
  \citenamefont {Levine}(1996)}]{SPAWN1}%
  \BibitemOpen
  \bibfield  {author} {\bibinfo {author} {\bibfnamefont {T.~J.}\ \bibnamefont
  {Mart\'{i}nez}}, \bibinfo {author} {\bibfnamefont {M.}~\bibnamefont
  {Ben-Nun}}, \ and\ \bibinfo {author} {\bibfnamefont {R.~D.}\ \bibnamefont
  {Levine}},\ }\href@noop {} {\bibfield  {journal} {\bibinfo  {journal} {J.
  Phys. Chem.}\ }\textbf {\bibinfo {volume} {100}},\ \bibinfo {pages} {7884}
  (\bibinfo {year} {1996})}\BibitemShut {NoStop}%
\bibitem [{\citenamefont {Ben-Nun}\ and\ \citenamefont
  {Mart\'{i}nez}(1998)}]{SPAWN2}%
  \BibitemOpen
  \bibfield  {author} {\bibinfo {author} {\bibfnamefont {M.}~\bibnamefont
  {Ben-Nun}}\ and\ \bibinfo {author} {\bibfnamefont {T.~J.}\ \bibnamefont
  {Mart\'{i}nez}},\ }\href@noop {} {\bibfield  {journal} {\bibinfo  {journal}
  {J. Chem. Phys.}\ }\textbf {\bibinfo {volume} {108}},\ \bibinfo {pages}
  {7244} (\bibinfo {year} {1998})}\BibitemShut {NoStop}%
\bibitem [{\citenamefont {Baer}(1975)}]{baer_cpl35}%
  \BibitemOpen
  \bibfield  {author} {\bibinfo {author} {\bibfnamefont {M.}~\bibnamefont
  {Baer}},\ }\href@noop {} {\bibfield  {journal} {\bibinfo  {journal} {Chem.
  Phys. Lett.}\ }\textbf {\bibinfo {volume} {35}},\ \bibinfo {pages} {112}
  (\bibinfo {year} {1975})}\BibitemShut {NoStop}%
\bibitem [{\citenamefont {Baer}(2002)}]{baer_phys_rep358}%
  \BibitemOpen
  \bibfield  {author} {\bibinfo {author} {\bibfnamefont {M.}~\bibnamefont
  {Baer}},\ }\href@noop {} {\bibfield  {journal} {\bibinfo  {journal} {Phys.
  Rep.}\ }\textbf {\bibinfo {volume} {358}},\ \bibinfo {pages} {75} (\bibinfo
  {year} {2002})}\BibitemShut {NoStop}%
\bibitem [{\citenamefont {Mukherjee}, \citenamefont {Mukherjee},\ and\
  \citenamefont {Adhikari}(2017)}]{soumya_jpca121}%
  \BibitemOpen
  \bibfield  {author} {\bibinfo {author} {\bibfnamefont {S.}~\bibnamefont
  {Mukherjee}}, \bibinfo {author} {\bibfnamefont {B.}~\bibnamefont
  {Mukherjee}}, \ and\ \bibinfo {author} {\bibfnamefont {S.}~\bibnamefont
  {Adhikari}},\ }\href@noop {} {\bibfield  {journal} {\bibinfo  {journal} {J.
  Phys. Chem. A}\ }\textbf {\bibinfo {volume} {121}},\ \bibinfo {pages} {6314}
  (\bibinfo {year} {2017})}\BibitemShut {NoStop}%
\bibitem [{\citenamefont {Mukherjee}\ \emph {et~al.}(2020)\citenamefont
  {Mukherjee}, \citenamefont {Naskar}, \citenamefont {Mukherjee}, \citenamefont
  {S.Ravi}, \citenamefont {Shamasundar}, \citenamefont {Mukhopadhyay},\ and\
  \citenamefont {Adhikari}}]{bijit_fh2}%
  \BibitemOpen
  \bibfield  {author} {\bibinfo {author} {\bibfnamefont {B.}~\bibnamefont
  {Mukherjee}}, \bibinfo {author} {\bibfnamefont {K.}~\bibnamefont {Naskar}},
  \bibinfo {author} {\bibfnamefont {S.}~\bibnamefont {Mukherjee}}, \bibinfo
  {author} {\bibnamefont {S.Ravi}}, \bibinfo {author} {\bibfnamefont {K.~R.}\
  \bibnamefont {Shamasundar}}, \bibinfo {author} {\bibfnamefont
  {D.}~\bibnamefont {Mukhopadhyay}}, \ and\ \bibinfo {author} {\bibfnamefont
  {S.}~\bibnamefont {Adhikari}},\ }\href@noop {} {\bibfield  {journal}
  {\bibinfo  {journal} {J. Chem. Phys.}\ }\textbf {\bibinfo {volume} {153}},\
  \bibinfo {pages} {174301} (\bibinfo {year} {2020})}\BibitemShut {NoStop}%
\bibitem [{\citenamefont {Mukherjee}\ \emph
  {et~al.}(2019{\natexlab{a}})\citenamefont {Mukherjee}, \citenamefont
  {Naskar}, \citenamefont {Mukherjee}, \citenamefont {Ghosh}, \citenamefont
  {Sahoo},\ and\ \citenamefont {Adhikari}}]{bijit_irpc38}%
  \BibitemOpen
  \bibfield  {author} {\bibinfo {author} {\bibfnamefont {B.}~\bibnamefont
  {Mukherjee}}, \bibinfo {author} {\bibfnamefont {K.}~\bibnamefont {Naskar}},
  \bibinfo {author} {\bibfnamefont {S.}~\bibnamefont {Mukherjee}}, \bibinfo
  {author} {\bibfnamefont {S.}~\bibnamefont {Ghosh}}, \bibinfo {author}
  {\bibfnamefont {T.}~\bibnamefont {Sahoo}}, \ and\ \bibinfo {author}
  {\bibfnamefont {S.}~\bibnamefont {Adhikari}},\ }\href@noop {} {\bibfield
  {journal} {\bibinfo  {journal} {Int. Rev. Phys. Chem.}\ }\textbf {\bibinfo
  {volume} {38}},\ \bibinfo {pages} {287} (\bibinfo {year}
  {2019}{\natexlab{a}})}\BibitemShut {NoStop}%
\bibitem [{\citenamefont {Naskar}\ \emph {et~al.}(2020)\citenamefont {Naskar},
  \citenamefont {Mukherjee}, \citenamefont {Mukherjee}, \citenamefont {Ravi},
  \citenamefont {Mukherjee}, \citenamefont {Sardar},\ and\ \citenamefont
  {Adhikari}}]{naskar_jctc}%
  \BibitemOpen
  \bibfield  {author} {\bibinfo {author} {\bibfnamefont {K.}~\bibnamefont
  {Naskar}}, \bibinfo {author} {\bibfnamefont {S.}~\bibnamefont {Mukherjee}},
  \bibinfo {author} {\bibfnamefont {B.}~\bibnamefont {Mukherjee}}, \bibinfo
  {author} {\bibfnamefont {S.}~\bibnamefont {Ravi}}, \bibinfo {author}
  {\bibfnamefont {S.}~\bibnamefont {Mukherjee}}, \bibinfo {author}
  {\bibfnamefont {S.}~\bibnamefont {Sardar}}, \ and\ \bibinfo {author}
  {\bibfnamefont {S.}~\bibnamefont {Adhikari}},\ }\href@noop {} {\bibfield
  {journal} {\bibinfo  {journal} {J. Chem. Theory Comput.}\ }\textbf {\bibinfo
  {volume} {16}},\ \bibinfo {pages} {1666} (\bibinfo {year}
  {2020})}\BibitemShut {NoStop}%
\bibitem [{\citenamefont {Baer}(1980)}]{baer_mp40}%
  \BibitemOpen
  \bibfield  {author} {\bibinfo {author} {\bibfnamefont {M.}~\bibnamefont
  {Baer}},\ }\href@noop {} {\bibfield  {journal} {\bibinfo  {journal} {Mol.
  Phys.}\ }\textbf {\bibinfo {volume} {40}},\ \bibinfo {pages} {1011} (\bibinfo
  {year} {1980})}\BibitemShut {NoStop}%
\bibitem [{\citenamefont {Baer}\ and\ \citenamefont
  {Englman}(1992)}]{baer_mp75}%
  \BibitemOpen
  \bibfield  {author} {\bibinfo {author} {\bibfnamefont {M.}~\bibnamefont
  {Baer}}\ and\ \bibinfo {author} {\bibfnamefont {R.}~\bibnamefont {Englman}},\
  }\href@noop {} {\bibfield  {journal} {\bibinfo  {journal} {Mol. Phys.}\
  }\textbf {\bibinfo {volume} {75}},\ \bibinfo {pages} {293} (\bibinfo {year}
  {1992})}\BibitemShut {NoStop}%
\bibitem [{\citenamefont {Kendrick}, \citenamefont {Mead},\ and\ \citenamefont
  {Truhlar}(2002)}]{kendrick_cp277}%
  \BibitemOpen
  \bibfield  {author} {\bibinfo {author} {\bibfnamefont {B.~K.}\ \bibnamefont
  {Kendrick}}, \bibinfo {author} {\bibfnamefont {C.~A.}\ \bibnamefont {Mead}},
  \ and\ \bibinfo {author} {\bibfnamefont {D.~G.}\ \bibnamefont {Truhlar}},\
  }\href@noop {} {\bibfield  {journal} {\bibinfo  {journal} {Chem. Phys.}\
  }\textbf {\bibinfo {volume} {277}},\ \bibinfo {pages} {31} (\bibinfo {year}
  {2002})}\BibitemShut {NoStop}%
\bibitem [{\citenamefont {Matsunaga}\ and\ \citenamefont
  {Yarkony}(1998)}]{yarkony_mp93}%
  \BibitemOpen
  \bibfield  {author} {\bibinfo {author} {\bibfnamefont {N.}~\bibnamefont
  {Matsunaga}}\ and\ \bibinfo {author} {\bibfnamefont {D.~R.}\ \bibnamefont
  {Yarkony}},\ }\href@noop {} {\bibfield  {journal} {\bibinfo  {journal} {Mol.
  Phys.}\ }\textbf {\bibinfo {volume} {93}},\ \bibinfo {pages} {79} (\bibinfo
  {year} {1998})}\BibitemShut {NoStop}%
\bibitem [{\citenamefont {Paul}\ \emph
  {et~al.}(2011{\natexlab{a}})\citenamefont {Paul}, \citenamefont {Ray},
  \citenamefont {Mukhopadhyay},\ and\ \citenamefont
  {Adhikari}}]{akpaul_cpl508}%
  \BibitemOpen
  \bibfield  {author} {\bibinfo {author} {\bibfnamefont {A.~K.}\ \bibnamefont
  {Paul}}, \bibinfo {author} {\bibfnamefont {S.}~\bibnamefont {Ray}}, \bibinfo
  {author} {\bibfnamefont {D.}~\bibnamefont {Mukhopadhyay}}, \ and\ \bibinfo
  {author} {\bibfnamefont {S.}~\bibnamefont {Adhikari}},\ }\href@noop {}
  {\bibfield  {journal} {\bibinfo  {journal} {Chem. Phys. Lett.}\ }\textbf
  {\bibinfo {volume} {508}},\ \bibinfo {pages} {300} (\bibinfo {year}
  {2011}{\natexlab{a}})}\BibitemShut {NoStop}%
\bibitem [{\citenamefont {Paul}\ \emph
  {et~al.}(2011{\natexlab{b}})\citenamefont {Paul}, \citenamefont {Ray},
  \citenamefont {Mukhopadhyay},\ and\ \citenamefont
  {Adhikari}}]{akpaul_jcp135}%
  \BibitemOpen
  \bibfield  {author} {\bibinfo {author} {\bibfnamefont {A.~K.}\ \bibnamefont
  {Paul}}, \bibinfo {author} {\bibfnamefont {S.}~\bibnamefont {Ray}}, \bibinfo
  {author} {\bibfnamefont {D.}~\bibnamefont {Mukhopadhyay}}, \ and\ \bibinfo
  {author} {\bibfnamefont {S.}~\bibnamefont {Adhikari}},\ }\href@noop {}
  {\bibfield  {journal} {\bibinfo  {journal} {J. Chem. Phys.}\ }\textbf
  {\bibinfo {volume} {135}},\ \bibinfo {pages} {034107} (\bibinfo {year}
  {2011}{\natexlab{b}})}\BibitemShut {NoStop}%
\bibitem [{\citenamefont {Mukherjee}\ \emph
  {et~al.}(2013{\natexlab{b}})\citenamefont {Mukherjee}, \citenamefont
  {Bandyopadhyay}, \citenamefont {Paul},\ and\ \citenamefont
  {Adhikari}}]{mukherjee_jpca117}%
  \BibitemOpen
  \bibfield  {author} {\bibinfo {author} {\bibfnamefont {S.}~\bibnamefont
  {Mukherjee}}, \bibinfo {author} {\bibfnamefont {S.}~\bibnamefont
  {Bandyopadhyay}}, \bibinfo {author} {\bibfnamefont {A.~K.}\ \bibnamefont
  {Paul}}, \ and\ \bibinfo {author} {\bibfnamefont {S.}~\bibnamefont
  {Adhikari}},\ }\href@noop {} {\bibfield  {journal} {\bibinfo  {journal} {J.
  Phys. Chem. A}\ }\textbf {\bibinfo {volume} {117}},\ \bibinfo {pages} {3475}
  (\bibinfo {year} {2013}{\natexlab{b}})}\BibitemShut {NoStop}%
\bibitem [{\citenamefont {Mukherjee}\ and\ \citenamefont
  {Adhikari}(2014)}]{mukherjee_cp440}%
  \BibitemOpen
  \bibfield  {author} {\bibinfo {author} {\bibfnamefont {S.}~\bibnamefont
  {Mukherjee}}\ and\ \bibinfo {author} {\bibfnamefont {S.}~\bibnamefont
  {Adhikari}},\ }\href@noop {} {\bibfield  {journal} {\bibinfo  {journal}
  {Chem. Phys.}\ }\textbf {\bibinfo {volume} {440}},\ \bibinfo {pages} {106}
  (\bibinfo {year} {2014})}\BibitemShut {NoStop}%
\bibitem [{\citenamefont {Mukherjee}, \citenamefont {Mukhopadhyay},\ and\
  \citenamefont {Adhikari}(2014)}]{mukherjee_jcp141}%
  \BibitemOpen
  \bibfield  {author} {\bibinfo {author} {\bibfnamefont {S.}~\bibnamefont
  {Mukherjee}}, \bibinfo {author} {\bibfnamefont {D.}~\bibnamefont
  {Mukhopadhyay}}, \ and\ \bibinfo {author} {\bibfnamefont {S.}~\bibnamefont
  {Adhikari}},\ }\href@noop {} {\bibfield  {journal} {\bibinfo  {journal} {J.
  Chem. Phys.}\ }\textbf {\bibinfo {volume} {141}},\ \bibinfo {pages} {204306}
  (\bibinfo {year} {2014})}\BibitemShut {NoStop}%
\bibitem [{\citenamefont {Sardar}\ \emph {et~al.}(2013)\citenamefont {Sardar},
  \citenamefont {Mukherjee}, \citenamefont {Paul},\ and\ \citenamefont
  {Adhikari}}]{sardar_cp416}%
  \BibitemOpen
  \bibfield  {author} {\bibinfo {author} {\bibfnamefont {S.}~\bibnamefont
  {Sardar}}, \bibinfo {author} {\bibfnamefont {S.}~\bibnamefont {Mukherjee}},
  \bibinfo {author} {\bibfnamefont {A.~K.}\ \bibnamefont {Paul}}, \ and\
  \bibinfo {author} {\bibfnamefont {S.}~\bibnamefont {Adhikari}},\ }\href@noop
  {} {\bibfield  {journal} {\bibinfo  {journal} {Chem. Phys.}\ }\textbf
  {\bibinfo {volume} {416}},\ \bibinfo {pages} {11} (\bibinfo {year}
  {2013})}\BibitemShut {NoStop}%
\bibitem [{\citenamefont {Mukherjee}\ \emph {et~al.}(2015)\citenamefont
  {Mukherjee}, \citenamefont {Mukherjee}, \citenamefont {Sardar},\ and\
  \citenamefont {Adhikari}}]{mukherjee_jcp143}%
  \BibitemOpen
  \bibfield  {author} {\bibinfo {author} {\bibfnamefont {S.}~\bibnamefont
  {Mukherjee}}, \bibinfo {author} {\bibfnamefont {B.}~\bibnamefont
  {Mukherjee}}, \bibinfo {author} {\bibfnamefont {S.}~\bibnamefont {Sardar}}, \
  and\ \bibinfo {author} {\bibfnamefont {S.}~\bibnamefont {Adhikari}},\
  }\href@noop {} {\bibfield  {journal} {\bibinfo  {journal} {J. Chem. Phys.}\
  }\textbf {\bibinfo {volume} {143}},\ \bibinfo {pages} {244307} (\bibinfo
  {year} {2015})}\BibitemShut {NoStop}%
\bibitem [{\citenamefont {Mukherjee}\ \emph {et~al.}(2017)\citenamefont
  {Mukherjee}, \citenamefont {Mukherjee}, \citenamefont {Sardar}, \citenamefont
  {Shamasundar},\ and\ \citenamefont {Adhikari}}]{bijit_mp115}%
  \BibitemOpen
  \bibfield  {author} {\bibinfo {author} {\bibfnamefont {B.}~\bibnamefont
  {Mukherjee}}, \bibinfo {author} {\bibfnamefont {S.}~\bibnamefont
  {Mukherjee}}, \bibinfo {author} {\bibfnamefont {S.}~\bibnamefont {Sardar}},
  \bibinfo {author} {\bibfnamefont {K.~R.}\ \bibnamefont {Shamasundar}}, \ and\
  \bibinfo {author} {\bibfnamefont {S.}~\bibnamefont {Adhikari}},\ }\href@noop
  {} {\bibfield  {journal} {\bibinfo  {journal} {Mol. Phys.}\ }\textbf
  {\bibinfo {volume} {115}},\ \bibinfo {pages} {2833} (\bibinfo {year}
  {2017})}\BibitemShut {NoStop}%
\bibitem [{\citenamefont {Mukherjee}\ \emph
  {et~al.}(2019{\natexlab{b}})\citenamefont {Mukherjee}, \citenamefont {Dutta},
  \citenamefont {Mukherjee}, \citenamefont {Sardar},\ and\ \citenamefont
  {Adhikari}}]{soumya_jcp150}%
  \BibitemOpen
  \bibfield  {author} {\bibinfo {author} {\bibfnamefont {S.}~\bibnamefont
  {Mukherjee}}, \bibinfo {author} {\bibfnamefont {J.}~\bibnamefont {Dutta}},
  \bibinfo {author} {\bibfnamefont {B.}~\bibnamefont {Mukherjee}}, \bibinfo
  {author} {\bibfnamefont {S.}~\bibnamefont {Sardar}}, \ and\ \bibinfo {author}
  {\bibfnamefont {S.}~\bibnamefont {Adhikari}},\ }\href@noop {} {\bibfield
  {journal} {\bibinfo  {journal} {J. Chem. Phys.}\ }\textbf {\bibinfo {volume}
  {150}},\ \bibinfo {pages} {064308} (\bibinfo {year}
  {2019}{\natexlab{b}})}\BibitemShut {NoStop}%
\bibitem [{\citenamefont {Werner}\ \emph {et~al.}(2012)\citenamefont {Werner},
  \citenamefont {Knowles}, \citenamefont {Knizia}, \citenamefont {Manby},\ and\
  \citenamefont {Sch{\ifmmode\ddot{u}\else\"{u}\fi}tz}}]{MOLPRO_brief}%
  \BibitemOpen
  \bibfield  {author} {\bibinfo {author} {\bibfnamefont {H.-J.}\ \bibnamefont
  {Werner}}, \bibinfo {author} {\bibfnamefont {P.~J.}\ \bibnamefont {Knowles}},
  \bibinfo {author} {\bibfnamefont {G.}~\bibnamefont {Knizia}}, \bibinfo
  {author} {\bibfnamefont {F.~R.}\ \bibnamefont {Manby}}, \ and\ \bibinfo
  {author} {\bibfnamefont {M.}~\bibnamefont
  {Sch{\ifmmode\ddot{u}\else\"{u}\fi}tz}},\ }\href@noop {} {\bibfield
  {journal} {\bibinfo  {journal} {WIREs Comput. Mol. Sci.}\ }\textbf {\bibinfo
  {volume} {2}},\ \bibinfo {pages} {242} (\bibinfo {year} {2012})}\BibitemShut
  {NoStop}%
\bibitem [{\citenamefont {Lischka}\ \emph {et~al.}(2017)\citenamefont
  {Lischka}, \citenamefont {Shepard}, \citenamefont {Shavitt}, \citenamefont
  {Pitzer}, \citenamefont {Dallos} \emph {et~al.}}]{columbus}%
  \BibitemOpen
  \bibfield  {author} {\bibinfo {author} {\bibfnamefont {H.}~\bibnamefont
  {Lischka}}, \bibinfo {author} {\bibfnamefont {R.}~\bibnamefont {Shepard}},
  \bibinfo {author} {\bibfnamefont {I.}~\bibnamefont {Shavitt}}, \bibinfo
  {author} {\bibfnamefont {R.~M.}\ \bibnamefont {Pitzer}}, \bibinfo {author}
  {\bibfnamefont {M.}~\bibnamefont {Dallos}},  \emph {et~al.},\ }\href@noop {}
  {\enquote {\bibinfo {title} {Columbus, an ab initio electronic structure
  program, release 7.0},}\ } (\bibinfo {year} {2017})\BibitemShut {NoStop}%
\bibitem [{\citenamefont {Schmidt}\ \emph {et~al.}(1993)\citenamefont
  {Schmidt}, \citenamefont {Baldridge}, \citenamefont {Boatz}, \citenamefont
  {Elbert}, \citenamefont {Gordon} \emph {et~al.}}]{gamess}%
  \BibitemOpen
  \bibfield  {author} {\bibinfo {author} {\bibfnamefont {M.~W.}\ \bibnamefont
  {Schmidt}}, \bibinfo {author} {\bibfnamefont {K.~K.}\ \bibnamefont
  {Baldridge}}, \bibinfo {author} {\bibfnamefont {J.~A.}\ \bibnamefont
  {Boatz}}, \bibinfo {author} {\bibfnamefont {S.~T.}\ \bibnamefont {Elbert}},
  \bibinfo {author} {\bibfnamefont {M.~S.}\ \bibnamefont {Gordon}},  \emph
  {et~al.},\ }\href@noop {} {\bibfield  {journal} {\bibinfo  {journal} {J.
  Comput. Chem.}\ }\textbf {\bibinfo {volume} {14}},\ \bibinfo {pages} {1347}
  (\bibinfo {year} {1993})}\BibitemShut {NoStop}%
\bibitem [{\citenamefont {Frisch}\ \emph {et~al.}(2016)\citenamefont {Frisch},
  \citenamefont {Trucks}, \citenamefont {Schlegel}, \citenamefont {Scuseria},
  \citenamefont {Robb} \emph {et~al.}}]{g16}%
  \BibitemOpen
  \bibfield  {author} {\bibinfo {author} {\bibfnamefont {M.~J.}\ \bibnamefont
  {Frisch}}, \bibinfo {author} {\bibfnamefont {G.~W.}\ \bibnamefont {Trucks}},
  \bibinfo {author} {\bibfnamefont {H.~B.}\ \bibnamefont {Schlegel}}, \bibinfo
  {author} {\bibfnamefont {G.~E.}\ \bibnamefont {Scuseria}}, \bibinfo {author}
  {\bibfnamefont {M.~A.}\ \bibnamefont {Robb}},  \emph {et~al.},\ }\href@noop
  {} {\enquote {\bibinfo {title} {Gaussian˜16 {R}evision {C}.01},}\ }
  (\bibinfo {year} {2016}),\ \bibinfo {note} {gaussian Inc. Wallingford
  CT}\BibitemShut {NoStop}%
\bibitem [{\citenamefont {Sah}\ \emph {et~al.}(2023)\citenamefont {Sah},
  \citenamefont {Mukherjee}, \citenamefont {Naskar}, \citenamefont {Hazra},\
  and\ \citenamefont {Adhikari}}]{mantu_ijqc_curl}%
  \BibitemOpen
  \bibfield  {author} {\bibinfo {author} {\bibfnamefont {M.~K.}\ \bibnamefont
  {Sah}}, \bibinfo {author} {\bibfnamefont {S.}~\bibnamefont {Mukherjee}},
  \bibinfo {author} {\bibfnamefont {K.}~\bibnamefont {Naskar}}, \bibinfo
  {author} {\bibfnamefont {S.}~\bibnamefont {Hazra}}, \ and\ \bibinfo {author}
  {\bibfnamefont {S.}~\bibnamefont {Adhikari}},\ }\href@noop {} {\bibfield
  {journal} {\bibinfo  {journal} {International Journal of Quantum Chemistry}\
  }\textbf {\bibinfo {volume} {123}},\ \bibinfo {pages} {e27212} (\bibinfo
  {year} {2023})}\BibitemShut {NoStop}%
\bibitem [{\citenamefont {Mukherjee}\ \emph
  {et~al.}(2018{\natexlab{b}})\citenamefont {Mukherjee}, \citenamefont
  {Mukherjee}, \citenamefont {Dutta}, \citenamefont {Sardar},\ and\
  \citenamefont {Adhikari}}]{soumya_acs}%
  \BibitemOpen
  \bibfield  {author} {\bibinfo {author} {\bibfnamefont {S.}~\bibnamefont
  {Mukherjee}}, \bibinfo {author} {\bibfnamefont {B.}~\bibnamefont
  {Mukherjee}}, \bibinfo {author} {\bibfnamefont {J.}~\bibnamefont {Dutta}},
  \bibinfo {author} {\bibfnamefont {S.}~\bibnamefont {Sardar}}, \ and\ \bibinfo
  {author} {\bibfnamefont {S.}~\bibnamefont {Adhikari}},\ }\href@noop {}
  {\bibfield  {journal} {\bibinfo  {journal} {ACS omega}\ }\textbf {\bibinfo
  {volume} {3}},\ \bibinfo {pages} {12465} (\bibinfo {year}
  {2018}{\natexlab{b}})}\BibitemShut {NoStop}%
\bibitem [{\citenamefont {Mukherjee}\ \emph
  {et~al.}(2019{\natexlab{c}})\citenamefont {Mukherjee}, \citenamefont
  {Mukherjee}, \citenamefont {Sardar},\ and\ \citenamefont
  {Adhikari}}]{soumya_comptc}%
  \BibitemOpen
  \bibfield  {author} {\bibinfo {author} {\bibfnamefont {S.}~\bibnamefont
  {Mukherjee}}, \bibinfo {author} {\bibfnamefont {B.}~\bibnamefont
  {Mukherjee}}, \bibinfo {author} {\bibfnamefont {S.}~\bibnamefont {Sardar}}, \
  and\ \bibinfo {author} {\bibfnamefont {S.}~\bibnamefont {Adhikari}},\
  }\href@noop {} {\bibfield  {journal} {\bibinfo  {journal} {Comp. Theor.
  Chem.}\ }\textbf {\bibinfo {volume} {1154}},\ \bibinfo {pages} {57} (\bibinfo
  {year} {2019}{\natexlab{c}})}\BibitemShut {NoStop}%
\bibitem [{\citenamefont {Dutta}\ \emph {et~al.}(2020)\citenamefont {Dutta},
  \citenamefont {Mukherjee}, \citenamefont {Naskar}, \citenamefont {Ghosh},
  \citenamefont {Mukherjee}, \citenamefont {Ravi},\ and\ \citenamefont
  {Adhikari}}]{joy_pccp}%
  \BibitemOpen
  \bibfield  {author} {\bibinfo {author} {\bibfnamefont {J.}~\bibnamefont
  {Dutta}}, \bibinfo {author} {\bibfnamefont {S.}~\bibnamefont {Mukherjee}},
  \bibinfo {author} {\bibfnamefont {K.}~\bibnamefont {Naskar}}, \bibinfo
  {author} {\bibfnamefont {S.}~\bibnamefont {Ghosh}}, \bibinfo {author}
  {\bibfnamefont {B.}~\bibnamefont {Mukherjee}}, \bibinfo {author}
  {\bibfnamefont {S.}~\bibnamefont {Ravi}}, \ and\ \bibinfo {author}
  {\bibfnamefont {S.}~\bibnamefont {Adhikari}},\ }\href@noop {} {\bibfield
  {journal} {\bibinfo  {journal} {Phys. Chem. Chem. Phys. (Perspective)}\
  }\textbf {\bibinfo {volume} {22}},\ \bibinfo {pages} {27496} (\bibinfo {year}
  {2020})}\BibitemShut {NoStop}%
\bibitem [{\citenamefont {Hazra}\ \emph {et~al.}(2022)\citenamefont {Hazra},
  \citenamefont {Mukherjee}, \citenamefont {Ravi}, \citenamefont {Sardar},\
  and\ \citenamefont {Adhikari}}]{saikat_pyra}%
  \BibitemOpen
  \bibfield  {author} {\bibinfo {author} {\bibfnamefont {S.}~\bibnamefont
  {Hazra}}, \bibinfo {author} {\bibfnamefont {S.}~\bibnamefont {Mukherjee}},
  \bibinfo {author} {\bibfnamefont {S.}~\bibnamefont {Ravi}}, \bibinfo {author}
  {\bibfnamefont {S.}~\bibnamefont {Sardar}}, \ and\ \bibinfo {author}
  {\bibfnamefont {S.}~\bibnamefont {Adhikari}},\ }\href@noop {} {\bibfield
  {journal} {\bibinfo  {journal} {ChemPhysChem}\ }\textbf {\bibinfo {volume}
  {23}},\ \bibinfo {pages} {e202200482} (\bibinfo {year} {2022})}\BibitemShut
  {NoStop}%
\bibitem [{\citenamefont {Ghosh}\ \emph {et~al.}(2017)\citenamefont {Ghosh},
  \citenamefont {Mukherjee}, \citenamefont {Mukherjee}, \citenamefont {Mandal},
  \citenamefont {Sharma}, \citenamefont {Chaudhury},\ and\ \citenamefont
  {Adhikari}}]{jcp-147-074105-SA}%
  \BibitemOpen
  \bibfield  {author} {\bibinfo {author} {\bibfnamefont {S.}~\bibnamefont
  {Ghosh}}, \bibinfo {author} {\bibfnamefont {S.}~\bibnamefont {Mukherjee}},
  \bibinfo {author} {\bibfnamefont {B.}~\bibnamefont {Mukherjee}}, \bibinfo
  {author} {\bibfnamefont {S.}~\bibnamefont {Mandal}}, \bibinfo {author}
  {\bibfnamefont {R.}~\bibnamefont {Sharma}}, \bibinfo {author} {\bibfnamefont
  {P.}~\bibnamefont {Chaudhury}}, \ and\ \bibinfo {author} {\bibfnamefont
  {S.}~\bibnamefont {Adhikari}},\ }\href@noop {} {\bibfield  {journal}
  {\bibinfo  {journal} {J. Chem. Phys.}\ }\textbf {\bibinfo {volume} {147}},\
  \bibinfo {pages} {074105/1} (\bibinfo {year} {2017})}\BibitemShut {NoStop}%
\bibitem [{\citenamefont {Jahn}\ and\ \citenamefont
  {Teller}(1937)}]{JT_prs161}%
  \BibitemOpen
  \bibfield  {author} {\bibinfo {author} {\bibfnamefont {H.~A.}\ \bibnamefont
  {Jahn}}\ and\ \bibinfo {author} {\bibfnamefont {E.}~\bibnamefont {Teller}},\
  }\href@noop {} {\bibfield  {journal} {\bibinfo  {journal} {Proc. Roy. Soc.
  (London)}\ }\textbf {\bibinfo {volume} {161}},\ \bibinfo {pages} {220}
  (\bibinfo {year} {1937})}\BibitemShut {NoStop}%
\bibitem [{\citenamefont {Bersuker}(2006)}]{bersuker_book}%
  \BibitemOpen
  \bibfield  {author} {\bibinfo {author} {\bibfnamefont {I.~B.}\ \bibnamefont
  {Bersuker}},\ }\href@noop {} {\emph {\bibinfo {title} {{\it The Jahn-Teller
  Effect}}}}\ (\bibinfo  {publisher} {Cambridge University Press},\ \bibinfo
  {year} {2006})\BibitemShut {NoStop}%
\bibitem [{\citenamefont {Garc\'{\i}a-Fern\'andez}\ \emph
  {et~al.}(2005)\citenamefont {Garc\'{\i}a-Fern\'andez}, \citenamefont
  {Bersuker}, \citenamefont {Aramburu}, \citenamefont {Barriuso},\ and\
  \citenamefont {Moreno}}]{Pablo_prb71}%
  \BibitemOpen
  \bibfield  {author} {\bibinfo {author} {\bibfnamefont {P.}~\bibnamefont
  {Garc\'{\i}a-Fern\'andez}}, \bibinfo {author} {\bibfnamefont {I.~B.}\
  \bibnamefont {Bersuker}}, \bibinfo {author} {\bibfnamefont {J.~A.}\
  \bibnamefont {Aramburu}}, \bibinfo {author} {\bibfnamefont {M.~T.}\
  \bibnamefont {Barriuso}}, \ and\ \bibinfo {author} {\bibfnamefont
  {M.}~\bibnamefont {Moreno}},\ }\href@noop {} {\bibfield  {journal} {\bibinfo
  {journal} {Phys. Rev. B}\ }\textbf {\bibinfo {volume} {71}},\ \bibinfo
  {pages} {184117} (\bibinfo {year} {2005})}\BibitemShut {NoStop}%
\bibitem [{\citenamefont {Renner}(1934)}]{renner_zp92}%
  \BibitemOpen
  \bibfield  {author} {\bibinfo {author} {\bibfnamefont {E.}~\bibnamefont
  {Renner}},\ }\href@noop {} {\bibfield  {journal} {\bibinfo  {journal} {Z.
  Phys.}\ }\textbf {\bibinfo {volume} {92}},\ \bibinfo {pages} {172} (\bibinfo
  {year} {1934})}\BibitemShut {NoStop}%
\bibitem [{\citenamefont {Pae}\ and\ \citenamefont
  {Hizhnyakov}(2017)}]{Hizhnyakov_jcp147}%
  \BibitemOpen
  \bibfield  {author} {\bibinfo {author} {\bibfnamefont {K.}~\bibnamefont
  {Pae}}\ and\ \bibinfo {author} {\bibfnamefont {V.}~\bibnamefont
  {Hizhnyakov}},\ }\href@noop {} {\bibfield  {journal} {\bibinfo  {journal} {J.
  Chem. Phys.}\ }\textbf {\bibinfo {volume} {147}},\ \bibinfo {pages} {084107}
  (\bibinfo {year} {2017})}\BibitemShut {NoStop}%
\bibitem [{\citenamefont {Mukherjee}\ \emph
  {et~al.}(2018{\natexlab{c}})\citenamefont {Mukherjee}, \citenamefont
  {Mukhopadhyay}, \citenamefont {Adhikari},\ and\ \citenamefont
  {Baer}}]{bijit_mp116}%
  \BibitemOpen
  \bibfield  {author} {\bibinfo {author} {\bibfnamefont {B.}~\bibnamefont
  {Mukherjee}}, \bibinfo {author} {\bibfnamefont {D.}~\bibnamefont
  {Mukhopadhyay}}, \bibinfo {author} {\bibfnamefont {S.}~\bibnamefont
  {Adhikari}}, \ and\ \bibinfo {author} {\bibfnamefont {M.}~\bibnamefont
  {Baer}},\ }\href@noop {} {\bibfield  {journal} {\bibinfo  {journal} {Mol.
  Phys.}\ }\textbf {\bibinfo {volume} {116}},\ \bibinfo {pages} {2435}
  (\bibinfo {year} {2018}{\natexlab{c}})}\BibitemShut {NoStop}%
\bibitem [{\citenamefont {Sah}\ \emph {et~al.}(2025)\citenamefont {Sah},
  \citenamefont {Mukherjee}, \citenamefont {Ravi},\ and\ \citenamefont
  {Adhikari}}]{mantu_tio6_jcp}%
  \BibitemOpen
  \bibfield  {author} {\bibinfo {author} {\bibfnamefont {M.~K.}\ \bibnamefont
  {Sah}}, \bibinfo {author} {\bibfnamefont {S.}~\bibnamefont {Mukherjee}},
  \bibinfo {author} {\bibfnamefont {S.}~\bibnamefont {Ravi}}, \ and\ \bibinfo
  {author} {\bibfnamefont {S.}~\bibnamefont {Adhikari}},\ }\href@noop {}
  {\bibfield  {journal} {\bibinfo  {journal} {The Journal of Chemical Physics}\
  }\textbf {\bibinfo {volume} {under review}} (\bibinfo {year}
  {2025})}\BibitemShut {NoStop}%
\bibitem [{\citenamefont {Naskar}\ \emph
  {et~al.}(2023{\natexlab{a}})\citenamefont {Naskar}, \citenamefont {Ravi},
  \citenamefont {Adhikari}, \citenamefont {Baer},\ and\ \citenamefont
  {Sathyamurthy}}]{naskar_jpca_heh2+}%
  \BibitemOpen
  \bibfield  {author} {\bibinfo {author} {\bibfnamefont {K.}~\bibnamefont
  {Naskar}}, \bibinfo {author} {\bibfnamefont {S.}~\bibnamefont {Ravi}},
  \bibinfo {author} {\bibfnamefont {S.}~\bibnamefont {Adhikari}}, \bibinfo
  {author} {\bibfnamefont {M.}~\bibnamefont {Baer}}, \ and\ \bibinfo {author}
  {\bibfnamefont {N.}~\bibnamefont {Sathyamurthy}},\ }\href@noop {} {\bibfield
  {journal} {\bibinfo  {journal} {J. Phys. Chem. A}\ }\textbf {\bibinfo
  {volume} {127}},\ \bibinfo {pages} {3832} (\bibinfo {year}
  {2023}{\natexlab{a}})}\BibitemShut {NoStop}%
\bibitem [{\citenamefont {Naskar}\ \emph
  {et~al.}(2023{\natexlab{b}})\citenamefont {Naskar}, \citenamefont {Ghosh},
  \citenamefont {Adhikari}, \citenamefont {Baer},\ and\ \citenamefont
  {Sathyamurthy}}]{koushik_heh2+_dyna}%
  \BibitemOpen
  \bibfield  {author} {\bibinfo {author} {\bibfnamefont {K.}~\bibnamefont
  {Naskar}}, \bibinfo {author} {\bibfnamefont {S.}~\bibnamefont {Ghosh}},
  \bibinfo {author} {\bibfnamefont {S.}~\bibnamefont {Adhikari}}, \bibinfo
  {author} {\bibfnamefont {M.}~\bibnamefont {Baer}}, \ and\ \bibinfo {author}
  {\bibfnamefont {N.}~\bibnamefont {Sathyamurthy}},\ }\href@noop {} {\bibfield
  {journal} {\bibinfo  {journal} {J. Chem. Phys.}\ }\textbf {\bibinfo {volume}
  {159}},\ \bibinfo {pages} {034302} (\bibinfo {year}
  {2023}{\natexlab{b}})}\BibitemShut {NoStop}%
\bibitem [{\citenamefont {Naskar}\ \emph {et~al.}(2024)\citenamefont {Naskar},
  \citenamefont {Mukherjee}, \citenamefont {Ghosh},\ and\ \citenamefont
  {Adhikari}}]{koushik_fh2_dyna}%
  \BibitemOpen
  \bibfield  {author} {\bibinfo {author} {\bibfnamefont {K.}~\bibnamefont
  {Naskar}}, \bibinfo {author} {\bibfnamefont {S.}~\bibnamefont {Mukherjee}},
  \bibinfo {author} {\bibfnamefont {S.}~\bibnamefont {Ghosh}}, \ and\ \bibinfo
  {author} {\bibfnamefont {S.}~\bibnamefont {Adhikari}},\ }\href@noop {}
  {\bibfield  {journal} {\bibinfo  {journal} {The Journal of Physical Chemistry
  A}\ }\textbf {\bibinfo {volume} {128}},\ \bibinfo {pages} {1438} (\bibinfo
  {year} {2024})}\BibitemShut {NoStop}%
\bibitem [{\citenamefont {Michelsen}, \citenamefont {Rettner},\ and\
  \citenamefont {Auerbach}(1992)}]{HA2}%
  \BibitemOpen
  \bibfield  {author} {\bibinfo {author} {\bibfnamefont {H.~A.}\ \bibnamefont
  {Michelsen}}, \bibinfo {author} {\bibfnamefont {C.~T.}\ \bibnamefont
  {Rettner}}, \ and\ \bibinfo {author} {\bibfnamefont {D.~J.}\ \bibnamefont
  {Auerbach}},\ }\href@noop {} {\bibfield  {journal} {\bibinfo  {journal}
  {Surf.\ Sci.}\ }\textbf {\bibinfo {volume} {272}},\ \bibinfo {pages} {65}
  (\bibinfo {year} {1992})}\BibitemShut {NoStop}%
\bibitem [{\citenamefont {Rettner}, \citenamefont {Auerbach},\ and\
  \citenamefont {Michelsen}(1992)}]{CTD}%
  \BibitemOpen
  \bibfield  {author} {\bibinfo {author} {\bibfnamefont {C.~T.}\ \bibnamefont
  {Rettner}}, \bibinfo {author} {\bibfnamefont {D.~J.}\ \bibnamefont
  {Auerbach}}, \ and\ \bibinfo {author} {\bibfnamefont {H.~A.}\ \bibnamefont
  {Michelsen}},\ }\href@noop {} {\bibfield  {journal} {\bibinfo  {journal}
  {Phys.\ Rev.\ Lett.}\ }\textbf {\bibinfo {volume} {68}},\ \bibinfo {pages}
  {1164} (\bibinfo {year} {1992})}\BibitemShut {NoStop}%
\bibitem [{\citenamefont {Rettner}, \citenamefont {Michelsen},\ and\
  \citenamefont {Auerbach}(1993{\natexlab{a}})}]{CTH}%
  \BibitemOpen
  \bibfield  {author} {\bibinfo {author} {\bibfnamefont {C.~T.}\ \bibnamefont
  {Rettner}}, \bibinfo {author} {\bibfnamefont {H.~A.}\ \bibnamefont
  {Michelsen}}, \ and\ \bibinfo {author} {\bibfnamefont {D.~J.}\ \bibnamefont
  {Auerbach}},\ }\href@noop {} {\bibfield  {journal} {\bibinfo  {journal}
  {Faraday Discuss.}\ }\textbf {\bibinfo {volume} {96}},\ \bibinfo {pages} {17}
  (\bibinfo {year} {1993}{\natexlab{a}})}\BibitemShut {NoStop}%
\bibitem [{\citenamefont {Rettner}, \citenamefont {Michelsen},\ and\
  \citenamefont {Auerbach}(1993{\natexlab{b}})}]{CTR2}%
  \BibitemOpen
  \bibfield  {author} {\bibinfo {author} {\bibfnamefont {C.~T.}\ \bibnamefont
  {Rettner}}, \bibinfo {author} {\bibfnamefont {H.~A.}\ \bibnamefont
  {Michelsen}}, \ and\ \bibinfo {author} {\bibfnamefont {D.~J.}\ \bibnamefont
  {Auerbach}},\ }\href@noop {} {\bibfield  {journal} {\bibinfo  {journal}
  {Chem. Phys.}\ }\textbf {\bibinfo {volume} {175}},\ \bibinfo {pages} {157}
  (\bibinfo {year} {1993}{\natexlab{b}})}\BibitemShut {NoStop}%
\bibitem [{\citenamefont {Rettner}, \citenamefont {Michelsen},\ and\
  \citenamefont {Auerbach}(1995)}]{CTR}%
  \BibitemOpen
  \bibfield  {author} {\bibinfo {author} {\bibfnamefont {C.~T.}\ \bibnamefont
  {Rettner}}, \bibinfo {author} {\bibfnamefont {H.~A.}\ \bibnamefont
  {Michelsen}}, \ and\ \bibinfo {author} {\bibfnamefont {D.~J.}\ \bibnamefont
  {Auerbach}},\ }\href@noop {} {\bibfield  {journal} {\bibinfo  {journal} {J.\
  Chem.\ Phys.}\ }\textbf {\bibinfo {volume} {102}},\ \bibinfo {pages} {4625}
  (\bibinfo {year} {1995})}\BibitemShut {NoStop}%
\bibitem [{\citenamefont {Michelsen}\ \emph {et~al.}(1993)\citenamefont
  {Michelsen}, \citenamefont {Rettner}, \citenamefont {Auerbach},\ and\
  \citenamefont {Zare}}]{HAM}%
  \BibitemOpen
  \bibfield  {author} {\bibinfo {author} {\bibfnamefont {H.~A.}\ \bibnamefont
  {Michelsen}}, \bibinfo {author} {\bibfnamefont {C.~T.}\ \bibnamefont
  {Rettner}}, \bibinfo {author} {\bibfnamefont {D.~J.}\ \bibnamefont
  {Auerbach}}, \ and\ \bibinfo {author} {\bibfnamefont {R.~N.}\ \bibnamefont
  {Zare}},\ }\href@noop {} {\bibfield  {journal} {\bibinfo  {journal} {J.\
  Chem.\ Phys.}\ }\textbf {\bibinfo {volume} {98}},\ \bibinfo {pages} {8294}
  (\bibinfo {year} {1993})}\BibitemShut {NoStop}%
\bibitem [{\citenamefont {Hou}\ \emph {et~al.}(1997)\citenamefont {Hou},
  \citenamefont {Gulding}, \citenamefont {Rettner}, \citenamefont {Wodtke},\
  and\ \citenamefont {Auerbach}}]{EXPNEW2}%
  \BibitemOpen
  \bibfield  {author} {\bibinfo {author} {\bibfnamefont {H.}~\bibnamefont
  {Hou}}, \bibinfo {author} {\bibfnamefont {S.~J.}\ \bibnamefont {Gulding}},
  \bibinfo {author} {\bibfnamefont {C.~T.}\ \bibnamefont {Rettner}}, \bibinfo
  {author} {\bibfnamefont {A.~M.}\ \bibnamefont {Wodtke}}, \ and\ \bibinfo
  {author} {\bibfnamefont {D.~J.}\ \bibnamefont {Auerbach}},\ }\href@noop {}
  {\bibfield  {journal} {\bibinfo  {journal} {Science.}\ }\textbf {\bibinfo
  {volume} {277}},\ \bibinfo {pages} {80} (\bibinfo {year} {1997})}\BibitemShut
  {NoStop}%
\bibitem [{\citenamefont {Murphy}\ and\ \citenamefont
  {Hodgson}(1998)}]{EXPNEW3}%
  \BibitemOpen
  \bibfield  {author} {\bibinfo {author} {\bibfnamefont {M.~J.}\ \bibnamefont
  {Murphy}}\ and\ \bibinfo {author} {\bibfnamefont {A.}~\bibnamefont
  {Hodgson}},\ }\href@noop {} {\bibfield  {journal} {\bibinfo  {journal} {J.
  Chem. Phys.}\ }\textbf {\bibinfo {volume} {108}},\ \bibinfo {pages} {4199}
  (\bibinfo {year} {1998})}\BibitemShut {NoStop}%
\bibitem [{\citenamefont {Kaufmann}\ \emph {et~al.}(2018)\citenamefont
  {Kaufmann}, \citenamefont {Shuai}, \citenamefont {Auerbach}, \citenamefont
  {Schwarzer},\ and\ \citenamefont {Wodtke}}]{Wodtke}%
  \BibitemOpen
  \bibfield  {author} {\bibinfo {author} {\bibfnamefont {S.}~\bibnamefont
  {Kaufmann}}, \bibinfo {author} {\bibfnamefont {Q.}~\bibnamefont {Shuai}},
  \bibinfo {author} {\bibfnamefont {D.~J.}\ \bibnamefont {Auerbach}}, \bibinfo
  {author} {\bibfnamefont {D.}~\bibnamefont {Schwarzer}}, \ and\ \bibinfo
  {author} {\bibfnamefont {A.~M.}\ \bibnamefont {Wodtke}},\ }\href@noop {}
  {\bibfield  {journal} {\bibinfo  {journal} {J. Chem. Phys.}\ }\textbf
  {\bibinfo {volume} {148}},\ \bibinfo {pages} {194703} (\bibinfo {year}
  {2018})}\BibitemShut {NoStop}%
\bibitem [{\citenamefont {Chadwick}\ and\ \citenamefont
  {Alexandrowicz}(2024)}]{Chadwick}%
  \BibitemOpen
  \bibfield  {author} {\bibinfo {author} {\bibfnamefont {H.}~\bibnamefont
  {Chadwick}}\ and\ \bibinfo {author} {\bibfnamefont {G.}~\bibnamefont
  {Alexandrowicz}},\ }\href@noop {} {\bibfield  {journal} {\bibinfo  {journal}
  {Faraday Discuss.}\ }\textbf {\bibinfo {volume} {251}},\ \bibinfo {pages}
  {76} (\bibinfo {year} {2024})}\BibitemShut {NoStop}%
\bibitem [{\citenamefont {Wang}, \citenamefont {Darling},\ and\ \citenamefont
  {Holloway}(2001)}]{HO1}%
  \BibitemOpen
  \bibfield  {author} {\bibinfo {author} {\bibfnamefont {Z.~S.}\ \bibnamefont
  {Wang}}, \bibinfo {author} {\bibfnamefont {G.~R.}\ \bibnamefont {Darling}}, \
  and\ \bibinfo {author} {\bibfnamefont {S.}~\bibnamefont {Holloway}},\
  }\href@noop {} {\bibfield  {journal} {\bibinfo  {journal} {Phys.\ Rev.\
  Lett.}\ }\textbf {\bibinfo {volume} {87}},\ \bibinfo {pages} {226102/1}
  (\bibinfo {year} {2001})}\BibitemShut {NoStop}%
\bibitem [{\citenamefont {D\'{i}az}\ \emph {et~al.}(2010)\citenamefont
  {D\'{i}az}, \citenamefont {Olsen}, \citenamefont {Auerbach},\ and\
  \citenamefont {Kroes}}]{CDGJK}%
  \BibitemOpen
  \bibfield  {author} {\bibinfo {author} {\bibfnamefont {C.}~\bibnamefont
  {D\'{i}az}}, \bibinfo {author} {\bibfnamefont {R.~A.}\ \bibnamefont {Olsen}},
  \bibinfo {author} {\bibfnamefont {D.~J.}\ \bibnamefont {Auerbach}}, \ and\
  \bibinfo {author} {\bibfnamefont {G.~J.}\ \bibnamefont {Kroes}},\ }\href@noop
  {} {\bibfield  {journal} {\bibinfo  {journal} {Phys. Chem. Chem. Phys.}\
  }\textbf {\bibinfo {volume} {12}},\ \bibinfo {pages} {6499} (\bibinfo {year}
  {2010})}\BibitemShut {NoStop}%
\bibitem [{\citenamefont {Hand}\ and\ \citenamefont {Haris}(1990)}]{HA1}%
  \BibitemOpen
  \bibfield  {author} {\bibinfo {author} {\bibfnamefont {M.}~\bibnamefont
  {Hand}}\ and\ \bibinfo {author} {\bibfnamefont {J.}~\bibnamefont {Haris}},\
  }\href@noop {} {\bibfield  {journal} {\bibinfo  {journal} {J.\ Chem.\ Phys.}\
  }\textbf {\bibinfo {volume} {92}},\ \bibinfo {pages} {7610} (\bibinfo {year}
  {1990})}\BibitemShut {NoStop}%
\bibitem [{\citenamefont {Dohle}\ and\ \citenamefont {Saalfrank}(1997)}]{SAL}%
  \BibitemOpen
  \bibfield  {author} {\bibinfo {author} {\bibfnamefont {M.}~\bibnamefont
  {Dohle}}\ and\ \bibinfo {author} {\bibfnamefont {P.}~\bibnamefont
  {Saalfrank}},\ }\href@noop {} {\bibfield  {journal} {\bibinfo  {journal}
  {Surf.\ Sci.}\ }\textbf {\bibinfo {volume} {373}},\ \bibinfo {pages} {95}
  (\bibinfo {year} {1997})}\BibitemShut {NoStop}%
\bibitem [{\citenamefont {Adhikari}\ and\ \citenamefont {Billing}(2000)}]{SAG}%
  \BibitemOpen
  \bibfield  {author} {\bibinfo {author} {\bibfnamefont {S.}~\bibnamefont
  {Adhikari}}\ and\ \bibinfo {author} {\bibfnamefont {G.~D.}\ \bibnamefont
  {Billing}},\ }\href@noop {} {\bibfield  {journal} {\bibinfo  {journal} {J.\
  Chem.\ Phys.}\ }\textbf {\bibinfo {volume} {112}},\ \bibinfo {pages} {3884}
  (\bibinfo {year} {2000})}\BibitemShut {NoStop}%
\bibitem [{\citenamefont {Sahoo}, \citenamefont {Sardar},\ and\ \citenamefont
  {Adhikari}(2011{\natexlab{a}})}]{TSS}%
  \BibitemOpen
  \bibfield  {author} {\bibinfo {author} {\bibfnamefont {T.}~\bibnamefont
  {Sahoo}}, \bibinfo {author} {\bibfnamefont {S.}~\bibnamefont {Sardar}}, \
  and\ \bibinfo {author} {\bibfnamefont {S.}~\bibnamefont {Adhikari}},\
  }\href@noop {} {\bibfield  {journal} {\bibinfo  {journal} {Phys. Chem. Chem.
  Phys.}\ }\textbf {\bibinfo {volume} {13}},\ \bibinfo {pages} {10100}
  (\bibinfo {year} {2011}{\natexlab{a}})}\BibitemShut {NoStop}%
\bibitem [{\citenamefont {Sahoo}\ \emph {et~al.}(2011)\citenamefont {Sahoo},
  \citenamefont {Sardar}, \citenamefont {Mondal}, \citenamefont {Sarkar},\ and\
  \citenamefont {Adhikari}}]{TSS1}%
  \BibitemOpen
  \bibfield  {author} {\bibinfo {author} {\bibfnamefont {T.}~\bibnamefont
  {Sahoo}}, \bibinfo {author} {\bibfnamefont {S.}~\bibnamefont {Sardar}},
  \bibinfo {author} {\bibfnamefont {P.}~\bibnamefont {Mondal}}, \bibinfo
  {author} {\bibfnamefont {B.}~\bibnamefont {Sarkar}}, \ and\ \bibinfo {author}
  {\bibfnamefont {S.}~\bibnamefont {Adhikari}},\ }\href@noop {} {\bibfield
  {journal} {\bibinfo  {journal} {J. Phys. Chem. A}\ }\textbf {\bibinfo
  {volume} {115}},\ \bibinfo {pages} {5256} (\bibinfo {year}
  {2011})}\BibitemShut {NoStop}%
\bibitem [{\citenamefont {Sahoo}, \citenamefont {Sardar},\ and\ \citenamefont
  {Adhikari}(2011{\natexlab{b}})}]{TSS2}%
  \BibitemOpen
  \bibfield  {author} {\bibinfo {author} {\bibfnamefont {T.}~\bibnamefont
  {Sahoo}}, \bibinfo {author} {\bibfnamefont {S.}~\bibnamefont {Sardar}}, \
  and\ \bibinfo {author} {\bibfnamefont {S.}~\bibnamefont {Adhikari}},\
  }\href@noop {} {\bibfield  {journal} {\bibinfo  {journal} {Phys. Scr.}\
  }\textbf {\bibinfo {volume} {84}},\ \bibinfo {pages} {028105/1} (\bibinfo
  {year} {2011}{\natexlab{b}})}\BibitemShut {NoStop}%
\bibitem [{\citenamefont {Sahoo}, \citenamefont {Mukherjee},\ and\
  \citenamefont {Adhikari}(2012)}]{TSSM}%
  \BibitemOpen
  \bibfield  {author} {\bibinfo {author} {\bibfnamefont {T.}~\bibnamefont
  {Sahoo}}, \bibinfo {author} {\bibfnamefont {S.}~\bibnamefont {Mukherjee}}, \
  and\ \bibinfo {author} {\bibfnamefont {S.}~\bibnamefont {Adhikari}},\
  }\href@noop {} {\bibfield  {journal} {\bibinfo  {journal} {J.\ Chem.\ Phys.}\
  }\textbf {\bibinfo {volume} {136}},\ \bibinfo {pages} {84306/1} (\bibinfo
  {year} {2012})}\BibitemShut {NoStop}%
\bibitem [{\citenamefont {Nattino}\ \emph {et~al.}(2014)\citenamefont
  {Nattino}, \citenamefont {Genova}, \citenamefont {Guijt}, \citenamefont
  {Muzas}, \citenamefont {D\'{i}az}, \citenamefont {Auerbach},\ and\
  \citenamefont {Kroes}}]{FNGJK1}%
  \BibitemOpen
  \bibfield  {author} {\bibinfo {author} {\bibfnamefont {F.}~\bibnamefont
  {Nattino}}, \bibinfo {author} {\bibfnamefont {A.}~\bibnamefont {Genova}},
  \bibinfo {author} {\bibfnamefont {M.}~\bibnamefont {Guijt}}, \bibinfo
  {author} {\bibfnamefont {A.~S.}\ \bibnamefont {Muzas}}, \bibinfo {author}
  {\bibfnamefont {C.}~\bibnamefont {D\'{i}az}}, \bibinfo {author}
  {\bibfnamefont {D.~J.}\ \bibnamefont {Auerbach}}, \ and\ \bibinfo {author}
  {\bibfnamefont {G.~J.}\ \bibnamefont {Kroes}},\ }\href@noop {} {\bibfield
  {journal} {\bibinfo  {journal} {J. Chem. Phys.}\ }\textbf {\bibinfo {volume}
  {141}},\ \bibinfo {pages} {124705} (\bibinfo {year} {2014})}\BibitemShut
  {NoStop}%
\bibitem [{\citenamefont {Feit}, \citenamefont {Fleck},\ and\ \citenamefont
  {Steiger}(1982)}]{SPO}%
  \BibitemOpen
  \bibfield  {author} {\bibinfo {author} {\bibfnamefont {M.~D.}\ \bibnamefont
  {Feit}}, \bibinfo {author} {\bibfnamefont {J.~A.}\ \bibnamefont {Fleck}}, \
  and\ \bibinfo {author} {\bibfnamefont {A.}~\bibnamefont {Steiger}},\
  }\href@noop {} {\bibfield  {journal} {\bibinfo  {journal} {J. Comput. Phys.}\
  }\textbf {\bibinfo {volume} {47}},\ \bibinfo {pages} {412} (\bibinfo {year}
  {1982})}\BibitemShut {NoStop}%
\bibitem [{\citenamefont {Wijzenbroek}\ and\ \citenamefont
  {Somers}(2012)}]{MWMF}%
  \BibitemOpen
  \bibfield  {author} {\bibinfo {author} {\bibfnamefont {M.}~\bibnamefont
  {Wijzenbroek}}\ and\ \bibinfo {author} {\bibfnamefont {M.~F.}\ \bibnamefont
  {Somers}},\ }\href@noop {} {\bibfield  {journal} {\bibinfo  {journal} {J.
  Chem. Phys.}\ }\textbf {\bibinfo {volume} {137}},\ \bibinfo {pages} {054703}
  (\bibinfo {year} {2012})}\BibitemShut {NoStop}%
\bibitem [{\citenamefont {Kroes}\ and\ \citenamefont
  {D\'{i}az}(2016)}]{QD-AIMD-H2-D2-Cu-review-2016}%
  \BibitemOpen
  \bibfield  {author} {\bibinfo {author} {\bibfnamefont {G.-J.}\ \bibnamefont
  {Kroes}}\ and\ \bibinfo {author} {\bibfnamefont {C.}~\bibnamefont
  {D\'{i}az}},\ }\href@noop {} {\bibfield  {journal} {\bibinfo  {journal}
  {Chem. Soc. Rev.}\ }\textbf {\bibinfo {volume} {45}},\ \bibinfo {pages}
  {3658} (\bibinfo {year} {2016})}\BibitemShut {NoStop}%
\bibitem [{\citenamefont {Zhu}\ \emph {et~al.}(2020)\citenamefont {Zhu},
  \citenamefont {Zhang}, \citenamefont {Zhang}, \citenamefont {Zhou},\ and\
  \citenamefont {Jiang}}]{H2-Cu-multiple}%
  \BibitemOpen
  \bibfield  {author} {\bibinfo {author} {\bibfnamefont {L.}~\bibnamefont
  {Zhu}}, \bibinfo {author} {\bibfnamefont {Y.}~\bibnamefont {Zhang}}, \bibinfo
  {author} {\bibfnamefont {L.}~\bibnamefont {Zhang}}, \bibinfo {author}
  {\bibfnamefont {X.}~\bibnamefont {Zhou}}, \ and\ \bibinfo {author}
  {\bibfnamefont {B.}~\bibnamefont {Jiang}},\ }\href@noop {} {\bibfield
  {journal} {\bibinfo  {journal} {Phys. Chem. Chem. Phys.}\ }\textbf {\bibinfo
  {volume} {22}},\ \bibinfo {pages} {13958} (\bibinfo {year}
  {2020})}\BibitemShut {NoStop}%
\bibitem [{\citenamefont {Spiering}, \citenamefont {Wijzenbroek},\ and\
  \citenamefont {Somers}(2018)}]{PSMFS}%
  \BibitemOpen
  \bibfield  {author} {\bibinfo {author} {\bibfnamefont {P.}~\bibnamefont
  {Spiering}}, \bibinfo {author} {\bibfnamefont {M.}~\bibnamefont
  {Wijzenbroek}}, \ and\ \bibinfo {author} {\bibfnamefont {M.~F.}\ \bibnamefont
  {Somers}},\ }\href@noop {} {\bibfield  {journal} {\bibinfo  {journal} {J.
  Chem. Phys.}\ }\textbf {\bibinfo {volume} {149}},\ \bibinfo {pages} {234702}
  (\bibinfo {year} {2018})}\BibitemShut {NoStop}%
\bibitem [{\citenamefont {Dutta}, \citenamefont {Mandal},\ and\ \citenamefont
  {Adhikari}(2022)}]{VV-Hartree-2021}%
  \BibitemOpen
  \bibfield  {author} {\bibinfo {author} {\bibfnamefont {J.}~\bibnamefont
  {Dutta}}, \bibinfo {author} {\bibfnamefont {S.}~\bibnamefont {Mandal}}, \
  and\ \bibinfo {author} {\bibfnamefont {S.}~\bibnamefont {Adhikari}},\
  }\href@noop {} {\bibfield  {journal} {\bibinfo  {journal} {Chem. Phys.}\
  }\textbf {\bibinfo {volume} {552}},\ \bibinfo {pages} {111371} (\bibinfo
  {year} {2022})}\BibitemShut {NoStop}%
\bibitem [{\citenamefont {Smits}\ and\ \citenamefont
  {Somers}(2021)}]{EAM-SCM-QC-Mark-2021}%
  \BibitemOpen
  \bibfield  {author} {\bibinfo {author} {\bibfnamefont {B.}~\bibnamefont
  {Smits}}\ and\ \bibinfo {author} {\bibfnamefont {M.~F.}\ \bibnamefont
  {Somers}},\ }\href@noop {} {\bibfield  {journal} {\bibinfo  {journal} {J.
  Chem. Phys.}\ }\textbf {\bibinfo {volume} {154}},\ \bibinfo {pages} {074710}
  (\bibinfo {year} {2021})}\BibitemShut {NoStop}%
\bibitem [{\citenamefont {Smeets}\ and\ \citenamefont
  {Kroes}(2021)}]{SRP-DF-new-H2-Cu-Kroes-2021}%
  \BibitemOpen
  \bibfield  {author} {\bibinfo {author} {\bibfnamefont {E.~W.~F.}\
  \bibnamefont {Smeets}}\ and\ \bibinfo {author} {\bibfnamefont {G.-J.}\
  \bibnamefont {Kroes}},\ }\href@noop {} {\bibfield  {journal} {\bibinfo
  {journal} {Phys. Chem. Chem. Phys.}\ }\textbf {\bibinfo {volume} {23}},\
  \bibinfo {pages} {7875} (\bibinfo {year} {2021})}\BibitemShut {NoStop}%
\bibitem [{\citenamefont {Kroes}(2021)}]{Kroes-Review-2021}%
  \BibitemOpen
  \bibfield  {author} {\bibinfo {author} {\bibfnamefont {G.-J.}\ \bibnamefont
  {Kroes}},\ }\href@noop {} {\bibfield  {journal} {\bibinfo  {journal} {Phys.
  Chem. Chem. Phys.}\ }\textbf {\bibinfo {volume} {23}},\ \bibinfo {pages}
  {8962} (\bibinfo {year} {2021})}\BibitemShut {NoStop}%
\bibitem [{\citenamefont {Dutta}\ \emph {et~al.}(2021)\citenamefont {Dutta},
  \citenamefont {Mandal}, \citenamefont {Adhikari}, \citenamefont {Spiering},
  \citenamefont {Meyer},\ and\ \citenamefont {Somers}}]{SAMARK}%
  \BibitemOpen
  \bibfield  {author} {\bibinfo {author} {\bibfnamefont {J.}~\bibnamefont
  {Dutta}}, \bibinfo {author} {\bibfnamefont {S.}~\bibnamefont {Mandal}},
  \bibinfo {author} {\bibfnamefont {S.}~\bibnamefont {Adhikari}}, \bibinfo
  {author} {\bibfnamefont {P.}~\bibnamefont {Spiering}}, \bibinfo {author}
  {\bibfnamefont {J.}~\bibnamefont {Meyer}}, \ and\ \bibinfo {author}
  {\bibfnamefont {M.~F.}\ \bibnamefont {Somers}},\ }\href@noop {} {\bibfield
  {journal} {\bibinfo  {journal} {J. Chem. Phys.}\ }\textbf {\bibinfo {volume}
  {154}},\ \bibinfo {pages} {104103} (\bibinfo {year} {2021})}\BibitemShut
  {NoStop}%
\bibitem [{\citenamefont {Dutta}\ \emph
  {et~al.}(2022{\natexlab{b}})\citenamefont {Dutta}, \citenamefont {Naskar},
  \citenamefont {Adhikari}, \citenamefont {Meyer},\ and\ \citenamefont
  {Somers}}]{Joy_D2_jcp_2022}%
  \BibitemOpen
  \bibfield  {author} {\bibinfo {author} {\bibfnamefont {J.}~\bibnamefont
  {Dutta}}, \bibinfo {author} {\bibfnamefont {K.}~\bibnamefont {Naskar}},
  \bibinfo {author} {\bibfnamefont {S.}~\bibnamefont {Adhikari}}, \bibinfo
  {author} {\bibfnamefont {J.}~\bibnamefont {Meyer}}, \ and\ \bibinfo {author}
  {\bibfnamefont {M.~F.}\ \bibnamefont {Somers}},\ }\href@noop {} {\bibfield
  {journal} {\bibinfo  {journal} {J. Chem. Phys.}\ }\textbf {\bibinfo {volume}
  {157}},\ \bibinfo {pages} {194112} (\bibinfo {year}
  {2022}{\natexlab{b}})}\BibitemShut {NoStop}%
\bibitem [{\citenamefont {Sah}\ \emph {et~al.}(2024)\citenamefont {Sah},
  \citenamefont {Naskar}, \citenamefont {Adhikari}, \citenamefont {Smits},
  \citenamefont {Meyer},\ and\ \citenamefont {Somers}}]{Mantu_jcp_2024}%
  \BibitemOpen
  \bibfield  {author} {\bibinfo {author} {\bibfnamefont {M.~K.}\ \bibnamefont
  {Sah}}, \bibinfo {author} {\bibfnamefont {K.}~\bibnamefont {Naskar}},
  \bibinfo {author} {\bibfnamefont {S.}~\bibnamefont {Adhikari}}, \bibinfo
  {author} {\bibfnamefont {B.}~\bibnamefont {Smits}}, \bibinfo {author}
  {\bibfnamefont {J.}~\bibnamefont {Meyer}}, \ and\ \bibinfo {author}
  {\bibfnamefont {M.~F.}\ \bibnamefont {Somers}},\ }\href@noop {} {\bibfield
  {journal} {\bibinfo  {journal} {J. Chem. Phys.}\ }\textbf {\bibinfo {volume}
  {161}},\ \bibinfo {pages} {014306} (\bibinfo {year} {2024})}\BibitemShut
  {NoStop}%
\bibitem [{\citenamefont {Smits}\ \emph {et~al.}(2024)\citenamefont {Smits},
  \citenamefont {Sah}, \citenamefont {Naskar}, \citenamefont {Adhikari},
  \citenamefont {Meyer},\ and\ \citenamefont {Somers}}]{Bauke_jcp_2024}%
  \BibitemOpen
  \bibfield  {author} {\bibinfo {author} {\bibfnamefont {B.}~\bibnamefont
  {Smits}}, \bibinfo {author} {\bibfnamefont {M.~K.}\ \bibnamefont {Sah}},
  \bibinfo {author} {\bibfnamefont {K.}~\bibnamefont {Naskar}}, \bibinfo
  {author} {\bibfnamefont {S.}~\bibnamefont {Adhikari}}, \bibinfo {author}
  {\bibfnamefont {J.}~\bibnamefont {Meyer}}, \ and\ \bibinfo {author}
  {\bibfnamefont {M.~F.}\ \bibnamefont {Somers}},\ }\href@noop {} {\bibfield
  {journal} {\bibinfo  {journal} {J. Chem. Phys.}\ }\textbf {\bibinfo {volume}
  {161}},\ \bibinfo {pages} {154702} (\bibinfo {year} {2024})}\BibitemShut
  {NoStop}%
\bibitem [{\citenamefont {Sah}\ \emph {et~al.}(2026)\citenamefont {Sah},
  \citenamefont {Adhikari}, \citenamefont {Smits}, \citenamefont {Meyer},\ and\
  \citenamefont {Somers}}]{Mantu_JPCC_2026}%
  \BibitemOpen
  \bibfield  {author} {\bibinfo {author} {\bibfnamefont {M.~K.}\ \bibnamefont
  {Sah}}, \bibinfo {author} {\bibfnamefont {S.}~\bibnamefont {Adhikari}},
  \bibinfo {author} {\bibfnamefont {B.}~\bibnamefont {Smits}}, \bibinfo
  {author} {\bibfnamefont {J.}~\bibnamefont {Meyer}}, \ and\ \bibinfo {author}
  {\bibfnamefont {M.~F.}\ \bibnamefont {Somers}},\ }\href@noop {} {\bibfield
  {journal} {\bibinfo  {journal} {J. Phys. Chem. C}\ }\textbf {\bibinfo
  {volume} {130}},\ \bibinfo {pages} {9010} (\bibinfo {year}
  {2026})}\BibitemShut {NoStop}%
\bibitem [{\citenamefont {Huang}\ \emph {et~al.}(1993)\citenamefont {Huang},
  \citenamefont {Zhu}, \citenamefont {Kouri},\ and\ \citenamefont
  {Hoffman}}]{QD-method-1993}%
  \BibitemOpen
  \bibfield  {author} {\bibinfo {author} {\bibfnamefont {Y.}~\bibnamefont
  {Huang}}, \bibinfo {author} {\bibfnamefont {W.}~\bibnamefont {Zhu}}, \bibinfo
  {author} {\bibfnamefont {D.~J.}\ \bibnamefont {Kouri}}, \ and\ \bibinfo
  {author} {\bibfnamefont {D.~K.}\ \bibnamefont {Hoffman}},\ }\href@noop {}
  {\bibfield  {journal} {\bibinfo  {journal} {Chem. Phys. Lett.}\ }\textbf
  {\bibinfo {volume} {206}},\ \bibinfo {pages} {96} (\bibinfo {year}
  {1993})}\BibitemShut {NoStop}%
\bibitem [{\citenamefont {Guo}, \citenamefont {Menzel},\ and\ \citenamefont
  {Jackson}(2018)}]{SRP-DF-24-CH4-step-Ni-2018}%
  \BibitemOpen
  \bibfield  {author} {\bibinfo {author} {\bibfnamefont {H.}~\bibnamefont
  {Guo}}, \bibinfo {author} {\bibfnamefont {J.~P.}\ \bibnamefont {Menzel}}, \
  and\ \bibinfo {author} {\bibfnamefont {B.}~\bibnamefont {Jackson}},\
  }\href@noop {} {\bibfield  {journal} {\bibinfo  {journal} {J. Chem. Phys.}\
  }\textbf {\bibinfo {volume} {149}},\ \bibinfo {pages} {244704} (\bibinfo
  {year} {2018})}\BibitemShut {NoStop}%
\bibitem [{\citenamefont {Guo}\ and\ \citenamefont
  {Jackson}(2019)}]{SRP-DF-23-CH4-step-Ni-2019}%
  \BibitemOpen
  \bibfield  {author} {\bibinfo {author} {\bibfnamefont {H.}~\bibnamefont
  {Guo}}\ and\ \bibinfo {author} {\bibfnamefont {B.}~\bibnamefont {Jackson}},\
  }\href@noop {} {\bibfield  {journal} {\bibinfo  {journal} {J. Chem. Phys.}\
  }\textbf {\bibinfo {volume} {150}},\ \bibinfo {pages} {204703} (\bibinfo
  {year} {2019})}\BibitemShut {NoStop}%
\bibitem [{\citenamefont {Mandal}\ \emph {et~al.}(2019)\citenamefont {Mandal},
  \citenamefont {Ghosh}, \citenamefont {Sardar},\ and\ \citenamefont
  {Adhikari}}]{SMIRPC}%
  \BibitemOpen
  \bibfield  {author} {\bibinfo {author} {\bibfnamefont {S.}~\bibnamefont
  {Mandal}}, \bibinfo {author} {\bibfnamefont {S.}~\bibnamefont {Ghosh}},
  \bibinfo {author} {\bibfnamefont {S.~S.}\ \bibnamefont {Sardar}}, \ and\
  \bibinfo {author} {\bibfnamefont {S.}~\bibnamefont {Adhikari}},\ }\href@noop
  {} {\bibfield  {journal} {\bibinfo  {journal} {Int. Rev. Phys. Chem.}\
  }\textbf {\bibinfo {volume} {37}},\ \bibinfo {pages} {607} (\bibinfo {year}
  {2019})}\BibitemShut {NoStop}%
\bibitem [{\citenamefont {D\'{i}az}\ \emph {et~al.}(2009)\citenamefont
  {D\'{i}az}, \citenamefont {Pijper}, \citenamefont {Olsen}, \citenamefont
  {Busnengo}, \citenamefont {Auerbach},\ and\ \citenamefont {Kroes}}]{SC1}%
  \BibitemOpen
  \bibfield  {author} {\bibinfo {author} {\bibfnamefont {C.}~\bibnamefont
  {D\'{i}az}}, \bibinfo {author} {\bibfnamefont {E.}~\bibnamefont {Pijper}},
  \bibinfo {author} {\bibfnamefont {R.~A.}\ \bibnamefont {Olsen}}, \bibinfo
  {author} {\bibfnamefont {H.~F.}\ \bibnamefont {Busnengo}}, \bibinfo {author}
  {\bibfnamefont {D.~J.}\ \bibnamefont {Auerbach}}, \ and\ \bibinfo {author}
  {\bibfnamefont {G.~J.}\ \bibnamefont {Kroes}},\ }\href@noop {} {\bibfield
  {journal} {\bibinfo  {journal} {Science}\ }\textbf {\bibinfo {volume}
  {326}},\ \bibinfo {pages} {832} (\bibinfo {year} {2009})}\BibitemShut
  {NoStop}%
\bibitem [{\citenamefont {Spiering}\ and\ \citenamefont
  {Meyer}(2018)}]{SRP-DF-8-H2-D2-Cu-LDFA-ODF-elhp-2018}%
  \BibitemOpen
  \bibfield  {author} {\bibinfo {author} {\bibfnamefont {P.}~\bibnamefont
  {Spiering}}\ and\ \bibinfo {author} {\bibfnamefont {J.}~\bibnamefont
  {Meyer}},\ }\href@noop {} {\bibfield  {journal} {\bibinfo  {journal} {J.
  Phys. Chem. Lett.}\ }\textbf {\bibinfo {volume} {9}},\ \bibinfo {pages}
  {1803} (\bibinfo {year} {2018})}\BibitemShut {NoStop}%
\bibitem [{\citenamefont {Jiang}, \citenamefont {Li},\ and\ \citenamefont
  {Guo}(2020)}]{HD-NNP-perspective-2020}%
  \BibitemOpen
  \bibfield  {author} {\bibinfo {author} {\bibfnamefont {B.}~\bibnamefont
  {Jiang}}, \bibinfo {author} {\bibfnamefont {J.}~\bibnamefont {Li}}, \ and\
  \bibinfo {author} {\bibfnamefont {H.}~\bibnamefont {Guo}},\ }\href@noop {}
  {\bibfield  {journal} {\bibinfo  {journal} {J. Phys. Chem. Lett.}\ }\textbf
  {\bibinfo {volume} {11}},\ \bibinfo {pages} {5120} (\bibinfo {year}
  {2020})}\BibitemShut {NoStop}%
\bibitem [{\citenamefont {Nattino}\ \emph {et~al.}(2012)\citenamefont
  {Nattino}, \citenamefont {D\'{i}az}, \citenamefont {Jackson},\ and\
  \citenamefont {Kroes}}]{FNGJK}%
  \BibitemOpen
  \bibfield  {author} {\bibinfo {author} {\bibfnamefont {F.}~\bibnamefont
  {Nattino}}, \bibinfo {author} {\bibfnamefont {C.}~\bibnamefont {D\'{i}az}},
  \bibinfo {author} {\bibfnamefont {B.}~\bibnamefont {Jackson}}, \ and\
  \bibinfo {author} {\bibfnamefont {G.~J.}\ \bibnamefont {Kroes}},\ }\href@noop
  {} {\bibfield  {journal} {\bibinfo  {journal} {Phy. Rev. Lett.}\ }\textbf
  {\bibinfo {volume} {108}},\ \bibinfo {pages} {236104/1} (\bibinfo {year}
  {2012})}\BibitemShut {NoStop}%
\bibitem [{\citenamefont {Jackson}, \citenamefont {Nattino},\ and\
  \citenamefont {Kroes}(2014)}]{PBE-19-CH4-on-Ni-AIMD-2014}%
  \BibitemOpen
  \bibfield  {author} {\bibinfo {author} {\bibfnamefont {B.}~\bibnamefont
  {Jackson}}, \bibinfo {author} {\bibfnamefont {F.}~\bibnamefont {Nattino}}, \
  and\ \bibinfo {author} {\bibfnamefont {G.-J.}\ \bibnamefont {Kroes}},\
  }\href@noop {} {\bibfield  {journal} {\bibinfo  {journal} {J. Chem. Phys.}\
  }\textbf {\bibinfo {volume} {141}},\ \bibinfo {pages} {054102} (\bibinfo
  {year} {2014})}\BibitemShut {NoStop}%
\bibitem [{\citenamefont {Kroes}, \citenamefont {Juaristi},\ and\ \citenamefont
  {Alducin}(2017)}]{SRP-DF-17-H2-Cu-GLO-AIMDEF-elhp-2017}%
  \BibitemOpen
  \bibfield  {author} {\bibinfo {author} {\bibfnamefont {G.-J.}\ \bibnamefont
  {Kroes}}, \bibinfo {author} {\bibfnamefont {J.~I.}\ \bibnamefont {Juaristi}},
  \ and\ \bibinfo {author} {\bibfnamefont {M.}~\bibnamefont {Alducin}},\
  }\href@noop {} {\bibfield  {journal} {\bibinfo  {journal} {J. Phys. Chem. C}\
  }\textbf {\bibinfo {volume} {121}},\ \bibinfo {pages} {13617} (\bibinfo
  {year} {2017})}\BibitemShut {NoStop}%
\bibitem [{\citenamefont {Tiwari}, \citenamefont {Nave},\ and\ \citenamefont
  {Jackson}(2010)}]{AKT1}%
  \BibitemOpen
  \bibfield  {author} {\bibinfo {author} {\bibfnamefont {A.~K.}\ \bibnamefont
  {Tiwari}}, \bibinfo {author} {\bibfnamefont {S.}~\bibnamefont {Nave}}, \ and\
  \bibinfo {author} {\bibfnamefont {B.}~\bibnamefont {Jackson}},\ }\href@noop
  {} {\bibfield  {journal} {\bibinfo  {journal} {J. Chem. Phys.}\ }\textbf
  {\bibinfo {volume} {132}},\ \bibinfo {pages} {134702} (\bibinfo {year}
  {2010})}\BibitemShut {NoStop}%
\bibitem [{\citenamefont {Tiwari}, \citenamefont {Nave},\ and\ \citenamefont
  {Jackson}(2009)}]{AKT}%
  \BibitemOpen
  \bibfield  {author} {\bibinfo {author} {\bibfnamefont {A.~K.}\ \bibnamefont
  {Tiwari}}, \bibinfo {author} {\bibfnamefont {S.}~\bibnamefont {Nave}}, \ and\
  \bibinfo {author} {\bibfnamefont {B.}~\bibnamefont {Jackson}},\ }\href@noop
  {} {\bibfield  {journal} {\bibinfo  {journal} {Phys.\ Rev.\ Lett.}\ }\textbf
  {\bibinfo {volume} {103}},\ \bibinfo {pages} {253201/1} (\bibinfo {year}
  {2009})}\BibitemShut {NoStop}%
\bibitem [{\citenamefont {Dohle}, \citenamefont {Saalfrank},\ and\
  \citenamefont {Uzer}(1998)}]{UZER1}%
  \BibitemOpen
  \bibfield  {author} {\bibinfo {author} {\bibfnamefont {M.}~\bibnamefont
  {Dohle}}, \bibinfo {author} {\bibfnamefont {P.}~\bibnamefont {Saalfrank}}, \
  and\ \bibinfo {author} {\bibfnamefont {T.}~\bibnamefont {Uzer}},\ }\href@noop
  {} {\bibfield  {journal} {\bibinfo  {journal} {Surf. Sci.}\ }\textbf
  {\bibinfo {volume} {409}},\ \bibinfo {pages} {37} (\bibinfo {year}
  {1998})}\BibitemShut {NoStop}%
\bibitem [{\citenamefont {Juaristi}\ \emph {et~al.}(2008)\citenamefont
  {Juaristi}, \citenamefont {Alducin}, \citenamefont {{n}o}, \citenamefont
  {Busnengo},\ and\ \citenamefont {Salin}}]{JIJ1}%
  \BibitemOpen
  \bibfield  {author} {\bibinfo {author} {\bibfnamefont {J.~I.}\ \bibnamefont
  {Juaristi}}, \bibinfo {author} {\bibfnamefont {M.}~\bibnamefont {Alducin}},
  \bibinfo {author} {\bibfnamefont {R.~D.~M.}\ \bibnamefont {{n}o}}, \bibinfo
  {author} {\bibfnamefont {H.~F.}\ \bibnamefont {Busnengo}}, \ and\ \bibinfo
  {author} {\bibfnamefont {A.}~\bibnamefont {Salin}},\ }\href@noop {}
  {\bibfield  {journal} {\bibinfo  {journal} {Phys. Rev. Lett.}\ }\textbf
  {\bibinfo {volume} {100}},\ \bibinfo {pages} {116102/1} (\bibinfo {year}
  {2008})}\BibitemShut {NoStop}%
\bibitem [{\citenamefont {Rittmeyer}, \citenamefont {Bukas},\ and\
  \citenamefont {Reuter}(2018)}]{Dynamical-models-review-2018}%
  \BibitemOpen
  \bibfield  {author} {\bibinfo {author} {\bibfnamefont {S.~P.}\ \bibnamefont
  {Rittmeyer}}, \bibinfo {author} {\bibfnamefont {V.~J.}\ \bibnamefont
  {Bukas}}, \ and\ \bibinfo {author} {\bibfnamefont {K.}~\bibnamefont
  {Reuter}},\ }\href@noop {} {\bibfield  {journal} {\bibinfo  {journal} {Adv.
  Phys.: X}\ }\textbf {\bibinfo {volume} {3}},\ \bibinfo {pages} {1381574}
  (\bibinfo {year} {2018})}\BibitemShut {NoStop}%
\bibitem [{\citenamefont {Shenvi}, \citenamefont {Roy},\ and\ \citenamefont
  {Tully}(2009{\natexlab{a}})}]{IESH-2009}%
  \BibitemOpen
  \bibfield  {author} {\bibinfo {author} {\bibfnamefont {N.}~\bibnamefont
  {Shenvi}}, \bibinfo {author} {\bibfnamefont {S.}~\bibnamefont {Roy}}, \ and\
  \bibinfo {author} {\bibfnamefont {J.~C.}\ \bibnamefont {Tully}},\ }\href@noop
  {} {\bibfield  {journal} {\bibinfo  {journal} {J. Chem. Phys.}\ }\textbf
  {\bibinfo {volume} {130}},\ \bibinfo {pages} {174107} (\bibinfo {year}
  {2009}{\natexlab{a}})}\BibitemShut {NoStop}%
\bibitem [{\citenamefont {Shenvi}, \citenamefont {Roy},\ and\ \citenamefont
  {Tully}(2009{\natexlab{b}})}]{JCTully-elhp1}%
  \BibitemOpen
  \bibfield  {author} {\bibinfo {author} {\bibfnamefont {N.}~\bibnamefont
  {Shenvi}}, \bibinfo {author} {\bibfnamefont {S.}~\bibnamefont {Roy}}, \ and\
  \bibinfo {author} {\bibfnamefont {J.~C.}\ \bibnamefont {Tully}},\ }\href@noop
  {} {\bibfield  {journal} {\bibinfo  {journal} {Science}\ }\textbf {\bibinfo
  {volume} {326}},\ \bibinfo {pages} {829} (\bibinfo {year}
  {2009}{\natexlab{b}})}\BibitemShut {NoStop}%
\bibitem [{\citenamefont {Mandal}\ \emph
  {et~al.}(2015{\natexlab{a}})\citenamefont {Mandal}, \citenamefont {Sahoo},
  \citenamefont {Ghosh},\ and\ \citenamefont {Adhikari}}]{SM-JTCC}%
  \BibitemOpen
  \bibfield  {author} {\bibinfo {author} {\bibfnamefont {S.}~\bibnamefont
  {Mandal}}, \bibinfo {author} {\bibfnamefont {T.}~\bibnamefont {Sahoo}},
  \bibinfo {author} {\bibfnamefont {S.}~\bibnamefont {Ghosh}}, \ and\ \bibinfo
  {author} {\bibfnamefont {S.}~\bibnamefont {Adhikari}},\ }\href@noop {}
  {\bibfield  {journal} {\bibinfo  {journal} {J. Theor. Comput. Chem.}\
  }\textbf {\bibinfo {volume} {14}},\ \bibinfo {pages} {1550028} (\bibinfo
  {year} {2015}{\natexlab{a}})}\BibitemShut {NoStop}%
\bibitem [{\citenamefont {Mandal}\ \emph
  {et~al.}(2015{\natexlab{b}})\citenamefont {Mandal}, \citenamefont {Sahoo},
  \citenamefont {Ghosh},\ and\ \citenamefont {Adhikari}}]{SM-MOLPHYS}%
  \BibitemOpen
  \bibfield  {author} {\bibinfo {author} {\bibfnamefont {S.}~\bibnamefont
  {Mandal}}, \bibinfo {author} {\bibfnamefont {T.}~\bibnamefont {Sahoo}},
  \bibinfo {author} {\bibfnamefont {S.}~\bibnamefont {Ghosh}}, \ and\ \bibinfo
  {author} {\bibfnamefont {S.}~\bibnamefont {Adhikari}},\ }\href@noop {}
  {\bibfield  {journal} {\bibinfo  {journal} {Mol. Phys.}\ }\textbf {\bibinfo
  {volume} {113}},\ \bibinfo {pages} {3042} (\bibinfo {year}
  {2015}{\natexlab{b}})}\BibitemShut {NoStop}%
\end{thebibliography}%


\begin{thebibliography}{115}%
\makeatletter
\providecommand \@ifxundefined [1]{%
 \@ifx{#1\undefined}
}%
\providecommand \@ifnum [1]{%
 \ifnum #1\expandafter \@firstoftwo
 \else \expandafter \@secondoftwo
 \fi
}%
\providecommand \@ifx [1]{%
 \ifx #1\expandafter \@firstoftwo
 \else \expandafter \@secondoftwo
 \fi
}%
\providecommand \natexlab [1]{#1}%
\providecommand \enquote  [1]{``#1''}%
\providecommand \bibnamefont  [1]{#1}%
\providecommand \bibfnamefont [1]{#1}%
\providecommand \citenamefont [1]{#1}%
\providecommand \href@noop [0]{\@secondoftwo}%
\providecommand \href [0]{\begingroup \@sanitize@url \@href}%
\providecommand \@href[1]{\@@startlink{#1}\@@href}%
\providecommand \@@href[1]{\endgroup#1\@@endlink}%
\providecommand \@sanitize@url [0]{\catcode `\\12\catcode `\$12\catcode
  `\&12\catcode `\#12\catcode `\^12\catcode `\_12\catcode `\%12\relax}%
\providecommand \@@startlink[1]{}%
\providecommand \@@endlink[0]{}%
\providecommand \url  [0]{\begingroup\@sanitize@url \@url }%
\providecommand \@url [1]{\endgroup\@href {#1}{\urlprefix }}%
\providecommand \urlprefix  [0]{URL }%
\providecommand \Eprint [0]{\href }%
\providecommand \doibase [0]{http://dx.doi.org/}%
\providecommand \selectlanguage [0]{\@gobble}%
\providecommand \bibinfo  [0]{\@secondoftwo}%
\providecommand \bibfield  [0]{\@secondoftwo}%
\providecommand \translation [1]{[#1]}%
\providecommand \BibitemOpen [0]{}%
\providecommand \bibitemStop [0]{}%
\providecommand \bibitemNoStop [0]{.\EOS\space}%
\providecommand \EOS [0]{\spacefactor3000\relax}%
\providecommand \BibitemShut  [1]{\csname bibitem#1\endcsname}%
\let\auto@bib@innerbib\@empty
\bibitem [{\citenamefont {Michelsen}, \citenamefont {Rettner},\ and\
  \citenamefont {Auerbach}(1992)}]{HA2}%
  \BibitemOpen
  \bibfield  {author} {\bibinfo {author} {\bibfnamefont {H.~A.}\ \bibnamefont
  {Michelsen}}, \bibinfo {author} {\bibfnamefont {C.~T.}\ \bibnamefont
  {Rettner}}, \ and\ \bibinfo {author} {\bibfnamefont {D.~J.}\ \bibnamefont
  {Auerbach}},\ }\href@noop {} {\bibfield  {journal} {\bibinfo  {journal}
  {Surf.\ Sci.}\ }\textbf {\bibinfo {volume} {272}},\ \bibinfo {pages} {65}
  (\bibinfo {year} {1992})}\BibitemShut {NoStop}%
\bibitem [{\citenamefont {Rettner}, \citenamefont {Auerbach},\ and\
  \citenamefont {Michelsen}(1992)}]{CTD}%
  \BibitemOpen
  \bibfield  {author} {\bibinfo {author} {\bibfnamefont {C.~T.}\ \bibnamefont
  {Rettner}}, \bibinfo {author} {\bibfnamefont {D.~J.}\ \bibnamefont
  {Auerbach}}, \ and\ \bibinfo {author} {\bibfnamefont {H.~A.}\ \bibnamefont
  {Michelsen}},\ }\href@noop {} {\bibfield  {journal} {\bibinfo  {journal}
  {Phys.\ Rev.\ Lett.}\ }\textbf {\bibinfo {volume} {68}},\ \bibinfo {pages}
  {1164} (\bibinfo {year} {1992})}\BibitemShut {NoStop}%
\bibitem [{\citenamefont {Rettner}, \citenamefont {Michelsen},\ and\
  \citenamefont {Auerbach}(1993{\natexlab{a}})}]{CTH}%
  \BibitemOpen
  \bibfield  {author} {\bibinfo {author} {\bibfnamefont {C.~T.}\ \bibnamefont
  {Rettner}}, \bibinfo {author} {\bibfnamefont {H.~A.}\ \bibnamefont
  {Michelsen}}, \ and\ \bibinfo {author} {\bibfnamefont {D.~J.}\ \bibnamefont
  {Auerbach}},\ }\href@noop {} {\bibfield  {journal} {\bibinfo  {journal}
  {Faraday Discuss.}\ }\textbf {\bibinfo {volume} {96}},\ \bibinfo {pages} {17}
  (\bibinfo {year} {1993}{\natexlab{a}})}\BibitemShut {NoStop}%
\bibitem [{\citenamefont {Rettner}, \citenamefont {Michelsen},\ and\
  \citenamefont {Auerbach}(1993{\natexlab{b}})}]{CTR2}%
  \BibitemOpen
  \bibfield  {author} {\bibinfo {author} {\bibfnamefont {C.~T.}\ \bibnamefont
  {Rettner}}, \bibinfo {author} {\bibfnamefont {H.~A.}\ \bibnamefont
  {Michelsen}}, \ and\ \bibinfo {author} {\bibfnamefont {D.~J.}\ \bibnamefont
  {Auerbach}},\ }\href@noop {} {\bibfield  {journal} {\bibinfo  {journal}
  {Chem. Phys.}\ }\textbf {\bibinfo {volume} {175}},\ \bibinfo {pages} {157}
  (\bibinfo {year} {1993}{\natexlab{b}})}\BibitemShut {NoStop}%
\bibitem [{\citenamefont {Rettner}, \citenamefont {Michelsen},\ and\
  \citenamefont {Auerbach}(1995)}]{CTR}%
  \BibitemOpen
  \bibfield  {author} {\bibinfo {author} {\bibfnamefont {C.~T.}\ \bibnamefont
  {Rettner}}, \bibinfo {author} {\bibfnamefont {H.~A.}\ \bibnamefont
  {Michelsen}}, \ and\ \bibinfo {author} {\bibfnamefont {D.~J.}\ \bibnamefont
  {Auerbach}},\ }\href@noop {} {\bibfield  {journal} {\bibinfo  {journal} {J.\
  Chem.\ Phys.}\ }\textbf {\bibinfo {volume} {102}},\ \bibinfo {pages} {4625}
  (\bibinfo {year} {1995})}\BibitemShut {NoStop}%
\bibitem [{\citenamefont {Michelsen}\ \emph {et~al.}(1993)\citenamefont
  {Michelsen}, \citenamefont {Rettner}, \citenamefont {Auerbach},\ and\
  \citenamefont {Zare}}]{HAM}%
  \BibitemOpen
  \bibfield  {author} {\bibinfo {author} {\bibfnamefont {H.~A.}\ \bibnamefont
  {Michelsen}}, \bibinfo {author} {\bibfnamefont {C.~T.}\ \bibnamefont
  {Rettner}}, \bibinfo {author} {\bibfnamefont {D.~J.}\ \bibnamefont
  {Auerbach}}, \ and\ \bibinfo {author} {\bibfnamefont {R.~N.}\ \bibnamefont
  {Zare}},\ }\href@noop {} {\bibfield  {journal} {\bibinfo  {journal} {J.\
  Chem.\ Phys.}\ }\textbf {\bibinfo {volume} {98}},\ \bibinfo {pages} {8294}
  (\bibinfo {year} {1993})}\BibitemShut {NoStop}%
\bibitem [{\citenamefont {Hou}\ \emph {et~al.}(1997)\citenamefont {Hou},
  \citenamefont {Gulding}, \citenamefont {Rettner}, \citenamefont {Wodtke},\
  and\ \citenamefont {Auerbach}}]{EXPNEW2}%
  \BibitemOpen
  \bibfield  {author} {\bibinfo {author} {\bibfnamefont {H.}~\bibnamefont
  {Hou}}, \bibinfo {author} {\bibfnamefont {S.~J.}\ \bibnamefont {Gulding}},
  \bibinfo {author} {\bibfnamefont {C.~T.}\ \bibnamefont {Rettner}}, \bibinfo
  {author} {\bibfnamefont {A.~M.}\ \bibnamefont {Wodtke}}, \ and\ \bibinfo
  {author} {\bibfnamefont {D.~J.}\ \bibnamefont {Auerbach}},\ }\href@noop {}
  {\bibfield  {journal} {\bibinfo  {journal} {Science.}\ }\textbf {\bibinfo
  {volume} {277}},\ \bibinfo {pages} {80} (\bibinfo {year} {1997})}\BibitemShut
  {NoStop}%
\bibitem [{\citenamefont {Murphy}\ and\ \citenamefont
  {Hodgson}(1998)}]{EXPNEW3}%
  \BibitemOpen
  \bibfield  {author} {\bibinfo {author} {\bibfnamefont {M.~J.}\ \bibnamefont
  {Murphy}}\ and\ \bibinfo {author} {\bibfnamefont {A.}~\bibnamefont
  {Hodgson}},\ }\href@noop {} {\bibfield  {journal} {\bibinfo  {journal} {J.
  Chem. Phys.}\ }\textbf {\bibinfo {volume} {108}},\ \bibinfo {pages} {4199}
  (\bibinfo {year} {1998})}\BibitemShut {NoStop}%
\bibitem [{\citenamefont {Kaufmann}\ \emph {et~al.}(2018)\citenamefont
  {Kaufmann}, \citenamefont {Shuai}, \citenamefont {Auerbach}, \citenamefont
  {Schwarzer},\ and\ \citenamefont {Wodtke}}]{Wodtke}%
  \BibitemOpen
  \bibfield  {author} {\bibinfo {author} {\bibfnamefont {S.}~\bibnamefont
  {Kaufmann}}, \bibinfo {author} {\bibfnamefont {Q.}~\bibnamefont {Shuai}},
  \bibinfo {author} {\bibfnamefont {D.~J.}\ \bibnamefont {Auerbach}}, \bibinfo
  {author} {\bibfnamefont {D.}~\bibnamefont {Schwarzer}}, \ and\ \bibinfo
  {author} {\bibfnamefont {A.~M.}\ \bibnamefont {Wodtke}},\ }\href@noop {}
  {\bibfield  {journal} {\bibinfo  {journal} {J. Chem. Phys.}\ }\textbf
  {\bibinfo {volume} {148}},\ \bibinfo {pages} {194703} (\bibinfo {year}
  {2018})}\BibitemShut {NoStop}%
\bibitem [{\citenamefont {Chadwick}\ and\ \citenamefont
  {Alexandrowicz}(2024)}]{Chadwick}%
  \BibitemOpen
  \bibfield  {author} {\bibinfo {author} {\bibfnamefont {H.}~\bibnamefont
  {Chadwick}}\ and\ \bibinfo {author} {\bibfnamefont {G.}~\bibnamefont
  {Alexandrowicz}},\ }\href@noop {} {\bibfield  {journal} {\bibinfo  {journal}
  {Faraday Discuss.}\ }\textbf {\bibinfo {volume} {251}},\ \bibinfo {pages}
  {76} (\bibinfo {year} {2024})}\BibitemShut {NoStop}%
\bibitem [{\citenamefont {Wang}, \citenamefont {Darling},\ and\ \citenamefont
  {Holloway}(2001)}]{HO1}%
  \BibitemOpen
  \bibfield  {author} {\bibinfo {author} {\bibfnamefont {Z.~S.}\ \bibnamefont
  {Wang}}, \bibinfo {author} {\bibfnamefont {G.~R.}\ \bibnamefont {Darling}}, \
  and\ \bibinfo {author} {\bibfnamefont {S.}~\bibnamefont {Holloway}},\
  }\href@noop {} {\bibfield  {journal} {\bibinfo  {journal} {Phys.\ Rev.\
  Lett.}\ }\textbf {\bibinfo {volume} {87}},\ \bibinfo {pages} {226102/1}
  (\bibinfo {year} {2001})}\BibitemShut {NoStop}%
\bibitem [{\citenamefont {D\'{i}az}\ \emph
  {et~al.}(2010{\natexlab{a}})\citenamefont {D\'{i}az}, \citenamefont {Olsen},
  \citenamefont {Auerbach},\ and\ \citenamefont {Kroes}}]{CDGJK}%
  \BibitemOpen
  \bibfield  {author} {\bibinfo {author} {\bibfnamefont {C.}~\bibnamefont
  {D\'{i}az}}, \bibinfo {author} {\bibfnamefont {R.~A.}\ \bibnamefont {Olsen}},
  \bibinfo {author} {\bibfnamefont {D.~J.}\ \bibnamefont {Auerbach}}, \ and\
  \bibinfo {author} {\bibfnamefont {G.~J.}\ \bibnamefont {Kroes}},\ }\href@noop
  {} {\bibfield  {journal} {\bibinfo  {journal} {Phys. Chem. Chem. Phys.}\
  }\textbf {\bibinfo {volume} {12}},\ \bibinfo {pages} {6499} (\bibinfo {year}
  {2010}{\natexlab{a}})}\BibitemShut {NoStop}%
\bibitem [{\citenamefont {Hand}\ and\ \citenamefont {Haris}(1990)}]{HA1}%
  \BibitemOpen
  \bibfield  {author} {\bibinfo {author} {\bibfnamefont {M.}~\bibnamefont
  {Hand}}\ and\ \bibinfo {author} {\bibfnamefont {J.}~\bibnamefont {Haris}},\
  }\href@noop {} {\bibfield  {journal} {\bibinfo  {journal} {J.\ Chem.\ Phys.}\
  }\textbf {\bibinfo {volume} {92}},\ \bibinfo {pages} {7610} (\bibinfo {year}
  {1990})}\BibitemShut {NoStop}%
\bibitem [{\citenamefont {Dohle}\ and\ \citenamefont {Saalfrank}(1997)}]{SAL}%
  \BibitemOpen
  \bibfield  {author} {\bibinfo {author} {\bibfnamefont {M.}~\bibnamefont
  {Dohle}}\ and\ \bibinfo {author} {\bibfnamefont {P.}~\bibnamefont
  {Saalfrank}},\ }\href@noop {} {\bibfield  {journal} {\bibinfo  {journal}
  {Surf.\ Sci.}\ }\textbf {\bibinfo {volume} {373}},\ \bibinfo {pages} {95}
  (\bibinfo {year} {1997})}\BibitemShut {NoStop}%
\bibitem [{\citenamefont {Adhikari}\ and\ \citenamefont {Billing}(2000)}]{SAG}%
  \BibitemOpen
  \bibfield  {author} {\bibinfo {author} {\bibfnamefont {S.}~\bibnamefont
  {Adhikari}}\ and\ \bibinfo {author} {\bibfnamefont {G.~D.}\ \bibnamefont
  {Billing}},\ }\href@noop {} {\bibfield  {journal} {\bibinfo  {journal} {J.\
  Chem.\ Phys.}\ }\textbf {\bibinfo {volume} {112}},\ \bibinfo {pages} {3884}
  (\bibinfo {year} {2000})}\BibitemShut {NoStop}%
\bibitem [{\citenamefont {Sahoo}, \citenamefont {Sardar},\ and\ \citenamefont
  {Adhikari}(2011{\natexlab{a}})}]{TSS}%
  \BibitemOpen
  \bibfield  {author} {\bibinfo {author} {\bibfnamefont {T.}~\bibnamefont
  {Sahoo}}, \bibinfo {author} {\bibfnamefont {S.}~\bibnamefont {Sardar}}, \
  and\ \bibinfo {author} {\bibfnamefont {S.}~\bibnamefont {Adhikari}},\
  }\href@noop {} {\bibfield  {journal} {\bibinfo  {journal} {Phys. Chem. Chem.
  Phys.}\ }\textbf {\bibinfo {volume} {13}},\ \bibinfo {pages} {10100}
  (\bibinfo {year} {2011}{\natexlab{a}})}\BibitemShut {NoStop}%
\bibitem [{\citenamefont {Sahoo}\ \emph {et~al.}(2011)\citenamefont {Sahoo},
  \citenamefont {Sardar}, \citenamefont {Mondal}, \citenamefont {Sarkar},\ and\
  \citenamefont {Adhikari}}]{TSS1}%
  \BibitemOpen
  \bibfield  {author} {\bibinfo {author} {\bibfnamefont {T.}~\bibnamefont
  {Sahoo}}, \bibinfo {author} {\bibfnamefont {S.}~\bibnamefont {Sardar}},
  \bibinfo {author} {\bibfnamefont {P.}~\bibnamefont {Mondal}}, \bibinfo
  {author} {\bibfnamefont {B.}~\bibnamefont {Sarkar}}, \ and\ \bibinfo {author}
  {\bibfnamefont {S.}~\bibnamefont {Adhikari}},\ }\href@noop {} {\bibfield
  {journal} {\bibinfo  {journal} {J. Phys. Chem. A}\ }\textbf {\bibinfo
  {volume} {115}},\ \bibinfo {pages} {5256} (\bibinfo {year}
  {2011})}\BibitemShut {NoStop}%
\bibitem [{\citenamefont {Sahoo}, \citenamefont {Sardar},\ and\ \citenamefont
  {Adhikari}(2011{\natexlab{b}})}]{TSS2}%
  \BibitemOpen
  \bibfield  {author} {\bibinfo {author} {\bibfnamefont {T.}~\bibnamefont
  {Sahoo}}, \bibinfo {author} {\bibfnamefont {S.}~\bibnamefont {Sardar}}, \
  and\ \bibinfo {author} {\bibfnamefont {S.}~\bibnamefont {Adhikari}},\
  }\href@noop {} {\bibfield  {journal} {\bibinfo  {journal} {Phys. Scr.}\
  }\textbf {\bibinfo {volume} {84}},\ \bibinfo {pages} {028105/1} (\bibinfo
  {year} {2011}{\natexlab{b}})}\BibitemShut {NoStop}%
\bibitem [{\citenamefont {Sahoo}, \citenamefont {Mukherjee},\ and\
  \citenamefont {Adhikari}(2012{\natexlab{a}})}]{TSSM}%
  \BibitemOpen
  \bibfield  {author} {\bibinfo {author} {\bibfnamefont {T.}~\bibnamefont
  {Sahoo}}, \bibinfo {author} {\bibfnamefont {S.}~\bibnamefont {Mukherjee}}, \
  and\ \bibinfo {author} {\bibfnamefont {S.}~\bibnamefont {Adhikari}},\
  }\href@noop {} {\bibfield  {journal} {\bibinfo  {journal} {J.\ Chem.\ Phys.}\
  }\textbf {\bibinfo {volume} {136}},\ \bibinfo {pages} {84306/1} (\bibinfo
  {year} {2012}{\natexlab{a}})}\BibitemShut {NoStop}%
\bibitem [{\citenamefont {Nattino}\ \emph {et~al.}(2014)\citenamefont
  {Nattino}, \citenamefont {Genova}, \citenamefont {Guijt}, \citenamefont
  {Muzas}, \citenamefont {D\'{i}az}, \citenamefont {Auerbach},\ and\
  \citenamefont {Kroes}}]{FNGJK1}%
  \BibitemOpen
  \bibfield  {author} {\bibinfo {author} {\bibfnamefont {F.}~\bibnamefont
  {Nattino}}, \bibinfo {author} {\bibfnamefont {A.}~\bibnamefont {Genova}},
  \bibinfo {author} {\bibfnamefont {M.}~\bibnamefont {Guijt}}, \bibinfo
  {author} {\bibfnamefont {A.~S.}\ \bibnamefont {Muzas}}, \bibinfo {author}
  {\bibfnamefont {C.}~\bibnamefont {D\'{i}az}}, \bibinfo {author}
  {\bibfnamefont {D.~J.}\ \bibnamefont {Auerbach}}, \ and\ \bibinfo {author}
  {\bibfnamefont {G.~J.}\ \bibnamefont {Kroes}},\ }\href@noop {} {\bibfield
  {journal} {\bibinfo  {journal} {J. Chem. Phys.}\ }\textbf {\bibinfo {volume}
  {141}},\ \bibinfo {pages} {124705} (\bibinfo {year} {2014})}\BibitemShut
  {NoStop}%
\bibitem [{\citenamefont {Feit}, \citenamefont {Fleck},\ and\ \citenamefont
  {Steiger}(1982)}]{SPO}%
  \BibitemOpen
  \bibfield  {author} {\bibinfo {author} {\bibfnamefont {M.~D.}\ \bibnamefont
  {Feit}}, \bibinfo {author} {\bibfnamefont {J.~A.}\ \bibnamefont {Fleck}}, \
  and\ \bibinfo {author} {\bibfnamefont {A.}~\bibnamefont {Steiger}},\
  }\href@noop {} {\bibfield  {journal} {\bibinfo  {journal} {J. Comput. Phys.}\
  }\textbf {\bibinfo {volume} {47}},\ \bibinfo {pages} {412} (\bibinfo {year}
  {1982})}\BibitemShut {NoStop}%
\bibitem [{\citenamefont {Wijzenbroek}\ and\ \citenamefont
  {Somers}(2012)}]{MWMF}%
  \BibitemOpen
  \bibfield  {author} {\bibinfo {author} {\bibfnamefont {M.}~\bibnamefont
  {Wijzenbroek}}\ and\ \bibinfo {author} {\bibfnamefont {M.~F.}\ \bibnamefont
  {Somers}},\ }\href@noop {} {\bibfield  {journal} {\bibinfo  {journal} {J.
  Chem. Phys.}\ }\textbf {\bibinfo {volume} {137}},\ \bibinfo {pages} {054703}
  (\bibinfo {year} {2012})}\BibitemShut {NoStop}%
\bibitem [{\citenamefont {Kroes}\ and\ \citenamefont
  {D\'{i}az}(2016)}]{QD-AIMD-H2-D2-Cu-review-2016}%
  \BibitemOpen
  \bibfield  {author} {\bibinfo {author} {\bibfnamefont {G.-J.}\ \bibnamefont
  {Kroes}}\ and\ \bibinfo {author} {\bibfnamefont {C.}~\bibnamefont
  {D\'{i}az}},\ }\href@noop {} {\bibfield  {journal} {\bibinfo  {journal}
  {Chem. Soc. Rev.}\ }\textbf {\bibinfo {volume} {45}},\ \bibinfo {pages}
  {3658} (\bibinfo {year} {2016})}\BibitemShut {NoStop}%
\bibitem [{\citenamefont {Zhu}\ \emph {et~al.}(2020)\citenamefont {Zhu},
  \citenamefont {Zhang}, \citenamefont {Zhang}, \citenamefont {Zhou},\ and\
  \citenamefont {Jiang}}]{H2-Cu-multiple}%
  \BibitemOpen
  \bibfield  {author} {\bibinfo {author} {\bibfnamefont {L.}~\bibnamefont
  {Zhu}}, \bibinfo {author} {\bibfnamefont {Y.}~\bibnamefont {Zhang}}, \bibinfo
  {author} {\bibfnamefont {L.}~\bibnamefont {Zhang}}, \bibinfo {author}
  {\bibfnamefont {X.}~\bibnamefont {Zhou}}, \ and\ \bibinfo {author}
  {\bibfnamefont {B.}~\bibnamefont {Jiang}},\ }\href@noop {} {\bibfield
  {journal} {\bibinfo  {journal} {Phys. Chem. Chem. Phys.}\ }\textbf {\bibinfo
  {volume} {22}},\ \bibinfo {pages} {13958} (\bibinfo {year}
  {2020})}\BibitemShut {NoStop}%
\bibitem [{\citenamefont {Spiering}, \citenamefont {Wijzenbroek},\ and\
  \citenamefont {Somers}(2018)}]{PSMFS}%
  \BibitemOpen
  \bibfield  {author} {\bibinfo {author} {\bibfnamefont {P.}~\bibnamefont
  {Spiering}}, \bibinfo {author} {\bibfnamefont {M.}~\bibnamefont
  {Wijzenbroek}}, \ and\ \bibinfo {author} {\bibfnamefont {M.~F.}\ \bibnamefont
  {Somers}},\ }\href@noop {} {\bibfield  {journal} {\bibinfo  {journal} {J.
  Chem. Phys.}\ }\textbf {\bibinfo {volume} {149}},\ \bibinfo {pages} {234702}
  (\bibinfo {year} {2018})}\BibitemShut {NoStop}%
\bibitem [{\citenamefont {Dutta}, \citenamefont {Mandal},\ and\ \citenamefont
  {Adhikari}(2022)}]{VV-Hartree-2021}%
  \BibitemOpen
  \bibfield  {author} {\bibinfo {author} {\bibfnamefont {J.}~\bibnamefont
  {Dutta}}, \bibinfo {author} {\bibfnamefont {S.}~\bibnamefont {Mandal}}, \
  and\ \bibinfo {author} {\bibfnamefont {S.}~\bibnamefont {Adhikari}},\
  }\href@noop {} {\bibfield  {journal} {\bibinfo  {journal} {Chem. Phys.}\
  }\textbf {\bibinfo {volume} {552}},\ \bibinfo {pages} {111371} (\bibinfo
  {year} {2022})}\BibitemShut {NoStop}%
\bibitem [{\citenamefont {Smits}\ and\ \citenamefont
  {Somers}(2021)}]{EAM-SCM-QC-Mark-2021}%
  \BibitemOpen
  \bibfield  {author} {\bibinfo {author} {\bibfnamefont {B.}~\bibnamefont
  {Smits}}\ and\ \bibinfo {author} {\bibfnamefont {M.~F.}\ \bibnamefont
  {Somers}},\ }\href@noop {} {\bibfield  {journal} {\bibinfo  {journal} {J.
  Chem. Phys.}\ }\textbf {\bibinfo {volume} {154}},\ \bibinfo {pages} {074710}
  (\bibinfo {year} {2021})}\BibitemShut {NoStop}%
\bibitem [{\citenamefont {Smeets}\ and\ \citenamefont
  {Kroes}(2021)}]{SRP-DF-new-H2-Cu-Kroes-2021}%
  \BibitemOpen
  \bibfield  {author} {\bibinfo {author} {\bibfnamefont {E.~W.~F.}\
  \bibnamefont {Smeets}}\ and\ \bibinfo {author} {\bibfnamefont {G.-J.}\
  \bibnamefont {Kroes}},\ }\href@noop {} {\bibfield  {journal} {\bibinfo
  {journal} {Phys. Chem. Chem. Phys.}\ }\textbf {\bibinfo {volume} {23}},\
  \bibinfo {pages} {7875} (\bibinfo {year} {2021})}\BibitemShut {NoStop}%
\bibitem [{\citenamefont {Kroes}(2021)}]{Kroes-Review-2021}%
  \BibitemOpen
  \bibfield  {author} {\bibinfo {author} {\bibfnamefont {G.-J.}\ \bibnamefont
  {Kroes}},\ }\href@noop {} {\bibfield  {journal} {\bibinfo  {journal} {Phys.
  Chem. Chem. Phys.}\ }\textbf {\bibinfo {volume} {23}},\ \bibinfo {pages}
  {8962} (\bibinfo {year} {2021})}\BibitemShut {NoStop}%
\bibitem [{\citenamefont {Dutta}\ \emph {et~al.}(2021)\citenamefont {Dutta},
  \citenamefont {Mandal}, \citenamefont {Adhikari}, \citenamefont {Spiering},
  \citenamefont {Meyer},\ and\ \citenamefont {Somers}}]{SAMARK}%
  \BibitemOpen
  \bibfield  {author} {\bibinfo {author} {\bibfnamefont {J.}~\bibnamefont
  {Dutta}}, \bibinfo {author} {\bibfnamefont {S.}~\bibnamefont {Mandal}},
  \bibinfo {author} {\bibfnamefont {S.}~\bibnamefont {Adhikari}}, \bibinfo
  {author} {\bibfnamefont {P.}~\bibnamefont {Spiering}}, \bibinfo {author}
  {\bibfnamefont {J.}~\bibnamefont {Meyer}}, \ and\ \bibinfo {author}
  {\bibfnamefont {M.~F.}\ \bibnamefont {Somers}},\ }\href@noop {} {\bibfield
  {journal} {\bibinfo  {journal} {J. Chem. Phys.}\ }\textbf {\bibinfo {volume}
  {154}},\ \bibinfo {pages} {104103} (\bibinfo {year} {2021})}\BibitemShut
  {NoStop}%
\bibitem [{\citenamefont {Dutta}\ \emph {et~al.}(2022)\citenamefont {Dutta},
  \citenamefont {Naskar}, \citenamefont {Adhikari}, \citenamefont {Meyer},\
  and\ \citenamefont {Somers}}]{Joy_D2_jcp_2022}%
  \BibitemOpen
  \bibfield  {author} {\bibinfo {author} {\bibfnamefont {J.}~\bibnamefont
  {Dutta}}, \bibinfo {author} {\bibfnamefont {K.}~\bibnamefont {Naskar}},
  \bibinfo {author} {\bibfnamefont {S.}~\bibnamefont {Adhikari}}, \bibinfo
  {author} {\bibfnamefont {J.}~\bibnamefont {Meyer}}, \ and\ \bibinfo {author}
  {\bibfnamefont {M.~F.}\ \bibnamefont {Somers}},\ }\href@noop {} {\bibfield
  {journal} {\bibinfo  {journal} {J. Chem. Phys.}\ }\textbf {\bibinfo {volume}
  {157}},\ \bibinfo {pages} {194112} (\bibinfo {year} {2022})}\BibitemShut
  {NoStop}%
\bibitem [{\citenamefont {Sah}\ \emph {et~al.}(2024)\citenamefont {Sah},
  \citenamefont {Naskar}, \citenamefont {Adhikari}, \citenamefont {Smits},
  \citenamefont {Meyer},\ and\ \citenamefont {Somers}}]{Mantu_jcp_2024}%
  \BibitemOpen
  \bibfield  {author} {\bibinfo {author} {\bibfnamefont {M.~K.}\ \bibnamefont
  {Sah}}, \bibinfo {author} {\bibfnamefont {K.}~\bibnamefont {Naskar}},
  \bibinfo {author} {\bibfnamefont {S.}~\bibnamefont {Adhikari}}, \bibinfo
  {author} {\bibfnamefont {B.}~\bibnamefont {Smits}}, \bibinfo {author}
  {\bibfnamefont {J.}~\bibnamefont {Meyer}}, \ and\ \bibinfo {author}
  {\bibfnamefont {M.~F.}\ \bibnamefont {Somers}},\ }\href@noop {} {\bibfield
  {journal} {\bibinfo  {journal} {J. Chem. Phys.}\ }\textbf {\bibinfo {volume}
  {161}},\ \bibinfo {pages} {014306} (\bibinfo {year} {2024})}\BibitemShut
  {NoStop}%
\bibitem [{\citenamefont {Smits}\ \emph {et~al.}(2024)\citenamefont {Smits},
  \citenamefont {Sah}, \citenamefont {Naskar}, \citenamefont {Adhikari},
  \citenamefont {Meyer},\ and\ \citenamefont {Somers}}]{Bauke_jcp_2024}%
  \BibitemOpen
  \bibfield  {author} {\bibinfo {author} {\bibfnamefont {B.}~\bibnamefont
  {Smits}}, \bibinfo {author} {\bibfnamefont {M.~K.}\ \bibnamefont {Sah}},
  \bibinfo {author} {\bibfnamefont {K.}~\bibnamefont {Naskar}}, \bibinfo
  {author} {\bibfnamefont {S.}~\bibnamefont {Adhikari}}, \bibinfo {author}
  {\bibfnamefont {J.}~\bibnamefont {Meyer}}, \ and\ \bibinfo {author}
  {\bibfnamefont {M.~F.}\ \bibnamefont {Somers}},\ }\href@noop {} {\bibfield
  {journal} {\bibinfo  {journal} {J. Chem. Phys.}\ }\textbf {\bibinfo {volume}
  {161}},\ \bibinfo {pages} {154702} (\bibinfo {year} {2024})}\BibitemShut
  {NoStop}%
\bibitem [{\citenamefont {Sah}\ \emph {et~al.}(2026)\citenamefont {Sah},
  \citenamefont {Adhikari}, \citenamefont {Smits}, \citenamefont {Meyer},\ and\
  \citenamefont {Somers}}]{Mantu_JPCC_2026}%
  \BibitemOpen
  \bibfield  {author} {\bibinfo {author} {\bibfnamefont {M.~K.}\ \bibnamefont
  {Sah}}, \bibinfo {author} {\bibfnamefont {S.}~\bibnamefont {Adhikari}},
  \bibinfo {author} {\bibfnamefont {B.}~\bibnamefont {Smits}}, \bibinfo
  {author} {\bibfnamefont {J.}~\bibnamefont {Meyer}}, \ and\ \bibinfo {author}
  {\bibfnamefont {M.~F.}\ \bibnamefont {Somers}},\ }\href@noop {} {\bibfield
  {journal} {\bibinfo  {journal} {J. Phys. Chem. C}\ }\textbf {\bibinfo
  {volume} {130}},\ \bibinfo {pages} {9010} (\bibinfo {year}
  {2026})}\BibitemShut {NoStop}%
\bibitem [{\citenamefont {D\'{i}az}\ \emph {et~al.}(2009)\citenamefont
  {D\'{i}az}, \citenamefont {Pijper}, \citenamefont {Olsen}, \citenamefont
  {Busnengo}, \citenamefont {Auerbach},\ and\ \citenamefont {Kroes}}]{SC1}%
  \BibitemOpen
  \bibfield  {author} {\bibinfo {author} {\bibfnamefont {C.}~\bibnamefont
  {D\'{i}az}}, \bibinfo {author} {\bibfnamefont {E.}~\bibnamefont {Pijper}},
  \bibinfo {author} {\bibfnamefont {R.~A.}\ \bibnamefont {Olsen}}, \bibinfo
  {author} {\bibfnamefont {H.~F.}\ \bibnamefont {Busnengo}}, \bibinfo {author}
  {\bibfnamefont {D.~J.}\ \bibnamefont {Auerbach}}, \ and\ \bibinfo {author}
  {\bibfnamefont {G.~J.}\ \bibnamefont {Kroes}},\ }\href@noop {} {\bibfield
  {journal} {\bibinfo  {journal} {Science}\ }\textbf {\bibinfo {volume}
  {326}},\ \bibinfo {pages} {832} (\bibinfo {year} {2009})}\BibitemShut
  {NoStop}%
\bibitem [{\citenamefont {Busnengo}, \citenamefont {Salin},\ and\ \citenamefont
  {Dong}(2000)}]{CRP}%
  \BibitemOpen
  \bibfield  {author} {\bibinfo {author} {\bibfnamefont {H.~F.}\ \bibnamefont
  {Busnengo}}, \bibinfo {author} {\bibfnamefont {A.}~\bibnamefont {Salin}}, \
  and\ \bibinfo {author} {\bibfnamefont {W.}~\bibnamefont {Dong}},\ }\href@noop
  {} {\bibfield  {journal} {\bibinfo  {journal} {J. Chem. Phys.}\ }\textbf
  {\bibinfo {volume} {112}},\ \bibinfo {pages} {7641} (\bibinfo {year}
  {2000})}\BibitemShut {NoStop}%
\bibitem [{\citenamefont {Nattino}\ \emph
  {et~al.}(2012{\natexlab{a}})\citenamefont {Nattino}, \citenamefont
  {D\'{i}az}, \citenamefont {Jackson},\ and\ \citenamefont {Kroes}}]{CRP3}%
  \BibitemOpen
  \bibfield  {author} {\bibinfo {author} {\bibfnamefont {F.}~\bibnamefont
  {Nattino}}, \bibinfo {author} {\bibfnamefont {C.}~\bibnamefont {D\'{i}az}},
  \bibinfo {author} {\bibfnamefont {B.}~\bibnamefont {Jackson}}, \ and\
  \bibinfo {author} {\bibfnamefont {G.-J.}\ \bibnamefont {Kroes}},\ }\href@noop
  {} {\bibfield  {journal} {\bibinfo  {journal} {Phys. Rev. Lett.}\ }\textbf
  {\bibinfo {volume} {108}},\ \bibinfo {pages} {236104} (\bibinfo {year}
  {2012}{\natexlab{a}})}\BibitemShut {NoStop}%
\bibitem [{\citenamefont {Kroes}(2015)}]{CRP4}%
  \BibitemOpen
  \bibfield  {author} {\bibinfo {author} {\bibfnamefont {G.-J.}\ \bibnamefont
  {Kroes}},\ }\href@noop {} {\bibfield  {journal} {\bibinfo  {journal} {J.
  Phys. Chem. Lett.}\ }\textbf {\bibinfo {volume} {6}},\ \bibinfo {pages}
  {4106} (\bibinfo {year} {2015})}\BibitemShut {NoStop}%
\bibitem [{\citenamefont {Zhou}\ \emph {et~al.}(2021)\citenamefont {Zhou},
  \citenamefont {Zhang}, \citenamefont {Yin}, \citenamefont {Hu},\ and\
  \citenamefont {Jiang}}]{NN-PES-perspective-2021}%
  \BibitemOpen
  \bibfield  {author} {\bibinfo {author} {\bibfnamefont {X.}~\bibnamefont
  {Zhou}}, \bibinfo {author} {\bibfnamefont {Y.}~\bibnamefont {Zhang}},
  \bibinfo {author} {\bibfnamefont {R.}~\bibnamefont {Yin}}, \bibinfo {author}
  {\bibfnamefont {C.}~\bibnamefont {Hu}}, \ and\ \bibinfo {author}
  {\bibfnamefont {B.}~\bibnamefont {Jiang}},\ }\href@noop {} {\bibfield
  {journal} {\bibinfo  {journal} {Chin. J. Chem.}\ }\textbf {\bibinfo {volume}
  {39}},\ \bibinfo {pages} {2917} (\bibinfo {year} {2021})}\BibitemShut
  {NoStop}%
\bibitem [{\citenamefont {Mandal}\ \emph {et~al.}(2019)\citenamefont {Mandal},
  \citenamefont {Ghosh}, \citenamefont {Sardar},\ and\ \citenamefont
  {Adhikari}}]{SMIRPC}%
  \BibitemOpen
  \bibfield  {author} {\bibinfo {author} {\bibfnamefont {S.}~\bibnamefont
  {Mandal}}, \bibinfo {author} {\bibfnamefont {S.}~\bibnamefont {Ghosh}},
  \bibinfo {author} {\bibfnamefont {S.~S.}\ \bibnamefont {Sardar}}, \ and\
  \bibinfo {author} {\bibfnamefont {S.}~\bibnamefont {Adhikari}},\ }\href@noop
  {} {\bibfield  {journal} {\bibinfo  {journal} {Int. Rev. Phys. Chem.}\
  }\textbf {\bibinfo {volume} {37}},\ \bibinfo {pages} {607} (\bibinfo {year}
  {2019})}\BibitemShut {NoStop}%
\bibitem [{\citenamefont {Huang}\ \emph {et~al.}(1993)\citenamefont {Huang},
  \citenamefont {Zhu}, \citenamefont {Kouri},\ and\ \citenamefont
  {Hoffman}}]{QD-method-1993}%
  \BibitemOpen
  \bibfield  {author} {\bibinfo {author} {\bibfnamefont {Y.}~\bibnamefont
  {Huang}}, \bibinfo {author} {\bibfnamefont {W.}~\bibnamefont {Zhu}}, \bibinfo
  {author} {\bibfnamefont {D.~J.}\ \bibnamefont {Kouri}}, \ and\ \bibinfo
  {author} {\bibfnamefont {D.~K.}\ \bibnamefont {Hoffman}},\ }\href@noop {}
  {\bibfield  {journal} {\bibinfo  {journal} {Chem. Phys. Lett.}\ }\textbf
  {\bibinfo {volume} {206}},\ \bibinfo {pages} {96} (\bibinfo {year}
  {1993})}\BibitemShut {NoStop}%
\bibitem [{\citenamefont {Guo}, \citenamefont {Menzel},\ and\ \citenamefont
  {Jackson}(2018)}]{SRP-DF-24-CH4-step-Ni-2018}%
  \BibitemOpen
  \bibfield  {author} {\bibinfo {author} {\bibfnamefont {H.}~\bibnamefont
  {Guo}}, \bibinfo {author} {\bibfnamefont {J.~P.}\ \bibnamefont {Menzel}}, \
  and\ \bibinfo {author} {\bibfnamefont {B.}~\bibnamefont {Jackson}},\
  }\href@noop {} {\bibfield  {journal} {\bibinfo  {journal} {J. Chem. Phys.}\
  }\textbf {\bibinfo {volume} {149}},\ \bibinfo {pages} {244704} (\bibinfo
  {year} {2018})}\BibitemShut {NoStop}%
\bibitem [{\citenamefont {Guo}\ and\ \citenamefont
  {Jackson}(2019)}]{SRP-DF-23-CH4-step-Ni-2019}%
  \BibitemOpen
  \bibfield  {author} {\bibinfo {author} {\bibfnamefont {H.}~\bibnamefont
  {Guo}}\ and\ \bibinfo {author} {\bibfnamefont {B.}~\bibnamefont {Jackson}},\
  }\href@noop {} {\bibfield  {journal} {\bibinfo  {journal} {J. Chem. Phys.}\
  }\textbf {\bibinfo {volume} {150}},\ \bibinfo {pages} {204703} (\bibinfo
  {year} {2019})}\BibitemShut {NoStop}%
\bibitem [{\citenamefont {Spiering}\ and\ \citenamefont
  {Meyer}(2018)}]{SRP-DF-8-H2-D2-Cu-LDFA-ODF-elhp-2018}%
  \BibitemOpen
  \bibfield  {author} {\bibinfo {author} {\bibfnamefont {P.}~\bibnamefont
  {Spiering}}\ and\ \bibinfo {author} {\bibfnamefont {J.}~\bibnamefont
  {Meyer}},\ }\href@noop {} {\bibfield  {journal} {\bibinfo  {journal} {J.
  Phys. Chem. Lett.}\ }\textbf {\bibinfo {volume} {9}},\ \bibinfo {pages}
  {1803} (\bibinfo {year} {2018})}\BibitemShut {NoStop}%
\bibitem [{\citenamefont {Jiang}, \citenamefont {Li},\ and\ \citenamefont
  {Guo}(2020)}]{HD-NNP-perspective-2020}%
  \BibitemOpen
  \bibfield  {author} {\bibinfo {author} {\bibfnamefont {B.}~\bibnamefont
  {Jiang}}, \bibinfo {author} {\bibfnamefont {J.}~\bibnamefont {Li}}, \ and\
  \bibinfo {author} {\bibfnamefont {H.}~\bibnamefont {Guo}},\ }\href@noop {}
  {\bibfield  {journal} {\bibinfo  {journal} {J. Phys. Chem. Lett.}\ }\textbf
  {\bibinfo {volume} {11}},\ \bibinfo {pages} {5120} (\bibinfo {year}
  {2020})}\BibitemShut {NoStop}%
\bibitem [{\citenamefont {Nattino}\ \emph
  {et~al.}(2012{\natexlab{b}})\citenamefont {Nattino}, \citenamefont
  {D\'{i}az}, \citenamefont {Jackson},\ and\ \citenamefont {Kroes}}]{FNGJK}%
  \BibitemOpen
  \bibfield  {author} {\bibinfo {author} {\bibfnamefont {F.}~\bibnamefont
  {Nattino}}, \bibinfo {author} {\bibfnamefont {C.}~\bibnamefont {D\'{i}az}},
  \bibinfo {author} {\bibfnamefont {B.}~\bibnamefont {Jackson}}, \ and\
  \bibinfo {author} {\bibfnamefont {G.~J.}\ \bibnamefont {Kroes}},\ }\href@noop
  {} {\bibfield  {journal} {\bibinfo  {journal} {Phy. Rev. Lett.}\ }\textbf
  {\bibinfo {volume} {108}},\ \bibinfo {pages} {236104/1} (\bibinfo {year}
  {2012}{\natexlab{b}})}\BibitemShut {NoStop}%
\bibitem [{\citenamefont {Jackson}, \citenamefont {Nattino},\ and\
  \citenamefont {Kroes}(2014)}]{PBE-19-CH4-on-Ni-AIMD-2014}%
  \BibitemOpen
  \bibfield  {author} {\bibinfo {author} {\bibfnamefont {B.}~\bibnamefont
  {Jackson}}, \bibinfo {author} {\bibfnamefont {F.}~\bibnamefont {Nattino}}, \
  and\ \bibinfo {author} {\bibfnamefont {G.-J.}\ \bibnamefont {Kroes}},\
  }\href@noop {} {\bibfield  {journal} {\bibinfo  {journal} {J. Chem. Phys.}\
  }\textbf {\bibinfo {volume} {141}},\ \bibinfo {pages} {054102} (\bibinfo
  {year} {2014})}\BibitemShut {NoStop}%
\bibitem [{\citenamefont {Kroes}, \citenamefont {Juaristi},\ and\ \citenamefont
  {Alducin}(2017)}]{SRP-DF-17-H2-Cu-GLO-AIMDEF-elhp-2017}%
  \BibitemOpen
  \bibfield  {author} {\bibinfo {author} {\bibfnamefont {G.-J.}\ \bibnamefont
  {Kroes}}, \bibinfo {author} {\bibfnamefont {J.~I.}\ \bibnamefont {Juaristi}},
  \ and\ \bibinfo {author} {\bibfnamefont {M.}~\bibnamefont {Alducin}},\
  }\href@noop {} {\bibfield  {journal} {\bibinfo  {journal} {J. Phys. Chem. C}\
  }\textbf {\bibinfo {volume} {121}},\ \bibinfo {pages} {13617} (\bibinfo
  {year} {2017})}\BibitemShut {NoStop}%
\bibitem [{\citenamefont {Kroes}\ and\ \citenamefont
  {Meyer}(2025)}]{Kroes_2025}%
  \BibitemOpen
  \bibfield  {author} {\bibinfo {author} {\bibfnamefont {G.-J.}\ \bibnamefont
  {Kroes}}\ and\ \bibinfo {author} {\bibfnamefont {J.}~\bibnamefont {Meyer}},\
  }\href@noop {} {\bibfield  {journal} {\bibinfo  {journal} {Chem. Sci.}\
  }\textbf {\bibinfo {volume} {16}},\ \bibinfo {pages} {480} (\bibinfo {year}
  {2025})}\BibitemShut {NoStop}%
\bibitem [{\citenamefont {Liu}, \citenamefont {Fu},\ and\ \citenamefont
  {Zhang}(2014)}]{Zhang-QD-H2-Cu111}%
  \BibitemOpen
  \bibfield  {author} {\bibinfo {author} {\bibfnamefont {T.}~\bibnamefont
  {Liu}}, \bibinfo {author} {\bibfnamefont {B.}~\bibnamefont {Fu}}, \ and\
  \bibinfo {author} {\bibfnamefont {D.~H.}\ \bibnamefont {Zhang}},\ }\href@noop
  {} {\bibfield  {journal} {\bibinfo  {journal} {J. Chem. Phys.}\ }\textbf
  {\bibinfo {volume} {141}},\ \bibinfo {pages} {194302} (\bibinfo {year}
  {2014})}\BibitemShut {NoStop}%
\bibitem [{\citenamefont {D\'{i}az}\ \emph
  {et~al.}(2010{\natexlab{b}})\citenamefont {D\'{i}az}, \citenamefont {Olsen},
  \citenamefont {Busnengo},\ and\ \citenamefont {Kroes}}]{H2-Cu111-CRP}%
  \BibitemOpen
  \bibfield  {author} {\bibinfo {author} {\bibfnamefont {C.}~\bibnamefont
  {D\'{i}az}}, \bibinfo {author} {\bibfnamefont {R.~A.}\ \bibnamefont {Olsen}},
  \bibinfo {author} {\bibfnamefont {H.~F.}\ \bibnamefont {Busnengo}}, \ and\
  \bibinfo {author} {\bibfnamefont {G.-J.}\ \bibnamefont {Kroes}},\ }\href@noop
  {} {\bibfield  {journal} {\bibinfo  {journal} {J. Phys. Chem. C.}\ }\textbf
  {\bibinfo {volume} {114}},\ \bibinfo {pages} {11192} (\bibinfo {year}
  {2010}{\natexlab{b}})}\BibitemShut {NoStop}%
\bibitem [{\citenamefont {Jiang}\ and\ \citenamefont
  {Guo}(2014)}]{6D-QD-HD-NNP-B-Jiang-H-Guo}%
  \BibitemOpen
  \bibfield  {author} {\bibinfo {author} {\bibfnamefont {B.}~\bibnamefont
  {Jiang}}\ and\ \bibinfo {author} {\bibfnamefont {H.}~\bibnamefont {Guo}},\
  }\href@noop {} {\bibfield  {journal} {\bibinfo  {journal} {Phys. Chem. Chem.
  Phys.}\ }\textbf {\bibinfo {volume} {16}},\ \bibinfo {pages} {24704}
  (\bibinfo {year} {2014})}\BibitemShut {NoStop}%
\bibitem [{\citenamefont {Wijzenbroek}\ \emph {et~al.}(2015)\citenamefont
  {Wijzenbroek}, \citenamefont {Klein}, \citenamefont {Smits}, \citenamefont
  {Somers},\ and\ \citenamefont
  {Kroes}}]{GGA-exchange+correlation-8-H2-Metal-2015}%
  \BibitemOpen
  \bibfield  {author} {\bibinfo {author} {\bibfnamefont {M.}~\bibnamefont
  {Wijzenbroek}}, \bibinfo {author} {\bibfnamefont {D.~M.}\ \bibnamefont
  {Klein}}, \bibinfo {author} {\bibfnamefont {B.}~\bibnamefont {Smits}},
  \bibinfo {author} {\bibfnamefont {M.~F.}\ \bibnamefont {Somers}}, \ and\
  \bibinfo {author} {\bibfnamefont {G.-J.}\ \bibnamefont {Kroes}},\ }\href@noop
  {} {\bibfield  {journal} {\bibinfo  {journal} {J. Phys. Chem. A}\ }\textbf
  {\bibinfo {volume} {119}},\ \bibinfo {pages} {12146} (\bibinfo {year}
  {2015})}\BibitemShut {NoStop}%
\bibitem [{\citenamefont {Chen}\ \emph {et~al.}(2018)\citenamefont {Chen},
  \citenamefont {Zhou}, \citenamefont {Zhang},\ and\ \citenamefont
  {Jiang}}]{GGA-exchange+correlation-2-CH3OH-Cu-2019}%
  \BibitemOpen
  \bibfield  {author} {\bibinfo {author} {\bibfnamefont {J.~L.}\ \bibnamefont
  {Chen}}, \bibinfo {author} {\bibfnamefont {X.~Y.}\ \bibnamefont {Zhou}},
  \bibinfo {author} {\bibfnamefont {Y.~L.}\ \bibnamefont {Zhang}}, \ and\
  \bibinfo {author} {\bibfnamefont {B.}~\bibnamefont {Jiang}},\ }\href@noop {}
  {\bibfield  {journal} {\bibinfo  {journal} {Nat Commun}\ }\textbf {\bibinfo
  {volume} {9}},\ \bibinfo {pages} {4039} (\bibinfo {year} {2018})}\BibitemShut
  {NoStop}%
\bibitem [{\citenamefont {Smeets}, \citenamefont {Voss},\ and\ \citenamefont
  {Kroes}(2019)}]{mGGA-1-H2-Cu-Ag-2019}%
  \BibitemOpen
  \bibfield  {author} {\bibinfo {author} {\bibfnamefont {E.~W.~F.}\
  \bibnamefont {Smeets}}, \bibinfo {author} {\bibfnamefont {J.}~\bibnamefont
  {Voss}}, \ and\ \bibinfo {author} {\bibfnamefont {G.-J.}\ \bibnamefont
  {Kroes}},\ }\href@noop {} {\bibfield  {journal} {\bibinfo  {journal} {J.
  Phys. Chem. A}\ }\textbf {\bibinfo {volume} {123}},\ \bibinfo {pages} {5395}
  (\bibinfo {year} {2019})}\BibitemShut {NoStop}%
\bibitem [{\citenamefont {Gonzalez-Lafont}, \citenamefont {Truong},\ and\
  \citenamefont {Truhlar}(1991)}]{SRP-origin}%
  \BibitemOpen
  \bibfield  {author} {\bibinfo {author} {\bibfnamefont {A.}~\bibnamefont
  {Gonzalez-Lafont}}, \bibinfo {author} {\bibfnamefont {T.~N.}\ \bibnamefont
  {Truong}}, \ and\ \bibinfo {author} {\bibfnamefont {D.~G.}\ \bibnamefont
  {Truhlar}},\ }\href@noop {} {\bibfield  {journal} {\bibinfo  {journal} {J.
  Phys. Chem.}\ }\textbf {\bibinfo {volume} {95}},\ \bibinfo {pages} {4618}
  (\bibinfo {year} {1991})}\BibitemShut {NoStop}%
\bibitem [{\citenamefont {Chuang}\ \emph {et~al.}(1999)\citenamefont {Chuang},
  \citenamefont {Radhakrishnan}, \citenamefont {Fast}, \citenamefont {Cramer},\
  and\ \citenamefont {Truhlar}}]{SRP-DFT-Truhlar}%
  \BibitemOpen
  \bibfield  {author} {\bibinfo {author} {\bibfnamefont {Y.-Y.}\ \bibnamefont
  {Chuang}}, \bibinfo {author} {\bibfnamefont {M.~L.}\ \bibnamefont
  {Radhakrishnan}}, \bibinfo {author} {\bibfnamefont {P.~L.}\ \bibnamefont
  {Fast}}, \bibinfo {author} {\bibfnamefont {C.~J.}\ \bibnamefont {Cramer}}, \
  and\ \bibinfo {author} {\bibfnamefont {D.~G.}\ \bibnamefont {Truhlar}},\
  }\href@noop {} {\bibfield  {journal} {\bibinfo  {journal} {J. Phys. Chem. A}\
  }\textbf {\bibinfo {volume} {103}},\ \bibinfo {pages} {4893} (\bibinfo {year}
  {1999})}\BibitemShut {NoStop}%
\bibitem [{\citenamefont {Mallikarjun~Sharada}\ \emph
  {et~al.}(2017)\citenamefont {Mallikarjun~Sharada}, \citenamefont {Bligaard},
  \citenamefont {Luntz}, \citenamefont {Kroes},\ and\ \citenamefont
  {Nørskov}}]{SBH10-SRP-DF-database}%
  \BibitemOpen
  \bibfield  {author} {\bibinfo {author} {\bibfnamefont {S.}~\bibnamefont
  {Mallikarjun~Sharada}}, \bibinfo {author} {\bibfnamefont {T.}~\bibnamefont
  {Bligaard}}, \bibinfo {author} {\bibfnamefont {A.~C.}\ \bibnamefont {Luntz}},
  \bibinfo {author} {\bibfnamefont {G.-J.}\ \bibnamefont {Kroes}}, \ and\
  \bibinfo {author} {\bibfnamefont {J.~K.}\ \bibnamefont {Nørskov}},\
  }\href@noop {} {\bibfield  {journal} {\bibinfo  {journal} {J. Phys. Chem. C}\
  }\textbf {\bibinfo {volume} {121}},\ \bibinfo {pages} {19807} (\bibinfo
  {year} {2017})}\BibitemShut {NoStop}%
\bibitem [{\citenamefont {Gerrits}\ \emph {et~al.}(2020)\citenamefont
  {Gerrits}, \citenamefont {Smeets}, \citenamefont {Vuckovic}, \citenamefont
  {Powell}, \citenamefont {Doblhoff-Dier},\ and\ \citenamefont
  {Kroes}}]{SRP-DF-38-limitation-2020}%
  \BibitemOpen
  \bibfield  {author} {\bibinfo {author} {\bibfnamefont {N.}~\bibnamefont
  {Gerrits}}, \bibinfo {author} {\bibfnamefont {E.~W.~F.}\ \bibnamefont
  {Smeets}}, \bibinfo {author} {\bibfnamefont {S.}~\bibnamefont {Vuckovic}},
  \bibinfo {author} {\bibfnamefont {A.~D.}\ \bibnamefont {Powell}}, \bibinfo
  {author} {\bibfnamefont {K.}~\bibnamefont {Doblhoff-Dier}}, \ and\ \bibinfo
  {author} {\bibfnamefont {G.-J.}\ \bibnamefont {Kroes}},\ }\href@noop {}
  {\bibfield  {journal} {\bibinfo  {journal} {J. Phys. Chem. Lett.}\ }\textbf
  {\bibinfo {volume} {11}},\ \bibinfo {pages} {10552} (\bibinfo {year}
  {2020})}\BibitemShut {NoStop}%
\bibitem [{\citenamefont {Yin}\ \emph {et~al.}(2018)\citenamefont {Yin},
  \citenamefont {Zhang}, \citenamefont {Libisch}, \citenamefont {Carter},
  \citenamefont {Guo},\ and\ \citenamefont {Jiang}}]{ECW-3-O2-Al-2018}%
  \BibitemOpen
  \bibfield  {author} {\bibinfo {author} {\bibfnamefont {R.}~\bibnamefont
  {Yin}}, \bibinfo {author} {\bibfnamefont {Y.}~\bibnamefont {Zhang}}, \bibinfo
  {author} {\bibfnamefont {F.}~\bibnamefont {Libisch}}, \bibinfo {author}
  {\bibfnamefont {E.~A.}\ \bibnamefont {Carter}}, \bibinfo {author}
  {\bibfnamefont {H.}~\bibnamefont {Guo}}, \ and\ \bibinfo {author}
  {\bibfnamefont {B.}~\bibnamefont {Jiang}},\ }\href@noop {} {\bibfield
  {journal} {\bibinfo  {journal} {J. Phys. Chem. Lett.}\ }\textbf {\bibinfo
  {volume} {9}},\ \bibinfo {pages} {3271} (\bibinfo {year} {2018})}\BibitemShut
  {NoStop}%
\bibitem [{\citenamefont {Doblhoff-Dier}\ \emph {et~al.}(2017)\citenamefont
  {Doblhoff-Dier}, \citenamefont {Meyer}, \citenamefont {Hoggan},\ and\
  \citenamefont {Kroes}}]{DMC-1-H2-Cu-2017}%
  \BibitemOpen
  \bibfield  {author} {\bibinfo {author} {\bibfnamefont {K.}~\bibnamefont
  {Doblhoff-Dier}}, \bibinfo {author} {\bibfnamefont {J.}~\bibnamefont
  {Meyer}}, \bibinfo {author} {\bibfnamefont {P.~E.}\ \bibnamefont {Hoggan}}, \
  and\ \bibinfo {author} {\bibfnamefont {G.-J.}\ \bibnamefont {Kroes}},\
  }\href@noop {} {\bibfield  {journal} {\bibinfo  {journal} {J. Chem. Theory
  Comput.}\ }\textbf {\bibinfo {volume} {13}},\ \bibinfo {pages} {3208}
  (\bibinfo {year} {2017})}\BibitemShut {NoStop}%
\bibitem [{\citenamefont {Powell}, \citenamefont {Kroes},\ and\ \citenamefont
  {Doblhoff-Dier}(2020)}]{DMC-3-2020}%
  \BibitemOpen
  \bibfield  {author} {\bibinfo {author} {\bibfnamefont {A.~D.}\ \bibnamefont
  {Powell}}, \bibinfo {author} {\bibfnamefont {G.-J.}\ \bibnamefont {Kroes}}, \
  and\ \bibinfo {author} {\bibfnamefont {K.}~\bibnamefont {Doblhoff-Dier}},\
  }\href@noop {} {\bibfield  {journal} {\bibinfo  {journal} {J. Chem. Phys.}\
  }\textbf {\bibinfo {volume} {153}},\ \bibinfo {pages} {224701} (\bibinfo
  {year} {2020})}\BibitemShut {NoStop}%
\bibitem [{\citenamefont {Bonfanti}\ \emph {et~al.}(2013)\citenamefont
  {Bonfanti}, \citenamefont {Somers}, \citenamefont {Díaz}, \citenamefont
  {Busnengo},\ and\ \citenamefont {Kroes}}]{Kroes-7D-QD}%
  \BibitemOpen
  \bibfield  {author} {\bibinfo {author} {\bibfnamefont {M.}~\bibnamefont
  {Bonfanti}}, \bibinfo {author} {\bibfnamefont {M.~F.}\ \bibnamefont
  {Somers}}, \bibinfo {author} {\bibfnamefont {C.}~\bibnamefont {Díaz}},
  \bibinfo {author} {\bibfnamefont {H.~F.}\ \bibnamefont {Busnengo}}, \ and\
  \bibinfo {author} {\bibfnamefont {G.-J.}\ \bibnamefont {Kroes}},\ }\href@noop
  {} {\bibfield  {journal} {\bibinfo  {journal} {Z. Phys. Chem.}\ }\textbf
  {\bibinfo {volume} {227}},\ \bibinfo {pages} {1397} (\bibinfo {year}
  {2013})}\BibitemShut {NoStop}%
\bibitem [{\citenamefont {Behler}(2015)}]{HD-NN-tutorial-review}%
  \BibitemOpen
  \bibfield  {author} {\bibinfo {author} {\bibfnamefont {J.}~\bibnamefont
  {Behler}},\ }\href@noop {} {\bibfield  {journal} {\bibinfo  {journal} {Int.
  J. Quantum Chem.}\ }\textbf {\bibinfo {volume} {115}},\ \bibinfo {pages}
  {1032} (\bibinfo {year} {2015})}\BibitemShut {NoStop}%
\bibitem [{\citenamefont {Behler}\ and\ \citenamefont
  {Parrinello}(2007)}]{AtNNmethod}%
  \BibitemOpen
  \bibfield  {author} {\bibinfo {author} {\bibfnamefont {J.}~\bibnamefont
  {Behler}}\ and\ \bibinfo {author} {\bibfnamefont {M.}~\bibnamefont
  {Parrinello}},\ }\href@noop {} {\bibfield  {journal} {\bibinfo  {journal}
  {Phys. Rev. Lett.}\ }\textbf {\bibinfo {volume} {98}},\ \bibinfo {pages}
  {146401} (\bibinfo {year} {2007})}\BibitemShut {NoStop}%
\bibitem [{\citenamefont {Liu}, \citenamefont {Fu},\ and\ \citenamefont
  {Zhang}(2019)}]{6D-QD-HD-NNP-D-Zhang-2}%
  \BibitemOpen
  \bibfield  {author} {\bibinfo {author} {\bibfnamefont {T.}~\bibnamefont
  {Liu}}, \bibinfo {author} {\bibfnamefont {B.}~\bibnamefont {Fu}}, \ and\
  \bibinfo {author} {\bibfnamefont {D.~H.}\ \bibnamefont {Zhang}},\ }\href@noop
  {} {\bibfield  {journal} {\bibinfo  {journal} {J. Chem. Phys.}\ }\textbf
  {\bibinfo {volume} {151}},\ \bibinfo {pages} {144707} (\bibinfo {year}
  {2019})}\BibitemShut {NoStop}%
\bibitem [{\citenamefont {Dohle}, \citenamefont {Saalfrank},\ and\
  \citenamefont {Uzer}(1998)}]{UZER1}%
  \BibitemOpen
  \bibfield  {author} {\bibinfo {author} {\bibfnamefont {M.}~\bibnamefont
  {Dohle}}, \bibinfo {author} {\bibfnamefont {P.}~\bibnamefont {Saalfrank}}, \
  and\ \bibinfo {author} {\bibfnamefont {T.}~\bibnamefont {Uzer}},\ }\href@noop
  {} {\bibfield  {journal} {\bibinfo  {journal} {Surf. Sci.}\ }\textbf
  {\bibinfo {volume} {409}},\ \bibinfo {pages} {37} (\bibinfo {year}
  {1998})}\BibitemShut {NoStop}%
\bibitem [{\citenamefont {Tully}(1980)}]{GLO-1}%
  \BibitemOpen
  \bibfield  {author} {\bibinfo {author} {\bibfnamefont {J.~C.}\ \bibnamefont
  {Tully}},\ }\href@noop {} {\bibfield  {journal} {\bibinfo  {journal} {J.
  Chem. Phys.}\ }\textbf {\bibinfo {volume} {73}},\ \bibinfo {pages} {1975}
  (\bibinfo {year} {1980})}\BibitemShut {NoStop}%
\bibitem [{\citenamefont {Adelman}\ and\ \citenamefont {Doll}(1976)}]{GLO-2}%
  \BibitemOpen
  \bibfield  {author} {\bibinfo {author} {\bibfnamefont {S.~A.}\ \bibnamefont
  {Adelman}}\ and\ \bibinfo {author} {\bibfnamefont {J.~D.}\ \bibnamefont
  {Doll}},\ }\href@noop {} {\bibfield  {journal} {\bibinfo  {journal} {J. Chem.
  Phys.}\ }\textbf {\bibinfo {volume} {64}},\ \bibinfo {pages} {2375} (\bibinfo
  {year} {1976})}\BibitemShut {NoStop}%
\bibitem [{\citenamefont {Busnengo}\ \emph {et~al.}(2005)\citenamefont
  {Busnengo}, \citenamefont {Di~C\'esare}, \citenamefont {Dong},\ and\
  \citenamefont {Salin}}]{CRP-H2-Pd-thermally-distorted-2005}%
  \BibitemOpen
  \bibfield  {author} {\bibinfo {author} {\bibfnamefont {H.~F.}\ \bibnamefont
  {Busnengo}}, \bibinfo {author} {\bibfnamefont {M.~A.}\ \bibnamefont
  {Di~C\'esare}}, \bibinfo {author} {\bibfnamefont {W.}~\bibnamefont {Dong}}, \
  and\ \bibinfo {author} {\bibfnamefont {A.}~\bibnamefont {Salin}},\
  }\href@noop {} {\bibfield  {journal} {\bibinfo  {journal} {Phys. Rev. B}\
  }\textbf {\bibinfo {volume} {72}},\ \bibinfo {pages} {125411} (\bibinfo
  {year} {2005})}\BibitemShut {NoStop}%
\bibitem [{\citenamefont {Zhou}\ and\ \citenamefont
  {Jiang}(2019)}]{SRP-DF-9-gas-surface-GLO-new-2019}%
  \BibitemOpen
  \bibfield  {author} {\bibinfo {author} {\bibfnamefont {X.}~\bibnamefont
  {Zhou}}\ and\ \bibinfo {author} {\bibfnamefont {B.}~\bibnamefont {Jiang}},\
  }\href@noop {} {\bibfield  {journal} {\bibinfo  {journal} {J. Chem. Phys.}\
  }\textbf {\bibinfo {volume} {150}},\ \bibinfo {pages} {024704} (\bibinfo
  {year} {2019})}\BibitemShut {NoStop}%
\bibitem [{\citenamefont {Tiwari}, \citenamefont {Nave},\ and\ \citenamefont
  {Jackson}(2010)}]{AKT1}%
  \BibitemOpen
  \bibfield  {author} {\bibinfo {author} {\bibfnamefont {A.~K.}\ \bibnamefont
  {Tiwari}}, \bibinfo {author} {\bibfnamefont {S.}~\bibnamefont {Nave}}, \ and\
  \bibinfo {author} {\bibfnamefont {B.}~\bibnamefont {Jackson}},\ }\href@noop
  {} {\bibfield  {journal} {\bibinfo  {journal} {J. Chem. Phys.}\ }\textbf
  {\bibinfo {volume} {132}},\ \bibinfo {pages} {134702} (\bibinfo {year}
  {2010})}\BibitemShut {NoStop}%
\bibitem [{\citenamefont {Tiwari}, \citenamefont {Nave},\ and\ \citenamefont
  {Jackson}(2009)}]{AKT}%
  \BibitemOpen
  \bibfield  {author} {\bibinfo {author} {\bibfnamefont {A.~K.}\ \bibnamefont
  {Tiwari}}, \bibinfo {author} {\bibfnamefont {S.}~\bibnamefont {Nave}}, \ and\
  \bibinfo {author} {\bibfnamefont {B.}~\bibnamefont {Jackson}},\ }\href@noop
  {} {\bibfield  {journal} {\bibinfo  {journal} {Phys.\ Rev.\ Lett.}\ }\textbf
  {\bibinfo {volume} {103}},\ \bibinfo {pages} {253201/1} (\bibinfo {year}
  {2009})}\BibitemShut {NoStop}%
\bibitem [{\citenamefont
  {Jackson}(2020)}]{CH4-trapping-1-Ni-Ir-Pt-Jackson-2020}%
  \BibitemOpen
  \bibfield  {author} {\bibinfo {author} {\bibfnamefont {B.}~\bibnamefont
  {Jackson}},\ }\href@noop {} {\bibfield  {journal} {\bibinfo  {journal} {J.
  Chem. Phys.}\ }\textbf {\bibinfo {volume} {153}},\ \bibinfo {pages} {034704}
  (\bibinfo {year} {2020})}\BibitemShut {NoStop}%
\bibitem [{\citenamefont {Jackson}(2021)}]{CH4-trapping-2-Ir-Jackson-2021}%
  \BibitemOpen
  \bibfield  {author} {\bibinfo {author} {\bibfnamefont {B.}~\bibnamefont
  {Jackson}},\ }\href@noop {} {\bibfield  {journal} {\bibinfo  {journal} {J.
  Chem. Phys.}\ }\textbf {\bibinfo {volume} {155}},\ \bibinfo {pages} {044705}
  (\bibinfo {year} {2021})}\BibitemShut {NoStop}%
\bibitem [{\citenamefont {Smits}\ and\ \citenamefont
  {Somers}(2022)}]{Bauke_jcp_2022}%
  \BibitemOpen
  \bibfield  {author} {\bibinfo {author} {\bibfnamefont {B.}~\bibnamefont
  {Smits}}\ and\ \bibinfo {author} {\bibfnamefont {M.~F.}\ \bibnamefont
  {Somers}},\ }\href@noop {} {\bibfield  {journal} {\bibinfo  {journal} {J.
  Chem. Phys.}\ }\textbf {\bibinfo {volume} {157}},\ \bibinfo {pages} {134704}
  (\bibinfo {year} {2022})}\BibitemShut {NoStop}%
\bibitem [{\citenamefont {Smits}\ and\ \citenamefont
  {Somers}(2023)}]{Bauke_jcp_2023}%
  \BibitemOpen
  \bibfield  {author} {\bibinfo {author} {\bibfnamefont {B.}~\bibnamefont
  {Smits}}\ and\ \bibinfo {author} {\bibfnamefont {M.~F.}\ \bibnamefont
  {Somers}},\ }\href@noop {} {\bibfield  {journal} {\bibinfo  {journal} {J.
  Chem. Phys.}\ }\textbf {\bibinfo {volume} {158}},\ \bibinfo {pages} {014704}
  (\bibinfo {year} {2023})}\BibitemShut {NoStop}%
\bibitem [{\citenamefont {Mandal}\ \emph
  {et~al.}(2015{\natexlab{a}})\citenamefont {Mandal}, \citenamefont {Sahoo},
  \citenamefont {Ghosh},\ and\ \citenamefont {Adhikari}}]{SM-JTCC}%
  \BibitemOpen
  \bibfield  {author} {\bibinfo {author} {\bibfnamefont {S.}~\bibnamefont
  {Mandal}}, \bibinfo {author} {\bibfnamefont {T.}~\bibnamefont {Sahoo}},
  \bibinfo {author} {\bibfnamefont {S.}~\bibnamefont {Ghosh}}, \ and\ \bibinfo
  {author} {\bibfnamefont {S.}~\bibnamefont {Adhikari}},\ }\href@noop {}
  {\bibfield  {journal} {\bibinfo  {journal} {J. Theor. Comput. Chem.}\
  }\textbf {\bibinfo {volume} {14}},\ \bibinfo {pages} {1550028} (\bibinfo
  {year} {2015}{\natexlab{a}})}\BibitemShut {NoStop}%
\bibitem [{\citenamefont {Mandal}\ \emph
  {et~al.}(2015{\natexlab{b}})\citenamefont {Mandal}, \citenamefont {Sahoo},
  \citenamefont {Ghosh},\ and\ \citenamefont {Adhikari}}]{SM-MOLPHYS}%
  \BibitemOpen
  \bibfield  {author} {\bibinfo {author} {\bibfnamefont {S.}~\bibnamefont
  {Mandal}}, \bibinfo {author} {\bibfnamefont {T.}~\bibnamefont {Sahoo}},
  \bibinfo {author} {\bibfnamefont {S.}~\bibnamefont {Ghosh}}, \ and\ \bibinfo
  {author} {\bibfnamefont {S.}~\bibnamefont {Adhikari}},\ }\href@noop {}
  {\bibfield  {journal} {\bibinfo  {journal} {Mol. Phys.}\ }\textbf {\bibinfo
  {volume} {113}},\ \bibinfo {pages} {3042} (\bibinfo {year}
  {2015}{\natexlab{b}})}\BibitemShut {NoStop}%
\bibitem [{\citenamefont {Kroes}\ and\ \citenamefont
  {Somers}(2005)}]{SPO-DVR-code-2}%
  \BibitemOpen
  \bibfield  {author} {\bibinfo {author} {\bibfnamefont {G.~J.}\ \bibnamefont
  {Kroes}}\ and\ \bibinfo {author} {\bibfnamefont {M.~F.}\ \bibnamefont
  {Somers}},\ }\href@noop {} {\bibfield  {journal} {\bibinfo  {journal} {J.
  Theor. Comput. Chem.}\ }\textbf {\bibinfo {volume} {04}},\ \bibinfo {pages}
  {493} (\bibinfo {year} {2005})}\BibitemShut {NoStop}%
\bibitem [{\citenamefont {Kindt}\ \emph {et~al.}(1998)\citenamefont {Kindt},
  \citenamefont {Tully}, \citenamefont {Head-Gordon},\ and\ \citenamefont
  {Gomez}}]{JCT}%
  \BibitemOpen
  \bibfield  {author} {\bibinfo {author} {\bibfnamefont {J.~T.}\ \bibnamefont
  {Kindt}}, \bibinfo {author} {\bibfnamefont {J.~C.}\ \bibnamefont {Tully}},
  \bibinfo {author} {\bibfnamefont {M.}~\bibnamefont {Head-Gordon}}, \ and\
  \bibinfo {author} {\bibfnamefont {M.~A.}\ \bibnamefont {Gomez}},\ }\href@noop
  {} {\bibfield  {journal} {\bibinfo  {journal} {J. Chem. Phys.}\ }\textbf
  {\bibinfo {volume} {109}},\ \bibinfo {pages} {3629} (\bibinfo {year}
  {1998})}\BibitemShut {NoStop}%
\bibitem [{\citenamefont {Tully}(2000)}]{JCTullyrev}%
  \BibitemOpen
  \bibfield  {author} {\bibinfo {author} {\bibfnamefont {J.~C.}\ \bibnamefont
  {Tully}},\ }\href@noop {} {\bibfield  {journal} {\bibinfo  {journal} {Annu.
  Rev. Phys. Chem.}\ }\textbf {\bibinfo {volume} {51}},\ \bibinfo {pages} {153}
  (\bibinfo {year} {2000})}\BibitemShut {NoStop}%
\bibitem [{\citenamefont {Saalfrank}(2006)}]{Saalfrankrev}%
  \BibitemOpen
  \bibfield  {author} {\bibinfo {author} {\bibfnamefont {P.}~\bibnamefont
  {Saalfrank}},\ }\href@noop {} {\bibfield  {journal} {\bibinfo  {journal}
  {Chem. Rev.}\ }\textbf {\bibinfo {volume} {106}},\ \bibinfo {pages} {4116}
  (\bibinfo {year} {2006})}\BibitemShut {NoStop}%
\bibitem [{\citenamefont {Wodtke}, \citenamefont {Matsiev},\ and\ \citenamefont
  {Auerbach}(2008)}]{Wodtkeelhp}%
  \BibitemOpen
  \bibfield  {author} {\bibinfo {author} {\bibfnamefont {A.~M.}\ \bibnamefont
  {Wodtke}}, \bibinfo {author} {\bibfnamefont {D.}~\bibnamefont {Matsiev}}, \
  and\ \bibinfo {author} {\bibfnamefont {D.~J.}\ \bibnamefont {Auerbach}},\
  }\href@noop {} {\bibfield  {journal} {\bibinfo  {journal} {Prog. Surf. Sci.}\
  }\textbf {\bibinfo {volume} {83}},\ \bibinfo {pages} {167} (\bibinfo {year}
  {2008})}\BibitemShut {NoStop}%
\bibitem [{\citenamefont {Shenvi}, \citenamefont {Roy},\ and\ \citenamefont
  {Tully}(2009{\natexlab{a}})}]{JCTully-elhp1}%
  \BibitemOpen
  \bibfield  {author} {\bibinfo {author} {\bibfnamefont {N.}~\bibnamefont
  {Shenvi}}, \bibinfo {author} {\bibfnamefont {S.}~\bibnamefont {Roy}}, \ and\
  \bibinfo {author} {\bibfnamefont {J.~C.}\ \bibnamefont {Tully}},\ }\href@noop
  {} {\bibfield  {journal} {\bibinfo  {journal} {Science}\ }\textbf {\bibinfo
  {volume} {326}},\ \bibinfo {pages} {829} (\bibinfo {year}
  {2009}{\natexlab{a}})}\BibitemShut {NoStop}%
\bibitem [{\citenamefont {Roy}, \citenamefont {Shenvi},\ and\ \citenamefont
  {Tully}(2009)}]{JCTully-elhp2}%
  \BibitemOpen
  \bibfield  {author} {\bibinfo {author} {\bibfnamefont {S.}~\bibnamefont
  {Roy}}, \bibinfo {author} {\bibfnamefont {N.~A.}\ \bibnamefont {Shenvi}}, \
  and\ \bibinfo {author} {\bibfnamefont {J.~C.}\ \bibnamefont {Tully}},\
  }\href@noop {} {\bibfield  {journal} {\bibinfo  {journal} {J. Chem. Phys.}\
  }\textbf {\bibinfo {volume} {130}},\ \bibinfo {pages} {174716} (\bibinfo
  {year} {2009})}\BibitemShut {NoStop}%
\bibitem [{\citenamefont {Bukas}\ \emph {et~al.}(2015)\citenamefont {Bukas},
  \citenamefont {Mitra}, \citenamefont {Meyer},\ and\ \citenamefont
  {Reuter}}]{O2-Pd100-JM1}%
  \BibitemOpen
  \bibfield  {author} {\bibinfo {author} {\bibfnamefont {V.~J.}\ \bibnamefont
  {Bukas}}, \bibinfo {author} {\bibfnamefont {S.}~\bibnamefont {Mitra}},
  \bibinfo {author} {\bibfnamefont {J.}~\bibnamefont {Meyer}}, \ and\ \bibinfo
  {author} {\bibfnamefont {K.}~\bibnamefont {Reuter}},\ }\href@noop {}
  {\bibfield  {journal} {\bibinfo  {journal} {J. Chem. Phys.}\ }\textbf
  {\bibinfo {volume} {143}},\ \bibinfo {pages} {034705} (\bibinfo {year}
  {2015})}\BibitemShut {NoStop}%
\bibitem [{\citenamefont {Tremblay}(2013)}]{Tremblay6}%
  \BibitemOpen
  \bibfield  {author} {\bibinfo {author} {\bibfnamefont {J.~C.}\ \bibnamefont
  {Tremblay}},\ }\href@noop {} {\bibfield  {journal} {\bibinfo  {journal} {J.
  Chem. Phys.}\ }\textbf {\bibinfo {volume} {138}},\ \bibinfo {pages} {244106}
  (\bibinfo {year} {2013})}\BibitemShut {NoStop}%
\bibitem [{\citenamefont {Juaristi}\ \emph {et~al.}(2008)\citenamefont
  {Juaristi}, \citenamefont {Alducin}, \citenamefont {{n}o}, \citenamefont
  {Busnengo},\ and\ \citenamefont {Salin}}]{JIJ1}%
  \BibitemOpen
  \bibfield  {author} {\bibinfo {author} {\bibfnamefont {J.~I.}\ \bibnamefont
  {Juaristi}}, \bibinfo {author} {\bibfnamefont {M.}~\bibnamefont {Alducin}},
  \bibinfo {author} {\bibfnamefont {R.~D.~M.}\ \bibnamefont {{n}o}}, \bibinfo
  {author} {\bibfnamefont {H.~F.}\ \bibnamefont {Busnengo}}, \ and\ \bibinfo
  {author} {\bibfnamefont {A.}~\bibnamefont {Salin}},\ }\href@noop {}
  {\bibfield  {journal} {\bibinfo  {journal} {Phys. Rev. Lett.}\ }\textbf
  {\bibinfo {volume} {100}},\ \bibinfo {pages} {116102/1} (\bibinfo {year}
  {2008})}\BibitemShut {NoStop}%
\bibitem [{\citenamefont {Shenvi}, \citenamefont {Roy},\ and\ \citenamefont
  {Tully}(2009{\natexlab{b}})}]{IESH-2009}%
  \BibitemOpen
  \bibfield  {author} {\bibinfo {author} {\bibfnamefont {N.}~\bibnamefont
  {Shenvi}}, \bibinfo {author} {\bibfnamefont {S.}~\bibnamefont {Roy}}, \ and\
  \bibinfo {author} {\bibfnamefont {J.~C.}\ \bibnamefont {Tully}},\ }\href@noop
  {} {\bibfield  {journal} {\bibinfo  {journal} {J. Chem. Phys.}\ }\textbf
  {\bibinfo {volume} {130}},\ \bibinfo {pages} {174107} (\bibinfo {year}
  {2009}{\natexlab{b}})}\BibitemShut {NoStop}%
\bibitem [{\citenamefont {Spiering}\ \emph {et~al.}(2019)\citenamefont
  {Spiering}, \citenamefont {Shakouri}, \citenamefont {Behler}, \citenamefont
  {Kroes},\ and\ \citenamefont {Meyer}}]{RPBE-1-N2-Ru-ODF-elhp-2019}%
  \BibitemOpen
  \bibfield  {author} {\bibinfo {author} {\bibfnamefont {P.}~\bibnamefont
  {Spiering}}, \bibinfo {author} {\bibfnamefont {K.}~\bibnamefont {Shakouri}},
  \bibinfo {author} {\bibfnamefont {J.}~\bibnamefont {Behler}}, \bibinfo
  {author} {\bibfnamefont {G.-J.}\ \bibnamefont {Kroes}}, \ and\ \bibinfo
  {author} {\bibfnamefont {J.}~\bibnamefont {Meyer}},\ }\href@noop {}
  {\bibfield  {journal} {\bibinfo  {journal} {J. Phys. Chem. Lett.}\ }\textbf
  {\bibinfo {volume} {10}},\ \bibinfo {pages} {2957} (\bibinfo {year}
  {2019})}\BibitemShut {NoStop}%
\bibitem [{\citenamefont {Maurer}\ \emph {et~al.}(2016)\citenamefont {Maurer},
  \citenamefont {Askerka}, \citenamefont {Batista},\ and\ \citenamefont
  {Tully}}]{ODF-5}%
  \BibitemOpen
  \bibfield  {author} {\bibinfo {author} {\bibfnamefont {R.~J.}\ \bibnamefont
  {Maurer}}, \bibinfo {author} {\bibfnamefont {M.}~\bibnamefont {Askerka}},
  \bibinfo {author} {\bibfnamefont {V.~S.}\ \bibnamefont {Batista}}, \ and\
  \bibinfo {author} {\bibfnamefont {J.~C.}\ \bibnamefont {Tully}},\ }\href@noop
  {} {\bibfield  {journal} {\bibinfo  {journal} {Phys. Rev. B}\ }\textbf
  {\bibinfo {volume} {94}},\ \bibinfo {pages} {115432} (\bibinfo {year}
  {2016})}\BibitemShut {NoStop}%
\bibitem [{\citenamefont {Rittmeyer}, \citenamefont {Bukas},\ and\
  \citenamefont {Reuter}(2018)}]{Dynamical-models-review-2018}%
  \BibitemOpen
  \bibfield  {author} {\bibinfo {author} {\bibfnamefont {S.~P.}\ \bibnamefont
  {Rittmeyer}}, \bibinfo {author} {\bibfnamefont {V.~J.}\ \bibnamefont
  {Bukas}}, \ and\ \bibinfo {author} {\bibfnamefont {K.}~\bibnamefont
  {Reuter}},\ }\href@noop {} {\bibfield  {journal} {\bibinfo  {journal} {Adv.
  Phys.: X}\ }\textbf {\bibinfo {volume} {3}},\ \bibinfo {pages} {1381574}
  (\bibinfo {year} {2018})}\BibitemShut {NoStop}%
\bibitem [{\citenamefont {Pe\~{n}a Torres}\ \emph {et~al.}(2019)\citenamefont
  {Pe\~{n}a Torres}, \citenamefont {Busnengo}, \citenamefont {Juaristi},
  \citenamefont {Larregaray},\ and\ \citenamefont
  {Crespos}}]{GGA-exchange+correlation-1-N2-W-elhp-2019}%
  \BibitemOpen
  \bibfield  {author} {\bibinfo {author} {\bibfnamefont {A.}~\bibnamefont
  {Pe\~{n}a Torres}}, \bibinfo {author} {\bibfnamefont {H.~F.}\ \bibnamefont
  {Busnengo}}, \bibinfo {author} {\bibfnamefont {J.~I.}\ \bibnamefont
  {Juaristi}}, \bibinfo {author} {\bibfnamefont {P.}~\bibnamefont
  {Larregaray}}, \ and\ \bibinfo {author} {\bibfnamefont {C.}~\bibnamefont
  {Crespos}},\ }\href@noop {} {\bibfield  {journal} {\bibinfo  {journal} {J.
  Phys. Chem. C}\ }\textbf {\bibinfo {volume} {123}},\ \bibinfo {pages} {2900}
  (\bibinfo {year} {2019})}\BibitemShut {NoStop}%
\bibitem [{\citenamefont {Luo}, \citenamefont {Zhou},\ and\ \citenamefont
  {Jiang}(2018)}]{PBE-6-CO2-surface-mode-elhp-2018}%
  \BibitemOpen
  \bibfield  {author} {\bibinfo {author} {\bibfnamefont {X.}~\bibnamefont
  {Luo}}, \bibinfo {author} {\bibfnamefont {X.}~\bibnamefont {Zhou}}, \ and\
  \bibinfo {author} {\bibfnamefont {B.}~\bibnamefont {Jiang}},\ }\href@noop {}
  {\bibfield  {journal} {\bibinfo  {journal} {J. Chem. Phys.}\ }\textbf
  {\bibinfo {volume} {148}},\ \bibinfo {pages} {174702} (\bibinfo {year}
  {2018})}\BibitemShut {NoStop}%
\bibitem [{\citenamefont {Novko}\ \emph {et~al.}(2016)\citenamefont {Novko},
  \citenamefont {Blanco-Rey}, \citenamefont {Alducin},\ and\ \citenamefont
  {Juaristi}}]{RPBE-2-H2-Pd-N-Ag-N2-Fe-AIMDEF-elhp-2016}%
  \BibitemOpen
  \bibfield  {author} {\bibinfo {author} {\bibfnamefont {D.}~\bibnamefont
  {Novko}}, \bibinfo {author} {\bibfnamefont {M.}~\bibnamefont {Blanco-Rey}},
  \bibinfo {author} {\bibfnamefont {M.}~\bibnamefont {Alducin}}, \ and\
  \bibinfo {author} {\bibfnamefont {J.~I.}\ \bibnamefont {Juaristi}},\
  }\href@noop {} {\bibfield  {journal} {\bibinfo  {journal} {Phys. Rev. B}\
  }\textbf {\bibinfo {volume} {93}},\ \bibinfo {pages} {245435} (\bibinfo
  {year} {2016})}\BibitemShut {NoStop}%
\bibitem [{\citenamefont {F\"uchsel}\ \emph {et~al.}(2019)\citenamefont
  {F\"uchsel}, \citenamefont {Zhou}, \citenamefont {Jiang}, \citenamefont
  {Juaristi}, \citenamefont {Alducin}, \citenamefont {Guo},\ and\ \citenamefont
  {Kroes}}]{GGA-exchange+correlation-5-HCl-Au-AIMD-2019}%
  \BibitemOpen
  \bibfield  {author} {\bibinfo {author} {\bibfnamefont {G.}~\bibnamefont
  {F\"uchsel}}, \bibinfo {author} {\bibfnamefont {X.}~\bibnamefont {Zhou}},
  \bibinfo {author} {\bibfnamefont {B.}~\bibnamefont {Jiang}}, \bibinfo
  {author} {\bibfnamefont {J.~I.~n.}\ \bibnamefont {Juaristi}}, \bibinfo
  {author} {\bibfnamefont {M.}~\bibnamefont {Alducin}}, \bibinfo {author}
  {\bibfnamefont {H.}~\bibnamefont {Guo}}, \ and\ \bibinfo {author}
  {\bibfnamefont {G.-J.}\ \bibnamefont {Kroes}},\ }\href@noop {} {\bibfield
  {journal} {\bibinfo  {journal} {J. Phys. Chem. C}\ }\textbf {\bibinfo
  {volume} {123}},\ \bibinfo {pages} {2287} (\bibinfo {year}
  {2019})}\BibitemShut {NoStop}%
\bibitem [{\citenamefont {Truong}, \citenamefont {Truhlar},\ and\ \citenamefont
  {Garrett}(1989)}]{TNT}%
  \BibitemOpen
  \bibfield  {author} {\bibinfo {author} {\bibfnamefont {T.~N.}\ \bibnamefont
  {Truong}}, \bibinfo {author} {\bibfnamefont {D.~G.}\ \bibnamefont {Truhlar}},
  \ and\ \bibinfo {author} {\bibfnamefont {B.~C.}\ \bibnamefont {Garrett}},\
  }\href@noop {} {\bibfield  {journal} {\bibinfo  {journal} {J.\ Phys.\ Chem.}\
  }\textbf {\bibinfo {volume} {93}},\ \bibinfo {pages} {8227} (\bibinfo {year}
  {1989})}\BibitemShut {NoStop}%
\bibitem [{\citenamefont {Olsen}\ \emph {et~al.}(1997)\citenamefont {Olsen},
  \citenamefont {Kroes}, \citenamefont {Lovvik},\ and\ \citenamefont
  {Baerends}}]{OLKR}%
  \BibitemOpen
  \bibfield  {author} {\bibinfo {author} {\bibfnamefont {R.~A.}\ \bibnamefont
  {Olsen}}, \bibinfo {author} {\bibfnamefont {G.~J.}\ \bibnamefont {Kroes}},
  \bibinfo {author} {\bibfnamefont {O.~M.}\ \bibnamefont {Lovvik}}, \ and\
  \bibinfo {author} {\bibfnamefont {E.~J.}\ \bibnamefont {Baerends}},\
  }\href@noop {} {\bibfield  {journal} {\bibinfo  {journal} {J. Chem. Phys.}\
  }\textbf {\bibinfo {volume} {107}},\ \bibinfo {pages} {10652} (\bibinfo
  {year} {1997})}\BibitemShut {NoStop}%
\bibitem [{\citenamefont {Pechukas}\ and\ \citenamefont {Light}(1966)}]{PPJS}%
  \BibitemOpen
  \bibfield  {author} {\bibinfo {author} {\bibfnamefont {P.}~\bibnamefont
  {Pechukas}}\ and\ \bibinfo {author} {\bibfnamefont {J.~C.}\ \bibnamefont
  {Light}},\ }\href@noop {} {\bibfield  {journal} {\bibinfo  {journal} {J.\
  Chem.\ Phys.}\ }\textbf {\bibinfo {volume} {44}},\ \bibinfo {pages} {3897}
  (\bibinfo {year} {1966})}\BibitemShut {NoStop}%
\bibitem [{\citenamefont {Mengel}\ and\ \citenamefont
  {Billing}(1997)}]{BILEH2}%
  \BibitemOpen
  \bibfield  {author} {\bibinfo {author} {\bibfnamefont {S.~K.}\ \bibnamefont
  {Mengel}}\ and\ \bibinfo {author} {\bibfnamefont {G.~D.}\ \bibnamefont
  {Billing}},\ }\href@noop {} {\bibfield  {journal} {\bibinfo  {journal} {J.
  Phys. Chem. B}\ }\textbf {\bibinfo {volume} {101}},\ \bibinfo {pages} {10781}
  (\bibinfo {year} {1997})}\BibitemShut {NoStop}%
\bibitem [{\citenamefont {Pines}(1963)}]{DPI}%
  \BibitemOpen
  \bibfield  {author} {\bibinfo {author} {\bibfnamefont {D.}~\bibnamefont
  {Pines}},\ }\href@noop {} {\emph {\bibinfo {title} {Elementary Excitations in
  Solids}}}\ (\bibinfo  {publisher} {Bejamin/Cummings, Reading},\ \bibinfo
  {year} {1963})\BibitemShut {NoStop}%
\bibitem [{\citenamefont {Billing}\ and\ \citenamefont
  {Mikkelsen}(1997)}]{GDB2}%
  \BibitemOpen
  \bibfield  {author} {\bibinfo {author} {\bibfnamefont {G.~D.}\ \bibnamefont
  {Billing}}\ and\ \bibinfo {author} {\bibfnamefont {K.~V.}\ \bibnamefont
  {Mikkelsen}},\ }\href@noop {} {\emph {\bibinfo {title} {Advance Molecular
  Dynamics and Chemical Kinetics}}}\ (\bibinfo {year} {1997})\BibitemShut
  {NoStop}%
\bibitem [{\citenamefont {Michelsen}\ and\ \citenamefont
  {Auerbach}(1991)}]{HAM1}%
  \BibitemOpen
  \bibfield  {author} {\bibinfo {author} {\bibfnamefont {H.~A.}\ \bibnamefont
  {Michelsen}}\ and\ \bibinfo {author} {\bibfnamefont {D.~J.}\ \bibnamefont
  {Auerbach}},\ }\href@noop {} {\bibfield  {journal} {\bibinfo  {journal} {J.
  Chem. Phys.}\ }\textbf {\bibinfo {volume} {94}},\ \bibinfo {pages} {7502}
  (\bibinfo {year} {1991})}\BibitemShut {NoStop}%
\bibitem [{\citenamefont {Lee}\ and\ \citenamefont {DePristo}(1996)}]{CYL2}%
  \BibitemOpen
  \bibfield  {author} {\bibinfo {author} {\bibfnamefont {C.~Y.}\ \bibnamefont
  {Lee}}\ and\ \bibinfo {author} {\bibfnamefont {A.~E.}\ \bibnamefont
  {DePristo}},\ }\href@noop {} {\bibfield  {journal} {\bibinfo  {journal} {J.
  Chem. Phys.}\ }\textbf {\bibinfo {volume} {85}},\ \bibinfo {pages} {4161}
  (\bibinfo {year} {1996})}\BibitemShut {NoStop}%
\bibitem [{\citenamefont {Watts}\ \emph {et~al.}(2001)\citenamefont {Watts},
  \citenamefont {Sitz}, \citenamefont {McCormack}, \citenamefont {Kroes},
  \citenamefont {Olsen}, \citenamefont {Groeneveld}, \citenamefont {Stralen},
  \citenamefont {Baerends},\ and\ \citenamefont {Mowrey}}]{GJ3}%
  \BibitemOpen
  \bibfield  {author} {\bibinfo {author} {\bibfnamefont {E.}~\bibnamefont
  {Watts}}, \bibinfo {author} {\bibfnamefont {G.~O.}\ \bibnamefont {Sitz}},
  \bibinfo {author} {\bibfnamefont {D.~A.}\ \bibnamefont {McCormack}}, \bibinfo
  {author} {\bibfnamefont {G.~J.}\ \bibnamefont {Kroes}}, \bibinfo {author}
  {\bibfnamefont {R.~A.}\ \bibnamefont {Olsen}}, \bibinfo {author}
  {\bibfnamefont {J.~A.}\ \bibnamefont {Groeneveld}}, \bibinfo {author}
  {\bibfnamefont {J.~N. P.~V.}\ \bibnamefont {Stralen}}, \bibinfo {author}
  {\bibfnamefont {E.~J.}\ \bibnamefont {Baerends}}, \ and\ \bibinfo {author}
  {\bibfnamefont {R.~C.}\ \bibnamefont {Mowrey}},\ }\href@noop {} {\bibfield
  {journal} {\bibinfo  {journal} {J.\ Chem.\ Phys.}\ }\textbf {\bibinfo
  {volume} {114}},\ \bibinfo {pages} {495} (\bibinfo {year}
  {2001})}\BibitemShut {NoStop}%
\bibitem [{\citenamefont {Sahoo}, \citenamefont {Mukherjee},\ and\
  \citenamefont {Adhikari}(2012{\natexlab{b}})}]{Tapas_JCP_2012}%
  \BibitemOpen
  \bibfield  {author} {\bibinfo {author} {\bibfnamefont {T.}~\bibnamefont
  {Sahoo}}, \bibinfo {author} {\bibfnamefont {S.}~\bibnamefont {Mukherjee}}, \
  and\ \bibinfo {author} {\bibfnamefont {S.}~\bibnamefont {Adhikari}},\
  }\href@noop {} {\bibfield  {journal} {\bibinfo  {journal} {The Journal of
  Chemical Physics}\ }\textbf {\bibinfo {volume} {136}},\ \bibinfo {pages}
  {084306} (\bibinfo {year} {2012}{\natexlab{b}})}\BibitemShut {NoStop}%
\bibitem [{\citenamefont {Hamza}\ and\ \citenamefont {Madix}(1985)}]{AVH}%
  \BibitemOpen
  \bibfield  {author} {\bibinfo {author} {\bibfnamefont {A.~V.}\ \bibnamefont
  {Hamza}}\ and\ \bibinfo {author} {\bibfnamefont {R.~J.}\ \bibnamefont
  {Madix}},\ }\href@noop {} {\bibfield  {journal} {\bibinfo  {journal} {J.
  Phys. Chem.}\ }\textbf {\bibinfo {volume} {89}},\ \bibinfo {pages} {5381}
  (\bibinfo {year} {1985})}\BibitemShut {NoStop}%
\bibitem [{\citenamefont {Sheng}\ \emph {et~al.}(2011)\citenamefont {Sheng},
  \citenamefont {Kramer}, \citenamefont {Cadien}, \citenamefont {Fujita},\ and\
  \citenamefont {Chen}}]{Sheng_2011}%
  \BibitemOpen
  \bibfield  {author} {\bibinfo {author} {\bibfnamefont {H.~W.}\ \bibnamefont
  {Sheng}}, \bibinfo {author} {\bibfnamefont {M.~J.}\ \bibnamefont {Kramer}},
  \bibinfo {author} {\bibfnamefont {A.}~\bibnamefont {Cadien}}, \bibinfo
  {author} {\bibfnamefont {T.}~\bibnamefont {Fujita}}, \ and\ \bibinfo {author}
  {\bibfnamefont {M.~W.}\ \bibnamefont {Chen}},\ }\href@noop {} {\bibfield
  {journal} {\bibinfo  {journal} {Phys. Rev. B}\ }\textbf {\bibinfo {volume}
  {83}},\ \bibinfo {pages} {134118} (\bibinfo {year} {2011})}\BibitemShut
  {NoStop}%
\bibitem [{\citenamefont {Togo}(2023)}]{togo_Firstprinciples_2023}%
  \BibitemOpen
  \bibfield  {author} {\bibinfo {author} {\bibfnamefont {A.}~\bibnamefont
  {Togo}},\ }\href@noop {} {\bibfield  {journal} {\bibinfo  {journal} {J. Phys.
  Soc. Jpn.}\ }\textbf {\bibinfo {volume} {92}},\ \bibinfo {pages} {012001}
  (\bibinfo {year} {2023})}\BibitemShut {NoStop}%
\bibitem [{\citenamefont {Togo}\ \emph {et~al.}(2023)\citenamefont {Togo},
  \citenamefont {Chaput}, \citenamefont {Tadano},\ and\ \citenamefont
  {Tanaka}}]{togo_Implementation_2023}%
  \BibitemOpen
  \bibfield  {author} {\bibinfo {author} {\bibfnamefont {A.}~\bibnamefont
  {Togo}}, \bibinfo {author} {\bibfnamefont {L.}~\bibnamefont {Chaput}},
  \bibinfo {author} {\bibfnamefont {T.}~\bibnamefont {Tadano}}, \ and\ \bibinfo
  {author} {\bibfnamefont {I.}~\bibnamefont {Tanaka}},\ }\href@noop {}
  {\bibfield  {journal} {\bibinfo  {journal} {J. Phys.: Condens. Matter}\
  }\textbf {\bibinfo {volume} {35}},\ \bibinfo {pages} {353001} (\bibinfo
  {year} {2023})}\BibitemShut {NoStop}%
\bibitem [{\citenamefont {Thompson}\ \emph {et~al.}(2022)\citenamefont
  {Thompson}, \citenamefont {Aktulga}, \citenamefont {Berger}, \citenamefont
  {Bolintineanu}, \citenamefont {Brown}, \citenamefont {Crozier}, \citenamefont
  {in~'t Veld}, \citenamefont {Kohlmeyer}, \citenamefont {Moore}, \citenamefont
  {Nguyen}, \citenamefont {Shan}, \citenamefont {Stevens}, \citenamefont
  {Tranchida}, \citenamefont {Trott},\ and\ \citenamefont
  {Plimpton}}]{LAMMPS_2022}%
  \BibitemOpen
  \bibfield  {author} {\bibinfo {author} {\bibfnamefont {A.~P.}\ \bibnamefont
  {Thompson}}, \bibinfo {author} {\bibfnamefont {H.~M.}\ \bibnamefont
  {Aktulga}}, \bibinfo {author} {\bibfnamefont {R.}~\bibnamefont {Berger}},
  \bibinfo {author} {\bibfnamefont {D.~S.}\ \bibnamefont {Bolintineanu}},
  \bibinfo {author} {\bibfnamefont {W.~M.}\ \bibnamefont {Brown}}, \bibinfo
  {author} {\bibfnamefont {P.~S.}\ \bibnamefont {Crozier}}, \bibinfo {author}
  {\bibfnamefont {P.~J.}\ \bibnamefont {in~'t Veld}}, \bibinfo {author}
  {\bibfnamefont {A.}~\bibnamefont {Kohlmeyer}}, \bibinfo {author}
  {\bibfnamefont {S.~G.}\ \bibnamefont {Moore}}, \bibinfo {author}
  {\bibfnamefont {T.~D.}\ \bibnamefont {Nguyen}}, \bibinfo {author}
  {\bibfnamefont {R.}~\bibnamefont {Shan}}, \bibinfo {author} {\bibfnamefont
  {M.~J.}\ \bibnamefont {Stevens}}, \bibinfo {author} {\bibfnamefont
  {J.}~\bibnamefont {Tranchida}}, \bibinfo {author} {\bibfnamefont
  {C.}~\bibnamefont {Trott}}, \ and\ \bibinfo {author} {\bibfnamefont {S.~J.}\
  \bibnamefont {Plimpton}},\ }\href@noop {} {\bibfield  {journal} {\bibinfo
  {journal} {Comput. Phys. Commun.}\ }\textbf {\bibinfo {volume} {271}},\
  \bibinfo {pages} {108171} (\bibinfo {year} {2022})}\BibitemShut {NoStop}%
\bibitem [{\citenamefont {Kresse}\ and\ \citenamefont
  {Hafner}(1993)}]{kresse_ab_1993}%
  \BibitemOpen
  \bibfield  {author} {\bibinfo {author} {\bibfnamefont {G.}~\bibnamefont
  {Kresse}}\ and\ \bibinfo {author} {\bibfnamefont {J.}~\bibnamefont
  {Hafner}},\ }\href@noop {} {\bibfield  {journal} {\bibinfo  {journal} {Phys.
  Rev. B}\ }\textbf {\bibinfo {volume} {47}},\ \bibinfo {pages} {558} (\bibinfo
  {year} {1993})}\BibitemShut {NoStop}%
\bibitem [{\citenamefont {Kresse}\ and\ \citenamefont
  {Furthm\"{u}ller}(1996{\natexlab{a}})}]{kresse_efficient_1996}%
  \BibitemOpen
  \bibfield  {author} {\bibinfo {author} {\bibfnamefont {G.}~\bibnamefont
  {Kresse}}\ and\ \bibinfo {author} {\bibfnamefont {J.}~\bibnamefont
  {Furthm\"{u}ller}},\ }\href@noop {} {\bibfield  {journal} {\bibinfo
  {journal} {Phys. Rev. B}\ }\textbf {\bibinfo {volume} {54}},\ \bibinfo
  {pages} {11169} (\bibinfo {year} {1996}{\natexlab{a}})}\BibitemShut {NoStop}%
\bibitem [{\citenamefont {Kresse}\ and\ \citenamefont
  {Furthm\"{u}ller}(1996{\natexlab{b}})}]{kresse_efficiency_1996}%
  \BibitemOpen
  \bibfield  {author} {\bibinfo {author} {\bibfnamefont {G.}~\bibnamefont
  {Kresse}}\ and\ \bibinfo {author} {\bibfnamefont {J.}~\bibnamefont
  {Furthm\"{u}ller}},\ }\href@noop {} {\bibfield  {journal} {\bibinfo
  {journal} {Comput. Mater. Sci.}\ }\textbf {\bibinfo {volume} {6}},\ \bibinfo
  {pages} {15} (\bibinfo {year} {1996}{\natexlab{b}})}\BibitemShut {NoStop}%
\end{thebibliography}%


\begin{thebibliography}{6}%
\makeatletter
\providecommand \@ifxundefined [1]{%
 \@ifx{#1\undefined}
}%
\providecommand \@ifnum [1]{%
 \ifnum #1\expandafter \@firstoftwo
 \else \expandafter \@secondoftwo
 \fi
}%
\providecommand \@ifx [1]{%
 \ifx #1\expandafter \@firstoftwo
 \else \expandafter \@secondoftwo
 \fi
}%
\providecommand \natexlab [1]{#1}%
\providecommand \enquote  [1]{``#1''}%
\providecommand \bibnamefont  [1]{#1}%
\providecommand \bibfnamefont [1]{#1}%
\providecommand \citenamefont [1]{#1}%
\providecommand \href@noop [0]{\@secondoftwo}%
\providecommand \href [0]{\begingroup \@sanitize@url \@href}%
\providecommand \@href[1]{\@@startlink{#1}\@@href}%
\providecommand \@@href[1]{\endgroup#1\@@endlink}%
\providecommand \@sanitize@url [0]{\catcode `\\12\catcode `\$12\catcode
  `\&12\catcode `\#12\catcode `\^12\catcode `\_12\catcode `\%12\relax}%
\providecommand \@@startlink[1]{}%
\providecommand \@@endlink[0]{}%
\providecommand \url  [0]{\begingroup\@sanitize@url \@url }%
\providecommand \@url [1]{\endgroup\@href {#1}{\urlprefix }}%
\providecommand \urlprefix  [0]{URL }%
\providecommand \Eprint [0]{\href }%
\providecommand \doibase [0]{http://dx.doi.org/}%
\providecommand \selectlanguage [0]{\@gobble}%
\providecommand \bibinfo  [0]{\@secondoftwo}%
\providecommand \bibfield  [0]{\@secondoftwo}%
\providecommand \translation [1]{[#1]}%
\providecommand \BibitemOpen [0]{}%
\providecommand \bibitemStop [0]{}%
\providecommand \bibitemNoStop [0]{.\EOS\space}%
\providecommand \EOS [0]{\spacefactor3000\relax}%
\providecommand \BibitemShut  [1]{\csname bibitem#1\endcsname}%
\let\auto@bib@innerbib\@empty
\bibitem [{\citenamefont {Jahn}\ and\ \citenamefont {Teller}(1937)}]{JT}%
  \BibitemOpen
  \bibfield  {author} {\bibinfo {author} {\bibfnamefont {H.~A.}\ \bibnamefont
  {Jahn}}\ and\ \bibinfo {author} {\bibfnamefont {E.}~\bibnamefont {Teller}},\
  }\href@noop {} {\bibfield  {journal} {\bibinfo  {journal} {Proc. R. Soc.
  London A}\ }\textbf {\bibinfo {volume} {161}},\ \bibinfo {pages} {220}
  (\bibinfo {year} {1937})}\BibitemShut {NoStop}%
\bibitem [{\citenamefont {Longuet-Higgins}(1975)}]{longuet-higgins_prsl344}%
  \BibitemOpen
  \bibfield  {author} {\bibinfo {author} {\bibfnamefont {H.~C.}\ \bibnamefont
  {Longuet-Higgins}},\ }\href@noop {} {\bibfield  {journal} {\bibinfo
  {journal} {Proc. R. Soc. London, Ser. A}\ }\textbf {\bibinfo {volume}
  {344}},\ \bibinfo {pages} {147} (\bibinfo {year} {1975})}\BibitemShut
  {NoStop}%
\bibitem [{\citenamefont {Klein}(1974)}]{klein_jcp61}%
  \BibitemOpen
  \bibfield  {author} {\bibinfo {author} {\bibfnamefont {D.~J.}\ \bibnamefont
  {Klein}},\ }\href@noop {} {\bibfield  {journal} {\bibinfo  {journal} {J.
  Chem. Phys.}\ }\textbf {\bibinfo {volume} {61}},\ \bibinfo {pages} {786}
  (\bibinfo {year} {1974})}\BibitemShut {NoStop}%
\bibitem [{\citenamefont {Bersuker}(2001)}]{JT-review-Bersuker}%
  \BibitemOpen
  \bibfield  {author} {\bibinfo {author} {\bibfnamefont {I.~B.}\ \bibnamefont
  {Bersuker}},\ }\href@noop {} {\bibfield  {journal} {\bibinfo  {journal}
  {Chem. Rev.}\ }\textbf {\bibinfo {volume} {101}},\ \bibinfo {pages} {1067}
  (\bibinfo {year} {2001})}\BibitemShut {NoStop}%
\bibitem [{\citenamefont {Jahn}(1937)}]{JT1}%
  \BibitemOpen
  \bibfield  {author} {\bibinfo {author} {\bibfnamefont {H.~A.}\ \bibnamefont
  {Jahn}},\ }\href@noop {} {\bibfield  {journal} {\bibinfo  {journal} {Proc. R.
  Soc. London A}\ }\textbf {\bibinfo {volume} {164}},\ \bibinfo {pages} {117}
  (\bibinfo {year} {1937})}\BibitemShut {NoStop}%
\bibitem [{\citenamefont {Mukherjee}\ \emph {et~al.}(2017)\citenamefont
  {Mukherjee}, \citenamefont {Mukherjee}, \citenamefont {Sardar}, \citenamefont
  {Shamasundar},\ and\ \citenamefont {Adhikari}}]{bijit_mp115}%
  \BibitemOpen
  \bibfield  {author} {\bibinfo {author} {\bibfnamefont {B.}~\bibnamefont
  {Mukherjee}}, \bibinfo {author} {\bibfnamefont {S.}~\bibnamefont
  {Mukherjee}}, \bibinfo {author} {\bibfnamefont {S.}~\bibnamefont {Sardar}},
  \bibinfo {author} {\bibfnamefont {K.~R.}\ \bibnamefont {Shamasundar}}, \ and\
  \bibinfo {author} {\bibfnamefont {S.}~\bibnamefont {Adhikari}},\ }\href@noop
  {} {\bibfield  {journal} {\bibinfo  {journal} {Mol. Phys.}\ }\textbf
  {\bibinfo {volume} {115}},\ \bibinfo {pages} {2833} (\bibinfo {year}
  {2017})}\BibitemShut {NoStop}%
\end{thebibliography}%


\begin{thebibliography}{13}%
\makeatletter
\providecommand \@ifxundefined [1]{%
 \@ifx{#1\undefined}
}%
\providecommand \@ifnum [1]{%
 \ifnum #1\expandafter \@firstoftwo
 \else \expandafter \@secondoftwo
 \fi
}%
\providecommand \@ifx [1]{%
 \ifx #1\expandafter \@firstoftwo
 \else \expandafter \@secondoftwo
 \fi
}%
\providecommand \natexlab [1]{#1}%
\providecommand \enquote  [1]{``#1''}%
\providecommand \bibnamefont  [1]{#1}%
\providecommand \bibfnamefont [1]{#1}%
\providecommand \citenamefont [1]{#1}%
\providecommand \href@noop [0]{\@secondoftwo}%
\providecommand \href [0]{\begingroup \@sanitize@url \@href}%
\providecommand \@href[1]{\@@startlink{#1}\@@href}%
\providecommand \@@href[1]{\endgroup#1\@@endlink}%
\providecommand \@sanitize@url [0]{\catcode `\\12\catcode `\$12\catcode
  `\&12\catcode `\#12\catcode `\^12\catcode `\_12\catcode `\%12\relax}%
\providecommand \@@startlink[1]{}%
\providecommand \@@endlink[0]{}%
\providecommand \url  [0]{\begingroup\@sanitize@url \@url }%
\providecommand \@url [1]{\endgroup\@href {#1}{\urlprefix }}%
\providecommand \urlprefix  [0]{URL }%
\providecommand \Eprint [0]{\href }%
\providecommand \doibase [0]{http://dx.doi.org/}%
\providecommand \selectlanguage [0]{\@gobble}%
\providecommand \bibinfo  [0]{\@secondoftwo}%
\providecommand \bibfield  [0]{\@secondoftwo}%
\providecommand \translation [1]{[#1]}%
\providecommand \BibitemOpen [0]{}%
\providecommand \bibitemStop [0]{}%
\providecommand \bibitemNoStop [0]{.\EOS\space}%
\providecommand \EOS [0]{\spacefactor3000\relax}%
\providecommand \BibitemShut  [1]{\csname bibitem#1\endcsname}%
\let\auto@bib@innerbib\@empty
\bibitem [{\citenamefont {Jahn}\ and\ \citenamefont {Teller}(1937)}]{JT:PRS}%
  \BibitemOpen
  \bibfield  {author} {\bibinfo {author} {\bibfnamefont {H.~A.}\ \bibnamefont
  {Jahn}}\ and\ \bibinfo {author} {\bibfnamefont {E.}~\bibnamefont {Teller}},\
  }\href@noop {} {\bibfield  {journal} {\bibinfo  {journal} {Proc. Roy. Soc.
  (London)}\ }\textbf {\bibinfo {volume} {161}},\ \bibinfo {pages} {220}
  (\bibinfo {year} {1937})}\BibitemShut {NoStop}%
\bibitem [{\citenamefont {Longuet-Higgins}(1961)}]{longuet-higgins}%
  \BibitemOpen
  \bibfield  {author} {\bibinfo {author} {\bibfnamefont {H.~C.}\ \bibnamefont
  {Longuet-Higgins}},\ }\href@noop {} {\emph {\bibinfo {title} {{\it Advances
  in Spectroscopy}}}},\ Vol.~\bibinfo {volume} {2}\ (\bibinfo  {publisher}
  {Interscience Publishers, Ltd.},\ \bibinfo {address} {New York and London},\
  \bibinfo {year} {1961})\ Chap.\ \bibinfo {chapter} {Some Recent Developments
  in the Theory of Molecular Energy Levels}\BibitemShut {NoStop}%
\bibitem [{\citenamefont {Longuet-Higgins}(1975)}]{longuet-higgins_prsl344}%
  \BibitemOpen
  \bibfield  {author} {\bibinfo {author} {\bibfnamefont {H.~C.}\ \bibnamefont
  {Longuet-Higgins}},\ }\href@noop {} {\bibfield  {journal} {\bibinfo
  {journal} {Proc. R. Soc. London, Ser. A}\ }\textbf {\bibinfo {volume}
  {344}},\ \bibinfo {pages} {147} (\bibinfo {year} {1975})}\BibitemShut
  {NoStop}%
\bibitem [{\citenamefont {Herzberg}\ and\ \citenamefont
  {Longuet-Higgins}(1963)}]{herzberg_dfs35}%
  \BibitemOpen
  \bibfield  {author} {\bibinfo {author} {\bibfnamefont {G.}~\bibnamefont
  {Herzberg}}\ and\ \bibinfo {author} {\bibfnamefont {H.~C.}\ \bibnamefont
  {Longuet-Higgins}},\ }\href@noop {} {\bibfield  {journal} {\bibinfo
  {journal} {Discuss Faraday Soc.}\ }\textbf {\bibinfo {volume} {35}},\
  \bibinfo {pages} {77} (\bibinfo {year} {1963})}\BibitemShut {NoStop}%
\bibitem [{\citenamefont {Vleck}(1939)}]{van-vleck}%
  \BibitemOpen
  \bibfield  {author} {\bibinfo {author} {\bibfnamefont {J.~H.~V.}\
  \bibnamefont {Vleck}},\ }\href@noop {} {\bibfield  {journal} {\bibinfo
  {journal} {J. Chem. Phys.}\ }\textbf {\bibinfo {volume} {7}},\ \bibinfo
  {pages} {72} (\bibinfo {year} {1939})}\BibitemShut {NoStop}%
\bibitem [{\citenamefont {Kovaleva}\ \emph {et~al.}(2016)\citenamefont
  {Kovaleva}, \citenamefont {Kugel}, \citenamefont {Pot\.{u}\u{c}ek},
  \citenamefont {Kusmartseva}, \citenamefont {Goryachev}, \citenamefont
  {Bryknar}, \citenamefont {Demikhov}, \citenamefont {Trepakov}, \citenamefont
  {Dejneka}, \citenamefont {Kusmartsev},\ and\ \citenamefont
  {Stoneham}}]{Natalia_JETP}%
  \BibitemOpen
  \bibfield  {author} {\bibinfo {author} {\bibfnamefont {N.~N.}\ \bibnamefont
  {Kovaleva}}, \bibinfo {author} {\bibfnamefont {K.~I.}\ \bibnamefont {Kugel}},
  \bibinfo {author} {\bibfnamefont {Z.}~\bibnamefont {Pot\.{u}\u{c}ek}},
  \bibinfo {author} {\bibfnamefont {O.~E.}\ \bibnamefont {Kusmartseva}},
  \bibinfo {author} {\bibfnamefont {N.~S.}\ \bibnamefont {Goryachev}}, \bibinfo
  {author} {\bibfnamefont {Z.}~\bibnamefont {Bryknar}}, \bibinfo {author}
  {\bibfnamefont {E.~I.}\ \bibnamefont {Demikhov}}, \bibinfo {author}
  {\bibfnamefont {V.~A.}\ \bibnamefont {Trepakov}}, \bibinfo {author}
  {\bibfnamefont {A.}~\bibnamefont {Dejneka}}, \bibinfo {author} {\bibfnamefont
  {F.}~\bibnamefont {Kusmartsev}}, \ and\ \bibinfo {author} {\bibfnamefont
  {A.~M.}\ \bibnamefont {Stoneham}},\ }\href@noop {} {\bibfield  {journal}
  {\bibinfo  {journal} {J. Exp. Theor. Phys.}\ }\textbf {\bibinfo {volume}
  {122}},\ \bibinfo {pages} {890} (\bibinfo {year} {2016})}\BibitemShut
  {NoStop}%
\bibitem [{\citenamefont {Dutta}\ \emph {et~al.}(2022)\citenamefont {Dutta},
  \citenamefont {Ravi}, \citenamefont {Mukherjee}, \citenamefont {Ojha},\ and\
  \citenamefont {Adhikari}}]{joy_jpca}%
  \BibitemOpen
  \bibfield  {author} {\bibinfo {author} {\bibfnamefont {J.}~\bibnamefont
  {Dutta}}, \bibinfo {author} {\bibfnamefont {S.}~\bibnamefont {Ravi}},
  \bibinfo {author} {\bibfnamefont {S.}~\bibnamefont {Mukherjee}}, \bibinfo
  {author} {\bibfnamefont {A.~K.}\ \bibnamefont {Ojha}}, \ and\ \bibinfo
  {author} {\bibfnamefont {S.}~\bibnamefont {Adhikari}},\ }\href@noop {}
  {\bibfield  {journal} {\bibinfo  {journal} {J. Phys. Chem. A}\ }\textbf
  {\bibinfo {volume} {126}},\ \bibinfo {pages} {691} (\bibinfo {year}
  {2022})}\BibitemShut {NoStop}%
\bibitem [{\citenamefont {Kugel}\ and\ \citenamefont
  {Khomskii}(1982)}]{Kugel-Khomskii}%
  \BibitemOpen
  \bibfield  {author} {\bibinfo {author} {\bibfnamefont {K.~I.}\ \bibnamefont
  {Kugel}}\ and\ \bibinfo {author} {\bibfnamefont {D.~I.}\ \bibnamefont
  {Khomskii}},\ }\href@noop {} {\bibfield  {journal} {\bibinfo  {journal} {Sov.
  Phys. Usp.}\ }\textbf {\bibinfo {volume} {25}},\ \bibinfo {pages} {231}
  (\bibinfo {year} {1982})}\BibitemShut {NoStop}%
\bibitem [{\citenamefont {Ole\'s}\ \emph {et~al.}(2005)\citenamefont {Ole\'s},
  \citenamefont {Khaliullin}, \citenamefont {Horsch},\ and\ \citenamefont
  {Feiner}}]{Oles-Feiner}%
  \BibitemOpen
  \bibfield  {author} {\bibinfo {author} {\bibfnamefont {A.~M.}\ \bibnamefont
  {Ole\'s}}, \bibinfo {author} {\bibfnamefont {G.}~\bibnamefont {Khaliullin}},
  \bibinfo {author} {\bibfnamefont {P.}~\bibnamefont {Horsch}}, \ and\ \bibinfo
  {author} {\bibfnamefont {L.~F.}\ \bibnamefont {Feiner}},\ }\href@noop {}
  {\bibfield  {journal} {\bibinfo  {journal} {Phys. Rev. B}\ }\textbf {\bibinfo
  {volume} {72}},\ \bibinfo {pages} {214431} (\bibinfo {year}
  {2005})}\BibitemShut {NoStop}%
\bibitem [{\citenamefont {Longuet-Higgins}\ \emph {et~al.}(1958)\citenamefont
  {Longuet-Higgins}, \citenamefont {\"Opik}, \citenamefont {Pryce},\ and\
  \citenamefont {Sack}}]{lh-opik}%
  \BibitemOpen
  \bibfield  {author} {\bibinfo {author} {\bibfnamefont {H.~C.}\ \bibnamefont
  {Longuet-Higgins}}, \bibinfo {author} {\bibfnamefont {U.}~\bibnamefont
  {\"Opik}}, \bibinfo {author} {\bibfnamefont {M.~H.~L.}\ \bibnamefont
  {Pryce}}, \ and\ \bibinfo {author} {\bibfnamefont {R.~A.}\ \bibnamefont
  {Sack}},\ }\href@noop {} {\bibfield  {journal} {\bibinfo  {journal} {Proc. R.
  Soc. London, Ser. A}\ }\textbf {\bibinfo {volume} {244}},\ \bibinfo {pages}
  {1} (\bibinfo {year} {1958})}\BibitemShut {NoStop}%
\bibitem [{\citenamefont {O'Brien}(1964)}]{O'Brien}%
  \BibitemOpen
  \bibfield  {author} {\bibinfo {author} {\bibfnamefont {M.~C.~M.}\
  \bibnamefont {O'Brien}},\ }\href@noop {} {\bibfield  {journal} {\bibinfo
  {journal} {Proc. Roy. Soc. A}\ }\textbf {\bibinfo {volume} {281}},\ \bibinfo
  {pages} {323} (\bibinfo {year} {1964})}\BibitemShut {NoStop}%
\bibitem [{\citenamefont {Dutta}, \citenamefont {Adhikari},\ and\ \citenamefont
  {Kovaleva}(2019)}]{joy_jcp}%
  \BibitemOpen
  \bibfield  {author} {\bibinfo {author} {\bibfnamefont {J.}~\bibnamefont
  {Dutta}}, \bibinfo {author} {\bibfnamefont {S.}~\bibnamefont {Adhikari}}, \
  and\ \bibinfo {author} {\bibfnamefont {N.}~\bibnamefont {Kovaleva}},\
  }\href@noop {} {\bibfield  {journal} {\bibinfo  {journal} {J. Chem. Phys.}\
  }\textbf {\bibinfo {volume} {150}},\ \bibinfo {pages} {064703} (\bibinfo
  {year} {2019})}\BibitemShut {NoStop}%
\bibitem [{\citenamefont {Slonczewski}(1963)}]{Slonczewski}%
  \BibitemOpen
  \bibfield  {author} {\bibinfo {author} {\bibfnamefont {J.~C.}\ \bibnamefont
  {Slonczewski}},\ }\href@noop {} {\bibfield  {journal} {\bibinfo  {journal}
  {Phys. Rev.}\ }\textbf {\bibinfo {volume} {131}},\ \bibinfo {pages} {1596}
  (\bibinfo {year} {1963})}\BibitemShut {NoStop}%
\end{thebibliography}%


\begin{thebibliography}{25}%
\makeatletter
\providecommand \@ifxundefined [1]{%
 \@ifx{#1\undefined}
}%
\providecommand \@ifnum [1]{%
 \ifnum #1\expandafter \@firstoftwo
 \else \expandafter \@secondoftwo
 \fi
}%
\providecommand \@ifx [1]{%
 \ifx #1\expandafter \@firstoftwo
 \else \expandafter \@secondoftwo
 \fi
}%
\providecommand \natexlab [1]{#1}%
\providecommand \enquote  [1]{``#1''}%
\providecommand \bibnamefont  [1]{#1}%
\providecommand \bibfnamefont [1]{#1}%
\providecommand \citenamefont [1]{#1}%
\providecommand \href@noop [0]{\@secondoftwo}%
\providecommand \href [0]{\begingroup \@sanitize@url \@href}%
\providecommand \@href[1]{\@@startlink{#1}\@@href}%
\providecommand \@@href[1]{\endgroup#1\@@endlink}%
\providecommand \@sanitize@url [0]{\catcode `\\12\catcode `\$12\catcode
  `\&12\catcode `\#12\catcode `\^12\catcode `\_12\catcode `\%12\relax}%
\providecommand \@@startlink[1]{}%
\providecommand \@@endlink[0]{}%
\providecommand \url  [0]{\begingroup\@sanitize@url \@url }%
\providecommand \@url [1]{\endgroup\@href {#1}{\urlprefix }}%
\providecommand \urlprefix  [0]{URL }%
\providecommand \Eprint [0]{\href }%
\providecommand \doibase [0]{http://dx.doi.org/}%
\providecommand \selectlanguage [0]{\@gobble}%
\providecommand \bibinfo  [0]{\@secondoftwo}%
\providecommand \bibfield  [0]{\@secondoftwo}%
\providecommand \translation [1]{[#1]}%
\providecommand \BibitemOpen [0]{}%
\providecommand \bibitemStop [0]{}%
\providecommand \bibitemNoStop [0]{.\EOS\space}%
\providecommand \EOS [0]{\spacefactor3000\relax}%
\providecommand \BibitemShut  [1]{\csname bibitem#1\endcsname}%
\let\auto@bib@innerbib\@empty
\bibitem [{\citenamefont {Hering}\ \emph {et~al.}(1980)\citenamefont {Hering},
  \citenamefont {Brooks}, \citenamefont {Curl}, \citenamefont {Judson},\ and\
  \citenamefont {Lowe}}]{hering_prl44}%
  \BibitemOpen
  \bibfield  {author} {\bibinfo {author} {\bibfnamefont {P.}~\bibnamefont
  {Hering}}, \bibinfo {author} {\bibfnamefont {P.~R.}\ \bibnamefont {Brooks}},
  \bibinfo {author} {\bibfnamefont {R.~F.}\ \bibnamefont {Curl}}, \bibinfo
  {author} {\bibfnamefont {R.~S.}\ \bibnamefont {Judson}}, \ and\ \bibinfo
  {author} {\bibfnamefont {R.~S.}\ \bibnamefont {Lowe}},\ }\href {\doibase
  10.1103/PhysRevLett.44.687} {\bibfield  {journal} {\bibinfo  {journal} {Phys.
  Rev. Lett.}\ }\textbf {\bibinfo {volume} {44}},\ \bibinfo {pages} {687}
  (\bibinfo {year} {1980})}\BibitemShut {NoStop}%
\bibitem [{\citenamefont {Zimmerman}, \citenamefont {Yuan},\ and\ \citenamefont
  {George}(1977)}]{zimmerman_jcp66}%
  \BibitemOpen
  \bibfield  {author} {\bibinfo {author} {\bibfnamefont {I.~H.}\ \bibnamefont
  {Zimmerman}}, \bibinfo {author} {\bibfnamefont {J.-M.}\ \bibnamefont {Yuan}},
  \ and\ \bibinfo {author} {\bibfnamefont {T.~F.}\ \bibnamefont {George}},\
  }\href@noop {} {\bibfield  {journal} {\bibinfo  {journal} {The Journal of
  Chemical Physics}\ }\textbf {\bibinfo {volume} {66}},\ \bibinfo {pages}
  {2638} (\bibinfo {year} {1977})}\BibitemShut {NoStop}%
\bibitem [{\citenamefont {George}\ \emph {et~al.}(1979)\citenamefont {George},
  \citenamefont {Zimmerman}, \citenamefont {DeVries}, \citenamefont {Yuan},
  \citenamefont {Lam}, \citenamefont {Bellum}, \citenamefont {Lee},
  \citenamefont {Slutsky},\ and\ \citenamefont {Lin}}]{george_book}%
  \BibitemOpen
  \bibfield  {author} {\bibinfo {author} {\bibfnamefont {T.~F.}\ \bibnamefont
  {George}}, \bibinfo {author} {\bibfnamefont {I.~H.}\ \bibnamefont
  {Zimmerman}}, \bibinfo {author} {\bibfnamefont {P.~L.}\ \bibnamefont
  {DeVries}}, \bibinfo {author} {\bibfnamefont {J.-M.}\ \bibnamefont {Yuan}},
  \bibinfo {author} {\bibfnamefont {K.-S.}\ \bibnamefont {Lam}}, \bibinfo
  {author} {\bibfnamefont {J.~C.}\ \bibnamefont {Bellum}}, \bibinfo {author}
  {\bibfnamefont {H.-W.}\ \bibnamefont {Lee}}, \bibinfo {author} {\bibfnamefont
  {M.~S.}\ \bibnamefont {Slutsky}}, \ and\ \bibinfo {author} {\bibfnamefont
  {J.}~\bibnamefont {Lin}},\ }\href@noop {} {\bibfield  {journal} {\bibinfo
  {journal} {Chemical and Biochemical Applications of Lasers, edited by CB
  Moore (Academic, New York, 1979)}\ }\textbf {\bibinfo {volume} {4}},\
  \bibinfo {pages} {253} (\bibinfo {year} {1979})}\BibitemShut {NoStop}%
\bibitem [{\citenamefont {Light}\ and\ \citenamefont
  {Altenberger-Siczek}(1979)}]{light_jcp70}%
  \BibitemOpen
  \bibfield  {author} {\bibinfo {author} {\bibfnamefont {J.~C.}\ \bibnamefont
  {Light}}\ and\ \bibinfo {author} {\bibfnamefont {A.}~\bibnamefont
  {Altenberger-Siczek}},\ }\href@noop {} {\bibfield  {journal} {\bibinfo
  {journal} {The Journal of Chemical Physics}\ }\textbf {\bibinfo {volume}
  {70}},\ \bibinfo {pages} {4108} (\bibinfo {year} {1979})}\BibitemShut
  {NoStop}%
\bibitem [{\citenamefont {Chu}(1981)}]{chu_jcp75}%
  \BibitemOpen
  \bibfield  {author} {\bibinfo {author} {\bibfnamefont {S.-I.}\ \bibnamefont
  {Chu}},\ }\href@noop {} {\bibfield  {journal} {\bibinfo  {journal} {The
  Journal of Chemical Physics}\ }\textbf {\bibinfo {volume} {75}},\ \bibinfo
  {pages} {2215} (\bibinfo {year} {1981})}\BibitemShut {NoStop}%
\bibitem [{\citenamefont {Zimmerman}, \citenamefont {Baer},\ and\ \citenamefont
  {George}(1983)}]{zimmerman_jcp87}%
  \BibitemOpen
  \bibfield  {author} {\bibinfo {author} {\bibfnamefont {I.~H.}\ \bibnamefont
  {Zimmerman}}, \bibinfo {author} {\bibfnamefont {M.}~\bibnamefont {Baer}}, \
  and\ \bibinfo {author} {\bibfnamefont {T.~F.}\ \bibnamefont {George}},\
  }\href@noop {} {\bibfield  {journal} {\bibinfo  {journal} {The Journal of
  Physical Chemistry}\ }\textbf {\bibinfo {volume} {87}},\ \bibinfo {pages}
  {1478} (\bibinfo {year} {1983})}\BibitemShut {NoStop}%
\bibitem [{\citenamefont {Last}\ \emph {et~al.}(1983)\citenamefont {Last},
  \citenamefont {Baer}, \citenamefont {Zimmerman},\ and\ \citenamefont
  {George}}]{last_cpl101}%
  \BibitemOpen
  \bibfield  {author} {\bibinfo {author} {\bibfnamefont {I.}~\bibnamefont
  {Last}}, \bibinfo {author} {\bibfnamefont {M.}~\bibnamefont {Baer}}, \bibinfo
  {author} {\bibfnamefont {I.~H.}\ \bibnamefont {Zimmerman}}, \ and\ \bibinfo
  {author} {\bibfnamefont {T.~F.}\ \bibnamefont {George}},\ }\href@noop {}
  {\bibfield  {journal} {\bibinfo  {journal} {Chemical physics letters}\
  }\textbf {\bibinfo {volume} {101}},\ \bibinfo {pages} {163} (\bibinfo {year}
  {1983})}\BibitemShut {NoStop}%
\bibitem [{\citenamefont {Last}\ and\ \citenamefont {Baer}(1985)}]{last_jcp82}%
  \BibitemOpen
  \bibfield  {author} {\bibinfo {author} {\bibfnamefont {I.}~\bibnamefont
  {Last}}\ and\ \bibinfo {author} {\bibfnamefont {M.}~\bibnamefont {Baer}},\
  }\href@noop {} {\bibfield  {journal} {\bibinfo  {journal} {The Journal of
  chemical physics}\ }\textbf {\bibinfo {volume} {82}},\ \bibinfo {pages}
  {4954} (\bibinfo {year} {1985})}\BibitemShut {NoStop}%
\bibitem [{\citenamefont {Manz}\ \emph {et~al.}(2000)\citenamefont {Manz},
  \citenamefont {Naundorf}, \citenamefont {Yamashita},\ and\ \citenamefont
  {Zhao}}]{manz_jcp113}%
  \BibitemOpen
  \bibfield  {author} {\bibinfo {author} {\bibfnamefont {J.}~\bibnamefont
  {Manz}}, \bibinfo {author} {\bibfnamefont {H.}~\bibnamefont {Naundorf}},
  \bibinfo {author} {\bibfnamefont {K.}~\bibnamefont {Yamashita}}, \ and\
  \bibinfo {author} {\bibfnamefont {Y.}~\bibnamefont {Zhao}},\ }\href@noop {}
  {\bibfield  {journal} {\bibinfo  {journal} {The Journal of Chemical Physics}\
  }\textbf {\bibinfo {volume} {113}},\ \bibinfo {pages} {8969} (\bibinfo {year}
  {2000})}\BibitemShut {NoStop}%
\bibitem [{\citenamefont {Barth}\ and\ \citenamefont
  {Manz}(2007)}]{barth_pra75}%
  \BibitemOpen
  \bibfield  {author} {\bibinfo {author} {\bibfnamefont {I.}~\bibnamefont
  {Barth}}\ and\ \bibinfo {author} {\bibfnamefont {J.}~\bibnamefont {Manz}},\
  }\href@noop {} {\bibfield  {journal} {\bibinfo  {journal} {Physical Review
  A—Atomic, Molecular, and Optical Physics}\ }\textbf {\bibinfo {volume}
  {75}},\ \bibinfo {pages} {012510} (\bibinfo {year} {2007})}\BibitemShut
  {NoStop}%
\bibitem [{\citenamefont {Yonehara}\ and\ \citenamefont
  {Takatsuka}(2008)}]{yonehara_jcp128}%
  \BibitemOpen
  \bibfield  {author} {\bibinfo {author} {\bibfnamefont {T.}~\bibnamefont
  {Yonehara}}\ and\ \bibinfo {author} {\bibfnamefont {K.}~\bibnamefont
  {Takatsuka}},\ }\href@noop {} {\bibfield  {journal} {\bibinfo  {journal} {The
  Journal of chemical physics}\ }\textbf {\bibinfo {volume} {128}} (\bibinfo
  {year} {2008})}\BibitemShut {NoStop}%
\bibitem [{\citenamefont {Sarkar}, \citenamefont {Adhikari},\ and\
  \citenamefont {Baer}(2007{\natexlab{a}})}]{sarkar_jcp126}%
  \BibitemOpen
  \bibfield  {author} {\bibinfo {author} {\bibfnamefont {B.}~\bibnamefont
  {Sarkar}}, \bibinfo {author} {\bibfnamefont {S.}~\bibnamefont {Adhikari}}, \
  and\ \bibinfo {author} {\bibfnamefont {M.}~\bibnamefont {Baer}},\ }\href@noop
  {} {\bibfield  {journal} {\bibinfo  {journal} {The Journal of chemical
  physics}\ }\textbf {\bibinfo {volume} {126}} (\bibinfo {year}
  {2007}{\natexlab{a}})}\BibitemShut {NoStop}%
\bibitem [{\citenamefont {Sarkar}, \citenamefont {Adhikari},\ and\
  \citenamefont {Baer}(2007{\natexlab{b}})}]{sarkar_jcp127_1}%
  \BibitemOpen
  \bibfield  {author} {\bibinfo {author} {\bibfnamefont {B.}~\bibnamefont
  {Sarkar}}, \bibinfo {author} {\bibfnamefont {S.}~\bibnamefont {Adhikari}}, \
  and\ \bibinfo {author} {\bibfnamefont {M.}~\bibnamefont {Baer}},\ }\href@noop
  {} {\bibfield  {journal} {\bibinfo  {journal} {The Journal of chemical
  physics}\ }\textbf {\bibinfo {volume} {127}} (\bibinfo {year}
  {2007}{\natexlab{b}})}\BibitemShut {NoStop}%
\bibitem [{\citenamefont {Sarkar}, \citenamefont {Adhikari},\ and\
  \citenamefont {Baer}(2007{\natexlab{c}})}]{sarkar_jcp127_2}%
  \BibitemOpen
  \bibfield  {author} {\bibinfo {author} {\bibfnamefont {B.}~\bibnamefont
  {Sarkar}}, \bibinfo {author} {\bibfnamefont {S.}~\bibnamefont {Adhikari}}, \
  and\ \bibinfo {author} {\bibfnamefont {M.}~\bibnamefont {Baer}},\ }\href@noop
  {} {\bibfield  {journal} {\bibinfo  {journal} {The Journal of chemical
  physics}\ }\textbf {\bibinfo {volume} {127}} (\bibinfo {year}
  {2007}{\natexlab{c}})}\BibitemShut {NoStop}%
\bibitem [{\citenamefont {Paul}, \citenamefont {Adhikari},\ and\ \citenamefont
  {Baer}(2010{\natexlab{a}})}]{paul_jcp132}%
  \BibitemOpen
  \bibfield  {author} {\bibinfo {author} {\bibfnamefont {A.~K.}\ \bibnamefont
  {Paul}}, \bibinfo {author} {\bibfnamefont {S.}~\bibnamefont {Adhikari}}, \
  and\ \bibinfo {author} {\bibfnamefont {M.}~\bibnamefont {Baer}},\ }\href@noop
  {} {\bibfield  {journal} {\bibinfo  {journal} {The Journal of chemical
  physics}\ }\textbf {\bibinfo {volume} {132}} (\bibinfo {year}
  {2010}{\natexlab{a}})}\BibitemShut {NoStop}%
\bibitem [{\citenamefont {Paul}\ \emph {et~al.}(2009)\citenamefont {Paul},
  \citenamefont {Adhikari}, \citenamefont {Mukhopadhyay}, \citenamefont
  {Hal{\'a}sz}, \citenamefont {Vib{\'o}k}, \citenamefont {Baer},\ and\
  \citenamefont {Baer}}]{paul_jpca113}%
  \BibitemOpen
  \bibfield  {author} {\bibinfo {author} {\bibfnamefont {A.~K.}\ \bibnamefont
  {Paul}}, \bibinfo {author} {\bibfnamefont {S.}~\bibnamefont {Adhikari}},
  \bibinfo {author} {\bibfnamefont {D.}~\bibnamefont {Mukhopadhyay}}, \bibinfo
  {author} {\bibfnamefont {G.~J.}\ \bibnamefont {Hal{\'a}sz}}, \bibinfo
  {author} {\bibfnamefont {A.}~\bibnamefont {Vib{\'o}k}}, \bibinfo {author}
  {\bibfnamefont {R.}~\bibnamefont {Baer}}, \ and\ \bibinfo {author}
  {\bibfnamefont {M.}~\bibnamefont {Baer}},\ }\href@noop {} {\bibfield
  {journal} {\bibinfo  {journal} {The Journal of Physical Chemistry A}\
  }\textbf {\bibinfo {volume} {113}},\ \bibinfo {pages} {7331} (\bibinfo {year}
  {2009})}\BibitemShut {NoStop}%
\bibitem [{\citenamefont {Paul}\ \emph {et~al.}(2010)\citenamefont {Paul},
  \citenamefont {Adhikari}, \citenamefont {Baer},\ and\ \citenamefont
  {Baer}}]{paul_pra81}%
  \BibitemOpen
  \bibfield  {author} {\bibinfo {author} {\bibfnamefont {A.~K.}\ \bibnamefont
  {Paul}}, \bibinfo {author} {\bibfnamefont {S.}~\bibnamefont {Adhikari}},
  \bibinfo {author} {\bibfnamefont {M.}~\bibnamefont {Baer}}, \ and\ \bibinfo
  {author} {\bibfnamefont {R.}~\bibnamefont {Baer}},\ }\href@noop {} {\bibfield
   {journal} {\bibinfo  {journal} {Physical Review A}\ }\textbf {\bibinfo
  {volume} {81}},\ \bibinfo {pages} {013412} (\bibinfo {year}
  {2010})}\BibitemShut {NoStop}%
\bibitem [{\citenamefont {Paul}, \citenamefont {Adhikari},\ and\ \citenamefont
  {Baer}(2010{\natexlab{b}})}]{PAUL_PR496}%
  \BibitemOpen
  \bibfield  {author} {\bibinfo {author} {\bibfnamefont {A.~K.}\ \bibnamefont
  {Paul}}, \bibinfo {author} {\bibfnamefont {S.}~\bibnamefont {Adhikari}}, \
  and\ \bibinfo {author} {\bibfnamefont {M.}~\bibnamefont {Baer}},\ }\href@noop
  {} {\bibfield  {journal} {\bibinfo  {journal} {Physics reports}\ }\textbf
  {\bibinfo {volume} {496}},\ \bibinfo {pages} {79} (\bibinfo {year}
  {2010}{\natexlab{b}})}\BibitemShut {NoStop}%
\bibitem [{\citenamefont {Gavrila}(1991)}]{gavrila_book}%
  \BibitemOpen
  \bibfield  {author} {\bibinfo {author} {\bibfnamefont {M.}~\bibnamefont
  {Gavrila}},\ }\href@noop {} {\  (\bibinfo {year} {1991})}\BibitemShut
  {NoStop}%
\bibitem [{\citenamefont {Corkum}\ \emph {et~al.}(1999)\citenamefont {Corkum},
  \citenamefont {Ellert}, \citenamefont {Mehendale}, \citenamefont {Dietrich},
  \citenamefont {Hankin}, \citenamefont {Aseyev}, \citenamefont {Rayner},\ and\
  \citenamefont {Villeneuve}}]{corkum_fd113}%
  \BibitemOpen
  \bibfield  {author} {\bibinfo {author} {\bibfnamefont {P.}~\bibnamefont
  {Corkum}}, \bibinfo {author} {\bibfnamefont {C.}~\bibnamefont {Ellert}},
  \bibinfo {author} {\bibfnamefont {M.}~\bibnamefont {Mehendale}}, \bibinfo
  {author} {\bibfnamefont {P.}~\bibnamefont {Dietrich}}, \bibinfo {author}
  {\bibfnamefont {S.}~\bibnamefont {Hankin}}, \bibinfo {author} {\bibfnamefont
  {S.}~\bibnamefont {Aseyev}}, \bibinfo {author} {\bibfnamefont
  {D.}~\bibnamefont {Rayner}}, \ and\ \bibinfo {author} {\bibfnamefont
  {D.}~\bibnamefont {Villeneuve}},\ }\href@noop {} {\bibfield  {journal}
  {\bibinfo  {journal} {Faraday Discussions}\ }\textbf {\bibinfo {volume}
  {113}},\ \bibinfo {pages} {47} (\bibinfo {year} {1999})}\BibitemShut
  {NoStop}%
\bibitem [{\citenamefont {Cohen-Tannoudji}(1998)}]{cohen_rmp70}%
  \BibitemOpen
  \bibfield  {author} {\bibinfo {author} {\bibfnamefont {C.~N.}\ \bibnamefont
  {Cohen-Tannoudji}},\ }\href@noop {} {\bibfield  {journal} {\bibinfo
  {journal} {Reviews of Modern Physics}\ }\textbf {\bibinfo {volume} {70}},\
  \bibinfo {pages} {707} (\bibinfo {year} {1998})}\BibitemShut {NoStop}%
\bibitem [{\citenamefont {Baer}(1975)}]{baer_cpl35}%
  \BibitemOpen
  \bibfield  {author} {\bibinfo {author} {\bibfnamefont {M.}~\bibnamefont
  {Baer}},\ }\href@noop {} {\bibfield  {journal} {\bibinfo  {journal} {Chem.
  Phys. Lett.}\ }\textbf {\bibinfo {volume} {35}},\ \bibinfo {pages} {112}
  (\bibinfo {year} {1975})}\BibitemShut {NoStop}%
\bibitem [{\citenamefont {Baer}\ \emph {et~al.}(2003)\citenamefont {Baer},
  \citenamefont {Kouri}, \citenamefont {Baer},\ and\ \citenamefont
  {Hoffman}}]{baer_jcp119}%
  \BibitemOpen
  \bibfield  {author} {\bibinfo {author} {\bibfnamefont {R.}~\bibnamefont
  {Baer}}, \bibinfo {author} {\bibfnamefont {D.~J.}\ \bibnamefont {Kouri}},
  \bibinfo {author} {\bibfnamefont {M.}~\bibnamefont {Baer}}, \ and\ \bibinfo
  {author} {\bibfnamefont {D.~K.}\ \bibnamefont {Hoffman}},\ }\href@noop {}
  {\bibfield  {journal} {\bibinfo  {journal} {The Journal of chemical physics}\
  }\textbf {\bibinfo {volume} {119}},\ \bibinfo {pages} {6998} (\bibinfo {year}
  {2003})}\BibitemShut {NoStop}%
\bibitem [{\citenamefont {Baer}(2006)}]{baer_book}%
  \BibitemOpen
  \bibfield  {author} {\bibinfo {author} {\bibfnamefont {M.}~\bibnamefont
  {Baer}},\ }\href@noop {} {\emph {\bibinfo {title} {{\it Beyond Born -
  Oppenheimer: Conical intersections and Electronic nonadiabatic coupling
  terms}}}}\ (\bibinfo  {publisher} {Wiley Interscience},\ \bibinfo {address}
  {New Jersey},\ \bibinfo {year} {2006})\BibitemShut {NoStop}%
\bibitem [{\citenamefont {Englman}, \citenamefont {Yahalom},\ and\
  \citenamefont {Baer}(2002)}]{englman_ijqc90}%
  \BibitemOpen
  \bibfield  {author} {\bibinfo {author} {\bibfnamefont {R.}~\bibnamefont
  {Englman}}, \bibinfo {author} {\bibfnamefont {A.}~\bibnamefont {Yahalom}}, \
  and\ \bibinfo {author} {\bibfnamefont {M.}~\bibnamefont {Baer}},\ }\href@noop
  {} {\bibfield  {journal} {\bibinfo  {journal} {International journal of
  quantum chemistry}\ }\textbf {\bibinfo {volume} {90}},\ \bibinfo {pages}
  {266} (\bibinfo {year} {2002})}\BibitemShut {NoStop}%
\end{thebibliography}%


\providecommand{\newblock}{}
\begin{thebibliography}{100}
\expandafter\ifx\csname url\endcsname\relax
  \def\url#1{{\tt #1}}\fi
\expandafter\ifx\csname urlprefix\endcsname\relax\def\urlprefix{URL }\fi
\providecommand{\eprint}[2][]{\url{#2}}

\bibitem{kdc_acp57}
K\"oppel H, Domcke W and Cederbaum L~S 1984 {\em Adv. Chem. Phys.\/} {\bf 57}
  59--246

\bibitem{koppel_jcp110}
Thiel A and K\"oppel H 1999 {\em J. Chem. Phys.\/} {\bf 110} 9371

\bibitem{koppel_jcp150}
K\"oppel H, Gronki J and Mahapatra S 2001 {\em J. Chem. Phys.\/} {\bf 150} 2377

\bibitem{koppel_mp104}
K\"oppel H and Schubert B 2006 {\em Mol. Phys.\/} {\bf 104} 1069

\bibitem{viel_jcp120}
Viel A and Eisfeld W 2004 {\em J. Chem. Phys.\/} {\bf 120} 4603--4613

\bibitem{bhatt_jpca118}
Bhattacharyya S, Opalka D, Poluyanov L~V and Domcke W 2014 {\em J. Phys. Chem.
  A\/} {\bf 118} 11962--11970

\bibitem{hickman_pccp20}
Hickman R~J, Lang R~A and Zeng T 2018 {\em Phys. Chem. Chem. Phys.\/} {\bf 20}
  12312--12322

\bibitem{issaka_pccp21}
Seidu I, Goel P, Wang X~G, Chen B, Wang X~B and Zeng T 2019 {\em Phys. Chem.
  Chem. Phys.\/} {\bf 21} 8679--8690

\bibitem{mukherjee_jpca117}
Mukherjee S, Bandyopadhyay S, Paul A~K and Adhikari S 2013 {\em J. Phys. Chem.
  A\/} {\bf 117} 3475--95

\bibitem{mukherjee_cp440}
Mukherjee S and Adhikari S 2014 {\em Chem. Phys.\/} {\bf 440} 106--18

\bibitem{sardar_cp416}
Sardar S, Mukherjee S, Paul A~K and Adhikari S 2013 {\em Chem. Phys.\/} {\bf
  416} 11--20

\bibitem{mukherjee_jcp143}
Mukherjee S, Mukherjee B, Sardar S and Adhikari S 2015 {\em J. Chem. Phys.\/}
  {\bf 143} 244307

\bibitem{soumya_jpca121}
Mukherjee S, Mukherjee B and Adhikari S 2017 {\em J. Phys. Chem. A\/} {\bf 121}
  6314--26

\bibitem{bijit_mp115}
Mukherjee B, Mukherjee S, Sardar S, Shamasundar K~R and Adhikari S 2017 {\em
  Mol. Phys.\/} {\bf 115} 2833--48

\bibitem{soumya_acs}
Mukherjee S, Mukherjee B, Dutta J, Sardar S and Adhikari S 2018 {\em ACS
  omega\/} {\bf 3} 12465--75

\bibitem{bijit_cp515}
Mukherjee B, Mukherjee S, Sardar S, Shamasundar K~R and Adhikari S 2018 {\em
  Chem. Phys.\/} {\bf 515} 350--9

\bibitem{soumya_jcp150}
Mukherjee S, Dutta J, Mukherjee B, Sardar S and Adhikari S 2019 {\em J. Chem.
  Phys.\/} {\bf 150} 064308

\bibitem{soumya_comptc}
Mukherjee S, Mukherjee B, Sardar S and Adhikari S 2019 {\em Comp. Theor.
  Chem.\/} {\bf 1154} 57--67

\bibitem{bijit_jpcs1148}
Mukherjee B, Ghosh S and Adhikari S 2018 {\em J. Phys.: Conf. Ser.\/} {\bf
  1148} 012001

\bibitem{bijit_irpc38}
Mukherjee B, Naskar K, Mukherjee S, Ghosh S, Sahoo T and Adhikari S 2019 {\em
  Int. Rev. Phys. Chem.\/} {\bf 38} 287--341

\bibitem{joy_pccp}
Dutta J, Mukherjee S, Naskar K, Ghosh S, Mukherjee B, Ravi S and Adhikari S
  2020 {\em Phys. Chem. Chem. Phys. (Perspective)\/} {\bf 22} 27496--27524

\bibitem{soumya_bz}
Mukherjee S, Ravi S, Naskar K, Sardar S and Adhikari S 2021 {\em J. Chem.
  Phys.\/} {\bf 154} 094306

\bibitem{soumya_tfb}
Mukherjee S, Ravi S, Dutta J, Sardar S and Adhikari S 2022 {\em Phys. Chem.
  Chem. Phys.\/} {\bf 24} 2185--2202

\bibitem{saikat_pyra}
Hazra S, Mukherjee S, Ravi S, Sardar S and Adhikari S 2022 {\em ChemPhysChem\/}
  {\bf 23} e202200482

\bibitem{JT_prs161}
Jahn H~A and Teller E 1937 {\em Proc. Roy. Soc. (London)\/} {\bf 161} 220--35

\bibitem{bersuker_book}
Bersuker I~B 2006 {\em {\it The Jahn-Teller Effect}\/} (Cambridge University
  Press)

\bibitem{Pablo_prb71}
Garc\'{\i}a-Fern\'andez P, Bersuker I~B, Aramburu J~A, Barriuso M~T and Moreno
  M 2005 {\em Phys. Rev. B\/} {\bf 71}(18) 184117

\bibitem{Pablo_prl96}
Garc\'{\i}a-Fern\'andez P, Bersuker I~B and Boggs J~E 2006 {\em Phys. Rev.
  Lett.\/} {\bf 96}(16) 163005

\bibitem{Hizhnyakov_jcp147}
Pae K and Hizhnyakov V 2017 {\em J. Chem. Phys.\/} {\bf 147} 084107

\bibitem{Hizhnyakov_cpl525}
Hizhnyakov V, Pae K and Vaikj\"arv T 2012 {\em Chem. Phys. Lett.\/} {\bf
  525-526} 64--68 ISSN 0009-2614

\bibitem{renner_zp92}
Renner E 1934 {\em Z. Phys.\/} {\bf 92} 172--93

\bibitem{saikat_cpc26}
Hazra S, Basu S, Mukherjee S, Ravi S, Sardar S and Adhikari S 2025 {\em
  ChemPhysChem\/} {\bf 26} e202500461

\bibitem{SSMP107}
Sardar S, Paul A~K and Adhikari S 2009 {\em Mol. Phys.\/} {\bf 107} 2467

\bibitem{SSJPCA118}
Khan B~A, Sardar S, Sarkar P and Adhikari S 2014 {\em J. Phys. Chem. A\/} {\bf
  118 (49)} 11451

\bibitem{SSPCCP13}
Sardar S, Puzari P and Adhikari S 2011 {\em Phys. Chem. Chem. Phys.\/} {\bf 13}
  15960

\bibitem{SSJTCC12}
Khan B~A, Sardar S, Sarkar P and Adhikari S 2013 {\em J. Theo. Com. Chem.\/}
  {\bf 12} 1350042

\bibitem{SSPCCP10}
Sardar S, Paul A~K, Mondal P, Sarkar B and Adhikari S 2008 {\em Phys. Chem.
  Chem. Phys.\/} {\bf 10} 6388

\bibitem{subhankar_ijqc111}
Sardar S, Paul A~K, Sharma R and Adhikari S 2011 {\em Int. J. Quantum Chem.\/}
  {\bf 111} 2741--59

\bibitem{subhankar_jcp130}
Sardar S, Paul A~K, Sharma R and Adhikari S 2009 {\em J. Chem. Phys.\/} {\bf
  130} 144302

\bibitem{souvik_irpc37}
Mandal S, Ghosh S, Sardar S and Adhikari S 2019 {\em Int. Rev. Phys. Chem.\/}
  {\bf 37} 607--700

\bibitem{GHH}
Herzberg G and Higgins H~C~L 1963 {\em Discuss. Faraday Soc.\/} {\bf 35} 77

\bibitem{PPP}
Pechukas P 1969 {\em Phys. Rev.\/} {\bf 181} 174

\bibitem{JC1}
Tully J~C and Preston R~K 1971 {\em J. Chem. Phys.\/} {\bf 55} 562

\bibitem{WHM}
Miller W~H and George T~F 1972 {\em J. Chem. Phys.\/} {\bf 56} 5637

\bibitem{YAR}
Yarkony D~R 1996 {\em Rev. Mod. Phys.\/} {\bf 68} 985

\bibitem{MBB}
Baer M {\em Beyond Born-Oppenheimer: Electronic Nonadiabatic Coupling Terms and
  Conical Intersections; Wiley-Interscience (2006)\/}

\bibitem{BS1}
Sarkar B and Adhikari S 2006 {\em J. Chem. Phys.\/} {\bf 124} 074101

\bibitem{GSM}
Schatz G and Ratner M~A {\em Quantum Mechanics in Chemistry; Prentice Hall,
  Englewood Cliffs: NJ (1993)\/}

\bibitem{SACPL309}
Adhikari S and Billing G~D 1999 {\em Chem. Phys. Lett.\/} {\bf 309} 249

\bibitem{SAJCP11148}
Adhikari S and Billing G~D 1999 {\em J. Chem. Phys.\/} {\bf 111} 48

\bibitem{SACP300}
Puzari P, Deshpande S~A and Adhikari S 2004 {\em Chem. Phys.\/} {\bf 300} 305

\bibitem{SAJCP113}
Adhikari S and Billing G~D 2000 {\em J. Chem. Phys.\/} {\bf 113} 1409

\bibitem{SACPL321}
Billing G~D and Adhikari S 2000 {\em Chem. Phys. Lett.\/} {\bf 321} 197

\bibitem{SAJCP118}
Barkakaty B and Adhikari S 2003 {\em J. Chem. Phys.\/} {\bf 118} 5302

\bibitem{SAJCP123}
Puzari P, Swathi R~S, Sarkar B and Adhikari S 2005 {\em J. Chem. Phys.\/} {\bf
  123} 134317

\bibitem{SAJCP125}
Puzari P, Sarkar B and Adhikari S 2006 {\em J. Chem. Phys.\/} {\bf 125} 194316

\bibitem{subhankar_jcs124}
Sardar S and Adhikari S 2012 {\em J. Chem. Sci.\/} {\bf 124} 51--8

\bibitem{SSCPL496}
Sardar S, Puzari P and Adhikari S 2010 {\em Chem. Phys. Lett.\/} {\bf 496} 341

\bibitem{BO}
Born M and Oppenheimer R 1927 {\em Ann. Phys.\/} {\bf 389} 457--484 ISSN
  0003-3804

\bibitem{BH}
Born M and Huang K 1954 {\em Dynamical Theory of Crystal Lattices\/} (Oxford
  University Press, Oxford)

\bibitem{hellmann}
Hellmann H 1937 {\em {\it Einfuring in die Quantenchemie}\/} (Germany: Franz
  Duetiche:Leipzig)

\bibitem{feynman_pr56}
Feynman R 1939 {\em Phys. Rev.\/} {\bf 56} 340--3

\bibitem{SSJCS122}
Sardar S, Paul A~K and Adhikari S 2010 {\em J. Chem. Sc.\/} {\bf 122} 491

\bibitem{PAM}
Dirac P~A~M 1930 {\em Proc. Cambridge Philos. Soc.\/} {\bf 26} 376

\bibitem{SAIJQC434}
Puzari P and Adhikari S 2004 {\em Int. J. Quant. Chem.\/} {\bf 98} 434

\bibitem{mahapatra_jcp110}
Mahapatra S, K\"oppel H and Cederbaum L 1999 {\em J. Chem. Phys.\/} {\bf 110}
  5691--701

\bibitem{jackels_jcp63}
Jackels C~F and Davidson E~R 1975 {\em J. Chem. Phys.\/} {\bf 63} 4672

\bibitem{kaldor_cpl185}
Kaldor U 1991 {\em Chem. Phys. Lett.\/} {\bf 185} 131--5

\bibitem{haller_jms111}
Haller E, K\"oppel H and Cederbaum L 1985 {\em J. Mol. Spec.\/} {\bf 111}
  377--97

\bibitem{schinke_cp346}
Schinke R, Grebenshchikov S~Y and Zhu H 2008 {\em Chem. Phys.\/} {\bf 346}
  99--114

\bibitem{joyeux_jcp119}
Joyeux M, Jost R,  and Lombardi M 2003 {\em J. Chem. Phys.\/} {\bf 119}
  5923--32

\bibitem{sanrey_jcp125}
Sanrey M and Joyeux M 2006 {\em J. Chem. Phys.\/} {\bf 125} 014304

\bibitem{takatsuka_jcp132}
Arasaki Y, Takatsuka K, Wang K and McKoy V 2010 {\em J. Chem. Phys.\/} {\bf
  132} 124307

\bibitem{takatsuka_pccp13}
Arasaki Y, Wang K, McKoy V and Takatsuka K 2011 {\em Phys. Chem. Chem. Phys.\/}
  {\bf 13} 8681--9

\bibitem{MOLPRO_brief}
Werner H~J, Knowles P~J, Knizia G, Manby F~R and
  Sch{\ifmmode\ddot{u}\else\"{u}\fi}tz M 2012 {\em WIREs Comput. Mol. Sci.\/}
  {\bf 2} 242--253 ISSN 1759-0876

\bibitem{weaver_jcp90}
Weaver A, Metz R~B, Bradforth S~E and Neumark D~M 1989 {\em J. Chem. Phys.\/}
  {\bf 90} 2070--1

\bibitem{naskar_jctc}
Naskar K, Mukherjee S, Mukherjee B, Ravi S, Mukherjee S, Sardar S and Adhikari
  S 2020 {\em J. Chem. Theory Comput.\/} {\bf 16} 1666--1680

\bibitem{mantu_ijqc_curl}
Sah M~K, Mukherjee S, Naskar K, Hazra S and Adhikari S 2023 {\em International
  Journal of Quantum Chemistry\/} {\bf 123} e27212

\bibitem{sardar_pccp13}
Sardar S, Puzari P and Adhikari S 2011 {\em Phys. Chem. Chem. Phys.\/} {\bf 13}
  15960

\bibitem{sardar_jcs124}
Sardar S and Adhikari S 2012 {\em J. Chem. Sci.\/} {\bf 124} 51

\bibitem{sardar_jpca118}
Khan B~A, Sardar S, Sarkar P and Adhikari S 2014 {\em J. Phys. Chem. A\/} {\bf
  118} 11451

\bibitem{jpca-119-12392-SA}
Ghosh S, Sahoo T, Adhikari S, Sharma R and Varandas A~J~C 2015 {\em J. Phys.
  Chem. A\/} {\bf 119} 12392--12403

\bibitem{jcp-147-074105-SA}
Ghosh S, Mukherjee S, Mukherjee B, Mandal S, Sharma R, Chaudhury P and Adhikari
  S 2017 {\em J. Chem. Phys.\/} {\bf 147} 074105/1--16

\bibitem{mandal_irpc37}
Mandal S, Ghosh S, Sardar S and Adhikari S 2018 {\em Int. Rev. Phys. Chem.\/}
  {\bf 37} 607--700

\bibitem{sadygov_jcp110}
Sadygov R~G and Yarkony D~R 1999 {\em J. Chem. Phys.\/} {\bf 110} 3639--42

\bibitem{vivie-rieldle_cp223}
de~Vivie-Riedle R, Gaus J, Bona\v{c}i\'c-Kouteck\'y V, Manz J, Reischl-Lenz B
  and Saalfrank P 1997 {\em Chem. Phys.\/} {\bf 223} 1--14

\bibitem{meiswinkel_cp144}
Meiswinkel R and K\"oppel H 1990 {\em Chem. Phys.\/} {\bf 144} 117--28

\bibitem{koizumi_prl83}
Koizumi H and Bersuker I~B 1999 {\em Phys. Rev. Lett.\/} {\bf 83} 3009--12

\bibitem{cocchini_jcp88}
Cocchini F, Upton T~H and Andreoni W 1988 {\em J. Chem. Phys.\/} {\bf 88}
  6068--77

\bibitem{hauser_jcp129}
Hauser A~W, Callegari C, Soldan P and Ernst W~E 2008 {\em J. Chem. Phys.\/}
  {\bf 129} 044307

\bibitem{hauser_cp375}
Hauser A~W, Callegari C, Soldan P and Ernst W~E 2010 {\em Chem. Phys.\/} {\bf
  375} 73

\bibitem{delacretaz_prl56}
Delacr\'etaz G, Grant E~R, Whetten R~L, W\"oste L and Zwanziger J~W 1986 {\em
  Phys. Rev. Lett.\/} {\bf 56} 2598--601

\bibitem{ernst_pst112}
Ernst W~E and Golonzka O 2004 {\em Phys. Scrp.\/} {\bf T112} 27

\bibitem{akpaul_jcp131}
Paul A~K, Sardar S, Sarkar B and Adhikari S 2009 {\em J. Chem. Phys.\/} {\bf
  131} 124312

\bibitem{baer_jcp66}
Top Z~H and Baer M 1977 {\em J. Chem. Phys.\/} {\bf 66} 1363

\bibitem{sarkar_jcp124}
Sarkar B and Adhikari S 2006 {\em J. Chem. Phys.\/} {\bf 124} 074101

\bibitem{stock_jcp103}
Stock G, Woywod C, Domcke W, Swinney T and Hudson B~S 1995 {\em J. Chem.
  Phys.\/} {\bf 103} 6851

\bibitem{domcke_jcp93}
Stock G and Domcke W 1990 {\em J. Chem. Phys.\/} {\bf 93} 5496

\bibitem{yamazaki_fdcs75}
Yamazaki I, Murao T, Yamanaka T and Yoshihara K 1983 {\em Faraday Discuss.
  Chem. Soc.\/} {\bf 75} 395--405

\bibitem{seidner_jcp96}
Seidner L, Stock G and Sobolewski A~L 1992 {\em J. Chem. Phys.\/} {\bf 96} 5298

\bibitem{domcke_jcp98}
Sobolewski A~L, Woywod C and Domcke W 1993 {\em J. Chem. Phys.\/} {\bf 98} 5627

\bibitem{guo_jrs4074}
Guo J~L, Liu C, Xie B~B, Zhao Y~Y, Pei K~M, Wang H~G, Zheng X, Ai Y~J, Chen
  X~B, Fang W~H and Yeung C~S 2012 {\em J. Raman Spectrosc.\/} {\bf 4074}

\bibitem{kanno_4}
Kanno M, Ito Y, Shimakura N, Koseki S, Kono H and Fujimura Y 2015 {\em Phys.
  Chem. Chem. Phys.\/} {\bf 17} 2012--2024

\bibitem{woywood_jcp100}
Woywod C, Domcke W, Sobolewski A~L and Werner H~J 1994 {\em J. Chem. Phys.\/}
  {\bf 100} 1400

\bibitem{worth_jcp109}
Worth G~A, Meyer H~D and Cederbaum L~S 1998 {\em J. Chem. Phys.\/} {\bf 109}
  3518--3529

\bibitem{Burghardt_jcp129}
Burghardt I, Giri K and Worth G~A 2008 {\em J. Chem. Phys.\/} {\bf 129} 174104

\bibitem{vendrell_jcp134}
Vendrell O and Meyer H~D 2011 {\em J. Chem. Phys.\/} {\bf 134} 044135

\bibitem{richings_15}
Richings G~W, Robertson C and Habershon S 2019 {\em J. Chem. Theory Comput.\/}
  {\bf 15} 857--870

\bibitem{herzberg_dfs35}
Herzberg G and Longuet-Higgins H~C 1963 {\em Discuss Faraday Soc.\/} {\bf 35}
  77--82

\bibitem{pancharatnam_pias44}
Pancharatnam S 1956 {\em Proc. Ind. Acad. Sci. A\/} {\bf 44} 247

\bibitem{berry_prca392}
Berry M~V 1984 {\em Proc. R. Soc., Lond. A\/} {\bf 392} 45

\bibitem{mantu_curl}
Sah M~K, Mukherjee S, Naskar K, Hazra S and Adhikari S 2023 {\em Int. J.
  Quantum Chem.\/} {\bf DOI: 10.1002/qua.27212}

\bibitem{raab_jcp110}
Raab A, Worth G~A, Meyer H~D and Cederbaum L~S 1999 {\em J. Chem. Phys.\/} {\bf
  110} 936--946

\bibitem{mahapatra_cp347}
Faraji S, K\"oppel H, Eisfeld W and Mahapatra S 2008 {\em Chem. Phys.\/} {\bf
  347} 110--9

\bibitem{faraji_jcp_struct}
Faraji S and Köppel H 2008 {\em The Journal of Chemical Physics\/} {\bf 129}
  074310

\bibitem{sardar_pccp}
Sardar S, Puzari P and Adhikari S 2011 {\em Phys. Chem. Chem. Phys.\/} {\bf
  13}(35) 15960--15972

\bibitem{eisfeld_jcp113}
Eisfeld W and Morokuma K 2000 {\em J. Chem. Phys.\/} {\bf 113} 5587--97

\bibitem{mayer_jcp100}
Mayer M, Cederbaum L~S and K\"oppel H 1994 {\em J. Chem. Phys.\/} {\bf 100}
  899--911

\bibitem{okumura_ps73}
Okumura M, Stanton J~F, Deev A and Sommar J 2006 {\em Phys. Scr.\/} {\bf 73}
  C64

\bibitem{stanton_jcp126}
Stanton J~F 2007 {\em J. Chem. Phys.\/} {\bf 126} 134309

\bibitem{nelson_jpc87}
Nelson H~H, Pasternack L and McDonald J~R 1983 {\em J. Phys. Chem.\/} {\bf 87}
  1286

\bibitem{ishiwata_jpc87}
Ishiwata T, Fujiwara I, Naruge Y, Obi K and Tanaka I 1983 {\em J. Phys.
  Chem.\/} {\bf 87} 1349

\bibitem{ishiwata_jcp82}
Ishiwata T, Tanaka I, Kawaguchi K and Hirota E 1985 {\em J. Chem. Phys.\/} {\bf
  82} 2196

\bibitem{hirota_jcp107}
Hirota E, Ishiwata T, Kawaguchi K, Fujitake M, Ohashi N and Tanaka I 1997 {\em
  J. Chem. Phys.\/} {\bf 107} 2829--38

\bibitem{weaver_jcp94}
Weaver A, Arnold D~W, Bradforth S~E and Neumark D~M 1991 {\em J. Chem. Phys.\/}
  {\bf 94} 1740--51

\bibitem{deev_jcp122}
Deev A, Stanton J~F and Okumura M 2005 {\em J. Chem. Phys.\/} {\bf 122} 224305

\bibitem{codd_jcp142}
Codd T, Chen M~W, Roudjane M, Stanton J~F and Miller T~A 2015 {\em J. Chem.
  Phys.\/} {\bf 142} 184305

\bibitem{scott_jcp}
Scott D~W, Messerly J~F, Todd S~S, Hossenlopp I~A, Osborn A and McCullough J~P
  1963 {\em The Journal of Chemical Physics\/} {\bf 38} 532--539

\bibitem{abraham_jcsb}
Abraham R~J and Cooper M~A 1968 {\em J. Chem. Soc. B\/} (0) 1094--1098

\bibitem{gerritsen_sapms}
Gerritsen J and MacLean C 1971 {\em Spectrochimica Acta Part A: Molecular
  Spectroscopy\/} {\bf 27} 1495--1503

\bibitem{clark_jesc}
Clark D, Kilcast D, Adams D and Musgrave W 1972 {\em Journal of Electron
  Spectroscopy and Related Phenomena\/} {\bf 1} 227--250

\bibitem{shirley_jacs}
Davis D~W, Shirley D~A and Thomas T~D 1972 {\em Journal of the American
  Chemical Society\/} {\bf 94} 6565--6575

\bibitem{gilbert_cpl}
GILBERT R and SANDORFY C 1974 {\em Chemical Physics Letters\/} {\bf 27}
  457--463

\bibitem{bieri_jesc}
Bieri G, Åsbrink L and {Von Niessen} W 1981 {\em Journal of Electron
  Spectroscopy and Related Phenomena\/} {\bf 23} 281--322

\bibitem{tsuchiya_cpl}
Tsuchiya Y, Takazawa K, Fujii M and Ito M 1991 {\em Chemical Physics Letters\/}
  {\bf 183} 107--112

\bibitem{tsuchiya_jcp}
Tsuchiya Y, Takazawa K, Fujii M and Ito M 1992 {\em The Journal of Physical
  Chemistry\/} {\bf 96} 99--104

\bibitem{philis_jpb}
Philis J, Bolovinos A, Andritsopoulos G, Pantos E and Tsekeris P 1981 {\em
  Journal of Physics B: Atomic and Molecular Physics\/} {\bf 14} 3621

\bibitem{goly_cp}
Goly T, Spoerel U and Stahl W 2002 {\em Chemical Physics\/} {\bf 283} 289--296

\bibitem{kwon_jcp_2003}
Kwon C~H, Kim H~L and Kim M~S 2003 {\em J. Chem. Phys\/} {\bf 118} 6327--6335

\bibitem{faraji_jcp_dyna}
Faraji S, Meyer H~D and Köppel H 2008 {\em The Journal of Chemical Physics\/}
  {\bf 129} 074311

\bibitem{kwon_jcp}
Kwon C~H, Kim H~L and Kim M~S 2002 {\em The Journal of Chemical Physics\/} {\bf
  116} 10361--10371

\bibitem{gilbert_cpl17}
Gilbert R, Sauvageau P and Sandorfy C 1972 {\em Chem. Phys. Lett.\/} {\bf 17}
  465--70

\bibitem{bieri_jesrp23}
Bieri G, {\AA}sbrink L and Niessen W~V 1981 {\em J. Electron Spectrosc. Relat.
  Phenom.\/} {\bf 23} 281--322

\bibitem{dujardin_jcp79}
Dujardin G, Leach S, Dutuit O, Govers T and Guyon P~M 1983 {\em J. Chem.
  Phys.\/} {\bf 79} 644--657

\bibitem{braden_jpca104}
Braden D~A and Hudson B~S 2000 {\em J. Phys. Chem. A\/} {\bf 104} 982--989

\bibitem{palmer_jms49}
Palmer M~H, Moyes W, Spiers M and Ridyard J~N~A 1978 {\em J. Mol. Struct.\/}
  {\bf 49} 105--23

\bibitem{mahapatra_pccp11}
Mondal T and Mahapatra S 2009 {\em Phys. Chem. Chem. Phys.\/} {\bf 11}
  10867--80

\bibitem{koppel_jcp89}
K\"oppel H, Cederbaum L~S and Domcke W 1988 {\em J. Chem. Phys.\/} {\bf 89}
  2023--40

\bibitem{Koppel_jcp117_1}
D\"oscher M, K\"oppel H and Szalay P~G 2002 {\em J. Chem. Phys\/} {\bf 117}
  2645--56

\bibitem{Koppel_jcp117_2}
K\"oppel H, D\"oscher M, B\^aldea I, Meyer H~D and Szalay P~G 2002 {\em J.
  Chem. Phys.\/} {\bf 117} 2657--71

\bibitem{johnson_jcp117}
Johnson P~M 2002 {\em J. Chem. Phys.\/} {\bf 117} 9991

\bibitem{mantu_jcp159_path}
Sah M~K, Mukherjee S, Saha S, Naskar K and Adhikari S 2023 {\em The Journal of
  Chemical Physics\/} {\bf 159} 244116

\bibitem{Baltzer_cp224}
Baltzer P, Karlsson L, Wannberg B, \"Ohrwall G, Holland D~M~P, MacDonald M~A,
  Hayes M~A and von Niesson W 1997 {\em Chem. Phys.\/} {\bf 224} 95--119

\bibitem{Koppel_aqc44}
K\"oppel H, B\^aldea I and Szalay P~G 2003 {\em Adv. Quantum Chem.\/} {\bf 44}
  199

\bibitem{mukherjee_jcp141}
Mukherjee S, Mukhopadhyay D and Adhikari S 2014 {\em J. Chem. Phys.\/} {\bf
  141} 204306

\end{thebibliography}


\begin{thebibliography}{199}%
\makeatletter
\providecommand \@ifxundefined [1]{%
 \@ifx{#1\undefined}
}%
\providecommand \@ifnum [1]{%
 \ifnum #1\expandafter \@firstoftwo
 \else \expandafter \@secondoftwo
 \fi
}%
\providecommand \@ifx [1]{%
 \ifx #1\expandafter \@firstoftwo
 \else \expandafter \@secondoftwo
 \fi
}%
\providecommand \natexlab [1]{#1}%
\providecommand \enquote  [1]{``#1''}%
\providecommand \bibnamefont  [1]{#1}%
\providecommand \bibfnamefont [1]{#1}%
\providecommand \citenamefont [1]{#1}%
\providecommand \href@noop [0]{\@secondoftwo}%
\providecommand \href [0]{\begingroup \@sanitize@url \@href}%
\providecommand \@href[1]{\@@startlink{#1}\@@href}%
\providecommand \@@href[1]{\endgroup#1\@@endlink}%
\providecommand \@sanitize@url [0]{\catcode `\\12\catcode `\$12\catcode
  `\&12\catcode `\#12\catcode `\^12\catcode `\_12\catcode `\%12\relax}%
\providecommand \@@startlink[1]{}%
\providecommand \@@endlink[0]{}%
\providecommand \url  [0]{\begingroup\@sanitize@url \@url }%
\providecommand \@url [1]{\endgroup\@href {#1}{\urlprefix }}%
\providecommand \urlprefix  [0]{URL }%
\providecommand \Eprint [0]{\href }%
\providecommand \doibase [0]{http://dx.doi.org/}%
\providecommand \selectlanguage [0]{\@gobble}%
\providecommand \bibinfo  [0]{\@secondoftwo}%
\providecommand \bibfield  [0]{\@secondoftwo}%
\providecommand \translation [1]{[#1]}%
\providecommand \BibitemOpen [0]{}%
\providecommand \bibitemStop [0]{}%
\providecommand \bibitemNoStop [0]{.\EOS\space}%
\providecommand \EOS [0]{\spacefactor3000\relax}%
\providecommand \BibitemShut  [1]{\csname bibitem#1\endcsname}%
\let\auto@bib@innerbib\@empty
\bibitem [{\citenamefont {Schafer}\ \emph {et~al.}(1970)\citenamefont
  {Schafer}, \citenamefont {Siska}, \citenamefont {Parson}, \citenamefont
  {Tully}, \citenamefont {Wong},\ and\ \citenamefont {Lee}}]{jcp-53-3385-LEE}%
  \BibitemOpen
  \bibfield  {author} {\bibinfo {author} {\bibfnamefont {T.~P.}\ \bibnamefont
  {Schafer}}, \bibinfo {author} {\bibfnamefont {P.~E.}\ \bibnamefont {Siska}},
  \bibinfo {author} {\bibfnamefont {J.~M.}\ \bibnamefont {Parson}}, \bibinfo
  {author} {\bibfnamefont {F.~P.}\ \bibnamefont {Tully}}, \bibinfo {author}
  {\bibfnamefont {Y.~C.}\ \bibnamefont {Wong}}, \ and\ \bibinfo {author}
  {\bibfnamefont {Y.~T.}\ \bibnamefont {Lee}},\ }\href@noop {} {\bibfield
  {journal} {\bibinfo  {journal} {J. Chem. Phys.}\ }\textbf {\bibinfo {volume}
  {53}},\ \bibinfo {pages} {3385} (\bibinfo {year} {1970})}\BibitemShut
  {NoStop}%
\bibitem [{\citenamefont {Neumark}\ \emph
  {et~al.}(1985{\natexlab{a}})\citenamefont {Neumark}, \citenamefont {Wodtke},
  \citenamefont {Robinson}, \citenamefont {Hayden},\ and\ \citenamefont
  {Lee}}]{jcp-82-3045-LEE}%
  \BibitemOpen
  \bibfield  {author} {\bibinfo {author} {\bibfnamefont {D.~M.}\ \bibnamefont
  {Neumark}}, \bibinfo {author} {\bibfnamefont {A.~M.}\ \bibnamefont {Wodtke}},
  \bibinfo {author} {\bibfnamefont {G.~N.}\ \bibnamefont {Robinson}}, \bibinfo
  {author} {\bibfnamefont {C.~C.}\ \bibnamefont {Hayden}}, \ and\ \bibinfo
  {author} {\bibfnamefont {Y.~T.}\ \bibnamefont {Lee}},\ }\href {\doibase
  http://dx.doi.org/10.1063/1.448254} {\bibfield  {journal} {\bibinfo
  {journal} {J. Chem. Phys.}\ }\textbf {\bibinfo {volume} {82}},\ \bibinfo
  {pages} {3045} (\bibinfo {year} {1985}{\natexlab{a}})}\BibitemShut {NoStop}%
\bibitem [{\citenamefont {Neumark}\ \emph
  {et~al.}(1985{\natexlab{b}})\citenamefont {Neumark}, \citenamefont {Wodtke},
  \citenamefont {Robinson}, \citenamefont {Hayden}, \citenamefont {Shobatake},
  \citenamefont {Sparks}, \citenamefont {Schafer},\ and\ \citenamefont
  {Lee}}]{jcp-82-3067-LEE}%
  \BibitemOpen
  \bibfield  {author} {\bibinfo {author} {\bibfnamefont {D.~M.}\ \bibnamefont
  {Neumark}}, \bibinfo {author} {\bibfnamefont {A.~M.}\ \bibnamefont {Wodtke}},
  \bibinfo {author} {\bibfnamefont {G.~N.}\ \bibnamefont {Robinson}}, \bibinfo
  {author} {\bibfnamefont {C.~C.}\ \bibnamefont {Hayden}}, \bibinfo {author}
  {\bibfnamefont {K.}~\bibnamefont {Shobatake}}, \bibinfo {author}
  {\bibfnamefont {R.~K.}\ \bibnamefont {Sparks}}, \bibinfo {author}
  {\bibfnamefont {T.~P.}\ \bibnamefont {Schafer}}, \ and\ \bibinfo {author}
  {\bibfnamefont {Y.~T.}\ \bibnamefont {Lee}},\ }\href@noop {} {\bibfield
  {journal} {\bibinfo  {journal} {J. Chem. Phys.}\ }\textbf {\bibinfo {volume}
  {82}},\ \bibinfo {pages} {3067} (\bibinfo {year}
  {1985}{\natexlab{b}})}\BibitemShut {NoStop}%
\bibitem [{\citenamefont {Faubel}\ \emph {et~al.}(1991)\citenamefont {Faubel},
  \citenamefont {Schlemmer}, \citenamefont {Sondermann},\ and\ \citenamefont
  {Toennies}}]{jcp-94-4676-TOENN}%
  \BibitemOpen
  \bibfield  {author} {\bibinfo {author} {\bibfnamefont {M.}~\bibnamefont
  {Faubel}}, \bibinfo {author} {\bibfnamefont {S.}~\bibnamefont {Schlemmer}},
  \bibinfo {author} {\bibfnamefont {F.}~\bibnamefont {Sondermann}}, \ and\
  \bibinfo {author} {\bibfnamefont {J.~P.}\ \bibnamefont {Toennies}},\
  }\href@noop {} {\bibfield  {journal} {\bibinfo  {journal} {J. Chem. Phys.}\
  }\textbf {\bibinfo {volume} {94}},\ \bibinfo {pages} {4676} (\bibinfo {year}
  {1991})}\BibitemShut {NoStop}%
\bibitem [{\citenamefont {Faubel}\ \emph {et~al.}(1994)\citenamefont {Faubel},
  \citenamefont {Rusin}, \citenamefont {Schlemmer}, \citenamefont {Sondermann},
  \citenamefont {Tappe},\ and\ \citenamefont {Toennies}}]{jcp-101-2106-TOENN}%
  \BibitemOpen
  \bibfield  {author} {\bibinfo {author} {\bibfnamefont {M.}~\bibnamefont
  {Faubel}}, \bibinfo {author} {\bibfnamefont {L.}~\bibnamefont {Rusin}},
  \bibinfo {author} {\bibfnamefont {S.}~\bibnamefont {Schlemmer}}, \bibinfo
  {author} {\bibfnamefont {F.}~\bibnamefont {Sondermann}}, \bibinfo {author}
  {\bibfnamefont {U.}~\bibnamefont {Tappe}}, \ and\ \bibinfo {author}
  {\bibfnamefont {J.~P.}\ \bibnamefont {Toennies}},\ }\href@noop {} {\bibfield
  {journal} {\bibinfo  {journal} {J. Chem. Phys.}\ }\textbf {\bibinfo {volume}
  {101}},\ \bibinfo {pages} {2106} (\bibinfo {year} {1994})}\BibitemShut
  {NoStop}%
\bibitem [{\citenamefont {Baer}\ \emph {et~al.}(1996)\citenamefont {Baer},
  \citenamefont {Faubel}, \citenamefont {Martinez‐Haya}, \citenamefont
  {Rusin}, \citenamefont {Tappe},\ and\ \citenamefont
  {Toennies}}]{jcp-104-2743-TOENN}%
  \BibitemOpen
  \bibfield  {author} {\bibinfo {author} {\bibfnamefont {M.}~\bibnamefont
  {Baer}}, \bibinfo {author} {\bibfnamefont {M.}~\bibnamefont {Faubel}},
  \bibinfo {author} {\bibfnamefont {B.}~\bibnamefont {Martinez‐Haya}},
  \bibinfo {author} {\bibfnamefont {L.~Y.}\ \bibnamefont {Rusin}}, \bibinfo
  {author} {\bibfnamefont {U.}~\bibnamefont {Tappe}}, \ and\ \bibinfo {author}
  {\bibfnamefont {J.~P.}\ \bibnamefont {Toennies}},\ }\href@noop {} {\bibfield
  {journal} {\bibinfo  {journal} {J. Chem. Phys.}\ }\textbf {\bibinfo {volume}
  {104}},\ \bibinfo {pages} {2743} (\bibinfo {year} {1996})}\BibitemShut
  {NoStop}%
\bibitem [{\citenamefont {Faubel}\ \emph {et~al.}(1997)\citenamefont {Faubel},
  \citenamefont {Martinez-Haya}, \citenamefont {Rusin}, \citenamefont {Tappe},\
  and\ \citenamefont {Toennies}}]{jpcA-101-6415-TOENN}%
  \BibitemOpen
  \bibfield  {author} {\bibinfo {author} {\bibfnamefont {M.}~\bibnamefont
  {Faubel}}, \bibinfo {author} {\bibfnamefont {B.}~\bibnamefont
  {Martinez-Haya}}, \bibinfo {author} {\bibfnamefont {L.~Y.}\ \bibnamefont
  {Rusin}}, \bibinfo {author} {\bibfnamefont {U.}~\bibnamefont {Tappe}}, \ and\
  \bibinfo {author} {\bibfnamefont {J.~P.}\ \bibnamefont {Toennies}},\
  }\href@noop {} {\bibfield  {journal} {\bibinfo  {journal} {J. Phys. Chem. A}\
  }\textbf {\bibinfo {volume} {101}},\ \bibinfo {pages} {6415} (\bibinfo {year}
  {1997})}\BibitemShut {NoStop}%
\bibitem [{\citenamefont {Baer}\ \emph {et~al.}(1998)\citenamefont {Baer},
  \citenamefont {Faubel}, \citenamefont {Martinez‐Haya}, \citenamefont
  {Rusin}, \citenamefont {Tappe},\ and\ \citenamefont
  {Toennies}}]{jcp-108-9694-TOENN}%
  \BibitemOpen
  \bibfield  {author} {\bibinfo {author} {\bibfnamefont {M.}~\bibnamefont
  {Baer}}, \bibinfo {author} {\bibfnamefont {M.}~\bibnamefont {Faubel}},
  \bibinfo {author} {\bibfnamefont {B.}~\bibnamefont {Martinez‐Haya}},
  \bibinfo {author} {\bibfnamefont {L.~Y.}\ \bibnamefont {Rusin}}, \bibinfo
  {author} {\bibfnamefont {U.}~\bibnamefont {Tappe}}, \ and\ \bibinfo {author}
  {\bibfnamefont {J.~P.}\ \bibnamefont {Toennies}},\ }\href@noop {} {\bibfield
  {journal} {\bibinfo  {journal} {J. Chem. Phys.}\ }\textbf {\bibinfo {volume}
  {108}},\ \bibinfo {pages} {9694} (\bibinfo {year} {1998})}\BibitemShut
  {NoStop}%
\bibitem [{\citenamefont {Wurzberg}\ and\ \citenamefont
  {Houston}(1980)}]{jcp-72-4811-HOUSTON}%
  \BibitemOpen
  \bibfield  {author} {\bibinfo {author} {\bibfnamefont {E.}~\bibnamefont
  {Wurzberg}}\ and\ \bibinfo {author} {\bibfnamefont {P.~L.}\ \bibnamefont
  {Houston}},\ }\href@noop {} {\bibfield  {journal} {\bibinfo  {journal} {J.
  Chem. Phys.}\ }\textbf {\bibinfo {volume} {72}},\ \bibinfo {pages} {4811}
  (\bibinfo {year} {1980})}\BibitemShut {NoStop}%
\bibitem [{\citenamefont {Heidner}\ \emph {et~al.}(1980)\citenamefont
  {Heidner}, \citenamefont {Bott}, \citenamefont {Gardner},\ and\ \citenamefont
  {Melzer}}]{jcp-72-4815-HEIDNER}%
  \BibitemOpen
  \bibfield  {author} {\bibinfo {author} {\bibfnamefont {R.~F.}\ \bibnamefont
  {Heidner}}, \bibinfo {author} {\bibfnamefont {J.~F.}\ \bibnamefont {Bott}},
  \bibinfo {author} {\bibfnamefont {C.~E.}\ \bibnamefont {Gardner}}, \ and\
  \bibinfo {author} {\bibfnamefont {J.~E.}\ \bibnamefont {Melzer}},\
  }\href@noop {} {\bibfield  {journal} {\bibinfo  {journal} {J. Chem. Phys.}\
  }\textbf {\bibinfo {volume} {72}},\ \bibinfo {pages} {4815} (\bibinfo {year}
  {1980})}\BibitemShut {NoStop}%
\bibitem [{\citenamefont {Stevens}, \citenamefont {Brune},\ and\ \citenamefont
  {Anderson}(1989)}]{jpc-93-4068-ANDERSON}%
  \BibitemOpen
  \bibfield  {author} {\bibinfo {author} {\bibfnamefont {P.~S.}\ \bibnamefont
  {Stevens}}, \bibinfo {author} {\bibfnamefont {W.~H.}\ \bibnamefont {Brune}},
  \ and\ \bibinfo {author} {\bibfnamefont {J.~G.}\ \bibnamefont {Anderson}},\
  }\href@noop {} {\bibfield  {journal} {\bibinfo  {journal} {J. Phys. Chem.}\
  }\textbf {\bibinfo {volume} {93}},\ \bibinfo {pages} {4068} (\bibinfo {year}
  {1989})}\BibitemShut {NoStop}%
\bibitem [{\citenamefont {Tizniti}\ \emph {et~al.}(2014)\citenamefont
  {Tizniti}, \citenamefont {Picard}, \citenamefont {Lique}, \citenamefont
  {Berteloite}, \citenamefont {Canosa}, \citenamefont {Alexander},\ and\
  \citenamefont {Sims}}]{nchem-6-141-SIMS}%
  \BibitemOpen
  \bibfield  {author} {\bibinfo {author} {\bibfnamefont {M.}~\bibnamefont
  {Tizniti}}, \bibinfo {author} {\bibfnamefont {S.~D.~L.}\ \bibnamefont
  {Picard}}, \bibinfo {author} {\bibfnamefont {F.}~\bibnamefont {Lique}},
  \bibinfo {author} {\bibfnamefont {C.}~\bibnamefont {Berteloite}}, \bibinfo
  {author} {\bibfnamefont {A.}~\bibnamefont {Canosa}}, \bibinfo {author}
  {\bibfnamefont {M.~H.}\ \bibnamefont {Alexander}}, \ and\ \bibinfo {author}
  {\bibfnamefont {I.~R.}\ \bibnamefont {Sims}},\ }\href {\doibase
  10.1038/NCHEM.1835} {\bibfield  {journal} {\bibinfo  {journal} {Nature
  Chem.}\ }\textbf {\bibinfo {volume} {6}},\ \bibinfo {pages} {141} (\bibinfo
  {year} {2014})}\BibitemShut {NoStop}%
\bibitem [{\citenamefont {Bedjanian}(2021{\natexlab{a}})}]{bedjaninan}%
  \BibitemOpen
  \bibfield  {author} {\bibinfo {author} {\bibfnamefont {Y.}~\bibnamefont
  {Bedjanian}},\ }\href {\doibase https://doi.org/10.1002/kin.21462} {\bibfield
   {journal} {\bibinfo  {journal} {Int. J. Chem. Kin.}\ }\textbf {\bibinfo
  {volume} {53}},\ \bibinfo {pages} {527} (\bibinfo {year}
  {2021}{\natexlab{a}})},\ \Eprint
  {http://arxiv.org/abs/https://onlinelibrary.wiley.com/doi/pdf/10.1002/kin.21462}
  {https://onlinelibrary.wiley.com/doi/pdf/10.1002/kin.21462} \BibitemShut
  {NoStop}%
\bibitem [{\citenamefont {Aquilanti}\ \emph {et~al.}(2005)\citenamefont
  {Aquilanti}, \citenamefont {Cavalli}, \citenamefont {De~Fazio}, \citenamefont
  {Volpi}, \citenamefont {Aguilar},\ and\ \citenamefont {Lucas}}]{aql2}%
  \BibitemOpen
  \bibfield  {author} {\bibinfo {author} {\bibfnamefont {V.}~\bibnamefont
  {Aquilanti}}, \bibinfo {author} {\bibfnamefont {S.}~\bibnamefont {Cavalli}},
  \bibinfo {author} {\bibfnamefont {D.}~\bibnamefont {De~Fazio}}, \bibinfo
  {author} {\bibfnamefont {A.}~\bibnamefont {Volpi}}, \bibinfo {author}
  {\bibfnamefont {A.}~\bibnamefont {Aguilar}}, \ and\ \bibinfo {author}
  {\bibfnamefont {J.~M.}\ \bibnamefont {Lucas}},\ }\href {\doibase
  10.1016/j.chemphys.2004.03.027} {\bibfield  {journal} {\bibinfo  {journal}
  {Chem. Phys.}\ }\textbf {\bibinfo {volume} {308}},\ \bibinfo {pages} {237}
  (\bibinfo {year} {2005})}\BibitemShut {NoStop}%
\bibitem [{\citenamefont {Chen}, \citenamefont {Sun},\ and\ \citenamefont
  {Zhang}(2015)}]{jcp-142-024303-ZHANG}%
  \BibitemOpen
  \bibfield  {author} {\bibinfo {author} {\bibfnamefont {J.}~\bibnamefont
  {Chen}}, \bibinfo {author} {\bibfnamefont {Z.}~\bibnamefont {Sun}}, \ and\
  \bibinfo {author} {\bibfnamefont {D.~H.}\ \bibnamefont {Zhang}},\ }\href@noop
  {} {\bibfield  {journal} {\bibinfo  {journal} {J. Chem. Phys.}\ }\textbf
  {\bibinfo {volume} {142}},\ \bibinfo {pages} {024303} (\bibinfo {year}
  {2015})}\BibitemShut {NoStop}%
\bibitem [{\citenamefont {Fu}, \citenamefont {Xu},\ and\ \citenamefont
  {Zhang}(2008)}]{fxz_pes}%
  \BibitemOpen
  \bibfield  {author} {\bibinfo {author} {\bibfnamefont {B.}~\bibnamefont
  {Fu}}, \bibinfo {author} {\bibfnamefont {X.}~\bibnamefont {Xu}}, \ and\
  \bibinfo {author} {\bibfnamefont {D.~H.}\ \bibnamefont {Zhang}},\ }\href
  {\doibase 10.1063/1.2955729} {\bibfield  {journal} {\bibinfo  {journal} {J.
  Chem. Phys.}\ }\textbf {\bibinfo {volume} {129}},\ \bibinfo {pages} {011103}
  (\bibinfo {year} {2008})}\BibitemShut {NoStop}%
\bibitem [{\citenamefont {Naskar}, \citenamefont {Ghosh},\ and\ \citenamefont
  {Adhikari}(2022{\natexlab{a}})}]{Naskar2022_JPCA_126}%
  \BibitemOpen
  \bibfield  {author} {\bibinfo {author} {\bibfnamefont {K.}~\bibnamefont
  {Naskar}}, \bibinfo {author} {\bibfnamefont {S.}~\bibnamefont {Ghosh}}, \
  and\ \bibinfo {author} {\bibfnamefont {S.}~\bibnamefont {Adhikari}},\ }\href
  {\doibase 10.1021/acs.jpca.2c01209} {\bibfield  {journal} {\bibinfo
  {journal} {J. Phys. Chem. A}\ }\textbf {\bibinfo {volume} {126}},\ \bibinfo
  {pages} {3311} (\bibinfo {year} {2022}{\natexlab{a}})}\BibitemShut {NoStop}%
\bibitem [{\citenamefont {Naskar}\ \emph {et~al.}(2024)\citenamefont {Naskar},
  \citenamefont {Mukherjee}, \citenamefont {Ghosh},\ and\ \citenamefont
  {Adhikari}}]{FH2_dynamics}%
  \BibitemOpen
  \bibfield  {author} {\bibinfo {author} {\bibfnamefont {K.}~\bibnamefont
  {Naskar}}, \bibinfo {author} {\bibfnamefont {S.}~\bibnamefont {Mukherjee}},
  \bibinfo {author} {\bibfnamefont {S.}~\bibnamefont {Ghosh}}, \ and\ \bibinfo
  {author} {\bibfnamefont {S.}~\bibnamefont {Adhikari}},\ }\href@noop {}
  {\bibfield  {journal} {\bibinfo  {journal} {J. Phys. Chem. A}\ }\textbf
  {\bibinfo {volume} {128}},\ \bibinfo {pages} {1438} (\bibinfo {year}
  {2024})}\BibitemShut {NoStop}%
\bibitem [{\citenamefont {Castillo}\ \emph
  {et~al.}(1996{\natexlab{a}})\citenamefont {Castillo}, \citenamefont
  {Manolopoulos}, \citenamefont {Stark},\ and\ \citenamefont
  {Werner}}]{werner_jcp104}%
  \BibitemOpen
  \bibfield  {author} {\bibinfo {author} {\bibfnamefont {J.~F.}\ \bibnamefont
  {Castillo}}, \bibinfo {author} {\bibfnamefont {D.~E.}\ \bibnamefont
  {Manolopoulos}}, \bibinfo {author} {\bibfnamefont {K.}~\bibnamefont {Stark}},
  \ and\ \bibinfo {author} {\bibfnamefont {H.-J.}\ \bibnamefont {Werner}},\
  }\href@noop {} {\bibfield  {journal} {\bibinfo  {journal} {J. Chem. Phys.}\
  }\textbf {\bibinfo {volume} {104}},\ \bibinfo {pages} {6531} (\bibinfo {year}
  {1996}{\natexlab{a}})}\BibitemShut {NoStop}%
\bibitem [{\citenamefont {Castillo}\ \emph
  {et~al.}(1998{\natexlab{a}})\citenamefont {Castillo}, \citenamefont {Hartke},
  \citenamefont {Werner}, \citenamefont {Aoiz}, \citenamefont {nares},\ and\
  \citenamefont {Mart\'inez-Haya}}]{werner_jcp109}%
  \BibitemOpen
  \bibfield  {author} {\bibinfo {author} {\bibfnamefont {J.~F.}\ \bibnamefont
  {Castillo}}, \bibinfo {author} {\bibfnamefont {B.}~\bibnamefont {Hartke}},
  \bibinfo {author} {\bibfnamefont {H.-J.}\ \bibnamefont {Werner}}, \bibinfo
  {author} {\bibfnamefont {F.~J.}\ \bibnamefont {Aoiz}}, \bibinfo {author}
  {\bibfnamefont {L.~B.}\ \bibnamefont {nares}}, \ and\ \bibinfo {author}
  {\bibfnamefont {B.}~\bibnamefont {Mart\'inez-Haya}},\ }\href@noop {}
  {\bibfield  {journal} {\bibinfo  {journal} {J. Chem. Phys.}\ }\textbf
  {\bibinfo {volume} {109}},\ \bibinfo {pages} {7224} (\bibinfo {year}
  {1998}{\natexlab{a}})}\BibitemShut {NoStop}%
\bibitem [{\citenamefont {Mukherjee}\ \emph
  {et~al.}(2020{\natexlab{a}})\citenamefont {Mukherjee}, \citenamefont
  {Naskar}, \citenamefont {Mukherjee}, \citenamefont {Ravi}, \citenamefont
  {Shamasundar}, \citenamefont {Mukhopadhyay},\ and\ \citenamefont
  {Adhikari}}]{FH2_Surf}%
  \BibitemOpen
  \bibfield  {author} {\bibinfo {author} {\bibfnamefont {B.}~\bibnamefont
  {Mukherjee}}, \bibinfo {author} {\bibfnamefont {K.}~\bibnamefont {Naskar}},
  \bibinfo {author} {\bibfnamefont {S.}~\bibnamefont {Mukherjee}}, \bibinfo
  {author} {\bibfnamefont {S.}~\bibnamefont {Ravi}}, \bibinfo {author}
  {\bibfnamefont {K.~R.}\ \bibnamefont {Shamasundar}}, \bibinfo {author}
  {\bibfnamefont {D.}~\bibnamefont {Mukhopadhyay}}, \ and\ \bibinfo {author}
  {\bibfnamefont {S.}~\bibnamefont {Adhikari}},\ }\href@noop {} {\bibfield
  {journal} {\bibinfo  {journal} {J. Chem. Phys.}\ }\textbf {\bibinfo {volume}
  {153}},\ \bibinfo {pages} {174301} (\bibinfo {year}
  {2020}{\natexlab{a}})}\BibitemShut {NoStop}%
\bibitem [{\citenamefont {Das}\ \emph {et~al.}(2011)\citenamefont {Das},
  \citenamefont {Mukhopadhyay}, \citenamefont {Adhikari},\ and\ \citenamefont
  {Baer}}]{cpl-517-92-ANITA}%
  \BibitemOpen
  \bibfield  {author} {\bibinfo {author} {\bibfnamefont {A.}~\bibnamefont
  {Das}}, \bibinfo {author} {\bibfnamefont {D.}~\bibnamefont {Mukhopadhyay}},
  \bibinfo {author} {\bibfnamefont {S.}~\bibnamefont {Adhikari}}, \ and\
  \bibinfo {author} {\bibfnamefont {M.}~\bibnamefont {Baer}},\ }\href@noop {}
  {\bibfield  {journal} {\bibinfo  {journal} {Chem. Phys. Lett.}\ }\textbf
  {\bibinfo {volume} {517}},\ \bibinfo {pages} {92} (\bibinfo {year}
  {2011})}\BibitemShut {NoStop}%
\bibitem [{\citenamefont {Oka}(1980)}]{OKA:PRL}%
  \BibitemOpen
  \bibfield  {author} {\bibinfo {author} {\bibfnamefont {T.}~\bibnamefont
  {Oka}},\ }\href@noop {} {\bibfield  {journal} {\bibinfo  {journal} {Phys.
  Rev. Lett.}\ }\textbf {\bibinfo {volume} {45}},\ \bibinfo {pages} {531}
  (\bibinfo {year} {1980})}\BibitemShut {NoStop}%
\bibitem [{\citenamefont {Mukherjee}, \citenamefont {Mukhopadhyay},\ and\
  \citenamefont {Adhikari}(2014{\natexlab{a}})}]{H3+_Surf}%
  \BibitemOpen
  \bibfield  {author} {\bibinfo {author} {\bibfnamefont {S.}~\bibnamefont
  {Mukherjee}}, \bibinfo {author} {\bibfnamefont {D.}~\bibnamefont
  {Mukhopadhyay}}, \ and\ \bibinfo {author} {\bibfnamefont {S.}~\bibnamefont
  {Adhikari}},\ }\href@noop {} {\bibfield  {journal} {\bibinfo  {journal} {J.
  Chem. Phys.}\ }\textbf {\bibinfo {volume} {141}},\ \bibinfo {pages} {204306}
  (\bibinfo {year} {2014}{\natexlab{a}})}\BibitemShut {NoStop}%
\bibitem [{\citenamefont {Fehsenfeld}\ \emph {et~al.}(1974)\citenamefont
  {Fehsenfeld}, \citenamefont {Albritton}, \citenamefont {Bush}, \citenamefont
  {Fournier}, \citenamefont {Govers},\ and\ \citenamefont
  {Fournier}}]{jcp-61-2150-FOUR}%
  \BibitemOpen
  \bibfield  {author} {\bibinfo {author} {\bibfnamefont {F.~C.}\ \bibnamefont
  {Fehsenfeld}}, \bibinfo {author} {\bibfnamefont {D.~L.}\ \bibnamefont
  {Albritton}}, \bibinfo {author} {\bibfnamefont {Y.~A.}\ \bibnamefont {Bush}},
  \bibinfo {author} {\bibfnamefont {P.~G.}\ \bibnamefont {Fournier}}, \bibinfo
  {author} {\bibfnamefont {T.~R.}\ \bibnamefont {Govers}}, \ and\ \bibinfo
  {author} {\bibfnamefont {J.}~\bibnamefont {Fournier}},\ }\href@noop {}
  {\bibfield  {journal} {\bibinfo  {journal} {J. Chem. Phys.}\ }\textbf
  {\bibinfo {volume} {61}},\ \bibinfo {pages} {2150} (\bibinfo {year}
  {1974})}\BibitemShut {NoStop}%
\bibitem [{\citenamefont {Villinger}, \citenamefont {Henchman},\ and\
  \citenamefont {Lindinger}(1982)}]{jcp-76-1590-HEN2}%
  \BibitemOpen
  \bibfield  {author} {\bibinfo {author} {\bibfnamefont {H.}~\bibnamefont
  {Villinger}}, \bibinfo {author} {\bibfnamefont {M.~J.}\ \bibnamefont
  {Henchman}}, \ and\ \bibinfo {author} {\bibfnamefont {W.}~\bibnamefont
  {Lindinger}},\ }\href@noop {} {\bibfield  {journal} {\bibinfo  {journal} {J.
  Chem. Phys.}\ }\textbf {\bibinfo {volume} {76}},\ \bibinfo {pages} {1590}
  (\bibinfo {year} {1982})}\BibitemShut {NoStop}%
\bibitem [{\citenamefont {Gerlich}(1992)}]{advcp-82-1-GER}%
  \BibitemOpen
  \bibfield  {author} {\bibinfo {author} {\bibfnamefont {D.}~\bibnamefont
  {Gerlich}},\ }\href@noop {} {\bibfield  {journal} {\bibinfo  {journal} {Adv.
  Chem. Phys.}\ }\textbf {\bibinfo {volume} {82}},\ \bibinfo {pages} {1}
  (\bibinfo {year} {1992})}\BibitemShut {NoStop}%
\bibitem [{\citenamefont {Gerlich}(1990)}]{Gerlich_1990}%
  \BibitemOpen
  \bibfield  {author} {\bibinfo {author} {\bibfnamefont {D.}~\bibnamefont
  {Gerlich}},\ }\href@noop {} {\bibfield  {journal} {\bibinfo  {journal} {J.
  Chem. Phys.}\ }\textbf {\bibinfo {volume} {92}},\ \bibinfo {pages} {2377}
  (\bibinfo {year} {1990})}\BibitemShut {NoStop}%
\bibitem [{\citenamefont {Honvault}\ \emph {et~al.}(2012)\citenamefont
  {Honvault}, \citenamefont {Jorfi}, \citenamefont {Gonz\'alez-Lezana},
  \citenamefont {Faure},\ and\ \citenamefont {Pagani}}]{HONV_2011_Er}%
  \BibitemOpen
  \bibfield  {author} {\bibinfo {author} {\bibfnamefont {P.}~\bibnamefont
  {Honvault}}, \bibinfo {author} {\bibfnamefont {M.}~\bibnamefont {Jorfi}},
  \bibinfo {author} {\bibfnamefont {T.}~\bibnamefont {Gonz\'alez-Lezana}},
  \bibinfo {author} {\bibfnamefont {A.}~\bibnamefont {Faure}}, \ and\ \bibinfo
  {author} {\bibfnamefont {L.}~\bibnamefont {Pagani}},\ }\href@noop {}
  {\bibfield  {journal} {\bibinfo  {journal} {Phys. Rev. Lett.}\ }\textbf
  {\bibinfo {volume} {108}},\ \bibinfo {pages} {109903} (\bibinfo {year}
  {2012})}\BibitemShut {NoStop}%
\bibitem [{\citenamefont {Hazra}\ \emph {et~al.}(2024)\citenamefont {Hazra},
  \citenamefont {Naskar}, \citenamefont {Adhikari},\ and\ \citenamefont
  {Varandas}}]{ortho-para_2024}%
  \BibitemOpen
  \bibfield  {author} {\bibinfo {author} {\bibfnamefont {S.}~\bibnamefont
  {Hazra}}, \bibinfo {author} {\bibfnamefont {K.}~\bibnamefont {Naskar}},
  \bibinfo {author} {\bibfnamefont {S.}~\bibnamefont {Adhikari}}, \ and\
  \bibinfo {author} {\bibfnamefont {A.~J.~C.}\ \bibnamefont {Varandas}},\
  }\href {\doibase 10.1021/acs.jpca.4c02243} {\bibfield  {journal} {\bibinfo
  {journal} {The Journal of Physical Chemistry A}\ }\textbf {\bibinfo {volume}
  {128}},\ \bibinfo {pages} {8833} (\bibinfo {year} {2024})}\BibitemShut
  {NoStop}%
\bibitem [{\citenamefont {Ghosh}\ \emph
  {et~al.}(2017{\natexlab{a}})\citenamefont {Ghosh}, \citenamefont {Mukherjee},
  \citenamefont {Mukherjee}, \citenamefont {Mandal}, \citenamefont {Sharma},
  \citenamefont {Chaudhury},\ and\ \citenamefont
  {Adhikari}}]{jcp-147-074105-SA}%
  \BibitemOpen
  \bibfield  {author} {\bibinfo {author} {\bibfnamefont {S.}~\bibnamefont
  {Ghosh}}, \bibinfo {author} {\bibfnamefont {S.}~\bibnamefont {Mukherjee}},
  \bibinfo {author} {\bibfnamefont {B.}~\bibnamefont {Mukherjee}}, \bibinfo
  {author} {\bibfnamefont {S.}~\bibnamefont {Mandal}}, \bibinfo {author}
  {\bibfnamefont {R.}~\bibnamefont {Sharma}}, \bibinfo {author} {\bibfnamefont
  {P.}~\bibnamefont {Chaudhury}}, \ and\ \bibinfo {author} {\bibfnamefont
  {S.}~\bibnamefont {Adhikari}},\ }\href@noop {} {\bibfield  {journal}
  {\bibinfo  {journal} {J. Chem. Phys.}\ }\textbf {\bibinfo {volume} {147}},\
  \bibinfo {pages} {074105/1} (\bibinfo {year}
  {2017}{\natexlab{a}})}\BibitemShut {NoStop}%
\bibitem [{\citenamefont {Ghosh}\ \emph {et~al.}(2021)\citenamefont {Ghosh},
  \citenamefont {Sahoo}, \citenamefont {Baer},\ and\ \citenamefont
  {Adhikari}}]{H3+_dynamics_2021}%
  \BibitemOpen
  \bibfield  {author} {\bibinfo {author} {\bibfnamefont {S.}~\bibnamefont
  {Ghosh}}, \bibinfo {author} {\bibfnamefont {T.}~\bibnamefont {Sahoo}},
  \bibinfo {author} {\bibfnamefont {M.}~\bibnamefont {Baer}}, \ and\ \bibinfo
  {author} {\bibfnamefont {S.}~\bibnamefont {Adhikari}},\ }\href@noop {}
  {\bibfield  {journal} {\bibinfo  {journal} {J. Phys. Chem. A}\ }\textbf
  {\bibinfo {volume} {125}},\ \bibinfo {pages} {731} (\bibinfo {year}
  {2021})}\BibitemShut {NoStop}%
\bibitem [{\citenamefont {Klein}\ and\ \citenamefont
  {Friedman}(1964)}]{Klein1964_JCP_41}%
  \BibitemOpen
  \bibfield  {author} {\bibinfo {author} {\bibfnamefont {F.~S.}\ \bibnamefont
  {Klein}}\ and\ \bibinfo {author} {\bibfnamefont {L.}~\bibnamefont
  {Friedman}},\ }\href {\doibase 10.1063/1.1726159} {\bibfield  {journal}
  {\bibinfo  {journal} {J. Chem. Phys.}\ }\textbf {\bibinfo {volume} {41}},\
  \bibinfo {pages} {1789} (\bibinfo {year} {1964})}\BibitemShut {NoStop}%
\bibitem [{\citenamefont {Chupka}\ and\ \citenamefont
  {Russell}(1968)}]{chupka1968_1}%
  \BibitemOpen
  \bibfield  {author} {\bibinfo {author} {\bibfnamefont {W.~A.}\ \bibnamefont
  {Chupka}}\ and\ \bibinfo {author} {\bibfnamefont {M.~E.}\ \bibnamefont
  {Russell}},\ }\href {\doibase 10.1063/1.1670068} {\bibfield  {journal}
  {\bibinfo  {journal} {J. Chem. Phys.}\ }\textbf {\bibinfo {volume} {49}},\
  \bibinfo {pages} {5426} (\bibinfo {year} {1968})}\BibitemShut {NoStop}%
\bibitem [{\citenamefont {Chupka}, \citenamefont {Berkowitz},\ and\
  \citenamefont {Russell}(1969)}]{chupka1968_2}%
  \BibitemOpen
  \bibfield  {author} {\bibinfo {author} {\bibfnamefont {W.~A.}\ \bibnamefont
  {Chupka}}, \bibinfo {author} {\bibfnamefont {J.}~\bibnamefont {Berkowitz}}, \
  and\ \bibinfo {author} {\bibfnamefont {M.}~\bibnamefont {Russell}},\
  }\href@noop {} {\bibfield  {journal} {\bibinfo  {journal} {6th Int. Conf. on
  the Physics of Electronic and Atomic Collisions: ICPEAC VI}\ ,\ \bibinfo
  {pages} {71}} (\bibinfo {year} {1969})}\BibitemShut {NoStop}%
\bibitem [{\citenamefont {Tang}\ \emph
  {et~al.}(2005{\natexlab{a}})\citenamefont {Tang}, \citenamefont {Xu},
  \citenamefont {Zhang}, \citenamefont {Hou}, \citenamefont {Chang},
  \citenamefont {Ng}, \citenamefont {Chiu}, \citenamefont {Dressler},\ and\
  \citenamefont {Levandier}}]{Tang2005_JCP_122}%
  \BibitemOpen
  \bibfield  {author} {\bibinfo {author} {\bibfnamefont {X.~N.}\ \bibnamefont
  {Tang}}, \bibinfo {author} {\bibfnamefont {H.}~\bibnamefont {Xu}}, \bibinfo
  {author} {\bibfnamefont {T.}~\bibnamefont {Zhang}}, \bibinfo {author}
  {\bibfnamefont {Y.}~\bibnamefont {Hou}}, \bibinfo {author} {\bibfnamefont
  {C.}~\bibnamefont {Chang}}, \bibinfo {author} {\bibfnamefont {C.~Y.}\
  \bibnamefont {Ng}}, \bibinfo {author} {\bibfnamefont {Y.}~\bibnamefont
  {Chiu}}, \bibinfo {author} {\bibfnamefont {R.~A.}\ \bibnamefont {Dressler}},
  \ and\ \bibinfo {author} {\bibfnamefont {D.~J.}\ \bibnamefont {Levandier}},\
  }\href {\doibase 10.1063/1.1883169} {\bibfield  {journal} {\bibinfo
  {journal} {J. Chem. Phys.}\ }\textbf {\bibinfo {volume} {122}},\ \bibinfo
  {pages} {164301} (\bibinfo {year} {2005}{\natexlab{a}})}\BibitemShut
  {NoStop}%
\bibitem [{\citenamefont {Turner}, \citenamefont {Dutuit},\ and\ \citenamefont
  {Lee}()}]{Turner2022_JCP_81}%
  \BibitemOpen
  \bibfield  {author} {\bibinfo {author} {\bibfnamefont {T.}~\bibnamefont
  {Turner}}, \bibinfo {author} {\bibfnamefont {O.}~\bibnamefont {Dutuit}}, \
  and\ \bibinfo {author} {\bibfnamefont {Y.~T.}\ \bibnamefont {Lee}},\
  }\href@noop {} {\ }\BibitemShut {NoStop}%
\bibitem [{\citenamefont {Maiti}\ \emph {et~al.}(2002)\citenamefont {Maiti},
  \citenamefont {Kalyanaraman}, \citenamefont {Panda},\ and\ \citenamefont
  {Sathyamurthy}}]{Maiti2002_JCP_117}%
  \BibitemOpen
  \bibfield  {author} {\bibinfo {author} {\bibfnamefont {B.}~\bibnamefont
  {Maiti}}, \bibinfo {author} {\bibfnamefont {C.}~\bibnamefont {Kalyanaraman}},
  \bibinfo {author} {\bibfnamefont {A.~N.}\ \bibnamefont {Panda}}, \ and\
  \bibinfo {author} {\bibfnamefont {N.}~\bibnamefont {Sathyamurthy}},\ }\href
  {\doibase 10.1063/1.1518002} {\bibfield  {journal} {\bibinfo  {journal} {J.
  Chem. Phys.}\ }\textbf {\bibinfo {volume} {117}},\ \bibinfo {pages} {9719}
  (\bibinfo {year} {2002})}\BibitemShut {NoStop}%
\bibitem [{\citenamefont {Kraemer}, \citenamefont {\v{S}pirko},\ and\
  \citenamefont {Bludsk\'{y}}(2002)}]{Kraemer2002_CP_276}%
  \BibitemOpen
  \bibfield  {author} {\bibinfo {author} {\bibfnamefont {W.~P.}\ \bibnamefont
  {Kraemer}}, \bibinfo {author} {\bibfnamefont {V.}~\bibnamefont {\v{S}pirko}},
  \ and\ \bibinfo {author} {\bibfnamefont {O.}~\bibnamefont {Bludsk\'{y}}},\
  }\href {\doibase 10.1016/S0301-0104(01)00694-2} {\bibfield  {journal}
  {\bibinfo  {journal} {Chem. Phys.}\ }\textbf {\bibinfo {volume} {276}},\
  \bibinfo {pages} {225} (\bibinfo {year} {2002})}\BibitemShut {NoStop}%
\bibitem [{\citenamefont {Mruga\l{}a}, \citenamefont {\v{S}pirko},\ and\
  \citenamefont {Kraemer}(2003)}]{Mrugala2003_JCP_118}%
  \BibitemOpen
  \bibfield  {author} {\bibinfo {author} {\bibfnamefont {F.}~\bibnamefont
  {Mruga\l{}a}}, \bibinfo {author} {\bibfnamefont {V.}~\bibnamefont
  {\v{S}pirko}}, \ and\ \bibinfo {author} {\bibfnamefont {W.~P.}\ \bibnamefont
  {Kraemer}},\ }\href {\doibase 10.1063/1.1573184} {\bibfield  {journal}
  {\bibinfo  {journal} {J. Chem. Phys.}\ }\textbf {\bibinfo {volume} {118}},\
  \bibinfo {pages} {10547} (\bibinfo {year} {2003})}\BibitemShut {NoStop}%
\bibitem [{\citenamefont {Tiwari}\ and\ \citenamefont
  {Sathyamurthy}(2005)}]{Tiwari2005_CPL_414}%
  \BibitemOpen
  \bibfield  {author} {\bibinfo {author} {\bibfnamefont {A.~K.}\ \bibnamefont
  {Tiwari}}\ and\ \bibinfo {author} {\bibfnamefont {N.}~\bibnamefont
  {Sathyamurthy}},\ }\href {\doibase 10.1016/j.cplett.2005.08.130} {\bibfield
  {journal} {\bibinfo  {journal} {Chem. Phys. Lett.}\ }\textbf {\bibinfo
  {volume} {414}},\ \bibinfo {pages} {509} (\bibinfo {year}
  {2005})}\BibitemShut {NoStop}%
\bibitem [{\citenamefont {Panda}\ and\ \citenamefont
  {Sathyamurthy}(2005)}]{Panda2005_JCP_122}%
  \BibitemOpen
  \bibfield  {author} {\bibinfo {author} {\bibfnamefont {A.~N.}\ \bibnamefont
  {Panda}}\ and\ \bibinfo {author} {\bibfnamefont {N.}~\bibnamefont
  {Sathyamurthy}},\ }\href {\doibase 10.1063/1.1839866} {\bibfield  {journal}
  {\bibinfo  {journal} {J. Chem. Phys.}\ }\textbf {\bibinfo {volume} {122}},\
  \bibinfo {pages} {054304} (\bibinfo {year} {2005})}\BibitemShut {NoStop}%
\bibitem [{\citenamefont {Panda}\ and\ \citenamefont
  {Althorpe}(2006)}]{Panda2006_CPL_419}%
  \BibitemOpen
  \bibfield  {author} {\bibinfo {author} {\bibfnamefont {A.~N.}\ \bibnamefont
  {Panda}}\ and\ \bibinfo {author} {\bibfnamefont {S.~C.}\ \bibnamefont
  {Althorpe}},\ }\href {\doibase 10.1016/j.cplett.2005.11.093} {\bibfield
  {journal} {\bibinfo  {journal} {Chem. Phys. Lett.}\ }\textbf {\bibinfo
  {volume} {419}},\ \bibinfo {pages} {245} (\bibinfo {year}
  {2006})}\BibitemShut {NoStop}%
\bibitem [{\citenamefont {Tang}\ \emph {et~al.}(2007)\citenamefont {Tang},
  \citenamefont {Houchins}, \citenamefont {Lau}, \citenamefont {Ng},
  \citenamefont {Dressler}, \citenamefont {Chiu}, \citenamefont {Chu},\ and\
  \citenamefont {Han}}]{Tang2007_JCP_127}%
  \BibitemOpen
  \bibfield  {author} {\bibinfo {author} {\bibfnamefont {X.}~\bibnamefont
  {Tang}}, \bibinfo {author} {\bibfnamefont {C.}~\bibnamefont {Houchins}},
  \bibinfo {author} {\bibfnamefont {K.-C.}\ \bibnamefont {Lau}}, \bibinfo
  {author} {\bibfnamefont {C.~Y.}\ \bibnamefont {Ng}}, \bibinfo {author}
  {\bibfnamefont {R.~A.}\ \bibnamefont {Dressler}}, \bibinfo {author}
  {\bibfnamefont {Y.-H.}\ \bibnamefont {Chiu}}, \bibinfo {author}
  {\bibfnamefont {T.-S.}\ \bibnamefont {Chu}}, \ and\ \bibinfo {author}
  {\bibfnamefont {K.-L.}\ \bibnamefont {Han}},\ }\href {\doibase
  10.1063/1.2800009} {\bibfield  {journal} {\bibinfo  {journal} {J. Chem.
  Phys.}\ }\textbf {\bibinfo {volume} {127}},\ \bibinfo {pages} {164318}
  (\bibinfo {year} {2007})}\BibitemShut {NoStop}%
\bibitem [{\citenamefont {Chupka}(1972)}]{chupka1972}%
  \BibitemOpen
  \bibfield  {author} {\bibinfo {author} {\bibfnamefont {W.~A.}\ \bibnamefont
  {Chupka}},\ }\href@noop {} {\bibfield  {journal} {\bibinfo  {journal}
  {Ion-Molecule Reactions, vol.1 Ed. J. L. Franklin, Chap. 3, Plenum, New York,
  p.}\ ,\ \bibinfo {pages} {72}} (\bibinfo {year} {1972})}\BibitemShut
  {NoStop}%
\bibitem [{\citenamefont {Schneider}\ \emph {et~al.}(1976)\citenamefont
  {Schneider}, \citenamefont {Havemann}, \citenamefont {Zülicke},
  \citenamefont {Pac{\'{a}}k}, \citenamefont {Birkinshaw},\ and\ \citenamefont
  {Herman}}]{CPL_37_1976_Schneider}%
  \BibitemOpen
  \bibfield  {author} {\bibinfo {author} {\bibfnamefont {F.}~\bibnamefont
  {Schneider}}, \bibinfo {author} {\bibfnamefont {U.}~\bibnamefont {Havemann}},
  \bibinfo {author} {\bibfnamefont {L.}~\bibnamefont {Zülicke}}, \bibinfo
  {author} {\bibfnamefont {V.}~\bibnamefont {Pac{\'{a}}k}}, \bibinfo {author}
  {\bibfnamefont {K.}~\bibnamefont {Birkinshaw}}, \ and\ \bibinfo {author}
  {\bibfnamefont {Z.}~\bibnamefont {Herman}},\ }\href {\doibase
  10.1016/0009-2614(76)80225-4} {\bibfield  {journal} {\bibinfo  {journal}
  {Chem. Phys. Lett.}\ }\textbf {\bibinfo {volume} {37}},\ \bibinfo {pages}
  {323} (\bibinfo {year} {1976})}\BibitemShut {NoStop}%
\bibitem [{\citenamefont {Pac{\'{a}}k}\ \emph {et~al.}(1977)\citenamefont
  {Pac{\'{a}}k}, \citenamefont {Havemann}, \citenamefont {Herman},
  \citenamefont {Schneider},\ and\ \citenamefont
  {Zülicke}}]{CPL_49_1977_Pacak}%
  \BibitemOpen
  \bibfield  {author} {\bibinfo {author} {\bibfnamefont {V.}~\bibnamefont
  {Pac{\'{a}}k}}, \bibinfo {author} {\bibfnamefont {U.}~\bibnamefont
  {Havemann}}, \bibinfo {author} {\bibfnamefont {Z.}~\bibnamefont {Herman}},
  \bibinfo {author} {\bibfnamefont {F.}~\bibnamefont {Schneider}}, \ and\
  \bibinfo {author} {\bibfnamefont {L.}~\bibnamefont {Zülicke}},\ }\href
  {\doibase 10.1016/0009-2614(77)80585-x} {\bibfield  {journal} {\bibinfo
  {journal} {Chem. Phys. Lett.}\ }\textbf {\bibinfo {volume} {49}},\ \bibinfo
  {pages} {273} (\bibinfo {year} {1977})}\BibitemShut {NoStop}%
\bibitem [{\citenamefont {Pollard}\ \emph {et~al.}(1991)\citenamefont
  {Pollard}, \citenamefont {Syage}, \citenamefont {Johnson},\ and\
  \citenamefont {Cohen}}]{JCP_94_1991_Pollard}%
  \BibitemOpen
  \bibfield  {author} {\bibinfo {author} {\bibfnamefont {J.~E.}\ \bibnamefont
  {Pollard}}, \bibinfo {author} {\bibfnamefont {J.~A.}\ \bibnamefont {Syage}},
  \bibinfo {author} {\bibfnamefont {L.~K.}\ \bibnamefont {Johnson}}, \ and\
  \bibinfo {author} {\bibfnamefont {R.~B.}\ \bibnamefont {Cohen}},\ }\href
  {\doibase 10.1063/1.460718} {\bibfield  {journal} {\bibinfo  {journal} {J.
  Chem. Phys.}\ }\textbf {\bibinfo {volume} {94}},\ \bibinfo {pages} {8615}
  (\bibinfo {year} {1991})}\BibitemShut {NoStop}%
\bibitem [{\citenamefont {Pollard}, \citenamefont {Johnson},\ and\
  \citenamefont {Cohen}(1991)}]{JCP_95_1991_Pollard}%
  \BibitemOpen
  \bibfield  {author} {\bibinfo {author} {\bibfnamefont {J.~E.}\ \bibnamefont
  {Pollard}}, \bibinfo {author} {\bibfnamefont {L.~K.}\ \bibnamefont
  {Johnson}}, \ and\ \bibinfo {author} {\bibfnamefont {R.~B.}\ \bibnamefont
  {Cohen}},\ }\href {\doibase 10.1063/1.461705} {\bibfield  {journal} {\bibinfo
   {journal} {J. Chem. Phys.}\ }\textbf {\bibinfo {volume} {95}},\ \bibinfo
  {pages} {4894} (\bibinfo {year} {1991})}\BibitemShut {NoStop}%
\bibitem [{\citenamefont {Pijkeren}, \citenamefont {Eck},\ and\ \citenamefont
  {Niehaus}(1983)}]{CPL_96_1983_Pijkeren}%
  \BibitemOpen
  \bibfield  {author} {\bibinfo {author} {\bibfnamefont {D.~V.}\ \bibnamefont
  {Pijkeren}}, \bibinfo {author} {\bibfnamefont {J.~V.}\ \bibnamefont {Eck}}, \
  and\ \bibinfo {author} {\bibfnamefont {A.}~\bibnamefont {Niehaus}},\ }\href
  {\doibase 10.1016/0009-2614(83)80108-0} {\bibfield  {journal} {\bibinfo
  {journal} {Chem. Phys. Lett.}\ }\textbf {\bibinfo {volume} {96}},\ \bibinfo
  {pages} {20} (\bibinfo {year} {1983})}\BibitemShut {NoStop}%
\bibitem [{\citenamefont {Herman}\ and\ \citenamefont
  {Koyano}(1987)}]{JCSFT2_83_1987_Herman}%
  \BibitemOpen
  \bibfield  {author} {\bibinfo {author} {\bibfnamefont {Z.}~\bibnamefont
  {Herman}}\ and\ \bibinfo {author} {\bibfnamefont {I.}~\bibnamefont
  {Koyano}},\ }\href {\doibase 10.1039/f29878300127} {\bibfield  {journal}
  {\bibinfo  {journal} {J. Chem. Soc. Faraday Trans. 2}\ }\textbf {\bibinfo
  {volume} {83}},\ \bibinfo {pages} {127} (\bibinfo {year} {1987})}\BibitemShut
  {NoStop}%
\bibitem [{\citenamefont {Baer}\ and\ \citenamefont
  {Nakamura}(1987)}]{Baer1987_JPC_91}%
  \BibitemOpen
  \bibfield  {author} {\bibinfo {author} {\bibfnamefont {M.}~\bibnamefont
  {Baer}}\ and\ \bibinfo {author} {\bibfnamefont {H.}~\bibnamefont
  {Nakamura}},\ }\href {\doibase 10.1021/j100305a024} {\bibfield  {journal}
  {\bibinfo  {journal} {J. Phys. Chem.}\ }\textbf {\bibinfo {volume} {91}},\
  \bibinfo {pages} {5503} (\bibinfo {year} {1987})}\BibitemShut {NoStop}%
\bibitem [{\citenamefont {Aquilanti}\ \emph {et~al.}(2000)\citenamefont
  {Aquilanti}, \citenamefont {Capecchi}, \citenamefont {Cavalli}, \citenamefont
  {De~Fazio}, \citenamefont {Palmieri}, \citenamefont {Puzzarini},
  \citenamefont {Aguilar}, \citenamefont {Gim\'{e}nez},\ and\ \citenamefont
  {Lucas}}]{Aquilanti2000_CPL_318}%
  \BibitemOpen
  \bibfield  {author} {\bibinfo {author} {\bibfnamefont {V.}~\bibnamefont
  {Aquilanti}}, \bibinfo {author} {\bibfnamefont {G.}~\bibnamefont {Capecchi}},
  \bibinfo {author} {\bibfnamefont {S.}~\bibnamefont {Cavalli}}, \bibinfo
  {author} {\bibfnamefont {D.}~\bibnamefont {De~Fazio}}, \bibinfo {author}
  {\bibfnamefont {P.}~\bibnamefont {Palmieri}}, \bibinfo {author}
  {\bibfnamefont {C.}~\bibnamefont {Puzzarini}}, \bibinfo {author}
  {\bibfnamefont {A.}~\bibnamefont {Aguilar}}, \bibinfo {author} {\bibfnamefont
  {X.}~\bibnamefont {Gim\'{e}nez}}, \ and\ \bibinfo {author} {\bibfnamefont
  {J.~M.}\ \bibnamefont {Lucas}},\ }\href {\doibase
  10.1016/S0009-2614(00)00067-1} {\bibfield  {journal} {\bibinfo  {journal}
  {Chem. Phys. Lett.}\ }\textbf {\bibinfo {volume} {318}},\ \bibinfo {pages}
  {619} (\bibinfo {year} {2000})}\BibitemShut {NoStop}%
\bibitem [{\citenamefont {Xu}\ \emph {et~al.}(2008)\citenamefont {Xu},
  \citenamefont {Liu}, \citenamefont {Luan}, \citenamefont {Zhang},\ and\
  \citenamefont {Zhang}}]{Xu2008_CPL_464}%
  \BibitemOpen
  \bibfield  {author} {\bibinfo {author} {\bibfnamefont {W.}~\bibnamefont
  {Xu}}, \bibinfo {author} {\bibfnamefont {X.}~\bibnamefont {Liu}}, \bibinfo
  {author} {\bibfnamefont {S.}~\bibnamefont {Luan}}, \bibinfo {author}
  {\bibfnamefont {Q.}~\bibnamefont {Zhang}}, \ and\ \bibinfo {author}
  {\bibfnamefont {P.}~\bibnamefont {Zhang}},\ }\href {\doibase
  10.1016/j.cplett.2008.08.011} {\bibfield  {journal} {\bibinfo  {journal}
  {Chem. Phys. Lett.}\ }\textbf {\bibinfo {volume} {464}},\ \bibinfo {pages}
  {92} (\bibinfo {year} {2008})}\BibitemShut {NoStop}%
\bibitem [{\citenamefont {De~Fazio}\ \emph {et~al.}(2012)\citenamefont
  {De~Fazio}, \citenamefont {de~Castro-Vitores}, \citenamefont {Aguado},
  \citenamefont {Aquilanti},\ and\ \citenamefont
  {Cavalli}}]{JCP_137_2012_Fazio}%
  \BibitemOpen
  \bibfield  {author} {\bibinfo {author} {\bibfnamefont {D.}~\bibnamefont
  {De~Fazio}}, \bibinfo {author} {\bibfnamefont {M.}~\bibnamefont
  {de~Castro-Vitores}}, \bibinfo {author} {\bibfnamefont {A.}~\bibnamefont
  {Aguado}}, \bibinfo {author} {\bibfnamefont {V.}~\bibnamefont {Aquilanti}}, \
  and\ \bibinfo {author} {\bibfnamefont {S.}~\bibnamefont {Cavalli}},\ }\href
  {\doibase 10.1063/1.4772651} {\bibfield  {journal} {\bibinfo  {journal} {J.
  Chem. Phys.}\ }\textbf {\bibinfo {volume} {137}},\ \bibinfo {pages} {244306}
  (\bibinfo {year} {2012})}\BibitemShut {NoStop}%
\bibitem [{\citenamefont {Kumar}, \citenamefont {Sathyamurthy},\ and\
  \citenamefont {Bhalla}(1993)}]{JCP_98_1993_Kumar}%
  \BibitemOpen
  \bibfield  {author} {\bibinfo {author} {\bibfnamefont {S.}~\bibnamefont
  {Kumar}}, \bibinfo {author} {\bibfnamefont {N.}~\bibnamefont {Sathyamurthy}},
  \ and\ \bibinfo {author} {\bibfnamefont {K.~C.}\ \bibnamefont {Bhalla}},\
  }\href {\doibase 10.1063/1.464972} {\bibfield  {journal} {\bibinfo  {journal}
  {J. Chem. Phys.}\ }\textbf {\bibinfo {volume} {98}},\ \bibinfo {pages} {4680}
  (\bibinfo {year} {1993})}\BibitemShut {NoStop}%
\bibitem [{\citenamefont {Turner}, \citenamefont {Dutuit},\ and\ \citenamefont
  {Lee}(1984{\natexlab{a}})}]{JCP_81_1984_Turner}%
  \BibitemOpen
  \bibfield  {author} {\bibinfo {author} {\bibfnamefont {T.}~\bibnamefont
  {Turner}}, \bibinfo {author} {\bibfnamefont {O.}~\bibnamefont {Dutuit}}, \
  and\ \bibinfo {author} {\bibfnamefont {Y.~T.}\ \bibnamefont {Lee}},\ }\href
  {\doibase 10.1063/1.448073} {\bibfield  {journal} {\bibinfo  {journal} {J.
  Chem. Phys.}\ }\textbf {\bibinfo {volume} {81}},\ \bibinfo {pages} {3475}
  (\bibinfo {year} {1984}{\natexlab{a}})}\BibitemShut {NoStop}%
\bibitem [{\citenamefont {Achtenhagen}, \citenamefont {Schweizer},\ and\
  \citenamefont {Gerlich}(1993)}]{Achtenhagen}%
  \BibitemOpen
  \bibfield  {author} {\bibinfo {author} {\bibfnamefont {M.}~\bibnamefont
  {Achtenhagen}}, \bibinfo {author} {\bibfnamefont {M.}~\bibnamefont
  {Schweizer}}, \ and\ \bibinfo {author} {\bibfnamefont {D.}~\bibnamefont
  {Gerlich}},\ }\href {\doibase 10.1063/1.464972} {\enquote {\bibinfo {title}
  {{He,H$_{2}^{+}$} dynamics revisited},}\ }\bibinfo {howpublished} {private
  communication reported by S. Kumar, N. Sathyamurthy, and K. C. Bhalla}
  (\bibinfo {year} {1993})\BibitemShut {NoStop}%
\bibitem [{\citenamefont {Naskar}\ \emph
  {et~al.}(2023{\natexlab{a}})\citenamefont {Naskar}, \citenamefont {Ravi},
  \citenamefont {Adhikari}, \citenamefont {Baer},\ and\ \citenamefont
  {Sathyamurthy}}]{HeH2_Surf}%
  \BibitemOpen
  \bibfield  {author} {\bibinfo {author} {\bibfnamefont {K.}~\bibnamefont
  {Naskar}}, \bibinfo {author} {\bibfnamefont {S.}~\bibnamefont {Ravi}},
  \bibinfo {author} {\bibfnamefont {S.}~\bibnamefont {Adhikari}}, \bibinfo
  {author} {\bibfnamefont {M.}~\bibnamefont {Baer}}, \ and\ \bibinfo {author}
  {\bibfnamefont {N.}~\bibnamefont {Sathyamurthy}},\ }\href@noop {} {\bibfield
  {journal} {\bibinfo  {journal} {J. Phys. Chem. A.}\ }\textbf {\bibinfo
  {volume} {127}},\ \bibinfo {pages} {3832} (\bibinfo {year}
  {2023}{\natexlab{a}})}\BibitemShut {NoStop}%
\bibitem [{\citenamefont {Naskar}\ \emph
  {et~al.}(2023{\natexlab{b}})\citenamefont {Naskar}, \citenamefont {Ghosh},
  \citenamefont {Adhikari}, \citenamefont {Baer},\ and\ \citenamefont
  {Sathyamurthy}}]{HeH2+_dynamics_2023}%
  \BibitemOpen
  \bibfield  {author} {\bibinfo {author} {\bibfnamefont {K.}~\bibnamefont
  {Naskar}}, \bibinfo {author} {\bibfnamefont {S.}~\bibnamefont {Ghosh}},
  \bibinfo {author} {\bibfnamefont {S.}~\bibnamefont {Adhikari}}, \bibinfo
  {author} {\bibfnamefont {M.}~\bibnamefont {Baer}}, \ and\ \bibinfo {author}
  {\bibfnamefont {N.}~\bibnamefont {Sathyamurthy}},\ }\href@noop {} {\bibfield
  {journal} {\bibinfo  {journal} {J. Chem. Phys.}\ }\textbf {\bibinfo {volume}
  {159}},\ \bibinfo {pages} {034302} (\bibinfo {year}
  {2023}{\natexlab{b}})}\BibitemShut {NoStop}%
\bibitem [{\citenamefont {Mondal}\ and\ \citenamefont
  {Mukhopadhyay}(2021)}]{IJQC_121_2021_Mondal}%
  \BibitemOpen
  \bibfield  {author} {\bibinfo {author} {\bibfnamefont {R.}~\bibnamefont
  {Mondal}}\ and\ \bibinfo {author} {\bibfnamefont {D.}~\bibnamefont
  {Mukhopadhyay}},\ }\href {\doibase 10.1002/qua.26552} {\bibfield  {journal}
  {\bibinfo  {journal} {Int. J. Quantum Chem.}\ }\textbf {\bibinfo {volume}
  {121}},\ \bibinfo {pages} {e26552} (\bibinfo {year} {2021})}\BibitemShut
  {NoStop}%
\bibitem [{\citenamefont {Gupta}\ \emph {et~al.}(2020)\citenamefont {Gupta},
  \citenamefont {Dhindhwal}, \citenamefont {Baer}, \citenamefont
  {Sathyamurthy}, \citenamefont {Ravi}, \citenamefont {Mukherjee},
  \citenamefont {Mukherjee},\ and\ \citenamefont
  {Adhikari}}]{MolPhys_118_2020_Gupta}%
  \BibitemOpen
  \bibfield  {author} {\bibinfo {author} {\bibfnamefont {A.~K.}\ \bibnamefont
  {Gupta}}, \bibinfo {author} {\bibfnamefont {V.}~\bibnamefont {Dhindhwal}},
  \bibinfo {author} {\bibfnamefont {M.}~\bibnamefont {Baer}}, \bibinfo {author}
  {\bibfnamefont {N.}~\bibnamefont {Sathyamurthy}}, \bibinfo {author}
  {\bibfnamefont {S.}~\bibnamefont {Ravi}}, \bibinfo {author} {\bibfnamefont
  {S.}~\bibnamefont {Mukherjee}}, \bibinfo {author} {\bibfnamefont
  {B.}~\bibnamefont {Mukherjee}}, \ and\ \bibinfo {author} {\bibfnamefont
  {S.}~\bibnamefont {Adhikari}},\ }\href {\doibase
  10.1080/00268976.2019.1683243} {\bibfield  {journal} {\bibinfo  {journal}
  {Mol. Phys.}\ }\textbf {\bibinfo {volume} {118}},\ \bibinfo {pages}
  {e1683243} (\bibinfo {year} {2020})}\BibitemShut {NoStop}%
\bibitem [{\citenamefont {Ravi}\ \emph
  {et~al.}(2021{\natexlab{a}})\citenamefont {Ravi}, \citenamefont {Mukherjee},
  \citenamefont {Mukherjee}, \citenamefont {Adhikari}, \citenamefont
  {Sathyamurthy},\ and\ \citenamefont {Baer}}]{MolPhys_119_2021_ravi}%
  \BibitemOpen
  \bibfield  {author} {\bibinfo {author} {\bibfnamefont {S.}~\bibnamefont
  {Ravi}}, \bibinfo {author} {\bibfnamefont {S.}~\bibnamefont {Mukherjee}},
  \bibinfo {author} {\bibfnamefont {B.}~\bibnamefont {Mukherjee}}, \bibinfo
  {author} {\bibfnamefont {S.}~\bibnamefont {Adhikari}}, \bibinfo {author}
  {\bibfnamefont {N.}~\bibnamefont {Sathyamurthy}}, \ and\ \bibinfo {author}
  {\bibfnamefont {M.}~\bibnamefont {Baer}},\ }\href {\doibase
  10.1080/00268976.2020.1811907} {\bibfield  {journal} {\bibinfo  {journal}
  {Mol. Phys.}\ }\textbf {\bibinfo {volume} {119}},\ \bibinfo {pages} {1}
  (\bibinfo {year} {2021}{\natexlab{a}})}\BibitemShut {NoStop}%
\bibitem [{\citenamefont {Kuppermann}\ and\ \citenamefont
  {Schatz}(1975)}]{Kuppermann_1975}%
  \BibitemOpen
  \bibfield  {author} {\bibinfo {author} {\bibfnamefont {A.}~\bibnamefont
  {Kuppermann}}\ and\ \bibinfo {author} {\bibfnamefont {G.~C.}\ \bibnamefont
  {Schatz}},\ }\href@noop {} {\bibfield  {journal} {\bibinfo  {journal} {The
  Journal of Chemical Physics}\ }\textbf {\bibinfo {volume} {62}},\ \bibinfo
  {pages} {2502} (\bibinfo {year} {1975})}\BibitemShut {NoStop}%
\bibitem [{\citenamefont {Walker}, \citenamefont {Stechel},\ and\ \citenamefont
  {Light}(1978)}]{Walker_1978}%
  \BibitemOpen
  \bibfield  {author} {\bibinfo {author} {\bibfnamefont {R.~B.}\ \bibnamefont
  {Walker}}, \bibinfo {author} {\bibfnamefont {E.~B.}\ \bibnamefont {Stechel}},
  \ and\ \bibinfo {author} {\bibfnamefont {J.~C.}\ \bibnamefont {Light}},\
  }\href@noop {} {\bibfield  {journal} {\bibinfo  {journal} {Journal of
  Chemical Physics}\ }\textbf {\bibinfo {volume} {69}},\ \bibinfo {pages}
  {2922} (\bibinfo {year} {1978})}\BibitemShut {NoStop}%
\bibitem [{\citenamefont {Charutz}, \citenamefont {Last},\ and\ \citenamefont
  {Baer}(1997)}]{Charutz_Baer_1997}%
  \BibitemOpen
  \bibfield  {author} {\bibinfo {author} {\bibfnamefont {D.~M.}\ \bibnamefont
  {Charutz}}, \bibinfo {author} {\bibfnamefont {I.}~\bibnamefont {Last}}, \
  and\ \bibinfo {author} {\bibfnamefont {M.}~\bibnamefont {Baer}},\ }\href@noop
  {} {\bibfield  {journal} {\bibinfo  {journal} {J. Chem. Phys.}\ }\textbf
  {\bibinfo {volume} {106}},\ \bibinfo {pages} {7654} (\bibinfo {year}
  {1997})}\BibitemShut {NoStop}%
\bibitem [{\citenamefont {Kendrick}(2001)}]{Kendrick_2001}%
  \BibitemOpen
  \bibfield  {author} {\bibinfo {author} {\bibfnamefont {B.~K.}\ \bibnamefont
  {Kendrick}},\ }\href@noop {} {\bibfield  {journal} {\bibinfo  {journal} {J.
  Chem. Phys.}\ }\textbf {\bibinfo {volume} {114}},\ \bibinfo {pages} {8796}
  (\bibinfo {year} {2001})}\BibitemShut {NoStop}%
\bibitem [{\citenamefont {Aoiz}, \citenamefont {BaÑares},\ and\ \citenamefont
  {Herrero}(2005)}]{Aoiz_2005}%
  \BibitemOpen
  \bibfield  {author} {\bibinfo {author} {\bibfnamefont {F.~J.}\ \bibnamefont
  {Aoiz}}, \bibinfo {author} {\bibfnamefont {L.}~\bibnamefont {BaÑares}}, \
  and\ \bibinfo {author} {\bibfnamefont {V.~J.}\ \bibnamefont {Herrero}},\
  }\href@noop {} {\bibfield  {journal} {\bibinfo  {journal} {Int. Rev. Phys.
  Chem.}\ }\textbf {\bibinfo {volume} {24}},\ \bibinfo {pages} {119} (\bibinfo
  {year} {2005})}\BibitemShut {NoStop}%
\bibitem [{\citenamefont {Adelman}\ \emph {et~al.}(1992)\citenamefont
  {Adelman}, \citenamefont {Shafer}, \citenamefont {Kliner},\ and\
  \citenamefont {Zare}}]{Zare_92}%
  \BibitemOpen
  \bibfield  {author} {\bibinfo {author} {\bibfnamefont {D.~E.}\ \bibnamefont
  {Adelman}}, \bibinfo {author} {\bibfnamefont {N.~E.}\ \bibnamefont {Shafer}},
  \bibinfo {author} {\bibfnamefont {D.~A.~V.}\ \bibnamefont {Kliner}}, \ and\
  \bibinfo {author} {\bibfnamefont {R.~N.}\ \bibnamefont {Zare}},\ }\href@noop
  {} {\bibfield  {journal} {\bibinfo  {journal} {J. Chem. Phys.}\ }\textbf
  {\bibinfo {volume} {97}},\ \bibinfo {pages} {7323} (\bibinfo {year}
  {1992})}\BibitemShut {NoStop}%
\bibitem [{\citenamefont {Nieh}\ and\ \citenamefont
  {Valentini}(1988)}]{Nieh_Valentini}%
  \BibitemOpen
  \bibfield  {author} {\bibinfo {author} {\bibfnamefont {J.-C.}\ \bibnamefont
  {Nieh}}\ and\ \bibinfo {author} {\bibfnamefont {J.~J.}\ \bibnamefont
  {Valentini}},\ }\href@noop {} {\bibfield  {journal} {\bibinfo  {journal}
  {Phys. Rev. Lett.}\ }\textbf {\bibinfo {volume} {60}},\ \bibinfo {pages}
  {519} (\bibinfo {year} {1988})}\BibitemShut {NoStop}%
\bibitem [{\citenamefont {Continetti}, \citenamefont {Balko},\ and\
  \citenamefont {Lee}(1990)}]{Y.T_Lee_90}%
  \BibitemOpen
  \bibfield  {author} {\bibinfo {author} {\bibfnamefont {R.~E.}\ \bibnamefont
  {Continetti}}, \bibinfo {author} {\bibfnamefont {B.~A.}\ \bibnamefont
  {Balko}}, \ and\ \bibinfo {author} {\bibfnamefont {Y.~T.}\ \bibnamefont
  {Lee}},\ }\href@noop {} {\bibfield  {journal} {\bibinfo  {journal} {J. Chem.
  Phys.}\ }\textbf {\bibinfo {volume} {93}},\ \bibinfo {pages} {5719} (\bibinfo
  {year} {1990})}\BibitemShut {NoStop}%
\bibitem [{\citenamefont {Ba{\~n}ares}\ \emph {et~al.}(1998)\citenamefont
  {Ba{\~n}ares}, \citenamefont {Aoiz}, \citenamefont {Herrero}, \citenamefont
  {D'Mello}, \citenamefont {Niederjohann}, \citenamefont {Seekamp-Rahn},
  \citenamefont {Wrede},\ and\ \citenamefont
  {Schnieder}}]{Banares_Exp_Theo_1998}%
  \BibitemOpen
  \bibfield  {author} {\bibinfo {author} {\bibfnamefont {L.}~\bibnamefont
  {Ba{\~n}ares}}, \bibinfo {author} {\bibfnamefont {F.~J.}\ \bibnamefont
  {Aoiz}}, \bibinfo {author} {\bibfnamefont {V.~J.}\ \bibnamefont {Herrero}},
  \bibinfo {author} {\bibfnamefont {M.~J.}\ \bibnamefont {D'Mello}}, \bibinfo
  {author} {\bibfnamefont {B.}~\bibnamefont {Niederjohann}}, \bibinfo {author}
  {\bibfnamefont {K.}~\bibnamefont {Seekamp-Rahn}}, \bibinfo {author}
  {\bibfnamefont {E.}~\bibnamefont {Wrede}}, \ and\ \bibinfo {author}
  {\bibfnamefont {L.}~\bibnamefont {Schnieder}},\ }\href@noop {} {\bibfield
  {journal} {\bibinfo  {journal} {J. Chem. Phys.}\ }\textbf {\bibinfo {volume}
  {108}},\ \bibinfo {pages} {6160} (\bibinfo {year} {1998})}\BibitemShut
  {NoStop}%
\bibitem [{\citenamefont {Bean}\ \emph {et~al.}(2002)\citenamefont {Bean},
  \citenamefont {Ayers}, \citenamefont {Fern{\'a}ndez-Alonso},\ and\
  \citenamefont {Zare}}]{Bean_Zare_DCS_2002}%
  \BibitemOpen
  \bibfield  {author} {\bibinfo {author} {\bibfnamefont {B.~D.}\ \bibnamefont
  {Bean}}, \bibinfo {author} {\bibfnamefont {J.~D.}\ \bibnamefont {Ayers}},
  \bibinfo {author} {\bibfnamefont {F.}~\bibnamefont {Fern{\'a}ndez-Alonso}}, \
  and\ \bibinfo {author} {\bibfnamefont {R.~N.}\ \bibnamefont {Zare}},\
  }\href@noop {} {\bibfield  {journal} {\bibinfo  {journal} {J. Chem. Phys.}\
  }\textbf {\bibinfo {volume} {116}},\ \bibinfo {pages} {6634} (\bibinfo {year}
  {2002})}\BibitemShut {NoStop}%
\bibitem [{\citenamefont {Mielke}\ \emph {et~al.}(2003)\citenamefont {Mielke},
  \citenamefont {Peterson}, \citenamefont {Schwenke}, \citenamefont {Garrett},
  \citenamefont {Truhlar}, \citenamefont {Michael}, \citenamefont {Su},\ and\
  \citenamefont {Sutherland}}]{Mielke_Sutherland_2003}%
  \BibitemOpen
  \bibfield  {author} {\bibinfo {author} {\bibfnamefont {S.~L.}\ \bibnamefont
  {Mielke}}, \bibinfo {author} {\bibfnamefont {K.~A.}\ \bibnamefont
  {Peterson}}, \bibinfo {author} {\bibfnamefont {D.~W.}\ \bibnamefont
  {Schwenke}}, \bibinfo {author} {\bibfnamefont {B.~C.}\ \bibnamefont
  {Garrett}}, \bibinfo {author} {\bibfnamefont {D.~G.}\ \bibnamefont
  {Truhlar}}, \bibinfo {author} {\bibfnamefont {J.~V.}\ \bibnamefont
  {Michael}}, \bibinfo {author} {\bibfnamefont {M.-C.}\ \bibnamefont {Su}}, \
  and\ \bibinfo {author} {\bibfnamefont {J.~W.}\ \bibnamefont {Sutherland}},\
  }\href@noop {} {\bibfield  {journal} {\bibinfo  {journal} {Phys. Rev. Lett.}\
  }\textbf {\bibinfo {volume} {91}},\ \bibinfo {pages} {063201} (\bibinfo
  {year} {2003})}\BibitemShut {NoStop}%
\bibitem [{\citenamefont {Dai}\ \emph {et~al.}(2003)\citenamefont {Dai},
  \citenamefont {Wang}, \citenamefont {Harich}, \citenamefont {Wang},
  \citenamefont {Yang}, \citenamefont {Chao},\ and\ \citenamefont
  {Skodje}}]{Dai_Wang_Science_2003}%
  \BibitemOpen
  \bibfield  {author} {\bibinfo {author} {\bibfnamefont {D.}~\bibnamefont
  {Dai}}, \bibinfo {author} {\bibfnamefont {C.~C.}\ \bibnamefont {Wang}},
  \bibinfo {author} {\bibfnamefont {S.~A.}\ \bibnamefont {Harich}}, \bibinfo
  {author} {\bibfnamefont {X.}~\bibnamefont {Wang}}, \bibinfo {author}
  {\bibfnamefont {X.}~\bibnamefont {Yang}}, \bibinfo {author} {\bibfnamefont
  {S.~D.}\ \bibnamefont {Chao}}, \ and\ \bibinfo {author} {\bibfnamefont
  {R.~T.}\ \bibnamefont {Skodje}},\ }\href@noop {} {\bibfield  {journal}
  {\bibinfo  {journal} {Science}\ }\textbf {\bibinfo {volume} {300}},\ \bibinfo
  {pages} {1730} (\bibinfo {year} {2003})}\BibitemShut {NoStop}%
\bibitem [{\citenamefont {Ausfelder}\ \emph {et~al.}(2004)\citenamefont
  {Ausfelder}, \citenamefont {Pomerantz}, \citenamefont {Zare}, \citenamefont
  {Althorpe}, \citenamefont {Aoiz}, \citenamefont {Ba{\~n}ares},\ and\
  \citenamefont {Castillo}}]{Althorpe_Zare_ICS_2004}%
  \BibitemOpen
  \bibfield  {author} {\bibinfo {author} {\bibfnamefont {F.}~\bibnamefont
  {Ausfelder}}, \bibinfo {author} {\bibfnamefont {A.~E.}\ \bibnamefont
  {Pomerantz}}, \bibinfo {author} {\bibfnamefont {R.~N.}\ \bibnamefont {Zare}},
  \bibinfo {author} {\bibfnamefont {S.~C.}\ \bibnamefont {Althorpe}}, \bibinfo
  {author} {\bibfnamefont {F.~J.}\ \bibnamefont {Aoiz}}, \bibinfo {author}
  {\bibfnamefont {L.}~\bibnamefont {Ba{\~n}ares}}, \ and\ \bibinfo {author}
  {\bibfnamefont {J.~F.}\ \bibnamefont {Castillo}},\ }\href@noop {} {\bibfield
  {journal} {\bibinfo  {journal} {J. Chem. Phys.}\ }\textbf {\bibinfo {volume}
  {120}},\ \bibinfo {pages} {3255} (\bibinfo {year} {2004})}\BibitemShut
  {NoStop}%
\bibitem [{\citenamefont {Pomerantz}\ \emph {et~al.}(2004)\citenamefont
  {Pomerantz}, \citenamefont {Ausfelder}, \citenamefont {Zare}, \citenamefont
  {Althorpe}, \citenamefont {Aoiz}, \citenamefont {Ba{\~n}ares},\ and\
  \citenamefont {Castillo}}]{Pomerantz_2004_ref_20}%
  \BibitemOpen
  \bibfield  {author} {\bibinfo {author} {\bibfnamefont {A.~E.}\ \bibnamefont
  {Pomerantz}}, \bibinfo {author} {\bibfnamefont {F.}~\bibnamefont
  {Ausfelder}}, \bibinfo {author} {\bibfnamefont {R.~N.}\ \bibnamefont {Zare}},
  \bibinfo {author} {\bibfnamefont {S.~C.}\ \bibnamefont {Althorpe}}, \bibinfo
  {author} {\bibfnamefont {F.~J.}\ \bibnamefont {Aoiz}}, \bibinfo {author}
  {\bibfnamefont {L.}~\bibnamefont {Ba{\~n}ares}}, \ and\ \bibinfo {author}
  {\bibfnamefont {J.~F.}\ \bibnamefont {Castillo}},\ }\href@noop {} {\bibfield
  {journal} {\bibinfo  {journal} {J. Chem. Phys.}\ }\textbf {\bibinfo {volume}
  {120}},\ \bibinfo {pages} {3244} (\bibinfo {year} {2004})}\BibitemShut
  {NoStop}%
\bibitem [{\citenamefont {Koszinowski}\ \emph {et~al.}(2005)\citenamefont
  {Koszinowski}, \citenamefont {Goldberg}, \citenamefont {Pomerantz},
  \citenamefont {Zare}, \citenamefont {Juanes-Marcos},\ and\ \citenamefont
  {Althorpe}}]{zare_2005}%
  \BibitemOpen
  \bibfield  {author} {\bibinfo {author} {\bibfnamefont {K.}~\bibnamefont
  {Koszinowski}}, \bibinfo {author} {\bibfnamefont {N.~T.}\ \bibnamefont
  {Goldberg}}, \bibinfo {author} {\bibfnamefont {A.~E.}\ \bibnamefont
  {Pomerantz}}, \bibinfo {author} {\bibfnamefont {R.~N.}\ \bibnamefont {Zare}},
  \bibinfo {author} {\bibfnamefont {J.~C.}\ \bibnamefont {Juanes-Marcos}}, \
  and\ \bibinfo {author} {\bibfnamefont {S.~C.}\ \bibnamefont {Althorpe}},\
  }\href@noop {} {\bibfield  {journal} {\bibinfo  {journal} {J. Chem. Phys.}\
  }\textbf {\bibinfo {volume} {123}},\ \bibinfo {pages} {054306} (\bibinfo
  {year} {2005})}\BibitemShut {NoStop}%
\bibitem [{\citenamefont {Hal{\'a}sz}\ \emph {et~al.}(2003)\citenamefont
  {Hal{\'a}sz}, \citenamefont {Vib{\'o}k}, \citenamefont {Mebel},\ and\
  \citenamefont {Baer}}]{Halasz_Baer2003}%
  \BibitemOpen
  \bibfield  {author} {\bibinfo {author} {\bibfnamefont {G.}~\bibnamefont
  {Hal{\'a}sz}}, \bibinfo {author} {\bibfnamefont {A.}~\bibnamefont
  {Vib{\'o}k}}, \bibinfo {author} {\bibfnamefont {A.~M.}\ \bibnamefont
  {Mebel}}, \ and\ \bibinfo {author} {\bibfnamefont {M.}~\bibnamefont {Baer}},\
  }\href@noop {} {\bibfield  {journal} {\bibinfo  {journal} {J. Chem. Phys.}\
  }\textbf {\bibinfo {volume} {118}},\ \bibinfo {pages} {3052} (\bibinfo {year}
  {2003})}\BibitemShut {NoStop}%
\bibitem [{\citenamefont {Baer}\ \emph {et~al.}(2004)\citenamefont {Baer},
  \citenamefont {Ve'Rtesi}, \citenamefont {Hal{\'a}sz}, \citenamefont {Vibok},\
  and\ \citenamefont {Suhai}}]{Baer_2004}%
  \BibitemOpen
  \bibfield  {author} {\bibinfo {author} {\bibfnamefont {M.}~\bibnamefont
  {Baer}}, \bibinfo {author} {\bibfnamefont {T.}~\bibnamefont {Ve'Rtesi}},
  \bibinfo {author} {\bibfnamefont {G.}~\bibnamefont {Hal{\'a}sz}}, \bibinfo
  {author} {\bibfnamefont {A.}~\bibnamefont {Vibok}}, \ and\ \bibinfo {author}
  {\bibfnamefont {S.}~\bibnamefont {Suhai}},\ }\href@noop {} {\bibfield
  {journal} {\bibinfo  {journal} {Faraday Discussions}\ }\textbf {\bibinfo
  {volume} {127}},\ \bibinfo {pages} {337} (\bibinfo {year}
  {2004})}\BibitemShut {NoStop}%
\bibitem [{\citenamefont {Ghosh}\ \emph {et~al.}(2025)\citenamefont {Ghosh},
  \citenamefont {Roy}, \citenamefont {Kumar}, \citenamefont {Naskar},
  \citenamefont {Ravi},\ and\ \citenamefont {Adhikari}}]{Ghosh_H3_2025}%
  \BibitemOpen
  \bibfield  {author} {\bibinfo {author} {\bibfnamefont {A.}~\bibnamefont
  {Ghosh}}, \bibinfo {author} {\bibfnamefont {A.}~\bibnamefont {Roy}}, \bibinfo
  {author} {\bibfnamefont {A.}~\bibnamefont {Kumar}}, \bibinfo {author}
  {\bibfnamefont {K.}~\bibnamefont {Naskar}}, \bibinfo {author} {\bibfnamefont
  {S.}~\bibnamefont {Ravi}}, \ and\ \bibinfo {author} {\bibfnamefont
  {S.}~\bibnamefont {Adhikari}},\ }\href@noop {} {\bibfield  {journal}
  {\bibinfo  {journal} {J. Phys. Chem. A}\ }\textbf {\bibinfo {volume} {129}},\
  \bibinfo {pages} {6315} (\bibinfo {year} {2025})}\BibitemShut {NoStop}%
\bibitem [{\citenamefont {Gao}\ \emph {et~al.}(2015)\citenamefont {Gao},
  \citenamefont {Sneha}, \citenamefont {Bouakline}, \citenamefont {Althorpe},\
  and\ \citenamefont {Zare}}]{Althorpe_2015_DCS}%
  \BibitemOpen
  \bibfield  {author} {\bibinfo {author} {\bibfnamefont {H.}~\bibnamefont
  {Gao}}, \bibinfo {author} {\bibfnamefont {M.}~\bibnamefont {Sneha}}, \bibinfo
  {author} {\bibfnamefont {F.}~\bibnamefont {Bouakline}}, \bibinfo {author}
  {\bibfnamefont {S.~C.}\ \bibnamefont {Althorpe}}, \ and\ \bibinfo {author}
  {\bibfnamefont {R.~N.}\ \bibnamefont {Zare}},\ }\href@noop {} {\bibfield
  {journal} {\bibinfo  {journal} {J. Phys. Chem. A}\ }\textbf {\bibinfo
  {volume} {119}},\ \bibinfo {pages} {12036} (\bibinfo {year}
  {2015})}\BibitemShut {NoStop}%
\bibitem [{\citenamefont {Jankunas}\ \emph {et~al.}(2013)\citenamefont
  {Jankunas}, \citenamefont {Sneha}, \citenamefont {Zare}, \citenamefont
  {Bouakline},\ and\ \citenamefont {Althorpe}}]{Jankunas_Althrope_2013}%
  \BibitemOpen
  \bibfield  {author} {\bibinfo {author} {\bibfnamefont {J.}~\bibnamefont
  {Jankunas}}, \bibinfo {author} {\bibfnamefont {M.}~\bibnamefont {Sneha}},
  \bibinfo {author} {\bibfnamefont {R.~N.}\ \bibnamefont {Zare}}, \bibinfo
  {author} {\bibfnamefont {F.}~\bibnamefont {Bouakline}}, \ and\ \bibinfo
  {author} {\bibfnamefont {S.~C.}\ \bibnamefont {Althorpe}},\ }\href@noop {}
  {\bibfield  {journal} {\bibinfo  {journal} {The Journal of Chemical Physics}\
  }\textbf {\bibinfo {volume} {138}},\ \bibinfo {pages} {094310} (\bibinfo
  {year} {2013})}\BibitemShut {NoStop}%
\bibitem [{\citenamefont {Varandas}\ \emph {et~al.}(1987)\citenamefont
  {Varandas}, \citenamefont {Brown}, \citenamefont {Mead}, \citenamefont
  {Truhlar},\ and\ \citenamefont {Blais}}]{DMBE_1987}%
  \BibitemOpen
  \bibfield  {author} {\bibinfo {author} {\bibfnamefont {A.~J.~C.}\
  \bibnamefont {Varandas}}, \bibinfo {author} {\bibfnamefont {F.~B.}\
  \bibnamefont {Brown}}, \bibinfo {author} {\bibfnamefont {C.~A.}\ \bibnamefont
  {Mead}}, \bibinfo {author} {\bibfnamefont {D.~G.}\ \bibnamefont {Truhlar}}, \
  and\ \bibinfo {author} {\bibfnamefont {N.~C.}\ \bibnamefont {Blais}},\
  }\href@noop {} {\bibfield  {journal} {\bibinfo  {journal} {J. Chem. Phys.}\
  }\textbf {\bibinfo {volume} {86}},\ \bibinfo {pages} {6258} (\bibinfo {year}
  {1987})}\BibitemShut {NoStop}%
\bibitem [{\citenamefont {Hazra}\ \emph {et~al.}(2026)\citenamefont {Hazra},
  \citenamefont {Roy}, \citenamefont {Ghosh}, \citenamefont {Naskar},\ and\
  \citenamefont {Adhikari}}]{Scat_Review}%
  \BibitemOpen
  \bibfield  {author} {\bibinfo {author} {\bibfnamefont {S.}~\bibnamefont
  {Hazra}}, \bibinfo {author} {\bibfnamefont {A.}~\bibnamefont {Roy}}, \bibinfo
  {author} {\bibfnamefont {A.}~\bibnamefont {Ghosh}}, \bibinfo {author}
  {\bibfnamefont {K.}~\bibnamefont {Naskar}}, \ and\ \bibinfo {author}
  {\bibfnamefont {S.}~\bibnamefont {Adhikari}},\ }\href {\doibase
  https://doi.org/10.1002/jcc.70464} {\bibfield  {journal} {\bibinfo  {journal}
  {Journal of Computational Chemistry}\ }\textbf {\bibinfo {volume} {47}},\
  \bibinfo {pages} {e70464} (\bibinfo {year} {2026})}\BibitemShut {NoStop}%
\bibitem [{\citenamefont {Neuhauser}\ and\ \citenamefont
  {Baer}(1990)}]{jcp-92-3419-BAER}%
  \BibitemOpen
  \bibfield  {author} {\bibinfo {author} {\bibfnamefont {D.}~\bibnamefont
  {Neuhauser}}\ and\ \bibinfo {author} {\bibfnamefont {M.}~\bibnamefont
  {Baer}},\ }\href@noop {} {\bibfield  {journal} {\bibinfo  {journal} {J. Chem.
  Phys.}\ }\textbf {\bibinfo {volume} {92}},\ \bibinfo {pages} {3419} (\bibinfo
  {year} {1990})}\BibitemShut {NoStop}%
\bibitem [{\citenamefont {Neuhasuer}\ and\ \citenamefont
  {Baer}(1989)}]{jcp-90-4351-BAER}%
  \BibitemOpen
  \bibfield  {author} {\bibinfo {author} {\bibfnamefont {D.}~\bibnamefont
  {Neuhasuer}}\ and\ \bibinfo {author} {\bibfnamefont {M.}~\bibnamefont
  {Baer}},\ }\href@noop {} {\bibfield  {journal} {\bibinfo  {journal} {J. Chem.
  Phys.}\ }\textbf {\bibinfo {volume} {90}},\ \bibinfo {pages} {4351} (\bibinfo
  {year} {1989})}\BibitemShut {NoStop}%
\bibitem [{\citenamefont {Zhang}\ and\ \citenamefont
  {Miller}(1989)}]{jcp-91-1528-MILL}%
  \BibitemOpen
  \bibfield  {author} {\bibinfo {author} {\bibfnamefont {J.~Z.~H.}\
  \bibnamefont {Zhang}}\ and\ \bibinfo {author} {\bibfnamefont {W.~H.}\
  \bibnamefont {Miller}},\ }\href@noop {} {\bibfield  {journal} {\bibinfo
  {journal} {J. Chem. Phys.}\ }\textbf {\bibinfo {volume} {91}},\ \bibinfo
  {pages} {1528} (\bibinfo {year} {1989})}\BibitemShut {NoStop}%
\bibitem [{\citenamefont {Webster}\ and\ \citenamefont
  {Light}(1986)}]{jcp-85-4744-LIT2}%
  \BibitemOpen
  \bibfield  {author} {\bibinfo {author} {\bibfnamefont {F.}~\bibnamefont
  {Webster}}\ and\ \bibinfo {author} {\bibfnamefont {J.~C.}\ \bibnamefont
  {Light}},\ }\href@noop {} {\bibfield  {journal} {\bibinfo  {journal} {J.
  Chem. Phys.}\ }\textbf {\bibinfo {volume} {85}},\ \bibinfo {pages} {4744}
  (\bibinfo {year} {1986})}\BibitemShut {NoStop}%
\bibitem [{\citenamefont {Webster}\ and\ \citenamefont
  {Light}(1989)}]{jcp-90-265-LIT2}%
  \BibitemOpen
  \bibfield  {author} {\bibinfo {author} {\bibfnamefont {F.}~\bibnamefont
  {Webster}}\ and\ \bibinfo {author} {\bibfnamefont {J.~C.}\ \bibnamefont
  {Light}},\ }\href@noop {} {\bibfield  {journal} {\bibinfo  {journal} {J.
  Chem. Phys.}\ }\textbf {\bibinfo {volume} {90}},\ \bibinfo {pages} {265}
  (\bibinfo {year} {1989})}\BibitemShut {NoStop}%
\bibitem [{\citenamefont {Baer}\ and\ \citenamefont
  {Kouri}(1971)}]{prA-4-1924-MBK2}%
  \BibitemOpen
  \bibfield  {author} {\bibinfo {author} {\bibfnamefont {M.}~\bibnamefont
  {Baer}}\ and\ \bibinfo {author} {\bibfnamefont {D.~J.}\ \bibnamefont
  {Kouri}},\ }\href@noop {} {\bibfield  {journal} {\bibinfo  {journal} {Phys.
  Rev. A}\ }\textbf {\bibinfo {volume} {4}},\ \bibinfo {pages} {1924} (\bibinfo
  {year} {1971})}\BibitemShut {NoStop}%
\bibitem [{\citenamefont {Sun}, \citenamefont {Guo},\ and\ \citenamefont
  {Zhang}(2010)}]{jcp-132-084112-GUO}%
  \BibitemOpen
  \bibfield  {author} {\bibinfo {author} {\bibfnamefont {Z.}~\bibnamefont
  {Sun}}, \bibinfo {author} {\bibfnamefont {H.}~\bibnamefont {Guo}}, \ and\
  \bibinfo {author} {\bibfnamefont {D.~H.}\ \bibnamefont {Zhang}},\ }\href@noop
  {} {\bibfield  {journal} {\bibinfo  {journal} {J. Chem. Phys.}\ }\textbf
  {\bibinfo {volume} {132}},\ \bibinfo {pages} {084112/1} (\bibinfo {year}
  {2010})}\BibitemShut {NoStop}%
\bibitem [{\citenamefont {Carrasco}\ and\ \citenamefont
  {Roncero}(2006)}]{jcp-125-054102-GOM}%
  \BibitemOpen
  \bibfield  {author} {\bibinfo {author} {\bibfnamefont {S.~G.}\ \bibnamefont
  {Carrasco}}\ and\ \bibinfo {author} {\bibfnamefont {O.}~\bibnamefont
  {Roncero}},\ }\href@noop {} {\bibfield  {journal} {\bibinfo  {journal} {J.
  Chem. Phys.}\ }\textbf {\bibinfo {volume} {125}},\ \bibinfo {pages}
  {054102/1} (\bibinfo {year} {2006})}\BibitemShut {NoStop}%
\bibitem [{\citenamefont {Zhang}, \citenamefont {Chu},\ and\ \citenamefont
  {Miller}(1988)}]{jcp-88-6233-MILL}%
  \BibitemOpen
  \bibfield  {author} {\bibinfo {author} {\bibfnamefont {J.~Z.~H.}\
  \bibnamefont {Zhang}}, \bibinfo {author} {\bibfnamefont {S.-I.}\ \bibnamefont
  {Chu}}, \ and\ \bibinfo {author} {\bibfnamefont {W.~H.}\ \bibnamefont
  {Miller}},\ }\href@noop {} {\bibfield  {journal} {\bibinfo  {journal} {J.
  Chem. Phys.}\ }\textbf {\bibinfo {volume} {88}},\ \bibinfo {pages} {6233}
  (\bibinfo {year} {1988})}\BibitemShut {NoStop}%
\bibitem [{\citenamefont {Zhu}, \citenamefont {Peng},\ and\ \citenamefont
  {Zhang}(1997)}]{jcp-106-1742-ZHA1}%
  \BibitemOpen
  \bibfield  {author} {\bibinfo {author} {\bibfnamefont {W.}~\bibnamefont
  {Zhu}}, \bibinfo {author} {\bibfnamefont {T.}~\bibnamefont {Peng}}, \ and\
  \bibinfo {author} {\bibfnamefont {J.~Z.~H.}\ \bibnamefont {Zhang}},\
  }\href@noop {} {\bibfield  {journal} {\bibinfo  {journal} {J. Chem. Phys.}\
  }\textbf {\bibinfo {volume} {106}},\ \bibinfo {pages} {1742} (\bibinfo {year}
  {1997})}\BibitemShut {NoStop}%
\bibitem [{\citenamefont {Kouri}\ \emph {et~al.}(1996)\citenamefont {Kouri},
  \citenamefont {Hoffman}, \citenamefont {Peng},\ and\ \citenamefont
  {Zhang}}]{cpl-262-519-ZHA1}%
  \BibitemOpen
  \bibfield  {author} {\bibinfo {author} {\bibfnamefont {D.~J.}\ \bibnamefont
  {Kouri}}, \bibinfo {author} {\bibfnamefont {D.~K.}\ \bibnamefont {Hoffman}},
  \bibinfo {author} {\bibfnamefont {T.}~\bibnamefont {Peng}}, \ and\ \bibinfo
  {author} {\bibfnamefont {J.~Z.~H.}\ \bibnamefont {Zhang}},\ }\href@noop {}
  {\bibfield  {journal} {\bibinfo  {journal} {Chem. Phys. Lett.}\ }\textbf
  {\bibinfo {volume} {262}},\ \bibinfo {pages} {519} (\bibinfo {year}
  {1996})}\BibitemShut {NoStop}%
\bibitem [{\citenamefont {Peng}\ and\ \citenamefont
  {Zhang}(1996)}]{jcp-105-6072-ZHA1}%
  \BibitemOpen
  \bibfield  {author} {\bibinfo {author} {\bibfnamefont {T.}~\bibnamefont
  {Peng}}\ and\ \bibinfo {author} {\bibfnamefont {J.~Z.~H.}\ \bibnamefont
  {Zhang}},\ }\href@noop {} {\bibfield  {journal} {\bibinfo  {journal} {J.
  Chem. Phys.}\ }\textbf {\bibinfo {volume} {105}},\ \bibinfo {pages} {6072}
  (\bibinfo {year} {1996})}\BibitemShut {NoStop}%
\bibitem [{\citenamefont {Althorpe}, \citenamefont {Kouri},\ and\ \citenamefont
  {Hoffmann}(1997{\natexlab{a}})}]{jcp-106-7629-ALT1}%
  \BibitemOpen
  \bibfield  {author} {\bibinfo {author} {\bibfnamefont {S.~C.}\ \bibnamefont
  {Althorpe}}, \bibinfo {author} {\bibfnamefont {D.~J.}\ \bibnamefont {Kouri}},
  \ and\ \bibinfo {author} {\bibfnamefont {D.~K.}\ \bibnamefont {Hoffmann}},\
  }\href@noop {} {\bibfield  {journal} {\bibinfo  {journal} {J. Chem. Phys.}\
  }\textbf {\bibinfo {volume} {106}},\ \bibinfo {pages} {7629} (\bibinfo {year}
  {1997}{\natexlab{a}})}\BibitemShut {NoStop}%
\bibitem [{\citenamefont {Althorpe}, \citenamefont {Kouri},\ and\ \citenamefont
  {Hoffmann}(1997{\natexlab{b}})}]{jcp-107-7816-ALT1}%
  \BibitemOpen
  \bibfield  {author} {\bibinfo {author} {\bibfnamefont {S.~C.}\ \bibnamefont
  {Althorpe}}, \bibinfo {author} {\bibfnamefont {D.~J.}\ \bibnamefont {Kouri}},
  \ and\ \bibinfo {author} {\bibfnamefont {D.~K.}\ \bibnamefont {Hoffmann}},\
  }\href@noop {} {\bibfield  {journal} {\bibinfo  {journal} {J. Chem. Phys.}\
  }\textbf {\bibinfo {volume} {107}},\ \bibinfo {pages} {7816} (\bibinfo {year}
  {1997}{\natexlab{b}})}\BibitemShut {NoStop}%
\bibitem [{\citenamefont {Johnson}(1983{\natexlab{a}})}]{jcp-79-1916-JOH}%
  \BibitemOpen
  \bibfield  {author} {\bibinfo {author} {\bibfnamefont {B.~R.}\ \bibnamefont
  {Johnson}},\ }\href@noop {} {\bibfield  {journal} {\bibinfo  {journal} {J.
  Chem. Phys.}\ }\textbf {\bibinfo {volume} {79}},\ \bibinfo {pages} {1916}
  (\bibinfo {year} {1983}{\natexlab{a}})}\BibitemShut {NoStop}%
\bibitem [{\citenamefont {Xu}, \citenamefont {Baer},\ and\ \citenamefont
  {Varandas}(2000)}]{Xu2000_JCP_112}%
  \BibitemOpen
  \bibfield  {author} {\bibinfo {author} {\bibfnamefont {Z.}~\bibnamefont
  {Xu}}, \bibinfo {author} {\bibfnamefont {M.}~\bibnamefont {Baer}}, \ and\
  \bibinfo {author} {\bibfnamefont {A.~J.~C.}\ \bibnamefont {Varandas}},\
  }\href@noop {} {\bibfield  {journal} {\bibinfo  {journal} {J. Chem. Phys.}\
  }\textbf {\bibinfo {volume} {112}},\ \bibinfo {pages} {2746} (\bibinfo {year}
  {2000})}\BibitemShut {NoStop}%
\bibitem [{\citenamefont {Adhikari}\ and\ \citenamefont
  {Varandas}(2013)}]{cpc-184-270-SA}%
  \BibitemOpen
  \bibfield  {author} {\bibinfo {author} {\bibfnamefont {S.}~\bibnamefont
  {Adhikari}}\ and\ \bibinfo {author} {\bibfnamefont {A.~J.~C.}\ \bibnamefont
  {Varandas}},\ }\href@noop {} {\bibfield  {journal} {\bibinfo  {journal}
  {Comput. Phys. Comm.}\ }\textbf {\bibinfo {volume} {184}},\ \bibinfo {pages}
  {270} (\bibinfo {year} {2013})}\BibitemShut {NoStop}%
\bibitem [{\citenamefont {Billing}\ and\ \citenamefont
  {Markovic}(1993)}]{jcp-99-2674-BILL}%
  \BibitemOpen
  \bibfield  {author} {\bibinfo {author} {\bibfnamefont {G.~D.}\ \bibnamefont
  {Billing}}\ and\ \bibinfo {author} {\bibfnamefont {N.}~\bibnamefont
  {Markovic}},\ }\href@noop {} {\bibfield  {journal} {\bibinfo  {journal} {J.
  Chem. Phys.}\ }\textbf {\bibinfo {volume} {99}},\ \bibinfo {pages} {2674}
  (\bibinfo {year} {1993})}\BibitemShut {NoStop}%
\bibitem [{\citenamefont {Billing}\ and\ \citenamefont
  {Markovic}(1994)}]{jcp-100-1085-BILL}%
  \BibitemOpen
  \bibfield  {author} {\bibinfo {author} {\bibfnamefont {G.~D.}\ \bibnamefont
  {Billing}}\ and\ \bibinfo {author} {\bibfnamefont {N.}~\bibnamefont
  {Markovic}},\ }\href@noop {} {\bibfield  {journal} {\bibinfo  {journal} {J.
  Chem. Phys.}\ }\textbf {\bibinfo {volume} {100}},\ \bibinfo {pages} {1085}
  (\bibinfo {year} {1994})}\BibitemShut {NoStop}%
\bibitem [{\citenamefont {Markovic}\ and\ \citenamefont
  {Billing}(1995)}]{cp-191-247-MAR}%
  \BibitemOpen
  \bibfield  {author} {\bibinfo {author} {\bibfnamefont {N.}~\bibnamefont
  {Markovic}}\ and\ \bibinfo {author} {\bibfnamefont {G.~D.}\ \bibnamefont
  {Billing}},\ }\href@noop {} {\bibfield  {journal} {\bibinfo  {journal} {Chem.
  Phys.}\ }\textbf {\bibinfo {volume} {191}},\ \bibinfo {pages} {247} (\bibinfo
  {year} {1995})}\BibitemShut {NoStop}%
\bibitem [{\citenamefont {Skouteris}\ \emph {et~al.}(1999)\citenamefont
  {Skouteris}, \citenamefont {Manolopoulos}, \citenamefont {Bian},
  \citenamefont {Werner}, \citenamefont {Lai},\ and\ \citenamefont
  {Liu}}]{science-286-1713-HYP2}%
  \BibitemOpen
  \bibfield  {author} {\bibinfo {author} {\bibfnamefont {D.}~\bibnamefont
  {Skouteris}}, \bibinfo {author} {\bibfnamefont {D.~E.}\ \bibnamefont
  {Manolopoulos}}, \bibinfo {author} {\bibfnamefont {W.}~\bibnamefont {Bian}},
  \bibinfo {author} {\bibfnamefont {W.~J.}\ \bibnamefont {Werner}}, \bibinfo
  {author} {\bibfnamefont {L.~H.}\ \bibnamefont {Lai}}, \ and\ \bibinfo
  {author} {\bibfnamefont {K.}~\bibnamefont {Liu}},\ }\href@noop {} {\bibfield
  {journal} {\bibinfo  {journal} {Science}\ }\textbf {\bibinfo {volume}
  {286}},\ \bibinfo {pages} {1713} (\bibinfo {year} {1999})}\BibitemShut
  {NoStop}%
\bibitem [{\citenamefont {Honvault}\ and\ \citenamefont
  {Launay}(1999{\natexlab{a}})}]{jcp-111-6665-HYP2}%
  \BibitemOpen
  \bibfield  {author} {\bibinfo {author} {\bibfnamefont {P.}~\bibnamefont
  {Honvault}}\ and\ \bibinfo {author} {\bibfnamefont {J.~M.}\ \bibnamefont
  {Launay}},\ }\href@noop {} {\bibfield  {journal} {\bibinfo  {journal} {J.
  Chem. Phys.}\ }\textbf {\bibinfo {volume} {111}},\ \bibinfo {pages} {6665}
  (\bibinfo {year} {1999}{\natexlab{a}})}\BibitemShut {NoStop}%
\bibitem [{\citenamefont {Lagana}\ \emph {et~al.}(2000)\citenamefont {Lagana},
  \citenamefont {Bolloni}, \citenamefont {Crocchianti},\ and\ \citenamefont
  {Parker}}]{cpl-324-466-HYP2}%
  \BibitemOpen
  \bibfield  {author} {\bibinfo {author} {\bibfnamefont {A.}~\bibnamefont
  {Lagana}}, \bibinfo {author} {\bibfnamefont {A.}~\bibnamefont {Bolloni}},
  \bibinfo {author} {\bibfnamefont {S.}~\bibnamefont {Crocchianti}}, \ and\
  \bibinfo {author} {\bibfnamefont {G.~A.}\ \bibnamefont {Parker}},\
  }\href@noop {} {\bibfield  {journal} {\bibinfo  {journal} {Chem. Phys.
  Lett.}\ }\textbf {\bibinfo {volume} {324}},\ \bibinfo {pages} {466} (\bibinfo
  {year} {2000})}\BibitemShut {NoStop}%
\bibitem [{\citenamefont {Sahoo}\ \emph {et~al.}(2014)\citenamefont {Sahoo},
  \citenamefont {Ghosh}, \citenamefont {Adhikari}, \citenamefont {Sharma},\
  and\ \citenamefont {Varandas}}]{jpcA-118-4837-SA}%
  \BibitemOpen
  \bibfield  {author} {\bibinfo {author} {\bibfnamefont {T.}~\bibnamefont
  {Sahoo}}, \bibinfo {author} {\bibfnamefont {S.}~\bibnamefont {Ghosh}},
  \bibinfo {author} {\bibfnamefont {S.}~\bibnamefont {Adhikari}}, \bibinfo
  {author} {\bibfnamefont {R.}~\bibnamefont {Sharma}}, \ and\ \bibinfo {author}
  {\bibfnamefont {A.~J.~C.}\ \bibnamefont {Varandas}},\ }\href@noop {}
  {\bibfield  {journal} {\bibinfo  {journal} {J. Phys. Chem. A}\ }\textbf
  {\bibinfo {volume} {118}},\ \bibinfo {pages} {4837} (\bibinfo {year}
  {2014})}\BibitemShut {NoStop}%
\bibitem [{\citenamefont {Kuppermann}\ and\ \citenamefont
  {Hipes}(1986)}]{jcp-84-5962-HYP1}%
  \BibitemOpen
  \bibfield  {author} {\bibinfo {author} {\bibfnamefont {A.}~\bibnamefont
  {Kuppermann}}\ and\ \bibinfo {author} {\bibfnamefont {P.~G.}\ \bibnamefont
  {Hipes}},\ }\href@noop {} {\bibfield  {journal} {\bibinfo  {journal} {J.
  Chem. Phys.}\ }\textbf {\bibinfo {volume} {84}},\ \bibinfo {pages} {5962}
  (\bibinfo {year} {1986})}\BibitemShut {NoStop}%
\bibitem [{\citenamefont {Sahoo}\ \emph {et~al.}(2015)\citenamefont {Sahoo},
  \citenamefont {Ghosh}, \citenamefont {Adhikari}, \citenamefont {Sharma},\
  and\ \citenamefont {Varandas}}]{jcp-142-024304-SA}%
  \BibitemOpen
  \bibfield  {author} {\bibinfo {author} {\bibfnamefont {T.}~\bibnamefont
  {Sahoo}}, \bibinfo {author} {\bibfnamefont {S.}~\bibnamefont {Ghosh}},
  \bibinfo {author} {\bibfnamefont {S.}~\bibnamefont {Adhikari}}, \bibinfo
  {author} {\bibfnamefont {R.}~\bibnamefont {Sharma}}, \ and\ \bibinfo {author}
  {\bibfnamefont {A.~J.~C.}\ \bibnamefont {Varandas}},\ }\href@noop {}
  {\bibfield  {journal} {\bibinfo  {journal} {J. Chem. Phys.}\ }\textbf
  {\bibinfo {volume} {142}},\ \bibinfo {pages} {024304/1} (\bibinfo {year}
  {2015})}\BibitemShut {NoStop}%
\bibitem [{\citenamefont {Ghosh}\ \emph
  {et~al.}(2017{\natexlab{b}})\citenamefont {Ghosh}, \citenamefont {Sharma},
  \citenamefont {Adhikari},\ and\ \citenamefont {Varandas}}]{cpl-675-85-SA}%
  \BibitemOpen
  \bibfield  {author} {\bibinfo {author} {\bibfnamefont {S.}~\bibnamefont
  {Ghosh}}, \bibinfo {author} {\bibfnamefont {R.}~\bibnamefont {Sharma}},
  \bibinfo {author} {\bibfnamefont {S.}~\bibnamefont {Adhikari}}, \ and\
  \bibinfo {author} {\bibfnamefont {A.~J.~C.}\ \bibnamefont {Varandas}},\
  }\href@noop {} {\bibfield  {journal} {\bibinfo  {journal} {Chem. Phys.
  Lett.}\ }\textbf {\bibinfo {volume} {675}},\ \bibinfo {pages} {85} (\bibinfo
  {year} {2017}{\natexlab{b}})}\BibitemShut {NoStop}%
\bibitem [{\citenamefont {Ghosh}\ \emph {et~al.}(2018)\citenamefont {Ghosh},
  \citenamefont {Sharma}, \citenamefont {Adhikari},\ and\ \citenamefont
  {Varandas}}]{pccp-20-478-SA}%
  \BibitemOpen
  \bibfield  {author} {\bibinfo {author} {\bibfnamefont {S.}~\bibnamefont
  {Ghosh}}, \bibinfo {author} {\bibfnamefont {R.}~\bibnamefont {Sharma}},
  \bibinfo {author} {\bibfnamefont {S.}~\bibnamefont {Adhikari}}, \ and\
  \bibinfo {author} {\bibfnamefont {A.~J.~C.}\ \bibnamefont {Varandas}},\
  }\href@noop {} {\bibfield  {journal} {\bibinfo  {journal} {Phys. Chem. Chem.
  Phys.}\ }\textbf {\bibinfo {volume} {20}},\ \bibinfo {pages} {478} (\bibinfo
  {year} {2018})}\BibitemShut {NoStop}%
\bibitem [{\citenamefont {Ghosh}\ \emph {et~al.}(2019)\citenamefont {Ghosh},
  \citenamefont {Sharma}, \citenamefont {Adhikari},\ and\ \citenamefont
  {Varandas}}]{pccp-21-20166-SA}%
  \BibitemOpen
  \bibfield  {author} {\bibinfo {author} {\bibfnamefont {S.}~\bibnamefont
  {Ghosh}}, \bibinfo {author} {\bibfnamefont {R.}~\bibnamefont {Sharma}},
  \bibinfo {author} {\bibfnamefont {S.}~\bibnamefont {Adhikari}}, \ and\
  \bibinfo {author} {\bibfnamefont {A.~J.~C.}\ \bibnamefont {Varandas}},\
  }\href@noop {} {\bibfield  {journal} {\bibinfo  {journal} {Phys. Chem. Chem.
  Phys.}\ }\textbf {\bibinfo {volume} {21}},\ \bibinfo {pages} {20166}
  (\bibinfo {year} {2019})}\BibitemShut {NoStop}%
\bibitem [{\citenamefont {Tang}\ \emph
  {et~al.}(2005{\natexlab{b}})\citenamefont {Tang}, \citenamefont {Xu},
  \citenamefont {Zhang}, \citenamefont {Hou}, \citenamefont {Chang},
  \citenamefont {Ng}, \citenamefont {Chiu}, \citenamefont {Dressler},\ and\
  \citenamefont {Levandier}}]{heh2+_dyna}%
  \BibitemOpen
  \bibfield  {author} {\bibinfo {author} {\bibfnamefont {X.~N.}\ \bibnamefont
  {Tang}}, \bibinfo {author} {\bibfnamefont {H.}~\bibnamefont {Xu}}, \bibinfo
  {author} {\bibfnamefont {T.}~\bibnamefont {Zhang}}, \bibinfo {author}
  {\bibfnamefont {Y.}~\bibnamefont {Hou}}, \bibinfo {author} {\bibfnamefont
  {C.}~\bibnamefont {Chang}}, \bibinfo {author} {\bibfnamefont {C.~Y.}\
  \bibnamefont {Ng}}, \bibinfo {author} {\bibfnamefont {Y.}~\bibnamefont
  {Chiu}}, \bibinfo {author} {\bibfnamefont {R.~A.}\ \bibnamefont {Dressler}},
  \ and\ \bibinfo {author} {\bibfnamefont {D.~J.}\ \bibnamefont {Levandier}},\
  }\href@noop {} {\bibfield  {journal} {\bibinfo  {journal} {J. Chem. Phys.}\
  }\textbf {\bibinfo {volume} {122}},\ \bibinfo {pages} {164301} (\bibinfo
  {year} {2005}{\natexlab{b}})}\BibitemShut {NoStop}%
\bibitem [{\citenamefont {Ghosh}\ \emph {et~al.}(2015)\citenamefont {Ghosh},
  \citenamefont {Sahoo}, \citenamefont {Adhikari}, \citenamefont {Sharma},\
  and\ \citenamefont {Varandas}}]{jpca-119-12392-SA}%
  \BibitemOpen
  \bibfield  {author} {\bibinfo {author} {\bibfnamefont {S.}~\bibnamefont
  {Ghosh}}, \bibinfo {author} {\bibfnamefont {T.}~\bibnamefont {Sahoo}},
  \bibinfo {author} {\bibfnamefont {S.}~\bibnamefont {Adhikari}}, \bibinfo
  {author} {\bibfnamefont {R.}~\bibnamefont {Sharma}}, \ and\ \bibinfo {author}
  {\bibfnamefont {A.~J.~C.}\ \bibnamefont {Varandas}},\ }\href@noop {}
  {\bibfield  {journal} {\bibinfo  {journal} {J. Phys. Chem. A}\ }\textbf
  {\bibinfo {volume} {119}},\ \bibinfo {pages} {12392} (\bibinfo {year}
  {2015})}\BibitemShut {NoStop}%
\bibitem [{\citenamefont {Born}\ and\ \citenamefont {Oppenheimer}(1927)}]{BO}%
  \BibitemOpen
  \bibfield  {author} {\bibinfo {author} {\bibfnamefont {M.}~\bibnamefont
  {Born}}\ and\ \bibinfo {author} {\bibfnamefont {R.}~\bibnamefont
  {Oppenheimer}},\ }\href@noop {} {\bibfield  {journal} {\bibinfo  {journal}
  {Ann. Phys.}\ }\textbf {\bibinfo {volume} {389}},\ \bibinfo {pages} {457}
  (\bibinfo {year} {1927})}\BibitemShut {NoStop}%
\bibitem [{\citenamefont {Born}\ and\ \citenamefont {Huang}(1954)}]{BH}%
  \BibitemOpen
  \bibfield  {author} {\bibinfo {author} {\bibfnamefont {M.}~\bibnamefont
  {Born}}\ and\ \bibinfo {author} {\bibfnamefont {K.}~\bibnamefont {Huang}},\
  }\href@noop {} {\emph {\bibinfo {title} {Dynamical Theory of Crystal
  Lattices}}}\ (\bibinfo  {publisher} {Oxford University Press, Oxford},\
  \bibinfo {year} {1954})\BibitemShut {NoStop}%
\bibitem [{\citenamefont
  {Baer}(2006{\natexlab{a}})}]{Baer2006_bbo_book_sect211}%
  \BibitemOpen
  \bibfield  {author} {\bibinfo {author} {\bibfnamefont {M.}~\bibnamefont
  {Baer}},\ }\enquote {\bibinfo {title} {{Beyond Born-Oppenheimer: Electronic
  Nonadiabatic Coupling Terms and Conical Intersections}},}\ \ (\bibinfo
  {publisher} {Wiley},\ \bibinfo {address} {Hoboken, NJ, USA},\ \bibinfo {year}
  {2006})\ Chap.\ \bibinfo {chapter} {2.1.1}\BibitemShut {NoStop}%
\bibitem [{\citenamefont {Clusius}(1941)}]{clusius1941einfuhrung}%
  \BibitemOpen
  \bibfield  {author} {\bibinfo {author} {\bibnamefont {Clusius}},\ }\href@noop
  {} {\enquote {\bibinfo {title} {Einf{\"u}hrung in die quantenchemie. von h.
  hellmann. 350 s., 43 abb., 35 tab. franz deuticke, leipzig u. wien 1937. pr.
  geh. rm. 20,-. geb. rm. 22,-},}\ } (\bibinfo {year} {1941})\BibitemShut
  {NoStop}%
\bibitem [{\citenamefont {Feynman}(1939)}]{feynman1939forces}%
  \BibitemOpen
  \bibfield  {author} {\bibinfo {author} {\bibfnamefont {R.~P.}\ \bibnamefont
  {Feynman}},\ }\href@noop {} {\bibfield  {journal} {\bibinfo  {journal} {Phys.
  Rev.}\ }\textbf {\bibinfo {volume} {56}},\ \bibinfo {pages} {340} (\bibinfo
  {year} {1939})}\BibitemShut {NoStop}%
\bibitem [{\citenamefont {Epstein}(1954)}]{epstein1954note}%
  \BibitemOpen
  \bibfield  {author} {\bibinfo {author} {\bibfnamefont {S.~T.}\ \bibnamefont
  {Epstein}},\ }\href@noop {} {\bibfield  {journal} {\bibinfo  {journal} {Am.
  J. Phys.}\ }\textbf {\bibinfo {volume} {22}},\ \bibinfo {pages} {613}
  (\bibinfo {year} {1954})}\BibitemShut {NoStop}%
\bibitem [{\citenamefont {Mukherjee}\ \emph {et~al.}(2024)\citenamefont
  {Mukherjee}, \citenamefont {Saha}, \citenamefont {Ghosh}, \citenamefont
  {Adhikari}, \citenamefont {Sathyamurthy},\ and\ \citenamefont
  {Baer}}]{mukherjee2024quasi}%
  \BibitemOpen
  \bibfield  {author} {\bibinfo {author} {\bibfnamefont {S.}~\bibnamefont
  {Mukherjee}}, \bibinfo {author} {\bibfnamefont {S.}~\bibnamefont {Saha}},
  \bibinfo {author} {\bibfnamefont {S.}~\bibnamefont {Ghosh}}, \bibinfo
  {author} {\bibfnamefont {S.}~\bibnamefont {Adhikari}}, \bibinfo {author}
  {\bibfnamefont {N.}~\bibnamefont {Sathyamurthy}}, \ and\ \bibinfo {author}
  {\bibfnamefont {M.}~\bibnamefont {Baer}},\ }\href@noop {} {\bibfield
  {journal} {\bibinfo  {journal} {J. Phys. Chem. A}\ }\textbf {\bibinfo
  {volume} {128}},\ \bibinfo {pages} {7691} (\bibinfo {year}
  {2024})}\BibitemShut {NoStop}%
\bibitem [{\citenamefont {Baer}(1975{\natexlab{a}})}]{Baer1975_CPL_35}%
  \BibitemOpen
  \bibfield  {author} {\bibinfo {author} {\bibfnamefont {M.}~\bibnamefont
  {Baer}},\ }\href@noop {} {\bibfield  {journal} {\bibinfo  {journal} {Chem.
  Phys. Lett.}\ }\textbf {\bibinfo {volume} {35}},\ \bibinfo {pages} {112}
  (\bibinfo {year} {1975}{\natexlab{a}})}\BibitemShut {NoStop}%
\bibitem [{\citenamefont {Top}\ and\ \citenamefont {Baer}(1977)}]{TB}%
  \BibitemOpen
  \bibfield  {author} {\bibinfo {author} {\bibfnamefont {Z.~H.}\ \bibnamefont
  {Top}}\ and\ \bibinfo {author} {\bibfnamefont {M.}~\bibnamefont {Baer}},\
  }\href@noop {} {\bibfield  {journal} {\bibinfo  {journal} {J. Chem. Phys.}\
  }\textbf {\bibinfo {volume} {66}},\ \bibinfo {pages} {1363} (\bibinfo {year}
  {1977})}\BibitemShut {NoStop}%
\bibitem [{\citenamefont {Sarkar}\ and\ \citenamefont
  {Adhikari}(2006)}]{Sarkar2006_JCP_124_EBO}%
  \BibitemOpen
  \bibfield  {author} {\bibinfo {author} {\bibfnamefont {B.}~\bibnamefont
  {Sarkar}}\ and\ \bibinfo {author} {\bibfnamefont {S.}~\bibnamefont
  {Adhikari}},\ }\href@noop {} {\bibfield  {journal} {\bibinfo  {journal} {J.
  Chem. Phys.}\ }\textbf {\bibinfo {volume} {124}},\ \bibinfo {pages} {074101}
  (\bibinfo {year} {2006})}\BibitemShut {NoStop}%
\bibitem [{\citenamefont {Mukherjee}\ \emph
  {et~al.}(2020{\natexlab{b}})\citenamefont {Mukherjee}, \citenamefont
  {Naskar}, \citenamefont {Mukherjee}, \citenamefont {Ravi}, \citenamefont
  {Shamasundar}, \citenamefont {Mukhopadhyay},\ and\ \citenamefont
  {Adhikari}}]{jcp-153-174301-SA}%
  \BibitemOpen
  \bibfield  {author} {\bibinfo {author} {\bibfnamefont {B.}~\bibnamefont
  {Mukherjee}}, \bibinfo {author} {\bibfnamefont {K.}~\bibnamefont {Naskar}},
  \bibinfo {author} {\bibfnamefont {S.}~\bibnamefont {Mukherjee}}, \bibinfo
  {author} {\bibfnamefont {S.}~\bibnamefont {Ravi}}, \bibinfo {author}
  {\bibfnamefont {K.~R.}\ \bibnamefont {Shamasundar}}, \bibinfo {author}
  {\bibfnamefont {D.}~\bibnamefont {Mukhopadhyay}}, \ and\ \bibinfo {author}
  {\bibfnamefont {S.}~\bibnamefont {Adhikari}},\ }\href@noop {} {\bibfield
  {journal} {\bibinfo  {journal} {J. Chem. Phys.}\ }\textbf {\bibinfo {volume}
  {153}},\ \bibinfo {pages} {174301} (\bibinfo {year}
  {2020}{\natexlab{b}})}\BibitemShut {NoStop}%
\bibitem [{\citenamefont {Werner}\ \emph {et~al.}(2010)\citenamefont {Werner},
  \citenamefont {Knowles}, \citenamefont {Knizia}, \citenamefont {Manby},
  \citenamefont {{Sch\"{u}tz}}, \citenamefont {Celani}, \citenamefont {Korona},
  \citenamefont {Lindh}, \citenamefont {Mitrushenkov}, \citenamefont {Rauhut},
  \citenamefont {Shamasundar}, \citenamefont {Adler}, \citenamefont {Amos},
  \citenamefont {Bernhardsson}, \citenamefont {Berning}, \citenamefont
  {Cooper}, \citenamefont {Deegan}, \citenamefont {Dobbyn}, \citenamefont
  {Eckert}, \citenamefont {Goll}, \citenamefont {Hampel}, \citenamefont
  {Hesselmann}, \citenamefont {Hetzer}, \citenamefont {Hrenar}, \citenamefont
  {Jansen}, \citenamefont {K\"oppl}, \citenamefont {Liu}, \citenamefont
  {Lloyd}, \citenamefont {Mata}, \citenamefont {May}, \citenamefont
  {McNicholas}, \citenamefont {Meyer}, \citenamefont {Mura}, \citenamefont
  {Nicklass}, \citenamefont {O'Neill}, \citenamefont {Palmieri}, \citenamefont
  {Peng}, \citenamefont {Pfl\"uger}, \citenamefont {Pitzer}, \citenamefont
  {Reiher}, \citenamefont {Shiozaki}, \citenamefont {Stoll}, \citenamefont
  {Stone}, \citenamefont {Tarroni}, \citenamefont {Thorsteinsson},\ and\
  \citenamefont {Wang}}]{MOL}%
  \BibitemOpen
  \bibfield  {author} {\bibinfo {author} {\bibfnamefont {H.-J.}\ \bibnamefont
  {Werner}}, \bibinfo {author} {\bibfnamefont {P.~J.}\ \bibnamefont {Knowles}},
  \bibinfo {author} {\bibfnamefont {G.}~\bibnamefont {Knizia}}, \bibinfo
  {author} {\bibfnamefont {F.~R.}\ \bibnamefont {Manby}}, \bibinfo {author}
  {\bibfnamefont {M.}~\bibnamefont {{Sch\"{u}tz}}}, \bibinfo {author}
  {\bibfnamefont {P.}~\bibnamefont {Celani}}, \bibinfo {author} {\bibfnamefont
  {T.}~\bibnamefont {Korona}}, \bibinfo {author} {\bibfnamefont
  {R.}~\bibnamefont {Lindh}}, \bibinfo {author} {\bibfnamefont
  {A.}~\bibnamefont {Mitrushenkov}}, \bibinfo {author} {\bibfnamefont
  {G.}~\bibnamefont {Rauhut}}, \bibinfo {author} {\bibfnamefont {K.~R.}\
  \bibnamefont {Shamasundar}}, \bibinfo {author} {\bibfnamefont {T.~B.}\
  \bibnamefont {Adler}}, \bibinfo {author} {\bibfnamefont {R.~D.}\ \bibnamefont
  {Amos}}, \bibinfo {author} {\bibfnamefont {A.}~\bibnamefont {Bernhardsson}},
  \bibinfo {author} {\bibfnamefont {A.}~\bibnamefont {Berning}}, \bibinfo
  {author} {\bibfnamefont {D.~L.}\ \bibnamefont {Cooper}}, \bibinfo {author}
  {\bibfnamefont {M.~J.~O.}\ \bibnamefont {Deegan}}, \bibinfo {author}
  {\bibfnamefont {A.~J.}\ \bibnamefont {Dobbyn}}, \bibinfo {author}
  {\bibfnamefont {F.}~\bibnamefont {Eckert}}, \bibinfo {author} {\bibfnamefont
  {E.}~\bibnamefont {Goll}}, \bibinfo {author} {\bibfnamefont {C.}~\bibnamefont
  {Hampel}}, \bibinfo {author} {\bibfnamefont {A.}~\bibnamefont {Hesselmann}},
  \bibinfo {author} {\bibfnamefont {G.}~\bibnamefont {Hetzer}}, \bibinfo
  {author} {\bibfnamefont {T.}~\bibnamefont {Hrenar}}, \bibinfo {author}
  {\bibfnamefont {G.}~\bibnamefont {Jansen}}, \bibinfo {author} {\bibfnamefont
  {C.}~\bibnamefont {K\"oppl}}, \bibinfo {author} {\bibfnamefont
  {Y.}~\bibnamefont {Liu}}, \bibinfo {author} {\bibfnamefont {A.~W.}\
  \bibnamefont {Lloyd}}, \bibinfo {author} {\bibfnamefont {R.~A.}\ \bibnamefont
  {Mata}}, \bibinfo {author} {\bibfnamefont {A.~J.}\ \bibnamefont {May}},
  \bibinfo {author} {\bibfnamefont {S.~J.}\ \bibnamefont {McNicholas}},
  \bibinfo {author} {\bibfnamefont {W.}~\bibnamefont {Meyer}}, \bibinfo
  {author} {\bibfnamefont {M.~E.}\ \bibnamefont {Mura}}, \bibinfo {author}
  {\bibfnamefont {A.}~\bibnamefont {Nicklass}}, \bibinfo {author}
  {\bibfnamefont {D.~P.}\ \bibnamefont {O'Neill}}, \bibinfo {author}
  {\bibfnamefont {P.}~\bibnamefont {Palmieri}}, \bibinfo {author}
  {\bibfnamefont {D.}~\bibnamefont {Peng}}, \bibinfo {author} {\bibfnamefont
  {K.}~\bibnamefont {Pfl\"uger}}, \bibinfo {author} {\bibfnamefont
  {R.}~\bibnamefont {Pitzer}}, \bibinfo {author} {\bibfnamefont
  {M.}~\bibnamefont {Reiher}}, \bibinfo {author} {\bibfnamefont
  {T.}~\bibnamefont {Shiozaki}}, \bibinfo {author} {\bibfnamefont
  {H.}~\bibnamefont {Stoll}}, \bibinfo {author} {\bibfnamefont {A.~J.}\
  \bibnamefont {Stone}}, \bibinfo {author} {\bibfnamefont {R.}~\bibnamefont
  {Tarroni}}, \bibinfo {author} {\bibfnamefont {T.}~\bibnamefont
  {Thorsteinsson}}, \ and\ \bibinfo {author} {\bibfnamefont {M.}~\bibnamefont
  {Wang}},\ }\href@noop {} {\enquote {\bibinfo {title} {Molpro, version 2010.1,
  a package of ab initio programs},}\ } (\bibinfo {year} {2010}),\ \bibinfo
  {note} {see http://www.molpro.net}\BibitemShut {NoStop}%
\bibitem [{\citenamefont {Yarkony}(1996)}]{Yarkony_jpc100}%
  \BibitemOpen
  \bibfield  {author} {\bibinfo {author} {\bibfnamefont {D.~R.}\ \bibnamefont
  {Yarkony}},\ }\href@noop {} {\bibfield  {journal} {\bibinfo  {journal} {J.
  Phys. Chem.}\ }\textbf {\bibinfo {volume} {100}},\ \bibinfo {pages} {17439}
  (\bibinfo {year} {1996})}\BibitemShut {NoStop}%
\bibitem [{\citenamefont {Alexander}, \citenamefont {Manolopoulos},\ and\
  \citenamefont {Werner}(2000{\natexlab{a}})}]{jcp-113-11084-WERNER}%
  \BibitemOpen
  \bibfield  {author} {\bibinfo {author} {\bibfnamefont {M.~H.}\ \bibnamefont
  {Alexander}}, \bibinfo {author} {\bibfnamefont {D.~E.}\ \bibnamefont
  {Manolopoulos}}, \ and\ \bibinfo {author} {\bibfnamefont {H.-J.}\
  \bibnamefont {Werner}},\ }\href@noop {} {\bibfield  {journal} {\bibinfo
  {journal} {J. Chem. Phys.}\ }\textbf {\bibinfo {volume} {113}},\ \bibinfo
  {pages} {11084} (\bibinfo {year} {2000}{\natexlab{a}})}\BibitemShut {NoStop}%
\bibitem [{\citenamefont {Baer}(2006{\natexlab{b}})}]{BaerBook2006}%
  \BibitemOpen
  \bibfield  {author} {\bibinfo {author} {\bibfnamefont {M.}~\bibnamefont
  {Baer}},\ }\href {\doibase 10.1002/0471780081} {\emph {\bibinfo {title}
  {{Beyond Born - Oppenheimer: Conical Intersections and Electronic
  Nonadiabatic Coupling Terms}}}}\ (\bibinfo {year} {2006})\BibitemShut
  {NoStop}%
\bibitem [{\citenamefont {Baer}(2002)}]{Baer2002_PhysRep}%
  \BibitemOpen
  \bibfield  {author} {\bibinfo {author} {\bibfnamefont {M.}~\bibnamefont
  {Baer}},\ }\href {\doibase 10.1016/S0370-1573(01)00052-7} {\bibfield
  {journal} {\bibinfo  {journal} {Phys. Rep.}\ }\textbf {\bibinfo {volume}
  {358}},\ \bibinfo {pages} {75} (\bibinfo {year} {2002})}\BibitemShut
  {NoStop}%
\bibitem [{\citenamefont {Baer}(1975{\natexlab{b}})}]{baer_cpl35}%
  \BibitemOpen
  \bibfield  {author} {\bibinfo {author} {\bibfnamefont {M.}~\bibnamefont
  {Baer}},\ }\href@noop {} {\bibfield  {journal} {\bibinfo  {journal} {Chem.
  Phys. Lett.}\ }\textbf {\bibinfo {volume} {35}},\ \bibinfo {pages} {112}
  (\bibinfo {year} {1975}{\natexlab{b}})}\BibitemShut {NoStop}%
\bibitem [{\citenamefont {Mukherjee}\ \emph {et~al.}(2019)\citenamefont
  {Mukherjee}, \citenamefont {Naskar}, \citenamefont {Mukherjee}, \citenamefont
  {Ghosh}, \citenamefont {Sahoo},\ and\ \citenamefont
  {Adhikari}}]{irpc-38-287-SA}%
  \BibitemOpen
  \bibfield  {author} {\bibinfo {author} {\bibfnamefont {B.}~\bibnamefont
  {Mukherjee}}, \bibinfo {author} {\bibfnamefont {K.}~\bibnamefont {Naskar}},
  \bibinfo {author} {\bibfnamefont {S.}~\bibnamefont {Mukherjee}}, \bibinfo
  {author} {\bibfnamefont {S.}~\bibnamefont {Ghosh}}, \bibinfo {author}
  {\bibfnamefont {T.}~\bibnamefont {Sahoo}}, \ and\ \bibinfo {author}
  {\bibfnamefont {S.}~\bibnamefont {Adhikari}},\ }\href@noop {} {\bibfield
  {journal} {\bibinfo  {journal} {Int. Rev. Phys. Chem.}\ }\textbf {\bibinfo
  {volume} {38}},\ \bibinfo {pages} {287} (\bibinfo {year} {2019})}\BibitemShut
  {NoStop}%
\bibitem [{\citenamefont {Dutta}\ \emph
  {et~al.}(2020{\natexlab{a}})\citenamefont {Dutta}, \citenamefont {Mukherjee},
  \citenamefont {Naskar}, \citenamefont {Ghosh}, \citenamefont {Mukherjee},
  \citenamefont {Ravi},\ and\ \citenamefont {Adhikari}}]{pccp-22-27496-SA}%
  \BibitemOpen
  \bibfield  {author} {\bibinfo {author} {\bibfnamefont {J.}~\bibnamefont
  {Dutta}}, \bibinfo {author} {\bibfnamefont {S.}~\bibnamefont {Mukherjee}},
  \bibinfo {author} {\bibfnamefont {K.}~\bibnamefont {Naskar}}, \bibinfo
  {author} {\bibfnamefont {S.}~\bibnamefont {Ghosh}}, \bibinfo {author}
  {\bibfnamefont {B.}~\bibnamefont {Mukherjee}}, \bibinfo {author}
  {\bibfnamefont {S.}~\bibnamefont {Ravi}}, \ and\ \bibinfo {author}
  {\bibfnamefont {S.}~\bibnamefont {Adhikari}},\ }\href@noop {} {\bibfield
  {journal} {\bibinfo  {journal} {Phys. Chem. Chem. Phys.}\ }\textbf {\bibinfo
  {volume} {22}},\ \bibinfo {pages} {27496} (\bibinfo {year}
  {2020}{\natexlab{a}})}\BibitemShut {NoStop}%
\bibitem [{\citenamefont {Naskar}\ \emph
  {et~al.}(2020{\natexlab{a}})\citenamefont {Naskar}, \citenamefont
  {Mukherjee}, \citenamefont {Mukherjee}, \citenamefont {Ravi}, \citenamefont
  {Mukherjee}, \citenamefont {Sardar},\ and\ \citenamefont
  {Adhikari}}]{jctc-16-1666-SA}%
  \BibitemOpen
  \bibfield  {author} {\bibinfo {author} {\bibfnamefont {K.}~\bibnamefont
  {Naskar}}, \bibinfo {author} {\bibfnamefont {S.}~\bibnamefont {Mukherjee}},
  \bibinfo {author} {\bibfnamefont {B.}~\bibnamefont {Mukherjee}}, \bibinfo
  {author} {\bibfnamefont {S.}~\bibnamefont {Ravi}}, \bibinfo {author}
  {\bibfnamefont {S.}~\bibnamefont {Mukherjee}}, \bibinfo {author}
  {\bibfnamefont {S.}~\bibnamefont {Sardar}}, \ and\ \bibinfo {author}
  {\bibfnamefont {S.}~\bibnamefont {Adhikari}},\ }\href@noop {} {\bibfield
  {journal} {\bibinfo  {journal} {J. Chem. Theo. Comput.}\ }\textbf {\bibinfo
  {volume} {16}},\ \bibinfo {pages} {1666} (\bibinfo {year}
  {2020}{\natexlab{a}})}\BibitemShut {NoStop}%
\bibitem [{\citenamefont {Johnson}(1980)}]{Johnson1980_JCP_73}%
  \BibitemOpen
  \bibfield  {author} {\bibinfo {author} {\bibfnamefont {B.~R.}\ \bibnamefont
  {Johnson}},\ }\href@noop {} {\bibfield  {journal} {\bibinfo  {journal} {J.
  Chem. Phys.}\ }\textbf {\bibinfo {volume} {73}},\ \bibinfo {pages} {5051}
  (\bibinfo {year} {1980})}\BibitemShut {NoStop}%
\bibitem [{\citenamefont {Kosloff}\ and\ \citenamefont
  {Kosloff}(1986)}]{jcomphy-52-35-DKRR}%
  \BibitemOpen
  \bibfield  {author} {\bibinfo {author} {\bibfnamefont {D.}~\bibnamefont
  {Kosloff}}\ and\ \bibinfo {author} {\bibfnamefont {R.}~\bibnamefont
  {Kosloff}},\ }\href@noop {} {\bibfield  {journal} {\bibinfo  {journal} {J.
  Comput. Phys.}\ }\textbf {\bibinfo {volume} {52}},\ \bibinfo {pages} {35}
  (\bibinfo {year} {1986})}\BibitemShut {NoStop}%
\bibitem [{\citenamefont {Park}\ and\ \citenamefont
  {Light}(1986)}]{jcp-85-5870-TJPR}%
  \BibitemOpen
  \bibfield  {author} {\bibinfo {author} {\bibfnamefont {T.~J.}\ \bibnamefont
  {Park}}\ and\ \bibinfo {author} {\bibfnamefont {J.~C.}\ \bibnamefont
  {Light}},\ }\href@noop {} {\bibfield  {journal} {\bibinfo  {journal} {J.
  Chem. Phys.}\ }\textbf {\bibinfo {volume} {85}},\ \bibinfo {pages} {5870}
  (\bibinfo {year} {1986})}\BibitemShut {NoStop}%
\bibitem [{\citenamefont {Aoiz}\ \emph {et~al.}(1996)\citenamefont {Aoiz},
  \citenamefont {Banares}, \citenamefont {Herrero}, \citenamefont {Stark},\
  and\ \citenamefont {Werner}}]{cpl-254-341-AOIZ}%
  \BibitemOpen
  \bibfield  {author} {\bibinfo {author} {\bibfnamefont {F.~J.}\ \bibnamefont
  {Aoiz}}, \bibinfo {author} {\bibfnamefont {L.}~\bibnamefont {Banares}},
  \bibinfo {author} {\bibfnamefont {V.}~\bibnamefont {Herrero}}, \bibinfo
  {author} {\bibfnamefont {K.}~\bibnamefont {Stark}}, \ and\ \bibinfo {author}
  {\bibfnamefont {H.-J.}\ \bibnamefont {Werner}},\ }\href@noop {} {\bibfield
  {journal} {\bibinfo  {journal} {Chem. Phys. Lett.}\ }\textbf {\bibinfo
  {volume} {254}},\ \bibinfo {pages} {341} (\bibinfo {year}
  {1996})}\BibitemShut {NoStop}%
\bibitem [{\citenamefont {Aoiz}\ \emph {et~al.}(1994)\citenamefont {Aoiz},
  \citenamefont {Banares}, \citenamefont {Herrero}, \citenamefont {Rabanos},
  \citenamefont {K.Stark},\ and\ \citenamefont
  {H.-J.Werner}}]{cpl-223-215-AOIZ}%
  \BibitemOpen
  \bibfield  {author} {\bibinfo {author} {\bibfnamefont {F.~J.}\ \bibnamefont
  {Aoiz}}, \bibinfo {author} {\bibfnamefont {L.}~\bibnamefont {Banares}},
  \bibinfo {author} {\bibfnamefont {V.~J.}\ \bibnamefont {Herrero}}, \bibinfo
  {author} {\bibfnamefont {V.}~\bibnamefont {Rabanos}}, \bibinfo {author}
  {\bibnamefont {K.Stark}}, \ and\ \bibinfo {author} {\bibnamefont
  {H.-J.Werner}},\ }\href@noop {} {\bibfield  {journal} {\bibinfo  {journal}
  {Chem. Phys. Lett.}\ }\textbf {\bibinfo {volume} {223}},\ \bibinfo {pages}
  {215} (\bibinfo {year} {1994})}\BibitemShut {NoStop}%
\bibitem [{\citenamefont {Stark}\ and\ \citenamefont
  {Werner}(1996{\natexlab{a}})}]{SW_jcp104}%
  \BibitemOpen
  \bibfield  {author} {\bibinfo {author} {\bibfnamefont {K.}~\bibnamefont
  {Stark}}\ and\ \bibinfo {author} {\bibfnamefont {H.-J.}\ \bibnamefont
  {Werner}},\ }\href@noop {} {\bibfield  {journal} {\bibinfo  {journal} {J.
  Chem. Phys.}\ }\textbf {\bibinfo {volume} {104}},\ \bibinfo {pages} {6515}
  (\bibinfo {year} {1996}{\natexlab{a}})}\BibitemShut {NoStop}%
\bibitem [{\citenamefont {Castillo}\ \emph
  {et~al.}(1996{\natexlab{b}})\citenamefont {Castillo}, \citenamefont
  {Manolopoulos}, \citenamefont {Stark},\ and\ \citenamefont
  {Werner}}]{jcp-104-6531-CAST}%
  \BibitemOpen
  \bibfield  {author} {\bibinfo {author} {\bibfnamefont {J.~F.}\ \bibnamefont
  {Castillo}}, \bibinfo {author} {\bibfnamefont {D.~E.}\ \bibnamefont
  {Manolopoulos}}, \bibinfo {author} {\bibfnamefont {K.}~\bibnamefont {Stark}},
  \ and\ \bibinfo {author} {\bibfnamefont {H.~J.}\ \bibnamefont {Werner}},\
  }\href@noop {} {\bibfield  {journal} {\bibinfo  {journal} {J. Chem. Phys.}\
  }\textbf {\bibinfo {volume} {104}},\ \bibinfo {pages} {6531} (\bibinfo {year}
  {1996}{\natexlab{b}})}\BibitemShut {NoStop}%
\bibitem [{\citenamefont {Hartke}\ and\ \citenamefont
  {Werner}(1997{\natexlab{a}})}]{HSW_cpl280}%
  \BibitemOpen
  \bibfield  {author} {\bibinfo {author} {\bibfnamefont {B.}~\bibnamefont
  {Hartke}}\ and\ \bibinfo {author} {\bibfnamefont {H.-J.}\ \bibnamefont
  {Werner}},\ }\href@noop {} {\bibfield  {journal} {\bibinfo  {journal} {Chem.
  Phys. Lett.}\ }\textbf {\bibinfo {volume} {280}},\ \bibinfo {pages} {430}
  (\bibinfo {year} {1997}{\natexlab{a}})}\BibitemShut {NoStop}%
\bibitem [{\citenamefont {Alexander}, \citenamefont {Manolopoulos},\ and\
  \citenamefont {Werner}(2000{\natexlab{b}})}]{ASW_jcp113}%
  \BibitemOpen
  \bibfield  {author} {\bibinfo {author} {\bibfnamefont {M.~H.}\ \bibnamefont
  {Alexander}}, \bibinfo {author} {\bibfnamefont {D.~E.}\ \bibnamefont
  {Manolopoulos}}, \ and\ \bibinfo {author} {\bibfnamefont {H.-J.}\
  \bibnamefont {Werner}},\ }\href@noop {} {\bibfield  {journal} {\bibinfo
  {journal} {J. Chem. Phys.}\ }\textbf {\bibinfo {volume} {113}},\ \bibinfo
  {pages} {11084} (\bibinfo {year} {2000}{\natexlab{b}})}\BibitemShut {NoStop}%
\bibitem [{\citenamefont {Li}\ \emph {et~al.}(2007)\citenamefont {Li},
  \citenamefont {Werner}, \citenamefont {Lique},\ and\ \citenamefont
  {Alexander}}]{Werner_jcp127}%
  \BibitemOpen
  \bibfield  {author} {\bibinfo {author} {\bibfnamefont {G.}~\bibnamefont
  {Li}}, \bibinfo {author} {\bibfnamefont {H.-J.}\ \bibnamefont {Werner}},
  \bibinfo {author} {\bibfnamefont {F.}~\bibnamefont {Lique}}, \ and\ \bibinfo
  {author} {\bibfnamefont {M.~H.}\ \bibnamefont {Alexander}},\ }\href@noop {}
  {\bibfield  {journal} {\bibinfo  {journal} {J. Chem. Phys.}\ }\textbf
  {\bibinfo {volume} {127}},\ \bibinfo {pages} {174302} (\bibinfo {year}
  {2007})}\BibitemShut {NoStop}%
\bibitem [{\citenamefont {Xu}, \citenamefont {Xie},\ and\ \citenamefont
  {Zhang}(2006)}]{cjcp-19-96-XU}%
  \BibitemOpen
  \bibfield  {author} {\bibinfo {author} {\bibfnamefont {C.}~\bibnamefont
  {Xu}}, \bibinfo {author} {\bibfnamefont {D.}~\bibnamefont {Xie}}, \ and\
  \bibinfo {author} {\bibfnamefont {D.~H.}\ \bibnamefont {Zhang}},\ }\href@noop
  {} {\bibfield  {journal} {\bibinfo  {journal} {Chin. J. Chem. Phys.}\
  }\textbf {\bibinfo {volume} {19}},\ \bibinfo {pages} {96} (\bibinfo {year}
  {2006})}\BibitemShut {NoStop}%
\bibitem [{\citenamefont {Li}, \citenamefont {Song},\ and\ \citenamefont
  {Varandas}(2015)}]{epjd-69-22-AJCV}%
  \BibitemOpen
  \bibfield  {author} {\bibinfo {author} {\bibfnamefont {Y.-Q.}\ \bibnamefont
  {Li}}, \bibinfo {author} {\bibfnamefont {Y.-Z.}\ \bibnamefont {Song}}, \ and\
  \bibinfo {author} {\bibfnamefont {A.~J.~C.}\ \bibnamefont {Varandas}},\
  }\href@noop {} {\bibfield  {journal} {\bibinfo  {journal} {Eur. Phys. J. D}\
  }\textbf {\bibinfo {volume} {69}},\ \bibinfo {pages} {22} (\bibinfo {year}
  {2015})}\BibitemShut {NoStop}%
\bibitem [{\citenamefont {Das}\ \emph {et~al.}(2012)\citenamefont {Das},
  \citenamefont {Mukhopadhyay}, \citenamefont {Adhikari},\ and\ \citenamefont
  {Baer}}]{ijqc-112-2561-ANITA}%
  \BibitemOpen
  \bibfield  {author} {\bibinfo {author} {\bibfnamefont {A.}~\bibnamefont
  {Das}}, \bibinfo {author} {\bibfnamefont {D.}~\bibnamefont {Mukhopadhyay}},
  \bibinfo {author} {\bibfnamefont {S.}~\bibnamefont {Adhikari}}, \ and\
  \bibinfo {author} {\bibfnamefont {M.}~\bibnamefont {Baer}},\ }\href@noop {}
  {\bibfield  {journal} {\bibinfo  {journal} {Int. J. Quant. Chem.}\ }\textbf
  {\bibinfo {volume} {112}},\ \bibinfo {pages} {2561} (\bibinfo {year}
  {2012})}\BibitemShut {NoStop}%
\bibitem [{\citenamefont {Dutta}\ \emph
  {et~al.}(2020{\natexlab{b}})\citenamefont {Dutta}, \citenamefont {Mukherjee},
  \citenamefont {Naskar}, \citenamefont {Ghosh}, \citenamefont {Mukherjee},
  \citenamefont {Ravi},\ and\ \citenamefont {Adhikari}}]{joy_pccp}%
  \BibitemOpen
  \bibfield  {author} {\bibinfo {author} {\bibfnamefont {J.}~\bibnamefont
  {Dutta}}, \bibinfo {author} {\bibfnamefont {S.}~\bibnamefont {Mukherjee}},
  \bibinfo {author} {\bibfnamefont {K.}~\bibnamefont {Naskar}}, \bibinfo
  {author} {\bibfnamefont {S.}~\bibnamefont {Ghosh}}, \bibinfo {author}
  {\bibfnamefont {B.}~\bibnamefont {Mukherjee}}, \bibinfo {author}
  {\bibfnamefont {S.}~\bibnamefont {Ravi}}, \ and\ \bibinfo {author}
  {\bibfnamefont {S.}~\bibnamefont {Adhikari}},\ }\href@noop {} {\bibfield
  {journal} {\bibinfo  {journal} {Phys. Chem. Chem. Phys. (Perspective)}\
  }\textbf {\bibinfo {volume} {22}},\ \bibinfo {pages} {27496} (\bibinfo {year}
  {2020}{\natexlab{b}})}\BibitemShut {NoStop}%
\bibitem [{\citenamefont {Preston}\ and\ \citenamefont
  {Tully}(1970)}]{jcp-54-4297-TUL}%
  \BibitemOpen
  \bibfield  {author} {\bibinfo {author} {\bibfnamefont {K.}~\bibnamefont
  {Preston}}\ and\ \bibinfo {author} {\bibfnamefont {J.~C.}\ \bibnamefont
  {Tully}},\ }\href@noop {} {\bibfield  {journal} {\bibinfo  {journal} {J.
  Chem. Phys.}\ }\textbf {\bibinfo {volume} {54}},\ \bibinfo {pages} {4297}
  (\bibinfo {year} {1970})}\BibitemShut {NoStop}%
\bibitem [{\citenamefont {Kamisaka}\ \emph
  {et~al.}(2002{\natexlab{a}})\citenamefont {Kamisaka}, \citenamefont {Bian},
  \citenamefont {Nobusada},\ and\ \citenamefont
  {Nakamura}}]{Kamisaka_JCP_2002}%
  \BibitemOpen
  \bibfield  {author} {\bibinfo {author} {\bibfnamefont {H.}~\bibnamefont
  {Kamisaka}}, \bibinfo {author} {\bibfnamefont {W.}~\bibnamefont {Bian}},
  \bibinfo {author} {\bibfnamefont {K.}~\bibnamefont {Nobusada}}, \ and\
  \bibinfo {author} {\bibfnamefont {H.}~\bibnamefont {Nakamura}},\ }\href@noop
  {} {\bibfield  {journal} {\bibinfo  {journal} {J. Chem. Phys.}\ }\textbf
  {\bibinfo {volume} {116}},\ \bibinfo {pages} {654} (\bibinfo {year}
  {2002}{\natexlab{a}})}\BibitemShut {NoStop}%
\bibitem [{\citenamefont {Viegas}, \citenamefont {Alijah},\ and\ \citenamefont
  {Varandas}(2007{\natexlab{a}})}]{jcp-126-074309-AJCV}%
  \BibitemOpen
  \bibfield  {author} {\bibinfo {author} {\bibfnamefont {L.~P.}\ \bibnamefont
  {Viegas}}, \bibinfo {author} {\bibfnamefont {A.}~\bibnamefont {Alijah}}, \
  and\ \bibinfo {author} {\bibfnamefont {A.~J.~C.}\ \bibnamefont {Varandas}},\
  }\href@noop {} {\bibfield  {journal} {\bibinfo  {journal} {J. Chem. Phys.}\
  }\textbf {\bibinfo {volume} {126}},\ \bibinfo {pages} {074309/1} (\bibinfo
  {year} {2007}{\natexlab{a}})}\BibitemShut {NoStop}%
\bibitem [{\citenamefont {Mukherjee}, \citenamefont {Mukhopadhyay},\ and\
  \citenamefont {Adhikari}(2014{\natexlab{b}})}]{jcp-141-204306-SA}%
  \BibitemOpen
  \bibfield  {author} {\bibinfo {author} {\bibfnamefont {S.}~\bibnamefont
  {Mukherjee}}, \bibinfo {author} {\bibfnamefont {D.}~\bibnamefont
  {Mukhopadhyay}}, \ and\ \bibinfo {author} {\bibfnamefont {S.}~\bibnamefont
  {Adhikari}},\ }\href@noop {} {\bibfield  {journal} {\bibinfo  {journal} {J.
  Chem. Phys.}\ }\textbf {\bibinfo {volume} {141}},\ \bibinfo {pages}
  {204306/1} (\bibinfo {year} {2014}{\natexlab{b}})}\BibitemShut {NoStop}%
\bibitem [{\citenamefont {Dunning}(1989)}]{dunning_basis}%
  \BibitemOpen
  \bibfield  {author} {\bibinfo {author} {\bibfnamefont {T.~H.}\ \bibnamefont
  {Dunning}},\ }\href@noop {} {\bibfield  {journal} {\bibinfo  {journal} {J.
  Chem. Phys.}\ }\textbf {\bibinfo {volume} {90}},\ \bibinfo {pages} {1007}
  (\bibinfo {year} {1989})}\BibitemShut {NoStop}%
\bibitem [{\citenamefont {Naskar}\ \emph
  {et~al.}(2020{\natexlab{b}})\citenamefont {Naskar}, \citenamefont
  {Mukherjee}, \citenamefont {Mukherjee}, \citenamefont {Ravi}, \citenamefont
  {Mukherjee}, \citenamefont {Sardar},\ and\ \citenamefont
  {Adhikari}}]{ADT_JCTC_16}%
  \BibitemOpen
  \bibfield  {author} {\bibinfo {author} {\bibfnamefont {K.}~\bibnamefont
  {Naskar}}, \bibinfo {author} {\bibfnamefont {S.}~\bibnamefont {Mukherjee}},
  \bibinfo {author} {\bibfnamefont {B.}~\bibnamefont {Mukherjee}}, \bibinfo
  {author} {\bibfnamefont {S.}~\bibnamefont {Ravi}}, \bibinfo {author}
  {\bibfnamefont {S.}~\bibnamefont {Mukherjee}}, \bibinfo {author}
  {\bibfnamefont {S.}~\bibnamefont {Sardar}}, \ and\ \bibinfo {author}
  {\bibfnamefont {S.}~\bibnamefont {Adhikari}},\ }\href@noop {} {\bibfield
  {journal} {\bibinfo  {journal} {J. Chem. Theo. and Comp.}\ }\textbf {\bibinfo
  {volume} {16}},\ \bibinfo {pages} {1666} (\bibinfo {year}
  {2020}{\natexlab{b}})}\BibitemShut {NoStop}%
\bibitem [{\citenamefont {Mukherjee}, \citenamefont {Mukhopadhyay},\ and\
  \citenamefont {Adhikari}(2014{\natexlab{c}})}]{mukherjee_jcp141}%
  \BibitemOpen
  \bibfield  {author} {\bibinfo {author} {\bibfnamefont {S.}~\bibnamefont
  {Mukherjee}}, \bibinfo {author} {\bibfnamefont {D.}~\bibnamefont
  {Mukhopadhyay}}, \ and\ \bibinfo {author} {\bibfnamefont {S.}~\bibnamefont
  {Adhikari}},\ }\href@noop {} {\bibfield  {journal} {\bibinfo  {journal} {J.
  Chem. Phys.}\ }\textbf {\bibinfo {volume} {141}},\ \bibinfo {pages} {204306}
  (\bibinfo {year} {2014}{\natexlab{c}})}\BibitemShut {NoStop}%
\bibitem [{\citenamefont {Russek}\ and\ \citenamefont
  {Furlan}(1989)}]{Russek1989}%
  \BibitemOpen
  \bibfield  {author} {\bibinfo {author} {\bibfnamefont {A.}~\bibnamefont
  {Russek}}\ and\ \bibinfo {author} {\bibfnamefont {R.~J.}\ \bibnamefont
  {Furlan}},\ }\href {\doibase 10.1103/PhysRevA.39.5034} {\bibfield  {journal}
  {\bibinfo  {journal} {Phys. Rev. A}\ }\textbf {\bibinfo {volume} {39}},\
  \bibinfo {pages} {5034} (\bibinfo {year} {1989})}\BibitemShut {NoStop}%
\bibitem [{\citenamefont {Furlan}, \citenamefont {Bent},\ and\ \citenamefont
  {Russek}(1990)}]{Furlan1990}%
  \BibitemOpen
  \bibfield  {author} {\bibinfo {author} {\bibfnamefont {R.~J.}\ \bibnamefont
  {Furlan}}, \bibinfo {author} {\bibfnamefont {G.}~\bibnamefont {Bent}}, \ and\
  \bibinfo {author} {\bibfnamefont {A.}~\bibnamefont {Russek}},\ }\href
  {\doibase 10.1063/1.458936} {\bibfield  {journal} {\bibinfo  {journal} {J.
  Chem. Phys.}\ }\textbf {\bibinfo {volume} {93}},\ \bibinfo {pages} {6676}
  (\bibinfo {year} {1990})}\BibitemShut {NoStop}%
\bibitem [{\citenamefont {Schneider}\ and\ \citenamefont
  {Z\"{u}licke}(1979)}]{Schneider1979}%
  \BibitemOpen
  \bibfield  {author} {\bibinfo {author} {\bibfnamefont {F.}~\bibnamefont
  {Schneider}}\ and\ \bibinfo {author} {\bibfnamefont {L.}~\bibnamefont
  {Z\"{u}licke}},\ }\href {\doibase 10.1016/0009-2614(79)85212-4} {\bibfield
  {journal} {\bibinfo  {journal} {Chem. Phys. Lett.}\ }\textbf {\bibinfo
  {volume} {67}},\ \bibinfo {pages} {491} (\bibinfo {year} {1979})}\BibitemShut
  {NoStop}%
\bibitem [{\citenamefont {Schneider}\ \emph {et~al.}(1983)\citenamefont
  {Schneider}, \citenamefont {Z\"{u}licke}, \citenamefont {Pol\'{a}k},\ and\
  \citenamefont {Vojt\'{i}k}}]{Schneider1983}%
  \BibitemOpen
  \bibfield  {author} {\bibinfo {author} {\bibfnamefont {F.}~\bibnamefont
  {Schneider}}, \bibinfo {author} {\bibfnamefont {L.}~\bibnamefont
  {Z\"{u}licke}}, \bibinfo {author} {\bibfnamefont {R.}~\bibnamefont
  {Pol\'{a}k}}, \ and\ \bibinfo {author} {\bibfnamefont {J.}~\bibnamefont
  {Vojt\'{i}k}},\ }\href {\doibase 10.1016/0301-0104(83)85040-X} {\bibfield
  {journal} {\bibinfo  {journal} {Chem. Phys.}\ }\textbf {\bibinfo {volume}
  {76}},\ \bibinfo {pages} {259} (\bibinfo {year} {1983})}\BibitemShut
  {NoStop}%
\bibitem [{\citenamefont {Aguado}, \citenamefont {Su\'{a}rez},\ and\
  \citenamefont {Paniagua}(1993)}]{Aguado1993}%
  \BibitemOpen
  \bibfield  {author} {\bibinfo {author} {\bibfnamefont {A.}~\bibnamefont
  {Aguado}}, \bibinfo {author} {\bibfnamefont {C.}~\bibnamefont {Su\'{a}rez}},
  \ and\ \bibinfo {author} {\bibfnamefont {M.}~\bibnamefont {Paniagua}},\
  }\href {\doibase 10.1063/1.464676} {\bibfield  {journal} {\bibinfo  {journal}
  {J. Chem. Phys.}\ }\textbf {\bibinfo {volume} {98}},\ \bibinfo {pages} {308}
  (\bibinfo {year} {1993})}\BibitemShut {NoStop}%
\bibitem [{\citenamefont {Sidis}(1996)}]{Sidis1996}%
  \BibitemOpen
  \bibfield  {author} {\bibinfo {author} {\bibfnamefont {V.}~\bibnamefont
  {Sidis}},\ }\href {\doibase 10.1016/0301-0104(96)00060-2} {\bibfield
  {journal} {\bibinfo  {journal} {Chem. Phys.}\ }\textbf {\bibinfo {volume}
  {209}},\ \bibinfo {pages} {313} (\bibinfo {year} {1996})}\BibitemShut
  {NoStop}%
\bibitem [{\citenamefont {Ravi}\ \emph
  {et~al.}(2021{\natexlab{b}})\citenamefont {Ravi}, \citenamefont {Mukherjee},
  \citenamefont {Mukherjee}, \citenamefont {Adhikari}, \citenamefont
  {Sathyamurthy},\ and\ \citenamefont {Baer}}]{Ravi2021_molphys_119}%
  \BibitemOpen
  \bibfield  {author} {\bibinfo {author} {\bibfnamefont {S.}~\bibnamefont
  {Ravi}}, \bibinfo {author} {\bibfnamefont {S.}~\bibnamefont {Mukherjee}},
  \bibinfo {author} {\bibfnamefont {B.}~\bibnamefont {Mukherjee}}, \bibinfo
  {author} {\bibfnamefont {S.}~\bibnamefont {Adhikari}}, \bibinfo {author}
  {\bibfnamefont {N.}~\bibnamefont {Sathyamurthy}}, \ and\ \bibinfo {author}
  {\bibfnamefont {M.}~\bibnamefont {Baer}},\ }\href {\doibase
  10.1080/00268976.2020.1811907} {\bibfield  {journal} {\bibinfo  {journal}
  {Mol. Phys.}\ }\textbf {\bibinfo {volume} {119}},\ \bibinfo {pages}
  {e1811907} (\bibinfo {year} {2021}{\natexlab{b}})}\BibitemShut {NoStop}%
\bibitem [{\citenamefont {Truhlar}\ and\ \citenamefont
  {Horowitz}(1978)}]{LSTH_1978}%
  \BibitemOpen
  \bibfield  {author} {\bibinfo {author} {\bibfnamefont {D.~G.}\ \bibnamefont
  {Truhlar}}\ and\ \bibinfo {author} {\bibfnamefont {C.~J.}\ \bibnamefont
  {Horowitz}},\ }\href@noop {} {\bibfield  {journal} {\bibinfo  {journal} {J.
  Chem. Phys.}\ }\textbf {\bibinfo {volume} {68}},\ \bibinfo {pages} {2466}
  (\bibinfo {year} {1978})}\BibitemShut {NoStop}%
\bibitem [{\citenamefont {Boothroyd}\ \emph {et~al.}(1996)\citenamefont
  {Boothroyd}, \citenamefont {Keogh}, \citenamefont {Martin},\ and\
  \citenamefont {Peterson}}]{BKMP2_1996}%
  \BibitemOpen
  \bibfield  {author} {\bibinfo {author} {\bibfnamefont {A.~I.}\ \bibnamefont
  {Boothroyd}}, \bibinfo {author} {\bibfnamefont {W.~J.}\ \bibnamefont
  {Keogh}}, \bibinfo {author} {\bibfnamefont {P.~G.}\ \bibnamefont {Martin}}, \
  and\ \bibinfo {author} {\bibfnamefont {M.~R.}\ \bibnamefont {Peterson}},\
  }\href@noop {} {\bibfield  {journal} {\bibinfo  {journal} {J. Chem. Phys.}\
  }\textbf {\bibinfo {volume} {104}},\ \bibinfo {pages} {7139} (\bibinfo {year}
  {1996})}\BibitemShut {NoStop}%
\bibitem [{\citenamefont {Mielke}, \citenamefont {Garrett},\ and\ \citenamefont
  {Peterson}(2002)}]{CCI_2002}%
  \BibitemOpen
  \bibfield  {author} {\bibinfo {author} {\bibfnamefont {S.~L.}\ \bibnamefont
  {Mielke}}, \bibinfo {author} {\bibfnamefont {B.~C.}\ \bibnamefont {Garrett}},
  \ and\ \bibinfo {author} {\bibfnamefont {K.~A.}\ \bibnamefont {Peterson}},\
  }\href@noop {} {\bibfield  {journal} {\bibinfo  {journal} {J. Chem. Phys.}\
  }\textbf {\bibinfo {volume} {116}},\ \bibinfo {pages} {4142} (\bibinfo {year}
  {2002})}\BibitemShut {NoStop}%
\bibitem [{\citenamefont {Abrol}\ and\ \citenamefont
  {Kuppermann}(2002)}]{Abrol_Kuppermann_2002}%
  \BibitemOpen
  \bibfield  {author} {\bibinfo {author} {\bibfnamefont {R.}~\bibnamefont
  {Abrol}}\ and\ \bibinfo {author} {\bibfnamefont {A.}~\bibnamefont
  {Kuppermann}},\ }\href@noop {} {\bibfield  {journal} {\bibinfo  {journal} {J.
  Chem. Phys.}\ }\textbf {\bibinfo {volume} {116}},\ \bibinfo {pages} {1035}
  (\bibinfo {year} {2002})}\BibitemShut {NoStop}%
\bibitem [{\citenamefont {Werner}\ \emph {et~al.}(2012)\citenamefont {Werner},
  \citenamefont {Knowles}, \citenamefont {Knizia}, \citenamefont {Manby},\ and\
  \citenamefont {Sch{\ifmmode\ddot{u}\else\"{u}\fi}tz}}]{MOLPRO_brief}%
  \BibitemOpen
  \bibfield  {author} {\bibinfo {author} {\bibfnamefont {H.-J.}\ \bibnamefont
  {Werner}}, \bibinfo {author} {\bibfnamefont {P.~J.}\ \bibnamefont {Knowles}},
  \bibinfo {author} {\bibfnamefont {G.}~\bibnamefont {Knizia}}, \bibinfo
  {author} {\bibfnamefont {F.~R.}\ \bibnamefont {Manby}}, \ and\ \bibinfo
  {author} {\bibfnamefont {M.}~\bibnamefont
  {Sch{\ifmmode\ddot{u}\else\"{u}\fi}tz}},\ }\href@noop {} {\bibfield
  {journal} {\bibinfo  {journal} {WIREs Comput. Mol. Sci.}\ }\textbf {\bibinfo
  {volume} {2}},\ \bibinfo {pages} {242} (\bibinfo {year} {2012})}\BibitemShut
  {NoStop}%
\bibitem [{\citenamefont {Werner}\ \emph {et~al.}(2020)\citenamefont {Werner},
  \citenamefont {Knowles}, \citenamefont {Manby}, \citenamefont {Black},
  \citenamefont {Doll}, \citenamefont {Heßelmann}, \citenamefont {Kats},
  \citenamefont {Köhn}, \citenamefont {Korona}, \citenamefont {Kreplin},
  \citenamefont {Ma}, \citenamefont {Miller}, \citenamefont {Mitrushchenkov},
  \citenamefont {Peterson}, \citenamefont {Polyak}, \citenamefont {Rauhut},\
  and\ \citenamefont {Sibaev}}]{MOLPRO_2020}%
  \BibitemOpen
  \bibfield  {author} {\bibinfo {author} {\bibfnamefont {H.-J.}\ \bibnamefont
  {Werner}}, \bibinfo {author} {\bibfnamefont {P.~J.}\ \bibnamefont {Knowles}},
  \bibinfo {author} {\bibfnamefont {F.~R.}\ \bibnamefont {Manby}}, \bibinfo
  {author} {\bibfnamefont {J.~A.}\ \bibnamefont {Black}}, \bibinfo {author}
  {\bibfnamefont {K.}~\bibnamefont {Doll}}, \bibinfo {author} {\bibfnamefont
  {A.}~\bibnamefont {Heßelmann}}, \bibinfo {author} {\bibfnamefont
  {D.}~\bibnamefont {Kats}}, \bibinfo {author} {\bibfnamefont {A.}~\bibnamefont
  {Köhn}}, \bibinfo {author} {\bibfnamefont {T.}~\bibnamefont {Korona}},
  \bibinfo {author} {\bibfnamefont {D.~A.}\ \bibnamefont {Kreplin}}, \bibinfo
  {author} {\bibfnamefont {Q.}~\bibnamefont {Ma}}, \bibinfo {author}
  {\bibfnamefont {I.}~\bibnamefont {Miller}, \bibfnamefont {Thomas~F.}},
  \bibinfo {author} {\bibfnamefont {A.}~\bibnamefont {Mitrushchenkov}},
  \bibinfo {author} {\bibfnamefont {K.~A.}\ \bibnamefont {Peterson}}, \bibinfo
  {author} {\bibfnamefont {I.}~\bibnamefont {Polyak}}, \bibinfo {author}
  {\bibfnamefont {G.}~\bibnamefont {Rauhut}}, \ and\ \bibinfo {author}
  {\bibfnamefont {M.}~\bibnamefont {Sibaev}},\ }\href@noop {} {\bibfield
  {journal} {\bibinfo  {journal} {J. Chem. Phys.}\ }\textbf {\bibinfo {volume}
  {152}},\ \bibinfo {pages} {144107} (\bibinfo {year} {2020})}\BibitemShut
  {NoStop}%
\bibitem [{\citenamefont {Naskar}, \citenamefont {Ghosh},\ and\ \citenamefont
  {Adhikari}(2022{\natexlab{b}})}]{FH2_adia_dynamics}%
  \BibitemOpen
  \bibfield  {author} {\bibinfo {author} {\bibfnamefont {K.}~\bibnamefont
  {Naskar}}, \bibinfo {author} {\bibfnamefont {S.}~\bibnamefont {Ghosh}}, \
  and\ \bibinfo {author} {\bibfnamefont {S.}~\bibnamefont {Adhikari}},\ }\href
  {\doibase 10.1021/acs.jpca.2c01209} {\bibfield  {journal} {\bibinfo
  {journal} {The Journal of Physical Chemistry A}\ }\textbf {\bibinfo {volume}
  {126}},\ \bibinfo {pages} {3311} (\bibinfo {year} {2022}{\natexlab{b}})},\
  \bibinfo {note} {pMID: 35594416}\BibitemShut {NoStop}%
\bibitem [{\citenamefont {Rosenman}\ \emph {et~al.}(1995)\citenamefont
  {Rosenman}, \citenamefont {Hochman-Kowal}, \citenamefont {Persky},\ and\
  \citenamefont {Baer}}]{jpc-99-16523-BAER}%
  \BibitemOpen
  \bibfield  {author} {\bibinfo {author} {\bibfnamefont {E.}~\bibnamefont
  {Rosenman}}, \bibinfo {author} {\bibfnamefont {S.}~\bibnamefont
  {Hochman-Kowal}}, \bibinfo {author} {\bibfnamefont {A.}~\bibnamefont
  {Persky}}, \ and\ \bibinfo {author} {\bibfnamefont {M.}~\bibnamefont
  {Baer}},\ }\href {\doibase https://doi.org/10.1021/j100045a007} {\bibfield
  {journal} {\bibinfo  {journal} {J. Phys. Chem.}\ }\textbf {\bibinfo {volume}
  {99}},\ \bibinfo {pages} {16523} (\bibinfo {year} {1995})}\BibitemShut
  {NoStop}%
\bibitem [{\citenamefont {Castillo}\ \emph
  {et~al.}(1998{\natexlab{b}})\citenamefont {Castillo}, \citenamefont {Hartke},
  \citenamefont {Werner}, \citenamefont {Aoiz}, \citenamefont {Ba\`nares},\
  and\ \citenamefont {Mart\'inez-Haya}}]{jcp-109-7224-WERNER}%
  \BibitemOpen
  \bibfield  {author} {\bibinfo {author} {\bibfnamefont {J.~F.}\ \bibnamefont
  {Castillo}}, \bibinfo {author} {\bibfnamefont {B.}~\bibnamefont {Hartke}},
  \bibinfo {author} {\bibfnamefont {H.-J.}\ \bibnamefont {Werner}}, \bibinfo
  {author} {\bibfnamefont {F.~J.}\ \bibnamefont {Aoiz}}, \bibinfo {author}
  {\bibfnamefont {L.}~\bibnamefont {Ba\`nares}}, \ and\ \bibinfo {author}
  {\bibfnamefont {B.}~\bibnamefont {Mart\'inez-Haya}},\ }\href {\doibase
  10.1063/1.477401} {\bibfield  {journal} {\bibinfo  {journal} {J. Chem.
  Phys.}\ }\textbf {\bibinfo {volume} {109}},\ \bibinfo {pages} {7224}
  (\bibinfo {year} {1998}{\natexlab{b}})}\BibitemShut {NoStop}%
\bibitem [{\citenamefont {Aoiz}, \citenamefont {Banares},\ and\ \citenamefont
  {Castillo}(1999)}]{jcp-111-4013-AOIZ}%
  \BibitemOpen
  \bibfield  {author} {\bibinfo {author} {\bibfnamefont {F.~J.}\ \bibnamefont
  {Aoiz}}, \bibinfo {author} {\bibfnamefont {L.}~\bibnamefont {Banares}}, \
  and\ \bibinfo {author} {\bibfnamefont {J.~F.}\ \bibnamefont {Castillo}},\
  }\href {\doibase http://dx.doi.org/10.1063/1.479703} {\bibfield  {journal}
  {\bibinfo  {journal} {J. Chem. Phys.}\ }\textbf {\bibinfo {volume} {111}},\
  \bibinfo {pages} {4013} (\bibinfo {year} {1999})}\BibitemShut {NoStop}%
\bibitem [{\citenamefont {Markovic}\ and\ \citenamefont
  {Billing}(2000)}]{mp-98-1771-BILLING}%
  \BibitemOpen
  \bibfield  {author} {\bibinfo {author} {\bibfnamefont {N.}~\bibnamefont
  {Markovic}}\ and\ \bibinfo {author} {\bibfnamefont {G.~D.}\ \bibnamefont
  {Billing}},\ }\href@noop {} {\bibfield  {journal} {\bibinfo  {journal}
  {Molec. Phys.}\ }\textbf {\bibinfo {volume} {98}},\ \bibinfo {pages} {1771}
  (\bibinfo {year} {2000})}\BibitemShut {NoStop}%
\bibitem [{\citenamefont {Muckerman}(1981)}]{tcap-6A-1-MUCKER}%
  \BibitemOpen
  \bibfield  {author} {\bibinfo {author} {\bibfnamefont {J.~T.}\ \bibnamefont
  {Muckerman}},\ }\href@noop {} {\emph {\bibinfo {title} {Theoretical Chemistry
  Advances and Perspectives}}},\ Vol.~\bibinfo {volume} {6A}\ (\bibinfo
  {publisher} {Academic Press, New York},\ \bibinfo {year} {1981})\ p.~\bibinfo
  {pages} {1}\BibitemShut {NoStop}%
\bibitem [{\citenamefont {Lynch}\ \emph {et~al.}(1991)\citenamefont {Lynch},
  \citenamefont {Steckler}, \citenamefont {Schwenke}, \citenamefont
  {Varandas},\ and\ \citenamefont {Truhlar}}]{jcp-94-7136-TRU}%
  \BibitemOpen
  \bibfield  {author} {\bibinfo {author} {\bibfnamefont {G.~C.}\ \bibnamefont
  {Lynch}}, \bibinfo {author} {\bibfnamefont {R.}~\bibnamefont {Steckler}},
  \bibinfo {author} {\bibfnamefont {D.~W.}\ \bibnamefont {Schwenke}}, \bibinfo
  {author} {\bibfnamefont {A.~J.~C.}\ \bibnamefont {Varandas}}, \ and\ \bibinfo
  {author} {\bibfnamefont {D.~G.}\ \bibnamefont {Truhlar}},\ }\href@noop {}
  {\bibfield  {journal} {\bibinfo  {journal} {J. Chem. Phys.}\ }\textbf
  {\bibinfo {volume} {94}},\ \bibinfo {pages} {7136} (\bibinfo {year}
  {1991})}\BibitemShut {NoStop}%
\bibitem [{\citenamefont {Stark}\ and\ \citenamefont
  {Werner}(1996{\natexlab{b}})}]{jcp-104-6515-WERNER}%
  \BibitemOpen
  \bibfield  {author} {\bibinfo {author} {\bibfnamefont {K.}~\bibnamefont
  {Stark}}\ and\ \bibinfo {author} {\bibfnamefont {H.~J.}\ \bibnamefont
  {Werner}},\ }\href@noop {} {\bibfield  {journal} {\bibinfo  {journal} {J.
  Chem. Phys.}\ }\textbf {\bibinfo {volume} {104}},\ \bibinfo {pages} {6515}
  (\bibinfo {year} {1996}{\natexlab{b}})}\BibitemShut {NoStop}%
\bibitem [{\citenamefont {Hartke}\ and\ \citenamefont
  {Werner}(1997{\natexlab{b}})}]{cpl-280-430-HARTKE}%
  \BibitemOpen
  \bibfield  {author} {\bibinfo {author} {\bibfnamefont {B.}~\bibnamefont
  {Hartke}}\ and\ \bibinfo {author} {\bibfnamefont {H.-J.}\ \bibnamefont
  {Werner}},\ }\href@noop {} {\bibfield  {journal} {\bibinfo  {journal} {Chem.
  Phys. Lett.}\ }\textbf {\bibinfo {volume} {280}},\ \bibinfo {pages} {430}
  (\bibinfo {year} {1997}{\natexlab{b}})}\BibitemShut {NoStop}%
\bibitem [{\citenamefont {Aoiz}\ \emph {et~al.}(1993)\citenamefont {Aoiz},
  \citenamefont {Herrero}, \citenamefont {Nogueira},\ and\ \citenamefont
  {R\'abanos}}]{cpl-204-359-AOIZ}%
  \BibitemOpen
  \bibfield  {author} {\bibinfo {author} {\bibfnamefont {F.~J.}\ \bibnamefont
  {Aoiz}}, \bibinfo {author} {\bibfnamefont {V.}~\bibnamefont {Herrero}},
  \bibinfo {author} {\bibfnamefont {M.}~\bibnamefont {Nogueira}}, \ and\
  \bibinfo {author} {\bibfnamefont {V.~S.}\ \bibnamefont {R\'abanos}},\ }\href
  {\doibase https://doi.org/10.1016/0009-2614(93)90022-S} {\bibfield  {journal}
  {\bibinfo  {journal} {Chem. Phys. Lett.}\ }\textbf {\bibinfo {volume}
  {204}},\ \bibinfo {pages} {359} (\bibinfo {year} {1993})}\BibitemShut
  {NoStop}%
\bibitem [{\citenamefont {Honvault}\ and\ \citenamefont
  {Launay}(1999{\natexlab{b}})}]{cpl-303-657-HON}%
  \BibitemOpen
  \bibfield  {author} {\bibinfo {author} {\bibfnamefont {P.}~\bibnamefont
  {Honvault}}\ and\ \bibinfo {author} {\bibfnamefont {J.-M.}\ \bibnamefont
  {Launay}},\ }\href@noop {} {\bibfield  {journal} {\bibinfo  {journal} {Chem.
  Phys. Lett.}\ }\textbf {\bibinfo {volume} {303}},\ \bibinfo {pages} {657}
  (\bibinfo {year} {1999}{\natexlab{b}})}\BibitemShut {NoStop}%
\bibitem [{\citenamefont {Dong}, \citenamefont {Lee},\ and\ \citenamefont
  {Liu}(2000)}]{jcp-113-3633-LIU}%
  \BibitemOpen
  \bibfield  {author} {\bibinfo {author} {\bibfnamefont {F.}~\bibnamefont
  {Dong}}, \bibinfo {author} {\bibfnamefont {S.-H.}\ \bibnamefont {Lee}}, \
  and\ \bibinfo {author} {\bibfnamefont {K.}~\bibnamefont {Liu}},\ }\href@noop
  {} {\bibfield  {journal} {\bibinfo  {journal} {J. Chem. Phys.}\ }\textbf
  {\bibinfo {volume} {113}},\ \bibinfo {pages} {3633} (\bibinfo {year}
  {2000})}\BibitemShut {NoStop}%
\bibitem [{\citenamefont {Bedjanian}(2021{\natexlab{b}})}]{bredjaninan}%
  \BibitemOpen
  \bibfield  {author} {\bibinfo {author} {\bibfnamefont {Y.}~\bibnamefont
  {Bedjanian}},\ }\href {\doibase https://doi.org/10.1002/kin.21462} {\bibfield
   {journal} {\bibinfo  {journal} {Int. J. Chem. Kin.}\ }\textbf {\bibinfo
  {volume} {53}},\ \bibinfo {pages} {527} (\bibinfo {year}
  {2021}{\natexlab{b}})},\ \Eprint
  {http://arxiv.org/abs/https://onlinelibrary.wiley.com/doi/pdf/10.1002/kin.21462}
  {https://onlinelibrary.wiley.com/doi/pdf/10.1002/kin.21462} \BibitemShut
  {NoStop}%
\bibitem [{\citenamefont {Rosenman}\ \emph {et~al.}(1996)\citenamefont
  {Rosenman}, \citenamefont {Hochman-Kowal}, \citenamefont {Persky},\ and\
  \citenamefont {Baer}}]{Rosenman1996_CPC_257}%
  \BibitemOpen
  \bibfield  {author} {\bibinfo {author} {\bibfnamefont {E.}~\bibnamefont
  {Rosenman}}, \bibinfo {author} {\bibfnamefont {S.}~\bibnamefont
  {Hochman-Kowal}}, \bibinfo {author} {\bibfnamefont {A.}~\bibnamefont
  {Persky}}, \ and\ \bibinfo {author} {\bibfnamefont {M.}~\bibnamefont
  {Baer}},\ }\href@noop {} {\bibfield  {journal} {\bibinfo  {journal} {Chem.
  Phys. Lett.}\ }\textbf {\bibinfo {volume} {257}},\ \bibinfo {pages} {421}
  (\bibinfo {year} {1996})}\BibitemShut {NoStop}%
\bibitem [{\citenamefont {Zhang}\ \emph {et~al.}(2003)\citenamefont {Zhang},
  \citenamefont {Xie}, \citenamefont {Han},\ and\ \citenamefont
  {Zhang}}]{jcp-119-12921-HAN}%
  \BibitemOpen
  \bibfield  {author} {\bibinfo {author} {\bibfnamefont {Y.}~\bibnamefont
  {Zhang}}, \bibinfo {author} {\bibfnamefont {T.-X.}\ \bibnamefont {Xie}},
  \bibinfo {author} {\bibfnamefont {K.~L.}\ \bibnamefont {Han}}, \ and\
  \bibinfo {author} {\bibfnamefont {J.~Z.~H.}\ \bibnamefont {Zhang}},\
  }\href@noop {} {\bibfield  {journal} {\bibinfo  {journal} {J. Chem. Phys.}\
  }\textbf {\bibinfo {volume} {119}},\ \bibinfo {pages} {12921} (\bibinfo
  {year} {2003})}\BibitemShut {NoStop}%
\bibitem [{\citenamefont {Naskar}\ \emph
  {et~al.}(2023{\natexlab{c}})\citenamefont {Naskar}, \citenamefont {Ravi},
  \citenamefont {Adhikari}, \citenamefont {Baer},\ and\ \citenamefont
  {Sathyamurthy}}]{Naskar2023_JPCA_127}%
  \BibitemOpen
  \bibfield  {author} {\bibinfo {author} {\bibfnamefont {K.}~\bibnamefont
  {Naskar}}, \bibinfo {author} {\bibfnamefont {S.}~\bibnamefont {Ravi}},
  \bibinfo {author} {\bibfnamefont {S.}~\bibnamefont {Adhikari}}, \bibinfo
  {author} {\bibfnamefont {M.}~\bibnamefont {Baer}}, \ and\ \bibinfo {author}
  {\bibfnamefont {N.}~\bibnamefont {Sathyamurthy}},\ }\href {\doibase
  10.1021/acs.jpca.3c01047} {\bibfield  {journal} {\bibinfo  {journal} {J.
  Phys. Chem. A}\ }\textbf {\bibinfo {volume} {127}},\ \bibinfo {pages} {3832}
  (\bibinfo {year} {2023}{\natexlab{c}})}\BibitemShut {NoStop}%
\bibitem [{\citenamefont {Viegas}, \citenamefont {Alijah},\ and\ \citenamefont
  {Varandas}(2007{\natexlab{b}})}]{viegas_jcp126}%
  \BibitemOpen
  \bibfield  {author} {\bibinfo {author} {\bibfnamefont {L.~P.}\ \bibnamefont
  {Viegas}}, \bibinfo {author} {\bibfnamefont {A.}~\bibnamefont {Alijah}}, \
  and\ \bibinfo {author} {\bibfnamefont {A.~J.~C.}\ \bibnamefont {Varandas}},\
  }\href@noop {} {\bibfield  {journal} {\bibinfo  {journal} {J. Chem. Phys.}\
  }\textbf {\bibinfo {volume} {126}},\ \bibinfo {pages} {074309} (\bibinfo
  {year} {2007}{\natexlab{b}})}\BibitemShut {NoStop}%
\bibitem [{\citenamefont {Chu}, \citenamefont {Varandas},\ and\ \citenamefont
  {Han}(2009)}]{cpl-471-222-HAN3}%
  \BibitemOpen
  \bibfield  {author} {\bibinfo {author} {\bibfnamefont {T.~S.}\ \bibnamefont
  {Chu}}, \bibinfo {author} {\bibfnamefont {A.~J.~C.}\ \bibnamefont
  {Varandas}}, \ and\ \bibinfo {author} {\bibfnamefont {K.~L.}\ \bibnamefont
  {Han}},\ }\href@noop {} {\bibfield  {journal} {\bibinfo  {journal} {Chem.
  Phys. Lett.}\ }\textbf {\bibinfo {volume} {471}},\ \bibinfo {pages} {222}
  (\bibinfo {year} {2009})}\BibitemShut {NoStop}%
\bibitem [{\citenamefont {Chu}\ and\ \citenamefont
  {Han}(2005)}]{jpcA-109-2050-HAN1}%
  \BibitemOpen
  \bibfield  {author} {\bibinfo {author} {\bibfnamefont {T.~S.}\ \bibnamefont
  {Chu}}\ and\ \bibinfo {author} {\bibfnamefont {K.~L.}\ \bibnamefont {Han}},\
  }\href@noop {} {\bibfield  {journal} {\bibinfo  {journal} {J. Phy. Chem. A}\
  }\textbf {\bibinfo {volume} {109}},\ \bibinfo {pages} {2050} (\bibinfo {year}
  {2005})}\BibitemShut {NoStop}%
\bibitem [{\citenamefont {Kamisaka}\ \emph
  {et~al.}(2002{\natexlab{b}})\citenamefont {Kamisaka}, \citenamefont {Bian},
  \citenamefont {Nobusada},\ and\ \citenamefont {Nakamura}}]{jcp-116-654-KAM}%
  \BibitemOpen
  \bibfield  {author} {\bibinfo {author} {\bibfnamefont {A.}~\bibnamefont
  {Kamisaka}}, \bibinfo {author} {\bibfnamefont {W.}~\bibnamefont {Bian}},
  \bibinfo {author} {\bibfnamefont {K.}~\bibnamefont {Nobusada}}, \ and\
  \bibinfo {author} {\bibfnamefont {H.}~\bibnamefont {Nakamura}},\ }\href@noop
  {} {\bibfield  {journal} {\bibinfo  {journal} {J. Chem. Phys.}\ }\textbf
  {\bibinfo {volume} {116}},\ \bibinfo {pages} {654} (\bibinfo {year}
  {2002}{\natexlab{b}})}\BibitemShut {NoStop}%
\bibitem [{\citenamefont {Schlier}, \citenamefont {Nowotny},\ and\
  \citenamefont {Teloy}(1987)}]{cp-111-401-SCH}%
  \BibitemOpen
  \bibfield  {author} {\bibinfo {author} {\bibfnamefont {C.}~\bibnamefont
  {Schlier}}, \bibinfo {author} {\bibfnamefont {U.}~\bibnamefont {Nowotny}}, \
  and\ \bibinfo {author} {\bibfnamefont {E.}~\bibnamefont {Teloy}},\
  }\href@noop {} {\bibfield  {journal} {\bibinfo  {journal} {Chem. Phys.}\
  }\textbf {\bibinfo {volume} {111}},\ \bibinfo {pages} {401} (\bibinfo {year}
  {1987})}\BibitemShut {NoStop}%
\bibitem [{\citenamefont {Kamisaka}\ \emph
  {et~al.}(2002{\natexlab{c}})\citenamefont {Kamisaka}, \citenamefont {Bian},
  \citenamefont {Nobusada},\ and\ \citenamefont {Nakamura}}]{KBNN}%
  \BibitemOpen
  \bibfield  {author} {\bibinfo {author} {\bibfnamefont {H.}~\bibnamefont
  {Kamisaka}}, \bibinfo {author} {\bibfnamefont {W.}~\bibnamefont {Bian}},
  \bibinfo {author} {\bibfnamefont {K.}~\bibnamefont {Nobusada}}, \ and\
  \bibinfo {author} {\bibfnamefont {H.}~\bibnamefont {Nakamura}},\ }\href@noop
  {} {\bibfield  {journal} {\bibinfo  {journal} {J. Chem. Phys.}\ }\textbf
  {\bibinfo {volume} {116}},\ \bibinfo {pages} {654} (\bibinfo {year}
  {2002}{\natexlab{c}})}\BibitemShut {NoStop}%
\bibitem [{\citenamefont {Last}, \citenamefont {Gilibert},\ and\ \citenamefont
  {Baer}(1997)}]{jcp-107-1451-LAST}%
  \BibitemOpen
  \bibfield  {author} {\bibinfo {author} {\bibfnamefont {I.}~\bibnamefont
  {Last}}, \bibinfo {author} {\bibfnamefont {M.}~\bibnamefont {Gilibert}}, \
  and\ \bibinfo {author} {\bibfnamefont {M.}~\bibnamefont {Baer}},\ }\href@noop
  {} {\bibfield  {journal} {\bibinfo  {journal} {J. Chem. Phys.}\ }\textbf
  {\bibinfo {volume} {107}},\ \bibinfo {pages} {1451} (\bibinfo {year}
  {1997})}\BibitemShut {NoStop}%
\bibitem [{\citenamefont {Karpas}, \citenamefont {Anicich},\ and\ \citenamefont
  {Huntress(Jr.)}(1979)}]{jcp-70-2877-KARPAS}%
  \BibitemOpen
  \bibfield  {author} {\bibinfo {author} {\bibfnamefont {Z.}~\bibnamefont
  {Karpas}}, \bibinfo {author} {\bibfnamefont {V.}~\bibnamefont {Anicich}}, \
  and\ \bibinfo {author} {\bibfnamefont {W.~T.}\ \bibnamefont
  {Huntress(Jr.)}},\ }\href@noop {} {\bibfield  {journal} {\bibinfo  {journal}
  {J. Chem. Phys.}\ }\textbf {\bibinfo {volume} {70}},\ \bibinfo {pages} {2877}
  (\bibinfo {year} {1979})}\BibitemShut {NoStop}%
\bibitem [{\citenamefont {Naskar}\ \emph
  {et~al.}(2023{\natexlab{d}})\citenamefont {Naskar}, \citenamefont {Ravi},
  \citenamefont {Adhikari}, \citenamefont {Baer},\ and\ \citenamefont
  {Sathyamurthy}}]{heh2+_surface_article}%
  \BibitemOpen
  \bibfield  {author} {\bibinfo {author} {\bibfnamefont {K.}~\bibnamefont
  {Naskar}}, \bibinfo {author} {\bibfnamefont {S.}~\bibnamefont {Ravi}},
  \bibinfo {author} {\bibfnamefont {S.}~\bibnamefont {Adhikari}}, \bibinfo
  {author} {\bibfnamefont {M.}~\bibnamefont {Baer}}, \ and\ \bibinfo {author}
  {\bibfnamefont {N.}~\bibnamefont {Sathyamurthy}},\ }\href {\doibase
  10.1021/acs.jpca.3c01047} {\bibfield  {journal} {\bibinfo  {journal} {J.
  Phys. Chem. A}\ }\textbf {\bibinfo {volume} {127}},\ \bibinfo {pages} {3832}
  (\bibinfo {year} {2023}{\natexlab{d}})}\BibitemShut {NoStop}%
\bibitem [{\citenamefont {Turner}, \citenamefont {Dutuit},\ and\ \citenamefont
  {Lee}(1984{\natexlab{b}})}]{Turner1984_JCP_81}%
  \BibitemOpen
  \bibfield  {author} {\bibinfo {author} {\bibfnamefont {T.}~\bibnamefont
  {Turner}}, \bibinfo {author} {\bibfnamefont {O.}~\bibnamefont {Dutuit}}, \
  and\ \bibinfo {author} {\bibfnamefont {Y.~T.}\ \bibnamefont {Lee}},\ }\href
  {\doibase 10.1063/1.448073} {\bibfield  {journal} {\bibinfo  {journal} {J.
  Chem. Phys.}\ }\textbf {\bibinfo {volume} {81}},\ \bibinfo {pages} {3475}
  (\bibinfo {year} {1984}{\natexlab{b}})}\BibitemShut {NoStop}%
\bibitem [{\citenamefont {Bouakline}, \citenamefont {Althorpe},\ and\
  \citenamefont {Pel{\'a}ez~Ruiz}(2008)}]{Althorpe_2008}%
  \BibitemOpen
  \bibfield  {author} {\bibinfo {author} {\bibfnamefont {F.}~\bibnamefont
  {Bouakline}}, \bibinfo {author} {\bibfnamefont {S.~C.}\ \bibnamefont
  {Althorpe}}, \ and\ \bibinfo {author} {\bibfnamefont {D.}~\bibnamefont
  {Pel{\'a}ez~Ruiz}},\ }\href@noop {} {\bibfield  {journal} {\bibinfo
  {journal} {J. Chem. Phys.}\ }\textbf {\bibinfo {volume} {128}},\ \bibinfo
  {pages} {124322} (\bibinfo {year} {2008})}\BibitemShut {NoStop}%
\bibitem [{\citenamefont {Launay}\ and\ \citenamefont
  {Dourneuf}(1989)}]{Launay_1989}%
  \BibitemOpen
  \bibfield  {author} {\bibinfo {author} {\bibfnamefont {J.}~\bibnamefont
  {Launay}}\ and\ \bibinfo {author} {\bibfnamefont {M.~L.}\ \bibnamefont
  {Dourneuf}},\ }\href@noop {} {\bibfield  {journal} {\bibinfo  {journal}
  {Chemical Physics Letters}\ }\textbf {\bibinfo {volume} {163}},\ \bibinfo
  {pages} {178} (\bibinfo {year} {1989})}\BibitemShut {NoStop}%
\bibitem [{\citenamefont {Johnson}(1983{\natexlab{b}})}]{Johnson1983_JCP_79}%
  \BibitemOpen
  \bibfield  {author} {\bibinfo {author} {\bibfnamefont {B.~R.}\ \bibnamefont
  {Johnson}},\ }\href@noop {} {\bibfield  {journal} {\bibinfo  {journal} {J.
  Chem. Phys.}\ }\textbf {\bibinfo {volume} {79}},\ \bibinfo {pages} {1916}
  (\bibinfo {year} {1983}{\natexlab{b}})}\BibitemShut {NoStop}%
\end{thebibliography}%


\begin{thebibliography}{47}%
\makeatletter
\providecommand \@ifxundefined [1]{%
 \@ifx{#1\undefined}
}%
\providecommand \@ifnum [1]{%
 \ifnum #1\expandafter \@firstoftwo
 \else \expandafter \@secondoftwo
 \fi
}%
\providecommand \@ifx [1]{%
 \ifx #1\expandafter \@firstoftwo
 \else \expandafter \@secondoftwo
 \fi
}%
\providecommand \natexlab [1]{#1}%
\providecommand \enquote  [1]{``#1''}%
\providecommand \bibnamefont  [1]{#1}%
\providecommand \bibfnamefont [1]{#1}%
\providecommand \citenamefont [1]{#1}%
\providecommand \href@noop [0]{\@secondoftwo}%
\providecommand \href [0]{\begingroup \@sanitize@url \@href}%
\providecommand \@href[1]{\@@startlink{#1}\@@href}%
\providecommand \@@href[1]{\endgroup#1\@@endlink}%
\providecommand \@sanitize@url [0]{\catcode `\\12\catcode `\$12\catcode
  `\&12\catcode `\#12\catcode `\^12\catcode `\_12\catcode `\%12\relax}%
\providecommand \@@startlink[1]{}%
\providecommand \@@endlink[0]{}%
\providecommand \url  [0]{\begingroup\@sanitize@url \@url }%
\providecommand \@url [1]{\endgroup\@href {#1}{\urlprefix }}%
\providecommand \urlprefix  [0]{URL }%
\providecommand \Eprint [0]{\href }%
\providecommand \doibase [0]{http://dx.doi.org/}%
\providecommand \selectlanguage [0]{\@gobble}%
\providecommand \bibinfo  [0]{\@secondoftwo}%
\providecommand \bibfield  [0]{\@secondoftwo}%
\providecommand \translation [1]{[#1]}%
\providecommand \BibitemOpen [0]{}%
\providecommand \bibitemStop [0]{}%
\providecommand \bibitemNoStop [0]{.\EOS\space}%
\providecommand \EOS [0]{\spacefactor3000\relax}%
\providecommand \BibitemShut  [1]{\csname bibitem#1\endcsname}%
\let\auto@bib@innerbib\@empty
\bibitem [{\citenamefont {Tully}\ and\ \citenamefont
  {Preston}(1971)}]{TULLY:JCP55}%
  \BibitemOpen
  \bibfield  {author} {\bibinfo {author} {\bibfnamefont {J.~C.}\ \bibnamefont
  {Tully}}\ and\ \bibinfo {author} {\bibfnamefont {R.~K.}\ \bibnamefont
  {Preston}},\ }\href@noop {} {\bibfield  {journal} {\bibinfo  {journal} {J.
  Chem. Phys.}\ }\textbf {\bibinfo {volume} {55}},\ \bibinfo {pages} {562}
  (\bibinfo {year} {1971})}\BibitemShut {NoStop}%
\bibitem [{\citenamefont {Bjerre}\ and\ \citenamefont
  {Nikitin}(1967)}]{nikitin_cpl1}%
  \BibitemOpen
  \bibfield  {author} {\bibinfo {author} {\bibfnamefont {A.}~\bibnamefont
  {Bjerre}}\ and\ \bibinfo {author} {\bibfnamefont {E.~E.}\ \bibnamefont
  {Nikitin}},\ }\href@noop {} {\bibfield  {journal} {\bibinfo  {journal} {Chem.
  Phys. Lett.}\ }\textbf {\bibinfo {volume} {1}},\ \bibinfo {pages} {179}
  (\bibinfo {year} {1967})}\BibitemShut {NoStop}%
\bibitem [{\citenamefont {Stine}\ and\ \citenamefont
  {Muckerman}(1976)}]{muckerman_jcp65}%
  \BibitemOpen
  \bibfield  {author} {\bibinfo {author} {\bibfnamefont {J.~R.}\ \bibnamefont
  {Stine}}\ and\ \bibinfo {author} {\bibfnamefont {J.~T.}\ \bibnamefont
  {Muckerman}},\ }\href@noop {} {\bibfield  {journal} {\bibinfo  {journal} {J
  Chem. Phys.}\ }\textbf {\bibinfo {volume} {65}},\ \bibinfo {pages} {3975}
  (\bibinfo {year} {1976})}\BibitemShut {NoStop}%
\bibitem [{\citenamefont {Tully}(1990)}]{tully_jcp93}%
  \BibitemOpen
  \bibfield  {author} {\bibinfo {author} {\bibfnamefont {J.~C.}\ \bibnamefont
  {Tully}},\ }\href@noop {} {\bibfield  {journal} {\bibinfo  {journal} {J.
  Chem. Phys.}\ }\textbf {\bibinfo {volume} {93}},\ \bibinfo {pages} {1061}
  (\bibinfo {year} {1990})}\BibitemShut {NoStop}%
\bibitem [{\citenamefont {McLafferty}(1983)}]{mclafferty_jcp79}%
  \BibitemOpen
  \bibfield  {author} {\bibinfo {author} {\bibfnamefont {F.}~\bibnamefont
  {McLafferty}},\ }\href@noop {} {\bibfield  {journal} {\bibinfo  {journal} {J.
  Chem. Phys.}\ }\textbf {\bibinfo {volume} {79}},\ \bibinfo {pages} {4922}
  (\bibinfo {year} {1983})}\BibitemShut {NoStop}%
\bibitem [{\citenamefont {Ali}\ and\ \citenamefont {Miller}(1983)}]{ali_jcp78}%
  \BibitemOpen
  \bibfield  {author} {\bibinfo {author} {\bibfnamefont {D.~P.}\ \bibnamefont
  {Ali}}\ and\ \bibinfo {author} {\bibfnamefont {W.~H.}\ \bibnamefont
  {Miller}},\ }\href@noop {} {\bibfield  {journal} {\bibinfo  {journal} {J.
  Chem. Phys.}\ }\textbf {\bibinfo {volume} {78}},\ \bibinfo {pages} {6640}
  (\bibinfo {year} {1983})}\BibitemShut {NoStop}%
\bibitem [{\citenamefont {Kuntz}(1991)}]{kuntz_jcp95}%
  \BibitemOpen
  \bibfield  {author} {\bibinfo {author} {\bibfnamefont {P.~J.}\ \bibnamefont
  {Kuntz}},\ }\href@noop {} {\bibfield  {journal} {\bibinfo  {journal} {J.
  Chem. Phys.}\ }\textbf {\bibinfo {volume} {95}},\ \bibinfo {pages} {141}
  (\bibinfo {year} {1991})}\BibitemShut {NoStop}%
\bibitem [{\citenamefont {Hammes-Schiffer}\ and\ \citenamefont
  {Tully}(1994)}]{tully_jcp101}%
  \BibitemOpen
  \bibfield  {author} {\bibinfo {author} {\bibfnamefont {S.}~\bibnamefont
  {Hammes-Schiffer}}\ and\ \bibinfo {author} {\bibfnamefont {J.~C.}\
  \bibnamefont {Tully}},\ }\href@noop {} {\bibfield  {journal} {\bibinfo
  {journal} {J. Chem. Phys.}\ }\textbf {\bibinfo {volume} {101}},\ \bibinfo
  {pages} {4657} (\bibinfo {year} {1994})}\BibitemShut {NoStop}%
\bibitem [{\citenamefont {Tapavicza}, \citenamefont {Tavernelli},\ and\
  \citenamefont {Rothlisberger}(2007)}]{rothlisberger_prl98}%
  \BibitemOpen
  \bibfield  {author} {\bibinfo {author} {\bibfnamefont {E.}~\bibnamefont
  {Tapavicza}}, \bibinfo {author} {\bibfnamefont {I.}~\bibnamefont
  {Tavernelli}}, \ and\ \bibinfo {author} {\bibfnamefont {U.}~\bibnamefont
  {Rothlisberger}},\ }\href@noop {} {\bibfield  {journal} {\bibinfo  {journal}
  {Phys. Rev. Lett.}\ }\textbf {\bibinfo {volume} {98}},\ \bibinfo {pages}
  {023001} (\bibinfo {year} {2007})}\BibitemShut {NoStop}%
\bibitem [{\citenamefont {Werner}\ \emph {et~al.}(2008)\citenamefont {Werner},
  \citenamefont {Mitric}, \citenamefont {Suzuki},\ and\ \citenamefont
  {Bonacic-Kouteck\'y}}]{werner_cp349}%
  \BibitemOpen
  \bibfield  {author} {\bibinfo {author} {\bibfnamefont {U.}~\bibnamefont
  {Werner}}, \bibinfo {author} {\bibfnamefont {R.}~\bibnamefont {Mitric}},
  \bibinfo {author} {\bibfnamefont {T.}~\bibnamefont {Suzuki}}, \ and\ \bibinfo
  {author} {\bibfnamefont {V.}~\bibnamefont {Bonacic-Kouteck\'y}},\ }\href@noop
  {} {\bibfield  {journal} {\bibinfo  {journal} {Chem. Phys.}\ }\textbf
  {\bibinfo {volume} {349}},\ \bibinfo {pages} {319} (\bibinfo {year}
  {2008})}\BibitemShut {NoStop}%
\bibitem [{\citenamefont {Ibele}\ \emph {et~al.}(2021)\citenamefont {Ibele},
  \citenamefont {Lassmann}, \citenamefont {Mart\'inez},\ and\ \citenamefont
  {Curchod}}]{martinez_jcp154}%
  \BibitemOpen
  \bibfield  {author} {\bibinfo {author} {\bibfnamefont {L.~M.}\ \bibnamefont
  {Ibele}}, \bibinfo {author} {\bibfnamefont {Y.}~\bibnamefont {Lassmann}},
  \bibinfo {author} {\bibfnamefont {T.~J.}\ \bibnamefont {Mart\'inez}}, \ and\
  \bibinfo {author} {\bibfnamefont {B.~F.~E.}\ \bibnamefont {Curchod}},\
  }\href@noop {} {\bibfield  {journal} {\bibinfo  {journal} {J. Chem. Phys.}\
  }\textbf {\bibinfo {volume} {154}},\ \bibinfo {pages} {104110/1} (\bibinfo
  {year} {2021})}\BibitemShut {NoStop}%
\bibitem [{\citenamefont {Perkins}\ \emph {et~al.}(2015)\citenamefont
  {Perkins}, \citenamefont {Herr\'aez-Aguilar}, \citenamefont {McCrudden},
  \citenamefont {Klos}, \citenamefont {Aoiz},\ and\ \citenamefont
  {Brouard}}]{klos_jcp142}%
  \BibitemOpen
  \bibfield  {author} {\bibinfo {author} {\bibfnamefont {T.}~\bibnamefont
  {Perkins}}, \bibinfo {author} {\bibfnamefont {D.}~\bibnamefont
  {Herr\'aez-Aguilar}}, \bibinfo {author} {\bibfnamefont {G.}~\bibnamefont
  {McCrudden}}, \bibinfo {author} {\bibfnamefont {J.}~\bibnamefont {Klos}},
  \bibinfo {author} {\bibfnamefont {F.~J.}\ \bibnamefont {Aoiz}}, \ and\
  \bibinfo {author} {\bibfnamefont {M.}~\bibnamefont {Brouard}},\ }\href@noop
  {} {\bibfield  {journal} {\bibinfo  {journal} {J. Chem. Phys.}\ }\textbf
  {\bibinfo {volume} {142}},\ \bibinfo {pages} {144307} (\bibinfo {year}
  {2015})}\BibitemShut {NoStop}%
\bibitem [{\citenamefont {Warshel}\ and\ \citenamefont
  {Karplus}(1975)}]{karplus_cpl32}%
  \BibitemOpen
  \bibfield  {author} {\bibinfo {author} {\bibfnamefont {A.}~\bibnamefont
  {Warshel}}\ and\ \bibinfo {author} {\bibfnamefont {M.}~\bibnamefont
  {Karplus}},\ }\href@noop {} {\bibfield  {journal} {\bibinfo  {journal} {Chem.
  Phys. Lett.}\ }\textbf {\bibinfo {volume} {32}},\ \bibinfo {pages} {11}
  (\bibinfo {year} {1975})}\BibitemShut {NoStop}%
\bibitem [{\citenamefont {Tachikawa}\ \emph {et~al.}(1997)\citenamefont
  {Tachikawa}, \citenamefont {Ohnishi}, \citenamefont {Hamabayashi},\ and\
  \citenamefont {Yoshida}}]{yoshida_jpca101}%
  \BibitemOpen
  \bibfield  {author} {\bibinfo {author} {\bibfnamefont {H.}~\bibnamefont
  {Tachikawa}}, \bibinfo {author} {\bibfnamefont {K.}~\bibnamefont {Ohnishi}},
  \bibinfo {author} {\bibfnamefont {T.}~\bibnamefont {Hamabayashi}}, \ and\
  \bibinfo {author} {\bibfnamefont {H.}~\bibnamefont {Yoshida}},\ }\href@noop
  {} {\bibfield  {journal} {\bibinfo  {journal} {J. Phys. Chem. A}\ }\textbf
  {\bibinfo {volume} {101}},\ \bibinfo {pages} {2229} (\bibinfo {year}
  {1997})}\BibitemShut {NoStop}%
\bibitem [{\citenamefont {Hu}\ \emph {et~al.}(2008)\citenamefont {Hu},
  \citenamefont {Lendvay}, \citenamefont {Maiti},\ and\ \citenamefont
  {Schatz}}]{schatz_jpca112}%
  \BibitemOpen
  \bibfield  {author} {\bibinfo {author} {\bibfnamefont {W.}~\bibnamefont
  {Hu}}, \bibinfo {author} {\bibfnamefont {G.}~\bibnamefont {Lendvay}},
  \bibinfo {author} {\bibfnamefont {B.}~\bibnamefont {Maiti}}, \ and\ \bibinfo
  {author} {\bibfnamefont {G.~C.}\ \bibnamefont {Schatz}},\ }\href@noop {}
  {\bibfield  {journal} {\bibinfo  {journal} {J. Phys. Chem. A}\ }\textbf
  {\bibinfo {volume} {112}},\ \bibinfo {pages} {2093} (\bibinfo {year}
  {2008})}\BibitemShut {NoStop}%
\bibitem [{\citenamefont {Carbogno}\ \emph {et~al.}(2010)\citenamefont
  {Carbogno}, \citenamefont {Behler}, \citenamefont {Reuter},\ and\
  \citenamefont {Gross}}]{gross_prb81}%
  \BibitemOpen
  \bibfield  {author} {\bibinfo {author} {\bibfnamefont {C.}~\bibnamefont
  {Carbogno}}, \bibinfo {author} {\bibfnamefont {J.}~\bibnamefont {Behler}},
  \bibinfo {author} {\bibfnamefont {K.}~\bibnamefont {Reuter}}, \ and\ \bibinfo
  {author} {\bibfnamefont {A.}~\bibnamefont {Gross}},\ }\href@noop {}
  {\bibfield  {journal} {\bibinfo  {journal} {Phys. Rev. B}\ }\textbf {\bibinfo
  {volume} {81}},\ \bibinfo {pages} {035410} (\bibinfo {year}
  {2010})}\BibitemShut {NoStop}%
\bibitem [{\citenamefont {Yamashita}\ and\ \citenamefont
  {Morokuma}(1990)}]{morokuma_cpl169}%
  \BibitemOpen
  \bibfield  {author} {\bibinfo {author} {\bibfnamefont {K.}~\bibnamefont
  {Yamashita}}\ and\ \bibinfo {author} {\bibfnamefont {K.}~\bibnamefont
  {Morokuma}},\ }\href@noop {} {\bibfield  {journal} {\bibinfo  {journal}
  {Chem. Phys. Lett.}\ }\textbf {\bibinfo {volume} {169}},\ \bibinfo {pages}
  {263} (\bibinfo {year} {1990})}\BibitemShut {NoStop}%
\bibitem [{\citenamefont {Gai}\ and\ \citenamefont {Voth}(1992)}]{voth_jap71}%
  \BibitemOpen
  \bibfield  {author} {\bibinfo {author} {\bibfnamefont {H.}~\bibnamefont
  {Gai}}\ and\ \bibinfo {author} {\bibfnamefont {G.~A.}\ \bibnamefont {Voth}},\
  }\href@noop {} {\bibfield  {journal} {\bibinfo  {journal} {J. Appl. Phys.}\
  }\textbf {\bibinfo {volume} {71}},\ \bibinfo {pages} {1415} (\bibinfo {year}
  {1992})}\BibitemShut {NoStop}%
\bibitem [{\citenamefont {Jones}, \citenamefont {Acocella},\ and\ \citenamefont
  {Zerbetto}(2008)}]{zerbetto_jpca112}%
  \BibitemOpen
  \bibfield  {author} {\bibinfo {author} {\bibfnamefont {G.~A.}\ \bibnamefont
  {Jones}}, \bibinfo {author} {\bibfnamefont {A.}~\bibnamefont {Acocella}}, \
  and\ \bibinfo {author} {\bibfnamefont {F.}~\bibnamefont {Zerbetto}},\
  }\href@noop {} {\bibfield  {journal} {\bibinfo  {journal} {J. Phys. Chem. A}\
  }\textbf {\bibinfo {volume} {112}},\ \bibinfo {pages} {9650} (\bibinfo {year}
  {2008})}\BibitemShut {NoStop}%
\bibitem [{\citenamefont {Mitric}, \citenamefont {Petersen},\ and\
  \citenamefont {Bonacic-Koutecky}(2009)}]{mitric_pra79}%
  \BibitemOpen
  \bibfield  {author} {\bibinfo {author} {\bibfnamefont {R.}~\bibnamefont
  {Mitric}}, \bibinfo {author} {\bibfnamefont {J.}~\bibnamefont {Petersen}}, \
  and\ \bibinfo {author} {\bibfnamefont {V.}~\bibnamefont {Bonacic-Koutecky}},\
  }\href@noop {} {\bibfield  {journal} {\bibinfo  {journal} {Phys. Rev. A}\
  }\textbf {\bibinfo {volume} {79}},\ \bibinfo {pages} {053416} (\bibinfo
  {year} {2009})}\BibitemShut {NoStop}%
\bibitem [{\citenamefont {Herman}\ and\ \citenamefont
  {Arce}(1994)}]{herman_cp183}%
  \BibitemOpen
  \bibfield  {author} {\bibinfo {author} {\bibfnamefont {M.~F.}\ \bibnamefont
  {Herman}}\ and\ \bibinfo {author} {\bibfnamefont {J.~C.}\ \bibnamefont
  {Arce}},\ }\href@noop {} {\bibfield  {journal} {\bibinfo  {journal} {Chem.
  Phys.}\ }\textbf {\bibinfo {volume} {183}},\ \bibinfo {pages} {335} (\bibinfo
  {year} {1994})}\BibitemShut {NoStop}%
\bibitem [{\citenamefont {Mukherjee}\ \emph {et~al.}(2022)\citenamefont
  {Mukherjee}, \citenamefont {Hazra}, \citenamefont {Ghosh}, \citenamefont
  {Mukherjee},\ and\ \citenamefont {Adhikari}}]{soumya_hopp}%
  \BibitemOpen
  \bibfield  {author} {\bibinfo {author} {\bibfnamefont {S.}~\bibnamefont
  {Mukherjee}}, \bibinfo {author} {\bibfnamefont {S.}~\bibnamefont {Hazra}},
  \bibinfo {author} {\bibfnamefont {S.}~\bibnamefont {Ghosh}}, \bibinfo
  {author} {\bibfnamefont {S.}~\bibnamefont {Mukherjee}}, \ and\ \bibinfo
  {author} {\bibfnamefont {S.}~\bibnamefont {Adhikari}},\ }\href@noop {}
  {\bibfield  {journal} {\bibinfo  {journal} {Chem. Phys.}\ }\textbf {\bibinfo
  {volume} {560}},\ \bibinfo {pages} {111588} (\bibinfo {year}
  {2022})}\BibitemShut {NoStop}%
\bibitem [{\citenamefont {Mukherjee}, \citenamefont {Mukhopadhyay},\ and\
  \citenamefont {Adhikari}(2014)}]{mukherjee_jcp141}%
  \BibitemOpen
  \bibfield  {author} {\bibinfo {author} {\bibfnamefont {S.}~\bibnamefont
  {Mukherjee}}, \bibinfo {author} {\bibfnamefont {D.}~\bibnamefont
  {Mukhopadhyay}}, \ and\ \bibinfo {author} {\bibfnamefont {S.}~\bibnamefont
  {Adhikari}},\ }\href@noop {} {\bibfield  {journal} {\bibinfo  {journal} {J.
  Chem. Phys.}\ }\textbf {\bibinfo {volume} {141}},\ \bibinfo {pages} {204306}
  (\bibinfo {year} {2014})}\BibitemShut {NoStop}%
\bibitem [{\citenamefont {Fabiano}, \citenamefont {Keal},\ and\ \citenamefont
  {Thiel}(2008)}]{thiel_cp349}%
  \BibitemOpen
  \bibfield  {author} {\bibinfo {author} {\bibfnamefont {E.}~\bibnamefont
  {Fabiano}}, \bibinfo {author} {\bibfnamefont {T.~W.}\ \bibnamefont {Keal}}, \
  and\ \bibinfo {author} {\bibfnamefont {W.}~\bibnamefont {Thiel}},\
  }\href@noop {} {\bibfield  {journal} {\bibinfo  {journal} {Chem. Phys.}\
  }\textbf {\bibinfo {volume} {349}},\ \bibinfo {pages} {334} (\bibinfo {year}
  {2008})}\BibitemShut {NoStop}%
\bibitem [{\citenamefont {Ghosh}\ \emph {et~al.}(2017)\citenamefont {Ghosh},
  \citenamefont {Mukherjee}, \citenamefont {Mukherjee}, \citenamefont {Mandal},
  \citenamefont {Sharma}, \citenamefont {Chaudhury},\ and\ \citenamefont
  {Adhikari}}]{jcp-147-074105-SA}%
  \BibitemOpen
  \bibfield  {author} {\bibinfo {author} {\bibfnamefont {S.}~\bibnamefont
  {Ghosh}}, \bibinfo {author} {\bibfnamefont {S.}~\bibnamefont {Mukherjee}},
  \bibinfo {author} {\bibfnamefont {B.}~\bibnamefont {Mukherjee}}, \bibinfo
  {author} {\bibfnamefont {S.}~\bibnamefont {Mandal}}, \bibinfo {author}
  {\bibfnamefont {R.}~\bibnamefont {Sharma}}, \bibinfo {author} {\bibfnamefont
  {P.}~\bibnamefont {Chaudhury}}, \ and\ \bibinfo {author} {\bibfnamefont
  {S.}~\bibnamefont {Adhikari}},\ }\href@noop {} {\bibfield  {journal}
  {\bibinfo  {journal} {J. Chem. Phys.}\ }\textbf {\bibinfo {volume} {147}},\
  \bibinfo {pages} {074105/1} (\bibinfo {year} {2017})}\BibitemShut {NoStop}%
\bibitem [{\citenamefont {Ghosh}\ \emph {et~al.}(2021)\citenamefont {Ghosh},
  \citenamefont {Sahoo}, \citenamefont {Baer},\ and\ \citenamefont
  {Adhikari}}]{ghosh_jpca125}%
  \BibitemOpen
  \bibfield  {author} {\bibinfo {author} {\bibfnamefont {S.}~\bibnamefont
  {Ghosh}}, \bibinfo {author} {\bibfnamefont {T.}~\bibnamefont {Sahoo}},
  \bibinfo {author} {\bibfnamefont {M.}~\bibnamefont {Baer}}, \ and\ \bibinfo
  {author} {\bibfnamefont {S.}~\bibnamefont {Adhikari}},\ }\href@noop {}
  {\bibfield  {journal} {\bibinfo  {journal} {J. Phys. Chem. A}\ }\textbf
  {\bibinfo {volume} {125}},\ \bibinfo {pages} {731} (\bibinfo {year}
  {2021})}\BibitemShut {NoStop}%
\bibitem [{\citenamefont {Schlier}, \citenamefont {Nowotny},\ and\
  \citenamefont {Teloy}(1987)}]{cp-111-401-SCH}%
  \BibitemOpen
  \bibfield  {author} {\bibinfo {author} {\bibfnamefont {C.}~\bibnamefont
  {Schlier}}, \bibinfo {author} {\bibfnamefont {U.}~\bibnamefont {Nowotny}}, \
  and\ \bibinfo {author} {\bibfnamefont {E.}~\bibnamefont {Teloy}},\
  }\href@noop {} {\bibfield  {journal} {\bibinfo  {journal} {Chem. Phys.}\
  }\textbf {\bibinfo {volume} {111}},\ \bibinfo {pages} {401} (\bibinfo {year}
  {1987})}\BibitemShut {NoStop}%
\bibitem [{\citenamefont {Karpas}, \citenamefont {Anicich},\ and\ \citenamefont
  {Huntress}(1979)}]{karpas_jcp70}%
  \BibitemOpen
  \bibfield  {author} {\bibinfo {author} {\bibfnamefont {Z.}~\bibnamefont
  {Karpas}}, \bibinfo {author} {\bibfnamefont {V.}~\bibnamefont {Anicich}}, \
  and\ \bibinfo {author} {\bibfnamefont {W.~T.~J.}\ \bibnamefont {Huntress}},\
  }\href@noop {} {\bibfield  {journal} {\bibinfo  {journal} {J. Chem. Phys.}\
  }\textbf {\bibinfo {volume} {70}},\ \bibinfo {pages} {2877} (\bibinfo {year}
  {1979})}\BibitemShut {NoStop}%
\bibitem [{\citenamefont {Mukherjee}\ \emph {et~al.}(2024)\citenamefont
  {Mukherjee}, \citenamefont {Saha}, \citenamefont {Ghosh}, \citenamefont
  {Adhikari}, \citenamefont {Sathyamurthy},\ and\ \citenamefont
  {Baer}}]{friction_jpca128}%
  \BibitemOpen
  \bibfield  {author} {\bibinfo {author} {\bibfnamefont {S.}~\bibnamefont
  {Mukherjee}}, \bibinfo {author} {\bibfnamefont {S.}~\bibnamefont {Saha}},
  \bibinfo {author} {\bibfnamefont {S.}~\bibnamefont {Ghosh}}, \bibinfo
  {author} {\bibfnamefont {S.}~\bibnamefont {Adhikari}}, \bibinfo {author}
  {\bibfnamefont {N.}~\bibnamefont {Sathyamurthy}}, \ and\ \bibinfo {author}
  {\bibfnamefont {M.}~\bibnamefont {Baer}},\ }\href@noop {} {\bibfield
  {journal} {\bibinfo  {journal} {J. Phys. Chem. A}\ }\textbf {\bibinfo
  {volume} {128}},\ \bibinfo {pages} {7691} (\bibinfo {year}
  {2024})}\BibitemShut {NoStop}%
\bibitem [{\citenamefont {Porter}, \citenamefont {Raff},\ and\ \citenamefont
  {Miller}(1975)}]{miller_jcp63}%
  \BibitemOpen
  \bibfield  {author} {\bibinfo {author} {\bibfnamefont {R.~N.}\ \bibnamefont
  {Porter}}, \bibinfo {author} {\bibfnamefont {L.~M.}\ \bibnamefont {Raff}}, \
  and\ \bibinfo {author} {\bibfnamefont {W.~H.}\ \bibnamefont {Miller}},\
  }\href@noop {} {\bibfield  {journal} {\bibinfo  {journal} {J. Chem. Phys.}\
  }\textbf {\bibinfo {volume} {63}},\ \bibinfo {pages} {2214} (\bibinfo {year}
  {1975})}\BibitemShut {NoStop}%
\bibitem [{\citenamefont {Johnson}(1983)}]{johnson_jcp79}%
  \BibitemOpen
  \bibfield  {author} {\bibinfo {author} {\bibfnamefont {B.~R.}\ \bibnamefont
  {Johnson}},\ }\href@noop {} {\bibfield  {journal} {\bibinfo  {journal} {J.
  Chem. Phys.}\ }\textbf {\bibinfo {volume} {79}},\ \bibinfo {pages} {1906}
  (\bibinfo {year} {1983})}\BibitemShut {NoStop}%
\bibitem [{\citenamefont {Billing}(2003)}]{Billing_book}%
  \BibitemOpen
  \bibfield  {author} {\bibinfo {author} {\bibfnamefont {G.~D.}\ \bibnamefont
  {Billing}},\ }\href@noop {} {\emph {\bibinfo {title} {The Quantum Classical
  Theory}}}\ (\bibinfo  {publisher} {Oxford University Press, Oxford},\
  \bibinfo {year} {2003})\BibitemShut {NoStop}%
\bibitem [{\citenamefont {Ravi}\ \emph {et~al.}(2020)\citenamefont {Ravi},
  \citenamefont {Mukherjee}, \citenamefont {Mukherjee}, \citenamefont
  {Adhikari}, \citenamefont {Sathyamurthy},\ and\ \citenamefont
  {Baer}}]{ravi_epjd}%
  \BibitemOpen
  \bibfield  {author} {\bibinfo {author} {\bibfnamefont {S.}~\bibnamefont
  {Ravi}}, \bibinfo {author} {\bibfnamefont {S.}~\bibnamefont {Mukherjee}},
  \bibinfo {author} {\bibfnamefont {B.}~\bibnamefont {Mukherjee}}, \bibinfo
  {author} {\bibfnamefont {S.}~\bibnamefont {Adhikari}}, \bibinfo {author}
  {\bibfnamefont {N.}~\bibnamefont {Sathyamurthy}}, \ and\ \bibinfo {author}
  {\bibfnamefont {M.}~\bibnamefont {Baer}},\ }\href@noop {} {\bibfield
  {journal} {\bibinfo  {journal} {The European Physical Journal D}\ }\textbf
  {\bibinfo {volume} {74}},\ \bibinfo {pages} {239} (\bibinfo {year}
  {2020})}\BibitemShut {NoStop}%
\bibitem [{\citenamefont {Ravi}\ \emph {et~al.}(2021)\citenamefont {Ravi},
  \citenamefont {Mukherjee}, \citenamefont {Mukherjee}, \citenamefont
  {Adhikari}, \citenamefont {Sathyamurthy},\ and\ \citenamefont
  {Baer}}]{ravi_mp119}%
  \BibitemOpen
  \bibfield  {author} {\bibinfo {author} {\bibfnamefont {S.}~\bibnamefont
  {Ravi}}, \bibinfo {author} {\bibfnamefont {S.}~\bibnamefont {Mukherjee}},
  \bibinfo {author} {\bibfnamefont {B.}~\bibnamefont {Mukherjee}}, \bibinfo
  {author} {\bibfnamefont {S.}~\bibnamefont {Adhikari}}, \bibinfo {author}
  {\bibfnamefont {N.}~\bibnamefont {Sathyamurthy}}, \ and\ \bibinfo {author}
  {\bibfnamefont {M.}~\bibnamefont {Baer}},\ }\href@noop {} {\bibfield
  {journal} {\bibinfo  {journal} {Molecular Physics}\ }\textbf {\bibinfo
  {volume} {119}},\ \bibinfo {pages} {e1811907} (\bibinfo {year}
  {2021})}\BibitemShut {NoStop}%
\bibitem [{\citenamefont {Goldstein}(1956)}]{goldstein_classical_book}%
  \BibitemOpen
  \bibfield  {author} {\bibinfo {author} {\bibfnamefont {H.}~\bibnamefont
  {Goldstein}},\ }\href@noop {} {\emph {\bibinfo {title} {Classical
  mechanics}}}\ (\bibinfo  {publisher} {Addison-Wesley Publishing Company:
  Cambridge, Massachusetts},\ \bibinfo {year} {1956})\BibitemShut {NoStop}%
\bibitem [{\citenamefont {Landau}\ and\ \citenamefont
  {Lifshitz}(1997)}]{landau_mechanics_book}%
  \BibitemOpen
  \bibfield  {author} {\bibinfo {author} {\bibfnamefont {L.~D.}\ \bibnamefont
  {Landau}}\ and\ \bibinfo {author} {\bibfnamefont {E.~M.}\ \bibnamefont
  {Lifshitz}},\ }\href@noop {} {\emph {\bibinfo {title} {Mechanics}}}\
  (\bibinfo  {publisher} {Pergamon Press: New York},\ \bibinfo {year}
  {1997})\BibitemShut {NoStop}%
\bibitem [{\citenamefont {Charutz}, \citenamefont {Baer},\ and\ \citenamefont
  {Baer}(1997)}]{baer_cpl265}%
  \BibitemOpen
  \bibfield  {author} {\bibinfo {author} {\bibfnamefont {D.~M.}\ \bibnamefont
  {Charutz}}, \bibinfo {author} {\bibfnamefont {R.}~\bibnamefont {Baer}}, \
  and\ \bibinfo {author} {\bibfnamefont {M.}~\bibnamefont {Baer}},\ }\href@noop
  {} {\bibfield  {journal} {\bibinfo  {journal} {Chem. Phys. Letts.}\ }\textbf
  {\bibinfo {volume} {265}},\ \bibinfo {pages} {629} (\bibinfo {year}
  {1997})}\BibitemShut {NoStop}%
\bibitem [{\citenamefont {Sarkar}\ and\ \citenamefont
  {Adhikari}(2006)}]{sarkar_jcp124}%
  \BibitemOpen
  \bibfield  {author} {\bibinfo {author} {\bibfnamefont {B.}~\bibnamefont
  {Sarkar}}\ and\ \bibinfo {author} {\bibfnamefont {S.}~\bibnamefont
  {Adhikari}},\ }\href@noop {} {\bibfield  {journal} {\bibinfo  {journal} {J.
  Chem. Phys.}\ }\textbf {\bibinfo {volume} {124}},\ \bibinfo {pages} {074101}
  (\bibinfo {year} {2006})}\BibitemShut {NoStop}%
\bibitem [{\citenamefont {Hopper}(1980)}]{hopper_jcp_73}%
  \BibitemOpen
  \bibfield  {author} {\bibinfo {author} {\bibfnamefont {D.~G.}\ \bibnamefont
  {Hopper}},\ }\href@noop {} {\bibfield  {journal} {\bibinfo  {journal} {The
  Journal of Chemical Physics}\ }\textbf {\bibinfo {volume} {73}},\ \bibinfo
  {pages} {3289} (\bibinfo {year} {1980})}\BibitemShut {NoStop}%
\bibitem [{\citenamefont {Kramers}(1940)}]{kramers_physica_7}%
  \BibitemOpen
  \bibfield  {author} {\bibinfo {author} {\bibfnamefont {H.~A.}\ \bibnamefont
  {Kramers}},\ }\href@noop {} {\bibfield  {journal} {\bibinfo  {journal}
  {physica}\ }\textbf {\bibinfo {volume} {7}},\ \bibinfo {pages} {284}
  (\bibinfo {year} {1940})}\BibitemShut {NoStop}%
\bibitem [{\citenamefont {Head-Gordon}\ and\ \citenamefont
  {Tully}(1995)}]{head_jcp103}%
  \BibitemOpen
  \bibfield  {author} {\bibinfo {author} {\bibfnamefont {M.}~\bibnamefont
  {Head-Gordon}}\ and\ \bibinfo {author} {\bibfnamefont {J.~C.}\ \bibnamefont
  {Tully}},\ }\href@noop {} {\bibfield  {journal} {\bibinfo  {journal} {The
  Journal of chemical physics}\ }\textbf {\bibinfo {volume} {103}},\ \bibinfo
  {pages} {10137} (\bibinfo {year} {1995})}\BibitemShut {NoStop}%
\bibitem [{\citenamefont {Spiering}\ \emph {et~al.}(2019)\citenamefont
  {Spiering}, \citenamefont {Shakouri}, \citenamefont {Behler}, \citenamefont
  {Kroes},\ and\ \citenamefont {Meyer}}]{spiering_jpcl10}%
  \BibitemOpen
  \bibfield  {author} {\bibinfo {author} {\bibfnamefont {P.}~\bibnamefont
  {Spiering}}, \bibinfo {author} {\bibfnamefont {K.}~\bibnamefont {Shakouri}},
  \bibinfo {author} {\bibfnamefont {J.}~\bibnamefont {Behler}}, \bibinfo
  {author} {\bibfnamefont {G.-J.}\ \bibnamefont {Kroes}}, \ and\ \bibinfo
  {author} {\bibfnamefont {J.}~\bibnamefont {Meyer}},\ }\href@noop {}
  {\bibfield  {journal} {\bibinfo  {journal} {The journal of physical chemistry
  letters}\ }\textbf {\bibinfo {volume} {10}},\ \bibinfo {pages} {2957}
  (\bibinfo {year} {2019})}\BibitemShut {NoStop}%
\bibitem [{\citenamefont {Smits}\ \emph {et~al.}(2024)\citenamefont {Smits},
  \citenamefont {Sah}, \citenamefont {Naskar}, \citenamefont {Adhikari},
  \citenamefont {Meyer},\ and\ \citenamefont {Somers}}]{smits_jcp161}%
  \BibitemOpen
  \bibfield  {author} {\bibinfo {author} {\bibfnamefont {B.}~\bibnamefont
  {Smits}}, \bibinfo {author} {\bibfnamefont {M.~K.}\ \bibnamefont {Sah}},
  \bibinfo {author} {\bibfnamefont {K.}~\bibnamefont {Naskar}}, \bibinfo
  {author} {\bibfnamefont {S.}~\bibnamefont {Adhikari}}, \bibinfo {author}
  {\bibfnamefont {J.}~\bibnamefont {Meyer}}, \ and\ \bibinfo {author}
  {\bibfnamefont {M.~F.}\ \bibnamefont {Somers}},\ }\href@noop {} {\bibfield
  {journal} {\bibinfo  {journal} {The Journal of Chemical Physics}\ }\textbf
  {\bibinfo {volume} {161}} (\bibinfo {year} {2024})}\BibitemShut {NoStop}%
\bibitem [{\citenamefont {Sah}\ \emph {et~al.}(2024)\citenamefont {Sah},
  \citenamefont {Naskar}, \citenamefont {Adhikari}, \citenamefont {Smits},
  \citenamefont {Meyer},\ and\ \citenamefont {Somers}}]{sah_jcp161}%
  \BibitemOpen
  \bibfield  {author} {\bibinfo {author} {\bibfnamefont {M.~K.}\ \bibnamefont
  {Sah}}, \bibinfo {author} {\bibfnamefont {K.}~\bibnamefont {Naskar}},
  \bibinfo {author} {\bibfnamefont {S.}~\bibnamefont {Adhikari}}, \bibinfo
  {author} {\bibfnamefont {B.}~\bibnamefont {Smits}}, \bibinfo {author}
  {\bibfnamefont {J.}~\bibnamefont {Meyer}}, \ and\ \bibinfo {author}
  {\bibfnamefont {M.~F.}\ \bibnamefont {Somers}},\ }\href@noop {} {\bibfield
  {journal} {\bibinfo  {journal} {The Journal of Chemical Physics}\ }\textbf
  {\bibinfo {volume} {161}} (\bibinfo {year} {2024})}\BibitemShut {NoStop}%
\bibitem [{\citenamefont {Saalfrank}(2006)}]{saalfrank_cr106}%
  \BibitemOpen
  \bibfield  {author} {\bibinfo {author} {\bibfnamefont {P.}~\bibnamefont
  {Saalfrank}},\ }\href@noop {} {\bibfield  {journal} {\bibinfo  {journal}
  {Chemical Reviews}\ }\textbf {\bibinfo {volume} {106}},\ \bibinfo {pages}
  {4116} (\bibinfo {year} {2006})}\BibitemShut {NoStop}%
\bibitem [{\citenamefont {Baer}\ \emph {et~al.}(2024)\citenamefont {Baer},
  \citenamefont {Mukherjee}, \citenamefont {Ravi}, \citenamefont {Adhikari},\
  and\ \citenamefont {Sathyamurthy}}]{baer_aqc89}%
  \BibitemOpen
  \bibfield  {author} {\bibinfo {author} {\bibfnamefont {M.}~\bibnamefont
  {Baer}}, \bibinfo {author} {\bibfnamefont {S.}~\bibnamefont {Mukherjee}},
  \bibinfo {author} {\bibfnamefont {S.}~\bibnamefont {Ravi}}, \bibinfo {author}
  {\bibfnamefont {S.}~\bibnamefont {Adhikari}}, \ and\ \bibinfo {author}
  {\bibfnamefont {N.}~\bibnamefont {Sathyamurthy}},\ }in\ \href@noop {} {\emph
  {\bibinfo {booktitle} {Advances in Quantum Chemistry}}},\ Vol.~\bibinfo
  {volume} {89}\ (\bibinfo  {publisher} {Elsevier},\ \bibinfo {year} {2024})\
  pp.\ \bibinfo {pages} {291--304}\BibitemShut {NoStop}%
\bibitem [{\citenamefont {Mukherjee}\ \emph {et~al.}(2015)\citenamefont
  {Mukherjee}, \citenamefont {Mukherjee}, \citenamefont {Sardar},\ and\
  \citenamefont {Adhikari}}]{mukherjee_jcp143}%
  \BibitemOpen
  \bibfield  {author} {\bibinfo {author} {\bibfnamefont {S.}~\bibnamefont
  {Mukherjee}}, \bibinfo {author} {\bibfnamefont {B.}~\bibnamefont
  {Mukherjee}}, \bibinfo {author} {\bibfnamefont {S.}~\bibnamefont {Sardar}}, \
  and\ \bibinfo {author} {\bibfnamefont {S.}~\bibnamefont {Adhikari}},\
  }\href@noop {} {\bibfield  {journal} {\bibinfo  {journal} {J. Chem. Phys.}\
  }\textbf {\bibinfo {volume} {143}},\ \bibinfo {pages} {244307} (\bibinfo
  {year} {2015})}\BibitemShut {NoStop}%
\end{thebibliography}%


\begin{thebibliography}{102}%
\makeatletter
\providecommand \@ifxundefined [1]{%
 \@ifx{#1\undefined}
}%
\providecommand \@ifnum [1]{%
 \ifnum #1\expandafter \@firstoftwo
 \else \expandafter \@secondoftwo
 \fi
}%
\providecommand \@ifx [1]{%
 \ifx #1\expandafter \@firstoftwo
 \else \expandafter \@secondoftwo
 \fi
}%
\providecommand \natexlab [1]{#1}%
\providecommand \enquote  [1]{``#1''}%
\providecommand \bibnamefont  [1]{#1}%
\providecommand \bibfnamefont [1]{#1}%
\providecommand \citenamefont [1]{#1}%
\providecommand \href@noop [0]{\@secondoftwo}%
\providecommand \href [0]{\begingroup \@sanitize@url \@href}%
\providecommand \@href[1]{\@@startlink{#1}\@@href}%
\providecommand \@@href[1]{\endgroup#1\@@endlink}%
\providecommand \@sanitize@url [0]{\catcode `\\12\catcode `\$12\catcode
  `\&12\catcode `\#12\catcode `\^12\catcode `\_12\catcode `\%12\relax}%
\providecommand \@@startlink[1]{}%
\providecommand \@@endlink[0]{}%
\providecommand \url  [0]{\begingroup\@sanitize@url \@url }%
\providecommand \@url [1]{\endgroup\@href {#1}{\urlprefix }}%
\providecommand \urlprefix  [0]{URL }%
\providecommand \Eprint [0]{\href }%
\providecommand \doibase [0]{http://dx.doi.org/}%
\providecommand \selectlanguage [0]{\@gobble}%
\providecommand \bibinfo  [0]{\@secondoftwo}%
\providecommand \bibfield  [0]{\@secondoftwo}%
\providecommand \translation [1]{[#1]}%
\providecommand \BibitemOpen [0]{}%
\providecommand \bibitemStop [0]{}%
\providecommand \bibitemNoStop [0]{.\EOS\space}%
\providecommand \EOS [0]{\spacefactor3000\relax}%
\providecommand \BibitemShut  [1]{\csname bibitem#1\endcsname}%
\let\auto@bib@innerbib\@empty
\bibitem [{\citenamefont {Tokura}\ and\ \citenamefont
  {Tomioka}(1999)}]{Tokura_1999}%
  \BibitemOpen
  \bibfield  {author} {\bibinfo {author} {\bibfnamefont {Y.}~\bibnamefont
  {Tokura}}\ and\ \bibinfo {author} {\bibfnamefont {Y.}~\bibnamefont
  {Tomioka}},\ }\href {\doibase https://doi.org/10.1016/S0304-8853(99)00352-2}
  {\bibfield  {journal} {\bibinfo  {journal} {Journal of Magnetism and Magnetic
  Materials}\ }\textbf {\bibinfo {volume} {200}},\ \bibinfo {pages} {1}
  (\bibinfo {year} {1999})}\BibitemShut {NoStop}%
\bibitem [{\citenamefont {Dagotto}, \citenamefont {Hotta},\ and\ \citenamefont
  {Moreo}(2001)}]{Dagotto_2001}%
  \BibitemOpen
  \bibfield  {author} {\bibinfo {author} {\bibfnamefont {E.}~\bibnamefont
  {Dagotto}}, \bibinfo {author} {\bibfnamefont {T.}~\bibnamefont {Hotta}}, \
  and\ \bibinfo {author} {\bibfnamefont {A.}~\bibnamefont {Moreo}},\ }\href
  {\doibase https://doi.org/10.1016/S0370-1573(00)00121-6} {\bibfield
  {journal} {\bibinfo  {journal} {Physics Reports}\ }\textbf {\bibinfo {volume}
  {344}},\ \bibinfo {pages} {1} (\bibinfo {year} {2001})}\BibitemShut {NoStop}%
\bibitem [{\citenamefont {Lines}\ and\ \citenamefont
  {Glass}(2001)}]{Lines_2001}%
  \BibitemOpen
  \bibfield  {author} {\bibinfo {author} {\bibfnamefont {M.}~\bibnamefont
  {Lines}}\ and\ \bibinfo {author} {\bibfnamefont {A.}~\bibnamefont {Glass}},\
  }\href {https://books.google.co.in/books?id=p6ruJH8C84kC} {\emph {\bibinfo
  {title} {Principles and Applications of Ferroelectrics and Related
  Materials}}},\ International series of monographs on physics\ (\bibinfo
  {publisher} {OUP Oxford},\ \bibinfo {year} {2001})\BibitemShut {NoStop}%
\bibitem [{Jaf(1971)}]{Jaffe_1971}%
  \BibitemOpen
  in\ \href {\doibase https://doi.org/10.1016/B978-0-12-379550-2.50001-6}
  {\emph {\bibinfo {booktitle} {Piezoelectric Ceramics}}},\ \bibinfo {editor}
  {edited by\ \bibinfo {editor} {\bibfnamefont {B.}~\bibnamefont {JAFFE}},
  \bibinfo {editor} {\bibfnamefont {W.~R.}\ \bibnamefont {COOK}}, \ and\
  \bibinfo {editor} {\bibfnamefont {H.}~\bibnamefont {JAFFE}}}\ (\bibinfo
  {publisher} {Academic Press},\ \bibinfo {year} {1971})\ p.~\bibinfo {pages}
  {ii}\BibitemShut {NoStop}%
\bibitem [{\citenamefont {Ramesh}\ and\ \citenamefont
  {Spaldin}(2007)}]{Ramesh_2007}%
  \BibitemOpen
  \bibfield  {author} {\bibinfo {author} {\bibfnamefont {R.}~\bibnamefont
  {Ramesh}}\ and\ \bibinfo {author} {\bibfnamefont {N.~A.}\ \bibnamefont
  {Spaldin}},\ }\href {\doibase 10.1038/nmat1805} {\bibfield  {journal}
  {\bibinfo  {journal} {Nature Materials}\ }\textbf {\bibinfo {volume} {6}},\
  \bibinfo {pages} {21–29} (\bibinfo {year} {2007})}\BibitemShut {NoStop}%
\bibitem [{\citenamefont {Jahn}\ and\ \citenamefont {Teller}(1937)}]{JT}%
  \BibitemOpen
  \bibfield  {author} {\bibinfo {author} {\bibfnamefont {H.~A.}\ \bibnamefont
  {Jahn}}\ and\ \bibinfo {author} {\bibfnamefont {E.}~\bibnamefont {Teller}},\
  }\href@noop {} {\bibfield  {journal} {\bibinfo  {journal} {Proc. R. Soc.
  London A}\ }\textbf {\bibinfo {volume} {161}},\ \bibinfo {pages} {220}
  (\bibinfo {year} {1937})}\BibitemShut {NoStop}%
\bibitem [{\citenamefont {Jahn}(1937)}]{JT1}%
  \BibitemOpen
  \bibfield  {author} {\bibinfo {author} {\bibfnamefont {H.~A.}\ \bibnamefont
  {Jahn}},\ }\href@noop {} {\bibfield  {journal} {\bibinfo  {journal} {Proc. R.
  Soc. London A}\ }\textbf {\bibinfo {volume} {164}},\ \bibinfo {pages} {117}
  (\bibinfo {year} {1937})}\BibitemShut {NoStop}%
\bibitem [{\citenamefont {Vleck}(1939{\natexlab{a}})}]{VanVleck1}%
  \BibitemOpen
  \bibfield  {author} {\bibinfo {author} {\bibfnamefont {V.}~\bibnamefont
  {Vleck}},\ }\href@noop {} {\bibfield  {journal} {\bibinfo  {journal} {J.
  Chem. Phys}\ }\textbf {\bibinfo {volume} {7}},\ \bibinfo {pages} {61}
  (\bibinfo {year} {1939}{\natexlab{a}})}\BibitemShut {NoStop}%
\bibitem [{\citenamefont {Vleck}(1939{\natexlab{b}})}]{VanVleck}%
  \BibitemOpen
  \bibfield  {author} {\bibinfo {author} {\bibfnamefont {V.}~\bibnamefont
  {Vleck}},\ }\href@noop {} {\bibfield  {journal} {\bibinfo  {journal} {J.
  Chem. Phys}\ }\textbf {\bibinfo {volume} {7}},\ \bibinfo {pages} {72}
  (\bibinfo {year} {1939}{\natexlab{b}})}\BibitemShut {NoStop}%
\bibitem [{\citenamefont {\"Opik}\ and\ \citenamefont {Pryce}(1957)}]{Opik}%
  \BibitemOpen
  \bibfield  {author} {\bibinfo {author} {\bibfnamefont {U.}~\bibnamefont
  {\"Opik}}\ and\ \bibinfo {author} {\bibfnamefont {M.~H.~L.}\ \bibnamefont
  {Pryce}},\ }\href@noop {} {\bibfield  {journal} {\bibinfo  {journal} {Proc.
  R. Soc. London, Ser. A}\ }\textbf {\bibinfo {volume} {238}},\ \bibinfo
  {pages} {425} (\bibinfo {year} {1957})}\BibitemShut {NoStop}%
\bibitem [{\citenamefont {Longuet-Higgins}(1961)}]{Longuet-Higgins}%
  \BibitemOpen
  \bibfield  {author} {\bibinfo {author} {\bibfnamefont {H.~C.}\ \bibnamefont
  {Longuet-Higgins}},\ }\href@noop {} {\emph {\bibinfo {title} {{\it Advances
  in Spectroscopy}}}},\ Vol.~\bibinfo {volume} {2}\ (\bibinfo  {publisher}
  {Interscience Publishers, Ltd.},\ \bibinfo {address} {New York and London},\
  \bibinfo {year} {1961})\ Chap.\ \bibinfo {chapter} {Some Recent Developments
  in the Theory of Molecular Energy Levels}\BibitemShut {NoStop}%
\bibitem [{\citenamefont {O'Brien}(1964)}]{O'Brien}%
  \BibitemOpen
  \bibfield  {author} {\bibinfo {author} {\bibfnamefont {M.~C.~M.}\
  \bibnamefont {O'Brien}},\ }\href@noop {} {\bibfield  {journal} {\bibinfo
  {journal} {Proc. Roy. Soc. A}\ }\textbf {\bibinfo {volume} {281}},\ \bibinfo
  {pages} {323} (\bibinfo {year} {1964})}\BibitemShut {NoStop}%
\bibitem [{\citenamefont {Bersuker}(1963)}]{Bersuker1}%
  \BibitemOpen
  \bibfield  {author} {\bibinfo {author} {\bibfnamefont {I.~B.}\ \bibnamefont
  {Bersuker}},\ }\href@noop {} {\bibfield  {journal} {\bibinfo  {journal} {Sov.
  Phys. JETP}\ }\textbf {\bibinfo {volume} {17}},\ \bibinfo {pages} {933}
  (\bibinfo {year} {1963})}\BibitemShut {NoStop}%
\bibitem [{\citenamefont {Bersuker}, \citenamefont {Vekhter},\ and\
  \citenamefont {Ogurtsov}(1975)}]{Sovphys}%
  \BibitemOpen
  \bibfield  {author} {\bibinfo {author} {\bibfnamefont {I.~B.}\ \bibnamefont
  {Bersuker}}, \bibinfo {author} {\bibfnamefont {B.~J.}\ \bibnamefont
  {Vekhter}}, \ and\ \bibinfo {author} {\bibfnamefont {I.~Y.}\ \bibnamefont
  {Ogurtsov}},\ }\href@noop {} {\bibfield  {journal} {\bibinfo  {journal} {Sov.
  Phys. Usp.}\ }\textbf {\bibinfo {volume} {18}},\ \bibinfo {pages} {569}
  (\bibinfo {year} {1975})}\BibitemShut {NoStop}%
\bibitem [{\citenamefont {Englman}\ and\ \citenamefont
  {Halperin}(1970)}]{Cooperative-JT-Englman-I}%
  \BibitemOpen
  \bibfield  {author} {\bibinfo {author} {\bibfnamefont {R.}~\bibnamefont
  {Englman}}\ and\ \bibinfo {author} {\bibfnamefont {B.}~\bibnamefont
  {Halperin}},\ }\href@noop {} {\bibfield  {journal} {\bibinfo  {journal}
  {Phys. Rev. B}\ }\textbf {\bibinfo {volume} {2}},\ \bibinfo {pages} {75}
  (\bibinfo {year} {1970})}\BibitemShut {NoStop}%
\bibitem [{\citenamefont {Englman}\ and\ \citenamefont
  {Halperin}(1971)}]{Cooperative-JT-Englman-II}%
  \BibitemOpen
  \bibfield  {author} {\bibinfo {author} {\bibfnamefont {R.}~\bibnamefont
  {Englman}}\ and\ \bibinfo {author} {\bibfnamefont {B.}~\bibnamefont
  {Halperin}},\ }\href@noop {} {\bibfield  {journal} {\bibinfo  {journal}
  {Phys. Rev. B}\ }\textbf {\bibinfo {volume} {3}},\ \bibinfo {pages} {1698}
  (\bibinfo {year} {1971})}\BibitemShut {NoStop}%
\bibitem [{\citenamefont {Slonczewski}(1963)}]{JCS}%
  \BibitemOpen
  \bibfield  {author} {\bibinfo {author} {\bibfnamefont {J.~C.}\ \bibnamefont
  {Slonczewski}},\ }\href@noop {} {\bibfield  {journal} {\bibinfo  {journal}
  {Phys. Rev}\ }\textbf {\bibinfo {volume} {131}},\ \bibinfo {pages} {1596}
  (\bibinfo {year} {1963})}\BibitemShut {NoStop}%
\bibitem [{\citenamefont {Englman}(1972)}]{REnglman}%
  \BibitemOpen
  \bibfield  {author} {\bibinfo {author} {\bibfnamefont {R.}~\bibnamefont
  {Englman}},\ }\href@noop {} {\emph {\bibinfo {title} {The Jahn-Teller Effect
  in Molecules and Crystals}}}\ (\bibinfo  {publisher} {Wiley},\ \bibinfo
  {address} {New York},\ \bibinfo {year} {1972})\BibitemShut {NoStop}%
\bibitem [{\citenamefont {Bersuker}\ and\ \citenamefont
  {Polinger}(1989)}]{IBersuker-VPolinger}%
  \BibitemOpen
  \bibfield  {author} {\bibinfo {author} {\bibfnamefont {I.~B.}\ \bibnamefont
  {Bersuker}}\ and\ \bibinfo {author} {\bibfnamefont {V.~Z.}\ \bibnamefont
  {Polinger}},\ }\href@noop {} {\emph {\bibinfo {title} {Vibronic Interactions
  in Molecules and Crystals}}}\ (\bibinfo  {publisher} {Springer},\ \bibinfo
  {address} {Berlin},\ \bibinfo {year} {1989})\BibitemShut {NoStop}%
\bibitem [{\citenamefont {Bersuker}(1996)}]{IBersukerbook}%
  \BibitemOpen
  \bibfield  {author} {\bibinfo {author} {\bibfnamefont {I.~B.}\ \bibnamefont
  {Bersuker}},\ }\href@noop {} {\emph {\bibinfo {title} {Electronic Structure
  and Properties of Transition Metal Compounds}}}\ (\bibinfo  {publisher} {J
  Wiley},\ \bibinfo {address} {New York},\ \bibinfo {year} {1996})\BibitemShut
  {NoStop}%
\bibitem [{\citenamefont {Bersuker}(2001)}]{JT-review-Bersuker}%
  \BibitemOpen
  \bibfield  {author} {\bibinfo {author} {\bibfnamefont {I.~B.}\ \bibnamefont
  {Bersuker}},\ }\href@noop {} {\bibfield  {journal} {\bibinfo  {journal}
  {Chem. Rev.}\ }\textbf {\bibinfo {volume} {101}},\ \bibinfo {pages} {1067}
  (\bibinfo {year} {2001})}\BibitemShut {NoStop}%
\bibitem [{\citenamefont {Bersuker}(2006)}]{IBB}%
  \BibitemOpen
  \bibfield  {author} {\bibinfo {author} {\bibfnamefont {I.~B.}\ \bibnamefont
  {Bersuker}},\ }\href@noop {} {\emph {\bibinfo {title} {The Jahn-Teller
  Effect}}}\ (\bibinfo  {publisher} {Cambridge University Press},\ \bibinfo
  {address} {Cambridge},\ \bibinfo {year} {2006})\BibitemShut {NoStop}%
\bibitem [{\citenamefont {Kugel}\ and\ \citenamefont
  {Khomskii}(1982)}]{Kugel-Khomskii}%
  \BibitemOpen
  \bibfield  {author} {\bibinfo {author} {\bibfnamefont {K.~I.}\ \bibnamefont
  {Kugel}}\ and\ \bibinfo {author} {\bibfnamefont {D.~I.}\ \bibnamefont
  {Khomskii}},\ }\href@noop {} {\bibfield  {journal} {\bibinfo  {journal} {Sov.
  Phys. Usp.}\ }\textbf {\bibinfo {volume} {25}},\ \bibinfo {pages} {231}
  (\bibinfo {year} {1982})}\BibitemShut {NoStop}%
\bibitem [{\citenamefont {Ole\'s}\ \emph {et~al.}(2005)\citenamefont {Ole\'s},
  \citenamefont {Khaliullin}, \citenamefont {Horsch},\ and\ \citenamefont
  {Feiner}}]{Oles-Feiner}%
  \BibitemOpen
  \bibfield  {author} {\bibinfo {author} {\bibfnamefont {A.~M.}\ \bibnamefont
  {Ole\'s}}, \bibinfo {author} {\bibfnamefont {G.}~\bibnamefont {Khaliullin}},
  \bibinfo {author} {\bibfnamefont {P.}~\bibnamefont {Horsch}}, \ and\ \bibinfo
  {author} {\bibfnamefont {L.~F.}\ \bibnamefont {Feiner}},\ }\href@noop {}
  {\bibfield  {journal} {\bibinfo  {journal} {Phys. Rev. B}\ }\textbf {\bibinfo
  {volume} {72}},\ \bibinfo {pages} {214431} (\bibinfo {year}
  {2005})}\BibitemShut {NoStop}%
\bibitem [{\citenamefont {Kovaleva}\ \emph {et~al.}(2004)\citenamefont
  {Kovaleva}, \citenamefont {Boris}, \citenamefont {Bernhard}, \citenamefont
  {Kulakov}, \citenamefont {Pimenov}, \citenamefont {Balbashov}, \citenamefont
  {Khaliullin},\ and\ \citenamefont {Keimer}}]{Natalia_JETP1}%
  \BibitemOpen
  \bibfield  {author} {\bibinfo {author} {\bibfnamefont {N.~N.}\ \bibnamefont
  {Kovaleva}}, \bibinfo {author} {\bibfnamefont {A.~V.}\ \bibnamefont {Boris}},
  \bibinfo {author} {\bibfnamefont {C.}~\bibnamefont {Bernhard}}, \bibinfo
  {author} {\bibfnamefont {A.}~\bibnamefont {Kulakov}}, \bibinfo {author}
  {\bibfnamefont {A.}~\bibnamefont {Pimenov}}, \bibinfo {author} {\bibfnamefont
  {A.~M.}\ \bibnamefont {Balbashov}}, \bibinfo {author} {\bibfnamefont
  {G.}~\bibnamefont {Khaliullin}}, \ and\ \bibinfo {author} {\bibfnamefont
  {B.}~\bibnamefont {Keimer}},\ }\href@noop {} {\bibfield  {journal} {\bibinfo
  {journal} {Phys. Rev. Lett.}\ }\textbf {\bibinfo {volume} {93}},\ \bibinfo
  {pages} {147204} (\bibinfo {year} {2004})}\BibitemShut {NoStop}%
\bibitem [{\citenamefont {Kovaleva}\ \emph {et~al.}(2010)\citenamefont
  {Kovaleva}, \citenamefont {Ole\'s}, \citenamefont {Balbashov}, \citenamefont
  {Maljuk}, \citenamefont {Argyriou}, \citenamefont {Khaliullin},\ and\
  \citenamefont {Keimer}}]{Natalia_JETP2}%
  \BibitemOpen
  \bibfield  {author} {\bibinfo {author} {\bibfnamefont {N.~N.}\ \bibnamefont
  {Kovaleva}}, \bibinfo {author} {\bibfnamefont {A.~M.}\ \bibnamefont
  {Ole\'s}}, \bibinfo {author} {\bibfnamefont {A.~M.}\ \bibnamefont
  {Balbashov}}, \bibinfo {author} {\bibfnamefont {A.}~\bibnamefont {Maljuk}},
  \bibinfo {author} {\bibfnamefont {D.~N.}\ \bibnamefont {Argyriou}}, \bibinfo
  {author} {\bibfnamefont {G.}~\bibnamefont {Khaliullin}}, \ and\ \bibinfo
  {author} {\bibfnamefont {B.}~\bibnamefont {Keimer}},\ }\href@noop {}
  {\bibfield  {journal} {\bibinfo  {journal} {Phys. Rev. B.}\ }\textbf
  {\bibinfo {volume} {81}},\ \bibinfo {pages} {235130} (\bibinfo {year}
  {2010})}\BibitemShut {NoStop}%
\bibitem [{\citenamefont {Mondal}\ \emph {et~al.}(2011)\citenamefont {Mondal},
  \citenamefont {Opalka}, \citenamefont {Poluyanov},\ and\ \citenamefont
  {Domcke}}]{domcke_cp387}%
  \BibitemOpen
  \bibfield  {author} {\bibinfo {author} {\bibfnamefont {P.}~\bibnamefont
  {Mondal}}, \bibinfo {author} {\bibfnamefont {D.}~\bibnamefont {Opalka}},
  \bibinfo {author} {\bibfnamefont {L.~V.}\ \bibnamefont {Poluyanov}}, \ and\
  \bibinfo {author} {\bibfnamefont {W.}~\bibnamefont {Domcke}},\ }\href@noop {}
  {\bibfield  {journal} {\bibinfo  {journal} {Chem. Phys.}\ }\textbf {\bibinfo
  {volume} {387}},\ \bibinfo {pages} {56} (\bibinfo {year} {2011})}\BibitemShut
  {NoStop}%
\bibitem [{\citenamefont {Solomonik}\ and\ \citenamefont
  {Mukhanov}(2012)}]{mukhanov_jsc53}%
  \BibitemOpen
  \bibfield  {author} {\bibinfo {author} {\bibfnamefont {V.~G.}\ \bibnamefont
  {Solomonik}}\ and\ \bibinfo {author} {\bibfnamefont {A.~A.}\ \bibnamefont
  {Mukhanov}},\ }\href@noop {} {\bibfield  {journal} {\bibinfo  {journal} {J.
  Struct. Chem.}\ }\textbf {\bibinfo {volume} {53}},\ \bibinfo {pages} {819}
  (\bibinfo {year} {2012})}\BibitemShut {NoStop}%
\bibitem [{\citenamefont {Yates}\ and\ \citenamefont
  {Pitzer}(1979)}]{pitzer_jcp70}%
  \BibitemOpen
  \bibfield  {author} {\bibinfo {author} {\bibfnamefont {J.~H.}\ \bibnamefont
  {Yates}}\ and\ \bibinfo {author} {\bibfnamefont {R.~M.}\ \bibnamefont
  {Pitzer}},\ }\href@noop {} {\bibfield  {journal} {\bibinfo  {journal} {J.
  Chem. Phys.}\ }\textbf {\bibinfo {volume} {70}},\ \bibinfo {pages} {4049}
  (\bibinfo {year} {1979})}\BibitemShut {NoStop}%
\bibitem [{\citenamefont {Solomonik}\ and\ \citenamefont
  {Mukhanov}(2014)}]{mukhanov_rjpca88}%
  \BibitemOpen
  \bibfield  {author} {\bibinfo {author} {\bibfnamefont {V.~G.}\ \bibnamefont
  {Solomonik}}\ and\ \bibinfo {author} {\bibfnamefont {A.~A.}\ \bibnamefont
  {Mukhanov}},\ }\href@noop {} {\bibfield  {journal} {\bibinfo  {journal}
  {Russ. J. Phys. Chem.}\ }\textbf {\bibinfo {volume} {88}},\ \bibinfo {pages}
  {85} (\bibinfo {year} {2014})}\BibitemShut {NoStop}%
\bibitem [{\citenamefont {Nandipati}\ \emph {et~al.}(2020)\citenamefont
  {Nandipati}, \citenamefont {Vasilyev}, \citenamefont {Navarkin},
  \citenamefont {Solomonik},\ and\ \citenamefont {Domcke}}]{domcke_jcp152}%
  \BibitemOpen
  \bibfield  {author} {\bibinfo {author} {\bibfnamefont {K.~R.}\ \bibnamefont
  {Nandipati}}, \bibinfo {author} {\bibfnamefont {O.~A.}\ \bibnamefont
  {Vasilyev}}, \bibinfo {author} {\bibfnamefont {I.~S.}\ \bibnamefont
  {Navarkin}}, \bibinfo {author} {\bibfnamefont {V.~G.}\ \bibnamefont
  {Solomonik}}, \ and\ \bibinfo {author} {\bibfnamefont {W.}~\bibnamefont
  {Domcke}},\ }\href@noop {} {\bibfield  {journal} {\bibinfo  {journal} {J.
  Chem. Phys.}\ }\textbf {\bibinfo {volume} {152}},\ \bibinfo {pages} {094304}
  (\bibinfo {year} {2020})}\BibitemShut {NoStop}%
\bibitem [{\citenamefont {Rodr\'{i}guez-Carvajal}\ \emph
  {et~al.}(1998)\citenamefont {Rodr\'{i}guez-Carvajal}, \citenamefont
  {Hennion}, \citenamefont {Moussa}, \citenamefont {Moudden}, \citenamefont
  {Pinsard},\ and\ \citenamefont {Revcolevschi}}]{Phase}%
  \BibitemOpen
  \bibfield  {author} {\bibinfo {author} {\bibfnamefont {J.}~\bibnamefont
  {Rodr\'{i}guez-Carvajal}}, \bibinfo {author} {\bibfnamefont {M.}~\bibnamefont
  {Hennion}}, \bibinfo {author} {\bibfnamefont {F.}~\bibnamefont {Moussa}},
  \bibinfo {author} {\bibfnamefont {A.~H.}\ \bibnamefont {Moudden}}, \bibinfo
  {author} {\bibfnamefont {L.}~\bibnamefont {Pinsard}}, \ and\ \bibinfo
  {author} {\bibfnamefont {A.}~\bibnamefont {Revcolevschi}},\ }\href@noop {}
  {\bibfield  {journal} {\bibinfo  {journal} {Phys. Rev. B}\ }\textbf {\bibinfo
  {volume} {57}},\ \bibinfo {pages} {R3189} (\bibinfo {year}
  {1998})}\BibitemShut {NoStop}%
\bibitem [{\citenamefont {Hirota}\ \emph {et~al.}(1996)\citenamefont {Hirota},
  \citenamefont {Kaneko}, \citenamefont {Nishizawa},\ and\ \citenamefont
  {Endoh}}]{Hirota}%
  \BibitemOpen
  \bibfield  {author} {\bibinfo {author} {\bibfnamefont {K.}~\bibnamefont
  {Hirota}}, \bibinfo {author} {\bibfnamefont {N.}~\bibnamefont {Kaneko}},
  \bibinfo {author} {\bibfnamefont {A.}~\bibnamefont {Nishizawa}}, \ and\
  \bibinfo {author} {\bibfnamefont {Y.}~\bibnamefont {Endoh}},\ }\href@noop {}
  {\bibfield  {journal} {\bibinfo  {journal} {J. Phys. Soc. Japan}\ }\textbf
  {\bibinfo {volume} {65}},\ \bibinfo {pages} {3736} (\bibinfo {year}
  {1996})}\BibitemShut {NoStop}%
\bibitem [{\citenamefont {Kovaleva}\ \emph {et~al.}(2016)\citenamefont
  {Kovaleva}, \citenamefont {Kugel}, \citenamefont {Pot\.{u}\u{c}ek},
  \citenamefont {Kusmartseva}, \citenamefont {Goryachev}, \citenamefont
  {Bryknar}, \citenamefont {Demikhov}, \citenamefont {Trepakov}, \citenamefont
  {Dejneka}, \citenamefont {Kusmartsev},\ and\ \citenamefont
  {Stoneham}}]{Natalia_JETP}%
  \BibitemOpen
  \bibfield  {author} {\bibinfo {author} {\bibfnamefont {N.~N.}\ \bibnamefont
  {Kovaleva}}, \bibinfo {author} {\bibfnamefont {K.~I.}\ \bibnamefont {Kugel}},
  \bibinfo {author} {\bibfnamefont {Z.}~\bibnamefont {Pot\.{u}\u{c}ek}},
  \bibinfo {author} {\bibfnamefont {O.~E.}\ \bibnamefont {Kusmartseva}},
  \bibinfo {author} {\bibfnamefont {N.~S.}\ \bibnamefont {Goryachev}}, \bibinfo
  {author} {\bibfnamefont {Z.}~\bibnamefont {Bryknar}}, \bibinfo {author}
  {\bibfnamefont {E.~I.}\ \bibnamefont {Demikhov}}, \bibinfo {author}
  {\bibfnamefont {V.~A.}\ \bibnamefont {Trepakov}}, \bibinfo {author}
  {\bibfnamefont {A.}~\bibnamefont {Dejneka}}, \bibinfo {author} {\bibfnamefont
  {F.}~\bibnamefont {Kusmartsev}}, \ and\ \bibinfo {author} {\bibfnamefont
  {A.~M.}\ \bibnamefont {Stoneham}},\ }\href@noop {} {\bibfield  {journal}
  {\bibinfo  {journal} {J. Exp. Theor. Phys.}\ }\textbf {\bibinfo {volume}
  {122}},\ \bibinfo {pages} {890} (\bibinfo {year} {2016})}\BibitemShut
  {NoStop}%
\bibitem [{\citenamefont {Dutta}, \citenamefont {Adhikari},\ and\ \citenamefont
  {Kovaleva}(2019)}]{joy_jcp}%
  \BibitemOpen
  \bibfield  {author} {\bibinfo {author} {\bibfnamefont {J.}~\bibnamefont
  {Dutta}}, \bibinfo {author} {\bibfnamefont {S.}~\bibnamefont {Adhikari}}, \
  and\ \bibinfo {author} {\bibfnamefont {N.}~\bibnamefont {Kovaleva}},\
  }\href@noop {} {\bibfield  {journal} {\bibinfo  {journal} {J. Chem. Phys.}\
  }\textbf {\bibinfo {volume} {150}},\ \bibinfo {pages} {064703} (\bibinfo
  {year} {2019})}\BibitemShut {NoStop}%
\bibitem [{\citenamefont {Kovaleva}\ \emph {et~al.}(2017)\citenamefont
  {Kovaleva}, \citenamefont {Kusmartseva}, \citenamefont {Kugel},\ and\
  \citenamefont {Kusmartsev}}]{MultiRaman}%
  \BibitemOpen
  \bibfield  {author} {\bibinfo {author} {\bibfnamefont {N.~N.}\ \bibnamefont
  {Kovaleva}}, \bibinfo {author} {\bibfnamefont {O.~E.}\ \bibnamefont
  {Kusmartseva}}, \bibinfo {author} {\bibfnamefont {K.~I.}\ \bibnamefont
  {Kugel}}, \ and\ \bibinfo {author} {\bibfnamefont {F.~V.}\ \bibnamefont
  {Kusmartsev}},\ }\href@noop {} {\bibfield  {journal} {\bibinfo  {journal} {J.
  Phys.: Conf. Ser.}\ }\textbf {\bibinfo {volume} {833}},\ \bibinfo {pages}
  {012005} (\bibinfo {year} {2017})}\BibitemShut {NoStop}%
\bibitem [{\citenamefont {Kovaleva}\ \emph {et~al.}(2002)\citenamefont
  {Kovaleva}, \citenamefont {Gavartin}, \citenamefont {Shluger}, \citenamefont
  {Boris},\ and\ \citenamefont {Stoneham}}]{Natalia_JETP3}%
  \BibitemOpen
  \bibfield  {author} {\bibinfo {author} {\bibfnamefont {N.~N.}\ \bibnamefont
  {Kovaleva}}, \bibinfo {author} {\bibfnamefont {J.~L.}\ \bibnamefont
  {Gavartin}}, \bibinfo {author} {\bibfnamefont {A.~L.}\ \bibnamefont
  {Shluger}}, \bibinfo {author} {\bibfnamefont {A.~V.}\ \bibnamefont {Boris}},
  \ and\ \bibinfo {author} {\bibfnamefont {A.~M.}\ \bibnamefont {Stoneham}},\
  }\href@noop {} {\bibfield  {journal} {\bibinfo  {journal} {J. Exp. Theor.
  Phys.}\ }\textbf {\bibinfo {volume} {94}},\ \bibinfo {pages} {178} (\bibinfo
  {year} {2002})}\BibitemShut {NoStop}%
\bibitem [{\citenamefont {Liehr}\ and\ \citenamefont
  {Ballhausen}(1958)}]{quadratic-coupling1}%
  \BibitemOpen
  \bibfield  {author} {\bibinfo {author} {\bibfnamefont {A.~D.}\ \bibnamefont
  {Liehr}}\ and\ \bibinfo {author} {\bibfnamefont {C.~J.}\ \bibnamefont
  {Ballhausen}},\ }\href@noop {} {\bibfield  {journal} {\bibinfo  {journal}
  {Ann. Phys. (N.Y.)}\ }\textbf {\bibinfo {volume} {3}},\ \bibinfo {pages}
  {304} (\bibinfo {year} {1958})}\BibitemShut {NoStop}%
\bibitem [{\citenamefont {Zwanziger}\ and\ \citenamefont
  {Grant}(1987)}]{quadratic-coupling-strong1}%
  \BibitemOpen
  \bibfield  {author} {\bibinfo {author} {\bibfnamefont {J.~W.}\ \bibnamefont
  {Zwanziger}}\ and\ \bibinfo {author} {\bibfnamefont {E.~R.}\ \bibnamefont
  {Grant}},\ }\href@noop {} {\bibfield  {journal} {\bibinfo  {journal} {J.
  Chem. Phys.}\ }\textbf {\bibinfo {volume} {87}},\ \bibinfo {pages} {2954}
  (\bibinfo {year} {1987})}\BibitemShut {NoStop}%
\bibitem [{\citenamefont {Sadygov}\ and\ \citenamefont
  {Yarkony}(1999)}]{quadratic-coupling-Li3}%
  \BibitemOpen
  \bibfield  {author} {\bibinfo {author} {\bibfnamefont {R.~G.}\ \bibnamefont
  {Sadygov}}\ and\ \bibinfo {author} {\bibfnamefont {D.~R.}\ \bibnamefont
  {Yarkony}},\ }\href@noop {} {\bibfield  {journal} {\bibinfo  {journal} {J.
  Chem. Phys.}\ }\textbf {\bibinfo {volume} {110}},\ \bibinfo {pages} {3639}
  (\bibinfo {year} {1999})}\BibitemShut {NoStop}%
\bibitem [{\citenamefont {Yarkony}(1999)}]{quadratic-coupling-Na3}%
  \BibitemOpen
  \bibfield  {author} {\bibinfo {author} {\bibfnamefont {D.~R.}\ \bibnamefont
  {Yarkony}},\ }\href@noop {} {\bibfield  {journal} {\bibinfo  {journal} {J.
  Chem. Phys.}\ }\textbf {\bibinfo {volume} {111}},\ \bibinfo {pages} {4906}
  (\bibinfo {year} {1999})}\BibitemShut {NoStop}%
\bibitem [{\citenamefont {Mukherjee}\ \emph {et~al.}(2013)\citenamefont
  {Mukherjee}, \citenamefont {Bandyopadhyay}, \citenamefont {Paul},\ and\
  \citenamefont {Adhikari}}]{mukherjee_jpca117}%
  \BibitemOpen
  \bibfield  {author} {\bibinfo {author} {\bibfnamefont {S.}~\bibnamefont
  {Mukherjee}}, \bibinfo {author} {\bibfnamefont {S.}~\bibnamefont
  {Bandyopadhyay}}, \bibinfo {author} {\bibfnamefont {A.~K.}\ \bibnamefont
  {Paul}}, \ and\ \bibinfo {author} {\bibfnamefont {S.}~\bibnamefont
  {Adhikari}},\ }\href@noop {} {\bibfield  {journal} {\bibinfo  {journal} {J.
  Phys. Chem. A}\ }\textbf {\bibinfo {volume} {117}},\ \bibinfo {pages} {3475}
  (\bibinfo {year} {2013})}\BibitemShut {NoStop}%
\bibitem [{\citenamefont {Koizumi}\ and\ \citenamefont
  {Bersuker}(1999)}]{quadratic-coupling-Exe}%
  \BibitemOpen
  \bibfield  {author} {\bibinfo {author} {\bibfnamefont {H.}~\bibnamefont
  {Koizumi}}\ and\ \bibinfo {author} {\bibfnamefont {I.~B.}\ \bibnamefont
  {Bersuker}},\ }\href@noop {} {\bibfield  {journal} {\bibinfo  {journal}
  {Phys. Rev. Lett.}\ }\textbf {\bibinfo {volume} {83}},\ \bibinfo {pages}
  {3009} (\bibinfo {year} {1999})}\BibitemShut {NoStop}%
\bibitem [{\citenamefont {Koizumi}\ \emph {et~al.}(2000)\citenamefont
  {Koizumi}, \citenamefont {Bersuker}, \citenamefont {Boggs},\ and\
  \citenamefont {Polinger}}]{quadratic-coupling-Txt2}%
  \BibitemOpen
  \bibfield  {author} {\bibinfo {author} {\bibfnamefont {H.}~\bibnamefont
  {Koizumi}}, \bibinfo {author} {\bibfnamefont {I.~B.}\ \bibnamefont
  {Bersuker}}, \bibinfo {author} {\bibfnamefont {J.~E.}\ \bibnamefont {Boggs}},
  \ and\ \bibinfo {author} {\bibfnamefont {V.~Z.}\ \bibnamefont {Polinger}},\
  }\href@noop {} {\bibfield  {journal} {\bibinfo  {journal} {J. Chem. Phys.}\
  }\textbf {\bibinfo {volume} {112}},\ \bibinfo {pages} {8470} (\bibinfo {year}
  {2000})}\BibitemShut {NoStop}%
\bibitem [{\citenamefont {Pae}\ and\ \citenamefont
  {Hizhnyakov}(2017)}]{quadratic-coupling-Exe-nonadiabatic}%
  \BibitemOpen
  \bibfield  {author} {\bibinfo {author} {\bibfnamefont {K.}~\bibnamefont
  {Pae}}\ and\ \bibinfo {author} {\bibfnamefont {V.}~\bibnamefont
  {Hizhnyakov}},\ }\href@noop {} {\bibfield  {journal} {\bibinfo  {journal} {J.
  Chem. Phys.}\ }\textbf {\bibinfo {volume} {147}},\ \bibinfo {pages} {084107}
  (\bibinfo {year} {2017})}\BibitemShut {NoStop}%
\bibitem [{\citenamefont {Garc\'{i}a-Fern\'{a}ndez}\ \emph
  {et~al.}(2005)\citenamefont {Garc\'{i}a-Fern\'{a}ndez}, \citenamefont
  {Bersuker}, \citenamefont {Aramburu}, \citenamefont {Barriuso},\ and\
  \citenamefont {Moreno}}]{warping-Exe}%
  \BibitemOpen
  \bibfield  {author} {\bibinfo {author} {\bibfnamefont {P.}~\bibnamefont
  {Garc\'{i}a-Fern\'{a}ndez}}, \bibinfo {author} {\bibfnamefont {I.~B.}\
  \bibnamefont {Bersuker}}, \bibinfo {author} {\bibfnamefont {J.~A.}\
  \bibnamefont {Aramburu}}, \bibinfo {author} {\bibfnamefont {M.~T.}\
  \bibnamefont {Barriuso}}, \ and\ \bibinfo {author} {\bibfnamefont
  {M.}~\bibnamefont {Moreno}},\ }\href@noop {} {\bibfield  {journal} {\bibinfo
  {journal} {Phys. Rev. B.}\ }\textbf {\bibinfo {volume} {71}},\ \bibinfo
  {pages} {184117} (\bibinfo {year} {2005})}\BibitemShut {NoStop}%
\bibitem [{\citenamefont {Hizhnyakov}, \citenamefont {Pae},\ and\ \citenamefont
  {Vaikj\"arv}(2012)}]{Hizhnyakov1}%
  \BibitemOpen
  \bibfield  {author} {\bibinfo {author} {\bibfnamefont {V.}~\bibnamefont
  {Hizhnyakov}}, \bibinfo {author} {\bibfnamefont {K.}~\bibnamefont {Pae}}, \
  and\ \bibinfo {author} {\bibfnamefont {T.}~\bibnamefont {Vaikj\"arv}},\
  }\href@noop {} {\bibfield  {journal} {\bibinfo  {journal} {Chem. Phys.
  Lett.}\ }\textbf {\bibinfo {volume} {525-526}},\ \bibinfo {pages} {64}
  (\bibinfo {year} {2012})}\BibitemShut {NoStop}%
\bibitem [{\citenamefont {Pae}\ and\ \citenamefont
  {Hizhnyakov}(2013)}]{Hizhnyakov2}%
  \BibitemOpen
  \bibfield  {author} {\bibinfo {author} {\bibfnamefont {K.}~\bibnamefont
  {Pae}}\ and\ \bibinfo {author} {\bibfnamefont {V.}~\bibnamefont
  {Hizhnyakov}},\ }\href@noop {} {\bibfield  {journal} {\bibinfo  {journal} {J.
  Chem. Phys.}\ }\textbf {\bibinfo {volume} {138}},\ \bibinfo {pages} {104103}
  (\bibinfo {year} {2013})}\BibitemShut {NoStop}%
\bibitem [{\citenamefont {Pae}\ and\ \citenamefont
  {Hizhnyakov}(2014)}]{Hizhnyakov3}%
  \BibitemOpen
  \bibfield  {author} {\bibinfo {author} {\bibfnamefont {K.}~\bibnamefont
  {Pae}}\ and\ \bibinfo {author} {\bibfnamefont {V.}~\bibnamefont
  {Hizhnyakov}},\ }\href@noop {} {\bibfield  {journal} {\bibinfo  {journal} {J.
  Chem. Phys.}\ }\textbf {\bibinfo {volume} {141}},\ \bibinfo {pages} {234113}
  (\bibinfo {year} {2014})}\BibitemShut {NoStop}%
\bibitem [{\citenamefont {Pae}\ and\ \citenamefont
  {Hizhnyakov}(2016)}]{Hizhnyakov4}%
  \BibitemOpen
  \bibfield  {author} {\bibinfo {author} {\bibfnamefont {K.}~\bibnamefont
  {Pae}}\ and\ \bibinfo {author} {\bibfnamefont {V.}~\bibnamefont
  {Hizhnyakov}},\ }\href@noop {} {\bibfield  {journal} {\bibinfo  {journal} {J.
  Chem. Phys.}\ }\textbf {\bibinfo {volume} {145}},\ \bibinfo {pages} {064108}
  (\bibinfo {year} {2016})}\BibitemShut {NoStop}%
\bibitem [{\citenamefont {K\"{o}ppel}\ \emph {et~al.}(2009)\citenamefont
  {K\"{o}ppel}, \citenamefont {Yarkony}, \citenamefont {Barentzen},\ and\
  \citenamefont {Eds.}}]{Bersuker-Yarkony-book}%
  \BibitemOpen
  \bibfield  {author} {\bibinfo {author} {\bibfnamefont {H.}~\bibnamefont
  {K\"{o}ppel}}, \bibinfo {author} {\bibfnamefont {D.~R.}\ \bibnamefont
  {Yarkony}}, \bibinfo {author} {\bibfnamefont {H.}~\bibnamefont {Barentzen}},
  \ and\ \bibinfo {author} {\bibnamefont {Eds.}},\ }\href@noop {} {\emph
  {\bibinfo {title} {The Jahn-Teller effect. Fundamentals and Implications for
  Physics and Chemistry}}}\ (\bibinfo  {publisher} {Springer Series of Chemical
  Physics; Springer},\ \bibinfo {address} {Heidelberg},\ \bibinfo {year}
  {2009})\BibitemShut {NoStop}%
\bibitem [{\citenamefont {Garcia-Fernandez}, \citenamefont {Bersuker},\ and\
  \citenamefont {Boggs}(2006)}]{Hidden-JT1}%
  \BibitemOpen
  \bibfield  {author} {\bibinfo {author} {\bibfnamefont {P.}~\bibnamefont
  {Garcia-Fernandez}}, \bibinfo {author} {\bibfnamefont {I.~B.}\ \bibnamefont
  {Bersuker}}, \ and\ \bibinfo {author} {\bibfnamefont {J.~E.}\ \bibnamefont
  {Boggs}},\ }\href@noop {} {\bibfield  {journal} {\bibinfo  {journal} {Phys.
  Rev. Lett.}\ }\textbf {\bibinfo {volume} {96}},\ \bibinfo {pages} {163005}
  (\bibinfo {year} {2006})}\BibitemShut {NoStop}%
\bibitem [{\citenamefont {Liu}\ \emph {et~al.}(2009)\citenamefont {Liu},
  \citenamefont {Bersuker}, \citenamefont {Zou},\ and\ \citenamefont
  {Boggs}}]{Hidden-JT-PJT}%
  \BibitemOpen
  \bibfield  {author} {\bibinfo {author} {\bibfnamefont {Y.}~\bibnamefont
  {Liu}}, \bibinfo {author} {\bibfnamefont {I.~B.}\ \bibnamefont {Bersuker}},
  \bibinfo {author} {\bibfnamefont {W.}~\bibnamefont {Zou}}, \ and\ \bibinfo
  {author} {\bibfnamefont {J.~E.}\ \bibnamefont {Boggs}},\ }\href@noop {}
  {\bibfield  {journal} {\bibinfo  {journal} {J. Chem. Theory Comput.}\
  }\textbf {\bibinfo {volume} {5}},\ \bibinfo {pages} {2679} (\bibinfo {year}
  {2009})}\BibitemShut {NoStop}%
\bibitem [{\citenamefont {Bersuker}(1980)}]{PJT0}%
  \BibitemOpen
  \bibfield  {author} {\bibinfo {author} {\bibfnamefont {I.~B.}\ \bibnamefont
  {Bersuker}},\ }\href@noop {} {\bibfield  {journal} {\bibinfo  {journal}
  {Nouv. J. Chim.}\ }\textbf {\bibinfo {volume} {4}},\ \bibinfo {pages} {139}
  (\bibinfo {year} {1980})}\BibitemShut {NoStop}%
\bibitem [{\citenamefont {Bersuker}, \citenamefont {Gorinchoi},\ and\
  \citenamefont {Polinger}(1984)}]{PJT1}%
  \BibitemOpen
  \bibfield  {author} {\bibinfo {author} {\bibfnamefont {I.~B.}\ \bibnamefont
  {Bersuker}}, \bibinfo {author} {\bibfnamefont {N.~N.}\ \bibnamefont
  {Gorinchoi}}, \ and\ \bibinfo {author} {\bibfnamefont {V.~Z.}\ \bibnamefont
  {Polinger}},\ }\href@noop {} {\bibfield  {journal} {\bibinfo  {journal}
  {Theoret. Chim. Acta}\ }\textbf {\bibinfo {volume} {66}},\ \bibinfo {pages}
  {161} (\bibinfo {year} {1984})}\BibitemShut {NoStop}%
\bibitem [{\citenamefont {Bersuker}(1988)}]{PJT2}%
  \BibitemOpen
  \bibfield  {author} {\bibinfo {author} {\bibfnamefont {I.~B.}\ \bibnamefont
  {Bersuker}},\ }\href@noop {} {\bibfield  {journal} {\bibinfo  {journal} {Pure
  Appl. Chem.}\ }\textbf {\bibinfo {volume} {60}},\ \bibinfo {pages} {1167}
  (\bibinfo {year} {1988})}\BibitemShut {NoStop}%
\bibitem [{\citenamefont {Bersuker}, \citenamefont {Gorinchoi},\ and\
  \citenamefont {Polinger}(1992)}]{PJT3}%
  \BibitemOpen
  \bibfield  {author} {\bibinfo {author} {\bibfnamefont {I.~B.}\ \bibnamefont
  {Bersuker}}, \bibinfo {author} {\bibfnamefont {N.~N.}\ \bibnamefont
  {Gorinchoi}}, \ and\ \bibinfo {author} {\bibfnamefont {V.~Z.}\ \bibnamefont
  {Polinger}},\ }\href@noop {} {\bibfield  {journal} {\bibinfo  {journal} {J.
  Mol. Struct.}\ }\textbf {\bibinfo {volume} {270}},\ \bibinfo {pages} {369}
  (\bibinfo {year} {1992})}\BibitemShut {NoStop}%
\bibitem [{\citenamefont {Bersuker}(2018)}]{Bersuker18112018}%
  \BibitemOpen
  \bibfield  {author} {\bibinfo {author} {\bibfnamefont {I.~B.}\ \bibnamefont
  {Bersuker}},\ }\href@noop {} {\bibfield  {journal} {\bibinfo  {journal}
  {Ferroelectrics}\ }\textbf {\bibinfo {volume} {536}},\ \bibinfo {pages} {1}
  (\bibinfo {year} {2018})}\BibitemShut {NoStop}%
\bibitem [{\citenamefont {Bersuker}(2021)}]{Bersuker_review_2021}%
  \BibitemOpen
  \bibfield  {author} {\bibinfo {author} {\bibfnamefont {I.~B.}\ \bibnamefont
  {Bersuker}},\ }\href@noop {} {\bibfield  {journal} {\bibinfo  {journal}
  {Chem. Rev.}\ }\textbf {\bibinfo {volume} {121}},\ \bibinfo {pages} {1463}
  (\bibinfo {year} {2021})}\BibitemShut {NoStop}%
\bibitem [{\citenamefont {Bersuker}(1966)}]{Bersuker_1966}%
  \BibitemOpen
  \bibfield  {author} {\bibinfo {author} {\bibfnamefont {I.~B.}\ \bibnamefont
  {Bersuker}},\ }\href@noop {} {\bibfield  {journal} {\bibinfo  {journal}
  {Phys. Lett.}\ }\textbf {\bibinfo {volume} {20}},\ \bibinfo {pages} {589}
  (\bibinfo {year} {1966})}\BibitemShut {NoStop}%
\bibitem [{\citenamefont {Kole\.{z}y\'{n}ski}\ and\ \citenamefont
  {Tkacz-\'{S}miech}(2005)}]{Kolezynski_2005}%
  \BibitemOpen
  \bibfield  {author} {\bibinfo {author} {\bibfnamefont {A.}~\bibnamefont
  {Kole\.{z}y\'{n}ski}}\ and\ \bibinfo {author} {\bibfnamefont
  {K.}~\bibnamefont {Tkacz-\'{S}miech}},\ }\href@noop {} {\bibfield  {journal}
  {\bibinfo  {journal} {Ferroelectrics}\ }\textbf {\bibinfo {volume} {314}},\
  \bibinfo {pages} {123} (\bibinfo {year} {2005})}\BibitemShut {NoStop}%
\bibitem [{\citenamefont {Liu}\ \emph {et~al.}(2013)\citenamefont {Liu},
  \citenamefont {Zhang}, \citenamefont {Liu}, \citenamefont {Wang},\ and\
  \citenamefont {Liu}}]{Qi-Jun_Liu_2013}%
  \BibitemOpen
  \bibfield  {author} {\bibinfo {author} {\bibfnamefont {Q.-J.}\ \bibnamefont
  {Liu}}, \bibinfo {author} {\bibfnamefont {N.-C.}\ \bibnamefont {Zhang}},
  \bibinfo {author} {\bibfnamefont {F.-S.}\ \bibnamefont {Liu}}, \bibinfo
  {author} {\bibfnamefont {H.-Y.}\ \bibnamefont {Wang}}, \ and\ \bibinfo
  {author} {\bibfnamefont {Z.-T.}\ \bibnamefont {Liu}},\ }\href@noop {}
  {\bibfield  {journal} {\bibinfo  {journal} {Opt. Mater.}\ }\textbf {\bibinfo
  {volume} {35}},\ \bibinfo {pages} {2629} (\bibinfo {year}
  {2013})}\BibitemShut {NoStop}%
\bibitem [{\citenamefont {Polinger}, \citenamefont {Garc\'{\i}a-Fern\'andez},\
  and\ \citenamefont {Bersuker}(2015)}]{bersuker_pbcm457}%
  \BibitemOpen
  \bibfield  {author} {\bibinfo {author} {\bibfnamefont {V.}~\bibnamefont
  {Polinger}}, \bibinfo {author} {\bibfnamefont {P.}~\bibnamefont
  {Garc\'{\i}a-Fern\'andez}}, \ and\ \bibinfo {author} {\bibfnamefont {I.~B.}\
  \bibnamefont {Bersuker}},\ }\href@noop {} {\bibfield  {journal} {\bibinfo
  {journal} {Physica B: Condensed Matter}\ }\textbf {\bibinfo {volume} {457}},\
  \bibinfo {pages} {296} (\bibinfo {year} {2015})}\BibitemShut {NoStop}%
\bibitem [{\citenamefont {Bersuker}(2015)}]{Bersuker_2015}%
  \BibitemOpen
  \bibfield  {author} {\bibinfo {author} {\bibfnamefont {I.~B.}\ \bibnamefont
  {Bersuker}},\ }\href@noop {} {\bibfield  {journal} {\bibinfo  {journal}
  {Appl. Phys. Lett.}\ }\textbf {\bibinfo {volume} {106}},\ \bibinfo {pages}
  {022903} (\bibinfo {year} {2015})}\BibitemShut {NoStop}%
\bibitem [{\citenamefont {Mirhosseini}\ \emph {et~al.}(2010)\citenamefont
  {Mirhosseini}, \citenamefont {Maznichenko}, \citenamefont {Abdelouahed},
  \citenamefont {Ostanin}, \citenamefont {Ernst}, \citenamefont {Mertig},\ and\
  \citenamefont {Henk}}]{Mirhosseini_phys_rev_2010}%
  \BibitemOpen
  \bibfield  {author} {\bibinfo {author} {\bibfnamefont {H.}~\bibnamefont
  {Mirhosseini}}, \bibinfo {author} {\bibfnamefont {I.~V.}\ \bibnamefont
  {Maznichenko}}, \bibinfo {author} {\bibfnamefont {S.}~\bibnamefont
  {Abdelouahed}}, \bibinfo {author} {\bibfnamefont {S.}~\bibnamefont
  {Ostanin}}, \bibinfo {author} {\bibfnamefont {A.}~\bibnamefont {Ernst}},
  \bibinfo {author} {\bibfnamefont {I.}~\bibnamefont {Mertig}}, \ and\ \bibinfo
  {author} {\bibfnamefont {J.}~\bibnamefont {Henk}},\ }\href {\doibase
  10.1103/PhysRevB.81.073406} {\bibfield  {journal} {\bibinfo  {journal} {Phys.
  Rev. B}\ }\textbf {\bibinfo {volume} {81}},\ \bibinfo {pages} {073406}
  (\bibinfo {year} {2010})}\BibitemShut {NoStop}%
\bibitem [{\citenamefont {Zhong}\ \emph {et~al.}(2015)\citenamefont {Zhong},
  \citenamefont {Si}, \citenamefont {Zhang}, \citenamefont {Yin}, \citenamefont
  {Yunoki},\ and\ \citenamefont {Held}}]{Zhong_Adv_Mat}%
  \BibitemOpen
  \bibfield  {author} {\bibinfo {author} {\bibfnamefont {Z.}~\bibnamefont
  {Zhong}}, \bibinfo {author} {\bibfnamefont {L.}~\bibnamefont {Si}}, \bibinfo
  {author} {\bibfnamefont {Q.}~\bibnamefont {Zhang}}, \bibinfo {author}
  {\bibfnamefont {W.-G.}\ \bibnamefont {Yin}}, \bibinfo {author} {\bibfnamefont
  {S.}~\bibnamefont {Yunoki}}, \ and\ \bibinfo {author} {\bibfnamefont
  {K.}~\bibnamefont {Held}},\ }\href@noop {} {\bibfield  {journal} {\bibinfo
  {journal} {Adv. Mater. Interfaces}\ }\textbf {\bibinfo {volume} {2}},\
  \bibinfo {pages} {1400445} (\bibinfo {year} {2015})}\BibitemShut {NoStop}%
\bibitem [{\citenamefont {Ali}\ \emph {et~al.}(2019)\citenamefont {Ali},
  \citenamefont {Khan}, \citenamefont {Ali}, \citenamefont {Khan},\ and\
  \citenamefont {Ahmad}}]{Ali_BFO_BTO}%
  \BibitemOpen
  \bibfield  {author} {\bibinfo {author} {\bibfnamefont {A.}~\bibnamefont
  {Ali}}, \bibinfo {author} {\bibfnamefont {I.}~\bibnamefont {Khan}}, \bibinfo
  {author} {\bibfnamefont {Z.}~\bibnamefont {Ali}}, \bibinfo {author}
  {\bibfnamefont {F.}~\bibnamefont {Khan}}, \ and\ \bibinfo {author}
  {\bibfnamefont {I.}~\bibnamefont {Ahmad}},\ }\href@noop {} {\bibfield
  {journal} {\bibinfo  {journal} {Int. J. Mod. Phys. B}\ }\textbf {\bibinfo
  {volume} {33}},\ \bibinfo {pages} {1950231} (\bibinfo {year}
  {2019})}\BibitemShut {NoStop}%
\bibitem [{\citenamefont {Pinczuk}\ \emph {et~al.}(1967)\citenamefont
  {Pinczuk}, \citenamefont {Taylor}, \citenamefont {Burstein},\ and\
  \citenamefont {Lefkowitz}}]{Pinczuki_1967}%
  \BibitemOpen
  \bibfield  {author} {\bibinfo {author} {\bibfnamefont {A.}~\bibnamefont
  {Pinczuk}}, \bibinfo {author} {\bibfnamefont {W.}~\bibnamefont {Taylor}},
  \bibinfo {author} {\bibfnamefont {E.}~\bibnamefont {Burstein}}, \ and\
  \bibinfo {author} {\bibfnamefont {I.}~\bibnamefont {Lefkowitz}},\ }\href@noop
  {} {\bibfield  {journal} {\bibinfo  {journal} {Solid State Commun.}\ }\textbf
  {\bibinfo {volume} {5}},\ \bibinfo {pages} {429} (\bibinfo {year}
  {1967})}\BibitemShut {NoStop}%
\bibitem [{\citenamefont {Turik}\ and\ \citenamefont
  {Shevchenko}(1979)}]{Turik_Shevchenko_1979}%
  \BibitemOpen
  \bibfield  {author} {\bibinfo {author} {\bibfnamefont {A.~V.}\ \bibnamefont
  {Turik}}\ and\ \bibinfo {author} {\bibfnamefont {N.~B.}\ \bibnamefont
  {Shevchenko}},\ }\href@noop {} {\bibfield  {journal} {\bibinfo  {journal}
  {Phys. Stat. Sol. (B)}\ }\textbf {\bibinfo {volume} {95}},\ \bibinfo {pages}
  {585} (\bibinfo {year} {1979})}\BibitemShut {NoStop}%
\bibitem [{\citenamefont {Cord}\ and\ \citenamefont
  {Courths}(1985)}]{R_Cord_1985}%
  \BibitemOpen
  \bibfield  {author} {\bibinfo {author} {\bibfnamefont {B.}~\bibnamefont
  {Cord}}\ and\ \bibinfo {author} {\bibfnamefont {R.}~\bibnamefont {Courths}},\
  }\href@noop {} {\bibfield  {journal} {\bibinfo  {journal} {Surf. Sci.}\
  }\textbf {\bibinfo {volume} {152-153}},\ \bibinfo {pages} {1141} (\bibinfo
  {year} {1985})}\BibitemShut {NoStop}%
\bibitem [{\citenamefont {Carley}\ \emph {et~al.}(1987)\citenamefont {Carley},
  \citenamefont {Chalker}, \citenamefont {Riviere},\ and\ \citenamefont
  {Roberts}}]{Roberts_1987}%
  \BibitemOpen
  \bibfield  {author} {\bibinfo {author} {\bibfnamefont {A.~F.}\ \bibnamefont
  {Carley}}, \bibinfo {author} {\bibfnamefont {P.~R.}\ \bibnamefont {Chalker}},
  \bibinfo {author} {\bibfnamefont {J.~C.}\ \bibnamefont {Riviere}}, \ and\
  \bibinfo {author} {\bibfnamefont {M.~W.}\ \bibnamefont {Roberts}},\
  }\href@noop {} {\bibfield  {journal} {\bibinfo  {journal} {J. Chem. Soc.
  Faraday Trans. 1}\ }\textbf {\bibinfo {volume} {83}},\ \bibinfo {pages} {351}
  (\bibinfo {year} {1987})}\BibitemShut {NoStop}%
\bibitem [{\citenamefont {Nasser}(2000)}]{Nasser_Expt_2000}%
  \BibitemOpen
  \bibfield  {author} {\bibinfo {author} {\bibfnamefont {S.~A.}\ \bibnamefont
  {Nasser}},\ }\href@noop {} {\bibfield  {journal} {\bibinfo  {journal} {Appl.
  Surf. Sci.}\ }\textbf {\bibinfo {volume} {157}},\ \bibinfo {pages} {14}
  (\bibinfo {year} {2000})}\BibitemShut {NoStop}%
\bibitem [{\citenamefont {Wan}, \citenamefont {Han},\ and\ \citenamefont
  {Zhu}(2008)}]{Wan_2008}%
  \BibitemOpen
  \bibfield  {author} {\bibinfo {author} {\bibfnamefont {F.}~\bibnamefont
  {Wan}}, \bibinfo {author} {\bibfnamefont {J.}~\bibnamefont {Han}}, \ and\
  \bibinfo {author} {\bibfnamefont {Z.}~\bibnamefont {Zhu}},\ }\href@noop {}
  {\bibfield  {journal} {\bibinfo  {journal} {Phys. Lett. A}\ }\textbf
  {\bibinfo {volume} {372}},\ \bibinfo {pages} {2137} (\bibinfo {year}
  {2008})}\BibitemShut {NoStop}%
\bibitem [{\citenamefont {Sah}\ \emph {et~al.}(2025)\citenamefont {Sah},
  \citenamefont {Mukherjee}, \citenamefont {Ravi},\ and\ \citenamefont
  {Adhikari}}]{Mantu_BTO}%
  \BibitemOpen
  \bibfield  {author} {\bibinfo {author} {\bibfnamefont {M.~K.}\ \bibnamefont
  {Sah}}, \bibinfo {author} {\bibfnamefont {S.}~\bibnamefont {Mukherjee}},
  \bibinfo {author} {\bibfnamefont {S.}~\bibnamefont {Ravi}}, \ and\ \bibinfo
  {author} {\bibfnamefont {S.}~\bibnamefont {Adhikari}},\ }\href@noop {}
  {\bibfield  {journal} {\bibinfo  {journal} {Physical Chemistry Chemical
  Physics}\ }\textbf {\bibinfo {volume} {27}},\ \bibinfo {pages} {15759}
  (\bibinfo {year} {2025})}\BibitemShut {NoStop}%
\bibitem [{\citenamefont {Longuet-Higgins}(1975)}]{longuet-higgins_prsl344}%
  \BibitemOpen
  \bibfield  {author} {\bibinfo {author} {\bibfnamefont {H.~C.}\ \bibnamefont
  {Longuet-Higgins}},\ }\href@noop {} {\bibfield  {journal} {\bibinfo
  {journal} {Proc. R. Soc. London, Ser. A}\ }\textbf {\bibinfo {volume}
  {344}},\ \bibinfo {pages} {147} (\bibinfo {year} {1975})}\BibitemShut
  {NoStop}%
\bibitem [{\citenamefont {Kumar~Sah}\ and\ \citenamefont
  {Adhikari}()}]{Mantu_Solid-state_Review}%
  \BibitemOpen
  \bibfield  {author} {\bibinfo {author} {\bibfnamefont {M.}~\bibnamefont
  {Kumar~Sah}}\ and\ \bibinfo {author} {\bibfnamefont {S.}~\bibnamefont
  {Adhikari}},\ }\href@noop {} {\bibfield  {journal} {\bibinfo  {journal} {J.
  Comput. Chem.}\ }\textbf {\bibinfo {volume} {46}},\ \bibinfo {pages}
  {e70239}}\BibitemShut {NoStop}%
\bibitem [{\citenamefont {Born}\ and\ \citenamefont {Oppenheimer}(1927)}]{BO}%
  \BibitemOpen
  \bibfield  {author} {\bibinfo {author} {\bibfnamefont {M.}~\bibnamefont
  {Born}}\ and\ \bibinfo {author} {\bibfnamefont {R.}~\bibnamefont
  {Oppenheimer}},\ }\href@noop {} {\bibfield  {journal} {\bibinfo  {journal}
  {Ann. Phys.}\ }\textbf {\bibinfo {volume} {389}},\ \bibinfo {pages} {457}
  (\bibinfo {year} {1927})}\BibitemShut {NoStop}%
\bibitem [{\citenamefont {Born}\ and\ \citenamefont {Huang}(1954)}]{BH}%
  \BibitemOpen
  \bibfield  {author} {\bibinfo {author} {\bibfnamefont {M.}~\bibnamefont
  {Born}}\ and\ \bibinfo {author} {\bibfnamefont {K.}~\bibnamefont {Huang}},\
  }\href@noop {} {\emph {\bibinfo {title} {Dynamical Theory of Crystal
  Lattices}}}\ (\bibinfo  {publisher} {Oxford University Press, Oxford},\
  \bibinfo {year} {1954})\BibitemShut {NoStop}%
\bibitem [{\citenamefont {Hellmann}(1937)}]{hellmann}%
  \BibitemOpen
  \bibfield  {author} {\bibinfo {author} {\bibfnamefont {H.}~\bibnamefont
  {Hellmann}},\ }\href@noop {} {\emph {\bibinfo {title} {{\it Einfuring in die
  Quantenchemie}}}}\ (\bibinfo  {publisher} {Franz Duetiche:Leipzig},\ \bibinfo
  {address} {Germany},\ \bibinfo {year} {1937})\BibitemShut {NoStop}%
\bibitem [{\citenamefont {Feynmann}\ and\ \citenamefont
  {Hibbs}(1965)}]{feynmann}%
  \BibitemOpen
  \bibfield  {author} {\bibinfo {author} {\bibfnamefont {R.~P.}\ \bibnamefont
  {Feynmann}}\ and\ \bibinfo {author} {\bibfnamefont {A.~R.}\ \bibnamefont
  {Hibbs}},\ }\href@noop {} {\emph {\bibinfo {title} {Quantum Mechanics and
  Path Integrals}}}\ (\bibinfo  {publisher} {McCraw-Hill, New York},\ \bibinfo
  {year} {1965})\BibitemShut {NoStop}%
\bibitem [{\citenamefont {Baer}(2002)}]{baer_phys_rep358}%
  \BibitemOpen
  \bibfield  {author} {\bibinfo {author} {\bibfnamefont {M.}~\bibnamefont
  {Baer}},\ }\href@noop {} {\bibfield  {journal} {\bibinfo  {journal} {Phys.
  Rep.}\ }\textbf {\bibinfo {volume} {358}},\ \bibinfo {pages} {75} (\bibinfo
  {year} {2002})}\BibitemShut {NoStop}%
\bibitem [{\citenamefont {Mukherjee}\ \emph {et~al.}(2019)\citenamefont
  {Mukherjee}, \citenamefont {Naskar}, \citenamefont {Mukherjee}, \citenamefont
  {Ghosh}, \citenamefont {Sahoo},\ and\ \citenamefont
  {Adhikari}}]{bijit_irpc38}%
  \BibitemOpen
  \bibfield  {author} {\bibinfo {author} {\bibfnamefont {B.}~\bibnamefont
  {Mukherjee}}, \bibinfo {author} {\bibfnamefont {K.}~\bibnamefont {Naskar}},
  \bibinfo {author} {\bibfnamefont {S.}~\bibnamefont {Mukherjee}}, \bibinfo
  {author} {\bibfnamefont {S.}~\bibnamefont {Ghosh}}, \bibinfo {author}
  {\bibfnamefont {T.}~\bibnamefont {Sahoo}}, \ and\ \bibinfo {author}
  {\bibfnamefont {S.}~\bibnamefont {Adhikari}},\ }\href@noop {} {\bibfield
  {journal} {\bibinfo  {journal} {Int. Rev. Phys. Chem.}\ }\textbf {\bibinfo
  {volume} {38}},\ \bibinfo {pages} {287} (\bibinfo {year} {2019})}\BibitemShut
  {NoStop}%
\bibitem [{\citenamefont {Hazra}\ \emph {et~al.}(2022)\citenamefont {Hazra},
  \citenamefont {Mukherjee}, \citenamefont {Ravi}, \citenamefont {Sardar},\
  and\ \citenamefont {Adhikari}}]{saikat_pyra}%
  \BibitemOpen
  \bibfield  {author} {\bibinfo {author} {\bibfnamefont {S.}~\bibnamefont
  {Hazra}}, \bibinfo {author} {\bibfnamefont {S.}~\bibnamefont {Mukherjee}},
  \bibinfo {author} {\bibfnamefont {S.}~\bibnamefont {Ravi}}, \bibinfo {author}
  {\bibfnamefont {S.}~\bibnamefont {Sardar}}, \ and\ \bibinfo {author}
  {\bibfnamefont {S.}~\bibnamefont {Adhikari}},\ }\href@noop {} {\bibfield
  {journal} {\bibinfo  {journal} {ChemPhysChem}\ }\textbf {\bibinfo {volume}
  {23}},\ \bibinfo {pages} {e202200482} (\bibinfo {year} {2022})}\BibitemShut
  {NoStop}%
\bibitem [{\citenamefont {Naskar}\ \emph {et~al.}(2020)\citenamefont {Naskar},
  \citenamefont {Mukherjee}, \citenamefont {Mukherjee}, \citenamefont {Ravi},
  \citenamefont {Mukherjee}, \citenamefont {Sardar},\ and\ \citenamefont
  {Adhikari}}]{naskar_jctc}%
  \BibitemOpen
  \bibfield  {author} {\bibinfo {author} {\bibfnamefont {K.}~\bibnamefont
  {Naskar}}, \bibinfo {author} {\bibfnamefont {S.}~\bibnamefont {Mukherjee}},
  \bibinfo {author} {\bibfnamefont {B.}~\bibnamefont {Mukherjee}}, \bibinfo
  {author} {\bibfnamefont {S.}~\bibnamefont {Ravi}}, \bibinfo {author}
  {\bibfnamefont {S.}~\bibnamefont {Mukherjee}}, \bibinfo {author}
  {\bibfnamefont {S.}~\bibnamefont {Sardar}}, \ and\ \bibinfo {author}
  {\bibfnamefont {S.}~\bibnamefont {Adhikari}},\ }\href@noop {} {\bibfield
  {journal} {\bibinfo  {journal} {J. Chem. Theory Comput.}\ }\textbf {\bibinfo
  {volume} {16}},\ \bibinfo {pages} {1666} (\bibinfo {year}
  {2020})}\BibitemShut {NoStop}%
\bibitem [{\citenamefont {Adhikari}\ and\ \citenamefont
  {Billing}(2000)}]{SAJCP113}%
  \BibitemOpen
  \bibfield  {author} {\bibinfo {author} {\bibfnamefont {S.}~\bibnamefont
  {Adhikari}}\ and\ \bibinfo {author} {\bibfnamefont {G.~D.}\ \bibnamefont
  {Billing}},\ }\href@noop {} {\bibfield  {journal} {\bibinfo  {journal} {J.
  Chem. Phys.}\ }\textbf {\bibinfo {volume} {113}},\ \bibinfo {pages} {1409}
  (\bibinfo {year} {2000})}\BibitemShut {NoStop}%
\bibitem [{\citenamefont {Puzari}, \citenamefont {Sarkar},\ and\ \citenamefont
  {Adhikari}(2004)}]{SAJCP121}%
  \BibitemOpen
  \bibfield  {author} {\bibinfo {author} {\bibfnamefont {P.}~\bibnamefont
  {Puzari}}, \bibinfo {author} {\bibfnamefont {B.}~\bibnamefont {Sarkar}}, \
  and\ \bibinfo {author} {\bibfnamefont {S.}~\bibnamefont {Adhikari}},\
  }\href@noop {} {\bibfield  {journal} {\bibinfo  {journal} {J. Chem. Phys.}\
  }\textbf {\bibinfo {volume} {121}},\ \bibinfo {pages} {707} (\bibinfo {year}
  {2004})}\BibitemShut {NoStop}%
\bibitem [{\citenamefont {Puzari}, \citenamefont {Sarkar},\ and\ \citenamefont
  {Adhikari}(2005)}]{SAIJQC105}%
  \BibitemOpen
  \bibfield  {author} {\bibinfo {author} {\bibfnamefont {P.}~\bibnamefont
  {Puzari}}, \bibinfo {author} {\bibfnamefont {B.}~\bibnamefont {Sarkar}}, \
  and\ \bibinfo {author} {\bibfnamefont {S.}~\bibnamefont {Adhikari}},\
  }\href@noop {} {\bibfield  {journal} {\bibinfo  {journal} {Int. J. Quant.
  Chem.}\ }\textbf {\bibinfo {volume} {105}},\ \bibinfo {pages} {209} (\bibinfo
  {year} {2005})}\BibitemShut {NoStop}%
\bibitem [{\citenamefont {Khan}\ \emph {et~al.}(2014)\citenamefont {Khan},
  \citenamefont {Sardar}, \citenamefont {Sarkar},\ and\ \citenamefont
  {Adhikari}}]{SSJPCA118}%
  \BibitemOpen
  \bibfield  {author} {\bibinfo {author} {\bibfnamefont {B.~A.}\ \bibnamefont
  {Khan}}, \bibinfo {author} {\bibfnamefont {S.}~\bibnamefont {Sardar}},
  \bibinfo {author} {\bibfnamefont {P.}~\bibnamefont {Sarkar}}, \ and\ \bibinfo
  {author} {\bibfnamefont {S.}~\bibnamefont {Adhikari}},\ }\href@noop {}
  {\bibfield  {journal} {\bibinfo  {journal} {J. Phys. Chem. A}\ }\textbf
  {\bibinfo {volume} {118 (49)}},\ \bibinfo {pages} {11451} (\bibinfo {year}
  {2014})}\BibitemShut {NoStop}%
\bibitem [{\citenamefont {Mandal}\ \emph {et~al.}(2019)\citenamefont {Mandal},
  \citenamefont {Ghosh}, \citenamefont {Sardar},\ and\ \citenamefont
  {Adhikari}}]{souvik_irpc37}%
  \BibitemOpen
  \bibfield  {author} {\bibinfo {author} {\bibfnamefont {S.}~\bibnamefont
  {Mandal}}, \bibinfo {author} {\bibfnamefont {S.}~\bibnamefont {Ghosh}},
  \bibinfo {author} {\bibfnamefont {S.}~\bibnamefont {Sardar}}, \ and\ \bibinfo
  {author} {\bibfnamefont {S.}~\bibnamefont {Adhikari}},\ }\href@noop {}
  {\bibfield  {journal} {\bibinfo  {journal} {Int. Rev. Phys. Chem.}\ }\textbf
  {\bibinfo {volume} {37}},\ \bibinfo {pages} {607} (\bibinfo {year}
  {2019})}\BibitemShut {NoStop}%
\bibitem [{\citenamefont {Dutta}\ \emph {et~al.}(2022)\citenamefont {Dutta},
  \citenamefont {Ravi}, \citenamefont {Mukherjee}, \citenamefont {Ojha},\ and\
  \citenamefont {Adhikari}}]{joy_jpca}%
  \BibitemOpen
  \bibfield  {author} {\bibinfo {author} {\bibfnamefont {J.}~\bibnamefont
  {Dutta}}, \bibinfo {author} {\bibfnamefont {S.}~\bibnamefont {Ravi}},
  \bibinfo {author} {\bibfnamefont {S.}~\bibnamefont {Mukherjee}}, \bibinfo
  {author} {\bibfnamefont {A.~K.}\ \bibnamefont {Ojha}}, \ and\ \bibinfo
  {author} {\bibfnamefont {S.}~\bibnamefont {Adhikari}},\ }\href@noop {}
  {\bibfield  {journal} {\bibinfo  {journal} {J. Phys. Chem. A}\ }\textbf
  {\bibinfo {volume} {126}},\ \bibinfo {pages} {691} (\bibinfo {year}
  {2022})}\BibitemShut {NoStop}%
\bibitem [{\citenamefont {Mukherjee}\ \emph {et~al.}(2022)\citenamefont
  {Mukherjee}, \citenamefont {Ravi}, \citenamefont {Dutta}, \citenamefont
  {Sardar},\ and\ \citenamefont {Adhikari}}]{soumya_tfb}%
  \BibitemOpen
  \bibfield  {author} {\bibinfo {author} {\bibfnamefont {S.}~\bibnamefont
  {Mukherjee}}, \bibinfo {author} {\bibfnamefont {S.}~\bibnamefont {Ravi}},
  \bibinfo {author} {\bibfnamefont {J.}~\bibnamefont {Dutta}}, \bibinfo
  {author} {\bibfnamefont {S.}~\bibnamefont {Sardar}}, \ and\ \bibinfo {author}
  {\bibfnamefont {S.}~\bibnamefont {Adhikari}},\ }\href@noop {} {\bibfield
  {journal} {\bibinfo  {journal} {Phys. Chem. Chem. Phys.}\ }\textbf {\bibinfo
  {volume} {24}},\ \bibinfo {pages} {2185} (\bibinfo {year}
  {2022})}\BibitemShut {NoStop}%
\bibitem [{\citenamefont {Bersuker}\ and\ \citenamefont
  {Polinger}(2020)}]{bersuker_cm5}%
  \BibitemOpen
  \bibfield  {author} {\bibinfo {author} {\bibfnamefont {I.~B.}\ \bibnamefont
  {Bersuker}}\ and\ \bibinfo {author} {\bibfnamefont {V.}~\bibnamefont
  {Polinger}},\ }\href@noop {} {\bibfield  {journal} {\bibinfo  {journal}
  {Condensed Matter}\ }\textbf {\bibinfo {volume} {5}} (\bibinfo {year}
  {2020})}\BibitemShut {NoStop}%
\bibitem [{\citenamefont {Baldini}\ \emph {et~al.}(2018)\citenamefont
  {Baldini}, \citenamefont {Kubacka}, \citenamefont {Mallett}, \citenamefont
  {Ma}, \citenamefont {Koohpayeh}, \citenamefont {Zhu}, \citenamefont
  {Bernhard}, \citenamefont {Johnson},\ and\ \citenamefont
  {Carbone}}]{Berhard-TbMnO}%
  \BibitemOpen
  \bibfield  {author} {\bibinfo {author} {\bibfnamefont {E.}~\bibnamefont
  {Baldini}}, \bibinfo {author} {\bibfnamefont {T.}~\bibnamefont {Kubacka}},
  \bibinfo {author} {\bibfnamefont {B.~P.~P.}\ \bibnamefont {Mallett}},
  \bibinfo {author} {\bibfnamefont {C.}~\bibnamefont {Ma}}, \bibinfo {author}
  {\bibfnamefont {S.~M.}\ \bibnamefont {Koohpayeh}}, \bibinfo {author}
  {\bibfnamefont {Y.~M.}\ \bibnamefont {Zhu}}, \bibinfo {author} {\bibfnamefont
  {C.}~\bibnamefont {Bernhard}}, \bibinfo {author} {\bibfnamefont {S.~L.}\
  \bibnamefont {Johnson}}, \ and\ \bibinfo {author} {\bibfnamefont
  {F.}~\bibnamefont {Carbone}},\ }\href@noop {} {\bibfield  {journal} {\bibinfo
   {journal} {Phys. Rev. B}\ }\textbf {\bibinfo {volume} {97}},\ \bibinfo
  {pages} {125149} (\bibinfo {year} {2018})}\BibitemShut {NoStop}%
\bibitem [{\citenamefont {Kotomin}\ \emph {et~al.}(2005)\citenamefont
  {Kotomin}, \citenamefont {Evarestov}, \citenamefont {Mastrikov},\ and\
  \citenamefont {Maier}}]{maier_pccp7}%
  \BibitemOpen
  \bibfield  {author} {\bibinfo {author} {\bibfnamefont {E.~A.}\ \bibnamefont
  {Kotomin}}, \bibinfo {author} {\bibfnamefont {R.~A.}\ \bibnamefont
  {Evarestov}}, \bibinfo {author} {\bibfnamefont {Y.~A.}\ \bibnamefont
  {Mastrikov}}, \ and\ \bibinfo {author} {\bibfnamefont {J.}~\bibnamefont
  {Maier}},\ }\href@noop {} {\bibfield  {journal} {\bibinfo  {journal} {Phys.
  Chem. Chem. Phys.}\ }\textbf {\bibinfo {volume} {7}},\ \bibinfo {pages}
  {2346} (\bibinfo {year} {2005})}\BibitemShut {NoStop}%
\bibitem [{\citenamefont {Evarestov}\ \emph {et~al.}(2005)\citenamefont
  {Evarestov}, \citenamefont {Kotomin}, \citenamefont {Mastrikov},
  \citenamefont {Gryaznov}, \citenamefont {Heifets},\ and\ \citenamefont
  {Maier}}]{maier_prb72}%
  \BibitemOpen
  \bibfield  {author} {\bibinfo {author} {\bibfnamefont {R.~A.}\ \bibnamefont
  {Evarestov}}, \bibinfo {author} {\bibfnamefont {E.~A.}\ \bibnamefont
  {Kotomin}}, \bibinfo {author} {\bibfnamefont {Y.~A.}\ \bibnamefont
  {Mastrikov}}, \bibinfo {author} {\bibfnamefont {D.}~\bibnamefont {Gryaznov}},
  \bibinfo {author} {\bibfnamefont {E.}~\bibnamefont {Heifets}}, \ and\
  \bibinfo {author} {\bibfnamefont {J.}~\bibnamefont {Maier}},\ }\href@noop {}
  {\bibfield  {journal} {\bibinfo  {journal} {Phys. Rev. B}\ }\textbf {\bibinfo
  {volume} {72}},\ \bibinfo {pages} {214411} (\bibinfo {year}
  {2005})}\BibitemShut {NoStop}%
\bibitem [{\citenamefont {Werner}\ \emph {et~al.}(2012)\citenamefont {Werner},
  \citenamefont {Knowles}, \citenamefont {Knizia}, \citenamefont {Manby},\ and\
  \citenamefont {Sch{\ifmmode\ddot{u}\else\"{u}\fi}tz}}]{MOLPRO_brief}%
  \BibitemOpen
  \bibfield  {author} {\bibinfo {author} {\bibfnamefont {H.-J.}\ \bibnamefont
  {Werner}}, \bibinfo {author} {\bibfnamefont {P.~J.}\ \bibnamefont {Knowles}},
  \bibinfo {author} {\bibfnamefont {G.}~\bibnamefont {Knizia}}, \bibinfo
  {author} {\bibfnamefont {F.~R.}\ \bibnamefont {Manby}}, \ and\ \bibinfo
  {author} {\bibfnamefont {M.}~\bibnamefont
  {Sch{\ifmmode\ddot{u}\else\"{u}\fi}tz}},\ }\href@noop {} {\bibfield
  {journal} {\bibinfo  {journal} {WIREs Comput. Mol. Sci.}\ }\textbf {\bibinfo
  {volume} {2}},\ \bibinfo {pages} {242} (\bibinfo {year} {2012})}\BibitemShut
  {NoStop}%
\bibitem [{\citenamefont {Kwei}\ \emph {et~al.}(1993)\citenamefont {Kwei},
  \citenamefont {Lawson}, \citenamefont {Billinge},\ and\ \citenamefont
  {Cheong}}]{Kwei_Bond_order_exp}%
  \BibitemOpen
  \bibfield  {author} {\bibinfo {author} {\bibfnamefont {G.~H.}\ \bibnamefont
  {Kwei}}, \bibinfo {author} {\bibfnamefont {A.~C.}\ \bibnamefont {Lawson}},
  \bibinfo {author} {\bibfnamefont {S.~J.~L.}\ \bibnamefont {Billinge}}, \ and\
  \bibinfo {author} {\bibfnamefont {S.~W.}\ \bibnamefont {Cheong}},\
  }\href@noop {} {\bibfield  {journal} {\bibinfo  {journal} {J. Phys. Chem.}\
  }\textbf {\bibinfo {volume} {97}},\ \bibinfo {pages} {2368} (\bibinfo {year}
  {1993})}\BibitemShut {NoStop}%
\bibitem [{\citenamefont {Moin}\ \emph {et~al.}(2024)\citenamefont {Moin},
  \citenamefont {Anwar}, \citenamefont {Babar}, \citenamefont {Thumu},\ and\
  \citenamefont {Ali}}]{Heliyon_bond_order_theor}%
  \BibitemOpen
  \bibfield  {author} {\bibinfo {author} {\bibfnamefont {M.}~\bibnamefont
  {Moin}}, \bibinfo {author} {\bibfnamefont {A.~W.}\ \bibnamefont {Anwar}},
  \bibinfo {author} {\bibfnamefont {M.}~\bibnamefont {Babar}}, \bibinfo
  {author} {\bibfnamefont {U.}~\bibnamefont {Thumu}}, \ and\ \bibinfo {author}
  {\bibfnamefont {A.}~\bibnamefont {Ali}},\ }\href@noop {} {\bibfield
  {journal} {\bibinfo  {journal} {Heliyon}\ }\textbf {\bibinfo {volume} {10}},\
  \bibinfo {pages} {e24607} (\bibinfo {year} {2024})}\BibitemShut {NoStop}%
\bibitem [{\citenamefont {Mahmoud}\ \emph {et~al.}(2014)\citenamefont
  {Mahmoud}, \citenamefont {Erba}, \citenamefont {El-Kelany}, \citenamefont
  {R\'erat},\ and\ \citenamefont {Orlando}}]{A_Mahmoud}%
  \BibitemOpen
  \bibfield  {author} {\bibinfo {author} {\bibfnamefont {A.}~\bibnamefont
  {Mahmoud}}, \bibinfo {author} {\bibfnamefont {A.}~\bibnamefont {Erba}},
  \bibinfo {author} {\bibfnamefont {K.~E.}\ \bibnamefont {El-Kelany}}, \bibinfo
  {author} {\bibfnamefont {M.}~\bibnamefont {R\'erat}}, \ and\ \bibinfo
  {author} {\bibfnamefont {R.}~\bibnamefont {Orlando}},\ }\href@noop {}
  {\bibfield  {journal} {\bibinfo  {journal} {Phys. Rev. B}\ }\textbf {\bibinfo
  {volume} {89}},\ \bibinfo {pages} {045103} (\bibinfo {year}
  {2014})}\BibitemShut {NoStop}%
\bibitem [{\citenamefont {Pertosa}, \citenamefont {Hollinger},\ and\
  \citenamefont {Michel-Calendini}(1978)}]{Pertosa_Theory_Spectra}%
  \BibitemOpen
  \bibfield  {author} {\bibinfo {author} {\bibfnamefont {P.}~\bibnamefont
  {Pertosa}}, \bibinfo {author} {\bibfnamefont {G.}~\bibnamefont {Hollinger}},
  \ and\ \bibinfo {author} {\bibfnamefont {F.~M.}\ \bibnamefont
  {Michel-Calendini}},\ }\href {\doibase 10.1103/PhysRevB.18.5177} {\bibfield
  {journal} {\bibinfo  {journal} {Phys. Rev. B}\ }\textbf {\bibinfo {volume}
  {18}},\ \bibinfo {pages} {5177} (\bibinfo {year} {1978})}\BibitemShut
  {NoStop}%
\bibitem [{\citenamefont {Merz}(1949)}]{Walter_spont_1949}%
  \BibitemOpen
  \bibfield  {author} {\bibinfo {author} {\bibfnamefont {W.~J.}\ \bibnamefont
  {Merz}},\ }\href {\doibase 10.1103/PhysRev.76.1221} {\bibfield  {journal}
  {\bibinfo  {journal} {Phys. Rev.}\ }\textbf {\bibinfo {volume} {76}},\
  \bibinfo {pages} {1221} (\bibinfo {year} {1949})}\BibitemShut {NoStop}%
\bibitem [{\citenamefont {Polinger}(2013)}]{Polinger_2013}%
  \BibitemOpen
  \bibfield  {author} {\bibinfo {author} {\bibfnamefont {V.}~\bibnamefont
  {Polinger}},\ }\href@noop {} {\bibfield  {journal} {\bibinfo  {journal} {J.
  Phys. Conf. Ser.}\ }\textbf {\bibinfo {volume} {428}},\ \bibinfo {pages}
  {012026} (\bibinfo {year} {2013})}\BibitemShut {NoStop}%
\end{thebibliography}%


\begin{thebibliography}{664}%
\makeatletter
\providecommand \@ifxundefined [1]{%
 \@ifx{#1\undefined}
}%
\providecommand \@ifnum [1]{%
 \ifnum #1\expandafter \@firstoftwo
 \else \expandafter \@secondoftwo
 \fi
}%
\providecommand \@ifx [1]{%
 \ifx #1\expandafter \@firstoftwo
 \else \expandafter \@secondoftwo
 \fi
}%
\providecommand \natexlab [1]{#1}%
\providecommand \enquote  [1]{``#1''}%
\providecommand \bibnamefont  [1]{#1}%
\providecommand \bibfnamefont [1]{#1}%
\providecommand \citenamefont [1]{#1}%
\providecommand \href@noop [0]{\@secondoftwo}%
\providecommand \href [0]{\begingroup \@sanitize@url \@href}%
\providecommand \@href[1]{\@@startlink{#1}\@@href}%
\providecommand \@@href[1]{\endgroup#1\@@endlink}%
\providecommand \@sanitize@url [0]{\catcode `\\12\catcode `\$12\catcode
  `\&12\catcode `\#12\catcode `\^12\catcode `\_12\catcode `\%12\relax}%
\providecommand \@@startlink[1]{}%
\providecommand \@@endlink[0]{}%
\providecommand \url  [0]{\begingroup\@sanitize@url \@url }%
\providecommand \@url [1]{\endgroup\@href {#1}{\urlprefix }}%
\providecommand \urlprefix  [0]{URL }%
\providecommand \Eprint [0]{\href }%
\providecommand \doibase [0]{http://dx.doi.org/}%
\providecommand \selectlanguage [0]{\@gobble}%
\providecommand \bibinfo  [0]{\@secondoftwo}%
\providecommand \bibfield  [0]{\@secondoftwo}%
\providecommand \translation [1]{[#1]}%
\providecommand \BibitemOpen [0]{}%
\providecommand \bibitemStop [0]{}%
\providecommand \bibitemNoStop [0]{.\EOS\space}%
\providecommand \EOS [0]{\spacefactor3000\relax}%
\providecommand \BibitemShut  [1]{\csname bibitem#1\endcsname}%
\let\auto@bib@innerbib\@empty
\bibitem [{\citenamefont {Longuet-Higgins}\ \emph {et~al.}(1958)\citenamefont
  {Longuet-Higgins}, \citenamefont {\"Opik}, \citenamefont {Pryce},\ and\
  \citenamefont {Sack}}]{lh-opik}%
  \BibitemOpen
  \bibfield  {author} {\bibinfo {author} {\bibfnamefont {H.~C.}\ \bibnamefont
  {Longuet-Higgins}}, \bibinfo {author} {\bibfnamefont {U.}~\bibnamefont
  {\"Opik}}, \bibinfo {author} {\bibfnamefont {M.~H.~L.}\ \bibnamefont
  {Pryce}}, \ and\ \bibinfo {author} {\bibfnamefont {R.~A.}\ \bibnamefont
  {Sack}},\ }\href@noop {} {\bibfield  {journal} {\bibinfo  {journal} {Proc. R.
  Soc. London, Ser. A}\ }\textbf {\bibinfo {volume} {244}},\ \bibinfo {pages}
  {1} (\bibinfo {year} {1958})}\BibitemShut {NoStop}%
\bibitem [{\citenamefont {Sarkar}\ and\ \citenamefont
  {Adhikari}(2006{\natexlab{a}})}]{sarkar_jcp124}%
  \BibitemOpen
  \bibfield  {author} {\bibinfo {author} {\bibfnamefont {B.}~\bibnamefont
  {Sarkar}}\ and\ \bibinfo {author} {\bibfnamefont {S.}~\bibnamefont
  {Adhikari}},\ }\href@noop {} {\bibfield  {journal} {\bibinfo  {journal} {J.
  Chem. Phys.}\ }\textbf {\bibinfo {volume} {124}},\ \bibinfo {pages} {074101}
  (\bibinfo {year} {2006}{\natexlab{a}})}\BibitemShut {NoStop}%
\bibitem [{\citenamefont {Sarkar}\ and\ \citenamefont
  {Adhikari}(2008)}]{sarkar_jpca112}%
  \BibitemOpen
  \bibfield  {author} {\bibinfo {author} {\bibfnamefont {B.}~\bibnamefont
  {Sarkar}}\ and\ \bibinfo {author} {\bibfnamefont {S.}~\bibnamefont
  {Adhikari}},\ }\href@noop {} {\bibfield  {journal} {\bibinfo  {journal} {J.
  Phys. Chem. A}\ }\textbf {\bibinfo {volume} {112}},\ \bibinfo {pages} {9868}
  (\bibinfo {year} {2008})}\BibitemShut {NoStop}%
\bibitem [{\citenamefont {Paul}, \citenamefont {Adhikari},\ and\ \citenamefont
  {Baer}(2010{\natexlab{a}})}]{PAUL_PR496}%
  \BibitemOpen
  \bibfield  {author} {\bibinfo {author} {\bibfnamefont {A.~K.}\ \bibnamefont
  {Paul}}, \bibinfo {author} {\bibfnamefont {S.}~\bibnamefont {Adhikari}}, \
  and\ \bibinfo {author} {\bibfnamefont {M.}~\bibnamefont {Baer}},\ }\href@noop
  {} {\bibfield  {journal} {\bibinfo  {journal} {Physics reports}\ }\textbf
  {\bibinfo {volume} {496}},\ \bibinfo {pages} {79} (\bibinfo {year}
  {2010}{\natexlab{a}})}\BibitemShut {NoStop}%
\bibitem [{\citenamefont {Mukherjee}\ \emph
  {et~al.}(2022{\natexlab{a}})\citenamefont {Mukherjee}, \citenamefont {Hazra},
  \citenamefont {Ghosh}, \citenamefont {Mukherjee},\ and\ \citenamefont
  {Adhikari}}]{soumya_hopp}%
  \BibitemOpen
  \bibfield  {author} {\bibinfo {author} {\bibfnamefont {S.}~\bibnamefont
  {Mukherjee}}, \bibinfo {author} {\bibfnamefont {S.}~\bibnamefont {Hazra}},
  \bibinfo {author} {\bibfnamefont {S.}~\bibnamefont {Ghosh}}, \bibinfo
  {author} {\bibfnamefont {S.}~\bibnamefont {Mukherjee}}, \ and\ \bibinfo
  {author} {\bibfnamefont {S.}~\bibnamefont {Adhikari}},\ }\href@noop {}
  {\bibfield  {journal} {\bibinfo  {journal} {Chem. Phys.}\ }\textbf {\bibinfo
  {volume} {560}},\ \bibinfo {pages} {111588} (\bibinfo {year}
  {2022}{\natexlab{a}})}\BibitemShut {NoStop}%
\bibitem [{\citenamefont {Born}\ and\ \citenamefont {Oppenheimer}(1927)}]{BO}%
  \BibitemOpen
  \bibfield  {author} {\bibinfo {author} {\bibfnamefont {M.}~\bibnamefont
  {Born}}\ and\ \bibinfo {author} {\bibfnamefont {R.}~\bibnamefont
  {Oppenheimer}},\ }\href@noop {} {\bibfield  {journal} {\bibinfo  {journal}
  {Ann. Phys.}\ }\textbf {\bibinfo {volume} {389}},\ \bibinfo {pages} {457}
  (\bibinfo {year} {1927})}\BibitemShut {NoStop}%
\bibitem [{\citenamefont {Born}\ and\ \citenamefont
  {Huang}(1954{\natexlab{a}})}]{BH}%
  \BibitemOpen
  \bibfield  {author} {\bibinfo {author} {\bibfnamefont {M.}~\bibnamefont
  {Born}}\ and\ \bibinfo {author} {\bibfnamefont {K.}~\bibnamefont {Huang}},\
  }\href@noop {} {\emph {\bibinfo {title} {Dynamical Theory of Crystal
  Lattices}}}\ (\bibinfo  {publisher} {Oxford University Press, Oxford},\
  \bibinfo {year} {1954})\BibitemShut {NoStop}%
\bibitem [{\citenamefont {Top}\ and\ \citenamefont
  {Baer}(1977{\natexlab{a}})}]{baer_jcp66}%
  \BibitemOpen
  \bibfield  {author} {\bibinfo {author} {\bibfnamefont {Z.~H.}\ \bibnamefont
  {Top}}\ and\ \bibinfo {author} {\bibfnamefont {M.}~\bibnamefont {Baer}},\
  }\href@noop {} {\bibfield  {journal} {\bibinfo  {journal} {J. Chem. Phys.}\
  }\textbf {\bibinfo {volume} {66}},\ \bibinfo {pages} {1363} (\bibinfo {year}
  {1977}{\natexlab{a}})}\BibitemShut {NoStop}%
\bibitem [{\citenamefont {Adhikari}\ and\ \citenamefont
  {Billing}(1999{\natexlab{a}})}]{adhikari_jcp111}%
  \BibitemOpen
  \bibfield  {author} {\bibinfo {author} {\bibfnamefont {S.}~\bibnamefont
  {Adhikari}}\ and\ \bibinfo {author} {\bibfnamefont {G.~D.}\ \bibnamefont
  {Billing}},\ }\href@noop {} {\bibfield  {journal} {\bibinfo  {journal} {J.
  Chem. Phys.}\ }\textbf {\bibinfo {volume} {111}},\ \bibinfo {pages} {40}
  (\bibinfo {year} {1999}{\natexlab{a}})}\BibitemShut {NoStop}%
\bibitem [{\citenamefont {Varandas}\ and\ \citenamefont
  {Xu}(2000)}]{varandas_jcp112}%
  \BibitemOpen
  \bibfield  {author} {\bibinfo {author} {\bibfnamefont {A.~J.~C.}\
  \bibnamefont {Varandas}}\ and\ \bibinfo {author} {\bibfnamefont {Z.~R.}\
  \bibnamefont {Xu}},\ }\href@noop {} {\bibfield  {journal} {\bibinfo
  {journal} {J. Chem. Phys.}\ }\textbf {\bibinfo {volume} {112}},\ \bibinfo
  {pages} {2121} (\bibinfo {year} {2000})}\BibitemShut {NoStop}%
\bibitem [{\citenamefont {Curchod}\ and\ \citenamefont
  {Mart\'inez}(2018)}]{martinez_chemrev}%
  \BibitemOpen
  \bibfield  {author} {\bibinfo {author} {\bibfnamefont {B.~F.~E.}\
  \bibnamefont {Curchod}}\ and\ \bibinfo {author} {\bibfnamefont {T.~J.}\
  \bibnamefont {Mart\'inez}},\ }\href@noop {} {\bibfield  {journal} {\bibinfo
  {journal} {Chem. Rev.}\ }\textbf {\bibinfo {volume} {118}},\ \bibinfo {pages}
  {3305} (\bibinfo {year} {2018})}\BibitemShut {NoStop}%
\bibitem [{\citenamefont {Mukherjee}\ \emph
  {et~al.}(2018{\natexlab{a}})\citenamefont {Mukherjee}, \citenamefont
  {Mukherjee}, \citenamefont {Sardar}, \citenamefont {Shamasundar},\ and\
  \citenamefont {Adhikari}}]{bijit_cp515}%
  \BibitemOpen
  \bibfield  {author} {\bibinfo {author} {\bibfnamefont {B.}~\bibnamefont
  {Mukherjee}}, \bibinfo {author} {\bibfnamefont {S.}~\bibnamefont
  {Mukherjee}}, \bibinfo {author} {\bibfnamefont {S.}~\bibnamefont {Sardar}},
  \bibinfo {author} {\bibfnamefont {K.~R.}\ \bibnamefont {Shamasundar}}, \ and\
  \bibinfo {author} {\bibfnamefont {S.}~\bibnamefont {Adhikari}},\ }\href@noop
  {} {\bibfield  {journal} {\bibinfo  {journal} {Chem. Phys.}\ }\textbf
  {\bibinfo {volume} {515}},\ \bibinfo {pages} {350} (\bibinfo {year}
  {2018}{\natexlab{a}})}\BibitemShut {NoStop}%
\bibitem [{\citenamefont {Mukherjee}\ \emph {et~al.}(2021)\citenamefont
  {Mukherjee}, \citenamefont {Ravi}, \citenamefont {Naskar}, \citenamefont
  {Sardar},\ and\ \citenamefont {Adhikari}}]{soumya_bz}%
  \BibitemOpen
  \bibfield  {author} {\bibinfo {author} {\bibfnamefont {S.}~\bibnamefont
  {Mukherjee}}, \bibinfo {author} {\bibfnamefont {S.}~\bibnamefont {Ravi}},
  \bibinfo {author} {\bibfnamefont {K.}~\bibnamefont {Naskar}}, \bibinfo
  {author} {\bibfnamefont {S.}~\bibnamefont {Sardar}}, \ and\ \bibinfo {author}
  {\bibfnamefont {S.}~\bibnamefont {Adhikari}},\ }\href@noop {} {\bibfield
  {journal} {\bibinfo  {journal} {J. Chem. Phys.}\ }\textbf {\bibinfo {volume}
  {154}},\ \bibinfo {pages} {094306} (\bibinfo {year} {2021})}\BibitemShut
  {NoStop}%
\bibitem [{\citenamefont {Mukherjee}\ \emph
  {et~al.}(2022{\natexlab{b}})\citenamefont {Mukherjee}, \citenamefont {Ravi},
  \citenamefont {Dutta}, \citenamefont {Sardar},\ and\ \citenamefont
  {Adhikari}}]{soumya_tfb}%
  \BibitemOpen
  \bibfield  {author} {\bibinfo {author} {\bibfnamefont {S.}~\bibnamefont
  {Mukherjee}}, \bibinfo {author} {\bibfnamefont {S.}~\bibnamefont {Ravi}},
  \bibinfo {author} {\bibfnamefont {J.}~\bibnamefont {Dutta}}, \bibinfo
  {author} {\bibfnamefont {S.}~\bibnamefont {Sardar}}, \ and\ \bibinfo {author}
  {\bibfnamefont {S.}~\bibnamefont {Adhikari}},\ }\href@noop {} {\bibfield
  {journal} {\bibinfo  {journal} {Phys. Chem. Chem. Phys.}\ }\textbf {\bibinfo
  {volume} {24}},\ \bibinfo {pages} {2185} (\bibinfo {year}
  {2022}{\natexlab{b}})}\BibitemShut {NoStop}%
\bibitem [{\citenamefont {Baer}, \citenamefont {Niedner-Schatterburg},\ and\
  \citenamefont {Toennies}(1989)}]{baer_jcp91}%
  \BibitemOpen
  \bibfield  {author} {\bibinfo {author} {\bibfnamefont {M.}~\bibnamefont
  {Baer}}, \bibinfo {author} {\bibfnamefont {G.}~\bibnamefont
  {Niedner-Schatterburg}}, \ and\ \bibinfo {author} {\bibfnamefont {J.~P.}\
  \bibnamefont {Toennies}},\ }\href@noop {} {\bibfield  {journal} {\bibinfo
  {journal} {J. Chem. Phys.}\ }\textbf {\bibinfo {volume} {91}},\ \bibinfo
  {pages} {4169} (\bibinfo {year} {1989})}\BibitemShut {NoStop}%
\bibitem [{\citenamefont {Grimbert}, \citenamefont {Sidis},\ and\ \citenamefont
  {Cobut}(1998)}]{grimbert_jcp108}%
  \BibitemOpen
  \bibfield  {author} {\bibinfo {author} {\bibfnamefont {D.}~\bibnamefont
  {Grimbert}}, \bibinfo {author} {\bibfnamefont {V.}~\bibnamefont {Sidis}}, \
  and\ \bibinfo {author} {\bibfnamefont {V.}~\bibnamefont {Cobut}},\
  }\href@noop {} {\bibfield  {journal} {\bibinfo  {journal} {J. Chem. Phys.}\
  }\textbf {\bibinfo {volume} {108}},\ \bibinfo {pages} {6331} (\bibinfo {year}
  {1998})}\BibitemShut {NoStop}%
\bibitem [{\citenamefont {Baer}\ \emph
  {et~al.}(1996{\natexlab{a}})\citenamefont {Baer}, \citenamefont {Charutz},
  \citenamefont {Kosloff},\ and\ \citenamefont {Baer}}]{baer_jcp105}%
  \BibitemOpen
  \bibfield  {author} {\bibinfo {author} {\bibfnamefont {R.}~\bibnamefont
  {Baer}}, \bibinfo {author} {\bibfnamefont {D.~M.}\ \bibnamefont {Charutz}},
  \bibinfo {author} {\bibfnamefont {R.}~\bibnamefont {Kosloff}}, \ and\
  \bibinfo {author} {\bibfnamefont {M.}~\bibnamefont {Baer}},\ }\href@noop {}
  {\bibfield  {journal} {\bibinfo  {journal} {J. Chem. Phys.}\ }\textbf
  {\bibinfo {volume} {105}},\ \bibinfo {pages} {9141} (\bibinfo {year}
  {1996}{\natexlab{a}})}\BibitemShut {NoStop}%
\bibitem [{\citenamefont {Baer}\ \emph {et~al.}(2000)\citenamefont {Baer},
  \citenamefont {Lin}, \citenamefont {Alijah}, \citenamefont {Adhikari},\ and\
  \citenamefont {Billing}}]{baer_pra62}%
  \BibitemOpen
  \bibfield  {author} {\bibinfo {author} {\bibfnamefont {M.}~\bibnamefont
  {Baer}}, \bibinfo {author} {\bibfnamefont {S.~H.}\ \bibnamefont {Lin}},
  \bibinfo {author} {\bibfnamefont {A.}~\bibnamefont {Alijah}}, \bibinfo
  {author} {\bibfnamefont {S.}~\bibnamefont {Adhikari}}, \ and\ \bibinfo
  {author} {\bibfnamefont {G.~D.}\ \bibnamefont {Billing}},\ }\href@noop {}
  {\bibfield  {journal} {\bibinfo  {journal} {Phys. Rev. A}\ }\textbf {\bibinfo
  {volume} {62}},\ \bibinfo {pages} {32506} (\bibinfo {year}
  {2000})}\BibitemShut {NoStop}%
\bibitem [{\citenamefont {Adhikari}\ \emph {et~al.}(2000)\citenamefont
  {Adhikari}, \citenamefont {Billing}, \citenamefont {Alijah}, \citenamefont
  {Lin},\ and\ \citenamefont {Baer}}]{adhikari_pra62}%
  \BibitemOpen
  \bibfield  {author} {\bibinfo {author} {\bibfnamefont {S.}~\bibnamefont
  {Adhikari}}, \bibinfo {author} {\bibfnamefont {G.~D.}\ \bibnamefont
  {Billing}}, \bibinfo {author} {\bibfnamefont {A.}~\bibnamefont {Alijah}},
  \bibinfo {author} {\bibfnamefont {S.~H.}\ \bibnamefont {Lin}}, \ and\
  \bibinfo {author} {\bibfnamefont {M.}~\bibnamefont {Baer}},\ }\href@noop {}
  {\bibfield  {journal} {\bibinfo  {journal} {Phys. Rev. A}\ }\textbf {\bibinfo
  {volume} {62}},\ \bibinfo {pages} {32507} (\bibinfo {year}
  {2000})}\BibitemShut {NoStop}%
\bibitem [{\citenamefont {Sarkar}\ and\ \citenamefont
  {Adhikari}(2009)}]{sarkar_ijqc109}%
  \BibitemOpen
  \bibfield  {author} {\bibinfo {author} {\bibfnamefont {B.}~\bibnamefont
  {Sarkar}}\ and\ \bibinfo {author} {\bibfnamefont {S.}~\bibnamefont
  {Adhikari}},\ }\href@noop {} {\bibfield  {journal} {\bibinfo  {journal} {Int.
  J. Quantum Chem.}\ }\textbf {\bibinfo {volume} {109}},\ \bibinfo {pages}
  {650} (\bibinfo {year} {2009})}\BibitemShut {NoStop}%
\bibitem [{\citenamefont {Paul}\ \emph
  {et~al.}(2009{\natexlab{a}})\citenamefont {Paul}, \citenamefont {Sardar},
  \citenamefont {Sarkar},\ and\ \citenamefont {Adhikari}}]{akpaul_jcp131}%
  \BibitemOpen
  \bibfield  {author} {\bibinfo {author} {\bibfnamefont {A.~K.}\ \bibnamefont
  {Paul}}, \bibinfo {author} {\bibfnamefont {S.}~\bibnamefont {Sardar}},
  \bibinfo {author} {\bibfnamefont {B.}~\bibnamefont {Sarkar}}, \ and\ \bibinfo
  {author} {\bibfnamefont {S.}~\bibnamefont {Adhikari}},\ }\href@noop {}
  {\bibfield  {journal} {\bibinfo  {journal} {J. Chem. Phys.}\ }\textbf
  {\bibinfo {volume} {131}},\ \bibinfo {pages} {124312} (\bibinfo {year}
  {2009}{\natexlab{a}})}\BibitemShut {NoStop}%
\bibitem [{\citenamefont {Dutta}\ \emph
  {et~al.}(2022{\natexlab{a}})\citenamefont {Dutta}, \citenamefont {Ravi},
  \citenamefont {Mukherjee}, \citenamefont {Ojha},\ and\ \citenamefont
  {Adhikari}}]{joy_jpca}%
  \BibitemOpen
  \bibfield  {author} {\bibinfo {author} {\bibfnamefont {J.}~\bibnamefont
  {Dutta}}, \bibinfo {author} {\bibfnamefont {S.}~\bibnamefont {Ravi}},
  \bibinfo {author} {\bibfnamefont {S.}~\bibnamefont {Mukherjee}}, \bibinfo
  {author} {\bibfnamefont {A.~K.}\ \bibnamefont {Ojha}}, \ and\ \bibinfo
  {author} {\bibfnamefont {S.}~\bibnamefont {Adhikari}},\ }\href@noop {}
  {\bibfield  {journal} {\bibinfo  {journal} {J. Phys. Chem. A}\ }\textbf
  {\bibinfo {volume} {126}},\ \bibinfo {pages} {691} (\bibinfo {year}
  {2022}{\natexlab{a}})}\BibitemShut {NoStop}%
\bibitem [{\citenamefont {Hellmann}(1937)}]{hellmann}%
  \BibitemOpen
  \bibfield  {author} {\bibinfo {author} {\bibfnamefont {H.}~\bibnamefont
  {Hellmann}},\ }\href@noop {} {\emph {\bibinfo {title} {{\it Einfuring in die
  Quantenchemie}}}}\ (\bibinfo  {publisher} {Franz Duetiche:Leipzig},\ \bibinfo
  {address} {Germany},\ \bibinfo {year} {1937})\BibitemShut {NoStop}%
\bibitem [{\citenamefont {Feynman}(1939{\natexlab{a}})}]{feynman_pr56}%
  \BibitemOpen
  \bibfield  {author} {\bibinfo {author} {\bibfnamefont {R.}~\bibnamefont
  {Feynman}},\ }\href@noop {} {\bibfield  {journal} {\bibinfo  {journal} {Phys.
  Rev.}\ }\textbf {\bibinfo {volume} {56}},\ \bibinfo {pages} {340} (\bibinfo
  {year} {1939}{\natexlab{a}})}\BibitemShut {NoStop}%
\bibitem [{\citenamefont {Mukherjee}\ \emph
  {et~al.}(2024{\natexlab{a}})\citenamefont {Mukherjee}, \citenamefont {Saha},
  \citenamefont {Ghosh}, \citenamefont {Adhikari}, \citenamefont
  {Sathyamurthy},\ and\ \citenamefont {Baer}}]{friction_jpca128}%
  \BibitemOpen
  \bibfield  {author} {\bibinfo {author} {\bibfnamefont {S.}~\bibnamefont
  {Mukherjee}}, \bibinfo {author} {\bibfnamefont {S.}~\bibnamefont {Saha}},
  \bibinfo {author} {\bibfnamefont {S.}~\bibnamefont {Ghosh}}, \bibinfo
  {author} {\bibfnamefont {S.}~\bibnamefont {Adhikari}}, \bibinfo {author}
  {\bibfnamefont {N.}~\bibnamefont {Sathyamurthy}}, \ and\ \bibinfo {author}
  {\bibfnamefont {M.}~\bibnamefont {Baer}},\ }\href@noop {} {\bibfield
  {journal} {\bibinfo  {journal} {J. Phys. Chem. A}\ }\textbf {\bibinfo
  {volume} {128}},\ \bibinfo {pages} {7691} (\bibinfo {year}
  {2024}{\natexlab{a}})}\BibitemShut {NoStop}%
\bibitem [{\citenamefont {Longuet-Higgins}(1961)}]{longuet-higgins}%
  \BibitemOpen
  \bibfield  {author} {\bibinfo {author} {\bibfnamefont {H.~C.}\ \bibnamefont
  {Longuet-Higgins}},\ }\href@noop {} {\emph {\bibinfo {title} {{\it Advances
  in Spectroscopy}}}},\ Vol.~\bibinfo {volume} {2}\ (\bibinfo  {publisher}
  {Interscience Publishers, Ltd.},\ \bibinfo {address} {New York and London},\
  \bibinfo {year} {1961})\ Chap.\ \bibinfo {chapter} {Some Recent Developments
  in the Theory of Molecular Energy Levels}\BibitemShut {NoStop}%
\bibitem [{\citenamefont {Longuet-Higgins}(1975)}]{longuet-higgins_prsl344}%
  \BibitemOpen
  \bibfield  {author} {\bibinfo {author} {\bibfnamefont {H.~C.}\ \bibnamefont
  {Longuet-Higgins}},\ }\href@noop {} {\bibfield  {journal} {\bibinfo
  {journal} {Proc. R. Soc. London, Ser. A}\ }\textbf {\bibinfo {volume}
  {344}},\ \bibinfo {pages} {147} (\bibinfo {year} {1975})}\BibitemShut
  {NoStop}%
\bibitem [{\citenamefont {Herzberg}\ and\ \citenamefont
  {Longuet-Higgins}(1963)}]{herzberg_dfs35}%
  \BibitemOpen
  \bibfield  {author} {\bibinfo {author} {\bibfnamefont {G.}~\bibnamefont
  {Herzberg}}\ and\ \bibinfo {author} {\bibfnamefont {H.~C.}\ \bibnamefont
  {Longuet-Higgins}},\ }\href@noop {} {\bibfield  {journal} {\bibinfo
  {journal} {Discuss Faraday Soc.}\ }\textbf {\bibinfo {volume} {35}},\
  \bibinfo {pages} {77} (\bibinfo {year} {1963})}\BibitemShut {NoStop}%
\bibitem [{\citenamefont {Baer}(2006{\natexlab{a}})}]{baer_book}%
  \BibitemOpen
  \bibfield  {author} {\bibinfo {author} {\bibfnamefont {M.}~\bibnamefont
  {Baer}},\ }\href@noop {} {\emph {\bibinfo {title} {{\it Beyond Born -
  Oppenheimer: Conical intersections and Electronic nonadiabatic coupling
  terms}}}}\ (\bibinfo  {publisher} {Wiley Interscience},\ \bibinfo {address}
  {New Jersey},\ \bibinfo {year} {2006})\BibitemShut {NoStop}%
\bibitem [{\citenamefont {Mead}\ and\ \citenamefont
  {Truhlar}(1982)}]{mead_jcp77}%
  \BibitemOpen
  \bibfield  {author} {\bibinfo {author} {\bibfnamefont {C.~A.}\ \bibnamefont
  {Mead}}\ and\ \bibinfo {author} {\bibfnamefont {D.~G.}\ \bibnamefont
  {Truhlar}},\ }\href@noop {} {\bibfield  {journal} {\bibinfo  {journal} {J.
  Chem. Phys.}\ }\textbf {\bibinfo {volume} {77}},\ \bibinfo {pages} {6090}
  (\bibinfo {year} {1982})}\BibitemShut {NoStop}%
\bibitem [{\citenamefont {Varandas}, \citenamefont {Tennyson},\ and\
  \citenamefont {Murrell}(1979)}]{varandas_cpl61}%
  \BibitemOpen
  \bibfield  {author} {\bibinfo {author} {\bibfnamefont {A.~J.~C.}\
  \bibnamefont {Varandas}}, \bibinfo {author} {\bibfnamefont {J.}~\bibnamefont
  {Tennyson}}, \ and\ \bibinfo {author} {\bibfnamefont {J.~N.}\ \bibnamefont
  {Murrell}},\ }\href@noop {} {\bibfield  {journal} {\bibinfo  {journal} {Chem.
  Phys. Lett.}\ }\textbf {\bibinfo {volume} {61}},\ \bibinfo {pages} {431}
  (\bibinfo {year} {1979})}\BibitemShut {NoStop}%
\bibitem [{\citenamefont {Mukherjee}\ \emph
  {et~al.}(2013{\natexlab{a}})\citenamefont {Mukherjee}, \citenamefont
  {Bandyopadhyay}, \citenamefont {Paul},\ and\ \citenamefont
  {Adhikari}}]{saikat_jpca117}%
  \BibitemOpen
  \bibfield  {author} {\bibinfo {author} {\bibfnamefont {S.}~\bibnamefont
  {Mukherjee}}, \bibinfo {author} {\bibfnamefont {S.}~\bibnamefont
  {Bandyopadhyay}}, \bibinfo {author} {\bibfnamefont {A.~K.}\ \bibnamefont
  {Paul}}, \ and\ \bibinfo {author} {\bibfnamefont {S.}~\bibnamefont
  {Adhikari}},\ }\href@noop {} {\bibfield  {journal} {\bibinfo  {journal} {J.
  Phys. Chem. A}\ }\textbf {\bibinfo {volume} {117}},\ \bibinfo {pages} {3475}
  (\bibinfo {year} {2013}{\natexlab{a}})}\BibitemShut {NoStop}%
\bibitem [{\citenamefont {Neumark}\ \emph
  {et~al.}(1985{\natexlab{a}})\citenamefont {Neumark}, \citenamefont {Wodtke},
  \citenamefont {Robinson}, \citenamefont {Hayden},\ and\ \citenamefont
  {Lee}}]{ytlee_jcp82_3045}%
  \BibitemOpen
  \bibfield  {author} {\bibinfo {author} {\bibfnamefont {D.~M.}\ \bibnamefont
  {Neumark}}, \bibinfo {author} {\bibfnamefont {A.~M.}\ \bibnamefont {Wodtke}},
  \bibinfo {author} {\bibfnamefont {G.~N.}\ \bibnamefont {Robinson}}, \bibinfo
  {author} {\bibfnamefont {C.~C.}\ \bibnamefont {Hayden}}, \ and\ \bibinfo
  {author} {\bibfnamefont {Y.~T.}\ \bibnamefont {Lee}},\ }\href@noop {}
  {\bibfield  {journal} {\bibinfo  {journal} {J. Chem. Phys.}\ }\textbf
  {\bibinfo {volume} {82}},\ \bibinfo {pages} {3045} (\bibinfo {year}
  {1985}{\natexlab{a}})}\BibitemShut {NoStop}%
\bibitem [{\citenamefont {Castillo}\ \emph
  {et~al.}(1998{\natexlab{a}})\citenamefont {Castillo}, \citenamefont {Hartke},
  \citenamefont {Werner}, \citenamefont {Aoiz}, \citenamefont {nares},\ and\
  \citenamefont {Mart\'inez-Haya}}]{werner_jcp109}%
  \BibitemOpen
  \bibfield  {author} {\bibinfo {author} {\bibfnamefont {J.~F.}\ \bibnamefont
  {Castillo}}, \bibinfo {author} {\bibfnamefont {B.}~\bibnamefont {Hartke}},
  \bibinfo {author} {\bibfnamefont {H.-J.}\ \bibnamefont {Werner}}, \bibinfo
  {author} {\bibfnamefont {F.~J.}\ \bibnamefont {Aoiz}}, \bibinfo {author}
  {\bibfnamefont {L.~B.}\ \bibnamefont {nares}}, \ and\ \bibinfo {author}
  {\bibfnamefont {B.}~\bibnamefont {Mart\'inez-Haya}},\ }\href@noop {}
  {\bibfield  {journal} {\bibinfo  {journal} {J. Chem. Phys.}\ }\textbf
  {\bibinfo {volume} {109}},\ \bibinfo {pages} {7224} (\bibinfo {year}
  {1998}{\natexlab{a}})}\BibitemShut {NoStop}%
\bibitem [{\citenamefont {Hobey}\ and\ \citenamefont
  {Mclachlan}(1960)}]{hobey_jcp33}%
  \BibitemOpen
  \bibfield  {author} {\bibinfo {author} {\bibfnamefont {W.~D.}\ \bibnamefont
  {Hobey}}\ and\ \bibinfo {author} {\bibfnamefont {A.~D.}\ \bibnamefont
  {Mclachlan}},\ }\href@noop {} {\bibfield  {journal} {\bibinfo  {journal} {J.
  Chem. Phys.}\ }\textbf {\bibinfo {volume} {33}},\ \bibinfo {pages} {1695}
  (\bibinfo {year} {1960})}\BibitemShut {NoStop}%
\bibitem [{\citenamefont {Lichten}(1967)}]{litchen_pr164}%
  \BibitemOpen
  \bibfield  {author} {\bibinfo {author} {\bibfnamefont {W.}~\bibnamefont
  {Lichten}},\ }\href@noop {} {\bibfield  {journal} {\bibinfo  {journal} {Phys.
  Rev.}\ }\textbf {\bibinfo {volume} {164}},\ \bibinfo {pages} {131} (\bibinfo
  {year} {1967})}\BibitemShut {NoStop}%
\bibitem [{\citenamefont {Smith}(1969)}]{smith_pr179}%
  \BibitemOpen
  \bibfield  {author} {\bibinfo {author} {\bibfnamefont {F.~T.}\ \bibnamefont
  {Smith}},\ }\href@noop {} {\bibfield  {journal} {\bibinfo  {journal} {Phys.
  Rev.}\ }\textbf {\bibinfo {volume} {179}},\ \bibinfo {pages} {111} (\bibinfo
  {year} {1969})}\BibitemShut {NoStop}%
\bibitem [{\citenamefont {Thiel}\ and\ \citenamefont
  {K\"oppel}(1999)}]{koppel_jcp110}%
  \BibitemOpen
  \bibfield  {author} {\bibinfo {author} {\bibfnamefont {A.}~\bibnamefont
  {Thiel}}\ and\ \bibinfo {author} {\bibfnamefont {H.}~\bibnamefont
  {K\"oppel}},\ }\href@noop {} {\bibfield  {journal} {\bibinfo  {journal} {J.
  Chem. Phys.}\ }\textbf {\bibinfo {volume} {110}},\ \bibinfo {pages} {9371}
  (\bibinfo {year} {1999})}\BibitemShut {NoStop}%
\bibitem [{\citenamefont {K\"oppel}, \citenamefont {Gronki},\ and\
  \citenamefont {Mahapatra}(2001)}]{koppel_jcp150}%
  \BibitemOpen
  \bibfield  {author} {\bibinfo {author} {\bibfnamefont {H.}~\bibnamefont
  {K\"oppel}}, \bibinfo {author} {\bibfnamefont {J.}~\bibnamefont {Gronki}}, \
  and\ \bibinfo {author} {\bibfnamefont {S.}~\bibnamefont {Mahapatra}},\
  }\href@noop {} {\bibfield  {journal} {\bibinfo  {journal} {J. Chem. Phys.}\
  }\textbf {\bibinfo {volume} {150}},\ \bibinfo {pages} {2377} (\bibinfo {year}
  {2001})}\BibitemShut {NoStop}%
\bibitem [{\citenamefont {Viel}\ and\ \citenamefont
  {Eisfeld}(2004)}]{viel_jcp120}%
  \BibitemOpen
  \bibfield  {author} {\bibinfo {author} {\bibfnamefont {A.}~\bibnamefont
  {Viel}}\ and\ \bibinfo {author} {\bibfnamefont {W.}~\bibnamefont {Eisfeld}},\
  }\href@noop {} {\bibfield  {journal} {\bibinfo  {journal} {J. Chem. Phys.}\
  }\textbf {\bibinfo {volume} {120}},\ \bibinfo {pages} {4603} (\bibinfo {year}
  {2004})}\BibitemShut {NoStop}%
\bibitem [{\citenamefont {Abedi}, \citenamefont {Maitra},\ and\ \citenamefont
  {Gross}(2010)}]{gross_prl105}%
  \BibitemOpen
  \bibfield  {author} {\bibinfo {author} {\bibfnamefont {A.}~\bibnamefont
  {Abedi}}, \bibinfo {author} {\bibfnamefont {N.~T.}\ \bibnamefont {Maitra}}, \
  and\ \bibinfo {author} {\bibfnamefont {E.~K.~U.}\ \bibnamefont {Gross}},\
  }\href@noop {} {\bibfield  {journal} {\bibinfo  {journal} {Phys. Rev. Lett.}\
  }\textbf {\bibinfo {volume} {105}},\ \bibinfo {pages} {123002} (\bibinfo
  {year} {2010})}\BibitemShut {NoStop}%
\bibitem [{\citenamefont {Abedi}, \citenamefont {Maitra},\ and\ \citenamefont
  {Gross}(2012)}]{gross_jcp137}%
  \BibitemOpen
  \bibfield  {author} {\bibinfo {author} {\bibfnamefont {A.}~\bibnamefont
  {Abedi}}, \bibinfo {author} {\bibfnamefont {N.~T.}\ \bibnamefont {Maitra}}, \
  and\ \bibinfo {author} {\bibfnamefont {E.~K.~U.}\ \bibnamefont {Gross}},\
  }\href@noop {} {\bibfield  {journal} {\bibinfo  {journal} {J. Chem. Phys.}\
  }\textbf {\bibinfo {volume} {137}},\ \bibinfo {pages} {22A530} (\bibinfo
  {year} {2012})}\BibitemShut {NoStop}%
\bibitem [{\citenamefont {Worth}\ \emph {et~al.}(2008)\citenamefont {Worth},
  \citenamefont {Meyer}, \citenamefont {K\"oppel}, \citenamefont {Cederbaum},\
  and\ \citenamefont {Burghardt}}]{worth_irpc27}%
  \BibitemOpen
  \bibfield  {author} {\bibinfo {author} {\bibfnamefont {G.~A.}\ \bibnamefont
  {Worth}}, \bibinfo {author} {\bibfnamefont {H.-D.}\ \bibnamefont {Meyer}},
  \bibinfo {author} {\bibfnamefont {H.}~\bibnamefont {K\"oppel}}, \bibinfo
  {author} {\bibfnamefont {L.~S.}\ \bibnamefont {Cederbaum}}, \ and\ \bibinfo
  {author} {\bibfnamefont {I.}~\bibnamefont {Burghardt}},\ }\href@noop {}
  {\bibfield  {journal} {\bibinfo  {journal} {Int. Rev. Phys. Chem.}\ }\textbf
  {\bibinfo {volume} {27}},\ \bibinfo {pages} {569} (\bibinfo {year}
  {2008})}\BibitemShut {NoStop}%
\bibitem [{\citenamefont {Richings}\ and\ \citenamefont
  {Habershon}(2018)}]{richings_jcp148}%
  \BibitemOpen
  \bibfield  {author} {\bibinfo {author} {\bibfnamefont {G.~W.}\ \bibnamefont
  {Richings}}\ and\ \bibinfo {author} {\bibfnamefont {S.}~\bibnamefont
  {Habershon}},\ }\href@noop {} {\bibfield  {journal} {\bibinfo  {journal} {J.
  Chem. Phys.}\ }\textbf {\bibinfo {volume} {148}},\ \bibinfo {pages} {134116}
  (\bibinfo {year} {2018})}\BibitemShut {NoStop}%
\bibitem [{\citenamefont {Richings}\ \emph {et~al.}(2015)\citenamefont
  {Richings}, \citenamefont {Polyak}, \citenamefont {Spinlove}, \citenamefont
  {Worth}, , \citenamefont {Burghardt},\ and\ \citenamefont
  {Lasorne}}]{worth_irpc34}%
  \BibitemOpen
  \bibfield  {author} {\bibinfo {author} {\bibfnamefont {G.~W.}\ \bibnamefont
  {Richings}}, \bibinfo {author} {\bibfnamefont {I.}~\bibnamefont {Polyak}},
  \bibinfo {author} {\bibfnamefont {K.}~\bibnamefont {Spinlove}}, \bibinfo
  {author} {\bibfnamefont {G.~A.}\ \bibnamefont {Worth}}, , \bibinfo {author}
  {\bibfnamefont {I.}~\bibnamefont {Burghardt}}, \ and\ \bibinfo {author}
  {\bibfnamefont {B.}~\bibnamefont {Lasorne}},\ }\href@noop {} {\bibfield
  {journal} {\bibinfo  {journal} {Int. Rev. Phys. Chem.}\ }\textbf {\bibinfo
  {volume} {34}},\ \bibinfo {pages} {269} (\bibinfo {year} {2015})}\BibitemShut
  {NoStop}%
\bibitem [{\citenamefont {Richings}\ and\ \citenamefont
  {Worth}(2015)}]{worth_jpca119}%
  \BibitemOpen
  \bibfield  {author} {\bibinfo {author} {\bibfnamefont {G.~W.}\ \bibnamefont
  {Richings}}\ and\ \bibinfo {author} {\bibfnamefont {G.~A.}\ \bibnamefont
  {Worth}},\ }\href@noop {} {\bibfield  {journal} {\bibinfo  {journal} {J.
  Phys. Chem. A}\ }\textbf {\bibinfo {volume} {119}},\ \bibinfo {pages} {12457}
  (\bibinfo {year} {2015})}\BibitemShut {NoStop}%
\bibitem [{\citenamefont {Mart\'{i}nez}, \citenamefont {Ben-Nun},\ and\
  \citenamefont {Levine}(1996)}]{SPAWN1}%
  \BibitemOpen
  \bibfield  {author} {\bibinfo {author} {\bibfnamefont {T.~J.}\ \bibnamefont
  {Mart\'{i}nez}}, \bibinfo {author} {\bibfnamefont {M.}~\bibnamefont
  {Ben-Nun}}, \ and\ \bibinfo {author} {\bibfnamefont {R.~D.}\ \bibnamefont
  {Levine}},\ }\href@noop {} {\bibfield  {journal} {\bibinfo  {journal} {J.
  Phys. Chem.}\ }\textbf {\bibinfo {volume} {100}},\ \bibinfo {pages} {7884}
  (\bibinfo {year} {1996})}\BibitemShut {NoStop}%
\bibitem [{\citenamefont {Ben-Nun}\ and\ \citenamefont
  {Mart\'{i}nez}(1998)}]{SPAWN2}%
  \BibitemOpen
  \bibfield  {author} {\bibinfo {author} {\bibfnamefont {M.}~\bibnamefont
  {Ben-Nun}}\ and\ \bibinfo {author} {\bibfnamefont {T.~J.}\ \bibnamefont
  {Mart\'{i}nez}},\ }\href@noop {} {\bibfield  {journal} {\bibinfo  {journal}
  {J. Chem. Phys.}\ }\textbf {\bibinfo {volume} {108}},\ \bibinfo {pages}
  {7244} (\bibinfo {year} {1998})}\BibitemShut {NoStop}%
\bibitem [{\citenamefont {Baer}(1975{\natexlab{a}})}]{baer_cpl35}%
  \BibitemOpen
  \bibfield  {author} {\bibinfo {author} {\bibfnamefont {M.}~\bibnamefont
  {Baer}},\ }\href@noop {} {\bibfield  {journal} {\bibinfo  {journal} {Chem.
  Phys. Lett.}\ }\textbf {\bibinfo {volume} {35}},\ \bibinfo {pages} {112}
  (\bibinfo {year} {1975}{\natexlab{a}})}\BibitemShut {NoStop}%
\bibitem [{\citenamefont {Baer}(2002{\natexlab{a}})}]{baer_phys_rep358}%
  \BibitemOpen
  \bibfield  {author} {\bibinfo {author} {\bibfnamefont {M.}~\bibnamefont
  {Baer}},\ }\href@noop {} {\bibfield  {journal} {\bibinfo  {journal} {Phys.
  Rep.}\ }\textbf {\bibinfo {volume} {358}},\ \bibinfo {pages} {75} (\bibinfo
  {year} {2002}{\natexlab{a}})}\BibitemShut {NoStop}%
\bibitem [{\citenamefont {Mukherjee}, \citenamefont {Mukherjee},\ and\
  \citenamefont {Adhikari}(2017)}]{soumya_jpca121}%
  \BibitemOpen
  \bibfield  {author} {\bibinfo {author} {\bibfnamefont {S.}~\bibnamefont
  {Mukherjee}}, \bibinfo {author} {\bibfnamefont {B.}~\bibnamefont
  {Mukherjee}}, \ and\ \bibinfo {author} {\bibfnamefont {S.}~\bibnamefont
  {Adhikari}},\ }\href@noop {} {\bibfield  {journal} {\bibinfo  {journal} {J.
  Phys. Chem. A}\ }\textbf {\bibinfo {volume} {121}},\ \bibinfo {pages} {6314}
  (\bibinfo {year} {2017})}\BibitemShut {NoStop}%
\bibitem [{\citenamefont {Mukherjee}\ \emph
  {et~al.}(2020{\natexlab{a}})\citenamefont {Mukherjee}, \citenamefont
  {Naskar}, \citenamefont {Mukherjee}, \citenamefont {S.Ravi}, \citenamefont
  {Shamasundar}, \citenamefont {Mukhopadhyay},\ and\ \citenamefont
  {Adhikari}}]{bijit_fh2}%
  \BibitemOpen
  \bibfield  {author} {\bibinfo {author} {\bibfnamefont {B.}~\bibnamefont
  {Mukherjee}}, \bibinfo {author} {\bibfnamefont {K.}~\bibnamefont {Naskar}},
  \bibinfo {author} {\bibfnamefont {S.}~\bibnamefont {Mukherjee}}, \bibinfo
  {author} {\bibnamefont {S.Ravi}}, \bibinfo {author} {\bibfnamefont {K.~R.}\
  \bibnamefont {Shamasundar}}, \bibinfo {author} {\bibfnamefont
  {D.}~\bibnamefont {Mukhopadhyay}}, \ and\ \bibinfo {author} {\bibfnamefont
  {S.}~\bibnamefont {Adhikari}},\ }\href@noop {} {\bibfield  {journal}
  {\bibinfo  {journal} {J. Chem. Phys.}\ }\textbf {\bibinfo {volume} {153}},\
  \bibinfo {pages} {174301} (\bibinfo {year} {2020}{\natexlab{a}})}\BibitemShut
  {NoStop}%
\bibitem [{\citenamefont {Mukherjee}\ \emph
  {et~al.}(2019{\natexlab{a}})\citenamefont {Mukherjee}, \citenamefont
  {Naskar}, \citenamefont {Mukherjee}, \citenamefont {Ghosh}, \citenamefont
  {Sahoo},\ and\ \citenamefont {Adhikari}}]{bijit_irpc38}%
  \BibitemOpen
  \bibfield  {author} {\bibinfo {author} {\bibfnamefont {B.}~\bibnamefont
  {Mukherjee}}, \bibinfo {author} {\bibfnamefont {K.}~\bibnamefont {Naskar}},
  \bibinfo {author} {\bibfnamefont {S.}~\bibnamefont {Mukherjee}}, \bibinfo
  {author} {\bibfnamefont {S.}~\bibnamefont {Ghosh}}, \bibinfo {author}
  {\bibfnamefont {T.}~\bibnamefont {Sahoo}}, \ and\ \bibinfo {author}
  {\bibfnamefont {S.}~\bibnamefont {Adhikari}},\ }\href@noop {} {\bibfield
  {journal} {\bibinfo  {journal} {Int. Rev. Phys. Chem.}\ }\textbf {\bibinfo
  {volume} {38}},\ \bibinfo {pages} {287} (\bibinfo {year}
  {2019}{\natexlab{a}})}\BibitemShut {NoStop}%
\bibitem [{\citenamefont {Naskar}\ \emph
  {et~al.}(2020{\natexlab{a}})\citenamefont {Naskar}, \citenamefont
  {Mukherjee}, \citenamefont {Mukherjee}, \citenamefont {Ravi}, \citenamefont
  {Mukherjee}, \citenamefont {Sardar},\ and\ \citenamefont
  {Adhikari}}]{naskar_jctc}%
  \BibitemOpen
  \bibfield  {author} {\bibinfo {author} {\bibfnamefont {K.}~\bibnamefont
  {Naskar}}, \bibinfo {author} {\bibfnamefont {S.}~\bibnamefont {Mukherjee}},
  \bibinfo {author} {\bibfnamefont {B.}~\bibnamefont {Mukherjee}}, \bibinfo
  {author} {\bibfnamefont {S.}~\bibnamefont {Ravi}}, \bibinfo {author}
  {\bibfnamefont {S.}~\bibnamefont {Mukherjee}}, \bibinfo {author}
  {\bibfnamefont {S.}~\bibnamefont {Sardar}}, \ and\ \bibinfo {author}
  {\bibfnamefont {S.}~\bibnamefont {Adhikari}},\ }\href@noop {} {\bibfield
  {journal} {\bibinfo  {journal} {J. Chem. Theory Comput.}\ }\textbf {\bibinfo
  {volume} {16}},\ \bibinfo {pages} {1666} (\bibinfo {year}
  {2020}{\natexlab{a}})}\BibitemShut {NoStop}%
\bibitem [{\citenamefont {Baer}(1980)}]{baer_mp40}%
  \BibitemOpen
  \bibfield  {author} {\bibinfo {author} {\bibfnamefont {M.}~\bibnamefont
  {Baer}},\ }\href@noop {} {\bibfield  {journal} {\bibinfo  {journal} {Mol.
  Phys.}\ }\textbf {\bibinfo {volume} {40}},\ \bibinfo {pages} {1011} (\bibinfo
  {year} {1980})}\BibitemShut {NoStop}%
\bibitem [{\citenamefont {Baer}\ and\ \citenamefont
  {Englman}(1992)}]{baer_mp75}%
  \BibitemOpen
  \bibfield  {author} {\bibinfo {author} {\bibfnamefont {M.}~\bibnamefont
  {Baer}}\ and\ \bibinfo {author} {\bibfnamefont {R.}~\bibnamefont {Englman}},\
  }\href@noop {} {\bibfield  {journal} {\bibinfo  {journal} {Mol. Phys.}\
  }\textbf {\bibinfo {volume} {75}},\ \bibinfo {pages} {293} (\bibinfo {year}
  {1992})}\BibitemShut {NoStop}%
\bibitem [{\citenamefont {Kendrick}, \citenamefont {Mead},\ and\ \citenamefont
  {Truhlar}(2002)}]{kendrick_cp277}%
  \BibitemOpen
  \bibfield  {author} {\bibinfo {author} {\bibfnamefont {B.~K.}\ \bibnamefont
  {Kendrick}}, \bibinfo {author} {\bibfnamefont {C.~A.}\ \bibnamefont {Mead}},
  \ and\ \bibinfo {author} {\bibfnamefont {D.~G.}\ \bibnamefont {Truhlar}},\
  }\href@noop {} {\bibfield  {journal} {\bibinfo  {journal} {Chem. Phys.}\
  }\textbf {\bibinfo {volume} {277}},\ \bibinfo {pages} {31} (\bibinfo {year}
  {2002})}\BibitemShut {NoStop}%
\bibitem [{\citenamefont {Matsunaga}\ and\ \citenamefont
  {Yarkony}(1998)}]{yarkony_mp93}%
  \BibitemOpen
  \bibfield  {author} {\bibinfo {author} {\bibfnamefont {N.}~\bibnamefont
  {Matsunaga}}\ and\ \bibinfo {author} {\bibfnamefont {D.~R.}\ \bibnamefont
  {Yarkony}},\ }\href@noop {} {\bibfield  {journal} {\bibinfo  {journal} {Mol.
  Phys.}\ }\textbf {\bibinfo {volume} {93}},\ \bibinfo {pages} {79} (\bibinfo
  {year} {1998})}\BibitemShut {NoStop}%
\bibitem [{\citenamefont {Paul}\ \emph
  {et~al.}(2011{\natexlab{a}})\citenamefont {Paul}, \citenamefont {Ray},
  \citenamefont {Mukhopadhyay},\ and\ \citenamefont
  {Adhikari}}]{akpaul_cpl508}%
  \BibitemOpen
  \bibfield  {author} {\bibinfo {author} {\bibfnamefont {A.~K.}\ \bibnamefont
  {Paul}}, \bibinfo {author} {\bibfnamefont {S.}~\bibnamefont {Ray}}, \bibinfo
  {author} {\bibfnamefont {D.}~\bibnamefont {Mukhopadhyay}}, \ and\ \bibinfo
  {author} {\bibfnamefont {S.}~\bibnamefont {Adhikari}},\ }\href@noop {}
  {\bibfield  {journal} {\bibinfo  {journal} {Chem. Phys. Lett.}\ }\textbf
  {\bibinfo {volume} {508}},\ \bibinfo {pages} {300} (\bibinfo {year}
  {2011}{\natexlab{a}})}\BibitemShut {NoStop}%
\bibitem [{\citenamefont {Paul}\ \emph
  {et~al.}(2011{\natexlab{b}})\citenamefont {Paul}, \citenamefont {Ray},
  \citenamefont {Mukhopadhyay},\ and\ \citenamefont
  {Adhikari}}]{akpaul_jcp135}%
  \BibitemOpen
  \bibfield  {author} {\bibinfo {author} {\bibfnamefont {A.~K.}\ \bibnamefont
  {Paul}}, \bibinfo {author} {\bibfnamefont {S.}~\bibnamefont {Ray}}, \bibinfo
  {author} {\bibfnamefont {D.}~\bibnamefont {Mukhopadhyay}}, \ and\ \bibinfo
  {author} {\bibfnamefont {S.}~\bibnamefont {Adhikari}},\ }\href@noop {}
  {\bibfield  {journal} {\bibinfo  {journal} {J. Chem. Phys.}\ }\textbf
  {\bibinfo {volume} {135}},\ \bibinfo {pages} {034107} (\bibinfo {year}
  {2011}{\natexlab{b}})}\BibitemShut {NoStop}%
\bibitem [{\citenamefont {Mukherjee}\ \emph
  {et~al.}(2013{\natexlab{b}})\citenamefont {Mukherjee}, \citenamefont
  {Bandyopadhyay}, \citenamefont {Paul},\ and\ \citenamefont
  {Adhikari}}]{mukherjee_jpca117}%
  \BibitemOpen
  \bibfield  {author} {\bibinfo {author} {\bibfnamefont {S.}~\bibnamefont
  {Mukherjee}}, \bibinfo {author} {\bibfnamefont {S.}~\bibnamefont
  {Bandyopadhyay}}, \bibinfo {author} {\bibfnamefont {A.~K.}\ \bibnamefont
  {Paul}}, \ and\ \bibinfo {author} {\bibfnamefont {S.}~\bibnamefont
  {Adhikari}},\ }\href@noop {} {\bibfield  {journal} {\bibinfo  {journal} {J.
  Phys. Chem. A}\ }\textbf {\bibinfo {volume} {117}},\ \bibinfo {pages} {3475}
  (\bibinfo {year} {2013}{\natexlab{b}})}\BibitemShut {NoStop}%
\bibitem [{\citenamefont {Mukherjee}\ and\ \citenamefont
  {Adhikari}(2014)}]{mukherjee_cp440}%
  \BibitemOpen
  \bibfield  {author} {\bibinfo {author} {\bibfnamefont {S.}~\bibnamefont
  {Mukherjee}}\ and\ \bibinfo {author} {\bibfnamefont {S.}~\bibnamefont
  {Adhikari}},\ }\href@noop {} {\bibfield  {journal} {\bibinfo  {journal}
  {Chem. Phys.}\ }\textbf {\bibinfo {volume} {440}},\ \bibinfo {pages} {106}
  (\bibinfo {year} {2014})}\BibitemShut {NoStop}%
\bibitem [{\citenamefont {Mukherjee}, \citenamefont {Mukhopadhyay},\ and\
  \citenamefont {Adhikari}(2014{\natexlab{a}})}]{mukherjee_jcp141}%
  \BibitemOpen
  \bibfield  {author} {\bibinfo {author} {\bibfnamefont {S.}~\bibnamefont
  {Mukherjee}}, \bibinfo {author} {\bibfnamefont {D.}~\bibnamefont
  {Mukhopadhyay}}, \ and\ \bibinfo {author} {\bibfnamefont {S.}~\bibnamefont
  {Adhikari}},\ }\href@noop {} {\bibfield  {journal} {\bibinfo  {journal} {J.
  Chem. Phys.}\ }\textbf {\bibinfo {volume} {141}},\ \bibinfo {pages} {204306}
  (\bibinfo {year} {2014}{\natexlab{a}})}\BibitemShut {NoStop}%
\bibitem [{\citenamefont {Sardar}\ \emph {et~al.}(2013)\citenamefont {Sardar},
  \citenamefont {Mukherjee}, \citenamefont {Paul},\ and\ \citenamefont
  {Adhikari}}]{sardar_cp416}%
  \BibitemOpen
  \bibfield  {author} {\bibinfo {author} {\bibfnamefont {S.}~\bibnamefont
  {Sardar}}, \bibinfo {author} {\bibfnamefont {S.}~\bibnamefont {Mukherjee}},
  \bibinfo {author} {\bibfnamefont {A.~K.}\ \bibnamefont {Paul}}, \ and\
  \bibinfo {author} {\bibfnamefont {S.}~\bibnamefont {Adhikari}},\ }\href@noop
  {} {\bibfield  {journal} {\bibinfo  {journal} {Chem. Phys.}\ }\textbf
  {\bibinfo {volume} {416}},\ \bibinfo {pages} {11} (\bibinfo {year}
  {2013})}\BibitemShut {NoStop}%
\bibitem [{\citenamefont {Mukherjee}\ \emph {et~al.}(2015)\citenamefont
  {Mukherjee}, \citenamefont {Mukherjee}, \citenamefont {Sardar},\ and\
  \citenamefont {Adhikari}}]{mukherjee_jcp143}%
  \BibitemOpen
  \bibfield  {author} {\bibinfo {author} {\bibfnamefont {S.}~\bibnamefont
  {Mukherjee}}, \bibinfo {author} {\bibfnamefont {B.}~\bibnamefont
  {Mukherjee}}, \bibinfo {author} {\bibfnamefont {S.}~\bibnamefont {Sardar}}, \
  and\ \bibinfo {author} {\bibfnamefont {S.}~\bibnamefont {Adhikari}},\
  }\href@noop {} {\bibfield  {journal} {\bibinfo  {journal} {J. Chem. Phys.}\
  }\textbf {\bibinfo {volume} {143}},\ \bibinfo {pages} {244307} (\bibinfo
  {year} {2015})}\BibitemShut {NoStop}%
\bibitem [{\citenamefont {Mukherjee}\ \emph {et~al.}(2017)\citenamefont
  {Mukherjee}, \citenamefont {Mukherjee}, \citenamefont {Sardar}, \citenamefont
  {Shamasundar},\ and\ \citenamefont {Adhikari}}]{bijit_mp115}%
  \BibitemOpen
  \bibfield  {author} {\bibinfo {author} {\bibfnamefont {B.}~\bibnamefont
  {Mukherjee}}, \bibinfo {author} {\bibfnamefont {S.}~\bibnamefont
  {Mukherjee}}, \bibinfo {author} {\bibfnamefont {S.}~\bibnamefont {Sardar}},
  \bibinfo {author} {\bibfnamefont {K.~R.}\ \bibnamefont {Shamasundar}}, \ and\
  \bibinfo {author} {\bibfnamefont {S.}~\bibnamefont {Adhikari}},\ }\href@noop
  {} {\bibfield  {journal} {\bibinfo  {journal} {Mol. Phys.}\ }\textbf
  {\bibinfo {volume} {115}},\ \bibinfo {pages} {2833} (\bibinfo {year}
  {2017})}\BibitemShut {NoStop}%
\bibitem [{\citenamefont {Mukherjee}\ \emph
  {et~al.}(2019{\natexlab{b}})\citenamefont {Mukherjee}, \citenamefont {Dutta},
  \citenamefont {Mukherjee}, \citenamefont {Sardar},\ and\ \citenamefont
  {Adhikari}}]{soumya_jcp150}%
  \BibitemOpen
  \bibfield  {author} {\bibinfo {author} {\bibfnamefont {S.}~\bibnamefont
  {Mukherjee}}, \bibinfo {author} {\bibfnamefont {J.}~\bibnamefont {Dutta}},
  \bibinfo {author} {\bibfnamefont {B.}~\bibnamefont {Mukherjee}}, \bibinfo
  {author} {\bibfnamefont {S.}~\bibnamefont {Sardar}}, \ and\ \bibinfo {author}
  {\bibfnamefont {S.}~\bibnamefont {Adhikari}},\ }\href@noop {} {\bibfield
  {journal} {\bibinfo  {journal} {J. Chem. Phys.}\ }\textbf {\bibinfo {volume}
  {150}},\ \bibinfo {pages} {064308} (\bibinfo {year}
  {2019}{\natexlab{b}})}\BibitemShut {NoStop}%
\bibitem [{\citenamefont {Werner}\ \emph {et~al.}(2012)\citenamefont {Werner},
  \citenamefont {Knowles}, \citenamefont {Knizia}, \citenamefont {Manby},\ and\
  \citenamefont {Sch{\ifmmode\ddot{u}\else\"{u}\fi}tz}}]{MOLPRO_brief}%
  \BibitemOpen
  \bibfield  {author} {\bibinfo {author} {\bibfnamefont {H.-J.}\ \bibnamefont
  {Werner}}, \bibinfo {author} {\bibfnamefont {P.~J.}\ \bibnamefont {Knowles}},
  \bibinfo {author} {\bibfnamefont {G.}~\bibnamefont {Knizia}}, \bibinfo
  {author} {\bibfnamefont {F.~R.}\ \bibnamefont {Manby}}, \ and\ \bibinfo
  {author} {\bibfnamefont {M.}~\bibnamefont
  {Sch{\ifmmode\ddot{u}\else\"{u}\fi}tz}},\ }\href@noop {} {\bibfield
  {journal} {\bibinfo  {journal} {WIREs Comput. Mol. Sci.}\ }\textbf {\bibinfo
  {volume} {2}},\ \bibinfo {pages} {242} (\bibinfo {year} {2012})}\BibitemShut
  {NoStop}%
\bibitem [{\citenamefont {Lischka}\ \emph {et~al.}(2017)\citenamefont
  {Lischka}, \citenamefont {Shepard}, \citenamefont {Shavitt}, \citenamefont
  {Pitzer}, \citenamefont {Dallos} \emph {et~al.}}]{columbus}%
  \BibitemOpen
  \bibfield  {author} {\bibinfo {author} {\bibfnamefont {H.}~\bibnamefont
  {Lischka}}, \bibinfo {author} {\bibfnamefont {R.}~\bibnamefont {Shepard}},
  \bibinfo {author} {\bibfnamefont {I.}~\bibnamefont {Shavitt}}, \bibinfo
  {author} {\bibfnamefont {R.~M.}\ \bibnamefont {Pitzer}}, \bibinfo {author}
  {\bibfnamefont {M.}~\bibnamefont {Dallos}},  \emph {et~al.},\ }\href@noop {}
  {\enquote {\bibinfo {title} {Columbus, an ab initio electronic structure
  program, release 7.0},}\ } (\bibinfo {year} {2017})\BibitemShut {NoStop}%
\bibitem [{\citenamefont {Schmidt}\ \emph {et~al.}(1993)\citenamefont
  {Schmidt}, \citenamefont {Baldridge}, \citenamefont {Boatz}, \citenamefont
  {Elbert}, \citenamefont {Gordon} \emph {et~al.}}]{gamess}%
  \BibitemOpen
  \bibfield  {author} {\bibinfo {author} {\bibfnamefont {M.~W.}\ \bibnamefont
  {Schmidt}}, \bibinfo {author} {\bibfnamefont {K.~K.}\ \bibnamefont
  {Baldridge}}, \bibinfo {author} {\bibfnamefont {J.~A.}\ \bibnamefont
  {Boatz}}, \bibinfo {author} {\bibfnamefont {S.~T.}\ \bibnamefont {Elbert}},
  \bibinfo {author} {\bibfnamefont {M.~S.}\ \bibnamefont {Gordon}},  \emph
  {et~al.},\ }\href@noop {} {\bibfield  {journal} {\bibinfo  {journal} {J.
  Comput. Chem.}\ }\textbf {\bibinfo {volume} {14}},\ \bibinfo {pages} {1347}
  (\bibinfo {year} {1993})}\BibitemShut {NoStop}%
\bibitem [{\citenamefont {Frisch}\ \emph {et~al.}(2016)\citenamefont {Frisch},
  \citenamefont {Trucks}, \citenamefont {Schlegel}, \citenamefont {Scuseria},
  \citenamefont {Robb} \emph {et~al.}}]{g16}%
  \BibitemOpen
  \bibfield  {author} {\bibinfo {author} {\bibfnamefont {M.~J.}\ \bibnamefont
  {Frisch}}, \bibinfo {author} {\bibfnamefont {G.~W.}\ \bibnamefont {Trucks}},
  \bibinfo {author} {\bibfnamefont {H.~B.}\ \bibnamefont {Schlegel}}, \bibinfo
  {author} {\bibfnamefont {G.~E.}\ \bibnamefont {Scuseria}}, \bibinfo {author}
  {\bibfnamefont {M.~A.}\ \bibnamefont {Robb}},  \emph {et~al.},\ }\href@noop
  {} {\enquote {\bibinfo {title} {Gaussian˜16 {R}evision {C}.01},}\ }
  (\bibinfo {year} {2016}),\ \bibinfo {note} {gaussian Inc. Wallingford
  CT}\BibitemShut {NoStop}%
\bibitem [{\citenamefont {Sah}\ \emph {et~al.}(2023{\natexlab{a}})\citenamefont
  {Sah}, \citenamefont {Mukherjee}, \citenamefont {Naskar}, \citenamefont
  {Hazra},\ and\ \citenamefont {Adhikari}}]{mantu_ijqc_curl}%
  \BibitemOpen
  \bibfield  {author} {\bibinfo {author} {\bibfnamefont {M.~K.}\ \bibnamefont
  {Sah}}, \bibinfo {author} {\bibfnamefont {S.}~\bibnamefont {Mukherjee}},
  \bibinfo {author} {\bibfnamefont {K.}~\bibnamefont {Naskar}}, \bibinfo
  {author} {\bibfnamefont {S.}~\bibnamefont {Hazra}}, \ and\ \bibinfo {author}
  {\bibfnamefont {S.}~\bibnamefont {Adhikari}},\ }\href@noop {} {\bibfield
  {journal} {\bibinfo  {journal} {International Journal of Quantum Chemistry}\
  }\textbf {\bibinfo {volume} {123}},\ \bibinfo {pages} {e27212} (\bibinfo
  {year} {2023}{\natexlab{a}})}\BibitemShut {NoStop}%
\bibitem [{\citenamefont {Mukherjee}\ \emph
  {et~al.}(2018{\natexlab{b}})\citenamefont {Mukherjee}, \citenamefont
  {Mukherjee}, \citenamefont {Dutta}, \citenamefont {Sardar},\ and\
  \citenamefont {Adhikari}}]{soumya_acs}%
  \BibitemOpen
  \bibfield  {author} {\bibinfo {author} {\bibfnamefont {S.}~\bibnamefont
  {Mukherjee}}, \bibinfo {author} {\bibfnamefont {B.}~\bibnamefont
  {Mukherjee}}, \bibinfo {author} {\bibfnamefont {J.}~\bibnamefont {Dutta}},
  \bibinfo {author} {\bibfnamefont {S.}~\bibnamefont {Sardar}}, \ and\ \bibinfo
  {author} {\bibfnamefont {S.}~\bibnamefont {Adhikari}},\ }\href@noop {}
  {\bibfield  {journal} {\bibinfo  {journal} {ACS omega}\ }\textbf {\bibinfo
  {volume} {3}},\ \bibinfo {pages} {12465} (\bibinfo {year}
  {2018}{\natexlab{b}})}\BibitemShut {NoStop}%
\bibitem [{\citenamefont {Mukherjee}\ \emph
  {et~al.}(2019{\natexlab{c}})\citenamefont {Mukherjee}, \citenamefont
  {Mukherjee}, \citenamefont {Sardar},\ and\ \citenamefont
  {Adhikari}}]{soumya_comptc}%
  \BibitemOpen
  \bibfield  {author} {\bibinfo {author} {\bibfnamefont {S.}~\bibnamefont
  {Mukherjee}}, \bibinfo {author} {\bibfnamefont {B.}~\bibnamefont
  {Mukherjee}}, \bibinfo {author} {\bibfnamefont {S.}~\bibnamefont {Sardar}}, \
  and\ \bibinfo {author} {\bibfnamefont {S.}~\bibnamefont {Adhikari}},\
  }\href@noop {} {\bibfield  {journal} {\bibinfo  {journal} {Comp. Theor.
  Chem.}\ }\textbf {\bibinfo {volume} {1154}},\ \bibinfo {pages} {57} (\bibinfo
  {year} {2019}{\natexlab{c}})}\BibitemShut {NoStop}%
\bibitem [{\citenamefont {Dutta}\ \emph
  {et~al.}(2020{\natexlab{a}})\citenamefont {Dutta}, \citenamefont {Mukherjee},
  \citenamefont {Naskar}, \citenamefont {Ghosh}, \citenamefont {Mukherjee},
  \citenamefont {Ravi},\ and\ \citenamefont {Adhikari}}]{joy_pccp}%
  \BibitemOpen
  \bibfield  {author} {\bibinfo {author} {\bibfnamefont {J.}~\bibnamefont
  {Dutta}}, \bibinfo {author} {\bibfnamefont {S.}~\bibnamefont {Mukherjee}},
  \bibinfo {author} {\bibfnamefont {K.}~\bibnamefont {Naskar}}, \bibinfo
  {author} {\bibfnamefont {S.}~\bibnamefont {Ghosh}}, \bibinfo {author}
  {\bibfnamefont {B.}~\bibnamefont {Mukherjee}}, \bibinfo {author}
  {\bibfnamefont {S.}~\bibnamefont {Ravi}}, \ and\ \bibinfo {author}
  {\bibfnamefont {S.}~\bibnamefont {Adhikari}},\ }\href@noop {} {\bibfield
  {journal} {\bibinfo  {journal} {Phys. Chem. Chem. Phys. (Perspective)}\
  }\textbf {\bibinfo {volume} {22}},\ \bibinfo {pages} {27496} (\bibinfo {year}
  {2020}{\natexlab{a}})}\BibitemShut {NoStop}%
\bibitem [{\citenamefont {Hazra}\ \emph {et~al.}(2022)\citenamefont {Hazra},
  \citenamefont {Mukherjee}, \citenamefont {Ravi}, \citenamefont {Sardar},\
  and\ \citenamefont {Adhikari}}]{saikat_pyra}%
  \BibitemOpen
  \bibfield  {author} {\bibinfo {author} {\bibfnamefont {S.}~\bibnamefont
  {Hazra}}, \bibinfo {author} {\bibfnamefont {S.}~\bibnamefont {Mukherjee}},
  \bibinfo {author} {\bibfnamefont {S.}~\bibnamefont {Ravi}}, \bibinfo {author}
  {\bibfnamefont {S.}~\bibnamefont {Sardar}}, \ and\ \bibinfo {author}
  {\bibfnamefont {S.}~\bibnamefont {Adhikari}},\ }\href@noop {} {\bibfield
  {journal} {\bibinfo  {journal} {ChemPhysChem}\ }\textbf {\bibinfo {volume}
  {23}},\ \bibinfo {pages} {e202200482} (\bibinfo {year} {2022})}\BibitemShut
  {NoStop}%
\bibitem [{\citenamefont {Ghosh}\ \emph
  {et~al.}(2017{\natexlab{a}})\citenamefont {Ghosh}, \citenamefont {Mukherjee},
  \citenamefont {Mukherjee}, \citenamefont {Mandal}, \citenamefont {Sharma},
  \citenamefont {Chaudhury},\ and\ \citenamefont
  {Adhikari}}]{jcp-147-074105-SA}%
  \BibitemOpen
  \bibfield  {author} {\bibinfo {author} {\bibfnamefont {S.}~\bibnamefont
  {Ghosh}}, \bibinfo {author} {\bibfnamefont {S.}~\bibnamefont {Mukherjee}},
  \bibinfo {author} {\bibfnamefont {B.}~\bibnamefont {Mukherjee}}, \bibinfo
  {author} {\bibfnamefont {S.}~\bibnamefont {Mandal}}, \bibinfo {author}
  {\bibfnamefont {R.}~\bibnamefont {Sharma}}, \bibinfo {author} {\bibfnamefont
  {P.}~\bibnamefont {Chaudhury}}, \ and\ \bibinfo {author} {\bibfnamefont
  {S.}~\bibnamefont {Adhikari}},\ }\href@noop {} {\bibfield  {journal}
  {\bibinfo  {journal} {J. Chem. Phys.}\ }\textbf {\bibinfo {volume} {147}},\
  \bibinfo {pages} {074105/1} (\bibinfo {year}
  {2017}{\natexlab{a}})}\BibitemShut {NoStop}%
\bibitem [{\citenamefont {Jahn}\ and\ \citenamefont
  {Teller}(1937{\natexlab{a}})}]{JT_prs161}%
  \BibitemOpen
  \bibfield  {author} {\bibinfo {author} {\bibfnamefont {H.~A.}\ \bibnamefont
  {Jahn}}\ and\ \bibinfo {author} {\bibfnamefont {E.}~\bibnamefont {Teller}},\
  }\href@noop {} {\bibfield  {journal} {\bibinfo  {journal} {Proc. Roy. Soc.
  (London)}\ }\textbf {\bibinfo {volume} {161}},\ \bibinfo {pages} {220}
  (\bibinfo {year} {1937}{\natexlab{a}})}\BibitemShut {NoStop}%
\bibitem [{\citenamefont {Bersuker}(2006{\natexlab{a}})}]{bersuker_book}%
  \BibitemOpen
  \bibfield  {author} {\bibinfo {author} {\bibfnamefont {I.~B.}\ \bibnamefont
  {Bersuker}},\ }\href@noop {} {\emph {\bibinfo {title} {{\it The Jahn-Teller
  Effect}}}}\ (\bibinfo  {publisher} {Cambridge University Press},\ \bibinfo
  {year} {2006})\BibitemShut {NoStop}%
\bibitem [{\citenamefont {Garc\'{\i}a-Fern\'andez}\ \emph
  {et~al.}(2005)\citenamefont {Garc\'{\i}a-Fern\'andez}, \citenamefont
  {Bersuker}, \citenamefont {Aramburu}, \citenamefont {Barriuso},\ and\
  \citenamefont {Moreno}}]{Pablo_prb71}%
  \BibitemOpen
  \bibfield  {author} {\bibinfo {author} {\bibfnamefont {P.}~\bibnamefont
  {Garc\'{\i}a-Fern\'andez}}, \bibinfo {author} {\bibfnamefont {I.~B.}\
  \bibnamefont {Bersuker}}, \bibinfo {author} {\bibfnamefont {J.~A.}\
  \bibnamefont {Aramburu}}, \bibinfo {author} {\bibfnamefont {M.~T.}\
  \bibnamefont {Barriuso}}, \ and\ \bibinfo {author} {\bibfnamefont
  {M.}~\bibnamefont {Moreno}},\ }\href@noop {} {\bibfield  {journal} {\bibinfo
  {journal} {Phys. Rev. B}\ }\textbf {\bibinfo {volume} {71}},\ \bibinfo
  {pages} {184117} (\bibinfo {year} {2005})}\BibitemShut {NoStop}%
\bibitem [{\citenamefont {Renner}(1934)}]{renner_zp92}%
  \BibitemOpen
  \bibfield  {author} {\bibinfo {author} {\bibfnamefont {E.}~\bibnamefont
  {Renner}},\ }\href@noop {} {\bibfield  {journal} {\bibinfo  {journal} {Z.
  Phys.}\ }\textbf {\bibinfo {volume} {92}},\ \bibinfo {pages} {172} (\bibinfo
  {year} {1934})}\BibitemShut {NoStop}%
\bibitem [{\citenamefont {Pae}\ and\ \citenamefont
  {Hizhnyakov}(2017{\natexlab{a}})}]{Hizhnyakov_jcp147}%
  \BibitemOpen
  \bibfield  {author} {\bibinfo {author} {\bibfnamefont {K.}~\bibnamefont
  {Pae}}\ and\ \bibinfo {author} {\bibfnamefont {V.}~\bibnamefont
  {Hizhnyakov}},\ }\href@noop {} {\bibfield  {journal} {\bibinfo  {journal} {J.
  Chem. Phys.}\ }\textbf {\bibinfo {volume} {147}},\ \bibinfo {pages} {084107}
  (\bibinfo {year} {2017}{\natexlab{a}})}\BibitemShut {NoStop}%
\bibitem [{\citenamefont {Mukherjee}\ \emph
  {et~al.}(2018{\natexlab{c}})\citenamefont {Mukherjee}, \citenamefont
  {Mukhopadhyay}, \citenamefont {Adhikari},\ and\ \citenamefont
  {Baer}}]{bijit_mp116}%
  \BibitemOpen
  \bibfield  {author} {\bibinfo {author} {\bibfnamefont {B.}~\bibnamefont
  {Mukherjee}}, \bibinfo {author} {\bibfnamefont {D.}~\bibnamefont
  {Mukhopadhyay}}, \bibinfo {author} {\bibfnamefont {S.}~\bibnamefont
  {Adhikari}}, \ and\ \bibinfo {author} {\bibfnamefont {M.}~\bibnamefont
  {Baer}},\ }\href@noop {} {\bibfield  {journal} {\bibinfo  {journal} {Mol.
  Phys.}\ }\textbf {\bibinfo {volume} {116}},\ \bibinfo {pages} {2435}
  (\bibinfo {year} {2018}{\natexlab{c}})}\BibitemShut {NoStop}%
\bibitem [{\citenamefont {Sah}\ \emph {et~al.}(2025{\natexlab{a}})\citenamefont
  {Sah}, \citenamefont {Mukherjee}, \citenamefont {Ravi},\ and\ \citenamefont
  {Adhikari}}]{mantu_tio6_jcp}%
  \BibitemOpen
  \bibfield  {author} {\bibinfo {author} {\bibfnamefont {M.~K.}\ \bibnamefont
  {Sah}}, \bibinfo {author} {\bibfnamefont {S.}~\bibnamefont {Mukherjee}},
  \bibinfo {author} {\bibfnamefont {S.}~\bibnamefont {Ravi}}, \ and\ \bibinfo
  {author} {\bibfnamefont {S.}~\bibnamefont {Adhikari}},\ }\href@noop {}
  {\bibfield  {journal} {\bibinfo  {journal} {The Journal of Chemical Physics}\
  }\textbf {\bibinfo {volume} {under review}} (\bibinfo {year}
  {2025}{\natexlab{a}})}\BibitemShut {NoStop}%
\bibitem [{\citenamefont {Naskar}\ \emph
  {et~al.}(2023{\natexlab{a}})\citenamefont {Naskar}, \citenamefont {Ravi},
  \citenamefont {Adhikari}, \citenamefont {Baer},\ and\ \citenamefont
  {Sathyamurthy}}]{naskar_jpca_heh2+}%
  \BibitemOpen
  \bibfield  {author} {\bibinfo {author} {\bibfnamefont {K.}~\bibnamefont
  {Naskar}}, \bibinfo {author} {\bibfnamefont {S.}~\bibnamefont {Ravi}},
  \bibinfo {author} {\bibfnamefont {S.}~\bibnamefont {Adhikari}}, \bibinfo
  {author} {\bibfnamefont {M.}~\bibnamefont {Baer}}, \ and\ \bibinfo {author}
  {\bibfnamefont {N.}~\bibnamefont {Sathyamurthy}},\ }\href@noop {} {\bibfield
  {journal} {\bibinfo  {journal} {J. Phys. Chem. A}\ }\textbf {\bibinfo
  {volume} {127}},\ \bibinfo {pages} {3832} (\bibinfo {year}
  {2023}{\natexlab{a}})}\BibitemShut {NoStop}%
\bibitem [{\citenamefont {Naskar}\ \emph
  {et~al.}(2023{\natexlab{b}})\citenamefont {Naskar}, \citenamefont {Ghosh},
  \citenamefont {Adhikari}, \citenamefont {Baer},\ and\ \citenamefont
  {Sathyamurthy}}]{koushik_heh2+_dyna}%
  \BibitemOpen
  \bibfield  {author} {\bibinfo {author} {\bibfnamefont {K.}~\bibnamefont
  {Naskar}}, \bibinfo {author} {\bibfnamefont {S.}~\bibnamefont {Ghosh}},
  \bibinfo {author} {\bibfnamefont {S.}~\bibnamefont {Adhikari}}, \bibinfo
  {author} {\bibfnamefont {M.}~\bibnamefont {Baer}}, \ and\ \bibinfo {author}
  {\bibfnamefont {N.}~\bibnamefont {Sathyamurthy}},\ }\href@noop {} {\bibfield
  {journal} {\bibinfo  {journal} {J. Chem. Phys.}\ }\textbf {\bibinfo {volume}
  {159}},\ \bibinfo {pages} {034302} (\bibinfo {year}
  {2023}{\natexlab{b}})}\BibitemShut {NoStop}%
\bibitem [{\citenamefont {Naskar}\ \emph
  {et~al.}(2024{\natexlab{a}})\citenamefont {Naskar}, \citenamefont
  {Mukherjee}, \citenamefont {Ghosh},\ and\ \citenamefont
  {Adhikari}}]{koushik_fh2_dyna}%
  \BibitemOpen
  \bibfield  {author} {\bibinfo {author} {\bibfnamefont {K.}~\bibnamefont
  {Naskar}}, \bibinfo {author} {\bibfnamefont {S.}~\bibnamefont {Mukherjee}},
  \bibinfo {author} {\bibfnamefont {S.}~\bibnamefont {Ghosh}}, \ and\ \bibinfo
  {author} {\bibfnamefont {S.}~\bibnamefont {Adhikari}},\ }\href@noop {}
  {\bibfield  {journal} {\bibinfo  {journal} {The Journal of Physical Chemistry
  A}\ }\textbf {\bibinfo {volume} {128}},\ \bibinfo {pages} {1438} (\bibinfo
  {year} {2024}{\natexlab{a}})}\BibitemShut {NoStop}%
\bibitem [{\citenamefont {Michelsen}, \citenamefont {Rettner},\ and\
  \citenamefont {Auerbach}(1992)}]{HA2}%
  \BibitemOpen
  \bibfield  {author} {\bibinfo {author} {\bibfnamefont {H.~A.}\ \bibnamefont
  {Michelsen}}, \bibinfo {author} {\bibfnamefont {C.~T.}\ \bibnamefont
  {Rettner}}, \ and\ \bibinfo {author} {\bibfnamefont {D.~J.}\ \bibnamefont
  {Auerbach}},\ }\href@noop {} {\bibfield  {journal} {\bibinfo  {journal}
  {Surf.\ Sci.}\ }\textbf {\bibinfo {volume} {272}},\ \bibinfo {pages} {65}
  (\bibinfo {year} {1992})}\BibitemShut {NoStop}%
\bibitem [{\citenamefont {Rettner}, \citenamefont {Auerbach},\ and\
  \citenamefont {Michelsen}(1992)}]{CTD}%
  \BibitemOpen
  \bibfield  {author} {\bibinfo {author} {\bibfnamefont {C.~T.}\ \bibnamefont
  {Rettner}}, \bibinfo {author} {\bibfnamefont {D.~J.}\ \bibnamefont
  {Auerbach}}, \ and\ \bibinfo {author} {\bibfnamefont {H.~A.}\ \bibnamefont
  {Michelsen}},\ }\href@noop {} {\bibfield  {journal} {\bibinfo  {journal}
  {Phys.\ Rev.\ Lett.}\ }\textbf {\bibinfo {volume} {68}},\ \bibinfo {pages}
  {1164} (\bibinfo {year} {1992})}\BibitemShut {NoStop}%
\bibitem [{\citenamefont {Rettner}, \citenamefont {Michelsen},\ and\
  \citenamefont {Auerbach}(1993{\natexlab{a}})}]{CTH}%
  \BibitemOpen
  \bibfield  {author} {\bibinfo {author} {\bibfnamefont {C.~T.}\ \bibnamefont
  {Rettner}}, \bibinfo {author} {\bibfnamefont {H.~A.}\ \bibnamefont
  {Michelsen}}, \ and\ \bibinfo {author} {\bibfnamefont {D.~J.}\ \bibnamefont
  {Auerbach}},\ }\href@noop {} {\bibfield  {journal} {\bibinfo  {journal}
  {Faraday Discuss.}\ }\textbf {\bibinfo {volume} {96}},\ \bibinfo {pages} {17}
  (\bibinfo {year} {1993}{\natexlab{a}})}\BibitemShut {NoStop}%
\bibitem [{\citenamefont {Rettner}, \citenamefont {Michelsen},\ and\
  \citenamefont {Auerbach}(1993{\natexlab{b}})}]{CTR2}%
  \BibitemOpen
  \bibfield  {author} {\bibinfo {author} {\bibfnamefont {C.~T.}\ \bibnamefont
  {Rettner}}, \bibinfo {author} {\bibfnamefont {H.~A.}\ \bibnamefont
  {Michelsen}}, \ and\ \bibinfo {author} {\bibfnamefont {D.~J.}\ \bibnamefont
  {Auerbach}},\ }\href@noop {} {\bibfield  {journal} {\bibinfo  {journal}
  {Chem. Phys.}\ }\textbf {\bibinfo {volume} {175}},\ \bibinfo {pages} {157}
  (\bibinfo {year} {1993}{\natexlab{b}})}\BibitemShut {NoStop}%
\bibitem [{\citenamefont {Rettner}, \citenamefont {Michelsen},\ and\
  \citenamefont {Auerbach}(1995)}]{CTR}%
  \BibitemOpen
  \bibfield  {author} {\bibinfo {author} {\bibfnamefont {C.~T.}\ \bibnamefont
  {Rettner}}, \bibinfo {author} {\bibfnamefont {H.~A.}\ \bibnamefont
  {Michelsen}}, \ and\ \bibinfo {author} {\bibfnamefont {D.~J.}\ \bibnamefont
  {Auerbach}},\ }\href@noop {} {\bibfield  {journal} {\bibinfo  {journal} {J.\
  Chem.\ Phys.}\ }\textbf {\bibinfo {volume} {102}},\ \bibinfo {pages} {4625}
  (\bibinfo {year} {1995})}\BibitemShut {NoStop}%
\bibitem [{\citenamefont {Michelsen}\ \emph {et~al.}(1993)\citenamefont
  {Michelsen}, \citenamefont {Rettner}, \citenamefont {Auerbach},\ and\
  \citenamefont {Zare}}]{HAM}%
  \BibitemOpen
  \bibfield  {author} {\bibinfo {author} {\bibfnamefont {H.~A.}\ \bibnamefont
  {Michelsen}}, \bibinfo {author} {\bibfnamefont {C.~T.}\ \bibnamefont
  {Rettner}}, \bibinfo {author} {\bibfnamefont {D.~J.}\ \bibnamefont
  {Auerbach}}, \ and\ \bibinfo {author} {\bibfnamefont {R.~N.}\ \bibnamefont
  {Zare}},\ }\href@noop {} {\bibfield  {journal} {\bibinfo  {journal} {J.\
  Chem.\ Phys.}\ }\textbf {\bibinfo {volume} {98}},\ \bibinfo {pages} {8294}
  (\bibinfo {year} {1993})}\BibitemShut {NoStop}%
\bibitem [{\citenamefont {Hou}\ \emph {et~al.}(1997)\citenamefont {Hou},
  \citenamefont {Gulding}, \citenamefont {Rettner}, \citenamefont {Wodtke},\
  and\ \citenamefont {Auerbach}}]{EXPNEW2}%
  \BibitemOpen
  \bibfield  {author} {\bibinfo {author} {\bibfnamefont {H.}~\bibnamefont
  {Hou}}, \bibinfo {author} {\bibfnamefont {S.~J.}\ \bibnamefont {Gulding}},
  \bibinfo {author} {\bibfnamefont {C.~T.}\ \bibnamefont {Rettner}}, \bibinfo
  {author} {\bibfnamefont {A.~M.}\ \bibnamefont {Wodtke}}, \ and\ \bibinfo
  {author} {\bibfnamefont {D.~J.}\ \bibnamefont {Auerbach}},\ }\href@noop {}
  {\bibfield  {journal} {\bibinfo  {journal} {Science.}\ }\textbf {\bibinfo
  {volume} {277}},\ \bibinfo {pages} {80} (\bibinfo {year} {1997})}\BibitemShut
  {NoStop}%
\bibitem [{\citenamefont {Murphy}\ and\ \citenamefont
  {Hodgson}(1998)}]{EXPNEW3}%
  \BibitemOpen
  \bibfield  {author} {\bibinfo {author} {\bibfnamefont {M.~J.}\ \bibnamefont
  {Murphy}}\ and\ \bibinfo {author} {\bibfnamefont {A.}~\bibnamefont
  {Hodgson}},\ }\href@noop {} {\bibfield  {journal} {\bibinfo  {journal} {J.
  Chem. Phys.}\ }\textbf {\bibinfo {volume} {108}},\ \bibinfo {pages} {4199}
  (\bibinfo {year} {1998})}\BibitemShut {NoStop}%
\bibitem [{\citenamefont {Kaufmann}\ \emph {et~al.}(2018)\citenamefont
  {Kaufmann}, \citenamefont {Shuai}, \citenamefont {Auerbach}, \citenamefont
  {Schwarzer},\ and\ \citenamefont {Wodtke}}]{Wodtke}%
  \BibitemOpen
  \bibfield  {author} {\bibinfo {author} {\bibfnamefont {S.}~\bibnamefont
  {Kaufmann}}, \bibinfo {author} {\bibfnamefont {Q.}~\bibnamefont {Shuai}},
  \bibinfo {author} {\bibfnamefont {D.~J.}\ \bibnamefont {Auerbach}}, \bibinfo
  {author} {\bibfnamefont {D.}~\bibnamefont {Schwarzer}}, \ and\ \bibinfo
  {author} {\bibfnamefont {A.~M.}\ \bibnamefont {Wodtke}},\ }\href@noop {}
  {\bibfield  {journal} {\bibinfo  {journal} {J. Chem. Phys.}\ }\textbf
  {\bibinfo {volume} {148}},\ \bibinfo {pages} {194703} (\bibinfo {year}
  {2018})}\BibitemShut {NoStop}%
\bibitem [{\citenamefont {Wang}, \citenamefont {Darling},\ and\ \citenamefont
  {Holloway}(2001)}]{HO1}%
  \BibitemOpen
  \bibfield  {author} {\bibinfo {author} {\bibfnamefont {Z.~S.}\ \bibnamefont
  {Wang}}, \bibinfo {author} {\bibfnamefont {G.~R.}\ \bibnamefont {Darling}}, \
  and\ \bibinfo {author} {\bibfnamefont {S.}~\bibnamefont {Holloway}},\
  }\href@noop {} {\bibfield  {journal} {\bibinfo  {journal} {Phys.\ Rev.\
  Lett.}\ }\textbf {\bibinfo {volume} {87}},\ \bibinfo {pages} {226102/1}
  (\bibinfo {year} {2001})}\BibitemShut {NoStop}%
\bibitem [{\citenamefont {D\'{i}az}\ \emph
  {et~al.}(2010{\natexlab{a}})\citenamefont {D\'{i}az}, \citenamefont {Olsen},
  \citenamefont {Auerbach},\ and\ \citenamefont {Kroes}}]{CDGJK}%
  \BibitemOpen
  \bibfield  {author} {\bibinfo {author} {\bibfnamefont {C.}~\bibnamefont
  {D\'{i}az}}, \bibinfo {author} {\bibfnamefont {R.~A.}\ \bibnamefont {Olsen}},
  \bibinfo {author} {\bibfnamefont {D.~J.}\ \bibnamefont {Auerbach}}, \ and\
  \bibinfo {author} {\bibfnamefont {G.~J.}\ \bibnamefont {Kroes}},\ }\href@noop
  {} {\bibfield  {journal} {\bibinfo  {journal} {Phys. Chem. Chem. Phys.}\
  }\textbf {\bibinfo {volume} {12}},\ \bibinfo {pages} {6499} (\bibinfo {year}
  {2010}{\natexlab{a}})}\BibitemShut {NoStop}%
\bibitem [{\citenamefont {Hand}\ and\ \citenamefont {Haris}(1990)}]{HA1}%
  \BibitemOpen
  \bibfield  {author} {\bibinfo {author} {\bibfnamefont {M.}~\bibnamefont
  {Hand}}\ and\ \bibinfo {author} {\bibfnamefont {J.}~\bibnamefont {Haris}},\
  }\href@noop {} {\bibfield  {journal} {\bibinfo  {journal} {J.\ Chem.\ Phys.}\
  }\textbf {\bibinfo {volume} {92}},\ \bibinfo {pages} {7610} (\bibinfo {year}
  {1990})}\BibitemShut {NoStop}%
\bibitem [{\citenamefont {Dohle}\ and\ \citenamefont {Saalfrank}(1997)}]{SAL}%
  \BibitemOpen
  \bibfield  {author} {\bibinfo {author} {\bibfnamefont {M.}~\bibnamefont
  {Dohle}}\ and\ \bibinfo {author} {\bibfnamefont {P.}~\bibnamefont
  {Saalfrank}},\ }\href@noop {} {\bibfield  {journal} {\bibinfo  {journal}
  {Surf.\ Sci.}\ }\textbf {\bibinfo {volume} {373}},\ \bibinfo {pages} {95}
  (\bibinfo {year} {1997})}\BibitemShut {NoStop}%
\bibitem [{\citenamefont {Adhikari}\ and\ \citenamefont
  {Billing}(2000{\natexlab{a}})}]{SAG}%
  \BibitemOpen
  \bibfield  {author} {\bibinfo {author} {\bibfnamefont {S.}~\bibnamefont
  {Adhikari}}\ and\ \bibinfo {author} {\bibfnamefont {G.~D.}\ \bibnamefont
  {Billing}},\ }\href@noop {} {\bibfield  {journal} {\bibinfo  {journal} {J.\
  Chem.\ Phys.}\ }\textbf {\bibinfo {volume} {112}},\ \bibinfo {pages} {3884}
  (\bibinfo {year} {2000}{\natexlab{a}})}\BibitemShut {NoStop}%
\bibitem [{\citenamefont {Sahoo}, \citenamefont {Sardar},\ and\ \citenamefont
  {Adhikari}(2011{\natexlab{a}})}]{TSS}%
  \BibitemOpen
  \bibfield  {author} {\bibinfo {author} {\bibfnamefont {T.}~\bibnamefont
  {Sahoo}}, \bibinfo {author} {\bibfnamefont {S.}~\bibnamefont {Sardar}}, \
  and\ \bibinfo {author} {\bibfnamefont {S.}~\bibnamefont {Adhikari}},\
  }\href@noop {} {\bibfield  {journal} {\bibinfo  {journal} {Phys. Chem. Chem.
  Phys.}\ }\textbf {\bibinfo {volume} {13}},\ \bibinfo {pages} {10100}
  (\bibinfo {year} {2011}{\natexlab{a}})}\BibitemShut {NoStop}%
\bibitem [{\citenamefont {Sahoo}\ \emph {et~al.}(2011)\citenamefont {Sahoo},
  \citenamefont {Sardar}, \citenamefont {Mondal}, \citenamefont {Sarkar},\ and\
  \citenamefont {Adhikari}}]{TSS1}%
  \BibitemOpen
  \bibfield  {author} {\bibinfo {author} {\bibfnamefont {T.}~\bibnamefont
  {Sahoo}}, \bibinfo {author} {\bibfnamefont {S.}~\bibnamefont {Sardar}},
  \bibinfo {author} {\bibfnamefont {P.}~\bibnamefont {Mondal}}, \bibinfo
  {author} {\bibfnamefont {B.}~\bibnamefont {Sarkar}}, \ and\ \bibinfo {author}
  {\bibfnamefont {S.}~\bibnamefont {Adhikari}},\ }\href@noop {} {\bibfield
  {journal} {\bibinfo  {journal} {J. Phys. Chem. A}\ }\textbf {\bibinfo
  {volume} {115}},\ \bibinfo {pages} {5256} (\bibinfo {year}
  {2011})}\BibitemShut {NoStop}%
\bibitem [{\citenamefont {Sahoo}, \citenamefont {Sardar},\ and\ \citenamefont
  {Adhikari}(2011{\natexlab{b}})}]{TSS2}%
  \BibitemOpen
  \bibfield  {author} {\bibinfo {author} {\bibfnamefont {T.}~\bibnamefont
  {Sahoo}}, \bibinfo {author} {\bibfnamefont {S.}~\bibnamefont {Sardar}}, \
  and\ \bibinfo {author} {\bibfnamefont {S.}~\bibnamefont {Adhikari}},\
  }\href@noop {} {\bibfield  {journal} {\bibinfo  {journal} {Phys. Scr.}\
  }\textbf {\bibinfo {volume} {84}},\ \bibinfo {pages} {028105/1} (\bibinfo
  {year} {2011}{\natexlab{b}})}\BibitemShut {NoStop}%
\bibitem [{\citenamefont {Sahoo}, \citenamefont {Mukherjee},\ and\
  \citenamefont {Adhikari}(2012)}]{TSSM}%
  \BibitemOpen
  \bibfield  {author} {\bibinfo {author} {\bibfnamefont {T.}~\bibnamefont
  {Sahoo}}, \bibinfo {author} {\bibfnamefont {S.}~\bibnamefont {Mukherjee}}, \
  and\ \bibinfo {author} {\bibfnamefont {S.}~\bibnamefont {Adhikari}},\
  }\href@noop {} {\bibfield  {journal} {\bibinfo  {journal} {J.\ Chem.\ Phys.}\
  }\textbf {\bibinfo {volume} {136}},\ \bibinfo {pages} {84306/1} (\bibinfo
  {year} {2012})}\BibitemShut {NoStop}%
\bibitem [{\citenamefont {Nattino}\ \emph {et~al.}(2014)\citenamefont
  {Nattino}, \citenamefont {Genova}, \citenamefont {Guijt}, \citenamefont
  {Muzas}, \citenamefont {D\'{i}az}, \citenamefont {Auerbach},\ and\
  \citenamefont {Kroes}}]{FNGJK1}%
  \BibitemOpen
  \bibfield  {author} {\bibinfo {author} {\bibfnamefont {F.}~\bibnamefont
  {Nattino}}, \bibinfo {author} {\bibfnamefont {A.}~\bibnamefont {Genova}},
  \bibinfo {author} {\bibfnamefont {M.}~\bibnamefont {Guijt}}, \bibinfo
  {author} {\bibfnamefont {A.~S.}\ \bibnamefont {Muzas}}, \bibinfo {author}
  {\bibfnamefont {C.}~\bibnamefont {D\'{i}az}}, \bibinfo {author}
  {\bibfnamefont {D.~J.}\ \bibnamefont {Auerbach}}, \ and\ \bibinfo {author}
  {\bibfnamefont {G.~J.}\ \bibnamefont {Kroes}},\ }\href@noop {} {\bibfield
  {journal} {\bibinfo  {journal} {J. Chem. Phys.}\ }\textbf {\bibinfo {volume}
  {141}},\ \bibinfo {pages} {124705} (\bibinfo {year} {2014})}\BibitemShut
  {NoStop}%
\bibitem [{\citenamefont {Feit}, \citenamefont {Fleck},\ and\ \citenamefont
  {Steiger}(1982)}]{SPO}%
  \BibitemOpen
  \bibfield  {author} {\bibinfo {author} {\bibfnamefont {M.~D.}\ \bibnamefont
  {Feit}}, \bibinfo {author} {\bibfnamefont {J.~A.}\ \bibnamefont {Fleck}}, \
  and\ \bibinfo {author} {\bibfnamefont {A.}~\bibnamefont {Steiger}},\
  }\href@noop {} {\bibfield  {journal} {\bibinfo  {journal} {J. Comput. Phys.}\
  }\textbf {\bibinfo {volume} {47}},\ \bibinfo {pages} {412} (\bibinfo {year}
  {1982})}\BibitemShut {NoStop}%
\bibitem [{\citenamefont {Wijzenbroek}\ and\ \citenamefont
  {Somers}(2012)}]{MWMF}%
  \BibitemOpen
  \bibfield  {author} {\bibinfo {author} {\bibfnamefont {M.}~\bibnamefont
  {Wijzenbroek}}\ and\ \bibinfo {author} {\bibfnamefont {M.~F.}\ \bibnamefont
  {Somers}},\ }\href@noop {} {\bibfield  {journal} {\bibinfo  {journal} {J.
  Chem. Phys.}\ }\textbf {\bibinfo {volume} {137}},\ \bibinfo {pages} {054703}
  (\bibinfo {year} {2012})}\BibitemShut {NoStop}%
\bibitem [{\citenamefont {Kroes}\ and\ \citenamefont
  {D\'{i}az}(2016)}]{QD-AIMD-H2-D2-Cu-review-2016}%
  \BibitemOpen
  \bibfield  {author} {\bibinfo {author} {\bibfnamefont {G.-J.}\ \bibnamefont
  {Kroes}}\ and\ \bibinfo {author} {\bibfnamefont {C.}~\bibnamefont
  {D\'{i}az}},\ }\href@noop {} {\bibfield  {journal} {\bibinfo  {journal}
  {Chem. Soc. Rev.}\ }\textbf {\bibinfo {volume} {45}},\ \bibinfo {pages}
  {3658} (\bibinfo {year} {2016})}\BibitemShut {NoStop}%
\bibitem [{\citenamefont {Zhu}\ \emph {et~al.}(2020)\citenamefont {Zhu},
  \citenamefont {Zhang}, \citenamefont {Zhang}, \citenamefont {Zhou},\ and\
  \citenamefont {Jiang}}]{H2-Cu-multiple}%
  \BibitemOpen
  \bibfield  {author} {\bibinfo {author} {\bibfnamefont {L.}~\bibnamefont
  {Zhu}}, \bibinfo {author} {\bibfnamefont {Y.}~\bibnamefont {Zhang}}, \bibinfo
  {author} {\bibfnamefont {L.}~\bibnamefont {Zhang}}, \bibinfo {author}
  {\bibfnamefont {X.}~\bibnamefont {Zhou}}, \ and\ \bibinfo {author}
  {\bibfnamefont {B.}~\bibnamefont {Jiang}},\ }\href@noop {} {\bibfield
  {journal} {\bibinfo  {journal} {Phys. Chem. Chem. Phys.}\ }\textbf {\bibinfo
  {volume} {22}},\ \bibinfo {pages} {13958} (\bibinfo {year}
  {2020})}\BibitemShut {NoStop}%
\bibitem [{\citenamefont {Spiering}, \citenamefont {Wijzenbroek},\ and\
  \citenamefont {Somers}(2018)}]{PSMFS}%
  \BibitemOpen
  \bibfield  {author} {\bibinfo {author} {\bibfnamefont {P.}~\bibnamefont
  {Spiering}}, \bibinfo {author} {\bibfnamefont {M.}~\bibnamefont
  {Wijzenbroek}}, \ and\ \bibinfo {author} {\bibfnamefont {M.~F.}\ \bibnamefont
  {Somers}},\ }\href@noop {} {\bibfield  {journal} {\bibinfo  {journal} {J.
  Chem. Phys.}\ }\textbf {\bibinfo {volume} {149}},\ \bibinfo {pages} {234702}
  (\bibinfo {year} {2018})}\BibitemShut {NoStop}%
\bibitem [{\citenamefont {Dutta}, \citenamefont {Mandal},\ and\ \citenamefont
  {Adhikari}(2022)}]{VV-Hartree-2021}%
  \BibitemOpen
  \bibfield  {author} {\bibinfo {author} {\bibfnamefont {J.}~\bibnamefont
  {Dutta}}, \bibinfo {author} {\bibfnamefont {S.}~\bibnamefont {Mandal}}, \
  and\ \bibinfo {author} {\bibfnamefont {S.}~\bibnamefont {Adhikari}},\
  }\href@noop {} {\bibfield  {journal} {\bibinfo  {journal} {Chem. Phys.}\
  }\textbf {\bibinfo {volume} {552}},\ \bibinfo {pages} {111371} (\bibinfo
  {year} {2022})}\BibitemShut {NoStop}%
\bibitem [{\citenamefont {Smits}\ and\ \citenamefont
  {Somers}(2021)}]{EAM-SCM-QC-Mark-2021}%
  \BibitemOpen
  \bibfield  {author} {\bibinfo {author} {\bibfnamefont {B.}~\bibnamefont
  {Smits}}\ and\ \bibinfo {author} {\bibfnamefont {M.~F.}\ \bibnamefont
  {Somers}},\ }\href@noop {} {\bibfield  {journal} {\bibinfo  {journal} {J.
  Chem. Phys.}\ }\textbf {\bibinfo {volume} {154}},\ \bibinfo {pages} {074710}
  (\bibinfo {year} {2021})}\BibitemShut {NoStop}%
\bibitem [{\citenamefont {Smeets}\ and\ \citenamefont
  {Kroes}(2021)}]{SRP-DF-new-H2-Cu-Kroes-2021}%
  \BibitemOpen
  \bibfield  {author} {\bibinfo {author} {\bibfnamefont {E.~W.~F.}\
  \bibnamefont {Smeets}}\ and\ \bibinfo {author} {\bibfnamefont {G.-J.}\
  \bibnamefont {Kroes}},\ }\href@noop {} {\bibfield  {journal} {\bibinfo
  {journal} {Phys. Chem. Chem. Phys.}\ }\textbf {\bibinfo {volume} {23}},\
  \bibinfo {pages} {7875} (\bibinfo {year} {2021})}\BibitemShut {NoStop}%
\bibitem [{\citenamefont {Kroes}(2021)}]{Kroes-Review-2021}%
  \BibitemOpen
  \bibfield  {author} {\bibinfo {author} {\bibfnamefont {G.-J.}\ \bibnamefont
  {Kroes}},\ }\href@noop {} {\bibfield  {journal} {\bibinfo  {journal} {Phys.
  Chem. Chem. Phys.}\ }\textbf {\bibinfo {volume} {23}},\ \bibinfo {pages}
  {8962} (\bibinfo {year} {2021})}\BibitemShut {NoStop}%
\bibitem [{\citenamefont {Dutta}\ \emph {et~al.}(2021)\citenamefont {Dutta},
  \citenamefont {Mandal}, \citenamefont {Adhikari}, \citenamefont {Spiering},
  \citenamefont {Meyer},\ and\ \citenamefont {Somers}}]{SAMARK}%
  \BibitemOpen
  \bibfield  {author} {\bibinfo {author} {\bibfnamefont {J.}~\bibnamefont
  {Dutta}}, \bibinfo {author} {\bibfnamefont {S.}~\bibnamefont {Mandal}},
  \bibinfo {author} {\bibfnamefont {S.}~\bibnamefont {Adhikari}}, \bibinfo
  {author} {\bibfnamefont {P.}~\bibnamefont {Spiering}}, \bibinfo {author}
  {\bibfnamefont {J.}~\bibnamefont {Meyer}}, \ and\ \bibinfo {author}
  {\bibfnamefont {M.~F.}\ \bibnamefont {Somers}},\ }\href@noop {} {\bibfield
  {journal} {\bibinfo  {journal} {J. Chem. Phys.}\ }\textbf {\bibinfo {volume}
  {154}},\ \bibinfo {pages} {104103} (\bibinfo {year} {2021})}\BibitemShut
  {NoStop}%
\bibitem [{\citenamefont {Dutta}\ \emph
  {et~al.}(2022{\natexlab{b}})\citenamefont {Dutta}, \citenamefont {Naskar},
  \citenamefont {Adhikari}, \citenamefont {Meyer},\ and\ \citenamefont
  {Somers}}]{Joy_D2_jcp_2022}%
  \BibitemOpen
  \bibfield  {author} {\bibinfo {author} {\bibfnamefont {J.}~\bibnamefont
  {Dutta}}, \bibinfo {author} {\bibfnamefont {K.}~\bibnamefont {Naskar}},
  \bibinfo {author} {\bibfnamefont {S.}~\bibnamefont {Adhikari}}, \bibinfo
  {author} {\bibfnamefont {J.}~\bibnamefont {Meyer}}, \ and\ \bibinfo {author}
  {\bibfnamefont {M.~F.}\ \bibnamefont {Somers}},\ }\href@noop {} {\bibfield
  {journal} {\bibinfo  {journal} {J. Chem. Phys.}\ }\textbf {\bibinfo {volume}
  {157}},\ \bibinfo {pages} {194112} (\bibinfo {year}
  {2022}{\natexlab{b}})}\BibitemShut {NoStop}%
\bibitem [{\citenamefont {Huang}\ \emph {et~al.}(1993)\citenamefont {Huang},
  \citenamefont {Zhu}, \citenamefont {Kouri},\ and\ \citenamefont
  {Hoffman}}]{QD-method-1993}%
  \BibitemOpen
  \bibfield  {author} {\bibinfo {author} {\bibfnamefont {Y.}~\bibnamefont
  {Huang}}, \bibinfo {author} {\bibfnamefont {W.}~\bibnamefont {Zhu}}, \bibinfo
  {author} {\bibfnamefont {D.~J.}\ \bibnamefont {Kouri}}, \ and\ \bibinfo
  {author} {\bibfnamefont {D.~K.}\ \bibnamefont {Hoffman}},\ }\href@noop {}
  {\bibfield  {journal} {\bibinfo  {journal} {Chem. Phys. Lett.}\ }\textbf
  {\bibinfo {volume} {206}},\ \bibinfo {pages} {96} (\bibinfo {year}
  {1993})}\BibitemShut {NoStop}%
\bibitem [{\citenamefont {Guo}, \citenamefont {Menzel},\ and\ \citenamefont
  {Jackson}(2018)}]{SRP-DF-24-CH4-step-Ni-2018}%
  \BibitemOpen
  \bibfield  {author} {\bibinfo {author} {\bibfnamefont {H.}~\bibnamefont
  {Guo}}, \bibinfo {author} {\bibfnamefont {J.~P.}\ \bibnamefont {Menzel}}, \
  and\ \bibinfo {author} {\bibfnamefont {B.}~\bibnamefont {Jackson}},\
  }\href@noop {} {\bibfield  {journal} {\bibinfo  {journal} {J. Chem. Phys.}\
  }\textbf {\bibinfo {volume} {149}},\ \bibinfo {pages} {244704} (\bibinfo
  {year} {2018})}\BibitemShut {NoStop}%
\bibitem [{\citenamefont {Guo}\ and\ \citenamefont
  {Jackson}(2019)}]{SRP-DF-23-CH4-step-Ni-2019}%
  \BibitemOpen
  \bibfield  {author} {\bibinfo {author} {\bibfnamefont {H.}~\bibnamefont
  {Guo}}\ and\ \bibinfo {author} {\bibfnamefont {B.}~\bibnamefont {Jackson}},\
  }\href@noop {} {\bibfield  {journal} {\bibinfo  {journal} {J. Chem. Phys.}\
  }\textbf {\bibinfo {volume} {150}},\ \bibinfo {pages} {204703} (\bibinfo
  {year} {2019})}\BibitemShut {NoStop}%
\bibitem [{\citenamefont {Mandal}\ \emph
  {et~al.}(2019{\natexlab{a}})\citenamefont {Mandal}, \citenamefont {Ghosh},
  \citenamefont {Sardar},\ and\ \citenamefont {Adhikari}}]{SMIRPC}%
  \BibitemOpen
  \bibfield  {author} {\bibinfo {author} {\bibfnamefont {S.}~\bibnamefont
  {Mandal}}, \bibinfo {author} {\bibfnamefont {S.}~\bibnamefont {Ghosh}},
  \bibinfo {author} {\bibfnamefont {S.~S.}\ \bibnamefont {Sardar}}, \ and\
  \bibinfo {author} {\bibfnamefont {S.}~\bibnamefont {Adhikari}},\ }\href@noop
  {} {\bibfield  {journal} {\bibinfo  {journal} {Int. Rev. Phys. Chem.}\
  }\textbf {\bibinfo {volume} {37}},\ \bibinfo {pages} {607} (\bibinfo {year}
  {2019}{\natexlab{a}})}\BibitemShut {NoStop}%
\bibitem [{\citenamefont {D\'{i}az}\ \emph {et~al.}(2009)\citenamefont
  {D\'{i}az}, \citenamefont {Pijper}, \citenamefont {Olsen}, \citenamefont
  {Busnengo}, \citenamefont {Auerbach},\ and\ \citenamefont {Kroes}}]{SC1}%
  \BibitemOpen
  \bibfield  {author} {\bibinfo {author} {\bibfnamefont {C.}~\bibnamefont
  {D\'{i}az}}, \bibinfo {author} {\bibfnamefont {E.}~\bibnamefont {Pijper}},
  \bibinfo {author} {\bibfnamefont {R.~A.}\ \bibnamefont {Olsen}}, \bibinfo
  {author} {\bibfnamefont {H.~F.}\ \bibnamefont {Busnengo}}, \bibinfo {author}
  {\bibfnamefont {D.~J.}\ \bibnamefont {Auerbach}}, \ and\ \bibinfo {author}
  {\bibfnamefont {G.~J.}\ \bibnamefont {Kroes}},\ }\href@noop {} {\bibfield
  {journal} {\bibinfo  {journal} {Science}\ }\textbf {\bibinfo {volume}
  {326}},\ \bibinfo {pages} {832} (\bibinfo {year} {2009})}\BibitemShut
  {NoStop}%
\bibitem [{\citenamefont {Spiering}\ and\ \citenamefont
  {Meyer}(2018)}]{SRP-DF-8-H2-D2-Cu-LDFA-ODF-elhp-2018}%
  \BibitemOpen
  \bibfield  {author} {\bibinfo {author} {\bibfnamefont {P.}~\bibnamefont
  {Spiering}}\ and\ \bibinfo {author} {\bibfnamefont {J.}~\bibnamefont
  {Meyer}},\ }\href@noop {} {\bibfield  {journal} {\bibinfo  {journal} {J.
  Phys. Chem. Lett.}\ }\textbf {\bibinfo {volume} {9}},\ \bibinfo {pages}
  {1803} (\bibinfo {year} {2018})}\BibitemShut {NoStop}%
\bibitem [{\citenamefont {Jiang}, \citenamefont {Li},\ and\ \citenamefont
  {Guo}(2020)}]{HD-NNP-perspective-2020}%
  \BibitemOpen
  \bibfield  {author} {\bibinfo {author} {\bibfnamefont {B.}~\bibnamefont
  {Jiang}}, \bibinfo {author} {\bibfnamefont {J.}~\bibnamefont {Li}}, \ and\
  \bibinfo {author} {\bibfnamefont {H.}~\bibnamefont {Guo}},\ }\href@noop {}
  {\bibfield  {journal} {\bibinfo  {journal} {J. Phys. Chem. Lett.}\ }\textbf
  {\bibinfo {volume} {11}},\ \bibinfo {pages} {5120} (\bibinfo {year}
  {2020})}\BibitemShut {NoStop}%
\bibitem [{\citenamefont {Nattino}\ \emph
  {et~al.}(2012{\natexlab{a}})\citenamefont {Nattino}, \citenamefont
  {D\'{i}az}, \citenamefont {Jackson},\ and\ \citenamefont {Kroes}}]{FNGJK}%
  \BibitemOpen
  \bibfield  {author} {\bibinfo {author} {\bibfnamefont {F.}~\bibnamefont
  {Nattino}}, \bibinfo {author} {\bibfnamefont {C.}~\bibnamefont {D\'{i}az}},
  \bibinfo {author} {\bibfnamefont {B.}~\bibnamefont {Jackson}}, \ and\
  \bibinfo {author} {\bibfnamefont {G.~J.}\ \bibnamefont {Kroes}},\ }\href@noop
  {} {\bibfield  {journal} {\bibinfo  {journal} {Phy. Rev. Lett.}\ }\textbf
  {\bibinfo {volume} {108}},\ \bibinfo {pages} {236104/1} (\bibinfo {year}
  {2012}{\natexlab{a}})}\BibitemShut {NoStop}%
\bibitem [{\citenamefont {Jackson}, \citenamefont {Nattino},\ and\
  \citenamefont {Kroes}(2014)}]{PBE-19-CH4-on-Ni-AIMD-2014}%
  \BibitemOpen
  \bibfield  {author} {\bibinfo {author} {\bibfnamefont {B.}~\bibnamefont
  {Jackson}}, \bibinfo {author} {\bibfnamefont {F.}~\bibnamefont {Nattino}}, \
  and\ \bibinfo {author} {\bibfnamefont {G.-J.}\ \bibnamefont {Kroes}},\
  }\href@noop {} {\bibfield  {journal} {\bibinfo  {journal} {J. Chem. Phys.}\
  }\textbf {\bibinfo {volume} {141}},\ \bibinfo {pages} {054102} (\bibinfo
  {year} {2014})}\BibitemShut {NoStop}%
\bibitem [{\citenamefont {Kroes}, \citenamefont {Juaristi},\ and\ \citenamefont
  {Alducin}(2017)}]{SRP-DF-17-H2-Cu-GLO-AIMDEF-elhp-2017}%
  \BibitemOpen
  \bibfield  {author} {\bibinfo {author} {\bibfnamefont {G.-J.}\ \bibnamefont
  {Kroes}}, \bibinfo {author} {\bibfnamefont {J.~I.}\ \bibnamefont {Juaristi}},
  \ and\ \bibinfo {author} {\bibfnamefont {M.}~\bibnamefont {Alducin}},\
  }\href@noop {} {\bibfield  {journal} {\bibinfo  {journal} {J. Phys. Chem. C}\
  }\textbf {\bibinfo {volume} {121}},\ \bibinfo {pages} {13617} (\bibinfo
  {year} {2017})}\BibitemShut {NoStop}%
\bibitem [{\citenamefont {Tiwari}, \citenamefont {Nave},\ and\ \citenamefont
  {Jackson}(2010)}]{AKT1}%
  \BibitemOpen
  \bibfield  {author} {\bibinfo {author} {\bibfnamefont {A.~K.}\ \bibnamefont
  {Tiwari}}, \bibinfo {author} {\bibfnamefont {S.}~\bibnamefont {Nave}}, \ and\
  \bibinfo {author} {\bibfnamefont {B.}~\bibnamefont {Jackson}},\ }\href@noop
  {} {\bibfield  {journal} {\bibinfo  {journal} {J. Chem. Phys.}\ }\textbf
  {\bibinfo {volume} {132}},\ \bibinfo {pages} {134702} (\bibinfo {year}
  {2010})}\BibitemShut {NoStop}%
\bibitem [{\citenamefont {Tiwari}, \citenamefont {Nave},\ and\ \citenamefont
  {Jackson}(2009)}]{AKT}%
  \BibitemOpen
  \bibfield  {author} {\bibinfo {author} {\bibfnamefont {A.~K.}\ \bibnamefont
  {Tiwari}}, \bibinfo {author} {\bibfnamefont {S.}~\bibnamefont {Nave}}, \ and\
  \bibinfo {author} {\bibfnamefont {B.}~\bibnamefont {Jackson}},\ }\href@noop
  {} {\bibfield  {journal} {\bibinfo  {journal} {Phys.\ Rev.\ Lett.}\ }\textbf
  {\bibinfo {volume} {103}},\ \bibinfo {pages} {253201/1} (\bibinfo {year}
  {2009})}\BibitemShut {NoStop}%
\bibitem [{\citenamefont {Dohle}, \citenamefont {Saalfrank},\ and\
  \citenamefont {Uzer}(1998)}]{UZER1}%
  \BibitemOpen
  \bibfield  {author} {\bibinfo {author} {\bibfnamefont {M.}~\bibnamefont
  {Dohle}}, \bibinfo {author} {\bibfnamefont {P.}~\bibnamefont {Saalfrank}}, \
  and\ \bibinfo {author} {\bibfnamefont {T.}~\bibnamefont {Uzer}},\ }\href@noop
  {} {\bibfield  {journal} {\bibinfo  {journal} {Surf. Sci.}\ }\textbf
  {\bibinfo {volume} {409}},\ \bibinfo {pages} {37} (\bibinfo {year}
  {1998})}\BibitemShut {NoStop}%
\bibitem [{\citenamefont {Juaristi}\ \emph {et~al.}(2008)\citenamefont
  {Juaristi}, \citenamefont {Alducin}, \citenamefont {{n}o}, \citenamefont
  {Busnengo},\ and\ \citenamefont {Salin}}]{JIJ1}%
  \BibitemOpen
  \bibfield  {author} {\bibinfo {author} {\bibfnamefont {J.~I.}\ \bibnamefont
  {Juaristi}}, \bibinfo {author} {\bibfnamefont {M.}~\bibnamefont {Alducin}},
  \bibinfo {author} {\bibfnamefont {R.~D.~M.}\ \bibnamefont {{n}o}}, \bibinfo
  {author} {\bibfnamefont {H.~F.}\ \bibnamefont {Busnengo}}, \ and\ \bibinfo
  {author} {\bibfnamefont {A.}~\bibnamefont {Salin}},\ }\href@noop {}
  {\bibfield  {journal} {\bibinfo  {journal} {Phys. Rev. Lett.}\ }\textbf
  {\bibinfo {volume} {100}},\ \bibinfo {pages} {116102/1} (\bibinfo {year}
  {2008})}\BibitemShut {NoStop}%
\bibitem [{\citenamefont {Rittmeyer}, \citenamefont {Bukas},\ and\
  \citenamefont {Reuter}(2018)}]{Dynamical-models-review-2018}%
  \BibitemOpen
  \bibfield  {author} {\bibinfo {author} {\bibfnamefont {S.~P.}\ \bibnamefont
  {Rittmeyer}}, \bibinfo {author} {\bibfnamefont {V.~J.}\ \bibnamefont
  {Bukas}}, \ and\ \bibinfo {author} {\bibfnamefont {K.}~\bibnamefont
  {Reuter}},\ }\href@noop {} {\bibfield  {journal} {\bibinfo  {journal} {Adv.
  Phys.: X}\ }\textbf {\bibinfo {volume} {3}},\ \bibinfo {pages} {1381574}
  (\bibinfo {year} {2018})}\BibitemShut {NoStop}%
\bibitem [{\citenamefont {Shenvi}, \citenamefont {Roy},\ and\ \citenamefont
  {Tully}(2009{\natexlab{a}})}]{IESH-2009}%
  \BibitemOpen
  \bibfield  {author} {\bibinfo {author} {\bibfnamefont {N.}~\bibnamefont
  {Shenvi}}, \bibinfo {author} {\bibfnamefont {S.}~\bibnamefont {Roy}}, \ and\
  \bibinfo {author} {\bibfnamefont {J.~C.}\ \bibnamefont {Tully}},\ }\href@noop
  {} {\bibfield  {journal} {\bibinfo  {journal} {J. Chem. Phys.}\ }\textbf
  {\bibinfo {volume} {130}},\ \bibinfo {pages} {174107} (\bibinfo {year}
  {2009}{\natexlab{a}})}\BibitemShut {NoStop}%
\bibitem [{\citenamefont {Shenvi}, \citenamefont {Roy},\ and\ \citenamefont
  {Tully}(2009{\natexlab{b}})}]{JCTully-elhp1}%
  \BibitemOpen
  \bibfield  {author} {\bibinfo {author} {\bibfnamefont {N.}~\bibnamefont
  {Shenvi}}, \bibinfo {author} {\bibfnamefont {S.}~\bibnamefont {Roy}}, \ and\
  \bibinfo {author} {\bibfnamefont {J.~C.}\ \bibnamefont {Tully}},\ }\href@noop
  {} {\bibfield  {journal} {\bibinfo  {journal} {Science}\ }\textbf {\bibinfo
  {volume} {326}},\ \bibinfo {pages} {829} (\bibinfo {year}
  {2009}{\natexlab{b}})}\BibitemShut {NoStop}%
\bibitem [{\citenamefont {Mandal}\ \emph
  {et~al.}(2015{\natexlab{a}})\citenamefont {Mandal}, \citenamefont {Sahoo},
  \citenamefont {Ghosh},\ and\ \citenamefont {Adhikari}}]{SM-JTCC}%
  \BibitemOpen
  \bibfield  {author} {\bibinfo {author} {\bibfnamefont {S.}~\bibnamefont
  {Mandal}}, \bibinfo {author} {\bibfnamefont {T.}~\bibnamefont {Sahoo}},
  \bibinfo {author} {\bibfnamefont {S.}~\bibnamefont {Ghosh}}, \ and\ \bibinfo
  {author} {\bibfnamefont {S.}~\bibnamefont {Adhikari}},\ }\href@noop {}
  {\bibfield  {journal} {\bibinfo  {journal} {J. Theor. Comput. Chem.}\
  }\textbf {\bibinfo {volume} {14}},\ \bibinfo {pages} {1550028} (\bibinfo
  {year} {2015}{\natexlab{a}})}\BibitemShut {NoStop}%
\bibitem [{\citenamefont {Mandal}\ \emph
  {et~al.}(2015{\natexlab{b}})\citenamefont {Mandal}, \citenamefont {Sahoo},
  \citenamefont {Ghosh},\ and\ \citenamefont {Adhikari}}]{SM-MOLPHYS}%
  \BibitemOpen
  \bibfield  {author} {\bibinfo {author} {\bibfnamefont {S.}~\bibnamefont
  {Mandal}}, \bibinfo {author} {\bibfnamefont {T.}~\bibnamefont {Sahoo}},
  \bibinfo {author} {\bibfnamefont {S.}~\bibnamefont {Ghosh}}, \ and\ \bibinfo
  {author} {\bibfnamefont {S.}~\bibnamefont {Adhikari}},\ }\href@noop {}
  {\bibfield  {journal} {\bibinfo  {journal} {Mol. Phys.}\ }\textbf {\bibinfo
  {volume} {113}},\ \bibinfo {pages} {3042} (\bibinfo {year}
  {2015}{\natexlab{b}})}\BibitemShut {NoStop}%
\bibitem [{\citenamefont {Jahn}\ and\ \citenamefont
  {Teller}(1937{\natexlab{b}})}]{JT}%
  \BibitemOpen
  \bibfield  {author} {\bibinfo {author} {\bibfnamefont {H.~A.}\ \bibnamefont
  {Jahn}}\ and\ \bibinfo {author} {\bibfnamefont {E.}~\bibnamefont {Teller}},\
  }\href@noop {} {\bibfield  {journal} {\bibinfo  {journal} {Proc. R. Soc.
  London A}\ }\textbf {\bibinfo {volume} {161}},\ \bibinfo {pages} {220}
  (\bibinfo {year} {1937}{\natexlab{b}})}\BibitemShut {NoStop}%
\bibitem [{\citenamefont {Klein}(1974)}]{klein_jcp61}%
  \BibitemOpen
  \bibfield  {author} {\bibinfo {author} {\bibfnamefont {D.~J.}\ \bibnamefont
  {Klein}},\ }\href@noop {} {\bibfield  {journal} {\bibinfo  {journal} {J.
  Chem. Phys.}\ }\textbf {\bibinfo {volume} {61}},\ \bibinfo {pages} {786}
  (\bibinfo {year} {1974})}\BibitemShut {NoStop}%
\bibitem [{\citenamefont {Bersuker}(2001)}]{JT-review-Bersuker}%
  \BibitemOpen
  \bibfield  {author} {\bibinfo {author} {\bibfnamefont {I.~B.}\ \bibnamefont
  {Bersuker}},\ }\href@noop {} {\bibfield  {journal} {\bibinfo  {journal}
  {Chem. Rev.}\ }\textbf {\bibinfo {volume} {101}},\ \bibinfo {pages} {1067}
  (\bibinfo {year} {2001})}\BibitemShut {NoStop}%
\bibitem [{\citenamefont {Jahn}(1937)}]{JT1}%
  \BibitemOpen
  \bibfield  {author} {\bibinfo {author} {\bibfnamefont {H.~A.}\ \bibnamefont
  {Jahn}},\ }\href@noop {} {\bibfield  {journal} {\bibinfo  {journal} {Proc. R.
  Soc. London A}\ }\textbf {\bibinfo {volume} {164}},\ \bibinfo {pages} {117}
  (\bibinfo {year} {1937})}\BibitemShut {NoStop}%
\bibitem [{\citenamefont {Jahn}\ and\ \citenamefont
  {Teller}(1937{\natexlab{c}})}]{JT:PRS}%
  \BibitemOpen
  \bibfield  {author} {\bibinfo {author} {\bibfnamefont {H.~A.}\ \bibnamefont
  {Jahn}}\ and\ \bibinfo {author} {\bibfnamefont {E.}~\bibnamefont {Teller}},\
  }\href@noop {} {\bibfield  {journal} {\bibinfo  {journal} {Proc. Roy. Soc.
  (London)}\ }\textbf {\bibinfo {volume} {161}},\ \bibinfo {pages} {220}
  (\bibinfo {year} {1937}{\natexlab{c}})}\BibitemShut {NoStop}%
\bibitem [{\citenamefont {Vleck}(1939{\natexlab{a}})}]{van-vleck}%
  \BibitemOpen
  \bibfield  {author} {\bibinfo {author} {\bibfnamefont {J.~H.~V.}\
  \bibnamefont {Vleck}},\ }\href@noop {} {\bibfield  {journal} {\bibinfo
  {journal} {J. Chem. Phys.}\ }\textbf {\bibinfo {volume} {7}},\ \bibinfo
  {pages} {72} (\bibinfo {year} {1939}{\natexlab{a}})}\BibitemShut {NoStop}%
\bibitem [{\citenamefont {Kovaleva}\ \emph {et~al.}(2016)\citenamefont
  {Kovaleva}, \citenamefont {Kugel}, \citenamefont {Pot\.{u}\u{c}ek},
  \citenamefont {Kusmartseva}, \citenamefont {Goryachev}, \citenamefont
  {Bryknar}, \citenamefont {Demikhov}, \citenamefont {Trepakov}, \citenamefont
  {Dejneka}, \citenamefont {Kusmartsev},\ and\ \citenamefont
  {Stoneham}}]{Natalia_JETP}%
  \BibitemOpen
  \bibfield  {author} {\bibinfo {author} {\bibfnamefont {N.~N.}\ \bibnamefont
  {Kovaleva}}, \bibinfo {author} {\bibfnamefont {K.~I.}\ \bibnamefont {Kugel}},
  \bibinfo {author} {\bibfnamefont {Z.}~\bibnamefont {Pot\.{u}\u{c}ek}},
  \bibinfo {author} {\bibfnamefont {O.~E.}\ \bibnamefont {Kusmartseva}},
  \bibinfo {author} {\bibfnamefont {N.~S.}\ \bibnamefont {Goryachev}}, \bibinfo
  {author} {\bibfnamefont {Z.}~\bibnamefont {Bryknar}}, \bibinfo {author}
  {\bibfnamefont {E.~I.}\ \bibnamefont {Demikhov}}, \bibinfo {author}
  {\bibfnamefont {V.~A.}\ \bibnamefont {Trepakov}}, \bibinfo {author}
  {\bibfnamefont {A.}~\bibnamefont {Dejneka}}, \bibinfo {author} {\bibfnamefont
  {F.}~\bibnamefont {Kusmartsev}}, \ and\ \bibinfo {author} {\bibfnamefont
  {A.~M.}\ \bibnamefont {Stoneham}},\ }\href@noop {} {\bibfield  {journal}
  {\bibinfo  {journal} {J. Exp. Theor. Phys.}\ }\textbf {\bibinfo {volume}
  {122}},\ \bibinfo {pages} {890} (\bibinfo {year} {2016})}\BibitemShut
  {NoStop}%
\bibitem [{\citenamefont {Kugel}\ and\ \citenamefont
  {Khomskii}(1982)}]{Kugel-Khomskii}%
  \BibitemOpen
  \bibfield  {author} {\bibinfo {author} {\bibfnamefont {K.~I.}\ \bibnamefont
  {Kugel}}\ and\ \bibinfo {author} {\bibfnamefont {D.~I.}\ \bibnamefont
  {Khomskii}},\ }\href@noop {} {\bibfield  {journal} {\bibinfo  {journal} {Sov.
  Phys. Usp.}\ }\textbf {\bibinfo {volume} {25}},\ \bibinfo {pages} {231}
  (\bibinfo {year} {1982})}\BibitemShut {NoStop}%
\bibitem [{\citenamefont {Ole\'s}\ \emph {et~al.}(2005)\citenamefont {Ole\'s},
  \citenamefont {Khaliullin}, \citenamefont {Horsch},\ and\ \citenamefont
  {Feiner}}]{Oles-Feiner}%
  \BibitemOpen
  \bibfield  {author} {\bibinfo {author} {\bibfnamefont {A.~M.}\ \bibnamefont
  {Ole\'s}}, \bibinfo {author} {\bibfnamefont {G.}~\bibnamefont {Khaliullin}},
  \bibinfo {author} {\bibfnamefont {P.}~\bibnamefont {Horsch}}, \ and\ \bibinfo
  {author} {\bibfnamefont {L.~F.}\ \bibnamefont {Feiner}},\ }\href@noop {}
  {\bibfield  {journal} {\bibinfo  {journal} {Phys. Rev. B}\ }\textbf {\bibinfo
  {volume} {72}},\ \bibinfo {pages} {214431} (\bibinfo {year}
  {2005})}\BibitemShut {NoStop}%
\bibitem [{\citenamefont {O'Brien}(1964)}]{O'Brien}%
  \BibitemOpen
  \bibfield  {author} {\bibinfo {author} {\bibfnamefont {M.~C.~M.}\
  \bibnamefont {O'Brien}},\ }\href@noop {} {\bibfield  {journal} {\bibinfo
  {journal} {Proc. Roy. Soc. A}\ }\textbf {\bibinfo {volume} {281}},\ \bibinfo
  {pages} {323} (\bibinfo {year} {1964})}\BibitemShut {NoStop}%
\bibitem [{\citenamefont {Dutta}, \citenamefont {Adhikari},\ and\ \citenamefont
  {Kovaleva}(2019)}]{joy_jcp}%
  \BibitemOpen
  \bibfield  {author} {\bibinfo {author} {\bibfnamefont {J.}~\bibnamefont
  {Dutta}}, \bibinfo {author} {\bibfnamefont {S.}~\bibnamefont {Adhikari}}, \
  and\ \bibinfo {author} {\bibfnamefont {N.}~\bibnamefont {Kovaleva}},\
  }\href@noop {} {\bibfield  {journal} {\bibinfo  {journal} {J. Chem. Phys.}\
  }\textbf {\bibinfo {volume} {150}},\ \bibinfo {pages} {064703} (\bibinfo
  {year} {2019})}\BibitemShut {NoStop}%
\bibitem [{\citenamefont {Slonczewski}(1963{\natexlab{a}})}]{Slonczewski}%
  \BibitemOpen
  \bibfield  {author} {\bibinfo {author} {\bibfnamefont {J.~C.}\ \bibnamefont
  {Slonczewski}},\ }\href@noop {} {\bibfield  {journal} {\bibinfo  {journal}
  {Phys. Rev.}\ }\textbf {\bibinfo {volume} {131}},\ \bibinfo {pages} {1596}
  (\bibinfo {year} {1963}{\natexlab{a}})}\BibitemShut {NoStop}%
\bibitem [{\citenamefont {Feynmann}\ and\ \citenamefont
  {Hibbs}(1965)}]{feynmann}%
  \BibitemOpen
  \bibfield  {author} {\bibinfo {author} {\bibfnamefont {R.~P.}\ \bibnamefont
  {Feynmann}}\ and\ \bibinfo {author} {\bibfnamefont {A.~R.}\ \bibnamefont
  {Hibbs}},\ }\href@noop {} {\emph {\bibinfo {title} {Quantum Mechanics and
  Path Integrals}}}\ (\bibinfo  {publisher} {McCraw-Hill, New York},\ \bibinfo
  {year} {1965})\BibitemShut {NoStop}%
\bibitem [{\citenamefont {Sah}\ \emph {et~al.}(2023{\natexlab{b}})\citenamefont
  {Sah}, \citenamefont {Mukherjee}, \citenamefont {Saha}, \citenamefont
  {Naskar},\ and\ \citenamefont {Adhikari}}]{mantu_jcp159_path}%
  \BibitemOpen
  \bibfield  {author} {\bibinfo {author} {\bibfnamefont {M.~K.}\ \bibnamefont
  {Sah}}, \bibinfo {author} {\bibfnamefont {S.}~\bibnamefont {Mukherjee}},
  \bibinfo {author} {\bibfnamefont {S.}~\bibnamefont {Saha}}, \bibinfo {author}
  {\bibfnamefont {K.}~\bibnamefont {Naskar}}, \ and\ \bibinfo {author}
  {\bibfnamefont {S.}~\bibnamefont {Adhikari}},\ }\href@noop {} {\bibfield
  {journal} {\bibinfo  {journal} {The Journal of Chemical Physics}\ }\textbf
  {\bibinfo {volume} {159}},\ \bibinfo {pages} {244116} (\bibinfo {year}
  {2023}{\natexlab{b}})}\BibitemShut {NoStop}%
\bibitem [{\citenamefont {Longuet-Higgins}(1963)}]{longuet-higgins_mp6}%
  \BibitemOpen
  \bibfield  {author} {\bibinfo {author} {\bibfnamefont {H.~C.}\ \bibnamefont
  {Longuet-Higgins}},\ }\href@noop {} {\bibfield  {journal} {\bibinfo
  {journal} {Mol. Phys.}\ }\textbf {\bibinfo {volume} {6}},\ \bibinfo {pages}
  {445} (\bibinfo {year} {1963})}\BibitemShut {NoStop}%
\bibitem [{\citenamefont {Bunker}\ and\ \citenamefont
  {Jensen}(2006)}]{bunker_book}%
  \BibitemOpen
  \bibfield  {author} {\bibinfo {author} {\bibfnamefont {P.~R.}\ \bibnamefont
  {Bunker}}\ and\ \bibinfo {author} {\bibfnamefont {P.}~\bibnamefont
  {Jensen}},\ }\href@noop {} {\emph {\bibinfo {title} {{\it Molecular symmetry
  and spectroscopy: 2nd edition}}}}\ (\bibinfo  {publisher} {NRC Research
  Press},\ \bibinfo {year} {2006})\BibitemShut {NoStop}%
\bibitem [{\citenamefont {Al-Jabour}\ \emph {et~al.}(2010)\citenamefont
  {Al-Jabour}, \citenamefont {Baer}, \citenamefont {Deeb}, \citenamefont
  {Leibscher}, \citenamefont {Manz}, \citenamefont {Xu},\ and\ \citenamefont
  {Zilberg}}]{al-jabour_jpca114}%
  \BibitemOpen
  \bibfield  {author} {\bibinfo {author} {\bibfnamefont {S.}~\bibnamefont
  {Al-Jabour}}, \bibinfo {author} {\bibfnamefont {M.}~\bibnamefont {Baer}},
  \bibinfo {author} {\bibfnamefont {O.}~\bibnamefont {Deeb}}, \bibinfo {author}
  {\bibfnamefont {M.}~\bibnamefont {Leibscher}}, \bibinfo {author}
  {\bibfnamefont {J.}~\bibnamefont {Manz}}, \bibinfo {author} {\bibfnamefont
  {X.}~\bibnamefont {Xu}}, \ and\ \bibinfo {author} {\bibfnamefont
  {S.}~\bibnamefont {Zilberg}},\ }\href@noop {} {\bibfield  {journal} {\bibinfo
   {journal} {J. Phys. Chem. A}\ }\textbf {\bibinfo {volume} {114}},\ \bibinfo
  {pages} {2991} (\bibinfo {year} {2010})}\BibitemShut {NoStop}%
\bibitem [{\citenamefont {Kettle}(2007)}]{kettle_book}%
  \BibitemOpen
  \bibfield  {author} {\bibinfo {author} {\bibfnamefont {S.~F.~A.}\
  \bibnamefont {Kettle}},\ }\href@noop {} {\emph {\bibinfo {title} {Symmetry
  and Structure: Readable Group Theory for Chemists: 3rd edition}}}\ (\bibinfo
  {publisher} {J. Wiley and Sons Inc: West Sussex},\ \bibinfo {address}
  {England},\ \bibinfo {year} {2007})\BibitemShut {NoStop}%
\bibitem [{\citenamefont {V\'ertesi}\ \emph {et~al.}(2004)\citenamefont
  {V\'ertesi}, \citenamefont {Vib\'ok}, \citenamefont {Hal\'asz},\ and\
  \citenamefont {Baer}}]{baer_jcp120}%
  \BibitemOpen
  \bibfield  {author} {\bibinfo {author} {\bibfnamefont {T.}~\bibnamefont
  {V\'ertesi}}, \bibinfo {author} {\bibfnamefont {A.}~\bibnamefont {Vib\'ok}},
  \bibinfo {author} {\bibfnamefont {G.~J.}\ \bibnamefont {Hal\'asz}}, \ and\
  \bibinfo {author} {\bibfnamefont {M.}~\bibnamefont {Baer}},\ }\href@noop {}
  {\bibfield  {journal} {\bibinfo  {journal} {J. Chem. Phys.}\ }\textbf
  {\bibinfo {volume} {120}},\ \bibinfo {pages} {2565} (\bibinfo {year}
  {2004})}\BibitemShut {NoStop}%
\bibitem [{\citenamefont {Cocchini}, \citenamefont {Upton},\ and\ \citenamefont
  {Andreoni}(1988)}]{cocchini_jcp88}%
  \BibitemOpen
  \bibfield  {author} {\bibinfo {author} {\bibfnamefont {F.}~\bibnamefont
  {Cocchini}}, \bibinfo {author} {\bibfnamefont {T.~H.}\ \bibnamefont {Upton}},
  \ and\ \bibinfo {author} {\bibfnamefont {W.}~\bibnamefont {Andreoni}},\
  }\href@noop {} {\bibfield  {journal} {\bibinfo  {journal} {J. Chem. Phys.}\
  }\textbf {\bibinfo {volume} {88}},\ \bibinfo {pages} {6068} (\bibinfo {year}
  {1988})}\BibitemShut {NoStop}%
\bibitem [{\citenamefont {Mebel}\ \emph {et~al.}(2002)\citenamefont {Mebel},
  \citenamefont {Hal\'asz}, \citenamefont {Vib\'ok}, \citenamefont {Alijah},\
  and\ \citenamefont {Baer}}]{baer_jcp117}%
  \BibitemOpen
  \bibfield  {author} {\bibinfo {author} {\bibfnamefont {A.~M.}\ \bibnamefont
  {Mebel}}, \bibinfo {author} {\bibfnamefont {G.~J.}\ \bibnamefont {Hal\'asz}},
  \bibinfo {author} {\bibfnamefont {A.}~\bibnamefont {Vib\'ok}}, \bibinfo
  {author} {\bibfnamefont {A.}~\bibnamefont {Alijah}}, \ and\ \bibinfo {author}
  {\bibfnamefont {M.}~\bibnamefont {Baer}},\ }\href@noop {} {\bibfield
  {journal} {\bibinfo  {journal} {J. Chem. Phys.}\ }\textbf {\bibinfo {volume}
  {117}},\ \bibinfo {pages} {991} (\bibinfo {year} {2002})}\BibitemShut
  {NoStop}%
\bibitem [{\citenamefont {K\"oppel}, \citenamefont {Domcke},\ and\
  \citenamefont {Cederbaum}(1984)}]{kdc_acp57}%
  \BibitemOpen
  \bibfield  {author} {\bibinfo {author} {\bibfnamefont {H.}~\bibnamefont
  {K\"oppel}}, \bibinfo {author} {\bibfnamefont {W.}~\bibnamefont {Domcke}}, \
  and\ \bibinfo {author} {\bibfnamefont {L.~S.}\ \bibnamefont {Cederbaum}},\
  }\href@noop {} {\bibfield  {journal} {\bibinfo  {journal} {Adv. Chem. Phys.}\
  }\textbf {\bibinfo {volume} {57}},\ \bibinfo {pages} {59} (\bibinfo {year}
  {1984})}\BibitemShut {NoStop}%
\bibitem [{\citenamefont {K\"oppel}\ and\ \citenamefont
  {Schubert}(2006)}]{koppel_mp104}%
  \BibitemOpen
  \bibfield  {author} {\bibinfo {author} {\bibfnamefont {H.}~\bibnamefont
  {K\"oppel}}\ and\ \bibinfo {author} {\bibfnamefont {B.}~\bibnamefont
  {Schubert}},\ }\href@noop {} {\bibfield  {journal} {\bibinfo  {journal} {Mol.
  Phys.}\ }\textbf {\bibinfo {volume} {104}},\ \bibinfo {pages} {1069}
  (\bibinfo {year} {2006})}\BibitemShut {NoStop}%
\bibitem [{\citenamefont {Bhattacharyya}\ \emph {et~al.}(2014)\citenamefont
  {Bhattacharyya}, \citenamefont {Opalka}, \citenamefont {Poluyanov},\ and\
  \citenamefont {Domcke}}]{bhatt_jpca118}%
  \BibitemOpen
  \bibfield  {author} {\bibinfo {author} {\bibfnamefont {S.}~\bibnamefont
  {Bhattacharyya}}, \bibinfo {author} {\bibfnamefont {D.}~\bibnamefont
  {Opalka}}, \bibinfo {author} {\bibfnamefont {L.~V.}\ \bibnamefont
  {Poluyanov}}, \ and\ \bibinfo {author} {\bibfnamefont {W.}~\bibnamefont
  {Domcke}},\ }\href@noop {} {\bibfield  {journal} {\bibinfo  {journal} {J.
  Phys. Chem. A}\ }\textbf {\bibinfo {volume} {118}},\ \bibinfo {pages} {11962}
  (\bibinfo {year} {2014})}\BibitemShut {NoStop}%
\bibitem [{\citenamefont {Hickman}, \citenamefont {Lang},\ and\ \citenamefont
  {Zeng}(2018)}]{hickman_pccp20}%
  \BibitemOpen
  \bibfield  {author} {\bibinfo {author} {\bibfnamefont {R.~J.}\ \bibnamefont
  {Hickman}}, \bibinfo {author} {\bibfnamefont {R.~A.}\ \bibnamefont {Lang}}, \
  and\ \bibinfo {author} {\bibfnamefont {T.}~\bibnamefont {Zeng}},\ }\href@noop
  {} {\bibfield  {journal} {\bibinfo  {journal} {Phys. Chem. Chem. Phys.}\
  }\textbf {\bibinfo {volume} {20}},\ \bibinfo {pages} {12312} (\bibinfo {year}
  {2018})}\BibitemShut {NoStop}%
\bibitem [{\citenamefont {Seidu}\ \emph {et~al.}(2019)\citenamefont {Seidu},
  \citenamefont {Goel}, \citenamefont {Wang}, \citenamefont {Chen},
  \citenamefont {Wang},\ and\ \citenamefont {Zeng}}]{issaka_pccp21}%
  \BibitemOpen
  \bibfield  {author} {\bibinfo {author} {\bibfnamefont {I.}~\bibnamefont
  {Seidu}}, \bibinfo {author} {\bibfnamefont {P.}~\bibnamefont {Goel}},
  \bibinfo {author} {\bibfnamefont {X.~G.}\ \bibnamefont {Wang}}, \bibinfo
  {author} {\bibfnamefont {B.}~\bibnamefont {Chen}}, \bibinfo {author}
  {\bibfnamefont {X.~B.}\ \bibnamefont {Wang}}, \ and\ \bibinfo {author}
  {\bibfnamefont {T.}~\bibnamefont {Zeng}},\ }\href@noop {} {\bibfield
  {journal} {\bibinfo  {journal} {Phys. Chem. Chem. Phys.}\ }\textbf {\bibinfo
  {volume} {21}},\ \bibinfo {pages} {8679} (\bibinfo {year}
  {2019})}\BibitemShut {NoStop}%
\bibitem [{\citenamefont {Gossel}, \citenamefont {Lacombe},\ and\ \citenamefont
  {Maitra}(2019)}]{maitra_jcp150}%
  \BibitemOpen
  \bibfield  {author} {\bibinfo {author} {\bibfnamefont {G.~H.}\ \bibnamefont
  {Gossel}}, \bibinfo {author} {\bibfnamefont {L.}~\bibnamefont {Lacombe}}, \
  and\ \bibinfo {author} {\bibfnamefont {N.~T.}\ \bibnamefont {Maitra}},\
  }\href@noop {} {\bibfield  {journal} {\bibinfo  {journal} {J. Chem. Phys.}\
  }\textbf {\bibinfo {volume} {150}},\ \bibinfo {pages} {154112} (\bibinfo
  {year} {2019})}\BibitemShut {NoStop}%
\bibitem [{\citenamefont {Curchod}\ and\ \citenamefont
  {Agostini}(2017)}]{agostini_jpcl8}%
  \BibitemOpen
  \bibfield  {author} {\bibinfo {author} {\bibfnamefont {B.~F.~E.}\
  \bibnamefont {Curchod}}\ and\ \bibinfo {author} {\bibfnamefont
  {F.}~\bibnamefont {Agostini}},\ }\href@noop {} {\bibfield  {journal}
  {\bibinfo  {journal} {J. Phys. Chem. Lett.}\ }\textbf {\bibinfo {volume}
  {8}},\ \bibinfo {pages} {831} (\bibinfo {year} {2017})}\BibitemShut {NoStop}%
\bibitem [{\citenamefont {Agostini}\ and\ \citenamefont
  {Curchod}(2018)}]{agostini_epjb91}%
  \BibitemOpen
  \bibfield  {author} {\bibinfo {author} {\bibfnamefont {F.}~\bibnamefont
  {Agostini}}\ and\ \bibinfo {author} {\bibfnamefont {B.~F.~E.}\ \bibnamefont
  {Curchod}},\ }\href@noop {} {\bibfield  {journal} {\bibinfo  {journal} {Eur.
  Phys. J. B}\ }\textbf {\bibinfo {volume} {91}},\ \bibinfo {pages} {141}
  (\bibinfo {year} {2018})}\BibitemShut {NoStop}%
\bibitem [{\citenamefont {Min}\ \emph {et~al.}(2017)\citenamefont {Min},
  \citenamefont {Agostini}, \citenamefont {Tavernelli},\ and\ \citenamefont
  {Gross}}]{gross_jpcl8}%
  \BibitemOpen
  \bibfield  {author} {\bibinfo {author} {\bibfnamefont {S.~K.}\ \bibnamefont
  {Min}}, \bibinfo {author} {\bibfnamefont {F.}~\bibnamefont {Agostini}},
  \bibinfo {author} {\bibfnamefont {I.}~\bibnamefont {Tavernelli}}, \ and\
  \bibinfo {author} {\bibfnamefont {E.~K.~U.}\ \bibnamefont {Gross}},\
  }\href@noop {} {\bibfield  {journal} {\bibinfo  {journal} {J. Phys. Chem.
  Lett.}\ }\textbf {\bibinfo {volume} {8}},\ \bibinfo {pages} {3048} (\bibinfo
  {year} {2017})}\BibitemShut {NoStop}%
\bibitem [{\citenamefont {Richings}\ and\ \citenamefont
  {Worth}(2017)}]{worth_cpl683}%
  \BibitemOpen
  \bibfield  {author} {\bibinfo {author} {\bibfnamefont {G.~W.}\ \bibnamefont
  {Richings}}\ and\ \bibinfo {author} {\bibfnamefont {G.~A.}\ \bibnamefont
  {Worth}},\ }\href@noop {} {\bibfield  {journal} {\bibinfo  {journal} {Chem.
  Phys. Lett.}\ }\textbf {\bibinfo {volume} {683}},\ \bibinfo {pages} {606}
  (\bibinfo {year} {2017})}\BibitemShut {NoStop}%
\bibitem [{\citenamefont {Born}\ and\ \citenamefont
  {Huang}(1954{\natexlab{b}})}]{takatsuka_book}%
  \BibitemOpen
  \bibfield  {author} {\bibinfo {author} {\bibfnamefont {M.}~\bibnamefont
  {Born}}\ and\ \bibinfo {author} {\bibfnamefont {K.}~\bibnamefont {Huang}},\
  }\href@noop {} {\emph {\bibinfo {title} {{\it Dynamical Theory of Crystal
  Lattices}}}}\ (\bibinfo  {publisher} {Oxford University Press},\ \bibinfo
  {address} {Oxford},\ \bibinfo {year} {1954})\BibitemShut {NoStop}%
\bibitem [{\citenamefont {Yonehara}, \citenamefont {Hanasaki},\ and\
  \citenamefont {Takatsuka}(2012)}]{takatsuka_rev}%
  \BibitemOpen
  \bibfield  {author} {\bibinfo {author} {\bibfnamefont {T.}~\bibnamefont
  {Yonehara}}, \bibinfo {author} {\bibfnamefont {K.}~\bibnamefont {Hanasaki}},
  \ and\ \bibinfo {author} {\bibfnamefont {K.}~\bibnamefont {Takatsuka}},\
  }\href@noop {} {\bibfield  {journal} {\bibinfo  {journal} {Chem. Rev.}\
  }\textbf {\bibinfo {volume} {112}},\ \bibinfo {pages} {499} (\bibinfo {year}
  {2012})}\BibitemShut {NoStop}%
\bibitem [{\citenamefont {Takatsuka}(2024)}]{takatsuka_jcp160}%
  \BibitemOpen
  \bibfield  {author} {\bibinfo {author} {\bibfnamefont {K.}~\bibnamefont
  {Takatsuka}},\ }\href@noop {} {\bibfield  {journal} {\bibinfo  {journal} {J.
  Chem. Phys.}\ }\textbf {\bibinfo {volume} {160}},\ \bibinfo {pages} {044112}
  (\bibinfo {year} {2024})}\BibitemShut {NoStop}%
\bibitem [{\citenamefont {\"Opik}\ and\ \citenamefont {Pryce}(1957)}]{opik}%
  \BibitemOpen
  \bibfield  {author} {\bibinfo {author} {\bibfnamefont {U.}~\bibnamefont
  {\"Opik}}\ and\ \bibinfo {author} {\bibfnamefont {M.~H.~L.}\ \bibnamefont
  {Pryce}},\ }\href@noop {} {\bibfield  {journal} {\bibinfo  {journal} {Proc.
  R. Soc. London, Ser. A}\ }\textbf {\bibinfo {volume} {238}},\ \bibinfo
  {pages} {425} (\bibinfo {year} {1957})}\BibitemShut {NoStop}%
\bibitem [{\citenamefont {Mahapatra}, \citenamefont {K\"oppel},\ and\
  \citenamefont {Cederbaum}(1999)}]{mahapatra_jcp110}%
  \BibitemOpen
  \bibfield  {author} {\bibinfo {author} {\bibfnamefont {S.}~\bibnamefont
  {Mahapatra}}, \bibinfo {author} {\bibfnamefont {H.}~\bibnamefont {K\"oppel}},
  \ and\ \bibinfo {author} {\bibfnamefont {L.}~\bibnamefont {Cederbaum}},\
  }\href@noop {} {\bibfield  {journal} {\bibinfo  {journal} {J. Chem. Phys.}\
  }\textbf {\bibinfo {volume} {110}},\ \bibinfo {pages} {5691} (\bibinfo {year}
  {1999})}\BibitemShut {NoStop}%
\bibitem [{\citenamefont {D\"oscher}, \citenamefont {K\"oppel},\ and\
  \citenamefont {Szalay}(2002)}]{Koppel_jcp117_1}%
  \BibitemOpen
  \bibfield  {author} {\bibinfo {author} {\bibfnamefont {M.}~\bibnamefont
  {D\"oscher}}, \bibinfo {author} {\bibfnamefont {H.}~\bibnamefont {K\"oppel}},
  \ and\ \bibinfo {author} {\bibfnamefont {P.~G.}\ \bibnamefont {Szalay}},\
  }\href@noop {} {\bibfield  {journal} {\bibinfo  {journal} {J. Chem. Phys}\
  }\textbf {\bibinfo {volume} {117}},\ \bibinfo {pages} {2645} (\bibinfo {year}
  {2002})}\BibitemShut {NoStop}%
\bibitem [{\citenamefont {Mondal}\ and\ \citenamefont
  {Mahapatra}(2009)}]{mahapatra_pccp11}%
  \BibitemOpen
  \bibfield  {author} {\bibinfo {author} {\bibfnamefont {T.}~\bibnamefont
  {Mondal}}\ and\ \bibinfo {author} {\bibfnamefont {S.}~\bibnamefont
  {Mahapatra}},\ }\href@noop {} {\bibfield  {journal} {\bibinfo  {journal}
  {Phys. Chem. Chem. Phys.}\ }\textbf {\bibinfo {volume} {11}},\ \bibinfo
  {pages} {10867} (\bibinfo {year} {2009})}\BibitemShut {NoStop}%
\bibitem [{\citenamefont {Faraji}\ \emph {et~al.}(2008)\citenamefont {Faraji},
  \citenamefont {K\"oppel}, \citenamefont {Eisfeld},\ and\ \citenamefont
  {Mahapatra}}]{mahapatra_cp347}%
  \BibitemOpen
  \bibfield  {author} {\bibinfo {author} {\bibfnamefont {S.}~\bibnamefont
  {Faraji}}, \bibinfo {author} {\bibfnamefont {H.}~\bibnamefont {K\"oppel}},
  \bibinfo {author} {\bibfnamefont {W.}~\bibnamefont {Eisfeld}}, \ and\
  \bibinfo {author} {\bibfnamefont {S.}~\bibnamefont {Mahapatra}},\ }\href@noop
  {} {\bibfield  {journal} {\bibinfo  {journal} {Chem. Phys.}\ }\textbf
  {\bibinfo {volume} {347}},\ \bibinfo {pages} {110} (\bibinfo {year}
  {2008})}\BibitemShut {NoStop}%
\bibitem [{\citenamefont {Hering}\ \emph {et~al.}(1980)\citenamefont {Hering},
  \citenamefont {Brooks}, \citenamefont {Curl}, \citenamefont {Judson},\ and\
  \citenamefont {Lowe}}]{hering_prl44}%
  \BibitemOpen
  \bibfield  {author} {\bibinfo {author} {\bibfnamefont {P.}~\bibnamefont
  {Hering}}, \bibinfo {author} {\bibfnamefont {P.~R.}\ \bibnamefont {Brooks}},
  \bibinfo {author} {\bibfnamefont {R.~F.}\ \bibnamefont {Curl}}, \bibinfo
  {author} {\bibfnamefont {R.~S.}\ \bibnamefont {Judson}}, \ and\ \bibinfo
  {author} {\bibfnamefont {R.~S.}\ \bibnamefont {Lowe}},\ }\href {\doibase
  10.1103/PhysRevLett.44.687} {\bibfield  {journal} {\bibinfo  {journal} {Phys.
  Rev. Lett.}\ }\textbf {\bibinfo {volume} {44}},\ \bibinfo {pages} {687}
  (\bibinfo {year} {1980})}\BibitemShut {NoStop}%
\bibitem [{\citenamefont {Zimmerman}, \citenamefont {Yuan},\ and\ \citenamefont
  {George}(1977)}]{zimmerman_jcp66}%
  \BibitemOpen
  \bibfield  {author} {\bibinfo {author} {\bibfnamefont {I.~H.}\ \bibnamefont
  {Zimmerman}}, \bibinfo {author} {\bibfnamefont {J.-M.}\ \bibnamefont {Yuan}},
  \ and\ \bibinfo {author} {\bibfnamefont {T.~F.}\ \bibnamefont {George}},\
  }\href@noop {} {\bibfield  {journal} {\bibinfo  {journal} {The Journal of
  Chemical Physics}\ }\textbf {\bibinfo {volume} {66}},\ \bibinfo {pages}
  {2638} (\bibinfo {year} {1977})}\BibitemShut {NoStop}%
\bibitem [{\citenamefont {George}\ \emph {et~al.}(1979)\citenamefont {George},
  \citenamefont {Zimmerman}, \citenamefont {DeVries}, \citenamefont {Yuan},
  \citenamefont {Lam}, \citenamefont {Bellum}, \citenamefont {Lee},
  \citenamefont {Slutsky},\ and\ \citenamefont {Lin}}]{george_book}%
  \BibitemOpen
  \bibfield  {author} {\bibinfo {author} {\bibfnamefont {T.~F.}\ \bibnamefont
  {George}}, \bibinfo {author} {\bibfnamefont {I.~H.}\ \bibnamefont
  {Zimmerman}}, \bibinfo {author} {\bibfnamefont {P.~L.}\ \bibnamefont
  {DeVries}}, \bibinfo {author} {\bibfnamefont {J.-M.}\ \bibnamefont {Yuan}},
  \bibinfo {author} {\bibfnamefont {K.-S.}\ \bibnamefont {Lam}}, \bibinfo
  {author} {\bibfnamefont {J.~C.}\ \bibnamefont {Bellum}}, \bibinfo {author}
  {\bibfnamefont {H.-W.}\ \bibnamefont {Lee}}, \bibinfo {author} {\bibfnamefont
  {M.~S.}\ \bibnamefont {Slutsky}}, \ and\ \bibinfo {author} {\bibfnamefont
  {J.}~\bibnamefont {Lin}},\ }\href@noop {} {\bibfield  {journal} {\bibinfo
  {journal} {Chemical and Biochemical Applications of Lasers, edited by CB
  Moore (Academic, New York, 1979)}\ }\textbf {\bibinfo {volume} {4}},\
  \bibinfo {pages} {253} (\bibinfo {year} {1979})}\BibitemShut {NoStop}%
\bibitem [{\citenamefont {Light}\ and\ \citenamefont
  {Altenberger-Siczek}(1979)}]{light_jcp70}%
  \BibitemOpen
  \bibfield  {author} {\bibinfo {author} {\bibfnamefont {J.~C.}\ \bibnamefont
  {Light}}\ and\ \bibinfo {author} {\bibfnamefont {A.}~\bibnamefont
  {Altenberger-Siczek}},\ }\href@noop {} {\bibfield  {journal} {\bibinfo
  {journal} {The Journal of Chemical Physics}\ }\textbf {\bibinfo {volume}
  {70}},\ \bibinfo {pages} {4108} (\bibinfo {year} {1979})}\BibitemShut
  {NoStop}%
\bibitem [{\citenamefont {Chu}(1981)}]{chu_jcp75}%
  \BibitemOpen
  \bibfield  {author} {\bibinfo {author} {\bibfnamefont {S.-I.}\ \bibnamefont
  {Chu}},\ }\href@noop {} {\bibfield  {journal} {\bibinfo  {journal} {The
  Journal of Chemical Physics}\ }\textbf {\bibinfo {volume} {75}},\ \bibinfo
  {pages} {2215} (\bibinfo {year} {1981})}\BibitemShut {NoStop}%
\bibitem [{\citenamefont {Zimmerman}, \citenamefont {Baer},\ and\ \citenamefont
  {George}(1983)}]{zimmerman_jcp87}%
  \BibitemOpen
  \bibfield  {author} {\bibinfo {author} {\bibfnamefont {I.~H.}\ \bibnamefont
  {Zimmerman}}, \bibinfo {author} {\bibfnamefont {M.}~\bibnamefont {Baer}}, \
  and\ \bibinfo {author} {\bibfnamefont {T.~F.}\ \bibnamefont {George}},\
  }\href@noop {} {\bibfield  {journal} {\bibinfo  {journal} {The Journal of
  Physical Chemistry}\ }\textbf {\bibinfo {volume} {87}},\ \bibinfo {pages}
  {1478} (\bibinfo {year} {1983})}\BibitemShut {NoStop}%
\bibitem [{\citenamefont {Last}\ \emph {et~al.}(1983)\citenamefont {Last},
  \citenamefont {Baer}, \citenamefont {Zimmerman},\ and\ \citenamefont
  {George}}]{last_cpl101}%
  \BibitemOpen
  \bibfield  {author} {\bibinfo {author} {\bibfnamefont {I.}~\bibnamefont
  {Last}}, \bibinfo {author} {\bibfnamefont {M.}~\bibnamefont {Baer}}, \bibinfo
  {author} {\bibfnamefont {I.~H.}\ \bibnamefont {Zimmerman}}, \ and\ \bibinfo
  {author} {\bibfnamefont {T.~F.}\ \bibnamefont {George}},\ }\href@noop {}
  {\bibfield  {journal} {\bibinfo  {journal} {Chemical physics letters}\
  }\textbf {\bibinfo {volume} {101}},\ \bibinfo {pages} {163} (\bibinfo {year}
  {1983})}\BibitemShut {NoStop}%
\bibitem [{\citenamefont {Last}\ and\ \citenamefont {Baer}(1985)}]{last_jcp82}%
  \BibitemOpen
  \bibfield  {author} {\bibinfo {author} {\bibfnamefont {I.}~\bibnamefont
  {Last}}\ and\ \bibinfo {author} {\bibfnamefont {M.}~\bibnamefont {Baer}},\
  }\href@noop {} {\bibfield  {journal} {\bibinfo  {journal} {The Journal of
  chemical physics}\ }\textbf {\bibinfo {volume} {82}},\ \bibinfo {pages}
  {4954} (\bibinfo {year} {1985})}\BibitemShut {NoStop}%
\bibitem [{\citenamefont {Manz}\ \emph {et~al.}(2000)\citenamefont {Manz},
  \citenamefont {Naundorf}, \citenamefont {Yamashita},\ and\ \citenamefont
  {Zhao}}]{manz_jcp113}%
  \BibitemOpen
  \bibfield  {author} {\bibinfo {author} {\bibfnamefont {J.}~\bibnamefont
  {Manz}}, \bibinfo {author} {\bibfnamefont {H.}~\bibnamefont {Naundorf}},
  \bibinfo {author} {\bibfnamefont {K.}~\bibnamefont {Yamashita}}, \ and\
  \bibinfo {author} {\bibfnamefont {Y.}~\bibnamefont {Zhao}},\ }\href@noop {}
  {\bibfield  {journal} {\bibinfo  {journal} {The Journal of Chemical Physics}\
  }\textbf {\bibinfo {volume} {113}},\ \bibinfo {pages} {8969} (\bibinfo {year}
  {2000})}\BibitemShut {NoStop}%
\bibitem [{\citenamefont {Barth}\ and\ \citenamefont
  {Manz}(2007)}]{barth_pra75}%
  \BibitemOpen
  \bibfield  {author} {\bibinfo {author} {\bibfnamefont {I.}~\bibnamefont
  {Barth}}\ and\ \bibinfo {author} {\bibfnamefont {J.}~\bibnamefont {Manz}},\
  }\href@noop {} {\bibfield  {journal} {\bibinfo  {journal} {Physical Review
  A—Atomic, Molecular, and Optical Physics}\ }\textbf {\bibinfo {volume}
  {75}},\ \bibinfo {pages} {012510} (\bibinfo {year} {2007})}\BibitemShut
  {NoStop}%
\bibitem [{\citenamefont {Yonehara}\ and\ \citenamefont
  {Takatsuka}(2008)}]{yonehara_jcp128}%
  \BibitemOpen
  \bibfield  {author} {\bibinfo {author} {\bibfnamefont {T.}~\bibnamefont
  {Yonehara}}\ and\ \bibinfo {author} {\bibfnamefont {K.}~\bibnamefont
  {Takatsuka}},\ }\href@noop {} {\bibfield  {journal} {\bibinfo  {journal} {The
  Journal of chemical physics}\ }\textbf {\bibinfo {volume} {128}} (\bibinfo
  {year} {2008})}\BibitemShut {NoStop}%
\bibitem [{\citenamefont {Sarkar}, \citenamefont {Adhikari},\ and\
  \citenamefont {Baer}(2007{\natexlab{a}})}]{sarkar_jcp126}%
  \BibitemOpen
  \bibfield  {author} {\bibinfo {author} {\bibfnamefont {B.}~\bibnamefont
  {Sarkar}}, \bibinfo {author} {\bibfnamefont {S.}~\bibnamefont {Adhikari}}, \
  and\ \bibinfo {author} {\bibfnamefont {M.}~\bibnamefont {Baer}},\ }\href@noop
  {} {\bibfield  {journal} {\bibinfo  {journal} {The Journal of chemical
  physics}\ }\textbf {\bibinfo {volume} {126}} (\bibinfo {year}
  {2007}{\natexlab{a}})}\BibitemShut {NoStop}%
\bibitem [{\citenamefont {Sarkar}, \citenamefont {Adhikari},\ and\
  \citenamefont {Baer}(2007{\natexlab{b}})}]{sarkar_jcp127_1}%
  \BibitemOpen
  \bibfield  {author} {\bibinfo {author} {\bibfnamefont {B.}~\bibnamefont
  {Sarkar}}, \bibinfo {author} {\bibfnamefont {S.}~\bibnamefont {Adhikari}}, \
  and\ \bibinfo {author} {\bibfnamefont {M.}~\bibnamefont {Baer}},\ }\href@noop
  {} {\bibfield  {journal} {\bibinfo  {journal} {The Journal of chemical
  physics}\ }\textbf {\bibinfo {volume} {127}} (\bibinfo {year}
  {2007}{\natexlab{b}})}\BibitemShut {NoStop}%
\bibitem [{\citenamefont {Sarkar}, \citenamefont {Adhikari},\ and\
  \citenamefont {Baer}(2007{\natexlab{c}})}]{sarkar_jcp127_2}%
  \BibitemOpen
  \bibfield  {author} {\bibinfo {author} {\bibfnamefont {B.}~\bibnamefont
  {Sarkar}}, \bibinfo {author} {\bibfnamefont {S.}~\bibnamefont {Adhikari}}, \
  and\ \bibinfo {author} {\bibfnamefont {M.}~\bibnamefont {Baer}},\ }\href@noop
  {} {\bibfield  {journal} {\bibinfo  {journal} {The Journal of chemical
  physics}\ }\textbf {\bibinfo {volume} {127}} (\bibinfo {year}
  {2007}{\natexlab{c}})}\BibitemShut {NoStop}%
\bibitem [{\citenamefont {Paul}, \citenamefont {Adhikari},\ and\ \citenamefont
  {Baer}(2010{\natexlab{b}})}]{paul_jcp132}%
  \BibitemOpen
  \bibfield  {author} {\bibinfo {author} {\bibfnamefont {A.~K.}\ \bibnamefont
  {Paul}}, \bibinfo {author} {\bibfnamefont {S.}~\bibnamefont {Adhikari}}, \
  and\ \bibinfo {author} {\bibfnamefont {M.}~\bibnamefont {Baer}},\ }\href@noop
  {} {\bibfield  {journal} {\bibinfo  {journal} {The Journal of chemical
  physics}\ }\textbf {\bibinfo {volume} {132}} (\bibinfo {year}
  {2010}{\natexlab{b}})}\BibitemShut {NoStop}%
\bibitem [{\citenamefont {Paul}\ \emph
  {et~al.}(2009{\natexlab{b}})\citenamefont {Paul}, \citenamefont {Adhikari},
  \citenamefont {Mukhopadhyay}, \citenamefont {Hal{\'a}sz}, \citenamefont
  {Vib{\'o}k}, \citenamefont {Baer},\ and\ \citenamefont
  {Baer}}]{paul_jpca113}%
  \BibitemOpen
  \bibfield  {author} {\bibinfo {author} {\bibfnamefont {A.~K.}\ \bibnamefont
  {Paul}}, \bibinfo {author} {\bibfnamefont {S.}~\bibnamefont {Adhikari}},
  \bibinfo {author} {\bibfnamefont {D.}~\bibnamefont {Mukhopadhyay}}, \bibinfo
  {author} {\bibfnamefont {G.~J.}\ \bibnamefont {Hal{\'a}sz}}, \bibinfo
  {author} {\bibfnamefont {A.}~\bibnamefont {Vib{\'o}k}}, \bibinfo {author}
  {\bibfnamefont {R.}~\bibnamefont {Baer}}, \ and\ \bibinfo {author}
  {\bibfnamefont {M.}~\bibnamefont {Baer}},\ }\href@noop {} {\bibfield
  {journal} {\bibinfo  {journal} {The Journal of Physical Chemistry A}\
  }\textbf {\bibinfo {volume} {113}},\ \bibinfo {pages} {7331} (\bibinfo {year}
  {2009}{\natexlab{b}})}\BibitemShut {NoStop}%
\bibitem [{\citenamefont {Paul}\ \emph {et~al.}(2010)\citenamefont {Paul},
  \citenamefont {Adhikari}, \citenamefont {Baer},\ and\ \citenamefont
  {Baer}}]{paul_pra81}%
  \BibitemOpen
  \bibfield  {author} {\bibinfo {author} {\bibfnamefont {A.~K.}\ \bibnamefont
  {Paul}}, \bibinfo {author} {\bibfnamefont {S.}~\bibnamefont {Adhikari}},
  \bibinfo {author} {\bibfnamefont {M.}~\bibnamefont {Baer}}, \ and\ \bibinfo
  {author} {\bibfnamefont {R.}~\bibnamefont {Baer}},\ }\href@noop {} {\bibfield
   {journal} {\bibinfo  {journal} {Physical Review A}\ }\textbf {\bibinfo
  {volume} {81}},\ \bibinfo {pages} {013412} (\bibinfo {year}
  {2010})}\BibitemShut {NoStop}%
\bibitem [{\citenamefont {Gavrila}(1991)}]{gavrila_book}%
  \BibitemOpen
  \bibfield  {author} {\bibinfo {author} {\bibfnamefont {M.}~\bibnamefont
  {Gavrila}},\ }\href@noop {} {\  (\bibinfo {year} {1991})}\BibitemShut
  {NoStop}%
\bibitem [{\citenamefont {Corkum}\ \emph {et~al.}(1999)\citenamefont {Corkum},
  \citenamefont {Ellert}, \citenamefont {Mehendale}, \citenamefont {Dietrich},
  \citenamefont {Hankin}, \citenamefont {Aseyev}, \citenamefont {Rayner},\ and\
  \citenamefont {Villeneuve}}]{corkum_fd113}%
  \BibitemOpen
  \bibfield  {author} {\bibinfo {author} {\bibfnamefont {P.}~\bibnamefont
  {Corkum}}, \bibinfo {author} {\bibfnamefont {C.}~\bibnamefont {Ellert}},
  \bibinfo {author} {\bibfnamefont {M.}~\bibnamefont {Mehendale}}, \bibinfo
  {author} {\bibfnamefont {P.}~\bibnamefont {Dietrich}}, \bibinfo {author}
  {\bibfnamefont {S.}~\bibnamefont {Hankin}}, \bibinfo {author} {\bibfnamefont
  {S.}~\bibnamefont {Aseyev}}, \bibinfo {author} {\bibfnamefont
  {D.}~\bibnamefont {Rayner}}, \ and\ \bibinfo {author} {\bibfnamefont
  {D.}~\bibnamefont {Villeneuve}},\ }\href@noop {} {\bibfield  {journal}
  {\bibinfo  {journal} {Faraday Discussions}\ }\textbf {\bibinfo {volume}
  {113}},\ \bibinfo {pages} {47} (\bibinfo {year} {1999})}\BibitemShut
  {NoStop}%
\bibitem [{\citenamefont {Cohen-Tannoudji}(1998)}]{cohen_rmp70}%
  \BibitemOpen
  \bibfield  {author} {\bibinfo {author} {\bibfnamefont {C.~N.}\ \bibnamefont
  {Cohen-Tannoudji}},\ }\href@noop {} {\bibfield  {journal} {\bibinfo
  {journal} {Reviews of Modern Physics}\ }\textbf {\bibinfo {volume} {70}},\
  \bibinfo {pages} {707} (\bibinfo {year} {1998})}\BibitemShut {NoStop}%
\bibitem [{\citenamefont {Baer}\ \emph {et~al.}(2003)\citenamefont {Baer},
  \citenamefont {Kouri}, \citenamefont {Baer},\ and\ \citenamefont
  {Hoffman}}]{baer_jcp119}%
  \BibitemOpen
  \bibfield  {author} {\bibinfo {author} {\bibfnamefont {R.}~\bibnamefont
  {Baer}}, \bibinfo {author} {\bibfnamefont {D.~J.}\ \bibnamefont {Kouri}},
  \bibinfo {author} {\bibfnamefont {M.}~\bibnamefont {Baer}}, \ and\ \bibinfo
  {author} {\bibfnamefont {D.~K.}\ \bibnamefont {Hoffman}},\ }\href@noop {}
  {\bibfield  {journal} {\bibinfo  {journal} {The Journal of chemical physics}\
  }\textbf {\bibinfo {volume} {119}},\ \bibinfo {pages} {6998} (\bibinfo {year}
  {2003})}\BibitemShut {NoStop}%
\bibitem [{\citenamefont {Englman}, \citenamefont {Yahalom},\ and\
  \citenamefont {Baer}(2002)}]{englman_ijqc90}%
  \BibitemOpen
  \bibfield  {author} {\bibinfo {author} {\bibfnamefont {R.}~\bibnamefont
  {Englman}}, \bibinfo {author} {\bibfnamefont {A.}~\bibnamefont {Yahalom}}, \
  and\ \bibinfo {author} {\bibfnamefont {M.}~\bibnamefont {Baer}},\ }\href@noop
  {} {\bibfield  {journal} {\bibinfo  {journal} {International journal of
  quantum chemistry}\ }\textbf {\bibinfo {volume} {90}},\ \bibinfo {pages}
  {266} (\bibinfo {year} {2002})}\BibitemShut {NoStop}%
\bibitem [{\citenamefont {Mukherjee}, \citenamefont {Ghosh},\ and\
  \citenamefont {Adhikari}(2018)}]{bijit_jpcs1148}%
  \BibitemOpen
  \bibfield  {author} {\bibinfo {author} {\bibfnamefont {B.}~\bibnamefont
  {Mukherjee}}, \bibinfo {author} {\bibfnamefont {S.}~\bibnamefont {Ghosh}}, \
  and\ \bibinfo {author} {\bibfnamefont {S.}~\bibnamefont {Adhikari}},\
  }\href@noop {} {\bibfield  {journal} {\bibinfo  {journal} {J. Phys.: Conf.
  Ser.}\ }\textbf {\bibinfo {volume} {1148}},\ \bibinfo {pages} {012001}
  (\bibinfo {year} {2018})}\BibitemShut {NoStop}%
\bibitem [{\citenamefont {Garc\'{\i}a-Fern\'andez}, \citenamefont {Bersuker},\
  and\ \citenamefont {Boggs}(2006)}]{Pablo_prl96}%
  \BibitemOpen
  \bibfield  {author} {\bibinfo {author} {\bibfnamefont {P.}~\bibnamefont
  {Garc\'{\i}a-Fern\'andez}}, \bibinfo {author} {\bibfnamefont {I.~B.}\
  \bibnamefont {Bersuker}}, \ and\ \bibinfo {author} {\bibfnamefont {J.~E.}\
  \bibnamefont {Boggs}},\ }\href@noop {} {\bibfield  {journal} {\bibinfo
  {journal} {Phys. Rev. Lett.}\ }\textbf {\bibinfo {volume} {96}},\ \bibinfo
  {pages} {163005} (\bibinfo {year} {2006})}\BibitemShut {NoStop}%
\bibitem [{\citenamefont {Hizhnyakov}, \citenamefont {Pae},\ and\ \citenamefont
  {Vaikj\"arv}(2012{\natexlab{a}})}]{Hizhnyakov_cpl525}%
  \BibitemOpen
  \bibfield  {author} {\bibinfo {author} {\bibfnamefont {V.}~\bibnamefont
  {Hizhnyakov}}, \bibinfo {author} {\bibfnamefont {K.}~\bibnamefont {Pae}}, \
  and\ \bibinfo {author} {\bibfnamefont {T.}~\bibnamefont {Vaikj\"arv}},\
  }\href@noop {} {\bibfield  {journal} {\bibinfo  {journal} {Chem. Phys.
  Lett.}\ }\textbf {\bibinfo {volume} {525-526}},\ \bibinfo {pages} {64}
  (\bibinfo {year} {2012}{\natexlab{a}})}\BibitemShut {NoStop}%
\bibitem [{\citenamefont {Hazra}\ \emph {et~al.}(2025)\citenamefont {Hazra},
  \citenamefont {Basu}, \citenamefont {Mukherjee}, \citenamefont {Ravi},
  \citenamefont {Sardar},\ and\ \citenamefont {Adhikari}}]{saikat_cpc26}%
  \BibitemOpen
  \bibfield  {author} {\bibinfo {author} {\bibfnamefont {S.}~\bibnamefont
  {Hazra}}, \bibinfo {author} {\bibfnamefont {S.}~\bibnamefont {Basu}},
  \bibinfo {author} {\bibfnamefont {S.}~\bibnamefont {Mukherjee}}, \bibinfo
  {author} {\bibfnamefont {S.}~\bibnamefont {Ravi}}, \bibinfo {author}
  {\bibfnamefont {S.}~\bibnamefont {Sardar}}, \ and\ \bibinfo {author}
  {\bibfnamefont {S.}~\bibnamefont {Adhikari}},\ }\href@noop {} {\bibfield
  {journal} {\bibinfo  {journal} {ChemPhysChem}\ }\textbf {\bibinfo {volume}
  {26}},\ \bibinfo {pages} {e202500461} (\bibinfo {year} {2025})}\BibitemShut
  {NoStop}%
\bibitem [{\citenamefont {Sardar}, \citenamefont {Paul},\ and\ \citenamefont
  {Adhikari}(2009)}]{SSMP107}%
  \BibitemOpen
  \bibfield  {author} {\bibinfo {author} {\bibfnamefont {S.}~\bibnamefont
  {Sardar}}, \bibinfo {author} {\bibfnamefont {A.~K.}\ \bibnamefont {Paul}}, \
  and\ \bibinfo {author} {\bibfnamefont {S.}~\bibnamefont {Adhikari}},\
  }\href@noop {} {\bibfield  {journal} {\bibinfo  {journal} {Mol. Phys.}\
  }\textbf {\bibinfo {volume} {107}},\ \bibinfo {pages} {2467} (\bibinfo {year}
  {2009})}\BibitemShut {NoStop}%
\bibitem [{\citenamefont {Khan}\ \emph
  {et~al.}(2014{\natexlab{a}})\citenamefont {Khan}, \citenamefont {Sardar},
  \citenamefont {Sarkar},\ and\ \citenamefont {Adhikari}}]{SSJPCA118}%
  \BibitemOpen
  \bibfield  {author} {\bibinfo {author} {\bibfnamefont {B.~A.}\ \bibnamefont
  {Khan}}, \bibinfo {author} {\bibfnamefont {S.}~\bibnamefont {Sardar}},
  \bibinfo {author} {\bibfnamefont {P.}~\bibnamefont {Sarkar}}, \ and\ \bibinfo
  {author} {\bibfnamefont {S.}~\bibnamefont {Adhikari}},\ }\href@noop {}
  {\bibfield  {journal} {\bibinfo  {journal} {J. Phys. Chem. A}\ }\textbf
  {\bibinfo {volume} {118 (49)}},\ \bibinfo {pages} {11451} (\bibinfo {year}
  {2014}{\natexlab{a}})}\BibitemShut {NoStop}%
\bibitem [{\citenamefont {Sardar}, \citenamefont {Puzari},\ and\ \citenamefont
  {Adhikari}(2011{\natexlab{a}})}]{SSPCCP13}%
  \BibitemOpen
  \bibfield  {author} {\bibinfo {author} {\bibfnamefont {S.}~\bibnamefont
  {Sardar}}, \bibinfo {author} {\bibfnamefont {P.}~\bibnamefont {Puzari}}, \
  and\ \bibinfo {author} {\bibfnamefont {S.}~\bibnamefont {Adhikari}},\
  }\href@noop {} {\bibfield  {journal} {\bibinfo  {journal} {Phys. Chem. Chem.
  Phys.}\ }\textbf {\bibinfo {volume} {13}},\ \bibinfo {pages} {15960}
  (\bibinfo {year} {2011}{\natexlab{a}})}\BibitemShut {NoStop}%
\bibitem [{\citenamefont {Khan}\ \emph {et~al.}(2013)\citenamefont {Khan},
  \citenamefont {Sardar}, \citenamefont {Sarkar},\ and\ \citenamefont
  {Adhikari}}]{SSJTCC12}%
  \BibitemOpen
  \bibfield  {author} {\bibinfo {author} {\bibfnamefont {B.~A.}\ \bibnamefont
  {Khan}}, \bibinfo {author} {\bibfnamefont {S.}~\bibnamefont {Sardar}},
  \bibinfo {author} {\bibfnamefont {P.}~\bibnamefont {Sarkar}}, \ and\ \bibinfo
  {author} {\bibfnamefont {S.}~\bibnamefont {Adhikari}},\ }\href@noop {}
  {\bibfield  {journal} {\bibinfo  {journal} {J. Theo. Com. Chem.}\ }\textbf
  {\bibinfo {volume} {12}},\ \bibinfo {pages} {1350042} (\bibinfo {year}
  {2013})}\BibitemShut {NoStop}%
\bibitem [{\citenamefont {Sardar}\ \emph {et~al.}(2008)\citenamefont {Sardar},
  \citenamefont {Paul}, \citenamefont {Mondal}, \citenamefont {Sarkar},\ and\
  \citenamefont {Adhikari}}]{SSPCCP10}%
  \BibitemOpen
  \bibfield  {author} {\bibinfo {author} {\bibfnamefont {S.}~\bibnamefont
  {Sardar}}, \bibinfo {author} {\bibfnamefont {A.~K.}\ \bibnamefont {Paul}},
  \bibinfo {author} {\bibfnamefont {P.}~\bibnamefont {Mondal}}, \bibinfo
  {author} {\bibfnamefont {B.}~\bibnamefont {Sarkar}}, \ and\ \bibinfo {author}
  {\bibfnamefont {S.}~\bibnamefont {Adhikari}},\ }\href@noop {} {\bibfield
  {journal} {\bibinfo  {journal} {Phys. Chem. Chem. Phys.}\ }\textbf {\bibinfo
  {volume} {10}},\ \bibinfo {pages} {6388} (\bibinfo {year}
  {2008})}\BibitemShut {NoStop}%
\bibitem [{\citenamefont {Sardar}\ \emph {et~al.}(2011)\citenamefont {Sardar},
  \citenamefont {Paul}, \citenamefont {Sharma},\ and\ \citenamefont
  {Adhikari}}]{subhankar_ijqc111}%
  \BibitemOpen
  \bibfield  {author} {\bibinfo {author} {\bibfnamefont {S.}~\bibnamefont
  {Sardar}}, \bibinfo {author} {\bibfnamefont {A.~K.}\ \bibnamefont {Paul}},
  \bibinfo {author} {\bibfnamefont {R.}~\bibnamefont {Sharma}}, \ and\ \bibinfo
  {author} {\bibfnamefont {S.}~\bibnamefont {Adhikari}},\ }\href@noop {}
  {\bibfield  {journal} {\bibinfo  {journal} {Int. J. Quantum Chem.}\ }\textbf
  {\bibinfo {volume} {111}},\ \bibinfo {pages} {2741} (\bibinfo {year}
  {2011})}\BibitemShut {NoStop}%
\bibitem [{\citenamefont {Sardar}\ \emph {et~al.}(2009)\citenamefont {Sardar},
  \citenamefont {Paul}, \citenamefont {Sharma},\ and\ \citenamefont
  {Adhikari}}]{subhankar_jcp130}%
  \BibitemOpen
  \bibfield  {author} {\bibinfo {author} {\bibfnamefont {S.}~\bibnamefont
  {Sardar}}, \bibinfo {author} {\bibfnamefont {A.~K.}\ \bibnamefont {Paul}},
  \bibinfo {author} {\bibfnamefont {R.}~\bibnamefont {Sharma}}, \ and\ \bibinfo
  {author} {\bibfnamefont {S.}~\bibnamefont {Adhikari}},\ }\href@noop {}
  {\bibfield  {journal} {\bibinfo  {journal} {J. Chem. Phys.}\ }\textbf
  {\bibinfo {volume} {130}},\ \bibinfo {pages} {144302} (\bibinfo {year}
  {2009})}\BibitemShut {NoStop}%
\bibitem [{\citenamefont {Mandal}\ \emph
  {et~al.}(2019{\natexlab{b}})\citenamefont {Mandal}, \citenamefont {Ghosh},
  \citenamefont {Sardar},\ and\ \citenamefont {Adhikari}}]{souvik_irpc37}%
  \BibitemOpen
  \bibfield  {author} {\bibinfo {author} {\bibfnamefont {S.}~\bibnamefont
  {Mandal}}, \bibinfo {author} {\bibfnamefont {S.}~\bibnamefont {Ghosh}},
  \bibinfo {author} {\bibfnamefont {S.}~\bibnamefont {Sardar}}, \ and\ \bibinfo
  {author} {\bibfnamefont {S.}~\bibnamefont {Adhikari}},\ }\href@noop {}
  {\bibfield  {journal} {\bibinfo  {journal} {Int. Rev. Phys. Chem.}\ }\textbf
  {\bibinfo {volume} {37}},\ \bibinfo {pages} {607} (\bibinfo {year}
  {2019}{\natexlab{b}})}\BibitemShut {NoStop}%
\bibitem [{\citenamefont {Herzberg}\ and\ \citenamefont {Higgins}(1963)}]{GHH}%
  \BibitemOpen
  \bibfield  {author} {\bibinfo {author} {\bibfnamefont {G.}~\bibnamefont
  {Herzberg}}\ and\ \bibinfo {author} {\bibfnamefont {H.~C.~L.}\ \bibnamefont
  {Higgins}},\ }\href@noop {} {\bibfield  {journal} {\bibinfo  {journal}
  {Discuss. Faraday Soc.}\ }\textbf {\bibinfo {volume} {35}},\ \bibinfo {pages}
  {77} (\bibinfo {year} {1963})}\BibitemShut {NoStop}%
\bibitem [{\citenamefont {Pechukas}(1969)}]{PPP}%
  \BibitemOpen
  \bibfield  {author} {\bibinfo {author} {\bibfnamefont {P.}~\bibnamefont
  {Pechukas}},\ }\href@noop {} {\bibfield  {journal} {\bibinfo  {journal}
  {Phys. Rev.}\ }\textbf {\bibinfo {volume} {181}},\ \bibinfo {pages} {174}
  (\bibinfo {year} {1969})}\BibitemShut {NoStop}%
\bibitem [{\citenamefont {Tully}\ and\ \citenamefont
  {Preston}(1971{\natexlab{a}})}]{JC1}%
  \BibitemOpen
  \bibfield  {author} {\bibinfo {author} {\bibfnamefont {J.~C.}\ \bibnamefont
  {Tully}}\ and\ \bibinfo {author} {\bibfnamefont {R.~K.}\ \bibnamefont
  {Preston}},\ }\href@noop {} {\bibfield  {journal} {\bibinfo  {journal} {J.
  Chem. Phys.}\ }\textbf {\bibinfo {volume} {55}},\ \bibinfo {pages} {562}
  (\bibinfo {year} {1971}{\natexlab{a}})}\BibitemShut {NoStop}%
\bibitem [{\citenamefont {Miller}\ and\ \citenamefont {George}(1972)}]{WHM}%
  \BibitemOpen
  \bibfield  {author} {\bibinfo {author} {\bibfnamefont {W.~H.}\ \bibnamefont
  {Miller}}\ and\ \bibinfo {author} {\bibfnamefont {T.~F.}\ \bibnamefont
  {George}},\ }\href@noop {} {\bibfield  {journal} {\bibinfo  {journal} {J.
  Chem. Phys.}\ }\textbf {\bibinfo {volume} {56}},\ \bibinfo {pages} {5637}
  (\bibinfo {year} {1972})}\BibitemShut {NoStop}%
\bibitem [{\citenamefont {Yarkony}(1996{\natexlab{a}})}]{YAR}%
  \BibitemOpen
  \bibfield  {author} {\bibinfo {author} {\bibfnamefont {D.~R.}\ \bibnamefont
  {Yarkony}},\ }\href@noop {} {\bibfield  {journal} {\bibinfo  {journal} {Rev.
  Mod. Phys.}\ }\textbf {\bibinfo {volume} {68}},\ \bibinfo {pages} {985}
  (\bibinfo {year} {1996}{\natexlab{a}})}\BibitemShut {NoStop}%
\bibitem [{\citenamefont {Baer}()}]{MBB}%
  \BibitemOpen
  \bibfield  {author} {\bibinfo {author} {\bibfnamefont {M.}~\bibnamefont
  {Baer}},\ }\href@noop {} {\emph {\bibinfo {title} {Beyond Born-Oppenheimer:
  Electronic Nonadiabatic Coupling Terms and Conical Intersections;
  Wiley-Interscience (2006)}}}\BibitemShut {NoStop}%
\bibitem [{\citenamefont {Sarkar}\ and\ \citenamefont
  {Adhikari}(2006{\natexlab{b}})}]{BS1}%
  \BibitemOpen
  \bibfield  {author} {\bibinfo {author} {\bibfnamefont {B.}~\bibnamefont
  {Sarkar}}\ and\ \bibinfo {author} {\bibfnamefont {S.}~\bibnamefont
  {Adhikari}},\ }\href@noop {} {\bibfield  {journal} {\bibinfo  {journal} {J.
  Chem. Phys.}\ }\textbf {\bibinfo {volume} {124}},\ \bibinfo {pages} {074101}
  (\bibinfo {year} {2006}{\natexlab{b}})}\BibitemShut {NoStop}%
\bibitem [{\citenamefont {Schatz}\ and\ \citenamefont {Ratner}()}]{GSM}%
  \BibitemOpen
  \bibfield  {author} {\bibinfo {author} {\bibfnamefont {G.}~\bibnamefont
  {Schatz}}\ and\ \bibinfo {author} {\bibfnamefont {M.~A.}\ \bibnamefont
  {Ratner}},\ }\href@noop {} {\emph {\bibinfo {title} {Quantum Mechanics in
  Chemistry; Prentice Hall, Englewood Cliffs: NJ (1993)}}}\BibitemShut
  {NoStop}%
\bibitem [{\citenamefont {Adhikari}\ and\ \citenamefont
  {Billing}(1999{\natexlab{b}})}]{SACPL309}%
  \BibitemOpen
  \bibfield  {author} {\bibinfo {author} {\bibfnamefont {S.}~\bibnamefont
  {Adhikari}}\ and\ \bibinfo {author} {\bibfnamefont {G.~D.}\ \bibnamefont
  {Billing}},\ }\href@noop {} {\bibfield  {journal} {\bibinfo  {journal} {Chem.
  Phys. Lett.}\ }\textbf {\bibinfo {volume} {309}},\ \bibinfo {pages} {249}
  (\bibinfo {year} {1999}{\natexlab{b}})}\BibitemShut {NoStop}%
\bibitem [{\citenamefont {Adhikari}\ and\ \citenamefont
  {Billing}(1999{\natexlab{c}})}]{SAJCP11148}%
  \BibitemOpen
  \bibfield  {author} {\bibinfo {author} {\bibfnamefont {S.}~\bibnamefont
  {Adhikari}}\ and\ \bibinfo {author} {\bibfnamefont {G.~D.}\ \bibnamefont
  {Billing}},\ }\href@noop {} {\bibfield  {journal} {\bibinfo  {journal} {J.
  Chem. Phys.}\ }\textbf {\bibinfo {volume} {111}},\ \bibinfo {pages} {48}
  (\bibinfo {year} {1999}{\natexlab{c}})}\BibitemShut {NoStop}%
\bibitem [{\citenamefont {Puzari}, \citenamefont {Deshpande},\ and\
  \citenamefont {Adhikari}(2004)}]{SACP300}%
  \BibitemOpen
  \bibfield  {author} {\bibinfo {author} {\bibfnamefont {P.}~\bibnamefont
  {Puzari}}, \bibinfo {author} {\bibfnamefont {S.~A.}\ \bibnamefont
  {Deshpande}}, \ and\ \bibinfo {author} {\bibfnamefont {S.}~\bibnamefont
  {Adhikari}},\ }\href@noop {} {\bibfield  {journal} {\bibinfo  {journal}
  {Chem. Phys.}\ }\textbf {\bibinfo {volume} {300}},\ \bibinfo {pages} {305}
  (\bibinfo {year} {2004})}\BibitemShut {NoStop}%
\bibitem [{\citenamefont {Adhikari}\ and\ \citenamefont
  {Billing}(2000{\natexlab{b}})}]{SAJCP113}%
  \BibitemOpen
  \bibfield  {author} {\bibinfo {author} {\bibfnamefont {S.}~\bibnamefont
  {Adhikari}}\ and\ \bibinfo {author} {\bibfnamefont {G.~D.}\ \bibnamefont
  {Billing}},\ }\href@noop {} {\bibfield  {journal} {\bibinfo  {journal} {J.
  Chem. Phys.}\ }\textbf {\bibinfo {volume} {113}},\ \bibinfo {pages} {1409}
  (\bibinfo {year} {2000}{\natexlab{b}})}\BibitemShut {NoStop}%
\bibitem [{\citenamefont {Billing}\ and\ \citenamefont
  {Adhikari}(2000)}]{SACPL321}%
  \BibitemOpen
  \bibfield  {author} {\bibinfo {author} {\bibfnamefont {G.~D.}\ \bibnamefont
  {Billing}}\ and\ \bibinfo {author} {\bibfnamefont {S.}~\bibnamefont
  {Adhikari}},\ }\href@noop {} {\bibfield  {journal} {\bibinfo  {journal}
  {Chem. Phys. Lett.}\ }\textbf {\bibinfo {volume} {321}},\ \bibinfo {pages}
  {197} (\bibinfo {year} {2000})}\BibitemShut {NoStop}%
\bibitem [{\citenamefont {Barkakaty}\ and\ \citenamefont
  {Adhikari}(2003)}]{SAJCP118}%
  \BibitemOpen
  \bibfield  {author} {\bibinfo {author} {\bibfnamefont {B.}~\bibnamefont
  {Barkakaty}}\ and\ \bibinfo {author} {\bibfnamefont {S.}~\bibnamefont
  {Adhikari}},\ }\href@noop {} {\bibfield  {journal} {\bibinfo  {journal} {J.
  Chem. Phys.}\ }\textbf {\bibinfo {volume} {118}},\ \bibinfo {pages} {5302}
  (\bibinfo {year} {2003})}\BibitemShut {NoStop}%
\bibitem [{\citenamefont {Puzari}\ \emph {et~al.}(2005)\citenamefont {Puzari},
  \citenamefont {Swathi}, \citenamefont {Sarkar},\ and\ \citenamefont
  {Adhikari}}]{SAJCP123}%
  \BibitemOpen
  \bibfield  {author} {\bibinfo {author} {\bibfnamefont {P.}~\bibnamefont
  {Puzari}}, \bibinfo {author} {\bibfnamefont {R.~S.}\ \bibnamefont {Swathi}},
  \bibinfo {author} {\bibfnamefont {B.}~\bibnamefont {Sarkar}}, \ and\ \bibinfo
  {author} {\bibfnamefont {S.}~\bibnamefont {Adhikari}},\ }\href@noop {}
  {\bibfield  {journal} {\bibinfo  {journal} {J. Chem. Phys.}\ }\textbf
  {\bibinfo {volume} {123}},\ \bibinfo {pages} {134317} (\bibinfo {year}
  {2005})}\BibitemShut {NoStop}%
\bibitem [{\citenamefont {Puzari}, \citenamefont {Sarkar},\ and\ \citenamefont
  {Adhikari}(2006)}]{SAJCP125}%
  \BibitemOpen
  \bibfield  {author} {\bibinfo {author} {\bibfnamefont {P.}~\bibnamefont
  {Puzari}}, \bibinfo {author} {\bibfnamefont {B.}~\bibnamefont {Sarkar}}, \
  and\ \bibinfo {author} {\bibfnamefont {S.}~\bibnamefont {Adhikari}},\
  }\href@noop {} {\bibfield  {journal} {\bibinfo  {journal} {J. Chem. Phys.}\
  }\textbf {\bibinfo {volume} {125}},\ \bibinfo {pages} {194316} (\bibinfo
  {year} {2006})}\BibitemShut {NoStop}%
\bibitem [{\citenamefont {Sardar}\ and\ \citenamefont
  {Adhikari}(2012{\natexlab{a}})}]{subhankar_jcs124}%
  \BibitemOpen
  \bibfield  {author} {\bibinfo {author} {\bibfnamefont {S.}~\bibnamefont
  {Sardar}}\ and\ \bibinfo {author} {\bibfnamefont {S.}~\bibnamefont
  {Adhikari}},\ }\href@noop {} {\bibfield  {journal} {\bibinfo  {journal} {J.
  Chem. Sci.}\ }\textbf {\bibinfo {volume} {124}},\ \bibinfo {pages} {51}
  (\bibinfo {year} {2012}{\natexlab{a}})}\BibitemShut {NoStop}%
\bibitem [{\citenamefont {Sardar}, \citenamefont {Puzari},\ and\ \citenamefont
  {Adhikari}(2010)}]{SSCPL496}%
  \BibitemOpen
  \bibfield  {author} {\bibinfo {author} {\bibfnamefont {S.}~\bibnamefont
  {Sardar}}, \bibinfo {author} {\bibfnamefont {P.}~\bibnamefont {Puzari}}, \
  and\ \bibinfo {author} {\bibfnamefont {S.}~\bibnamefont {Adhikari}},\
  }\href@noop {} {\bibfield  {journal} {\bibinfo  {journal} {Chem. Phys.
  Lett.}\ }\textbf {\bibinfo {volume} {496}},\ \bibinfo {pages} {341} (\bibinfo
  {year} {2010})}\BibitemShut {NoStop}%
\bibitem [{\citenamefont {Sardar}, \citenamefont {Paul},\ and\ \citenamefont
  {Adhikari}(2010)}]{SSJCS122}%
  \BibitemOpen
  \bibfield  {author} {\bibinfo {author} {\bibfnamefont {S.}~\bibnamefont
  {Sardar}}, \bibinfo {author} {\bibfnamefont {A.~K.}\ \bibnamefont {Paul}}, \
  and\ \bibinfo {author} {\bibfnamefont {S.}~\bibnamefont {Adhikari}},\
  }\href@noop {} {\bibfield  {journal} {\bibinfo  {journal} {J. Chem. Sc.}\
  }\textbf {\bibinfo {volume} {122}},\ \bibinfo {pages} {491} (\bibinfo {year}
  {2010})}\BibitemShut {NoStop}%
\bibitem [{\citenamefont {Dirac}(1930)}]{PAM}%
  \BibitemOpen
  \bibfield  {author} {\bibinfo {author} {\bibfnamefont {P.~A.~M.}\
  \bibnamefont {Dirac}},\ }\href@noop {} {\bibfield  {journal} {\bibinfo
  {journal} {Proc. Cambridge Philos. Soc.}\ }\textbf {\bibinfo {volume} {26}},\
  \bibinfo {pages} {376} (\bibinfo {year} {1930})}\BibitemShut {NoStop}%
\bibitem [{\citenamefont {Puzari}\ and\ \citenamefont
  {Adhikari}(2004)}]{SAIJQC434}%
  \BibitemOpen
  \bibfield  {author} {\bibinfo {author} {\bibfnamefont {P.}~\bibnamefont
  {Puzari}}\ and\ \bibinfo {author} {\bibfnamefont {S.}~\bibnamefont
  {Adhikari}},\ }\href@noop {} {\bibfield  {journal} {\bibinfo  {journal} {Int.
  J. Quant. Chem.}\ }\textbf {\bibinfo {volume} {98}},\ \bibinfo {pages} {434}
  (\bibinfo {year} {2004})}\BibitemShut {NoStop}%
\bibitem [{\citenamefont {Jackels}\ and\ \citenamefont
  {Davidson}(1975)}]{jackels_jcp63}%
  \BibitemOpen
  \bibfield  {author} {\bibinfo {author} {\bibfnamefont {C.~F.}\ \bibnamefont
  {Jackels}}\ and\ \bibinfo {author} {\bibfnamefont {E.~R.}\ \bibnamefont
  {Davidson}},\ }\href@noop {} {\bibfield  {journal} {\bibinfo  {journal} {J.
  Chem. Phys.}\ }\textbf {\bibinfo {volume} {63}},\ \bibinfo {pages} {4672}
  (\bibinfo {year} {1975})}\BibitemShut {NoStop}%
\bibitem [{\citenamefont {Kaldor}(1991)}]{kaldor_cpl185}%
  \BibitemOpen
  \bibfield  {author} {\bibinfo {author} {\bibfnamefont {U.}~\bibnamefont
  {Kaldor}},\ }\href@noop {} {\bibfield  {journal} {\bibinfo  {journal} {Chem.
  Phys. Lett.}\ }\textbf {\bibinfo {volume} {185}},\ \bibinfo {pages} {131}
  (\bibinfo {year} {1991})}\BibitemShut {NoStop}%
\bibitem [{\citenamefont {Haller}, \citenamefont {K\"oppel},\ and\
  \citenamefont {Cederbaum}(1985)}]{haller_jms111}%
  \BibitemOpen
  \bibfield  {author} {\bibinfo {author} {\bibfnamefont {E.}~\bibnamefont
  {Haller}}, \bibinfo {author} {\bibfnamefont {H.}~\bibnamefont {K\"oppel}}, \
  and\ \bibinfo {author} {\bibfnamefont {L.}~\bibnamefont {Cederbaum}},\
  }\href@noop {} {\bibfield  {journal} {\bibinfo  {journal} {J. Mol. Spec.}\
  }\textbf {\bibinfo {volume} {111}},\ \bibinfo {pages} {377} (\bibinfo {year}
  {1985})}\BibitemShut {NoStop}%
\bibitem [{\citenamefont {Schinke}, \citenamefont {Grebenshchikov},\ and\
  \citenamefont {Zhu}(2008)}]{schinke_cp346}%
  \BibitemOpen
  \bibfield  {author} {\bibinfo {author} {\bibfnamefont {R.}~\bibnamefont
  {Schinke}}, \bibinfo {author} {\bibfnamefont {S.~Y.}\ \bibnamefont
  {Grebenshchikov}}, \ and\ \bibinfo {author} {\bibfnamefont {H.}~\bibnamefont
  {Zhu}},\ }\href@noop {} {\bibfield  {journal} {\bibinfo  {journal} {Chem.
  Phys.}\ }\textbf {\bibinfo {volume} {346}},\ \bibinfo {pages} {99} (\bibinfo
  {year} {2008})}\BibitemShut {NoStop}%
\bibitem [{\citenamefont {Joyeux}\ \emph {et~al.}(2003)\citenamefont {Joyeux},
  \citenamefont {Jost}, ,\ and\ \citenamefont {Lombardi}}]{joyeux_jcp119}%
  \BibitemOpen
  \bibfield  {author} {\bibinfo {author} {\bibfnamefont {M.}~\bibnamefont
  {Joyeux}}, \bibinfo {author} {\bibfnamefont {R.}~\bibnamefont {Jost}}, , \
  and\ \bibinfo {author} {\bibfnamefont {M.}~\bibnamefont {Lombardi}},\
  }\href@noop {} {\bibfield  {journal} {\bibinfo  {journal} {J. Chem. Phys.}\
  }\textbf {\bibinfo {volume} {119}},\ \bibinfo {pages} {5923} (\bibinfo {year}
  {2003})}\BibitemShut {NoStop}%
\bibitem [{\citenamefont {Sanrey}\ and\ \citenamefont
  {Joyeux}(2006)}]{sanrey_jcp125}%
  \BibitemOpen
  \bibfield  {author} {\bibinfo {author} {\bibfnamefont {M.}~\bibnamefont
  {Sanrey}}\ and\ \bibinfo {author} {\bibfnamefont {M.}~\bibnamefont
  {Joyeux}},\ }\href@noop {} {\bibfield  {journal} {\bibinfo  {journal} {J.
  Chem. Phys.}\ }\textbf {\bibinfo {volume} {125}},\ \bibinfo {pages} {014304}
  (\bibinfo {year} {2006})}\BibitemShut {NoStop}%
\bibitem [{\citenamefont {Arasaki}\ \emph {et~al.}(2010)\citenamefont
  {Arasaki}, \citenamefont {Takatsuka}, \citenamefont {Wang},\ and\
  \citenamefont {McKoy}}]{takatsuka_jcp132}%
  \BibitemOpen
  \bibfield  {author} {\bibinfo {author} {\bibfnamefont {Y.}~\bibnamefont
  {Arasaki}}, \bibinfo {author} {\bibfnamefont {K.}~\bibnamefont {Takatsuka}},
  \bibinfo {author} {\bibfnamefont {K.}~\bibnamefont {Wang}}, \ and\ \bibinfo
  {author} {\bibfnamefont {V.}~\bibnamefont {McKoy}},\ }\href@noop {}
  {\bibfield  {journal} {\bibinfo  {journal} {J. Chem. Phys.}\ }\textbf
  {\bibinfo {volume} {132}},\ \bibinfo {pages} {124307} (\bibinfo {year}
  {2010})}\BibitemShut {NoStop}%
\bibitem [{\citenamefont {Arasaki}\ \emph {et~al.}(2011)\citenamefont
  {Arasaki}, \citenamefont {Wang}, \citenamefont {McKoy},\ and\ \citenamefont
  {Takatsuka}}]{takatsuka_pccp13}%
  \BibitemOpen
  \bibfield  {author} {\bibinfo {author} {\bibfnamefont {Y.}~\bibnamefont
  {Arasaki}}, \bibinfo {author} {\bibfnamefont {K.}~\bibnamefont {Wang}},
  \bibinfo {author} {\bibfnamefont {V.}~\bibnamefont {McKoy}}, \ and\ \bibinfo
  {author} {\bibfnamefont {K.}~\bibnamefont {Takatsuka}},\ }\href@noop {}
  {\bibfield  {journal} {\bibinfo  {journal} {Phys. Chem. Chem. Phys.}\
  }\textbf {\bibinfo {volume} {13}},\ \bibinfo {pages} {8681} (\bibinfo {year}
  {2011})}\BibitemShut {NoStop}%
\bibitem [{\citenamefont {Weaver}\ \emph {et~al.}(1989)\citenamefont {Weaver},
  \citenamefont {Metz}, \citenamefont {Bradforth},\ and\ \citenamefont
  {Neumark}}]{weaver_jcp90}%
  \BibitemOpen
  \bibfield  {author} {\bibinfo {author} {\bibfnamefont {A.}~\bibnamefont
  {Weaver}}, \bibinfo {author} {\bibfnamefont {R.~B.}\ \bibnamefont {Metz}},
  \bibinfo {author} {\bibfnamefont {S.~E.}\ \bibnamefont {Bradforth}}, \ and\
  \bibinfo {author} {\bibfnamefont {D.~M.}\ \bibnamefont {Neumark}},\
  }\href@noop {} {\bibfield  {journal} {\bibinfo  {journal} {J. Chem. Phys.}\
  }\textbf {\bibinfo {volume} {90}},\ \bibinfo {pages} {2070} (\bibinfo {year}
  {1989})}\BibitemShut {NoStop}%
\bibitem [{\citenamefont {Sardar}, \citenamefont {Puzari},\ and\ \citenamefont
  {Adhikari}(2011{\natexlab{b}})}]{sardar_pccp13}%
  \BibitemOpen
  \bibfield  {author} {\bibinfo {author} {\bibfnamefont {S.}~\bibnamefont
  {Sardar}}, \bibinfo {author} {\bibfnamefont {P.}~\bibnamefont {Puzari}}, \
  and\ \bibinfo {author} {\bibfnamefont {S.}~\bibnamefont {Adhikari}},\
  }\href@noop {} {\bibfield  {journal} {\bibinfo  {journal} {Phys. Chem. Chem.
  Phys.}\ }\textbf {\bibinfo {volume} {13}},\ \bibinfo {pages} {15960}
  (\bibinfo {year} {2011}{\natexlab{b}})}\BibitemShut {NoStop}%
\bibitem [{\citenamefont {Sardar}\ and\ \citenamefont
  {Adhikari}(2012{\natexlab{b}})}]{sardar_jcs124}%
  \BibitemOpen
  \bibfield  {author} {\bibinfo {author} {\bibfnamefont {S.}~\bibnamefont
  {Sardar}}\ and\ \bibinfo {author} {\bibfnamefont {S.}~\bibnamefont
  {Adhikari}},\ }\href@noop {} {\bibfield  {journal} {\bibinfo  {journal} {J.
  Chem. Sci.}\ }\textbf {\bibinfo {volume} {124}},\ \bibinfo {pages} {51}
  (\bibinfo {year} {2012}{\natexlab{b}})}\BibitemShut {NoStop}%
\bibitem [{\citenamefont {Khan}\ \emph
  {et~al.}(2014{\natexlab{b}})\citenamefont {Khan}, \citenamefont {Sardar},
  \citenamefont {Sarkar},\ and\ \citenamefont {Adhikari}}]{sardar_jpca118}%
  \BibitemOpen
  \bibfield  {author} {\bibinfo {author} {\bibfnamefont {B.~A.}\ \bibnamefont
  {Khan}}, \bibinfo {author} {\bibfnamefont {S.}~\bibnamefont {Sardar}},
  \bibinfo {author} {\bibfnamefont {P.}~\bibnamefont {Sarkar}}, \ and\ \bibinfo
  {author} {\bibfnamefont {S.}~\bibnamefont {Adhikari}},\ }\href@noop {}
  {\bibfield  {journal} {\bibinfo  {journal} {J. Phys. Chem. A}\ }\textbf
  {\bibinfo {volume} {118}},\ \bibinfo {pages} {11451} (\bibinfo {year}
  {2014}{\natexlab{b}})}\BibitemShut {NoStop}%
\bibitem [{\citenamefont {Ghosh}\ \emph {et~al.}(2015)\citenamefont {Ghosh},
  \citenamefont {Sahoo}, \citenamefont {Adhikari}, \citenamefont {Sharma},\
  and\ \citenamefont {Varandas}}]{jpca-119-12392-SA}%
  \BibitemOpen
  \bibfield  {author} {\bibinfo {author} {\bibfnamefont {S.}~\bibnamefont
  {Ghosh}}, \bibinfo {author} {\bibfnamefont {T.}~\bibnamefont {Sahoo}},
  \bibinfo {author} {\bibfnamefont {S.}~\bibnamefont {Adhikari}}, \bibinfo
  {author} {\bibfnamefont {R.}~\bibnamefont {Sharma}}, \ and\ \bibinfo {author}
  {\bibfnamefont {A.~J.~C.}\ \bibnamefont {Varandas}},\ }\href@noop {}
  {\bibfield  {journal} {\bibinfo  {journal} {J. Phys. Chem. A}\ }\textbf
  {\bibinfo {volume} {119}},\ \bibinfo {pages} {12392} (\bibinfo {year}
  {2015})}\BibitemShut {NoStop}%
\bibitem [{\citenamefont {Mandal}\ \emph {et~al.}(2018)\citenamefont {Mandal},
  \citenamefont {Ghosh}, \citenamefont {Sardar},\ and\ \citenamefont
  {Adhikari}}]{mandal_irpc37}%
  \BibitemOpen
  \bibfield  {author} {\bibinfo {author} {\bibfnamefont {S.}~\bibnamefont
  {Mandal}}, \bibinfo {author} {\bibfnamefont {S.}~\bibnamefont {Ghosh}},
  \bibinfo {author} {\bibfnamefont {S.}~\bibnamefont {Sardar}}, \ and\ \bibinfo
  {author} {\bibfnamefont {S.}~\bibnamefont {Adhikari}},\ }\href@noop {}
  {\bibfield  {journal} {\bibinfo  {journal} {Int. Rev. Phys. Chem.}\ }\textbf
  {\bibinfo {volume} {37}},\ \bibinfo {pages} {607} (\bibinfo {year}
  {2018})}\BibitemShut {NoStop}%
\bibitem [{\citenamefont {Sadygov}\ and\ \citenamefont
  {Yarkony}(1999{\natexlab{a}})}]{sadygov_jcp110}%
  \BibitemOpen
  \bibfield  {author} {\bibinfo {author} {\bibfnamefont {R.~G.}\ \bibnamefont
  {Sadygov}}\ and\ \bibinfo {author} {\bibfnamefont {D.~R.}\ \bibnamefont
  {Yarkony}},\ }\href@noop {} {\bibfield  {journal} {\bibinfo  {journal} {J.
  Chem. Phys.}\ }\textbf {\bibinfo {volume} {110}},\ \bibinfo {pages} {3639}
  (\bibinfo {year} {1999}{\natexlab{a}})}\BibitemShut {NoStop}%
\bibitem [{\citenamefont {de~Vivie-Riedle}\ \emph {et~al.}(1997)\citenamefont
  {de~Vivie-Riedle}, \citenamefont {Gaus}, \citenamefont
  {Bona\v{c}i\'c-Kouteck\'y}, \citenamefont {Manz}, \citenamefont
  {Reischl-Lenz},\ and\ \citenamefont {Saalfrank}}]{vivie-rieldle_cp223}%
  \BibitemOpen
  \bibfield  {author} {\bibinfo {author} {\bibfnamefont {R.}~\bibnamefont
  {de~Vivie-Riedle}}, \bibinfo {author} {\bibfnamefont {J.}~\bibnamefont
  {Gaus}}, \bibinfo {author} {\bibfnamefont {V.}~\bibnamefont
  {Bona\v{c}i\'c-Kouteck\'y}}, \bibinfo {author} {\bibfnamefont
  {J.}~\bibnamefont {Manz}}, \bibinfo {author} {\bibfnamefont {B.}~\bibnamefont
  {Reischl-Lenz}}, \ and\ \bibinfo {author} {\bibfnamefont {P.}~\bibnamefont
  {Saalfrank}},\ }\href@noop {} {\bibfield  {journal} {\bibinfo  {journal}
  {Chem. Phys.}\ }\textbf {\bibinfo {volume} {223}},\ \bibinfo {pages} {1}
  (\bibinfo {year} {1997})}\BibitemShut {NoStop}%
\bibitem [{\citenamefont {Meiswinkel}\ and\ \citenamefont
  {K\"oppel}(1990)}]{meiswinkel_cp144}%
  \BibitemOpen
  \bibfield  {author} {\bibinfo {author} {\bibfnamefont {R.}~\bibnamefont
  {Meiswinkel}}\ and\ \bibinfo {author} {\bibfnamefont {H.}~\bibnamefont
  {K\"oppel}},\ }\href@noop {} {\bibfield  {journal} {\bibinfo  {journal}
  {Chem. Phys.}\ }\textbf {\bibinfo {volume} {144}},\ \bibinfo {pages} {117}
  (\bibinfo {year} {1990})}\BibitemShut {NoStop}%
\bibitem [{\citenamefont {Koizumi}\ and\ \citenamefont
  {Bersuker}(1999{\natexlab{a}})}]{koizumi_prl83}%
  \BibitemOpen
  \bibfield  {author} {\bibinfo {author} {\bibfnamefont {H.}~\bibnamefont
  {Koizumi}}\ and\ \bibinfo {author} {\bibfnamefont {I.~B.}\ \bibnamefont
  {Bersuker}},\ }\href@noop {} {\bibfield  {journal} {\bibinfo  {journal}
  {Phys. Rev. Lett.}\ }\textbf {\bibinfo {volume} {83}},\ \bibinfo {pages}
  {3009} (\bibinfo {year} {1999}{\natexlab{a}})}\BibitemShut {NoStop}%
\bibitem [{\citenamefont {Hauser}\ \emph {et~al.}(2008)\citenamefont {Hauser},
  \citenamefont {Callegari}, \citenamefont {Soldan},\ and\ \citenamefont
  {Ernst}}]{hauser_jcp129}%
  \BibitemOpen
  \bibfield  {author} {\bibinfo {author} {\bibfnamefont {A.~W.}\ \bibnamefont
  {Hauser}}, \bibinfo {author} {\bibfnamefont {C.}~\bibnamefont {Callegari}},
  \bibinfo {author} {\bibfnamefont {P.}~\bibnamefont {Soldan}}, \ and\ \bibinfo
  {author} {\bibfnamefont {W.~E.}\ \bibnamefont {Ernst}},\ }\href@noop {}
  {\bibfield  {journal} {\bibinfo  {journal} {J. Chem. Phys.}\ }\textbf
  {\bibinfo {volume} {129}},\ \bibinfo {pages} {044307} (\bibinfo {year}
  {2008})}\BibitemShut {NoStop}%
\bibitem [{\citenamefont {Hauser}\ \emph {et~al.}(2010)\citenamefont {Hauser},
  \citenamefont {Callegari}, \citenamefont {Soldan},\ and\ \citenamefont
  {Ernst}}]{hauser_cp375}%
  \BibitemOpen
  \bibfield  {author} {\bibinfo {author} {\bibfnamefont {A.~W.}\ \bibnamefont
  {Hauser}}, \bibinfo {author} {\bibfnamefont {C.}~\bibnamefont {Callegari}},
  \bibinfo {author} {\bibfnamefont {P.}~\bibnamefont {Soldan}}, \ and\ \bibinfo
  {author} {\bibfnamefont {W.~E.}\ \bibnamefont {Ernst}},\ }\href@noop {}
  {\bibfield  {journal} {\bibinfo  {journal} {Chem. Phys.}\ }\textbf {\bibinfo
  {volume} {375}},\ \bibinfo {pages} {73} (\bibinfo {year} {2010})}\BibitemShut
  {NoStop}%
\bibitem [{\citenamefont {Delacr\'etaz}\ \emph {et~al.}(1986)\citenamefont
  {Delacr\'etaz}, \citenamefont {Grant}, \citenamefont {Whetten}, \citenamefont
  {W\"oste},\ and\ \citenamefont {Zwanziger}}]{delacretaz_prl56}%
  \BibitemOpen
  \bibfield  {author} {\bibinfo {author} {\bibfnamefont {G.}~\bibnamefont
  {Delacr\'etaz}}, \bibinfo {author} {\bibfnamefont {E.~R.}\ \bibnamefont
  {Grant}}, \bibinfo {author} {\bibfnamefont {R.~L.}\ \bibnamefont {Whetten}},
  \bibinfo {author} {\bibfnamefont {L.}~\bibnamefont {W\"oste}}, \ and\
  \bibinfo {author} {\bibfnamefont {J.~W.}\ \bibnamefont {Zwanziger}},\
  }\href@noop {} {\bibfield  {journal} {\bibinfo  {journal} {Phys. Rev. Lett.}\
  }\textbf {\bibinfo {volume} {56}},\ \bibinfo {pages} {2598} (\bibinfo {year}
  {1986})}\BibitemShut {NoStop}%
\bibitem [{\citenamefont {Ernst}\ and\ \citenamefont
  {Golonzka}(2004)}]{ernst_pst112}%
  \BibitemOpen
  \bibfield  {author} {\bibinfo {author} {\bibfnamefont {W.~E.}\ \bibnamefont
  {Ernst}}\ and\ \bibinfo {author} {\bibfnamefont {O.}~\bibnamefont
  {Golonzka}},\ }\href@noop {} {\bibfield  {journal} {\bibinfo  {journal}
  {Phys. Scrp.}\ }\textbf {\bibinfo {volume} {T112}},\ \bibinfo {pages} {27}
  (\bibinfo {year} {2004})}\BibitemShut {NoStop}%
\bibitem [{\citenamefont {Stock}\ \emph {et~al.}(1995)\citenamefont {Stock},
  \citenamefont {Woywod}, \citenamefont {Domcke}, \citenamefont {Swinney},\
  and\ \citenamefont {Hudson}}]{stock_jcp103}%
  \BibitemOpen
  \bibfield  {author} {\bibinfo {author} {\bibfnamefont {G.}~\bibnamefont
  {Stock}}, \bibinfo {author} {\bibfnamefont {C.}~\bibnamefont {Woywod}},
  \bibinfo {author} {\bibfnamefont {W.}~\bibnamefont {Domcke}}, \bibinfo
  {author} {\bibfnamefont {T.}~\bibnamefont {Swinney}}, \ and\ \bibinfo
  {author} {\bibfnamefont {B.~S.}\ \bibnamefont {Hudson}},\ }\href@noop {}
  {\bibfield  {journal} {\bibinfo  {journal} {J. Chem. Phys.}\ }\textbf
  {\bibinfo {volume} {103}},\ \bibinfo {pages} {6851} (\bibinfo {year}
  {1995})}\BibitemShut {NoStop}%
\bibitem [{\citenamefont {Stock}\ and\ \citenamefont
  {Domcke}(1990)}]{domcke_jcp93}%
  \BibitemOpen
  \bibfield  {author} {\bibinfo {author} {\bibfnamefont {G.}~\bibnamefont
  {Stock}}\ and\ \bibinfo {author} {\bibfnamefont {W.}~\bibnamefont {Domcke}},\
  }\href@noop {} {\bibfield  {journal} {\bibinfo  {journal} {J. Chem. Phys.}\
  }\textbf {\bibinfo {volume} {93}},\ \bibinfo {pages} {5496} (\bibinfo {year}
  {1990})}\BibitemShut {NoStop}%
\bibitem [{\citenamefont {Yamazaki}\ \emph {et~al.}(1983)\citenamefont
  {Yamazaki}, \citenamefont {Murao}, \citenamefont {Yamanaka},\ and\
  \citenamefont {Yoshihara}}]{yamazaki_fdcs75}%
  \BibitemOpen
  \bibfield  {author} {\bibinfo {author} {\bibfnamefont {I.}~\bibnamefont
  {Yamazaki}}, \bibinfo {author} {\bibfnamefont {T.}~\bibnamefont {Murao}},
  \bibinfo {author} {\bibfnamefont {T.}~\bibnamefont {Yamanaka}}, \ and\
  \bibinfo {author} {\bibfnamefont {K.}~\bibnamefont {Yoshihara}},\ }\href@noop
  {} {\bibfield  {journal} {\bibinfo  {journal} {Faraday Discuss. Chem. Soc.}\
  }\textbf {\bibinfo {volume} {75}},\ \bibinfo {pages} {395} (\bibinfo {year}
  {1983})}\BibitemShut {NoStop}%
\bibitem [{\citenamefont {Seidner}, \citenamefont {Stock},\ and\ \citenamefont
  {Sobolewski}(1992)}]{seidner_jcp96}%
  \BibitemOpen
  \bibfield  {author} {\bibinfo {author} {\bibfnamefont {L.}~\bibnamefont
  {Seidner}}, \bibinfo {author} {\bibfnamefont {G.}~\bibnamefont {Stock}}, \
  and\ \bibinfo {author} {\bibfnamefont {A.~L.}\ \bibnamefont {Sobolewski}},\
  }\href@noop {} {\bibfield  {journal} {\bibinfo  {journal} {J. Chem. Phys.}\
  }\textbf {\bibinfo {volume} {96}},\ \bibinfo {pages} {5298} (\bibinfo {year}
  {1992})}\BibitemShut {NoStop}%
\bibitem [{\citenamefont {Sobolewski}, \citenamefont {Woywod},\ and\
  \citenamefont {Domcke}(1993)}]{domcke_jcp98}%
  \BibitemOpen
  \bibfield  {author} {\bibinfo {author} {\bibfnamefont {A.~L.}\ \bibnamefont
  {Sobolewski}}, \bibinfo {author} {\bibfnamefont {C.}~\bibnamefont {Woywod}},
  \ and\ \bibinfo {author} {\bibfnamefont {W.}~\bibnamefont {Domcke}},\
  }\href@noop {} {\bibfield  {journal} {\bibinfo  {journal} {J. Chem. Phys.}\
  }\textbf {\bibinfo {volume} {98}},\ \bibinfo {pages} {5627} (\bibinfo {year}
  {1993})}\BibitemShut {NoStop}%
\bibitem [{\citenamefont {Guo}\ \emph {et~al.}(2012)\citenamefont {Guo},
  \citenamefont {Liu}, \citenamefont {Xie}, \citenamefont {Zhao}, \citenamefont
  {Pei}, \citenamefont {Wang}, \citenamefont {Zheng}, \citenamefont {Ai},
  \citenamefont {Chen}, \citenamefont {Fang},\ and\ \citenamefont
  {Yeung}}]{guo_jrs4074}%
  \BibitemOpen
  \bibfield  {author} {\bibinfo {author} {\bibfnamefont {J.~L.}\ \bibnamefont
  {Guo}}, \bibinfo {author} {\bibfnamefont {C.}~\bibnamefont {Liu}}, \bibinfo
  {author} {\bibfnamefont {B.~B.}\ \bibnamefont {Xie}}, \bibinfo {author}
  {\bibfnamefont {Y.~Y.}\ \bibnamefont {Zhao}}, \bibinfo {author}
  {\bibfnamefont {K.~M.}\ \bibnamefont {Pei}}, \bibinfo {author} {\bibfnamefont
  {H.~G.}\ \bibnamefont {Wang}}, \bibinfo {author} {\bibfnamefont
  {X.}~\bibnamefont {Zheng}}, \bibinfo {author} {\bibfnamefont {Y.~J.}\
  \bibnamefont {Ai}}, \bibinfo {author} {\bibfnamefont {X.~B.}\ \bibnamefont
  {Chen}}, \bibinfo {author} {\bibfnamefont {W.~H.}\ \bibnamefont {Fang}}, \
  and\ \bibinfo {author} {\bibfnamefont {C.~S.}\ \bibnamefont {Yeung}},\ }\href
  {\doibase 10.1002/jrs.4074} {\bibfield  {journal} {\bibinfo  {journal} {J.
  Raman Spectrosc.}\ }\textbf {\bibinfo {volume} {4074}} (\bibinfo {year}
  {2012}),\ 10.1002/jrs.4074}\BibitemShut {NoStop}%
\bibitem [{\citenamefont {Kanno}\ \emph {et~al.}(2015)\citenamefont {Kanno},
  \citenamefont {Ito}, \citenamefont {Shimakura}, \citenamefont {Koseki},
  \citenamefont {Kono},\ and\ \citenamefont {Fujimura}}]{kanno_4}%
  \BibitemOpen
  \bibfield  {author} {\bibinfo {author} {\bibfnamefont {M.}~\bibnamefont
  {Kanno}}, \bibinfo {author} {\bibfnamefont {Y.}~\bibnamefont {Ito}}, \bibinfo
  {author} {\bibfnamefont {N.}~\bibnamefont {Shimakura}}, \bibinfo {author}
  {\bibfnamefont {S.}~\bibnamefont {Koseki}}, \bibinfo {author} {\bibfnamefont
  {H.}~\bibnamefont {Kono}}, \ and\ \bibinfo {author} {\bibfnamefont
  {Y.}~\bibnamefont {Fujimura}},\ }\href@noop {} {\bibfield  {journal}
  {\bibinfo  {journal} {Phys. Chem. Chem. Phys.}\ }\textbf {\bibinfo {volume}
  {17}},\ \bibinfo {pages} {2012} (\bibinfo {year} {2015})}\BibitemShut
  {NoStop}%
\bibitem [{\citenamefont {Woywod}\ \emph {et~al.}(1994)\citenamefont {Woywod},
  \citenamefont {Domcke}, \citenamefont {Sobolewski},\ and\ \citenamefont
  {Werner}}]{woywood_jcp100}%
  \BibitemOpen
  \bibfield  {author} {\bibinfo {author} {\bibfnamefont {C.}~\bibnamefont
  {Woywod}}, \bibinfo {author} {\bibfnamefont {W.}~\bibnamefont {Domcke}},
  \bibinfo {author} {\bibfnamefont {A.~L.}\ \bibnamefont {Sobolewski}}, \ and\
  \bibinfo {author} {\bibfnamefont {H.~J.}\ \bibnamefont {Werner}},\
  }\href@noop {} {\bibfield  {journal} {\bibinfo  {journal} {J. Chem. Phys.}\
  }\textbf {\bibinfo {volume} {100}},\ \bibinfo {pages} {1400} (\bibinfo {year}
  {1994})}\BibitemShut {NoStop}%
\bibitem [{\citenamefont {Worth}, \citenamefont {Meyer},\ and\ \citenamefont
  {Cederbaum}(1998)}]{worth_jcp109}%
  \BibitemOpen
  \bibfield  {author} {\bibinfo {author} {\bibfnamefont {G.~A.}\ \bibnamefont
  {Worth}}, \bibinfo {author} {\bibfnamefont {H.~D.}\ \bibnamefont {Meyer}}, \
  and\ \bibinfo {author} {\bibfnamefont {L.~S.}\ \bibnamefont {Cederbaum}},\
  }\href {\doibase 10.1063/1.476947} {\bibfield  {journal} {\bibinfo  {journal}
  {J. Chem. Phys.}\ }\textbf {\bibinfo {volume} {109}},\ \bibinfo {pages}
  {3518} (\bibinfo {year} {1998})}\BibitemShut {NoStop}%
\bibitem [{\citenamefont {Burghardt}, \citenamefont {Giri},\ and\ \citenamefont
  {Worth}(2008)}]{Burghardt_jcp129}%
  \BibitemOpen
  \bibfield  {author} {\bibinfo {author} {\bibfnamefont {I.}~\bibnamefont
  {Burghardt}}, \bibinfo {author} {\bibfnamefont {K.}~\bibnamefont {Giri}}, \
  and\ \bibinfo {author} {\bibfnamefont {G.~A.}\ \bibnamefont {Worth}},\ }\href
  {\doibase 10.1063/1.2996349} {\bibfield  {journal} {\bibinfo  {journal} {J.
  Chem. Phys.}\ }\textbf {\bibinfo {volume} {129}},\ \bibinfo {pages} {174104}
  (\bibinfo {year} {2008})}\BibitemShut {NoStop}%
\bibitem [{\citenamefont {Vendrell}\ and\ \citenamefont
  {Meyer}(2011)}]{vendrell_jcp134}%
  \BibitemOpen
  \bibfield  {author} {\bibinfo {author} {\bibfnamefont {O.}~\bibnamefont
  {Vendrell}}\ and\ \bibinfo {author} {\bibfnamefont {H.~D.}\ \bibnamefont
  {Meyer}},\ }\href {\doibase 10.1063/1.3535541} {\bibfield  {journal}
  {\bibinfo  {journal} {J. Chem. Phys.}\ }\textbf {\bibinfo {volume} {134}},\
  \bibinfo {pages} {044135} (\bibinfo {year} {2011})}\BibitemShut {NoStop}%
\bibitem [{\citenamefont {Richings}, \citenamefont {Robertson},\ and\
  \citenamefont {Habershon}(2019)}]{richings_15}%
  \BibitemOpen
  \bibfield  {author} {\bibinfo {author} {\bibfnamefont {G.~W.}\ \bibnamefont
  {Richings}}, \bibinfo {author} {\bibfnamefont {C.}~\bibnamefont {Robertson}},
  \ and\ \bibinfo {author} {\bibfnamefont {S.}~\bibnamefont {Habershon}},\
  }\href {\doibase 10.1021/acs.jctc.8b00819} {\bibfield  {journal} {\bibinfo
  {journal} {J. Chem. Theory Comput.}\ }\textbf {\bibinfo {volume} {15}},\
  \bibinfo {pages} {857} (\bibinfo {year} {2019})}\BibitemShut {NoStop}%
\bibitem [{\citenamefont {Pancharatnam}(1956)}]{pancharatnam_pias44}%
  \BibitemOpen
  \bibfield  {author} {\bibinfo {author} {\bibfnamefont {S.}~\bibnamefont
  {Pancharatnam}},\ }\href@noop {} {\bibfield  {journal} {\bibinfo  {journal}
  {Proc. Ind. Acad. Sci. A}\ }\textbf {\bibinfo {volume} {44}},\ \bibinfo
  {pages} {247} (\bibinfo {year} {1956})}\BibitemShut {NoStop}%
\bibitem [{\citenamefont {Berry}(1984)}]{berry_prca392}%
  \BibitemOpen
  \bibfield  {author} {\bibinfo {author} {\bibfnamefont {M.~V.}\ \bibnamefont
  {Berry}},\ }\href@noop {} {\bibfield  {journal} {\bibinfo  {journal} {Proc.
  R. Soc., Lond. A}\ }\textbf {\bibinfo {volume} {392}},\ \bibinfo {pages} {45}
  (\bibinfo {year} {1984})}\BibitemShut {NoStop}%
\bibitem [{\citenamefont {Sah}\ \emph {et~al.}(2023{\natexlab{c}})\citenamefont
  {Sah}, \citenamefont {Mukherjee}, \citenamefont {Naskar}, \citenamefont
  {Hazra},\ and\ \citenamefont {Adhikari}}]{mantu_curl}%
  \BibitemOpen
  \bibfield  {author} {\bibinfo {author} {\bibfnamefont {M.~K.}\ \bibnamefont
  {Sah}}, \bibinfo {author} {\bibfnamefont {S.}~\bibnamefont {Mukherjee}},
  \bibinfo {author} {\bibfnamefont {K.}~\bibnamefont {Naskar}}, \bibinfo
  {author} {\bibfnamefont {S.}~\bibnamefont {Hazra}}, \ and\ \bibinfo {author}
  {\bibfnamefont {S.}~\bibnamefont {Adhikari}},\ }\href@noop {} {\bibfield
  {journal} {\bibinfo  {journal} {Int. J. Quantum Chem.}\ }\textbf {\bibinfo
  {volume} {DOI: 10.1002/qua.27212}} (\bibinfo {year}
  {2023}{\natexlab{c}})}\BibitemShut {NoStop}%
\bibitem [{\citenamefont {Raab}\ \emph {et~al.}(1999)\citenamefont {Raab},
  \citenamefont {Worth}, \citenamefont {Meyer},\ and\ \citenamefont
  {Cederbaum}}]{raab_jcp110}%
  \BibitemOpen
  \bibfield  {author} {\bibinfo {author} {\bibfnamefont {A.}~\bibnamefont
  {Raab}}, \bibinfo {author} {\bibfnamefont {G.~A.}\ \bibnamefont {Worth}},
  \bibinfo {author} {\bibfnamefont {H.-D.}\ \bibnamefont {Meyer}}, \ and\
  \bibinfo {author} {\bibfnamefont {L.~S.}\ \bibnamefont {Cederbaum}},\ }\href
  {\doibase 10.1063/1.478061} {\bibfield  {journal} {\bibinfo  {journal} {J.
  Chem. Phys.}\ }\textbf {\bibinfo {volume} {110}},\ \bibinfo {pages} {936}
  (\bibinfo {year} {1999})}\BibitemShut {NoStop}%
\bibitem [{\citenamefont {Faraji}\ and\ \citenamefont
  {Köppel}(2008)}]{faraji_jcp_struct}%
  \BibitemOpen
  \bibfield  {author} {\bibinfo {author} {\bibfnamefont {S.}~\bibnamefont
  {Faraji}}\ and\ \bibinfo {author} {\bibfnamefont {H.}~\bibnamefont
  {Köppel}},\ }\href@noop {} {\bibfield  {journal} {\bibinfo  {journal} {The
  Journal of Chemical Physics}\ }\textbf {\bibinfo {volume} {129}},\ \bibinfo
  {pages} {074310} (\bibinfo {year} {2008})}\BibitemShut {NoStop}%
\bibitem [{\citenamefont {Sardar}, \citenamefont {Puzari},\ and\ \citenamefont
  {Adhikari}(2011{\natexlab{c}})}]{sardar_pccp}%
  \BibitemOpen
  \bibfield  {author} {\bibinfo {author} {\bibfnamefont {S.}~\bibnamefont
  {Sardar}}, \bibinfo {author} {\bibfnamefont {P.}~\bibnamefont {Puzari}}, \
  and\ \bibinfo {author} {\bibfnamefont {S.}~\bibnamefont {Adhikari}},\
  }\href@noop {} {\bibfield  {journal} {\bibinfo  {journal} {Phys. Chem. Chem.
  Phys.}\ }\textbf {\bibinfo {volume} {13}},\ \bibinfo {pages} {15960}
  (\bibinfo {year} {2011}{\natexlab{c}})}\BibitemShut {NoStop}%
\bibitem [{\citenamefont {Eisfeld}\ and\ \citenamefont
  {Morokuma}(2000)}]{eisfeld_jcp113}%
  \BibitemOpen
  \bibfield  {author} {\bibinfo {author} {\bibfnamefont {W.}~\bibnamefont
  {Eisfeld}}\ and\ \bibinfo {author} {\bibfnamefont {K.}~\bibnamefont
  {Morokuma}},\ }\href@noop {} {\bibfield  {journal} {\bibinfo  {journal} {J.
  Chem. Phys.}\ }\textbf {\bibinfo {volume} {113}},\ \bibinfo {pages} {5587}
  (\bibinfo {year} {2000})}\BibitemShut {NoStop}%
\bibitem [{\citenamefont {Mayer}, \citenamefont {Cederbaum},\ and\
  \citenamefont {K\"oppel}(1994)}]{mayer_jcp100}%
  \BibitemOpen
  \bibfield  {author} {\bibinfo {author} {\bibfnamefont {M.}~\bibnamefont
  {Mayer}}, \bibinfo {author} {\bibfnamefont {L.~S.}\ \bibnamefont
  {Cederbaum}}, \ and\ \bibinfo {author} {\bibfnamefont {H.}~\bibnamefont
  {K\"oppel}},\ }\href@noop {} {\bibfield  {journal} {\bibinfo  {journal} {J.
  Chem. Phys.}\ }\textbf {\bibinfo {volume} {100}},\ \bibinfo {pages} {899}
  (\bibinfo {year} {1994})}\BibitemShut {NoStop}%
\bibitem [{\citenamefont {Okumura}\ \emph {et~al.}(2006)\citenamefont
  {Okumura}, \citenamefont {Stanton}, \citenamefont {Deev},\ and\ \citenamefont
  {Sommar}}]{okumura_ps73}%
  \BibitemOpen
  \bibfield  {author} {\bibinfo {author} {\bibfnamefont {M.}~\bibnamefont
  {Okumura}}, \bibinfo {author} {\bibfnamefont {J.~F.}\ \bibnamefont
  {Stanton}}, \bibinfo {author} {\bibfnamefont {A.}~\bibnamefont {Deev}}, \
  and\ \bibinfo {author} {\bibfnamefont {J.}~\bibnamefont {Sommar}},\
  }\href@noop {} {\bibfield  {journal} {\bibinfo  {journal} {Phys. Scr.}\
  }\textbf {\bibinfo {volume} {73}},\ \bibinfo {pages} {C64} (\bibinfo {year}
  {2006})}\BibitemShut {NoStop}%
\bibitem [{\citenamefont {Stanton}(2007)}]{stanton_jcp126}%
  \BibitemOpen
  \bibfield  {author} {\bibinfo {author} {\bibfnamefont {J.~F.}\ \bibnamefont
  {Stanton}},\ }\href@noop {} {\bibfield  {journal} {\bibinfo  {journal} {J.
  Chem. Phys.}\ }\textbf {\bibinfo {volume} {126}},\ \bibinfo {pages} {134309}
  (\bibinfo {year} {2007})}\BibitemShut {NoStop}%
\bibitem [{\citenamefont {Nelson}, \citenamefont {Pasternack},\ and\
  \citenamefont {McDonald}(1983)}]{nelson_jpc87}%
  \BibitemOpen
  \bibfield  {author} {\bibinfo {author} {\bibfnamefont {H.~H.}\ \bibnamefont
  {Nelson}}, \bibinfo {author} {\bibfnamefont {L.}~\bibnamefont {Pasternack}},
  \ and\ \bibinfo {author} {\bibfnamefont {J.~R.}\ \bibnamefont {McDonald}},\
  }\href@noop {} {\bibfield  {journal} {\bibinfo  {journal} {J. Phys. Chem.}\
  }\textbf {\bibinfo {volume} {87}},\ \bibinfo {pages} {1286} (\bibinfo {year}
  {1983})}\BibitemShut {NoStop}%
\bibitem [{\citenamefont {Ishiwata}\ \emph {et~al.}(1983)\citenamefont
  {Ishiwata}, \citenamefont {Fujiwara}, \citenamefont {Naruge}, \citenamefont
  {Obi},\ and\ \citenamefont {Tanaka}}]{ishiwata_jpc87}%
  \BibitemOpen
  \bibfield  {author} {\bibinfo {author} {\bibfnamefont {T.}~\bibnamefont
  {Ishiwata}}, \bibinfo {author} {\bibfnamefont {I.}~\bibnamefont {Fujiwara}},
  \bibinfo {author} {\bibfnamefont {Y.}~\bibnamefont {Naruge}}, \bibinfo
  {author} {\bibfnamefont {K.}~\bibnamefont {Obi}}, \ and\ \bibinfo {author}
  {\bibfnamefont {I.}~\bibnamefont {Tanaka}},\ }\href@noop {} {\bibfield
  {journal} {\bibinfo  {journal} {J. Phys. Chem.}\ }\textbf {\bibinfo {volume}
  {87}},\ \bibinfo {pages} {1349} (\bibinfo {year} {1983})}\BibitemShut
  {NoStop}%
\bibitem [{\citenamefont {Ishiwata}\ \emph {et~al.}(1985)\citenamefont
  {Ishiwata}, \citenamefont {Tanaka}, \citenamefont {Kawaguchi},\ and\
  \citenamefont {Hirota}}]{ishiwata_jcp82}%
  \BibitemOpen
  \bibfield  {author} {\bibinfo {author} {\bibfnamefont {T.}~\bibnamefont
  {Ishiwata}}, \bibinfo {author} {\bibfnamefont {I.}~\bibnamefont {Tanaka}},
  \bibinfo {author} {\bibfnamefont {K.}~\bibnamefont {Kawaguchi}}, \ and\
  \bibinfo {author} {\bibfnamefont {E.}~\bibnamefont {Hirota}},\ }\href@noop {}
  {\bibfield  {journal} {\bibinfo  {journal} {J. Chem. Phys.}\ }\textbf
  {\bibinfo {volume} {82}},\ \bibinfo {pages} {2196} (\bibinfo {year}
  {1985})}\BibitemShut {NoStop}%
\bibitem [{\citenamefont {Hirota}\ \emph {et~al.}(1997)\citenamefont {Hirota},
  \citenamefont {Ishiwata}, \citenamefont {Kawaguchi}, \citenamefont
  {Fujitake}, \citenamefont {Ohashi},\ and\ \citenamefont
  {Tanaka}}]{hirota_jcp107}%
  \BibitemOpen
  \bibfield  {author} {\bibinfo {author} {\bibfnamefont {E.}~\bibnamefont
  {Hirota}}, \bibinfo {author} {\bibfnamefont {T.}~\bibnamefont {Ishiwata}},
  \bibinfo {author} {\bibfnamefont {K.}~\bibnamefont {Kawaguchi}}, \bibinfo
  {author} {\bibfnamefont {M.}~\bibnamefont {Fujitake}}, \bibinfo {author}
  {\bibfnamefont {N.}~\bibnamefont {Ohashi}}, \ and\ \bibinfo {author}
  {\bibfnamefont {I.}~\bibnamefont {Tanaka}},\ }\href@noop {} {\bibfield
  {journal} {\bibinfo  {journal} {J. Chem. Phys.}\ }\textbf {\bibinfo {volume}
  {107}},\ \bibinfo {pages} {2829} (\bibinfo {year} {1997})}\BibitemShut
  {NoStop}%
\bibitem [{\citenamefont {Weaver}\ \emph {et~al.}(1991)\citenamefont {Weaver},
  \citenamefont {Arnold}, \citenamefont {Bradforth},\ and\ \citenamefont
  {Neumark}}]{weaver_jcp94}%
  \BibitemOpen
  \bibfield  {author} {\bibinfo {author} {\bibfnamefont {A.}~\bibnamefont
  {Weaver}}, \bibinfo {author} {\bibfnamefont {D.~W.}\ \bibnamefont {Arnold}},
  \bibinfo {author} {\bibfnamefont {S.~E.}\ \bibnamefont {Bradforth}}, \ and\
  \bibinfo {author} {\bibfnamefont {D.~M.}\ \bibnamefont {Neumark}},\
  }\href@noop {} {\bibfield  {journal} {\bibinfo  {journal} {J. Chem. Phys.}\
  }\textbf {\bibinfo {volume} {94}},\ \bibinfo {pages} {1740} (\bibinfo {year}
  {1991})}\BibitemShut {NoStop}%
\bibitem [{\citenamefont {Deev}, \citenamefont {Stanton},\ and\ \citenamefont
  {Okumura}(2005)}]{deev_jcp122}%
  \BibitemOpen
  \bibfield  {author} {\bibinfo {author} {\bibfnamefont {A.}~\bibnamefont
  {Deev}}, \bibinfo {author} {\bibfnamefont {J.~F.}\ \bibnamefont {Stanton}}, \
  and\ \bibinfo {author} {\bibfnamefont {M.}~\bibnamefont {Okumura}},\
  }\href@noop {} {\bibfield  {journal} {\bibinfo  {journal} {J. Chem. Phys.}\
  }\textbf {\bibinfo {volume} {122}},\ \bibinfo {pages} {224305} (\bibinfo
  {year} {2005})}\BibitemShut {NoStop}%
\bibitem [{\citenamefont {Codd}\ \emph {et~al.}(2015)\citenamefont {Codd},
  \citenamefont {Chen}, \citenamefont {Roudjane}, \citenamefont {Stanton},\
  and\ \citenamefont {Miller}}]{codd_jcp142}%
  \BibitemOpen
  \bibfield  {author} {\bibinfo {author} {\bibfnamefont {T.}~\bibnamefont
  {Codd}}, \bibinfo {author} {\bibfnamefont {M.-W.}\ \bibnamefont {Chen}},
  \bibinfo {author} {\bibfnamefont {M.}~\bibnamefont {Roudjane}}, \bibinfo
  {author} {\bibfnamefont {J.~F.}\ \bibnamefont {Stanton}}, \ and\ \bibinfo
  {author} {\bibfnamefont {T.~A.}\ \bibnamefont {Miller}},\ }\href@noop {}
  {\bibfield  {journal} {\bibinfo  {journal} {J. Chem. Phys.}\ }\textbf
  {\bibinfo {volume} {142}},\ \bibinfo {pages} {184305} (\bibinfo {year}
  {2015})}\BibitemShut {NoStop}%
\bibitem [{\citenamefont {Scott}\ \emph {et~al.}(1963)\citenamefont {Scott},
  \citenamefont {Messerly}, \citenamefont {Todd}, \citenamefont {Hossenlopp},
  \citenamefont {Osborn},\ and\ \citenamefont {McCullough}}]{scott_jcp}%
  \BibitemOpen
  \bibfield  {author} {\bibinfo {author} {\bibfnamefont {D.~W.}\ \bibnamefont
  {Scott}}, \bibinfo {author} {\bibfnamefont {J.~F.}\ \bibnamefont {Messerly}},
  \bibinfo {author} {\bibfnamefont {S.~S.}\ \bibnamefont {Todd}}, \bibinfo
  {author} {\bibfnamefont {I.~A.}\ \bibnamefont {Hossenlopp}}, \bibinfo
  {author} {\bibfnamefont {A.}~\bibnamefont {Osborn}}, \ and\ \bibinfo {author}
  {\bibfnamefont {J.~P.}\ \bibnamefont {McCullough}},\ }\href@noop {}
  {\bibfield  {journal} {\bibinfo  {journal} {The Journal of Chemical Physics}\
  }\textbf {\bibinfo {volume} {38}},\ \bibinfo {pages} {532} (\bibinfo {year}
  {1963})}\BibitemShut {NoStop}%
\bibitem [{\citenamefont {Abraham}\ and\ \citenamefont
  {Cooper}(1968)}]{abraham_jcsb}%
  \BibitemOpen
  \bibfield  {author} {\bibinfo {author} {\bibfnamefont {R.~J.}\ \bibnamefont
  {Abraham}}\ and\ \bibinfo {author} {\bibfnamefont {M.~A.}\ \bibnamefont
  {Cooper}},\ }\href@noop {} {\bibfield  {journal} {\bibinfo  {journal} {J.
  Chem. Soc. B}\ ,\ \bibinfo {pages} {1094}} (\bibinfo {year}
  {1968})}\BibitemShut {NoStop}%
\bibitem [{\citenamefont {Gerritsen}\ and\ \citenamefont
  {MacLean}(1971)}]{gerritsen_sapms}%
  \BibitemOpen
  \bibfield  {author} {\bibinfo {author} {\bibfnamefont {J.}~\bibnamefont
  {Gerritsen}}\ and\ \bibinfo {author} {\bibfnamefont {C.}~\bibnamefont
  {MacLean}},\ }\href@noop {} {\bibfield  {journal} {\bibinfo  {journal}
  {Spectrochimica Acta Part A: Molecular Spectroscopy}\ }\textbf {\bibinfo
  {volume} {27}},\ \bibinfo {pages} {1495} (\bibinfo {year}
  {1971})}\BibitemShut {NoStop}%
\bibitem [{\citenamefont {Clark}\ \emph {et~al.}(1972)\citenamefont {Clark},
  \citenamefont {Kilcast}, \citenamefont {Adams},\ and\ \citenamefont
  {Musgrave}}]{clark_jesc}%
  \BibitemOpen
  \bibfield  {author} {\bibinfo {author} {\bibfnamefont {D.}~\bibnamefont
  {Clark}}, \bibinfo {author} {\bibfnamefont {D.}~\bibnamefont {Kilcast}},
  \bibinfo {author} {\bibfnamefont {D.}~\bibnamefont {Adams}}, \ and\ \bibinfo
  {author} {\bibfnamefont {W.}~\bibnamefont {Musgrave}},\ }\href@noop {}
  {\bibfield  {journal} {\bibinfo  {journal} {Journal of Electron Spectroscopy
  and Related Phenomena}\ }\textbf {\bibinfo {volume} {1}},\ \bibinfo {pages}
  {227} (\bibinfo {year} {1972})}\BibitemShut {NoStop}%
\bibitem [{\citenamefont {Davis}, \citenamefont {Shirley},\ and\ \citenamefont
  {Thomas}(1972)}]{shirley_jacs}%
  \BibitemOpen
  \bibfield  {author} {\bibinfo {author} {\bibfnamefont {D.~W.}\ \bibnamefont
  {Davis}}, \bibinfo {author} {\bibfnamefont {D.~A.}\ \bibnamefont {Shirley}},
  \ and\ \bibinfo {author} {\bibfnamefont {T.~D.}\ \bibnamefont {Thomas}},\
  }\href@noop {} {\bibfield  {journal} {\bibinfo  {journal} {Journal of the
  American Chemical Society}\ }\textbf {\bibinfo {volume} {94}},\ \bibinfo
  {pages} {6565} (\bibinfo {year} {1972})}\BibitemShut {NoStop}%
\bibitem [{\citenamefont {GILBERT}\ and\ \citenamefont
  {SANDORFY}(1974)}]{gilbert_cpl}%
  \BibitemOpen
  \bibfield  {author} {\bibinfo {author} {\bibfnamefont {R.}~\bibnamefont
  {GILBERT}}\ and\ \bibinfo {author} {\bibfnamefont {C.}~\bibnamefont
  {SANDORFY}},\ }\href@noop {} {\bibfield  {journal} {\bibinfo  {journal}
  {Chemical Physics Letters}\ }\textbf {\bibinfo {volume} {27}},\ \bibinfo
  {pages} {457} (\bibinfo {year} {1974})}\BibitemShut {NoStop}%
\bibitem [{\citenamefont {Bieri}, \citenamefont {Åsbrink},\ and\ \citenamefont
  {{Von Niessen}}(1981)}]{bieri_jesc}%
  \BibitemOpen
  \bibfield  {author} {\bibinfo {author} {\bibfnamefont {G.}~\bibnamefont
  {Bieri}}, \bibinfo {author} {\bibfnamefont {L.}~\bibnamefont {Åsbrink}}, \
  and\ \bibinfo {author} {\bibfnamefont {W.}~\bibnamefont {{Von Niessen}}},\
  }\href@noop {} {\bibfield  {journal} {\bibinfo  {journal} {Journal of
  Electron Spectroscopy and Related Phenomena}\ }\textbf {\bibinfo {volume}
  {23}},\ \bibinfo {pages} {281} (\bibinfo {year} {1981})}\BibitemShut
  {NoStop}%
\bibitem [{\citenamefont {Tsuchiya}\ \emph {et~al.}(1991)\citenamefont
  {Tsuchiya}, \citenamefont {Takazawa}, \citenamefont {Fujii},\ and\
  \citenamefont {Ito}}]{tsuchiya_cpl}%
  \BibitemOpen
  \bibfield  {author} {\bibinfo {author} {\bibfnamefont {Y.}~\bibnamefont
  {Tsuchiya}}, \bibinfo {author} {\bibfnamefont {K.}~\bibnamefont {Takazawa}},
  \bibinfo {author} {\bibfnamefont {M.}~\bibnamefont {Fujii}}, \ and\ \bibinfo
  {author} {\bibfnamefont {M.}~\bibnamefont {Ito}},\ }\href@noop {} {\bibfield
  {journal} {\bibinfo  {journal} {Chemical Physics Letters}\ }\textbf {\bibinfo
  {volume} {183}},\ \bibinfo {pages} {107} (\bibinfo {year}
  {1991})}\BibitemShut {NoStop}%
\bibitem [{\citenamefont {Tsuchiya}\ \emph {et~al.}(1992)\citenamefont
  {Tsuchiya}, \citenamefont {Takazawa}, \citenamefont {Fujii},\ and\
  \citenamefont {Ito}}]{tsuchiya_jcp}%
  \BibitemOpen
  \bibfield  {author} {\bibinfo {author} {\bibfnamefont {Y.}~\bibnamefont
  {Tsuchiya}}, \bibinfo {author} {\bibfnamefont {K.}~\bibnamefont {Takazawa}},
  \bibinfo {author} {\bibfnamefont {M.}~\bibnamefont {Fujii}}, \ and\ \bibinfo
  {author} {\bibfnamefont {M.}~\bibnamefont {Ito}},\ }\href@noop {} {\bibfield
  {journal} {\bibinfo  {journal} {The Journal of Physical Chemistry}\ }\textbf
  {\bibinfo {volume} {96}},\ \bibinfo {pages} {99} (\bibinfo {year}
  {1992})}\BibitemShut {NoStop}%
\bibitem [{\citenamefont {Philis}\ \emph {et~al.}(1981)\citenamefont {Philis},
  \citenamefont {Bolovinos}, \citenamefont {Andritsopoulos}, \citenamefont
  {Pantos},\ and\ \citenamefont {Tsekeris}}]{philis_jpb}%
  \BibitemOpen
  \bibfield  {author} {\bibinfo {author} {\bibfnamefont {J.}~\bibnamefont
  {Philis}}, \bibinfo {author} {\bibfnamefont {A.}~\bibnamefont {Bolovinos}},
  \bibinfo {author} {\bibfnamefont {G.}~\bibnamefont {Andritsopoulos}},
  \bibinfo {author} {\bibfnamefont {E.}~\bibnamefont {Pantos}}, \ and\ \bibinfo
  {author} {\bibfnamefont {P.}~\bibnamefont {Tsekeris}},\ }\href@noop {}
  {\bibfield  {journal} {\bibinfo  {journal} {Journal of Physics B: Atomic and
  Molecular Physics}\ }\textbf {\bibinfo {volume} {14}},\ \bibinfo {pages}
  {3621} (\bibinfo {year} {1981})}\BibitemShut {NoStop}%
\bibitem [{\citenamefont {Goly}, \citenamefont {Spoerel},\ and\ \citenamefont
  {Stahl}(2002)}]{goly_cp}%
  \BibitemOpen
  \bibfield  {author} {\bibinfo {author} {\bibfnamefont {T.}~\bibnamefont
  {Goly}}, \bibinfo {author} {\bibfnamefont {U.}~\bibnamefont {Spoerel}}, \
  and\ \bibinfo {author} {\bibfnamefont {W.}~\bibnamefont {Stahl}},\
  }\href@noop {} {\bibfield  {journal} {\bibinfo  {journal} {Chemical Physics}\
  }\textbf {\bibinfo {volume} {283}},\ \bibinfo {pages} {289} (\bibinfo {year}
  {2002})}\BibitemShut {NoStop}%
\bibitem [{\citenamefont {Kwon}, \citenamefont {Kim},\ and\ \citenamefont
  {Kim}(2003)}]{kwon_jcp_2003}%
  \BibitemOpen
  \bibfield  {author} {\bibinfo {author} {\bibfnamefont {C.~H.}\ \bibnamefont
  {Kwon}}, \bibinfo {author} {\bibfnamefont {H.~L.}\ \bibnamefont {Kim}}, \
  and\ \bibinfo {author} {\bibfnamefont {M.~S.}\ \bibnamefont {Kim}},\
  }\href@noop {} {\bibfield  {journal} {\bibinfo  {journal} {J. Chem. Phys}\
  }\textbf {\bibinfo {volume} {118}},\ \bibinfo {pages} {6327} (\bibinfo {year}
  {2003})}\BibitemShut {NoStop}%
\bibitem [{\citenamefont {Faraji}, \citenamefont {Meyer},\ and\ \citenamefont
  {Köppel}(2008)}]{faraji_jcp_dyna}%
  \BibitemOpen
  \bibfield  {author} {\bibinfo {author} {\bibfnamefont {S.}~\bibnamefont
  {Faraji}}, \bibinfo {author} {\bibfnamefont {H.-D.}\ \bibnamefont {Meyer}}, \
  and\ \bibinfo {author} {\bibfnamefont {H.}~\bibnamefont {Köppel}},\
  }\href@noop {} {\bibfield  {journal} {\bibinfo  {journal} {The Journal of
  Chemical Physics}\ }\textbf {\bibinfo {volume} {129}},\ \bibinfo {pages}
  {074311} (\bibinfo {year} {2008})}\BibitemShut {NoStop}%
\bibitem [{\citenamefont {Kwon}, \citenamefont {Kim},\ and\ \citenamefont
  {Kim}(2002)}]{kwon_jcp}%
  \BibitemOpen
  \bibfield  {author} {\bibinfo {author} {\bibfnamefont {C.~H.}\ \bibnamefont
  {Kwon}}, \bibinfo {author} {\bibfnamefont {H.~L.}\ \bibnamefont {Kim}}, \
  and\ \bibinfo {author} {\bibfnamefont {M.~S.}\ \bibnamefont {Kim}},\
  }\href@noop {} {\bibfield  {journal} {\bibinfo  {journal} {The Journal of
  Chemical Physics}\ }\textbf {\bibinfo {volume} {116}},\ \bibinfo {pages}
  {10361} (\bibinfo {year} {2002})}\BibitemShut {NoStop}%
\bibitem [{\citenamefont {Gilbert}, \citenamefont {Sauvageau},\ and\
  \citenamefont {Sandorfy}(1972)}]{gilbert_cpl17}%
  \BibitemOpen
  \bibfield  {author} {\bibinfo {author} {\bibfnamefont {R.}~\bibnamefont
  {Gilbert}}, \bibinfo {author} {\bibfnamefont {P.}~\bibnamefont {Sauvageau}},
  \ and\ \bibinfo {author} {\bibfnamefont {C.}~\bibnamefont {Sandorfy}},\
  }\href@noop {} {\bibfield  {journal} {\bibinfo  {journal} {Chem. Phys.
  Lett.}\ }\textbf {\bibinfo {volume} {17}},\ \bibinfo {pages} {465} (\bibinfo
  {year} {1972})}\BibitemShut {NoStop}%
\bibitem [{\citenamefont {Bieri}, \citenamefont {{\AA}sbrink},\ and\
  \citenamefont {Niessen}(1981)}]{bieri_jesrp23}%
  \BibitemOpen
  \bibfield  {author} {\bibinfo {author} {\bibfnamefont {G.}~\bibnamefont
  {Bieri}}, \bibinfo {author} {\bibfnamefont {L.}~\bibnamefont {{\AA}sbrink}},
  \ and\ \bibinfo {author} {\bibfnamefont {W.~V.}\ \bibnamefont {Niessen}},\
  }\href@noop {} {\bibfield  {journal} {\bibinfo  {journal} {J. Electron
  Spectrosc. Relat. Phenom.}\ }\textbf {\bibinfo {volume} {23}},\ \bibinfo
  {pages} {281} (\bibinfo {year} {1981})}\BibitemShut {NoStop}%
\bibitem [{\citenamefont {Dujardin}\ \emph {et~al.}(1983)\citenamefont
  {Dujardin}, \citenamefont {Leach}, \citenamefont {Dutuit}, \citenamefont
  {Govers},\ and\ \citenamefont {Guyon}}]{dujardin_jcp79}%
  \BibitemOpen
  \bibfield  {author} {\bibinfo {author} {\bibfnamefont {G.}~\bibnamefont
  {Dujardin}}, \bibinfo {author} {\bibfnamefont {S.}~\bibnamefont {Leach}},
  \bibinfo {author} {\bibfnamefont {O.}~\bibnamefont {Dutuit}}, \bibinfo
  {author} {\bibfnamefont {T.}~\bibnamefont {Govers}}, \ and\ \bibinfo {author}
  {\bibfnamefont {P.~M.}\ \bibnamefont {Guyon}},\ }\href@noop {} {\bibfield
  {journal} {\bibinfo  {journal} {J. Chem. Phys.}\ }\textbf {\bibinfo {volume}
  {79}},\ \bibinfo {pages} {644} (\bibinfo {year} {1983})}\BibitemShut
  {NoStop}%
\bibitem [{\citenamefont {Braden}\ and\ \citenamefont
  {Hudson}(2000)}]{braden_jpca104}%
  \BibitemOpen
  \bibfield  {author} {\bibinfo {author} {\bibfnamefont {D.~A.}\ \bibnamefont
  {Braden}}\ and\ \bibinfo {author} {\bibfnamefont {B.~S.}\ \bibnamefont
  {Hudson}},\ }\href@noop {} {\bibfield  {journal} {\bibinfo  {journal} {J.
  Phys. Chem. A}\ }\textbf {\bibinfo {volume} {104}},\ \bibinfo {pages} {982}
  (\bibinfo {year} {2000})}\BibitemShut {NoStop}%
\bibitem [{\citenamefont {Palmer}\ \emph {et~al.}(1978)\citenamefont {Palmer},
  \citenamefont {Moyes}, \citenamefont {Spiers},\ and\ \citenamefont
  {Ridyard}}]{palmer_jms49}%
  \BibitemOpen
  \bibfield  {author} {\bibinfo {author} {\bibfnamefont {M.~H.}\ \bibnamefont
  {Palmer}}, \bibinfo {author} {\bibfnamefont {W.}~\bibnamefont {Moyes}},
  \bibinfo {author} {\bibfnamefont {M.}~\bibnamefont {Spiers}}, \ and\ \bibinfo
  {author} {\bibfnamefont {J.~N.~A.}\ \bibnamefont {Ridyard}},\ }\href@noop {}
  {\bibfield  {journal} {\bibinfo  {journal} {J. Mol. Struct.}\ }\textbf
  {\bibinfo {volume} {49}},\ \bibinfo {pages} {105} (\bibinfo {year}
  {1978})}\BibitemShut {NoStop}%
\bibitem [{\citenamefont {K\"oppel}, \citenamefont {Cederbaum},\ and\
  \citenamefont {Domcke}(1988)}]{koppel_jcp89}%
  \BibitemOpen
  \bibfield  {author} {\bibinfo {author} {\bibfnamefont {H.}~\bibnamefont
  {K\"oppel}}, \bibinfo {author} {\bibfnamefont {L.~S.}\ \bibnamefont
  {Cederbaum}}, \ and\ \bibinfo {author} {\bibfnamefont {W.}~\bibnamefont
  {Domcke}},\ }\href@noop {} {\bibfield  {journal} {\bibinfo  {journal} {J.
  Chem. Phys.}\ }\textbf {\bibinfo {volume} {89}},\ \bibinfo {pages} {2023}
  (\bibinfo {year} {1988})}\BibitemShut {NoStop}%
\bibitem [{\citenamefont {K\"oppel}\ \emph {et~al.}(2002)\citenamefont
  {K\"oppel}, \citenamefont {D\"oscher}, \citenamefont {B\^aldea},
  \citenamefont {Meyer},\ and\ \citenamefont {Szalay}}]{Koppel_jcp117_2}%
  \BibitemOpen
  \bibfield  {author} {\bibinfo {author} {\bibfnamefont {H.}~\bibnamefont
  {K\"oppel}}, \bibinfo {author} {\bibfnamefont {M.}~\bibnamefont {D\"oscher}},
  \bibinfo {author} {\bibfnamefont {I.}~\bibnamefont {B\^aldea}}, \bibinfo
  {author} {\bibfnamefont {H.~D.}\ \bibnamefont {Meyer}}, \ and\ \bibinfo
  {author} {\bibfnamefont {P.~G.}\ \bibnamefont {Szalay}},\ }\href@noop {}
  {\bibfield  {journal} {\bibinfo  {journal} {J. Chem. Phys.}\ }\textbf
  {\bibinfo {volume} {117}},\ \bibinfo {pages} {2657} (\bibinfo {year}
  {2002})}\BibitemShut {NoStop}%
\bibitem [{\citenamefont {Johnson}(2002)}]{johnson_jcp117}%
  \BibitemOpen
  \bibfield  {author} {\bibinfo {author} {\bibfnamefont {P.~M.}\ \bibnamefont
  {Johnson}},\ }\href@noop {} {\bibfield  {journal} {\bibinfo  {journal} {J.
  Chem. Phys.}\ }\textbf {\bibinfo {volume} {117}},\ \bibinfo {pages} {9991}
  (\bibinfo {year} {2002})}\BibitemShut {NoStop}%
\bibitem [{\citenamefont {Baltzer}\ \emph {et~al.}(1997)\citenamefont
  {Baltzer}, \citenamefont {Karlsson}, \citenamefont {Wannberg}, \citenamefont
  {\"Ohrwall}, \citenamefont {Holland}, \citenamefont {MacDonald},
  \citenamefont {Hayes},\ and\ \citenamefont {von Niesson}}]{Baltzer_cp224}%
  \BibitemOpen
  \bibfield  {author} {\bibinfo {author} {\bibfnamefont {P.}~\bibnamefont
  {Baltzer}}, \bibinfo {author} {\bibfnamefont {L.}~\bibnamefont {Karlsson}},
  \bibinfo {author} {\bibfnamefont {B.}~\bibnamefont {Wannberg}}, \bibinfo
  {author} {\bibfnamefont {G.}~\bibnamefont {\"Ohrwall}}, \bibinfo {author}
  {\bibfnamefont {D.~M.~P.}\ \bibnamefont {Holland}}, \bibinfo {author}
  {\bibfnamefont {M.~A.}\ \bibnamefont {MacDonald}}, \bibinfo {author}
  {\bibfnamefont {M.~A.}\ \bibnamefont {Hayes}}, \ and\ \bibinfo {author}
  {\bibfnamefont {W.}~\bibnamefont {von Niesson}},\ }\href@noop {} {\bibfield
  {journal} {\bibinfo  {journal} {Chem. Phys.}\ }\textbf {\bibinfo {volume}
  {224}},\ \bibinfo {pages} {95} (\bibinfo {year} {1997})}\BibitemShut
  {NoStop}%
\bibitem [{\citenamefont {K\"oppel}, \citenamefont {B\^aldea},\ and\
  \citenamefont {Szalay}(2003)}]{Koppel_aqc44}%
  \BibitemOpen
  \bibfield  {author} {\bibinfo {author} {\bibfnamefont {H.}~\bibnamefont
  {K\"oppel}}, \bibinfo {author} {\bibfnamefont {I.}~\bibnamefont {B\^aldea}},
  \ and\ \bibinfo {author} {\bibfnamefont {P.~G.}\ \bibnamefont {Szalay}},\
  }\href@noop {} {\bibfield  {journal} {\bibinfo  {journal} {Adv. Quantum
  Chem.}\ }\textbf {\bibinfo {volume} {44}},\ \bibinfo {pages} {199} (\bibinfo
  {year} {2003})}\BibitemShut {NoStop}%
\bibitem [{\citenamefont {Schafer}\ \emph {et~al.}(1970)\citenamefont
  {Schafer}, \citenamefont {Siska}, \citenamefont {Parson}, \citenamefont
  {Tully}, \citenamefont {Wong},\ and\ \citenamefont {Lee}}]{jcp-53-3385-LEE}%
  \BibitemOpen
  \bibfield  {author} {\bibinfo {author} {\bibfnamefont {T.~P.}\ \bibnamefont
  {Schafer}}, \bibinfo {author} {\bibfnamefont {P.~E.}\ \bibnamefont {Siska}},
  \bibinfo {author} {\bibfnamefont {J.~M.}\ \bibnamefont {Parson}}, \bibinfo
  {author} {\bibfnamefont {F.~P.}\ \bibnamefont {Tully}}, \bibinfo {author}
  {\bibfnamefont {Y.~C.}\ \bibnamefont {Wong}}, \ and\ \bibinfo {author}
  {\bibfnamefont {Y.~T.}\ \bibnamefont {Lee}},\ }\href@noop {} {\bibfield
  {journal} {\bibinfo  {journal} {J. Chem. Phys.}\ }\textbf {\bibinfo {volume}
  {53}},\ \bibinfo {pages} {3385} (\bibinfo {year} {1970})}\BibitemShut
  {NoStop}%
\bibitem [{\citenamefont {Neumark}\ \emph
  {et~al.}(1985{\natexlab{b}})\citenamefont {Neumark}, \citenamefont {Wodtke},
  \citenamefont {Robinson}, \citenamefont {Hayden},\ and\ \citenamefont
  {Lee}}]{jcp-82-3045-LEE}%
  \BibitemOpen
  \bibfield  {author} {\bibinfo {author} {\bibfnamefont {D.~M.}\ \bibnamefont
  {Neumark}}, \bibinfo {author} {\bibfnamefont {A.~M.}\ \bibnamefont {Wodtke}},
  \bibinfo {author} {\bibfnamefont {G.~N.}\ \bibnamefont {Robinson}}, \bibinfo
  {author} {\bibfnamefont {C.~C.}\ \bibnamefont {Hayden}}, \ and\ \bibinfo
  {author} {\bibfnamefont {Y.~T.}\ \bibnamefont {Lee}},\ }\href {\doibase
  http://dx.doi.org/10.1063/1.448254} {\bibfield  {journal} {\bibinfo
  {journal} {J. Chem. Phys.}\ }\textbf {\bibinfo {volume} {82}},\ \bibinfo
  {pages} {3045} (\bibinfo {year} {1985}{\natexlab{b}})}\BibitemShut {NoStop}%
\bibitem [{\citenamefont {Neumark}\ \emph
  {et~al.}(1985{\natexlab{c}})\citenamefont {Neumark}, \citenamefont {Wodtke},
  \citenamefont {Robinson}, \citenamefont {Hayden}, \citenamefont {Shobatake},
  \citenamefont {Sparks}, \citenamefont {Schafer},\ and\ \citenamefont
  {Lee}}]{jcp-82-3067-LEE}%
  \BibitemOpen
  \bibfield  {author} {\bibinfo {author} {\bibfnamefont {D.~M.}\ \bibnamefont
  {Neumark}}, \bibinfo {author} {\bibfnamefont {A.~M.}\ \bibnamefont {Wodtke}},
  \bibinfo {author} {\bibfnamefont {G.~N.}\ \bibnamefont {Robinson}}, \bibinfo
  {author} {\bibfnamefont {C.~C.}\ \bibnamefont {Hayden}}, \bibinfo {author}
  {\bibfnamefont {K.}~\bibnamefont {Shobatake}}, \bibinfo {author}
  {\bibfnamefont {R.~K.}\ \bibnamefont {Sparks}}, \bibinfo {author}
  {\bibfnamefont {T.~P.}\ \bibnamefont {Schafer}}, \ and\ \bibinfo {author}
  {\bibfnamefont {Y.~T.}\ \bibnamefont {Lee}},\ }\href@noop {} {\bibfield
  {journal} {\bibinfo  {journal} {J. Chem. Phys.}\ }\textbf {\bibinfo {volume}
  {82}},\ \bibinfo {pages} {3067} (\bibinfo {year}
  {1985}{\natexlab{c}})}\BibitemShut {NoStop}%
\bibitem [{\citenamefont {Faubel}\ \emph {et~al.}(1991)\citenamefont {Faubel},
  \citenamefont {Schlemmer}, \citenamefont {Sondermann},\ and\ \citenamefont
  {Toennies}}]{jcp-94-4676-TOENN}%
  \BibitemOpen
  \bibfield  {author} {\bibinfo {author} {\bibfnamefont {M.}~\bibnamefont
  {Faubel}}, \bibinfo {author} {\bibfnamefont {S.}~\bibnamefont {Schlemmer}},
  \bibinfo {author} {\bibfnamefont {F.}~\bibnamefont {Sondermann}}, \ and\
  \bibinfo {author} {\bibfnamefont {J.~P.}\ \bibnamefont {Toennies}},\
  }\href@noop {} {\bibfield  {journal} {\bibinfo  {journal} {J. Chem. Phys.}\
  }\textbf {\bibinfo {volume} {94}},\ \bibinfo {pages} {4676} (\bibinfo {year}
  {1991})}\BibitemShut {NoStop}%
\bibitem [{\citenamefont {Faubel}\ \emph {et~al.}(1994)\citenamefont {Faubel},
  \citenamefont {Rusin}, \citenamefont {Schlemmer}, \citenamefont {Sondermann},
  \citenamefont {Tappe},\ and\ \citenamefont {Toennies}}]{jcp-101-2106-TOENN}%
  \BibitemOpen
  \bibfield  {author} {\bibinfo {author} {\bibfnamefont {M.}~\bibnamefont
  {Faubel}}, \bibinfo {author} {\bibfnamefont {L.}~\bibnamefont {Rusin}},
  \bibinfo {author} {\bibfnamefont {S.}~\bibnamefont {Schlemmer}}, \bibinfo
  {author} {\bibfnamefont {F.}~\bibnamefont {Sondermann}}, \bibinfo {author}
  {\bibfnamefont {U.}~\bibnamefont {Tappe}}, \ and\ \bibinfo {author}
  {\bibfnamefont {J.~P.}\ \bibnamefont {Toennies}},\ }\href@noop {} {\bibfield
  {journal} {\bibinfo  {journal} {J. Chem. Phys.}\ }\textbf {\bibinfo {volume}
  {101}},\ \bibinfo {pages} {2106} (\bibinfo {year} {1994})}\BibitemShut
  {NoStop}%
\bibitem [{\citenamefont {Baer}\ \emph
  {et~al.}(1996{\natexlab{b}})\citenamefont {Baer}, \citenamefont {Faubel},
  \citenamefont {Martinez‐Haya}, \citenamefont {Rusin}, \citenamefont
  {Tappe},\ and\ \citenamefont {Toennies}}]{jcp-104-2743-TOENN}%
  \BibitemOpen
  \bibfield  {author} {\bibinfo {author} {\bibfnamefont {M.}~\bibnamefont
  {Baer}}, \bibinfo {author} {\bibfnamefont {M.}~\bibnamefont {Faubel}},
  \bibinfo {author} {\bibfnamefont {B.}~\bibnamefont {Martinez‐Haya}},
  \bibinfo {author} {\bibfnamefont {L.~Y.}\ \bibnamefont {Rusin}}, \bibinfo
  {author} {\bibfnamefont {U.}~\bibnamefont {Tappe}}, \ and\ \bibinfo {author}
  {\bibfnamefont {J.~P.}\ \bibnamefont {Toennies}},\ }\href@noop {} {\bibfield
  {journal} {\bibinfo  {journal} {J. Chem. Phys.}\ }\textbf {\bibinfo {volume}
  {104}},\ \bibinfo {pages} {2743} (\bibinfo {year}
  {1996}{\natexlab{b}})}\BibitemShut {NoStop}%
\bibitem [{\citenamefont {Faubel}\ \emph {et~al.}(1997)\citenamefont {Faubel},
  \citenamefont {Martinez-Haya}, \citenamefont {Rusin}, \citenamefont {Tappe},\
  and\ \citenamefont {Toennies}}]{jpcA-101-6415-TOENN}%
  \BibitemOpen
  \bibfield  {author} {\bibinfo {author} {\bibfnamefont {M.}~\bibnamefont
  {Faubel}}, \bibinfo {author} {\bibfnamefont {B.}~\bibnamefont
  {Martinez-Haya}}, \bibinfo {author} {\bibfnamefont {L.~Y.}\ \bibnamefont
  {Rusin}}, \bibinfo {author} {\bibfnamefont {U.}~\bibnamefont {Tappe}}, \ and\
  \bibinfo {author} {\bibfnamefont {J.~P.}\ \bibnamefont {Toennies}},\
  }\href@noop {} {\bibfield  {journal} {\bibinfo  {journal} {J. Phys. Chem. A}\
  }\textbf {\bibinfo {volume} {101}},\ \bibinfo {pages} {6415} (\bibinfo {year}
  {1997})}\BibitemShut {NoStop}%
\bibitem [{\citenamefont {Baer}\ \emph {et~al.}(1998)\citenamefont {Baer},
  \citenamefont {Faubel}, \citenamefont {Martinez‐Haya}, \citenamefont
  {Rusin}, \citenamefont {Tappe},\ and\ \citenamefont
  {Toennies}}]{jcp-108-9694-TOENN}%
  \BibitemOpen
  \bibfield  {author} {\bibinfo {author} {\bibfnamefont {M.}~\bibnamefont
  {Baer}}, \bibinfo {author} {\bibfnamefont {M.}~\bibnamefont {Faubel}},
  \bibinfo {author} {\bibfnamefont {B.}~\bibnamefont {Martinez‐Haya}},
  \bibinfo {author} {\bibfnamefont {L.~Y.}\ \bibnamefont {Rusin}}, \bibinfo
  {author} {\bibfnamefont {U.}~\bibnamefont {Tappe}}, \ and\ \bibinfo {author}
  {\bibfnamefont {J.~P.}\ \bibnamefont {Toennies}},\ }\href@noop {} {\bibfield
  {journal} {\bibinfo  {journal} {J. Chem. Phys.}\ }\textbf {\bibinfo {volume}
  {108}},\ \bibinfo {pages} {9694} (\bibinfo {year} {1998})}\BibitemShut
  {NoStop}%
\bibitem [{\citenamefont {Wurzberg}\ and\ \citenamefont
  {Houston}(1980)}]{jcp-72-4811-HOUSTON}%
  \BibitemOpen
  \bibfield  {author} {\bibinfo {author} {\bibfnamefont {E.}~\bibnamefont
  {Wurzberg}}\ and\ \bibinfo {author} {\bibfnamefont {P.~L.}\ \bibnamefont
  {Houston}},\ }\href@noop {} {\bibfield  {journal} {\bibinfo  {journal} {J.
  Chem. Phys.}\ }\textbf {\bibinfo {volume} {72}},\ \bibinfo {pages} {4811}
  (\bibinfo {year} {1980})}\BibitemShut {NoStop}%
\bibitem [{\citenamefont {Heidner}\ \emph {et~al.}(1980)\citenamefont
  {Heidner}, \citenamefont {Bott}, \citenamefont {Gardner},\ and\ \citenamefont
  {Melzer}}]{jcp-72-4815-HEIDNER}%
  \BibitemOpen
  \bibfield  {author} {\bibinfo {author} {\bibfnamefont {R.~F.}\ \bibnamefont
  {Heidner}}, \bibinfo {author} {\bibfnamefont {J.~F.}\ \bibnamefont {Bott}},
  \bibinfo {author} {\bibfnamefont {C.~E.}\ \bibnamefont {Gardner}}, \ and\
  \bibinfo {author} {\bibfnamefont {J.~E.}\ \bibnamefont {Melzer}},\
  }\href@noop {} {\bibfield  {journal} {\bibinfo  {journal} {J. Chem. Phys.}\
  }\textbf {\bibinfo {volume} {72}},\ \bibinfo {pages} {4815} (\bibinfo {year}
  {1980})}\BibitemShut {NoStop}%
\bibitem [{\citenamefont {Stevens}, \citenamefont {Brune},\ and\ \citenamefont
  {Anderson}(1989)}]{jpc-93-4068-ANDERSON}%
  \BibitemOpen
  \bibfield  {author} {\bibinfo {author} {\bibfnamefont {P.~S.}\ \bibnamefont
  {Stevens}}, \bibinfo {author} {\bibfnamefont {W.~H.}\ \bibnamefont {Brune}},
  \ and\ \bibinfo {author} {\bibfnamefont {J.~G.}\ \bibnamefont {Anderson}},\
  }\href@noop {} {\bibfield  {journal} {\bibinfo  {journal} {J. Phys. Chem.}\
  }\textbf {\bibinfo {volume} {93}},\ \bibinfo {pages} {4068} (\bibinfo {year}
  {1989})}\BibitemShut {NoStop}%
\bibitem [{\citenamefont {Tizniti}\ \emph {et~al.}(2014)\citenamefont
  {Tizniti}, \citenamefont {Picard}, \citenamefont {Lique}, \citenamefont
  {Berteloite}, \citenamefont {Canosa}, \citenamefont {Alexander},\ and\
  \citenamefont {Sims}}]{nchem-6-141-SIMS}%
  \BibitemOpen
  \bibfield  {author} {\bibinfo {author} {\bibfnamefont {M.}~\bibnamefont
  {Tizniti}}, \bibinfo {author} {\bibfnamefont {S.~D.~L.}\ \bibnamefont
  {Picard}}, \bibinfo {author} {\bibfnamefont {F.}~\bibnamefont {Lique}},
  \bibinfo {author} {\bibfnamefont {C.}~\bibnamefont {Berteloite}}, \bibinfo
  {author} {\bibfnamefont {A.}~\bibnamefont {Canosa}}, \bibinfo {author}
  {\bibfnamefont {M.~H.}\ \bibnamefont {Alexander}}, \ and\ \bibinfo {author}
  {\bibfnamefont {I.~R.}\ \bibnamefont {Sims}},\ }\href {\doibase
  10.1038/NCHEM.1835} {\bibfield  {journal} {\bibinfo  {journal} {Nature
  Chem.}\ }\textbf {\bibinfo {volume} {6}},\ \bibinfo {pages} {141} (\bibinfo
  {year} {2014})}\BibitemShut {NoStop}%
\bibitem [{\citenamefont {Bedjanian}(2021{\natexlab{a}})}]{bedjaninan}%
  \BibitemOpen
  \bibfield  {author} {\bibinfo {author} {\bibfnamefont {Y.}~\bibnamefont
  {Bedjanian}},\ }\href {\doibase https://doi.org/10.1002/kin.21462} {\bibfield
   {journal} {\bibinfo  {journal} {Int. J. Chem. Kin.}\ }\textbf {\bibinfo
  {volume} {53}},\ \bibinfo {pages} {527} (\bibinfo {year}
  {2021}{\natexlab{a}})},\ \Eprint
  {http://arxiv.org/abs/https://onlinelibrary.wiley.com/doi/pdf/10.1002/kin.21462}
  {https://onlinelibrary.wiley.com/doi/pdf/10.1002/kin.21462} \BibitemShut
  {NoStop}%
\bibitem [{\citenamefont {Aquilanti}\ \emph {et~al.}(2005)\citenamefont
  {Aquilanti}, \citenamefont {Cavalli}, \citenamefont {De~Fazio}, \citenamefont
  {Volpi}, \citenamefont {Aguilar},\ and\ \citenamefont {Lucas}}]{aql2}%
  \BibitemOpen
  \bibfield  {author} {\bibinfo {author} {\bibfnamefont {V.}~\bibnamefont
  {Aquilanti}}, \bibinfo {author} {\bibfnamefont {S.}~\bibnamefont {Cavalli}},
  \bibinfo {author} {\bibfnamefont {D.}~\bibnamefont {De~Fazio}}, \bibinfo
  {author} {\bibfnamefont {A.}~\bibnamefont {Volpi}}, \bibinfo {author}
  {\bibfnamefont {A.}~\bibnamefont {Aguilar}}, \ and\ \bibinfo {author}
  {\bibfnamefont {J.~M.}\ \bibnamefont {Lucas}},\ }\href {\doibase
  10.1016/j.chemphys.2004.03.027} {\bibfield  {journal} {\bibinfo  {journal}
  {Chem. Phys.}\ }\textbf {\bibinfo {volume} {308}},\ \bibinfo {pages} {237}
  (\bibinfo {year} {2005})}\BibitemShut {NoStop}%
\bibitem [{\citenamefont {Chen}, \citenamefont {Sun},\ and\ \citenamefont
  {Zhang}(2015)}]{jcp-142-024303-ZHANG}%
  \BibitemOpen
  \bibfield  {author} {\bibinfo {author} {\bibfnamefont {J.}~\bibnamefont
  {Chen}}, \bibinfo {author} {\bibfnamefont {Z.}~\bibnamefont {Sun}}, \ and\
  \bibinfo {author} {\bibfnamefont {D.~H.}\ \bibnamefont {Zhang}},\ }\href@noop
  {} {\bibfield  {journal} {\bibinfo  {journal} {J. Chem. Phys.}\ }\textbf
  {\bibinfo {volume} {142}},\ \bibinfo {pages} {024303} (\bibinfo {year}
  {2015})}\BibitemShut {NoStop}%
\bibitem [{\citenamefont {Fu}, \citenamefont {Xu},\ and\ \citenamefont
  {Zhang}(2008)}]{fxz_pes}%
  \BibitemOpen
  \bibfield  {author} {\bibinfo {author} {\bibfnamefont {B.}~\bibnamefont
  {Fu}}, \bibinfo {author} {\bibfnamefont {X.}~\bibnamefont {Xu}}, \ and\
  \bibinfo {author} {\bibfnamefont {D.~H.}\ \bibnamefont {Zhang}},\ }\href
  {\doibase 10.1063/1.2955729} {\bibfield  {journal} {\bibinfo  {journal} {J.
  Chem. Phys.}\ }\textbf {\bibinfo {volume} {129}},\ \bibinfo {pages} {011103}
  (\bibinfo {year} {2008})}\BibitemShut {NoStop}%
\bibitem [{\citenamefont {Naskar}, \citenamefont {Ghosh},\ and\ \citenamefont
  {Adhikari}(2022{\natexlab{a}})}]{Naskar2022_JPCA_126}%
  \BibitemOpen
  \bibfield  {author} {\bibinfo {author} {\bibfnamefont {K.}~\bibnamefont
  {Naskar}}, \bibinfo {author} {\bibfnamefont {S.}~\bibnamefont {Ghosh}}, \
  and\ \bibinfo {author} {\bibfnamefont {S.}~\bibnamefont {Adhikari}},\ }\href
  {\doibase 10.1021/acs.jpca.2c01209} {\bibfield  {journal} {\bibinfo
  {journal} {J. Phys. Chem. A}\ }\textbf {\bibinfo {volume} {126}},\ \bibinfo
  {pages} {3311} (\bibinfo {year} {2022}{\natexlab{a}})}\BibitemShut {NoStop}%
\bibitem [{\citenamefont {Naskar}\ \emph
  {et~al.}(2024{\natexlab{b}})\citenamefont {Naskar}, \citenamefont
  {Mukherjee}, \citenamefont {Ghosh},\ and\ \citenamefont
  {Adhikari}}]{FH2_dynamics}%
  \BibitemOpen
  \bibfield  {author} {\bibinfo {author} {\bibfnamefont {K.}~\bibnamefont
  {Naskar}}, \bibinfo {author} {\bibfnamefont {S.}~\bibnamefont {Mukherjee}},
  \bibinfo {author} {\bibfnamefont {S.}~\bibnamefont {Ghosh}}, \ and\ \bibinfo
  {author} {\bibfnamefont {S.}~\bibnamefont {Adhikari}},\ }\href@noop {}
  {\bibfield  {journal} {\bibinfo  {journal} {J. Phys. Chem. A}\ }\textbf
  {\bibinfo {volume} {128}},\ \bibinfo {pages} {1438} (\bibinfo {year}
  {2024}{\natexlab{b}})}\BibitemShut {NoStop}%
\bibitem [{\citenamefont {Castillo}\ \emph
  {et~al.}(1996{\natexlab{a}})\citenamefont {Castillo}, \citenamefont
  {Manolopoulos}, \citenamefont {Stark},\ and\ \citenamefont
  {Werner}}]{werner_jcp104}%
  \BibitemOpen
  \bibfield  {author} {\bibinfo {author} {\bibfnamefont {J.~F.}\ \bibnamefont
  {Castillo}}, \bibinfo {author} {\bibfnamefont {D.~E.}\ \bibnamefont
  {Manolopoulos}}, \bibinfo {author} {\bibfnamefont {K.}~\bibnamefont {Stark}},
  \ and\ \bibinfo {author} {\bibfnamefont {H.-J.}\ \bibnamefont {Werner}},\
  }\href@noop {} {\bibfield  {journal} {\bibinfo  {journal} {J. Chem. Phys.}\
  }\textbf {\bibinfo {volume} {104}},\ \bibinfo {pages} {6531} (\bibinfo {year}
  {1996}{\natexlab{a}})}\BibitemShut {NoStop}%
\bibitem [{\citenamefont {Mukherjee}\ \emph
  {et~al.}(2020{\natexlab{b}})\citenamefont {Mukherjee}, \citenamefont
  {Naskar}, \citenamefont {Mukherjee}, \citenamefont {Ravi}, \citenamefont
  {Shamasundar}, \citenamefont {Mukhopadhyay},\ and\ \citenamefont
  {Adhikari}}]{FH2_Surf}%
  \BibitemOpen
  \bibfield  {author} {\bibinfo {author} {\bibfnamefont {B.}~\bibnamefont
  {Mukherjee}}, \bibinfo {author} {\bibfnamefont {K.}~\bibnamefont {Naskar}},
  \bibinfo {author} {\bibfnamefont {S.}~\bibnamefont {Mukherjee}}, \bibinfo
  {author} {\bibfnamefont {S.}~\bibnamefont {Ravi}}, \bibinfo {author}
  {\bibfnamefont {K.~R.}\ \bibnamefont {Shamasundar}}, \bibinfo {author}
  {\bibfnamefont {D.}~\bibnamefont {Mukhopadhyay}}, \ and\ \bibinfo {author}
  {\bibfnamefont {S.}~\bibnamefont {Adhikari}},\ }\href@noop {} {\bibfield
  {journal} {\bibinfo  {journal} {J. Chem. Phys.}\ }\textbf {\bibinfo {volume}
  {153}},\ \bibinfo {pages} {174301} (\bibinfo {year}
  {2020}{\natexlab{b}})}\BibitemShut {NoStop}%
\bibitem [{\citenamefont {Das}\ \emph {et~al.}(2011)\citenamefont {Das},
  \citenamefont {Mukhopadhyay}, \citenamefont {Adhikari},\ and\ \citenamefont
  {Baer}}]{cpl-517-92-ANITA}%
  \BibitemOpen
  \bibfield  {author} {\bibinfo {author} {\bibfnamefont {A.}~\bibnamefont
  {Das}}, \bibinfo {author} {\bibfnamefont {D.}~\bibnamefont {Mukhopadhyay}},
  \bibinfo {author} {\bibfnamefont {S.}~\bibnamefont {Adhikari}}, \ and\
  \bibinfo {author} {\bibfnamefont {M.}~\bibnamefont {Baer}},\ }\href@noop {}
  {\bibfield  {journal} {\bibinfo  {journal} {Chem. Phys. Lett.}\ }\textbf
  {\bibinfo {volume} {517}},\ \bibinfo {pages} {92} (\bibinfo {year}
  {2011})}\BibitemShut {NoStop}%
\bibitem [{\citenamefont {Oka}(1980)}]{OKA:PRL}%
  \BibitemOpen
  \bibfield  {author} {\bibinfo {author} {\bibfnamefont {T.}~\bibnamefont
  {Oka}},\ }\href@noop {} {\bibfield  {journal} {\bibinfo  {journal} {Phys.
  Rev. Lett.}\ }\textbf {\bibinfo {volume} {45}},\ \bibinfo {pages} {531}
  (\bibinfo {year} {1980})}\BibitemShut {NoStop}%
\bibitem [{\citenamefont {Mukherjee}, \citenamefont {Mukhopadhyay},\ and\
  \citenamefont {Adhikari}(2014{\natexlab{b}})}]{H3+_Surf}%
  \BibitemOpen
  \bibfield  {author} {\bibinfo {author} {\bibfnamefont {S.}~\bibnamefont
  {Mukherjee}}, \bibinfo {author} {\bibfnamefont {D.}~\bibnamefont
  {Mukhopadhyay}}, \ and\ \bibinfo {author} {\bibfnamefont {S.}~\bibnamefont
  {Adhikari}},\ }\href@noop {} {\bibfield  {journal} {\bibinfo  {journal} {J.
  Chem. Phys.}\ }\textbf {\bibinfo {volume} {141}},\ \bibinfo {pages} {204306}
  (\bibinfo {year} {2014}{\natexlab{b}})}\BibitemShut {NoStop}%
\bibitem [{\citenamefont {Fehsenfeld}\ \emph {et~al.}(1974)\citenamefont
  {Fehsenfeld}, \citenamefont {Albritton}, \citenamefont {Bush}, \citenamefont
  {Fournier}, \citenamefont {Govers},\ and\ \citenamefont
  {Fournier}}]{jcp-61-2150-FOUR}%
  \BibitemOpen
  \bibfield  {author} {\bibinfo {author} {\bibfnamefont {F.~C.}\ \bibnamefont
  {Fehsenfeld}}, \bibinfo {author} {\bibfnamefont {D.~L.}\ \bibnamefont
  {Albritton}}, \bibinfo {author} {\bibfnamefont {Y.~A.}\ \bibnamefont {Bush}},
  \bibinfo {author} {\bibfnamefont {P.~G.}\ \bibnamefont {Fournier}}, \bibinfo
  {author} {\bibfnamefont {T.~R.}\ \bibnamefont {Govers}}, \ and\ \bibinfo
  {author} {\bibfnamefont {J.}~\bibnamefont {Fournier}},\ }\href@noop {}
  {\bibfield  {journal} {\bibinfo  {journal} {J. Chem. Phys.}\ }\textbf
  {\bibinfo {volume} {61}},\ \bibinfo {pages} {2150} (\bibinfo {year}
  {1974})}\BibitemShut {NoStop}%
\bibitem [{\citenamefont {Villinger}, \citenamefont {Henchman},\ and\
  \citenamefont {Lindinger}(1982)}]{jcp-76-1590-HEN2}%
  \BibitemOpen
  \bibfield  {author} {\bibinfo {author} {\bibfnamefont {H.}~\bibnamefont
  {Villinger}}, \bibinfo {author} {\bibfnamefont {M.~J.}\ \bibnamefont
  {Henchman}}, \ and\ \bibinfo {author} {\bibfnamefont {W.}~\bibnamefont
  {Lindinger}},\ }\href@noop {} {\bibfield  {journal} {\bibinfo  {journal} {J.
  Chem. Phys.}\ }\textbf {\bibinfo {volume} {76}},\ \bibinfo {pages} {1590}
  (\bibinfo {year} {1982})}\BibitemShut {NoStop}%
\bibitem [{\citenamefont {Gerlich}(1992)}]{advcp-82-1-GER}%
  \BibitemOpen
  \bibfield  {author} {\bibinfo {author} {\bibfnamefont {D.}~\bibnamefont
  {Gerlich}},\ }\href@noop {} {\bibfield  {journal} {\bibinfo  {journal} {Adv.
  Chem. Phys.}\ }\textbf {\bibinfo {volume} {82}},\ \bibinfo {pages} {1}
  (\bibinfo {year} {1992})}\BibitemShut {NoStop}%
\bibitem [{\citenamefont {Gerlich}(1990)}]{Gerlich_1990}%
  \BibitemOpen
  \bibfield  {author} {\bibinfo {author} {\bibfnamefont {D.}~\bibnamefont
  {Gerlich}},\ }\href@noop {} {\bibfield  {journal} {\bibinfo  {journal} {J.
  Chem. Phys.}\ }\textbf {\bibinfo {volume} {92}},\ \bibinfo {pages} {2377}
  (\bibinfo {year} {1990})}\BibitemShut {NoStop}%
\bibitem [{\citenamefont {Honvault}\ \emph {et~al.}(2012)\citenamefont
  {Honvault}, \citenamefont {Jorfi}, \citenamefont {Gonz\'alez-Lezana},
  \citenamefont {Faure},\ and\ \citenamefont {Pagani}}]{HONV_2011_Er}%
  \BibitemOpen
  \bibfield  {author} {\bibinfo {author} {\bibfnamefont {P.}~\bibnamefont
  {Honvault}}, \bibinfo {author} {\bibfnamefont {M.}~\bibnamefont {Jorfi}},
  \bibinfo {author} {\bibfnamefont {T.}~\bibnamefont {Gonz\'alez-Lezana}},
  \bibinfo {author} {\bibfnamefont {A.}~\bibnamefont {Faure}}, \ and\ \bibinfo
  {author} {\bibfnamefont {L.}~\bibnamefont {Pagani}},\ }\href@noop {}
  {\bibfield  {journal} {\bibinfo  {journal} {Phys. Rev. Lett.}\ }\textbf
  {\bibinfo {volume} {108}},\ \bibinfo {pages} {109903} (\bibinfo {year}
  {2012})}\BibitemShut {NoStop}%
\bibitem [{\citenamefont {Hazra}\ \emph {et~al.}(2024)\citenamefont {Hazra},
  \citenamefont {Naskar}, \citenamefont {Adhikari},\ and\ \citenamefont
  {Varandas}}]{ortho-para_2024}%
  \BibitemOpen
  \bibfield  {author} {\bibinfo {author} {\bibfnamefont {S.}~\bibnamefont
  {Hazra}}, \bibinfo {author} {\bibfnamefont {K.}~\bibnamefont {Naskar}},
  \bibinfo {author} {\bibfnamefont {S.}~\bibnamefont {Adhikari}}, \ and\
  \bibinfo {author} {\bibfnamefont {A.~J.~C.}\ \bibnamefont {Varandas}},\
  }\href {\doibase 10.1021/acs.jpca.4c02243} {\bibfield  {journal} {\bibinfo
  {journal} {The Journal of Physical Chemistry A}\ }\textbf {\bibinfo {volume}
  {128}},\ \bibinfo {pages} {8833} (\bibinfo {year} {2024})}\BibitemShut
  {NoStop}%
\bibitem [{\citenamefont {Ghosh}\ \emph
  {et~al.}(2021{\natexlab{a}})\citenamefont {Ghosh}, \citenamefont {Sahoo},
  \citenamefont {Baer},\ and\ \citenamefont {Adhikari}}]{H3+_dynamics_2021}%
  \BibitemOpen
  \bibfield  {author} {\bibinfo {author} {\bibfnamefont {S.}~\bibnamefont
  {Ghosh}}, \bibinfo {author} {\bibfnamefont {T.}~\bibnamefont {Sahoo}},
  \bibinfo {author} {\bibfnamefont {M.}~\bibnamefont {Baer}}, \ and\ \bibinfo
  {author} {\bibfnamefont {S.}~\bibnamefont {Adhikari}},\ }\href@noop {}
  {\bibfield  {journal} {\bibinfo  {journal} {J. Phys. Chem. A}\ }\textbf
  {\bibinfo {volume} {125}},\ \bibinfo {pages} {731} (\bibinfo {year}
  {2021}{\natexlab{a}})}\BibitemShut {NoStop}%
\bibitem [{\citenamefont {Klein}\ and\ \citenamefont
  {Friedman}(1964)}]{Klein1964_JCP_41}%
  \BibitemOpen
  \bibfield  {author} {\bibinfo {author} {\bibfnamefont {F.~S.}\ \bibnamefont
  {Klein}}\ and\ \bibinfo {author} {\bibfnamefont {L.}~\bibnamefont
  {Friedman}},\ }\href {\doibase 10.1063/1.1726159} {\bibfield  {journal}
  {\bibinfo  {journal} {J. Chem. Phys.}\ }\textbf {\bibinfo {volume} {41}},\
  \bibinfo {pages} {1789} (\bibinfo {year} {1964})}\BibitemShut {NoStop}%
\bibitem [{\citenamefont {Chupka}\ and\ \citenamefont
  {Russell}(1968)}]{chupka1968_1}%
  \BibitemOpen
  \bibfield  {author} {\bibinfo {author} {\bibfnamefont {W.~A.}\ \bibnamefont
  {Chupka}}\ and\ \bibinfo {author} {\bibfnamefont {M.~E.}\ \bibnamefont
  {Russell}},\ }\href {\doibase 10.1063/1.1670068} {\bibfield  {journal}
  {\bibinfo  {journal} {J. Chem. Phys.}\ }\textbf {\bibinfo {volume} {49}},\
  \bibinfo {pages} {5426} (\bibinfo {year} {1968})}\BibitemShut {NoStop}%
\bibitem [{\citenamefont {Chupka}, \citenamefont {Berkowitz},\ and\
  \citenamefont {Russell}(1969)}]{chupka1968_2}%
  \BibitemOpen
  \bibfield  {author} {\bibinfo {author} {\bibfnamefont {W.~A.}\ \bibnamefont
  {Chupka}}, \bibinfo {author} {\bibfnamefont {J.}~\bibnamefont {Berkowitz}}, \
  and\ \bibinfo {author} {\bibfnamefont {M.}~\bibnamefont {Russell}},\
  }\href@noop {} {\bibfield  {journal} {\bibinfo  {journal} {6th Int. Conf. on
  the Physics of Electronic and Atomic Collisions: ICPEAC VI}\ ,\ \bibinfo
  {pages} {71}} (\bibinfo {year} {1969})}\BibitemShut {NoStop}%
\bibitem [{\citenamefont {Tang}\ \emph
  {et~al.}(2005{\natexlab{a}})\citenamefont {Tang}, \citenamefont {Xu},
  \citenamefont {Zhang}, \citenamefont {Hou}, \citenamefont {Chang},
  \citenamefont {Ng}, \citenamefont {Chiu}, \citenamefont {Dressler},\ and\
  \citenamefont {Levandier}}]{Tang2005_JCP_122}%
  \BibitemOpen
  \bibfield  {author} {\bibinfo {author} {\bibfnamefont {X.~N.}\ \bibnamefont
  {Tang}}, \bibinfo {author} {\bibfnamefont {H.}~\bibnamefont {Xu}}, \bibinfo
  {author} {\bibfnamefont {T.}~\bibnamefont {Zhang}}, \bibinfo {author}
  {\bibfnamefont {Y.}~\bibnamefont {Hou}}, \bibinfo {author} {\bibfnamefont
  {C.}~\bibnamefont {Chang}}, \bibinfo {author} {\bibfnamefont {C.~Y.}\
  \bibnamefont {Ng}}, \bibinfo {author} {\bibfnamefont {Y.}~\bibnamefont
  {Chiu}}, \bibinfo {author} {\bibfnamefont {R.~A.}\ \bibnamefont {Dressler}},
  \ and\ \bibinfo {author} {\bibfnamefont {D.~J.}\ \bibnamefont {Levandier}},\
  }\href {\doibase 10.1063/1.1883169} {\bibfield  {journal} {\bibinfo
  {journal} {J. Chem. Phys.}\ }\textbf {\bibinfo {volume} {122}},\ \bibinfo
  {pages} {164301} (\bibinfo {year} {2005}{\natexlab{a}})}\BibitemShut
  {NoStop}%
\bibitem [{\citenamefont {Turner}, \citenamefont {Dutuit},\ and\ \citenamefont
  {Lee}()}]{Turner2022_JCP_81}%
  \BibitemOpen
  \bibfield  {author} {\bibinfo {author} {\bibfnamefont {T.}~\bibnamefont
  {Turner}}, \bibinfo {author} {\bibfnamefont {O.}~\bibnamefont {Dutuit}}, \
  and\ \bibinfo {author} {\bibfnamefont {Y.~T.}\ \bibnamefont {Lee}},\
  }\href@noop {} {\ }\BibitemShut {NoStop}%
\bibitem [{\citenamefont {Maiti}\ \emph {et~al.}(2002)\citenamefont {Maiti},
  \citenamefont {Kalyanaraman}, \citenamefont {Panda},\ and\ \citenamefont
  {Sathyamurthy}}]{Maiti2002_JCP_117}%
  \BibitemOpen
  \bibfield  {author} {\bibinfo {author} {\bibfnamefont {B.}~\bibnamefont
  {Maiti}}, \bibinfo {author} {\bibfnamefont {C.}~\bibnamefont {Kalyanaraman}},
  \bibinfo {author} {\bibfnamefont {A.~N.}\ \bibnamefont {Panda}}, \ and\
  \bibinfo {author} {\bibfnamefont {N.}~\bibnamefont {Sathyamurthy}},\ }\href
  {\doibase 10.1063/1.1518002} {\bibfield  {journal} {\bibinfo  {journal} {J.
  Chem. Phys.}\ }\textbf {\bibinfo {volume} {117}},\ \bibinfo {pages} {9719}
  (\bibinfo {year} {2002})}\BibitemShut {NoStop}%
\bibitem [{\citenamefont {Kraemer}, \citenamefont {\v{S}pirko},\ and\
  \citenamefont {Bludsk\'{y}}(2002)}]{Kraemer2002_CP_276}%
  \BibitemOpen
  \bibfield  {author} {\bibinfo {author} {\bibfnamefont {W.~P.}\ \bibnamefont
  {Kraemer}}, \bibinfo {author} {\bibfnamefont {V.}~\bibnamefont {\v{S}pirko}},
  \ and\ \bibinfo {author} {\bibfnamefont {O.}~\bibnamefont {Bludsk\'{y}}},\
  }\href {\doibase 10.1016/S0301-0104(01)00694-2} {\bibfield  {journal}
  {\bibinfo  {journal} {Chem. Phys.}\ }\textbf {\bibinfo {volume} {276}},\
  \bibinfo {pages} {225} (\bibinfo {year} {2002})}\BibitemShut {NoStop}%
\bibitem [{\citenamefont {Mruga\l{}a}, \citenamefont {\v{S}pirko},\ and\
  \citenamefont {Kraemer}(2003)}]{Mrugala2003_JCP_118}%
  \BibitemOpen
  \bibfield  {author} {\bibinfo {author} {\bibfnamefont {F.}~\bibnamefont
  {Mruga\l{}a}}, \bibinfo {author} {\bibfnamefont {V.}~\bibnamefont
  {\v{S}pirko}}, \ and\ \bibinfo {author} {\bibfnamefont {W.~P.}\ \bibnamefont
  {Kraemer}},\ }\href {\doibase 10.1063/1.1573184} {\bibfield  {journal}
  {\bibinfo  {journal} {J. Chem. Phys.}\ }\textbf {\bibinfo {volume} {118}},\
  \bibinfo {pages} {10547} (\bibinfo {year} {2003})}\BibitemShut {NoStop}%
\bibitem [{\citenamefont {Tiwari}\ and\ \citenamefont
  {Sathyamurthy}(2005)}]{Tiwari2005_CPL_414}%
  \BibitemOpen
  \bibfield  {author} {\bibinfo {author} {\bibfnamefont {A.~K.}\ \bibnamefont
  {Tiwari}}\ and\ \bibinfo {author} {\bibfnamefont {N.}~\bibnamefont
  {Sathyamurthy}},\ }\href {\doibase 10.1016/j.cplett.2005.08.130} {\bibfield
  {journal} {\bibinfo  {journal} {Chem. Phys. Lett.}\ }\textbf {\bibinfo
  {volume} {414}},\ \bibinfo {pages} {509} (\bibinfo {year}
  {2005})}\BibitemShut {NoStop}%
\bibitem [{\citenamefont {Panda}\ and\ \citenamefont
  {Sathyamurthy}(2005)}]{Panda2005_JCP_122}%
  \BibitemOpen
  \bibfield  {author} {\bibinfo {author} {\bibfnamefont {A.~N.}\ \bibnamefont
  {Panda}}\ and\ \bibinfo {author} {\bibfnamefont {N.}~\bibnamefont
  {Sathyamurthy}},\ }\href {\doibase 10.1063/1.1839866} {\bibfield  {journal}
  {\bibinfo  {journal} {J. Chem. Phys.}\ }\textbf {\bibinfo {volume} {122}},\
  \bibinfo {pages} {054304} (\bibinfo {year} {2005})}\BibitemShut {NoStop}%
\bibitem [{\citenamefont {Panda}\ and\ \citenamefont
  {Althorpe}(2006)}]{Panda2006_CPL_419}%
  \BibitemOpen
  \bibfield  {author} {\bibinfo {author} {\bibfnamefont {A.~N.}\ \bibnamefont
  {Panda}}\ and\ \bibinfo {author} {\bibfnamefont {S.~C.}\ \bibnamefont
  {Althorpe}},\ }\href {\doibase 10.1016/j.cplett.2005.11.093} {\bibfield
  {journal} {\bibinfo  {journal} {Chem. Phys. Lett.}\ }\textbf {\bibinfo
  {volume} {419}},\ \bibinfo {pages} {245} (\bibinfo {year}
  {2006})}\BibitemShut {NoStop}%
\bibitem [{\citenamefont {Tang}\ \emph {et~al.}(2007)\citenamefont {Tang},
  \citenamefont {Houchins}, \citenamefont {Lau}, \citenamefont {Ng},
  \citenamefont {Dressler}, \citenamefont {Chiu}, \citenamefont {Chu},\ and\
  \citenamefont {Han}}]{Tang2007_JCP_127}%
  \BibitemOpen
  \bibfield  {author} {\bibinfo {author} {\bibfnamefont {X.}~\bibnamefont
  {Tang}}, \bibinfo {author} {\bibfnamefont {C.}~\bibnamefont {Houchins}},
  \bibinfo {author} {\bibfnamefont {K.-C.}\ \bibnamefont {Lau}}, \bibinfo
  {author} {\bibfnamefont {C.~Y.}\ \bibnamefont {Ng}}, \bibinfo {author}
  {\bibfnamefont {R.~A.}\ \bibnamefont {Dressler}}, \bibinfo {author}
  {\bibfnamefont {Y.-H.}\ \bibnamefont {Chiu}}, \bibinfo {author}
  {\bibfnamefont {T.-S.}\ \bibnamefont {Chu}}, \ and\ \bibinfo {author}
  {\bibfnamefont {K.-L.}\ \bibnamefont {Han}},\ }\href {\doibase
  10.1063/1.2800009} {\bibfield  {journal} {\bibinfo  {journal} {J. Chem.
  Phys.}\ }\textbf {\bibinfo {volume} {127}},\ \bibinfo {pages} {164318}
  (\bibinfo {year} {2007})}\BibitemShut {NoStop}%
\bibitem [{\citenamefont {Chupka}(1972)}]{chupka1972}%
  \BibitemOpen
  \bibfield  {author} {\bibinfo {author} {\bibfnamefont {W.~A.}\ \bibnamefont
  {Chupka}},\ }\href@noop {} {\bibfield  {journal} {\bibinfo  {journal}
  {Ion-Molecule Reactions, vol.1 Ed. J. L. Franklin, Chap. 3, Plenum, New York,
  p.}\ ,\ \bibinfo {pages} {72}} (\bibinfo {year} {1972})}\BibitemShut
  {NoStop}%
\bibitem [{\citenamefont {Schneider}\ \emph {et~al.}(1976)\citenamefont
  {Schneider}, \citenamefont {Havemann}, \citenamefont {Zülicke},
  \citenamefont {Pac{\'{a}}k}, \citenamefont {Birkinshaw},\ and\ \citenamefont
  {Herman}}]{CPL_37_1976_Schneider}%
  \BibitemOpen
  \bibfield  {author} {\bibinfo {author} {\bibfnamefont {F.}~\bibnamefont
  {Schneider}}, \bibinfo {author} {\bibfnamefont {U.}~\bibnamefont {Havemann}},
  \bibinfo {author} {\bibfnamefont {L.}~\bibnamefont {Zülicke}}, \bibinfo
  {author} {\bibfnamefont {V.}~\bibnamefont {Pac{\'{a}}k}}, \bibinfo {author}
  {\bibfnamefont {K.}~\bibnamefont {Birkinshaw}}, \ and\ \bibinfo {author}
  {\bibfnamefont {Z.}~\bibnamefont {Herman}},\ }\href {\doibase
  10.1016/0009-2614(76)80225-4} {\bibfield  {journal} {\bibinfo  {journal}
  {Chem. Phys. Lett.}\ }\textbf {\bibinfo {volume} {37}},\ \bibinfo {pages}
  {323} (\bibinfo {year} {1976})}\BibitemShut {NoStop}%
\bibitem [{\citenamefont {Pac{\'{a}}k}\ \emph {et~al.}(1977)\citenamefont
  {Pac{\'{a}}k}, \citenamefont {Havemann}, \citenamefont {Herman},
  \citenamefont {Schneider},\ and\ \citenamefont
  {Zülicke}}]{CPL_49_1977_Pacak}%
  \BibitemOpen
  \bibfield  {author} {\bibinfo {author} {\bibfnamefont {V.}~\bibnamefont
  {Pac{\'{a}}k}}, \bibinfo {author} {\bibfnamefont {U.}~\bibnamefont
  {Havemann}}, \bibinfo {author} {\bibfnamefont {Z.}~\bibnamefont {Herman}},
  \bibinfo {author} {\bibfnamefont {F.}~\bibnamefont {Schneider}}, \ and\
  \bibinfo {author} {\bibfnamefont {L.}~\bibnamefont {Zülicke}},\ }\href
  {\doibase 10.1016/0009-2614(77)80585-x} {\bibfield  {journal} {\bibinfo
  {journal} {Chem. Phys. Lett.}\ }\textbf {\bibinfo {volume} {49}},\ \bibinfo
  {pages} {273} (\bibinfo {year} {1977})}\BibitemShut {NoStop}%
\bibitem [{\citenamefont {Pollard}\ \emph {et~al.}(1991)\citenamefont
  {Pollard}, \citenamefont {Syage}, \citenamefont {Johnson},\ and\
  \citenamefont {Cohen}}]{JCP_94_1991_Pollard}%
  \BibitemOpen
  \bibfield  {author} {\bibinfo {author} {\bibfnamefont {J.~E.}\ \bibnamefont
  {Pollard}}, \bibinfo {author} {\bibfnamefont {J.~A.}\ \bibnamefont {Syage}},
  \bibinfo {author} {\bibfnamefont {L.~K.}\ \bibnamefont {Johnson}}, \ and\
  \bibinfo {author} {\bibfnamefont {R.~B.}\ \bibnamefont {Cohen}},\ }\href
  {\doibase 10.1063/1.460718} {\bibfield  {journal} {\bibinfo  {journal} {J.
  Chem. Phys.}\ }\textbf {\bibinfo {volume} {94}},\ \bibinfo {pages} {8615}
  (\bibinfo {year} {1991})}\BibitemShut {NoStop}%
\bibitem [{\citenamefont {Pollard}, \citenamefont {Johnson},\ and\
  \citenamefont {Cohen}(1991)}]{JCP_95_1991_Pollard}%
  \BibitemOpen
  \bibfield  {author} {\bibinfo {author} {\bibfnamefont {J.~E.}\ \bibnamefont
  {Pollard}}, \bibinfo {author} {\bibfnamefont {L.~K.}\ \bibnamefont
  {Johnson}}, \ and\ \bibinfo {author} {\bibfnamefont {R.~B.}\ \bibnamefont
  {Cohen}},\ }\href {\doibase 10.1063/1.461705} {\bibfield  {journal} {\bibinfo
   {journal} {J. Chem. Phys.}\ }\textbf {\bibinfo {volume} {95}},\ \bibinfo
  {pages} {4894} (\bibinfo {year} {1991})}\BibitemShut {NoStop}%
\bibitem [{\citenamefont {Pijkeren}, \citenamefont {Eck},\ and\ \citenamefont
  {Niehaus}(1983)}]{CPL_96_1983_Pijkeren}%
  \BibitemOpen
  \bibfield  {author} {\bibinfo {author} {\bibfnamefont {D.~V.}\ \bibnamefont
  {Pijkeren}}, \bibinfo {author} {\bibfnamefont {J.~V.}\ \bibnamefont {Eck}}, \
  and\ \bibinfo {author} {\bibfnamefont {A.}~\bibnamefont {Niehaus}},\ }\href
  {\doibase 10.1016/0009-2614(83)80108-0} {\bibfield  {journal} {\bibinfo
  {journal} {Chem. Phys. Lett.}\ }\textbf {\bibinfo {volume} {96}},\ \bibinfo
  {pages} {20} (\bibinfo {year} {1983})}\BibitemShut {NoStop}%
\bibitem [{\citenamefont {Herman}\ and\ \citenamefont
  {Koyano}(1987)}]{JCSFT2_83_1987_Herman}%
  \BibitemOpen
  \bibfield  {author} {\bibinfo {author} {\bibfnamefont {Z.}~\bibnamefont
  {Herman}}\ and\ \bibinfo {author} {\bibfnamefont {I.}~\bibnamefont
  {Koyano}},\ }\href {\doibase 10.1039/f29878300127} {\bibfield  {journal}
  {\bibinfo  {journal} {J. Chem. Soc. Faraday Trans. 2}\ }\textbf {\bibinfo
  {volume} {83}},\ \bibinfo {pages} {127} (\bibinfo {year} {1987})}\BibitemShut
  {NoStop}%
\bibitem [{\citenamefont {Baer}\ and\ \citenamefont
  {Nakamura}(1987)}]{Baer1987_JPC_91}%
  \BibitemOpen
  \bibfield  {author} {\bibinfo {author} {\bibfnamefont {M.}~\bibnamefont
  {Baer}}\ and\ \bibinfo {author} {\bibfnamefont {H.}~\bibnamefont
  {Nakamura}},\ }\href {\doibase 10.1021/j100305a024} {\bibfield  {journal}
  {\bibinfo  {journal} {J. Phys. Chem.}\ }\textbf {\bibinfo {volume} {91}},\
  \bibinfo {pages} {5503} (\bibinfo {year} {1987})}\BibitemShut {NoStop}%
\bibitem [{\citenamefont {Aquilanti}\ \emph {et~al.}(2000)\citenamefont
  {Aquilanti}, \citenamefont {Capecchi}, \citenamefont {Cavalli}, \citenamefont
  {De~Fazio}, \citenamefont {Palmieri}, \citenamefont {Puzzarini},
  \citenamefont {Aguilar}, \citenamefont {Gim\'{e}nez},\ and\ \citenamefont
  {Lucas}}]{Aquilanti2000_CPL_318}%
  \BibitemOpen
  \bibfield  {author} {\bibinfo {author} {\bibfnamefont {V.}~\bibnamefont
  {Aquilanti}}, \bibinfo {author} {\bibfnamefont {G.}~\bibnamefont {Capecchi}},
  \bibinfo {author} {\bibfnamefont {S.}~\bibnamefont {Cavalli}}, \bibinfo
  {author} {\bibfnamefont {D.}~\bibnamefont {De~Fazio}}, \bibinfo {author}
  {\bibfnamefont {P.}~\bibnamefont {Palmieri}}, \bibinfo {author}
  {\bibfnamefont {C.}~\bibnamefont {Puzzarini}}, \bibinfo {author}
  {\bibfnamefont {A.}~\bibnamefont {Aguilar}}, \bibinfo {author} {\bibfnamefont
  {X.}~\bibnamefont {Gim\'{e}nez}}, \ and\ \bibinfo {author} {\bibfnamefont
  {J.~M.}\ \bibnamefont {Lucas}},\ }\href {\doibase
  10.1016/S0009-2614(00)00067-1} {\bibfield  {journal} {\bibinfo  {journal}
  {Chem. Phys. Lett.}\ }\textbf {\bibinfo {volume} {318}},\ \bibinfo {pages}
  {619} (\bibinfo {year} {2000})}\BibitemShut {NoStop}%
\bibitem [{\citenamefont {Xu}\ \emph {et~al.}(2008)\citenamefont {Xu},
  \citenamefont {Liu}, \citenamefont {Luan}, \citenamefont {Zhang},\ and\
  \citenamefont {Zhang}}]{Xu2008_CPL_464}%
  \BibitemOpen
  \bibfield  {author} {\bibinfo {author} {\bibfnamefont {W.}~\bibnamefont
  {Xu}}, \bibinfo {author} {\bibfnamefont {X.}~\bibnamefont {Liu}}, \bibinfo
  {author} {\bibfnamefont {S.}~\bibnamefont {Luan}}, \bibinfo {author}
  {\bibfnamefont {Q.}~\bibnamefont {Zhang}}, \ and\ \bibinfo {author}
  {\bibfnamefont {P.}~\bibnamefont {Zhang}},\ }\href {\doibase
  10.1016/j.cplett.2008.08.011} {\bibfield  {journal} {\bibinfo  {journal}
  {Chem. Phys. Lett.}\ }\textbf {\bibinfo {volume} {464}},\ \bibinfo {pages}
  {92} (\bibinfo {year} {2008})}\BibitemShut {NoStop}%
\bibitem [{\citenamefont {De~Fazio}\ \emph {et~al.}(2012)\citenamefont
  {De~Fazio}, \citenamefont {de~Castro-Vitores}, \citenamefont {Aguado},
  \citenamefont {Aquilanti},\ and\ \citenamefont
  {Cavalli}}]{JCP_137_2012_Fazio}%
  \BibitemOpen
  \bibfield  {author} {\bibinfo {author} {\bibfnamefont {D.}~\bibnamefont
  {De~Fazio}}, \bibinfo {author} {\bibfnamefont {M.}~\bibnamefont
  {de~Castro-Vitores}}, \bibinfo {author} {\bibfnamefont {A.}~\bibnamefont
  {Aguado}}, \bibinfo {author} {\bibfnamefont {V.}~\bibnamefont {Aquilanti}}, \
  and\ \bibinfo {author} {\bibfnamefont {S.}~\bibnamefont {Cavalli}},\ }\href
  {\doibase 10.1063/1.4772651} {\bibfield  {journal} {\bibinfo  {journal} {J.
  Chem. Phys.}\ }\textbf {\bibinfo {volume} {137}},\ \bibinfo {pages} {244306}
  (\bibinfo {year} {2012})}\BibitemShut {NoStop}%
\bibitem [{\citenamefont {Kumar}, \citenamefont {Sathyamurthy},\ and\
  \citenamefont {Bhalla}(1993)}]{JCP_98_1993_Kumar}%
  \BibitemOpen
  \bibfield  {author} {\bibinfo {author} {\bibfnamefont {S.}~\bibnamefont
  {Kumar}}, \bibinfo {author} {\bibfnamefont {N.}~\bibnamefont {Sathyamurthy}},
  \ and\ \bibinfo {author} {\bibfnamefont {K.~C.}\ \bibnamefont {Bhalla}},\
  }\href {\doibase 10.1063/1.464972} {\bibfield  {journal} {\bibinfo  {journal}
  {J. Chem. Phys.}\ }\textbf {\bibinfo {volume} {98}},\ \bibinfo {pages} {4680}
  (\bibinfo {year} {1993})}\BibitemShut {NoStop}%
\bibitem [{\citenamefont {Turner}, \citenamefont {Dutuit},\ and\ \citenamefont
  {Lee}(1984{\natexlab{a}})}]{JCP_81_1984_Turner}%
  \BibitemOpen
  \bibfield  {author} {\bibinfo {author} {\bibfnamefont {T.}~\bibnamefont
  {Turner}}, \bibinfo {author} {\bibfnamefont {O.}~\bibnamefont {Dutuit}}, \
  and\ \bibinfo {author} {\bibfnamefont {Y.~T.}\ \bibnamefont {Lee}},\ }\href
  {\doibase 10.1063/1.448073} {\bibfield  {journal} {\bibinfo  {journal} {J.
  Chem. Phys.}\ }\textbf {\bibinfo {volume} {81}},\ \bibinfo {pages} {3475}
  (\bibinfo {year} {1984}{\natexlab{a}})}\BibitemShut {NoStop}%
\bibitem [{\citenamefont {Achtenhagen}, \citenamefont {Schweizer},\ and\
  \citenamefont {Gerlich}(1993)}]{Achtenhagen}%
  \BibitemOpen
  \bibfield  {author} {\bibinfo {author} {\bibfnamefont {M.}~\bibnamefont
  {Achtenhagen}}, \bibinfo {author} {\bibfnamefont {M.}~\bibnamefont
  {Schweizer}}, \ and\ \bibinfo {author} {\bibfnamefont {D.}~\bibnamefont
  {Gerlich}},\ }\href {\doibase 10.1063/1.464972} {\enquote {\bibinfo {title}
  {{He,H$_{2}^{+}$} dynamics revisited},}\ }\bibinfo {howpublished} {private
  communication reported by S. Kumar, N. Sathyamurthy, and K. C. Bhalla}
  (\bibinfo {year} {1993})\BibitemShut {NoStop}%
\bibitem [{\citenamefont {Naskar}\ \emph
  {et~al.}(2023{\natexlab{c}})\citenamefont {Naskar}, \citenamefont {Ravi},
  \citenamefont {Adhikari}, \citenamefont {Baer},\ and\ \citenamefont
  {Sathyamurthy}}]{HeH2_Surf}%
  \BibitemOpen
  \bibfield  {author} {\bibinfo {author} {\bibfnamefont {K.}~\bibnamefont
  {Naskar}}, \bibinfo {author} {\bibfnamefont {S.}~\bibnamefont {Ravi}},
  \bibinfo {author} {\bibfnamefont {S.}~\bibnamefont {Adhikari}}, \bibinfo
  {author} {\bibfnamefont {M.}~\bibnamefont {Baer}}, \ and\ \bibinfo {author}
  {\bibfnamefont {N.}~\bibnamefont {Sathyamurthy}},\ }\href@noop {} {\bibfield
  {journal} {\bibinfo  {journal} {J. Phys. Chem. A.}\ }\textbf {\bibinfo
  {volume} {127}},\ \bibinfo {pages} {3832} (\bibinfo {year}
  {2023}{\natexlab{c}})}\BibitemShut {NoStop}%
\bibitem [{\citenamefont {Naskar}\ \emph
  {et~al.}(2023{\natexlab{d}})\citenamefont {Naskar}, \citenamefont {Ghosh},
  \citenamefont {Adhikari}, \citenamefont {Baer},\ and\ \citenamefont
  {Sathyamurthy}}]{HeH2+_dynamics_2023}%
  \BibitemOpen
  \bibfield  {author} {\bibinfo {author} {\bibfnamefont {K.}~\bibnamefont
  {Naskar}}, \bibinfo {author} {\bibfnamefont {S.}~\bibnamefont {Ghosh}},
  \bibinfo {author} {\bibfnamefont {S.}~\bibnamefont {Adhikari}}, \bibinfo
  {author} {\bibfnamefont {M.}~\bibnamefont {Baer}}, \ and\ \bibinfo {author}
  {\bibfnamefont {N.}~\bibnamefont {Sathyamurthy}},\ }\href@noop {} {\bibfield
  {journal} {\bibinfo  {journal} {J. Chem. Phys.}\ }\textbf {\bibinfo {volume}
  {159}},\ \bibinfo {pages} {034302} (\bibinfo {year}
  {2023}{\natexlab{d}})}\BibitemShut {NoStop}%
\bibitem [{\citenamefont {Mondal}\ and\ \citenamefont
  {Mukhopadhyay}(2021)}]{IJQC_121_2021_Mondal}%
  \BibitemOpen
  \bibfield  {author} {\bibinfo {author} {\bibfnamefont {R.}~\bibnamefont
  {Mondal}}\ and\ \bibinfo {author} {\bibfnamefont {D.}~\bibnamefont
  {Mukhopadhyay}},\ }\href {\doibase 10.1002/qua.26552} {\bibfield  {journal}
  {\bibinfo  {journal} {Int. J. Quantum Chem.}\ }\textbf {\bibinfo {volume}
  {121}},\ \bibinfo {pages} {e26552} (\bibinfo {year} {2021})}\BibitemShut
  {NoStop}%
\bibitem [{\citenamefont {Gupta}\ \emph {et~al.}(2020)\citenamefont {Gupta},
  \citenamefont {Dhindhwal}, \citenamefont {Baer}, \citenamefont
  {Sathyamurthy}, \citenamefont {Ravi}, \citenamefont {Mukherjee},
  \citenamefont {Mukherjee},\ and\ \citenamefont
  {Adhikari}}]{MolPhys_118_2020_Gupta}%
  \BibitemOpen
  \bibfield  {author} {\bibinfo {author} {\bibfnamefont {A.~K.}\ \bibnamefont
  {Gupta}}, \bibinfo {author} {\bibfnamefont {V.}~\bibnamefont {Dhindhwal}},
  \bibinfo {author} {\bibfnamefont {M.}~\bibnamefont {Baer}}, \bibinfo {author}
  {\bibfnamefont {N.}~\bibnamefont {Sathyamurthy}}, \bibinfo {author}
  {\bibfnamefont {S.}~\bibnamefont {Ravi}}, \bibinfo {author} {\bibfnamefont
  {S.}~\bibnamefont {Mukherjee}}, \bibinfo {author} {\bibfnamefont
  {B.}~\bibnamefont {Mukherjee}}, \ and\ \bibinfo {author} {\bibfnamefont
  {S.}~\bibnamefont {Adhikari}},\ }\href {\doibase
  10.1080/00268976.2019.1683243} {\bibfield  {journal} {\bibinfo  {journal}
  {Mol. Phys.}\ }\textbf {\bibinfo {volume} {118}},\ \bibinfo {pages}
  {e1683243} (\bibinfo {year} {2020})}\BibitemShut {NoStop}%
\bibitem [{\citenamefont {Ravi}\ \emph
  {et~al.}(2021{\natexlab{a}})\citenamefont {Ravi}, \citenamefont {Mukherjee},
  \citenamefont {Mukherjee}, \citenamefont {Adhikari}, \citenamefont
  {Sathyamurthy},\ and\ \citenamefont {Baer}}]{MolPhys_119_2021_ravi}%
  \BibitemOpen
  \bibfield  {author} {\bibinfo {author} {\bibfnamefont {S.}~\bibnamefont
  {Ravi}}, \bibinfo {author} {\bibfnamefont {S.}~\bibnamefont {Mukherjee}},
  \bibinfo {author} {\bibfnamefont {B.}~\bibnamefont {Mukherjee}}, \bibinfo
  {author} {\bibfnamefont {S.}~\bibnamefont {Adhikari}}, \bibinfo {author}
  {\bibfnamefont {N.}~\bibnamefont {Sathyamurthy}}, \ and\ \bibinfo {author}
  {\bibfnamefont {M.}~\bibnamefont {Baer}},\ }\href {\doibase
  10.1080/00268976.2020.1811907} {\bibfield  {journal} {\bibinfo  {journal}
  {Mol. Phys.}\ }\textbf {\bibinfo {volume} {119}},\ \bibinfo {pages} {1}
  (\bibinfo {year} {2021}{\natexlab{a}})}\BibitemShut {NoStop}%
\bibitem [{\citenamefont {Kuppermann}\ and\ \citenamefont
  {Schatz}(1975)}]{Kuppermann_1975}%
  \BibitemOpen
  \bibfield  {author} {\bibinfo {author} {\bibfnamefont {A.}~\bibnamefont
  {Kuppermann}}\ and\ \bibinfo {author} {\bibfnamefont {G.~C.}\ \bibnamefont
  {Schatz}},\ }\href@noop {} {\bibfield  {journal} {\bibinfo  {journal} {The
  Journal of Chemical Physics}\ }\textbf {\bibinfo {volume} {62}},\ \bibinfo
  {pages} {2502} (\bibinfo {year} {1975})}\BibitemShut {NoStop}%
\bibitem [{\citenamefont {Walker}, \citenamefont {Stechel},\ and\ \citenamefont
  {Light}(1978)}]{Walker_1978}%
  \BibitemOpen
  \bibfield  {author} {\bibinfo {author} {\bibfnamefont {R.~B.}\ \bibnamefont
  {Walker}}, \bibinfo {author} {\bibfnamefont {E.~B.}\ \bibnamefont {Stechel}},
  \ and\ \bibinfo {author} {\bibfnamefont {J.~C.}\ \bibnamefont {Light}},\
  }\href@noop {} {\bibfield  {journal} {\bibinfo  {journal} {Journal of
  Chemical Physics}\ }\textbf {\bibinfo {volume} {69}},\ \bibinfo {pages}
  {2922} (\bibinfo {year} {1978})}\BibitemShut {NoStop}%
\bibitem [{\citenamefont {Charutz}, \citenamefont {Last},\ and\ \citenamefont
  {Baer}(1997)}]{Charutz_Baer_1997}%
  \BibitemOpen
  \bibfield  {author} {\bibinfo {author} {\bibfnamefont {D.~M.}\ \bibnamefont
  {Charutz}}, \bibinfo {author} {\bibfnamefont {I.}~\bibnamefont {Last}}, \
  and\ \bibinfo {author} {\bibfnamefont {M.}~\bibnamefont {Baer}},\ }\href@noop
  {} {\bibfield  {journal} {\bibinfo  {journal} {J. Chem. Phys.}\ }\textbf
  {\bibinfo {volume} {106}},\ \bibinfo {pages} {7654} (\bibinfo {year}
  {1997})}\BibitemShut {NoStop}%
\bibitem [{\citenamefont {Kendrick}(2001)}]{Kendrick_2001}%
  \BibitemOpen
  \bibfield  {author} {\bibinfo {author} {\bibfnamefont {B.~K.}\ \bibnamefont
  {Kendrick}},\ }\href@noop {} {\bibfield  {journal} {\bibinfo  {journal} {J.
  Chem. Phys.}\ }\textbf {\bibinfo {volume} {114}},\ \bibinfo {pages} {8796}
  (\bibinfo {year} {2001})}\BibitemShut {NoStop}%
\bibitem [{\citenamefont {Aoiz}, \citenamefont {BaÑares},\ and\ \citenamefont
  {Herrero}(2005)}]{Aoiz_2005}%
  \BibitemOpen
  \bibfield  {author} {\bibinfo {author} {\bibfnamefont {F.~J.}\ \bibnamefont
  {Aoiz}}, \bibinfo {author} {\bibfnamefont {L.}~\bibnamefont {BaÑares}}, \
  and\ \bibinfo {author} {\bibfnamefont {V.~J.}\ \bibnamefont {Herrero}},\
  }\href@noop {} {\bibfield  {journal} {\bibinfo  {journal} {Int. Rev. Phys.
  Chem.}\ }\textbf {\bibinfo {volume} {24}},\ \bibinfo {pages} {119} (\bibinfo
  {year} {2005})}\BibitemShut {NoStop}%
\bibitem [{\citenamefont {Adelman}\ \emph {et~al.}(1992)\citenamefont
  {Adelman}, \citenamefont {Shafer}, \citenamefont {Kliner},\ and\
  \citenamefont {Zare}}]{Zare_92}%
  \BibitemOpen
  \bibfield  {author} {\bibinfo {author} {\bibfnamefont {D.~E.}\ \bibnamefont
  {Adelman}}, \bibinfo {author} {\bibfnamefont {N.~E.}\ \bibnamefont {Shafer}},
  \bibinfo {author} {\bibfnamefont {D.~A.~V.}\ \bibnamefont {Kliner}}, \ and\
  \bibinfo {author} {\bibfnamefont {R.~N.}\ \bibnamefont {Zare}},\ }\href@noop
  {} {\bibfield  {journal} {\bibinfo  {journal} {J. Chem. Phys.}\ }\textbf
  {\bibinfo {volume} {97}},\ \bibinfo {pages} {7323} (\bibinfo {year}
  {1992})}\BibitemShut {NoStop}%
\bibitem [{\citenamefont {Nieh}\ and\ \citenamefont
  {Valentini}(1988)}]{Nieh_Valentini}%
  \BibitemOpen
  \bibfield  {author} {\bibinfo {author} {\bibfnamefont {J.-C.}\ \bibnamefont
  {Nieh}}\ and\ \bibinfo {author} {\bibfnamefont {J.~J.}\ \bibnamefont
  {Valentini}},\ }\href@noop {} {\bibfield  {journal} {\bibinfo  {journal}
  {Phys. Rev. Lett.}\ }\textbf {\bibinfo {volume} {60}},\ \bibinfo {pages}
  {519} (\bibinfo {year} {1988})}\BibitemShut {NoStop}%
\bibitem [{\citenamefont {Continetti}, \citenamefont {Balko},\ and\
  \citenamefont {Lee}(1990)}]{Y.T_Lee_90}%
  \BibitemOpen
  \bibfield  {author} {\bibinfo {author} {\bibfnamefont {R.~E.}\ \bibnamefont
  {Continetti}}, \bibinfo {author} {\bibfnamefont {B.~A.}\ \bibnamefont
  {Balko}}, \ and\ \bibinfo {author} {\bibfnamefont {Y.~T.}\ \bibnamefont
  {Lee}},\ }\href@noop {} {\bibfield  {journal} {\bibinfo  {journal} {J. Chem.
  Phys.}\ }\textbf {\bibinfo {volume} {93}},\ \bibinfo {pages} {5719} (\bibinfo
  {year} {1990})}\BibitemShut {NoStop}%
\bibitem [{\citenamefont {Ba{\~n}ares}\ \emph {et~al.}(1998)\citenamefont
  {Ba{\~n}ares}, \citenamefont {Aoiz}, \citenamefont {Herrero}, \citenamefont
  {D'Mello}, \citenamefont {Niederjohann}, \citenamefont {Seekamp-Rahn},
  \citenamefont {Wrede},\ and\ \citenamefont
  {Schnieder}}]{Banares_Exp_Theo_1998}%
  \BibitemOpen
  \bibfield  {author} {\bibinfo {author} {\bibfnamefont {L.}~\bibnamefont
  {Ba{\~n}ares}}, \bibinfo {author} {\bibfnamefont {F.~J.}\ \bibnamefont
  {Aoiz}}, \bibinfo {author} {\bibfnamefont {V.~J.}\ \bibnamefont {Herrero}},
  \bibinfo {author} {\bibfnamefont {M.~J.}\ \bibnamefont {D'Mello}}, \bibinfo
  {author} {\bibfnamefont {B.}~\bibnamefont {Niederjohann}}, \bibinfo {author}
  {\bibfnamefont {K.}~\bibnamefont {Seekamp-Rahn}}, \bibinfo {author}
  {\bibfnamefont {E.}~\bibnamefont {Wrede}}, \ and\ \bibinfo {author}
  {\bibfnamefont {L.}~\bibnamefont {Schnieder}},\ }\href@noop {} {\bibfield
  {journal} {\bibinfo  {journal} {J. Chem. Phys.}\ }\textbf {\bibinfo {volume}
  {108}},\ \bibinfo {pages} {6160} (\bibinfo {year} {1998})}\BibitemShut
  {NoStop}%
\bibitem [{\citenamefont {Bean}\ \emph {et~al.}(2002)\citenamefont {Bean},
  \citenamefont {Ayers}, \citenamefont {Fern{\'a}ndez-Alonso},\ and\
  \citenamefont {Zare}}]{Bean_Zare_DCS_2002}%
  \BibitemOpen
  \bibfield  {author} {\bibinfo {author} {\bibfnamefont {B.~D.}\ \bibnamefont
  {Bean}}, \bibinfo {author} {\bibfnamefont {J.~D.}\ \bibnamefont {Ayers}},
  \bibinfo {author} {\bibfnamefont {F.}~\bibnamefont {Fern{\'a}ndez-Alonso}}, \
  and\ \bibinfo {author} {\bibfnamefont {R.~N.}\ \bibnamefont {Zare}},\
  }\href@noop {} {\bibfield  {journal} {\bibinfo  {journal} {J. Chem. Phys.}\
  }\textbf {\bibinfo {volume} {116}},\ \bibinfo {pages} {6634} (\bibinfo {year}
  {2002})}\BibitemShut {NoStop}%
\bibitem [{\citenamefont {Mielke}\ \emph {et~al.}(2003)\citenamefont {Mielke},
  \citenamefont {Peterson}, \citenamefont {Schwenke}, \citenamefont {Garrett},
  \citenamefont {Truhlar}, \citenamefont {Michael}, \citenamefont {Su},\ and\
  \citenamefont {Sutherland}}]{Mielke_Sutherland_2003}%
  \BibitemOpen
  \bibfield  {author} {\bibinfo {author} {\bibfnamefont {S.~L.}\ \bibnamefont
  {Mielke}}, \bibinfo {author} {\bibfnamefont {K.~A.}\ \bibnamefont
  {Peterson}}, \bibinfo {author} {\bibfnamefont {D.~W.}\ \bibnamefont
  {Schwenke}}, \bibinfo {author} {\bibfnamefont {B.~C.}\ \bibnamefont
  {Garrett}}, \bibinfo {author} {\bibfnamefont {D.~G.}\ \bibnamefont
  {Truhlar}}, \bibinfo {author} {\bibfnamefont {J.~V.}\ \bibnamefont
  {Michael}}, \bibinfo {author} {\bibfnamefont {M.-C.}\ \bibnamefont {Su}}, \
  and\ \bibinfo {author} {\bibfnamefont {J.~W.}\ \bibnamefont {Sutherland}},\
  }\href@noop {} {\bibfield  {journal} {\bibinfo  {journal} {Phys. Rev. Lett.}\
  }\textbf {\bibinfo {volume} {91}},\ \bibinfo {pages} {063201} (\bibinfo
  {year} {2003})}\BibitemShut {NoStop}%
\bibitem [{\citenamefont {Dai}\ \emph {et~al.}(2003)\citenamefont {Dai},
  \citenamefont {Wang}, \citenamefont {Harich}, \citenamefont {Wang},
  \citenamefont {Yang}, \citenamefont {Chao},\ and\ \citenamefont
  {Skodje}}]{Dai_Wang_Science_2003}%
  \BibitemOpen
  \bibfield  {author} {\bibinfo {author} {\bibfnamefont {D.}~\bibnamefont
  {Dai}}, \bibinfo {author} {\bibfnamefont {C.~C.}\ \bibnamefont {Wang}},
  \bibinfo {author} {\bibfnamefont {S.~A.}\ \bibnamefont {Harich}}, \bibinfo
  {author} {\bibfnamefont {X.}~\bibnamefont {Wang}}, \bibinfo {author}
  {\bibfnamefont {X.}~\bibnamefont {Yang}}, \bibinfo {author} {\bibfnamefont
  {S.~D.}\ \bibnamefont {Chao}}, \ and\ \bibinfo {author} {\bibfnamefont
  {R.~T.}\ \bibnamefont {Skodje}},\ }\href@noop {} {\bibfield  {journal}
  {\bibinfo  {journal} {Science}\ }\textbf {\bibinfo {volume} {300}},\ \bibinfo
  {pages} {1730} (\bibinfo {year} {2003})}\BibitemShut {NoStop}%
\bibitem [{\citenamefont {Ausfelder}\ \emph {et~al.}(2004)\citenamefont
  {Ausfelder}, \citenamefont {Pomerantz}, \citenamefont {Zare}, \citenamefont
  {Althorpe}, \citenamefont {Aoiz}, \citenamefont {Ba{\~n}ares},\ and\
  \citenamefont {Castillo}}]{Althorpe_Zare_ICS_2004}%
  \BibitemOpen
  \bibfield  {author} {\bibinfo {author} {\bibfnamefont {F.}~\bibnamefont
  {Ausfelder}}, \bibinfo {author} {\bibfnamefont {A.~E.}\ \bibnamefont
  {Pomerantz}}, \bibinfo {author} {\bibfnamefont {R.~N.}\ \bibnamefont {Zare}},
  \bibinfo {author} {\bibfnamefont {S.~C.}\ \bibnamefont {Althorpe}}, \bibinfo
  {author} {\bibfnamefont {F.~J.}\ \bibnamefont {Aoiz}}, \bibinfo {author}
  {\bibfnamefont {L.}~\bibnamefont {Ba{\~n}ares}}, \ and\ \bibinfo {author}
  {\bibfnamefont {J.~F.}\ \bibnamefont {Castillo}},\ }\href@noop {} {\bibfield
  {journal} {\bibinfo  {journal} {J. Chem. Phys.}\ }\textbf {\bibinfo {volume}
  {120}},\ \bibinfo {pages} {3255} (\bibinfo {year} {2004})}\BibitemShut
  {NoStop}%
\bibitem [{\citenamefont {Pomerantz}\ \emph {et~al.}(2004)\citenamefont
  {Pomerantz}, \citenamefont {Ausfelder}, \citenamefont {Zare}, \citenamefont
  {Althorpe}, \citenamefont {Aoiz}, \citenamefont {Ba{\~n}ares},\ and\
  \citenamefont {Castillo}}]{Pomerantz_2004_ref_20}%
  \BibitemOpen
  \bibfield  {author} {\bibinfo {author} {\bibfnamefont {A.~E.}\ \bibnamefont
  {Pomerantz}}, \bibinfo {author} {\bibfnamefont {F.}~\bibnamefont
  {Ausfelder}}, \bibinfo {author} {\bibfnamefont {R.~N.}\ \bibnamefont {Zare}},
  \bibinfo {author} {\bibfnamefont {S.~C.}\ \bibnamefont {Althorpe}}, \bibinfo
  {author} {\bibfnamefont {F.~J.}\ \bibnamefont {Aoiz}}, \bibinfo {author}
  {\bibfnamefont {L.}~\bibnamefont {Ba{\~n}ares}}, \ and\ \bibinfo {author}
  {\bibfnamefont {J.~F.}\ \bibnamefont {Castillo}},\ }\href@noop {} {\bibfield
  {journal} {\bibinfo  {journal} {J. Chem. Phys.}\ }\textbf {\bibinfo {volume}
  {120}},\ \bibinfo {pages} {3244} (\bibinfo {year} {2004})}\BibitemShut
  {NoStop}%
\bibitem [{\citenamefont {Koszinowski}\ \emph {et~al.}(2005)\citenamefont
  {Koszinowski}, \citenamefont {Goldberg}, \citenamefont {Pomerantz},
  \citenamefont {Zare}, \citenamefont {Juanes-Marcos},\ and\ \citenamefont
  {Althorpe}}]{zare_2005}%
  \BibitemOpen
  \bibfield  {author} {\bibinfo {author} {\bibfnamefont {K.}~\bibnamefont
  {Koszinowski}}, \bibinfo {author} {\bibfnamefont {N.~T.}\ \bibnamefont
  {Goldberg}}, \bibinfo {author} {\bibfnamefont {A.~E.}\ \bibnamefont
  {Pomerantz}}, \bibinfo {author} {\bibfnamefont {R.~N.}\ \bibnamefont {Zare}},
  \bibinfo {author} {\bibfnamefont {J.~C.}\ \bibnamefont {Juanes-Marcos}}, \
  and\ \bibinfo {author} {\bibfnamefont {S.~C.}\ \bibnamefont {Althorpe}},\
  }\href@noop {} {\bibfield  {journal} {\bibinfo  {journal} {J. Chem. Phys.}\
  }\textbf {\bibinfo {volume} {123}},\ \bibinfo {pages} {054306} (\bibinfo
  {year} {2005})}\BibitemShut {NoStop}%
\bibitem [{\citenamefont {Hal{\'a}sz}\ \emph {et~al.}(2003)\citenamefont
  {Hal{\'a}sz}, \citenamefont {Vib{\'o}k}, \citenamefont {Mebel},\ and\
  \citenamefont {Baer}}]{Halasz_Baer2003}%
  \BibitemOpen
  \bibfield  {author} {\bibinfo {author} {\bibfnamefont {G.}~\bibnamefont
  {Hal{\'a}sz}}, \bibinfo {author} {\bibfnamefont {A.}~\bibnamefont
  {Vib{\'o}k}}, \bibinfo {author} {\bibfnamefont {A.~M.}\ \bibnamefont
  {Mebel}}, \ and\ \bibinfo {author} {\bibfnamefont {M.}~\bibnamefont {Baer}},\
  }\href@noop {} {\bibfield  {journal} {\bibinfo  {journal} {J. Chem. Phys.}\
  }\textbf {\bibinfo {volume} {118}},\ \bibinfo {pages} {3052} (\bibinfo {year}
  {2003})}\BibitemShut {NoStop}%
\bibitem [{\citenamefont {Baer}\ \emph {et~al.}(2004)\citenamefont {Baer},
  \citenamefont {Ve'Rtesi}, \citenamefont {Hal{\'a}sz}, \citenamefont {Vibok},\
  and\ \citenamefont {Suhai}}]{Baer_2004}%
  \BibitemOpen
  \bibfield  {author} {\bibinfo {author} {\bibfnamefont {M.}~\bibnamefont
  {Baer}}, \bibinfo {author} {\bibfnamefont {T.}~\bibnamefont {Ve'Rtesi}},
  \bibinfo {author} {\bibfnamefont {G.}~\bibnamefont {Hal{\'a}sz}}, \bibinfo
  {author} {\bibfnamefont {A.}~\bibnamefont {Vibok}}, \ and\ \bibinfo {author}
  {\bibfnamefont {S.}~\bibnamefont {Suhai}},\ }\href@noop {} {\bibfield
  {journal} {\bibinfo  {journal} {Faraday Discussions}\ }\textbf {\bibinfo
  {volume} {127}},\ \bibinfo {pages} {337} (\bibinfo {year}
  {2004})}\BibitemShut {NoStop}%
\bibitem [{\citenamefont {Ghosh}\ \emph {et~al.}(2025)\citenamefont {Ghosh},
  \citenamefont {Roy}, \citenamefont {Kumar}, \citenamefont {Naskar},
  \citenamefont {Ravi},\ and\ \citenamefont {Adhikari}}]{Ghosh_H3_2025}%
  \BibitemOpen
  \bibfield  {author} {\bibinfo {author} {\bibfnamefont {A.}~\bibnamefont
  {Ghosh}}, \bibinfo {author} {\bibfnamefont {A.}~\bibnamefont {Roy}}, \bibinfo
  {author} {\bibfnamefont {A.}~\bibnamefont {Kumar}}, \bibinfo {author}
  {\bibfnamefont {K.}~\bibnamefont {Naskar}}, \bibinfo {author} {\bibfnamefont
  {S.}~\bibnamefont {Ravi}}, \ and\ \bibinfo {author} {\bibfnamefont
  {S.}~\bibnamefont {Adhikari}},\ }\href@noop {} {\bibfield  {journal}
  {\bibinfo  {journal} {J. Phys. Chem. A}\ }\textbf {\bibinfo {volume} {129}},\
  \bibinfo {pages} {6315} (\bibinfo {year} {2025})}\BibitemShut {NoStop}%
\bibitem [{\citenamefont {Gao}\ \emph {et~al.}(2015)\citenamefont {Gao},
  \citenamefont {Sneha}, \citenamefont {Bouakline}, \citenamefont {Althorpe},\
  and\ \citenamefont {Zare}}]{Althorpe_2015_DCS}%
  \BibitemOpen
  \bibfield  {author} {\bibinfo {author} {\bibfnamefont {H.}~\bibnamefont
  {Gao}}, \bibinfo {author} {\bibfnamefont {M.}~\bibnamefont {Sneha}}, \bibinfo
  {author} {\bibfnamefont {F.}~\bibnamefont {Bouakline}}, \bibinfo {author}
  {\bibfnamefont {S.~C.}\ \bibnamefont {Althorpe}}, \ and\ \bibinfo {author}
  {\bibfnamefont {R.~N.}\ \bibnamefont {Zare}},\ }\href@noop {} {\bibfield
  {journal} {\bibinfo  {journal} {J. Phys. Chem. A}\ }\textbf {\bibinfo
  {volume} {119}},\ \bibinfo {pages} {12036} (\bibinfo {year}
  {2015})}\BibitemShut {NoStop}%
\bibitem [{\citenamefont {Jankunas}\ \emph {et~al.}(2013)\citenamefont
  {Jankunas}, \citenamefont {Sneha}, \citenamefont {Zare}, \citenamefont
  {Bouakline},\ and\ \citenamefont {Althorpe}}]{Jankunas_Althrope_2013}%
  \BibitemOpen
  \bibfield  {author} {\bibinfo {author} {\bibfnamefont {J.}~\bibnamefont
  {Jankunas}}, \bibinfo {author} {\bibfnamefont {M.}~\bibnamefont {Sneha}},
  \bibinfo {author} {\bibfnamefont {R.~N.}\ \bibnamefont {Zare}}, \bibinfo
  {author} {\bibfnamefont {F.}~\bibnamefont {Bouakline}}, \ and\ \bibinfo
  {author} {\bibfnamefont {S.~C.}\ \bibnamefont {Althorpe}},\ }\href@noop {}
  {\bibfield  {journal} {\bibinfo  {journal} {The Journal of Chemical Physics}\
  }\textbf {\bibinfo {volume} {138}},\ \bibinfo {pages} {094310} (\bibinfo
  {year} {2013})}\BibitemShut {NoStop}%
\bibitem [{\citenamefont {Varandas}\ \emph {et~al.}(1987)\citenamefont
  {Varandas}, \citenamefont {Brown}, \citenamefont {Mead}, \citenamefont
  {Truhlar},\ and\ \citenamefont {Blais}}]{DMBE_1987}%
  \BibitemOpen
  \bibfield  {author} {\bibinfo {author} {\bibfnamefont {A.~J.~C.}\
  \bibnamefont {Varandas}}, \bibinfo {author} {\bibfnamefont {F.~B.}\
  \bibnamefont {Brown}}, \bibinfo {author} {\bibfnamefont {C.~A.}\ \bibnamefont
  {Mead}}, \bibinfo {author} {\bibfnamefont {D.~G.}\ \bibnamefont {Truhlar}}, \
  and\ \bibinfo {author} {\bibfnamefont {N.~C.}\ \bibnamefont {Blais}},\
  }\href@noop {} {\bibfield  {journal} {\bibinfo  {journal} {J. Chem. Phys.}\
  }\textbf {\bibinfo {volume} {86}},\ \bibinfo {pages} {6258} (\bibinfo {year}
  {1987})}\BibitemShut {NoStop}%
\bibitem [{\citenamefont {Neuhauser}\ and\ \citenamefont
  {Baer}(1990)}]{jcp-92-3419-BAER}%
  \BibitemOpen
  \bibfield  {author} {\bibinfo {author} {\bibfnamefont {D.}~\bibnamefont
  {Neuhauser}}\ and\ \bibinfo {author} {\bibfnamefont {M.}~\bibnamefont
  {Baer}},\ }\href@noop {} {\bibfield  {journal} {\bibinfo  {journal} {J. Chem.
  Phys.}\ }\textbf {\bibinfo {volume} {92}},\ \bibinfo {pages} {3419} (\bibinfo
  {year} {1990})}\BibitemShut {NoStop}%
\bibitem [{\citenamefont {Neuhasuer}\ and\ \citenamefont
  {Baer}(1989)}]{jcp-90-4351-BAER}%
  \BibitemOpen
  \bibfield  {author} {\bibinfo {author} {\bibfnamefont {D.}~\bibnamefont
  {Neuhasuer}}\ and\ \bibinfo {author} {\bibfnamefont {M.}~\bibnamefont
  {Baer}},\ }\href@noop {} {\bibfield  {journal} {\bibinfo  {journal} {J. Chem.
  Phys.}\ }\textbf {\bibinfo {volume} {90}},\ \bibinfo {pages} {4351} (\bibinfo
  {year} {1989})}\BibitemShut {NoStop}%
\bibitem [{\citenamefont {Zhang}\ and\ \citenamefont
  {Miller}(1989)}]{jcp-91-1528-MILL}%
  \BibitemOpen
  \bibfield  {author} {\bibinfo {author} {\bibfnamefont {J.~Z.~H.}\
  \bibnamefont {Zhang}}\ and\ \bibinfo {author} {\bibfnamefont {W.~H.}\
  \bibnamefont {Miller}},\ }\href@noop {} {\bibfield  {journal} {\bibinfo
  {journal} {J. Chem. Phys.}\ }\textbf {\bibinfo {volume} {91}},\ \bibinfo
  {pages} {1528} (\bibinfo {year} {1989})}\BibitemShut {NoStop}%
\bibitem [{\citenamefont {Webster}\ and\ \citenamefont
  {Light}(1986)}]{jcp-85-4744-LIT2}%
  \BibitemOpen
  \bibfield  {author} {\bibinfo {author} {\bibfnamefont {F.}~\bibnamefont
  {Webster}}\ and\ \bibinfo {author} {\bibfnamefont {J.~C.}\ \bibnamefont
  {Light}},\ }\href@noop {} {\bibfield  {journal} {\bibinfo  {journal} {J.
  Chem. Phys.}\ }\textbf {\bibinfo {volume} {85}},\ \bibinfo {pages} {4744}
  (\bibinfo {year} {1986})}\BibitemShut {NoStop}%
\bibitem [{\citenamefont {Webster}\ and\ \citenamefont
  {Light}(1989)}]{jcp-90-265-LIT2}%
  \BibitemOpen
  \bibfield  {author} {\bibinfo {author} {\bibfnamefont {F.}~\bibnamefont
  {Webster}}\ and\ \bibinfo {author} {\bibfnamefont {J.~C.}\ \bibnamefont
  {Light}},\ }\href@noop {} {\bibfield  {journal} {\bibinfo  {journal} {J.
  Chem. Phys.}\ }\textbf {\bibinfo {volume} {90}},\ \bibinfo {pages} {265}
  (\bibinfo {year} {1989})}\BibitemShut {NoStop}%
\bibitem [{\citenamefont {Baer}\ and\ \citenamefont
  {Kouri}(1971)}]{prA-4-1924-MBK2}%
  \BibitemOpen
  \bibfield  {author} {\bibinfo {author} {\bibfnamefont {M.}~\bibnamefont
  {Baer}}\ and\ \bibinfo {author} {\bibfnamefont {D.~J.}\ \bibnamefont
  {Kouri}},\ }\href@noop {} {\bibfield  {journal} {\bibinfo  {journal} {Phys.
  Rev. A}\ }\textbf {\bibinfo {volume} {4}},\ \bibinfo {pages} {1924} (\bibinfo
  {year} {1971})}\BibitemShut {NoStop}%
\bibitem [{\citenamefont {Sun}, \citenamefont {Guo},\ and\ \citenamefont
  {Zhang}(2010)}]{jcp-132-084112-GUO}%
  \BibitemOpen
  \bibfield  {author} {\bibinfo {author} {\bibfnamefont {Z.}~\bibnamefont
  {Sun}}, \bibinfo {author} {\bibfnamefont {H.}~\bibnamefont {Guo}}, \ and\
  \bibinfo {author} {\bibfnamefont {D.~H.}\ \bibnamefont {Zhang}},\ }\href@noop
  {} {\bibfield  {journal} {\bibinfo  {journal} {J. Chem. Phys.}\ }\textbf
  {\bibinfo {volume} {132}},\ \bibinfo {pages} {084112/1} (\bibinfo {year}
  {2010})}\BibitemShut {NoStop}%
\bibitem [{\citenamefont {Carrasco}\ and\ \citenamefont
  {Roncero}(2006)}]{jcp-125-054102-GOM}%
  \BibitemOpen
  \bibfield  {author} {\bibinfo {author} {\bibfnamefont {S.~G.}\ \bibnamefont
  {Carrasco}}\ and\ \bibinfo {author} {\bibfnamefont {O.}~\bibnamefont
  {Roncero}},\ }\href@noop {} {\bibfield  {journal} {\bibinfo  {journal} {J.
  Chem. Phys.}\ }\textbf {\bibinfo {volume} {125}},\ \bibinfo {pages}
  {054102/1} (\bibinfo {year} {2006})}\BibitemShut {NoStop}%
\bibitem [{\citenamefont {Zhang}, \citenamefont {Chu},\ and\ \citenamefont
  {Miller}(1988)}]{jcp-88-6233-MILL}%
  \BibitemOpen
  \bibfield  {author} {\bibinfo {author} {\bibfnamefont {J.~Z.~H.}\
  \bibnamefont {Zhang}}, \bibinfo {author} {\bibfnamefont {S.-I.}\ \bibnamefont
  {Chu}}, \ and\ \bibinfo {author} {\bibfnamefont {W.~H.}\ \bibnamefont
  {Miller}},\ }\href@noop {} {\bibfield  {journal} {\bibinfo  {journal} {J.
  Chem. Phys.}\ }\textbf {\bibinfo {volume} {88}},\ \bibinfo {pages} {6233}
  (\bibinfo {year} {1988})}\BibitemShut {NoStop}%
\bibitem [{\citenamefont {Zhu}, \citenamefont {Peng},\ and\ \citenamefont
  {Zhang}(1997)}]{jcp-106-1742-ZHA1}%
  \BibitemOpen
  \bibfield  {author} {\bibinfo {author} {\bibfnamefont {W.}~\bibnamefont
  {Zhu}}, \bibinfo {author} {\bibfnamefont {T.}~\bibnamefont {Peng}}, \ and\
  \bibinfo {author} {\bibfnamefont {J.~Z.~H.}\ \bibnamefont {Zhang}},\
  }\href@noop {} {\bibfield  {journal} {\bibinfo  {journal} {J. Chem. Phys.}\
  }\textbf {\bibinfo {volume} {106}},\ \bibinfo {pages} {1742} (\bibinfo {year}
  {1997})}\BibitemShut {NoStop}%
\bibitem [{\citenamefont {Kouri}\ \emph {et~al.}(1996)\citenamefont {Kouri},
  \citenamefont {Hoffman}, \citenamefont {Peng},\ and\ \citenamefont
  {Zhang}}]{cpl-262-519-ZHA1}%
  \BibitemOpen
  \bibfield  {author} {\bibinfo {author} {\bibfnamefont {D.~J.}\ \bibnamefont
  {Kouri}}, \bibinfo {author} {\bibfnamefont {D.~K.}\ \bibnamefont {Hoffman}},
  \bibinfo {author} {\bibfnamefont {T.}~\bibnamefont {Peng}}, \ and\ \bibinfo
  {author} {\bibfnamefont {J.~Z.~H.}\ \bibnamefont {Zhang}},\ }\href@noop {}
  {\bibfield  {journal} {\bibinfo  {journal} {Chem. Phys. Lett.}\ }\textbf
  {\bibinfo {volume} {262}},\ \bibinfo {pages} {519} (\bibinfo {year}
  {1996})}\BibitemShut {NoStop}%
\bibitem [{\citenamefont {Peng}\ and\ \citenamefont
  {Zhang}(1996)}]{jcp-105-6072-ZHA1}%
  \BibitemOpen
  \bibfield  {author} {\bibinfo {author} {\bibfnamefont {T.}~\bibnamefont
  {Peng}}\ and\ \bibinfo {author} {\bibfnamefont {J.~Z.~H.}\ \bibnamefont
  {Zhang}},\ }\href@noop {} {\bibfield  {journal} {\bibinfo  {journal} {J.
  Chem. Phys.}\ }\textbf {\bibinfo {volume} {105}},\ \bibinfo {pages} {6072}
  (\bibinfo {year} {1996})}\BibitemShut {NoStop}%
\bibitem [{\citenamefont {Althorpe}, \citenamefont {Kouri},\ and\ \citenamefont
  {Hoffmann}(1997{\natexlab{a}})}]{jcp-106-7629-ALT1}%
  \BibitemOpen
  \bibfield  {author} {\bibinfo {author} {\bibfnamefont {S.~C.}\ \bibnamefont
  {Althorpe}}, \bibinfo {author} {\bibfnamefont {D.~J.}\ \bibnamefont {Kouri}},
  \ and\ \bibinfo {author} {\bibfnamefont {D.~K.}\ \bibnamefont {Hoffmann}},\
  }\href@noop {} {\bibfield  {journal} {\bibinfo  {journal} {J. Chem. Phys.}\
  }\textbf {\bibinfo {volume} {106}},\ \bibinfo {pages} {7629} (\bibinfo {year}
  {1997}{\natexlab{a}})}\BibitemShut {NoStop}%
\bibitem [{\citenamefont {Althorpe}, \citenamefont {Kouri},\ and\ \citenamefont
  {Hoffmann}(1997{\natexlab{b}})}]{jcp-107-7816-ALT1}%
  \BibitemOpen
  \bibfield  {author} {\bibinfo {author} {\bibfnamefont {S.~C.}\ \bibnamefont
  {Althorpe}}, \bibinfo {author} {\bibfnamefont {D.~J.}\ \bibnamefont {Kouri}},
  \ and\ \bibinfo {author} {\bibfnamefont {D.~K.}\ \bibnamefont {Hoffmann}},\
  }\href@noop {} {\bibfield  {journal} {\bibinfo  {journal} {J. Chem. Phys.}\
  }\textbf {\bibinfo {volume} {107}},\ \bibinfo {pages} {7816} (\bibinfo {year}
  {1997}{\natexlab{b}})}\BibitemShut {NoStop}%
\bibitem [{\citenamefont {Johnson}(1983{\natexlab{a}})}]{jcp-79-1916-JOH}%
  \BibitemOpen
  \bibfield  {author} {\bibinfo {author} {\bibfnamefont {B.~R.}\ \bibnamefont
  {Johnson}},\ }\href@noop {} {\bibfield  {journal} {\bibinfo  {journal} {J.
  Chem. Phys.}\ }\textbf {\bibinfo {volume} {79}},\ \bibinfo {pages} {1916}
  (\bibinfo {year} {1983}{\natexlab{a}})}\BibitemShut {NoStop}%
\bibitem [{\citenamefont {Xu}, \citenamefont {Baer},\ and\ \citenamefont
  {Varandas}(2000)}]{Xu2000_JCP_112}%
  \BibitemOpen
  \bibfield  {author} {\bibinfo {author} {\bibfnamefont {Z.}~\bibnamefont
  {Xu}}, \bibinfo {author} {\bibfnamefont {M.}~\bibnamefont {Baer}}, \ and\
  \bibinfo {author} {\bibfnamefont {A.~J.~C.}\ \bibnamefont {Varandas}},\
  }\href@noop {} {\bibfield  {journal} {\bibinfo  {journal} {J. Chem. Phys.}\
  }\textbf {\bibinfo {volume} {112}},\ \bibinfo {pages} {2746} (\bibinfo {year}
  {2000})}\BibitemShut {NoStop}%
\bibitem [{\citenamefont {Adhikari}\ and\ \citenamefont
  {Varandas}(2013)}]{cpc-184-270-SA}%
  \BibitemOpen
  \bibfield  {author} {\bibinfo {author} {\bibfnamefont {S.}~\bibnamefont
  {Adhikari}}\ and\ \bibinfo {author} {\bibfnamefont {A.~J.~C.}\ \bibnamefont
  {Varandas}},\ }\href@noop {} {\bibfield  {journal} {\bibinfo  {journal}
  {Comput. Phys. Comm.}\ }\textbf {\bibinfo {volume} {184}},\ \bibinfo {pages}
  {270} (\bibinfo {year} {2013})}\BibitemShut {NoStop}%
\bibitem [{\citenamefont {Billing}\ and\ \citenamefont
  {Markovic}(1993)}]{jcp-99-2674-BILL}%
  \BibitemOpen
  \bibfield  {author} {\bibinfo {author} {\bibfnamefont {G.~D.}\ \bibnamefont
  {Billing}}\ and\ \bibinfo {author} {\bibfnamefont {N.}~\bibnamefont
  {Markovic}},\ }\href@noop {} {\bibfield  {journal} {\bibinfo  {journal} {J.
  Chem. Phys.}\ }\textbf {\bibinfo {volume} {99}},\ \bibinfo {pages} {2674}
  (\bibinfo {year} {1993})}\BibitemShut {NoStop}%
\bibitem [{\citenamefont {Billing}\ and\ \citenamefont
  {Markovic}(1994)}]{jcp-100-1085-BILL}%
  \BibitemOpen
  \bibfield  {author} {\bibinfo {author} {\bibfnamefont {G.~D.}\ \bibnamefont
  {Billing}}\ and\ \bibinfo {author} {\bibfnamefont {N.}~\bibnamefont
  {Markovic}},\ }\href@noop {} {\bibfield  {journal} {\bibinfo  {journal} {J.
  Chem. Phys.}\ }\textbf {\bibinfo {volume} {100}},\ \bibinfo {pages} {1085}
  (\bibinfo {year} {1994})}\BibitemShut {NoStop}%
\bibitem [{\citenamefont {Markovic}\ and\ \citenamefont
  {Billing}(1995)}]{cp-191-247-MAR}%
  \BibitemOpen
  \bibfield  {author} {\bibinfo {author} {\bibfnamefont {N.}~\bibnamefont
  {Markovic}}\ and\ \bibinfo {author} {\bibfnamefont {G.~D.}\ \bibnamefont
  {Billing}},\ }\href@noop {} {\bibfield  {journal} {\bibinfo  {journal} {Chem.
  Phys.}\ }\textbf {\bibinfo {volume} {191}},\ \bibinfo {pages} {247} (\bibinfo
  {year} {1995})}\BibitemShut {NoStop}%
\bibitem [{\citenamefont {Skouteris}\ \emph {et~al.}(1999)\citenamefont
  {Skouteris}, \citenamefont {Manolopoulos}, \citenamefont {Bian},
  \citenamefont {Werner}, \citenamefont {Lai},\ and\ \citenamefont
  {Liu}}]{science-286-1713-HYP2}%
  \BibitemOpen
  \bibfield  {author} {\bibinfo {author} {\bibfnamefont {D.}~\bibnamefont
  {Skouteris}}, \bibinfo {author} {\bibfnamefont {D.~E.}\ \bibnamefont
  {Manolopoulos}}, \bibinfo {author} {\bibfnamefont {W.}~\bibnamefont {Bian}},
  \bibinfo {author} {\bibfnamefont {W.~J.}\ \bibnamefont {Werner}}, \bibinfo
  {author} {\bibfnamefont {L.~H.}\ \bibnamefont {Lai}}, \ and\ \bibinfo
  {author} {\bibfnamefont {K.}~\bibnamefont {Liu}},\ }\href@noop {} {\bibfield
  {journal} {\bibinfo  {journal} {Science}\ }\textbf {\bibinfo {volume}
  {286}},\ \bibinfo {pages} {1713} (\bibinfo {year} {1999})}\BibitemShut
  {NoStop}%
\bibitem [{\citenamefont {Honvault}\ and\ \citenamefont
  {Launay}(1999{\natexlab{a}})}]{jcp-111-6665-HYP2}%
  \BibitemOpen
  \bibfield  {author} {\bibinfo {author} {\bibfnamefont {P.}~\bibnamefont
  {Honvault}}\ and\ \bibinfo {author} {\bibfnamefont {J.~M.}\ \bibnamefont
  {Launay}},\ }\href@noop {} {\bibfield  {journal} {\bibinfo  {journal} {J.
  Chem. Phys.}\ }\textbf {\bibinfo {volume} {111}},\ \bibinfo {pages} {6665}
  (\bibinfo {year} {1999}{\natexlab{a}})}\BibitemShut {NoStop}%
\bibitem [{\citenamefont {Lagana}\ \emph {et~al.}(2000)\citenamefont {Lagana},
  \citenamefont {Bolloni}, \citenamefont {Crocchianti},\ and\ \citenamefont
  {Parker}}]{cpl-324-466-HYP2}%
  \BibitemOpen
  \bibfield  {author} {\bibinfo {author} {\bibfnamefont {A.}~\bibnamefont
  {Lagana}}, \bibinfo {author} {\bibfnamefont {A.}~\bibnamefont {Bolloni}},
  \bibinfo {author} {\bibfnamefont {S.}~\bibnamefont {Crocchianti}}, \ and\
  \bibinfo {author} {\bibfnamefont {G.~A.}\ \bibnamefont {Parker}},\
  }\href@noop {} {\bibfield  {journal} {\bibinfo  {journal} {Chem. Phys.
  Lett.}\ }\textbf {\bibinfo {volume} {324}},\ \bibinfo {pages} {466} (\bibinfo
  {year} {2000})}\BibitemShut {NoStop}%
\bibitem [{\citenamefont {Sahoo}\ \emph {et~al.}(2014)\citenamefont {Sahoo},
  \citenamefont {Ghosh}, \citenamefont {Adhikari}, \citenamefont {Sharma},\
  and\ \citenamefont {Varandas}}]{jpcA-118-4837-SA}%
  \BibitemOpen
  \bibfield  {author} {\bibinfo {author} {\bibfnamefont {T.}~\bibnamefont
  {Sahoo}}, \bibinfo {author} {\bibfnamefont {S.}~\bibnamefont {Ghosh}},
  \bibinfo {author} {\bibfnamefont {S.}~\bibnamefont {Adhikari}}, \bibinfo
  {author} {\bibfnamefont {R.}~\bibnamefont {Sharma}}, \ and\ \bibinfo {author}
  {\bibfnamefont {A.~J.~C.}\ \bibnamefont {Varandas}},\ }\href@noop {}
  {\bibfield  {journal} {\bibinfo  {journal} {J. Phys. Chem. A}\ }\textbf
  {\bibinfo {volume} {118}},\ \bibinfo {pages} {4837} (\bibinfo {year}
  {2014})}\BibitemShut {NoStop}%
\bibitem [{\citenamefont {Kuppermann}\ and\ \citenamefont
  {Hipes}(1986)}]{jcp-84-5962-HYP1}%
  \BibitemOpen
  \bibfield  {author} {\bibinfo {author} {\bibfnamefont {A.}~\bibnamefont
  {Kuppermann}}\ and\ \bibinfo {author} {\bibfnamefont {P.~G.}\ \bibnamefont
  {Hipes}},\ }\href@noop {} {\bibfield  {journal} {\bibinfo  {journal} {J.
  Chem. Phys.}\ }\textbf {\bibinfo {volume} {84}},\ \bibinfo {pages} {5962}
  (\bibinfo {year} {1986})}\BibitemShut {NoStop}%
\bibitem [{\citenamefont {Sahoo}\ \emph {et~al.}(2015)\citenamefont {Sahoo},
  \citenamefont {Ghosh}, \citenamefont {Adhikari}, \citenamefont {Sharma},\
  and\ \citenamefont {Varandas}}]{jcp-142-024304-SA}%
  \BibitemOpen
  \bibfield  {author} {\bibinfo {author} {\bibfnamefont {T.}~\bibnamefont
  {Sahoo}}, \bibinfo {author} {\bibfnamefont {S.}~\bibnamefont {Ghosh}},
  \bibinfo {author} {\bibfnamefont {S.}~\bibnamefont {Adhikari}}, \bibinfo
  {author} {\bibfnamefont {R.}~\bibnamefont {Sharma}}, \ and\ \bibinfo {author}
  {\bibfnamefont {A.~J.~C.}\ \bibnamefont {Varandas}},\ }\href@noop {}
  {\bibfield  {journal} {\bibinfo  {journal} {J. Chem. Phys.}\ }\textbf
  {\bibinfo {volume} {142}},\ \bibinfo {pages} {024304/1} (\bibinfo {year}
  {2015})}\BibitemShut {NoStop}%
\bibitem [{\citenamefont {Ghosh}\ \emph
  {et~al.}(2017{\natexlab{b}})\citenamefont {Ghosh}, \citenamefont {Sharma},
  \citenamefont {Adhikari},\ and\ \citenamefont {Varandas}}]{cpl-675-85-SA}%
  \BibitemOpen
  \bibfield  {author} {\bibinfo {author} {\bibfnamefont {S.}~\bibnamefont
  {Ghosh}}, \bibinfo {author} {\bibfnamefont {R.}~\bibnamefont {Sharma}},
  \bibinfo {author} {\bibfnamefont {S.}~\bibnamefont {Adhikari}}, \ and\
  \bibinfo {author} {\bibfnamefont {A.~J.~C.}\ \bibnamefont {Varandas}},\
  }\href@noop {} {\bibfield  {journal} {\bibinfo  {journal} {Chem. Phys.
  Lett.}\ }\textbf {\bibinfo {volume} {675}},\ \bibinfo {pages} {85} (\bibinfo
  {year} {2017}{\natexlab{b}})}\BibitemShut {NoStop}%
\bibitem [{\citenamefont {Ghosh}\ \emph {et~al.}(2018)\citenamefont {Ghosh},
  \citenamefont {Sharma}, \citenamefont {Adhikari},\ and\ \citenamefont
  {Varandas}}]{pccp-20-478-SA}%
  \BibitemOpen
  \bibfield  {author} {\bibinfo {author} {\bibfnamefont {S.}~\bibnamefont
  {Ghosh}}, \bibinfo {author} {\bibfnamefont {R.}~\bibnamefont {Sharma}},
  \bibinfo {author} {\bibfnamefont {S.}~\bibnamefont {Adhikari}}, \ and\
  \bibinfo {author} {\bibfnamefont {A.~J.~C.}\ \bibnamefont {Varandas}},\
  }\href@noop {} {\bibfield  {journal} {\bibinfo  {journal} {Phys. Chem. Chem.
  Phys.}\ }\textbf {\bibinfo {volume} {20}},\ \bibinfo {pages} {478} (\bibinfo
  {year} {2018})}\BibitemShut {NoStop}%
\bibitem [{\citenamefont {Ghosh}\ \emph {et~al.}(2019)\citenamefont {Ghosh},
  \citenamefont {Sharma}, \citenamefont {Adhikari},\ and\ \citenamefont
  {Varandas}}]{pccp-21-20166-SA}%
  \BibitemOpen
  \bibfield  {author} {\bibinfo {author} {\bibfnamefont {S.}~\bibnamefont
  {Ghosh}}, \bibinfo {author} {\bibfnamefont {R.}~\bibnamefont {Sharma}},
  \bibinfo {author} {\bibfnamefont {S.}~\bibnamefont {Adhikari}}, \ and\
  \bibinfo {author} {\bibfnamefont {A.~J.~C.}\ \bibnamefont {Varandas}},\
  }\href@noop {} {\bibfield  {journal} {\bibinfo  {journal} {Phys. Chem. Chem.
  Phys.}\ }\textbf {\bibinfo {volume} {21}},\ \bibinfo {pages} {20166}
  (\bibinfo {year} {2019})}\BibitemShut {NoStop}%
\bibitem [{\citenamefont {Tang}\ \emph
  {et~al.}(2005{\natexlab{b}})\citenamefont {Tang}, \citenamefont {Xu},
  \citenamefont {Zhang}, \citenamefont {Hou}, \citenamefont {Chang},
  \citenamefont {Ng}, \citenamefont {Chiu}, \citenamefont {Dressler},\ and\
  \citenamefont {Levandier}}]{heh2+_dyna}%
  \BibitemOpen
  \bibfield  {author} {\bibinfo {author} {\bibfnamefont {X.~N.}\ \bibnamefont
  {Tang}}, \bibinfo {author} {\bibfnamefont {H.}~\bibnamefont {Xu}}, \bibinfo
  {author} {\bibfnamefont {T.}~\bibnamefont {Zhang}}, \bibinfo {author}
  {\bibfnamefont {Y.}~\bibnamefont {Hou}}, \bibinfo {author} {\bibfnamefont
  {C.}~\bibnamefont {Chang}}, \bibinfo {author} {\bibfnamefont {C.~Y.}\
  \bibnamefont {Ng}}, \bibinfo {author} {\bibfnamefont {Y.}~\bibnamefont
  {Chiu}}, \bibinfo {author} {\bibfnamefont {R.~A.}\ \bibnamefont {Dressler}},
  \ and\ \bibinfo {author} {\bibfnamefont {D.~J.}\ \bibnamefont {Levandier}},\
  }\href@noop {} {\bibfield  {journal} {\bibinfo  {journal} {J. Chem. Phys.}\
  }\textbf {\bibinfo {volume} {122}},\ \bibinfo {pages} {164301} (\bibinfo
  {year} {2005}{\natexlab{b}})}\BibitemShut {NoStop}%
\bibitem [{\citenamefont
  {Baer}(2006{\natexlab{b}})}]{Baer2006_bbo_book_sect211}%
  \BibitemOpen
  \bibfield  {author} {\bibinfo {author} {\bibfnamefont {M.}~\bibnamefont
  {Baer}},\ }\enquote {\bibinfo {title} {{Beyond Born-Oppenheimer: Electronic
  Nonadiabatic Coupling Terms and Conical Intersections}},}\ \ (\bibinfo
  {publisher} {Wiley},\ \bibinfo {address} {Hoboken, NJ, USA},\ \bibinfo {year}
  {2006})\ Chap.\ \bibinfo {chapter} {2.1.1}\BibitemShut {NoStop}%
\bibitem [{\citenamefont {Clusius}(1941)}]{clusius1941einfuhrung}%
  \BibitemOpen
  \bibfield  {author} {\bibinfo {author} {\bibnamefont {Clusius}},\ }\href@noop
  {} {\enquote {\bibinfo {title} {Einf{\"u}hrung in die quantenchemie. von h.
  hellmann. 350 s., 43 abb., 35 tab. franz deuticke, leipzig u. wien 1937. pr.
  geh. rm. 20,-. geb. rm. 22,-},}\ } (\bibinfo {year} {1941})\BibitemShut
  {NoStop}%
\bibitem [{\citenamefont {Feynman}(1939{\natexlab{b}})}]{feynman1939forces}%
  \BibitemOpen
  \bibfield  {author} {\bibinfo {author} {\bibfnamefont {R.~P.}\ \bibnamefont
  {Feynman}},\ }\href@noop {} {\bibfield  {journal} {\bibinfo  {journal} {Phys.
  Rev.}\ }\textbf {\bibinfo {volume} {56}},\ \bibinfo {pages} {340} (\bibinfo
  {year} {1939}{\natexlab{b}})}\BibitemShut {NoStop}%
\bibitem [{\citenamefont {Epstein}(1954)}]{epstein1954note}%
  \BibitemOpen
  \bibfield  {author} {\bibinfo {author} {\bibfnamefont {S.~T.}\ \bibnamefont
  {Epstein}},\ }\href@noop {} {\bibfield  {journal} {\bibinfo  {journal} {Am.
  J. Phys.}\ }\textbf {\bibinfo {volume} {22}},\ \bibinfo {pages} {613}
  (\bibinfo {year} {1954})}\BibitemShut {NoStop}%
\bibitem [{\citenamefont {Mukherjee}\ \emph
  {et~al.}(2024{\natexlab{b}})\citenamefont {Mukherjee}, \citenamefont {Saha},
  \citenamefont {Ghosh}, \citenamefont {Adhikari}, \citenamefont
  {Sathyamurthy},\ and\ \citenamefont {Baer}}]{mukherjee2024quasi}%
  \BibitemOpen
  \bibfield  {author} {\bibinfo {author} {\bibfnamefont {S.}~\bibnamefont
  {Mukherjee}}, \bibinfo {author} {\bibfnamefont {S.}~\bibnamefont {Saha}},
  \bibinfo {author} {\bibfnamefont {S.}~\bibnamefont {Ghosh}}, \bibinfo
  {author} {\bibfnamefont {S.}~\bibnamefont {Adhikari}}, \bibinfo {author}
  {\bibfnamefont {N.}~\bibnamefont {Sathyamurthy}}, \ and\ \bibinfo {author}
  {\bibfnamefont {M.}~\bibnamefont {Baer}},\ }\href@noop {} {\bibfield
  {journal} {\bibinfo  {journal} {J. Phys. Chem. A}\ }\textbf {\bibinfo
  {volume} {128}},\ \bibinfo {pages} {7691} (\bibinfo {year}
  {2024}{\natexlab{b}})}\BibitemShut {NoStop}%
\bibitem [{\citenamefont {Baer}(1975{\natexlab{b}})}]{Baer1975_CPL_35}%
  \BibitemOpen
  \bibfield  {author} {\bibinfo {author} {\bibfnamefont {M.}~\bibnamefont
  {Baer}},\ }\href@noop {} {\bibfield  {journal} {\bibinfo  {journal} {Chem.
  Phys. Lett.}\ }\textbf {\bibinfo {volume} {35}},\ \bibinfo {pages} {112}
  (\bibinfo {year} {1975}{\natexlab{b}})}\BibitemShut {NoStop}%
\bibitem [{\citenamefont {Top}\ and\ \citenamefont
  {Baer}(1977{\natexlab{b}})}]{TB}%
  \BibitemOpen
  \bibfield  {author} {\bibinfo {author} {\bibfnamefont {Z.~H.}\ \bibnamefont
  {Top}}\ and\ \bibinfo {author} {\bibfnamefont {M.}~\bibnamefont {Baer}},\
  }\href@noop {} {\bibfield  {journal} {\bibinfo  {journal} {J. Chem. Phys.}\
  }\textbf {\bibinfo {volume} {66}},\ \bibinfo {pages} {1363} (\bibinfo {year}
  {1977}{\natexlab{b}})}\BibitemShut {NoStop}%
\bibitem [{\citenamefont {Sarkar}\ and\ \citenamefont
  {Adhikari}(2006{\natexlab{c}})}]{Sarkar2006_JCP_124_EBO}%
  \BibitemOpen
  \bibfield  {author} {\bibinfo {author} {\bibfnamefont {B.}~\bibnamefont
  {Sarkar}}\ and\ \bibinfo {author} {\bibfnamefont {S.}~\bibnamefont
  {Adhikari}},\ }\href@noop {} {\bibfield  {journal} {\bibinfo  {journal} {J.
  Chem. Phys.}\ }\textbf {\bibinfo {volume} {124}},\ \bibinfo {pages} {074101}
  (\bibinfo {year} {2006}{\natexlab{c}})}\BibitemShut {NoStop}%
\bibitem [{\citenamefont {Mukherjee}\ \emph
  {et~al.}(2020{\natexlab{c}})\citenamefont {Mukherjee}, \citenamefont
  {Naskar}, \citenamefont {Mukherjee}, \citenamefont {Ravi}, \citenamefont
  {Shamasundar}, \citenamefont {Mukhopadhyay},\ and\ \citenamefont
  {Adhikari}}]{jcp-153-174301-SA}%
  \BibitemOpen
  \bibfield  {author} {\bibinfo {author} {\bibfnamefont {B.}~\bibnamefont
  {Mukherjee}}, \bibinfo {author} {\bibfnamefont {K.}~\bibnamefont {Naskar}},
  \bibinfo {author} {\bibfnamefont {S.}~\bibnamefont {Mukherjee}}, \bibinfo
  {author} {\bibfnamefont {S.}~\bibnamefont {Ravi}}, \bibinfo {author}
  {\bibfnamefont {K.~R.}\ \bibnamefont {Shamasundar}}, \bibinfo {author}
  {\bibfnamefont {D.}~\bibnamefont {Mukhopadhyay}}, \ and\ \bibinfo {author}
  {\bibfnamefont {S.}~\bibnamefont {Adhikari}},\ }\href@noop {} {\bibfield
  {journal} {\bibinfo  {journal} {J. Chem. Phys.}\ }\textbf {\bibinfo {volume}
  {153}},\ \bibinfo {pages} {174301} (\bibinfo {year}
  {2020}{\natexlab{c}})}\BibitemShut {NoStop}%
\bibitem [{\citenamefont {Werner}\ \emph {et~al.}(2010)\citenamefont {Werner},
  \citenamefont {Knowles}, \citenamefont {Knizia}, \citenamefont {Manby},
  \citenamefont {{Sch\"{u}tz}}, \citenamefont {Celani}, \citenamefont {Korona},
  \citenamefont {Lindh}, \citenamefont {Mitrushenkov}, \citenamefont {Rauhut},
  \citenamefont {Shamasundar}, \citenamefont {Adler}, \citenamefont {Amos},
  \citenamefont {Bernhardsson}, \citenamefont {Berning}, \citenamefont
  {Cooper}, \citenamefont {Deegan}, \citenamefont {Dobbyn}, \citenamefont
  {Eckert}, \citenamefont {Goll}, \citenamefont {Hampel}, \citenamefont
  {Hesselmann}, \citenamefont {Hetzer}, \citenamefont {Hrenar}, \citenamefont
  {Jansen}, \citenamefont {K\"oppl}, \citenamefont {Liu}, \citenamefont
  {Lloyd}, \citenamefont {Mata}, \citenamefont {May}, \citenamefont
  {McNicholas}, \citenamefont {Meyer}, \citenamefont {Mura}, \citenamefont
  {Nicklass}, \citenamefont {O'Neill}, \citenamefont {Palmieri}, \citenamefont
  {Peng}, \citenamefont {Pfl\"uger}, \citenamefont {Pitzer}, \citenamefont
  {Reiher}, \citenamefont {Shiozaki}, \citenamefont {Stoll}, \citenamefont
  {Stone}, \citenamefont {Tarroni}, \citenamefont {Thorsteinsson},\ and\
  \citenamefont {Wang}}]{MOL}%
  \BibitemOpen
  \bibfield  {author} {\bibinfo {author} {\bibfnamefont {H.-J.}\ \bibnamefont
  {Werner}}, \bibinfo {author} {\bibfnamefont {P.~J.}\ \bibnamefont {Knowles}},
  \bibinfo {author} {\bibfnamefont {G.}~\bibnamefont {Knizia}}, \bibinfo
  {author} {\bibfnamefont {F.~R.}\ \bibnamefont {Manby}}, \bibinfo {author}
  {\bibfnamefont {M.}~\bibnamefont {{Sch\"{u}tz}}}, \bibinfo {author}
  {\bibfnamefont {P.}~\bibnamefont {Celani}}, \bibinfo {author} {\bibfnamefont
  {T.}~\bibnamefont {Korona}}, \bibinfo {author} {\bibfnamefont
  {R.}~\bibnamefont {Lindh}}, \bibinfo {author} {\bibfnamefont
  {A.}~\bibnamefont {Mitrushenkov}}, \bibinfo {author} {\bibfnamefont
  {G.}~\bibnamefont {Rauhut}}, \bibinfo {author} {\bibfnamefont {K.~R.}\
  \bibnamefont {Shamasundar}}, \bibinfo {author} {\bibfnamefont {T.~B.}\
  \bibnamefont {Adler}}, \bibinfo {author} {\bibfnamefont {R.~D.}\ \bibnamefont
  {Amos}}, \bibinfo {author} {\bibfnamefont {A.}~\bibnamefont {Bernhardsson}},
  \bibinfo {author} {\bibfnamefont {A.}~\bibnamefont {Berning}}, \bibinfo
  {author} {\bibfnamefont {D.~L.}\ \bibnamefont {Cooper}}, \bibinfo {author}
  {\bibfnamefont {M.~J.~O.}\ \bibnamefont {Deegan}}, \bibinfo {author}
  {\bibfnamefont {A.~J.}\ \bibnamefont {Dobbyn}}, \bibinfo {author}
  {\bibfnamefont {F.}~\bibnamefont {Eckert}}, \bibinfo {author} {\bibfnamefont
  {E.}~\bibnamefont {Goll}}, \bibinfo {author} {\bibfnamefont {C.}~\bibnamefont
  {Hampel}}, \bibinfo {author} {\bibfnamefont {A.}~\bibnamefont {Hesselmann}},
  \bibinfo {author} {\bibfnamefont {G.}~\bibnamefont {Hetzer}}, \bibinfo
  {author} {\bibfnamefont {T.}~\bibnamefont {Hrenar}}, \bibinfo {author}
  {\bibfnamefont {G.}~\bibnamefont {Jansen}}, \bibinfo {author} {\bibfnamefont
  {C.}~\bibnamefont {K\"oppl}}, \bibinfo {author} {\bibfnamefont
  {Y.}~\bibnamefont {Liu}}, \bibinfo {author} {\bibfnamefont {A.~W.}\
  \bibnamefont {Lloyd}}, \bibinfo {author} {\bibfnamefont {R.~A.}\ \bibnamefont
  {Mata}}, \bibinfo {author} {\bibfnamefont {A.~J.}\ \bibnamefont {May}},
  \bibinfo {author} {\bibfnamefont {S.~J.}\ \bibnamefont {McNicholas}},
  \bibinfo {author} {\bibfnamefont {W.}~\bibnamefont {Meyer}}, \bibinfo
  {author} {\bibfnamefont {M.~E.}\ \bibnamefont {Mura}}, \bibinfo {author}
  {\bibfnamefont {A.}~\bibnamefont {Nicklass}}, \bibinfo {author}
  {\bibfnamefont {D.~P.}\ \bibnamefont {O'Neill}}, \bibinfo {author}
  {\bibfnamefont {P.}~\bibnamefont {Palmieri}}, \bibinfo {author}
  {\bibfnamefont {D.}~\bibnamefont {Peng}}, \bibinfo {author} {\bibfnamefont
  {K.}~\bibnamefont {Pfl\"uger}}, \bibinfo {author} {\bibfnamefont
  {R.}~\bibnamefont {Pitzer}}, \bibinfo {author} {\bibfnamefont
  {M.}~\bibnamefont {Reiher}}, \bibinfo {author} {\bibfnamefont
  {T.}~\bibnamefont {Shiozaki}}, \bibinfo {author} {\bibfnamefont
  {H.}~\bibnamefont {Stoll}}, \bibinfo {author} {\bibfnamefont {A.~J.}\
  \bibnamefont {Stone}}, \bibinfo {author} {\bibfnamefont {R.}~\bibnamefont
  {Tarroni}}, \bibinfo {author} {\bibfnamefont {T.}~\bibnamefont
  {Thorsteinsson}}, \ and\ \bibinfo {author} {\bibfnamefont {M.}~\bibnamefont
  {Wang}},\ }\href@noop {} {\enquote {\bibinfo {title} {Molpro, version 2010.1,
  a package of ab initio programs},}\ } (\bibinfo {year} {2010}),\ \bibinfo
  {note} {see http://www.molpro.net}\BibitemShut {NoStop}%
\bibitem [{\citenamefont {Yarkony}(1996{\natexlab{b}})}]{Yarkony_jpc100}%
  \BibitemOpen
  \bibfield  {author} {\bibinfo {author} {\bibfnamefont {D.~R.}\ \bibnamefont
  {Yarkony}},\ }\href@noop {} {\bibfield  {journal} {\bibinfo  {journal} {J.
  Phys. Chem.}\ }\textbf {\bibinfo {volume} {100}},\ \bibinfo {pages} {17439}
  (\bibinfo {year} {1996}{\natexlab{b}})}\BibitemShut {NoStop}%
\bibitem [{\citenamefont {Alexander}, \citenamefont {Manolopoulos},\ and\
  \citenamefont {Werner}(2000{\natexlab{a}})}]{jcp-113-11084-WERNER}%
  \BibitemOpen
  \bibfield  {author} {\bibinfo {author} {\bibfnamefont {M.~H.}\ \bibnamefont
  {Alexander}}, \bibinfo {author} {\bibfnamefont {D.~E.}\ \bibnamefont
  {Manolopoulos}}, \ and\ \bibinfo {author} {\bibfnamefont {H.-J.}\
  \bibnamefont {Werner}},\ }\href@noop {} {\bibfield  {journal} {\bibinfo
  {journal} {J. Chem. Phys.}\ }\textbf {\bibinfo {volume} {113}},\ \bibinfo
  {pages} {11084} (\bibinfo {year} {2000}{\natexlab{a}})}\BibitemShut {NoStop}%
\bibitem [{\citenamefont {Baer}(2006{\natexlab{c}})}]{BaerBook2006}%
  \BibitemOpen
  \bibfield  {author} {\bibinfo {author} {\bibfnamefont {M.}~\bibnamefont
  {Baer}},\ }\href {\doibase 10.1002/0471780081} {\emph {\bibinfo {title}
  {{Beyond Born - Oppenheimer: Conical Intersections and Electronic
  Nonadiabatic Coupling Terms}}}}\ (\bibinfo {year} {2006})\BibitemShut
  {NoStop}%
\bibitem [{\citenamefont {Baer}(2002{\natexlab{b}})}]{Baer2002_PhysRep}%
  \BibitemOpen
  \bibfield  {author} {\bibinfo {author} {\bibfnamefont {M.}~\bibnamefont
  {Baer}},\ }\href {\doibase 10.1016/S0370-1573(01)00052-7} {\bibfield
  {journal} {\bibinfo  {journal} {Phys. Rep.}\ }\textbf {\bibinfo {volume}
  {358}},\ \bibinfo {pages} {75} (\bibinfo {year}
  {2002}{\natexlab{b}})}\BibitemShut {NoStop}%
\bibitem [{\citenamefont {Mukherjee}\ \emph
  {et~al.}(2019{\natexlab{d}})\citenamefont {Mukherjee}, \citenamefont
  {Naskar}, \citenamefont {Mukherjee}, \citenamefont {Ghosh}, \citenamefont
  {Sahoo},\ and\ \citenamefont {Adhikari}}]{irpc-38-287-SA}%
  \BibitemOpen
  \bibfield  {author} {\bibinfo {author} {\bibfnamefont {B.}~\bibnamefont
  {Mukherjee}}, \bibinfo {author} {\bibfnamefont {K.}~\bibnamefont {Naskar}},
  \bibinfo {author} {\bibfnamefont {S.}~\bibnamefont {Mukherjee}}, \bibinfo
  {author} {\bibfnamefont {S.}~\bibnamefont {Ghosh}}, \bibinfo {author}
  {\bibfnamefont {T.}~\bibnamefont {Sahoo}}, \ and\ \bibinfo {author}
  {\bibfnamefont {S.}~\bibnamefont {Adhikari}},\ }\href@noop {} {\bibfield
  {journal} {\bibinfo  {journal} {Int. Rev. Phys. Chem.}\ }\textbf {\bibinfo
  {volume} {38}},\ \bibinfo {pages} {287} (\bibinfo {year}
  {2019}{\natexlab{d}})}\BibitemShut {NoStop}%
\bibitem [{\citenamefont {Dutta}\ \emph
  {et~al.}(2020{\natexlab{b}})\citenamefont {Dutta}, \citenamefont {Mukherjee},
  \citenamefont {Naskar}, \citenamefont {Ghosh}, \citenamefont {Mukherjee},
  \citenamefont {Ravi},\ and\ \citenamefont {Adhikari}}]{pccp-22-27496-SA}%
  \BibitemOpen
  \bibfield  {author} {\bibinfo {author} {\bibfnamefont {J.}~\bibnamefont
  {Dutta}}, \bibinfo {author} {\bibfnamefont {S.}~\bibnamefont {Mukherjee}},
  \bibinfo {author} {\bibfnamefont {K.}~\bibnamefont {Naskar}}, \bibinfo
  {author} {\bibfnamefont {S.}~\bibnamefont {Ghosh}}, \bibinfo {author}
  {\bibfnamefont {B.}~\bibnamefont {Mukherjee}}, \bibinfo {author}
  {\bibfnamefont {S.}~\bibnamefont {Ravi}}, \ and\ \bibinfo {author}
  {\bibfnamefont {S.}~\bibnamefont {Adhikari}},\ }\href@noop {} {\bibfield
  {journal} {\bibinfo  {journal} {Phys. Chem. Chem. Phys.}\ }\textbf {\bibinfo
  {volume} {22}},\ \bibinfo {pages} {27496} (\bibinfo {year}
  {2020}{\natexlab{b}})}\BibitemShut {NoStop}%
\bibitem [{\citenamefont {Naskar}\ \emph
  {et~al.}(2020{\natexlab{b}})\citenamefont {Naskar}, \citenamefont
  {Mukherjee}, \citenamefont {Mukherjee}, \citenamefont {Ravi}, \citenamefont
  {Mukherjee}, \citenamefont {Sardar},\ and\ \citenamefont
  {Adhikari}}]{jctc-16-1666-SA}%
  \BibitemOpen
  \bibfield  {author} {\bibinfo {author} {\bibfnamefont {K.}~\bibnamefont
  {Naskar}}, \bibinfo {author} {\bibfnamefont {S.}~\bibnamefont {Mukherjee}},
  \bibinfo {author} {\bibfnamefont {B.}~\bibnamefont {Mukherjee}}, \bibinfo
  {author} {\bibfnamefont {S.}~\bibnamefont {Ravi}}, \bibinfo {author}
  {\bibfnamefont {S.}~\bibnamefont {Mukherjee}}, \bibinfo {author}
  {\bibfnamefont {S.}~\bibnamefont {Sardar}}, \ and\ \bibinfo {author}
  {\bibfnamefont {S.}~\bibnamefont {Adhikari}},\ }\href@noop {} {\bibfield
  {journal} {\bibinfo  {journal} {J. Chem. Theo. Comput.}\ }\textbf {\bibinfo
  {volume} {16}},\ \bibinfo {pages} {1666} (\bibinfo {year}
  {2020}{\natexlab{b}})}\BibitemShut {NoStop}%
\bibitem [{\citenamefont {Johnson}(1980)}]{Johnson1980_JCP_73}%
  \BibitemOpen
  \bibfield  {author} {\bibinfo {author} {\bibfnamefont {B.~R.}\ \bibnamefont
  {Johnson}},\ }\href@noop {} {\bibfield  {journal} {\bibinfo  {journal} {J.
  Chem. Phys.}\ }\textbf {\bibinfo {volume} {73}},\ \bibinfo {pages} {5051}
  (\bibinfo {year} {1980})}\BibitemShut {NoStop}%
\bibitem [{\citenamefont {Kosloff}\ and\ \citenamefont
  {Kosloff}(1986)}]{jcomphy-52-35-DKRR}%
  \BibitemOpen
  \bibfield  {author} {\bibinfo {author} {\bibfnamefont {D.}~\bibnamefont
  {Kosloff}}\ and\ \bibinfo {author} {\bibfnamefont {R.}~\bibnamefont
  {Kosloff}},\ }\href@noop {} {\bibfield  {journal} {\bibinfo  {journal} {J.
  Comput. Phys.}\ }\textbf {\bibinfo {volume} {52}},\ \bibinfo {pages} {35}
  (\bibinfo {year} {1986})}\BibitemShut {NoStop}%
\bibitem [{\citenamefont {Park}\ and\ \citenamefont
  {Light}(1986)}]{jcp-85-5870-TJPR}%
  \BibitemOpen
  \bibfield  {author} {\bibinfo {author} {\bibfnamefont {T.~J.}\ \bibnamefont
  {Park}}\ and\ \bibinfo {author} {\bibfnamefont {J.~C.}\ \bibnamefont
  {Light}},\ }\href@noop {} {\bibfield  {journal} {\bibinfo  {journal} {J.
  Chem. Phys.}\ }\textbf {\bibinfo {volume} {85}},\ \bibinfo {pages} {5870}
  (\bibinfo {year} {1986})}\BibitemShut {NoStop}%
\bibitem [{\citenamefont {Aoiz}\ \emph {et~al.}(1996)\citenamefont {Aoiz},
  \citenamefont {Banares}, \citenamefont {Herrero}, \citenamefont {Stark},\
  and\ \citenamefont {Werner}}]{cpl-254-341-AOIZ}%
  \BibitemOpen
  \bibfield  {author} {\bibinfo {author} {\bibfnamefont {F.~J.}\ \bibnamefont
  {Aoiz}}, \bibinfo {author} {\bibfnamefont {L.}~\bibnamefont {Banares}},
  \bibinfo {author} {\bibfnamefont {V.}~\bibnamefont {Herrero}}, \bibinfo
  {author} {\bibfnamefont {K.}~\bibnamefont {Stark}}, \ and\ \bibinfo {author}
  {\bibfnamefont {H.-J.}\ \bibnamefont {Werner}},\ }\href@noop {} {\bibfield
  {journal} {\bibinfo  {journal} {Chem. Phys. Lett.}\ }\textbf {\bibinfo
  {volume} {254}},\ \bibinfo {pages} {341} (\bibinfo {year}
  {1996})}\BibitemShut {NoStop}%
\bibitem [{\citenamefont {Aoiz}\ \emph {et~al.}(1994)\citenamefont {Aoiz},
  \citenamefont {Banares}, \citenamefont {Herrero}, \citenamefont {Rabanos},
  \citenamefont {K.Stark},\ and\ \citenamefont
  {H.-J.Werner}}]{cpl-223-215-AOIZ}%
  \BibitemOpen
  \bibfield  {author} {\bibinfo {author} {\bibfnamefont {F.~J.}\ \bibnamefont
  {Aoiz}}, \bibinfo {author} {\bibfnamefont {L.}~\bibnamefont {Banares}},
  \bibinfo {author} {\bibfnamefont {V.~J.}\ \bibnamefont {Herrero}}, \bibinfo
  {author} {\bibfnamefont {V.}~\bibnamefont {Rabanos}}, \bibinfo {author}
  {\bibnamefont {K.Stark}}, \ and\ \bibinfo {author} {\bibnamefont
  {H.-J.Werner}},\ }\href@noop {} {\bibfield  {journal} {\bibinfo  {journal}
  {Chem. Phys. Lett.}\ }\textbf {\bibinfo {volume} {223}},\ \bibinfo {pages}
  {215} (\bibinfo {year} {1994})}\BibitemShut {NoStop}%
\bibitem [{\citenamefont {Stark}\ and\ \citenamefont
  {Werner}(1996{\natexlab{a}})}]{SW_jcp104}%
  \BibitemOpen
  \bibfield  {author} {\bibinfo {author} {\bibfnamefont {K.}~\bibnamefont
  {Stark}}\ and\ \bibinfo {author} {\bibfnamefont {H.-J.}\ \bibnamefont
  {Werner}},\ }\href@noop {} {\bibfield  {journal} {\bibinfo  {journal} {J.
  Chem. Phys.}\ }\textbf {\bibinfo {volume} {104}},\ \bibinfo {pages} {6515}
  (\bibinfo {year} {1996}{\natexlab{a}})}\BibitemShut {NoStop}%
\bibitem [{\citenamefont {Castillo}\ \emph
  {et~al.}(1996{\natexlab{b}})\citenamefont {Castillo}, \citenamefont
  {Manolopoulos}, \citenamefont {Stark},\ and\ \citenamefont
  {Werner}}]{jcp-104-6531-CAST}%
  \BibitemOpen
  \bibfield  {author} {\bibinfo {author} {\bibfnamefont {J.~F.}\ \bibnamefont
  {Castillo}}, \bibinfo {author} {\bibfnamefont {D.~E.}\ \bibnamefont
  {Manolopoulos}}, \bibinfo {author} {\bibfnamefont {K.}~\bibnamefont {Stark}},
  \ and\ \bibinfo {author} {\bibfnamefont {H.~J.}\ \bibnamefont {Werner}},\
  }\href@noop {} {\bibfield  {journal} {\bibinfo  {journal} {J. Chem. Phys.}\
  }\textbf {\bibinfo {volume} {104}},\ \bibinfo {pages} {6531} (\bibinfo {year}
  {1996}{\natexlab{b}})}\BibitemShut {NoStop}%
\bibitem [{\citenamefont {Hartke}\ and\ \citenamefont
  {Werner}(1997{\natexlab{a}})}]{HSW_cpl280}%
  \BibitemOpen
  \bibfield  {author} {\bibinfo {author} {\bibfnamefont {B.}~\bibnamefont
  {Hartke}}\ and\ \bibinfo {author} {\bibfnamefont {H.-J.}\ \bibnamefont
  {Werner}},\ }\href@noop {} {\bibfield  {journal} {\bibinfo  {journal} {Chem.
  Phys. Lett.}\ }\textbf {\bibinfo {volume} {280}},\ \bibinfo {pages} {430}
  (\bibinfo {year} {1997}{\natexlab{a}})}\BibitemShut {NoStop}%
\bibitem [{\citenamefont {Alexander}, \citenamefont {Manolopoulos},\ and\
  \citenamefont {Werner}(2000{\natexlab{b}})}]{ASW_jcp113}%
  \BibitemOpen
  \bibfield  {author} {\bibinfo {author} {\bibfnamefont {M.~H.}\ \bibnamefont
  {Alexander}}, \bibinfo {author} {\bibfnamefont {D.~E.}\ \bibnamefont
  {Manolopoulos}}, \ and\ \bibinfo {author} {\bibfnamefont {H.-J.}\
  \bibnamefont {Werner}},\ }\href@noop {} {\bibfield  {journal} {\bibinfo
  {journal} {J. Chem. Phys.}\ }\textbf {\bibinfo {volume} {113}},\ \bibinfo
  {pages} {11084} (\bibinfo {year} {2000}{\natexlab{b}})}\BibitemShut {NoStop}%
\bibitem [{\citenamefont {Li}\ \emph {et~al.}(2007)\citenamefont {Li},
  \citenamefont {Werner}, \citenamefont {Lique},\ and\ \citenamefont
  {Alexander}}]{Werner_jcp127}%
  \BibitemOpen
  \bibfield  {author} {\bibinfo {author} {\bibfnamefont {G.}~\bibnamefont
  {Li}}, \bibinfo {author} {\bibfnamefont {H.-J.}\ \bibnamefont {Werner}},
  \bibinfo {author} {\bibfnamefont {F.}~\bibnamefont {Lique}}, \ and\ \bibinfo
  {author} {\bibfnamefont {M.~H.}\ \bibnamefont {Alexander}},\ }\href@noop {}
  {\bibfield  {journal} {\bibinfo  {journal} {J. Chem. Phys.}\ }\textbf
  {\bibinfo {volume} {127}},\ \bibinfo {pages} {174302} (\bibinfo {year}
  {2007})}\BibitemShut {NoStop}%
\bibitem [{\citenamefont {Xu}, \citenamefont {Xie},\ and\ \citenamefont
  {Zhang}(2006)}]{cjcp-19-96-XU}%
  \BibitemOpen
  \bibfield  {author} {\bibinfo {author} {\bibfnamefont {C.}~\bibnamefont
  {Xu}}, \bibinfo {author} {\bibfnamefont {D.}~\bibnamefont {Xie}}, \ and\
  \bibinfo {author} {\bibfnamefont {D.~H.}\ \bibnamefont {Zhang}},\ }\href@noop
  {} {\bibfield  {journal} {\bibinfo  {journal} {Chin. J. Chem. Phys.}\
  }\textbf {\bibinfo {volume} {19}},\ \bibinfo {pages} {96} (\bibinfo {year}
  {2006})}\BibitemShut {NoStop}%
\bibitem [{\citenamefont {Li}, \citenamefont {Song},\ and\ \citenamefont
  {Varandas}(2015)}]{epjd-69-22-AJCV}%
  \BibitemOpen
  \bibfield  {author} {\bibinfo {author} {\bibfnamefont {Y.-Q.}\ \bibnamefont
  {Li}}, \bibinfo {author} {\bibfnamefont {Y.-Z.}\ \bibnamefont {Song}}, \ and\
  \bibinfo {author} {\bibfnamefont {A.~J.~C.}\ \bibnamefont {Varandas}},\
  }\href@noop {} {\bibfield  {journal} {\bibinfo  {journal} {Eur. Phys. J. D}\
  }\textbf {\bibinfo {volume} {69}},\ \bibinfo {pages} {22} (\bibinfo {year}
  {2015})}\BibitemShut {NoStop}%
\bibitem [{\citenamefont {Das}\ \emph {et~al.}(2012)\citenamefont {Das},
  \citenamefont {Mukhopadhyay}, \citenamefont {Adhikari},\ and\ \citenamefont
  {Baer}}]{ijqc-112-2561-ANITA}%
  \BibitemOpen
  \bibfield  {author} {\bibinfo {author} {\bibfnamefont {A.}~\bibnamefont
  {Das}}, \bibinfo {author} {\bibfnamefont {D.}~\bibnamefont {Mukhopadhyay}},
  \bibinfo {author} {\bibfnamefont {S.}~\bibnamefont {Adhikari}}, \ and\
  \bibinfo {author} {\bibfnamefont {M.}~\bibnamefont {Baer}},\ }\href@noop {}
  {\bibfield  {journal} {\bibinfo  {journal} {Int. J. Quant. Chem.}\ }\textbf
  {\bibinfo {volume} {112}},\ \bibinfo {pages} {2561} (\bibinfo {year}
  {2012})}\BibitemShut {NoStop}%
\bibitem [{\citenamefont {Preston}\ and\ \citenamefont
  {Tully}(1970)}]{jcp-54-4297-TUL}%
  \BibitemOpen
  \bibfield  {author} {\bibinfo {author} {\bibfnamefont {K.}~\bibnamefont
  {Preston}}\ and\ \bibinfo {author} {\bibfnamefont {J.~C.}\ \bibnamefont
  {Tully}},\ }\href@noop {} {\bibfield  {journal} {\bibinfo  {journal} {J.
  Chem. Phys.}\ }\textbf {\bibinfo {volume} {54}},\ \bibinfo {pages} {4297}
  (\bibinfo {year} {1970})}\BibitemShut {NoStop}%
\bibitem [{\citenamefont {Kamisaka}\ \emph
  {et~al.}(2002{\natexlab{a}})\citenamefont {Kamisaka}, \citenamefont {Bian},
  \citenamefont {Nobusada},\ and\ \citenamefont
  {Nakamura}}]{Kamisaka_JCP_2002}%
  \BibitemOpen
  \bibfield  {author} {\bibinfo {author} {\bibfnamefont {H.}~\bibnamefont
  {Kamisaka}}, \bibinfo {author} {\bibfnamefont {W.}~\bibnamefont {Bian}},
  \bibinfo {author} {\bibfnamefont {K.}~\bibnamefont {Nobusada}}, \ and\
  \bibinfo {author} {\bibfnamefont {H.}~\bibnamefont {Nakamura}},\ }\href@noop
  {} {\bibfield  {journal} {\bibinfo  {journal} {J. Chem. Phys.}\ }\textbf
  {\bibinfo {volume} {116}},\ \bibinfo {pages} {654} (\bibinfo {year}
  {2002}{\natexlab{a}})}\BibitemShut {NoStop}%
\bibitem [{\citenamefont {Viegas}, \citenamefont {Alijah},\ and\ \citenamefont
  {Varandas}(2007{\natexlab{a}})}]{jcp-126-074309-AJCV}%
  \BibitemOpen
  \bibfield  {author} {\bibinfo {author} {\bibfnamefont {L.~P.}\ \bibnamefont
  {Viegas}}, \bibinfo {author} {\bibfnamefont {A.}~\bibnamefont {Alijah}}, \
  and\ \bibinfo {author} {\bibfnamefont {A.~J.~C.}\ \bibnamefont {Varandas}},\
  }\href@noop {} {\bibfield  {journal} {\bibinfo  {journal} {J. Chem. Phys.}\
  }\textbf {\bibinfo {volume} {126}},\ \bibinfo {pages} {074309/1} (\bibinfo
  {year} {2007}{\natexlab{a}})}\BibitemShut {NoStop}%
\bibitem [{\citenamefont {Mukherjee}, \citenamefont {Mukhopadhyay},\ and\
  \citenamefont {Adhikari}(2014{\natexlab{c}})}]{jcp-141-204306-SA}%
  \BibitemOpen
  \bibfield  {author} {\bibinfo {author} {\bibfnamefont {S.}~\bibnamefont
  {Mukherjee}}, \bibinfo {author} {\bibfnamefont {D.}~\bibnamefont
  {Mukhopadhyay}}, \ and\ \bibinfo {author} {\bibfnamefont {S.}~\bibnamefont
  {Adhikari}},\ }\href@noop {} {\bibfield  {journal} {\bibinfo  {journal} {J.
  Chem. Phys.}\ }\textbf {\bibinfo {volume} {141}},\ \bibinfo {pages}
  {204306/1} (\bibinfo {year} {2014}{\natexlab{c}})}\BibitemShut {NoStop}%
\bibitem [{\citenamefont {Dunning}(1989)}]{dunning_basis}%
  \BibitemOpen
  \bibfield  {author} {\bibinfo {author} {\bibfnamefont {T.~H.}\ \bibnamefont
  {Dunning}},\ }\href@noop {} {\bibfield  {journal} {\bibinfo  {journal} {J.
  Chem. Phys.}\ }\textbf {\bibinfo {volume} {90}},\ \bibinfo {pages} {1007}
  (\bibinfo {year} {1989})}\BibitemShut {NoStop}%
\bibitem [{\citenamefont {Naskar}\ \emph
  {et~al.}(2020{\natexlab{c}})\citenamefont {Naskar}, \citenamefont
  {Mukherjee}, \citenamefont {Mukherjee}, \citenamefont {Ravi}, \citenamefont
  {Mukherjee}, \citenamefont {Sardar},\ and\ \citenamefont
  {Adhikari}}]{ADT_JCTC_16}%
  \BibitemOpen
  \bibfield  {author} {\bibinfo {author} {\bibfnamefont {K.}~\bibnamefont
  {Naskar}}, \bibinfo {author} {\bibfnamefont {S.}~\bibnamefont {Mukherjee}},
  \bibinfo {author} {\bibfnamefont {B.}~\bibnamefont {Mukherjee}}, \bibinfo
  {author} {\bibfnamefont {S.}~\bibnamefont {Ravi}}, \bibinfo {author}
  {\bibfnamefont {S.}~\bibnamefont {Mukherjee}}, \bibinfo {author}
  {\bibfnamefont {S.}~\bibnamefont {Sardar}}, \ and\ \bibinfo {author}
  {\bibfnamefont {S.}~\bibnamefont {Adhikari}},\ }\href@noop {} {\bibfield
  {journal} {\bibinfo  {journal} {J. Chem. Theo. and Comp.}\ }\textbf {\bibinfo
  {volume} {16}},\ \bibinfo {pages} {1666} (\bibinfo {year}
  {2020}{\natexlab{c}})}\BibitemShut {NoStop}%
\bibitem [{\citenamefont {Russek}\ and\ \citenamefont
  {Furlan}(1989)}]{Russek1989}%
  \BibitemOpen
  \bibfield  {author} {\bibinfo {author} {\bibfnamefont {A.}~\bibnamefont
  {Russek}}\ and\ \bibinfo {author} {\bibfnamefont {R.~J.}\ \bibnamefont
  {Furlan}},\ }\href {\doibase 10.1103/PhysRevA.39.5034} {\bibfield  {journal}
  {\bibinfo  {journal} {Phys. Rev. A}\ }\textbf {\bibinfo {volume} {39}},\
  \bibinfo {pages} {5034} (\bibinfo {year} {1989})}\BibitemShut {NoStop}%
\bibitem [{\citenamefont {Furlan}, \citenamefont {Bent},\ and\ \citenamefont
  {Russek}(1990)}]{Furlan1990}%
  \BibitemOpen
  \bibfield  {author} {\bibinfo {author} {\bibfnamefont {R.~J.}\ \bibnamefont
  {Furlan}}, \bibinfo {author} {\bibfnamefont {G.}~\bibnamefont {Bent}}, \ and\
  \bibinfo {author} {\bibfnamefont {A.}~\bibnamefont {Russek}},\ }\href
  {\doibase 10.1063/1.458936} {\bibfield  {journal} {\bibinfo  {journal} {J.
  Chem. Phys.}\ }\textbf {\bibinfo {volume} {93}},\ \bibinfo {pages} {6676}
  (\bibinfo {year} {1990})}\BibitemShut {NoStop}%
\bibitem [{\citenamefont {Schneider}\ and\ \citenamefont
  {Z\"{u}licke}(1979)}]{Schneider1979}%
  \BibitemOpen
  \bibfield  {author} {\bibinfo {author} {\bibfnamefont {F.}~\bibnamefont
  {Schneider}}\ and\ \bibinfo {author} {\bibfnamefont {L.}~\bibnamefont
  {Z\"{u}licke}},\ }\href {\doibase 10.1016/0009-2614(79)85212-4} {\bibfield
  {journal} {\bibinfo  {journal} {Chem. Phys. Lett.}\ }\textbf {\bibinfo
  {volume} {67}},\ \bibinfo {pages} {491} (\bibinfo {year} {1979})}\BibitemShut
  {NoStop}%
\bibitem [{\citenamefont {Schneider}\ \emph {et~al.}(1983)\citenamefont
  {Schneider}, \citenamefont {Z\"{u}licke}, \citenamefont {Pol\'{a}k},\ and\
  \citenamefont {Vojt\'{i}k}}]{Schneider1983}%
  \BibitemOpen
  \bibfield  {author} {\bibinfo {author} {\bibfnamefont {F.}~\bibnamefont
  {Schneider}}, \bibinfo {author} {\bibfnamefont {L.}~\bibnamefont
  {Z\"{u}licke}}, \bibinfo {author} {\bibfnamefont {R.}~\bibnamefont
  {Pol\'{a}k}}, \ and\ \bibinfo {author} {\bibfnamefont {J.}~\bibnamefont
  {Vojt\'{i}k}},\ }\href {\doibase 10.1016/0301-0104(83)85040-X} {\bibfield
  {journal} {\bibinfo  {journal} {Chem. Phys.}\ }\textbf {\bibinfo {volume}
  {76}},\ \bibinfo {pages} {259} (\bibinfo {year} {1983})}\BibitemShut
  {NoStop}%
\bibitem [{\citenamefont {Aguado}, \citenamefont {Su\'{a}rez},\ and\
  \citenamefont {Paniagua}(1993)}]{Aguado1993}%
  \BibitemOpen
  \bibfield  {author} {\bibinfo {author} {\bibfnamefont {A.}~\bibnamefont
  {Aguado}}, \bibinfo {author} {\bibfnamefont {C.}~\bibnamefont {Su\'{a}rez}},
  \ and\ \bibinfo {author} {\bibfnamefont {M.}~\bibnamefont {Paniagua}},\
  }\href {\doibase 10.1063/1.464676} {\bibfield  {journal} {\bibinfo  {journal}
  {J. Chem. Phys.}\ }\textbf {\bibinfo {volume} {98}},\ \bibinfo {pages} {308}
  (\bibinfo {year} {1993})}\BibitemShut {NoStop}%
\bibitem [{\citenamefont {Sidis}(1996)}]{Sidis1996}%
  \BibitemOpen
  \bibfield  {author} {\bibinfo {author} {\bibfnamefont {V.}~\bibnamefont
  {Sidis}},\ }\href {\doibase 10.1016/0301-0104(96)00060-2} {\bibfield
  {journal} {\bibinfo  {journal} {Chem. Phys.}\ }\textbf {\bibinfo {volume}
  {209}},\ \bibinfo {pages} {313} (\bibinfo {year} {1996})}\BibitemShut
  {NoStop}%
\bibitem [{\citenamefont {Ravi}\ \emph
  {et~al.}(2021{\natexlab{b}})\citenamefont {Ravi}, \citenamefont {Mukherjee},
  \citenamefont {Mukherjee}, \citenamefont {Adhikari}, \citenamefont
  {Sathyamurthy},\ and\ \citenamefont {Baer}}]{Ravi2021_molphys_119}%
  \BibitemOpen
  \bibfield  {author} {\bibinfo {author} {\bibfnamefont {S.}~\bibnamefont
  {Ravi}}, \bibinfo {author} {\bibfnamefont {S.}~\bibnamefont {Mukherjee}},
  \bibinfo {author} {\bibfnamefont {B.}~\bibnamefont {Mukherjee}}, \bibinfo
  {author} {\bibfnamefont {S.}~\bibnamefont {Adhikari}}, \bibinfo {author}
  {\bibfnamefont {N.}~\bibnamefont {Sathyamurthy}}, \ and\ \bibinfo {author}
  {\bibfnamefont {M.}~\bibnamefont {Baer}},\ }\href {\doibase
  10.1080/00268976.2020.1811907} {\bibfield  {journal} {\bibinfo  {journal}
  {Mol. Phys.}\ }\textbf {\bibinfo {volume} {119}},\ \bibinfo {pages}
  {e1811907} (\bibinfo {year} {2021}{\natexlab{b}})}\BibitemShut {NoStop}%
\bibitem [{\citenamefont {Truhlar}\ and\ \citenamefont
  {Horowitz}(1978)}]{LSTH_1978}%
  \BibitemOpen
  \bibfield  {author} {\bibinfo {author} {\bibfnamefont {D.~G.}\ \bibnamefont
  {Truhlar}}\ and\ \bibinfo {author} {\bibfnamefont {C.~J.}\ \bibnamefont
  {Horowitz}},\ }\href@noop {} {\bibfield  {journal} {\bibinfo  {journal} {J.
  Chem. Phys.}\ }\textbf {\bibinfo {volume} {68}},\ \bibinfo {pages} {2466}
  (\bibinfo {year} {1978})}\BibitemShut {NoStop}%
\bibitem [{\citenamefont {Boothroyd}\ \emph {et~al.}(1996)\citenamefont
  {Boothroyd}, \citenamefont {Keogh}, \citenamefont {Martin},\ and\
  \citenamefont {Peterson}}]{BKMP2_1996}%
  \BibitemOpen
  \bibfield  {author} {\bibinfo {author} {\bibfnamefont {A.~I.}\ \bibnamefont
  {Boothroyd}}, \bibinfo {author} {\bibfnamefont {W.~J.}\ \bibnamefont
  {Keogh}}, \bibinfo {author} {\bibfnamefont {P.~G.}\ \bibnamefont {Martin}}, \
  and\ \bibinfo {author} {\bibfnamefont {M.~R.}\ \bibnamefont {Peterson}},\
  }\href@noop {} {\bibfield  {journal} {\bibinfo  {journal} {J. Chem. Phys.}\
  }\textbf {\bibinfo {volume} {104}},\ \bibinfo {pages} {7139} (\bibinfo {year}
  {1996})}\BibitemShut {NoStop}%
\bibitem [{\citenamefont {Mielke}, \citenamefont {Garrett},\ and\ \citenamefont
  {Peterson}(2002)}]{CCI_2002}%
  \BibitemOpen
  \bibfield  {author} {\bibinfo {author} {\bibfnamefont {S.~L.}\ \bibnamefont
  {Mielke}}, \bibinfo {author} {\bibfnamefont {B.~C.}\ \bibnamefont {Garrett}},
  \ and\ \bibinfo {author} {\bibfnamefont {K.~A.}\ \bibnamefont {Peterson}},\
  }\href@noop {} {\bibfield  {journal} {\bibinfo  {journal} {J. Chem. Phys.}\
  }\textbf {\bibinfo {volume} {116}},\ \bibinfo {pages} {4142} (\bibinfo {year}
  {2002})}\BibitemShut {NoStop}%
\bibitem [{\citenamefont {Abrol}\ and\ \citenamefont
  {Kuppermann}(2002)}]{Abrol_Kuppermann_2002}%
  \BibitemOpen
  \bibfield  {author} {\bibinfo {author} {\bibfnamefont {R.}~\bibnamefont
  {Abrol}}\ and\ \bibinfo {author} {\bibfnamefont {A.}~\bibnamefont
  {Kuppermann}},\ }\href@noop {} {\bibfield  {journal} {\bibinfo  {journal} {J.
  Chem. Phys.}\ }\textbf {\bibinfo {volume} {116}},\ \bibinfo {pages} {1035}
  (\bibinfo {year} {2002})}\BibitemShut {NoStop}%
\bibitem [{\citenamefont {Werner}\ \emph {et~al.}(2020)\citenamefont {Werner},
  \citenamefont {Knowles}, \citenamefont {Manby}, \citenamefont {Black},
  \citenamefont {Doll}, \citenamefont {Heßelmann}, \citenamefont {Kats},
  \citenamefont {Köhn}, \citenamefont {Korona}, \citenamefont {Kreplin},
  \citenamefont {Ma}, \citenamefont {Miller}, \citenamefont {Mitrushchenkov},
  \citenamefont {Peterson}, \citenamefont {Polyak}, \citenamefont {Rauhut},\
  and\ \citenamefont {Sibaev}}]{MOLPRO_2020}%
  \BibitemOpen
  \bibfield  {author} {\bibinfo {author} {\bibfnamefont {H.-J.}\ \bibnamefont
  {Werner}}, \bibinfo {author} {\bibfnamefont {P.~J.}\ \bibnamefont {Knowles}},
  \bibinfo {author} {\bibfnamefont {F.~R.}\ \bibnamefont {Manby}}, \bibinfo
  {author} {\bibfnamefont {J.~A.}\ \bibnamefont {Black}}, \bibinfo {author}
  {\bibfnamefont {K.}~\bibnamefont {Doll}}, \bibinfo {author} {\bibfnamefont
  {A.}~\bibnamefont {Heßelmann}}, \bibinfo {author} {\bibfnamefont
  {D.}~\bibnamefont {Kats}}, \bibinfo {author} {\bibfnamefont {A.}~\bibnamefont
  {Köhn}}, \bibinfo {author} {\bibfnamefont {T.}~\bibnamefont {Korona}},
  \bibinfo {author} {\bibfnamefont {D.~A.}\ \bibnamefont {Kreplin}}, \bibinfo
  {author} {\bibfnamefont {Q.}~\bibnamefont {Ma}}, \bibinfo {author}
  {\bibfnamefont {I.}~\bibnamefont {Miller}, \bibfnamefont {Thomas~F.}},
  \bibinfo {author} {\bibfnamefont {A.}~\bibnamefont {Mitrushchenkov}},
  \bibinfo {author} {\bibfnamefont {K.~A.}\ \bibnamefont {Peterson}}, \bibinfo
  {author} {\bibfnamefont {I.}~\bibnamefont {Polyak}}, \bibinfo {author}
  {\bibfnamefont {G.}~\bibnamefont {Rauhut}}, \ and\ \bibinfo {author}
  {\bibfnamefont {M.}~\bibnamefont {Sibaev}},\ }\href@noop {} {\bibfield
  {journal} {\bibinfo  {journal} {J. Chem. Phys.}\ }\textbf {\bibinfo {volume}
  {152}},\ \bibinfo {pages} {144107} (\bibinfo {year} {2020})}\BibitemShut
  {NoStop}%
\bibitem [{\citenamefont {Naskar}, \citenamefont {Ghosh},\ and\ \citenamefont
  {Adhikari}(2022{\natexlab{b}})}]{FH2_adia_dynamics}%
  \BibitemOpen
  \bibfield  {author} {\bibinfo {author} {\bibfnamefont {K.}~\bibnamefont
  {Naskar}}, \bibinfo {author} {\bibfnamefont {S.}~\bibnamefont {Ghosh}}, \
  and\ \bibinfo {author} {\bibfnamefont {S.}~\bibnamefont {Adhikari}},\ }\href
  {\doibase 10.1021/acs.jpca.2c01209} {\bibfield  {journal} {\bibinfo
  {journal} {The Journal of Physical Chemistry A}\ }\textbf {\bibinfo {volume}
  {126}},\ \bibinfo {pages} {3311} (\bibinfo {year} {2022}{\natexlab{b}})},\
  \bibinfo {note} {pMID: 35594416}\BibitemShut {NoStop}%
\bibitem [{\citenamefont {Rosenman}\ \emph {et~al.}(1995)\citenamefont
  {Rosenman}, \citenamefont {Hochman-Kowal}, \citenamefont {Persky},\ and\
  \citenamefont {Baer}}]{jpc-99-16523-BAER}%
  \BibitemOpen
  \bibfield  {author} {\bibinfo {author} {\bibfnamefont {E.}~\bibnamefont
  {Rosenman}}, \bibinfo {author} {\bibfnamefont {S.}~\bibnamefont
  {Hochman-Kowal}}, \bibinfo {author} {\bibfnamefont {A.}~\bibnamefont
  {Persky}}, \ and\ \bibinfo {author} {\bibfnamefont {M.}~\bibnamefont
  {Baer}},\ }\href {\doibase https://doi.org/10.1021/j100045a007} {\bibfield
  {journal} {\bibinfo  {journal} {J. Phys. Chem.}\ }\textbf {\bibinfo {volume}
  {99}},\ \bibinfo {pages} {16523} (\bibinfo {year} {1995})}\BibitemShut
  {NoStop}%
\bibitem [{\citenamefont {Castillo}\ \emph
  {et~al.}(1998{\natexlab{b}})\citenamefont {Castillo}, \citenamefont {Hartke},
  \citenamefont {Werner}, \citenamefont {Aoiz}, \citenamefont {Ba\`nares},\
  and\ \citenamefont {Mart\'inez-Haya}}]{jcp-109-7224-WERNER}%
  \BibitemOpen
  \bibfield  {author} {\bibinfo {author} {\bibfnamefont {J.~F.}\ \bibnamefont
  {Castillo}}, \bibinfo {author} {\bibfnamefont {B.}~\bibnamefont {Hartke}},
  \bibinfo {author} {\bibfnamefont {H.-J.}\ \bibnamefont {Werner}}, \bibinfo
  {author} {\bibfnamefont {F.~J.}\ \bibnamefont {Aoiz}}, \bibinfo {author}
  {\bibfnamefont {L.}~\bibnamefont {Ba\`nares}}, \ and\ \bibinfo {author}
  {\bibfnamefont {B.}~\bibnamefont {Mart\'inez-Haya}},\ }\href {\doibase
  10.1063/1.477401} {\bibfield  {journal} {\bibinfo  {journal} {J. Chem.
  Phys.}\ }\textbf {\bibinfo {volume} {109}},\ \bibinfo {pages} {7224}
  (\bibinfo {year} {1998}{\natexlab{b}})}\BibitemShut {NoStop}%
\bibitem [{\citenamefont {Aoiz}, \citenamefont {Banares},\ and\ \citenamefont
  {Castillo}(1999)}]{jcp-111-4013-AOIZ}%
  \BibitemOpen
  \bibfield  {author} {\bibinfo {author} {\bibfnamefont {F.~J.}\ \bibnamefont
  {Aoiz}}, \bibinfo {author} {\bibfnamefont {L.}~\bibnamefont {Banares}}, \
  and\ \bibinfo {author} {\bibfnamefont {J.~F.}\ \bibnamefont {Castillo}},\
  }\href {\doibase http://dx.doi.org/10.1063/1.479703} {\bibfield  {journal}
  {\bibinfo  {journal} {J. Chem. Phys.}\ }\textbf {\bibinfo {volume} {111}},\
  \bibinfo {pages} {4013} (\bibinfo {year} {1999})}\BibitemShut {NoStop}%
\bibitem [{\citenamefont {Markovic}\ and\ \citenamefont
  {Billing}(2000)}]{mp-98-1771-BILLING}%
  \BibitemOpen
  \bibfield  {author} {\bibinfo {author} {\bibfnamefont {N.}~\bibnamefont
  {Markovic}}\ and\ \bibinfo {author} {\bibfnamefont {G.~D.}\ \bibnamefont
  {Billing}},\ }\href@noop {} {\bibfield  {journal} {\bibinfo  {journal}
  {Molec. Phys.}\ }\textbf {\bibinfo {volume} {98}},\ \bibinfo {pages} {1771}
  (\bibinfo {year} {2000})}\BibitemShut {NoStop}%
\bibitem [{\citenamefont {Muckerman}(1981)}]{tcap-6A-1-MUCKER}%
  \BibitemOpen
  \bibfield  {author} {\bibinfo {author} {\bibfnamefont {J.~T.}\ \bibnamefont
  {Muckerman}},\ }\href@noop {} {\emph {\bibinfo {title} {Theoretical Chemistry
  Advances and Perspectives}}},\ Vol.~\bibinfo {volume} {6A}\ (\bibinfo
  {publisher} {Academic Press, New York},\ \bibinfo {year} {1981})\ p.~\bibinfo
  {pages} {1}\BibitemShut {NoStop}%
\bibitem [{\citenamefont {Lynch}\ \emph {et~al.}(1991)\citenamefont {Lynch},
  \citenamefont {Steckler}, \citenamefont {Schwenke}, \citenamefont
  {Varandas},\ and\ \citenamefont {Truhlar}}]{jcp-94-7136-TRU}%
  \BibitemOpen
  \bibfield  {author} {\bibinfo {author} {\bibfnamefont {G.~C.}\ \bibnamefont
  {Lynch}}, \bibinfo {author} {\bibfnamefont {R.}~\bibnamefont {Steckler}},
  \bibinfo {author} {\bibfnamefont {D.~W.}\ \bibnamefont {Schwenke}}, \bibinfo
  {author} {\bibfnamefont {A.~J.~C.}\ \bibnamefont {Varandas}}, \ and\ \bibinfo
  {author} {\bibfnamefont {D.~G.}\ \bibnamefont {Truhlar}},\ }\href@noop {}
  {\bibfield  {journal} {\bibinfo  {journal} {J. Chem. Phys.}\ }\textbf
  {\bibinfo {volume} {94}},\ \bibinfo {pages} {7136} (\bibinfo {year}
  {1991})}\BibitemShut {NoStop}%
\bibitem [{\citenamefont {Stark}\ and\ \citenamefont
  {Werner}(1996{\natexlab{b}})}]{jcp-104-6515-WERNER}%
  \BibitemOpen
  \bibfield  {author} {\bibinfo {author} {\bibfnamefont {K.}~\bibnamefont
  {Stark}}\ and\ \bibinfo {author} {\bibfnamefont {H.~J.}\ \bibnamefont
  {Werner}},\ }\href@noop {} {\bibfield  {journal} {\bibinfo  {journal} {J.
  Chem. Phys.}\ }\textbf {\bibinfo {volume} {104}},\ \bibinfo {pages} {6515}
  (\bibinfo {year} {1996}{\natexlab{b}})}\BibitemShut {NoStop}%
\bibitem [{\citenamefont {Hartke}\ and\ \citenamefont
  {Werner}(1997{\natexlab{b}})}]{cpl-280-430-HARTKE}%
  \BibitemOpen
  \bibfield  {author} {\bibinfo {author} {\bibfnamefont {B.}~\bibnamefont
  {Hartke}}\ and\ \bibinfo {author} {\bibfnamefont {H.-J.}\ \bibnamefont
  {Werner}},\ }\href@noop {} {\bibfield  {journal} {\bibinfo  {journal} {Chem.
  Phys. Lett.}\ }\textbf {\bibinfo {volume} {280}},\ \bibinfo {pages} {430}
  (\bibinfo {year} {1997}{\natexlab{b}})}\BibitemShut {NoStop}%
\bibitem [{\citenamefont {Aoiz}\ \emph {et~al.}(1993)\citenamefont {Aoiz},
  \citenamefont {Herrero}, \citenamefont {Nogueira},\ and\ \citenamefont
  {R\'abanos}}]{cpl-204-359-AOIZ}%
  \BibitemOpen
  \bibfield  {author} {\bibinfo {author} {\bibfnamefont {F.~J.}\ \bibnamefont
  {Aoiz}}, \bibinfo {author} {\bibfnamefont {V.}~\bibnamefont {Herrero}},
  \bibinfo {author} {\bibfnamefont {M.}~\bibnamefont {Nogueira}}, \ and\
  \bibinfo {author} {\bibfnamefont {V.~S.}\ \bibnamefont {R\'abanos}},\ }\href
  {\doibase https://doi.org/10.1016/0009-2614(93)90022-S} {\bibfield  {journal}
  {\bibinfo  {journal} {Chem. Phys. Lett.}\ }\textbf {\bibinfo {volume}
  {204}},\ \bibinfo {pages} {359} (\bibinfo {year} {1993})}\BibitemShut
  {NoStop}%
\bibitem [{\citenamefont {Honvault}\ and\ \citenamefont
  {Launay}(1999{\natexlab{b}})}]{cpl-303-657-HON}%
  \BibitemOpen
  \bibfield  {author} {\bibinfo {author} {\bibfnamefont {P.}~\bibnamefont
  {Honvault}}\ and\ \bibinfo {author} {\bibfnamefont {J.-M.}\ \bibnamefont
  {Launay}},\ }\href@noop {} {\bibfield  {journal} {\bibinfo  {journal} {Chem.
  Phys. Lett.}\ }\textbf {\bibinfo {volume} {303}},\ \bibinfo {pages} {657}
  (\bibinfo {year} {1999}{\natexlab{b}})}\BibitemShut {NoStop}%
\bibitem [{\citenamefont {Dong}, \citenamefont {Lee},\ and\ \citenamefont
  {Liu}(2000)}]{jcp-113-3633-LIU}%
  \BibitemOpen
  \bibfield  {author} {\bibinfo {author} {\bibfnamefont {F.}~\bibnamefont
  {Dong}}, \bibinfo {author} {\bibfnamefont {S.-H.}\ \bibnamefont {Lee}}, \
  and\ \bibinfo {author} {\bibfnamefont {K.}~\bibnamefont {Liu}},\ }\href@noop
  {} {\bibfield  {journal} {\bibinfo  {journal} {J. Chem. Phys.}\ }\textbf
  {\bibinfo {volume} {113}},\ \bibinfo {pages} {3633} (\bibinfo {year}
  {2000})}\BibitemShut {NoStop}%
\bibitem [{\citenamefont {Bedjanian}(2021{\natexlab{b}})}]{bredjaninan}%
  \BibitemOpen
  \bibfield  {author} {\bibinfo {author} {\bibfnamefont {Y.}~\bibnamefont
  {Bedjanian}},\ }\href {\doibase https://doi.org/10.1002/kin.21462} {\bibfield
   {journal} {\bibinfo  {journal} {Int. J. Chem. Kin.}\ }\textbf {\bibinfo
  {volume} {53}},\ \bibinfo {pages} {527} (\bibinfo {year}
  {2021}{\natexlab{b}})},\ \Eprint
  {http://arxiv.org/abs/https://onlinelibrary.wiley.com/doi/pdf/10.1002/kin.21462}
  {https://onlinelibrary.wiley.com/doi/pdf/10.1002/kin.21462} \BibitemShut
  {NoStop}%
\bibitem [{\citenamefont {Rosenman}\ \emph {et~al.}(1996)\citenamefont
  {Rosenman}, \citenamefont {Hochman-Kowal}, \citenamefont {Persky},\ and\
  \citenamefont {Baer}}]{Rosenman1996_CPC_257}%
  \BibitemOpen
  \bibfield  {author} {\bibinfo {author} {\bibfnamefont {E.}~\bibnamefont
  {Rosenman}}, \bibinfo {author} {\bibfnamefont {S.}~\bibnamefont
  {Hochman-Kowal}}, \bibinfo {author} {\bibfnamefont {A.}~\bibnamefont
  {Persky}}, \ and\ \bibinfo {author} {\bibfnamefont {M.}~\bibnamefont
  {Baer}},\ }\href@noop {} {\bibfield  {journal} {\bibinfo  {journal} {Chem.
  Phys. Lett.}\ }\textbf {\bibinfo {volume} {257}},\ \bibinfo {pages} {421}
  (\bibinfo {year} {1996})}\BibitemShut {NoStop}%
\bibitem [{\citenamefont {Zhang}\ \emph {et~al.}(2003)\citenamefont {Zhang},
  \citenamefont {Xie}, \citenamefont {Han},\ and\ \citenamefont
  {Zhang}}]{jcp-119-12921-HAN}%
  \BibitemOpen
  \bibfield  {author} {\bibinfo {author} {\bibfnamefont {Y.}~\bibnamefont
  {Zhang}}, \bibinfo {author} {\bibfnamefont {T.-X.}\ \bibnamefont {Xie}},
  \bibinfo {author} {\bibfnamefont {K.~L.}\ \bibnamefont {Han}}, \ and\
  \bibinfo {author} {\bibfnamefont {J.~Z.~H.}\ \bibnamefont {Zhang}},\
  }\href@noop {} {\bibfield  {journal} {\bibinfo  {journal} {J. Chem. Phys.}\
  }\textbf {\bibinfo {volume} {119}},\ \bibinfo {pages} {12921} (\bibinfo
  {year} {2003})}\BibitemShut {NoStop}%
\bibitem [{\citenamefont {Naskar}\ \emph
  {et~al.}(2023{\natexlab{e}})\citenamefont {Naskar}, \citenamefont {Ravi},
  \citenamefont {Adhikari}, \citenamefont {Baer},\ and\ \citenamefont
  {Sathyamurthy}}]{Naskar2023_JPCA_127}%
  \BibitemOpen
  \bibfield  {author} {\bibinfo {author} {\bibfnamefont {K.}~\bibnamefont
  {Naskar}}, \bibinfo {author} {\bibfnamefont {S.}~\bibnamefont {Ravi}},
  \bibinfo {author} {\bibfnamefont {S.}~\bibnamefont {Adhikari}}, \bibinfo
  {author} {\bibfnamefont {M.}~\bibnamefont {Baer}}, \ and\ \bibinfo {author}
  {\bibfnamefont {N.}~\bibnamefont {Sathyamurthy}},\ }\href {\doibase
  10.1021/acs.jpca.3c01047} {\bibfield  {journal} {\bibinfo  {journal} {J.
  Phys. Chem. A}\ }\textbf {\bibinfo {volume} {127}},\ \bibinfo {pages} {3832}
  (\bibinfo {year} {2023}{\natexlab{e}})}\BibitemShut {NoStop}%
\bibitem [{\citenamefont {Viegas}, \citenamefont {Alijah},\ and\ \citenamefont
  {Varandas}(2007{\natexlab{b}})}]{viegas_jcp126}%
  \BibitemOpen
  \bibfield  {author} {\bibinfo {author} {\bibfnamefont {L.~P.}\ \bibnamefont
  {Viegas}}, \bibinfo {author} {\bibfnamefont {A.}~\bibnamefont {Alijah}}, \
  and\ \bibinfo {author} {\bibfnamefont {A.~J.~C.}\ \bibnamefont {Varandas}},\
  }\href@noop {} {\bibfield  {journal} {\bibinfo  {journal} {J. Chem. Phys.}\
  }\textbf {\bibinfo {volume} {126}},\ \bibinfo {pages} {074309} (\bibinfo
  {year} {2007}{\natexlab{b}})}\BibitemShut {NoStop}%
\bibitem [{\citenamefont {Chu}, \citenamefont {Varandas},\ and\ \citenamefont
  {Han}(2009)}]{cpl-471-222-HAN3}%
  \BibitemOpen
  \bibfield  {author} {\bibinfo {author} {\bibfnamefont {T.~S.}\ \bibnamefont
  {Chu}}, \bibinfo {author} {\bibfnamefont {A.~J.~C.}\ \bibnamefont
  {Varandas}}, \ and\ \bibinfo {author} {\bibfnamefont {K.~L.}\ \bibnamefont
  {Han}},\ }\href@noop {} {\bibfield  {journal} {\bibinfo  {journal} {Chem.
  Phys. Lett.}\ }\textbf {\bibinfo {volume} {471}},\ \bibinfo {pages} {222}
  (\bibinfo {year} {2009})}\BibitemShut {NoStop}%
\bibitem [{\citenamefont {Chu}\ and\ \citenamefont
  {Han}(2005)}]{jpcA-109-2050-HAN1}%
  \BibitemOpen
  \bibfield  {author} {\bibinfo {author} {\bibfnamefont {T.~S.}\ \bibnamefont
  {Chu}}\ and\ \bibinfo {author} {\bibfnamefont {K.~L.}\ \bibnamefont {Han}},\
  }\href@noop {} {\bibfield  {journal} {\bibinfo  {journal} {J. Phy. Chem. A}\
  }\textbf {\bibinfo {volume} {109}},\ \bibinfo {pages} {2050} (\bibinfo {year}
  {2005})}\BibitemShut {NoStop}%
\bibitem [{\citenamefont {Kamisaka}\ \emph
  {et~al.}(2002{\natexlab{b}})\citenamefont {Kamisaka}, \citenamefont {Bian},
  \citenamefont {Nobusada},\ and\ \citenamefont {Nakamura}}]{jcp-116-654-KAM}%
  \BibitemOpen
  \bibfield  {author} {\bibinfo {author} {\bibfnamefont {A.}~\bibnamefont
  {Kamisaka}}, \bibinfo {author} {\bibfnamefont {W.}~\bibnamefont {Bian}},
  \bibinfo {author} {\bibfnamefont {K.}~\bibnamefont {Nobusada}}, \ and\
  \bibinfo {author} {\bibfnamefont {H.}~\bibnamefont {Nakamura}},\ }\href@noop
  {} {\bibfield  {journal} {\bibinfo  {journal} {J. Chem. Phys.}\ }\textbf
  {\bibinfo {volume} {116}},\ \bibinfo {pages} {654} (\bibinfo {year}
  {2002}{\natexlab{b}})}\BibitemShut {NoStop}%
\bibitem [{\citenamefont {Schlier}, \citenamefont {Nowotny},\ and\
  \citenamefont {Teloy}(1987)}]{cp-111-401-SCH}%
  \BibitemOpen
  \bibfield  {author} {\bibinfo {author} {\bibfnamefont {C.}~\bibnamefont
  {Schlier}}, \bibinfo {author} {\bibfnamefont {U.}~\bibnamefont {Nowotny}}, \
  and\ \bibinfo {author} {\bibfnamefont {E.}~\bibnamefont {Teloy}},\
  }\href@noop {} {\bibfield  {journal} {\bibinfo  {journal} {Chem. Phys.}\
  }\textbf {\bibinfo {volume} {111}},\ \bibinfo {pages} {401} (\bibinfo {year}
  {1987})}\BibitemShut {NoStop}%
\bibitem [{\citenamefont {Kamisaka}\ \emph
  {et~al.}(2002{\natexlab{c}})\citenamefont {Kamisaka}, \citenamefont {Bian},
  \citenamefont {Nobusada},\ and\ \citenamefont {Nakamura}}]{KBNN}%
  \BibitemOpen
  \bibfield  {author} {\bibinfo {author} {\bibfnamefont {H.}~\bibnamefont
  {Kamisaka}}, \bibinfo {author} {\bibfnamefont {W.}~\bibnamefont {Bian}},
  \bibinfo {author} {\bibfnamefont {K.}~\bibnamefont {Nobusada}}, \ and\
  \bibinfo {author} {\bibfnamefont {H.}~\bibnamefont {Nakamura}},\ }\href@noop
  {} {\bibfield  {journal} {\bibinfo  {journal} {J. Chem. Phys.}\ }\textbf
  {\bibinfo {volume} {116}},\ \bibinfo {pages} {654} (\bibinfo {year}
  {2002}{\natexlab{c}})}\BibitemShut {NoStop}%
\bibitem [{\citenamefont {Last}, \citenamefont {Gilibert},\ and\ \citenamefont
  {Baer}(1997)}]{jcp-107-1451-LAST}%
  \BibitemOpen
  \bibfield  {author} {\bibinfo {author} {\bibfnamefont {I.}~\bibnamefont
  {Last}}, \bibinfo {author} {\bibfnamefont {M.}~\bibnamefont {Gilibert}}, \
  and\ \bibinfo {author} {\bibfnamefont {M.}~\bibnamefont {Baer}},\ }\href@noop
  {} {\bibfield  {journal} {\bibinfo  {journal} {J. Chem. Phys.}\ }\textbf
  {\bibinfo {volume} {107}},\ \bibinfo {pages} {1451} (\bibinfo {year}
  {1997})}\BibitemShut {NoStop}%
\bibitem [{\citenamefont {Karpas}, \citenamefont {Anicich},\ and\ \citenamefont
  {Huntress(Jr.)}(1979)}]{jcp-70-2877-KARPAS}%
  \BibitemOpen
  \bibfield  {author} {\bibinfo {author} {\bibfnamefont {Z.}~\bibnamefont
  {Karpas}}, \bibinfo {author} {\bibfnamefont {V.}~\bibnamefont {Anicich}}, \
  and\ \bibinfo {author} {\bibfnamefont {W.~T.}\ \bibnamefont
  {Huntress(Jr.)}},\ }\href@noop {} {\bibfield  {journal} {\bibinfo  {journal}
  {J. Chem. Phys.}\ }\textbf {\bibinfo {volume} {70}},\ \bibinfo {pages} {2877}
  (\bibinfo {year} {1979})}\BibitemShut {NoStop}%
\bibitem [{\citenamefont {Naskar}\ \emph
  {et~al.}(2023{\natexlab{f}})\citenamefont {Naskar}, \citenamefont {Ravi},
  \citenamefont {Adhikari}, \citenamefont {Baer},\ and\ \citenamefont
  {Sathyamurthy}}]{heh2+_surface_article}%
  \BibitemOpen
  \bibfield  {author} {\bibinfo {author} {\bibfnamefont {K.}~\bibnamefont
  {Naskar}}, \bibinfo {author} {\bibfnamefont {S.}~\bibnamefont {Ravi}},
  \bibinfo {author} {\bibfnamefont {S.}~\bibnamefont {Adhikari}}, \bibinfo
  {author} {\bibfnamefont {M.}~\bibnamefont {Baer}}, \ and\ \bibinfo {author}
  {\bibfnamefont {N.}~\bibnamefont {Sathyamurthy}},\ }\href {\doibase
  10.1021/acs.jpca.3c01047} {\bibfield  {journal} {\bibinfo  {journal} {J.
  Phys. Chem. A}\ }\textbf {\bibinfo {volume} {127}},\ \bibinfo {pages} {3832}
  (\bibinfo {year} {2023}{\natexlab{f}})}\BibitemShut {NoStop}%
\bibitem [{\citenamefont {Turner}, \citenamefont {Dutuit},\ and\ \citenamefont
  {Lee}(1984{\natexlab{b}})}]{Turner1984_JCP_81}%
  \BibitemOpen
  \bibfield  {author} {\bibinfo {author} {\bibfnamefont {T.}~\bibnamefont
  {Turner}}, \bibinfo {author} {\bibfnamefont {O.}~\bibnamefont {Dutuit}}, \
  and\ \bibinfo {author} {\bibfnamefont {Y.~T.}\ \bibnamefont {Lee}},\ }\href
  {\doibase 10.1063/1.448073} {\bibfield  {journal} {\bibinfo  {journal} {J.
  Chem. Phys.}\ }\textbf {\bibinfo {volume} {81}},\ \bibinfo {pages} {3475}
  (\bibinfo {year} {1984}{\natexlab{b}})}\BibitemShut {NoStop}%
\bibitem [{\citenamefont {Bouakline}, \citenamefont {Althorpe},\ and\
  \citenamefont {Pel{\'a}ez~Ruiz}(2008)}]{Althorpe_2008}%
  \BibitemOpen
  \bibfield  {author} {\bibinfo {author} {\bibfnamefont {F.}~\bibnamefont
  {Bouakline}}, \bibinfo {author} {\bibfnamefont {S.~C.}\ \bibnamefont
  {Althorpe}}, \ and\ \bibinfo {author} {\bibfnamefont {D.}~\bibnamefont
  {Pel{\'a}ez~Ruiz}},\ }\href@noop {} {\bibfield  {journal} {\bibinfo
  {journal} {J. Chem. Phys.}\ }\textbf {\bibinfo {volume} {128}},\ \bibinfo
  {pages} {124322} (\bibinfo {year} {2008})}\BibitemShut {NoStop}%
\bibitem [{\citenamefont {Launay}\ and\ \citenamefont
  {Dourneuf}(1989)}]{Launay_1989}%
  \BibitemOpen
  \bibfield  {author} {\bibinfo {author} {\bibfnamefont {J.}~\bibnamefont
  {Launay}}\ and\ \bibinfo {author} {\bibfnamefont {M.~L.}\ \bibnamefont
  {Dourneuf}},\ }\href@noop {} {\bibfield  {journal} {\bibinfo  {journal}
  {Chemical Physics Letters}\ }\textbf {\bibinfo {volume} {163}},\ \bibinfo
  {pages} {178} (\bibinfo {year} {1989})}\BibitemShut {NoStop}%
\bibitem [{\citenamefont {Johnson}(1983{\natexlab{b}})}]{Johnson1983_JCP_79}%
  \BibitemOpen
  \bibfield  {author} {\bibinfo {author} {\bibfnamefont {B.~R.}\ \bibnamefont
  {Johnson}},\ }\href@noop {} {\bibfield  {journal} {\bibinfo  {journal} {J.
  Chem. Phys.}\ }\textbf {\bibinfo {volume} {79}},\ \bibinfo {pages} {1916}
  (\bibinfo {year} {1983}{\natexlab{b}})}\BibitemShut {NoStop}%
\bibitem [{\citenamefont {Tully}\ and\ \citenamefont
  {Preston}(1971{\natexlab{b}})}]{TULLY:JCP55}%
  \BibitemOpen
  \bibfield  {author} {\bibinfo {author} {\bibfnamefont {J.~C.}\ \bibnamefont
  {Tully}}\ and\ \bibinfo {author} {\bibfnamefont {R.~K.}\ \bibnamefont
  {Preston}},\ }\href@noop {} {\bibfield  {journal} {\bibinfo  {journal} {J.
  Chem. Phys.}\ }\textbf {\bibinfo {volume} {55}},\ \bibinfo {pages} {562}
  (\bibinfo {year} {1971}{\natexlab{b}})}\BibitemShut {NoStop}%
\bibitem [{\citenamefont {Bjerre}\ and\ \citenamefont
  {Nikitin}(1967)}]{nikitin_cpl1}%
  \BibitemOpen
  \bibfield  {author} {\bibinfo {author} {\bibfnamefont {A.}~\bibnamefont
  {Bjerre}}\ and\ \bibinfo {author} {\bibfnamefont {E.~E.}\ \bibnamefont
  {Nikitin}},\ }\href@noop {} {\bibfield  {journal} {\bibinfo  {journal} {Chem.
  Phys. Lett.}\ }\textbf {\bibinfo {volume} {1}},\ \bibinfo {pages} {179}
  (\bibinfo {year} {1967})}\BibitemShut {NoStop}%
\bibitem [{\citenamefont {Stine}\ and\ \citenamefont
  {Muckerman}(1976)}]{muckerman_jcp65}%
  \BibitemOpen
  \bibfield  {author} {\bibinfo {author} {\bibfnamefont {J.~R.}\ \bibnamefont
  {Stine}}\ and\ \bibinfo {author} {\bibfnamefont {J.~T.}\ \bibnamefont
  {Muckerman}},\ }\href@noop {} {\bibfield  {journal} {\bibinfo  {journal} {J
  Chem. Phys.}\ }\textbf {\bibinfo {volume} {65}},\ \bibinfo {pages} {3975}
  (\bibinfo {year} {1976})}\BibitemShut {NoStop}%
\bibitem [{\citenamefont {Tully}(1990)}]{tully_jcp93}%
  \BibitemOpen
  \bibfield  {author} {\bibinfo {author} {\bibfnamefont {J.~C.}\ \bibnamefont
  {Tully}},\ }\href@noop {} {\bibfield  {journal} {\bibinfo  {journal} {J.
  Chem. Phys.}\ }\textbf {\bibinfo {volume} {93}},\ \bibinfo {pages} {1061}
  (\bibinfo {year} {1990})}\BibitemShut {NoStop}%
\bibitem [{\citenamefont {McLafferty}(1983)}]{mclafferty_jcp79}%
  \BibitemOpen
  \bibfield  {author} {\bibinfo {author} {\bibfnamefont {F.}~\bibnamefont
  {McLafferty}},\ }\href@noop {} {\bibfield  {journal} {\bibinfo  {journal} {J.
  Chem. Phys.}\ }\textbf {\bibinfo {volume} {79}},\ \bibinfo {pages} {4922}
  (\bibinfo {year} {1983})}\BibitemShut {NoStop}%
\bibitem [{\citenamefont {Ali}\ and\ \citenamefont {Miller}(1983)}]{ali_jcp78}%
  \BibitemOpen
  \bibfield  {author} {\bibinfo {author} {\bibfnamefont {D.~P.}\ \bibnamefont
  {Ali}}\ and\ \bibinfo {author} {\bibfnamefont {W.~H.}\ \bibnamefont
  {Miller}},\ }\href@noop {} {\bibfield  {journal} {\bibinfo  {journal} {J.
  Chem. Phys.}\ }\textbf {\bibinfo {volume} {78}},\ \bibinfo {pages} {6640}
  (\bibinfo {year} {1983})}\BibitemShut {NoStop}%
\bibitem [{\citenamefont {Kuntz}(1991)}]{kuntz_jcp95}%
  \BibitemOpen
  \bibfield  {author} {\bibinfo {author} {\bibfnamefont {P.~J.}\ \bibnamefont
  {Kuntz}},\ }\href@noop {} {\bibfield  {journal} {\bibinfo  {journal} {J.
  Chem. Phys.}\ }\textbf {\bibinfo {volume} {95}},\ \bibinfo {pages} {141}
  (\bibinfo {year} {1991})}\BibitemShut {NoStop}%
\bibitem [{\citenamefont {Hammes-Schiffer}\ and\ \citenamefont
  {Tully}(1994)}]{tully_jcp101}%
  \BibitemOpen
  \bibfield  {author} {\bibinfo {author} {\bibfnamefont {S.}~\bibnamefont
  {Hammes-Schiffer}}\ and\ \bibinfo {author} {\bibfnamefont {J.~C.}\
  \bibnamefont {Tully}},\ }\href@noop {} {\bibfield  {journal} {\bibinfo
  {journal} {J. Chem. Phys.}\ }\textbf {\bibinfo {volume} {101}},\ \bibinfo
  {pages} {4657} (\bibinfo {year} {1994})}\BibitemShut {NoStop}%
\bibitem [{\citenamefont {Tapavicza}, \citenamefont {Tavernelli},\ and\
  \citenamefont {Rothlisberger}(2007)}]{rothlisberger_prl98}%
  \BibitemOpen
  \bibfield  {author} {\bibinfo {author} {\bibfnamefont {E.}~\bibnamefont
  {Tapavicza}}, \bibinfo {author} {\bibfnamefont {I.}~\bibnamefont
  {Tavernelli}}, \ and\ \bibinfo {author} {\bibfnamefont {U.}~\bibnamefont
  {Rothlisberger}},\ }\href@noop {} {\bibfield  {journal} {\bibinfo  {journal}
  {Phys. Rev. Lett.}\ }\textbf {\bibinfo {volume} {98}},\ \bibinfo {pages}
  {023001} (\bibinfo {year} {2007})}\BibitemShut {NoStop}%
\bibitem [{\citenamefont {Werner}\ \emph {et~al.}(2008)\citenamefont {Werner},
  \citenamefont {Mitric}, \citenamefont {Suzuki},\ and\ \citenamefont
  {Bonacic-Kouteck\'y}}]{werner_cp349}%
  \BibitemOpen
  \bibfield  {author} {\bibinfo {author} {\bibfnamefont {U.}~\bibnamefont
  {Werner}}, \bibinfo {author} {\bibfnamefont {R.}~\bibnamefont {Mitric}},
  \bibinfo {author} {\bibfnamefont {T.}~\bibnamefont {Suzuki}}, \ and\ \bibinfo
  {author} {\bibfnamefont {V.}~\bibnamefont {Bonacic-Kouteck\'y}},\ }\href@noop
  {} {\bibfield  {journal} {\bibinfo  {journal} {Chem. Phys.}\ }\textbf
  {\bibinfo {volume} {349}},\ \bibinfo {pages} {319} (\bibinfo {year}
  {2008})}\BibitemShut {NoStop}%
\bibitem [{\citenamefont {Ibele}\ \emph {et~al.}(2021)\citenamefont {Ibele},
  \citenamefont {Lassmann}, \citenamefont {Mart\'inez},\ and\ \citenamefont
  {Curchod}}]{martinez_jcp154}%
  \BibitemOpen
  \bibfield  {author} {\bibinfo {author} {\bibfnamefont {L.~M.}\ \bibnamefont
  {Ibele}}, \bibinfo {author} {\bibfnamefont {Y.}~\bibnamefont {Lassmann}},
  \bibinfo {author} {\bibfnamefont {T.~J.}\ \bibnamefont {Mart\'inez}}, \ and\
  \bibinfo {author} {\bibfnamefont {B.~F.~E.}\ \bibnamefont {Curchod}},\
  }\href@noop {} {\bibfield  {journal} {\bibinfo  {journal} {J. Chem. Phys.}\
  }\textbf {\bibinfo {volume} {154}},\ \bibinfo {pages} {104110/1} (\bibinfo
  {year} {2021})}\BibitemShut {NoStop}%
\bibitem [{\citenamefont {Perkins}\ \emph {et~al.}(2015)\citenamefont
  {Perkins}, \citenamefont {Herr\'aez-Aguilar}, \citenamefont {McCrudden},
  \citenamefont {Klos}, \citenamefont {Aoiz},\ and\ \citenamefont
  {Brouard}}]{klos_jcp142}%
  \BibitemOpen
  \bibfield  {author} {\bibinfo {author} {\bibfnamefont {T.}~\bibnamefont
  {Perkins}}, \bibinfo {author} {\bibfnamefont {D.}~\bibnamefont
  {Herr\'aez-Aguilar}}, \bibinfo {author} {\bibfnamefont {G.}~\bibnamefont
  {McCrudden}}, \bibinfo {author} {\bibfnamefont {J.}~\bibnamefont {Klos}},
  \bibinfo {author} {\bibfnamefont {F.~J.}\ \bibnamefont {Aoiz}}, \ and\
  \bibinfo {author} {\bibfnamefont {M.}~\bibnamefont {Brouard}},\ }\href@noop
  {} {\bibfield  {journal} {\bibinfo  {journal} {J. Chem. Phys.}\ }\textbf
  {\bibinfo {volume} {142}},\ \bibinfo {pages} {144307} (\bibinfo {year}
  {2015})}\BibitemShut {NoStop}%
\bibitem [{\citenamefont {Warshel}\ and\ \citenamefont
  {Karplus}(1975)}]{karplus_cpl32}%
  \BibitemOpen
  \bibfield  {author} {\bibinfo {author} {\bibfnamefont {A.}~\bibnamefont
  {Warshel}}\ and\ \bibinfo {author} {\bibfnamefont {M.}~\bibnamefont
  {Karplus}},\ }\href@noop {} {\bibfield  {journal} {\bibinfo  {journal} {Chem.
  Phys. Lett.}\ }\textbf {\bibinfo {volume} {32}},\ \bibinfo {pages} {11}
  (\bibinfo {year} {1975})}\BibitemShut {NoStop}%
\bibitem [{\citenamefont {Tachikawa}\ \emph {et~al.}(1997)\citenamefont
  {Tachikawa}, \citenamefont {Ohnishi}, \citenamefont {Hamabayashi},\ and\
  \citenamefont {Yoshida}}]{yoshida_jpca101}%
  \BibitemOpen
  \bibfield  {author} {\bibinfo {author} {\bibfnamefont {H.}~\bibnamefont
  {Tachikawa}}, \bibinfo {author} {\bibfnamefont {K.}~\bibnamefont {Ohnishi}},
  \bibinfo {author} {\bibfnamefont {T.}~\bibnamefont {Hamabayashi}}, \ and\
  \bibinfo {author} {\bibfnamefont {H.}~\bibnamefont {Yoshida}},\ }\href@noop
  {} {\bibfield  {journal} {\bibinfo  {journal} {J. Phys. Chem. A}\ }\textbf
  {\bibinfo {volume} {101}},\ \bibinfo {pages} {2229} (\bibinfo {year}
  {1997})}\BibitemShut {NoStop}%
\bibitem [{\citenamefont {Hu}\ \emph {et~al.}(2008)\citenamefont {Hu},
  \citenamefont {Lendvay}, \citenamefont {Maiti},\ and\ \citenamefont
  {Schatz}}]{schatz_jpca112}%
  \BibitemOpen
  \bibfield  {author} {\bibinfo {author} {\bibfnamefont {W.}~\bibnamefont
  {Hu}}, \bibinfo {author} {\bibfnamefont {G.}~\bibnamefont {Lendvay}},
  \bibinfo {author} {\bibfnamefont {B.}~\bibnamefont {Maiti}}, \ and\ \bibinfo
  {author} {\bibfnamefont {G.~C.}\ \bibnamefont {Schatz}},\ }\href@noop {}
  {\bibfield  {journal} {\bibinfo  {journal} {J. Phys. Chem. A}\ }\textbf
  {\bibinfo {volume} {112}},\ \bibinfo {pages} {2093} (\bibinfo {year}
  {2008})}\BibitemShut {NoStop}%
\bibitem [{\citenamefont {Carbogno}\ \emph {et~al.}(2010)\citenamefont
  {Carbogno}, \citenamefont {Behler}, \citenamefont {Reuter},\ and\
  \citenamefont {Gross}}]{gross_prb81}%
  \BibitemOpen
  \bibfield  {author} {\bibinfo {author} {\bibfnamefont {C.}~\bibnamefont
  {Carbogno}}, \bibinfo {author} {\bibfnamefont {J.}~\bibnamefont {Behler}},
  \bibinfo {author} {\bibfnamefont {K.}~\bibnamefont {Reuter}}, \ and\ \bibinfo
  {author} {\bibfnamefont {A.}~\bibnamefont {Gross}},\ }\href@noop {}
  {\bibfield  {journal} {\bibinfo  {journal} {Phys. Rev. B}\ }\textbf {\bibinfo
  {volume} {81}},\ \bibinfo {pages} {035410} (\bibinfo {year}
  {2010})}\BibitemShut {NoStop}%
\bibitem [{\citenamefont {Yamashita}\ and\ \citenamefont
  {Morokuma}(1990)}]{morokuma_cpl169}%
  \BibitemOpen
  \bibfield  {author} {\bibinfo {author} {\bibfnamefont {K.}~\bibnamefont
  {Yamashita}}\ and\ \bibinfo {author} {\bibfnamefont {K.}~\bibnamefont
  {Morokuma}},\ }\href@noop {} {\bibfield  {journal} {\bibinfo  {journal}
  {Chem. Phys. Lett.}\ }\textbf {\bibinfo {volume} {169}},\ \bibinfo {pages}
  {263} (\bibinfo {year} {1990})}\BibitemShut {NoStop}%
\bibitem [{\citenamefont {Gai}\ and\ \citenamefont {Voth}(1992)}]{voth_jap71}%
  \BibitemOpen
  \bibfield  {author} {\bibinfo {author} {\bibfnamefont {H.}~\bibnamefont
  {Gai}}\ and\ \bibinfo {author} {\bibfnamefont {G.~A.}\ \bibnamefont {Voth}},\
  }\href@noop {} {\bibfield  {journal} {\bibinfo  {journal} {J. Appl. Phys.}\
  }\textbf {\bibinfo {volume} {71}},\ \bibinfo {pages} {1415} (\bibinfo {year}
  {1992})}\BibitemShut {NoStop}%
\bibitem [{\citenamefont {Jones}, \citenamefont {Acocella},\ and\ \citenamefont
  {Zerbetto}(2008)}]{zerbetto_jpca112}%
  \BibitemOpen
  \bibfield  {author} {\bibinfo {author} {\bibfnamefont {G.~A.}\ \bibnamefont
  {Jones}}, \bibinfo {author} {\bibfnamefont {A.}~\bibnamefont {Acocella}}, \
  and\ \bibinfo {author} {\bibfnamefont {F.}~\bibnamefont {Zerbetto}},\
  }\href@noop {} {\bibfield  {journal} {\bibinfo  {journal} {J. Phys. Chem. A}\
  }\textbf {\bibinfo {volume} {112}},\ \bibinfo {pages} {9650} (\bibinfo {year}
  {2008})}\BibitemShut {NoStop}%
\bibitem [{\citenamefont {Mitric}, \citenamefont {Petersen},\ and\
  \citenamefont {Bonacic-Koutecky}(2009)}]{mitric_pra79}%
  \BibitemOpen
  \bibfield  {author} {\bibinfo {author} {\bibfnamefont {R.}~\bibnamefont
  {Mitric}}, \bibinfo {author} {\bibfnamefont {J.}~\bibnamefont {Petersen}}, \
  and\ \bibinfo {author} {\bibfnamefont {V.}~\bibnamefont {Bonacic-Koutecky}},\
  }\href@noop {} {\bibfield  {journal} {\bibinfo  {journal} {Phys. Rev. A}\
  }\textbf {\bibinfo {volume} {79}},\ \bibinfo {pages} {053416} (\bibinfo
  {year} {2009})}\BibitemShut {NoStop}%
\bibitem [{\citenamefont {Herman}\ and\ \citenamefont
  {Arce}(1994)}]{herman_cp183}%
  \BibitemOpen
  \bibfield  {author} {\bibinfo {author} {\bibfnamefont {M.~F.}\ \bibnamefont
  {Herman}}\ and\ \bibinfo {author} {\bibfnamefont {J.~C.}\ \bibnamefont
  {Arce}},\ }\href@noop {} {\bibfield  {journal} {\bibinfo  {journal} {Chem.
  Phys.}\ }\textbf {\bibinfo {volume} {183}},\ \bibinfo {pages} {335} (\bibinfo
  {year} {1994})}\BibitemShut {NoStop}%
\bibitem [{\citenamefont {Fabiano}, \citenamefont {Keal},\ and\ \citenamefont
  {Thiel}(2008)}]{thiel_cp349}%
  \BibitemOpen
  \bibfield  {author} {\bibinfo {author} {\bibfnamefont {E.}~\bibnamefont
  {Fabiano}}, \bibinfo {author} {\bibfnamefont {T.~W.}\ \bibnamefont {Keal}}, \
  and\ \bibinfo {author} {\bibfnamefont {W.}~\bibnamefont {Thiel}},\
  }\href@noop {} {\bibfield  {journal} {\bibinfo  {journal} {Chem. Phys.}\
  }\textbf {\bibinfo {volume} {349}},\ \bibinfo {pages} {334} (\bibinfo {year}
  {2008})}\BibitemShut {NoStop}%
\bibitem [{\citenamefont {Ghosh}\ \emph
  {et~al.}(2021{\natexlab{b}})\citenamefont {Ghosh}, \citenamefont {Sahoo},
  \citenamefont {Baer},\ and\ \citenamefont {Adhikari}}]{ghosh_jpca125}%
  \BibitemOpen
  \bibfield  {author} {\bibinfo {author} {\bibfnamefont {S.}~\bibnamefont
  {Ghosh}}, \bibinfo {author} {\bibfnamefont {T.}~\bibnamefont {Sahoo}},
  \bibinfo {author} {\bibfnamefont {M.}~\bibnamefont {Baer}}, \ and\ \bibinfo
  {author} {\bibfnamefont {S.}~\bibnamefont {Adhikari}},\ }\href@noop {}
  {\bibfield  {journal} {\bibinfo  {journal} {J. Phys. Chem. A}\ }\textbf
  {\bibinfo {volume} {125}},\ \bibinfo {pages} {731} (\bibinfo {year}
  {2021}{\natexlab{b}})}\BibitemShut {NoStop}%
\bibitem [{\citenamefont {Karpas}, \citenamefont {Anicich},\ and\ \citenamefont
  {Huntress}(1979)}]{karpas_jcp70}%
  \BibitemOpen
  \bibfield  {author} {\bibinfo {author} {\bibfnamefont {Z.}~\bibnamefont
  {Karpas}}, \bibinfo {author} {\bibfnamefont {V.}~\bibnamefont {Anicich}}, \
  and\ \bibinfo {author} {\bibfnamefont {W.~T.~J.}\ \bibnamefont {Huntress}},\
  }\href@noop {} {\bibfield  {journal} {\bibinfo  {journal} {J. Chem. Phys.}\
  }\textbf {\bibinfo {volume} {70}},\ \bibinfo {pages} {2877} (\bibinfo {year}
  {1979})}\BibitemShut {NoStop}%
\bibitem [{\citenamefont {Porter}, \citenamefont {Raff},\ and\ \citenamefont
  {Miller}(1975)}]{miller_jcp63}%
  \BibitemOpen
  \bibfield  {author} {\bibinfo {author} {\bibfnamefont {R.~N.}\ \bibnamefont
  {Porter}}, \bibinfo {author} {\bibfnamefont {L.~M.}\ \bibnamefont {Raff}}, \
  and\ \bibinfo {author} {\bibfnamefont {W.~H.}\ \bibnamefont {Miller}},\
  }\href@noop {} {\bibfield  {journal} {\bibinfo  {journal} {J. Chem. Phys.}\
  }\textbf {\bibinfo {volume} {63}},\ \bibinfo {pages} {2214} (\bibinfo {year}
  {1975})}\BibitemShut {NoStop}%
\bibitem [{\citenamefont {Johnson}(1983{\natexlab{c}})}]{johnson_jcp79}%
  \BibitemOpen
  \bibfield  {author} {\bibinfo {author} {\bibfnamefont {B.~R.}\ \bibnamefont
  {Johnson}},\ }\href@noop {} {\bibfield  {journal} {\bibinfo  {journal} {J.
  Chem. Phys.}\ }\textbf {\bibinfo {volume} {79}},\ \bibinfo {pages} {1906}
  (\bibinfo {year} {1983}{\natexlab{c}})}\BibitemShut {NoStop}%
\bibitem [{\citenamefont {Billing}(2003)}]{Billing_book}%
  \BibitemOpen
  \bibfield  {author} {\bibinfo {author} {\bibfnamefont {G.~D.}\ \bibnamefont
  {Billing}},\ }\href@noop {} {\emph {\bibinfo {title} {The Quantum Classical
  Theory}}}\ (\bibinfo  {publisher} {Oxford University Press, Oxford},\
  \bibinfo {year} {2003})\BibitemShut {NoStop}%
\bibitem [{\citenamefont {Ravi}\ \emph {et~al.}(2020)\citenamefont {Ravi},
  \citenamefont {Mukherjee}, \citenamefont {Mukherjee}, \citenamefont
  {Adhikari}, \citenamefont {Sathyamurthy},\ and\ \citenamefont
  {Baer}}]{ravi_epjd}%
  \BibitemOpen
  \bibfield  {author} {\bibinfo {author} {\bibfnamefont {S.}~\bibnamefont
  {Ravi}}, \bibinfo {author} {\bibfnamefont {S.}~\bibnamefont {Mukherjee}},
  \bibinfo {author} {\bibfnamefont {B.}~\bibnamefont {Mukherjee}}, \bibinfo
  {author} {\bibfnamefont {S.}~\bibnamefont {Adhikari}}, \bibinfo {author}
  {\bibfnamefont {N.}~\bibnamefont {Sathyamurthy}}, \ and\ \bibinfo {author}
  {\bibfnamefont {M.}~\bibnamefont {Baer}},\ }\href@noop {} {\bibfield
  {journal} {\bibinfo  {journal} {The European Physical Journal D}\ }\textbf
  {\bibinfo {volume} {74}},\ \bibinfo {pages} {239} (\bibinfo {year}
  {2020})}\BibitemShut {NoStop}%
\bibitem [{\citenamefont {Ravi}\ \emph
  {et~al.}(2021{\natexlab{c}})\citenamefont {Ravi}, \citenamefont {Mukherjee},
  \citenamefont {Mukherjee}, \citenamefont {Adhikari}, \citenamefont
  {Sathyamurthy},\ and\ \citenamefont {Baer}}]{ravi_mp119}%
  \BibitemOpen
  \bibfield  {author} {\bibinfo {author} {\bibfnamefont {S.}~\bibnamefont
  {Ravi}}, \bibinfo {author} {\bibfnamefont {S.}~\bibnamefont {Mukherjee}},
  \bibinfo {author} {\bibfnamefont {B.}~\bibnamefont {Mukherjee}}, \bibinfo
  {author} {\bibfnamefont {S.}~\bibnamefont {Adhikari}}, \bibinfo {author}
  {\bibfnamefont {N.}~\bibnamefont {Sathyamurthy}}, \ and\ \bibinfo {author}
  {\bibfnamefont {M.}~\bibnamefont {Baer}},\ }\href@noop {} {\bibfield
  {journal} {\bibinfo  {journal} {Molecular Physics}\ }\textbf {\bibinfo
  {volume} {119}},\ \bibinfo {pages} {e1811907} (\bibinfo {year}
  {2021}{\natexlab{c}})}\BibitemShut {NoStop}%
\bibitem [{\citenamefont {Goldstein}(1956)}]{goldstein_classical_book}%
  \BibitemOpen
  \bibfield  {author} {\bibinfo {author} {\bibfnamefont {H.}~\bibnamefont
  {Goldstein}},\ }\href@noop {} {\emph {\bibinfo {title} {Classical
  mechanics}}}\ (\bibinfo  {publisher} {Addison-Wesley Publishing Company:
  Cambridge, Massachusetts},\ \bibinfo {year} {1956})\BibitemShut {NoStop}%
\bibitem [{\citenamefont {Landau}\ and\ \citenamefont
  {Lifshitz}(1997)}]{landau_mechanics_book}%
  \BibitemOpen
  \bibfield  {author} {\bibinfo {author} {\bibfnamefont {L.~D.}\ \bibnamefont
  {Landau}}\ and\ \bibinfo {author} {\bibfnamefont {E.~M.}\ \bibnamefont
  {Lifshitz}},\ }\href@noop {} {\emph {\bibinfo {title} {Mechanics}}}\
  (\bibinfo  {publisher} {Pergamon Press: New York},\ \bibinfo {year}
  {1997})\BibitemShut {NoStop}%
\bibitem [{\citenamefont {Charutz}, \citenamefont {Baer},\ and\ \citenamefont
  {Baer}(1997)}]{baer_cpl265}%
  \BibitemOpen
  \bibfield  {author} {\bibinfo {author} {\bibfnamefont {D.~M.}\ \bibnamefont
  {Charutz}}, \bibinfo {author} {\bibfnamefont {R.}~\bibnamefont {Baer}}, \
  and\ \bibinfo {author} {\bibfnamefont {M.}~\bibnamefont {Baer}},\ }\href@noop
  {} {\bibfield  {journal} {\bibinfo  {journal} {Chem. Phys. Letts.}\ }\textbf
  {\bibinfo {volume} {265}},\ \bibinfo {pages} {629} (\bibinfo {year}
  {1997})}\BibitemShut {NoStop}%
\bibitem [{\citenamefont {Hopper}(1980)}]{hopper_jcp_73}%
  \BibitemOpen
  \bibfield  {author} {\bibinfo {author} {\bibfnamefont {D.~G.}\ \bibnamefont
  {Hopper}},\ }\href@noop {} {\bibfield  {journal} {\bibinfo  {journal} {The
  Journal of Chemical Physics}\ }\textbf {\bibinfo {volume} {73}},\ \bibinfo
  {pages} {3289} (\bibinfo {year} {1980})}\BibitemShut {NoStop}%
\bibitem [{\citenamefont {Kramers}(1940)}]{kramers_physica_7}%
  \BibitemOpen
  \bibfield  {author} {\bibinfo {author} {\bibfnamefont {H.~A.}\ \bibnamefont
  {Kramers}},\ }\href@noop {} {\bibfield  {journal} {\bibinfo  {journal}
  {physica}\ }\textbf {\bibinfo {volume} {7}},\ \bibinfo {pages} {284}
  (\bibinfo {year} {1940})}\BibitemShut {NoStop}%
\bibitem [{\citenamefont {Head-Gordon}\ and\ \citenamefont
  {Tully}(1995)}]{head_jcp103}%
  \BibitemOpen
  \bibfield  {author} {\bibinfo {author} {\bibfnamefont {M.}~\bibnamefont
  {Head-Gordon}}\ and\ \bibinfo {author} {\bibfnamefont {J.~C.}\ \bibnamefont
  {Tully}},\ }\href@noop {} {\bibfield  {journal} {\bibinfo  {journal} {The
  Journal of chemical physics}\ }\textbf {\bibinfo {volume} {103}},\ \bibinfo
  {pages} {10137} (\bibinfo {year} {1995})}\BibitemShut {NoStop}%
\bibitem [{\citenamefont {Spiering}\ \emph
  {et~al.}(2019{\natexlab{a}})\citenamefont {Spiering}, \citenamefont
  {Shakouri}, \citenamefont {Behler}, \citenamefont {Kroes},\ and\
  \citenamefont {Meyer}}]{spiering_jpcl10}%
  \BibitemOpen
  \bibfield  {author} {\bibinfo {author} {\bibfnamefont {P.}~\bibnamefont
  {Spiering}}, \bibinfo {author} {\bibfnamefont {K.}~\bibnamefont {Shakouri}},
  \bibinfo {author} {\bibfnamefont {J.}~\bibnamefont {Behler}}, \bibinfo
  {author} {\bibfnamefont {G.-J.}\ \bibnamefont {Kroes}}, \ and\ \bibinfo
  {author} {\bibfnamefont {J.}~\bibnamefont {Meyer}},\ }\href@noop {}
  {\bibfield  {journal} {\bibinfo  {journal} {The journal of physical chemistry
  letters}\ }\textbf {\bibinfo {volume} {10}},\ \bibinfo {pages} {2957}
  (\bibinfo {year} {2019}{\natexlab{a}})}\BibitemShut {NoStop}%
\bibitem [{\citenamefont {Smits}\ \emph {et~al.}(2024)\citenamefont {Smits},
  \citenamefont {Sah}, \citenamefont {Naskar}, \citenamefont {Adhikari},
  \citenamefont {Meyer},\ and\ \citenamefont {Somers}}]{smits_jcp161}%
  \BibitemOpen
  \bibfield  {author} {\bibinfo {author} {\bibfnamefont {B.}~\bibnamefont
  {Smits}}, \bibinfo {author} {\bibfnamefont {M.~K.}\ \bibnamefont {Sah}},
  \bibinfo {author} {\bibfnamefont {K.}~\bibnamefont {Naskar}}, \bibinfo
  {author} {\bibfnamefont {S.}~\bibnamefont {Adhikari}}, \bibinfo {author}
  {\bibfnamefont {J.}~\bibnamefont {Meyer}}, \ and\ \bibinfo {author}
  {\bibfnamefont {M.~F.}\ \bibnamefont {Somers}},\ }\href@noop {} {\bibfield
  {journal} {\bibinfo  {journal} {The Journal of Chemical Physics}\ }\textbf
  {\bibinfo {volume} {161}} (\bibinfo {year} {2024})}\BibitemShut {NoStop}%
\bibitem [{\citenamefont {Sah}\ \emph {et~al.}(2024)\citenamefont {Sah},
  \citenamefont {Naskar}, \citenamefont {Adhikari}, \citenamefont {Smits},
  \citenamefont {Meyer},\ and\ \citenamefont {Somers}}]{sah_jcp161}%
  \BibitemOpen
  \bibfield  {author} {\bibinfo {author} {\bibfnamefont {M.~K.}\ \bibnamefont
  {Sah}}, \bibinfo {author} {\bibfnamefont {K.}~\bibnamefont {Naskar}},
  \bibinfo {author} {\bibfnamefont {S.}~\bibnamefont {Adhikari}}, \bibinfo
  {author} {\bibfnamefont {B.}~\bibnamefont {Smits}}, \bibinfo {author}
  {\bibfnamefont {J.}~\bibnamefont {Meyer}}, \ and\ \bibinfo {author}
  {\bibfnamefont {M.~F.}\ \bibnamefont {Somers}},\ }\href@noop {} {\bibfield
  {journal} {\bibinfo  {journal} {The Journal of Chemical Physics}\ }\textbf
  {\bibinfo {volume} {161}} (\bibinfo {year} {2024})}\BibitemShut {NoStop}%
\bibitem [{\citenamefont {Saalfrank}(2006{\natexlab{a}})}]{saalfrank_cr106}%
  \BibitemOpen
  \bibfield  {author} {\bibinfo {author} {\bibfnamefont {P.}~\bibnamefont
  {Saalfrank}},\ }\href@noop {} {\bibfield  {journal} {\bibinfo  {journal}
  {Chemical Reviews}\ }\textbf {\bibinfo {volume} {106}},\ \bibinfo {pages}
  {4116} (\bibinfo {year} {2006}{\natexlab{a}})}\BibitemShut {NoStop}%
\bibitem [{\citenamefont {Baer}\ \emph {et~al.}(2024)\citenamefont {Baer},
  \citenamefont {Mukherjee}, \citenamefont {Ravi}, \citenamefont {Adhikari},\
  and\ \citenamefont {Sathyamurthy}}]{baer_aqc89}%
  \BibitemOpen
  \bibfield  {author} {\bibinfo {author} {\bibfnamefont {M.}~\bibnamefont
  {Baer}}, \bibinfo {author} {\bibfnamefont {S.}~\bibnamefont {Mukherjee}},
  \bibinfo {author} {\bibfnamefont {S.}~\bibnamefont {Ravi}}, \bibinfo {author}
  {\bibfnamefont {S.}~\bibnamefont {Adhikari}}, \ and\ \bibinfo {author}
  {\bibfnamefont {N.}~\bibnamefont {Sathyamurthy}},\ }in\ \href@noop {} {\emph
  {\bibinfo {booktitle} {Advances in Quantum Chemistry}}},\ Vol.~\bibinfo
  {volume} {89}\ (\bibinfo  {publisher} {Elsevier},\ \bibinfo {year} {2024})\
  pp.\ \bibinfo {pages} {291--304}\BibitemShut {NoStop}%
\bibitem [{\citenamefont {Tokura}\ and\ \citenamefont
  {Tomioka}(1999)}]{Tokura_1999}%
  \BibitemOpen
  \bibfield  {author} {\bibinfo {author} {\bibfnamefont {Y.}~\bibnamefont
  {Tokura}}\ and\ \bibinfo {author} {\bibfnamefont {Y.}~\bibnamefont
  {Tomioka}},\ }\href {\doibase https://doi.org/10.1016/S0304-8853(99)00352-2}
  {\bibfield  {journal} {\bibinfo  {journal} {Journal of Magnetism and Magnetic
  Materials}\ }\textbf {\bibinfo {volume} {200}},\ \bibinfo {pages} {1}
  (\bibinfo {year} {1999})}\BibitemShut {NoStop}%
\bibitem [{\citenamefont {Dagotto}, \citenamefont {Hotta},\ and\ \citenamefont
  {Moreo}(2001)}]{Dagotto_2001}%
  \BibitemOpen
  \bibfield  {author} {\bibinfo {author} {\bibfnamefont {E.}~\bibnamefont
  {Dagotto}}, \bibinfo {author} {\bibfnamefont {T.}~\bibnamefont {Hotta}}, \
  and\ \bibinfo {author} {\bibfnamefont {A.}~\bibnamefont {Moreo}},\ }\href
  {\doibase https://doi.org/10.1016/S0370-1573(00)00121-6} {\bibfield
  {journal} {\bibinfo  {journal} {Physics Reports}\ }\textbf {\bibinfo {volume}
  {344}},\ \bibinfo {pages} {1} (\bibinfo {year} {2001})}\BibitemShut {NoStop}%
\bibitem [{\citenamefont {Lines}\ and\ \citenamefont
  {Glass}(2001)}]{Lines_2001}%
  \BibitemOpen
  \bibfield  {author} {\bibinfo {author} {\bibfnamefont {M.}~\bibnamefont
  {Lines}}\ and\ \bibinfo {author} {\bibfnamefont {A.}~\bibnamefont {Glass}},\
  }\href {https://books.google.co.in/books?id=p6ruJH8C84kC} {\emph {\bibinfo
  {title} {Principles and Applications of Ferroelectrics and Related
  Materials}}},\ International series of monographs on physics\ (\bibinfo
  {publisher} {OUP Oxford},\ \bibinfo {year} {2001})\BibitemShut {NoStop}%
\bibitem [{Jaf(1971)}]{Jaffe_1971}%
  \BibitemOpen
  in\ \href {\doibase https://doi.org/10.1016/B978-0-12-379550-2.50001-6}
  {\emph {\bibinfo {booktitle} {Piezoelectric Ceramics}}},\ \bibinfo {editor}
  {edited by\ \bibinfo {editor} {\bibfnamefont {B.}~\bibnamefont {JAFFE}},
  \bibinfo {editor} {\bibfnamefont {W.~R.}\ \bibnamefont {COOK}}, \ and\
  \bibinfo {editor} {\bibfnamefont {H.}~\bibnamefont {JAFFE}}}\ (\bibinfo
  {publisher} {Academic Press},\ \bibinfo {year} {1971})\ p.~\bibinfo {pages}
  {ii}\BibitemShut {NoStop}%
\bibitem [{\citenamefont {Ramesh}\ and\ \citenamefont
  {Spaldin}(2007)}]{Ramesh_2007}%
  \BibitemOpen
  \bibfield  {author} {\bibinfo {author} {\bibfnamefont {R.}~\bibnamefont
  {Ramesh}}\ and\ \bibinfo {author} {\bibfnamefont {N.~A.}\ \bibnamefont
  {Spaldin}},\ }\href {\doibase 10.1038/nmat1805} {\bibfield  {journal}
  {\bibinfo  {journal} {Nature Materials}\ }\textbf {\bibinfo {volume} {6}},\
  \bibinfo {pages} {21–29} (\bibinfo {year} {2007})}\BibitemShut {NoStop}%
\bibitem [{\citenamefont {Vleck}(1939{\natexlab{b}})}]{VanVleck1}%
  \BibitemOpen
  \bibfield  {author} {\bibinfo {author} {\bibfnamefont {V.}~\bibnamefont
  {Vleck}},\ }\href@noop {} {\bibfield  {journal} {\bibinfo  {journal} {J.
  Chem. Phys}\ }\textbf {\bibinfo {volume} {7}},\ \bibinfo {pages} {61}
  (\bibinfo {year} {1939}{\natexlab{b}})}\BibitemShut {NoStop}%
\bibitem [{\citenamefont {Vleck}(1939{\natexlab{c}})}]{VanVleck}%
  \BibitemOpen
  \bibfield  {author} {\bibinfo {author} {\bibfnamefont {V.}~\bibnamefont
  {Vleck}},\ }\href@noop {} {\bibfield  {journal} {\bibinfo  {journal} {J.
  Chem. Phys}\ }\textbf {\bibinfo {volume} {7}},\ \bibinfo {pages} {72}
  (\bibinfo {year} {1939}{\natexlab{c}})}\BibitemShut {NoStop}%
\bibitem [{\citenamefont {Kovaleva}\ \emph {et~al.}(2004)\citenamefont
  {Kovaleva}, \citenamefont {Boris}, \citenamefont {Bernhard}, \citenamefont
  {Kulakov}, \citenamefont {Pimenov}, \citenamefont {Balbashov}, \citenamefont
  {Khaliullin},\ and\ \citenamefont {Keimer}}]{Natalia_JETP1}%
  \BibitemOpen
  \bibfield  {author} {\bibinfo {author} {\bibfnamefont {N.~N.}\ \bibnamefont
  {Kovaleva}}, \bibinfo {author} {\bibfnamefont {A.~V.}\ \bibnamefont {Boris}},
  \bibinfo {author} {\bibfnamefont {C.}~\bibnamefont {Bernhard}}, \bibinfo
  {author} {\bibfnamefont {A.}~\bibnamefont {Kulakov}}, \bibinfo {author}
  {\bibfnamefont {A.}~\bibnamefont {Pimenov}}, \bibinfo {author} {\bibfnamefont
  {A.~M.}\ \bibnamefont {Balbashov}}, \bibinfo {author} {\bibfnamefont
  {G.}~\bibnamefont {Khaliullin}}, \ and\ \bibinfo {author} {\bibfnamefont
  {B.}~\bibnamefont {Keimer}},\ }\href@noop {} {\bibfield  {journal} {\bibinfo
  {journal} {Phys. Rev. Lett.}\ }\textbf {\bibinfo {volume} {93}},\ \bibinfo
  {pages} {147204} (\bibinfo {year} {2004})}\BibitemShut {NoStop}%
\bibitem [{\citenamefont {Kovaleva}\ \emph {et~al.}(2010)\citenamefont
  {Kovaleva}, \citenamefont {Ole\'s}, \citenamefont {Balbashov}, \citenamefont
  {Maljuk}, \citenamefont {Argyriou}, \citenamefont {Khaliullin},\ and\
  \citenamefont {Keimer}}]{Natalia_JETP2}%
  \BibitemOpen
  \bibfield  {author} {\bibinfo {author} {\bibfnamefont {N.~N.}\ \bibnamefont
  {Kovaleva}}, \bibinfo {author} {\bibfnamefont {A.~M.}\ \bibnamefont
  {Ole\'s}}, \bibinfo {author} {\bibfnamefont {A.~M.}\ \bibnamefont
  {Balbashov}}, \bibinfo {author} {\bibfnamefont {A.}~\bibnamefont {Maljuk}},
  \bibinfo {author} {\bibfnamefont {D.~N.}\ \bibnamefont {Argyriou}}, \bibinfo
  {author} {\bibfnamefont {G.}~\bibnamefont {Khaliullin}}, \ and\ \bibinfo
  {author} {\bibfnamefont {B.}~\bibnamefont {Keimer}},\ }\href@noop {}
  {\bibfield  {journal} {\bibinfo  {journal} {Phys. Rev. B.}\ }\textbf
  {\bibinfo {volume} {81}},\ \bibinfo {pages} {235130} (\bibinfo {year}
  {2010})}\BibitemShut {NoStop}%
\bibitem [{\citenamefont {Mondal}\ \emph {et~al.}(2011)\citenamefont {Mondal},
  \citenamefont {Opalka}, \citenamefont {Poluyanov},\ and\ \citenamefont
  {Domcke}}]{domcke_cp387}%
  \BibitemOpen
  \bibfield  {author} {\bibinfo {author} {\bibfnamefont {P.}~\bibnamefont
  {Mondal}}, \bibinfo {author} {\bibfnamefont {D.}~\bibnamefont {Opalka}},
  \bibinfo {author} {\bibfnamefont {L.~V.}\ \bibnamefont {Poluyanov}}, \ and\
  \bibinfo {author} {\bibfnamefont {W.}~\bibnamefont {Domcke}},\ }\href@noop {}
  {\bibfield  {journal} {\bibinfo  {journal} {Chem. Phys.}\ }\textbf {\bibinfo
  {volume} {387}},\ \bibinfo {pages} {56} (\bibinfo {year} {2011})}\BibitemShut
  {NoStop}%
\bibitem [{\citenamefont {Solomonik}\ and\ \citenamefont
  {Mukhanov}(2012)}]{mukhanov_jsc53}%
  \BibitemOpen
  \bibfield  {author} {\bibinfo {author} {\bibfnamefont {V.~G.}\ \bibnamefont
  {Solomonik}}\ and\ \bibinfo {author} {\bibfnamefont {A.~A.}\ \bibnamefont
  {Mukhanov}},\ }\href@noop {} {\bibfield  {journal} {\bibinfo  {journal} {J.
  Struct. Chem.}\ }\textbf {\bibinfo {volume} {53}},\ \bibinfo {pages} {819}
  (\bibinfo {year} {2012})}\BibitemShut {NoStop}%
\bibitem [{\citenamefont {Yates}\ and\ \citenamefont
  {Pitzer}(1979)}]{pitzer_jcp70}%
  \BibitemOpen
  \bibfield  {author} {\bibinfo {author} {\bibfnamefont {J.~H.}\ \bibnamefont
  {Yates}}\ and\ \bibinfo {author} {\bibfnamefont {R.~M.}\ \bibnamefont
  {Pitzer}},\ }\href@noop {} {\bibfield  {journal} {\bibinfo  {journal} {J.
  Chem. Phys.}\ }\textbf {\bibinfo {volume} {70}},\ \bibinfo {pages} {4049}
  (\bibinfo {year} {1979})}\BibitemShut {NoStop}%
\bibitem [{\citenamefont {Solomonik}\ and\ \citenamefont
  {Mukhanov}(2014)}]{mukhanov_rjpca88}%
  \BibitemOpen
  \bibfield  {author} {\bibinfo {author} {\bibfnamefont {V.~G.}\ \bibnamefont
  {Solomonik}}\ and\ \bibinfo {author} {\bibfnamefont {A.~A.}\ \bibnamefont
  {Mukhanov}},\ }\href@noop {} {\bibfield  {journal} {\bibinfo  {journal}
  {Russ. J. Phys. Chem.}\ }\textbf {\bibinfo {volume} {88}},\ \bibinfo {pages}
  {85} (\bibinfo {year} {2014})}\BibitemShut {NoStop}%
\bibitem [{\citenamefont {Nandipati}\ \emph {et~al.}(2020)\citenamefont
  {Nandipati}, \citenamefont {Vasilyev}, \citenamefont {Navarkin},
  \citenamefont {Solomonik},\ and\ \citenamefont {Domcke}}]{domcke_jcp152}%
  \BibitemOpen
  \bibfield  {author} {\bibinfo {author} {\bibfnamefont {K.~R.}\ \bibnamefont
  {Nandipati}}, \bibinfo {author} {\bibfnamefont {O.~A.}\ \bibnamefont
  {Vasilyev}}, \bibinfo {author} {\bibfnamefont {I.~S.}\ \bibnamefont
  {Navarkin}}, \bibinfo {author} {\bibfnamefont {V.~G.}\ \bibnamefont
  {Solomonik}}, \ and\ \bibinfo {author} {\bibfnamefont {W.}~\bibnamefont
  {Domcke}},\ }\href@noop {} {\bibfield  {journal} {\bibinfo  {journal} {J.
  Chem. Phys.}\ }\textbf {\bibinfo {volume} {152}},\ \bibinfo {pages} {094304}
  (\bibinfo {year} {2020})}\BibitemShut {NoStop}%
\bibitem [{\citenamefont {Rodr\'{i}guez-Carvajal}\ \emph
  {et~al.}(1998)\citenamefont {Rodr\'{i}guez-Carvajal}, \citenamefont
  {Hennion}, \citenamefont {Moussa}, \citenamefont {Moudden}, \citenamefont
  {Pinsard},\ and\ \citenamefont {Revcolevschi}}]{Phase}%
  \BibitemOpen
  \bibfield  {author} {\bibinfo {author} {\bibfnamefont {J.}~\bibnamefont
  {Rodr\'{i}guez-Carvajal}}, \bibinfo {author} {\bibfnamefont {M.}~\bibnamefont
  {Hennion}}, \bibinfo {author} {\bibfnamefont {F.}~\bibnamefont {Moussa}},
  \bibinfo {author} {\bibfnamefont {A.~H.}\ \bibnamefont {Moudden}}, \bibinfo
  {author} {\bibfnamefont {L.}~\bibnamefont {Pinsard}}, \ and\ \bibinfo
  {author} {\bibfnamefont {A.}~\bibnamefont {Revcolevschi}},\ }\href@noop {}
  {\bibfield  {journal} {\bibinfo  {journal} {Phys. Rev. B}\ }\textbf {\bibinfo
  {volume} {57}},\ \bibinfo {pages} {R3189} (\bibinfo {year}
  {1998})}\BibitemShut {NoStop}%
\bibitem [{\citenamefont {Hirota}\ \emph {et~al.}(1996)\citenamefont {Hirota},
  \citenamefont {Kaneko}, \citenamefont {Nishizawa},\ and\ \citenamefont
  {Endoh}}]{Hirota}%
  \BibitemOpen
  \bibfield  {author} {\bibinfo {author} {\bibfnamefont {K.}~\bibnamefont
  {Hirota}}, \bibinfo {author} {\bibfnamefont {N.}~\bibnamefont {Kaneko}},
  \bibinfo {author} {\bibfnamefont {A.}~\bibnamefont {Nishizawa}}, \ and\
  \bibinfo {author} {\bibfnamefont {Y.}~\bibnamefont {Endoh}},\ }\href@noop {}
  {\bibfield  {journal} {\bibinfo  {journal} {J. Phys. Soc. Japan}\ }\textbf
  {\bibinfo {volume} {65}},\ \bibinfo {pages} {3736} (\bibinfo {year}
  {1996})}\BibitemShut {NoStop}%
\bibitem [{\citenamefont {Kovaleva}\ \emph {et~al.}(2017)\citenamefont
  {Kovaleva}, \citenamefont {Kusmartseva}, \citenamefont {Kugel},\ and\
  \citenamefont {Kusmartsev}}]{MultiRaman}%
  \BibitemOpen
  \bibfield  {author} {\bibinfo {author} {\bibfnamefont {N.~N.}\ \bibnamefont
  {Kovaleva}}, \bibinfo {author} {\bibfnamefont {O.~E.}\ \bibnamefont
  {Kusmartseva}}, \bibinfo {author} {\bibfnamefont {K.~I.}\ \bibnamefont
  {Kugel}}, \ and\ \bibinfo {author} {\bibfnamefont {F.~V.}\ \bibnamefont
  {Kusmartsev}},\ }\href@noop {} {\bibfield  {journal} {\bibinfo  {journal} {J.
  Phys.: Conf. Ser.}\ }\textbf {\bibinfo {volume} {833}},\ \bibinfo {pages}
  {012005} (\bibinfo {year} {2017})}\BibitemShut {NoStop}%
\bibitem [{\citenamefont {Kovaleva}\ \emph {et~al.}(2002)\citenamefont
  {Kovaleva}, \citenamefont {Gavartin}, \citenamefont {Shluger}, \citenamefont
  {Boris},\ and\ \citenamefont {Stoneham}}]{Natalia_JETP3}%
  \BibitemOpen
  \bibfield  {author} {\bibinfo {author} {\bibfnamefont {N.~N.}\ \bibnamefont
  {Kovaleva}}, \bibinfo {author} {\bibfnamefont {J.~L.}\ \bibnamefont
  {Gavartin}}, \bibinfo {author} {\bibfnamefont {A.~L.}\ \bibnamefont
  {Shluger}}, \bibinfo {author} {\bibfnamefont {A.~V.}\ \bibnamefont {Boris}},
  \ and\ \bibinfo {author} {\bibfnamefont {A.~M.}\ \bibnamefont {Stoneham}},\
  }\href@noop {} {\bibfield  {journal} {\bibinfo  {journal} {J. Exp. Theor.
  Phys.}\ }\textbf {\bibinfo {volume} {94}},\ \bibinfo {pages} {178} (\bibinfo
  {year} {2002})}\BibitemShut {NoStop}%
\bibitem [{\citenamefont {Bersuker}(1963)}]{Bersuker1}%
  \BibitemOpen
  \bibfield  {author} {\bibinfo {author} {\bibfnamefont {I.~B.}\ \bibnamefont
  {Bersuker}},\ }\href@noop {} {\bibfield  {journal} {\bibinfo  {journal} {Sov.
  Phys. JETP}\ }\textbf {\bibinfo {volume} {17}},\ \bibinfo {pages} {933}
  (\bibinfo {year} {1963})}\BibitemShut {NoStop}%
\bibitem [{\citenamefont {Liehr}\ and\ \citenamefont
  {Ballhausen}(1958)}]{quadratic-coupling1}%
  \BibitemOpen
  \bibfield  {author} {\bibinfo {author} {\bibfnamefont {A.~D.}\ \bibnamefont
  {Liehr}}\ and\ \bibinfo {author} {\bibfnamefont {C.~J.}\ \bibnamefont
  {Ballhausen}},\ }\href@noop {} {\bibfield  {journal} {\bibinfo  {journal}
  {Ann. Phys. (N.Y.)}\ }\textbf {\bibinfo {volume} {3}},\ \bibinfo {pages}
  {304} (\bibinfo {year} {1958})}\BibitemShut {NoStop}%
\bibitem [{\citenamefont {Zwanziger}\ and\ \citenamefont
  {Grant}(1987)}]{quadratic-coupling-strong1}%
  \BibitemOpen
  \bibfield  {author} {\bibinfo {author} {\bibfnamefont {J.~W.}\ \bibnamefont
  {Zwanziger}}\ and\ \bibinfo {author} {\bibfnamefont {E.~R.}\ \bibnamefont
  {Grant}},\ }\href@noop {} {\bibfield  {journal} {\bibinfo  {journal} {J.
  Chem. Phys.}\ }\textbf {\bibinfo {volume} {87}},\ \bibinfo {pages} {2954}
  (\bibinfo {year} {1987})}\BibitemShut {NoStop}%
\bibitem [{\citenamefont {Sadygov}\ and\ \citenamefont
  {Yarkony}(1999{\natexlab{b}})}]{quadratic-coupling-Li3}%
  \BibitemOpen
  \bibfield  {author} {\bibinfo {author} {\bibfnamefont {R.~G.}\ \bibnamefont
  {Sadygov}}\ and\ \bibinfo {author} {\bibfnamefont {D.~R.}\ \bibnamefont
  {Yarkony}},\ }\href@noop {} {\bibfield  {journal} {\bibinfo  {journal} {J.
  Chem. Phys.}\ }\textbf {\bibinfo {volume} {110}},\ \bibinfo {pages} {3639}
  (\bibinfo {year} {1999}{\natexlab{b}})}\BibitemShut {NoStop}%
\bibitem [{\citenamefont {Yarkony}(1999)}]{quadratic-coupling-Na3}%
  \BibitemOpen
  \bibfield  {author} {\bibinfo {author} {\bibfnamefont {D.~R.}\ \bibnamefont
  {Yarkony}},\ }\href@noop {} {\bibfield  {journal} {\bibinfo  {journal} {J.
  Chem. Phys.}\ }\textbf {\bibinfo {volume} {111}},\ \bibinfo {pages} {4906}
  (\bibinfo {year} {1999})}\BibitemShut {NoStop}%
\bibitem [{\citenamefont {Koizumi}\ and\ \citenamefont
  {Bersuker}(1999{\natexlab{b}})}]{quadratic-coupling-Exe}%
  \BibitemOpen
  \bibfield  {author} {\bibinfo {author} {\bibfnamefont {H.}~\bibnamefont
  {Koizumi}}\ and\ \bibinfo {author} {\bibfnamefont {I.~B.}\ \bibnamefont
  {Bersuker}},\ }\href@noop {} {\bibfield  {journal} {\bibinfo  {journal}
  {Phys. Rev. Lett.}\ }\textbf {\bibinfo {volume} {83}},\ \bibinfo {pages}
  {3009} (\bibinfo {year} {1999}{\natexlab{b}})}\BibitemShut {NoStop}%
\bibitem [{\citenamefont {Koizumi}\ \emph {et~al.}(2000)\citenamefont
  {Koizumi}, \citenamefont {Bersuker}, \citenamefont {Boggs},\ and\
  \citenamefont {Polinger}}]{quadratic-coupling-Txt2}%
  \BibitemOpen
  \bibfield  {author} {\bibinfo {author} {\bibfnamefont {H.}~\bibnamefont
  {Koizumi}}, \bibinfo {author} {\bibfnamefont {I.~B.}\ \bibnamefont
  {Bersuker}}, \bibinfo {author} {\bibfnamefont {J.~E.}\ \bibnamefont {Boggs}},
  \ and\ \bibinfo {author} {\bibfnamefont {V.~Z.}\ \bibnamefont {Polinger}},\
  }\href@noop {} {\bibfield  {journal} {\bibinfo  {journal} {J. Chem. Phys.}\
  }\textbf {\bibinfo {volume} {112}},\ \bibinfo {pages} {8470} (\bibinfo {year}
  {2000})}\BibitemShut {NoStop}%
\bibitem [{\citenamefont {Pae}\ and\ \citenamefont
  {Hizhnyakov}(2017{\natexlab{b}})}]{quadratic-coupling-Exe-nonadiabatic}%
  \BibitemOpen
  \bibfield  {author} {\bibinfo {author} {\bibfnamefont {K.}~\bibnamefont
  {Pae}}\ and\ \bibinfo {author} {\bibfnamefont {V.}~\bibnamefont
  {Hizhnyakov}},\ }\href@noop {} {\bibfield  {journal} {\bibinfo  {journal} {J.
  Chem. Phys.}\ }\textbf {\bibinfo {volume} {147}},\ \bibinfo {pages} {084107}
  (\bibinfo {year} {2017}{\natexlab{b}})}\BibitemShut {NoStop}%
\bibitem [{\citenamefont {Garc\'{i}a-Fern\'{a}ndez}\ \emph
  {et~al.}(2005)\citenamefont {Garc\'{i}a-Fern\'{a}ndez}, \citenamefont
  {Bersuker}, \citenamefont {Aramburu}, \citenamefont {Barriuso},\ and\
  \citenamefont {Moreno}}]{warping-Exe}%
  \BibitemOpen
  \bibfield  {author} {\bibinfo {author} {\bibfnamefont {P.}~\bibnamefont
  {Garc\'{i}a-Fern\'{a}ndez}}, \bibinfo {author} {\bibfnamefont {I.~B.}\
  \bibnamefont {Bersuker}}, \bibinfo {author} {\bibfnamefont {J.~A.}\
  \bibnamefont {Aramburu}}, \bibinfo {author} {\bibfnamefont {M.~T.}\
  \bibnamefont {Barriuso}}, \ and\ \bibinfo {author} {\bibfnamefont
  {M.}~\bibnamefont {Moreno}},\ }\href@noop {} {\bibfield  {journal} {\bibinfo
  {journal} {Phys. Rev. B.}\ }\textbf {\bibinfo {volume} {71}},\ \bibinfo
  {pages} {184117} (\bibinfo {year} {2005})}\BibitemShut {NoStop}%
\bibitem [{\citenamefont {Slonczewski}(1963{\natexlab{b}})}]{JCS}%
  \BibitemOpen
  \bibfield  {author} {\bibinfo {author} {\bibfnamefont {J.~C.}\ \bibnamefont
  {Slonczewski}},\ }\href@noop {} {\bibfield  {journal} {\bibinfo  {journal}
  {Phys. Rev}\ }\textbf {\bibinfo {volume} {131}},\ \bibinfo {pages} {1596}
  (\bibinfo {year} {1963}{\natexlab{b}})}\BibitemShut {NoStop}%
\bibitem [{\citenamefont {Hizhnyakov}, \citenamefont {Pae},\ and\ \citenamefont
  {Vaikj\"arv}(2012{\natexlab{b}})}]{Hizhnyakov1}%
  \BibitemOpen
  \bibfield  {author} {\bibinfo {author} {\bibfnamefont {V.}~\bibnamefont
  {Hizhnyakov}}, \bibinfo {author} {\bibfnamefont {K.}~\bibnamefont {Pae}}, \
  and\ \bibinfo {author} {\bibfnamefont {T.}~\bibnamefont {Vaikj\"arv}},\
  }\href@noop {} {\bibfield  {journal} {\bibinfo  {journal} {Chem. Phys.
  Lett.}\ }\textbf {\bibinfo {volume} {525-526}},\ \bibinfo {pages} {64}
  (\bibinfo {year} {2012}{\natexlab{b}})}\BibitemShut {NoStop}%
\bibitem [{\citenamefont {Pae}\ and\ \citenamefont
  {Hizhnyakov}(2013)}]{Hizhnyakov2}%
  \BibitemOpen
  \bibfield  {author} {\bibinfo {author} {\bibfnamefont {K.}~\bibnamefont
  {Pae}}\ and\ \bibinfo {author} {\bibfnamefont {V.}~\bibnamefont
  {Hizhnyakov}},\ }\href@noop {} {\bibfield  {journal} {\bibinfo  {journal} {J.
  Chem. Phys.}\ }\textbf {\bibinfo {volume} {138}},\ \bibinfo {pages} {104103}
  (\bibinfo {year} {2013})}\BibitemShut {NoStop}%
\bibitem [{\citenamefont {Pae}\ and\ \citenamefont
  {Hizhnyakov}(2014)}]{Hizhnyakov3}%
  \BibitemOpen
  \bibfield  {author} {\bibinfo {author} {\bibfnamefont {K.}~\bibnamefont
  {Pae}}\ and\ \bibinfo {author} {\bibfnamefont {V.}~\bibnamefont
  {Hizhnyakov}},\ }\href@noop {} {\bibfield  {journal} {\bibinfo  {journal} {J.
  Chem. Phys.}\ }\textbf {\bibinfo {volume} {141}},\ \bibinfo {pages} {234113}
  (\bibinfo {year} {2014})}\BibitemShut {NoStop}%
\bibitem [{\citenamefont {Pae}\ and\ \citenamefont
  {Hizhnyakov}(2016)}]{Hizhnyakov4}%
  \BibitemOpen
  \bibfield  {author} {\bibinfo {author} {\bibfnamefont {K.}~\bibnamefont
  {Pae}}\ and\ \bibinfo {author} {\bibfnamefont {V.}~\bibnamefont
  {Hizhnyakov}},\ }\href@noop {} {\bibfield  {journal} {\bibinfo  {journal} {J.
  Chem. Phys.}\ }\textbf {\bibinfo {volume} {145}},\ \bibinfo {pages} {064108}
  (\bibinfo {year} {2016})}\BibitemShut {NoStop}%
\bibitem [{\citenamefont {K\"{o}ppel}\ \emph {et~al.}(2009)\citenamefont
  {K\"{o}ppel}, \citenamefont {Yarkony}, \citenamefont {Barentzen},\ and\
  \citenamefont {Eds.}}]{Bersuker-Yarkony-book}%
  \BibitemOpen
  \bibfield  {author} {\bibinfo {author} {\bibfnamefont {H.}~\bibnamefont
  {K\"{o}ppel}}, \bibinfo {author} {\bibfnamefont {D.~R.}\ \bibnamefont
  {Yarkony}}, \bibinfo {author} {\bibfnamefont {H.}~\bibnamefont {Barentzen}},
  \ and\ \bibinfo {author} {\bibnamefont {Eds.}},\ }\href@noop {} {\emph
  {\bibinfo {title} {The Jahn-Teller effect. Fundamentals and Implications for
  Physics and Chemistry}}}\ (\bibinfo  {publisher} {Springer Series of Chemical
  Physics; Springer},\ \bibinfo {address} {Heidelberg},\ \bibinfo {year}
  {2009})\BibitemShut {NoStop}%
\bibitem [{\citenamefont {Garcia-Fernandez}, \citenamefont {Bersuker},\ and\
  \citenamefont {Boggs}(2006)}]{Hidden-JT1}%
  \BibitemOpen
  \bibfield  {author} {\bibinfo {author} {\bibfnamefont {P.}~\bibnamefont
  {Garcia-Fernandez}}, \bibinfo {author} {\bibfnamefont {I.~B.}\ \bibnamefont
  {Bersuker}}, \ and\ \bibinfo {author} {\bibfnamefont {J.~E.}\ \bibnamefont
  {Boggs}},\ }\href@noop {} {\bibfield  {journal} {\bibinfo  {journal} {Phys.
  Rev. Lett.}\ }\textbf {\bibinfo {volume} {96}},\ \bibinfo {pages} {163005}
  (\bibinfo {year} {2006})}\BibitemShut {NoStop}%
\bibitem [{\citenamefont {Liu}\ \emph {et~al.}(2009)\citenamefont {Liu},
  \citenamefont {Bersuker}, \citenamefont {Zou},\ and\ \citenamefont
  {Boggs}}]{Hidden-JT-PJT}%
  \BibitemOpen
  \bibfield  {author} {\bibinfo {author} {\bibfnamefont {Y.}~\bibnamefont
  {Liu}}, \bibinfo {author} {\bibfnamefont {I.~B.}\ \bibnamefont {Bersuker}},
  \bibinfo {author} {\bibfnamefont {W.}~\bibnamefont {Zou}}, \ and\ \bibinfo
  {author} {\bibfnamefont {J.~E.}\ \bibnamefont {Boggs}},\ }\href@noop {}
  {\bibfield  {journal} {\bibinfo  {journal} {J. Chem. Theory Comput.}\
  }\textbf {\bibinfo {volume} {5}},\ \bibinfo {pages} {2679} (\bibinfo {year}
  {2009})}\BibitemShut {NoStop}%
\bibitem [{\citenamefont {Bersuker}\ and\ \citenamefont
  {Polinger}(1989)}]{IBersuker-VPolinger}%
  \BibitemOpen
  \bibfield  {author} {\bibinfo {author} {\bibfnamefont {I.~B.}\ \bibnamefont
  {Bersuker}}\ and\ \bibinfo {author} {\bibfnamefont {V.~Z.}\ \bibnamefont
  {Polinger}},\ }\href@noop {} {\emph {\bibinfo {title} {Vibronic Interactions
  in Molecules and Crystals}}}\ (\bibinfo  {publisher} {Springer},\ \bibinfo
  {address} {Berlin},\ \bibinfo {year} {1989})\BibitemShut {NoStop}%
\bibitem [{\citenamefont {Bersuker}(2006{\natexlab{b}})}]{IBB}%
  \BibitemOpen
  \bibfield  {author} {\bibinfo {author} {\bibfnamefont {I.~B.}\ \bibnamefont
  {Bersuker}},\ }\href@noop {} {\emph {\bibinfo {title} {The Jahn-Teller
  Effect}}}\ (\bibinfo  {publisher} {Cambridge University Press},\ \bibinfo
  {address} {Cambridge},\ \bibinfo {year} {2006})\BibitemShut {NoStop}%
\bibitem [{\citenamefont {Bersuker}(1980)}]{PJT0}%
  \BibitemOpen
  \bibfield  {author} {\bibinfo {author} {\bibfnamefont {I.~B.}\ \bibnamefont
  {Bersuker}},\ }\href@noop {} {\bibfield  {journal} {\bibinfo  {journal}
  {Nouv. J. Chim.}\ }\textbf {\bibinfo {volume} {4}},\ \bibinfo {pages} {139}
  (\bibinfo {year} {1980})}\BibitemShut {NoStop}%
\bibitem [{\citenamefont {Bersuker}, \citenamefont {Gorinchoi},\ and\
  \citenamefont {Polinger}(1984)}]{PJT1}%
  \BibitemOpen
  \bibfield  {author} {\bibinfo {author} {\bibfnamefont {I.~B.}\ \bibnamefont
  {Bersuker}}, \bibinfo {author} {\bibfnamefont {N.~N.}\ \bibnamefont
  {Gorinchoi}}, \ and\ \bibinfo {author} {\bibfnamefont {V.~Z.}\ \bibnamefont
  {Polinger}},\ }\href@noop {} {\bibfield  {journal} {\bibinfo  {journal}
  {Theoret. Chim. Acta}\ }\textbf {\bibinfo {volume} {66}},\ \bibinfo {pages}
  {161} (\bibinfo {year} {1984})}\BibitemShut {NoStop}%
\bibitem [{\citenamefont {Bersuker}(1988)}]{PJT2}%
  \BibitemOpen
  \bibfield  {author} {\bibinfo {author} {\bibfnamefont {I.~B.}\ \bibnamefont
  {Bersuker}},\ }\href@noop {} {\bibfield  {journal} {\bibinfo  {journal} {Pure
  Appl. Chem.}\ }\textbf {\bibinfo {volume} {60}},\ \bibinfo {pages} {1167}
  (\bibinfo {year} {1988})}\BibitemShut {NoStop}%
\bibitem [{\citenamefont {Bersuker}, \citenamefont {Gorinchoi},\ and\
  \citenamefont {Polinger}(1992)}]{PJT3}%
  \BibitemOpen
  \bibfield  {author} {\bibinfo {author} {\bibfnamefont {I.~B.}\ \bibnamefont
  {Bersuker}}, \bibinfo {author} {\bibfnamefont {N.~N.}\ \bibnamefont
  {Gorinchoi}}, \ and\ \bibinfo {author} {\bibfnamefont {V.~Z.}\ \bibnamefont
  {Polinger}},\ }\href@noop {} {\bibfield  {journal} {\bibinfo  {journal} {J.
  Mol. Struct.}\ }\textbf {\bibinfo {volume} {270}},\ \bibinfo {pages} {369}
  (\bibinfo {year} {1992})}\BibitemShut {NoStop}%
\bibitem [{\citenamefont {Bersuker}(2018)}]{Bersuker18112018}%
  \BibitemOpen
  \bibfield  {author} {\bibinfo {author} {\bibfnamefont {I.~B.}\ \bibnamefont
  {Bersuker}},\ }\href@noop {} {\bibfield  {journal} {\bibinfo  {journal}
  {Ferroelectrics}\ }\textbf {\bibinfo {volume} {536}},\ \bibinfo {pages} {1}
  (\bibinfo {year} {2018})}\BibitemShut {NoStop}%
\bibitem [{\citenamefont {Bersuker}(2021)}]{Bersuker_review_2021}%
  \BibitemOpen
  \bibfield  {author} {\bibinfo {author} {\bibfnamefont {I.~B.}\ \bibnamefont
  {Bersuker}},\ }\href@noop {} {\bibfield  {journal} {\bibinfo  {journal}
  {Chem. Rev.}\ }\textbf {\bibinfo {volume} {121}},\ \bibinfo {pages} {1463}
  (\bibinfo {year} {2021})}\BibitemShut {NoStop}%
\bibitem [{\citenamefont {Bersuker}(1966)}]{Bersuker_1966}%
  \BibitemOpen
  \bibfield  {author} {\bibinfo {author} {\bibfnamefont {I.~B.}\ \bibnamefont
  {Bersuker}},\ }\href@noop {} {\bibfield  {journal} {\bibinfo  {journal}
  {Phys. Lett.}\ }\textbf {\bibinfo {volume} {20}},\ \bibinfo {pages} {589}
  (\bibinfo {year} {1966})}\BibitemShut {NoStop}%
\bibitem [{\citenamefont {Kole\.{z}y\'{n}ski}\ and\ \citenamefont
  {Tkacz-\'{S}miech}(2005)}]{Kolezynski_2005}%
  \BibitemOpen
  \bibfield  {author} {\bibinfo {author} {\bibfnamefont {A.}~\bibnamefont
  {Kole\.{z}y\'{n}ski}}\ and\ \bibinfo {author} {\bibfnamefont
  {K.}~\bibnamefont {Tkacz-\'{S}miech}},\ }\href@noop {} {\bibfield  {journal}
  {\bibinfo  {journal} {Ferroelectrics}\ }\textbf {\bibinfo {volume} {314}},\
  \bibinfo {pages} {123} (\bibinfo {year} {2005})}\BibitemShut {NoStop}%
\bibitem [{\citenamefont {Liu}\ \emph {et~al.}(2013)\citenamefont {Liu},
  \citenamefont {Zhang}, \citenamefont {Liu}, \citenamefont {Wang},\ and\
  \citenamefont {Liu}}]{Qi-Jun_Liu_2013}%
  \BibitemOpen
  \bibfield  {author} {\bibinfo {author} {\bibfnamefont {Q.-J.}\ \bibnamefont
  {Liu}}, \bibinfo {author} {\bibfnamefont {N.-C.}\ \bibnamefont {Zhang}},
  \bibinfo {author} {\bibfnamefont {F.-S.}\ \bibnamefont {Liu}}, \bibinfo
  {author} {\bibfnamefont {H.-Y.}\ \bibnamefont {Wang}}, \ and\ \bibinfo
  {author} {\bibfnamefont {Z.-T.}\ \bibnamefont {Liu}},\ }\href@noop {}
  {\bibfield  {journal} {\bibinfo  {journal} {Opt. Mater.}\ }\textbf {\bibinfo
  {volume} {35}},\ \bibinfo {pages} {2629} (\bibinfo {year}
  {2013})}\BibitemShut {NoStop}%
\bibitem [{\citenamefont {Polinger}, \citenamefont {Garc\'{\i}a-Fern\'andez},\
  and\ \citenamefont {Bersuker}(2015)}]{bersuker_pbcm457}%
  \BibitemOpen
  \bibfield  {author} {\bibinfo {author} {\bibfnamefont {V.}~\bibnamefont
  {Polinger}}, \bibinfo {author} {\bibfnamefont {P.}~\bibnamefont
  {Garc\'{\i}a-Fern\'andez}}, \ and\ \bibinfo {author} {\bibfnamefont {I.~B.}\
  \bibnamefont {Bersuker}},\ }\href@noop {} {\bibfield  {journal} {\bibinfo
  {journal} {Physica B: Condensed Matter}\ }\textbf {\bibinfo {volume} {457}},\
  \bibinfo {pages} {296} (\bibinfo {year} {2015})}\BibitemShut {NoStop}%
\bibitem [{\citenamefont {Bersuker}(2015)}]{Bersuker_2015}%
  \BibitemOpen
  \bibfield  {author} {\bibinfo {author} {\bibfnamefont {I.~B.}\ \bibnamefont
  {Bersuker}},\ }\href@noop {} {\bibfield  {journal} {\bibinfo  {journal}
  {Appl. Phys. Lett.}\ }\textbf {\bibinfo {volume} {106}},\ \bibinfo {pages}
  {022903} (\bibinfo {year} {2015})}\BibitemShut {NoStop}%
\bibitem [{\citenamefont {Mirhosseini}\ \emph {et~al.}(2010)\citenamefont
  {Mirhosseini}, \citenamefont {Maznichenko}, \citenamefont {Abdelouahed},
  \citenamefont {Ostanin}, \citenamefont {Ernst}, \citenamefont {Mertig},\ and\
  \citenamefont {Henk}}]{Mirhosseini_phys_rev_2010}%
  \BibitemOpen
  \bibfield  {author} {\bibinfo {author} {\bibfnamefont {H.}~\bibnamefont
  {Mirhosseini}}, \bibinfo {author} {\bibfnamefont {I.~V.}\ \bibnamefont
  {Maznichenko}}, \bibinfo {author} {\bibfnamefont {S.}~\bibnamefont
  {Abdelouahed}}, \bibinfo {author} {\bibfnamefont {S.}~\bibnamefont
  {Ostanin}}, \bibinfo {author} {\bibfnamefont {A.}~\bibnamefont {Ernst}},
  \bibinfo {author} {\bibfnamefont {I.}~\bibnamefont {Mertig}}, \ and\ \bibinfo
  {author} {\bibfnamefont {J.}~\bibnamefont {Henk}},\ }\href {\doibase
  10.1103/PhysRevB.81.073406} {\bibfield  {journal} {\bibinfo  {journal} {Phys.
  Rev. B}\ }\textbf {\bibinfo {volume} {81}},\ \bibinfo {pages} {073406}
  (\bibinfo {year} {2010})}\BibitemShut {NoStop}%
\bibitem [{\citenamefont {Zhong}\ \emph {et~al.}(2015)\citenamefont {Zhong},
  \citenamefont {Si}, \citenamefont {Zhang}, \citenamefont {Yin}, \citenamefont
  {Yunoki},\ and\ \citenamefont {Held}}]{Zhong_Adv_Mat}%
  \BibitemOpen
  \bibfield  {author} {\bibinfo {author} {\bibfnamefont {Z.}~\bibnamefont
  {Zhong}}, \bibinfo {author} {\bibfnamefont {L.}~\bibnamefont {Si}}, \bibinfo
  {author} {\bibfnamefont {Q.}~\bibnamefont {Zhang}}, \bibinfo {author}
  {\bibfnamefont {W.-G.}\ \bibnamefont {Yin}}, \bibinfo {author} {\bibfnamefont
  {S.}~\bibnamefont {Yunoki}}, \ and\ \bibinfo {author} {\bibfnamefont
  {K.}~\bibnamefont {Held}},\ }\href@noop {} {\bibfield  {journal} {\bibinfo
  {journal} {Adv. Mater. Interfaces}\ }\textbf {\bibinfo {volume} {2}},\
  \bibinfo {pages} {1400445} (\bibinfo {year} {2015})}\BibitemShut {NoStop}%
\bibitem [{\citenamefont {Ali}\ \emph {et~al.}(2019)\citenamefont {Ali},
  \citenamefont {Khan}, \citenamefont {Ali}, \citenamefont {Khan},\ and\
  \citenamefont {Ahmad}}]{Ali_BFO_BTO}%
  \BibitemOpen
  \bibfield  {author} {\bibinfo {author} {\bibfnamefont {A.}~\bibnamefont
  {Ali}}, \bibinfo {author} {\bibfnamefont {I.}~\bibnamefont {Khan}}, \bibinfo
  {author} {\bibfnamefont {Z.}~\bibnamefont {Ali}}, \bibinfo {author}
  {\bibfnamefont {F.}~\bibnamefont {Khan}}, \ and\ \bibinfo {author}
  {\bibfnamefont {I.}~\bibnamefont {Ahmad}},\ }\href@noop {} {\bibfield
  {journal} {\bibinfo  {journal} {Int. J. Mod. Phys. B}\ }\textbf {\bibinfo
  {volume} {33}},\ \bibinfo {pages} {1950231} (\bibinfo {year}
  {2019})}\BibitemShut {NoStop}%
\bibitem [{\citenamefont {Pinczuk}\ \emph {et~al.}(1967)\citenamefont
  {Pinczuk}, \citenamefont {Taylor}, \citenamefont {Burstein},\ and\
  \citenamefont {Lefkowitz}}]{Pinczuki_1967}%
  \BibitemOpen
  \bibfield  {author} {\bibinfo {author} {\bibfnamefont {A.}~\bibnamefont
  {Pinczuk}}, \bibinfo {author} {\bibfnamefont {W.}~\bibnamefont {Taylor}},
  \bibinfo {author} {\bibfnamefont {E.}~\bibnamefont {Burstein}}, \ and\
  \bibinfo {author} {\bibfnamefont {I.}~\bibnamefont {Lefkowitz}},\ }\href@noop
  {} {\bibfield  {journal} {\bibinfo  {journal} {Solid State Commun.}\ }\textbf
  {\bibinfo {volume} {5}},\ \bibinfo {pages} {429} (\bibinfo {year}
  {1967})}\BibitemShut {NoStop}%
\bibitem [{\citenamefont {Turik}\ and\ \citenamefont
  {Shevchenko}(1979)}]{Turik_Shevchenko_1979}%
  \BibitemOpen
  \bibfield  {author} {\bibinfo {author} {\bibfnamefont {A.~V.}\ \bibnamefont
  {Turik}}\ and\ \bibinfo {author} {\bibfnamefont {N.~B.}\ \bibnamefont
  {Shevchenko}},\ }\href@noop {} {\bibfield  {journal} {\bibinfo  {journal}
  {Phys. Stat. Sol. (B)}\ }\textbf {\bibinfo {volume} {95}},\ \bibinfo {pages}
  {585} (\bibinfo {year} {1979})}\BibitemShut {NoStop}%
\bibitem [{\citenamefont {Cord}\ and\ \citenamefont
  {Courths}(1985)}]{R_Cord_1985}%
  \BibitemOpen
  \bibfield  {author} {\bibinfo {author} {\bibfnamefont {B.}~\bibnamefont
  {Cord}}\ and\ \bibinfo {author} {\bibfnamefont {R.}~\bibnamefont {Courths}},\
  }\href@noop {} {\bibfield  {journal} {\bibinfo  {journal} {Surf. Sci.}\
  }\textbf {\bibinfo {volume} {152-153}},\ \bibinfo {pages} {1141} (\bibinfo
  {year} {1985})}\BibitemShut {NoStop}%
\bibitem [{\citenamefont {Carley}\ \emph {et~al.}(1987)\citenamefont {Carley},
  \citenamefont {Chalker}, \citenamefont {Riviere},\ and\ \citenamefont
  {Roberts}}]{Roberts_1987}%
  \BibitemOpen
  \bibfield  {author} {\bibinfo {author} {\bibfnamefont {A.~F.}\ \bibnamefont
  {Carley}}, \bibinfo {author} {\bibfnamefont {P.~R.}\ \bibnamefont {Chalker}},
  \bibinfo {author} {\bibfnamefont {J.~C.}\ \bibnamefont {Riviere}}, \ and\
  \bibinfo {author} {\bibfnamefont {M.~W.}\ \bibnamefont {Roberts}},\
  }\href@noop {} {\bibfield  {journal} {\bibinfo  {journal} {J. Chem. Soc.
  Faraday Trans. 1}\ }\textbf {\bibinfo {volume} {83}},\ \bibinfo {pages} {351}
  (\bibinfo {year} {1987})}\BibitemShut {NoStop}%
\bibitem [{\citenamefont {Nasser}(2000)}]{Nasser_Expt_2000}%
  \BibitemOpen
  \bibfield  {author} {\bibinfo {author} {\bibfnamefont {S.~A.}\ \bibnamefont
  {Nasser}},\ }\href@noop {} {\bibfield  {journal} {\bibinfo  {journal} {Appl.
  Surf. Sci.}\ }\textbf {\bibinfo {volume} {157}},\ \bibinfo {pages} {14}
  (\bibinfo {year} {2000})}\BibitemShut {NoStop}%
\bibitem [{\citenamefont {Wan}, \citenamefont {Han},\ and\ \citenamefont
  {Zhu}(2008)}]{Wan_2008}%
  \BibitemOpen
  \bibfield  {author} {\bibinfo {author} {\bibfnamefont {F.}~\bibnamefont
  {Wan}}, \bibinfo {author} {\bibfnamefont {J.}~\bibnamefont {Han}}, \ and\
  \bibinfo {author} {\bibfnamefont {Z.}~\bibnamefont {Zhu}},\ }\href@noop {}
  {\bibfield  {journal} {\bibinfo  {journal} {Phys. Lett. A}\ }\textbf
  {\bibinfo {volume} {372}},\ \bibinfo {pages} {2137} (\bibinfo {year}
  {2008})}\BibitemShut {NoStop}%
\bibitem [{\citenamefont {Sah}\ \emph {et~al.}(2025{\natexlab{b}})\citenamefont
  {Sah}, \citenamefont {Mukherjee}, \citenamefont {Ravi},\ and\ \citenamefont
  {Adhikari}}]{Mantu_BTO}%
  \BibitemOpen
  \bibfield  {author} {\bibinfo {author} {\bibfnamefont {M.~K.}\ \bibnamefont
  {Sah}}, \bibinfo {author} {\bibfnamefont {S.}~\bibnamefont {Mukherjee}},
  \bibinfo {author} {\bibfnamefont {S.}~\bibnamefont {Ravi}}, \ and\ \bibinfo
  {author} {\bibfnamefont {S.}~\bibnamefont {Adhikari}},\ }\href
  {https://arxiv.org/abs/2505.02143} {\bibfield  {journal} {\bibinfo  {journal}
  {arXiv}\ }\textbf {\bibinfo {volume} {2505.02143}} (\bibinfo {year}
  {2025}{\natexlab{b}})}\BibitemShut {NoStop}%
\bibitem [{\citenamefont {Kumar~Sah}\ and\ \citenamefont
  {Adhikari}()}]{Mantu_Solid-state_Review}%
  \BibitemOpen
  \bibfield  {author} {\bibinfo {author} {\bibfnamefont {M.}~\bibnamefont
  {Kumar~Sah}}\ and\ \bibinfo {author} {\bibfnamefont {S.}~\bibnamefont
  {Adhikari}},\ }\href@noop {} {\bibfield  {journal} {\bibinfo  {journal} {J.
  Comput. Chem.}\ }\textbf {\bibinfo {volume} {46}},\ \bibinfo {pages}
  {e70239}}\BibitemShut {NoStop}%
\bibitem [{\citenamefont {Puzari}, \citenamefont {Sarkar},\ and\ \citenamefont
  {Adhikari}(2004)}]{SAJCP121}%
  \BibitemOpen
  \bibfield  {author} {\bibinfo {author} {\bibfnamefont {P.}~\bibnamefont
  {Puzari}}, \bibinfo {author} {\bibfnamefont {B.}~\bibnamefont {Sarkar}}, \
  and\ \bibinfo {author} {\bibfnamefont {S.}~\bibnamefont {Adhikari}},\
  }\href@noop {} {\bibfield  {journal} {\bibinfo  {journal} {J. Chem. Phys.}\
  }\textbf {\bibinfo {volume} {121}},\ \bibinfo {pages} {707} (\bibinfo {year}
  {2004})}\BibitemShut {NoStop}%
\bibitem [{\citenamefont {Puzari}, \citenamefont {Sarkar},\ and\ \citenamefont
  {Adhikari}(2005)}]{SAIJQC105}%
  \BibitemOpen
  \bibfield  {author} {\bibinfo {author} {\bibfnamefont {P.}~\bibnamefont
  {Puzari}}, \bibinfo {author} {\bibfnamefont {B.}~\bibnamefont {Sarkar}}, \
  and\ \bibinfo {author} {\bibfnamefont {S.}~\bibnamefont {Adhikari}},\
  }\href@noop {} {\bibfield  {journal} {\bibinfo  {journal} {Int. J. Quant.
  Chem.}\ }\textbf {\bibinfo {volume} {105}},\ \bibinfo {pages} {209} (\bibinfo
  {year} {2005})}\BibitemShut {NoStop}%
\bibitem [{\citenamefont {Bersuker}\ and\ \citenamefont
  {Polinger}(2020)}]{bersuker_cm5}%
  \BibitemOpen
  \bibfield  {author} {\bibinfo {author} {\bibfnamefont {I.~B.}\ \bibnamefont
  {Bersuker}}\ and\ \bibinfo {author} {\bibfnamefont {V.}~\bibnamefont
  {Polinger}},\ }\href@noop {} {\bibfield  {journal} {\bibinfo  {journal}
  {Condensed Matter}\ }\textbf {\bibinfo {volume} {5}} (\bibinfo {year}
  {2020})}\BibitemShut {NoStop}%
\bibitem [{\citenamefont {Baldini}\ \emph {et~al.}(2018)\citenamefont
  {Baldini}, \citenamefont {Kubacka}, \citenamefont {Mallett}, \citenamefont
  {Ma}, \citenamefont {Koohpayeh}, \citenamefont {Zhu}, \citenamefont
  {Bernhard}, \citenamefont {Johnson},\ and\ \citenamefont
  {Carbone}}]{Berhard-TbMnO}%
  \BibitemOpen
  \bibfield  {author} {\bibinfo {author} {\bibfnamefont {E.}~\bibnamefont
  {Baldini}}, \bibinfo {author} {\bibfnamefont {T.}~\bibnamefont {Kubacka}},
  \bibinfo {author} {\bibfnamefont {B.~P.~P.}\ \bibnamefont {Mallett}},
  \bibinfo {author} {\bibfnamefont {C.}~\bibnamefont {Ma}}, \bibinfo {author}
  {\bibfnamefont {S.~M.}\ \bibnamefont {Koohpayeh}}, \bibinfo {author}
  {\bibfnamefont {Y.~M.}\ \bibnamefont {Zhu}}, \bibinfo {author} {\bibfnamefont
  {C.}~\bibnamefont {Bernhard}}, \bibinfo {author} {\bibfnamefont {S.~L.}\
  \bibnamefont {Johnson}}, \ and\ \bibinfo {author} {\bibfnamefont
  {F.}~\bibnamefont {Carbone}},\ }\href@noop {} {\bibfield  {journal} {\bibinfo
   {journal} {Phys. Rev. B}\ }\textbf {\bibinfo {volume} {97}},\ \bibinfo
  {pages} {125149} (\bibinfo {year} {2018})}\BibitemShut {NoStop}%
\bibitem [{\citenamefont {Kotomin}\ \emph {et~al.}(2005)\citenamefont
  {Kotomin}, \citenamefont {Evarestov}, \citenamefont {Mastrikov},\ and\
  \citenamefont {Maier}}]{maier_pccp7}%
  \BibitemOpen
  \bibfield  {author} {\bibinfo {author} {\bibfnamefont {E.~A.}\ \bibnamefont
  {Kotomin}}, \bibinfo {author} {\bibfnamefont {R.~A.}\ \bibnamefont
  {Evarestov}}, \bibinfo {author} {\bibfnamefont {Y.~A.}\ \bibnamefont
  {Mastrikov}}, \ and\ \bibinfo {author} {\bibfnamefont {J.}~\bibnamefont
  {Maier}},\ }\href@noop {} {\bibfield  {journal} {\bibinfo  {journal} {Phys.
  Chem. Chem. Phys.}\ }\textbf {\bibinfo {volume} {7}},\ \bibinfo {pages}
  {2346} (\bibinfo {year} {2005})}\BibitemShut {NoStop}%
\bibitem [{\citenamefont {Evarestov}\ \emph {et~al.}(2005)\citenamefont
  {Evarestov}, \citenamefont {Kotomin}, \citenamefont {Mastrikov},
  \citenamefont {Gryaznov}, \citenamefont {Heifets},\ and\ \citenamefont
  {Maier}}]{maier_prb72}%
  \BibitemOpen
  \bibfield  {author} {\bibinfo {author} {\bibfnamefont {R.~A.}\ \bibnamefont
  {Evarestov}}, \bibinfo {author} {\bibfnamefont {E.~A.}\ \bibnamefont
  {Kotomin}}, \bibinfo {author} {\bibfnamefont {Y.~A.}\ \bibnamefont
  {Mastrikov}}, \bibinfo {author} {\bibfnamefont {D.}~\bibnamefont {Gryaznov}},
  \bibinfo {author} {\bibfnamefont {E.}~\bibnamefont {Heifets}}, \ and\
  \bibinfo {author} {\bibfnamefont {J.}~\bibnamefont {Maier}},\ }\href@noop {}
  {\bibfield  {journal} {\bibinfo  {journal} {Phys. Rev. B}\ }\textbf {\bibinfo
  {volume} {72}},\ \bibinfo {pages} {214411} (\bibinfo {year}
  {2005})}\BibitemShut {NoStop}%
\bibitem [{\citenamefont {Kwei}\ \emph {et~al.}(1993)\citenamefont {Kwei},
  \citenamefont {Lawson}, \citenamefont {Billinge},\ and\ \citenamefont
  {Cheong}}]{Kwei_Bond_order_exp}%
  \BibitemOpen
  \bibfield  {author} {\bibinfo {author} {\bibfnamefont {G.~H.}\ \bibnamefont
  {Kwei}}, \bibinfo {author} {\bibfnamefont {A.~C.}\ \bibnamefont {Lawson}},
  \bibinfo {author} {\bibfnamefont {S.~J.~L.}\ \bibnamefont {Billinge}}, \ and\
  \bibinfo {author} {\bibfnamefont {S.~W.}\ \bibnamefont {Cheong}},\
  }\href@noop {} {\bibfield  {journal} {\bibinfo  {journal} {J. Phys. Chem.}\
  }\textbf {\bibinfo {volume} {97}},\ \bibinfo {pages} {2368} (\bibinfo {year}
  {1993})}\BibitemShut {NoStop}%
\bibitem [{\citenamefont {Moin}\ \emph {et~al.}(2024)\citenamefont {Moin},
  \citenamefont {Anwar}, \citenamefont {Babar}, \citenamefont {Thumu},\ and\
  \citenamefont {Ali}}]{Heliyon_bond_order_theor}%
  \BibitemOpen
  \bibfield  {author} {\bibinfo {author} {\bibfnamefont {M.}~\bibnamefont
  {Moin}}, \bibinfo {author} {\bibfnamefont {A.~W.}\ \bibnamefont {Anwar}},
  \bibinfo {author} {\bibfnamefont {M.}~\bibnamefont {Babar}}, \bibinfo
  {author} {\bibfnamefont {U.}~\bibnamefont {Thumu}}, \ and\ \bibinfo {author}
  {\bibfnamefont {A.}~\bibnamefont {Ali}},\ }\href@noop {} {\bibfield
  {journal} {\bibinfo  {journal} {Heliyon}\ }\textbf {\bibinfo {volume} {10}},\
  \bibinfo {pages} {e24607} (\bibinfo {year} {2024})}\BibitemShut {NoStop}%
\bibitem [{\citenamefont {Mahmoud}\ \emph {et~al.}(2014)\citenamefont
  {Mahmoud}, \citenamefont {Erba}, \citenamefont {El-Kelany}, \citenamefont
  {R\'erat},\ and\ \citenamefont {Orlando}}]{A_Mahmoud}%
  \BibitemOpen
  \bibfield  {author} {\bibinfo {author} {\bibfnamefont {A.}~\bibnamefont
  {Mahmoud}}, \bibinfo {author} {\bibfnamefont {A.}~\bibnamefont {Erba}},
  \bibinfo {author} {\bibfnamefont {K.~E.}\ \bibnamefont {El-Kelany}}, \bibinfo
  {author} {\bibfnamefont {M.}~\bibnamefont {R\'erat}}, \ and\ \bibinfo
  {author} {\bibfnamefont {R.}~\bibnamefont {Orlando}},\ }\href@noop {}
  {\bibfield  {journal} {\bibinfo  {journal} {Phys. Rev. B}\ }\textbf {\bibinfo
  {volume} {89}},\ \bibinfo {pages} {045103} (\bibinfo {year}
  {2014})}\BibitemShut {NoStop}%
\bibitem [{\citenamefont {Pertosa}, \citenamefont {Hollinger},\ and\
  \citenamefont {Michel-Calendini}(1978)}]{Pertosa_Theory_Spectra}%
  \BibitemOpen
  \bibfield  {author} {\bibinfo {author} {\bibfnamefont {P.}~\bibnamefont
  {Pertosa}}, \bibinfo {author} {\bibfnamefont {G.}~\bibnamefont {Hollinger}},
  \ and\ \bibinfo {author} {\bibfnamefont {F.~M.}\ \bibnamefont
  {Michel-Calendini}},\ }\href {\doibase 10.1103/PhysRevB.18.5177} {\bibfield
  {journal} {\bibinfo  {journal} {Phys. Rev. B}\ }\textbf {\bibinfo {volume}
  {18}},\ \bibinfo {pages} {5177} (\bibinfo {year} {1978})}\BibitemShut
  {NoStop}%
\bibitem [{\citenamefont {Merz}(1949)}]{Walter_spont_1949}%
  \BibitemOpen
  \bibfield  {author} {\bibinfo {author} {\bibfnamefont {W.~J.}\ \bibnamefont
  {Merz}},\ }\href {\doibase 10.1103/PhysRev.76.1221} {\bibfield  {journal}
  {\bibinfo  {journal} {Phys. Rev.}\ }\textbf {\bibinfo {volume} {76}},\
  \bibinfo {pages} {1221} (\bibinfo {year} {1949})}\BibitemShut {NoStop}%
\bibitem [{\citenamefont {Polinger}(2013)}]{Polinger_2013}%
  \BibitemOpen
  \bibfield  {author} {\bibinfo {author} {\bibfnamefont {V.}~\bibnamefont
  {Polinger}},\ }\href@noop {} {\bibfield  {journal} {\bibinfo  {journal} {J.
  Phys. Conf. Ser.}\ }\textbf {\bibinfo {volume} {428}},\ \bibinfo {pages}
  {012026} (\bibinfo {year} {2013})}\BibitemShut {NoStop}%
\bibitem [{\citenamefont {Busnengo}, \citenamefont {Salin},\ and\ \citenamefont
  {Dong}(2000)}]{CRP}%
  \BibitemOpen
  \bibfield  {author} {\bibinfo {author} {\bibfnamefont {H.~F.}\ \bibnamefont
  {Busnengo}}, \bibinfo {author} {\bibfnamefont {A.}~\bibnamefont {Salin}}, \
  and\ \bibinfo {author} {\bibfnamefont {W.}~\bibnamefont {Dong}},\ }\href@noop
  {} {\bibfield  {journal} {\bibinfo  {journal} {J. Chem. Phys.}\ }\textbf
  {\bibinfo {volume} {112}},\ \bibinfo {pages} {7641} (\bibinfo {year}
  {2000})}\BibitemShut {NoStop}%
\bibitem [{\citenamefont {Nattino}\ \emph
  {et~al.}(2012{\natexlab{b}})\citenamefont {Nattino}, \citenamefont
  {D\'{i}az}, \citenamefont {Jackson},\ and\ \citenamefont {Kroes}}]{CRP3}%
  \BibitemOpen
  \bibfield  {author} {\bibinfo {author} {\bibfnamefont {F.}~\bibnamefont
  {Nattino}}, \bibinfo {author} {\bibfnamefont {C.}~\bibnamefont {D\'{i}az}},
  \bibinfo {author} {\bibfnamefont {B.}~\bibnamefont {Jackson}}, \ and\
  \bibinfo {author} {\bibfnamefont {G.-J.}\ \bibnamefont {Kroes}},\ }\href@noop
  {} {\bibfield  {journal} {\bibinfo  {journal} {Phys. Rev. Lett.}\ }\textbf
  {\bibinfo {volume} {108}},\ \bibinfo {pages} {236104} (\bibinfo {year}
  {2012}{\natexlab{b}})}\BibitemShut {NoStop}%
\bibitem [{\citenamefont {Kroes}(2015)}]{CRP4}%
  \BibitemOpen
  \bibfield  {author} {\bibinfo {author} {\bibfnamefont {G.-J.}\ \bibnamefont
  {Kroes}},\ }\href@noop {} {\bibfield  {journal} {\bibinfo  {journal} {J.
  Phys. Chem. Lett.}\ }\textbf {\bibinfo {volume} {6}},\ \bibinfo {pages}
  {4106} (\bibinfo {year} {2015})}\BibitemShut {NoStop}%
\bibitem [{\citenamefont {Zhou}\ \emph {et~al.}(2021)\citenamefont {Zhou},
  \citenamefont {Zhang}, \citenamefont {Yin}, \citenamefont {Hu},\ and\
  \citenamefont {Jiang}}]{NN-PES-perspective-2021}%
  \BibitemOpen
  \bibfield  {author} {\bibinfo {author} {\bibfnamefont {X.}~\bibnamefont
  {Zhou}}, \bibinfo {author} {\bibfnamefont {Y.}~\bibnamefont {Zhang}},
  \bibinfo {author} {\bibfnamefont {R.}~\bibnamefont {Yin}}, \bibinfo {author}
  {\bibfnamefont {C.}~\bibnamefont {Hu}}, \ and\ \bibinfo {author}
  {\bibfnamefont {B.}~\bibnamefont {Jiang}},\ }\href@noop {} {\bibfield
  {journal} {\bibinfo  {journal} {Chin. J. Chem.}\ }\textbf {\bibinfo {volume}
  {39}},\ \bibinfo {pages} {2917} (\bibinfo {year} {2021})}\BibitemShut
  {NoStop}%
\bibitem [{\citenamefont {Liu}, \citenamefont {Fu},\ and\ \citenamefont
  {Zhang}(2014)}]{Zhang-QD-H2-Cu111}%
  \BibitemOpen
  \bibfield  {author} {\bibinfo {author} {\bibfnamefont {T.}~\bibnamefont
  {Liu}}, \bibinfo {author} {\bibfnamefont {B.}~\bibnamefont {Fu}}, \ and\
  \bibinfo {author} {\bibfnamefont {D.~H.}\ \bibnamefont {Zhang}},\ }\href@noop
  {} {\bibfield  {journal} {\bibinfo  {journal} {J. Chem. Phys.}\ }\textbf
  {\bibinfo {volume} {141}},\ \bibinfo {pages} {194302} (\bibinfo {year}
  {2014})}\BibitemShut {NoStop}%
\bibitem [{\citenamefont {D\'{i}az}\ \emph
  {et~al.}(2010{\natexlab{b}})\citenamefont {D\'{i}az}, \citenamefont {Olsen},
  \citenamefont {Busnengo},\ and\ \citenamefont {Kroes}}]{H2-Cu111-CRP}%
  \BibitemOpen
  \bibfield  {author} {\bibinfo {author} {\bibfnamefont {C.}~\bibnamefont
  {D\'{i}az}}, \bibinfo {author} {\bibfnamefont {R.~A.}\ \bibnamefont {Olsen}},
  \bibinfo {author} {\bibfnamefont {H.~F.}\ \bibnamefont {Busnengo}}, \ and\
  \bibinfo {author} {\bibfnamefont {G.-J.}\ \bibnamefont {Kroes}},\ }\href@noop
  {} {\bibfield  {journal} {\bibinfo  {journal} {J. Phys. Chem. C.}\ }\textbf
  {\bibinfo {volume} {114}},\ \bibinfo {pages} {11192} (\bibinfo {year}
  {2010}{\natexlab{b}})}\BibitemShut {NoStop}%
\bibitem [{\citenamefont {Jiang}\ and\ \citenamefont
  {Guo}(2014)}]{6D-QD-HD-NNP-B-Jiang-H-Guo}%
  \BibitemOpen
  \bibfield  {author} {\bibinfo {author} {\bibfnamefont {B.}~\bibnamefont
  {Jiang}}\ and\ \bibinfo {author} {\bibfnamefont {H.}~\bibnamefont {Guo}},\
  }\href@noop {} {\bibfield  {journal} {\bibinfo  {journal} {Phys. Chem. Chem.
  Phys.}\ }\textbf {\bibinfo {volume} {16}},\ \bibinfo {pages} {24704}
  (\bibinfo {year} {2014})}\BibitemShut {NoStop}%
\bibitem [{\citenamefont {Wijzenbroek}\ \emph {et~al.}(2015)\citenamefont
  {Wijzenbroek}, \citenamefont {Klein}, \citenamefont {Smits}, \citenamefont
  {Somers},\ and\ \citenamefont
  {Kroes}}]{GGA-exchange+correlation-8-H2-Metal-2015}%
  \BibitemOpen
  \bibfield  {author} {\bibinfo {author} {\bibfnamefont {M.}~\bibnamefont
  {Wijzenbroek}}, \bibinfo {author} {\bibfnamefont {D.~M.}\ \bibnamefont
  {Klein}}, \bibinfo {author} {\bibfnamefont {B.}~\bibnamefont {Smits}},
  \bibinfo {author} {\bibfnamefont {M.~F.}\ \bibnamefont {Somers}}, \ and\
  \bibinfo {author} {\bibfnamefont {G.-J.}\ \bibnamefont {Kroes}},\ }\href@noop
  {} {\bibfield  {journal} {\bibinfo  {journal} {J. Phys. Chem. A}\ }\textbf
  {\bibinfo {volume} {119}},\ \bibinfo {pages} {12146} (\bibinfo {year}
  {2015})}\BibitemShut {NoStop}%
\bibitem [{\citenamefont {Chen}\ \emph {et~al.}(2018)\citenamefont {Chen},
  \citenamefont {Zhou}, \citenamefont {Zhang},\ and\ \citenamefont
  {Jiang}}]{GGA-exchange+correlation-2-CH3OH-Cu-2019}%
  \BibitemOpen
  \bibfield  {author} {\bibinfo {author} {\bibfnamefont {J.~L.}\ \bibnamefont
  {Chen}}, \bibinfo {author} {\bibfnamefont {X.~Y.}\ \bibnamefont {Zhou}},
  \bibinfo {author} {\bibfnamefont {Y.~L.}\ \bibnamefont {Zhang}}, \ and\
  \bibinfo {author} {\bibfnamefont {B.}~\bibnamefont {Jiang}},\ }\href@noop {}
  {\bibfield  {journal} {\bibinfo  {journal} {Nat Commun}\ }\textbf {\bibinfo
  {volume} {9}},\ \bibinfo {pages} {4039} (\bibinfo {year} {2018})}\BibitemShut
  {NoStop}%
\bibitem [{\citenamefont {Smeets}, \citenamefont {Voss},\ and\ \citenamefont
  {Kroes}(2019)}]{mGGA-1-H2-Cu-Ag-2019}%
  \BibitemOpen
  \bibfield  {author} {\bibinfo {author} {\bibfnamefont {E.~W.~F.}\
  \bibnamefont {Smeets}}, \bibinfo {author} {\bibfnamefont {J.}~\bibnamefont
  {Voss}}, \ and\ \bibinfo {author} {\bibfnamefont {G.-J.}\ \bibnamefont
  {Kroes}},\ }\href@noop {} {\bibfield  {journal} {\bibinfo  {journal} {J.
  Phys. Chem. A}\ }\textbf {\bibinfo {volume} {123}},\ \bibinfo {pages} {5395}
  (\bibinfo {year} {2019})}\BibitemShut {NoStop}%
\bibitem [{\citenamefont {Gonzalez-Lafont}, \citenamefont {Truong},\ and\
  \citenamefont {Truhlar}(1991)}]{SRP-origin}%
  \BibitemOpen
  \bibfield  {author} {\bibinfo {author} {\bibfnamefont {A.}~\bibnamefont
  {Gonzalez-Lafont}}, \bibinfo {author} {\bibfnamefont {T.~N.}\ \bibnamefont
  {Truong}}, \ and\ \bibinfo {author} {\bibfnamefont {D.~G.}\ \bibnamefont
  {Truhlar}},\ }\href@noop {} {\bibfield  {journal} {\bibinfo  {journal} {J.
  Phys. Chem.}\ }\textbf {\bibinfo {volume} {95}},\ \bibinfo {pages} {4618}
  (\bibinfo {year} {1991})}\BibitemShut {NoStop}%
\bibitem [{\citenamefont {Chuang}\ \emph {et~al.}(1999)\citenamefont {Chuang},
  \citenamefont {Radhakrishnan}, \citenamefont {Fast}, \citenamefont {Cramer},\
  and\ \citenamefont {Truhlar}}]{SRP-DFT-Truhlar}%
  \BibitemOpen
  \bibfield  {author} {\bibinfo {author} {\bibfnamefont {Y.-Y.}\ \bibnamefont
  {Chuang}}, \bibinfo {author} {\bibfnamefont {M.~L.}\ \bibnamefont
  {Radhakrishnan}}, \bibinfo {author} {\bibfnamefont {P.~L.}\ \bibnamefont
  {Fast}}, \bibinfo {author} {\bibfnamefont {C.~J.}\ \bibnamefont {Cramer}}, \
  and\ \bibinfo {author} {\bibfnamefont {D.~G.}\ \bibnamefont {Truhlar}},\
  }\href@noop {} {\bibfield  {journal} {\bibinfo  {journal} {J. Phys. Chem. A}\
  }\textbf {\bibinfo {volume} {103}},\ \bibinfo {pages} {4893} (\bibinfo {year}
  {1999})}\BibitemShut {NoStop}%
\bibitem [{\citenamefont {Mallikarjun~Sharada}\ \emph
  {et~al.}(2017)\citenamefont {Mallikarjun~Sharada}, \citenamefont {Bligaard},
  \citenamefont {Luntz}, \citenamefont {Kroes},\ and\ \citenamefont
  {Nørskov}}]{SBH10-SRP-DF-database}%
  \BibitemOpen
  \bibfield  {author} {\bibinfo {author} {\bibfnamefont {S.}~\bibnamefont
  {Mallikarjun~Sharada}}, \bibinfo {author} {\bibfnamefont {T.}~\bibnamefont
  {Bligaard}}, \bibinfo {author} {\bibfnamefont {A.~C.}\ \bibnamefont {Luntz}},
  \bibinfo {author} {\bibfnamefont {G.-J.}\ \bibnamefont {Kroes}}, \ and\
  \bibinfo {author} {\bibfnamefont {J.~K.}\ \bibnamefont {Nørskov}},\
  }\href@noop {} {\bibfield  {journal} {\bibinfo  {journal} {J. Phys. Chem. C}\
  }\textbf {\bibinfo {volume} {121}},\ \bibinfo {pages} {19807} (\bibinfo
  {year} {2017})}\BibitemShut {NoStop}%
\bibitem [{\citenamefont {Gerrits}\ \emph {et~al.}(2020)\citenamefont
  {Gerrits}, \citenamefont {Smeets}, \citenamefont {Vuckovic}, \citenamefont
  {Powell}, \citenamefont {Doblhoff-Dier},\ and\ \citenamefont
  {Kroes}}]{SRP-DF-38-limitation-2020}%
  \BibitemOpen
  \bibfield  {author} {\bibinfo {author} {\bibfnamefont {N.}~\bibnamefont
  {Gerrits}}, \bibinfo {author} {\bibfnamefont {E.~W.~F.}\ \bibnamefont
  {Smeets}}, \bibinfo {author} {\bibfnamefont {S.}~\bibnamefont {Vuckovic}},
  \bibinfo {author} {\bibfnamefont {A.~D.}\ \bibnamefont {Powell}}, \bibinfo
  {author} {\bibfnamefont {K.}~\bibnamefont {Doblhoff-Dier}}, \ and\ \bibinfo
  {author} {\bibfnamefont {G.-J.}\ \bibnamefont {Kroes}},\ }\href@noop {}
  {\bibfield  {journal} {\bibinfo  {journal} {J. Phys. Chem. Lett.}\ }\textbf
  {\bibinfo {volume} {11}},\ \bibinfo {pages} {10552} (\bibinfo {year}
  {2020})}\BibitemShut {NoStop}%
\bibitem [{\citenamefont {Yin}\ \emph {et~al.}(2018)\citenamefont {Yin},
  \citenamefont {Zhang}, \citenamefont {Libisch}, \citenamefont {Carter},
  \citenamefont {Guo},\ and\ \citenamefont {Jiang}}]{ECW-3-O2-Al-2018}%
  \BibitemOpen
  \bibfield  {author} {\bibinfo {author} {\bibfnamefont {R.}~\bibnamefont
  {Yin}}, \bibinfo {author} {\bibfnamefont {Y.}~\bibnamefont {Zhang}}, \bibinfo
  {author} {\bibfnamefont {F.}~\bibnamefont {Libisch}}, \bibinfo {author}
  {\bibfnamefont {E.~A.}\ \bibnamefont {Carter}}, \bibinfo {author}
  {\bibfnamefont {H.}~\bibnamefont {Guo}}, \ and\ \bibinfo {author}
  {\bibfnamefont {B.}~\bibnamefont {Jiang}},\ }\href@noop {} {\bibfield
  {journal} {\bibinfo  {journal} {J. Phys. Chem. Lett.}\ }\textbf {\bibinfo
  {volume} {9}},\ \bibinfo {pages} {3271} (\bibinfo {year} {2018})}\BibitemShut
  {NoStop}%
\bibitem [{\citenamefont {Doblhoff-Dier}\ \emph {et~al.}(2017)\citenamefont
  {Doblhoff-Dier}, \citenamefont {Meyer}, \citenamefont {Hoggan},\ and\
  \citenamefont {Kroes}}]{DMC-1-H2-Cu-2017}%
  \BibitemOpen
  \bibfield  {author} {\bibinfo {author} {\bibfnamefont {K.}~\bibnamefont
  {Doblhoff-Dier}}, \bibinfo {author} {\bibfnamefont {J.}~\bibnamefont
  {Meyer}}, \bibinfo {author} {\bibfnamefont {P.~E.}\ \bibnamefont {Hoggan}}, \
  and\ \bibinfo {author} {\bibfnamefont {G.-J.}\ \bibnamefont {Kroes}},\
  }\href@noop {} {\bibfield  {journal} {\bibinfo  {journal} {J. Chem. Theory
  Comput.}\ }\textbf {\bibinfo {volume} {13}},\ \bibinfo {pages} {3208}
  (\bibinfo {year} {2017})}\BibitemShut {NoStop}%
\bibitem [{\citenamefont {Powell}, \citenamefont {Kroes},\ and\ \citenamefont
  {Doblhoff-Dier}(2020)}]{DMC-3-2020}%
  \BibitemOpen
  \bibfield  {author} {\bibinfo {author} {\bibfnamefont {A.~D.}\ \bibnamefont
  {Powell}}, \bibinfo {author} {\bibfnamefont {G.-J.}\ \bibnamefont {Kroes}}, \
  and\ \bibinfo {author} {\bibfnamefont {K.}~\bibnamefont {Doblhoff-Dier}},\
  }\href@noop {} {\bibfield  {journal} {\bibinfo  {journal} {J. Chem. Phys.}\
  }\textbf {\bibinfo {volume} {153}},\ \bibinfo {pages} {224701} (\bibinfo
  {year} {2020})}\BibitemShut {NoStop}%
\bibitem [{\citenamefont {Bonfanti}\ \emph {et~al.}(2013)\citenamefont
  {Bonfanti}, \citenamefont {Somers}, \citenamefont {Díaz}, \citenamefont
  {Busnengo},\ and\ \citenamefont {Kroes}}]{Kroes-7D-QD}%
  \BibitemOpen
  \bibfield  {author} {\bibinfo {author} {\bibfnamefont {M.}~\bibnamefont
  {Bonfanti}}, \bibinfo {author} {\bibfnamefont {M.~F.}\ \bibnamefont
  {Somers}}, \bibinfo {author} {\bibfnamefont {C.}~\bibnamefont {Díaz}},
  \bibinfo {author} {\bibfnamefont {H.~F.}\ \bibnamefont {Busnengo}}, \ and\
  \bibinfo {author} {\bibfnamefont {G.-J.}\ \bibnamefont {Kroes}},\ }\href@noop
  {} {\bibfield  {journal} {\bibinfo  {journal} {Z. Phys. Chem.}\ }\textbf
  {\bibinfo {volume} {227}},\ \bibinfo {pages} {1397} (\bibinfo {year}
  {2013})}\BibitemShut {NoStop}%
\bibitem [{\citenamefont {Behler}(2015)}]{HD-NN-tutorial-review}%
  \BibitemOpen
  \bibfield  {author} {\bibinfo {author} {\bibfnamefont {J.}~\bibnamefont
  {Behler}},\ }\href@noop {} {\bibfield  {journal} {\bibinfo  {journal} {Int.
  J. Quantum Chem.}\ }\textbf {\bibinfo {volume} {115}},\ \bibinfo {pages}
  {1032} (\bibinfo {year} {2015})}\BibitemShut {NoStop}%
\bibitem [{\citenamefont {Behler}\ and\ \citenamefont
  {Parrinello}(2007)}]{AtNNmethod}%
  \BibitemOpen
  \bibfield  {author} {\bibinfo {author} {\bibfnamefont {J.}~\bibnamefont
  {Behler}}\ and\ \bibinfo {author} {\bibfnamefont {M.}~\bibnamefont
  {Parrinello}},\ }\href@noop {} {\bibfield  {journal} {\bibinfo  {journal}
  {Phys. Rev. Lett.}\ }\textbf {\bibinfo {volume} {98}},\ \bibinfo {pages}
  {146401} (\bibinfo {year} {2007})}\BibitemShut {NoStop}%
\bibitem [{\citenamefont {Liu}, \citenamefont {Fu},\ and\ \citenamefont
  {Zhang}(2019)}]{6D-QD-HD-NNP-D-Zhang-2}%
  \BibitemOpen
  \bibfield  {author} {\bibinfo {author} {\bibfnamefont {T.}~\bibnamefont
  {Liu}}, \bibinfo {author} {\bibfnamefont {B.}~\bibnamefont {Fu}}, \ and\
  \bibinfo {author} {\bibfnamefont {D.~H.}\ \bibnamefont {Zhang}},\ }\href@noop
  {} {\bibfield  {journal} {\bibinfo  {journal} {J. Chem. Phys.}\ }\textbf
  {\bibinfo {volume} {151}},\ \bibinfo {pages} {144707} (\bibinfo {year}
  {2019})}\BibitemShut {NoStop}%
\bibitem [{\citenamefont {Tully}(1980)}]{GLO-1}%
  \BibitemOpen
  \bibfield  {author} {\bibinfo {author} {\bibfnamefont {J.~C.}\ \bibnamefont
  {Tully}},\ }\href@noop {} {\bibfield  {journal} {\bibinfo  {journal} {J.
  Chem. Phys.}\ }\textbf {\bibinfo {volume} {73}},\ \bibinfo {pages} {1975}
  (\bibinfo {year} {1980})}\BibitemShut {NoStop}%
\bibitem [{\citenamefont {Adelman}\ and\ \citenamefont {Doll}(1976)}]{GLO-2}%
  \BibitemOpen
  \bibfield  {author} {\bibinfo {author} {\bibfnamefont {S.~A.}\ \bibnamefont
  {Adelman}}\ and\ \bibinfo {author} {\bibfnamefont {J.~D.}\ \bibnamefont
  {Doll}},\ }\href@noop {} {\bibfield  {journal} {\bibinfo  {journal} {J. Chem.
  Phys.}\ }\textbf {\bibinfo {volume} {64}},\ \bibinfo {pages} {2375} (\bibinfo
  {year} {1976})}\BibitemShut {NoStop}%
\bibitem [{\citenamefont {Busnengo}\ \emph {et~al.}(2005)\citenamefont
  {Busnengo}, \citenamefont {Di~C\'esare}, \citenamefont {Dong},\ and\
  \citenamefont {Salin}}]{CRP-H2-Pd-thermally-distorted-2005}%
  \BibitemOpen
  \bibfield  {author} {\bibinfo {author} {\bibfnamefont {H.~F.}\ \bibnamefont
  {Busnengo}}, \bibinfo {author} {\bibfnamefont {M.~A.}\ \bibnamefont
  {Di~C\'esare}}, \bibinfo {author} {\bibfnamefont {W.}~\bibnamefont {Dong}}, \
  and\ \bibinfo {author} {\bibfnamefont {A.}~\bibnamefont {Salin}},\
  }\href@noop {} {\bibfield  {journal} {\bibinfo  {journal} {Phys. Rev. B}\
  }\textbf {\bibinfo {volume} {72}},\ \bibinfo {pages} {125411} (\bibinfo
  {year} {2005})}\BibitemShut {NoStop}%
\bibitem [{\citenamefont {Zhou}\ and\ \citenamefont
  {Jiang}(2019)}]{SRP-DF-9-gas-surface-GLO-new-2019}%
  \BibitemOpen
  \bibfield  {author} {\bibinfo {author} {\bibfnamefont {X.}~\bibnamefont
  {Zhou}}\ and\ \bibinfo {author} {\bibfnamefont {B.}~\bibnamefont {Jiang}},\
  }\href@noop {} {\bibfield  {journal} {\bibinfo  {journal} {J. Chem. Phys.}\
  }\textbf {\bibinfo {volume} {150}},\ \bibinfo {pages} {024704} (\bibinfo
  {year} {2019})}\BibitemShut {NoStop}%
\bibitem [{\citenamefont
  {Jackson}(2020)}]{CH4-trapping-1-Ni-Ir-Pt-Jackson-2020}%
  \BibitemOpen
  \bibfield  {author} {\bibinfo {author} {\bibfnamefont {B.}~\bibnamefont
  {Jackson}},\ }\href@noop {} {\bibfield  {journal} {\bibinfo  {journal} {J.
  Chem. Phys.}\ }\textbf {\bibinfo {volume} {153}},\ \bibinfo {pages} {034704}
  (\bibinfo {year} {2020})}\BibitemShut {NoStop}%
\bibitem [{\citenamefont {Jackson}(2021)}]{CH4-trapping-2-Ir-Jackson-2021}%
  \BibitemOpen
  \bibfield  {author} {\bibinfo {author} {\bibfnamefont {B.}~\bibnamefont
  {Jackson}},\ }\href@noop {} {\bibfield  {journal} {\bibinfo  {journal} {J.
  Chem. Phys.}\ }\textbf {\bibinfo {volume} {155}},\ \bibinfo {pages} {044705}
  (\bibinfo {year} {2021})}\BibitemShut {NoStop}%
\bibitem [{\citenamefont {Smits}\ and\ \citenamefont
  {Somers}(2022)}]{Bauke_jcp_2022}%
  \BibitemOpen
  \bibfield  {author} {\bibinfo {author} {\bibfnamefont {B.}~\bibnamefont
  {Smits}}\ and\ \bibinfo {author} {\bibfnamefont {M.~F.}\ \bibnamefont
  {Somers}},\ }\href@noop {} {\bibfield  {journal} {\bibinfo  {journal} {J.
  Chem. Phys.}\ }\textbf {\bibinfo {volume} {157}},\ \bibinfo {pages} {134704}
  (\bibinfo {year} {2022})}\BibitemShut {NoStop}%
\bibitem [{\citenamefont {Smits}\ and\ \citenamefont
  {Somers}(2023)}]{Bauke_jcp_2023}%
  \BibitemOpen
  \bibfield  {author} {\bibinfo {author} {\bibfnamefont {B.}~\bibnamefont
  {Smits}}\ and\ \bibinfo {author} {\bibfnamefont {M.~F.}\ \bibnamefont
  {Somers}},\ }\href@noop {} {\bibfield  {journal} {\bibinfo  {journal} {J.
  Chem. Phys.}\ }\textbf {\bibinfo {volume} {158}},\ \bibinfo {pages} {014704}
  (\bibinfo {year} {2023})}\BibitemShut {NoStop}%
\bibitem [{\citenamefont {Kroes}\ and\ \citenamefont
  {Somers}(2005)}]{SPO-DVR-code-2}%
  \BibitemOpen
  \bibfield  {author} {\bibinfo {author} {\bibfnamefont {G.~J.}\ \bibnamefont
  {Kroes}}\ and\ \bibinfo {author} {\bibfnamefont {M.~F.}\ \bibnamefont
  {Somers}},\ }\href@noop {} {\bibfield  {journal} {\bibinfo  {journal} {J.
  Theor. Comput. Chem.}\ }\textbf {\bibinfo {volume} {04}},\ \bibinfo {pages}
  {493} (\bibinfo {year} {2005})}\BibitemShut {NoStop}%
\bibitem [{\citenamefont {Kindt}\ \emph {et~al.}(1998)\citenamefont {Kindt},
  \citenamefont {Tully}, \citenamefont {Head-Gordon},\ and\ \citenamefont
  {Gomez}}]{JCT}%
  \BibitemOpen
  \bibfield  {author} {\bibinfo {author} {\bibfnamefont {J.~T.}\ \bibnamefont
  {Kindt}}, \bibinfo {author} {\bibfnamefont {J.~C.}\ \bibnamefont {Tully}},
  \bibinfo {author} {\bibfnamefont {M.}~\bibnamefont {Head-Gordon}}, \ and\
  \bibinfo {author} {\bibfnamefont {M.~A.}\ \bibnamefont {Gomez}},\ }\href@noop
  {} {\bibfield  {journal} {\bibinfo  {journal} {J. Chem. Phys.}\ }\textbf
  {\bibinfo {volume} {109}},\ \bibinfo {pages} {3629} (\bibinfo {year}
  {1998})}\BibitemShut {NoStop}%
\bibitem [{\citenamefont {Tully}(2000)}]{JCTullyrev}%
  \BibitemOpen
  \bibfield  {author} {\bibinfo {author} {\bibfnamefont {J.~C.}\ \bibnamefont
  {Tully}},\ }\href@noop {} {\bibfield  {journal} {\bibinfo  {journal} {Annu.
  Rev. Phys. Chem.}\ }\textbf {\bibinfo {volume} {51}},\ \bibinfo {pages} {153}
  (\bibinfo {year} {2000})}\BibitemShut {NoStop}%
\bibitem [{\citenamefont {Saalfrank}(2006{\natexlab{b}})}]{Saalfrankrev}%
  \BibitemOpen
  \bibfield  {author} {\bibinfo {author} {\bibfnamefont {P.}~\bibnamefont
  {Saalfrank}},\ }\href@noop {} {\bibfield  {journal} {\bibinfo  {journal}
  {Chem. Rev.}\ }\textbf {\bibinfo {volume} {106}},\ \bibinfo {pages} {4116}
  (\bibinfo {year} {2006}{\natexlab{b}})}\BibitemShut {NoStop}%
\bibitem [{\citenamefont {Wodtke}, \citenamefont {Matsiev},\ and\ \citenamefont
  {Auerbach}(2008)}]{Wodtkeelhp}%
  \BibitemOpen
  \bibfield  {author} {\bibinfo {author} {\bibfnamefont {A.~M.}\ \bibnamefont
  {Wodtke}}, \bibinfo {author} {\bibfnamefont {D.}~\bibnamefont {Matsiev}}, \
  and\ \bibinfo {author} {\bibfnamefont {D.~J.}\ \bibnamefont {Auerbach}},\
  }\href@noop {} {\bibfield  {journal} {\bibinfo  {journal} {Prog. Surf. Sci.}\
  }\textbf {\bibinfo {volume} {83}},\ \bibinfo {pages} {167} (\bibinfo {year}
  {2008})}\BibitemShut {NoStop}%
\bibitem [{\citenamefont {Roy}, \citenamefont {Shenvi},\ and\ \citenamefont
  {Tully}(2009)}]{JCTully-elhp2}%
  \BibitemOpen
  \bibfield  {author} {\bibinfo {author} {\bibfnamefont {S.}~\bibnamefont
  {Roy}}, \bibinfo {author} {\bibfnamefont {N.~A.}\ \bibnamefont {Shenvi}}, \
  and\ \bibinfo {author} {\bibfnamefont {J.~C.}\ \bibnamefont {Tully}},\
  }\href@noop {} {\bibfield  {journal} {\bibinfo  {journal} {J. Chem. Phys.}\
  }\textbf {\bibinfo {volume} {130}},\ \bibinfo {pages} {174716} (\bibinfo
  {year} {2009})}\BibitemShut {NoStop}%
\bibitem [{\citenamefont {Bukas}\ \emph {et~al.}(2015)\citenamefont {Bukas},
  \citenamefont {Mitra}, \citenamefont {Meyer},\ and\ \citenamefont
  {Reuter}}]{O2-Pd100-JM1}%
  \BibitemOpen
  \bibfield  {author} {\bibinfo {author} {\bibfnamefont {V.~J.}\ \bibnamefont
  {Bukas}}, \bibinfo {author} {\bibfnamefont {S.}~\bibnamefont {Mitra}},
  \bibinfo {author} {\bibfnamefont {J.}~\bibnamefont {Meyer}}, \ and\ \bibinfo
  {author} {\bibfnamefont {K.}~\bibnamefont {Reuter}},\ }\href@noop {}
  {\bibfield  {journal} {\bibinfo  {journal} {J. Chem. Phys.}\ }\textbf
  {\bibinfo {volume} {143}},\ \bibinfo {pages} {034705} (\bibinfo {year}
  {2015})}\BibitemShut {NoStop}%
\bibitem [{\citenamefont {Tremblay}(2013)}]{Tremblay6}%
  \BibitemOpen
  \bibfield  {author} {\bibinfo {author} {\bibfnamefont {J.~C.}\ \bibnamefont
  {Tremblay}},\ }\href@noop {} {\bibfield  {journal} {\bibinfo  {journal} {J.
  Chem. Phys.}\ }\textbf {\bibinfo {volume} {138}},\ \bibinfo {pages} {244106}
  (\bibinfo {year} {2013})}\BibitemShut {NoStop}%
\bibitem [{\citenamefont {Spiering}\ \emph
  {et~al.}(2019{\natexlab{b}})\citenamefont {Spiering}, \citenamefont
  {Shakouri}, \citenamefont {Behler}, \citenamefont {Kroes},\ and\
  \citenamefont {Meyer}}]{RPBE-1-N2-Ru-ODF-elhp-2019}%
  \BibitemOpen
  \bibfield  {author} {\bibinfo {author} {\bibfnamefont {P.}~\bibnamefont
  {Spiering}}, \bibinfo {author} {\bibfnamefont {K.}~\bibnamefont {Shakouri}},
  \bibinfo {author} {\bibfnamefont {J.}~\bibnamefont {Behler}}, \bibinfo
  {author} {\bibfnamefont {G.-J.}\ \bibnamefont {Kroes}}, \ and\ \bibinfo
  {author} {\bibfnamefont {J.}~\bibnamefont {Meyer}},\ }\href@noop {}
  {\bibfield  {journal} {\bibinfo  {journal} {J. Phys. Chem. Lett.}\ }\textbf
  {\bibinfo {volume} {10}},\ \bibinfo {pages} {2957} (\bibinfo {year}
  {2019}{\natexlab{b}})}\BibitemShut {NoStop}%
\bibitem [{\citenamefont {Maurer}\ \emph {et~al.}(2016)\citenamefont {Maurer},
  \citenamefont {Askerka}, \citenamefont {Batista},\ and\ \citenamefont
  {Tully}}]{ODF-5}%
  \BibitemOpen
  \bibfield  {author} {\bibinfo {author} {\bibfnamefont {R.~J.}\ \bibnamefont
  {Maurer}}, \bibinfo {author} {\bibfnamefont {M.}~\bibnamefont {Askerka}},
  \bibinfo {author} {\bibfnamefont {V.~S.}\ \bibnamefont {Batista}}, \ and\
  \bibinfo {author} {\bibfnamefont {J.~C.}\ \bibnamefont {Tully}},\ }\href@noop
  {} {\bibfield  {journal} {\bibinfo  {journal} {Phys. Rev. B}\ }\textbf
  {\bibinfo {volume} {94}},\ \bibinfo {pages} {115432} (\bibinfo {year}
  {2016})}\BibitemShut {NoStop}%
\bibitem [{\citenamefont {Pe\~{n}a Torres}\ \emph {et~al.}(2019)\citenamefont
  {Pe\~{n}a Torres}, \citenamefont {Busnengo}, \citenamefont {Juaristi},
  \citenamefont {Larregaray},\ and\ \citenamefont
  {Crespos}}]{GGA-exchange+correlation-1-N2-W-elhp-2019}%
  \BibitemOpen
  \bibfield  {author} {\bibinfo {author} {\bibfnamefont {A.}~\bibnamefont
  {Pe\~{n}a Torres}}, \bibinfo {author} {\bibfnamefont {H.~F.}\ \bibnamefont
  {Busnengo}}, \bibinfo {author} {\bibfnamefont {J.~I.}\ \bibnamefont
  {Juaristi}}, \bibinfo {author} {\bibfnamefont {P.}~\bibnamefont
  {Larregaray}}, \ and\ \bibinfo {author} {\bibfnamefont {C.}~\bibnamefont
  {Crespos}},\ }\href@noop {} {\bibfield  {journal} {\bibinfo  {journal} {J.
  Phys. Chem. C}\ }\textbf {\bibinfo {volume} {123}},\ \bibinfo {pages} {2900}
  (\bibinfo {year} {2019})}\BibitemShut {NoStop}%
\bibitem [{\citenamefont {Luo}, \citenamefont {Zhou},\ and\ \citenamefont
  {Jiang}(2018)}]{PBE-6-CO2-surface-mode-elhp-2018}%
  \BibitemOpen
  \bibfield  {author} {\bibinfo {author} {\bibfnamefont {X.}~\bibnamefont
  {Luo}}, \bibinfo {author} {\bibfnamefont {X.}~\bibnamefont {Zhou}}, \ and\
  \bibinfo {author} {\bibfnamefont {B.}~\bibnamefont {Jiang}},\ }\href@noop {}
  {\bibfield  {journal} {\bibinfo  {journal} {J. Chem. Phys.}\ }\textbf
  {\bibinfo {volume} {148}},\ \bibinfo {pages} {174702} (\bibinfo {year}
  {2018})}\BibitemShut {NoStop}%
\bibitem [{\citenamefont {Novko}\ \emph {et~al.}(2016)\citenamefont {Novko},
  \citenamefont {Blanco-Rey}, \citenamefont {Alducin},\ and\ \citenamefont
  {Juaristi}}]{RPBE-2-H2-Pd-N-Ag-N2-Fe-AIMDEF-elhp-2016}%
  \BibitemOpen
  \bibfield  {author} {\bibinfo {author} {\bibfnamefont {D.}~\bibnamefont
  {Novko}}, \bibinfo {author} {\bibfnamefont {M.}~\bibnamefont {Blanco-Rey}},
  \bibinfo {author} {\bibfnamefont {M.}~\bibnamefont {Alducin}}, \ and\
  \bibinfo {author} {\bibfnamefont {J.~I.}\ \bibnamefont {Juaristi}},\
  }\href@noop {} {\bibfield  {journal} {\bibinfo  {journal} {Phys. Rev. B}\
  }\textbf {\bibinfo {volume} {93}},\ \bibinfo {pages} {245435} (\bibinfo
  {year} {2016})}\BibitemShut {NoStop}%
\bibitem [{\citenamefont {F\"uchsel}\ \emph {et~al.}(2019)\citenamefont
  {F\"uchsel}, \citenamefont {Zhou}, \citenamefont {Jiang}, \citenamefont
  {Juaristi}, \citenamefont {Alducin}, \citenamefont {Guo},\ and\ \citenamefont
  {Kroes}}]{GGA-exchange+correlation-5-HCl-Au-AIMD-2019}%
  \BibitemOpen
  \bibfield  {author} {\bibinfo {author} {\bibfnamefont {G.}~\bibnamefont
  {F\"uchsel}}, \bibinfo {author} {\bibfnamefont {X.}~\bibnamefont {Zhou}},
  \bibinfo {author} {\bibfnamefont {B.}~\bibnamefont {Jiang}}, \bibinfo
  {author} {\bibfnamefont {J.~I.~n.}\ \bibnamefont {Juaristi}}, \bibinfo
  {author} {\bibfnamefont {M.}~\bibnamefont {Alducin}}, \bibinfo {author}
  {\bibfnamefont {H.}~\bibnamefont {Guo}}, \ and\ \bibinfo {author}
  {\bibfnamefont {G.-J.}\ \bibnamefont {Kroes}},\ }\href@noop {} {\bibfield
  {journal} {\bibinfo  {journal} {J. Phys. Chem. C}\ }\textbf {\bibinfo
  {volume} {123}},\ \bibinfo {pages} {2287} (\bibinfo {year}
  {2019})}\BibitemShut {NoStop}%
\bibitem [{\citenamefont {Truong}, \citenamefont {Truhlar},\ and\ \citenamefont
  {Garrett}(1989)}]{TNT}%
  \BibitemOpen
  \bibfield  {author} {\bibinfo {author} {\bibfnamefont {T.~N.}\ \bibnamefont
  {Truong}}, \bibinfo {author} {\bibfnamefont {D.~G.}\ \bibnamefont {Truhlar}},
  \ and\ \bibinfo {author} {\bibfnamefont {B.~C.}\ \bibnamefont {Garrett}},\
  }\href@noop {} {\bibfield  {journal} {\bibinfo  {journal} {J.\ Phys.\ Chem.}\
  }\textbf {\bibinfo {volume} {93}},\ \bibinfo {pages} {8227} (\bibinfo {year}
  {1989})}\BibitemShut {NoStop}%
\bibitem [{\citenamefont {Olsen}\ \emph {et~al.}(1997)\citenamefont {Olsen},
  \citenamefont {Kroes}, \citenamefont {Lovvik},\ and\ \citenamefont
  {Baerends}}]{OLKR}%
  \BibitemOpen
  \bibfield  {author} {\bibinfo {author} {\bibfnamefont {R.~A.}\ \bibnamefont
  {Olsen}}, \bibinfo {author} {\bibfnamefont {G.~J.}\ \bibnamefont {Kroes}},
  \bibinfo {author} {\bibfnamefont {O.~M.}\ \bibnamefont {Lovvik}}, \ and\
  \bibinfo {author} {\bibfnamefont {E.~J.}\ \bibnamefont {Baerends}},\
  }\href@noop {} {\bibfield  {journal} {\bibinfo  {journal} {J. Chem. Phys.}\
  }\textbf {\bibinfo {volume} {107}},\ \bibinfo {pages} {10652} (\bibinfo
  {year} {1997})}\BibitemShut {NoStop}%
\bibitem [{\citenamefont {Pechukas}\ and\ \citenamefont {Light}(1966)}]{PPJS}%
  \BibitemOpen
  \bibfield  {author} {\bibinfo {author} {\bibfnamefont {P.}~\bibnamefont
  {Pechukas}}\ and\ \bibinfo {author} {\bibfnamefont {J.~C.}\ \bibnamefont
  {Light}},\ }\href@noop {} {\bibfield  {journal} {\bibinfo  {journal} {J.\
  Chem.\ Phys.}\ }\textbf {\bibinfo {volume} {44}},\ \bibinfo {pages} {3897}
  (\bibinfo {year} {1966})}\BibitemShut {NoStop}%
\bibitem [{\citenamefont {Mengel}\ and\ \citenamefont
  {Billing}(1997)}]{BILEH2}%
  \BibitemOpen
  \bibfield  {author} {\bibinfo {author} {\bibfnamefont {S.~K.}\ \bibnamefont
  {Mengel}}\ and\ \bibinfo {author} {\bibfnamefont {G.~D.}\ \bibnamefont
  {Billing}},\ }\href@noop {} {\bibfield  {journal} {\bibinfo  {journal} {J.
  Phys. Chem. B}\ }\textbf {\bibinfo {volume} {101}},\ \bibinfo {pages} {10781}
  (\bibinfo {year} {1997})}\BibitemShut {NoStop}%
\bibitem [{\citenamefont {Pines}(1963)}]{DPI}%
  \BibitemOpen
  \bibfield  {author} {\bibinfo {author} {\bibfnamefont {D.}~\bibnamefont
  {Pines}},\ }\href@noop {} {\emph {\bibinfo {title} {Elementary Excitations in
  Solids}}}\ (\bibinfo  {publisher} {Bejamin/Cummings, Reading},\ \bibinfo
  {year} {1963})\BibitemShut {NoStop}%
\bibitem [{\citenamefont {Billing}\ and\ \citenamefont
  {Mikkelsen}(1997)}]{GDB2}%
  \BibitemOpen
  \bibfield  {author} {\bibinfo {author} {\bibfnamefont {G.~D.}\ \bibnamefont
  {Billing}}\ and\ \bibinfo {author} {\bibfnamefont {K.~V.}\ \bibnamefont
  {Mikkelsen}},\ }\href@noop {} {\emph {\bibinfo {title} {Advance Molecular
  Dynamics and Chemical Kinetics}}}\ (\bibinfo {year} {1997})\BibitemShut
  {NoStop}%
\bibitem [{\citenamefont {Michelsen}\ and\ \citenamefont
  {Auerbach}(1991)}]{HAM1}%
  \BibitemOpen
  \bibfield  {author} {\bibinfo {author} {\bibfnamefont {H.~A.}\ \bibnamefont
  {Michelsen}}\ and\ \bibinfo {author} {\bibfnamefont {D.~J.}\ \bibnamefont
  {Auerbach}},\ }\href@noop {} {\bibfield  {journal} {\bibinfo  {journal} {J.
  Chem. Phys.}\ }\textbf {\bibinfo {volume} {94}},\ \bibinfo {pages} {7502}
  (\bibinfo {year} {1991})}\BibitemShut {NoStop}%
\bibitem [{\citenamefont {Lee}\ and\ \citenamefont {DePristo}(1996)}]{CYL2}%
  \BibitemOpen
  \bibfield  {author} {\bibinfo {author} {\bibfnamefont {C.~Y.}\ \bibnamefont
  {Lee}}\ and\ \bibinfo {author} {\bibfnamefont {A.~E.}\ \bibnamefont
  {DePristo}},\ }\href@noop {} {\bibfield  {journal} {\bibinfo  {journal} {J.
  Chem. Phys.}\ }\textbf {\bibinfo {volume} {85}},\ \bibinfo {pages} {4161}
  (\bibinfo {year} {1996})}\BibitemShut {NoStop}%
\bibitem [{\citenamefont {Watts}\ \emph {et~al.}(2001)\citenamefont {Watts},
  \citenamefont {Sitz}, \citenamefont {McCormack}, \citenamefont {Kroes},
  \citenamefont {Olsen}, \citenamefont {Groeneveld}, \citenamefont {Stralen},
  \citenamefont {Baerends},\ and\ \citenamefont {Mowrey}}]{GJ3}%
  \BibitemOpen
  \bibfield  {author} {\bibinfo {author} {\bibfnamefont {E.}~\bibnamefont
  {Watts}}, \bibinfo {author} {\bibfnamefont {G.~O.}\ \bibnamefont {Sitz}},
  \bibinfo {author} {\bibfnamefont {D.~A.}\ \bibnamefont {McCormack}}, \bibinfo
  {author} {\bibfnamefont {G.~J.}\ \bibnamefont {Kroes}}, \bibinfo {author}
  {\bibfnamefont {R.~A.}\ \bibnamefont {Olsen}}, \bibinfo {author}
  {\bibfnamefont {J.~A.}\ \bibnamefont {Groeneveld}}, \bibinfo {author}
  {\bibfnamefont {J.~N. P.~V.}\ \bibnamefont {Stralen}}, \bibinfo {author}
  {\bibfnamefont {E.~J.}\ \bibnamefont {Baerends}}, \ and\ \bibinfo {author}
  {\bibfnamefont {R.~C.}\ \bibnamefont {Mowrey}},\ }\href@noop {} {\bibfield
  {journal} {\bibinfo  {journal} {J.\ Chem.\ Phys.}\ }\textbf {\bibinfo
  {volume} {114}},\ \bibinfo {pages} {495} (\bibinfo {year}
  {2001})}\BibitemShut {NoStop}%
\bibitem [{\citenamefont {Hamza}\ and\ \citenamefont {Madix}(1985)}]{AVH}%
  \BibitemOpen
  \bibfield  {author} {\bibinfo {author} {\bibfnamefont {A.~V.}\ \bibnamefont
  {Hamza}}\ and\ \bibinfo {author} {\bibfnamefont {R.~J.}\ \bibnamefont
  {Madix}},\ }\href@noop {} {\bibfield  {journal} {\bibinfo  {journal} {J.
  Phys. Chem.}\ }\textbf {\bibinfo {volume} {89}},\ \bibinfo {pages} {5381}
  (\bibinfo {year} {1985})}\BibitemShut {NoStop}%
\bibitem [{\citenamefont {Kreyszig}(2025)}]{kreyszig}%
  \BibitemOpen
  \bibfield  {author} {\bibinfo {author} {\bibfnamefont {E.}~\bibnamefont
  {Kreyszig}},\ }\href@noop {} {\emph {\bibinfo {title} {Advanced Engineering
  Mathematics, International Adaptation}}}\ (\bibinfo  {publisher} {john wiley
  \& sons},\ \bibinfo {year} {2025})\BibitemShut {NoStop}%
\bibitem [{\citenamefont {Shankaraiah}(2023)}]{Shankaraiah}%
  \BibitemOpen
  \bibfield  {author} {\bibinfo {author} {\bibfnamefont {G.}~\bibnamefont
  {Shankaraiah}},\ }\href@noop {} {\bibfield  {journal} {\bibinfo  {journal}
  {International Journal of Research Publication and Reviews}\ }\textbf
  {\bibinfo {volume} {4}},\ \bibinfo {pages} {167} (\bibinfo {year}
  {2023})}\BibitemShut {NoStop}%
\bibitem [{\citenamefont {Alexander}, \citenamefont {Werner},\ and\
  \citenamefont {Manolopoulos}(1998)}]{jcp-109-5710-WERNER}%
  \BibitemOpen
  \bibfield  {author} {\bibinfo {author} {\bibfnamefont {M.~H.}\ \bibnamefont
  {Alexander}}, \bibinfo {author} {\bibfnamefont {H.}~\bibnamefont {Werner}}, \
  and\ \bibinfo {author} {\bibfnamefont {D.~E.}\ \bibnamefont {Manolopoulos}},\
  }\href@noop {} {\bibfield  {journal} {\bibinfo  {journal} {J. Chem. Phys.}\
  }\textbf {\bibinfo {volume} {109}},\ \bibinfo {pages} {5710} (\bibinfo {year}
  {1998})}\BibitemShut {NoStop}%
\bibitem [{\citenamefont {Arthurs}\ and\ \citenamefont {Dalgarno}(1960)}]{ADS}%
  \BibitemOpen
  \bibfield  {author} {\bibinfo {author} {\bibfnamefont {A.}~\bibnamefont
  {Arthurs}}\ and\ \bibinfo {author} {\bibfnamefont {A.}~\bibnamefont
  {Dalgarno}},\ }\href@noop {} {\bibfield  {journal} {\bibinfo  {journal}
  {Proc. R. Soc. Lond. Ser. A}\ }\textbf {\bibinfo {volume} {256}},\ \bibinfo
  {pages} {540} (\bibinfo {year} {1960})}\BibitemShut {NoStop}%
\bibitem [{\citenamefont {Gradshteyn}\ and\ \citenamefont {Ryzhik}()}]{ISG}%
  \BibitemOpen
  \bibfield  {author} {\bibinfo {author} {\bibfnamefont {I.~S.}\ \bibnamefont
  {Gradshteyn}}\ and\ \bibinfo {author} {\bibfnamefont {I.~M.}\ \bibnamefont
  {Ryzhik}},\ }\href@noop {} {\emph {\bibinfo {title} {Table of Integrals
  Series and Products; Academic Press, New York (1965)}}}\BibitemShut {NoStop}%
\bibitem [{\citenamefont {Billing}(1995)}]{GD1}%
  \BibitemOpen
  \bibfield  {author} {\bibinfo {author} {\bibfnamefont {G.~D.}\ \bibnamefont
  {Billing}},\ }\href@noop {} {\bibfield  {journal} {\bibinfo  {journal} {J.\
  Phys.\ Chem.}\ }\textbf {\bibinfo {volume} {99}},\ \bibinfo {pages} {15378}
  (\bibinfo {year} {1995})}\BibitemShut {NoStop}%
\end{thebibliography}%


\end{appendices}

\chapter*{Biography}\label{sec:bio}
\addcontentsline{toc}{chapter}{Biography}
\vspace{-0.5cm}
\section*{Dr. Soumya Mukherjee}

\noindent
Dr. Soumya Mukherjee earned his B.Sc. in Chemistry and M.Sc. in Physical Chemistry (specialization) from the University of Calcutta, India. Later on, he persuaded his PhD from Indian Association for the Cultivation of Science, Kolkata under the supervision of Professor Satrajit Adhikari in the area of Quantum Chemistry and Quantum Dynamics. He is now serving as Assistant Professor in the Department of Chemistry, Maulana Azad National Institute of Technology (MANIT), Bhopal. During his PhD tenure, Dr. Mukherjee published close to twenty-five (25) articles in various internationally reputed peer-reviewed journals. On the other hand, as an independent researcher along with his collaborators, close to ten (10) articles are published so far. In MANIT, Bhopal, Dr. Mukherjee’s research group is mainly involved on two different areas, (a) development of Complex Beyond Born-Oppenheimer (CBBO) theory, and (b) computational atmospheric and medicinal chemistry. He formulated analytical expressions of various adiabatic-to-diabatic transformation (ADT) quantities (like ADT equations and diabatic PESs matrices) for Abelian and non-Abelian cases considering six situations (different combinations of real/complex adiabatic PESs/spin-orbit coupling/ADT matrix) as well as implemented for representative systems, like C$_6$H$_6^+$ and F+H$_2$ system. 

\section*{Professor Satrajit Adhikari}
\noindent
Professor Satrajit Adhikari earned his B.Sc. in Chemistry and M.Sc. 
in Physical Chemistry (specialization) from the University of 
Burdwan, India, before completing his Ph.D. in Quantum Dynamics 
from IACS/Jadavpur University in 1996 under the supervision of Professor 
Sankar P. Bhattacharya. He pursued postdoctoral research at the 
University of Copenhagen, Denmark (1996–1999, with Professor G. D. 
Billing), followed by a Visiting Scientist position at the Fritz 
Haber Research Center, Israel, in 2000. He began his independent academic career 
as Senior Lecturer at BITS Pilani, Rajasthan (2000–2001) and then, joined the 
Indian Institute of Technology Guwahati as an Assistant Professor 
(2001–2007). In 2007, he moved to the Indian Association for the Cultivation of Science (IACS), Kolkata as Associate Professor and thereafter, promoted to Professor (2012), and finally, become Senior Professor since 2016 onwards. He has been invited as Adjuct Professor (2018-2020) and then, become Adjunct Research Professor (2021–present) at IISER Kolkata. He has served as Visiting Professor over dozens of well known Universities (University of Coimbra, Portugal; Hebrew University of Jerusalem, Israel; Leiden University, Netherland; Berlin Free University, University of Potsdam and University of Heidelberg, Germany; Qufu Normal University, China, etc.) Professor Adhikari's research interest covers the areas, namely, multi-state multi-mode quantum dynamics methodology, Beyond Born-Oppenheimer (BBO) theory and its application on spectroscopy, reactive scattering and solid state phenomena as well as finite temperature molecule-surface scattering process.  His work has resolved several longstanding conceptual and numerical challenges, which lead to the development of accurate and practical algorithms for computing reactive scattering cross sections as well as photodetachment and photoabsorption spectra of molecules using {\it ab initio} methods. Professor Adhikari along with his collaborators could able to publish nearly 180 articles in various internationally reputed journals. He has delivered more than forty (40) lectures in various internationally well-known conference/workshop series. He is an elected Fellow of the National Academy of 
Sciences, India (NASI) in 2019, and Fellow of the Indian Academy of 
Sciences (IASc) in 2024.

\end{document}